\documentclass[12pt]{article}

\usepackage{color,rotate,graphics,epsf,amssymb,latexsym}

\usepackage{hyperref}

\catcode`\@=11
\def \be  {\begin{equation}}
\def \ee  {\end{equation}}
\def \beq  {\begin{equation}}
\def \eeq  {\end{equation}}
\def \ba  {\begin{eqnarray}}
\def \ea  {\end{eqnarray}}
\def \baa {\begin{eqnarray*}}
\def \eaa {\end{eqnarray*}}
\def \lab #1 {\label{#1}}

\def \matrix #1 {\left(\begin{array}{cc} #1 \end{array}\right)}

\def \res{\mathop{\rm res}\nolimits}

\newcommand{\insertfig}[2]{\mbox{\epsfxsize=#1cm \epsfbox{#2.eps}}}
\newcommand{\Asym}{\mathop{\mbox{\bf A}}}
\newcommand{\Sym}{\mathop{\mbox{\bf S}}}
\newcommand{\ft}[2]{{\textstyle\frac{#1}{#2}}}

\newcommand{\as}{\ifmmode\alpha_{\rm s}\else{$\alpha_{\rm s}$}\fi}
\newcommand{\bit}[1]{\mbox{\boldmath$#1$}}

\newcommand{\CoA}{\mbox{CoA}}
\newcommand{\CeA}{\mbox{CeA}}
\newcommand{\SSA}{\mbox{SSA}}

\font\cmss=cmss12 
\def\1{\hbox{{1}\kern-.25em\hbox{l}}}
\def\bfZ{\relax{\hbox{\cmss Z\kern-.4em Z}}}

\font\cmss=cmss12 
\def\inbar{\,\vrule height1.5ex width.4pt depth0pt}
\def\IC{\relax\hbox{$\inbar\kern-.3em{\rm C}$}}
\def\IZ{\relax{\hbox{\cmss Z\kern-.4em Z}}}
\def\IR{{\hbox{{\rm I}\kern-.2em\hbox{\rm R}}}}

\def\IP{{\hbox{{\rm I}\kern-.2em\hbox{\rm P}}}}
\def\II{\hbox{{1}\kern-.25em\hbox{l}}}

\def\numberbysection{\@addtoreset{equation}{section}
                     \def\theequation{\thesection.\arabic{equation}}}
\numberbysection

\begin{document}

\begin{titlepage}
\begin{flushright}
\begin{tabular}{l}
JLAB--THY--04--34
\end{tabular}
\end{flushright}

\centerline{\large \bf Unraveling hadron structure with generalized parton distributions}

\vspace{10mm}

\centerline{\sc A.V. Belitsky}

\vspace{5mm}

\centerline{\it Department of Physics and Astronomy, Arizona State University}
\centerline{\it Tempe, AZ 85287-1504, USA}

\vspace{5mm}

\centerline{\sc A.V. Radyushkin%
                \footnote{Also at Bogoliubov Laboratory of Theoretical Physics, JINR, Dubna,
                          Russia.}
}

\vspace{5mm}

\centerline{\it Physics Department, Old Dominion University}
\centerline{\it Norfolk, VA 23529, USA}

\vspace{5mm}

\centerline{\it Theory Group, Jefferson Laboratory,}
\centerline{\it Newport News, VA 23606, USA}

\vspace{10mm}

\centerline{\bf Abstract}

\vspace{10mm}

The generalized parton distributions, introduced nearly a decade ago, have emerged
as a universal tool to describe hadrons in terms of quark and gluonic degrees of
freedom. They combine  the features of form factors, parton densities and
distribution amplitudes---the functions used for a long time in studies of hadronic
structure. Generalized parton distributions are analogous to the phase-space Wigner
quasi-probability function of non-relativistic quantum mechanics which encodes full
information on a quantum-mechanical system. We give an extensive review of main
achievements in the development of this formalism. We discuss physical interpretation
and basic properties of generalized parton distributions, their modeling and QCD
evolution in the leading and next-to-leading orders. We describe how these functions
enter a wide class of exclusive reactions, such as electro- and photo-production of
photons, lepton pairs, or mesons. The theory of these processes requires and implies
full control over diverse corrections and thus we outline the progress in handling
higher-order and higher-twist effects. We catalogue corresponding results and present
diverse techniques for their derivations. Subsequently, we address observables that
are sensitive to different characteristics of the nucleon structure in terms of
generalized parton distributions. The ultimate goal of the GPD approach is to provide
a three-dimensional spatial picture of the nucleon, direct measurement of the quark
orbital angular momentum, and various inter- and multi-parton correlations.

\vspace{10mm}

\centerline{\sl Dedicated to Anatoly V. Efremov on occasion of his 70th anniversary}

\end{titlepage}

\tableofcontents

\section{Preface}

The concept of generalized parton distributions (GPDs) was developed as a modern tool to
deliver a detailed description of the microscopic structure of hadrons in terms of their
elementary constituents \cite{MulDitRobGeyHor98,Ji96a,Rad1996a,Rad1996b,Ji96b,Rad1997a}.
The need for GPDs is dictated by the present-day situation in hadronic physics. The
fundamental particles which form hadrons are long known to be quarks and gluons, whose
interactions are described by the Lagrangian of quantum chromodynamics (QCD). However,
this knowledge is not sufficient at the moment to perform reliable, fully quantitative
calculations starting from first principles, since the precise mechanism of hadron
formation from the underlying quark and gluonic degrees of freedom is not completely
known and quantifiable to our dissatisfaction. We have to resort to hints from
experimental measurements in order to understand how QCD works and ultimately resolve
``the long-distance problem'' of QCD. The standard way out of these complications is to
use certain phenomenological functions. The well-known examples include form factors,
parton densities, and distribution amplitudes. The new functions---generalized parton
distributions---are  hybrids of these ``old'' functions which, in their turn, are the
limiting cases of the ``new'' ones. There are several existing reviews dedicated to
thorough discussion of diverse aspect of GPDs
\cite{Ji:1998pc,Radyushkin:2000uy,Goeke:2001tz,Diehl:2003ny}. Though the subject is
common, these reviews differ in the extent of the covered material, in the emphasis
on particular techniques and approaches to the GPD studies, and in the degree of
detail in the presentation of computations. The aim of the present endeavor is to
give a review of the theory of generalized parton distributions, with discussion of
calculation machinery sufficiently detailed to enable an interested reader to derive
many results without heavy use of original publications as well as to provide a
self-consistent compendium of analytical formulas.

The review is organized as follows. In Section \ref{Introduction}, we discuss general
quantum-mechanical aspects of investigating the structure of matter emphasizing
similarities of studies on atomic, nuclear and quark/gluon levels. In particular, we
address  the issue of interpreting the experimentally observable quantities in terms
of charge, magnetization, etc., distributions, and discuss Wigner distributions, the
closest quantum-mechanical analog of the generalized parton distributions. In Section
\ref{SectionClassification}, we give the classification of GPDs according to the
quantum numbers of the composite operators which represent them, and by the type of
the hadronic matrix elements involved. The consideration in that section is limited
to the leading-twist distributions. The twist-three functions are addressed later
in subsection \ref{TwistThreeGPDs}. Section \ref{Evolution} is devoted to the evolution
of generalized parton distributions. We discuss evolution kernels for the light-cone
operators in the coordinate representation, their transformation into the momentum
space, diagonalization of the mixing matrix, and interrelation between ``inclusive''
and  ``exclusive'' kernels. The use of the conformal symmetry allows to understand
many features of the evolution equations on a deeper level. This is also a topic of
a detailed discussion in Section \ref{Evolution}. The subject of  Section \ref{Compton}
is the virtual Compton amplitude, the major building block of the electron-hadron
Compton scattering process, potentially the cleanest source of information about GPDs.
We present a thorough study of the Compton amplitude including its gauge-invariant
tensor decomposition, one-loop corrections to the short-distance coefficient function
and power suppressed contributions, including twist-three and target mass effects.
In Section \ref{Phenomenology}, we give a comprehensive discussion of a few exclusive
reactions where GPDs can be measured or accessed, including deeply virtual Compton
scattering, doubly deeply virtual Compton scattering and hard exclusive meson
production. In the concluding Section \ref{Outlook}, we outline a few directions,
where significant improvements in the theory of GPDs can be made. A number of appendices
serve as a technical supplement to the material presented in the main text.

\section{Unraveling layers of matter: from atoms to partons}
\label{Introduction}

The wisdom goes back to the ancient Greeks who philosophized that matter
consists of tiny particles---atoms. However, the subatomic structure of
matter remained an unsolved puzzle till the beginning of the 20th century
when radioactivity was discovered and used by Rutherford in his seminal
experiments on large-angle scattering of $\alpha$-particles off atoms where
the outcome suggested that the atom bears a localized core---the nucleus.
On the other hand, electron beams were found to pass through atoms with no
or very little deflection forcing Lenard to hypothesize that atoms have
``wide empty spaces inside''. The $\alpha$-particles scattered more frequently
than their ``cousins'' $\beta$-particles, which produced very low, barely
observable rates due to tiny cross sections and, available at the time,
luminosity of the beam. Similar experiments but rather with light sources or
room-temperature neutrons are exploited nowadays to study the crystal's lattice.
If a crystal is placed in front of a source of visible light, the object will
merely leave a shadow on a screen behind it and one will not be able to detect
elementary building blocks which form it. Obviously, the visible light, having
the wavelength $\lambda_{\rm light} \sim 0.4 - 0.7 \ \mu{\rm m}$, cannot do
the job and resolve the internal structure of a crystal. The size of an individual
atom, say, hydrogen, is of order $R_{\rm atom} \sim \left( \alpha_{\rm em} m_e
\right)^{-1} \sim \left( 10 \ {\rm KeV} \right)^{- 1}$. Therefore, to ``see''
individual atoms in crystals one has to use light sources with a comparable or
smaller wavelength $\lambda_\gamma \leq R_{\rm atom}$, or equivalently, of energy
$E_\gamma \geq R^{-1}_{\rm atom}$. To do this kind of ``nano-photography''
a beam of X-rays is needed. To go further into the structure of atoms one has
to resort to even more energetic probes.

After the discovery of the nucleus' building blocks---the nucleons (i.e., protons
and neutrons)---the attention has been shifted to the extensive study of these
``elementary'' particles. However, their elementarity has been questioned since
Stern's experiments in 1932 \cite{IsrSimSte37} which measured magnitude of the
proton's magnetic moment about three times larger than that of the expected.
Hofstadter's experiments \cite{HofMcA55} with elastic electron scattering off
nucleons, $e N \to e' N'$, which accessed proton's electromagnetic form factors
revealed the deviation from those of a point-like object and demonstrated for
the first time that the nucleon has a spatial extent of order of one femtometer.
To probe femtometer scales, we rely on scattering experiments with high-energy
lepton beams, where the point-like nature of leptons does not represent an extra
uncertainty in the interpretation of the data contrary to beams of composite
particles. Inelastic lepton-nucleon scattering experiments, $\ell N \to \ell' X$,
conducted at Stanford Linear Accelerator Center \cite{FreKenTay69} fulfilled this
goal which led to the ultimate discovery of a new layer of matter by observing
events with the transfer of a large momentum from the electron to the proton.
If the proton would be a hard ball or a diffuse distribution of matter, such kind
of scattering would be improbable. An analogy to the latter is from the early
days of atomic physics when, according to the Thompson's model, the atom was
thought to consist of negatively charged electrons embedded into a jelly-like
medium of positive charge. Thus the results of the experiment were explained by
conjecturing the existence of point-like constituents inside the proton which
absorb a highly virtual $\gamma^\ast$-quantum emitted by the electron in the
course of the scattering process.

An earlier theoretical description of deeply inelastic events resulted into the
scaling hypothesis by Bjorken \cite{Bjo69}, stating that cross sections measured
there must not depend on dimensionful parameters, like particle virtualities.
Further development resulted into the formulation of the Feynman's parton model
\cite{Fey69,Fey71}. According to the parton  model, the hadron at high energy
can be viewed as a composition of a number of constituents---the partons---which
behave as an incoherent bunch of quanta at small space-time separations, and
the interaction of the probe happens with one of them. Current algebra analysis
of the deeply inelastic cross section had favored the quarks to be fermions with
spin one-half \cite{CalGro68}. It was quite tempting to identify these partons
with Gell-Mann and Zweig's quarks. This experimental discovery, which triggered
theoretical minds to reconsider the contemporary physical concepts, was analogous
to, if not greater than, the Rutherford's experiments in atomic physics. The
Feynman's partons were naturally identified with quark and gluon degrees of
freedom described by a nonabelian gauge field theory of strong interactions with
unbroken $SU (3)$ group---Quantum Chromodynamics (QCD).

The path which led to this fundamental finding was rather long. The analysis
of divergences of charge renormalization resulted in the discovery of the asymptotic
freedom \cite{Hoo73,GroWil73a,Pol73} in nonabelian gauge theories. The vacuum
in general is a polarizable medium of virtual particles. In abelian gauge theories
the former is dielectric, i.e., the medium screens the charge, and it becomes
weaker at finite distances---a lump of positive charge pulls negative charges
from the medium toward itself which partially neutralizes it causing its weakening
at a   distance. If the charge is point-like, i.e., has zero radius, it would be
screened completely  at finite distances. This is a well-known zero charge problem
\cite{LanPom55,Pom55}. For a finite-size charge the interaction becomes stronger
at smaller distances and inevitably it reaches the magnitude where the perturbation
theory becomes inapplicable. Of course, this inconsistency of the resummed theory
is achieved only at asymptotically high energy, so that this triviality of
interaction is irrelevant as soon as one discusses processes at Earth energies.
Contrary, when nonabelian fields are present the response of spin-one quanta is
paramagnetic \cite{Nie81,Hug81} so that effective coupling decreases with rising
energy.

The finding of the vanishing nonabelian charge at small distances has put into its
spot the concept of the gauged color group. Quantum Chromodynamics, baptized by
Gell-Mann and discussed earlier in the works by Fritzsch, Gell-Mann and Leutwyler
\cite{GelFri71}, has emerged as the theory of strong interactions. Its Lagrangian
\begin{equation}
\label{QCDLagrangian}
{\cal L}_{\scriptscriptstyle\rm QCD} (z)
=
\ft{i}2 \bar\psi (z)
\left(
\stackrel{{}_\rightarrow}{\cal D}_\mu - \stackrel{{}_\leftarrow}{\cal D}_\mu
\right)
\gamma^\mu \psi (z)
-
\ft14 F_{\mu\nu}^a (z) F^{\mu\nu}_a (z)
\,
\end{equation}
describes the  minimal interaction between the quarks and the glue, introduced
through the covariant derivatives
\begin{equation}
\stackrel{{}_\rightarrow}{\cal D}_\mu
\, = \,
\stackrel{{}_\rightarrow}{\partial}_\mu
- \, i g \, t^a A_\mu^a (z)
\, , \qquad
\stackrel{{}_\leftarrow}{\cal D}_\mu
\, = \,
\stackrel{{}_\leftarrow}{\partial}_\mu
+ \, i g \, t^a A_\mu^a (z)
\, .
\end{equation}
QCD predicts the desired Bjorken scaling at short distances which is, however,
broken by logarithmic corrections computable from  first principles
\cite{GroWil73b,GeoPol74}. It took quite a while before experiments, having
a sufficiently large window in the momentum transfer, had found a confirmation
for the scaling violation \cite{Gro79} intrinsic to renormalizable field theories.

The short-distance structure of the strong interaction dynamics was established,
and another even more fundamental issue arose: long distance behavior of the
theory. No free quarks have been  observed. Instead, when emitted in a given
process they conspire to form a color neutral object and fragment into observable
hadrons. When the process involves a hard momentum transfer they form jets.
This was yet another indirect evidence in favor of the quark-gluon substructure
of hadrons. The studies of jets explicitly demonstrated that they carry genuine
properties, computable from the first principles, of their predecessors---quarks
and gluons \cite{EllStiWeb96}. This unusual property of the color gauge theory
is the infrared slavery of hadron's constituents, or confinement, which has been
proven to be the most outstanding problem of strong interaction physics. No
efficient analytical tool has been developed so far to tackle the task. We have,
however, a rough physical picture according to which quarks inside a hadron are
bound by a string of glue. Once the quarks are separated far enough so that the
stretching force overwhelms the string tension, the latter breaks down only by
means of a creation of a quark-antiquark pair from the vacuum at its ends. The
quarks never show up free. With the development of lattice techniques, there is
a growing compelling evidence toward this description.

The nucleon represents a relativistic multi-particle quantum system in a bound
state whose dynamics is driven by strong interactions. The exhaustive exploration
of its structure and confining dynamics of its constituents is a cherished goal
of strong-interaction physics. This is the subject of the present review within
the framework of generalized parton distributions---a set of hadron
characteristics probing hadrons from a variety of different angles. This
approach is phenomenological and uses experimental data as an input in
theoretical considerations in order to draw solid conclusions about the internal
structure of the proton and its siblings.

\subsection{Quantum mechanical observables}

Before we move on to a detailed analysis of tools and observables used to unravel
the nucleon's content, let us draw a few close analogies from nonrelativistic
quantum mechanics. We hope that this will be extremely instructive and insightful
for the main subject of our analysis.

A quantum mechanical system is determined by its wave function $\psi$. Acquiring
the latter from theoretical considerations, like solving the Schr\"odinger equation,
or from experimental measurements, allows one to predict any physical observable
for a given system. The bulk of experimentally accessible quantities is sensitive
only to the absolute value of the wave function via measurements of cross sections.
The phase of the wave function is thus essentially unattainable there. To circumvent
the difficulty,  one has to measure correlations of wave functions or, more generally,
the density matrix $\varrho$. For a pure state,
\begin{equation}
\varrho (\bit{r}_1, \bit{r}_2) = \psi^\ast (\bit{r}_1) \psi (\bit{r}_2)
\, ,
\end{equation}
and thus the wave function is known once the density matrix is determined. If one is
able to measure the density matrix by means of the interference of a test system with
a reference source, possessing a priori known characteristics, this will then serve
the purpose of reconstructing the missing phase of the wave function. Succeeding
in this endeavor, complete information on a quantum mechanical system can be acquired.

\subsubsection{Atomic form factor and momentum density}

Let us recall a few physical observables which are conventionally used to
probe a quantum mechanical system and measure its wave function.

First, the spatial distribution of matter (or charge) in a system can be probed
through elastic scattering of electrons, photons, or neutrons, etc. The physical
quantity that one measures is the elastic form---also called structure---factor
$F (\bit{\Delta})$ which depends on the three-momentum transfer to the system
$\bit{\Delta}$. It is expressed (upon removing a point-like core due to scattering
off the atomic nucleus) as a Fourier transform of the absolute value squared of
the atomic wave function $\rho (\bit{r}) = |\psi (\bit{r})|^2$ with respect to
the position three-vector $\bit{r} = (r^x, r^y, r^z)$;
\begin{equation}
\rho (\bit{r})
=
\int d^3 \bit{\Delta} \,
{\rm e}^{i \bit{\scriptstyle \Delta} \cdot \bit{\scriptstyle r} /\hbar}
F (\bit{\Delta})
\, .
\end{equation}
One can easily compute the tail of the form factor for the scattering on a bound
state in a Coulomb potential, $V(\bit{r}) \sim 1/|\bit{r}|$, at large momentum
transfers. In the momentum space the latter behaves as $V(\bit{p}) \sim 1/\bit{p}^2$.
Then the solution to the Schr\"odinger equation,
\begin{equation}
\label{SchrodingerEq}
i \frac{d}{dt} \psi (\bit{r})
=
\left( H_0 + V (\bit{r}) \right) \psi (\bit{r})
\, ,
\end{equation}
where $H_0 = \bit{p}^2/(2 m)$ and $V (\bit{r})$ are the free-particle Hamiltonian
and the interaction potential, respectively, is given by
$$
\psi (\bit{p})
=
\frac{
\int d^3 \bit{p}' \, V (\bit{p} - \bit{p}') \psi (\bit{p}')
}{
E - \bit{p}^2/(2 m)
}
\ \,
\stackrel{|\bit{\scriptstyle p}| \to \infty}{\approx}
\
{\rm const.}
\frac{\psi (\bit{r} = 0)}{\bit{p}^4}
\, .
$$
Here all the quantities are Fourier transformed into the momentum space. In
the evaluation of the second equality we took into account that the kinetic
energy obviously dominates and we also neglected the dependence on $\bit{p}'$
in the argument of the potential compared to the large momentum $\bit{p}$.
The resulting integral is a Fourier transform of the wave function  at the
origin. Thus,
$$
F (\bit{\Delta})
=
\int d^3 \bit{p} \ \psi^\ast (\bit{p}) \psi (\bit{p} + \bit{\Delta})
\stackrel{|\bit{\scriptstyle \Delta}| \to \infty}{\approx}
\psi (\bit{\Delta}) \int d^3 \bit{p} \ \psi^\ast (\bit{p})
\approx
{\rm const.} \frac{| \psi (\bit{r} = 0) |^2}{\bit{\Delta}^4}
\, .
$$

The second approach to study the properties of a quantum mechanical system
is designed to measure the population of its constituents with a given momentum,
or the momentum distribution, through knock-out scattering. The momentum
density is expressed in terms of the absolute value of the momentum-space
wave function $n (\bit{p}) = |\psi (\bit{p})|^2$. An example of a
nonrelativistic quantum system where this quantity is measured is superfluid
helium (see, e.g., Ref.\ \cite{Cep95}). In the neutron-scattering experiment,
the distribution of angles and energies of a neutron beam passing through a
sample of helium is recorded, while the state of the liquid helium after
the scattering is not detected. At high momentum transfer, when a neutron
scatters off single atoms, the impulse approximation holds and the cross
section $\sigma (p^z)$ as a function of momentum $p^z$ is expressed via the
relation
$$
\sigma (p^z) \sim \int d p^x \, d p^y \, n (\bit{p})
$$
in terms of the Fourier transform of a single-particle density matrix
\begin{equation}
n (\bit{p})
=
\int \frac{d^3 \bit{r}_1}{(2 \pi \hbar)^3} \frac{d^3 \bit{r}_2}{(2 \pi \hbar)^3}
{\rm e}^{
i \bit{\scriptstyle p}
\cdot
(\bit{\scriptstyle r}_1 - \bit{\scriptstyle r}_2)/\hbar}
\varrho (\bit{r}_1, \bit{r}_2)
\, .
\end{equation}
For a pure state, it reduces to the square of the coordinate space wave
function, $\varrho (\bit{r}_1, \bit{r}_2) = \psi^\ast (\bit{r}_1) \psi
(\bit{r}_2)$. The momentum distribution of liquid helium $n (\bit{p})$
at $\bit{p} = 0$ determines the Bose-Einstein condensate. Another
well-known example of nonrelativistic momentum density is the nucleon
distributions in a nucleus measured through quasielastic electron scattering.

\subsubsection{Wigner distribution}
\label{WignerFunction}

As we emphasized, the observables outlined above probe only the diagonal elements
of the density matrix either in the coordinate or momentum space. Thus, one is
unable to get complete information on the quantum-mechanical state. The off-diagonal
elements of the density matrix remain unknown and one lacks phases of the wave
function. The question arises whether there exists an experimental observable which,
on the one hand, is sensitive to the phase structure of the wave function and, on
the other hand, has an intuitive interpretation. The answer to this question turns
out to be positive.

An equivalent to the density matrix approach and a classically more
comprehensive description of a quantum mechanical system is achieved by
means of a partially Fourier-transformed density matrix known as the
Wigner quasi-probability distribution \cite{Wig32,HilOcoScuWig84},
\begin{equation}
\label{QMWigner}
W (\bit{r}, \bit{p})
=
\int \frac{d^3 \bit{R}}{(2 \pi \hbar)^3} \,
{\rm e}^{- i \bit{\scriptstyle p} \cdot \bit{\scriptstyle R}/\hbar}
\psi^\ast \left( \bit{r} - \ft12 \bit{R} \right)
\psi \left( \bit{r} + \ft12 \bit{R} \right)
\, ,
\end{equation}
where $\bit{r}$ and $\bit{p}$ are $c$-numbers, not operators. Because it
involves wave functions at different spatial separations, it obviously
possesses complete knowledge on their imaginary phases, just like the
density matrix does. It has the following important features.

\begin{itemize}

\item The Wigner distribution is advantageous over the density matrix
since it is real:
\begin{equation}
W^\ast (\bit{r} , \bit{p}) = W (\bit{r} , \bit{p})
\, .
\end{equation}

\item It has a certain classical limit and thus can be thought of as an
extension of an intuitive classical concept to quantum physics. Namely,
in classical physics the state of a particle is specified by its
position $\bit{r}$ and momentum $\bit{p}$. In a gas of classical identical
particles, the single-particle properties are described by a phase-space
distribution $f (\bit{r}, \bit{p})$ representing the density of particles at
a phase-space point $(\bit{r},\bit{p})$. Time dependence of the distribution
is governed by the Boltzmann equation (or Liouville equation, in case the
particles are not interacting). In quantum mechanics, the position and
momentum operators do not commute; hence, in principle one cannot talk
about the joint momentum and position distribution of particles. Indeed,
a quantum-mechanical wave function depends either on spatial coordinates
or on the momenta, but never both, unless they are not reciprocal variables.

The Wigner distribution can, to some extent, be thought of as a distribution
of a quantum particle in the position and momentum spaces simultaneously.
The nonclassical feature of the Wigner function is that it cannot be fully
treated as a bona fide joint density of probability since it possesses
patches of ``negative probability" due to interference of quantum phases.
The quantum-mechanical uncertainty principle restricts the amount of
localization that a Wigner distribution can have. Therefore, the uncertainty
principle leads to a ``fuzzy'' phase-space description of the system
compared to the ``sharp'' determination of its momentum or coordinate observables
separately. In the classical limit, $W(\bit{r}, \bit{p})$ reduces to the usual
phase-space distribution and hence becomes positive definite. The Wigner
quasi-probability provides an appealing opportunity to describe a quantum
state using the classical concept of the phase space. The advantage of the
Wigner function is that it is real and it contains complete information about
the interference.

\item One can calculate any dynamical quantity $\hat{{\cal O}} (\hat{\bit{r}},
\hat{\bit{p}})$ by performing averages of the Wigner distribution (as if it
were a classical distribution) with the symbol of the operator ${\cal O}
(\bit{r}, \bit{p})$, where the momentum and the coordinates are ordered
according to the Weyl association rule \cite{HilOcoScuWig84},
\begin{equation}
\langle \hat{{\cal O}} \rangle
=
\int d^3 \bit{p} \, d^3 \bit{r} \, W (\bit{r}, \bit{p})
{\cal O}_{\rm Weyl} ( \bit{r}, \bit{p} )
\, .
\end{equation}

\item The marginal projections of the Wigner function lead to the familiar
space and momentum probability densities, namely;
\begin{equation}
\int d^3 \bit{p} \, W (\bit{r}, \bit{p})
=
\rho (\bit{r})
\, , \qquad
\int \frac{d^3 \bit{r}}{(2 \pi \hbar)^3} \, W (\bit{r}, \bit{p})
=
n (\bit{p})
\, .
\end{equation}

\end{itemize}

\subsubsection{Wigner distribution of a quantum oscillator}

\begin{figure}[t]
\unitlength1mm
\begin{center}
\mbox{
\begin{picture}(0,40)(100,0)
\put(20,0){\insertfig{5}{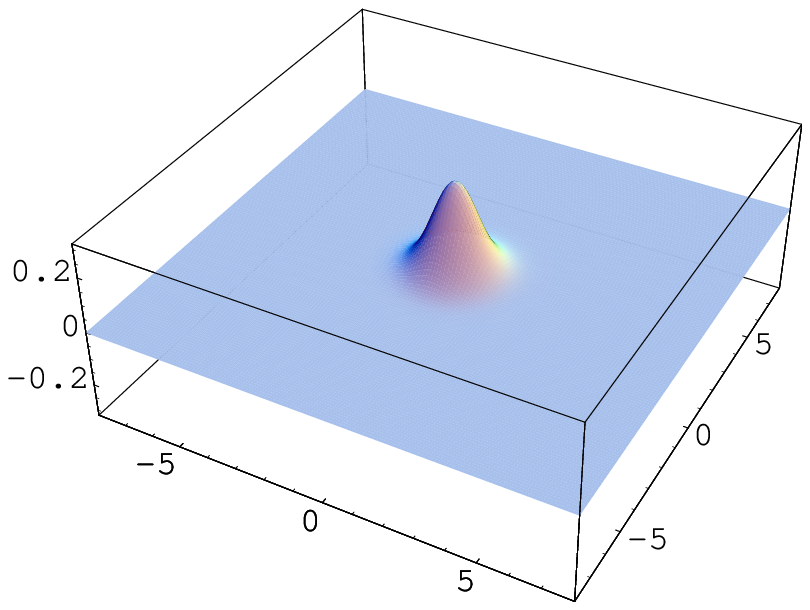}}
\put(75,0){\insertfig{5}{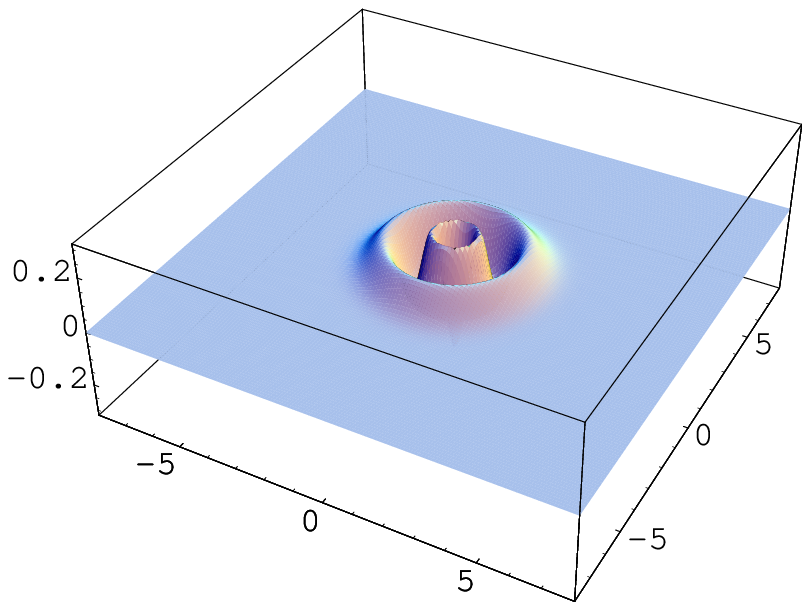}}
\put(130,0){\insertfig{5}{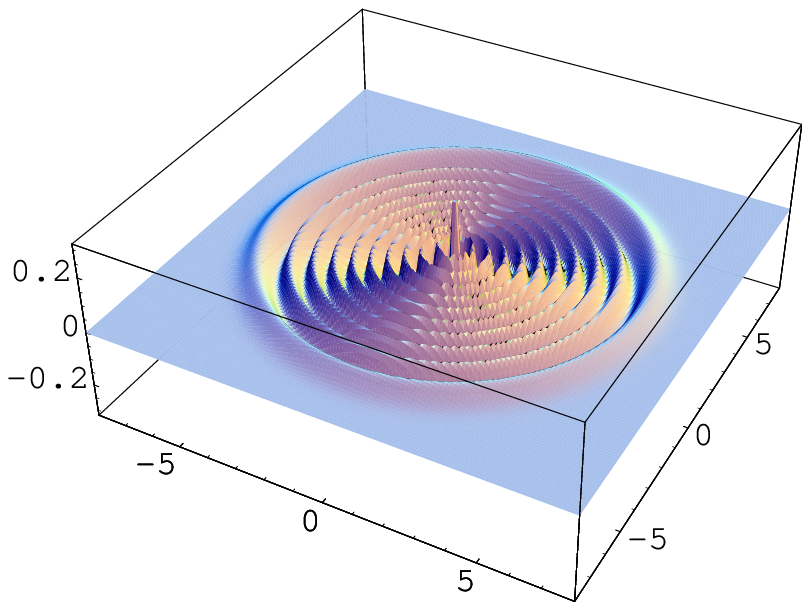}}
\end{picture}
}
\end{center}
\caption{\label{WignerQMO} The Wigner function for lowest-lying quantum states
of the harmonic oscillator, $n = 0$, $n = 2$ and higher level $n = 10$. The
peak for $n = 0$ state is the most probable phase-space point of a particle at
rest. The most probable orbit of the quantum oscillator is shown by the outermost
circular orbit in phase space from the solution of classical equations of motion
for the classical oscillator, see $n = 10$.}
\end{figure}

Let us illustrate the concept of the Wigner distribution using as an example
the simplest system: the one-dimensional harmonic oscillator. The Wigner
distribution for the $n$th excited state of the one-dimensional harmonic
oscillator of energy $E_n = \hbar \omega \left( n + \ft12 \right)$ is \cite{Gro46}
\begin{equation}
W_n (r, p)
=
\frac{(-1)^n}{\pi\hbar}
{\rm e}^{- 2 H / (\hbar\omega)}
L_n \left( \frac{4 H}{\hbar\omega} \right)
\, ,
\end{equation}
where $H$ stands for the classical Hamiltonian
$$
H(r, p) = \frac{p^2}{2 m} + \frac{m \omega^2 r^2}{2}
\, ,
$$
and $L_n$ is the $n$th order Laguerre polynomial. Figure \ref{WignerQMO}
gives a graphical demonstration of a few lowest-order Wigner distributions.

In the classical limit, the Wigner distribution is expected to become
a classical phase-space distribution. For systems which are statistical
ensembles, the limit $\hbar \rightarrow 0$ is often well behaved. For
example, for an ensemble of harmonic oscillators at finite temperature,
the Wigner distribution becomes the classical Boltzmann distribution as
$\hbar \rightarrow 0$, see, e.g., \cite{DraHab98}. For a single-particle
state, discussed here, the limit is more subtle. In the quasiclassical
limit---vanishing Planck constant and large quantum numbers---the Wigner
distribution of the harmonic oscillator turns into a mathematical
distribution which resides on classical trajectories, $E_\infty = n \hbar
\omega = {\rm fixed}$
\begin{equation}
\label{QClassWigner}
\lim_{ \hbar \to 0, \, n \to \infty }
W_n (r, p)
\sim
\delta \Big( H (r, p) - E_\infty \Big)
\, ,
\end{equation}
and implies that in the quasiclassical limit most of oscillations with $H (r, p)
< E_\infty$ nearly average to zero after integration with a smooth test function.
This result can be easily understood from the semiclassical form of the wave
function,
\begin{equation}
\psi (r) = C (r) {\rm e}^{i S (r)/\hbar}
\, ,
\end{equation}
with the classical action $S$. Substituting this into Eq.\ (\ref{QMWigner}),
reduced to the one-dimensional case, and expanding $S$ to first order in $r$,
one gets the quasiclassical Wigner distribution
\begin{equation}
W (r, p) = |C|^2 \delta \left( p - \frac{\partial S (r)}{\partial r} \right)
\, ,
\end{equation}
where the argument of the delta function describes a family of classical
trajectories, cf.\ Eq.\ (\ref{QClassWigner}).

The Wigner function incorporates quantum mechanical effects and at the
same time can be viewed as an analogue of the classical phase-space
distribution, encoding the distribution of observable quantities.

\subsubsection{Experimental access to Wigner distributions}
\label{QMWignerExperiment}

\begin{figure}[t]
\unitlength1mm
\begin{center}
\mbox{
\begin{picture}(0,45)(80,0)
\put(30,0){\insertfig{9}{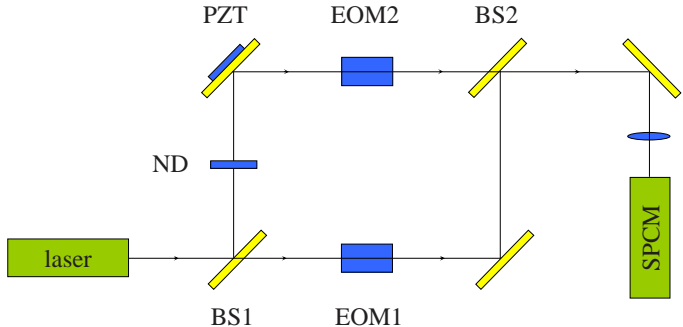}}
\end{picture}
}
\end{center}
\caption{\label{MachSender} Mach-Zender interferometric scheme for the measurement
of the quantum-mechanical Wigner distribution of a light mode. BS1 and BS2 denote
the low-reflection beam splitters. The quantum state is prepared using the neutral
density filter ND and a mirror mounted on a piezoelectric translator PZT. The
electro-optic modulators EOM1 and EOM2 control, respectively, the amplitude and
the phase of the point at which the Wigner function is measured. The signal field
is focused on a single photoncounting module SPCM.}
\end{figure}

The usefulness of the Wigner distribution concept from the viewpoint of
phenomenological applications is driven by the possibility of its experimental
measurement. To this end, one has to be able to access the correlations of
wave functions. The quantum-mechanical Wigner distribution is indeed a measurable
observable. The actual measurement has been performed for a simplest physical
quantum system---the quantum state of a light mode (a pulse of laser light of a
given frequency)---employing ideas of Vogel and Risken \cite{VogRis89}. It was
accessed via the method of homodyne tomography \cite{SmiBacRayFar93}, which is
based on measuring marginal observables and subsequent reconstruction by the
inverse Radon transformation (see, \cite{Leo97} for a review and Section
\ref{Sec-DD} below). The same Wigner distribution was accessed also by a more
direct measurement by means of the photon counting techniques based on a
Mach-Zender interferometric scheme \cite{BanRadWodKra99}; see Fig.\ \ref{MachSender}.
As we will see later in this paper, the experimental setup to measure the nucleon's
content via the concept of generalized parton distribution (GPDs) has very close
resemblance of this setup.

\subsection{Nucleon observables}

Both marginal projections of the Wigner dis\-tri\-bution in quantum
mechanics---the charge distribution and the momentum density---have their
analogues in QCD. Moreover, they share to some extent the probabilistic
interpretation of their non-relativistic ``cousins'', though not literally,
since they are plagued by complications due to the relativistic nature of the
problem and the settings where these observables are measured. In particular,
exploring the internal structure of the proton or neutron two methods are
conventionally used: (i) elastic reactions which measure form factors, and
(ii) inelastic scattering experiments which accesses momentum distributions
of nucleon's constituents.

Both approaches are complementary, but bear similar drawbacks. The form
factors do not yield direct information about the velocity of the constituents,
whereas the momentum distribution does not give information on their spatial
location. Quantum-mechanically, the phase structure is totally washed out as
well. As we emphasized in the previous section, more complete description of
the microscopic state lies in the correlation between the momentum and coordinate
space distributions: information of where a particle is located and, at the
same time, with what velocity it travels. However, the quantum-mechanical
interpretation will be obviously affected not only by the relativistic effects
we mentioned before but also by accommodation of the intrinsic interference
phenomena. Nonetheless, the notion of correlated position and momentum
distributions of quarks and gluons is very interesting, and it is clear that
the physics of a phase-space distribution must be very rich. Let us start by
reviewing in what circumstances the relativistic form factors and parton
distributions survive the nonrelativistic statistical interpretation.

\subsubsection{Nucleon electromagnetic form factors}

In this subsection, we re-examine  interpretation of electromagnetic form factors
of the proton regarding intrinsic ambiguities associated with  this interpretation.
The electromagnetic form factors are defined by the matrix element between the
nucleon states with different four-momenta of the quark electromagnetic current,
\begin{equation}
\label{QuarkEMcurrent}
j^\mu (x)
=
\sum_{q = u, d, \dots} Q_q \bar\psi_q (x) \gamma^\mu \psi_q (x)
\, ,
\end{equation}
where the quark charges take values
$$
Q_u = Q_c = \frac{2}{3} \, , \qquad Q_d = Q_s = - \frac{1}{3}
\, .
$$
Because the nucleon is a spin one-half particle, the matrix element is parametrized
by two form factors,
\begin{equation}
\label{JintermsofF12}
\langle p_2 | j^\mu (0) |p_1 \rangle
=
\bar u (p_2)
\left\{ \gamma^\mu F_1( \Delta^2)
+
\frac{i \sigma^{\nu\mu} \Delta_\nu}{2M_N}
F_2 (\Delta^2)
\right\}
u (p_1)
\, ,
\end{equation}
known as Dirac $F_1$ and Pauli $F_2$ form factors. They depend on the momentum
transfer\footnote{Usually, a convention with the opposite sign is used for the
momentum transfer in form factor  definitions, $\Delta = p_2 - p_1$. The motivation
is clear: the initial hadron gets the momentum transfer $p_2-p_1$ from the
electron. However, the most important class of physical processes involving GPDs
is deeply virtual photon or meson electroproduction, in which a spacelike virtual
photon converts into a real photon or a meson, requiring a  positive energy transfer
$E_1 - E_2$ from the initial hadron to the virtual photon. To avoid  writing
\textsl{positive}  quantities as \textsl{minus negative} ones, we will stick to
the convention $\Delta = p_1 - p_2$ throughout the paper.} $\Delta = p_1 - p_2$.
Our conventions for the Dirac matrices and the nucleon bispinor $u (p)$ are
summarized in Appendix \ref{conventions}.

At zero momentum transfer,  the form factors are normalized to static
properties of the nucleon. They can be inferred by studying the
zero-recoil limit of the matrix element, $p = p_1 \to p_2$. Introducing
the operators of charge and the magnetic moment at a time slice $t = 0$
(they do not change in time due to the current conservation)
\begin{equation}
Q
\equiv
\int d^3 \bit{r} \, j^0 (\bit{r})
\, , \qquad
\bit{\mu}
\equiv
\int d^3 \bit{r} \, [ \bit{r} \times \bit{j} (\bit{r}) ]
\, ,
\end{equation}
one finds the normalization
\begin{equation}
\label{FFnormalization}
\frac{ \langle p | Q | p \rangle }{ \langle p | p \rangle }
=
F_1 (0)
\, , \qquad
\frac{ \langle p | \bit{\mu} | p \rangle }{ \langle p | p \rangle }
=
\frac{\bit{s}}{M_N}
\left( F_1 (0) + F_2 (0) \right)
\, ,
\end{equation}
where the three-vector of spin $\bit{s} = \ft12 w^\ast \bit{\sigma} w$ is expressed
in terms of Weyl spinors $w$ (see Appendix \ref{Spinors}). The above relation for
the magnetic moment can be easily derived by introducing a Fourier transform of the
matrix element of spatial current, and expanding the exponential to the linear term
in the momentum transfer $\bit{\Delta}$;
$$
\int d^3 \bit{r} \,
{\rm e}^{- i \bit{\scriptstyle\Delta} \cdot \bit{\scriptstyle r}}
\langle p_2 | \bit{j} (\bit{r}) | p_1 \rangle
\approx
\int d^3 \bit{r} \, (1 - i \bit{\Delta} \cdot \bit{r})
\langle p_2 | \bit{j} (\bit{r}) | p_1 \rangle
$$
Making use of the identity $(\bit{r} \cdot \bit{\Delta}) \bit{j} = [[ \bit{r} \times
\bit{j} ] \times \bit{\Delta}]$ and the condition of the steady electromagnetic current
$\bit{\nabla} \cdot \bit{j} = 0$, one compares the matrix element to the expression
in term of form factors to find the equality (\ref{FFnormalization}).

It was realized long ago that the physical interpretation of the nucleon
form factors in terms of charge and magnetization distributions is obscured
by relativistic effects \cite{YenRavLev57}. Consider a system of size $R$ and
mass $M$. In a relativistic quantum theory, the system cannot be localized to
a precision better than its Compton wavelength $1/M$ (starting from this
section, we set $\hbar = 1$). As a consequence, the static size of the system
cannot be defined to a precision better than $1/M$. If $R \gg 1/M$, which is
the case for all nonrelativistic systems, this is not a significant constraint.
One can probe the internal structure of such a system with a wavelength
$\lambda \sim 1/|\bit{\Delta}|$ comparable to or even much smaller than $R$,
but still large enough compared to $1/M$ so that the probe does not induce an
appreciable recoil in the test system. A familiar example is the hydrogen atom
for which $R_{\rm atom} M_{\rm atom} \sim M_{\rm atom} /(m_e \alpha_{\rm em})
\sim 10^5$, and the form factor can be measured through electron scattering
with momentum transfer $|\bit{\Delta}| \ll M_{\rm atom}$.

When the probing wavelength is comparable to $1/M$, the form factors are
no longer determined by the internal structure alone. They also contain the
dynamical effects of Lorentz boosts because the initial and final nucleons
have different momenta due to non-negligible recoil. In a relativistic quantum
theory, the boost operators involve nontrivial dynamical effects which result
in the nucleon wave function being different in different frames (in the
instant form of quantization). Therefore in the region $|\bit{\Delta}| \sim M$,
the physical interpretation of  form factors is complicated because of the
entanglement of the internal and the center-of-mass motion in relativistic
dynamics. In the limit $|\bit{\Delta}| \gg M$, form factors depend crucially
on the physical mechanism producing the overall change of the nucleon momentum.
The structural effect involved is a very small part of the total nucleon wave
function since it corresponds to a few lowest Fock states only.

For the nucleon, $M_N R_N \sim 4$. Although much less certain than in the case
of the hydrogen atom, it still seems sensible to have a rest-frame picture for
the electromagnetic form factors, as long as one keeps in mind that equally
justified definitions of the nucleon sizes can differ by the effects of order
$1/M_N (R_N M_N)$. The form factors at $|\bit{\Delta}| \ge M_N \sim 1 \,
{\rm GeV}$ cannot be interpreted solely as information gained about the internal
structure of the nucleon.

To further clarify the uncertainty involved in the interpretation of the
electromagnetic form factors, let us review an explanation offered originally
by Sachs \cite{Sac60} and recently re-examined in Ref.\ \cite{BelJiYua03}.

\subsubsection{Form factors in the Breit frame and charge distributions}
\label{BreitFrameFFs}

To establish the notion of a (charge) distribution, one needs to create a
wave packet representing a static nucleon localized at $\bit{R}$,
\begin{equation}
\label{LocalizedNucleon}
| \bit{R} \rangle
=
\int \frac{d^3\bit{p}}{(2\pi)^3} \,
{\rm e}^{i \bit{\scriptstyle p} \cdot \bit{\scriptstyle R}} \,
{\mit\Psi} (\bit{p})
| \bit{p} \rangle
\, ,
\end{equation}
where ${\mit\Psi} (\bit{p})$ is the momentum space profile. The plane wave
state $|\bit{p}\rangle$ is normalized in a relativistic-invariant manner, as
defined in Appendix \ref{conventions}, so that
$$
\int \frac{d^3\bit{p}}{(2 \pi)^3}
\, (2 E_{\bit{\scriptstyle p}}) |{\mit\Psi} (\bit{p})|^2
=
1
$$
in order to have $\langle \bit{R} | \bit{R} \rangle = 1$. The wave packet
${\mit\Psi}$ is not an eigenstate of the free Hamiltonian. Therefore, as
time progresses, the wave packet will spread. The characteristic dispersion
time is proportional to
$$
t_{\rm char}
\sim
\int d^3\bit{p} \, (M_N/\bit{p}^2) |{\mit\Psi} (\bit{p})|^2
\, .
$$
It is long for a nonrelativistic system. But for a relativistic particle,
the spread could happen much faster compared to the characteristic
time scale of a weakly-interacting probe. The actual form of the wave
packet profile, however, does not matter since we want to capture
structural information of the particle itself and not of the auxiliary
wave packet.

\begin{figure}[t]
\unitlength1mm
\begin{center}
\mbox{
\begin{picture}(0,65)(30,0)
\put(0,0){\insertfig{6}{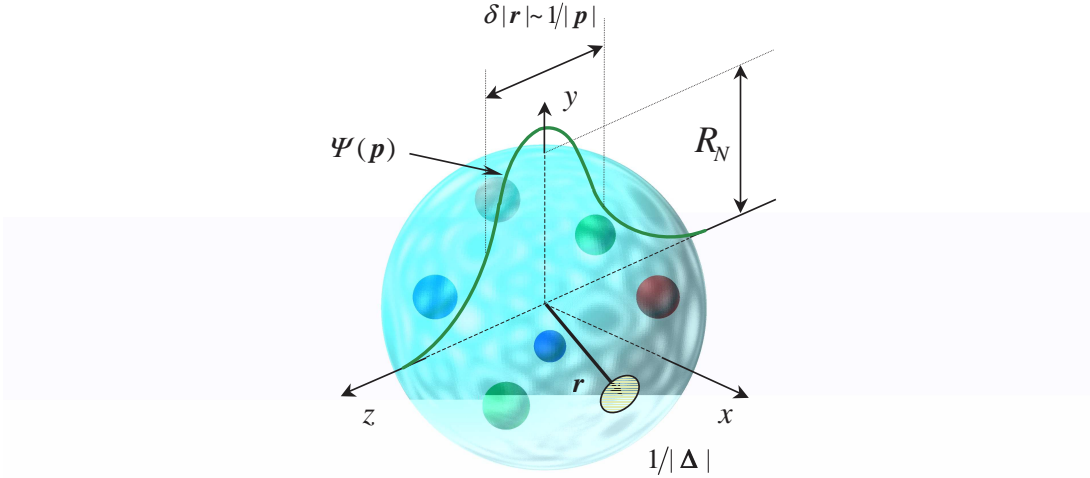}}
\end{picture}
}
\end{center}
\caption{\label{PhysicsFF} Localization of the nucleon with a wave packet.}
\end{figure}

Having localized the wave packet at $\bit{R} = 0$, we can use Eq.\
(\ref{LocalizedNucleon}) to calculate, for example, the charge
distribution in the wave packet,
\begin{equation}
\rho (\bit{r})
=
\langle \bit{R} = 0| j^0 (\bit{r}) | \bit{R} = 0 \rangle
\, ,
\end{equation}
where $\bit{r}$ measures the relative distance to the center, $\bit{R} = 0$,
see Fig.\ \ref{PhysicsFF}. Taking its Fourier transform, one gets
\begin{eqnarray}
F (\bit{\Delta})
\!\!\!&\equiv&\!\!\!
\int d^3 \bit{r} \,
{\rm e}^{- i \bit{\scriptstyle \Delta} \cdot \bit{\scriptstyle r}}
\rho (\bit{r})
\nonumber \\
&=&\!\!\!
\int
\frac{
d^3 \bit{p}
}{
(2\pi)^3
}
{\mit\Psi}^* \left( \bit{p} - \ft12 \bit{\Delta} \right)
{\mit\Psi} \left( \bit{p} + \ft12 \bit{\Delta} \right)
\left\langle \bit{p} - \ft12 \bit{\Delta} \right|
j^0 (0)
\left| \bit{p} + \ft12 \bit{\Delta} \right\rangle
\, ,
\end{eqnarray}
where we have changed the momentum integration variables, with $\bit{p}$
representing now the average momentum of the initial and final nucleons.
It is important to emphasize that the resolution momentum $\bit{\Delta}$
is now linked to the difference in the initial and final state momenta.
In non-relativistic quantum systems, because of large masses involved, the
momentum transfer causes little disturbance in velocity, and hence the
initial and final states have practically the same internal wave
functions. In relativistic systems, this is the origin of the difficulty
in interpreting the form factor: we do not have a matrix element involving
the same nucleon state before and after the interaction.

To have a situation free from ambiguities, several conditions have to be
imposed on momenta to avoid dangerous regions where the interpretation can
be affected by relativity, wave nature of the microscopic objects or the
wave packet used for the localization. Namely (see Fig.\ \ref{PhysicsFF}):

\begin{itemize}
\item To remove the effects of the wave packet, the necessary condition
on ${\mit\Psi} (\bit{p})$ is that the coordinate-space size of the
wave packet must be much smaller than the size of the test system,
the nucleon,
$$
\delta |\bit{r}| \ll R_N
\, .
$$
\item The probing wavelength, or the resolution scale, must be large compared
with the size of the wave packet, since one is not interested in details
of the wave packet itself:
$$
\delta |\bit{r}| \ll 1/|\bit{\Delta}|
\, .
$$
\item The size of the wave packet must be larger than the Compton
wavelength of the proton so as to be insensitive to the wave nature
of the proton. This results in
$$
\delta |\bit{r}| \gg 1/M_N
\, .
$$
The corresponding restriction on momenta allowed in the wave packet is
$|\bit{p}| \ll M_N$.
\end{itemize}
Therefore, the combined constraint on the wave packet profile is
\begin{equation}
\label{CombinedConstraint}
1/R_N \ll |\bit{\Delta}| \ll |\bit{p}| \ll M_N
\, ,
\end{equation}
where the size of the wave packet is $\delta |\bit{r}| \sim 1/|\bit{p}|$.
It is easy to see that the available window in the case of the proton is
very narrow since $R_N M_N \sim 1/4$.

The first two conditions allow us to ignore the $\bit{\Delta}$ dependence
in ${\mit\Psi}$ so that ${\mit\Psi} \left( \bit{p} \pm \ft12 \bit{\Delta}
\right) \approx {\mit\Psi} (\bit{p})$ and, therefore,
\begin{equation}
F (\bit{\Delta})
=
\int \frac{d^3 \bit{p}}{(2\pi)^3}
| {\mit\Psi} (\bit{p}) |^2
\left\langle \bit{p} - \ft12 \bit{\Delta} \right|
j_0 (0)
\left| \bit{p} + \ft12 \bit{\Delta} \right\rangle
\, .
\end{equation}
The extreme limit of the last inequality in Eq.\ (\ref{CombinedConstraint}) yields
a wave packet with a zero-momentum nucleon
\begin{equation}
| {\mit\Psi} (\bit{p}) |^2
=
\frac{(2\pi)^3}{2 M_N} \delta^{(3)} ( \bit{p} )
\, .
\end{equation}
Thus, one gets,
\begin{equation}
\label{FFF}
2 M_N F (\bit{\Delta})
=
\left\langle -\ft12 \bit{\Delta} \right|
j_0 (0)
\left|  \ft12 \bit{\Delta} \right\rangle
\, .
\end{equation}
This is the matrix element of the charge density in the Breit frame. The latter
has a unique property of the absence of the energy transfer from the incoming
to the outgoing nucleons, so that the hadronic four-momenta read
\begin{equation}
\label{BreitFrameMomenta}
p_1 = \left( E,  \ft12 \bit{\Delta} \right)
\, , \qquad
p_2 = \left( E, - \ft12 \bit{\Delta} \right)
\, .
\end{equation}
Thus the form factor (\ref{FFF}) is related (up to a factor) to the Sachs
electric form factor as
$$
F (\bit{\Delta}) = G_E ( - \bit{\Delta}^2) w_2^\ast w_1
$$
which is a superposition of the Dirac and Pauli form factors
\begin{equation}
G_E (- \bit{\Delta}^2) = F_1
( - \bit{\Delta}^2) - \frac{\bit{\Delta}^2}{4 M_N^2} F_2
( - \bit{\Delta}^2)
\, .
\end{equation}
The Weyl spinors
$w_2^\ast$, $w_1$ are the remnants of the Dirac bispinors. Hence, we
arrive at a textbook interpretation of $G_E$ as a Fourier transform of
the nucleon charge distribution normalized to the electric charge
(\ref{FFnormalization}). Likewise, the magnetic form factor $G_M$ is
a Fourier transform of the magnetization distribution
\begin{eqnarray}
\left\langle -\ft12 \bit{\Delta} \right|
\bit{j} (0)
\left|  \ft12 \bit{\Delta} \right\rangle
=
2 i [ \bit{\Delta} \times \bit{s} ]
G_M (- \bit{\Delta}^2)
\, ,
\end{eqnarray}
where
\begin{equation}
G_M (- \bit{\Delta}^2)
=
F_1 ( - \bit{\Delta}^2) + F_2 ( - \bit{\Delta}^2)
\, .
\end{equation}

\begin{figure}[t]
\unitlength1mm
\begin{center}
\mbox{
\begin{picture}(0,30)(25,0)
\put(0,0){\insertfig{5}{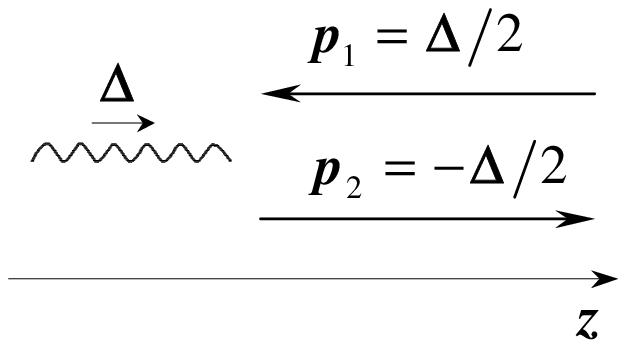}}
\end{picture}
}
\end{center}
\caption{\label{BreitFrame} Breit frame for the $\gamma^\ast p \to p'$
process.}
\end{figure}

Several times throughout our discussion we quoted  the proton radius.
What was meant by this is the root-mean-square radius, defined as
\begin{equation}
\label{3DprotonRadius}
R_N^2
=
\langle \bit{r}^2 \rangle_E
\equiv
\int d^3 \bit{r} \, \bit{r}^2 \rho (\bit{r})
=
- 6 \frac{\partial G_E (- \bit{\Delta}^2)}{\partial \bit{\Delta}^2}
\, ,
\end{equation}
and related to the slope of the electric form factor $G_E$. In the literature,
one can find other definitions of the proton charge radius, for instance,
$\langle \bit{r}^2 \rangle_D$, which is related to the slope of the Dirac form
factor $F_1$, rather than $G_E$.

The helicity of the proton is well defined in the Breit frame since the
incoming and outgoing particles' momenta are collinear to the same
direction. In this reference frame, the helicity non-flip transition is
determined by the electric form factor $G_E$, while the helicity-flip
transition is determined by the magnetic one, $G_M$. Of course, these
assignments are frame dependent since by a Lorentz boost one can reverse
the direction of the proton motion and thus changes the sign of proton's
helicity.

As we already emphasized before, the exceptional role of the Breit frame
is the absence of the energy exchange, distinguishing this unique frame
by the possibility of an unambiguous definition of the spatial Fourier
transform of the transition matrix elements of the electromagnetic current.
Of course, charge and magnetization distributions  defined in such a way
are not equivalent to the rest-frame densities. This is  due to the presence
of the Lorentz contraction effects along the direction $\bit{\Delta}$ when
$\bit{\Delta}^2 \gg 4 M_N^2$, which makes the nucleon look like a pancake.
If one attempts to compute a form factor boosted to the rest frame of the
nucleon, the Lorentz transformation tends to cut off its the momentum
dependence at high momentum transfer \cite{LicPag70} so that one effectively
finds $F \left( \Delta^2 \right) \to F \left( \Delta^2/\sqrt{1 - \Delta^2/4
M_N^2} \right)$. However, the procedure is not unique and thus it merely
results in yet another definition of what one means by the charge distribution.
Several types of prescriptions for extracting the static charge distributions
from available data on form factors have been used in recent analyses (see
\cite{Kel02}). Another physically-motivated approach is to find a relation
between the Fourier transform of a static charge density and the form
factor of a model obtained by boosting the wave function. For instance,
one can obtain such relations in the context of the Skyrme model \cite{Ji91}.
Different models lead to relations which differ by corrections of order $1/M_N$.

As we will establish later in Sect.\ \ref{FFinBjorkenFrame}, relativistic
corrections and Lorentz contraction effects are found to disappear in the
infinite momentum frame. In this frame,  the nucleon has an infinitely
large effective mass; hence, for physics in the transverse dimensions,
we are back to the nonrelativistic case. In particular, one can localize
the nucleon in the transverse coordinate space with no recoil corrections
at all. The Dirac form factor $F_1$ is found to be related to the charge
distribution in transverse plane, with information along the longitudinal
$z$-direction integrated out. The price one pays for eliminating the
relativistic corrections is one spatial dimension.

\subsubsection{Nucleon structure functions}

Let us discuss now the   observables probing the nucleon content through
inelastic processes, namely, the lepton-hadron deeply inelastic scattering.
Right before the incoming lepton hits the target, it fluctuates into a
lepton and a photon, $e (k) \to e (k') \gamma^\ast (q)$; the latter interacts
with the target $| p \rangle$ fragmenting into a number of hadrons in the
final state $| n \rangle$ with the total momentum $P_n = \sum_{k = 1}^N p_k$.
Since the electromagnetism is very weak compared to strong interactions
involved in the reaction dynamics, one can restrict the analysis to a
single-photon exchange.

\begin{figure}[t]
\unitlength1mm
\begin{center}
\mbox{
\begin{picture}(0,28)(25,0)
\put(0,0){\insertfig{5}{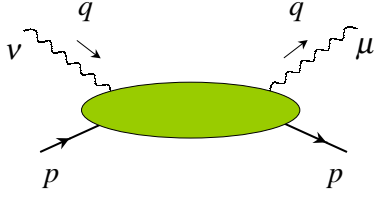}}
\end{picture}
}
\end{center}
\caption{\label{DeepInelasticTensor} Hadronic tensor of deep inelastic
scattering cross section determining the imaginary part of the forward
Compton scattering amplitude $\gamma^\ast (q) N (p) \to \gamma^\ast (q)
N (p)$.}
\end{figure}

The amplitude of this process is described by the current-to-current coupling,
\begin{equation}
{\cal A}_n
=
L_\mu (k, k') \langle n | j^\mu (0) | p \rangle \, ,
\end{equation}
of the hadronic transition amplitude $\langle n | j_\mu | p \rangle$ via
the local quark electromagnetic current (\ref{QuarkEMcurrent}) and the
leptonic current
$$
L^\mu (k, k') = \frac{i}{q^2} \bar u (k') \gamma^\mu u (k)
\, .
$$
The measurement is totally inclusive with respect to the final states and only
the scattered lepton is detected. Neglecting the phase-space factor, the
cross section reads:
\begin{equation}
\label{DIS-Compton}
\sigma_{\rm DIS}
=
\frac{\alpha_{\rm em}^2}{4 \pi}
\sum_n \left| {\cal A}_n \right|^2 (2 \pi)^4 \delta^{(4)} (p + q - P_n)
=
\frac{\alpha_{\rm em}^2}{q^4} L^\dagger_\mu L_\nu W^{\mu\nu}
\, ,
\end{equation}
where, using the completeness condition (\ref{CompletenessAppendix}) in summation
over the final states, we introduce the hadronic tensor
\begin{equation}
\label{HadronicTensorDIS}
W^{\mu\nu}
=
\frac{1}{4 \pi}
\int d^4 z \, {\rm e}^{i q \cdot z}
\langle p |
j^\mu (z) j^\nu (0)
| p \rangle
\, ,
\end{equation}
represented diagrammatically in Fig.\ \ref{DeepInelasticTensor} with the shown
assignments of indices and particle's momenta.

The decomposition of the hadronic tensor in independent Lorentz tensors
introduces the structure functions. For the spin one-half target the most
general form of $W^{\mu\nu}$ reads
\begin{eqnarray}
\label{DISstructurefunctions}
W^{\mu\nu}
\!\!\!&=&\!\!\!
- \left( g^{\mu\nu} - \frac{q^\mu q^\nu}{q^2} \right)
F_1 ( x_{\rm B}, {\cal Q}^2)
+
\frac{1}{p \cdot q}
\left(
p^\mu - \frac{p \cdot q}{q^2} q^\mu
\right)
\left(
p^\nu - \frac{p \cdot q}{q^2} q^\nu
\right)
F_2 ( x_{\rm B}, {\cal Q}^2)
\nonumber\\
&&
-
\frac{i}{p \cdot q} \varepsilon^{\mu\nu\rho\sigma} q_\rho s_\sigma
g_1 ( x_{\rm B}, {\cal Q}^2)
-
\frac{i}{p \cdot q} \varepsilon^{\mu\nu\rho\sigma} q_\rho
\left(
s_\sigma - \frac{s \cdot q}{p \cdot q} p_\sigma
\right)
g_2 ( x_{\rm B}, {\cal Q}^2)
\, .
\end{eqnarray}
The un- and polarized structure functions $F_i$ and $g_i$, respectively, depend
on  two variables: space-like momentum of the probe and the Bjorken variable,
\begin{equation}
\label{QcalAndBjorken}
q^2 \equiv - {\cal Q}^2
\, , \qquad
x_{\rm B} \equiv \frac{{\cal Q}^2}{2 p \cdot q}
\, .
\end{equation}
The hadronic tensor is related via the optical theorem---analogous to that
discussed in the Appendix \ref{OpticalTheorem} for the case of the vacuum
expectation value of two electromagnetic currents---to the imaginary part
\begin{equation}
W^{\mu\nu}
=
\frac{1}{2 \pi} \Im{\rm m} T^{\mu\nu}
\, ,
\end{equation}
of the forward Compton scattering amplitude
\begin{equation}
T^{\mu\nu}
=
\ i \! \int d^4 z \, {\rm e}^{i q \cdot z}
\langle p |
T \left\{
j^\mu (z) j^\nu (0)
\right\}
| p \rangle
\, ,
\end{equation}
determined by the chronological product of quark electromagnetic currents
(\ref{QuarkEMcurrent}).

In the deeply inelastic reaction, the photon interacting with the target acts as
a probe. The resolution is set by the inverse of the photon virtuality ${\cal Q}^2$.
Thus, the nucleon is probed with the resolution $\approx ( 0.2 \, {\rm fm} )/
({\cal Q} \, {\rm \, in \, GeV} )$. Since the nucleon's size is $R_N \sim 1 \,
{\rm fm}$, one concludes that for ${\cal Q}^2$ of order of a few GeV$^2$, the
photon penetrates the nucleon interior and interacts with its constituents. This
is the regime which initiates a hard scattering: the electron has to pass close
to one of the partons, i.e., at the distance $z^2 \sim 1/{\cal Q}^2$, to exchange
a photon of virtuality ${\cal Q}^2$. The structure functions have  a very important
property of scaling, i.e., ${\cal Q}^2$-independence, in the Bjorken limit;
$$
{\cal Q}^2 \to \infty
\, , \qquad
x_{\rm B} = \ \mbox{fixed}
\, .
$$
Analogous kinematical conditions will be imposed on other reactions
discussed in the present review which warrant the light-cone distance
dominance in scattering amplitudes and, hence, legitimate the application
of powerful methods of perturbative QCD.

\subsubsection{Infrared safety}

Let us study the question whether the hadronic tensor is indeed sensitive to
the structural information about  the nucleon. The argument is generic and
does not rely on the details of the interaction \cite{Ste00}; we have only to
keep in mind that the QCD coupling constant is growing at large space-time
scales. Thus we replace  QCD perturbation theory with a quantum-mechanical
one applying it to the Schr\"odinger equation with a scattering potential $V$
(Eq.\ (\ref{SchrodingerEq})). The solution of this equation will be expressed
in terms of asymptotic states at the infinite past where the interaction is
switched off adiabatically. The free-particle equation, $V = 0$, gives
$$
\psi_0 (t) = \exp (- i t E_m) | \omega_m \rangle
$$
with the stationary states which form a complete basis $\sum_m | \omega_m
\rangle \langle \omega_m | = 1$. The solution of the full equation is
expanded in free-particle states at the time $t$,
\begin{equation}
\psi (t) = \sum_m | \omega_m \rangle S_m
\, .
\end{equation}
$S_m$ is the evolution operator of the system from the unperturbed state
$\psi (- \infty) = | \omega_0 \rangle$, where $V (- \infty) = 0$, to an
observed state $| \omega_m \rangle$ at the time $t$ via a successive
sequence of interactions and the free propagation through intermediate states.
\begin{equation}
\label{QMscatteringAmplitude}
S_m
=
\int\limits_{- \infty}^{\infty} d \tau_m
{\rm e}^{- i ( E_{m - 1} - E_m ) \tau_m} V_{m - 1}
\int\limits_{- \infty}^{\tau_m} d \tau_{m - 1}
{\rm e}^{- i ( E_{m - 2} - E_{m - 1} ) \tau_{m - 1}} V_{m - 2}
\dots
V_1
\int\limits_{- \infty}^{\tau_2} d \tau_1 \
{\rm e}^{- i ( E_0 - E_1 ) \tau_2}
\, .
\end{equation}
The summation in repeated indices runs over states which are eigenfunctions of
the free equations of motion,  $V_m = \langle m | V | m \rangle$ is a matrix
element of a perturbation---the interaction potential---between the states
corresponding to  free particle propagation. The combined phase of the exponentials
$$
{\rm phase} = \sum_{n = 1}^m E_n (\tau_n - \tau_{n - 1})
\, .
$$
 is the free particle action: recall that $S = - \int d \tau H$.
At this point, one immediately concludes:
\begin{itemize}
\item  As long as the total phase---the free particle action---is time
dependent, there is no sensitivity to long-time scales where the coupling
constant grows. In this case, cancellations of long-time contributions occur
due to the oscillatory nature of the integrand.
\item However, if the phase is stationary, which corresponds to a stationary
classical action, particles travel along their classical trajectories and
the amplitude becomes sensitive to large-time dynamics.
\end{itemize}
Since $\alpha_s$ is large at large space-time scales, one cannot reliably
evaluate the amplitude using the  perturbative expansion, and the effects of
quark confinement are relevant. In the opposite situation, called ``the
infrared safety'' the quark confinement is not relevant, and the amplitude
can be computed from the perturbation theory alone to the leading power
accuracy in the momentum transfer $q^\mu$. This gives  an intuitive illustration
of the Coleman-Norton theorem \cite{ColNor65}, which is an essential part in
all the proofs of factorization theorems in QCD. Within the field-theoretical
framework, the above intuitive argument can be presented in a more rigorous
fashion in the form of the Landau equations for singularities of Feynman graphs.

However, there is only a very limited number of completely infrared safe
quantities. They include the  hadronic width of the electroweak $Z$-boson,
jet and  total inclusive cross sections in the electron-positron annihilation
into hadrons $e^+ e^- \to X$ \cite{EllStiWeb96}. In the latter case one has
\begin{equation}
\label{TotalXsection}
\sigma_{\rm tot} (q^2)
= \frac{1}{2 \pi} \sum_n \
\langle 0 | j_\mu (0) | n \rangle
\langle n | j^\mu (0) | 0 \rangle
\ (2 \pi)^4
\delta^{(4)} \left( \sum_i^n p_i - q \right)
\, ,
\end{equation}
where the  summation is performed with respect to the $n$-particle final states
having the total momentum $\sum_i^n p_i$. The cross section (\ref{TotalXsection})
is expressed, according to the optical theorem (see Appendix \ref{OpticalTheorem}),
in terms of the absorptive part of the vacuum polarization by
\begin{equation}
\label{InclusiveJetXsection}
\sigma_{\rm tot} (q^2) = \frac{1}{\pi} \Im{\rm m}
\ i \! \int d^4 z \, {\rm e}^{i q \cdot z}
\langle 0 |
T \left\{
j_\mu (z) j^\mu (0)
\right\}
| 0 \rangle \, ,
\end{equation}
The argument given above suggests that this quantity will be infrared safe.
Indeed, once the pair of quark and antiquark is created by the source $j_\mu$,
they travel back-to-back with the speed of light in the opposite directions
and cannot reassemble into a physical state absorbed by the $j^\dagger_\mu
= j^\mu$. Since there are no classical trajectories which would allow this
process, the observable is not sensitive to the infrared physics.

The discussion of the vacuum polarization in Minkowski space is complicated by
its non-analytic behavior in the form of branch cuts in $q^2$. They correspond
to the production thresholds of hadrons. On the other hand, the QCD description
involves quarks and gluons which are produced and propagate to infinite
distances. Of course, there is an implicit assumption here that the latter will
inevitably fragment into hadrons. A reliable theoretical evaluation can be
done only for the $q^2$ values away from the production of physical states. To
achieve this one has to use the dispersion relation which is the crucial element
of the optical theorem. Making the Wick rotation and going to Euclidean space
$- q^2 = {\cal Q}^2$  gives
\begin{equation}
\label{DispersionRele+e-}
i \! \int d^4 z \, {\rm e}^{i q \cdot z}
\langle 0 |
T \left\{
j_\mu (z) j^\mu (0)
\right\}
| 0 \rangle
= \int_0^\infty  d M^2 \frac{\sigma_{\rm tot} (M^2)}{{\cal Q}^2 + M^2} \, ,
\end{equation}
where the left-hand side is now an analytic function of ${\cal Q}^2$. In the deep
Euclidean region, ${\cal Q}_\mu \to \infty$, the distances probed in the current
correlator are very small, $z^\mu \to 0$, due to the Heisenberg uncertainty
principle, so one can safely use the QCD perturbation theory. When the
virtualities are not asymptotically large, one should include  power corrections
using  the local operator product expansion in short-distance
singularities:
\begin{equation}
j_\mu (z) j^\mu (0) = C_0 (z^2)   \1 + \sum_d C_d (z^2) {\cal O}_d (0)
\, ,
\end{equation}
where the first term is a purely perturbative contribution and the sum runs
over dimension-$d$ local operators.  For example, in the massless case the
lowest operator is ${\cal O}_4 (0) = F_{\mu\nu}^a (0) F^{\mu\nu}_a (0)$. After
the Fourier transformation, this series is translated into the power series
expansion in $1/{\cal Q}^2$. Matching both sides in Eq.\ (\ref{DispersionRele+e-})
one finds that the weighted integral of the physical cross section $\sigma_{\rm tot}
(q^2)$ is related to the quantities computed using the partonic language. Therefore,
the integral over the physical resonance spectrum corresponds to the integrals over
the quark continuum. This is the idea of the quark-hadron duality which states
that the physical cross section coincides in average with the partonic one.

The hadronic tensor, discussed in the previous section in relation to deeply
inelastic scattering, has a more complicated structure than the total $e^+e^-$
to hadrons cross section. The question is whether it can be expressed through
the same chronological product of two electromagnetic currents, albeit, in
different matrix elements? This is considered in the next section.

\subsubsection{Incoherence and scale separation}

A hard scattering cross section having at least one hadron in the initial
state (like the deeply inelastic scattering cross section) cannot be infrared
safe since in the preparation of the asymptotic hadron state its  constituents
have  strongly interacted among themselves for a long time in a bound state.
Obviously, the wave functions of quarks in a bound state differs from what it
would be if they were free, and this difference inevitably affects the cross
section.

\begin{figure}[t]
\unitlength1mm
\begin{center}
\mbox{
\begin{picture}(0,55)(85,0)
\put(55,0){\insertfig{6}{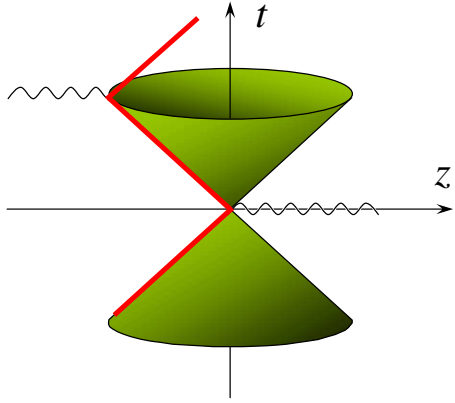}}
\end{picture}
}
\end{center}
\caption{\label{SpaceTimeTrajectory} Pinch surface: space-time trajectory of the
struck quark in the hadron.}
\end{figure}

Since the number of partons that carry the bulk of the hadron momentum
is small, the photon usually  will ``see'' only one parton per collision.
The probability for coherent scattering on an $n$-parton configuration
is suppressed by $n$th power of the photon virtuality,
\begin{equation}
\label{MultiPartScatDIS}
{\cal P}_n
\sim
\left( \frac{| \delta \bit{z}_\perp |^2 }{ \pi R_N^2 } \right)^n
\sim
\frac{1}{\left( {\cal Q}^2 \pi R_N^2 \right)^n}
\, ,
\end{equation}
where $\pi R_N^2$ is the transverse area of the nucleon. These
power-suppressed corrections go under the name of higher twists. Thus, in
leading approximation at high ${\cal Q}^2$ one can restrict all considerations
to the photon scattering on a single parton. This is known as handbag
approximation. Compared to the inclusive annihilation mentioned above, the
underlying physical picture for the forward Compton scattering on a quark
admits a classical trajectory, see Fig.\ \ref{SpaceTimeTrajectory}. A quark
taken from the hadron absorbs the virtual photon at $t = 0$ and, as a result,
accelerates. Then it re-emits a photon at later $t > 0$ and falls in the same
momentum state. After the energy is freed into the final state the parton
merges back into the parent hadron. As we already discussed above, the process
is not infrared safe and depends on the quark binding inside the nucleon.
The points of absorption and emission are separated by a light-like
distance. The character of relevant distances in the Compton amplitude
is a consequence of deep Euclidean kinematics, ${\cal Q}^2 \to \infty$. Large
virtualities, ${\cal Q}^2$, and energies, $\nu \equiv p \cdot q$, at fixed Bjorken
variable $x_{\rm B}$, probe short-distance and time structure of the
process, respectively. To derive the relevant distances in deeply
inelastic scattering,  let us switch  to a reference frame where the target
proton is at rest and the virtual photon's three-momentum points in the
direction opposite to the $z$-axis. Then
\begin{equation}
q^\mu
=
\left(
\frac{{\cal Q}^2}{2 M x_{\rm B}} ,
0,
0,
- \frac{{\cal Q}^2}{2 M x_{\rm B}} \sqrt{1 + 4 M^2 x^2_{\rm B}/{\cal Q}^2}
\
\right) \, .
\end{equation}
When ${\cal Q}^2$ is large, the  light-cone components of the momentum transfer
(see Appendix
\ref{LCvectors} for conventions) can be approximated by
\begin{equation}
q^- \sim {\cal Q}^2 / \left( M x_{\rm B} \right)
\, , \qquad
q^+ \sim M x_{\rm B}
\, .
\end{equation}
The integrand in Eq.\ (\ref{DIS-Compton}) is an oscillatory function and thus
gives vanishing result unless the distances involved are
\begin{equation}
z^- \sim 1 / \left( M x_{\rm B} \right)
\, , \qquad
z^+ \sim M x_{\rm B}/ {\cal Q}^2
\, .
\end{equation}
Therefore, provided transverse separations $z_\perp$ are small, the deeply
inelastic scattering probes strong interaction dynamics close to the
light-cone $z^2 \approx 0$, and we can neglect the dependence on all
coordinate components except for $z^-$. The latter is called the Ioffe
time \cite{Iof69} and has the meaning of the longitudinal distance probed
in the process. Its Fourier conjugate variable is the fraction
$x$ of the nucleon momentum carried by a parton interacting with the probe. In the
lowest order approximation,  $x = x_{\rm B}$.

The hard subprocess occupies a very small space-time volume. On the other
hand, the scales involved in the formation of the hadron nonperturbative
wave function are much larger, of order of a typical hadronic scale,
$1\ {\rm GeV}$. Hence, it is quite likely that the two scales are uncorrelated
and will not interfere. Thus, although the process depends on the hadronic
state from which a given constituent has come, this is basically irrelevant
for the  hard interactions. Moreover, all final state interactions
cancel in the deeply inelastic process. This is exhibited by the relation
(\ref{DIS-Compton}) following from the optical theorem. Thus, there is no
sensitivity to the soft final-state interaction and all information about long-distance
physics is encoded into a function which reflects the internal structure
of the proton, the so-called quark (gluon) distribution. The quantum
mechanical incoherence property of physics at different scales results
in the factorization property of the structure functions
(\ref{DISstructurefunctions})
\begin{equation}
\label{Factorization}
F_i \left( x_{\rm B}, {\cal Q}^2 \right)
=
\int_{x_{\rm B}}^1 \frac{d x}{x} \,
C_i \left( x_{\rm B}/x , {\cal Q}^2/\mu^2 \right)
\,
q ( x ; \mu^2 )
\, .
\end{equation}
Here $q$ is the quark distribution, which depends on  the momentum
fraction $x$ of the parent proton, and $C$ is a perturbatively computable
short-distance quark-photon cross section. This is the basis for the
predictive power of perturbative QCD.

The factorized expression for the structure functions has several
fundamental properties. The parton distributions are universal objects
and connect otherwise unrelated processes, like deeply inelastic
scattering and Drell-Yan production of lepton pairs in proton-proton
collisions $p p \to \ell^+ \ell^- X$, etc. Second, the formula implies
that the momentum scale dependence of leading term in $1/{\cal Q}^2$
expansion can be computed from  first principles. This will be the
subject of Section \ref{Evolution}.

\subsubsection{QCD parton distributions}
\label{QCDPartonDistributions}

\begin{figure}[t]
\begin{center}
\mbox{
\begin{picture}(0,80)(225,0)
\put(0,1){\insertfig{16}{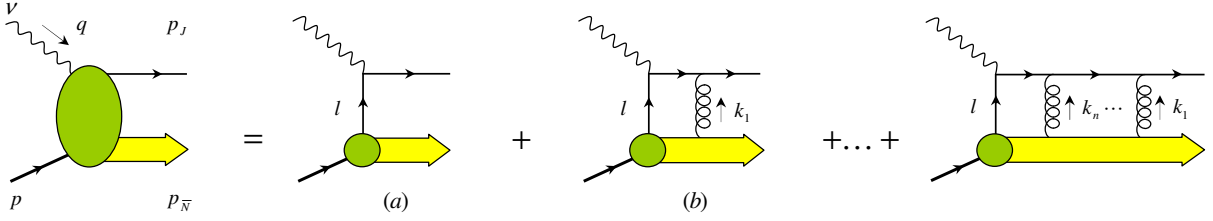}}
\end{picture}
}
\end{center}
\caption{\label{LinkDIS} Multi-gluon attachments to the struck quark in
deeply inelastic scattering which factorize to form the path-ordered
exponential.}
\end{figure}

Let us establish now  the operator content of QCD parton distributions.
The starting point of the analysis is  the expression (\ref{HadronicTensorDIS})
for the hadronic tensor. Inserting the complete set of hadronic states,
$\sum_n | n \rangle \langle n | = 1$, in the form (\ref{CompletenessExact}),
with a state $| n \rangle$ consisting of $n$ particles with the total momentum
$P_n = \sum_{k = 1}^n p_k$ and certain quantum numbers, one gets
\begin{equation}
\label{DIStensorOne}
W^{\mu\nu}
=
\frac{1}{4 \pi}
\sum_n \int \prod_{k = 1}^n
\frac{d^3 \bit{p}_k}{2 E_{\bit{\scriptstyle p}_k}(2 \pi)^3}
(2 \pi)^4 \delta^{(4)} \left( \sum_{l = 1}^n p_l - p - q \right)
\langle p | j^\mu (0) | p_1, \cdots, p_n \rangle
\langle p_1, \cdots, p_n | j^\nu (0) | p \rangle \, ,
\end{equation}
where the summation over $n$ implies the summation over the multiplicity
and quantum numbers of particles populating  the final state. For the
photon scattering on a single parton with momentum $\ell$, we separate
the final state into the current fragmentation and the target
fragmentation regions. The first one consists of an outgoing struck
quark which forms a jet with momentum $p_J$ and the second one is formed
by target spectators with momentum $P_{\bar{n}}$, $P_n = p_J +
P_{\bar{n}}$. The tree scattering amplitude corresponding to Fig.\
\ref{LinkDIS} (a) is given by
\begin{equation}
\label{TreeAmplitude}
\langle p_J, \bar{n} | j^\nu (0) | p \rangle_{(0)}
=
\bar u (p_J) \gamma^\nu \langle \bar{n} | \psi(0) | p \rangle \, ,
\end{equation}
where $\bar u$ is the Dirac spinor of the scattered quark. Substituting
Eq.\ (\ref{TreeAmplitude}) into (\ref{DIStensorOne}) we get the structure
function $F_1 (x_{\rm B}, Q^2)$;
\begin{eqnarray}
\label{PDFnongaugeinv}
F_1 (x_{\rm B}, {\cal Q}^2)
\!\!\!&=&\!\!\!
\frac{1}{2} \int^{{\cal Q}^2}
\!\! d^2 \bit{k}_\perp \, f^q (x_{\rm B}, \bit{k}_\perp)
\\
&=&\!\!\!
\frac{1}{4} \int^{Q^2} \!\! d^2 \bit{k}_\perp
\int \frac{d z^-}{2 \pi} \frac{d^2 \bit{z}_\perp}{(2\pi)^2}
{\rm e}^{
i x_{\rm B} p^+ z^-
-
i \bit{\scriptstyle k}_\perp \cdot \bit{\scriptstyle z}_\perp
}
\langle p |
\bar \psi (0, \bit{0})
\gamma^+
\psi ( z^-, \bit{z}_\perp )
| p \rangle
\,  . \nonumber
\end{eqnarray}
Here, $f^q (x_{\rm B}, \bit{k}_\perp)$ is the  transverse momentum-dependent
parton distribution in the tree-level approximation, with $x p^+ n^{\ast\mu}$
and $k^\mu_\perp$ being the light-cone and  transverse components of the quark
momentum $k^\mu$,
$$
k^\mu \approx  x p^+ n^{\ast\mu} + k^\mu_\perp
\, ,
$$
as explained in Appendix \ref{LCvectors}. Note, that the parton distribution
(\ref{PDFnongaugeinv}) is not gauge invariant as it stands, since the quark
fields are not at the same space-time point.

The gauge invariance of the parton distributions is restored after one performs
resummmation over the   multiple rescatterings of the observed final state jet
with the target fragments \cite{BroHoyMarPeiSan01,Col02,BelJiYua02}, see Fig.\
\ref{LinkDIS}. It is sufficient to consider the contribution from a single-gluon
exchange in Fig.\ \ref{LinkDIS} (b). It involves all the subtleties, the all-order
result being a straightforward generalization. The one-gluon amplitude is given by
\begin{equation}
\label{OneGluonAmplitude}
\langle p_J, \bar{n} | j^\nu (0) | p \rangle_{(1)}
=
g \bar u (p_J) \int \frac{d^4 k_1}{(2 \pi)^4}
\langle \bar n |
{\not\!\! A} (k_1) S (p_J - k_1) \gamma^\nu \psi(0)
| p \rangle \, ,
\end{equation}
with the free quark propagator $S (k)$.
It is  convenient to use now a frame similar to the Drell-Yan frame
defined in Appendix \ref{DYframe}. By looking at the light-cone expansion
of the quark propagator
\begin{equation}
S (p_J - k)
\approx
-
\frac{
p_J^- \gamma^+ + \not\!\bit{k}_\perp
}{
2 p^-_J k^+ + \bit{k}_\perp^2 - i0
} \, ,
\end{equation}
one immediately notices that the scaling contribution in the Bjorken
limit $p_{J -} \to \infty$ arises actually from two regions rather than
one.   They correspond to
\begin{itemize}
\item the extraction of the large $p^-_J$ component both from the numerator
and the denominator  (usually only this  term is considered);
\item scaling terms surviving the limit $k^+ \to 0$ before $p^-_J$ is sent
to infinity.
\end{itemize}
Obviously, the latter contribution is anomalous. Thus, we find for the
amplitude of the one-gluon exchange \cite{BelJiYua02},
\begin{eqnarray}
\label{Numerator}
\bar u (p_J)
\langle \bar{n} |
{\not\!\! A} (k) S (p_J - k) \gamma_\nu \psi (0)
| p \rangle
\!\!\!&\approx&\!\!\! - \frac{1}{k^+ - i 0}
\bar u (p_J) \gamma_\nu
\langle \bar{n} | A^+ (k_1) \psi (0) | p \rangle
\nonumber\\
&+&\!\!\! \bar u (p_J)
\frac{
\bit{\gamma}^\alpha_\perp {\not\! \bit{k}}_\perp \gamma_\nu
}{
\bit{k}_\perp^2 - i 0
}
\langle \bar{n} |
\bit{A}^\alpha_\perp (k^+ = 0, \bit{k}_\perp ) \psi (0)
| p \rangle
\, .
\end{eqnarray}
The first term comes from the expansion of the eikonal phase acquired by a
fast-propagating charge \cite{EfrRad80,ColSop82,LabSte85}. The second term
is something else. Its Fourier transform sets the argument of the gauge field
at $z^- = \infty$ \cite{BelJiYua02}. To see the effect of this term we make use
of the following representation of the transverse field \cite{BoeMulPij03}
\begin{equation}
\bit{A}^\alpha_\perp (\infty , \bit{z}_\perp)
=
- \bit{\nabla}^\alpha_\perp \int_{\bit{\scriptstyle z}_\perp}^\infty
d \bit{z}'_\perp \cdot \bit{A}_\perp (\infty , \bit{z}'_\perp)
\, ,
\end{equation}
with $\bit{\nabla}^\alpha_\perp = \partial/\partial\bit{z}^\alpha_\perp$.
The Fourier transform acquires a factor of transverse momentum
$\bit{\nabla}^\alpha_\perp \to i \bit{k}^\alpha_\perp$, and one sees that
the transverse propagator cancels and only the contour integral of the
gauge field in the transverse direction is left. Resumming to all orders,
one finds the complete result for the amplitude \cite{BelJiYua02}:
\begin{equation}
\label{FullAmplitude}
\langle p_J, \bar{n} | j^\nu (z) | p \rangle
=
\bar u (p_J) \gamma^\nu
\langle \bar{n} |
{\mit\Psi}_{\infty} (z_-, \bit{z}_\perp)
| p \rangle \, ,
\end{equation}
where we introduced a ``gauge-invariant'' quark field ${\mit\Psi}_{\infty}$.
Compared to the elementary quark field, it is augmented by the gauge links
attached to it,
\begin{equation}
\label{GaugeInvariantQuark}
{\mit\Psi}_{\infty} (z_-, \bit{z}_\perp)
\equiv
[\infty, \bit{\infty}; \infty, \bit{z}_\perp]
[\infty, \bit{z}_\perp; z^-, \bit{z}_\perp]
\, \psi (z_-, \bit{z}_\perp)
\, .
\end{equation}
One gauge link runs along the light cone
\begin{equation}
{}[\infty, \bit{z}_\perp; z^-, \bit{z}_\perp]
=
P \exp
\left(
i g \int_{z^-}^\infty d z'^- A^+ (z'^- , \bit{z}_\perp)
\right)
\, ,
\end{equation}
and then the integration path continues in the transverse direction producing
another gauge link
\begin{equation}
{}[\infty, \bit{\infty}; \infty, \bit{z}_\perp]
=
P \exp
\left(
- i g \int_{\bit{\scriptstyle z}_\perp}^\infty d \bit{z}'_\perp
\cdot
\bit{A}_\perp ( \infty , \bit{z}'_\perp )
\right)
\, .
\end{equation}

Multiplying (\ref{FullAmplitude}) by its complex conjugate, we obtain  the
expression for the gauge invariant transverse \mbox{momentum-dependent} parton
distribution \cite{BelJiYua02}
\begin{equation}
\label{DIStransversePDF}
f^q (x, \bit{k}_\perp)
=
\frac{1}{2}
\int \frac{d z^-}{2 \pi} \frac{d^2 \bit{z}}{(2 \pi)^2}
{\rm e}^{
i x p^+ z^-
-
i \bit{\scriptstyle k}_\perp \cdot \bit{\scriptstyle z}_\perp
}
\langle p |
\bar{\mit\Psi}_{\infty} (0, \bit{0})
\gamma^+
\, {\mit\Psi}_{\infty} (z^-, \bit{z}_\perp)
| p \rangle
\, .
\end{equation}
If there is an initial rather than the final state interaction, the direction
of the Wilson lines is changed: they come from the past to the interaction
point. Hence, in a generic situation, the quark field entering the distribution
\begin{equation}
{\mit\Psi}_{\infty \, {\rm sgn}(z^0)}
(z^-, \bit{z}_\perp)
\equiv
[\infty \cdot {\rm sgn} (z^0) , \bit{\infty} ;
\infty \cdot {\rm sgn} (z^0) , \bit{z}_\perp]
[\infty \cdot {\rm sgn} (z^0) , \bit{z}_\perp ;
z^-, \bit{z}_\perp]
\psi (z^-, \bit{z}_\perp)
\,
\end{equation}
acquires the phase in which the direction where the Wilson lines are pointing
is determined by the physical process in question. It is reflected by
${\rm sgn}(z_0)$ in the integration limit. In deeply inelastic and Drell-Yan
type processes  $z_0 > 0$ and $z_0 < 0$, respectively. The sign of $z_0$
depends on whether interactions generating the eikonal phase occur after or
before the struck quark is hit by the probe. In this formulation, the purely
collinear parton distributions are the integrals of the more general
distributions depending also on the transverse momentum
\begin{eqnarray}
\label{CollinearQuarkPDF}
f^q (x, \mu^2)
\!\!\!&=&\!\!\!
\int^{\mu^2} d^2 \bit{k}_\perp f^q (x, \bit{k}_\perp)
\nonumber\\
&=&\!\!\!
\frac{1}{2} \int \frac{d z^-}{2 \pi}
{\rm e}^{i x p^+ z^-}
\langle p |
\bar\psi (0, \bit{0})
\gamma^+
[0, \bit{0}; z^-, \bit{0}]
\, \psi (z^-, \bit{0})
| p \rangle
\,  .
\end{eqnarray}
We schematically exhibited the appearance of the cutoff scale dependence
of the transverse momentum integrals.

For positive $x$, the  function $f^q (x, \mu^2)$ can be interpreted as the
quark density $q (x, \mu^2)$, while for negative  momentum fractions $x$ it
is understood as the (minus) antiquark density:
\begin{equation}
\label{UnPolQuarkDensity}
f^q (x, \mu^2) = q (x, \mu^2) \theta (x) - \bar q (-x, \mu^2) \theta (-x) \, .
\end{equation}
In the Bjorken limit,  the structure functions are expressed in terms of quark
distributions. For instance, the proton polarization-independent structure
functions are
\begin{eqnarray}
\label{ForwardCallanGrossRel}
F_1 (x_{\rm B}, {\cal Q}^2)
\!\!\!&=&\!\!\!
\frac{1}{2 x_{\rm B}} F_2 (x_{\rm B}, {\cal Q}^2)
\\
&=&\!\!\!
\frac{1}{2}
\sum_q Q_q^2 \left( q (x_{\rm B}, {\cal Q}^2) + \bar q ( x_{\rm B}, {\cal Q}^2) \right)
\, , \nonumber
\end{eqnarray}
The proportionality of $F_1$  and $F_2$ structure functions is  a consequence of
the fact that the spin of quarks is one-half \cite{CalGro68}. It is instructive to
introduce a specific combination of $F_1$ and $F_2$, the longitudinal structure
function
\begin{equation}
F_L \equiv F_2 - 2 x_{\rm B} F_1
\, .
\end{equation}
For an unpolarized target, the structure functions are related to the
absorption cross sections of transversely and longitudinally polarized
photons
\begin{eqnarray*}
\sigma^{\gamma^\ast}_T
\!\!\!&=&\!\!\!
\frac{2 \pi^2 \alpha_{\rm em}^2}{K M_N}
\sum_{\lambda = \uparrow\downarrow}
\varepsilon_\mu^{\lambda \ast} W^{\mu\nu} \varepsilon_\nu^\lambda
\stackrel{{\cal Q}^2 \to \infty}{=}
\frac{4 \pi^2 \alpha_{\rm em}^2}{K M_N}
F_1 (x_{\rm B})
\, , \\
\sigma^{\gamma^\ast}_L
\!\!\!&=&\!\!\!
\frac{4 \pi^2 \alpha_{\rm em}^2}{K M_N} \
\varepsilon_{L \mu} W^{\mu\nu} \varepsilon_{L \nu}
\stackrel{{\cal Q}^2 \to \infty}{=}
\frac{4 \pi^2 \alpha_{\rm em}^2}{K M_N}
\frac{1}{2 x_{\rm B}}F_L (x_{\rm B})
\, ,
\end{eqnarray*}
respectively, with  $K$ being the photon flux factor. There are two
widely used conventions for it, due to Gilman \cite{Gil68}
\begin{equation}
\label{Gilman}
K = |\bit{q}|
=
\frac{p \cdot q}{M_N} \sqrt{1 + 4 M_N^2 x_{\rm B}^2/{\cal Q}^2}
\, ,
\end{equation}
and another one due to Hand \cite{Han63}
\begin{equation}
\label{Hand}
K = \frac{M_R^2 - M_N^2}{2 M_N} = \frac{p \cdot q}{M_N} (1 - x_{\rm B})
\, ,
\end{equation}
where $M_R$ is the mass of the intermediate state. In the Bjorken limit,
$F_1 \gg F_L$ for spin-one-half constituents, while for scalar partons the
inequality is reversed, $F_1 \ll F_L$. As is well known, the experimental
verification of the first inequality established the firm foundation for
identification of partons with Gell-Mann--Zweig quarks.

Analogous relations hold for the polarized structure function $g_1$, the
only difference is that spin-weighted rather than spin averaged quark
parton species enter the leading order prediction,
\begin{equation}
g_1 (x_{\rm B}, {\cal Q}^2)
=
\frac{1}{2} \sum_q Q_q^2
\left(
\Delta q (x_{\rm B}, {\cal Q}^2) + \Delta \bar q (x_{\rm B}, {\cal Q}^2)
\right)
\, .
\end{equation}
The definition of $\Delta q $, $\Delta \bar q$ differs  from Eq.\
(\ref{CollinearQuarkPDF}) by the presence of the chiral matrix, i.e.,
$\gamma^+ \to \gamma^+ \gamma^5$. We will discuss these in more detail
in subsequent sections.

\subsubsection{Parton distributions as momentum densities in the Bjorken frame}
\label{PDFsBjorkenFrame}

Let us turn to the physical picture of the deep-inelastic event. A very
intuitive interpretation of parton distributions arises in the so-called
Bjorken frame, where the proton moves along the $z$-axis and the photon
moves in the $x,y$  plane:
\begin{equation}
p^\mu
=
\left(
{\cal P} + \frac{M_N^2}{4 {\cal P}} , 0 , 0 , {\cal P}  - \frac{M_N^2}{4 {\cal P} }
\right)
\, , \qquad
q^{\mu} = (\omega , q^x, q^y, 0)
\, .
\end{equation}
Sending the proton's momentum to infinity,
\begin{equation}
{\cal P}  \to \infty
\, ,
\end{equation}
defines the infinite-momentum frame in which the energy of the virtual photon
vanishes
\begin{equation}
\omega
=
\frac{p \cdot q}{{\cal P} + M_N^2/(4 {\cal P} )}
\simeq
\frac{{\cal Q}^2}{2 x_{\rm B} {\cal P} }
\to 0
\, .
\end{equation}
Thus, there is no energy exchange between the lepton and the probed quark. The
photon momentum  becomes purely transverse: ${\cal Q}^2 \to \bit{q}_\perp^2$.
Thus, in the Bjorken frame the virtual photon is absorbed over a very short time
in a very small spatial area:
\begin{itemize}
\item A typical interaction time of constituents is inversely proportional
to the energy deficit of a given fluctuation of a particle with the energy
$E_0$ and three-momentum $\bit{k}_0$ into two partons with energies $E_{1,2}$
and three-momenta $\bit{k}_{1,2}$, where $\bit{k}_i = (\bit{k}_{\perp i},
x_i p^z)$. Namely,
\begin{equation}
\label{TimeDilation}
\delta t
\sim
\frac{1}{\delta E}
=
\frac{1}{E_0 - E_1 - E_2}
\sim
\frac{
x_0 x_1 x_2 p_z
}{
x_1 x_2 \bit{k}^2_{\perp 0}
-
x_0 x_2 \bit{k}^2_{\perp 1}
-
x_0 x_1 \bit{k}^2_{\perp 2}
}
\to
\infty
\,
\end{equation}
as $p_z \to \infty$. Therefore, one can treat partons as almost free in
the infinite momentum frame due to the time dilation (\ref{TimeDilation}).
During the time of transiting the target, the virtual photon ``sees" the
nucleon's constituents in a frozen state which is thus describable by an
instantaneous distribution of partons. Here it is instructive to make an
analogy with the X-ray crystallography. Recall that an X-ray scattered off
atoms reveals the crystal's structure owing to the fact that the oscillations
of atoms in the lattice sites can be neglected. Atoms can be considered at
rest during the time when X-rays pass through the crystal.
\item The highly virtual photon probes the transverse distance
\begin{equation}
\delta | \bit{z}_\perp |
\sim
\frac{1}{|\bit{q}_\perp|} = \frac{1}{{\cal Q}}
\end{equation}
in the
longitudinally Lorentz-contracted hadron.
\end{itemize}

To acquire a physically sensible picture for the parton distributions in the
Bjorken frame, one should use the light-like gauge $A^+ = 0$. In this physical
gauge, the gauge link in the definition of the collinear parton distribution
function disappears. In the formalism of the light-cone quantization reviewed
in Appendix \ref{LightConeQCD} (see \cite{BroPauPin98}), the leading-twist
parton distributions are defined by the ``good'' components of the field
operators $\bar\psi \gamma^+ \psi = \bar\psi_+ \gamma^+ \psi_+$. Notice that
the good components of quark and gluon fields behave as free only at the
light-cone time $z^+ = 0$, which is where these operators enter the definition
of parton distributions. This reflects the physics of the parton model: partons
are treated as free just during the short time when they are ``seen'' by a probe
like that of a highly virtual photon in a hard process. Notice also that the
parton states obtained by creation operators acting on the vacuum are defined
with the reference to the time of this hard interaction, rather than with that
to the far past or far future. The reason is obvious: we do not treat partons
as observable particles, since they are not in light of the quark confinement.
Substituting Eq.\ (\ref{FieldsCreatAnnihOps}) into (\ref{CollinearQuarkPDF}) gives
\begin{eqnarray}
\label{PartonicContentPDFs}
f^q (x)|_{x > 0}
\!\!\!&=&\!\!\!
\frac{1}{2 x}
\sum_{\lambda = \uparrow\downarrow}
\int \frac{d^2 \bit{k}_\perp}{(2 \pi)^3}
\frac{
\langle p |
b_\lambda^\dagger (x p^+, \bit{k}_\perp) b_\lambda ( x p^+, \bit{k}_\perp)
| p \rangle
}{
\langle p | p \rangle
}
\, , \nonumber \\
f^q (x)|_{x < 0}
\!\!\!&=&\!\!\!
\frac{- 1}{2 x}
\sum_{\lambda = \uparrow\downarrow}
\int \frac{d^2 \bit{k}_\perp}{(2 \pi)^3}
\frac{
\langle p |
d_\lambda^\dagger (x p_+, \bit{k}_\perp) d_\lambda ( x p_+, \bit{k}_\perp)
| p \rangle
}{
\langle p | p \rangle
}
\, ,
\end{eqnarray}
where $b^\dagger$ and $d^\dagger$ are the creation operators of a quark and
an anti-quark, respectively, with longitudinal momentum $k^+ \equiv x p^+$ and
transverse momentum $\bit{k}_\perp$, and $| p \rangle$ is a hadronic state with
a definite light-cone energy $p^+$ normalized by  $\langle p | p \rangle = 2 p^+
(2 \pi)^3 \delta^{(3)} (0)$. For positive momentum fractions, the parton
distribution is identical to the number of quarks with a given momentum
$x p^+$ in a fast-moving proton, while for negative $x$ it measures the number
of antiquarks. Thus, interpretation of parton distribution functions as parton
number densities \cite{Fey71} is obvious in the infinite momentum frame. The
positivity of parton distributions is also obvious, since in the light-cone
quantization one can cast them in the form,
$$
f^q (x)
=
\frac{1}{\sqrt{2}}
\sum_n \left| \langle n | \psi_+ (0) | p \rangle \right|^2
\delta ((1 - x) p^+ - P^+_n )
\, ,
$$
where $P_n = p_1 + p_2 + \dots + p_n$ is the momentum of intermediate particle
states.

\subsubsection{Parton distributions in the rest frame}

The definition of the parton distribution (\ref{CollinearQuarkPDF}) is boost
independent. One can check that the $z$-boost independence translates into the
reparametrization invariance of  parton distributions with respect to the
rescaling of light-cone vectors. Namely, the transformation
\begin{equation}
n^\mu \to \varrho \, n^\mu
\, , \qquad
n^{\ast \mu} \to \varrho^{- 1} \, n^{\ast \mu}
\,
\end{equation}
leaves the parton distributions intact. As a result, we can use parton distributions
in any frame. However, a clear separation of quarks from antiquarks for positive and
negative momentum fractions $x$, respectively, as well as the probabilistic parton
picture will hold only in the infinite momentum frame, since in any other frame the
partons will be strongly interacting.

By transforming all momenta to the rest frame, we see that the parton distribution
is determined by a measurement on the space-time surface $z^+ = 0$, i.e., the
measurement is not an instantaneous event. The variable $x$ is intrinsic to the
light-cone operator: it is a Fourier conjugate to the separation of quark fields
on the light cone. It maps into the momentum fraction of quarks only in the
formalism of light-cone quantization which is equivalent to the Feynman parton
model in the infinite momentum frame. It is easy to see that in the rest frame of
the nucleon, the Feynman variable $x$ is just a special combination of the off-shell
energy $k^0$ and momentum $k^z$,
$$
x = \frac{k^0 + k^z}{M_N}
\, .
$$
In other words, the parton distribution is the distribution of quarks projected along
a special direction in the four-dimensional energy-momentum space. The quarks with
different $k^0$ and $k^z$ can have the same $x$, and furthermore, the quarks are
naturally off-shell.

\subsection{Quark phase-space distribution}
\label{WignerDistrQuarks}

To parallel the description of a quantum state in nonrelativistic quantum mechanics
by means of the Wigner distribution, we introduce a similar concept within the
field-theoretical context of QCD \cite{Bel03,Ji03,BelJiYua03}. We generalize the
concept of phase-space distributions to the case of relativistic quarks and gluons
in the proton by  promoting the  wave functions of quantum-mechanical states to
the field operators acting on the Hilbert space of quantum states. We introduce
the (equal light-cone time) Wigner operator defining it by \cite{Ji03,BelJiYua03}
\begin{equation}
{\cal W}^q ( k^+ , \bit{k}_\perp ; \bit{r} )
=
\int \frac{d z^-}{2 \pi} \frac{d^2 \bit{z}_\perp}{(2 \pi)^2}
\,
{\rm e}^{
-
i k^+ z^-
+
i \bit{\scriptstyle k}_\perp \cdot \bit{\scriptstyle z}_\perp
}
\bar{\mit\Psi}_{\infty} (\bit{r} + z )
\gamma^+
{\mit\Psi}_{\infty} ( \bit{r} - z )
\, .
\end{equation}
We restricted the definition by  taking the correlation at equal light-cone time
since only such correlations are accessible in high-energy experiments. Here
$\bit{r}$ is the quark position and $k^\mu = (k^+, k^- = 0, \bit{k}_\perp)$ is
the momentum conjugate to the space-time separation $z^\mu = (z^+ = 0, z^-,
\bit{z}_\perp)$. Physically $k^\mu$ is the sum of incoming and outgoing partons'
momenta. The parton phase-space distribution is determined by the matrix element
of this operator sandwiched between the states of a localized proton
(\ref{LocalizedNucleon}), namely
\begin{equation}
W^q (x, \bit{k}_\perp ; \bit{r})
=
\langle \bit{R} = 0 |
{\cal W}^q ( k^+ , \bit{k}_\perp ; \bit{r} )
| \bit{R} = 0 \rangle
\, .
\end{equation}
Its Fourier transform with respect to the three-dimensional coordinate
$\bit{r}$ determines the generalized momentum-space correlation function
\begin{equation}
2 M_N \int \frac{d^3 \bit{r}}{(2 \pi)^3} \,
{\rm e}^{i \bit{\scriptstyle \Delta} \cdot \bit{\scriptstyle r}}
W^q \left( x, \bit{k}_\perp ; \bit{r} \right)
\equiv
{\cal F}^q \left( x, \bit{k}_\perp ; \eta, \bit{\Delta}_\perp \right)
\, .
\end{equation}
In the Breit frame\footnote{The limitation of this procedure will be
discussed below in Section \ref{GPDsAsWigner}.}, where the incoming
and outgoing momenta are given by Eq.\ (\ref{BreitFrameMomenta}), we
have\footnote{An analogous $\bit{k}_\perp$ unintegrated function was
introduced for gluons in Ref.\ \cite{MarRys01}, without any reference
to the Breit frame.}
\begin{eqnarray}
\label{GeneralCorrelation}
{\cal F}^q \left( x, \bit{k}_\perp ; \eta, \bit{\Delta}^2_\perp \right)
=
\int \frac{d z^-}{2 \pi} \frac{d^2 \bit{z}_\perp}{(2 \pi)^2} \,
{\rm e}^{
i x z^- p^+ - i \bit{\scriptstyle k}_\perp \cdot \bit{\scriptstyle z}_\perp
}
\langle -\ft12 \bit{\Delta} |
\bar {\mit\Psi}_{\infty} ( - z ) \gamma^+ {\mit\Psi}_{\infty} ( z )
| \ft12 \bit{\Delta} \rangle
\, .
\end{eqnarray}
Since there is an axial symmetry of the high-energy scattering with
the $z$-axis directed along the momentum of the incoming proton, it is
natural to separate the longitudinal and transverse kinematical
variables. The longitudinal variables include the ``external'' parameter of  skewness
$\eta$ and the ``internal'' Feynman momentum fraction $x$, defined as
\begin{equation}
\label{SkewnessInBreit}
\eta \equiv \frac{\Delta^+}{p^+} = \frac{\Delta^z}{2E}
\, , \qquad
x \equiv \frac{k^+}{p^+} = \frac{k^0 + k^z}{2 E}
\, ,
\end{equation}
respectively. The rightmost expressions in each of the two definitions hold in the
above ``brick wall" frame. Here the incoming (outgoing) proton energy is $E =
\sqrt{M_N^2 + \bit{\Delta}^2/4}$. We use twice the average proton momentum $p$
and the momentum transfer $\Delta$,
\begin{equation}
p = p_1 + p_2
\, , \qquad
\Delta = p_1 - p_2
\, .
\end{equation}

Obviously, Eq.\ (\ref{GeneralCorrelation}) is a generating function of all known
high-energy observables of the proton based on generalized parton distributions.
The Wigner distributions are related to:

\begin{itemize}
\item  the generalized parton distributions (GPDs) by a Fourier transformation and a
marginal projection over the transverse parton momentum:
\begin{equation}
\label{GPDviaWigner}
F^q \left( x, \eta, \bit{\Delta}^2_\perp \right)
=
\int d^2 \bit{k}_\perp
{\cal F}^q \left( x, \bit{k}_\perp ; \eta, \bit{\Delta}^2_\perp \right)
\, .
\end{equation}
\item the unintegrated forward parton distributions via a simple reduction
\begin{equation}
f^q (x, \bit{k}_\perp)
=
4 {\cal F}^q \left( x, 2 \bit{k}_\perp ; \eta = 0, \bit{\Delta}^2_\perp = 0 \right)
\, .
\end{equation}
The factor of 2 accompanying $\bit{k}$ in the argument of $\mathcal{F}$ is a
consequence of the reduction of the vector $k^\mu$ to the forward limit where
it becomes twice the quark momentum.
\end{itemize}

The function $W (x; \bit{r})$ is not a probability density. It is a three-dimensional
quasi-probability distribution of quarks with a selected Feynman variable $x$. We
remind the reader again that the light-cone momentum is understood here in its
rest-frame sense as merely a special combination of the off-shell energy and
momentum along $z$: $x \sim k^0 + k^z$. A few special limiting cases of the
parameter space endow the reduced Wigner functions with the density of probability
property. They are:
\begin{itemize}
\item The Mellin moments with respect to the Feynman variable $x$ of the Wigner
distribution yield various form factors. The first moment gives the electromagnetic
form factors discussed above, the second moment corresponds to gravitational form
factors (both discussed in detail below), and higher moments generate form factors
of higher-spin probes.
\item The $r^z$-integrated skewless GPD  (set to $\eta \sim p^z = 0$)  yields a
reduced Wigner distribution \cite{Bel03,Ji03} that coincides with the two-dimensional
density $q (x; \bit{r}_\perp)$ which is the Soper's impact-parameter parton
distribution \cite{Sop77}. Integrating it  over $\bit{r}_\perp$ gives  the usual
Feynman parton distribution.
\end{itemize}

The Wigner function gives a classical realization of phase-space distributions
in regions where its magnitude is larger than typical quantum fluctuations. For
the QCD case, the semiclassical picture arises most naturally for a large nucleus
with radius much larger than its Compton wavelength, $R_A \gg \lambda_A \sim 1/M_A$.

Unfortunately, the generic Wigner function (\ref{GeneralCorrelation}) is not
accessible experimentally, however, its first marginal projection (\ref{GPDviaWigner})
is. The rest of our discussion will be dedicated solely to the development of
the theory and phenomenology of GPDs.

\subsection{Exclusive versus inclusive processes}

Deeply inelastic scattering experiments discussed in the preceding paragraphs,
and other inclusive reactions, have been used for over the last three decades
as one of the major sources of information on the complicated long distance
dynamics of hadron constituents. However, information on the nucleon's
parton structure obtained in this way can only be used to access inclusive
properties of hadrons, like parton distributions. It is insufficient to
constrain the detailed picture of the hadron wave function, which is the
probability amplitude describing the hadron as a superposition of partonic
states, e.g., for the proton,
\begin{equation}
| p \rangle
= \psi_{uud} | u u d \rangle
+ \psi_{uudg} | u u d g \rangle
+ \psi_{uudgg} | u u d g g \rangle
+ \psi_{uud\bar{q}q} | u u d \bar{q} q \rangle
 + \cdots \, ,
\end{equation}
with precise assignment of particular  spin, flavor, longitudinal momentum
fractions and transverse momenta to each of the hadron constituents bound
inside the hadron. For instance, the quark distribution addressed above is
given (schematically) by the absolute value squared of the proton wave function,
\begin{equation}
f^q (x)
=
\int d x' \int d^2 \bit{k} \ d^2 {\bit{k}}^\prime
\
| \psi^\ast_{uud}
\left( x, 1 - x - x', x'; \bit{k} ,
{\bit{k}}^\prime , - \bit{k} - {\bit{k}}^\prime \right) |^2
\, ,
\end{equation}
(see Section \ref{SectionDistributionAmplitudes} for a more  precise formula) with
longitudinal momentum fractions $x$ of the struck and spectator quarks as well as
their transverse momenta kept explicit. In $f^q(x)$, all the transverse momenta of
the partons are integrated out. This obviously simplifies the theoretical description,
but the result lacks a plethora of important features of the strong interaction
physics.

There is a large class of hadronic reactions where one gets a more direct access
to the wave functions. Exclusive processes are defined as scattering reactions
where the kinematical parameters  of all initial and final state particles are
specified. From the experimental point of view, this implies that all outgoing
particles are detected. The reactions include elastic lepton-hadron scattering
giving information about  hadronic  form factors, hadron decays, hadron-hadron
scattering, leptoproduction of photons and mesons off hadrons.

Exclusive rates drop much faster with momentum transfer than their inclusive
counterparts. Contrary to deeply inelastic scattering, exclusive reaction
rates depend on the state of the hadronic system prior as well as after the
hard collision. It is a consequence of the coherent scattering of hadron
constituents by an external probe such that partons can form a single outgoing
hadron rather than a jet of particles. Because on the higher sensitivity to
the long-distance effects these processes on the one hand can shed more light
on the quark confinement but on the other hand they are more challenging for
theoretical analyses.

\section{Classification and properties of GPDs}
\label{SectionClassification}

The reduced phase-space distributions of partons (\ref{GPDviaWigner}) in the nucleon
are Fourier transforms of  generalized parton distributions. GPDs may be treated as
a formal generalization of the matrix elements of non-local light-cone operators
to the case of off-forward kinematics. They arise in a number of exclusive and
diffractive processes, to the detailed analysis of which we dedicate several
subsequent sections of this review.

In the present section, we give the classification of generalized parton
distributions according to the quantum numbers of the operators they are
represented by as well as by the hadron states these operators are sandwiched
in. We consider first the leading-twist distributions. The twist-three functions
will be addressed in Section \ref{TwistThreeGPDs}.

\subsection{Twist-two operators}
\label{ClassificationTwistTwo}

The leading terms in the asymptotic regime for hard processes are given by
operators of the lowest twist. In deeply inelastic scattering, such operators
are associated with contributions  exhibiting  Bjorken scaling. Formally, the
geometric twist is defined as dimension in mass units minus the Lorentz spin
of the operator
\begin{equation}
\tau = d - s
\, .
\end{equation}
A spin-$s$ tensor transforms as an irreducible representation of the Lorentz
group. The maximal spin for a given number of Lorentz indices is achieved when
they are all symmetrized. The irreducibility implies that the reduction to
lower-spin tensors is not possible: as a consequence, the contraction of any pair
of indices with the metric tensor gives zero. Thus, the Lorentz structure has
to be traceless, and its construction is straightforward. For instance, the
twist-two spin-$j$ operator built from scalar fields in a non-gauge theory is
given by
\begin{equation}
\label{MaxSpinWilson}
{\cal R}^{2,\phi\phi}_{\mu_1 \dots \mu_j} (0)
=
\Sym_{\mu_1 \dots \mu_j}
\phi^\dagger (0)
i \! \stackrel{{}_\leftrightarrow}{\partial}_{\mu_1}
\dots
i \! \stackrel{{}_\leftrightarrow}{\partial}_{\mu_j}
\phi (0)
\, ,
\end{equation}
where the left-right derivative is
\begin{equation}
\stackrel{{}_\leftrightarrow}{\partial}_\mu
\ \equiv \
\stackrel{{}_\rightarrow}{\partial}_\mu
-
\stackrel{{}_\leftarrow}{\partial}_\mu
\, .
\end{equation}
In ${\cal R}^{\tau, aa}$, the superscript $\tau$ stands for the twist of the
operator and $a$ specifies the particle content: QCD quarks $a = q$ and gluons
$a = g$ or scalars $a = \phi$ (the latter are used  for demonstration purposes
only).  The operation $\Sym$ denotes the symmetrization of the corresponding
Lorentz indices and trace subtraction. For example, in the case of a two-index
tensor it is
$$
\Sym_{\mu_1 \mu_2} t_{\mu_1 \mu_2}
=
\frac{1}{2!}
\left(
t_{\mu_1 \mu_2}
+
t_{\mu_2 \mu_1}
-
\frac{1}{2} g_{\mu_1 \mu_2} g^{\nu_1 \nu_2} t_{\nu_1\nu_2}
\right)
\, .
$$
A constructive all-order definition of this operation is given below in Section
\ref{HarmonicPolynomials}. In case of scalar fields $\phi$, the dimension of the
operator with $j$ derivatives is $d = 2 d_\phi + j$. Since the dimension of
$\phi$ is $d_\phi = 1$, we have $\tau = 2$. The towers of twist-two spin-$j$
operators can be conveniently summed into a non-local form and one can also
define the notion of twist for non-local light-ray operators in terms of Taylor
expansion,
\begin{equation}
\label{NonlocalScalarOper}
{\cal O}^{\phi\phi} ( - z^- , z^- )
=
\phi^\dagger ( - z^- ) \phi ( z^- )
=
\sum_{j = 0}^\infty \frac{(- i z^-)^j}{j!}
n^{\mu_1} \dots n^{\mu_j}
{\cal R}^{2, \phi\phi}_{\mu_1 \dots \mu_j} (0)
\, .
\end{equation}
When fermions and gauge bosons enter the game,  one has to separate first the
different spin projections contained in a given field operator. This can be
done in a most concise manner in the light-cone formalism (see Appendix
\ref{LightConeQCD}). Below, we outline the construction separately for quark
and gluon bilocal operators.

\subsubsection{Quark operators}

The four-component fermion field $\psi$, by means of the projection operators
\cite{KogSop70}
\begin{equation}
\label{FermionLCdec}
\psi = \psi_+ + \psi_-
\, , \qquad
\psi_\pm \equiv {\mit\Pi}^{\pm} \psi
\, , \qquad
{\mit\Pi}^{\pm} \equiv \ft12 \gamma^{\mp} \gamma^{\pm}
\, ,
\end{equation}
is decomposed in two terms $\psi_\pm$ that have the following light-cone spin:
\begin{equation}
\label{QuarkLCspin}
{\mit\Sigma}^{-+} \psi_{\pm} = \pm \ft12\, \psi_{\pm} \, .
\end{equation}
Here ${\mit\Sigma}^{\mu\nu} \equiv \ft{1}4 [\gamma^\mu , \gamma^\nu]$ is the
spin tensor. Since the canonical dimension of the fermion field is $d_q = 3/2$,
one finds that the $\psi_+$-component has the twist $\tau_q = 1$ since its
spin is $s_q = 1/2$. Similarly, for the $\psi_-$-component one finds $s_q = -
1/2$ and thus $\tau_q = 2$. Therefore, only the $\psi_+$ component enters a
nonlocal operator of the minimal twist. Going from the four-component to Weyl
spinors, the chiral projection of the ``good'' components yields two independent
fields
\begin{equation}
\psi_{+ \uparrow} = \ft12 (1 + \gamma^5 ) \psi_+
\, , \qquad
\psi_{+ \downarrow} = \ft12 (1 - \gamma^5 ) \psi_+
\, .
\end{equation}
The right $\psi_{+ \uparrow}$ and left $\psi_{+ \downarrow}$ projections of the
``good'' light-cone spinors possess a single nonvanishing component only,
which describes a state with a definite helicity. The helicity operator is
conventionally defined as (see Appendix \ref{Spinors})
\begin{equation}
\label{HelicityMatrix}
h \equiv
\ft12 \bar{\bit{e}}^i_\perp {\bit{e}}^j_\perp {\mit\Sigma}^{ij}
= i {\mit\Sigma}^{12}
\, ,
\qquad
\bit{e}^i_\perp
=
( \bar{\bit{e}}^i_\perp )^\ast
=
(1, i)
\, ,
\end{equation}
with eigenvalues designating their helicity
\begin{equation}
h \, \psi_{+ \uparrow} = \ft12 \psi_{+ \uparrow}
\, , \qquad
h \, \psi_{+ \downarrow} = - \ft12 \psi_{+ \downarrow}
\, .
\end{equation}

The counting of independent bilocal quark operators is especially transparent
within this formulation, since left and right fields each contain a single
non-vanishing component only. Namely, the fields
$$
\psi_{+ \uparrow} = \left( { \lambda_\uparrow \atop 0 } \right)
\, , \qquad
\psi_{+ \downarrow} = \left( { 0 \atop \bar\chi_\downarrow } \right)
\, ,
$$
with one-component two-dimensional Weyl spinors
$$
\lambda_\uparrow \sim \left( \, {1 \atop 0} \right)
\, , \qquad
\bar\chi_\downarrow \sim \left( \, {0 \atop 1} \, \right)
\, ,
$$
realize the $\left( 0, \ft12 \right)$ and $\left( \ft12, 0 \right)$
representations of the Lorentz group, respectively. Let us use them to build
all distinct helicity combinations for the two-particle leading twist
operators, i.e., $\left( \ft12, 0 \right) \otimes \left( \ft12, 0 \right) =
\left( 0, 0 \right) \oplus \left( 1, 0 \right)$ and $\left( \ft12, 0 \right)
\otimes \left( 0, \ft12 \right) = \left( \ft12, \ft12 \right)$. We can build
four distinct combinations and put them back into a covariant four-dimensional
form
\begin{eqnarray*}
\lambda^\ast_\uparrow \lambda^{\phantom{\ast}}_\uparrow
\pm
\bar\chi^\ast_\downarrow \bar\chi^{\phantom{\ast}}_\downarrow
=
\frac{1}{\sqrt{2}}
\bar\psi
\left\{
\!\!
\begin{array}{c}
\gamma^+
\\
\gamma^+ \gamma^5
\end{array}
\!\!
\right\}
\psi
\, , \qquad
\lambda^\ast_\uparrow \bar\chi^{\phantom{\ast}}_\downarrow
\pm
\bar\chi^\ast_\downarrow \lambda^{\phantom{\ast}}_\uparrow
=
\frac{i}{\sqrt{2}}
\bar\psi
\left\{
\!\!
\begin{array}{r}
i \sigma^{+2}
\\
\sigma^{+1}
\end{array}
\!\!
\right\}
\psi
\, .
\end{eqnarray*}
The latter two operators are merely two components of a two-dimensional vector.
Notice that since $\sigma^{\mu\nu}$ and $\sigma^{\mu\nu} \gamma^5$ are not
independent, the latter vertex will not appear as a possible Dirac structure.
As an outcome of this simple analysis we can introduce the following three
leading-twist bilocal quark operators:
\begin{eqnarray}
\label{QuarkLightRayOperators}
{\cal O}^{qq} (z^-_1, z^-_2)
\!\!\!&=&\!\!\!
\bar\psi (z^-_1)
[z^-_1 , z^-_2] \gamma^+
\psi (z^-_2)
\, , \\
\label{QuarkLightRayOperatorsOdd}
\widetilde {\cal O}^{qq} (z^-_1, z^-_2)
\!\!\!&=&\!\!\!
\bar\psi (z^-_1)
[z^-_1 , z^-_2] \gamma^+ \gamma^5
\psi (z^-_2)
\, , \\
{\cal T}^{qq}_\mu (z^-_1, z^-_2)
\!\!\!&=&\!\!\!
\bar\psi (z^-_1)
[z^-_1 , z^-_2] \sigma^{+ \perp}_{\phantom{+} \mu}
\psi (z^-_2)
\, .
\end{eqnarray}
The gauge link is added here to make the operator explicitly gauge invariant if other
than the  light-cone gauge $A^+=0$ is  used. As discussed in the introduction,
the  path-ordered exponentials appear in physical observables as a consequence
of the final (DIS) or initial (DY) state interactions.

\subsubsection{Gluon operators}

For a gauge field, in order to clearly identify its different spin components,
one has to project the Lorentz indices of the strength tensor $F^{\mu\nu}$
onto the longitudinal light-cone directions and the transverse space with the
two-dimensional metric $g^\perp_{\mu\nu}$. The light-cone spin assignments for
different projections of $F^{\mu\nu}$ are as follows
\begin{equation}
\label{GluonLCspin}
{\mit\Sigma}^{-+} F^{\pm \mu}_{\phantom{+} \, \perp} = \pm F^{\pm
\mu}_{\phantom{+} \, \perp} \, , \qquad {\mit\Sigma}^{-+}  F^{+ -} =
{\mit\Sigma}^{-+}  F^{\mu\nu}_{\perp\perp} = 0 \, ,
\end{equation}
where
$$
{\mit\Sigma}^{\mu\nu} F^{\rho\sigma}
=
g^{\mu\rho} F^{\nu\sigma} - g^{\nu\rho} F^{\mu\sigma}
-
g^{\mu\sigma} F^{\nu\rho} + g^{\nu\sigma} F^{\mu\rho}
\, .
$$
Since the canonical dimension of the strength tensor $F^{\mu\nu}$ is $d_g = 2$,
the $F^{+ \mu}_{\phantom{+} \, \perp}$-component possesses the twist $\tau_g = 1$
since its spin is $s_g = 1$. For $F^{+ -}$ and $F^{\mu\nu}_{\perp\perp}$-components
one gets $s_g = 0$ and therefore $\tau_g = 2$, while the $F^{- \mu}_{\phantom{+}
\, \perp}$-component $s_g = - 1$ and $\tau_g = 3$. Consequently, the minimal twist
is associated with the $F^{+ \mu}_{\phantom{+} \, \perp}$-component only.

The product of two vectors, each transforming as $\left( \ft12, \ft12 \right)$, can
be decomposed into irreducible representations of the Lorentz group as $\left( \ft12,
\ft12 \right) \otimes \left( \ft12, \ft12 \right) = \left( 0, 0 \right) \oplus
\left( \left( 1, 0 \right) \oplus \left( 0, 1 \right) \right) \oplus \left( 1, 1
\right)$. In terms of tensors this reads
\begin{eqnarray*}
g^{\mu\alpha}_\perp g^{\nu\beta}_\perp
=
\ft12
g^{\mu\nu}_\perp g^{\alpha\beta}_\perp
+
\ft12
\varepsilon^{\mu\nu}_\perp \varepsilon^{\alpha\beta}_\perp
+
\tau^{\mu\nu;\rho\sigma}_\perp \tau^{\alpha\beta;}_\perp{}_{\rho\sigma}
\, ,
\end{eqnarray*}
with the two-dimensional tensors being
\begin{eqnarray}
g^\perp_{\mu\nu}
\!\!\!&\equiv&\!\!\!
g_{\mu\nu} - n_\mu n^\ast_\nu
- n_\nu n^\ast_\mu
\, , \\
\varepsilon^\perp_{\mu\nu}
\!\!\!&\equiv&\!\!\!
\varepsilon^{\alpha\beta\rho\sigma}
g^\perp_{\alpha\mu} g^\perp_{\beta\nu} n^\ast_\rho n_\sigma
\, , \\
\tau_{\mu\nu;\rho\sigma}^\perp
\!\!\!&\equiv&\!\!\!
\ft12
g_{\mu\rho}^\perp g_{\nu\sigma}^\perp
+
\ft12
g_{\mu\sigma}^\perp g_{\nu\rho}^\perp
-
\ft12
g_{\mu\nu}^\perp g_{\rho\sigma}^\perp
\, .
\end{eqnarray}
The totally antisymmetric tensor is normalized as $\varepsilon^{0123} = 1$.
The totally symmetric and traceless in two pair of indices tensor
$\tau^\perp_{\mu\nu; \rho\sigma}$ possesses the properties
\begin{eqnarray*}
\tau^\perp_{\mu\nu;\rho\sigma} \tau^\perp_{\mu\nu;\rho'\sigma'}
=
\tau^\perp_{\rho\sigma;\rho'\sigma'}
\, , \qquad
\tau^\perp_{\mu\nu;\rho\sigma} = \tau^\perp_{\rho\sigma;\mu\nu}
\, , \qquad
\tau^\perp_{\mu\mu;\rho\sigma} = 0
\, , \qquad
\tau^\perp_{\mu\nu;\mu\nu} = 2
\, .
\end{eqnarray*}
Hence, there are three independent Lorentz structures for twist-two
two-particle gluonic operators.

This can be easily understood from counting two-gluon helicity states in the
formalism of the light-cone quantization. To this end, we build the holomorphic
and anti-holomorphic combinations of the gauge potential in the light-cone gauge
\begin{equation}
A_\perp \equiv A^x + i A^y \, , \qquad \bar A_\perp \equiv A^x - i A^y
\, .
\end{equation}
They describe  left and right circular polarizations with helicity $- 1$
and $+ 1$, respectively. Namely
\begin{equation}
h \, A_\perp = - A_\perp
\, , \qquad
h \, \bar{A}_\perp = \bar{A}_\perp
\, .
\end{equation}
Both of them are scalars with respect to the light-cone spin operator
${\mit\Sigma}^{-+}$, i.e., ${\mit\Sigma}^{-+} A = {\mit\Sigma}^{-+} \bar A = 0$.
Since the light-cone derivative carries one unit of spin, so will the derivatives
of the (anti-) holomorphic potentials. In the covariant form, they are related
to the gluon field-strength tensors
\begin{eqnarray*}
\partial^+ A_\perp
=
F^{+ x} - i \widetilde F^{+ x}
=
i \left( F^{+ y} - i \widetilde F^{+ y} \right)
\, , \qquad
\partial^+ \bar A_\perp
=
F^{+ x} + i \widetilde F^{+ x}
=
- i \left( F^{+ y} + i \widetilde F^{+ y} \right)
\, .
\end{eqnarray*}
For further convenience, we introduce the gluon negative- and positive-helicity
covariant fields
\begin{equation}
F^{\ + \mu}_{[-] \, \perp}
=
F^{+ \mu}_{\phantom{+} \, \perp}
-
i \widetilde F^{+ \mu}_{\phantom{+} \, \perp}
\, , \qquad
F^{\ + \mu}_{[+] \, \perp}
=
F^{+ \mu}_{\phantom{+} \, \perp}
+
i \widetilde F^{+ \mu}_{\phantom{+} \, \perp}
\, ,
\end{equation}
respectively. Here the dual gluon field strength is defined
as $\widetilde F^{\mu\nu} \equiv \ft12 \varepsilon^{\mu\nu\rho\sigma}
F_{\rho\sigma}$, so that $\widetilde F^+{}^\mu_\perp = \varepsilon^{\mu\nu}_\perp
F^+{}^\perp_\nu$ for two-dimensional transverse indices. From the positive-
and negative-helicity operators we build the even- and odd-parity combinations
$$
\left(
F^{\ + \mu}_{[-] \, \perp} F^{\ + \nu}_{[+] \, \perp}
\pm
F^{\ + \mu}_{[+] \, \perp} F^{\ + \nu}_{[-] \, \perp}
\right)
g^\perp_{\mu\nu}
=
4
\left\{
\!
\begin{array}{r}
g^\perp_{\mu\nu} \\
i \varepsilon^\perp_{\mu\nu}
\end{array}
\!
\right\}
F^{+ \mu}_{\phantom{+} \, \perp} F^{\nu +}_{\, \perp}
\, ,
$$
as well as the maximal-helicity operators
$$
\left(
F^{\ + \mu}_{[+] \, \perp} F^{\ + \nu}_{[+] \, \perp}
\pm
F^{\ + \mu}_{[-] \, \perp} F^{\ + \nu}_{[-] \, \perp}
\right)
=
4
\tau^{\mu\nu; \rho\sigma}_\perp
\left\{
\!
\begin{array}{r}
F^{+ \perp}_{\phantom{+} \, \rho} F^{\perp +}_{\, \sigma}
\\
i \widetilde F^{+ \perp}_{\phantom{+} \, \rho} F^{\perp +}_{\, \sigma}
\end{array}
\!
\right\}
\, .
$$
According to this nomenclature we introduce three leading-twist gluonic
operators:
\begin{eqnarray}
\label{GluonLightRayOperators}
{\cal O}^{gg} (z^-_1, z^-_2)
\!\!\!&=&\!\!\!
F^{+ \mu}_a (z^-_1)
[z^-_1 , z^-_2]^{ab} g_{\mu\nu}
F^{\nu +}_b (z^-_2)
\, , \\
\widetilde {\cal O}^{gg} (z^-_1, z^-_2)
\!\!\!&=&\!\!\!
F^{+ \mu}_a (z^-_1)
[z^-_1 , z^-_2]^{ab} i \varepsilon^\perp_{\mu\nu}
F^{\nu +}_b (z^-_2)
\, , \\
{\cal T}^{gg}_{\mu\nu} (z^-_1, z^-_2)
\!\!\!&=&\!\!\!
F^{+ \rho}_a (z^-_1)
[z^-_1 , z^-_2]^{ab} \tau^\perp_{\mu\nu ; \rho\sigma}
F^{\sigma +}_b (z^-_2)
\, ,
\label{GluonLightRayOperatorMaximalHelicity}
\end{eqnarray}
augmented by the Wison line in the adjoint representation
$$
[z^-_1 , z^-_2]^{ab}
=
P \exp
\left(
g \int_{z^-_2}^{z^-_1} d z'^- f^{abc} A_c^+ (z'^- , \bit{0}_\perp)
\right)
\, ,
$$
to respect gauge invariance. The links obey the following obvious properties
$$
[z^-_1 , z^-_2]^{ab}
=
[z^-_2 , z^-_1]^{ba}
\, , \qquad
[z^-_1 , z^-_2]^{ab}
=
[z^-_1 , z^-_3]^{ac}
[z^-_3 , z^-_2]^{cb}
\, .
$$

The construction we have just presented can be extended to higher twist
multi-particle operators which have the largest number of constituent
fields out of the set of all possible operators of twist-$N$ for a given
Lorentz spin. Namely, these operators are constructed from $N$ ``good''
fields, $\psi_+$ and $F^{+\mu}_{\phantom{+} \, \perp}$ living on the light
cone. Let us reiterate that the twist of such nonlocal operators equals
the number of elementary fields involved, and they are known in QCD as
quasipartonic operators \cite{BukFroLipKur85}.

\subsection{Operator matrix elements and GPDs}

The leading-twist generalized parton distributions arise as coefficients in the
decomposition of the off-forward hadronic matrix elements of the bilocal operators
introduced in the previous section. For the scalar operator (\ref{NonlocalScalarOper}),
one gets\footnote{The presence of the factor $p^+$ is a consequence of the $z$-boost
invariance of the generalized parton distributions, as will be explained in Section
\ref{BoostInvar}.}
\begin{equation}
\label{DefScalarGPD}
F^\phi (x, \eta, \Delta^2)
=
p^+ \int \frac{d z^-}{2 \pi} {\rm e}^{i x z^- p^+}
\langle p_2| {\cal O}^{\phi\phi} (- z^- , z^-) | p_1 \rangle
\, .
\end{equation}
The variable $x$, just like for the usual parton distribution functions, is the
Fourier conjugate to the Ioffe time. In terms of the outgoing $k_1$ and incoming
$k_2$ parton's momenta one can write $x$ as
$$
x  = \frac{k_1^+ + k_2^+}{p^+}
\ .
$$
GPDs also depend on the invariant $t$-channel momentum transfer $\Delta^2 =
(p_1 - p_2)^2$ and skewness
$$
\eta = \frac{\Delta^+}{p^+}
\, .
$$
We recall that we define the momentum transfer as $\Delta = p_1 - p_2$,
with the sign opposite to the usual convention used for form factors. This
choice\footnote{Notice that in a large number of papers $\eta$ is defined with
an opposite sign.} guarantees that $\eta$ is positive for exclusive deeply
virtual lepton-hadron scattering processes. Thus, the plus components of the
incoming and outgoing parton's momenta are
$$
k_1^+ = \frac{x + \eta}{2} \, p^+
\qquad   {\rm and} \qquad
k_2^+ = \frac{x - \eta}{2} \, p^+
\, ,
$$
respectively. This parametrization corresponds to the symmetric conventions of
Refs.\ \cite{MulDitRobGeyHor98} and \cite{Ji96a}. Another parametrization used
in the literature, corresponds to nonforward parton distributions \cite{Rad1997a}.
In this case, the parton momenta are measured in units of the incoming hadron's
momentum $p_1$. Then
$$
k_1^+ = X p_1^+
\qquad  {\rm and} \qquad
k_2^+ = (X - \zeta) \, p_1^+
\, .
$$
The parameters $X, \zeta$ are related to  $x, \eta$ by
$$
\eta = \frac{\zeta}{2 - \zeta}
\, , \qquad
x = \frac{X - \zeta/2}{1 - \zeta/2}
\, , \qquad
X = \frac{x + \eta}{1 + \eta}
\, , \qquad
X - \zeta = \frac{x - \eta}{1 + \eta}
\, .
$$
We will not use these variables in our presentation (see Ref.\ \cite{Rad1997a}
and Ref.\ \cite{Ji:1998pc} for more details).

\subsubsection{Time-ordering and support}
\label{TimeOrderingSupport}

Let us discuss first the issue of whether the operators defining  GPDs
have to be chronologically or normal ordered. Generally since they enter
as a part of a Feynman diagram describing the physical process, all fields
are time-ordered including those present in the two-particle operators
${\cal O}^{\phi\phi} (- z^- , z^-)$ determining the hadronic function in
question. Argumentation given below does not depend on the spin of constituent
or the target, so  we can use the scalar-field operator (\ref{NonlocalScalarOper})
to this end. We thus define
\begin{equation}
F^\phi (x, \eta, \Delta^2)
=
p^+ \int \frac{d z^-}{2 \pi} {\rm e}^{i x z^- p^+}
\langle p_2|
T \{\phi^\dagger ( - z^- ) \phi ( z^- )\}
| p_1 \rangle
\, .
\label{Fx}
\end{equation}
Our goal is to demonstrate that since the elementary fields enter the operator
at equal light-cone time $z^+ = 0$, the usual chronological ordering can be
omitted \cite{LanPolSh071,Jaf83,Rad1997a,DieGou98}. Notice that it is not even
required that the fields have to be separated by a light-like distance: the
argument holds even for the transverse-momentum dependent functions. The function
$F^\phi (x, \eta, \Delta^2)$  can be treated as a projection
\begin{equation}
F^\phi (x, \eta , \Delta^2)
=
\int_{- \infty}^\infty dk^+ \, d^2 \bit{k}_\perp \,
\delta \left( x - \frac{k^+}{p^+} \right)
\int_{- \infty}^{\infty} d k^- \, {\cal A} (k^+, \bit{k})
\, ,
\end{equation}
of a general Green function
\begin{eqnarray*}
{\cal A} (k) = \int \frac{d^4 z}{(2 \pi)^4}
\,
{\rm e}^{i k \cdot z}
\,
\langle p_2| T  \{\phi^\dagger ( - z ) \phi ( z )\} | p_1 \rangle
\,
\end{eqnarray*}
for (off-shell) parton scattering on a hadron, $\phi (- k_1) + h (p_1) \to
\phi (- k_2) + h (p_2)$, with   $z^\mu$ being here a four-dimensional vector
with all components nonvanishing. The next step is to demonstrate that the
integral over the $k^-$ component of the Green function ${\cal A}$ is identical
to the integral over its discontinuity,
\begin{eqnarray*}
\int_{- \infty}^{\infty} d k^- \, {\cal A} (k^+, \bit{k})
=
\int_{- \infty}^\infty dk^-
\,
{\rm disc}_{k^-}
{\cal A} (k^+, \bit{k})
\, .
\end{eqnarray*}

Thus we have to incorporate  the analytical properties of ${\cal A}$ as a function
of Mandelstam variables $s = (p_1 - k_1)^2$ and $u = (p_2 + k_1)^2$ and parton
virtualities $k_1^2 = (k - \Delta)^2/4$ and $k_2^2 = (k + \Delta)^2/4$.
As usual, we  will assume that the singularity structure of the non-perturbative
matrix element coincides with that of the matrix element given in terms of Feynman
diagrams, i.e., in perturbation theory. Then  ${\cal A}$ has no poles when
$\Re{\rm e} \, k_i^2 < 0$ (for ``negative virtualities'')  and no production
thresholds in the corresponding channels for negative Mandelstam invariants,
$\Re{\rm e} ( s, u ) < 0$.

\begin{figure}[t]
\begin{center}
\mbox{
\begin{picture}(0,80)(150,0)
\put(0,1){\insertfig{11}{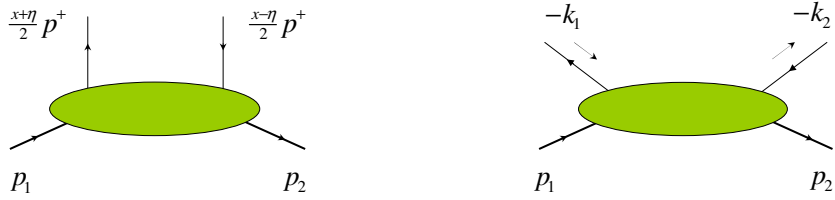}}
\end{picture}
}
\end{center}
\caption{\label{GPDgreenFunc} Longitudinal momentum flow in GPDs (left). GPDs
as an off-shell parton-hadron scattering amplitude (right).}
\end{figure}

To make correspondence between  the $s-,u-,k_1^2-$ and  $k_2^2-$singularities
of the Green function ${\cal A}$ and singularities in the $k^-$ plane, one should
express   $k^-$ in terms of these invariants:
\begin{eqnarray}
k^-_a
\!\!\!&=&\!\!\!
\frac{4 k_1^2 + ( \bit{k}_\perp + \bit{\Delta}_\perp )^2}{2 (k^+ + \Delta^+)}
-
\Delta^-
\, , \qquad
k^-_b
=
\frac{4 s + \bit{k}_\perp^2}{2 (k^+ - p^+)}
+ p^-
\, , \\
k^-_c
\!\!\!&=&\!\!\!
\frac{4 k_2^2 + ( \bit{k}_\perp - \bit{\Delta}_\perp )^2}{2 (k^+ - \Delta^+)}
+
\Delta^-
\, , \qquad
k^-_d
=
\frac{4 u + \bit{k}_\perp^2}{2 (k^+ + p^+)}
- p^-
\, ,
\end{eqnarray}
where $k^+ = x p^+$ and $\Delta^+ = \eta p^+$ are related to the external
variables $x$ and $\eta$. The $k^-$ arises in the denominators of propagators
with the causal Feynman prescription $k^2 + i 0$ and, therefore, produce
singularities of Feynman integrals in the complex plane while the prescription
determines the position of the contour bypassing them,
$$
\int_{- \infty}^\infty d k^- {\cal F} ( k^- + i 0 \cdot k^+_p)
\, ,
$$
with $p = a, b, c, d$ and $k^+_p$ being the denominators in above $k^-_p$.
Considering different regions of the momentum space, the singularities $k^-_p$
migrate in the complex $k^-$ plane and so will or will not contribute to the
$k^-$ contour integral. As we can see from the denominators of $k^-$, the sign
change occurs at $x = - 1 , - \eta , \eta , 1$ and thus four regions have to be
discussed (i) $|x| > 1$, (ii) $\eta < x < 1$, (iii) $- 1 < x < - \eta$, and
(iv) $|x| < \eta $. Let us discuss a couple of regions.
\begin{itemize}
\item In the region $|x| > 1$, all singularities are situated on the same
side of $\Im{\rm m} k^-$, since the denominators of $k^-_{a,b,c,d}$ are
either all positive for $x > 1$, or all negative for $x < -1$. Therefore, we
can close the integration contour in half-plane free of singularities and
assuming that ${\cal A}$ vanishes fast enough as $|k^-| \to \infty$ we can
drop the integral over the contour at infinity and get zero for the GPD.
Thus \cite{Rad1997a,FraFreGuzStr98,DieGou98},
\begin{equation}
\label{VanishGPDoutside}
F^\phi (x, \eta, \Delta^2) = 0 \, , \qquad |x| > 1
\, .
\end{equation}
\item In case $\eta < x < 1$, the $s$ singularities are in the upper half
plane, while the rest are in the lower. So we wrap the integration contour
about the $s$-channel production threshold which results into the difference
of integrals evaluated along the upper and lower sides of the branches
of the cut, and result into the discontinuity
$$
\int_{- \infty}^{\infty} d k^- \, {\cal A} (k^+, \bit{k})
=
\int_{- \infty}^\infty dk^-
{\rm disc}_s
\int \frac{d^4 z}{(2 \pi)^4}
{\rm e}^{i k \cdot z}
\langle p_2| T {\cal O}^{\phi\phi} (- z , z) | p_1 \rangle
\, .
$$
This is just what we wanted. The derivation works for any $\eta$ including
the forward limit $\eta=0$. In this sense, GPDs in this region are analogous
to the usual parton densities.
\end{itemize}

The other regions are considered in a similar way \cite{DieGou98}. In particular,
in the  $\, - 1 < x < - \eta$ region, one gets the  $u$-discontinuity in the
final formula instead of the $s$-discontinuity, and GPDs there are analogous to
the antiquark densities. The $|x| < \eta $ region is specific to the nonforward
kinematics: it disappears in the forward limit. The GPDs in this region correspond
to discontinuities in $k_1^2$ and $u$ or $k_2^2$ and $s$. They can be interpreted
as the (generalized) distribution amplitude describing the sharing of the longitudinal
momentum transfer $\eta p^+$ within the quark-antiquark pair emitted by the initial
(or absorbed by the final) hadron. The support properties of GPDs can be also studied
using their relation to double distributions, see Section \ref{DDs}.

\subsubsection{Counting GPDs}
\label{CountingGPDs}

For the operator matrix element of the transition of a spin-$s_1$ hadron
into a spin-$s_2$ hadron, one can have $(2s_1 + 1) \times (2s_2 + 1)$ hadron
helicity combinations. Notice that GPDs can initiate helicity-flip transitions.
This is possible due to the nonvanishing momentum exchange between the initial
and final states, which can produce  a non-zero orbital angular momentum
compensating the helicity deficit between the initial and final state
polarizations. In case of parton helicity non-flip transitions we have one
($n = 1$) operator of even and one of odd parity, i.e., $\lambda^\ast_\uparrow
\lambda^{\phantom{\ast}}_\uparrow \pm \bar\chi^\ast_\downarrow
\bar\chi^{\phantom{\ast}}_\downarrow$, respectively. On the other
hand, there are two operators ($n = 2$), $\lambda^\ast_\uparrow
\bar\chi^{\phantom{\ast}}_\downarrow$ and $\bar\chi^\ast_\downarrow
\lambda^{\phantom{\ast}}_\uparrow$ for the helicity-flip transitions
(recall the open Lorentz index in the maximal-helicity operators). To
get the total number of amplitudes, we have to multiply the number of
hadron-helicity combinations by the number of quark operators which
initiate the transition $n \times (2s_1 + 1) \times (2s_2 + 1)$. Not
all of the amplitudes are independent, however. It turns out that the
subsequent reduction works differently for generalized parton distributions
compared to the ordinary parton densities. Only the application of the
spatial-parity inversion results in a reduction of their number, contrary
to the case of forward parton distributions, where the time reversal provides
further relations between superficially independent functions. The number of
independent parton densities equals the number of independent helicity-conserving
helicity amplitudes allowed by parity and time reversal invariance, see Section
\ref{TimeReversalHermiticity} below. The number of generalized parton distributions
is obtained by counting helicity amplitudes under the constraints due to the
spatial parity only. The time-reversal invariance, on the contrary, determines the
overall phases and important symmetry properties of separate generalized parton
distributions but does not relate them to each other.

In the following few sections, we will discuss only flavor-diagonal GPDs, i.e.,
the transitions not changing the internal quantum numbers of the incoming
hadron in the final state. We will  introduce the number of independent GPDs
on a case-by-case basis. The general strategy can be found in Refs.\
\cite{Ji96b,JiLeb00,Die01,Hag04,CheJi04}. The flavor-nondiagonal
functions will be addressed later in Section \ref{FlavorChangingGPDs}.

\subsubsection{Boost invariance of GPDs}
\label{BoostInvar}

As explained in Appendix \ref{LCvectors}, and was mentioned in the introduction,
Lorentz symmetry implies that GPDs do not change under the reparametrization of
the light-cone vectors
$$
n^{\ast \mu} \to \varrho n^{\ast\mu}
\, , \qquad
n^\mu \to \varrho^{- 1} n^\mu
\, ,
$$
which implies in turn that they are also invariant under
\begin{equation}
k^- \to \varrho k^-
\, , \qquad
k^+ \to \varrho^{-1} k^+  \, .
\end{equation}
One easily observes that indeed the definition of GPDs given above have this
symmetry.

\subsubsection{Spin-zero hadrons}

For a spin-zero target there are two pairs of twist-two quark and gluon GPDs,
which arise from the operator matrix elements of the even-parity and
maximal-helicity operators introduced in Section \ref{ClassificationTwistTwo}.
Namely,
\begin{eqnarray}
\label{spin-zero-GPDs}
\langle p_2| {\cal O}^{qq} (- z^-, z^-) | p_1 \rangle
\!\!\!&=&\!\!\!
p^+
\int_{- 1}^1 d x \, {\rm e}^{- i x p^+ z^-}
H^q (x, \eta, \Delta^2)
\, , \\
\langle p_2| {\cal T}^{qq}_\mu (- z^-, z^-) | p_1 \rangle
\!\!\!&=&\!\!\!
p^+
\int_{- 1}^1 d x \, {\rm e}^{- i x p^+ z^-}
H^q_T (x, \eta, \Delta^2) \frac{\Delta^\perp_\mu}{2 M}
\, , \\
\langle p_2| {\cal O}^{gg} (- z^-, z^-) | p_1 \rangle
\!\!\!&=&\!\!\!
\frac{1}{4} ( p^+ )^2
\int_{- 1}^1 d x \, {\rm e}^{- i x p^+ z^-}
H^g (x, \eta, \Delta^2)
\, , \\
\label{spin-zero-GPDs2}
\langle p_2| {\cal T}^{gg}_{\mu\nu} (- z^-, z^-) | p_1 \rangle
\!\!\!&=&\!\!\!
\frac{1}{4} ( p^+ )^2
\int_{- 1}^1 d x \, {\rm e}^{- i x p^+ z^-}
H^g_T (x, \eta, \Delta^2)
\tau^\perp_{\mu\nu, \rho\sigma} \frac{\Delta^\rho \Delta^\sigma}{4 M^2}
\, .
\end{eqnarray}
The GPDs defined by the aligned parton-helicity operators are allowed due to nonzero
orbital angular momentum between the initial and final state hadrons. They die out
with vanishing momentum transfer $\Delta = 0$.

\subsubsection{Spin-one-half hadrons}
\label{GPDsSpinOneHalf}

In case of a spin-one-half target it is convenient to express the expectation
values of local operators in terms of spinor bilinears
\begin{equation}
\label{Def-ForFac}
\begin{array}{ll}
b = \bar u (p_2) u (p_1)
\, , \qquad
&
\widetilde b = \bar u (p_2) \gamma^5 u (p_1)
\, , \\
h^{\mu} = \bar u (p_2) \gamma^\mu u (p_1)
\, , \qquad
&
\widetilde{h}^\mu = \bar u (p_2) \gamma^\mu \gamma^5 u (p_1)
\, , \\
t^{\mu\nu} = \bar u (p_2) i \sigma^{\mu\nu} u (p_1)
\, , \qquad
&
\widetilde{t}^{\mu\nu} = \bar u (p_2) i \sigma^{\mu\nu} \gamma^5 u (p_1)
\, .
\end{array}
\end{equation}
Obviously, the dual tensor bilinear $\widetilde t^{\mu\nu}$ is obtained from
$t^{\mu\nu}$ by contraction with the totally antisymmetric $\varepsilon$-tensor
and can, therefore, be eliminated from the list of independent operators.
Furthermore, equations of motion show that in each parity sector there are
relations between the structures
(\ref{Def-ForFac})
\begin{equation}
\label{GordonIdentities}
\begin{array}{cc}
p^\mu b = ( M_{H_2} + M_{H_1} ) h^\mu  - t^{\nu\mu} \Delta_\nu
\, , \qquad
&
\Delta^\mu b = - ( M_{H_2} - M_{H_1} ) h^\mu  - t^{\nu\mu} p_\nu
\, , \\
\Delta^\mu \widetilde b
= - ( M_{H_2} + M_{H_1} ) \widetilde h^\mu - \widetilde t^{\nu\mu} p_\nu
\, , \qquad
&
p^\mu \widetilde b
= ( M_{H_2} - M_{H_1} ) \widetilde h^\mu  - \widetilde t^{\nu\mu} \Delta_\nu
\, ,
\nonumber
\end{array}
\end{equation}
where $M_{H_{1/2}}$ is the mass of the incoming/outgoing hadron. The ultimate
result of this consideration suggests that matrix elements for spin-one-half
target can be expressed in terms of a few Dirac bilinears chosen to be
\begin{eqnarray}
\label{DiracBilinearsSpinOneHalf}
h^\mu
\, , \qquad
e^\mu = \frac{t^{\nu\mu} \Delta_\nu}{M_{H_2} + M_{H_1}}
\, , \qquad
\widetilde h^\mu
\, , \qquad
\widetilde e^\mu = - \frac{\Delta^\mu \widetilde b} {M_{H_2} + M_{H_1}}
\, .
\end{eqnarray}
For further use, we give the light-cone projection of two of Eqs.\
(\ref{GordonIdentities}), namely,
\begin{equation}
\label{BilinearSubst}
b = (M_{H_2} + M_{H_1}) \frac{h^+  - e^+}{p^+}
\, , \qquad
\widetilde b = - (M_{H_2} + M_{H_1}) \frac{\widetilde e^+}{\Delta^+}
\, .
\end{equation}
Thus, for the vector and axial-vector operators there are two independent Dirac
structures \cite{Ji96a}, while for the maximal-helicity quark operator there
are four structures \cite{Die01}. The decomposition of the  matrix elements for
particular operators yields
\begin{eqnarray}
\label{vectorGPDnucleon}
\langle p_2 | {\cal O}^{qq} (- z^-, z^-) | p_1 \rangle
\!\!\!&=&\!\!\!
\int_{- 1}^{1} d x \ {\rm e}^{- i x p^+ z^- }
\left\{
h^+ H^q (x, \eta, \Delta^2) + e^+ E^q (x, \eta, \Delta^2)
\right\}
\, , \\
\label{axialGPDnucleon}
\langle p_2 | \widetilde {\cal O}^{qq} (- z^-, z^-) | p_1 \rangle
\!\!\!&=&\!\!\!
\int_{- 1}^{1} d x \ {\rm e}^{- i x p^+ z^- }
\left\{
\tilde h^+ \widetilde H^q (x, \eta, \Delta^2)
+
\tilde e^+ \widetilde E^q (x, \eta, \Delta^2)
\right\}
\, , \\
\label{transvGPDnucleon}
\langle p_2 | {\cal T}^{qq}_\mu (- z^-, z^-) | p_1 \rangle
\!\!\!&=&\!\!\!
\int_{- 1}^{1} d x \ {\rm e}^{- i x p^+ z^- }
\bigg\{
t^{+ \perp}_{\phantom{+} \mu} H^q_T (x, \eta, \Delta^2)
+
\frac{p^+ e^\perp_\mu}{M_N} \widetilde H^q_T (x, \eta, \Delta^2)
\\
&-&\!\!\!
\frac{1}{2 M_N}
\left( \Delta^\perp_\mu h^+ - \Delta^+ h^\perp_\mu \right)
E^q_T (x, \eta, \Delta^2)
-
\frac{p^+ h^\perp_\mu}{2 M_N} \widetilde E^q_T (x, \eta, \Delta^2)
\bigg\}
\, . \nonumber
\end{eqnarray}
The Dirac bilinears used here are those  introduced in Eq.\ (\ref{Def-ForFac}).
Note, that compared to Ref.\ \cite{Die01}, we have dropped the terms proportional
to transverse components of $p^\mu$ since we have in mind a DIS-type frame
where $p^\mu_\perp = 0$, see Appendix \ref{LCvectors}.

For the gluonic GPDs, we have a basically identical  parametrization
\begin{eqnarray}
\label{GluonGPDnucleon}
\langle p_2 | {\cal O}^{gg} (- z^-, z^-) | p_1 \rangle
\!\!\!&=&\!\!\!
\frac{1}{4} p^+
\int_{- 1}^{1} d x \
{\rm e}^{- i x p^+ z^-}
\left\{
h^+ H^g (x, \eta, \Delta^2) + e^+ E^g (x, \eta, \Delta^2)
\right\}
\, , \\
\langle p_2 |
\widetilde {\cal O}^{gg} (- z^-, z^-)
| p_1 \rangle
\!\!\!&=&\!\!\!
\frac{1}{4} p^+
\int_{- 1}^{1} d x \
{\rm e}^{- i x p^+ z^-}
\left\{
\tilde h^+ \widetilde H^g (x, \eta, \Delta^2)
+
\tilde e^+ \widetilde E^g (x, \eta, \Delta^2)
\right\}
\, , \\
\label{TGluonGPDnucleon}
\langle p_2 |
{\cal T}^{gg}_{\mu\nu} (- z^-, z^-)
| p_1 \rangle
\!\!\!&=&\!\!\!
\frac{1}{4} p^+
\int_{- 1}^{1} d x \ {\rm e}^{- i x p^+ z^- }
\bigg\{
t^{+ \perp}_{\phantom{+} \sigma} H^g_T (x, \eta, \Delta^2)
+
\frac{p^+ e^\perp_\sigma}{M_N} \widetilde H^g_T (x, \eta, \Delta^2)
\\
&-&\!\!\!
\frac{1}{2 M_N}
\left( \Delta^\perp_\sigma h^+ - \Delta^+ h^\perp_\sigma \right)
E^g_T (x, \eta, \Delta^2)
-
\frac{p^+ h^\perp_\sigma}{2 M_N} \widetilde E^g_T (x, \eta, \Delta^2)
\bigg\}
\tau^\perp_{\mu\nu;}{}^{\sigma\rho} \frac{- \Delta_\rho}{2 M_N}
\, . \nonumber
\end{eqnarray}
The decomposition for even and odd parity operators given here has been introduced
in Ref.\ \cite{Ji96b}, while the maximal-helicity gluon sector was addressed
in full in \cite{Die01}  (see also \cite{JiHoo98,BelMul00d} for an earlier
discussion).

For further reference, we introduce a target-independent, boost-invariant form
of GPDs. They formally have the same form as the parametrizations for the scalar
target (\ref{spin-zero-GPDs}), for instance for the parity-even sector,
\begin{eqnarray}
\label{TargetIndepQuarkFirst}
\langle p_2 | {\cal O}^{qq} (- z^- , z^-) | p_1 \rangle
\!\!\!&=&\!\!\!
p^+
\int_{-1}^1 dx \, {\rm e}^{- i x p^+ z^-} F^q (x, \eta, \Delta^2)
\, , \\
\label{TargetIndepGluonFirst}
\langle p_2 | {\cal O}^{gg} (- z^- , z^-) | p_1 \rangle
\!\!\!&=&\!\!\!
\frac{(p^+)^2}{4}
\int_{-1}^1 dx \, {\rm e}^{- i x p^+ z^-} F^g (x, \eta, \Delta^2)
\, ,
\end{eqnarray}
however, the involved functions $F^a$ $(a = q, g)$ admit subsequent expansion
in Dirac bilinears,
\begin{eqnarray*}
F^a (x , \eta, \Delta^2)
\!\!\!&=&\!\!\!
\frac{h^+}{p^+} H^a (x, \eta, \Delta^2)
+
\frac{e^+}{p^+} E^a (x, \eta, \Delta^2)
\, .
\end{eqnarray*}
The odd parity and the maximal-helicity functions are deduced from the above
by simple substitutions of operators and corresponding GPD functions.

\subsubsection{A comment on gluonic matrix elements}

Calculating  Feynman diagrams, corresponding to a given process, one needs to work
with  gauge potentials rather than with the gauge covariant field strength tensors.
The conversion formula is given by \cite{Rad1996b,Rad1997a}
\begin{eqnarray}
&&\int\frac{d z^-}{2 \pi} {\rm e}^{i x z^- p^+}
\langle p_2 |
A_a^{\mu} \left( - z^- \right) A_b^{\nu} \left( z^- \right)
| p_1 \rangle
=
\frac{\delta_{ab}}{N_c^2 - 1}
\frac{1}{(\eta - x - i 0)(x + \eta - i 0)}
\\
&&\qquad\qquad\qquad\qquad\qquad\times
\left\{
\ft12
g_{\perp}^{\mu \nu}
F^g (x, \eta , \Delta^2)
-
\ft{i}2 \varepsilon^{\mu\nu}_\perp \widetilde F^g (x, \eta , \Delta^2)
+
\tau^{\mu\nu;\rho\sigma}_\perp F^g_{T \rho\sigma} (x, \eta , \Delta^2)
\right\}
\, , \nonumber
\end{eqnarray}
in terms of the target-independent boost invariant gluonic GPDs, introduced in
the preceding section.

\subsubsection{Spin-one hadrons}

Finally, let us address GPDs for spin-one hadrons, deuteron being the most
obvious suspect. We take the polarization vectors for the incoming and outgoing
deuteron as $\varepsilon^\mu_1 \equiv \varepsilon^\mu (p_1)$ and $\varepsilon^\mu_2
\equiv \varepsilon^\mu (p_2)$, respectively. We need to parametrize the non-local
vector and axial-vector operators. Representing the tensor structures as
$\varepsilon^\ast_{2 \mu} (V^{\mu\nu} , A^{\mu\nu}) \varepsilon_{1 \nu}$ in
these two cases, one should keep only the tensors which do not vanish when
contracted with $\varepsilon_{1 \mu}$, $\varepsilon_{2 \nu}$, given the
orthogonality conditions $\varepsilon_1 \cdot p_1 = \varepsilon_2 \cdot p_2 = 0$
for the polarization vectors. With the constraints from parity invariance, one
finds that $V^{\mu\nu}$ contains five linear independent tensor structures
\cite{BerCanDiePir01}
\begin{eqnarray*}
g_{\beta \alpha}
\, , \quad
p_{\beta} n_{\alpha}
\, , \quad
n_{\beta} p'_{\alpha}
\, , \quad
p_{\beta} p'_{\alpha}
\, , \quad
n_{\beta} n_{\alpha}
\, .
\end{eqnarray*}
Similarly, the $A^{\mu\nu}$ tensors are linear combinations of the seven tensor
structures
\begin{eqnarray*}
&&
\varepsilon_{\mu \nu \beta \alpha} \, p^{\mu}p'^{\nu}
\, , \quad
\varepsilon_{\mu \nu \beta \alpha} \, n^{\mu}p^{\nu}
\, , \quad
\varepsilon_{\mu \nu \beta \alpha} \, n^{\mu}p'^{\nu}
\, , \quad
\varepsilon_{\mu \nu \rho \beta} \, p^{\mu}p'^{\nu} n^{\rho} n_{\alpha}
\, , \quad
\varepsilon_{\mu \nu \rho \beta} \, p^{\mu}p'^{\nu} n^{\rho} p'_{\alpha}
\, , \nonumber\\
&&
\qquad\qquad\qquad\qquad\qquad
\varepsilon_{\mu \nu \rho \alpha} \, p^{\mu}p'^{\nu} n^{\rho} n_{\beta}
\, , \quad
\varepsilon_{\mu \nu \rho \alpha} \, p^{\mu}p'^{\nu} n^{\rho} p_{\beta}
\, .
\end{eqnarray*}
Using the Schouten identity, see Appendix \ref{conventions}, one can show
that only four out of these seven are linearly independent.

Thus, the generalized parton distributions for a deuteron target are
defined as \cite{BerCanDiePir01,CanPir03}
\begin{eqnarray}
\label{SpinOneGPDs}
\langle p_2 | {\cal O}^{qq} (- z^-, z^-) | p_1 \rangle
\!\!\!&=&\!\!\!
\int_{- 1}^{1} d x \ {\rm e}^{- i x p^+ z^- }
\bigg\{
-
p^+ (\varepsilon_2^\ast \cdot \varepsilon_1) H_1^q (x, \eta, \Delta^2)
\\
&+&\!\!\!
\Big(
\varepsilon_1^+ (\varepsilon_2^\ast \cdot p)
+
\varepsilon_2^{+ \ast} (\varepsilon_1 \cdot p)
\Big)
H_2^q (x, \eta, \Delta^2)
-
\frac{p^+}{2 M^2_D}
(\varepsilon_2^\ast \cdot p) (\varepsilon_1 \cdot p)
H_3^q (x, \eta, \Delta^2)
\nonumber\\
&&{\hspace{-4cm}}
+
\Big(
\varepsilon_1^+ (\varepsilon_2^\ast \cdot p)
-
\varepsilon_2^{+ \ast} (\varepsilon_1 \cdot p)
\Big)
H_4^q (x, \eta, \Delta^2)
+
\frac{1}{p^+}
\left(
4 M_D^2 \varepsilon_1^+ \varepsilon_2^{+ \ast}
+
\frac{1}{3}
(p^+)^2 ( \varepsilon_2^\ast \cdot \varepsilon_1 )
\right)
H_5^q (x, \eta, \Delta^2)
\bigg\}
\, , \nonumber\\
\langle p_2 | \widetilde {\cal O}^{qq} (- z^-, z^-) | p_1 \rangle
\!\!\!&=&\!\!\!
\int_{- 1}^{1} d x \, {\rm e}^{- i x p^+ z^- }
\bigg\{
M_D^2 \varepsilon_{2 \mu}^\ast \varepsilon_{1 \nu}
\widetilde H_1^q (x, \eta, \Delta^2)
\\
&-&\!\!\!
\Delta_\mu
\Big(
\varepsilon_{1 \nu} (\varepsilon_2^\ast \cdot p)
+
\varepsilon_{2 \nu}^\ast (\varepsilon_1 \cdot p)
\Big)
\widetilde H_2^q (x, \eta, \Delta^2)
\nonumber\\
&&{\hspace{-4cm}}
-
\Delta_\mu
\Big(
\varepsilon_{1 \nu} (\varepsilon_2^\ast \cdot p)
-
\varepsilon_{2 \nu}^\ast (\varepsilon_1 \cdot p)
\Big)
\widetilde H_3^q (x, \eta, \Delta^2)
-
\Delta_\mu
\Big(
\varepsilon_{1 \nu} \varepsilon_2^{+ \ast}
+
\varepsilon_{2 \nu}^\ast \varepsilon_1^+
\Big)
\widetilde H_4^q (x, \eta, \Delta^2)
\bigg\}
\frac{i}{M_D^2} \varepsilon^{+ \rho \mu \nu} p_\rho
\, . \nonumber
\end{eqnarray}
Finally, the gluon distributions in the deuteron are parametrized analogously
to quark GPDs. They can be deduced from the above definitions by replacing the
bilocal quark operators by the gluon ones, i.e., ${\cal O}^{qq} \to {\cal O}^{gg}$,
changing the GPDs on the right-hand side, $H^q$ to $\ft14 p^+ H^g$, as was done
in Eq.\ (\ref{GluonGPDnucleon}). The extra factor of $p^+$ is needed to have a
boost-independent definition of GPDs. Note that  $H_i^g$'s are even and
$\widetilde H_i^g$'s are  odd in momentum fraction $x$. Other symmetry properties
will be established below.

\subsubsection{Implications of time reversal and hermiticity}
\label{TimeReversalHermiticity}

Now we are in a position to discuss the implications of the time-reversal
invariance on the symmetry properties of GPDs. The time-reversal operation
involves the complex conjugation of states. This can be easily observed in
a simple example of the non-relativistic Schr\"odinger equation where the
time derivative is accompanied by the imaginary unity. Thus the time-reversal
operator ${\cal T}$, defined on the Hilbert space of particle states, induces
a complex conjugation on any $c$-function standing to its right, ${\cal T} f
= f^\ast {\cal T}$. As a consequence, it is an antiunitary operation \cite{SteBook}
\begin{equation}
\label{ChargeConjState}
\langle {\mit\Omega}_2 | {\cal T}^\dagger {\cal T} | {\mit\Omega}_1 \rangle
=
\langle {\mit\Omega}_2 | {\mit\Omega}_1 \rangle^\ast
=
\langle {\mit\Omega}_1 | {\mit\Omega}_2 \rangle
\, .
\end{equation}
This equality can be easily verified by expanding the state $| {\mit\Omega}_2
\rangle$ in a complete basis of states. This is to be contrasted with the action
of a unitary operator ${\cal U}$ which obeys $\langle {\mit\Omega}_1 |
{\cal U}^\dagger {\cal U} | {\mit\Omega}_2 \rangle = \langle {\mit\Omega}_1 |
{\mit\Omega}_2 \rangle$, for instance the operator of spatial parity ${\cal U}
= {\cal P}$.

Consider the matrix element of the $\bit{z}_\perp$-dependent bilocal operator
$$
{\cal O} (- z^- , - \bit{z}_\perp ; z^- , \bit{z}_\perp )
=
{\mit\Phi}^\dagger_\infty (- z^- , - \bit{z}_\perp)
{\mit\Gamma}
{\mit\Phi}_\infty (z^- , \bit{z}_\perp)
$$
built by sandwiching the Dirac/Lorentz structure ${\mit\Gamma}$ between the
elementary fields ${\mit\Phi}$ with the Wilson lines attached to them, in
order to make the operator gauge invariant. Assuming  covariant gauges in
which the gauge potential is local and vanishes at infinity, we can write
$$
{\mit\Phi}_\infty (z^-, \bit{z}_\perp)
=
{}[ \infty , \bit{z}_\perp ; z^-, \bit{z}_\perp ]
{\mit\Phi} (z^-, \bit{z}_\perp)
\, ,
$$
similarly to the notation introduced in Eq.\ (\ref{GaugeInvariantQuark}).
For the  operator matrix element, the complex-conjugated version of Eq.\
(\ref{ChargeConjState}) implies
\begin{eqnarray}
\label{ChargeConjOper}
\langle {\mit\Omega}_2 |
{\cal O} (- z^-, - \bit{z}_\perp ; z^- , \bit{z}_\perp)
| {\mit\Omega}_1 \rangle
\!\!\!&=&\!\!\!
\langle {\mit\Omega}_2 |
{\cal T}^\dagger {\cal T} {\cal O}
(- z^-, - \bit{z}_\perp ; z^- , \bit{z}_\perp)
| {\mit\Omega}_1 \rangle^\ast
\\
&=&\!\!\!
\langle {\mit\Omega}_2 |
{\cal T}^\dagger {\cal T}
{\mit\Phi}^\dagger_\infty (- z^- , - \bit{z}_\perp)
{\cal T}^\dagger
{\mit\Gamma}^\ast
{\cal T}
{\mit\Phi}_\infty (z^- , \bit{z}_\perp)
{\cal T}^\dagger {\cal T}
| {\mit\Omega}_1 \rangle^\ast
\, . \nonumber
\end{eqnarray}
In the second line,  we ``commuted'' ${\cal T}$ through ${\mit\Gamma}$ and
inserted {\sl two} more antiunitary operations. They induce {\sl two} extra
complex conjugations compensating each other in the total result.  Note, that
the light-ray operator lives on the light-cone. Since the light-cone coordinate
is a combination of the usual time and coordinate, and the latter will not be
transformed under the time-reversal, thus (inconveniently) changing the minus
coordinate into the plus one, it is instructive to augment the above equation
(\ref{ChargeConjOper}) by inserting (three times) the unitary operation of
spatial parity inversion ${\cal P}^\dagger {\cal P} = 1$. Thus, it is convenient
to discuss the transformation of the light-cone correlation functions with
respect to the combined time and spatial parity reversal. These  discrete
operations act in the following way on the quark
\begin{eqnarray*}
{\cal T} \psi (z^0 , \bit{z}) {\cal T}^\dagger
\!\!\!&=&\!\!\!
\eta_T T \psi (- z^0 , \bit{z})
\, , \qquad
{\cal P} \psi (z^0 , \bit{z}) {\cal P}^\dagger
=
\eta_P \gamma^0 \psi (z^0 , - \bit{z})
\, , \\
{\cal T} \bar\psi (z^0 , \bit{z}) {\cal T}^\dagger
\!\!\!&=&\!\!\!
\eta^\ast_T \bar\psi (- z^0 , \bit{z}) T
\, , \qquad
{\cal P} \bar\psi (z^0 , \bit{z}) {\cal P}^\dagger
=
\eta^\ast_P \bar\psi (z^0 , - \bit{z}) \gamma^0
\, ,
\end{eqnarray*}
and gluon fields
\begin{eqnarray*}
{\cal T} A^\mu_a (z^0 , \bit{z}) {\cal T}^\dagger
\!\!\!&=&\!\!\!
A^a_\mu (- z^0 , \bit{z})
\, , \qquad
{\cal P} A^\mu_a (z^0 , \bit{z}) {\cal P}^\dagger
=
A^a_\mu (z^0 , - \bit{z})
\, .
\end{eqnarray*}
Here $T = - i \gamma^5 C$ where $C$ is the charge-conjugation matrix from
Appendix \ref{conventions} and $|\eta_i|^2 = 1$. Finally, the combined
reversal of the time and spatial parity of a single-particle state with
the four-momentum $p^\mu$ and spin four-vector $s^\mu$ is
\begin{equation}
\label{TimeReversalForState}
{\cal T} {\cal P} | p^\mu , s^\mu \rangle
=
{\rm e}^{i \chi (s)} | p^\mu , - s^\mu \rangle
\, ,
\end{equation}
where the phase factor $\chi (s) = \chi_0 + \pi (s - s^z) $ depends on the
intrinsic parity $\chi_0$ of the state, its spin $s$ and spin projection
$s^z$ on the quantization axis. Hermiticity, on the other hand, leads to
\begin{equation}
\langle {\mit\Omega}_2 |
{\mit\Phi}^\dagger_\infty (- z^-, - \bit{z}_\perp)
{\mit\Gamma}
{\mit\Phi}_\infty (z^-, \bit{z}_\perp)
| {\mit\Omega}_1 \rangle^\ast
=
\langle {\mit\Omega}_1 |
{\mit\Phi}^\dagger_{- \infty} (z^-, \bit{z}_\perp)
{\mit\Gamma}^\dagger
{\mit\Phi}_{- \infty} (- z^-, \bit{z}_\perp)
| {\mit\Omega}_2 \rangle
\, .
\end{equation}

As a simple demonstration of the implications of the above transformations
for corresponding GPDs, let us see what is the result of their action on
the bilocal light-cone quark vector operator (in the light-cone gauge), with
${\mit\Phi} = \psi$ and ${\mit\Gamma} = \gamma^0 \gamma^+$, sandwiched between
the hadronic states. The space-time reversal gives
\begin{eqnarray}
\langle p_2, s_2 | \bar\psi (- z^-) \gamma^+ \psi (z^-) | s_1 , p_1 \rangle
\!\!\!&=&\!\!\!
{\rm e}^{i [ \chi (s_2) - \chi (s_1) ]}
\langle p_2, - s_2 | \bar\psi (- z^-)
T^\dagger \gamma^0 \left( \gamma^+ \right)^\ast T \gamma^0
\psi (z^-) | - s_1 , p_1 \rangle^\ast
\nonumber\\
&=&\!\!\!
{\rm e}^{i [ \chi (s_2) - \chi (s_1) ]}
\langle p_1, - s_1 | \bar\psi (- z^-)
C \left( \gamma^+ \right)^T C
\psi (z^-) | - s_2 , p_2 \rangle
\nonumber\\
&=&\!\!\!
{\rm e}^{i [ \chi (s_2) - \chi (s_1) ]}
\langle p_1, - s_2 | \bar\psi (- z^-) \gamma^+ \psi (z^-) | - s_2 , p_2 \rangle
\, .
\label{TimeSpaceParityRules}
\end{eqnarray}
On the other hand,  the hermiticity immediately implies that
\begin{equation}
\langle p_2, s_2 | \bar\psi (- z^-) \gamma^+ \psi (z^-) | s_1 , p_1 \rangle^\ast
=
\langle p_1, s_1 | \bar\psi (z^-) \gamma^+ \psi (- z^-) | s_2 , p_2 \rangle
\, .
\end{equation}
These transformations lead to the following conditions on GPDs \cite{Ji:1998pc,Die01}.
\begin{itemize}
\item Spin-zero target: The time-reversal and hermiticity result in
\begin{equation}
H^q (x , \eta , \Delta^2) = H^q (x, - \eta, \Delta^2)
\, , \qquad
( H^q )^\ast (x , \eta , \Delta^2) = H^q (x, - \eta, \Delta^2)
\, ,
\end{equation}
respectively. The time reversal changes the sign of $\eta$ because the initial
and final state are interchanged. Thus, the GPDs are real functions and they are
even under the sign change of the skewness parameter. The last property has
profound implications on the functional dependence of GPDs on the skewness,
which we will discuss below in Section \ref{Polynomiality}.
\item Spin-one-half target:
\begin{equation}
\label{time}
F^q (x, \eta, \Delta^2) = F^q (x, - \eta, \Delta^2)
\end{equation}
for $F = H$, $\widetilde H$, $E$, $\widetilde E$, $H_T$, $\widetilde H_T$,
$E_T$, and
\begin{equation}
\label{special-time}
\widetilde{E}_T^q (x, \eta,  \Delta^2)
=
- \widetilde{E}_T^q (x, - \eta,  \Delta^2)
\, .
\end{equation}
Notice that the spin-dependent phases $\chi (s_i)$ in Eq.\ (\ref{TimeSpaceParityRules})
cancel with analogous factors arising from the transformation of hadronic wave functions
in the parametrization of the matrix elements when one converts the Dirac bispinors of
negative spin $u_{- s}$ to the one with positive spin $u_s$. Taking the complex conjugates
of Eqs.\ (\ref{vectorGPDnucleon}) -- (\ref{transvGPDnucleon}) gives, on the other hand,
\begin{equation}
\label{hermit}
( F^q )^\ast (x, \eta, \Delta^2) = F^q (x, - \eta,  \Delta^2)
\, ,
\end{equation}
for all distributions except $\widetilde{E}_T^q$. For the latter, one gets
\begin{equation}
\label{special-hermit}
( \widetilde{E}_T^q )^\ast (x, \eta,  \Delta^2)
=
- \widetilde{E}_T^q (x, - \eta,  \Delta^2)
\, .
\end{equation}
Taking these constraints together we see that all eight distributions are
required to be real-valued as a consequence of the time-reversal invariance.
In other words, this symmetry fixes the phases of the distributions, but does
not require any linear combination of them to be zero.

The result that $E_T^q$ and $\widetilde{E}_T^q$ have opposite behavior under
time reversal, (see Eqs.~(\ref{time}) and (\ref{special-time})), could have
been anticipated from the inspection of the tensors that multiply these GPDs in
their definition (\ref{transvGPDnucleon}). Namely, $\Delta$ changes sign under
$p_1 \leftrightarrow p_2$ but $p$ does not. As we have seen, this does not
constrain either of these distributions to be zero. It is interesting to note
that this situation changes if instead of the bilocal quark-antiquark operator
in Eq.\ (\ref{transvGPDnucleon}) one considers the local one,
\begin{equation}
\label{local-matrix}
\langle p_2 |
\bar\psi (0) \, i \sigma^{\mu\nu} \, \psi(0)\, | p_1 \rangle
\, .
\end{equation}
The time-reversal invariance does imply now that the form factor multiplying the
fourth bilinear, $\bar{u} (p_2) (\gamma^\mu p^\nu - p^\mu \gamma^\mu) u (p_1)$,
must vanish. Since the result must not depend on $\eta$, one finds
\begin{equation}
\label{moment}
\int_{-1}^1 dx\, \widetilde{E}_T^q(x, \eta,  \Delta^2) = 0 .
\end{equation}

Thus, one finds that, by time reversal invariance, there are only three independent
form factors for the  local matrix element (\ref{local-matrix}) but four independent
generalized quark distributions to describe the bilocal matrix element
(\ref{transvGPDnucleon}). In other words, the first moment of $\widetilde{E}_T^q$
is zero by the time reversal symmetry, but  its higher moments, $\int dx\, x^n \,
\widetilde{E}_T^q(x,\eta,  \Delta^2)$ with $n > 0$ are nonzero. In fact, the
higher moments correspond to local matrix elements as in (\ref{local-matrix}) but
contain additional derivatives. The corresponding Lorentz tensors have rank larger
than two,  and allow more than three independent form factors.
\item Spin-one target: combining hermiticity and time-reversal conditions one finds
that all nine GPDs are real. However, their behavior under time reversal is not
uniform and one gets \cite{BerCanDiePir01}:
\begin{eqnarray}
H^q_i (x, \eta, \Delta^2)
\!\!\!&=&\!\!\!
\phantom{-}
H^q_i (x, - \eta, \Delta^2)
\, ,
\qquad\quad i = 1, 2, 3, 5
\, , \nonumber\\
H^q_4 (x, \eta, \Delta^2)
\!\!\!&=&\!\!\!
- H^q_4(x, - \eta, \Delta^2)
\, , \nonumber\\
\widetilde H^q_i (x, \eta, \Delta^2)
\!\!\!&=&\!\!\!
\phantom{-}
\widetilde H^q_i (x, - \eta, \Delta^2)
\, ,
\qquad\quad
i = 1, 2, 4
\, , \nonumber\\
\widetilde H^q_3 (x, \eta, \Delta^2)
\!\!\!&=&\!\!\!
- \widetilde H^q_3 (x, - \eta, \Delta^2)
\, .
\label{TimeSpinOne}
\end{eqnarray}
As a consequence of the antisymmetry of GPDs $H_4^q$ and $\widetilde{H}_3^q$ under
the reversal of the skewness parameter, one finds the sum rules \cite{BerCanDiePir01}
\begin{equation}
\int_{- 1}^1 dx \, H_4^q (x, \eta, \Delta^2)
=
\int_{- 1}^1 dx \, \widetilde{H}_3^q (x, \eta, \Delta^2)
=
0 \, .
\end{equation}
The behavior of gluon GPDs  under the time reversal is the same as in
(\ref{TimeSpinOne}) for the corresponding quark distributions.
\end{itemize}

Let us reiterate the findings of this section. For non-forward kinematics the
time-reversal invariance fixes the overall phases of generalized parton
distributions and determines their symmetry properties with respect to the
skewness sign change, but it does not reduce the number of GPDs deduced from
the generic counting of allowed structures in the matrix elements based on the
number of non-vanishing parton-hadron helicity amplitudes.

\subsection{Forward limit}

Generalized parton distributions satisfy a number of remarkable constraints
which make them partially ``known'' in certain kinematical regions. Indeed,
we observe a close analogy of the operator definition of GPDs with that for the
conventional parton densities discussed in the introduction. Obviously, the
former will reduce to the latter when we set the $t$-channel momentum transfer
to be zero, i.e., take $\Delta = 0$. Let us analyze this reduction for targets
of different spins.

\subsubsection{Spin-zero hadrons}

The parton distributions in spin-zero targets are defined by a single function
for each operator matrix element. The leading twist densities are
\begin{eqnarray}
\langle p | {\cal O}^{qq} (0 , z^-) | p \rangle
\!\!\!&=&\!\!\!
2 p^+ \int_{0}^{1} d x
\left\{
q (x) {\rm e}^{- i x p^+ z^-}
-
\bar q (x) {\rm e}^{i x p^+ z^-}
\right\}
\, , \\
\langle p | {\cal O}^{gg} (0 , z^-) | p \rangle
\!\!\!&=&\!\!\!
(p^+)^2 \int_{0}^{1} d x \,
\left\{
{\rm e}^{- i x p^+ z^+} + {\rm e}^{i x p^+ z^-}
\right\}
x g (x)
\, .
\end{eqnarray}
Notice that for practical purposes the quark $q (x)$ and antiquark $\bar{q} (x)$
distributions, both defined for positive momentum fractions $x > 0$, can be combined
together into a single functions $f_q (x)$ which ``lives'' in the interval $- 1 \leq
x \leq 1$ as demonstrated in Eq.\ (\ref{UnPolQuarkDensity}). The quark distributions are
measurable in high-energy experiments like deeply inelastic scattering and Drell-Yan
lepton pair production. In case of the pion target, the fits of  parton distributions
to available experimental data can be found in Refs.\ \cite{GluReyVogSch99,SutMarRobSti92}.
Other (pseudo)scalar targets arise if nuclei are taken into consideration as possible
targets. The spin-zero GPDs (\ref{spin-zero-GPDs}) -- (\ref{spin-zero-GPDs2}) yield,
via the forward limiting procedure $\Delta \to 0$ such that $p_1 = p_2 \equiv p$, the
conventional parton distributions,
$$
H^q (x, 0, 0) = f^q (x) = q (x) \theta (x) - \bar{q} (- x) \theta (- x)
\, .
$$
Analogous relations hold for gluons with obvious substitutions of $q (x)$ by
$x g (x)$ and $\bar{q} (- x)$ by $x g (- x)$.

\subsubsection{Spin-one-half hadrons}

For non-zero spin targets---spin-one-half being the first nontrivial example---one
is allowed for polarized quark $\Delta q$ and antiquark $\Delta \bar q$ densities
in addition to unpolarized distributions $q(x)$ and $\bar q (x)$ due to the existence
of the spin pseudovector $s^\mu$. More than this, the set of parton distributions
for spin-one-half targets takes the advantage of the entire variety of the available
twist-two operators introduced in Section \ref{ClassificationTwistTwo} and leads to
the introduction of the chiral-odd quark densities $\delta q (x)$, which do not show
up though in conventional deeply inelastic lepton-hadron experiments. The complete set
of the operator matrix elements is parametrized as follows,
\begin{eqnarray}
\label{ForwardUnpolarizedPDF}
\langle p | {\cal O}^{qq} (0 , z^-) | p \rangle
\!\!\!&=&\!\!\!
2 p^+ \int_{0}^{1} d x
\left\{
q (x) {\rm e}^{- i x p^+ z^-}
-
\bar q (x) {\rm e}^{i x p^+ z^-}
\right\}
\, , \\
\langle p | \widetilde {\cal O}^{qq} (0 , z^-) | p \rangle
\!\!\!&=&\!\!\!
2 s^+ \int_{0}^{1} d x
\left\{
\Delta q (x) {\rm e}^{- i x p^+ z^-}
+
\Delta \bar q (x) {\rm e}^{i x p^+ z^-}
\right\}
\, , \\
\langle p | {\cal T}_\mu^{qq} (0 , z^-) | p \rangle
\!\!\!&=&\!\!\!
2 \widetilde s^\perp_\mu \int_{0}^{1} d x
\left\{
\delta q (x) {\rm e}^{- i x p^+ z^-}
-
\delta \bar q (x) {\rm e}^{i x p^+ z^-}
\right\}
\, ,
\label{ForwardTransversity}
\end{eqnarray}
making use of the nucleon polarization vector $s^\mu$. Here we defined also
the ``dual'' spin vector $\widetilde s^\perp_\mu = i \varepsilon^\perp_{\mu\nu}
s^\nu_\perp$ since the conventional definition of the transversity distribution
involves the Dirac matrix $\sigma^{+\perp} \gamma^5$ rather than $\sigma^{+\perp}$
used in our Eq.\ (\ref{ForwardTransversity}). Thus, the forward limit provides a
restriction on the GPDs $H$ and $\widetilde H$:
\begin{eqnarray}
\label{ForwardLimit}
H (x, 0, 0)
\!\!\!&=&\!\!\!
f^q (x)
=
q (x) \theta (x) - \bar q (- x) \theta (-x)
\, , \\
\widetilde H (x, 0, 0)
\!\!\!&=&\!\!\!
\Delta f^q (x)
=
\Delta q (x) \theta (x) + \Delta \bar q (- x) \theta (-x)
\, ,
\end{eqnarray}
where we introduced again the functions $(f_q, \Delta f_q)$ having support
for positive and negative momentum fractions. As it is clear from these
conventions, the antiquark distributions are determined by the obvious
relations $f^q (- x) = - \bar{q} (x)$ and $\Delta f^q (-x) = \Delta \bar{q} (x)$
for the chiral-even parity-even and odd cases, respectively. They have to
be complemented by the equation
$$
\delta \bar q (x) = - \delta f^q (- x)
$$
for the chiral odd density. No constraints arise on GPDs $E$ and $\widetilde E$,
since their Dirac structure vanishes as $\Delta \to 0$.

Recalling the definitions of the gluonic distributions,
\begin{eqnarray}
\label{NonpolGluonDensity}
\langle p | {\cal O}^{gg} (0 , z^-) | p \rangle
\!\!\!&=&\!\!\!
(p^+)^2 \int_{0}^{1} d x \,
\left\{
{\rm e}^{- i x p^+ z^+} + {\rm e}^{i x p^+ z^-}
\right\}
x g (x)
\, ,
\\
\langle p | \widetilde {\cal O}^{gg} (0 , z^-) | p \rangle
\!\!\!&=&\!\!\!
s^+ p^+
\int_{0}^{1} d x \,
\left\{
{\rm e}^{- i x p^+ z^-} - {\rm e}^{i x p^+ z^-}
\right\}
x \Delta g (x)
\, ,
\label{PolGluonDensity}
\end{eqnarray}
we get for gluon GPDs
\begin{eqnarray}
H^g (x, 0, 0)
\!\!\!&=&\!\!\!
x g (x) \theta (x) - x g (- x) \theta (-x)
\, , \\
\widetilde H^g (x, 0, 0)
\!\!\!&=&\!\!\!
x \Delta g (x) \theta (x) + x \Delta g (- x) \theta (-x)
\, ,
\end{eqnarray}
and no reduction formula for the maximal-helicity gluon sector since the
forward operator matrix elements of the operator $\mathcal{T}^g_{\mu\nu}$
between the spin-one-half hadronic states vanishes identically by means to
quark-hadron helicity conservation. In the off-forward case it is lifted
due to the injection of a non-zero orbital momentum of parton via the
nonvanishing momentum transfer $\Delta$.

The unpolarized quark distribution is customarily separated into valence and
sea contributions. Namely, for unpolarized functions one writes
\begin{equation}
f^q (x) = f^{q, \rm val} (x) + f^{q, \rm sea} (x)
\, ,
\end{equation}
with
$$
f^{q, \rm val} (x)
=
q^{\rm val} (x) \theta (x)
\equiv
\left\{ q (x) - \bar q (x) \right\} \theta (x)
\, , \qquad
f^{q, \rm sea} (x) = \bar q (x) \theta (x) - \bar q (-x) \theta (- x)
\, .
$$
Analogously, one defines valence and sea quark distributions in the polarized
case
\begin{equation}
\Delta f^q (x) = \Delta f^{q, \rm val} (x) + \Delta f^{q, \rm sea} (x)
\, ,
\end{equation}
with
$$
\Delta f^{q, \rm val} (x)
=
\Delta q^{\rm val} (x) \theta (x)
\equiv
\left\{ \Delta q (x) - \Delta \bar q (x) \right\} \theta (x)
\, , \qquad
\Delta f^{q, \rm sea} (x)
=
\bar \Delta q (x) \theta (x) + \Delta \bar q (-x) \theta (- x)
\, .
$$
Except for transversity distribution $\delta q (x)$, the nucleon parton distributions
are well constrained by experimental data in a wide range of momentum fractions $x$.
The fits to world data can be found in Refs.\
\cite{PumStuHusLaiNadTun02,MarRobStiTho02,GluReyVog98} and
\cite{GehSti96,GluReyStrVog01,BluBot02,LeaSidSta02,HirKumSai03} for the unpolarized
and polarized parton densities, respectively.

\subsubsection{Spin-one hadrons}
\label{ForwardSpinOneHadrons}

Finally, let us comment on the forward limit of the spin-one GPDs. Recall, that in
the parton model there are three independent structure functions in deep inelastic
scattering off a deuteron, i.e., $F_1$, $b_1$, $g_1$, whose probabilistic
interpretation in terms of quark densities reads \cite{HooJafMan89}
\begin{eqnarray}
F_1(x)
\!\!\!&=&\!\!\!
\ft{1}{6}
\sum_q Q_q^2
\left[
q^1 (x) + q^{-1} (x) + q^0 (x)
+
\bar{q}^1 (x) + \bar{q}^{-1} (x) + \bar{q}^0 (x)
\right]
\, , \nonumber \\
b_1(x)
\!\!\!&=&\!\!\!
\ft{1}{4} \sum_q Q_q^2
\left[
2 q^0 (x) - q^{1} (x) - q^{-1} (x)
+
2 \bar{q}^0 (x) - \bar{q}^{1} (x) - \bar{q}^{-1} (x)
\right]
\, , \nonumber \\
g_1 (x)
\!\!\!&=&\!\!\!
\ft{1}{2} \sum_q Q_q^2
\left[
q^1_\uparrow (x) - q^{-1}_\uparrow (x)
+
\bar{q}^1_\uparrow (x) - \bar{q}^{-1}_\uparrow (x)
\right]
\, .
\label{parton-model}
\end{eqnarray}
Here $q_{\uparrow (\downarrow)}^\lambda (x)$ represents the probability to find
a quark with momentum fraction $x$ and positive (negative) helicity in a deuteron
target of helicity $\lambda$. The unpolarized quark densities $q^\lambda$ are
defined as $q^{\lambda}(x) = q_{\uparrow}^\lambda(x) + q_\downarrow^\lambda(x)$.
From parity considerations, one has $q^\lambda_{\uparrow} = q^{-\lambda}_{\downarrow}$.
The densities for antiquarks are defined in a similar way. Note, that the probabilistic
interpretation for $F_1$ and $g_1$ is similar to that in the spin-one-half
case. However, the function $b_1$ does not appear for spin-one-half targets. In the
forward limit, the only structures in Eq.~(\ref{SpinOneGPDs}) that survive are those
proportional to $H_1$, $H_5$ and $\widetilde{H}_1$, because in that limit we have
$\Delta=0$ and $\varepsilon_1 \cdot p = \varepsilon_2 \cdot p = 0$. This yields
\begin{eqnarray}
\label{forward}
\begin{array}{ll}
H_1^q (x,0,0)
=
\ft{1}{3}
\left[
q^1 (x) + q^{-1} (x) + q^0 (x)
\right] \theta (x)
&\!\!\!
-
\ft{1}{3}
\left[
\bar{q}^1 (- x) + \bar{q}^{-1} (- x) + \bar{q}^0 (- x)
\right] \theta (- x)
\, , \\
& \\
H_5^q (x,0,0)
=
\ft{1}{2}
\left[
2 q^0 (x) - q^{1} (x) - q^{-1} (x)
\right] \theta (x)
&\!\!\!
-
\ft{1}{2}
\left[
2 \bar{q}^0 (- x) - \bar{q}^{1} (- x) - \bar{q}^{-1} (- x)
\right] \theta (- x)
\, , \\
& \\
\widetilde H_1^q (x,0,0)
=
\left[
q^1_\uparrow (x) - q^{-1}_\uparrow (x)
\right] \theta (x)
&\!\!\!
+
\left[
\bar{q}^1_\uparrow (- x) - \bar{q}^{-1}_\uparrow (- x)
\right] \theta (- x)
\, .
\end{array}
\end{eqnarray}
Note that the sum rule \cite{BerCanDiePir01}
\begin{eqnarray}
0
=
\int_{-1}^1 dx \, H_5^q (x,0,0)
=
\ft12 \int_{0}^1 dx \,
\left[
2 q^0(x) - q^{1}(x) - q^{-1}(x)
-
2 \bar{q}^0 (x) + \bar{q}^{1} (x) + \bar{q}^{-1} (x)
\right]
\, ,
\end{eqnarray}
corresponding to the parton model result $\int_0^1 b_1(x) = 0$ of Refs.\
\cite{EfrTer82,CloKum90,SchSzyTer99}, obtained under the assumption
that the quark sea $q - \bar{q}$ does not contribute to this integral,
and the vanishing of the integral
\begin{equation}
\int_{- 1}^1 dx \, \widetilde{H}_4^q (x, 0, 0) = 0
\, ,
\end{equation}
follow from the fact that the local limit of the matrix element of
the light-cone operator cannot involve the ``auxiliary'' light-like vector
$n^\nu$ in its parametrization.

\subsection{Form factors}

Another interesting limit of GPDs reduces them to the hadronic form factors.
We already discussed form factors in the  introductory section. Here we elaborate
on this issue in greater detail, starting from spin-zero hadrons and finishing
with the spin-one deuteron. Form factors arise as coefficients in the expansion
of matrix elements of  local currents in Dirac/Lorentz structures. It is useful
to define form factors for each particular quark flavor. The lowest Lorentz-spin
local operators are the electromagnetic current
\begin{equation}
j_q^\mu (z) = \bar\psi_q (z) \gamma^\mu \psi_q (z)
\, ,
\end{equation}
and the axial-vector current available through weak interaction of
leptons with hadrons
\begin{equation}
j_{q}^{5\mu} (z) = \bar\psi_q (z) \gamma^\mu \gamma^5 \psi_q (z)
\, .
\end{equation}
The tensor chiral-odd current is not directly measurable with any
known probe.

\subsubsection{Spin-zero hadrons}

A spin-zero hadron has only one  form factor
\begin{equation}
\langle p_2 | j_q^\mu (0) | p_1 \rangle
=
p^\mu F^q (\Delta^2)
\, .
\end{equation}
The reduction formula, obviously, is given by
\begin{equation}
\int_{-1}^1 dx \, H^q (x, \eta , \Delta^2)
=
F^q (\Delta^2)
\, .
\end{equation}
The disappearance of the skewness dependence on the right-hand side of this sum
rule,  as well as for other targets  discussed below, will be explained in Section
\ref{Polynomiality}. The well-known example of a spin-zero ``target'' is the
charged pion, in which case
\begin{equation}
\label{EMformfactorPion}
F_{\pi^+} = Q_u F^u - Q_d F^d
\, .
\end{equation}

\subsubsection{Spin-one-half hadrons}
\label{Spin12FFs}

Starting with spin-one-half hadrons, the matrix elements of the
axial-vector current are also non-zero. Analogously to the proton
matrix elements of the electromagetic current (\ref{QuarkEMcurrent}),
we define form factors for each specie of quarks separately
\begin{equation}
\label{vectorFFs}
\langle p_2 | j^\mu_q (0) |p_1 \rangle
=
h^\mu F^q_1 ( \Delta^2) + e^\mu F^q_2 (\Delta^2)
\, ,
\end{equation}
and analogously for the axial-vector current
\begin{equation}
\label{axialFFs}
\langle p_2 | j^{5\mu}_q (0) |p_1 \rangle
=
\tilde h^\mu G^q_A ( \Delta^2) + \tilde e^\mu G^q_P ( \Delta^2)
\, ,
\end{equation}
with axial $G_A$ and pseudoscalar $G_P$ form factors multiplied by the Dirac
bilinears, introduced in Section \ref{GPDsSpinOneHalf}. Note, that in the
case of axial current one can also write a term $\Delta_\nu \bar u (p_2)
\sigma^{\nu\mu} \gamma^5 u (p_1)$ which is similar to the electric dipole
moment of the nucleon. However, it is forbidden by the time-reversal
invariance.

Thus, the first moments of the twist-two GPDs are equal to the corresponding
quark  form factors in the nucleon
\begin{eqnarray}
\label{SumRule}
&&\int_{- 1}^{1} d x \, H^q (x, \eta, \Delta^2) = F^q_1 (\Delta^2)
\, , \qquad
\int_{- 1}^{1} d x \, E^q (x, \eta, \Delta^2) = F^q_2 (\Delta^2)
\, , \\
&&\int_{- 1}^{1} d x \, \widetilde H^q (x, \eta, \Delta^2) = G^q_A (\Delta^2)
\, , \qquad
\int_{- 1}^{1} d x \, \widetilde E^q (x, \eta, \Delta^2) = G^q_P (\Delta^2)
\, , \nonumber
\end{eqnarray}
i.e., Dirac, Pauli, axial, and pseudoscalar form factors, respectively.
The conventional proton $F_i^p$ and neutron $F_i^p$ electromagnetic form
factors are given in terms of quark ones
\begin{equation}
F_i^p = Q_u F_i^u + Q_d F_i^d
\, , \qquad
F_i^n = Q_d F_i^u + Q_u F_i^d
\, ,
\end{equation}
where in the second relation we used the isospin symmetry to express quark
operator matrix elements in the neutron in terms of the ones in the proton,
$\langle n | \bar{d} d | n \rangle = \langle p | \bar{u} u | p \rangle$ and
$\langle n | \bar{u} u | n \rangle = \langle p | \bar{d} d | p \rangle$.
Analogously we define decompositions for other form factors. The normalization
of the proton and neutron form factors at zero recoil are given in Table
\ref{FFProtonnormalization}.

\begin{table}[t]
\begin{center}
\begin{tabular}{lcccc}
\hline
$\quad i$ & $F_1^i (0)$ & $F_2^i (0)$ & $G_A^{(3)i} (0)$ & $G_P^{(3)i} (0)$
\phantom{$\bigg($}
\\ \hline\hline
(p)roton  & 1 & $\kappa^p = 1.79$   & $g_A^{(3)p} = 1.267$   & $4 g_A^{(3)p} (0) M_N^2/m_\pi^2$
\phantom{$\bigg($}
\\ \hline
(n)eutron & 0 & $\kappa^n = - 1.93$ & $g_A^{(3)n} = - 1.267$ & $4 g_A^{(3)n} (0) M_N^2/m_\pi^2$
\phantom{$\bigg($}
\\ \hline
\end{tabular}
\end{center}
\caption{\label{FFProtonnormalization} Normalization of nucleon electromagnetic
and isovector axial and pseudoscalar form factors. An exact isospin symmetry ($M_N
= M_p = M_n$) is implied for the axial form factors such that the isovector axial
constants coincide up to an overall sign, $g_A^{(3)} = g_A^{(3)p} = - g_A^{(3)n}$.}
\end{table}

Thus, the valence $u$- and $d$-quark form factors \footnote{The name reflects the
relation of these particular combination of hadronic form factors at zero recoil
to the valence-quark sum rules for forward parton densities.} can be extracted from
the proton and neutron form factors via the formulas
\begin{equation}
F^{u, \rm val}_i (\Delta^2)
=
2 F_i^p (\Delta^2) + F_i^n (\Delta^2)
\, , \qquad
F^{d, \rm val}_i (\Delta^2)
=
F_i^p (\Delta^2) + 2 F_i^n (\Delta^2)
\, .
\end{equation}
The latter are known from experimental measurements accessing the Sachs electric
and magnetic form factors, introduced earlier in Section \ref{BreitFrameFFs}.
The latter are related to the above Dirac and Pauli form factors by a set of
covariant equations
\begin{equation}
\label{DefinitionsOfSachsFFs}
G_E^i (\Delta^2)
=
F_1^i (\Delta^2) + \frac{\Delta^2}{4 M^2} F_2^i (\Delta^2)
\, , \qquad
G_M^i (\Delta^2)
=
F_1^i (\Delta^2) + F_2^i (\Delta^2)
\, .
\end{equation}
and can be pa\-ra\-met\-rized by dipole formulas in the small-$\Delta^2$ region
\begin{eqnarray}
\label{dipoleFF}
G_E^p (\Delta^2)
= \frac{1}{1 + \kappa_p} G_M^p (\Delta^2)
\!\!\!&=&\!\!\! \frac{1}{\kappa_n} G_M^n (\Delta^2)
\equiv G_D (\Delta^2)
= \left( 1 - \frac{\Delta^2}{0.71 \, {\rm GeV}^2} \right)^{- 2}
\, , \nonumber\\
G_E^n (\Delta^2)
\!\!\!&=&\!\!\! 0 \, .
\end{eqnarray}
Here we introduced the nucleon dipole form factor $G_D$, which is used as a
first order approximation for the $\Delta^2$ dependence. Available experimental
data are  rather well fitted by this simple form with certain deviations at
higher momentum transfers $|\Delta^2| > 2 \, {\rm GeV}^2$. Note that we have
set the neutron electric form factor equal to zero since at small $|\Delta^2|$
it vanishes as a first power of $\Delta^2$ with $-1/6$ times the neutron charge
radius $r_n^2$ as a coefficient, which is quite small, $r_n^2 \approx - 0.113\
{\rm fm}^2$. A modern summary of experimental results can be found in Ref.\ \cite{Gao03}.

The parametrizations for the sea-quark form factors, dubbed for their connection
to the first moment of the sea quark parton distributions, are guided by the counting
rules, discussed later in Section \ref{QuarkCountRules}. A plausible model ansatz
then reads in terms of Sachs form factors
\begin{equation}
\label{SQFF}
G_E^{\rm sea }(\Delta^2)
= \frac{1}{1 + \kappa_{\rm sea}} G_M^{\rm sea} (\Delta^2)
= \left( 1 - \frac{\Delta^2}{m_{\rm sea}^2} \right)^{- 3},
\end{equation}
with yet another mass cutoff $m_{\rm sea}$. The slope of $F_1^{\rm sea}$ is given
by $B_{\rm sea} = \partial F_1^{\rm sea} / \partial \Delta^2 |_{\Delta^2 = 0} = 3
/ m^2_{\rm sea} - \kappa_{\rm sea}/4M^2$.

The axial form factors are given analogously by
\begin{equation}
G^{i, \rm val}_A (\Delta^2)
=
\left( 1 - \frac{\Delta^2}{1.1 \, {\rm GeV}^2} \right)^{- 2}
\, ,
\qquad
G^{\rm sea}_A (\Delta^2)
=
\left( 1 - \frac{B_A}{3} \Delta^2\right)^{- 3}
\, ,
\end{equation}
where $i$ runs over $u$- and $d$-quark species. The first of these relations is
driven by experimental data, while the latter is merely a rough model. The slope
$B_A$ is considered as a free parameter, which should be fixed from experimental
data. Last but not least, the isovector pseudoscalar form factor, having the flavor
structure for the proton $G_P^{(3)} = \langle p | \bar{u} u - \bar{d} d | p \rangle$,
is dominated by the pion-pole contribution,
\begin{eqnarray}
\label{IsovectorPsedoscalarFF}
G_P^{(3)} (\Delta^2)
\approx
\frac{4 g_A^{(3)} M^2_N}{m_\pi^2 - \Delta^2}
\, ,
\end{eqnarray}
valid at $|\Delta| \sim m_\pi \approx 0.14 \ \mbox{GeV}$ and the residue of the
pole being proportional to the triplet axial constant $g_A^{(3)}$.

\subsubsection{Spin-one hadrons}

The conventional form factor decomposition of the vector and axial
currents for spin-1 hadrons \cite{Fre92,ArnCarGro81} is given by
\begin{eqnarray}
\langle p_2 | j_q^\mu (0) | p_1 \rangle
\!\!\!&=&\!\!\!
-
G^q_1 ( \Delta^2 ) ( \varepsilon_2^* \cdot \varepsilon_1) p^\mu
\\
&&\!\!\!
+
G^q_2 ( \Delta^2 )
\Big(
\varepsilon_1^\mu ( \varepsilon_2^* \cdot p )
+
\varepsilon_2^{* \mu} ( \varepsilon_1 \cdot p)
\Big)
-
G^q_3 ( \Delta^2 ) (\varepsilon_1 \cdot p) (\varepsilon_2^* \cdot p )
\frac{p^\mu}{2 M^2_D}
\, , \nonumber\\
\langle p_2 | j_q^{5 \mu} (0) | p_1 \rangle
\!\!\!&=&\!\!\!
i \widetilde G^q_1 ( \Delta^2 ) \varepsilon^{\mu\nu\rho\sigma}
\varepsilon^*_{2 \nu} \varepsilon_{1\rho} p_\sigma
+
i \widetilde G^q_2 ( \Delta^2 )
\varepsilon^{\mu\nu\rho\sigma}
\Delta_\nu p_\rho
\frac{
\varepsilon_{1 \sigma} (\varepsilon_2^* \cdot p)
+
\varepsilon^*_{2 \sigma} (\varepsilon_1 \cdot p)
}{
M^2_D
}
\, ,
\end{eqnarray}
The matrix elements are defined here flavor by flavor; to get form factors for
a particular hadron, one must weight the flavor components with appropriate
electromagnetic or weak charges and sum over flavors.

Integrating the matrix elements defining the twist-two GPDs in the deuteron
target we obtain the sum rules
\begin{eqnarray}
\int_{-1}^1 d x \, H^q_i (x, \eta, \Delta^2)
\!\!\!&=&\!\!\!
G^q_i (\Delta^2)
\, ,
\qquad\quad
i = 1,2,3
\, , \nonumber\\
\int_{-1}^1 d x \, \widetilde H^q_i (x, \eta, \Delta^2)
\!\!\!&=&\!\!\!
\widetilde G^q_i (\Delta^2)
\, ,
\qquad\quad
i = 1, 2
\, , \nonumber\\
\int_{-1}^1 d x \, H^q_4 (x, \eta, \Delta^2)
\!\!\!&=&\!\!\!
\int_{-1}^1 d x \, \widetilde H^q_3 (x, \eta, \Delta^2)
= 0
\, , \nonumber\\
\int_{-1}^1 d x \, H^q_5 (x, \eta, \Delta^2)
\!\!\!&=&\!\!\!
\int_{-1}^1 d x \, \widetilde H^q_4 (x, \eta, \Delta^2)
= 0
\, .
\end{eqnarray}
For $H_4$, $\widetilde H_3$ and $H_5$, $\widetilde H_4$, the integrals do not
correspond to form factors of the local vector or axial currents and therefore
vanish. In the case of $H_4$ and $\widetilde H_3$ this is due to the time reversal
constraints, whereas the definitions of $H_5$ and $\widetilde{H}_4$ involve the
tensor $n^\mu n^\nu/(p^+)^2$, whose analogue cannot appear in the decomposition
of local currents due to Lorentz invariance. Some of them we already established
in Section \ref{ForwardSpinOneHadrons}.

The set of form factors $G_1$, $G_2,$ and $G_3$ may be expressed in terms of
the charge monopole, magnetic dipole, and charge quadrupole form factors:
\begin{eqnarray}
\label{Rel-among-FF}
G_C(\Delta^2)
\!\!\!&=&\!\!\!
\left( 1 + \frac{2\tau}{3} \right) G_1(\Delta^2)
-
\frac{2\tau}{3}
\left[
G_2(\Delta^2) - (1 + \tau) G_3(\Delta^2)
\right]
\, ,
\nonumber\\
G_M(\Delta^2)
\!\!\!&=&\!\!\!
G_2(\Delta^2)
\, , \qquad
G_Q(\Delta^2)
=
G_1(\Delta^2) - G_2(\Delta^2) + (1 + \tau) G_3(\Delta^2)
\, ,
\nonumber
\end{eqnarray}
with $\tau \equiv - \Delta^2/(4 M^2)$. The normalization of these form
factors is given by
\begin{eqnarray}
\label{Def-StaPro}
G_C(0)=1
\, , \qquad
G_M(0) = \mu_d = 1.714
\, , \qquad
G_Q(0)= Q_d = 25.83
\, .
\end{eqnarray}
A parameterization for the deuteron form factors inspired by the quark counting
rules for large $|\Delta^2|$ can be found in Ref.\ \cite{KobSya95}:
\begin{eqnarray}
G_C(\Delta^2) &\!\!\!=\!\!\!&
\frac{G_D^2 \left( \ft14 \Delta^2 \right)}{1 + 2\tau}
\left[
\left( 1 - \frac{2}{3}\tau \right) g_{00}^+
+
\frac{8}{3} \sqrt{2 \tau} g_{+0}^+
-
\frac{2}{3}(1 - 2 \tau) g_{+-}^+
\right]
\, , \nonumber\\
G_M(\Delta^2)
\!\!\!&=&\!\!\!
\frac{G_D^2 \left(\ft14 \Delta^2 \right)}{1 + 2\tau}
\left[
2 g_{00}^+
-
\frac{2 (1 - 2\tau)}{\sqrt{2 \tau}} g_{+0}^+
-
2 g_{+-}^+
\right]
\, , \\
G_Q(\Delta^2)
\!\!\!&=&\!\!\!
\frac{G_D^2 \left( \ft14 \Delta^2 \right)}{1 + 2\tau}
\left[
-
g_{00}^+
+
\sqrt{\frac{2}{\tau}} g_{+0}^+
-
\frac{1 + \tau}{\tau} g_{+-}^+
\right]
\, .
\nonumber
\end{eqnarray}
The function $G_D \left( \Delta^2 \right)$ here is the standard dipole parameterization
for the nucleon form factor (\ref{dipoleFF}). The helicity amplitudes in the infinite
momentum frame are given by \cite{BroHil92}
\begin{eqnarray}
g_{00}^+ = \sum_{i=1}^4 \frac{a_i}{\alpha_i^2-\Delta^2}, \quad
g_{+0}^+ = \sqrt{-\Delta^2} \sum_{i=1}^4 \frac{b_i}{\beta_i^2-\Delta^2}, \quad
g_{+-}^+ = -\Delta^2 \sum_{i=1}^4 \frac{c_i}{\gamma_i^2-\Delta^2}.
\end{eqnarray}
The counting rules for large $|\Delta^2|$ predict the following behavior
\begin{eqnarray}
g_{00}^+ \sim (-\Delta)^{-2}, \quad g_{+0}^+ \sim (-\Delta)^{-3},
\quad
g_{+-}^+ \sim (-\Delta)^{-4},
\end{eqnarray}
which gives together with the static properties (\ref{Def-StaPro}) six constraints
for twenty four fitting parameters:
\begin{equation}
\label{Con-FitParDFF0}
\sum_{i=1}^4 \frac{a_i}{\alpha_i^2} =1, \
\sum_{i=1}^4 \frac{b_i}{\beta_i^2} =\frac{2 -\mu_d}{2\sqrt{2}M_d}, \
\sum_{i=1}^4 \frac{c_i}{\gamma_i^2} =\frac{1-\mu_d - Q_d}{4M_d^2},\
\sum_{i=1}^4 b_i= \sum_{i=1}^4 c_i=\sum_{i=1}^4 c_i \gamma_i^2  =0.
\nonumber
\end{equation}
To reduce this set to twelve parameters, one may introduce for each group
${\alpha_i}$, ${\beta_i}$, and ${\gamma_i}$ the algebraic relations:
\begin{eqnarray}
\label{Con-FitParDFF}
\alpha_i^2 = \alpha_1^2 + \left(\alpha_4^2-\alpha_1^2\right)
\frac{i-1}{3}\quad \mbox{for}
\quad i=1,\dots,4.
\end{eqnarray}
We quote here the fitting parameters from Ref.\ \cite{Abbetal00}. They are given
in Table \ref{Tab-DFF}.

\begin{table}[t]
\vspace{-0.5cm}
\begin{center}
\begin{tabular}{rcccc}
\hline
$i$ & 1 & 2 & 3 & 4
\\ \hline\hline
$a_i\ [{\rm fm}^{-2}]$
& $1.57057$ & $12.23792$ & $-42.04576$ & Eq.\ (\ref{Con-FitParDFF0})
\\ \hline
$b_i\ [{\rm fm}^{-1}]$
& $0.07043$ & $0.14443$ & Eq.\ (\ref{Con-FitParDFF0}) & Eq.\ (\ref{Con-FitParDFF0})
\\ \hline
$c_i$ \phantom{$[{\rm fm}^{-1}]$}
& $-$0.16577 & Eq.\ (\ref{Con-FitParDFF0}) & Eq.\ (\ref{Con-FitParDFF0}) &
Eq.\ (\ref{Con-FitParDFF0})
\\ \hline
$\alpha_i^2\ [{\rm fm}^{-2}]$
& $1.52501$ & Eq.\ (\ref{Con-FitParDFF}) & Eq.\ (\ref{Con-FitParDFF}) & $23.20415$
\\ \hline
$\beta_i^2\ [{\rm fm}^{-2}]$
& $43.67795$ & Eq.\ (\ref{Con-FitParDFF}) & Eq.\ (\ref{Con-FitParDFF}) & $2.80716$
\\ \hline
$\gamma_i^2\ [{\rm fm}^{-2}]$  & $1.87055$
& Eq.\ (\ref{Con-FitParDFF}) & Eq.\ (\ref{Con-FitParDFF}) & $41.1294$
\\ \hline
\end{tabular}
\end{center}
\caption{Sets of fitting parameters for the deuteron electromagnetic form factors.}
\label{Tab-DFF}
\end{table}

\subsection{Polynomiality and skewness dependence}
\label{Polynomiality}

Having discussed the simplest limiting cases of GPDs, we are going to address
now their generic properties, like the polynomiality in the skewness parameter
$\eta$. To properly address the issue, the formalism of Wilson operators is
especially handy. The Taylor expansion of non-local light-cone operators
results in a tower of local operators of increasing Lorentz spin. Namely, for
the even-parity quark and gluon operators we have an infinite series decomposition
\begin{eqnarray}
\label{QnonlocalOper}
{\cal O}^{qq} (- z^- , z^-)
\!\!\!&=&\!\!\!
\sum_{j = 0}^\infty
\frac{1}{j!} (- i z^-)^j
n^{\mu_1} n^{\mu_2} \dots n^{\mu_{j + 1}}
{\cal R}^{2,qq}_{\mu_1 \mu_2 \dots \mu_{j + 1}}
\, , \\
\label{GnonlocalOper}
{\cal O}^{gg} (- z^- , z^-)
\!\!\!&=&\!\!\!
\sum_{j = 0}^\infty
\frac{1}{j!} (- i z^-)^j
n^{\mu_1} n^{\mu_2} \dots n^{\mu_{j + 2}}
{\cal R}^{2,gg}_{\mu_1 \mu_2 \dots \mu_{j + 2}}
\, ,
\end{eqnarray}
in terms of the twist-two local operators
\begin{eqnarray}
\label{QuarkLocalOperator}
{\cal R}^{2,qq}_{\mu_1 \mu_2 \dots \mu_j}
\!\!\!&=&\!\!\!
\Sym_{\mu_1 \mu_2 \dots \mu_j}
\bar\psi
\gamma_{\mu_1}
i\!\stackrel{{}_\leftrightarrow}{\cal D}_{\mu_2}
i\!\stackrel{{}_\leftrightarrow}{\cal D}_{\mu_3}
\dots
i\!\stackrel{{}_\leftrightarrow}{\cal D}_{\mu_j}
\psi \, , \\
\label{GluonLocalOperator}
{\cal R}^{2,gg}_{\mu_1 \mu_2 \dots \mu_j}
\!\!\!&=&\!\!\!
\Sym_{\mu_1 \mu_2 \dots \mu_j}
F_{\mu_1 \nu}
i\!\stackrel{{}_\leftrightarrow}{\cal D}_{\mu_2}
i\!\stackrel{{}_\leftrightarrow}{\cal D}_{\mu_3}
\dots
i\!\stackrel{{}_\leftrightarrow}{\cal D}_{\mu_{j - 1}}
F^\nu{}_{\mu_j}
\, ,
\end{eqnarray}
constructed from the elementary fields and the covariant left-right derivative
acting on them,
\begin{equation}
\stackrel{{}_\leftrightarrow}{\cal D}_\mu
\ \equiv \
\stackrel{{}_\rightarrow}{\cal D}_\mu
-
\stackrel{{}_\leftarrow}{\cal D}_\mu
\, .
\end{equation}
As before, ${\Sym}$ is the operation of symmetrization and trace subtraction.
Analogous Taylor expansion hold for other twist-two quark and gluon operators.
The matrix elements of the local operators listed  above are related to the Mellin
moments of generalized parton distributions introduced in the preceding section.
For $j = 2$, the operators coincide with the quark and gluon parts of the classical,
symmetric energy-momentum tensor (see Appendix \ref{SymmetricEMtensor}), namely
\begin{eqnarray}
\label{QuarkEMtensor}
{\mit\Theta}^q_{\mu_1 \mu_2}
\!\!\!&=&\!\!\!
\ft{1}{2}
{\cal R}^{2,qq}_{\mu_1 \mu_2}
\equiv
\frac{i}{4}
\bar\psi
\left\{
\gamma_{\mu_1} \! \stackrel{{}_\leftrightarrow}{\cal D}_{\mu_2}
+
\, \gamma_{\mu_2} \! \stackrel{{}_\leftrightarrow}{\cal D}_{\mu_1}
-
\frac{1}{2} g_{\mu_1 \mu_2} \!
\stackrel{{}_\leftrightarrow}{\not\!\!{\cal D}} \,
\right\}
\psi \, , \\
\label{GluonEMtensor}
{\mit\Theta}^g_{\mu_1 \mu_2}
\!\!\!&=&\!\!\!
\phantom{\ft{1}{2}}
{\cal R}^{2,gg}_{\mu_1 \mu_2}
\equiv
\frac{1}{2}
\left\{
F_{\mu_1 \nu}^a F_{a \, \mu_2}^\nu
+
F_{\mu_2 \nu}^a F_{a \, \mu_1}^\nu
-
\frac{1}{2} g_{\mu_1 \mu_2}
F_{\mu \nu}^a F^{\nu\mu}_a
\right\}
\, .
\end{eqnarray}

When the index of the Dirac matrix in the quark bilocal operator
(\ref{QuarkLightRayOperators}) is not contracted with the lightlike vector
$n^\mu$, the operator ceases to have a definite geometric twist-two. In
the subsequent discussion, we will need a non-local form of the twist-two
quark operator which results from the extraction of the leading component
of the bilocal quark operator with a free Lorentz index of the Dirac matrix.
This form can be found with the help of an  identity \cite{BalBra89}, which
can be verified by Taylor expanding the light-ray operator,
\begin{eqnarray}
\label{TwistTwoOpenIndex}
{\cal R}^{2, qq}_{\mu} (- z^- , z^-)
\!\!\!&=&\!\!\!
\int_{-1}^1 du \,
\frac{\partial}{\partial z^\mu}
\bar\psi (- u z) {\not\!z} [- u z , u z] \psi (u z)
\\
&=&\!\!\!
\sum_{j = 0}^\infty
\frac{1}{j!} (- i z^-)^j
n^{\mu_1} \dots n^{\mu_j}
{\cal R}^{2,qq}_{\mu \mu_1 \dots \mu_j}
\, , \nonumber
\end{eqnarray}
with $z^\mu = z^- n^\mu$ on the right-hand side of the equality. Obviously,
$$
{\cal O}^{qq} (- z^- , z^-) = n^\mu {\cal R}^{2, qq}_\mu ( - z^-, z^- )
\, .
$$
The generalization to gluonic operators is straightforward. Having introduced the
local Wilson operators, we are in a position to discuss the polynomiality property
of Mellin momentns of GPDs, which are related to the hadronic matrix elements of
these operators.

\subsubsection{Spin-zero hadrons}

As usual, it is instructive to start with a spin-zero target. The parametrization
for the matrix elements of the completely symmetrized and traceless local Wilson
operators, introduced in the preceding section, sandwiched between the corresponding
states reads
\begin{eqnarray}
\label{LorentzVscal}
\!\!\!\!\!\!\!\!\!
\langle p_2 | {\cal R}^{2,qq}_{\mu_1 \dots \mu_j} | p_1 \rangle
\!\!\!&=&\!\!\!
\Sym_{\mu_1 \dots \mu_j}
\left\{
p_{\mu_1} \dots p_{\mu_j} H^q_{j, 0}
+
\Delta_{\mu_1} p_{\mu_2} \dots p_{\mu_j} H^q_{j, 1}
+
\dots
+
\Delta_{\mu_1} \dots \Delta_{\mu_j} H^q_{j , j}
\right\}
\, , \\
\!\!\!\!\!\!\!\!\!
\langle p_2 | \widetilde {\cal R}^{2,qq}_{\mu_1 \dots \mu_j} | p_1 \rangle
\!\!\!&=&\!\!\!
0
\, ,
\end{eqnarray}
in terms of the form factors $H_{j,k} (\Delta^2)$, whose dependence on the squared
momentum transfer $Q^2$ was not displayed for brevity. The vanishing of the matrix
element of the odd parity operator is a simple consequence of the fact that one
cannot form a ``twist-two'' axial-vector, i.e., involving only collinear
momenta, for a spin-zero hadron. Thus, only the vector operators develop non-zero
expectation values. For the Lorentz spin $j \geq 1$, there are exactly $j + 1$
form factors $H_{j,k}$ as had been established in Refs.\ \cite{Ji:1998pc,JiLeb00}.
Further, it is convenient to introduce a generating function of the coefficients
$H_{jk}$. One can see that, after contraction of the open Lorentz indices in Eq.\
(\ref{LorentzVscal}) with the light-cone vectors $n_\mu$ and summation over resulting
Wilson operators as in Eqs.\ (\ref{QnonlocalOper}) and (\ref{GnonlocalOper}), one
immediately establishes a relation to the Mellin moments of GPDs
\begin{eqnarray}
\label{FFversusGPDs}
H^q_{j, k} (\Delta^2)
= \frac{1}{k!} \left. \frac{d^k}{d \eta^k} \right|_{\eta = 0}
\int_{-1}^{1} dx\; x^{j - 1} H^q (x, \eta , \Delta^2)
\, ,
\end{eqnarray}
where $0 \le k \le j$, $1 \le j$. This equation demonstrates that any given
$x^{j - 1}$ moment of the GPD $H^q (x, \eta, \Delta^2)$ is a polynomial in
skewness $\eta$ of order $j$. Since this GPD is an even function of skewness,
as a consequence of the time-reversal invariance, only the Lorentz structures
even in $\Delta$ survive in Eq.\ (\ref{LorentzVscal}).

Later on in Section \ref{Compton} of this review, we will analyze the operator
product expansion for the off-forward Compton amplitude, which gives access
to GPDs. This will require the consideration of hadronic matrix elements of
non-local twist-two operators with an open Lorentz index, like Eq.\
(\ref{TwistTwoOpenIndex}), and their parametrization in terms of twist-two
GPDs. Having the decomposition (\ref{LorentzVscal}) and the relation
(\ref{FFversusGPDs}) at hand, the problem of designing the corresponding
operator matrix elements can be easily solved. To this end, we have to project
both sides of Eq.\ (\ref{LorentzVscal}) with the product of the light-cone
vectors $n_{\mu_2} \dots n_{\mu_j}$, leaving the index $\mu_1$ intact. The
obvious relation
\begin{eqnarray}
\label{ScalarContraction}
n_{\mu_1} \dots n_{\mu_j} \Sym_{\mu \mu_1 \dots \mu_j}
\Delta^\mu \dots \Delta^{\mu_k} p^{\mu_{k+1}} \dots p^{\mu_j}
=
\frac{( p^+ )^j}{j + 1}
\left\{
(j + 1 - k) p^\mu \eta^{k}
+
k \Delta^\mu \eta^{k-1}
\right\}
\, ,
\end{eqnarray}
is then translates into an equation for the off-forward matrix element of Wilson
operators,
\begin{equation}
\label{Def-Tw2-Sca}
n^{\mu_1} \dots n^{\mu_j}
\langle p_2 | {\cal R}^{2,qq}_{\mu \mu_1 \dots \mu_j} | p_1 \rangle
=
p_\mu ( p^+ )^{j}
\left( 1 - \frac{\eta}{j+1} \frac{d}{d\eta} \right) H^q_{j + 1} (\eta)
+
\Delta_\mu ( p^+ )^{j} \frac{1}{j + 1} \frac{d}{d\eta} H^q_{j + 1} (\eta)
\, .
\end{equation}
Here the coefficients in front of the two terms on the right-hand side of
Eq.\ (\ref{ScalarContraction}) become differential operators in skewness
acting on $H_{j + 1}(\eta)$. The latter is related to the Mellin moments
of GPDs,
\begin{eqnarray}
\label{SemiMomentsGPDs}
H^q_j (\eta)
=
\int_{-1}^{1} dx\; x^{j - 1} H^q (x, \eta)
=
\sum_{k = 0}^{j} \eta^k \ H^q_{j,  k}
\, ,
\end{eqnarray}
The moments $H^q_{j + 1} (\eta)$ can be found directly from the $+$ component
of the operator ${\cal R}^{2, qq}_+$. Obviously,
$$
n^\mu n^{\mu_1} \dots n^{\mu_j}
\langle p_2 | {\cal R}^{2,qq}_{\mu \mu_1 \dots \mu_j} | p_1 \rangle
=
( p^+ )^{j + 1} \ H^q_{j + 1} (\eta)
\, .
$$
As a consequence of symmetrization, Eq.\ (\ref{Def-Tw2-Sca}) contains a
Wandzura-Wilczek term proportional to $\Delta_{\mu}^\perp = \Delta_\mu -
\eta p_\mu$, which effectively enters as a twist-three contribution to
the scattering amplitude, as we will demonstrate later in Section
\ref{TwistThreeGPDs}. Finally, the matrix element of the twist-two
light-ray operator (\ref{TwistTwoOpenIndex}) can be obtained in a
straightforward manner by resummation:
\begin{eqnarray}
\label{ZeroSpinOpenIndexTw2}
\langle p_2 | {\cal R}^{2,qq}_\mu ( - z^- , z^- )| p_1 \rangle
=
\int_{-1}^1 d x \, {\rm e}^{- i z^- p^+ x}
\left(
p_\mu H^q (x, \eta)
+
\Delta_\mu^\perp
\int_{-1}^{1} dy \, W_2 (x, y) \frac{d}{d\eta} H^q (x, \eta)
\right)
, \ \
\end{eqnarray}
where the kernel $W_2(x, y)$  is given by
\begin{equation}
\label{W2kernel}
W_2(x, y)
=
\frac{1}{y} \theta(x) \theta(y - x) - \frac{1}{x} \theta(- x) \theta(x - y)
\, .
\end{equation}

\subsubsection{Spin-one-half hadrons}
\label{PolynomSpin12GPDs}

In parallel to the discussion in the previous section, the nucleon matrix
element of twist-two quark operators are parametrized via a set of ``form
factors'' as follows\footnote{Do not confuse these $A_{j,k}$ and $B_{j,k}$
form factors with the ones introduced in Ref.\ \cite{Ji96a}.}
\begin{eqnarray}
\label{Decomposition}
\langle p_2 | {\cal R}^{2,qq}_{\mu_1 \dots \mu_j} | p_1 \rangle
\!\!\!&=&\!\!\!
\Sym_{\rho \mu_1 \dots \mu_j}
h_{\mu_1}
\left\{ p_{\mu_2} \dots p_{\mu_j} A^q_{j, 0}
+
\cdots
+ \Delta_{\mu_2} \dots \Delta_{\mu_j} A^q_{j, j - 1}
\right\}
\\
&+&\!\!\!
\Sym_{\mu_1 \dots \mu_j}
\frac{b}{2 M_N}
\left\{
p_{\mu_1} \dots p_{\mu_j} B^q_{j, 0}
+
\Delta_{\mu_1} p_{\mu_2} \dots p_{\mu_j} B^q_{j, 1}
+
\dots
+
\Delta_{\mu_1} \dots \Delta_{\mu_j} B^q_{j , j}
\right\}
\, , \nonumber
\end{eqnarray}
where the structure in the second line is identical to Eq.\ (\ref{LorentzVscal}).
Analogous relations hold for the parity odd and chiral odd operator matrix elements,
though the form factor counting comes out different \cite{JiLeb00,Die01,Hag04,CheJi04}.
We used the Gordon identities (\ref{GordonIdentities}) in the above equation to reduce
the tensor structure involving $t^{\mu\nu}$ to the other two Dirac bilinears. This is
useful since one can use many results derived for the scalar target. As we discussed
above, there is an important constraint on the matrix elements (\ref{Decomposition})
imposed by the time-reversal invariance. Namely, the coefficients accompanying Lorentz
structures with odd powers of the momentum transfer $\Delta$ should vanish \cite{Ji:1998pc},
so that
$$
A^q_{j, 2n + 1} = B^q_{j, 2n + 1} = 0
\, .
$$
The nonvanishing form factors $A_{j,k}$ and $B_{j,k}$ are related to the moments of GPDs
\begin{equation}
\label{ABgpds}
A^q_j (\eta)
=
\int_{-1}^{1} dx\; x^{j - 1} A^q (x, \eta)
=
\sum_{k = 0}^{j - 1} \eta^k A^q_{j, k}
\, , \qquad
B^q_j (\eta)
=
\int_{-1}^{1} dx\; x^{j - 1} B^q (x, \eta)
=
\sum_{k = 0}^{j} \eta^k B^q_{j, k}
\, .
\end{equation}

Another equivalent representation for the matrix elements of twist-two operators
widely used in the literature is achieved by reduction of the scalar Dirac bilinear
$b$ accompanied by at least one vector $p_\mu$ using the Gordon identity back to
the vector $n^\mu$ and tensor $t^{\mu\nu}$ Dirac bilinears. This gives
\begin{eqnarray}
\label{GPDmomentsFFHandE}
\langle p_2 | {\cal R}^{2,qq}_{\mu_1 \dots \mu_j} | p_1 \rangle
\!\!\!&=&\!\!\! \Sym_{\mu_1 \dots \mu_j}
h_{\mu_1}
\left\{
p_{\mu_2} \dots p_{\mu_j} H^q_{j, 0} (\Delta^2)
+
\dots
+
\Delta_{\mu_2} \dots \Delta_{\mu_j} H^q_{j, j - 1} (\Delta^2)
\right\}
\nonumber\\
&+&\!\!\!
\Sym_{\mu_1 \dots \mu_j}
e_{\mu_1}
\left\{
p_{\mu_2} \dots p_{\mu_j} E^q_{j, 0} (\Delta^2)
+
\dots
+
\Delta_{\mu_2} \dots \Delta_{\mu_j} E^q_{j,j - 1} (\Delta^2)
\right\}
\nonumber\\
&+&\!\!\!
\Sym_{\mu_1 \dots \mu_j}
\frac{b}{2 M_N}
\Delta_{\mu_1} \dots \Delta_{\mu_j} D^q_{j} (\Delta^2)
\, ,
\end{eqnarray}
where
\begin{equation}
\label{FormFactorsABvsHE}
H^q_{j, k} = A^q_{j, k} + B^q_{j, k}
\, , \qquad
E^q_{j, k} = - B^q_{j, k}
\, , \qquad
D^q_j = B^q_{j, j}
\, .
\end{equation}
The reduced matrix elements $H^q$, $E^q$ and $D^q$ are related to the moments of
parity-even quark generalized parton distributions \cite{MulDitRobGeyHor98,Ji96a,Rad1997a}
via
\begin{eqnarray}
\label{PolynomialityNucleon}
\int_{-1}^1 d x \, x^{j - 1} H^q (x, \eta, \Delta^2)
\!\!\!&=&\!\!\! \sum_{k = 0}^{j - 1} \eta^k
H^q_{j, k} (\Delta^2) + \eta^j D^q_{j} (\Delta^2)
\, , \nonumber\\
\int_{-1}^1 d x \, x^{j - 1} E^q (x, \eta, \Delta^2)
\!\!\!&=&\!\!\! \sum_{k = 0}^{j - 1} \eta^k
E^q_{j, k} (\Delta^2) - \eta^j D^q_{j} (\Delta^2)
\, .
\end{eqnarray}
For gluon GPDs, the modification of the above formulas is very straightforward.
The parametrization of the parity-even gluon matrix elements in terms of form
factors $H^g$, $E^g$ and $D^g$ reads in complete analogy to (\ref{GPDmomentsFFHandE})
\begin{eqnarray}
\label{GPDmomentsFFHandEgluon}
\langle p_2 | {\cal R}^{2,gg}_{\mu_1 \dots \mu_j} | p_1 \rangle
\!\!\!&=&\!\!\!
\ft12
\Sym_{\mu_1 \dots \mu_j}
h_{\mu_1}
\left\{
p_{\mu_2} \dots p_{\mu_{j - 1}} H^g_{j, 0} (\Delta^2)
+
\dots
+
\Delta_{\mu_2} \dots \Delta_{\mu_{j - 1}} H^g_{j, j - 2} (\Delta^2)
\right\} p_{\mu_j}
\nonumber\\
&+&\!\!\!
\ft12
\Sym_{\mu_1 \dots \mu_j}
e_{\mu_1}
\left\{
p_{\mu_2} \dots p_{\mu_{j - 1}} E^g_{j, 0} (\Delta^2)
+
\dots
+
\Delta_{\mu_2} \dots \Delta_{\mu_{j - 1}} E^g_{j,j - 2} (\Delta^2)
\right\} p_{\mu_j}
\nonumber\\
&+&\!\!\!
\ft12
\Sym_{\mu_1 \dots \mu_j}
\frac{b}{2 M_N}
\Delta_{\mu_1} \dots \Delta_{\mu_j} D^g_{j} (\Delta^2)
\, ,
\end{eqnarray}
with form factors related to moments of gluon GPDs (\ref{GluonGPDnucleon}) via
\begin{eqnarray}
\label{PolynomialityNucleonGluon}
\frac{1}{2} \int_{-1}^1 d x \, x^{j - 2} H^g (x, \eta, \Delta^2)
\!\!\!&=&\!\!\! \sum_{k = 0}^{j - 2} \eta^k
H^g_{j, k} (\Delta^2) + \eta^j D^g_{j} (\Delta^2)
\, , \nonumber\\
\frac{1}{2} \int_{-1}^1 d x \, x^{j - 2} E^g (x, \eta, \Delta^2)
\!\!\!&=&\!\!\! \sum_{k = 0}^{j - 2} \eta^k
E^g_{j, k} (\Delta^2) - \eta^j D^g_{j} (\Delta^2)
\, .
\end{eqnarray}
The reduction of the value of the Lorentz spin in the exponential of the Mellin
integrand on the left-hand side of these equations by one unit $j \to j - 1$ is
an obvious consequence of the increased spin of gluon operators compared to the
quark ones. Notice that while quark GPDs do not possess definite symmetry properties
under the exchange $x \to - x$, the gluon GPDs do. Namely, due to the fact that
gluon and antigluons are the same particles, one finds
\begin{equation}
\begin{array}{ll}
H^g (- x, \eta, \Delta^2) = H^g (x, \eta, \Delta^2)
\, , \qquad
&
E^g (- x, \eta, \Delta^2) = E^g (x, \eta, \Delta^2)
\, , \\[-2mm]
& \\
\widetilde{H}^g (- x, \eta, \Delta^2) = - \widetilde{H}^g (x, \eta, \Delta^2)
\, , \qquad
&
\widetilde{E}^g (- x, \eta, \Delta^2) = - \widetilde{E}^g (x, \eta, \Delta^2)
\, ,
\end{array}
\end{equation}
for parity-even and parity-odd GPDs, respectively. This charge conjugation
property allows one to reduce the integral in Eq.\ (\ref{PolynomialityNucleonGluon})
to the domain $0 \leq x \leq 1$.

In establishing the relation between the nucleon matrix elements of the twist-two
operator with the open Lorentz index (\ref{TwistTwoOpenIndex}) and the nucleon
form factors, the only difference from the case of spin-zero target arises in the
implementation of the symmetrization procedure, namely,
\begin{eqnarray*}
n_{\mu_1} \dots n_{\mu_j} \Sym_{\mu \mu_1 \dots \mu_j}
h^\mu \Delta^{\mu_1} \dots \Delta^{\mu_k} p^{\mu_{k + 1}} \dots p^{\mu_j}
=
\frac{( p^+ )^{j - 1}}{j + 1}
\left\{
h^\mu \eta^{k} p^+
+
(j - k) p^\mu h^+ \eta^{k}
+
k \Delta^\mu h^+ \eta^{k - 1}
\right\}
\, , \!\!\!\!\!\!
\end{eqnarray*}
and provides a new structure
\begin{eqnarray}
\label{Def-Tw2-h}
n^{\mu_1} \dots n^{\mu_j}
\langle p_2 | {\cal R}^{2,qq}_{\mu \mu_1 \dots \mu_j} | p_1 \rangle
\!\!\!&=&\!\!\!
\frac{( p^+ )^{j - 1}}{j + 1}
\left(
(j + 1) p_\mu h^+
+
h_\mu - p_\mu h^+ + \Delta_\mu^\perp h^+
\frac{d}{d\eta}
\right)
A_{j + 1}(\eta)
\\
&+&\!\!\!
\frac{b}{2 M_N}
\left\{
p_\mu ( p^+ )^{j}
\left( 1 - \frac{\eta}{j+1} \frac{d}{d\eta} \right)
+
\Delta_\mu ( p^+ )^{j} \frac{1}{j + 1} \frac{d}{d\eta}
\right\} B^q_{j + 1} (\eta)
\, . \nonumber
\end{eqnarray}
It is easy to perform the  inverse Mellin transformation to establish the
corresponding relations to GPDs. To this end, one should match the
integrand of the momentum fraction integrals to the moment form $x^j f (x)$.
The procedure yields the equation
\begin{eqnarray}
\langle p_2 | {\cal R}^{2,qq}_\mu ( - z^- , z^- ) | p_1 \rangle
\!\!\!&=&\!\!\!
\int_{-1}^1 dx \, {\rm e}^{- i x z^- p^+} \int_0^1 dy \,
\Bigg\{
\left( h_\mu - \frac{h^+}{p^+} p_\mu \right)
W_2 (x, y) \left( H + E \right) (y, \eta)
\nonumber\\
&+&\!\!\!
\left(
p_\mu \delta (x - y) + \Delta_\mu^\perp W_2 (x, y)
\frac{d}{d \eta}
\right)
\left(
\frac{h^+}{p^+} H + \frac{e^+}{p^+} E
\right) (y, \eta)
\Bigg\}
\, ,
\label{TwistTwoOpenIndexWW}
\end{eqnarray}
where we returned to the conventional basis of Dirac structures by means of
the identity (\ref{BilinearSubst}) and conventional GPDs, related to the ones
defined in Eq.\ (\ref{ABgpds}) by
\begin{eqnarray}
\label{SPDsSubst}
\quad A^q = H^q + E^q ,
\quad B^q = - E^q
\, .
\end{eqnarray}
It is also useful to note that the identity $2 M_N ( p^+ h^\mu- p^\mu h^+ ) =
( p^+ t^{\nu \mu} - p^\mu t^{\nu +} ) \Delta_\nu$ allows to reduce the structure
accompanying $E$ in the first line of Eq.\ (\ref{TwistTwoOpenIndexWW}) to the
expected tensor form (\ref{vectorGPDnucleon}). Contracting the expression
(\ref{TwistTwoOpenIndexWW}) with the vector $n_\mu$ we obtain conventional
definitions of twist-two GPDs.

Since there are no further subtleties related to spin-one targets, we will not
discuss polynomiality properties of their GPDs here.

\subsection{GPDs and the proton spin puzzle}
\label{ProtonSpinPuzzle}

One of the motivations to study GPDs is their profound relation to the distribution
of angular momentum of quarks and gluons in the proton. As is by now well known,
the EMC measurement of the fraction of the proton spin carried by quarks, $\ft12
\Delta{\mit\Sigma}$,
\begin{equation}
\sum_q \langle p |j^{5 \mu}_q (0) | p \rangle
=
\Delta {\mit\Sigma} s^\mu
\, ,
\end{equation}
in polarized deeply inelastic scattering demonstrated that, instead of the much
anticipated result $\Delta{\mit\Sigma}^{\rm QM} = 1$, which is a prediction of the naive
quark model of the nucleon, the experimental finding was only about 20\% of this. This
``spin crisis'' triggered an enormous flood of theoretical and experimental activity.
Theoretical resolution of the spin puzzle was sought via many complementary ways including
the anomalous contributions due to gluons \cite{AltRos88,CarColMul88,EfrTer88}, large
negative polarization of strange quarks, significant role of the orbital angular momentum of
constituents \cite{JafMan90,Ji96a}, etc.\ (see Refs.\ \cite{AnsEfrLea95,LamRey00,FilJi01,Bas04}
for reviews). The experimental verification of the EMC finding is coming from several
sources, including deeply inelastic scattering and leptonic decays of hyperons.

\subsubsection{Axial coupling from polarized deeply inelastic scattering}
\label{AxialCouplingSU3}

\begin{table}[t]
\begin{center}
{\begin{tabular}{ccccc}
\multicolumn{5}{c}{} \\ [6pt]
\hline
$Q^2({\rm GeV}^2)$
&
${\mit\Gamma}_1^p$
&
${\mit\Gamma}_1^n$
&
${\mit\Gamma}_1^p - {\mit\Gamma}_1^n$
&
Ref.
\\
\hline
5
&
$0.118 \pm 0.008$
&
$-0.058 \pm 0.009$
&
$0.176 \pm 0.007$
&
\protect\cite{E155}
\\
10
&
$0.120 \pm 0.016$
&
$-0.078 \pm 0.021$
&
$0.198 \pm 0.023$
&
\protect\cite{SMC98,SMC98a}
\\
5
&
$0.121 \pm 0.018$
&
$-0.075 \pm 0.021$
&
$0.174 + 0.024 - 0.012$
&
\protect\cite{SMC98a}
\\
\hline
\end{tabular}}
\end{center}
\caption{\label{GammaPNdata} Experimental data on ${\mit\Gamma}_1^p$,
${\mit\Gamma}_1^n$, and Bjorken sum rule ${\mit\Gamma}_1^p - {\mit\Gamma}_1^n$
in the $\overline{\rm MS}$ scheme.}
\end{table}
The measurement of the polarized spin-dependent structure function $g_1$,
see Eq.\ (\ref{DISstructurefunctions}), gives access to the helicity-dependent
quark distribution functions, and thus to the fraction of hadron spin carried
by quarks deduced from their first Mellin moment,
\begin{equation}
\label{HelicitySumRule}
{\mit\Gamma}_1
=
\int_0^1 dx \, g_1 (x)
=
\frac{1}{2} \sum_q Q_q^2 \int_0^1 dx \,
\left( \Delta q (x) + \Delta \bar q (x) \right)
\equiv
\frac{1}{2} \sum_q Q_q^2 \left( \Delta q + \Delta \bar q \right)
\, .
\end{equation}
We have omitted the QCD radiative corrections to this parton model formula and
did not display the dependence of all quantities on the virtuality of the
resolving photon. The measurement of the structure functions on the proton and
neutron targets allows one to measure the following combinations of the axial
constants
\begin{equation}
{\mit\Gamma}_1^p
=
\frac{1}{9}
\left(
\frac{3}{4} g_A^{(3)} + \frac{1}{4} g_A^{(8)} + g_A^{(0)}
\right)
\, , \qquad
{\mit\Gamma}_1^n
=
\frac{1}{9}
\left(
- \frac{3}{4} g_A^{(3)} + \frac{1}{4} g_A^{(8)} + g_A^{(0)}
\right)
\, ,
\end{equation}
which are superpositions of the first moments of polarized quark densities,
\begin{equation}
\label{AxialConstants}
\begin{array}{rll}
g_A^{(0)}
&=
( \Delta u +  \Delta \bar u )
+
( \Delta d +  \Delta \bar d )
+
( \Delta s +  \Delta \bar s )
& \!\!\!\!\!\!\!
\equiv \Delta {\mit\Sigma}
\, , \\
g_A^{(3)}
&=
( \Delta u +  \Delta \bar u )
-
( \Delta u +  \Delta \bar u )
\, ,
&
\\
g_A^{(8)}
&=
( \Delta u +  \Delta \bar u )
+
( \Delta d +  \Delta \bar d )
- 2
( \Delta s +  \Delta \bar s )
\, .
&
\\
\end{array}
\end{equation}
The difference between the  first moments of the proton and neutron
structure functions is fixed by the Bjorken sum rule to the isovector
axial coupling,
\begin{equation}
\label{BjorkenSumRule}
{\mit\Gamma}_1^p
-
{\mit\Gamma}_1^n
=
\frac{1}{6} g_A^{(3)}
\, .
\end{equation}
This equation is a consequence of the current algebra and isospin symmetry.
Recent experimental data for the integrals ${\mit\Gamma}_1^p$ and ${\mit\Gamma}_1^n$
are summarized in Table \ref{GammaPNdata}. The deeply inelastic scattering data
alone do not allow to extract all axial constants since the octet and singlet
constants enter in the same combination into the proton and neutron sum rules.
Thus one has to resort to other sources of information in order to separate
different flavor contributions. Such knowledge is provided by hyperon decays
complemented by the use of the $SU(3)$ flavor symmetry, as discussed next.

\subsubsection{Axial coupling from semileptonic decays of hyperons}
\label{OctetOctetOctet}

Semileptonic decays of hyperons $B \to B' e^- \bar\nu_e$ measure the
transition matrix element of the charged weak current,
\begin{equation}
\label{BtoBtransition}
\langle B | \bar\psi_q \gamma^\mu (1 - \gamma^5) \psi_{q'} | B' \rangle
\, ,
\qquad
\psi_q
\equiv
\left(
\begin{array}{c}
u \\
d \\
s
\end{array}
\right)
\end{equation}
The approximate $SU (3)$ flavor symmetry of strong interactions allows one
to relate the above matrix elements among each other so that one deduces the
flavor-diagonal transitions, required for the spin sum rule, in terms of
experimentally measured flavor-changing decays. Let us give a few details on
the derivation of the $SU(3)$-flavor relations, since we will use the very same
identities for flavor-changing GPDs introduced in Section \ref{FlavorChangingGPDs}.

The $SU(3)$ baryon octet is represented conventionally by a matrix labeled
with a row index $a$ and a column index $b$,
\begin{equation}
B^a{}_b
=
\left(
\begin{array}{ccc}
\ft{1}{\sqrt{6}} {\mit\Lambda} + \ft{1}{\sqrt{2}} {\mit\Sigma}^0
&
{\mit\Sigma}^+
&
p
\\
{\mit\Sigma}^-
&
\ft{1}{\sqrt{6}} {\mit\Lambda} - \ft{1}{\sqrt{2}} {\mit\Sigma}^0
&
n
\\
{\mit\Xi}^-
&
{\mit\Xi}^0
&
- \ft{2}{\sqrt{6}} {\mit\Lambda}
\end{array}
\right)
\, ,
\end{equation}
so that, e.g., the proton is the element $B^1{}_3 = p$. The Lagrangian
describing the transition $B \to B'$ within the baryon octet via the octet
current has two $SU(3)$ invariant couplings, since there are two distinct
$\bit{8}$'s in the product of two, $\bit{8} \otimes \bit{8} = \bit{1} \oplus
\bit{8} \oplus \bit{8} \oplus \bit{10} \oplus \bit{\overline{10\!}\,} \oplus
\bit{27}$. This yields the symmetric ($D$) and antisymmetric ($F$) invariants
\cite{Swa63}
\begin{eqnarray}
\label{EightToEightLagrange}
{\cal L}_{\bit{\scriptstyle 8} \to \bit{\scriptstyle 8}}
\!\!\!&=&\!\!\!
D \, {\rm tr} \{ \bar B , B \} M
-
F \, {\rm tr} [ \bar B , B ] M
\nonumber\\
&=&\!\!\!
(D - F) \, \bar B^a{}_b M^c{}_a B^b{}_c
+
(D + F) \, \bar B^b{}_c M^c{}_a B^a{}_b
\, .
\end{eqnarray}
Here $\bar B \equiv B^\dagger$ is a hermitian conjugated matrix, e.g.,
$\bar{B}^3{}_1 = \bar{p}$, and the matrix $M^a{}_b$ describes the octet
of local vector ${\mit\Gamma} = \gamma^\mu$ and axial ${\mit\Gamma} =
\gamma^\mu \gamma^5$ currents,
\begin{equation}
M^{q'}{}_q = \bar\psi_q {\mit\Gamma} \psi_{q'}
-
\frac{1}{3} \delta_{q q'}
\sum_{q''} \bar\psi_{q''} {\mit\Gamma} \psi_{q''}
\,
\end{equation}
with corresponding two inequivalent sets of $D$ and $F$ constants, $F_V$, $D_V$
and $F_A$, $D_A$, respectively. Actually, these are not constants at all but
rather form factors due to the nonzero recoil $\Delta = p_1 - p_2$ in the
transition matrix element. There are as many of them as there are independent
structures in the decomposition of the matrix elements (\ref{BtoBtransition}) in
terms of Dirac bilinears, as, e.g., in Eqs.\ (\ref{vectorFFs}) and (\ref{axialFFs}),
with possible extra terms. However, since the recoil momentum in these decays is
negligible compared to the hyperon masses, $|\bit{\Delta}| \ll M$, one can use
the static approximation and take all form factors as constants at $\bit{\Delta}
= 0$. This also implies that the limitation of the consideration to the vector
and axial charges is legitimate. Since the $SU(3)$ vector current is a generator
of $SU(3)$ transformations, its matrix elements are fixed uniquely by the $SU(3)$
group structure constants $f^{abc}$, such that $F_V = 1$ and $D_V = 0$. These are
the matrix elements of the axial currents that are measured in these decays,
$F_A = F$, $D_A = D$.

For the transitions governed by the $M^u{}_d = \bar{d} u$ current, the Lagrangian
gives
\begin{eqnarray}
{\cal L}_{\bit{\scriptstyle 8} \to \bit{\scriptstyle 8}}
\!\!\!&=&\!\!\!
\sqrt{\ft23} D \, \bar{\mit\Lambda} (\bar d u) {\mit\Sigma}^-
+
\sqrt{\ft23} D \, \bar {\mit\Sigma}^+ (\bar d u) {\mit\Lambda}
\\
&+&\!\!\!
(D + F) \, \bar p (\bar d u) n
+
\sqrt{2} F \, \bar {\mit\Sigma}^0 (\bar d u) {\mit\Sigma}^-
-
\sqrt{2} F \, \bar {\mit\Sigma}^+ (\bar d u) {\mit\Sigma}^0
+
(D - F) \,
\bar {\mit\Xi}^0 (\bar d u) {\mit\Xi}^-
+ \dots
\, , \nonumber
\label{ExampleUDtransition}
\end{eqnarray}
with ellipsis standing for other flavor transitions. One can immediately read off the
strengths for the corresponding transitions
\begin{eqnarray}
\langle {\mit\Lambda} | \bar{u} d | {\mit\Sigma}^- \rangle
=
\langle {\mit\Sigma}^+ | \bar{u} d | {\mit\Lambda} \rangle
=
\sqrt{\ft23} D
\, , \qquad\quad \ \,
&&
\langle p | \bar{u} d | n \rangle
= D + F
\, , \\
\langle {\mit\Sigma}^0 | \bar{u} d | {\mit\Sigma}^- \rangle
=
-
\langle {\mit\Sigma}^+ | \bar{u} d | {\mit\Sigma}^0 \rangle
=
\sqrt{2} F
\, , \qquad
&&
\langle {\mit\Xi}^0 | \bar{u} d | {\mit\Xi}^- \rangle
=
D - F
\, . \nonumber
\end{eqnarray}
Extracting the proton-to-proton transitions from the above Lagrangian,
we get
\begin{equation}
\label{PtoPtrans}
{\cal L}_{\bit{\scriptstyle 8} \to \bit{\scriptstyle 8}}
=
\bar{p}
\left\{
\frac{1}{6} (3 F - D) \left( \bar{u} u + \bar{d} d - 2 \bar{s} s \right)
+
\frac{1}{2} (D + F) \left( \bar{u} u - \bar{d} d \right)
\right\} p
+ \dots
\, ,
\end{equation}
and, as a result\footnote{Since the  numerical factors appearing in the above
Lagrangian can be confusing, let  us give a schematic derivation of these
relations. The hadronic state is created from the vacuum by a creation operator
$| H \rangle = b^\dagger_H | 0 \rangle$,  and in the quark/hadron field $\psi$
only the annihilation operator is needed $\psi = b$ (we did not display the
quark/hadron wave function). Thus, we compute the matrix element of the
(H)eisenberg operator in perturbation theory defined, of course, in the
(D)irac representation with the  Lagrangian (\protect\ref{PtoPtrans}):
\begin{eqnarray*}
\langle p | \bar{u} u - \bar{d} d | p \rangle_{\rm H}
\!\!\!&=&\!\!\!
\langle p | :( \bar{u} u - \bar{d} d ):
{\cal L}_{\bit{\scriptstyle 8} \to \bit{\scriptstyle 8}} | p \rangle_{\rm D}
=
\langle p | :( \bar{u} u - \bar{d} d ):
\ft{1}{2} (D + F) :\bar{p} \left( \bar{u} u - \bar{d} d \right) p:
| p \rangle_{\rm D}
\\
&=&\!\!\!
\ft{1}{2} (D + F)
\langle 0 | b_p b^\dagger_p
:( \bar{u} u - \bar{d} d ):
:( \bar{u} u - \bar{d} d ): b_p b^\dagger_p
| 0 \rangle_{\rm D}
=
\ft{1}{2} (D + F)
\langle 0 | : \bar{u} u : : \bar{u} u :
+
: \bar{d} d : : \bar{d} d : | 0 \rangle_{\rm D}
\\
&=&\!\!\!
D + F
\, ,
\end{eqnarray*}
where  the operator inside $: :$ is subject to normal ordering. In the
last but one term, we have kept only pairing of quarks of the
same flavor.  We use normalization $\langle 0 | : \bar{u}
u : : \bar{u} u : | 0 \rangle_{\rm D} = \langle 0 | : \bar{d} d : : \bar{d}
d : | 0 \rangle_{\rm D} = 1$.},
\begin{equation}
\langle p | \bar{u} u - \bar{d} d | p \rangle
=
D + F
\, , \qquad
\langle p | \bar{u} u + \bar{d} d  - 2 \bar{s} s | p \rangle
=
3 F - D
\, .
\end{equation}
Solving these equations for the $F$ and $D$ couplings in terms of  matrix elements,
we obtain as a result the $SU(3)$ relations for the octet-to-octet transitions, for
instance,
\begin{eqnarray}
\langle p | \bar{u} u - \bar{d} d | p \rangle
=
\langle n | \bar{d} u | p \rangle
\, .
\end{eqnarray}
Thus, certain combinations of $D$
and $F$ decay constants are related to the triplet and octet axial
constants (\ref{AxialConstants}) via
\begin{equation}
g_A^{(3)} = F + D
\, , \qquad
g_A^{(8)} = 3 F - D
\, .
\end{equation}
The experimental data for a few decay modes are given in Table
\ref{HyperonDecays}, wherefrom we conclude that $D = 0.79 \pm 0.07$
and $F = 0.47 \pm 0.10$. Thus, the triplet axial constant parametrizing
the strength of the neutron decay  \cite{Pdg02}
$$
g_A^{(3)} = 1.2670 \pm 0.0035
$$
is in a  wonderful agreement with the Bjorken sum rule, see Eq.\
(\ref{BjorkenSumRule}) and Table \ref{GammaPNdata}. The octet
coupling is extracted from other hyperon decays, summarized in Table
\ref{HyperonDecays}, resulting in
$$
g_A^{(8)} = 0.58 \pm 0.03
\, .
$$
\begin{table}[t]
\begin{center}
{\begin{tabular}{llll}
\multicolumn{4}{l}{} \\ [6pt]
\hline
Axial transition
&
Matrix element
&
Strength
&
Experimental value
\\
\hline
$n \to p$
&
$\langle p | \bar{u}d | n \rangle$
&
$D + F$
&
$1.2670 \pm 0.0035$
\\
${\mit\Xi}^- \to {\mit\Lambda}^0$
&
$\langle {\mit\Lambda}^0 | \bar{u}s | {\mit\Xi}^- \rangle$
&
$\frac{1}{\sqrt{6}} (3 F - D)$
&
$0.25 \pm 0.05$
\\
${\mit\Sigma}^- \to n$
&
$\langle n | \bar{u}s | {\mit\Sigma}^- \rangle$
&
$D - F$
&
$0.340 \pm 0.017$
\\
${\mit\Lambda} \to p$
&
$\langle p | \bar{u}s | {\mit\Lambda}^0 \rangle$
&
$- \frac{1}{\sqrt{6}} (3 F + D)$
&
$- 0.718 \pm 0.015$
\\
\hline
\end{tabular}}
\end{center}
\caption{\label{HyperonDecays} Experimental data on hyperon decays
\cite{Pdg02} which measure the strength of axial transitions.}
\end{table}

Taking into account the QCD radiative corrections, which affect the coefficients
with which the axial constants enter the sum rule (\ref{HelicitySumRule}) by
multiplicative factors \cite{Lar93} on the right-hand side, one extracts the
singlet axial constant and thus the fraction of the nucleon spin carried by the
quarks to be \cite{BluBot02}
$$
\Delta {\mit\Sigma} = 0.29 \pm 0.10
$$
at $Q^2 = 4 \, {\rm GeV}^2$ in the $\overline{\rm MS}$ scheme. At the same time the
values extracted by experimental collaborations vary from $\Delta {\mit\Sigma} =
0.23 \pm 0.04 \pm 0.06$ at $Q^2 = 5 \, {\rm GeV}^2$ for SLAC E155 \cite{E155} to
$\Delta {\mit\Sigma} = 0.19 \pm 0.05 \pm 0.04$ at $Q^2 = 1 \, {\rm GeV}^2$ for SMC
\cite{SMC98,SMC98a}. As we emphasized before, this value is in sharp contradiction
with the expectations of the naive quark model of hadrons.

\subsubsection{Angular momentum of proton's constituents}

Thus, the spin carried by quarks cannot account for the whole spin of the proton.
The question arises  as to where the other sources of the hadron spin are resided.
To resolve successfully the issue, recall that the tensor of the angular momentum
is expressed solely in terms of the improved energy-momentum tensor (see Appendix
\ref{SymmetricEMtensor})
$$
M^{\mu\nu; \rho}
=
x^\mu {\mit\Theta}^{\rho\nu} - x^\nu {\mit\Theta}^{\rho\mu}
\, .
$$
The charge
$$
\mathbb{M}^{\mu\nu} = \int d^3 \bit{z} \, M^{\mu\nu; 0} (\bit{z})
$$
generates Lorentz transformations and obeys the standard $SO(3,1)$ algebra (see
Eq.\ (\ref{SO41ConformalAlgebra}) in Appendix \ref{ConformalSymAppendix}). The
three-vector of angular momentum determined from this charge as coordinate
moment of the three-vector of the momentum flow $\bit{\mit\Theta} = \left(
{\mit\Theta}^{0 1}, {\mit\Theta}^{0 2}, {\mit\Theta}^{0 3} \right)$,
\begin{equation}
\label{JfromMmunu}
J^i
=
\frac{1}{2} \varepsilon^{ijk} \mathbb{M}^{jk}
=
\int d^3 \bit{z} \, [ \bit{z} \times \bit{\mit\Theta} (\bit z) ]^i
\end{equation}
can be rewritten in the form \cite{JafMan90,Ji96a}
\begin{equation}
\bit{J}
= \int d^3 \bit{z} \,
\left\{
\frac{1}{2}
\psi^\dagger (\bit{z}) \bit{\mit\Sigma} \psi (\bit{z})
-
\psi^\dagger (\bit{z}) [ \bit{z} \times i \bit{\cal D} ] \psi (\bit{z})
+
[ \bit{E} (\bit{z}) \times \bit{B} (\bit{z}) ]
\right\}
\, ,
\end{equation}
which appears to have the meaning of the quark spin and orbital angular momentum
with covariant momentum $i \bit{\cal D}$ for the first two terms, respectively,
and the gluon angular momentum for the last one, expressed in terms of the Poynting
vector $[ \bit{E} \times \bit{B}]$ of chromoelectric $\bit{E}$ and chromomagnetic
$\bit{B}$ fields. Let us note, that, for a gauge particle, a further decomposition
of its angular momentum into its spin and orbital components is not possible. A
hand waving argument goes as follows \cite{LanLif82}: In quantum mechanics, the wave
function of a spin-$s$ particle is a symmetric rank-$2s$ spinor having $(2s + 1)$
components, which transform into each other under the rotation of the coordinate
system. The orbital wave function is related to the coordinate dependence of wave
functions and is given by the spherical harmonics of order $L$ for the angular
momentum $L$ of the system. Therefore, in order to distinguish clearly between the
spin and orbital momentum, the spin and coordinate properties must be independent. As
it is obvious, this condition is not fulfilled for gauge particles whose description
in terms of field operators inevitably involves a gauge condition. For instance,
in the Coulomb gauge, the gluon wave function is given by the three-potential
$\mbox{\boldmath$A$}^a (x)$ equivalent to the second rank spinor which, however,
is a subject to the gauge-fixing condition $\bit{\nabla} \cdot \mbox{\boldmath$A$}^a
(x) = 0$. As a result, the coordinate dependence of the vector cannot be
independently defined for each of its components and leads to the inability to
separate in a gauge-invariant manner the spin and orbital degrees of freedom. For
suggestions of a break-up of the gluon angular momentum into its spin and orbital
parts within the context of the QCD parton model picture, we refer the reader to
Refs.\ \cite{BasJaf98,Jaf00}. Within this framework, the gluon spin can be
identified with the first moment of the polarized gluon density $\Delta g (x)$
(see Eq.\ (\ref{PolGluonDensity})) in the light-cone gauge and then promoted to a
gauge-invariant operator matrix element making use of relations such as Eq.\
(\ref{FromPotentialToStrength}). Analogous consideration can applied to the gluon
orbital part, however, the matrix element of the corresponding operator does not
show up in any physical processes and thus presumably lack any phenomenological
significance \cite{Jaf00}.

For a proton at rest, with its spin vector $\bit{s}$ pointing in the positive
$z$-direction or for a proton moving along $z$ and in the helicity-up state,
the matrix element of the third component of the angular-momentum operator
determines the total proton spin in terms of separate contributions from
spin and orbital angular momentum of its constituents \cite{Ji96a,Ji97b}
\begin{equation}
\label{SpinSumRule}
\langle p, s^z | J^z | p, s^z \rangle
=
\frac{1}{2} \Delta {\mit\Sigma} (\mu^2) + \sum_q L^q (\mu^2) + J^g (\mu^2)
\equiv
\sum_q J^q (\mu^2) + J^g (\mu^2)
=
\frac{1}{2}
\, ,
\end{equation}
where the meaning  of different terms is self-explanatory. We introduced
an explicit renormalization scale dependence for separate terms which are
not conserved in distinction to the total sum which has zero
anomalous dimension and, hence,  is scale-invariant.

\subsubsection{Gravitational form factors}

There are no known processes where the forward matrix element of the angular
momentum operator (\ref{SpinSumRule}) can be measured. The angular momentum
three-vector is a coordinate moment of the momentum three-vector. A similar
situation occurred previously in case of electromagnetic form factors where the
nucleon magnetic moment, which corresponds to the coordinate moment of the electric
current, was given by  zero-recoil limit of the non-forward matrix element of
the electomagnetic current. Similarly, the form factors in the off-forward
matrix element of the energy-momentum tensor will allow to access the coordinate
moment of  the three-momentum operator. Since graviton couples to the symmetric
energy-momentum tensor (see Appendix \ref{SymmetricEMtensor}) these gravitational
form factors would be ``measurable'' in elastic graviton-nucleon scattering.
Obviously, such an experiment is not feasible. As we will discuss later in this
paper, one can access them in light-cone dominated Compton scattering. This was
the seminal observation in Ref.\ \cite{Ji96a}. The quark (\ref{QuarkEMtensor}) and
gluon (\ref{GluonEMtensor}) components of the QCD energy-momentum tensor  have
the following form factor decomposition \cite{Pag66,Ji96a}
\begin{eqnarray}
\label{GravitFormFactorDecomp}
\langle p_2 | {\mit\Theta}^{a, \mu\nu} | p_1 \rangle
=
\ft12
\Bigg(
H^a ( \Delta^2 ) p^{ \{ \mu } h^{ \nu \} }
+
E^a (\Delta^2)
p^{ \{ \mu } e^{\nu \} }
\!\!\!&+&\!\!\!
D^a ( \Delta^2 )
\frac{\Delta^\mu \Delta^\nu - g^{\mu\nu} \Delta^2}{2 M_N} b
\Bigg)
\nonumber\\
&\pm&\!\!\!
\widetilde D ( \Delta^2 ) M_N g^{\mu\nu} b
\, ,
\end{eqnarray}
where $a = q, g$ and the symmetrization of indices is defined as $\{ \mu\nu \}
= \ft12 (\mu\nu + \nu\mu)$. The form factors introduced here are related to
the ones arising in the decomposition of the twist-two vector operators
(\ref{GPDmomentsFFHandE}) via
\begin{equation}
H^a = H_{2, 0}^a
\, , \qquad
E^a = E_{2, 0}^a
\, , \qquad
D^a = D_2^a
\, .
\end{equation}
Note that contrary to the electromagentic form factors, the form factors of the
quark and gluon parts of the energy-momentum tensor separately depend on the
resolution $\mu$ (not displayed explicitly) of the hard probe they are measured
with. This is a consequence of their non-conservation, which is reflected
in the presence of an extra form factor $\widetilde D$. It disappears in the form
factor decomposition of the total energy-momentum tensor ${\mit\Theta}^{\mu\nu} =
{\mit\Theta}^{q, \mu\nu} + {\mit\Theta}^{g, \mu\nu}$. In the following,  we discuss
only matrix elements of the total energy-momentum tensor and thus eliminate the
superscript $a = q, g$ from the form factors. Due to the conservation of the total
energy-momentum tensor, all of its form factors are scale independent. Finally, let
us add for completeness the tensor decomposition of the matrix element of a spin-zero
target,
\begin{equation}
\label{PionGravitFFs}
\langle p_2| {\mit\Theta}^{\mu\nu} |p_1 \rangle
=
\ft12
\left( p^\mu p^\nu \theta_2 (\Delta^2)
+
(g^{\mu\nu}\Delta^2 - \Delta^\mu \Delta^\nu) \theta_1 (\Delta^2)
\right)
\, .
\end{equation}

\begin{figure}[t]
\unitlength1mm
\begin{center}
\mbox{
\begin{picture}(0,33)(27,0)
\put(0,0){\insertfig{5}{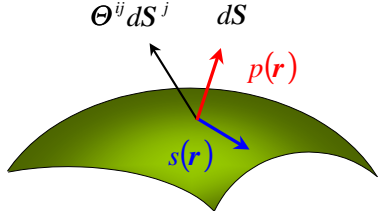}}
\end{picture}
}
\end{center}
\caption{\label{Pressure} Pressure and shear on the nucleon's surface,
considered as a continuous medium.}
\end{figure}

In order to uncover the physical interpretation of the form factors in the
parametrization (\ref{GravitFormFactorDecomp}), one performs the same analysis
with wave packets as was done for the matrix element of the quark electromagnetic
current in Section \ref{BreitFrameFFs}. As a result the Breit-frame gravitational
form factors naturally appear to be the Fourier transforms of spatial characteristics
of the nucleon \cite{BelJi02,BelMul02,Pol02,PolShu02}.
\begin{itemize}
\item The time component ${\mit\Theta}^{00}$ is the energy density and
thus its matrix element between the proton states localized at $\bit{R} =
\bit{0}$ measures the mass distribution inside the nucleon. Its is related
to the gravito-electric form factor \cite{BelJi02,BelMul02,Pol02,PolShu02},
\begin{equation}
\label{TimeTimeEMelements}
\int d^3 \bit{r} \,
{\rm e}^{- i \bit{\scriptstyle \Delta} \cdot \bit{\scriptstyle r}}
\langle \bit{0} | {\mit\Theta}^{00} (\bit{r}) | \bit{0} \rangle
=
M_N \sqrt{1 + \frac{\bit{\Delta}^2}{4 M_N^2}} w_2^\ast w_1
\left(
H - \frac{\bit{\Delta}^2}{4 M_N^2} ( E - D )
\right) (- \bit{\Delta}^2)
\, .
\end{equation}
Note that the mass distribution of constituents is different from their charge
distribution due to the presence of electromagnetically neutral gluons inside
hadrons which do not contribute to the electromagnetic form factors. At zero recoil,
$\bit{\Delta} = 0$, the matrix element of the energy-momentum tensor gives the
nucleon's momentum fractions carried by its constituents\footnote{Note however
that the contributing time components of the energy momentum tensor in Eq.\
(\ref{TimeTimeEMelements}) do contain interaction dependent terms in the light-cone
gauge $A^+ = 0$ and thus differ in interpretation from the momentum sum rule
(\ref{MomSumRule}) which involves the matrix element of the interaction-independent
light-cone projection of energy-momentum tensor $\langle p | {\mit\Theta}^{++} | p
\rangle$. See Ref.\ \cite{Jaf00} for a thorough discussion on this issue.} which is
obviously
\begin{equation}
\label{TotalMomentumSumRule}
H (0)
=
\sum_{a = q,g} H^a (0)
=
\sum_{a = q,g} \int_{- 1}^1 dx \, x \, f^a (x) = 1
\, .
\end{equation}
The separate components of this sum
\begin{equation}
\label{MomSumRule}
P^a = \int_{- 1}^1 dx \, x \, f^a (x)
\, ,
\end{equation}
define the fraction of hadron's momentum carried by the $a$-parton and
are measurable in deeply inelastic scattering.
\item The time-spatial component of the energy-momentum tensor $\bit{\mit\Theta}
\equiv ( {\mit\Theta}^{01}, {\mit\Theta}^{02}, {\mit\Theta}^{03})$ defines the
three-dimensional distribution of the orbital angular momentum
and is related to the nucleon gravito-magnetic form factor
\begin{equation}
\int d^3 \bit{r} \,
{\rm e}^{- i \bit{\scriptstyle \Delta} \cdot \bit{\scriptstyle r}}
\langle \bit{0} | {\mit\Theta}^{0k} (\bit{r}) | \bit{0} \rangle
=
\ft{i}4 \sqrt{1 + \frac{\bit{\Delta}^2}{4 M_N^2}}
w_2^\ast [\bit{\Delta} \times \bit{\sigma}]^k w_1
\left( H + E \right) ( - \bit{\Delta}^2 )
\, .
\end{equation}
To clearly see the physical content, we expand both sides of this equation up to
terms linear in the momentum transfer $\bit{\Delta}$ and finds
\begin{equation}
\langle \bit{0} | \bit{J} | \bit{0} \rangle
=
\int d^3 \bit{r} \,
\langle \bit{0} | [\bit{r} \times \bit{\mit\Theta} (\bit{r})] | \bit{0} \rangle
=
\frac{\bit{s}}{2} \left( H (0) + E (0) \right)
\, ,
\end{equation}
cf.\ (\ref{JfromMmunu}). Thus, the gravito-magnetic form factor is normalized
at zero recoil to the fractions of total nucleon spin carried by the angular
momentum of quarks and gluons \cite{Ji96a},
\begin{equation}
\label{SpinRule}
J^a
=
\frac{1}{2} \left( H^a ( 0 ) + E^a ( 0 ) \right)
\, ,
\end{equation}
with total angular momentum being, of course, the nucleon spin
$$
\sum_{q} J^q + J^g = \ft12
\, .
$$
This result, together with Eq.\ (\ref{TotalMomentumSumRule}) immediately implies
that the total gravitomagnetic moment is zero \cite{OkuKob62,KorZak70,BroHwaMaSch00}
(cf.\ \cite{Ter99})
\begin{equation}
\label{TotalGravitoMagneticMoment}
\sum_q E^q (0) + E^g (0) = 0
\, .
\end{equation}
\item Finally, the spatial components of the energy-momentum tensor in the
Breit frame allow one to separate the remaining form factor $D (\Delta^2)$,
\begin{equation}
\label{SpaceSpaceEM}
\int d^3 \bit{r} \,
{\rm e}^{- i \bit{\scriptstyle \Delta} \cdot \bit{\scriptstyle r}}
\langle \bit{0} | {\mit\Theta}^{jk} (\bit{r}) | \bit{0} \rangle
=
\frac{\Delta^j \Delta^k - \bit{\Delta}^2 \delta^{jk}}{4 M_N}
\sqrt{1 + \frac{\bit{\Delta}^2}{4 M_N^2}}
w_2^\ast w_1
D ( - \bit{\Delta}^2 )
\, .
\end{equation}
Considering the nucleon as a continuous medium of partons, the stress
tensor ${\mit\Theta}^{ij} (\bit{r})$ is a density of flow of the $i$-th
component of the momentum in the direction of the $j$-th axis through the
unit area in a unit time interval. In other words,  it characterizes
the force ${\mit\Theta}^{ij} d S^j$ experienced by quarks in an infinitesimal
surface element $d S^j$ at a distance $\bit{r}$ from the center of the
nucleon. We can introduce the following parametrization of the total static
stress tensor
\begin{equation}
\label{EMinShearPressure}
\langle \bit{0} | {\mit\Theta}^{jk} (\bit{r}) | \bit{0} \rangle
=
s (\bit{r})
\left( \frac{r^i r^j}{\bit{r}^2} - \frac{1}{3} \delta^{ij} \right)
+
p (\bit{r}) \delta^{ij}
\, ,
\end{equation}
in terms of the scalar functions which describe shear the $s (\bit{r})$ and pressure
$p (\bit{r})$, see Fig.\ \ref{Pressure}. They are related to each other by conservation
of the total energy-momentum tensor. The function $p(\bit{r})$ can be interpreted as
the radial distribution of the ``pressure" inside the nucleon. Thus a detailed
distribution of forces can be extracted from the $\Delta^2$-dependence of $D (\Delta^2)$.
Namely, substituting Eq.\ (\ref{EMinShearPressure}) into Eq.\ (\ref{SpaceSpaceEM}) one
finds that this form factor is related to the Fourier transform of the shear
\begin{equation}
\label{ShearForces}
\sqrt{1 + \frac{\bit{\Delta}^2}{4 M_N^2}}
\bit{\Delta}^2 D (- \bit{\Delta}^2) w_2^\ast w_1
=
6 M_N
\int d^3 \bit{r} \,
{\rm e}^{- i \bit{\scriptstyle \Delta} \cdot \bit{\scriptstyle r}}
s (\bit{r})
\left( \frac{r^i r^j}{\bit{r}^2} - \frac{1}{3} \delta^{ij} \right)
\frac{\Delta^j \Delta^k}{\bit{\Delta}^2}
\, .
\end{equation}
\end{itemize}

By taking the light-cone projection of the energy-momentum tensor and
comparing it to the polynomial form of the moments of generalized parton
distributions (\ref{PolynomialityNucleon}), one finds the following
relations
\begin{eqnarray}
\label{SecondMomentsHE}
&&\int_{-1}^1 dx \, x \, H^q (x, \eta, \Delta^2)
=
H^q (\Delta^2) + \eta^2 D^q (\Delta^2)
\, , \quad
\int_{-1}^1 dx \, x \, E^q (x, \eta, \Delta^2)
=
E^q (\Delta^2) - \eta^2 D^q (\Delta^2)
\, , \nonumber\\
&&\int_0^1 dx \, H^g (x, \eta, \Delta^2)
=
H^g (\Delta^2) + \eta^2 D^g (\Delta^2)
\, , \quad\ \
\int_0^1 dx \, E^g (x, \eta, \Delta^2)
=
E^g (\Delta^2) - \eta^2 D^g (\Delta^2)
\, . \nonumber\\
\end{eqnarray}

As it is clear from our presentation, the $\Delta^2$-dependence of the GPDs
provides us with a very detailed spatial images of the nucleon. So far, we
have addressed just their lowest two moments. As we found, the first moment
corresponds to electromagnetic and weak form factors, while the second is
expressed through the gravitational form factors. These two moments give
charge, magnetization, mass and  angular momentum distributions inside the
proton. Higher moments of GPDs correspond to coupling of the higher-spin
probes to the proton and shed light on its more complicated  characteristics.

\subsection{Wave functions and distribution amplitudes}
\label{SectionDistributionAmplitudes}

\begin{figure}[t]
\begin{center}
\hspace{0cm}
\mbox{
\begin{picture}(0,90)(60,0)
\put(0,0){\insertfig{4}{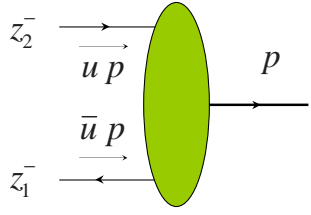}}
\end{picture}
}
\end{center}
\caption{\label{DistributionAmplitude} Two-particle distribution amplitude.
}
\end{figure}

As should be anticipated from our previous discussion and graphical representation
of GPDs in Fig.\ \ref{GPDgreenFunc}, in case when the incoming or outgoing momentum
is set to zero, a GPD reduces to a function of a single variable. This function
formally resembles the meson distribution amplitude. A meson distribution amplitude
(DA)  actually parametrizes a vacuum-to-hadron or hadron-to-vacuum matrix element
of a non-local light-cone operator\footnote{Here we do not display the gauge links.
Their presence is tacitly implied.}
\begin{equation}
\label{GenericDistributionAmplitude}
\langle p |
\bar \psi (z_1^-) \gamma^+ \psi (z_2^-)
| 0 \rangle
= - i p^+ f_M
\int_{0}^{1} d u \,
{\rm e}^{i p^+ ( u z_1^- + \bar{u} z_2^- )}
\phi_M (u)
\, .
\end{equation}
Expressing DA in the light-cone formalism in terms of creation and annihilation
operators, as shown in Fig.\ \ref{DistributionAmplitude},
\begin{equation}
\label{PartonicContentDAs}
i f_M \phi_M (u)
=
\frac{1}{\sqrt{u \bar{u}}}
\sum_{\lambda = \uparrow\downarrow}
\int \frac{d^2 \bit{k}_\perp}{(2 \pi)^3}
\frac{
\langle p |
b_\lambda^\dagger (u p^+ , \bit{k}_\perp)
d_{- \lambda}^\dagger (\bar{u} p^+ , - \bit{k}_\perp)
| 0 \rangle
}{\langle p | p \rangle}
\, ,
\end{equation}
one can see that the function is obviously not a density.

The discussion of distribution amplitudes and their $\bit{k}_\perp$-unintegrated
generalization (wave functions) has  double  purpose. First,  distribution
amplitudes enter as building blocks into numerous exclusive processes via
factorization theorems together with GPDs (see Section \ref{Sec-Int}). Thus
they should be treated as another unknown factors in the description of these
processes, and they should be understood in order to use those particular
reactions for extraction of GPDs. Second, the light-cone wave functions are
genuinely fundamental objects encoding information about hadronic structure on
the amplitude level. They can be used to represent a number of hadronic observable
in their terms. Formally, this reduces the problem of describing hadronic processes
to the computation of the wave functions from first principles. In the following
few sections, we provide first a short classification of meson distribution
amplitudes, to be used later on in the process of description of hard exclusive
meson production. Next, we outline the theory of baryon wave functions with
non-zero orbital angular momentum, and use them in the study of the overlap
representation of GPDs.

\subsubsection{Meson distribution amplitudes}
\label{MesonDA}

Leading twist meson distribution amplitudes (\ref{GenericDistributionAmplitude}),
which will appear in our discussion of phenomenology of hard exclusive meson
production, are classified below in few important cases by the spin and parity
of hadrons involved. Let us provide a compendium of results.

\noindent $\bullet$ Pseudoscalar distribution amplitudes:
\begin{eqnarray}
\label{LeadingTwistPionDA}
\langle \pi^+ (p) | \bar u (z_1^-) \gamma^+ \gamma^5 d (z_2^-) | 0 \rangle
\!\!\!&=&\!\!\!
- i p^+ f^+_\pi \int_0^1 du \, {\rm e}^{i p^+ ( u z_1^- + \bar u z_2^- )}
\phi_\pi (u)
\, , \\
\langle K^+ (p) | \bar u (z_1^-) \gamma^+ \gamma^5 s (z_2^-) | 0 \rangle
\!\!\!&=&\!\!\!
- i p^+ f^+_K \int_0^1 du \, {\rm e}^{i p^+ ( u z_1^- + \bar u z_2^- )}
\phi_K (u)
\, .
\end{eqnarray}
Notice that as a consequence of the $SU(2)$ and charge symmetry we have the
relations between matrix elements involving kaons
\begin{eqnarray}
\langle K^+ (p) | \bar u (z_1^-) \gamma^+ \gamma^5 s (z_2^-) | 0 \rangle
\!\!\!&=&\!\!\!
\langle K^- (p) | \bar s (z_1^-) \gamma^+ \gamma^5 u (z_2^-) | 0 \rangle
\\
\!\!\!&=&\!\!\!
\langle K^0 (p) | \bar d (z_1^-) \gamma^+ \gamma^5 s (z_2^-) | 0 \rangle
=
\langle \bar K^0 (p) | \bar s (z_1^-) \gamma^+ \gamma^5 d (z_2^-) | 0 \rangle
\, . \nonumber
\end{eqnarray}
\noindent $\bullet$ Distribution amplitudes of longitudinally polarized vector mesons:
\begin{eqnarray}
\langle \rho^+_L (p) |
\bar u (z_1^-) \gamma^+ d (z_2^-)
| 0 \rangle
\!\!\!&=&\!\!\!
p^+ f^+_\rho \int_0^1 du \, {\rm e}^{i p^+ ( u z_1^- + \bar u z_2^- )}
\phi_\rho (u)
\, , \\
\label{RhoMesonDA}
\langle \rho^0_L (p) |
\frac{1}{\sqrt{2}}
\left(
\bar u (z_1^-) \gamma^+ u (z_2^-)
-
\bar d (z_1^-) \gamma^+ d (z_2^-)
\right)
| 0 \rangle
\!\!\!&=&\!\!\!
p^+ f^0_\rho \int_0^1 du \, {\rm e}^{i p^+ ( u z_1^- + \bar u z_2^- )}
\phi_\rho (u)
\, , \nonumber\\
\langle \omega^0_L (p) |
\frac{1}{\sqrt{2}}
\left(
\bar u (z_1^-) \gamma^+ u (z_2^-)
+
\bar d (z_1^-) \gamma^+ d (z_2^-)
\right)
| 0 \rangle
\!\!\!&=&\!\!\!
p^+ f^0_\omega \int_0^1 du \, {\rm e}^{i p^+ ( u z_1^- + \bar u z_2^- )}
\phi_\omega (u)
\, . \nonumber
\end{eqnarray}
A convenient partial-wave expansion of distribution amplitudes is accomplished
in terms of Gegenbauer polynomials, the eigenfunctions of QCD evolution equations
(this will be explained in the next chapter),
\begin{equation}
\phi_M (u) = 6 u \bar u \sum_{n = 0}^\infty a_n^M C_n^{3/2} (2 u - 1)
\, , \qquad
a_0^M = 1
\, .
\end{equation}
The decay constants and hadronic parameters $a_n^M$ for different mesons are
summarized in Table \ref{MesonDAparameters}.

\begin{table}[t]
\begin{center}
\begin{tabular}{cccccc}
\hline
param. $\backslash$ M   & $\pi^+$ & $K^-$           & $\rho^+$        & $\rho^0$      & $\omega$ \\
\hline\hline
$f_M$, MeV              & 132     & $159 \pm 1.3$   & $195\pm 7$      & $216\pm 5$    & $195$    \\
\hline
$a_1^M$                 & 0       & $0.10 \pm 0.12$ & 0               &  0            & 0   \\
\hline
$a_2^M$                 & 0.19    & -               & $0.18 \pm 0.10$ & $-0.1\pm 0.2$ & -  \\
\hline
$a_4^M$                 & -0.13   & -               & -               & -             & -  \\
\hline
\end{tabular}
\end{center}
\caption{\label{MesonDAparameters} Decay constants \cite{Pdg02} and magnitude
of the first nontrivial harmonics in the partial wave expansion of mesonic
distribution amplitudes at the scale $\mu^2 \approx 1 \, {\rm GeV}^2$ for
$\pi^+$ \cite{BakPasSchSte04} (also $a_{2j} \sim 10^{-3}$ for $i = 3,4,5$ was
given there), $K^-$ \cite{BraLen04,KodMan04} (however, cf.\ \cite{Ball:2003sc}),
$\rho^+$ \cite{BalBra96}, $\rho^0$ \cite{Clerbaux:2000hb}.
}
\end{table}

\subsubsection{Nucleon wave function with orbital momentum}
\label{NucleonWaveFunctions}

The Fock decomposition of a hadronic state is made in components with increasing
number of constituents populating the hadron. It starts  with the minimally possible
configuration, which is a quark-antiquark pair for a meson or three quarks for a
baryon. Below, we limit our discussion to the lowest three-quark component of the
nucleon wave functions and classify them with respect to quark orbital angular
momentum \cite{JiMaYua02}. These wave functions are used both to unravel information
contained in matrix elements and to provide a nonperturbative description of nucleon
observables where the orbital momentum plays a significant role, like for instance
the Pauli form factor \cite{BroDre80,JaiRal93,BroHwaMaSch00,BelJiYua02b}, or the
hadron helicity-flip GPD $E^q$ \cite{BroHwaMaSch00,JiMaYua02} encoding the magnitude
of angular momentum carried by the quarks, etc. We consider below the proton state
with helicity-up and introduce the parametrization of the vacuum-to-proton matrix
elements at equal light-cone time $z^+ = 0$ in terms of coordinate-space wave
function with quark orbital angular momentum $L^z$. The following three-quark
spin and orbital momentum configurations are possible in order to build up
the nucleon state with the projection $s^z_p = 1/2$ of its spin:
\begin{equation}
\begin{array}{llll}
s^z_p = 1/2: \qquad
& \uparrow\downarrow\uparrow     &= \ \sum_q s_q^z = 1/2 \, , & \qquad L^z = 0 \, , \\
& \downarrow\downarrow\uparrow   &= \ \sum_q s_q^z = - 1/2 \, , & \qquad L^z = 1 \, , \\
& \uparrow\uparrow\uparrow       &= \ \sum_q s_q^z = 3/2 \, , & \qquad L^z = - 1 \, , \\
& \downarrow\downarrow\downarrow &= \ \sum_q s_q^z = - 3/2 \, , & \qquad L^z = 2 \, .
\end{array}
\end{equation}
They correspond to the following set of coordinate-space wave functions parametrizing
three-quark operators with the quantum numbers of the proton \cite{JiMaYua02}.

\noindent $\bullet$ $L^z = 0$:
\begin{eqnarray}
\label{NucleonWFLzero}
&&
\!\!\!\!\!\!\!\!\!
\langle 0 |
\frac{\varepsilon^{abc}}{\sqrt{6}}
\left(
u^a_{+ \uparrow} (z_1)
C \gamma^+
u^b_{+ \downarrow} (z_2)
\right)
d^c_{+ \uparrow} (z_3)
| p_\uparrow \rangle
=
\psi_1 (z_1, z_2, z_3) p^+ u_{+ \uparrow} (p)
\, , \\
&&
\!\!\!\!\!\!\!\!\!
\langle 0 |
\frac{\varepsilon^{abc}}{\sqrt{6}}
\left(
u^a_{+ \uparrow} (z_1)
C i \sigma^{+ i}
u^b_{+ \uparrow} (z_2)
\right)
d^c_{+ \downarrow} (z_3)
| p_\uparrow \rangle
=
\Big(
\psi_1 (z_1, z_3, z_2)
+
\psi_1 (z_2, z_3, z_1)
\Big)
p^+ \gamma^i u_{+ \uparrow} (p)
\, . \nonumber
\end{eqnarray}
\noindent $\bullet$ $L^z = 1$:
\begin{eqnarray}
&&
\!\!\!\!\!\!\!\!\!
\langle 0 |
\frac{\varepsilon^{abc}}{\sqrt{6}}
\left(
u^a_{+ \uparrow} (z_1)
C \gamma^+
u^b_{+ \downarrow} (z_2)
\right)
d^c_{+ \downarrow} (z_3)
| p_\uparrow \rangle
=
- i
\Big(\!
\not\!\!\bit{\nabla}_{\! 1\perp} \psi_3 (z_1, z_2, z_3) \,
+
\not\!\!\bit{\nabla}_{\! 2\perp} \psi_4 (z_1, z_2, z_3)
\Big)
p^+ u_{+ \uparrow} (p)
\, , \nonumber\\
&&
\!\!\!\!\!\!\!\!\!
\langle 0 |
\frac{\varepsilon^{abc}}{\sqrt{6}}
\left(
u^a_{+ \downarrow} (z_1)
C i \sigma^{+ i}
u^b_{+ \downarrow} (z_2)
\right)
d^c_{+ \uparrow} (z_3)
| p_\uparrow \rangle
\\
&&\qquad\qquad\qquad\qquad\qquad
\!\!\!\!\!\!\!\!\!
= i
\bigg(
\nabla_{\! + 1}^i
\Big(
\psi_4 (z_3, z_1, z_2)
-
\psi_3 (z_3, z_1, z_2)
-
\psi_3 (z_3, z_2, z_1)
\Big)
\nonumber\\
&&\qquad\qquad\qquad\qquad\qquad\quad\
\!\!\!\!\!\!\!\!\!
+
\nabla_{\! + 2}^i
\Big(
\psi_4 (z_3, z_2, z_1)
-
\psi_3 (z_3, z_2, z_1)
-
\psi_3 (z_3, z_1, z_2)
\Big)
\bigg)
p^+ u_{+ \uparrow} (p)
\, . \nonumber
\end{eqnarray}
\noindent $\bullet$ $L^z = - 1$:
\begin{eqnarray}
&&
\langle 0 |
\frac{\varepsilon^{abc}}{\sqrt{6}}
\left(
u^a_{+ \uparrow} (z_1)
C i \sigma^{+ i}
u^b_{+ \uparrow} (z_2)
\right)
d^c_{+ \uparrow} (z_3)
| p_\uparrow \rangle
\\
&&\qquad
= i
\bigg(
\nabla_{\! - 1}^i
\Big(
\psi_5 (z_1, z_3, z_2)
-
\psi_5 (z_1, z_2, z_3)
\Big)
+
\nabla_{\! - 2}^i
\Big(
\psi_5 (z_2, z_3, z_1)
-
\psi_5 (z_2, z_1, z_3)
\Big)
\bigg)
p^+ u_{+ \uparrow} (p)
\, , \nonumber
\end{eqnarray}
\noindent $\bullet$ $L^z = 2$:
\begin{eqnarray}
\label{NucleonWFLtwo}
&&
\langle 0 |
\frac{\varepsilon^{abc}}{\sqrt{6}}
\left(
u^a_{+ \downarrow} (z_1)
C i \sigma^{+ i}
u^b_{+ \downarrow} (z_2)
\right)
d^c_{+ \downarrow} (z_3)
| p_\uparrow \rangle
\\
&&\qquad
= 2 i^2 \Sym_{ij}
\bigg(
\nabla_{\! + 1}^i
\nabla_{\! \perp 2}^j
\Big(
\psi_6 (z_1, z_2, z_3)
+
\psi_6 (z_2, z_1, z_3)
-
\psi_6 (z_1, z_3, z_2)
-
\psi_6 (z_2, z_3, z_1)
\Big)
\nonumber\\
&&\qquad\qquad\qquad\qquad\qquad
+
\nabla_{\! + 1}^i
\nabla_{\! \perp 1}^j
\psi_6 (z_1, z_3, z_2)
+
\nabla_{\! + 2}^i
\nabla_{\! \perp 2}^j
\psi_6 (z_2, z_1, z_3)
\bigg)
p^+ u_{+ \uparrow} (p)
\, , \nonumber
\end{eqnarray}
where in the last equality we introduced the symmetrization and trace subtraction
operation for two-dimensional indices $\Sym_{ij} t_{ij} \equiv \ft12 (t_{ij} + t_{ji}
- \delta_{ij} t_{kk})$. The charge conjugation matrix $C$ used above is defined in
Eq.\ (\ref{Gamma5ChargeConj}) and $\bit{\nabla}_\perp \equiv \partial / \partial
\bit{z}_\perp$ is the transverse derivative. We also introduced the holomorphic and
antiholomorphic derivatives
\begin{equation}
\nabla_-^i \equiv \nabla^i - i \varepsilon^{ij} \nabla^j
=
\bit{e}^i_\perp \frac{\partial}{\partial z_\perp}
\, , \qquad
\nabla_+^i \equiv \nabla^i + i \varepsilon^{ij} \nabla^j
=
\bar {\bit e}^i_\perp \frac{\partial}{\partial {\bar z}_\perp}
\end{equation}
with respect to holomorphic $z_\perp \equiv z^x + i z^y$ and antiholomorphic
$\bar z_\perp \equiv z^x - i z^y$ coordinates. The two-dimensional vectors
involved here were previously introduced in Eq.\ (\ref{HelicityMatrix}).

The Fourier transform of the coordinate-space wave functions leads to the
conventional light-cone wave functions,
\begin{equation}
\label{FourierTransfMomCoord}
\psi (z_1, z_2, z_3)
=
\int [d x] [d^2 \bit{k}_\perp]
\, \psi (\kappa_1, \kappa_2, \kappa_3)
\exp\left(
- i p^+ \sum_{i = 1}^3 x_i z^-_i
+ i \sum_{i = 1}^3 \bit{k}_{i\perp} \cdot \bit{z}_{i\perp}
\right)
\, ,
\end{equation}
where we used a short-hand notation for the argument of the momentum-space wave
function
\begin{equation}
\kappa_i \equiv \left( x_i , \bit{k}_{i\perp} \right)
=
\left( x_i , k_i^x , k_i^y \right)
\, .
\end{equation}
We have taken into account that the wave functions depend on the momentum variables
via the  light-cone momentum fractions $x_i = k_i^+/p^+$ and transverse momenta
$\bit{k}_{i \perp}$ with respect to the momentum of the parent hadron. This is an
immediate consequence of the Lorentz invariance on the light cone. The integration
measures are\footnote{Our conventions for the integration measures as well as the
relation between the coordinate and momentum space wave functions differ from the
ones accepted in Ref.\ \cite{Diehl:2003ny} where the standard Brodsky-Lepage
normalization \cite{BroLep79} was assumed.}
\begin{equation}
[d x]
\equiv
\delta \left( x_1 + x_2 + x_3 - 1 \right)
\prod_{i = 1}^3 d x_i
\, , \quad
[d^2 \bit{k}_\perp]
\equiv
(2 \pi)^3
\delta^{(2)}
\left(
\bit{k}_{1\perp} + \bit{k}_{2\perp} + \bit{k}_{3\perp}
\right)
\prod_{i = 1}^3 \,
\frac{d^2 \bit{k}_{i \perp}}{(2 \pi)^3}
\, .
\end{equation}

Using the definitions of the coordinate-space wave functions, one can easily find
the nucleon state in terms of components possessing different angular momentum,
\begin{equation}
\label{NucleonState}
| p_\uparrow \rangle
=
| p_\uparrow \rangle_{L = 0}
+
| p_\uparrow \rangle_{L = 1}
+
| p_\uparrow \rangle_{L = - 1}
+
| p_\uparrow \rangle_{L = 2}
\, ,
\end{equation}
by evaluating the matrix elements with the nucleon state (\ref{NucleonWFLzero}) --
(\ref{NucleonWFLtwo}) at zero light-cone time, see Appendix \ref{LightConeQCD}. A
simple calculation yields the nucleon states expressed via the momentum space wave
functions defined in Eq.\ (\ref{FourierTransfMomCoord}),
\begin{eqnarray}
\label{L0ProtonState}
| p_\uparrow \rangle_{L = 0}
\!\!\!&=&\!\!\!
\frac{1}{2}
\int \frac{[d x] [d^2 \bit{k}_\perp]}{\sqrt{x_1 x_2 x_3}} \,
\psi_1 (\kappa_1, \kappa_2, \kappa_3)
\nonumber\\
&\times&\!\!\!
\frac{\varepsilon^{abc}}{\sqrt{6}} \,
b^{a \dagger}_u (\kappa_1, \uparrow)
\left\{
b^{b \dagger}_u (\kappa_2, \downarrow)
b^{c \dagger}_d (\kappa_3, \uparrow)
-
b^{b \dagger}_d (\kappa_2, \downarrow)
b^{c \dagger}_u (\kappa_3, \uparrow)
\right\}
| 0 \rangle
\, , \\
| p_\uparrow \rangle_{L = 1}
\!\!\!&=&\!\!\!
\frac{1}{2}
\int \frac{[d x] [d^2 \bit{k}_\perp]}{\sqrt{x_1 x_2 x_3}} \,
\left\{
k_{1 \perp} \psi_3 (\kappa_1, \kappa_2, \kappa_3)
+
k_{2 \perp} \psi_4 (\kappa_1, \kappa_2, \kappa_3)
\right\}
\nonumber\\
&\times&\!\!\!
\frac{\varepsilon^{abc}}{\sqrt{6}} \,
\left\{
b^{a \dagger}_u (\kappa_1, \uparrow)
b^{b \dagger}_u (\kappa_2, \downarrow)
b^{c \dagger}_d (\kappa_3, \downarrow)
-
b^{a \dagger}_d (\kappa_1, \uparrow)
b^{b \dagger}_u (\kappa_2, \downarrow)
b^{c \dagger}_u (\kappa_3, \downarrow)
\right\}
| 0 \rangle
\, , \\
| p_\uparrow \rangle_{L = - 1}
\!\!\!&=&\!\!\!
\frac{1}{2}
\int \frac{[d x] [d^2 \bit{k}_\perp]}{\sqrt{x_1 x_2 x_3}} \,
\bar k_{1 \perp} \psi_5 (\kappa_1, \kappa_2, \kappa_3)
\nonumber\\
&\times&\!\!\!
\frac{\varepsilon^{abc}}{\sqrt{6}} \,
b^{a \dagger}_u (\kappa_1, \uparrow)
\left\{
b^{b \dagger}_d (\kappa_2, \uparrow)
b^{c \dagger}_u (\kappa_3, \uparrow)
-
b^{b \dagger}_u (\kappa_2, \uparrow)
b^{c \dagger}_d (\kappa_3, \uparrow)
\right\}
| 0 \rangle
\, , \\
| p_\uparrow \rangle_{L = 2}
\!\!\!&=&\!\!\!
\frac{1}{2}
\int \frac{[d x] [d^2 \bit{k}_\perp]}{\sqrt{x_1 x_2 x_3}} \,
k_{1 \perp} k_{3 \perp} \psi_6 (\kappa_1, \kappa_2, \kappa_3)
\nonumber\\
&\times&\!\!\!
\frac{\varepsilon^{abc}}{\sqrt{6}} \,
b^{a \dagger}_u (\kappa_1, \downarrow)
\left\{
b^{b \dagger}_d (\kappa_2, \downarrow)
b^{c \dagger}_u (\kappa_3, \downarrow)
-
b^{b \dagger}_u (\kappa_2, \downarrow)
b^{c \dagger}_d (\kappa_3, \downarrow)
\right\}
| 0 \rangle
\, .
\label{Lm1ProtonState}
\end{eqnarray}
Here we introduced the holomorphic and antiholomorphic momenta
\begin{equation}
k_{i \perp} \equiv k_i^x + i \, k_i^y
\, , \qquad
\bar k_{i \perp} \equiv k_i^x - i \, k_i^y
\end{equation}
respectively. The state with zero orbital angular momentum is normalized as follows
\begin{equation}
{_{L = 0} \langle} p_{2 \uparrow} | p_{1 \uparrow} \rangle_{L = 0}
=
4 p_1^+ (2 \pi)^3 \delta ( p_1^+ - p_2^+ )
\delta^{(2)} (\bit{p}_{1 \perp} - \bit{p}_{2 \perp})
\int [dx] [d^2 \bit{k}_\perp]
|\psi_1 (\kappa_1, \kappa_2, \kappa_3)|^2
\, .
\end{equation}
The proton state with helicity-down can be obtained from the above Eqs.\
(\ref{L0ProtonState}) -- (\ref{Lm1ProtonState}) making use of the Jacob-Wick
transformation $\mathcal{Y}$,
\begin{equation}
\label{JacWicTransform}
\mathcal{Y} | p_{s^z} \rangle
\equiv
{\rm e}^{- i \pi J^y} \mathcal{P} | p_{s^z} \rangle
=
(- 1)^{s - s^z} | p_{- s^z} \rangle
\end{equation}
which consists of a sequence of the spatial reflection $\mathcal{P}$ (see Section
\ref{TimeReversalHermiticity}) and the spatial rotation $J^y$ (\ref{JfromMmunu})
by $180^\circ$ around the $y$-axis \cite{JacWic59}.

A few comments are in order concerning the properties of wave functions. As we know
from Section \ref{TimeReversalHermiticity}, under the reversal of spatial parity, the
three-components of the proton momentum change their sign, so that $(p^0, \bit{p})$
becomes $(p^0, - \bit{p})$. Under the time reversal, the momentum changes in the same
way. Therefore, under combined time-reversal and parity transformations, the momentum
of the proton is left intact (see Eq.\ (\ref{TimeReversalForState})). The same is true
for the quark momenta. On the other hand, the helicity changes sign under the combined
parity and time reversal, and hence the proton state $|p_\uparrow \rangle$ is
transformed into  $|p_\downarrow \rangle$. Similar changes occur for the quark states.
As we mentioned before, the time reversal acting on $c$-numbers replaces them with
their complex conjugates. Thus, under the combined reflection of space and time, a
positive-helicity proton state becomes a negative-helicity state $|p_\downarrow\rangle$,
with complex-conjugated wave function; flipped quark helicities, and $k^x \pm ik^y$
becoming $k^x\mp ik^y$. The transformed state is also accompanied by the phase factors
$(-1)^{s - s^z}$ (see Eq.\ (\ref{TimeReversalForState})). Comparing the resulting
expression with the ones calculated by means of the Jacob-Wick transformation
(\ref{JacWicTransform}), which does not involve the complex conjugation, we find that
all of the wave function amplitudes must be real,
\begin{equation}
\psi_i^\ast = \psi_i
\, .
\end{equation}
This  result, however, holds provided the gauge condition imposed on the gluon field
is also invariant under the discrete symmetries. In the light-cone gauge, the
condition $A^+=0$ does not fix the gauge completely, one must  specify additional
boundary conditions for the gauge field (see Appendix \ref{LightConeQCD}). This
additional gauge fixing might not be invariant under the combined parity and time
reversal. For example, the advanced and retarded boundary conditions transform into
each other under the time reversal transformation; however, the principal value
prescription is self-conjugate. In the former two cases, the wave function no longer
obey the reality condition.

Finally, it is instructive to establish relations between the wave functions
with non-zero orbital angular momentum and higher-twist distribution amplitudes,
which involve the ``bad'' component of the quark field. This can be done for the
twist-four distribution amplitudes where a complete classification is available
\cite{BraFriMahSte00}. The bridge is built by integrating the momentum-space wave
functions over the transverse momentum with appropriate weight factors. The
definitions of the twist-four distribution amplitudes may be found in Ref.\
\cite{BraFriMahSte00}. For reference purposes, we present here the expression
for the leading twist-three proton's distribution amplitude
\begin{eqnarray}
&&\langle 0 |
\frac{\varepsilon^{abc}}{\sqrt{6}}
\left(
u^a_{+ \uparrow} (z^-_1)
C \gamma^+
u^b_{+ \downarrow} (z^-_2)
\right)
d^c_{+ \uparrow} (z^-_3)
| p_\uparrow \rangle
\\
&&\qquad
= p^+ u_{+ \uparrow} (p)
\int [d x]
\exp \left( - i p^+ \sum_{i = 1}^3 x_i z_i^- \right)
{\mit\Phi}_3 (x_1, x_2, x_3)
\, . \nonumber
\end{eqnarray}
and the  complete set of twist-four distribution amplitudes\footnote{Note that we have
changed the overall normalization of these distribution amplitude as compared to the
original definition in Ref.\ \cite{BraFriMahSte00}, namely our ${\mit\Phi}$ is ${\mit\Phi}
= \frac{-1}{2 \sqrt{6}} {\mit\Phi}_{\mbox{\tiny\protect\cite{BraFriMahSte00}}}$.}
\begin{eqnarray}
\label{TwistFour1}
&&\langle 0 |
\frac{\varepsilon^{abc}}{\sqrt{6}}
\left(
u^a_{+ \uparrow} (z^-_1)
C \gamma^+
u^b_{+ \downarrow} (z^-_2)
\right)
\gamma^-
d^c_{- \uparrow} (z^-_3)
| p_\uparrow \rangle
\\
&&\qquad
= M_N
u_{+ \downarrow} (p)
\int [d x]
\exp \left( - i p^+ \sum_{i = 1}^3 x_i z_i^- \right)
{\mit\Phi}_4 (x_1, x_2, x_3)
\, , \nonumber\\
&&\langle 0 |
\frac{\varepsilon^{abc}}{\sqrt{6}}
\left(
u^a_{+ \uparrow} (z^-_1)
C \gamma^i_\perp
u^b_{- \downarrow} (z^-_2)
\right)
\gamma^i_\perp
d^c_{+ \downarrow} (z^-_3)
| p_\uparrow \rangle
\\
&&\qquad
= - M_N
u_{+ \uparrow} (p)
\int [d x]
\exp \left( - i p^+ \sum_{i = 1}^3 x_i z_i^- \right)
{\mit\Psi}_4 (x_1, x_2, x_3)
\, , \nonumber\\
\label{TwistFour3}
&&\langle 0 |
\frac{\varepsilon^{abc}}{\sqrt{6}}
\left(
u^a_{- \uparrow} (z^-_1)
C
u^b_{+ \uparrow} (z^-_2)
\right)
d^c_{+ \uparrow} (z^-_3)
| p_\uparrow \rangle
\\
&&\qquad
= - \frac{M_N}{2}
u_{+ \uparrow} (p)
\int [d x]
\exp \left( - i p^+ \sum_{i = 1}^3 x_i z_i^- \right)
{\mit\Xi}_4 (x_1, x_2, x_3)
\, . \nonumber
\end{eqnarray}
Neglecting the dynamical quark-gluon correlations, the desired connection between
Eqs.\ (\ref{NucleonWFLzero}) -- (\ref{NucleonWFLtwo}) and Eqs.\ (\ref{TwistFour1})
-- (\ref{TwistFour3}) is found by solving the free Dirac equation ${\not\!\partial}
\psi = 0$ for the ``bad'' components,
\begin{equation}
\psi_- (z^-, \bit{z}_\perp)
= \frac{i}{2}
\int \frac{d z'^-}{2 \pi} \int \frac{d x}{x}
{\rm e}^{i x (z^- - z'^-)}
\gamma^+ \bit{\gamma}_\perp \cdot\!\bit{\nabla}_\perp
\psi_+ (z'^-, \bit{z}_\perp)
\, ,
\end{equation}
where the prescription on the pole in $x$ has to be chosen according to the boundary
condition on the gauge field (see Appendix \ref{LightConeQCD}). Using the expressions
for nucleon state (\ref{NucleonState}), we finally get the representation for the
distribution amplitudes in terms of the wave functions \cite{BelJiYua02b}
\begin{eqnarray}
{\mit\Phi}_3 (x_1, x_2, x_3)
\!\!\!&=&\!\!\!
\int [d^2 \bit{k}_\perp] \,
\psi_1 (\kappa_1, \kappa_2, \kappa_3)
\, , \\
{\mit\Phi}_4 (x_2, x_1, x_3)
\!\!\!&=&\!\!\!
\frac{1}{M_N x_3}
\int [d^2 \bit{k}_\perp] \,
\bit{k}_{3\perp} \cdot
\Big(
\bit{k}_{1\perp}
\psi_3 (\kappa_1, \kappa_2, \kappa_3)
+
\bit{k}_{2\perp}
\psi_4 (\kappa_1, \kappa_2, \kappa_3)
\Big)
\, , \\
{\mit\Psi}_4 (x_1, x_2, x_3)
\!\!\!&=&\!\!\!
\frac{1}{M_N x_2}
\int [d^2 \bit{k}_\perp] \,
\bit{k}_{2\perp} \cdot
\Big(
\bit{k}_{1\perp}
\psi_3 (\kappa_1, \kappa_2, \kappa_3)
+
\bit{k}_{2\perp}
\psi_4 (\kappa_1, \kappa_2, \kappa_3)
\Big)
\, , \\
{\mit\Xi}_4 (x_1, x_2, x_3)
\!\!\!&=&\!\!\!
\frac{1}{M_N x_1}
\int [d^2 \bit{k}_\perp] \,
\bit{k}_{1\perp} \cdot
\bigg(
\bit{k}_{1\perp}
\Big(
\psi_5 (\kappa_1, \kappa_3, \kappa_2)
-
\psi_5 (\kappa_1, \kappa_2, \kappa_3)
\Big)
\\
&&\qquad\qquad\qquad\quad\ \ \ \
+ \bit{k}_{2\perp}
\Big(
\psi_5 (\kappa_2, \kappa_3, \kappa_1)
-
\psi_5 (\kappa_2, \kappa_1, \kappa_3)
\Big)
\bigg)
\, . \nonumber
\end{eqnarray}
Note the interchange of the first and second arguments in ${\mit\Phi}_4$!

\subsubsection{Partonic content of GPDs}
\label{PartonicContentGPDs}

\begin{figure}[t]
\begin{center}
\hspace{0cm}
\mbox{
\begin{picture}(0,100)(145,0)
\put(-40,-20){\insertfig{13}{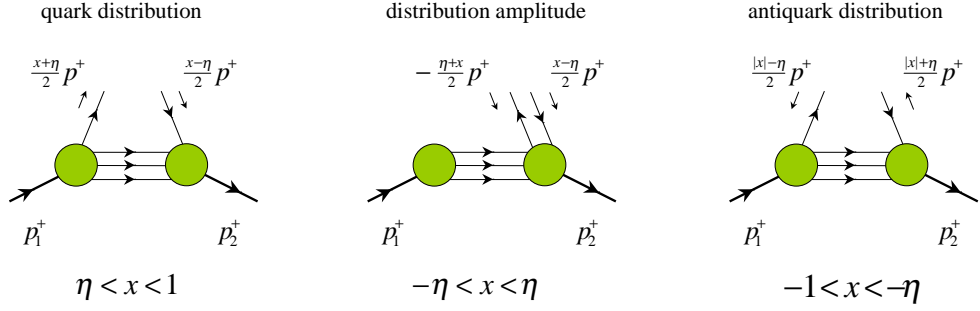}}
\end{picture}
}
\end{center}
\caption{\label{PartonicContent} Partonic content of generalized parton
distributions in different regions of the momentum-fraction space.}
\end{figure}

To get a clear physical understanding of  the structure of GPDs with respect to
the interplay of the Feynman momentum fraction and the skewness dependence, it
is instructive to use the light-cone approaches, like we did for the parton
distributions in Eq.\ (\ref{PartonicContentPDFs}) and distribution amplitudes
in Eq.\ (\ref{PartonicContentDAs}). In the light-cone quantization formalism, the
even-parity generalized parton distribution of a quark can be  written in terms
of creation and annihilation operators. In the region $x > \eta$, it has the
representation \cite{Ji:1998pc,GolMar98,DieFelJakKro00}
\begin{eqnarray}
\label{DISregion1GPDs}
F^q (x, \eta, \Delta^2)|_{x > \eta}
\!\!\!&=&\!\!\!
\frac{1}{\sqrt{x^2 - \eta^2}}
\sum_{\lambda = \uparrow\downarrow}
\int \frac{d^2 \bit{k}_\perp}{(2 \pi)^3}
\\
&\times&\!\!\!
\frac{
\langle p_2 |
b_\lambda^\dagger
\left( \ft12 (x - \eta) p^+, \bit{k}_\perp - \ft12 \bit{\Delta}_\perp \right)
b_\lambda
\left( \ft12 (x + \eta) p^+, \bit{k}_\perp + \ft12 \bit{\Delta}_\perp \right)
| p_1 \rangle
}{\langle p | p \rangle}
\, ,   \nonumber
\end{eqnarray}
where the normalization condition $\langle p | p \rangle = 2 p^+ (2 \pi)^3
\delta^{(3)} (0)$ is implied. This representation  has the form resembling the
expression for the quark distribution, cf.\ (\ref{PartonicContentPDFs}), and
coincides with   it when $p_1 = p_2$. In the region  $x < - \eta$, the GPD has
the form analogous to the antiquark distribution,
\begin{eqnarray}
F^q (x, \eta, \Delta^2)|_{x < - \eta}
\!\!\!&=&\!\!\!
- \frac{1}{\sqrt{x^2 - \eta^2}}
\sum_{\lambda = \uparrow\downarrow}
\int \frac{d^2 \bit{k}_\perp}{(2 \pi)^3}
\\
&\times&\!\!\!
\frac{
\langle p_2 |
d_\lambda^\dagger
\left( \ft12 (|x| - \eta) p^+, \bit{k}_\perp - \ft12 \bit{\Delta}_\perp \right)
d_\lambda
\left( \ft12 (|x| + \eta) p^+, \bit{k}_\perp + \ft12 \bit{\Delta}_\perp \right)
| p_1 \rangle
}{\langle p | p \rangle}
\, . \nonumber
\end{eqnarray}
Finally, in the central region $|x| < \eta$
\begin{eqnarray}
\label{ExclusivePartonicGPDs}
F^q (x, \eta, \Delta^2)|_{|x| < \eta}
\!\!\!&=&\!\!\!
- \frac{1}{\sqrt{\eta^2 - x^2}}
\sum_{\lambda = \uparrow\downarrow}
\int \frac{d^2 \bit{k}_\perp}{(2 \pi)^3}
\\
&\times&\!\!\!
\frac{
\langle p_2 |
b_\lambda^\dagger
\left( \ft12 (x - \eta) p^+, \bit{k}_\perp - \ft12 \bit{\Delta}_\perp \right)
d_{- \lambda}^\dagger
\left( - \ft12 (x + \eta) p^+, - \bit{k}_\perp - \ft12 \bit{\Delta}_\perp \right)
| p_1 \rangle
}{\langle p | p \rangle}
\, , \nonumber
\end{eqnarray}
we recover the form analogous to that encountered in the distribution amplitude
(\ref{PartonicContentDAs}). These representations are graphically illustrated
in Fig.\ \ref{PartonicContent}. Thus, in different regions of the momentum-fraction
space, GPDs share common properties with conventional inclusive parton densities
for $|x| > \eta$, and exclusive distribution amplitudes for $|x| < \eta$, which
illustrates their hybrid nature with respect to the  longitudinal momentum.

\subsubsection{Overlap representation of GPDs}
\label{OverlapGPDs}

In order to exhibit further their partonic content, the generalized parton
distribution can be represented in terms of the light-cone wave functions.
Restricting ourselves only to the lowest Fock component of the nucleon wave
function---i.e., that containing the minimal number of constituents---one can
express a GPD in terms of wave functions only in the ``inclusive'' domain
$|x| > \eta$. The ``exclusive'' central  region obviously requires the
interference of hadron wave functions which differ by two constituents, as
can be seen from the middle graph in Fig.\ \ref{PartonicContent}, and thus
involve three- and five-particle light-cone wave functions. In the absence
of a complete classification scheme for the latter, we will not discuss this
sector at all. For a generic representation in all regions of the momentum
fraction space, we refer to Refs.\ \cite{BroDieHwa00,DieFelJakKro00}.

Let us give a sample formula for the $d$-quark\footnote{The only reason to
chose the $d$-quark is that the  expression is more concise in this case, since
there is only one $d$-quark in the lowest Fock component of the proton wave
function.} helicity nonflip combination of GPDs via  the wave functions introduced
in Section \ref{NucleonWaveFunctions} for $\bit{\Delta}_\perp = 0$. Using the
Dirac bilinears in the light-cone frame given in Eq.\ (\ref{DiracBilinLCframe}),
we get the expression in terms of the wave function with zero orbital angular
momentum $L^z = 0$, thus neglecting contributions with $|L^z| > 0$,
\begin{eqnarray}
\label{HelNonFlipOverlapGPD}
&&\!\!\!\!\!\!\!\!\!\!\!
H^d (x, \eta, \Delta^2_{\rm min})
-
\frac{\eta^2}{1 - \eta^2} E^d (x, \eta, \Delta^2_{\rm min})
=
\int [dx] [d^2 \bit{k}_\perp]
\\
&&\!\!\!\!\!\!\!\!
\times
\Bigg\{
\delta \left( x_3 - x \right)
\psi_1^\ast\!
\left(
\frac{x_1}{1 - \eta} ,
\frac{x_2}{1 - \eta} ,
\frac{x_3 - \eta}{1 - \eta} ,
\bit{k}_{1 \perp} ,
\bit{k}_{2 \perp} ,
\bit{k}_{3 \perp}
\!\right)
\!
\psi_1\!
\left(
\frac{x_1}{1 + \eta} ,
\frac{x_2}{1 + \eta} ,
\frac{x_3 + \eta}{1 + \eta} ,
\bit{k}_{1 \perp} ,
\bit{k}_{2 \perp} ,
\bit{k}_{3 \perp}
\!\right)
\nonumber\\
&&\!\!\!\!
+ \delta \left( x_2 - x \right)
\psi_1^\ast\!
\left(
\frac{x_1}{1 - \eta} ,
\frac{x_2 - \eta}{1 - \eta} ,
\frac{x_3}{1 - \eta} ,
\bit{k}_{1 \perp} ,
\bit{k}_{3 \perp} ,
\bit{k}_{3 \perp}
\!\right)
\!
\psi_1\!
\left(
\frac{x_1}{1 + \eta} , \frac{x_2 + \eta}{1 + \eta} , \frac{x_3}{1 + \eta} ,
\bit{k}_{1 \perp} ,
\bit{k}_{2 \perp} ,
\bit{k}_{3 \perp}
\!\right)
\!\Bigg\}
. \nonumber
\end{eqnarray}
The additional terms with the overlap of $\psi_{3,4}$ with $\psi_5$ are quite analogous
to the displayed ones, however, they enter suppressed by a typical factor $\langle
\bit{k}_\perp^2 \rangle/M_N^2$, where $\langle \bit{k}_\perp^2 \rangle^{-1/2} \approx
300 \, {\rm MeV}$ is the average transverse momentum of quarks in the hadron wave
function. Notice that for $\bit{\Delta}_\perp = 0$, the invariant momentum transfer
squared $\Delta^2$ does not vanish but rather takes some minimal value
\begin{equation}
\label{MinMomDeltaSquared}
\Delta^2_{\rm min} = - \frac{4 \eta^2 M_N^2}{1 - \eta^2}
\, .
\end{equation}
When $\eta = 0$, one gets the conventional overlap representation of Feynman's
parton distributions. In case $\bit{\Delta}_\perp \neq 0$, the formula has to be
modified as follows. Consider a reference frame where the incoming hadron has no
transverse momentum $\bit{p}_{\perp 1} = 0$, while the outgoing proton carries
a non-vanishing transverse momentum equal to the total momentum transferred
$\bit{p}_{\perp 2} = - \bit{\Delta}_\perp$. Then, the arguments of the final-state
wave functions change to
\begin{equation}
\bit{k}_i \to \bit{k}_i - \frac{1 - x_i}{1 - \eta} \bit{\Delta}_\perp
\end{equation}
for the active parton, i.e., those with $i = 3$ and $i = 2$ for the first and
second term of Eq. (\ref{HelNonFlipOverlapGPD}), respectively, and
\begin{equation}
\label{PassiveTransMom}
\bit{k}_i \to \bit{k}_i + \frac{x_i}{1 - \eta} \bit{\Delta}_\perp
\, ,
\end{equation}
for the remaining  passive partons. This shift is well-known from the Drell-Yan
formula for elastic form factors \cite{Drell:1969km}. The representation in the
region $x < - \eta$ is  obtained by reversing the overall sign and replacing
$x \to - x$ in the argument of the $\delta$-function.

The hadron helicity flip amplitude expressed in terms of the GPD $E$ can be
represented in terms of the interference of the spherically symmetric wave
function $\psi_1$ and those with non-zero orbital angular momentum, i.e.,
$E \sim c_1 \psi_1^\ast \psi_3 + c_2 \psi_1^\ast \psi_4 + c_3 \psi_5^\ast \psi_6
+ \ {\rm h.c.}$ \cite{JiMaYua02}.

\subsection{Double distributions}
\label{DDs}

\begin{figure}[t]
\begin{center}
\hspace{0cm}
\mbox{
\begin{picture}(0,110)(190,0)
\put(0,0){\insertfig{13}{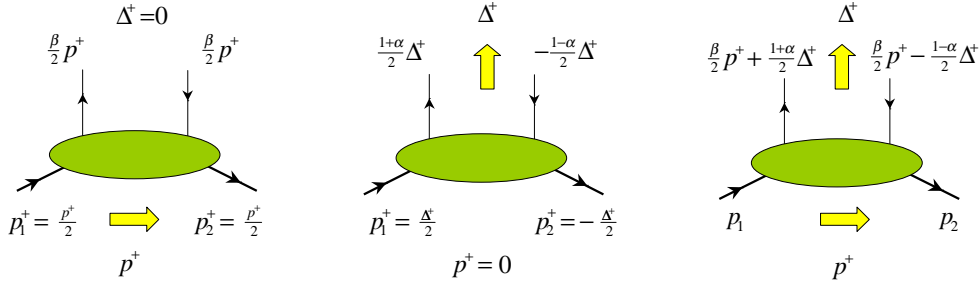}}
\end{picture}
}
\end{center}
\caption{\label{DDparametrization} Momentum flow in a double distribution.}
\end{figure}

Another complimentary and instructive way to look at GPDs is to treat them as
kinematic ``hybrids'' of the usual parton densities $q (\beta)$ and distribution
amplitudes $\phi (\alpha)$. Indeed, $q (\beta)$ corresponds to the forward limit
$\Delta = 0$ of the generalized parton distributions, when the momentum $p$ flows
only in the $s$-channel and the outgoing parton carries the momentum $\beta p^+/2$
of its parent hadron with momentum $p^+/2$ (see Fig.\ \ref{DDparametrization}). On
the other hand, if we take $p = 0$, the momentum $\Delta$ flows in the $t$-channel
only and is shared in fractions $(1 + \alpha) \Delta^+/2$ and $(1 - \alpha)
\Delta^+/2$ between the partons (both momenta treated as outgoing). In general case,
we deal with superposition of two momentum fluxes; then the plus components $k^+_i$
of the parton momenta $k_i$ can be represented as
\begin{equation}
\label{PartonMomentumDDsSymmetric}
k_1^+ = \beta \,\frac{ p^+}{2} \, + \frac{1 + \alpha}{2}\, \Delta^+
\, , \qquad
k_2^+ =  \beta \, \frac{ p^+}{2} \, - \frac{1 - \alpha}{2} \,\Delta^+
\, .
\end{equation}
This decomposition  corresponds to the following parametrization for the matrix element
of the composite operator (\ref{NonlocalScalarOper}) constructed out of scalar field
operators $\phi$, \cite{MulDitRobGeyHor98,Rad1996a,Rad1996b,Rad1997a,Radyushkin:1998es}.
\begin{eqnarray}
\label{ScalarDD}
\langle p_2 |
{\cal O}^{\phi\phi} (- z^- , z^-)
| p_1 \rangle
\!\!\!&=&\!\!\!
\int_{\mit\Omega} d \beta \, d \alpha \,
{\rm e}^{- i \beta z^- p^+ - i \alpha z^- \Delta^+}
f^\phi (\beta, \alpha, \Delta^2)
\nonumber\\
\!\!\!&=&\!\!\!
\int_{- 1}^{1} d x \, {\rm e}^{- i x z^- p^+}
F^\phi (x, \eta, \Delta^2) \, .
\end{eqnarray}
In the second line we have given for comparison the parametrization in terms of a
GPD, see Eq.\ (\ref{DefScalarGPD}).

\begin{figure}[t]
\begin{center}
\hspace{0cm}
\mbox{
\begin{picture}(0,140)(70,0)
\put(0,0){\insertfig{5}{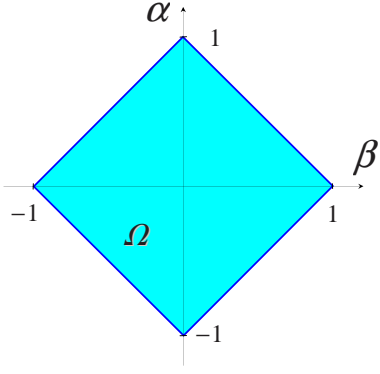}}
\end{picture}
}
\end{center}
\caption{\label{SupportDDs} Support region of double distributions.}
\end{figure}

The support area for $f^\phi (\beta, \alpha, \Delta^2)$ is shown in Fig.\
\ref{SupportDDs}. For any Feynman diagram, the spectral constraint $|\beta| +
|\alpha| \leq 1$ can be proved in the $\alpha$-representation
\cite{MulDitRobGeyHor98,Rad1997a} using the approach of Ref.\ \cite{Rad83}.
Hence,
\begin{equation}
\int_{\mit\Omega} d \beta \, d \alpha
=
\int_{-1}^{1} d \beta \int_{- 1 + |\beta|}^{1 - |\beta|} d \alpha
\, .
\end{equation}

Comparing Eq. (\ref{CollinearQuarkPDF}) with the $\Delta =0$ limit of the DD
definition (\ref{ScalarDD}) one can find the ``reduction formulas'' relating
the double distribution $f (\beta, \alpha, \Delta^2 = 0)$ to the scalar parton
distribution
\begin{eqnarray}
&&
\int_{- 1 + \beta}^{1 - \beta}
d \alpha \, f^\phi (\beta, \alpha, \Delta^2 = 0)
=
f^\phi (\beta)
\, .
\label{eq:redfsym}
\end{eqnarray}
Hence, the  positive-$\beta$ and negative-$\beta$ components of the double
distribution  $f^\phi (\beta , \alpha , \Delta^2 = 0)$ can be treated as
nonforward generalizations of quark and antiquark densities, respectively.
The usual ``forward'' densities $q (\beta)$ and $\bar q (\beta)$ are
thus given by integrating $f^\phi (\beta, \alpha , \Delta^2 = 0)$  over
vertical lines $\beta = {\rm const} $ for $\beta > 0$ and $\beta < 0$,
respectively.

Due to hermiticity and time-reversal invariance properties of nonforward
matrix elements, established in Section \ref{TimeReversalHermiticity}, the
DDs are even functions \cite{ManPilWei97} of the variable $\alpha$,
which is conjugate to
the skewness $\eta$
\begin{equation}
\label{TimeReversalDDs}
f^\phi ( \beta , \alpha , \Delta^2) = f^\phi ( \beta , - \alpha , \Delta^2)
\, .
\end{equation}

Equation (\ref{ScalarDD}) allows one to establish a relation between the GPD
$F^\phi (x, \eta, \Delta^2)$ and the DD $f^\phi (\beta, \alpha, \Delta^2)$. The
former is given as an integral of the latter, namely,
\begin{equation}
\label{nontodouble}
F^\phi (x, \eta, \Delta^2)
= \int_{- 1}^1 d \beta \int_{- 1 + |\beta|}^{1 - |\beta|} d \alpha \,
\delta ( \beta + \eta \alpha - x ) f^\phi ( \beta, \alpha, \Delta^2 )
\, .
\end{equation}
The $\alpha$-integration in the DD can be done by means of the delta-function
with the result
\begin{equation}
\label{RedFor-wro}
F^\phi (x, \eta)
=
\int_{- 1}^{1} \frac{d \beta}{\eta} \, {\mit\Xi} (\beta| x, \eta)
f^\phi \left( \beta, \frac{x - \beta}{\eta} \right)
\, ,
\end{equation}
where the step function defining the integration limits reads
\begin{eqnarray}
{\mit\Xi} (\beta| x, \eta)
= \theta (x > \eta)
\theta \left(
\ft{x + \eta}{1 + \eta} > \beta > \ft{x - \eta}{1 - \eta}
\right)
\!\!\!&+&\!\!\! \theta (- \eta > x)
\theta \left(
\ft{x + \eta}{1 - \eta} > \beta > \ft{x - \eta}{1 + \eta}
\right) \nonumber\\
&+&\!\!\! \theta (\eta > x > - \eta)
\theta \left(
\ft{x + \eta}{1 + \eta} > \beta > \ft{x - \eta}{1 + \eta}
\right)
\, .
\label{Limitsinbeta}
\end{eqnarray}

In this derivation, DDs are the starting point, while GPDs are obtained from them
by an integration. However, even if one starts directly with GPDs, the latter possess
{\sl the polynomiality  property}  which can be naturally incorporated only within
the formalism of double distributions. Namely,  the $x^j$ moment of $F^\phi (x, \eta,
\Delta^2)$ {\sl must be} a  $j$th order  polynomial of $\eta$. This restriction
on the interplay between $x$ and $\eta$ dependence of $F^\phi (x,\eta, \Delta^2)$
follows \cite{Ji:1998pc} from  the simple fact that the Lorentz indices $\mu_1
\ldots \mu_j$ of the  nonforward matrix element of a local twist-two operator
${\cal R}^{2, \phi\phi}_{ \mu_1 \ldots \mu_j}$ can be carried either by $p_\mu$
or by $\Delta_\mu$, as we clarified in Section \ref{Polynomiality}. As a result,
\begin{eqnarray}
n^{\mu_1} \dots n^{\mu_j}
\langle p_2 | {\cal R}^{2, \phi\phi}_{\mu_1 \dots \mu_j} | p_1 \rangle
\!\!\!&=&\!\!\!
n^{\mu_1} \dots n^{\mu_j}
\left\{
p_{\mu_1} \dots p_{\mu_j} F^\phi_{j,0}
+
\Delta_{\mu_1} \dots p_{\mu_j} F^\phi_{j,1}
+
\dots
+
\Delta_{\mu_1} \dots \Delta_{\mu_j} F^\phi_{j,j}
\right\}
\nonumber\\
&=&\!\!\!
\sum_{k = 0}^j (p^+)^{j-k}(\Delta^+)^k F^\phi_{j, k} =
(p^+)^j
\sum_{k = 0}^j \eta^k F^\phi_{j, k}
\, .
\label{poly}
\end{eqnarray}
The derivation of GPDs from  DDs automatically satisfies the polynomiality
condition (\ref{poly}), since
\begin{eqnarray}
\int_{- 1}^1 d x\, x^j F^\phi (x,\eta)
=
\int_{\mit\Omega} d \beta \, d \alpha\, (\beta + \eta \alpha)^j
f^\phi (\beta, \alpha)
=
\sum_{k = 0}^j
\eta^k
\left( { j \atop k } \right)
\int_{\mit\Omega} d \beta \, d \alpha\, \beta^{j - k} \alpha^k
f^\phi (\beta, \alpha)
\, .
\end{eqnarray}
Thus, the coefficients $F^\phi_{j,k}$ are given by
\begin{eqnarray}
F^\phi_{j,k}
=
\left( { j \atop k } \right)
\int_{\mit\Omega} d \beta \, d \alpha\, \beta^{j - k} \alpha^k
f^\phi (\beta, \alpha)
\equiv
\left( { j \atop k } \right) f^\phi_{j,k}
\, ,
\label{eq:Hjk}
\end{eqnarray}
i.e., they are proportional to the combined $\beta^{j - k} \alpha^k$ moments of the
DD $f^\phi (\beta, \alpha)$. The symmetry and support properties of DDs dictate a
nontrivial interplay between $j$ and $k$ dependence of $F^\phi_{j,k}$'s which should
be respected in the process of modeling GPDs.  After this observation, the use of
DDs is a natural step in building consistent parametrizations of GPDs.

\subsubsection{Gauge transformation of double distributions}
\label{spin0DD}

So far we have dealt with the  matrix elements of a scalar operator of a spinless
field (\ref{ScalarDD}). However, in the case of a vector twist-two operator
${\mathcal R}^2_\mu$ built either from spin-zero ($\phi^\dagger \partial^\mu \phi$)
or spin-one-half ($\bar \psi \gamma^{\mu} \psi$) fields, there is an open vector
index $\mu$. As a result, there appear extra Lorentz structures in the decomposition
of the matrix element which one should keep to avoid  violation of the polynomiality
condition for GPDs. This complication was studied in Ref.\ \cite{PolWei99}, where it
was proposed to add an extra term, the so-called D-term \cite{PolWei99} concentrated
in the central region $|x| \le \eta$ only. Such contributions may arise from the
$t$-channel exchange of mesonic-like states.

To illustrate the problem, let us consider the parametrization of the vector
quark operator ${\mathcal R}^{2, qq}_{\mu}(- z , z)$ in terms of double
distributions. Its matrix element has two components
\begin{equation}
\label{Twist2DecomPD}
\langle p_2 | {\cal R}^{2, qq}_{\mu} (- z^- , z^-) | p_1 \rangle
=
f^q (z^-) \, p_\mu \, + g^q (z^-) \, \Delta_\mu
\, .
\end{equation}
Multiplying ${\cal R}^{2, qq}_\mu$ with $n^\mu$ one immediately finds
\begin{eqnarray}
\langle p_2 | {\cal O}^{qq} (- z^- , z^-) | p_1 \rangle
\!\!\!&=&\!\!\!
\int_{\mit\Omega} d \beta \, d \alpha\,
{\rm e}^{- i z^- ( \beta p^+ +\alpha \Delta^+  )}
\left\{ p^+ f^q (\beta, \alpha) + \Delta^+ \, g^q (\beta, \alpha) \right\}
\, ,
\label{eq:DDfg}
\end{eqnarray}
where we introduced two double distributions corresponding to two structures
in Eq.\ (\ref{Twist2DecomPD}). The function $f^q (\beta, \alpha)$ is even in
$\alpha$, just as DD for the scalar operator, while $g^q (\beta, \alpha)$ is
odd in $\alpha$ because $\Delta$ changes sign under the reversal of time. The
corresponding  GPD can be written as
\begin{equation}
\label{DefDDfg}
H^q (x, \eta) =
\int_{\mit\Omega} d \beta \, d \alpha\,
\delta(\beta + \eta \alpha - x)
\left\{ f^q (\beta, \alpha) + \eta \, g^q (\beta, \alpha) \right\}
\, .
\end{equation}
Note,  that the $\alpha$-symmetry properties of $f^q (\beta, \alpha)$ and
$g^q (\beta, \alpha)$ guarantee that $H^q (x, \eta)$ is even in $\eta$. The
$x^j$ moments of $H^q (x, \eta)$ are now given by
\begin{eqnarray}
\int_{- 1}^1 d x \, x^j H^q (x, \eta)
& \equiv &
\sum_{k = 0}^{j+1}\eta^k H^q_{j+1,k} \\
&=&
f^q_{j,0} +  \sum_{k = 1}^{j} \eta^k \frac{j!}{k! (j-k+1)!}
\left[ (j - k + 1) f^q_{j,k} + k g^q_{j,k - 1} \right]
+
\eta^{j+1} g^q_{j,j}
\, , \nonumber
\label{eq:HFGjk}
\end{eqnarray}
where $f^q_{j,k}$ and  $g^q_{j,k}$, similarly to (\ref{eq:Hjk}), are defined as
$\beta^{j-k} \alpha^k$ moments of the relevant DD. The $j + 1$ subscript in
element $H^q_{j + 1,k}$ indicates the total number of vector indices of the
corresponding local operator and also the highest possible power $\eta^{j+1}$
of $\eta$ in the expansion of the $x^j$ moment of $H^q (x, \eta)$. The
$\eta^{j+1}$ term  can be obtained only from the DD $g^q (\beta, \alpha)$.
Alternatively, the zeroth power of $\eta$ can be obtained only from the DD
$f^q (\beta, \alpha)$. The $\eta^k$ powers with $1 \leq k \leq j$ come both
from the $f^q (\beta, \alpha)$ and $g^q (\beta, \alpha)$ DDs in the combination
$(j - k + 1) f^q_{j,k} + k g^q_{j,k-1}$. As a result, the coefficients $H^q_{j+1,k}$
do not change under the transformation
\begin{eqnarray}
f^q_{j,k} \to f^q_{j,k} + k \lambda^q_{j,k}
\, , \qquad
g^q_{j,k - 1} \to g^q_{j,k - 1} + (j - k + 1) \lambda^q_{j,k}
\, .
\end{eqnarray}
This means that separation of $H^q_{j+1,k}$ coefficients into $f^q$ and $g^q$
parts is not unambiguous \cite{PolWei99,Ter01}. However, in the double distribution
representation (\ref{eq:DDfg}), $f^q$ and $g^q$ contributions look like quite
distinct terms since they enter with different factors: $p^+$ and $\Delta^+$,
respectively. One can get rid of the latter integrating (\ref{eq:DDfg}) by parts.
Assuming that DDs vanish at the boundaries of the support region $\mit\Omega$, we
obtain
\begin{eqnarray}
z^- \langle p_2 | {\cal O}^{qq} (- z^- , z^-) | p_1 \rangle
=
- i
\int_{\mit\Omega} d \beta \, d \alpha\
{\rm e}^{- i z^- (\beta p^+  + \alpha\Delta^+  )}
\left\{
\frac{\partial}{\partial \beta} f^q (\beta, \alpha)
+
\frac{\partial}{\partial \alpha} g^q (\beta, \alpha)
\right\}
\, .
\end{eqnarray}
Thus, the functions $f^q$ and $g^q$ are defined modulo a ``gauge transformation''
\cite{Ter01}
\begin{equation}
f^q (\beta, \alpha)
\to
f^q (\beta, \alpha) + \frac{\partial}{\partial \alpha} \lambda^q ( \beta , \alpha )
\, , \qquad
g^q (\beta, \alpha)
\to
g^q (\beta, \alpha) - \frac{\partial}{\partial \beta} \lambda^q ( \beta , \alpha )
\, ,
\end{equation}
with $\lambda^q ( \beta, \alpha )$ being an odd function of $\alpha$, vanishing at
the boundaries of the DD support region\footnote{This restriction can be omitted,
see Ref.\ \cite{Tib04} for the analysis of the situation when DD does not vanish on
the boundaries of the support region.}. Since the $f^q_{j,0}$ term in (\ref{eq:HFGjk})
is unambiguously coming from an $f^q$-type DD, while the $g^q_{j,j}$ term only comes
from a $g^q$-type one, in general, the $\lambda^q$-transformation cannot completely
eliminate either one of the two DDs. But there are two  possibilities of ``almost
complete'' elimination.

\noindent $\bullet$ First, one can associate all the coefficients $H^q_{j+1,k}$
except the $k=j+1$ one with an $f^q$-type DD treating them as $\left( { j \atop
k } \right) f^q_{j,k}$, and consider only the last coefficient $H^q_{j+1,j+1} =
g^g_{j,j}$ as coming from a $g^q$-type contribution. This decomposition corresponds
to \cite{PolWei99}
\begin{eqnarray}
\label{RedFor-PW}
H^q (x, \eta) =
\int_{\mit\Omega} d \beta \, d \alpha\,
\delta(\beta + \eta \alpha - x) \,
f^q_D (\beta, \alpha)
+
{\rm sgn}(\eta) \, D^q (x/\eta)
\, ,
\end{eqnarray}
where $f^q_D (\beta, \alpha)$ satisfies
\begin{eqnarray}
\left. \frac {\partial f^q_D (\beta, \alpha)}{\partial \beta}
=
\frac {\partial f^q (\beta, \alpha)}{\partial \beta}
+
\left. \frac {\partial }{\partial \alpha}
\right[ g^q (\beta, \alpha) - \delta (\beta) D^q (\alpha) \right]
\, ,
\end{eqnarray}
and the $D$-term \cite{PolWei99} is given by
\begin{eqnarray}
\int_{- 1}^1 d \alpha \, \alpha^j \, D^q (\alpha)
=
g^q_{j,j}
\qquad {\rm or} \qquad
D^q (\alpha) = \int_{- 1+|\alpha|}^{1-|\alpha|} d\beta \, g^q (\beta,\alpha)
\, .
\end{eqnarray}
Since $D^q (\alpha)$ has the  $|\alpha| \le 1$ support, the $D$-term  in GPD is
entirely concentrated in the central region $|x| \leq |\eta|$. Obviously, $D^q
(\alpha)$ is an antisymmetric function $D^q (\alpha) = - D^q (\alpha) $ since
$g^q (\beta,\alpha)$ is odd in $\alpha$.

\noindent $\bullet$ Alternatively, we can decide that only the lowest
coefficient $H^q_{j,0}$ is related to an $f^q$-type contribution (given by
$f^q_{j,0}$) and treat all other $H^q_{j,k}$ coefficients as $\left(
{j \atop k-1 } \right) g^q_{j,k-1}$, thus  associating them with a $g^q$-type
DD. In this case \cite{Tib04}
\begin{eqnarray}
\label{RedFor-DY}
H^q (x, \eta)
=
f^q (x)
+
\eta \int_{\mit\Omega} d \beta \, d \alpha\, \delta(\beta + \eta \alpha - x)
g^q_F (\beta, \alpha)
\, ,
\end{eqnarray}
where now $g^q_F (\beta, \alpha)$ is a solution of
\begin{eqnarray}
\left. \frac{\partial g_F^q (\beta, \alpha)}{\partial \alpha}
=
\frac {\partial g^q_F (\beta, \alpha)}{\partial \alpha}
-
\left. \frac{\partial}{\partial \beta}
\right[ f^q (\beta, \alpha) - \delta (\alpha) f^q (\beta) \right]
\, ,
\end{eqnarray}
and $f^q (\beta)$ is the forward parton distribution
\begin{eqnarray}
\label{ForwardPDFfromDDs}
f^q (\beta)
=
\int_{-1+|\beta|}^{1-|\beta|} d\alpha \, f^q (\beta,\alpha)
\, .
\end{eqnarray}

\noindent $\bullet$ Another possibility \cite{BelKirMulSch00c} is the so-called
one-component DD. To understand the idea behind it, notice that,  in case of
scalar quarks, the local operators $\phi^\dagger \partial^\mu \partial^{\mu_1}
\ldots \partial^{\mu_j} \phi$ that appear in the expansion of the vector bilocal
operator $\phi^\dagger (0) \partial^{\mu} \phi (z^-)$ are the same as in the
expansion of the scalar bilocal operator $\phi^\dagger (0) \phi (z^-)$. The only
difference is that the index $\mu$ is treated as external in the vector case.
Thus, the matrix element of the local operator with $j$ derivatives in the
expansion of the scalar operator coincides with the  matrix element of the
vector-type local operator which has $j - 1$ ``expansion'' derivatives. Evidently,
both operators can be parametrized by the same DD $f^\phi (\beta, \alpha)$ of
the scalar operator. One should just realize that in the vector case one should
take the $x^{j-1}$ moment of $F^\phi (x, \eta)$ (or $x^j$ moment of  $F^\phi
(x, \eta)/x$) to get matrix element of the local operator with total $j$
derivatives. This gives
\begin{equation}
\label{AnalogueHscalar}
F^{\phi} (x, \eta)
=
x \int_{\mit\Omega} d \beta \, d \alpha\, \delta (\beta + \eta \alpha - x)
f^\phi (\beta, \alpha)
\, .
\end{equation}
Due to the delta function, we have $x = \beta + \eta \alpha$.

Obviously, one can use this definition also for a vector bilocal operator built
from spin-one-half fields. The parametrization of the scalar matrix elements yields
\begin{eqnarray}
\label{Def-cor-DD}
\langle p_2 | {\cal R}^{2, qq}_\mu ( - z^-, z^- ) | p_1 \rangle
=
\int_{\mit\Omega} d \beta \, d \alpha \,
\left( \beta p_\mu + \alpha \Delta_\mu \right)
h^q (\beta, \alpha)
\
{\rm e}^{- i \beta z^- p^+ - i \alpha z^- \Delta^+}
\, ,
\end{eqnarray}
where the DD $h^q (\beta, \alpha)$ can be defined through the coefficients
$H^q_{j,k}$ of the $\eta^k$ expansion (\ref{eq:HFGjk}) of the vector operator
\begin{equation}
\label{eq:one-DD}
H^q_{j, k}
\equiv
\left( {j \atop k} \right)
\int_{\mit\Omega} d \beta \, d \alpha\, \beta^{j-k} \alpha^{k} h^q (\beta, \alpha)
\, .
\end{equation}
The parametrization (\ref{eq:one-DD}) corresponds to the choice in (\ref{DefDDfg})
$$
f^q_h(\beta, \alpha) = \beta h^q (\beta, \alpha)
\, , \qquad
g^q_h (\beta, \alpha) = \alpha h^q (\beta, \alpha)
\ .
$$
The $\lambda^q$-transformation producing this result is a solution to the equation
\begin{equation}
\left (
\beta \frac{\partial}{\partial \beta} + \alpha \frac{\partial}{\partial \alpha}
\right )
\lambda^q (\beta, \alpha)
=
- \alpha f^q (\beta, \alpha) + \beta g^q (\beta, \alpha)
\, ,
\end{equation}
which is
\begin{equation}
\lambda^q (\beta, \alpha)
=
\int_1^{\infty} dt \,
\bigg[
\alpha  f^q (t \beta, t \alpha) - \beta g^q (t\beta, t\alpha)
\bigg]
\, .
\end{equation}
The $D$-term  and the forward parton density are related to $h^q(\beta, \alpha)$ by
\begin{eqnarray}
D^q (\alpha)
=
\alpha \, \int_{-1+|\alpha|}^{1-|\alpha|} d \beta \,
h^q (\beta,\alpha)
\, , \qquad
f^q (\beta)
=
\beta
\int_{-1+|\beta|}^{1-|\beta|} d\alpha \, h^q (\beta,\alpha)
\, .
\end{eqnarray}

\noindent $\bullet$ Finally, in Ref.\ \cite{Pobylitsa:2002vi} a DD was
introduced, which embeds a prefactor $(1 - x)^{2s}$ from a GPD of spin-$s$
parton in front of the DD integral. The relation of this DD to those we
discussed above is not straightforward.

\subsubsection{Radon tomography of double distributions with GPDs}
\label{Sec-DD}

\begin{figure}[t]
\begin{center}
\hspace{0cm}
\mbox{
\begin{picture}(0,110)(90,0)
\put(0,0){\insertfig{6}{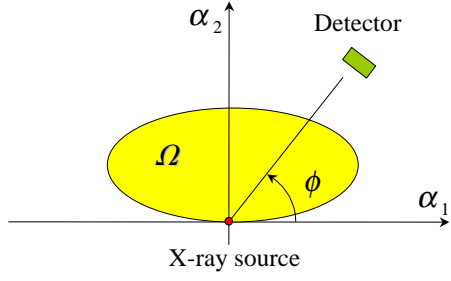}}
\end{picture}
}
\end{center}
\caption{\label{Tomography} The Radon transformation and the analogue of the
setup of the X-ray computed tomography. The X-ray source is at the center of
coordinates and the detector measures the intensity $I$ of the beam after its
passes through the object confined within the area ${\mit\Omega}$.}
\end{figure}

The relation (\ref{nontodouble}) is known as the Radon transformation, since
it is a one-dimensional slice of a two-dimensional function \cite{Leo97},
\begin{equation}
{\cal R} (I , \bit{n})
=
\int_{\mit\Omega} d^2 \bit{\alpha} \,
\delta (I - \bit{n} \cdot \bit{\alpha})
\, \tau (\bit{\alpha})
\, ,
\end{equation}
along the direction specified by a two-dimensional unit vector $\bit{n}$ (see Fig.\
\ref{Tomography})
$$
\bit{n} = ( \cos \phi , \sin \phi )
\, .
$$
For a function $\tau (\bit{\alpha})$ with the same support ${\mit\Omega}$ as for
the double distribution and the angle $\phi$ fixed by the skewness $\eta$,
$$
\phi = \arctan \eta
\, .
$$
we arrive at the above relation (\ref{nontodouble}) between the GPD
$$
F (x, \eta) = \cos \phi \ {\cal R} ( x \cos \phi, \bit{n})
\, ,
$$
and the double distribution
$$
\tau \left( \bit{\alpha} = ( \beta, \alpha ) \right) = f (\beta, \alpha)
\, .
$$

An explicit formula for the inverse Radon transform of the DD $h^q (\beta, \alpha)$ in terms
of $H^q (x,\eta)$ can be obtained by incorporating the single-DD representation for the
twist-two operator ${\cal R}^{2, qq}_\mu$ (\ref{Def-cor-DD}) \cite{BelKirMulSch00c} and is
achieved in a few easy steps. The first step consists in equating the right-hand sides
of Eqs.\ (\ref{ZeroSpinOpenIndexTw2}) and (\ref{Def-cor-DD}) and projecting the Lorentz
index $\mu$ onto the transverse plane such that it does not interfere with longitudinal
integrations performed at subsequent stages. Next, one Fourier transforms both sides of the
resulting equality with respect to the skewness $\eta = \Delta^+/p^+$. Integrating out the
arising delta-function accompanying the DD, one gets a factor of $|z^- p^+|^{-1}$. Since
the last step will consist in the Fourier transformation with respect to $z^-$, it is
instructive to use the following Fourier representation for $|z^- p^+|$:
\begin{equation}
|a|
=
-
\frac{1}{\pi} {\rm PV} \int_{- \infty}^\infty \frac{d\tau}{\tau^2} \,
{\rm e}^{i \tau a}
\, ,
\end{equation}
where  the principal value prescription is understood as \cite{GelShi64}
\begin{eqnarray}
{\rm PV} \int_{- \infty}^\infty \frac{d\tau}{\tau^2} \, {\cal F} (\tau)
\!\!\!&\equiv&\!\!\!
\frac{1}{2}
\int_{- \infty}^\infty d \tau \, {\cal F} (\tau)
\left\{
\frac{1}{(\tau + i 0)^2}
+
\frac{1}{(\tau - i 0)^2}
\right\}
\\
&=&\!\!\!
\int_{- \infty}^\infty \frac{d \tau}{\tau^2}\,
\{
{\cal F} ( \tau ) + {\cal F} (- \tau) - 2 {\cal F} (0)
\}
\, . \nonumber
\end{eqnarray}
Carrying out the advertised Fourier transform with respect to the light-cone
separation $z^-$, we find
\begin{eqnarray*}
\alpha h^q (\beta, \alpha)
=
-
\frac{1}{2\pi^2}
\int_{- \infty}^\infty d x
\int_{- \infty}^\infty d \eta \;
{\rm PV} \,\frac{1}{(\beta + \eta \alpha - x)^2}
\int_{- \infty}^\infty d x^\prime \, W_2 (x, x^\prime)
\frac{d}{d\eta}
H^q (x^\prime,\eta)
\, ,
\end{eqnarray*}
where $W_2$ kernel was introduced in Eq.\ (\ref{W2kernel}). Finally, we perform the
$x$ integration and integrate by parts with respect to $\eta$. Dropping the surface
term, we recover the result \cite{BelKirMulSch00c}
\begin{eqnarray}
\label{Res-Inv}
h^q (\beta, \alpha)
=
\frac{1}{2\pi^2} \int_{- \infty}^\infty d\eta \int_{- \infty}^\infty \frac{dx}{x}\,
{\rm PV}\,
\left\{
\frac{1}{(\beta + \eta \alpha)^2}
-
\frac{1}{(\beta + \eta \alpha - x)^2}
\right\}
H^q (x, \eta)
\, .
\end{eqnarray}
This form of the inverse Radon transformation is similar to that given in
Ref. \cite{GelShiVil66}. Note that the inversion kernel in Eq.(\ref{Res-Inv})
compensates the singularity of the $1/x$ factor, so that the integral over $x$
exists even if  the function $H^q (x, \eta)/x$ has a non-integrable singularity
at $x=0$. In case when $H^q (x, \eta)/x$ is integrable, the first term in the
integrand of Eq. (\ref{Res-Inv}) can be safely neglected, being proportional to
$\delta (\beta) \delta (\alpha)$ (see Eq.\ (\ref{AnalogueHscalar})).

According to Eq. (\ref{RedFor-PW}), the Radon transform of the $f_D^q (\beta,
\alpha)$ (\ref{RedFor-PW}) is given by  $(H^q - D^q)$. The inverse transformation
gives
\begin{eqnarray}
f^q_D (\beta, \alpha)
=
- \frac{1}{2\pi^2}
\int_{- \infty}^\infty d\eta \int_{- \infty}^\infty d x \
{\rm PV} \frac{1}{(\beta + \eta \alpha - x)^2}
\left\{
H^q (x, \eta) - {\rm sgn} (\eta) D^q \left( \frac{x}{\eta} \right)
\right\}
\, .
\end{eqnarray}
Here the term $1/(\beta + \eta \alpha)^2$ was omitted as explained above. Since
both $H^q (x, \eta)$  and  $D^q (x/\eta)$ can be expressed through $h^q (\beta,
\alpha)$, one can also write $f^q_D (\beta, \alpha)$ in terms of $h^q (\beta,
\alpha)$. The explicit expression was given in Ref.\ \cite{BelKirMulSch00c}:
\begin{eqnarray}
\label{ConToPW}
f^q_D (\beta, \alpha) = \beta h^q (\beta, \alpha)
-
\frac{d}{d\alpha} \alpha \int d \beta^\prime \, W_2 (\beta, \beta^\prime)
\beta^\prime h^q (\beta^\prime, \alpha)
\, .
\end{eqnarray}
In this sense, both $f^q_D$ and $D^q$ functions in (\ref{RedFor-PW}) can be
treated as different projections of the same function $h^q$. Also, it is easy
to recognize
\begin{eqnarray}
\lambda_h^q (\beta,\alpha) = - \alpha \int d\beta^\prime \,
W_2 (\beta, \beta^\prime)
\beta^\prime h^q (\beta^\prime, \alpha)
\end{eqnarray}
as the $\lambda^q$-transformation from  the one-component DD representation
(\ref{Def-cor-DD}) to (\ref{RedFor-PW}).

\subsubsection{Nucleon double distributions}
\label{SectionSingleCompDDnucleon}

In the present section, for reference purposes, we give a single component
representation of DDs for the spin-one-half target. We introduce the
parametrization using the decomposition in terms of Dirac bilinears $h^\mu$
and $b$ for the unpolarized sector,
\begin{eqnarray}
\label{SingleCompDDnucleonEven}
&&\langle p_2 | \mathcal{O}^{qq} (- z^- , z^-) | p_1 \rangle
\\
&&\qquad\qquad
=
\int_{\mit\Omega} d \beta \, d \alpha \,
{\rm e}^{- i \beta z^- p^+ - i \alpha z^- \Delta^+}
\left\{
h^+ h^q_A (\beta, \alpha, \Delta^2)
+
\frac{b}{2 M_N} (\beta p^+ + \alpha \Delta^+) h^q_B (\beta, \alpha, \Delta^2)
\right\}
\, , \nonumber
\end{eqnarray}
and $\widetilde{h}^\mu$ and $\widetilde{b}$ for the polarized one,
\begin{eqnarray}
\label{SingleCompDDnucleonOdd}
&&\langle p_2 | \widetilde{\mathcal{O}}^{qq} (- z^- , z^-) | p_1 \rangle
\\
&&\qquad\qquad
=
\int_{\mit\Omega} d \beta \, d \alpha \,
{\rm e}^{- i \beta z^- p^+ - i \alpha z^- \Delta^+}
\left\{
\widetilde{h}^+ \widetilde{h}^q_A (\beta, \alpha, \Delta^2)
+
\frac{\widetilde{b}}{2 M_N} (\beta p^+ + \alpha \Delta^+)
\widetilde{h}^q_B (\beta, \alpha, \Delta^2)
\right\}
\, . \nonumber
\end{eqnarray}
For gluons, the parametrization in terms of the same bilinears can be obtained from
these by simple substitutions: $\mathcal{O}^{qq} \to \mathcal{O}^{gg}$ on the
left-hand side and $h^q_A \to \ft14 p^+ h^g_A$, $h^q_B \to \ft14 (\beta p^+ + \alpha
\Delta^+)h^g_B$ on the right-hand side of Eq.\ (\ref{SingleCompDDnucleonEven}). Similar
replacements are done for the gluon equivalent of Eq.\ (\ref{SingleCompDDnucleonOdd})
with symbols dressed by tildes.

The moments of DDs $h^q_A$ and $h^q_B$ from Eq.\ (\ref{SingleCompDDnucleonEven}) are
related to the form factors and GPDs introduced in Eqs.\ (\ref{Decomposition}) and
(\ref{ABgpds}). Namely,
\begin{eqnarray}
\label{Def-GPD-mom-B}
B^q_{j, k}
\!\!\!&=&\!\!\!
\frac{1}{k!}
\left. \frac{\partial^k}{\partial\eta^k} \right|_{\eta = 0}
\int_{-1}^{1} dx\; x^{j - 1} B^q (x,\eta)
= \left({j \atop k}\right) \int_{{\mit\Omega}}
d\beta \, d\alpha \,  \beta^{j - k} \alpha^{k} h^q_B (\beta, \alpha)
\\
\label{Def-GPD-mom-A}
A_{j,  k}^q
\!\!\!&=&\!\!\! \frac{1}{k!}
\left. \frac{\partial^k}{\partial\eta^k} \right|_{\eta = 0}
\int_{-1}^{1} dx\; x^{j - 1} A^q (x,\eta)
= \left({j-1 \atop k}\right) \int_{{\mit\Omega}} d\beta \, d\alpha \,
\beta^{j-k-1} \alpha^{k} h_A^q (\beta,\alpha)
\, ,
\end{eqnarray}
where $j \geq 1$ and the index $k$ varies in the interval $0 \le k \le j$ for
$B^q$ and $0 \le k \le j - 1$ for $A^q$ structure. This is an immediate consequence
of the polynomiality condition: the $j$-th moment of the functions $A^a (x, \eta)$
and $B^q (x, \eta)$ are polynomials of order $(j - 1)$ and $j$ in $\eta$,
respectively. Matching the parametrization (\ref{SingleCompDDnucleonEven}) into
the one in terms of GPDs $A^q (x, \eta)$ and $B^q (x, \eta)$ introduced in Section
\ref{PolynomSpin12GPDs} one immediately finds
\begin{eqnarray}
\label{RelAtohA}
A^q (x, \eta)
\!\!\!&=&\!\!\!
\int_{\mit\Omega} d \beta \, d \alpha\,
\delta(\beta + \eta \alpha - x)
h^q_A (\beta, \alpha)
\, , \\
B^q (x, \eta)
\!\!\!&=&\!\!\!
x \int_{\mit\Omega} d \beta \, d \alpha\,
\delta(\beta + \eta \alpha - x)
h^q_B (\beta, \alpha)
\, ,
\end{eqnarray}
and analogous relations for the odd parity DDs (\ref{SingleCompDDnucleonOdd}). The
relation of GPDs $A^q$ and $B^q$ in terms of the standard $H^q$ and $E^q$ can be
found in Eq.\ (\ref{SPDsSubst}). Finally, for the axial channel in the analogous
parametrization with Dirac bilinears $\widetilde{h}^\mu$ and $\widetilde b$ as in
Eq.\ (\ref{Decomposition}), the GPDs $\tilde A^q$ and $\tilde B^q$ are identical
to the conventional $\widetilde H^q$ and $\widetilde E^q$.

\subsection{Analytic properties of DDs and GPDs}
\label{AnalyticPropertiesDDsAndGPDs}

The formalism of DDs also allows one to easily establish some important properties
of skewed distributions. Notice  that due to the cusp at the upper corner of the DD
support rhombus, the length of the integration line depends nonanalytically on $x$
for $x = \pm \eta $. Hence, unless the double distribution identically vanishes in a
finite region around the upper corner of the DD support rhombus, the $x$-dependence
of the relevant GPDs {\sl must be nonanalytic} at the border points $x = \pm \eta $.
Still, the length of the integration line is a continuous function of $x$. As a
result, if the double distribution $f(\beta, \alpha, \Delta^2)$ is not too singular
for small $\beta$, the skewed distribution $H(x, \eta, \Delta^2)$ is continuous at
the nonanalyticity points $x = \pm \eta $. Because of the  $1/(\xi \pm x)$ factors
present in hard amplitudes entering exclusive processes, as discussed later in
Section \ref{Phenomenology}, this property is crucial for perturbative factorization
of exclusive amplitudes involving GPDs, when the external kinematics sets $\eta = \xi$.
These features are manifested in model ans{\" a}tze that will be discussed in Section
\ref{LongDynamics}.

In principle, we cannot exclude the possibility that the functions $f (\beta,
\alpha, \Delta^2)$ have  singular terms at $\beta = 0$ proportional to $\delta
(\beta)$ or its derivative(s). Such terms have no projection  onto the usual
parton densities. We will denote them by $f_{\rm ex} (\beta, \alpha; \Delta^2)$.
They may be interpreted as coming from the $t$-channel meson-exchange type
contributions. In this case, the partons just share the plus-component of the
momentum transfer $\Delta$: information about the initial hadron momentum is
lost if the exchanged particle of mass $m_M$ can be described by a pole
propagator $1/(\Delta^2 - m_M^2)$. Hence, the meson-exchange contributions to
a double distribution may look like
\begin{equation}
f_{\rm ex}^+ (\beta, \alpha, \Delta^2)
\sim
\delta (\beta) \,
\frac{\varphi_M^+ (\alpha)}{m_M^2 - \Delta^2}
\, , \qquad
f_{\rm ex}^- (\beta, \alpha, \Delta^2)
\sim
\delta ' (\beta) \,
\frac{\varphi^-_M (\alpha)}{m_M^2 - \Delta^2}
\, ,
\end{equation}
etc., where $\varphi_M^{\pm} (\alpha)$ are the functions\footnote{These distribution
amplitudes are expressed via the one introduced in Section \ref{MesonDA} by means of
the change of the argument $\varphi_M (\alpha) = \phi_M \left( \ft{1 + \alpha}{2}
\right)$.} related to the distribution amplitudes of the relevant mesons. The two
examples given above above correspond to $\beta$-even ($C$-odd) and $\beta$-odd
($C$-even) parts of the double distribution $f (\beta, \alpha, \Delta^2)$. As an
obvious consequence of the time-reversal symmetry of DDs (\ref{TimeReversalDDs}),
the functions $\varphi_M^{\pm}(\alpha)$ for singular contributions $f_M^{\pm} (\beta,
\alpha, \Delta^2)$ are even functions $\varphi_M^{\pm} (\alpha) = \varphi_M^{\pm}
(-\alpha)$ of $\alpha$ both for $\beta$-even and $\beta$-odd components. If the
amplitude $\varphi_M (\alpha)$ vanishes at the end-points $\alpha = \pm 1$, the
$H_{\rm ex} (x,\eta , \Delta^2)$ part of GPD vanishes at $x = \pm \eta $. The total
function $H (x,\eta ; \Delta^2)$ is then continuous at the nonanalyticity points
$x = \pm \, \eta $. In the $C$-even case, one can get a continuous GPD in this case
only if the derivative $\varphi^\prime (\alpha)$ vanishes at the end points.

\subsection{Impact-parameter parton distributions}

In our presentation, GPDs were introduced as generalizations of the concept of
the non-relativistic Wigner function to the quantum phase-space quasi-probability
distribution in the nucleon. Thus, the partonic Wigner distributions are not
endowed with the density of probability interpretation: they intrinsically have
patches of ``negative probability'' due to the fact that they represent the
interference of two amplitudes, one for the outgoing parton from the initial
nucleon and another for the incoming parton in the final-state nucleon.

In this section, we demonstrate that in a particular case of zero value of the
skewness parameter, the generalized parton distributions acquire a well-defined
probability interpretation in the infinite-momentum frame similar to the
conventional collinear parton distribution functions. Moreover, the resulting
picture does not suffer from the presence of large relativistic corrections
inevitably affecting the rest-frame description. The price for this advantage
is the lack of one spatial dimension: instead one has a picture in two (transverse)
spatial dimensions and one (longitudinal) momentum direction. Note, that the
orthogonality of momentum and spatial degrees of freedom allows one to circumvent
constraints from the quantum mechanical uncertainty principle.

The probability interpretation of the skewness-independent function $H (x, \eta = 0,
\bit{\Delta}^2_\perp)$ was available, though not widely known, for a long time
\cite{Sop77}, where an equivalent function was treated as a two-dimensional
Fourier transform of the impact-parameter dependent parton distribution
$f (x, \bit{r}_\perp)$. Recently, this concept was revived in a series of papers
in Refs. \cite{Burkardt:2000za,Burkardt:2002hr,Burkardt:2002ks}.

\subsubsection{Electromagnetic form factors in the Bjorken frame}
\label{FFinBjorkenFrame}

Let us use the Bjorken frame adopted for discussion of form factors. In this case,
the average momentum of initial and outgoing nucleon can be chosen to have only
the time and $z$ components, while the momentum transfer can be made purely
transverse. To get an explicit form, one can take the general result from
Appendix \ref{LCvectors} where one sets $\eta = 0$ and use the reparametrization
invariance to choose $\varrho = \sqrt{2} {\cal P}$, so that the incoming and
outgoing momenta take the form
\begin{equation}
\label{MomentaBjorken}
p^\mu_{1, 2}
=
\left(
{\cal P} + \frac{M_N^2 + \ft14 \bit{\Delta}^2_\perp}{4 {\cal P}} ,
\pm \ft12 \bit{\Delta}_\perp ,
{\cal P} - \frac{M_N^2 + \ft14 \bit{\Delta}^2_\perp}{4 {\cal P}}
\right)
\, , \qquad
\Delta^\mu = (0 , \bit{\Delta}_\perp , 0)
\, .
\end{equation}

Since our aim is to give a transverse-space interpretation of form factors, we
have to localize the nucleon in the coordinate space with an appropriately
chosen two-dimensional wave packet, which possesses a definite $z$-component of
the three-momentum,
\begin{equation}
\label{ImpactParameterState}
| p^z , \bit{R}_\perp \rangle
=
\int \frac{d^2 \bit{p}_\perp}{(2 \pi)^2}
\frac{
{\rm e}^{i \bit{\scriptstyle p}_\perp \cdot \bit{\scriptstyle R}_\perp}
}{
\sqrt{2 E_{\bit{\scriptstyle p}}}
}
{\mit\Psi} (\bit{p}_\perp)
| p^z , \bit{p}_\perp \rangle
\, .
\end{equation}
Here the two-dimensional profile is related to the three-dimensional one,
discussed in the introduction, by
$$
{\mit\Psi} (\bit{p}) = 2 \pi \delta
(p^z - p'^z)
\frac{
{\mit\Psi} (\bit{p}_\perp)
}{
\sqrt{2 E_{\bit{\scriptstyle p}}}
}
\, .
$$
The mixed state (\ref{ImpactParameterState}) is normalized as
\begin{equation}
\langle p'^z , \bit{R}_\perp | p^z , \bit{R}_\perp \rangle
= 2 \pi \delta ( p'^z - p^z )
\, , \qquad
\int \frac{d^2 \bit{p}_\perp}{(2 \pi)^2}
| {\mit\Psi} (\bit{p}_\perp) |^2 = 1
\, .
\end{equation}
Let us put this state in the center of two-dimensional plane
and compute the Fourier transform of its  transverse charge density
\begin{eqnarray}
&&
\int d^2 \bit{r}_\perp \,
{\rm e}^{- i \bit{\scriptstyle \Delta}_\perp \cdot \bit{\scriptstyle r}_\perp}
\rho_\perp (\bit{r}_\perp)
\equiv
\int d^2 \bit{r}_\perp \,
{\rm e}^{- i \bit{\scriptstyle \Delta}_\perp \cdot \bit{\scriptstyle r}_\perp}
\langle p^z_2 , \bit{R}_\perp = 0 |
j^0 (\bit{r}_\perp)
| p^z_1 , \bit{R}_\perp = 0 \rangle
\\
&&\qquad
=
\int \frac{d^2 \bit{p}_\perp}{(2 \pi)^2}
\frac{
{\mit\Psi}^\ast \left( \bit{p}_\perp - \ft12 \bit{\Delta}_\perp \right)
}{
\sqrt{2 E_{\bit{\scriptstyle p_2}}}
}
\frac{
{\mit\Psi} \left( \bit{p}_\perp + \ft12 \bit{\Delta}_\perp \right)
}{
\sqrt{2 E_{\bit{\scriptstyle p_1}}}
}
\langle p^z_2 , \bit{p}_\perp - \ft12 \bit{\Delta}_\perp |
j^0 (0)
| p^z_1 , \bit{p}_\perp + \ft12 \bit{\Delta}_\perp \rangle
\, . \nonumber
\end{eqnarray}

To get rid of extraneous effects of the auxiliary wave packet and also to be
insensitive to all kinds of corrections, the following conditions should be
fulfilled (some of them are analogous to those discussed in the first section
dedicated to three-dimensional interpretation of form factors):
\begin{itemize}
\item We require that the  spatial
size of the wave packet  is not detectable, so it has to be smaller than the
resolution scale set by the inverse momentum transfer from the initial
to the final state. By
uncertainty principle,  this implies that typical momenta in the wave
packet are larger than the momentum of the probe
$$
| \bit{p}_\perp | \gg | \bit{\Delta}_\perp |
\, .
$$
Then one can neglect $\bit{\Delta}_\perp$ in the wave packet profiles
(but not in matrix elements, which are sensitive to small changes
in the momentum), ${\mit\Psi} \left( \bit{p}_\perp \pm \ft12
\bit{\Delta}_\perp \right) \approx {\mit\Psi} ( \bit{p}_\perp )$.
\item In order to  suppress relativistic corrections, and also to
give the proton a well-defined longitudinal momentum,
we need to avoid the energy exchange from the initial to
the final state. To this end,   we have to require
$$
p^z_i \gg | \bit{p}_\perp |
\, ,
$$
which results in a condition on the resolution scale \cite{Burkardt:2000za}
\begin{equation}
p^z_i \gg | \bit{\Delta}_\perp |
\, .
\end{equation}
\end{itemize}
In the infinite-momentum frame ${\cal P} \to \infty$, this will not be
a restriction at all, and one can achieve simulteneously a very precise
localization, since the ``effective'' Compton wave length $\sim 1/\sqrt{M_N^2
+ {\cal P}^2}$ contains the longitudinal momentum ${\cal P}$
\cite{Burkardt:2000za}), and  get rid of the relativistic corrections
(since $p^0_{1,2} \simeq p^z_{1,2} \simeq {\cal P}$).

For the helicity-non-flip transitions,
the matrix element of the electromagnetic current can be easily evaluated in
the infinite-momentum frame
$$
\langle p^z_2 , \bit{p}_\perp - \ft12 \bit{\Delta}_\perp |
j^0 (0)
| p^z_1 , \bit{p}_\perp + \ft12 \bit{\Delta}_\perp \rangle
=
2 {\cal P} F_1 ( - \bit{\Delta}_\perp^2 )
\, .
$$
This demonstrates that, in contrast to the three-dimensional picture, the
two-dimensional charge density in the infinite-momentum frame gives a
well-defined, free from relativistic effects interpretation of the Fourier
transform of the electromagenitc form factors as a two-dimensional charge
density,
\begin{equation}
\int d^2 \bit{r}_\perp \,
{\rm e}^{- i \bit{\scriptstyle \Delta}_\perp \cdot \bit{\scriptstyle r}_\perp}
\rho_\perp (\bit{r}_\perp)
=
F_1 (- \bit{\Delta}^2_\perp)
\, .
\end{equation}
Note also the  combinations of form factors entering the helicity-flip (non-flip)
transitions in the Breit frame, discussed in Section \ref{BreitFrameFFs}, differ
from those in the infinite-momentum frame as discussed at length below. This is
a mere consequence of the distinction between the light-cone and conventional
helicities.

The two-dimensional character of the dynamics probed by  form factors in the
Bjorken frame is reflected in the difference between the root-mean-square radii
of the three-dimensional charge densities, determined by form factors in the
Breit frame, see Eq.\ (\ref{3DprotonRadius}). Namely, in the infinite momentum
frame
\begin{equation}
\langle \bit{r}^2_\perp \rangle_D \equiv \int d^2 \bit{r}_\perp \,
\bit{r}^2_\perp \rho_\perp (\bit{r}_\perp)
=
- 4 \frac{\partial F_1 (- \bit{\Delta}^2_\perp)}{\partial \bit{\Delta}^2_\perp}
\, .
\end{equation}
Therefore, when one defines the charge radius in the Breit frame through
the Dirac form factor, rather than the electric form factor, one gets for the ratio
of the three-dimensional versus the two-dimensional radii
\begin{equation}
\frac{\langle \bit{r}^2 \rangle_D}{\langle \bit{r}^2_\perp \rangle_D}
=
\frac{3}{2}
\, .
\end{equation}

\subsubsection{Skewless GPDs as impact-parameter distributions}
\label{SkewlessGPDsImpactParton}

\begin{figure}[t]
\begin{center}
\mbox{
\begin{picture}(0,90)(90,0)
\put(0,-30){\insertfig{6}{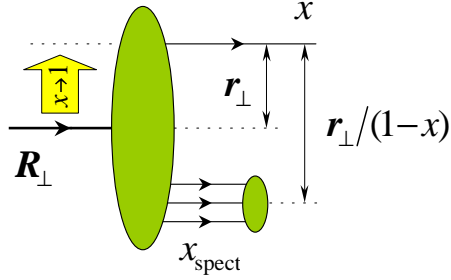}}
\end{picture}
}
\end{center}
\vspace{1cm}
\caption{\label{SkewlessImpactPS} Distances in the impact parameter space: the
transverse center-of-momentum $\bit{R}_\perp$ is a reference point from where the
distance to all partons is measured dressed by their momentum fractions. The
distance from the active parton to the spectator system is $\bit{r}_\perp^s
= \bit{r}_\perp/(1 - x)$.
}
\end{figure}

For the purpose of discussing high-energy observables, like conventional and
generalized parton distributions, the light-cone coordinates are more appropriate
than the energy-momentum variables. The former mix the transverse and longitudinal
momenta unless one takes the infinite momentum limit, as we have already demonstrated
in the previous subsection. For high-energy probes, these are the light-cone components
of momenta which are measured. Therefore, they are more suitable for designing a
viable formalism. Rewriting the relativistically-invariant momentum element in the
light-cone coordinates $d^3 \bit{p}/(2 E_{\bit{\scriptstyle p}}) = d p^+ d^2
\bit{p}_\perp/(2 p^+)$ with a definite $p^+$ instead of a definite $p^z$, we
change accordingly the wave-packet profile function and introduce the state
\begin{equation}
\label{LCmixedState}
| p^+ , \bit{R}_\perp \rangle
=
\int \frac{d^2 \bit{p}_\perp}{(2 \pi)^2}
{\rm e}^{i \bit{\scriptstyle p}_\perp \cdot \bit{\scriptstyle R}_\perp}
{\mit\Psi} (\bit{p}_\perp)
| p^+ , \bit{p}_\perp \rangle
\, .
\end{equation}
Here we do not display the helicity of the state, it will be restored later when it
will be needed for the presentation. This mixed state is normalized relativistically,
\begin{equation}
\langle p^{\prime +}, \bit{R}_\perp | p^+ , \bit{R}_\perp \rangle
= 2 p^+ (2 \pi) \delta \left( p^+ - p^{\prime +} \right)
\, , \qquad
\int\frac{d^2 \bit{p}_\perp}{(2 \pi)^2}
| {\mit\Psi} (\bit{p}_\perp) |^2 = 1
\, .
\end{equation}
It is localized in the sense that its transverse center of momentum is at
$\bit{R}_\perp$. For a state with total momentum $p^+$,  the transverse center of
momentum is defined as \cite{Burkardt:2000za,Burkardt:2002hr,Burkardt:2002ks}.
\begin{equation}
\bit{R}_\perp
\equiv
\frac{1}{p^+}
\int d z^- \, d^2 \bit{z}_\perp \, \bit{z}_\perp
{\mit\Theta}^{++} (z^+ = 0, z^-, \bit{z}_\perp)
\, ,
\end{equation}
where ${\mit\Theta}^{\mu \nu}$ is the energy momentum tensor. The eigenvalues of
the operator $\bit{R}_\perp$ on the state (\ref{LCmixedState}) have an intuitive
partonic representation in the light-cone gauge. It is obtained by expressing
${\mit\Theta}^{++}$ in terms of the light-cone creation and annihilation operators
and noticing that, after integrating over $z^-$, only the terms that are diagonal
in Fock space contribute, yielding
\begin{equation}
\label{Rperp}
\bit{R}_\perp =
\frac{
\sum_i k^+_i \bit{r}_{\perp,i}
}{
p^+
}
=
\sum_i x_i \bit{r}_{\perp,i}
\, .
\end{equation}
The summation in Eq.\ (\ref{Rperp}) is over all partons in the hadron and
$x_i \equiv k^+_i/p^+$ is the longitudinal momentum fraction carried by the
$i$-th parton in the infinite momentum frame. The quantity $\bit{R}_\perp$
is invariant under the transverse boost $\mathbb{M}^{+\perp}$ (see Appendix
\ref{ConformalSymAppendix}). This result should not be surprising, since the
momentum fractions $x_i $ play here the role very similar to that
of the mass fractions $m_i/\sum_i m_i$ in the expression for the center of
mass in non-relativistic quantum mechanics. Working with the localized
state $| p^+, \bit{R}_\perp \rangle$ is quite analogous to using the
center-of-mass frame in nonrelativistic physics, the main difference being
the change of the weight factors from the mass ratios to the longitudinal
momentum fractions in the expression for the transverse center of momentum.
This is related to the properties of the Galilean subgroup of transverse boosts.
Note finally that $\bit{r}_\perp$ measures the distance from the active quark
to the hadron's center of momentum (see Fig.\ \ref{SkewlessImpactPS}), while
the distance from the active quark to the center of momentum of spectators is
given by $\bit{r}^s_\perp = \bit{r}_\perp/(1 - x)$.

The objects we are going to discuss in this section are the matrix elements
between the state (\ref{LCmixedState}) of the partially Fourier transformed
bilocal quark operators
\begin{equation}
\label{FourierTransBilocal}
{\cal O}^{qq} (x, \bit{r}_\perp)
\equiv
\int \frac{d z^-}{2 \pi} {\rm e}^{i x p^+ z^-}
{\cal O}^{qq} ( - z^- , \bit{r}_\perp ; z^- , \bit{r}_\perp )
\, ,
\end{equation}
(with the change ${\cal O}^{qq} \to \widetilde {\cal O}^{qq}$ for the parity-odd
operators). The first condition applied in the preceding section to the profile
function remains unchanged. However, since now we have a fixed $p^+$ rather
than fixed $p^z = (p^+ - p^-)/\sqrt{2}$, the latter can become negative, $p^z <
0$, if the transverse momentum is too large so that $p^+ \ll p^- = (\bit{p}_\perp^2
+ M_N^2)/(2p^+)$. Then the desired interpretation of a proton moving with a
well-defined longitudinal momentum is spoiled. This  imposes a restriction on
the momenta in the wave packet $|\bit{p}_\perp| \ll p^+$ and as a result on the
attainable transverse localization, $|\bit{\Delta}_\perp| \ll p^+$. This is
analogous to the second condition of  the preceding section, but now applied to
the plus momentum. Thus, taking the proton with $p^+ \to \infty$, brings us to
the situation discussed there. In the Bjorken frame, where the momenta are given
by Eq.\ (\ref{MomentaBjorken}), $p^+ = {\cal P}/\sqrt{2}$.

To investigate the spin structure of generalized parton distributions, it is
useful to represent them in a form similar to that of helicity amplitudes. Since
we are dealing here with matrix elements involving two independent proton momenta,
some comments are in order regarding the choice of helicity states for the protons.
Note, that in the definitions of the distributions one singles out a direction
that defines the light-cone coordinates (in a physical process, where these
distributions appear, this direction is provided by the hard probe, such as the
virtual photon in deeply virtual Compton scattering). It is also useful to utilize
this light-cone direction for defining the spin states for the protons with momenta
$p_1$ and $p_2$. This leads to the concept of light-cone helicity states
\cite{KogSop70}. Using the light-cone spinors from Appendix \ref{Spinors} and
results for Dirac bilinears (\ref{DiracBilinLCframe}) derived there, one can easily
find for helicity-non-flip and -flip transitions \cite{Burkardt:2002hr,Diehl:2002he}
\begin{eqnarray}
\label{ImpactRepresentZeroSkewness}
H^q (x, \eta = 0, - \bit{\Delta}_\perp^2)
\!\!\!&=&\!\!\!
\int d^2 \bit{r}_\perp \,
{\rm e}^{- i \bit{\scriptstyle \Delta}_\perp \cdot \bit{\scriptstyle r}_\perp}
\langle \ft12 p^+ , \bit{0}_\perp , \uparrow |
{\cal O}^{qq} ( x, \bit{r}_\perp )
| \ft12 p^+ , \bit{0}_\perp , \uparrow \rangle
\, , \nonumber\\
- \frac{\Delta_\perp}{2 M_N}
E^q (x, \eta = 0, - \bit{\Delta}_\perp^2)
\!\!\!&=&\!\!\!
\int d^2 \bit{r}_\perp \,
{\rm e}^{- i \bit{\scriptstyle \Delta}_\perp \cdot \bit{\scriptstyle r}_\perp}
\langle \ft12 p^+ , \bit{0}_\perp , \downarrow |
{\cal O}^{qq} ( x, \bit{r}_\perp )
| \ft12 p^+ , \bit{0}_\perp , \uparrow \rangle
\, , \\
\widetilde{H}^q (x, \eta = 0, - \bit{\Delta}_\perp^2)
\!\!\!&=&\!\!\!
\int d^2 \bit{r}_\perp \,
{\rm e}^{- i \bit{\scriptstyle \Delta}_\perp \cdot \bit{\scriptstyle r}_\perp}
\langle \ft12 p^+ , \bit{0}_\perp , \uparrow |
\widetilde {\cal O}^{qq} ( x, \bit{r}_\perp )
| \ft12 p^+ , \bit{0}_\perp , \uparrow \rangle
\, , \nonumber
\end{eqnarray}
where, obviously, $p_1^+ = p_2^+ = p^+/2$ and the holomorphic transverse momentum
is defined as usual, $\Delta_\perp = \Delta^x + i \Delta^y$. Note, that we have
not given the expression for $\widetilde E$, since it decouples in the skewless
limit.

\begin{figure}[t]
\begin{center}
\mbox{
\begin{picture}(0,100)(110,0)
\put(0,-30){\insertfig{7}{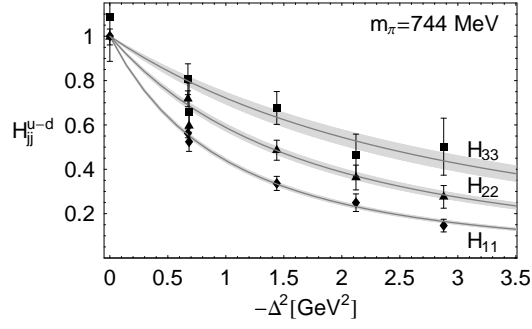}}
\end{picture}
}
\end{center}
\vspace{1cm}
\caption{\label{LatticeMomentsGPDs} Lattice calculation \cite{LHPC03} of the
lowest three form factors $H_{jj}$ with $j = 1, 2, 3$ in Mellin moments of
GPDs, see Eq.\ (\ref{GPDmomentsFFHandE}).
}
\vspace{-0.5cm}
\end{figure}

The fact that the distribution of partons in the impact parameter space is measured
with respect to the transverse center of momentum (\ref{Rperp}) has very profound
consequences on the functional form of the impact-parameter dependent distributions
and, as a result, that of GPDs. When the active parton carries almost the entire
momentum of the parent hadron, its momentum fraction $x$ approaches one, while for
all spectator partons $x_i \to 0$,  since the sum of all momentum fractions cannot
exceed unity, $\sum_{\rm spect.} x_i = 1 - x$. Therefore,  when $x = 1$, the center
of momentum
$$
\bit{R}_\perp = x \bit{r} + \sum_{\rm spect.} x_i \bit{r}_i
\,
$$
sits right on top of the active quark, i.e.,  $\bit{R}_\perp = \bit{r}$ (see
Fig.\ \ref{SkewlessImpactPS}). Thus, the transverse profile narrows down and
ultimately becomes concentrated at the position of the active quark $\sim
\delta^{(2)} (\bit{r})$ \cite{Burkardt:2002hr,Burkardt:2004bv}. In other words,
the transverse width of the impact-parameter dependent parton distribution
\begin{equation}
\langle \bit{r}_\perp^2 (x) \rangle
\equiv
\left.
\left( \int d^2 \bit{r}_\perp \, \bit{r}_\perp^2 f^q (x, \bit{r}_\perp) \right)
\right/
\left( \int d^2 \bit{r}_\perp \, f^q (x, \bit{r}_\perp) \right)
\, ,
\end{equation}
should vanish as $x \rightarrow 1$. As we can see, the distance between the
active parton and spectators in this asymptotic limit grows as $\bit{r}_\perp^s
= \bit{r}_\perp/(1 - x)$. Of course, this growth is stopped by confinement
effects,  so that the fast-moving active quark will not be separated from
spectators  by more than $|\bit{r}_\perp^s| \sim 1/m_h$, where $m_h$ is some
typical hadronic scale.  Going away from the asymptotic limit $x \to 1$, the
transverse distribution broadens out. In momentum space, such a picture implies
that GPDs Fourier conjugated to the impact-parameter parton distributions
must become $\Delta^2$-independent when the momentum fraction $x$ approaches
unity, i.e., the slope of skewless GPDs should approach zero when $x \to 1$
\cite{Burkardt:2002hr,Burkardt:2004bv,Yua03}. Thus, the transverse size is
strongly correlated with the longitudinal momentum. This behavior was confirmed
by lattice calculations of Mellin moments  of GPDs. It was observed \cite{LHPC03}
that, the higher the moment, the slower is its  fall-off with increasing momentum
transfer, see Fig.\ \ref{LatticeMomentsGPDs}.

Thus, one gets a very intuitive interpretation of skewless GPDs as Fourier
transforms of impact-parameter dependent parton distributions, e.g.,
\begin{equation}
f^q_H (x, \bit{r}_\perp)
=
\int \frac{d^2 \bit{\Delta}_\perp}{(2 \pi)^2}
{\rm e}^{i \bit{\scriptstyle\Delta}_\perp \cdot \bit{\scriptstyle r}_\perp}
H^q (x, 0, - \bit{\Delta}^2)
=
\langle p^+ , \bit{0}_\perp , \uparrow |
{\cal O}^{qq} ( x, \bit{r}_\perp )
| p^+ , \bit{0}_\perp , \uparrow \rangle
\, ,
\label{FourierImpactRepresentation}
\end{equation}
which are simultaneous probabilities that a parton is at a certain distance
$\bit{r}_\perp$ from the hadron's center-of-mass and carries a fraction $x$ of the
parent hadron's momentum $p$ \cite{Sop77}. These functions are positive definite,
which can be seen by expressing them  in terms of the creation/annihilation operators
in the formalism of the light-cone quantization
\begin{equation}
f^q_H (x, \bit{r}_\perp) |_{x > 0}
=
\frac{1}{2 x p^+ \delta (0^+)}
\sum_{\lambda = \uparrow\downarrow}
\langle
\ft{1}{2} p^+ , \bit{0}_\perp , \uparrow
|
b^\dagger_\lambda ( \ft{x}{2} p^+, \bit{r}_\perp)
b_\lambda ( \ft{x}{2} p^+, \bit{r}_\perp)
|
\ft{1}{2} p^+ , \bit{0}_\perp , \uparrow
\rangle
\, .
\end{equation}
The partially Fourier transformed annihilation operator used here is given by
$$
b_\lambda (k^+, \bit{r}_\perp)
=
\int
\frac{d^2 \bit{k}_\perp}{(2 \pi)^3}
{\rm e}^{i \bit{\scriptstyle k}_\perp \cdot \bit{\scriptstyle r}_\perp}
b_\lambda (k^+, \bit{k}_\perp)
\, .
$$
The helicity-flip distributions can be brought to the diagonal form---to endow
them with a probabilistic interpretation---by going to the transverse-spin basis
$$
| \pm \!\perp \rangle = \frac{1}{\sqrt{2}}
(
| \uparrow \rangle \pm | \downarrow \rangle
)
\, ,
$$
where
$$
\langle \downarrow | {\cal O} | \uparrow \rangle
=
\langle + \!\perp | {\cal O} | + \!\perp \rangle
-
\langle - \!\perp | {\cal O} | - \!\perp \rangle
\, .
$$

\begin{figure}[t]
\begin{center}
\mbox{
\begin{picture}(0,148)(215,0)
\put(-15,-18){\insertfig{3.9}{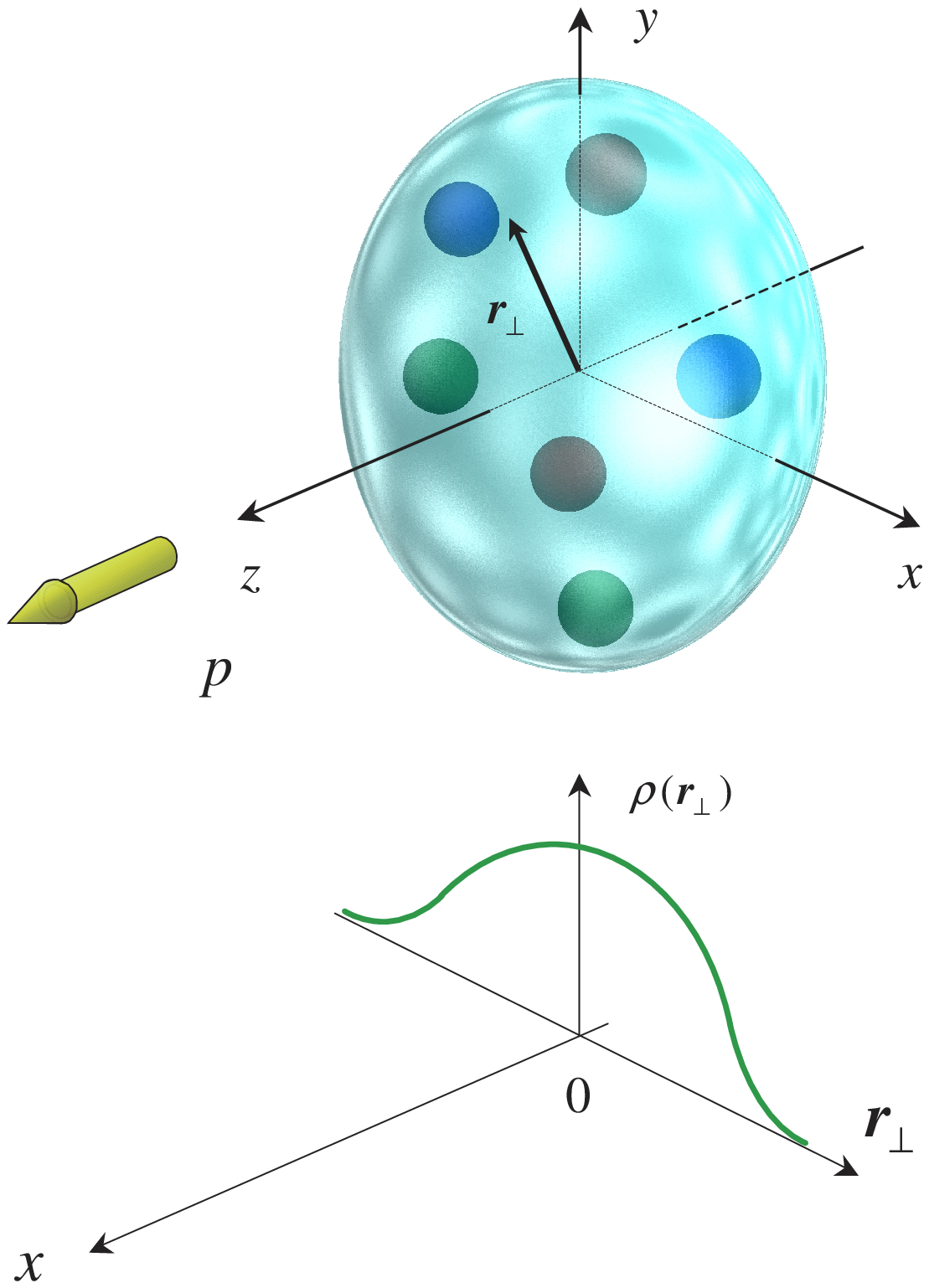}}
\put(155,-19){\insertfig{3.9}{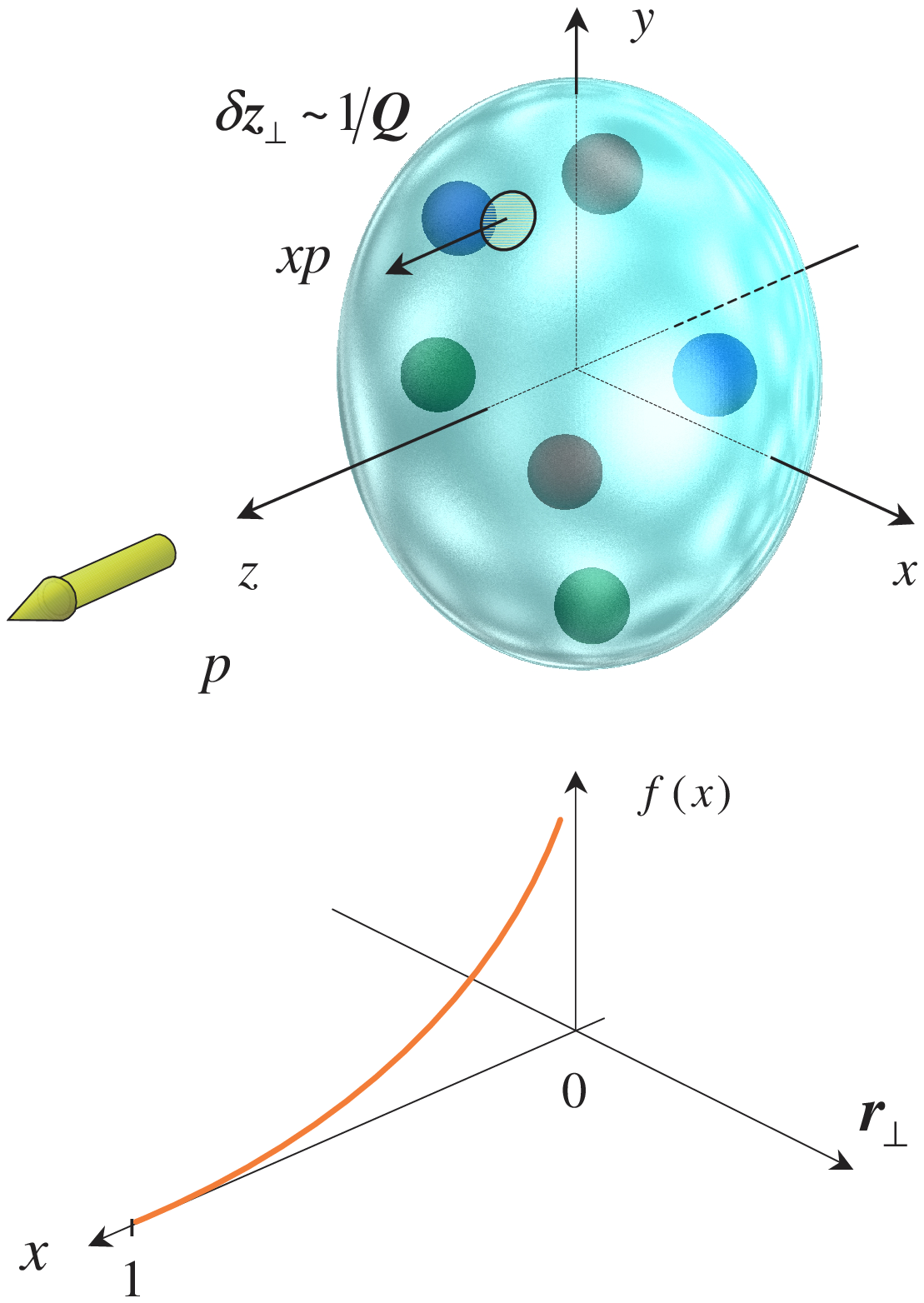}}
\put(325,-34){\insertfig{4}{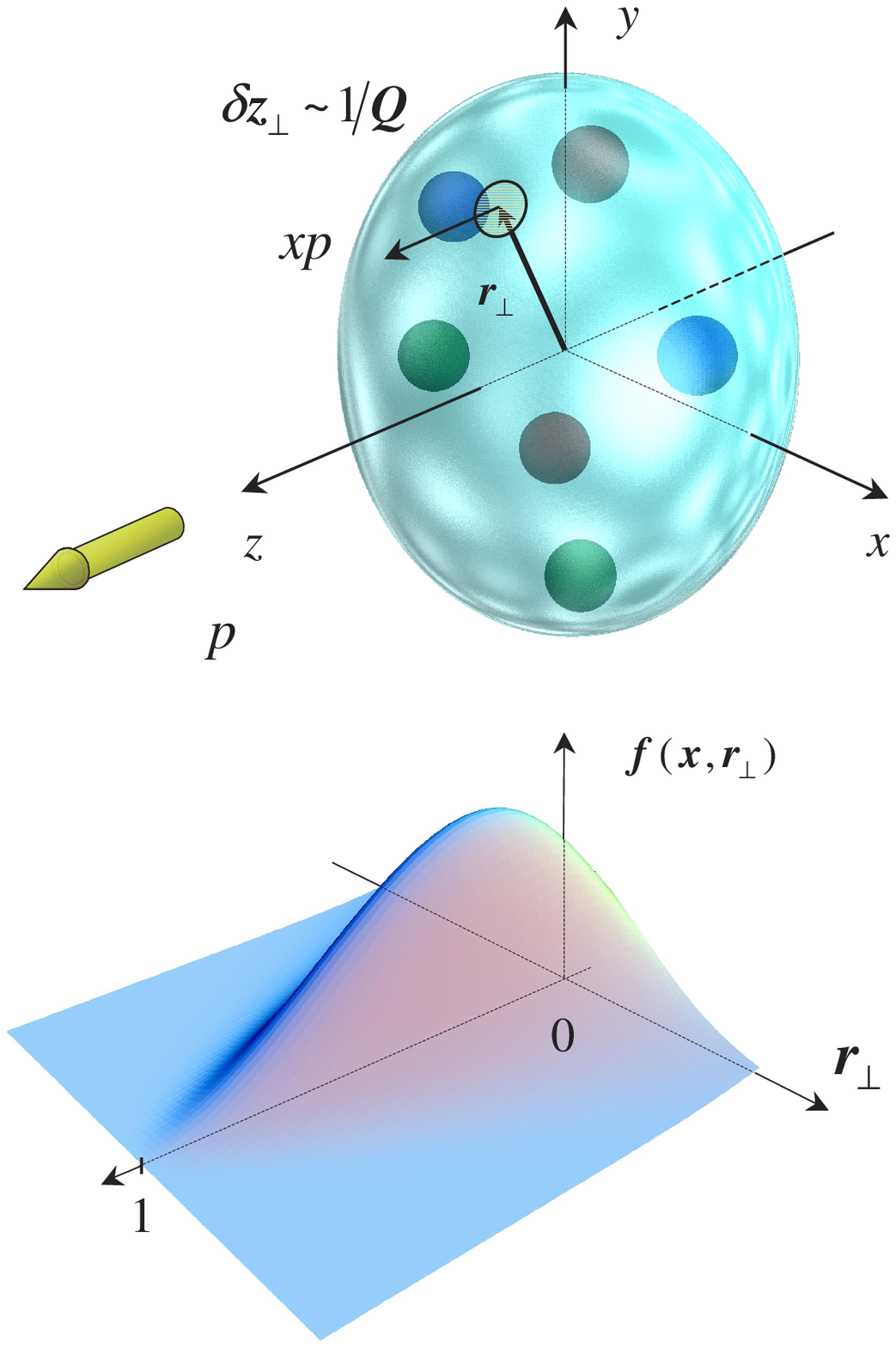}}
\end{picture}
}
\end{center}
\vspace{1cm}
\caption{\label{PartonContent} Probabilistic interpretation of form factors,
parton densities and generalized parton distributions at $\eta = 0$ in the
infinite momentum frame $p_z \to \infty$.}
\vspace{-0.5cm}
\end{figure}

Thus, GPDs regain a probabilistic interpretation once one sets skewness
$\eta = 0$. The function $f^q (x, \bit{r}_\perp)$ is a three-dimensional hybrid
distribution in one-dimensional momentum and two-dimensional coordinate
space. After  integrating over $\bit{r}_\perp$, one recovers the Feynman parton
distribution. On the other hand, after integration over $x$ one gets the
impact-parameter space distribution which is the Fourier transform of the elastic
form factor \cite{RalPir02,Diehl:2002he,BelMul02,Pol02,Fre02}. The pictorial
comparison of paron distributions, form factors and skewless GPDs and information
one accesses by studying these functions is demonstrated in Fig.\ \ref{PartonContent}.
Notice that the right-most graph demonstrates the phenomenon of shrinking of the
width of the impact-parameter distribution with increasing $x$ and the vanishing
of the height of the peak, just due to the fact that the probability to observe a
single quark carrying the whole momentum of the proton is negligible.

\subsubsection{GPDs in impact-parameter space}
\label{GPDsInImpactParameterSpace}

So far we have discussed GPDs for zero skewness and observed that they regain
the probabilistic interpretation analogous to the Feynman parton distributions.
Once one lifts the restriction of absence of the longitudinal momentum transfer
in the $t$-channel, GPDs cease to be densities, rather, they become interference
functions or quasiprobabilities according to their identification with Wigner
distributions. There is a number of new features and interesting subtleties which
appear in the impact parameter representation for $\eta \neq 0$, and now we are
going to discuss them \cite{Diehl:2002he}.

Since our goal is to evaluate the matrix element of the non-local operator
(\ref{FourierTransBilocal}) sandwiched between the mixed states (\ref{LCmixedState}),
we need to know the dependence on the transverse momenta with arbitrary
$\bit{p}_{1 \perp}$ and $\bit{p}_{2 \perp}$. As it was already indicated by the
evaluation of the Dirac bilinears in Appendix \ref{Spinors}, these momenta cease to
enter in the combination $\bit{\Delta}_\perp = \bit{p}_{1 \perp}- \bit{p}_{2 \perp}$,
unless one chooses the ``transverse'' Breit frame $\bit{p}_{1 \perp} = -
\bit{p}_{2 \perp} = \ft12 \bit{\Delta}_\perp$. Since $\bit{p}_{i \perp}$'s enter
as integration variables in (\ref{LCmixedState}), we cannot pick this reference
frame from the start,  and are allowed to impose any conditions only after the
integrations have been performed. The Lorentz-covariant combinations of Dirac
bilinears  analyzed in Appendix \ref{Spinors},  are found to depend on the
combination\footnote{Notice that for zero skewness, $p_1^+ = p_2^+$, this
combinations collapses to $\Delta_\perp/p_1^+$ and thus momentum-space matrix
elements do depend solely on $\Delta_\perp$ making the analysis of the previous
section legitimate.} $p_{1 \perp}/p_1^+ - p_{1 \perp}/p_1^+$ (see, e.g., Eq.\
(\ref{DiracBilinLCframe})). This is not an accident but a direct  consequence of
the transverse Lorentz-boost invariance of GPDs. These transformations leave the
plus-momentum invariant and change  the transverse momentum in proportion to a
parameter $\bit{v}_\perp$,
\begin{equation}
p^+ \to p^+
\, , \qquad
\bit{p}_\perp \to \bit{p}_\perp - p^+ \bit{v}_\perp
\, .
\end{equation}
As a consequence of this invariance, for generically defined unconstrained momenta
$\bit{p}_{1 \perp}$ and $\bit{p}_{2 \perp}$, a given GPD is a function
\begin{equation}
\label{GPDsWithArbirtP12tr}
F^q (x, \eta , \bit{D}_\perp)
=
\langle p_2 | {\cal O}^{qq} (x, \bit{0}_\perp) | p_2 \rangle
\end{equation}
of the momentum fraction $x$, skewness $\eta$ and the Lorentz invariant
combination of transverse momenta\footnote{Note that our vector $\bit{D}_\perp$
differs from the one introduced originally in Ref.\ \cite{Diehl:2002he} by a
global factor.}
\begin{equation}
\bit{D}_\perp
\equiv
\frac{2}{p^+}
\left(
\bit{p}_{1 \perp} p_2^+
-
\bit{p}_{2 \perp} p_1^+
\right)
=
(1 - \eta) \bit{p}_{1 \perp} - (1 + \eta) \bit{p}_{2 \perp}
\, .
\end{equation}
The four-dimensional momentum transfer squared is re-expressed in terms of
$\bit{D}_\perp$ via
\begin{equation}
\label{FourDeltaAsTwoDelta}
\Delta^2
=
\Delta^2_{\rm min} - \frac{\bit{D}_\perp^2}{1 - \eta^2}
\, ,
\end{equation}
where the minimum momentum transfer squared is defined in Eq.\ (\ref{MinMomDeltaSquared})
and formally coincides with the transverse Breit-frame expression (\ref{TranverseDeltaBreit}).
Obviously $\bit{D}_\perp \stackrel{\rm Breit}{=} \bit{\Delta}_\perp$.

\begin{figure}[t]
\begin{center}
\mbox{
\begin{picture}(0,75)(110,0)
\put(0,-30){\insertfig{8}{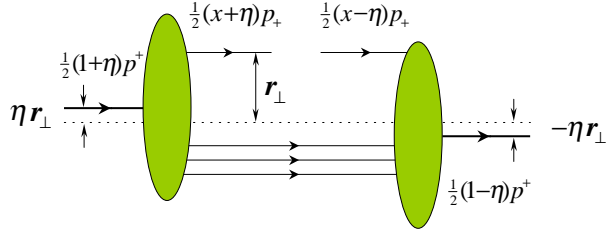}}
\end{picture}
}
\end{center}
\vspace{1cm}
\caption{\label{GPDimpactFullSkewness} Interpretation of GPDs in the impact
parameter space for non-zero skewness. The distributions of partons in
transverse plane is measured with respect to the two-hadron center of momentum.
Each of the hadrons is shifted away from it in the opposite directions linearly
proportional to the longitudinal deficit $\eta$.}
\end{figure}

Sandwiching the Fourier transformed light-cone operator (\ref{FourierTransBilocal})
between the mixed states (\ref{LCmixedState}) and changing the integration variables
from $\bit{p}_{1 \perp}$ and $\bit{p}_{2 \perp}$ to $\bit{D}_\perp$ and $\bit{p}_\perp$,
one immediately finds
\begin{eqnarray}
\label{ImpactParCaculation}
&&
\langle p^+_2 , \bit{r}_{2 \perp} |
{\cal O}^{qq} (x, \bit{0}_\perp)
| p^+_1 , \bit{r}_{1 \perp} \rangle
=
\int \frac{d^2 \bit{D}_\perp}{(2 \pi)^2}
{\rm e}^{- i \bit{\scriptstyle D}_\perp \cdot \bit{\scriptstyle r}_{1 \perp}/(1 - \eta)}
F^q (x, \eta, \bit{D}_\perp)
\\
&&\qquad
\times
\int \frac{d^2 \bit{p}_\perp}{(2 \pi)^2}
{\rm e}^{
i \bit{\scriptstyle p}_\perp \cdot
\left[
(1 - \eta) \bit{\scriptstyle r}_{2 \perp}
-
(1 + \eta) \bit{\scriptstyle r}_{1 \perp}
\right]
}
{\mit\Psi}^\ast \Big( (1 - \eta) \bit{p}_\perp \Big)
{\mit\Psi}^\ast \Big( (1 + \eta) \bit{p}_\perp + \bit{D}_\perp/(1 - \eta) \Big)
\, , \nonumber
\end{eqnarray}
where $p_1^+ = (1 + \eta)p^+/2$ and $p_2^+ = (1 - \eta)p^+/2$. Requiring that the
second exponential reduces to unity, one introduces
\begin{equation}
\label{HadronPositionsImpact}
\bit{r}_\perp
= - \frac{\bit{r}_{1 \perp}}{1 - \eta} = - \frac{\bit{r}_{2 \perp}}{1 + \eta}
\, .
\end{equation}
We impose the normalization condition on the wave packet
$$
{\cal N}_{\scriptscriptstyle\mit\Psi}
=
\int \frac{d^2 \bit{p}_\perp}{(2 \pi)^2}
{\mit\Psi}^\ast
\Big( (1 - \eta) \bit{p}_\perp - \ft12 \bit{D}_\perp/(1 + \eta) \Big)
{\mit\Psi}^\ast
\Big( (1 + \eta) \bit{p}_\perp + \ft12 \bit{D}_\perp/(1 - \eta) \Big)
= 1
\, ,
$$
implying that it is a slowly varying function of its argument. Notice that if
one chooses a ``nailed down'' proton state in the coordinate space, which
corresponds to ${\mit\Psi} = 1$, the  normalization factor becomes divergent
${\cal N}_{\scriptscriptstyle{\mit\Psi} = 1} = \delta^{(2)} (\bit{r}_\perp = 0)$.
Then, making use of
(\ref{HadronPositionsImpact}),
one immediately finds from Eq.\ (\ref{ImpactParCaculation})
\begin{equation}
\langle p^+_2 , -\eta \bit{r}_\perp |
{\cal O}^{qq} (x, \bit{r}_\perp)
| p^+_1 , \eta \bit{r}_\perp \rangle
=
\int \frac{d^2 \bit{D}_\perp}{(2 \pi)^2}
{\rm e}^{i \bit{\scriptstyle D}_\perp \cdot \bit{\scriptstyle r}_\perp}
F^q (x, \eta, \bit{D}_\perp)
\, ,
\end{equation}
where we have used $\langle \bit{r}_{2 \perp} - \bit{r}_\perp | {\cal O}
(\bit{0}_\perp) | \bit{r}_{1 \perp} - \bit{r}_\perp \rangle = \langle
\bit{r}_{2 \perp} | {\cal O} (\bit{r}_\perp)| \bit{r}_{1 \perp} \rangle$
resulting from translation invariance in the impact parameter space. Now,
one can safely impose the transverse Breit frame condition: the functional
dependence of GPDs on that condition  will not change. A very important
lesson one learns \cite{Diehl:2002he} is that  the presence of a non-zero
longitudinal momentum exchange in the $t$-channel affects the transverse
separation of incoming and outgoing hadrons (see Fig.\ \ref{GPDimpactFullSkewness}).
As a result, the overlap between these states decreases with increasing skewness.
It is interesting to notice that, for the partons, also, the transverse shift
depends only on the skewness but not on the momentum fraction $x$. Therefore,
information on the transverse localization of partons is not washed out in
quantities integrated over $x$ with an arbitrary weight depending on
longitudinal momenta. It is exactly the form in which  GPDs arise in a number
of high energy processes discussed later in this review.

Incorporating Eq.\ (\ref{DiracBilinLCframe}) to evaluate the light-cone helicity
matrix elements, one finds the following set of equations \cite{Diehl:2002he}
\begin{eqnarray}
\label{ImpactSkewnessDependent}
\sqrt{1 - \eta^2}
\bigg(
H (x, \eta , - \bit{\Delta}_\perp^2)
\,
-
\!\!\!&
\displaystyle\frac{\eta^2}{1 - \eta^2}
&\!\!\!
E
(x, \eta , - \bit{\Delta}_\perp^2)
\bigg)
\\
&=&\!\!\!\!\!\!\!\!\!\!
\int d^2 \bit{r}_\perp \,
{\rm e}^{- i \bit{\scriptstyle \Delta}_\perp \cdot \bit{\scriptstyle r}_\perp}
\langle p^+_2 , - \eta \bit{r}_\perp , \uparrow |
{\cal O}^{qq} (x, \bit{r}_\perp)
| p^+_1 , \eta \bit{r}_\perp , \uparrow \rangle
\, , \nonumber\\
- \frac{\Delta_\perp}{2 M_N \sqrt{1 - \eta^2}}
E (x, \eta , - \bit{\Delta}_\perp^2)
\!\!\!\!\!\!\!\!\!\!&=&\!\!\!\!\!\!\!\!\!\!
\int d^2 \bit{r}_\perp \,
{\rm e}^{- i \bit{\scriptstyle \Delta}_\perp \cdot \bit{\scriptstyle r}_\perp}
\langle p^+_2 , - \eta \bit{r}_\perp , \downarrow |
{\cal O}^{qq} (x, \bit{r}_\perp)
| p^+_1 , \eta \bit{r}_\perp , \uparrow \rangle
\, , \nonumber\\
\sqrt{1 - \eta^2}
\bigg(
\widetilde H (x, \eta , - \bit{\Delta}_\perp^2)
\,
-
\!\!\!&
\displaystyle\frac{\eta^2}{1 - \eta^2}
&\!\!\!
{\widetilde E} (x, \eta , - \bit{\Delta}_\perp^2)
\bigg)
\nonumber\\
&=&\!\!\!\!\!\!\!\!\!\!
\int d^2 \bit{r}_\perp \,
{\rm e}^{- i \bit{\scriptstyle \Delta}_\perp \cdot \bit{\scriptstyle r}_\perp}
\langle p^+_2 , - \eta \bit{r}_\perp , \uparrow |
\widetilde {\cal O}^{qq} (x, \bit{r}_\perp)
| p^+_1 , \eta \bit{r}_\perp , \uparrow \rangle
\, , \nonumber\\
- \frac{\eta \Delta_\perp}{2 M_N \sqrt{1 - \eta^2}}
\widetilde E (x, \eta , - \bit{\Delta}_\perp^2)
\!\!\!\!\!\!\!\!\!\!&=&\!\!\!\!\!\!\!\!\!\!
\int d^2 \bit{r}_\perp \,
{\rm e}^{- i \bit{\scriptstyle \Delta}_\perp \cdot \bit{\scriptstyle r}_\perp}
\langle p^+_2 , - \eta \bit{r}_\perp , \downarrow |
\widetilde {\cal O}^{qq} (x , \bit{r}_\perp)
| p^+_1 , \eta \bit{r}_\perp , \uparrow \rangle
\, . \nonumber
\end{eqnarray}
A more correct form for the third argument of GPDs in Eq.\
(\ref{GPDsWithArbirtP12tr}) is to write it as the four-dimensional momentum
transfer via Eq.\ (\ref{FourDeltaAsTwoDelta}). We hope that our shorter
notation will not lead to  confusion. Note, finally, that the zero-skewness
limit of Eqs.\ (\ref{ImpactRepresentZeroSkewness})  reproduces the results
of the previous section.

\subsection{Positivity constraints on GPDs}
\label{Positivity}

To carry out realistic modeling of generalized parton distributions, the latter
have to obey several first-principle constraints, like the reduction to the
usual parton densities and form factors, and have to fulfill the polynomiality
condition in the skewness parameter, as we established in previous sections.
Another nontrivial property of GPDs is reflected by so-called positivity constraints.
The terminology is introduced by analogy with inequalities satisfied by forward
parton densities. As was demonstrated in Eq.\ (\ref{PartonicContentPDFs}), the latter
are expressed as a diagonal matrix element of the number-of-particles operator,
$\langle p | b^\dagger b | p \rangle$, or as a norm $|{\cal A}|^2 = | b | p \rangle|^2$,
of the state ${\cal A} = b | p \rangle$ created with the action of the quark
annihilation operator $b$  on the hadronic state $ |\, p \rangle$. As a result,
the parton density is explicitly positive definite and can be treated as a probability.
On the other hand, generalized parton distributions correspond to non-diagonal
matrix elements, for instance, $\langle p_2 | b^\dagger b | p_1 \rangle$
in one of the inclusive regions (\ref{DISregion1GPDs}), and therefore they
describe the overlap $({\cal A}_2^* {\cal A}_1)$ of the state ${\cal A}_2 = b
| p_2 \rangle$ with ${\cal A}_1 =  b | p_1 \rangle$, rather than the norm. The
Cauchy inequality $|{\cal A}^*_2 {\cal A}_1| \leq |{\cal A}_2||{\cal A}_1|$ will be
the basic tool for getting positivity constraints on GPDs. When non-diagonal matrix
elements are parametrized by several GPDs, e.g., by $H$ and $E$ for spin-one-half
hadrons, the positivity constraints restrict possible values of particular combinations
of $H (x, \eta, \Delta^2)$ and $E (x, \eta, \Delta^2)$, with the bounds set in terms
of the forward parton densities $q (y)$ taken at particular values of its argument $y$
depending on both $x$ and $\eta$. Moreover, one can build the states ${\cal A}_{1,2}$ as
arbitrary superpositions of $b|p\rangle$ type states and this implies that, in principle,
there is an infinite set of positivity constraints. This presents a serious challenge for
building consistent models of GPDs.

\subsubsection{Inequalities relating GPDs and ordinary distributions}

\begin{figure}[t]
\begin{center}
\hspace{0cm}
\mbox{
\begin{picture}(0,140)(70,0)
\put(0,0){\insertfig{5}{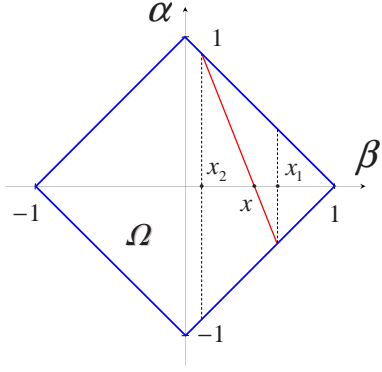}}
\end{picture}
}
\end{center}
\caption{\label{fig:posgpd} Integration line producing $H(x,\eta, \Delta^2)$.}
\end{figure}

As a first demonstration, let us address inequalities which arise between GPDs in
the inclusive domain and standard forward parton densities. In case when $x > \eta $,
the integration contour over DDs producing $H^q (x, \eta, - \Delta^2)$ (see Fig.\
\ref{fig:posgpd}) is inside the region between two vertical lines generating the
usual parton distributions $q (x_2)$ and $q (x_1)$ being functions of the momentum
fractions
$$
x_1 = \frac{x + \eta}{1 + \eta}
\, , \qquad
x_2
= \frac{x - \eta}{1- \eta}
\, ,
$$
cf.\ Eq. (\ref{Limitsinbeta}).
Then
\begin{eqnarray}
H^q (x,\eta, \Delta^2)|_{x \geq \eta}
=
\int_{x_2 }^{x_1}
\frac{d \beta}{\eta} f^q \left( \beta, \frac{\beta - x}{\eta} \right)
\, .
\end{eqnarray}
The combinations $x_1$, $x_2$ have a very simple interpretation: they measure the
momentum of the initial or final parton in units of the momentum of the relevant
hadron. Assuming a monotonic decrease of the double distribution $f^q (\beta,
\alpha)$ in the $\beta$-direction and a universal normalized profile in the
$\alpha$-direction, we conclude that $H^q (x, \eta, \Delta^2 = 0)$ is larger than
$q (x_1)$ but smaller than $q (x_2)$, $q (x_1) < H^q(x, \eta, \Delta^2 = 0) <
q (x_1)$. Moreover, if the slope of parton densities $q (x)$ gets steeper when
$x \to 0$, which is the case for any $x^{- a}$ behavior with $a > 0$, then we may
expect that $H^q (x, \eta, \Delta^2 = 0)$ is smaller than the average of $q (x_1)$
and $q (x_2)$:
$$
H^q (x, \eta, \Delta^2 = 0)
\leq
\frac{1}{2} \Big[ q (x_1) + q (x_2) \Big]
\, .
$$
An inequality of this kind was originally proposed in Refs.\ \cite{Martin:1997wy,Ji:1998pc}.
The basic idea used in the derivation is to treat GPDs as nondiagonal matrix elements
of the form $\langle {\mit\Psi}_{\rm out} | {\mit\Psi}_{\rm in} \rangle $ and then to
use the Schwartz inequality
$$
\langle {\mit\Psi}_{\rm out} | {\mit\Psi}_{\rm in} \rangle
\leq
\frac{1}{2}
\Big[
\langle {\mit\Psi}_{\rm in} | {\mit\Psi}_{\rm in} \rangle
+
\langle {\mit\Psi}_{\rm out} | {\mit\Psi}_{\rm out} \rangle
\Big]
\, ,
$$
interpreting the diagonal matrix elements in terms of the forward parton densities.
One step further into strengthening the constraint \cite{Pire:1998nw} was to use
the form involving the geometric average
$$
\langle {\mit\Psi}_{\rm out} | {\mit\Psi}_{\rm in} \rangle
\leq
\sqrt{|{\mit\Psi}_{\rm in}| |{\mit\Psi}_{\rm out}|}
\, .
$$
To this end, the GPDs are written as
\begin{equation}
H(x, \eta)
=
\sum_S \langle {\mit\Psi}_{\rm out} (x, \eta; S) | {\mit\Psi}_{\rm in} (x, \eta; S) \rangle
\, ,
\end{equation}
where $| {\mit\Psi}_{\rm in} (x, \eta; S) \rangle$ describes the probability amplitude
that the incoming nucleon with momentum $\ft12 (1 + \eta) p$ converts into a parton
with momentum $\ft12 (x + \eta) p$ and spectators $S$, while $\langle {\mit\Psi}_{\rm out}
(x,\eta; S)|$ describes the recombination of the spectator system $S$ and a parton with
momentum $\ft12 (x - \eta) p$ into the outgoing hadron with momentum $\ft12 (1 -  \eta) p$.
Then the Cauchy-Schwartz inequality states
\begin{equation}
\left|
\sum_S \langle {\mit\Psi}_{\rm out} (x, \eta; S) | {\mit\Psi}_{\rm in} (x, \eta; S) \rangle
\right|^2
\leq
\sum_S
\langle {\mit\Psi}_{\rm in} (x, \eta; S) | {\mit\Psi}_{\rm in} (x, \eta; S) \rangle
\sum_{S'}
\langle {\rm\Psi}_{\rm out} (x, \eta; S') | {\mit\Psi}_{\rm out} (x, \eta; S') \rangle
\, .
\label{eq:simpleineq}
\end{equation}

The forward matrix elements give the usual parton densities. For instance, for a
spinless target,
\begin{equation}
\sum_S \,
\langle {\mit\Psi}_{\rm in} (x, \eta; S) |
{\mit\Psi}_{\rm in} (x, \eta; S) \rangle
=
q (x_1)
 \qquad  ,  \qquad
\sum_S \,
\langle {\mit\Psi}_{\rm out} (x, \eta; S) |
 {\mit\Psi}_{\rm out} (x, \eta; S) \rangle
=
q (x_2)
\, .
\end{equation}
As a result, one obtains for the quark distributions
\cite{Pire:1998nw,Radyushkin:1998es,Pobylitsa:2002iu}
\begin{equation}
H^q (x, \eta, \Delta^2)
\leq
\sqrt{q (x_1) q (x_2)}
\, .
\label{eq:ineqquark}
\end{equation}
Similarly, for the gluon distributions, taking into account extra factors of
$x$ present in their definitions, gives the  inequality
\begin{equation}
H^g (x, \eta, \Delta^2) \leq \sqrt{(1 - \eta^2) x_1 x_2 g(x_1) g (x_2)}
\, .
\end{equation}

For unpolarized quarks in the nucleon, inequalities involve both $H$ and $E$ GPDs.
In particular, (\ref{eq:ineqquark}) was modified in Ref.\ \cite{Diehl:2000xz} into
\begin{equation}
\left| H - \frac{\eta^2}{1 - \eta^2} E \right|
\leq
\sqrt{\frac{q (x_1) q (x_2)}{1 - \eta^2}}
\, .
\label{eq:ineqquarknucleon}
\end{equation}
The same inequality, with the change $H \to \widetilde H$, $E \to \widetilde E$,
holds for polarized quark GPDs \cite{Pobylitsa:2002gw}. The inequality
(\ref{eq:ineqquarknucleon}) was enhanced in \cite{Pobylitsa:2001nt} to
\begin{equation}
\left( H (x, \eta, \Delta^2) - \frac{\eta^2}{1 - \eta^2} E (x, \eta, \Delta^2) \right)^2
+
\left( \frac{|\bit{\Delta}_\perp|}{2 M_N (1 - \eta^2)} E (x, \eta, \Delta^2) \right)^2
\leq
\frac{q (x_1) q (x_2)}{1 - \eta^2}
\, .
\end{equation}
This relation also gives a constraint on the helicity-flip GPD $E$ itself
\cite{Pobylitsa:2001nt,Diehl:2000xz}:
\begin{equation}
\frac{|\bit{\Delta}_\perp|}{2 M_N \sqrt{1 - \eta^2}} \left| E (x, \eta, \Delta^2) \right|
<
\sqrt{q (x_{1}) q(x_{2})}
\, .
\label{eq:Eqineq}
\end{equation}

\subsubsection{Inequalities in impact parameter representation}

All inequalities discussed in the previous subsection compare $\Delta^2$-dependent
GPDs $F^q (x, \eta, \Delta^2)$ with forward parton densities $q (x)$ which have no
$\Delta^2$-dependence. Since GPDs are expected to decrease with increasing $|\Delta^2|$,
the inequalities at lowest attainable $\Delta^2$ are the strongest. In fact, it is
possible to construct inequalities in which GPDs are compared to other GPDs which are
also decreasing with $|\Delta^2|$. The simplest example is obtained by taking $\eta = 0$
and  switching to the impact parameter representation (\ref{FourierImpactRepresentation}).
Then  an analogue of (\ref{eq:Eqineq}) is
\cite{Burkardt:2001ni},
\begin{equation}
\left| \bit{\nabla}_\perp f^q_E (x, 0, \bit{r}_\perp) \right|
< 2 M_N f^q_H (x, 0, \bit{r}_\perp)
\, ,
\label{eq:bEHineq}
\end{equation}
where the subscript $F = H, E$ indicates the origin of the impact parameter-dependent
parton distribution from the corresponding GPD (see Section \ref{SkewlessGPDsImpactParton}).
For discussion of other inequalities in the $\eta=0$ case see  Refs.\
\cite{Burkardt:2002sf,Burkardt:2002hr,Burkardt:2003ck,Burkardt:2000za}. There exist
also inequalities \cite{Diehl:2002he} between skewed ($\eta \neq 0$) and skewless
($\eta=0$) $\bit{r}_\perp$-dependent impact parameter distributions:
\begin{equation}
\sqrt{1 - \eta^2} |f^q_H (x, \eta, \bit{r}_\perp)|
\leq
\sqrt{
f^q_H \left( x_1, 0, \frac{\bit{r}_\perp}{1 + \eta} \right)
f^q_H \left( x_2, 0, \frac{\bit{r}_\perp}{1 - \eta} \right)
}
\, ,
\label{eq:HedHlessineq}
\end{equation}
where the third argument of the impact-parameter distributions correspond to the
distance between the probed quark and the transverse center-of-momentum of the hadron
\cite{Diehl:2002he}, as we demonstrated in Section \ref{GPDsInImpactParameterSpace}.

\subsubsection{General inequalities}

The derivation of inequalities is usually based on the positivity of the norm
of a Hilbert vector given by a superposition of  states like $b (x_i) |h (p_i,
\lambda_i) \rangle$, where $b (x_i)$ is the annihilation operator of the quark
with the momentum $x_i p_i$ and $| h (p_i, \lambda_i) \rangle$ is a hadronic
state with momentum $p_i$ and helicity $\lambda_i$. Choosing independent weights
$c_i$ for each state gives \cite{Pobylitsa:2002iu}
\begin{equation}
\left|
\sum\limits_i c_i b (x_i) |h (p_i, \lambda_i) \rangle
\right|^2
\geq 0
\label{norm-positivity}
\, .
\end{equation}
In the original example (\ref{eq:simpleineq}) only two states ${\mit\Psi}_{\rm in}$
and ${\mit\Psi}_{\rm out}$ were incorporated into the superposition. Taking a
general superposition, one may hope to derive the most generic inequalities
\cite{Pobylitsa:2002iu}. To this end, one constructs a general quark-hadron
superposition. It is given by a four-dimensional integral involving integration
over the hadron three-momentum and the quark longitudinal momentum. Hence, the
squared norm is an eight-dimensional integral. One integration is removed by the
longitudinal momentum conservation. As explained in \cite{Pobylitsa:2002iu}, by
using the impact parameter representation for GPDs, the remaining seven-dimensional
integral can be reduced to a two-dimensional one with respect to variables $x$ and
$\eta $:
\begin{equation}
\sum\limits_{\lambda _1 \lambda_2}
\
\int\limits_{-1}^{1} d \eta
\int\limits_{\eta}^{1}
\frac{d x}{(1 - x)^5}
\ p_{\lambda_2}^\ast (y_2) \, p_{\lambda_1} (y_1)
\, f_{\lambda_2 \lambda_1}
\left(
x, \eta , \frac{y_1 y_2}{1 - x} \bit{r}_\perp
\right) \geq 0
\, ,  \label{main-ineq}
\end{equation}
where $f_{\lambda_2 \lambda_1}$ is a generic notation for the impact parameter
representation of quark GPDs in the helicity basis. The arguments of the $p$-functions
$$
y_1 = \frac{1 - x}{1 + \eta}
\, , \qquad
y_2 = \frac{1 - x}{1 - \eta}
$$
correspond to the light-cone momentum of the spectator system measured in units of
the initial or final hadron, respectively. Since the functions $p_\lambda$ are
arbitrary, this relation generates  an infinite set of inequalities. In case of
spin-zero hadrons, the $\bit{r}_\perp$-space functions $f_{\lambda_2 \lambda_1}$
are simply $f^q = f^q_H$, while for spin-one-half case, they were obtained in Eq.\
(\ref{ImpactSkewnessDependent}) so that their matrix takes the form
\begin{equation}
f^q_{\lambda_2 \lambda_1} (x, \eta , - \bit{\Delta}_{\perp })
=
\sqrt{1 - \eta^2}
\left(
\begin{array}{cc}
H^q - \displaystyle\frac{\eta^2}{1 - \eta^2} E^q
& \quad
\displaystyle\frac{\bar\Delta}{2 M_N (1 - \eta^2)} E^q
\\
- \displaystyle\frac{\Delta}{2 M_N (1 - \eta^2)} E^q
& \quad
H^q - \displaystyle\frac{\eta^2}{1 - \eta^2} E^q
\end{array}
\right)_{\lambda_2 \lambda_1}
\, .
\label{F-H-E-nucleon}
\end{equation}
By choosing appropriate functions $p_\lambda$ in (\ref{main-ineq}) one can obtain
various inequalities \cite{Pobylitsa:2002iu}, including those discussed earlier.
In particular, taking $p(z) = c_1 \delta (z - y_1)+ c_2 \delta (z - y_2)$ and
optimizing the resulting inequality with respect to $c_1$ and $c_2$ gives Eq.\
(\ref{eq:HedHlessineq}). Furthermore, taking $c_i = d_i \exp(- p_{i \perp}
\bit{\Delta}_\perp)$ and optimizing with respect to $d_1$ and $d_2$ one obtains Eq.\
(\ref{eq:ineqquark}). It was checked in Refs.\ \cite{Pobylitsa:2002iu,Pire:1998nw}
that inequalities obtained in this way are not destroyed by one-loop evolution,
which will be discussed in Section \ref{Evolution}.

In Ref. \cite{Pobylitsa:2002vi} it was demonstrated that the general positivity
constraint (\ref{main-ineq}) is equivalent to the following representation for
GPDs in the impact-parameter space
\begin{equation}
f^q (x, \eta , \bit{r}_\perp )
=
(1 - x)^2
\sum\limits_n
Q_n \left( y_1, (1 - \eta ) \bit{r}_\perp \right)
Q_n \left( y_2, (1 + \eta ) \bit{r}_\perp \right)
\, ,
\label{pos-representation}
\end{equation}
valid in the inclusive domain $|x| < \eta$. Recalling that the momentum-fraction
interpretation of parameters $y_1$, $y_2$, we can interpret  the functions $Q_n$
as the $\bit{r}_\perp$-space transforms of objects similar to the light-cone wave
functions (see the explanation in Ref.\ \cite{Diehl:2003ny} after Eq.\ (198)). In
other words, Eq.\ (\ref{pos-representation}) has the structure of the generalized
overlap representation, cf.\ Section \ref{OverlapGPDs}. The prefactor $(1 - x)^2$
is specific for spin-one-half partons. For a generic spin-$s$ parton, it is
$(1 - x)^{2s + 1}$ \cite{Pobylitsa:2002vi}.

The positivity constraints are, of course, satisfied by perturbative diagrammatic
representation. In particular, for a toy scalar model---scalar quarks of mass
$m_\phi$ coupled to a scalar meson having the mass $m_M$---the contribution of the
triangle diagram corresponding to the matrix element of the bilocal operator
(\ref{NonlocalScalarOper}) takes the form \cite{Pobylitsa:2002vw}
\begin{equation}
f^\phi (x, \eta, \bit{r}_{\perp})
=
\frac{1 - x}{4 \pi}
\,
V_0 (y_1, (1 - \eta) \bit{r}_\perp) \, V_0 (y_2, (1 + \eta) \bit{r}_\perp)
\, ,
\label{Bbf}
\end{equation}
where the generalized impact-parameter light cone wave functions
\begin{eqnarray}
V_0 (y, \bit{r}_\perp)
=
\frac{1}{4 \pi y}
\int_0^{\infty} \frac{d\sigma}{\sigma}
\exp
\left(
-
\frac{ \bit{r}_\perp^2}{4 \sigma y^2}
-
\sigma \left( m_\phi^2 - y(1 - y) m_M^2 \right)
\right)
\, ,
\end{eqnarray}
can be concisely expressed through the McDonald function $K_0$. Feynman diagrams
also satisfy the polynomiality conditions. Thus, a possible  strategy for model
building is to construct GPDs by superposition of expressions corresponding to
perturbative diagrams \cite{MukMusPauRad03,Pobylitsa:2002vi} or having similar
structure \cite{Pobylitsa:2002vw}.

The positivity constraints provide a very sensitive test for self-consistency of
model expressions for GPDs. For instance, the triangle diagram for the bilocal
vector operator $\phi^\dagger (- z^-) \partial^\mu \phi (z^-)$ in the scalar toy
model \cite{MukMusPauRad03} has both the $p^\mu$ term corresponding to $F$-type
DD and the $\Delta^\mu$ part contributing to $G$-type DD. As observed in
\cite{Tiburzi:2002kr}, a GPD built from the $F$-type part alone violates even the
simplest positivity condition of Eq.\ (\ref{eq:ineqquark}) type. At the moment, it
seems that the most promising way to satisfy both the polynomiality conditions
and the infinite set of the positivity constraints is to combine the generic
representation (\ref{pos-representation}) with the formalism of double distributions.
Only a first step \cite{MukMusPauRad03,Pobylitsa:2002vi,Pobylitsa:2002vw} has been
made in this direction so far.

\subsection{Chiral properties of GPDs}
\label{ChiPTGPDs}

Although the dependence of GPDs on the momentum fraction variables $x$ and $\eta$
cannot be handled analytically from first principles, but only through elaborate
phenomenological modeling, the dependence on the transverse momentum transfer is
quite a different story. The small-$\Delta^2$ dependence can be revealed through
an effective field theory description at scales of order of the chiral symmetry
breaking. The conventional approach uses Goldstone bosons of the spontaneously
broken chiral symmetry as independent degrees of freedom within the framework
of chiral perturbation theory ($\chi$PT). The construction of effective operators
is performed by means of the standard power counting of $\chi$PT, which relies
on the fact that the Goldstone bosons do not interact at zero momentum.

In the present section, we review the application of $\chi$PT to the pion and
nucleon GPDs. We limit ourselves to the $SU(2)$ $\chi$PT. We consider single-nucleon
systems and, in order to have a consistent power counting, we use the formalism of
the heavy baryon $\chi$PT. It treats the nucleon as a non-relativistic infinitely
heavy particle \cite{JenMan91}. We present the results at  one loop level of
$\chi$PT. In this way, we compute the leading non-analytic corrections of the type
$|\Delta^2| \ln |\Delta^2|$ to GPDs, where $M_N^2 \gg |\Delta^2| \sim m_\pi^2$. Such
corrections are universal and allow us to get an insight into the structure of GPDs
at small momentum transfer. Note, that ultraviolet divergences generated by one-loop
diagrams  should  be absorbed into the coefficients of next-to-leading order
counterterms in the chiral expansion, as this domain is not described correctly
by low-momentum meson Lagrangian.

The calculational procedure is rather straightforward. First, one constructs
composite operators from the nucleon and pion fields which match the quantum
numbers of the operators built on the quark-gluon level, e.g., local Wilson
operators (\ref{QuarkLocalOperator}) -- (\ref{GluonLocalOperator}) or their
light-cone generating functions (\ref{QuarkLightRayOperators}) --
(\ref{GluonLightRayOperators}). The  operators made of nucleon and pion
fields are added to the effective Lagrangian \cite{ArnSav01,CheJi01}. Then one
uses the heavy-baryon chiral perturbation theory \cite{JenMan91} (for a
review, see \cite{BerKaiMei95}) for computation of the tree-level and one-loop
contributions. In our case, one needs the leading order pion-nucleon Lagrangian
\begin{equation}
\label{LOchiralLagrangian}
{\cal L}_{\scriptscriptstyle\rm LO} = \bar N_v
\left\{
i \, v \cdot {\cal D}
+ 2 g_A S \cdot {\cal A}
\right\}
N_v +
\frac{f_\pi^2}{8}
{\rm tr}
\left\{
\partial^\mu {\mit\Sigma} \ \partial_\mu {\mit\Sigma}^\dagger
\right\}
+ \lambda \,
{\rm tr}
\left\{ M_q
\left(
{\mit\Sigma} + {\mit\Sigma}^\dagger
\right)
\right\}
\, .
\end{equation}
Here, $N_v$ is the heavy-nucleon field operator, depending on the four-velocity
$v^\mu$ and normalized by $N_v (0) | p \rangle = u_v (p) | 0 \rangle$. The relation of
the heavy-nucleon spinor $u_v$ to the usual one $u$ is given in Appendix \ref{Spinors},
along with other useful relations required to perform calculations. The notation
$S_\mu \equiv \frac{i}{2} \sigma_{\mu\nu} \gamma_5 v^\nu$ is used for the
Pauli-Lubanski spin vector. The nonlinear chiral ${\mit\Sigma}$ field is constructed
from the triplet of the pion fields
\begin{equation}
{\mit\Sigma} \equiv \xi^2
=
\exp \left( \frac{i \sqrt{2}}{f_\pi} \pi^a \tau^a \right)
\, ,
\end{equation}
where $f_\pi = 132 \ {\rm MeV}$ is the pion decay constant and $\tau^a$ are the
flavor Pauli matrices. The axial-vector and vector (hidden in the covariant
derivative ${\cal D}_\mu \equiv \partial_\mu + {\cal V}_\mu$) pion potentials
are given by expressions
\begin{equation}
{\cal V}_\mu
\equiv
\frac{1}{2}
\left(
\xi^\dagger \partial_\mu \xi
+
\xi \partial_\mu \xi^\dagger
\right)
\, ,
\qquad
{\cal A}_\mu
\equiv
\frac{i}{2}
\left(
\xi^\dagger \partial_\mu \xi
-
\xi \partial_\mu \xi^\dagger
\right)
\, .
\end{equation}
The field ${\mit\Sigma}$ and the coupling $\lambda$, that occur in the effective
Lagrangian, are subject to the chiral counting rules:
\begin{equation}
\label{Chcount}
{\mit\Sigma} \, \sim \, {\cal O} ( \delta^0 )
\, , \quad
\partial^\mu {\mit\Sigma}
\, \sim \, {\cal O} ( \delta^1 )
\, , \quad
\lambda \sim \, {\cal O} ( \delta^2 )
\, ,
\end{equation}
with $\delta$ being a small parameter of the chiral expansion. In our subsequent
consideration it will be the $t$-channel momentum transfer $\Delta^\mu$. From
(\ref{Chcount}), one finds that the leading order effective Lagrangian is of the
order of $\Delta^2$. For the purposes of the present discussion, we will need
just the first non-trivial term in the expansion of the potentials in the pion
fields, namely,
$$
{\cal A}_\mu = - \frac{1}{\sqrt{2} f_\pi} \partial_\mu \,
\pi^a \tau^a + \dots
\, , \qquad  {\rm and} \qquad
{\cal V}_\mu = \frac{i}{(\sqrt{2} f_\pi)^2}
\varepsilon_{abc} \pi^a \partial_\mu \pi^b \tau^c
+
\dots \, ,
$$
respectively. In the mass term, the coefficient $\lambda$, accompanying the quark
mass matrix $M_q = {\rm diag} (m_u, m_d)$, is related to the quark condensate via
$\lambda = - \frac{1}{2} \langle \bar \psi \psi \rangle$.

\subsubsection{Pion GPDs}

We start with pion GPDs \cite{KivPol01}. Since the isosinglet and isotriplet
GPDs possess different chiral properties, it is instructive to study them
separately. These functions are related to the following combinations of
flavor-specific GPDs (\ref{spin-zero-GPDs}) as
\begin{equation}
H^{(0)}_\pi = H^u_\pi + H^d_\pi
\, , \qquad
H^{(1)}_\pi = H^u_\pi - H^d_\pi
\, .
\end{equation}
The isospin symmetry relates GPDs of different pion species among each other,
\begin{equation}
H^{u + d}_{\pi^+} = H^{u + d}_{\pi^-} = H^{u + d}_{\pi^0}
\, , \qquad
H^{u - d}_{\pi^+} = - H^{u - d}_{\pi^-}
\, , \qquad
H^{u - d}_{\pi^0} = 0
\, .
\end{equation}
These combinations have definite transformation properties under the time-reversal
and charge conjugation symmetries
\begin{equation}
H^{(I)}_\pi (x , \eta) = H^{(I)}_\pi (x , - \eta)
\, , \qquad
H^{(I)}_\pi (x , \eta) = (- 1)^{1 - I} H^{(I)}_\pi (- x , \eta)
\, .
\end{equation}

Before we compute their chiral behavior
in low momentum transfer as well as the pion mass, let us present a few exact
results based on their Goldstone nature. Namely, from the soft-pion theorems it
was established in Ref.\ \cite{Pol98} that
\begin{equation}
\label{SoftPion}
H^{(0)}_\pi (x, \eta = 1 , \Delta^2 = 0) = 0
\, , \qquad
H^{(1)}_\pi (x, \eta = 1 , \Delta^2 = 0) = \phi_\pi \left( \frac{1 + x}{2} \right)
\, .
\end{equation}
for isoscalar $I = 0$ and isovector $I = 1$ sectors. On the other hand, since the
$x$-moment of GPDs should be a polynomial in $\eta$ with the highest power of
$\eta$ no larger than two (see Eq.\ (\ref{FFversusGPDs})), one finds \cite{Pol98}
that
\begin{equation}
\label{PionGPD}
\int_{- 1}^1 dx \, x H^{(0)}_\pi (x, \eta , \Delta^2 = 0)
=
(1 - \eta^2) \sum_q P^q
\, ,
\end{equation}
where $P^q$ is the fraction of the pion's momentum carried by the flavor-$q$
quark. Notice that due to the chiral nature of the pion, the second moment
depends on a single rather than two independent non-perturbative parameters as
it is the case in the generic case of a spinless target. As we established in
Eqs.\ (\ref{TotalMomentumSumRule}) and (\ref{MomSumRule}), the value of $P^q$
is related to the inclusive parton distributions through
\begin{equation}
P^q
=
\int_{- 1}^1 dx \, x f^q (x)
=
\int_0^1 dx \, x \left( q (x) + \bar q (x) \right)
\, .
\end{equation}
This quantity depends on the resolution scale, which is not displayed. The sum
rule for the pion GPD (\ref{PionGPD}) imposes a constraint on the pion $D$-term,
$$
D^q (\alpha)
=
- \frac{5}{4} P^q (1 - \alpha^2)
\left\{ C^{3/2}_1 (\alpha) + \dots \right\}
\, ,
$$
thus fixing the magnitude of the leading term exactly.

Now we are in a position to figure out the dependence on the soft momentum
transfer and $m_\pi$ using $\chi$PT. First notice, that different light-cone
components of the $t$-channel momentum transfers have different scaling in
$\chi$PT. Namely, $\Delta^+$ is of order one since it forms the scaling variable
$\eta$, while the soft expansion parameter will be the transverse momentum
transfer $\delta = \bit{\Delta}_\perp$. Making use of this power counting
procedure, we have to construct effective hadronic operators made from the
non-linear pion field ${\mit\Sigma}$ and its derivatives, which will substitute
the partonic operator (\ref{QuarkLightRayOperators}) parametrizing GPDs at low
energy scales. Depending on the type of the light-cone component of the
derivative acting on ${\mit\Sigma}$, one finds the following scaling rules
in the infinite momentum frame:
\begin{equation}
\label{ChiPTbuilding}
\partial^+ {\mit\Sigma} \sim {\cal O} ( \bit{\Delta}_\perp^0 )
\, , \quad
\partial^- {\mit\Sigma} \sim {\cal O} ( \bit{\Delta}_\perp^2 )
\, , \quad
\bit{\nabla}_\perp {\mit\Sigma} \sim {\cal O} ( \bit{\Delta}_\perp )
\, .
\end{equation}

Using the building blocks (\ref{ChiPTbuilding}), one derives effective field
theory operators in terms of the Goldstone degrees of freedom by matching the
quantum numbers of the twist-two operators (\ref{QuarkLightRayOperators}).
The analysis, which exhibits the symmetry properties of the composite pion
operators can be done in terms of local Wilson operators. To the leading order
in the chiral expansion, one finds that the lowest spin-one operator in the
isoscalar sector vanishes identically---as a consequence of the charge
conjugation property of the nonlinear pion field $\tau^2 {\mit\Sigma}^\dagger
\tau^2 = {\mit\Sigma}^T$ and unitarity---so that the first nontrivial isoscalar
operator has spin two,
\begin{equation}
\label{SpinOneChiPT}
{\cal O}^{(0) \pi}_{\mu}
=
{\rm tr}
\left\{
{\mit\Sigma} \ \partial_\mu {\mit\Sigma}^\dagger
\right\}
=
0
\, , \qquad
{\cal O}^{(0)\pi}_{\mu_1 \mu_2}
=
f_\pi^2 a^{\pi}_2 \Sym_{\mu_1 \mu_2}
{\rm tr}
\left\{
\partial_{\mu_1} {\mit\Sigma} \ \partial_{\mu_2} {\mit\Sigma}^\dagger\
\right\}
\, ,
\end{equation}
On the other hand, even the spin-two operator in the isovector sector is zero,
\begin{equation}
\label{SpinTwoChiPT}
\Sym_{\mu_1 \mu_2}
{\rm tr} \left\{
\tau^a
\partial_{\mu_1} {\mit\Sigma} \ \partial_{\mu_2} {\mit\Sigma}^\dagger
\right\} = 0
\, .
\end{equation}
Actually, the origin of this equation is more general than the chiral limit. On
the parton level is corresponds to the vanishing of the first Mellin moment of
the isovector combination of GPDs $H^{u-d} (x, \eta, \Delta^2)$ (see Eq.\
(\ref{spin-zero-GPDs})) which is an even function in $x$ as a consequence of
the isospin symmetry of the quark operator and its odd parity under the charge
conjugation. One can further classify all high-spin local pion operators according
to this scheme \cite{BelJi02}. It is much more convenient however to deal with
non-local pion operators, which effectively resum the entire series of Wilson
operators. Namely, relying on general symmetry considerations one can introduce
the operator \cite{KivPol01}:
\begin{eqnarray}
\label{NonLocalPionOperator}
{\cal O}^{\pi} (z^-)
\!\!\!&=&\!\!\!
\frac{f_\pi^2}{8}\ \int_{-1}^1 d \beta \int_{-1 + |\beta|}^{1 - |\beta|} d \alpha
\, f (\beta , \alpha)
\Bigg\{
{\mit\Sigma}
\left( ( \alpha + \beta ) z^- \right)
i \! \stackrel{{}_\leftrightarrow}{\partial}{\!}^+
{\mit\Sigma}^\dagger
\left( ( \alpha - \beta ) z^- \right)
\nonumber\\
&&\qquad\qquad\qquad\qquad\qquad\qquad \,
+
{\mit\Sigma}^\dagger
\left( ( \alpha + \beta ) z^- \right)
i \! \stackrel{{}_\leftrightarrow}{\partial}{\!}^+
{\mit\Sigma}
\left( ( \alpha - \beta ) z^- \right)
\Bigg\}
\, ,
\end{eqnarray}
which is an $SU (2)$ matrix. Its trace with $\tau^A$, ${\rm tr} \, \tau^A
{\cal O}^{\pi}$, and expansion in local operators can be shown to yield only
the operators with required symmetry and structure, with (\ref{SpinOneChiPT})
and (\ref{SpinTwoChiPT}) being the first terms in Taylor series in powers of
the Ioffe time $z^-$. In the above definition $f (\beta, \alpha)$ is a
generating function of a tower of low-energy constants. It is a double
distribution, discussed in the preceding section, though to the lowest order
in the chiral expansion. The low-energy constants match the partonic degrees
of freedom to the pionic: their values are not determined within the effective
field theory and rather considered as phenomenological parameters. There is a
number of pion operators with more than two ${\mit\Sigma}$ fields since their
insertion does not alter the twist of the composite operator, however, they do
not contribute to the one-loop matrix elements we are interested in here and will
be totally omitted.

The pion GPDs are defined in terms of (\ref{NonLocalPionOperator}) as a matrix
element of the hadronic operator
\begin{equation}
\int
\frac{d z^-}{2 \pi} \, {\rm e}^{- i x p^+ z^- }
\langle \pi^b (p_2) |
{\rm tr} \, \tau^C {\cal O}^{\pi} (z^-)
| \pi^a (p_1) \rangle
=
\ft{1}{2} {\rm tr} ( \tau^C \tau^a \tau^b ) \
H^{(I)}_\pi (x, \eta, \Delta)
\, ,
\end{equation}
where $\tau^A = (\1 , \tau^a)$ and the index $A$ runs over the values $A = 0,
1, 2, 3$. The above formula introduces both the isovector (of isospin $I[a] = 1$)
and isoscalar (of isospin $I[0] = 0$) GPDs.

Notice that the first moment of the double distribution $f^{(I)}$ is matched to
known nonperturbative parameters
\begin{eqnarray}
\int_{\mit\Omega} d\alpha d\beta \ f^{(0)} (\beta, \alpha) \beta
\!\!\!&=&\!\!\!
\sum_q P^q = a_2^\pi
\, \\
\int_{\mit\Omega} d\alpha d\beta \ f^{(1)} (\beta, \alpha)
\!\!\!&=&\!\!\!
1
\, .
\end{eqnarray}
As it is clear from the context, the low-energy constant $a_2^{\pi}$ is related
to the momentum fraction of the pion carried by quarks and gluons forming it. For
an on-shell pion at low normalization point $\mu^2 \approx 0.3 \ {\rm GeV}^2$ one
has \cite{GluReyVogSch99} $a^{\pi,q}_2 \approx 0.7$ and respectively $a^{\pi,g}_2
\approx 0.3$ due to the momentum conservation. The interpretation of the generating
functions $f^{(I)} (\beta, \alpha)$ as DDs implies  that these functions depend on
the factorization scale $\mu$. The functional dependence on  this parameter is
described by evolution equations to be discussed in the following sections. For the
sake of brevity, we do not write this argument explicitly.

\begin{figure}[t]
\begin{center}
\hspace{0cm}
\mbox{
\begin{picture}(0,80)(110,0)
\put(0,0){\insertfig{7}{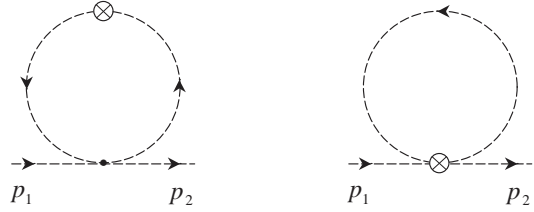}}
\end{picture}
}
\end{center}
\caption{\label{PionOneLoop} Chiral loops generating non-analytic behavior
of pion GPDs. The four-pion vertex in the leftmost diagram originates from
the expansion of effective Lagrangian ${\cal L}_{\scriptscriptstyle\rm LO}$.}
\end{figure}

Computation of the Feynman diagrams in Fig. \ref{PionOneLoop} yields the
following expressions for pion GPDs with leading non-analytic corrections
included \cite{KivPol01,DieManSch05}
\begin{eqnarray}
H^{(0)}_\pi (x, \eta, \Delta^2)
\!\!\!&=&\!\!\!
{\mit\Phi}^{(0)} \left( \frac{x}{\eta}, \frac{1}{\eta} \right)
+
\frac{
\theta\left( \eta - |x| \right)
}{
2 (4 \pi f_\pi)^2
}\
\left( m_\pi^2 - 2 \Delta^2 \right)
\int_{-1}^1 d \sigma \,  \ln m_\pi^2 (\sigma)
\
{\mit\Phi}^{(0)} \left( \frac{x}{\eta} , \sigma \right)
\, ,
\nonumber\\
H^{(1)}_\pi (x, \eta, \Delta^2)
\!\!\!&=&\!\!\!
\left(
1
-
2 \frac{m_\pi^2 \ln m_\pi^2}{(4 \pi f_\pi)^2}
\right)
{\mit\Phi}^{(1)} \left( \frac{x}{\eta}, \frac{1}{\eta} \right)
\nonumber\\
&+&\!\!\!
\frac{ \theta \left( \eta - |x| \right)}{(4 \pi f_\pi)^2 \eta}
\int_{-1}^1 d \sigma
\, m_\pi^2 (\sigma) \ln m_\pi^2 (\sigma)
\
\frac{\partial}{\partial\sigma} {\mit\Phi}^{(1)}
\left( \frac{x}{\eta}, \sigma \right)
\, ,
\label{ChiPTpionGPDs}
\end{eqnarray}
where
$$
m^2_\pi (\sigma) \equiv m^2_\pi - (1 - \sigma^2) \frac{\Delta^2}{4}
$$
and
$$
{\mit\Phi}^{(I)} (u, v)
=
\int_{\mit\Omega} d \alpha d \beta \,
f^{(I)} (\beta , \alpha)
\left\{
v \delta (u - \alpha - v \beta) - (1 - I) \delta (u - \alpha - \beta)
\right\}
\, ,
$$
with the integration going over the same simplex ${\mit\Omega}$ as for conventional
DDs. Note that the argument of the chiral logarithms can be  made dimensionless by
introducing a chiral cut-off $\Lambda_\chi^2$, the dependence on which cancels
between these one-loop expressions and the next-to-leading order counterterms in
the derivative expansion of the effective chiral Lagrangian
${\cal L}_{\scriptscriptstyle\rm NLO}$ \cite{GasLeu84}.

Let us discuss a few limiting cases of Eq.\ (\ref{ChiPTpionGPDs}).
\begin{itemize}
\item Conventional form factors are found by forming the first two moments of
pion GPDs,
\begin{eqnarray}
\int_{-1}^1 dx \ H^{(1)}_\pi (x, \eta, \Delta^2)
\!\!\!&=&\!\!\!
2 F_{\pi^+} (\Delta^2)
\, , \\
\int_{-1}^1 dx \ x\ H^{(0)}_\pi (x, \eta, \Delta^2)
\!\!\!&=&\!\!\!
\theta_2 (\Delta^2) - \eta^2 \theta_1 (\Delta^2)
\, ,
\end{eqnarray}
with the pion electromagnetic and gravitational form factors defined in Eqs.\
(\ref{EMformfactorPion}) and (\ref{PionGravitFFs}), respectively. Integrating
the one-loop expression (\ref{ChiPTpionGPDs}), we thus obtain the well-known
results for the leading non-analytic contribution to the pion electromagnetic
\cite{GasLeu84} and gravitational form factors \cite{DonLeu91}
\begin{eqnarray}
F_\pi (\Delta^2)
\!\!\!&=&\!\!\!
1 - 2 \frac{m_\pi^2 \ln m_\pi^2}{(4\pi f_\pi)^2}
+
\frac{1}{ (4 \pi f_\pi)^2}
\int_{- 1}^1 d \sigma
\, m_\pi^2 (\sigma) \ln  m_\pi^2 (\sigma)
\, , \\
\theta_2 (\Delta^2)
\!\!\!&=&\!\!\!
P^q
\, , \\
\theta_1 (\Delta^2)
\!\!\!&=&\!\!\!
P^q
\left(
1
+
\frac{m_\pi^2 - 2 \Delta^2}{2 (4\pi f_\pi)^2}
\int_{- 1}^1 d \sigma \ (1 - \sigma^2) \
\ln m_\pi^2 (\sigma)
\right) \, ,
\end{eqnarray}
respectively.
\item Parton densities are found by taking the skewless limit. Note, however, that
the limit $\eta \to 0$ and the chiral limit $m_\pi \to 0$ do not commute in Eq.\
(\ref{ChiPTpionGPDs}). To reproduce the results obtained recently in Refs.\
\cite{ArnSav01,CheJi01}, we have to send $\eta \to 0$ {\sl before} $m_\pi \to 0$.
This yields
\begin{equation}
q(x)
=
q_0 (x) \left(1 - 2 \frac{m_\pi^2 \ln m_\pi^2}{(4\pi f_\pi)^2}\right)
+
2 \delta(x)\ \frac{m_\pi^2 \ln m_\pi^2}{(4\pi f_\pi)^2}
\, ,
\end{equation}
with $q_0 (x)$ being the quark distribution in the chiral limit.
\end{itemize}

\subsubsection{Nucleon GPDs}

Undoubtedly, the most profound manifestation of the chiral symmetry in the
dynamics of nucleon GPDs is the appearance of the pion pole in the isovector
combination $\widetilde{E}^{u - d} = \widetilde{E}^u - \widetilde{E}^d$.
Analogously to the behavior of the psedoscalar nucleon form factor introduced
in Eq.\ (\ref{IsovectorPsedoscalarFF}), the quark-antiquark pair determining
the operator content of GPD $\widetilde{E}$ can fluctuate into an off-shell
pion and due to its Goldstone nature $\widetilde{E}^{u - d}$ develops this
transition develops a pole in $\Delta^2$. The explicit chiral symmetry breaking
proportional to $m_q$ gives a small mass to the pion so that the pole is shifted
from zero to the physical value $\Delta^2 = m_\pi^2$. At the pole, the single
pion exchange dominates over other resonances with the same quantum numbers,
such that it is possible to have a legitimate approximation for $|\Delta^2| \leq
m_\pi^2$. Continuing to the physical region of space-like $\Delta^2$, one writes
for the proton \cite{ManPilWei97,PenPolGoe00}
\begin{equation}
\label{PionPoleInEtilde}
\widetilde{E}^{u - d} (x, \eta, \Delta^2 \to m_\pi^2)
=
\frac{\theta (\eta - |x|)}{2 \eta} \phi_\pi \left( \frac{x + \eta}{2\eta} \right)
\frac{4 g_A^{(3)}M_N^2}{m_\pi^2 - \Delta^2}
\, ,
\end{equation}
where the bound state of the quark-antiquark pair is determined by the pion
leading twist distribution amplitude (\ref{LeadingTwistPionDA}). By means of the
Goldberger-Treiman, the pion-nucleon coupling constant $g_{\pi NN}$ was reduced
to the isovector axial constant $g_A^{(3)}$, $f_\pi g_{\pi NN} = \sqrt{2} M_N
g_A^{(3)}$. Using the isospin symmetry one can finally write for specific flavor
components
\begin{equation}
\widetilde{E}^u = - \widetilde{E}^d = \frac{1}{2} \widetilde{E}^{u - d}
\, .
\end{equation}

Now we turn to the chiral expansion of other nucleon GPDs \cite{BelJi02}. Under
the chiral $SU (2)_L \otimes SU (2)_R$ group, the quark operators determining their
content transform under either isovector $(3,1) \oplus (1, 3)$ or isoscalar $(1, 1)$
representations. These correspond to the flavor non-singlet ${\cal R}^{2, qq, a}$ and
singlet ${\cal R}^{2, qq, 0}$ quark operators, respectively,
\begin{equation}
\label{QuarkOperator}
{\cal R}^{2, qq, A}_{\mu_1 \mu_2 \dots \mu_j}
= \Sym_{\mu_1 \mu_2 \dots \mu_j}
\bar\psi
\tau^A
\gamma_{\mu_1}
i\!\stackrel{{}_\leftrightarrow}{\cal D}_{\mu_1}
i\!\stackrel{{}_\leftrightarrow}{\cal D}_{\mu_2}
\dots
i\!\stackrel{{}_\leftrightarrow}{\cal D}_{\mu_j}
\psi \, .
\end{equation}
For the purpose of a unique translation of the Lorentz-covariant matrix elements
in Eq.\ (\ref{GPDmomentsFFHandE}) into non-relativistic ones,  it is is convenient
to use the Breit reference frame \cite{BerKaiKamMei92,BerFeaHemMei98}, as we
discussed at length in Section \ref{BreitFrameFFs}. Exploiting the results from
the Appendix \ref{Spinors}, the decomposition (\ref{GPDmomentsFFHandE}) transforms
in the Breit frame into the equation
\begin{eqnarray}
&&\!\!\!\!\!\!\!\!\!\!\!
\langle p_2 | {\cal R}^{2, qq, A}_{\mu_1 \dots \mu_j} | p_1 \rangle
= \! \Sym_{\mu_1 \dots \mu_j}\!\!
\bar u_v (p_2) \tau^A u_v (p_1) v_{\mu_1} \!
\left\{
(2 M_N)^{j - 1}
G^E_{j, 0} v_{\mu_2} \dots v_{\mu_j} \!
+
\dots
+
G^E_{j, j-1} \Delta_{\mu_2} \dots \Delta_{\mu_j}
\right\}
\nonumber\\
&&\qquad
+\!
\Sym_{\mu_1 \dots \mu_j}
\frac{\bar u_v (p_2) \tau^A [S_{\mu_1}, S \cdot \Delta] u_v (p_1)}{M_N}
\left\{
(2 M_N)^{j - 1}
G^M_{j, 0} v_{\mu_2} \dots v_{\mu_j}
+
\dots
+
G^M_{j, j-1} \Delta_{\mu_2} \dots \Delta_{\mu_j}
\right\}
\!\!\!
\nonumber\\
&&\qquad
+
\Sym_{\mu_1 \dots \mu_j}
\frac{\bar u_v (p_2) \tau^A u_v (p_1)}{2 M_N}
\Delta_{\mu_1} \dots \Delta_{\mu_j} D_{j} \, ,
\label{DefIsoSpinMatrElem}
\end{eqnarray}
where for brevity we did not display the dependence of the form factors $G^E_{jk}$,
$G^M_{jk}$ and $D_j$ on the momentum transfer $\Delta^2$ and dropped higher-order
corrections in the ratio $\Delta^2/M_N^2$. Obviously due to the constraint from
the time-reversal invariance (see Section \ref{TimeReversalHermiticity}) the form
factors accompanying the odd power of the momentum transfer $\Delta_\mu$ has to be
set to zero. The electric- and magnetic-type form factors $G^E_{jk}$ and $G^M_{jk}$
are related to the previously introduced $H_{jk}$ and $E_{jk}$ in Eq.\
(\ref{GPDmomentsFFHandE}) via
\begin{equation}
G^E_{jk} = H_{jk} + \frac{\Delta^2}{4 M_N^2} E_{jk}
\, , \qquad
G^M_{jk} = H_{jk} + E_{jk}
\, .
\end{equation}
Note, that $G^E_{1,0}$ and $G^M_{1,0}$ correspond to the standard Sachs nucleon
electromagnetic form factors (\ref{DefinitionsOfSachsFFs}),
\begin{equation}
G^E_{1,0} = G_E
\, , \qquad
G^M_{1,0} = G_M
\, .
\end{equation}
They have been extensively discussed in the literature \cite{BerKaiKamMei92,BerFeaHemMei98},
and therefore we will not address their chiral properties here.

To proceed with the analysis of chiral loops, let us complement the basis of pion
operators, discussed in the preceding section, by adding hadronic twist-two
operators bilinear in the heavy nucleon fields. The nucleon operators admit the
form that mimics the tensor decomposition of the off-forward matrix elements
(\ref{DefIsoSpinMatrElem}). The leading operators contributing to isoscalar and
isovector combinations of $G^E_{j,0}$ and $G^M_{j,0}$ are
\begin{eqnarray}
\label{BaryonOperators}
{\cal O}^{N, A}_{\mu_1 \mu_2 \dots \mu_j}
\!\!\!&=&\!\!\!
a^N_j (2 M_N)^{j - 1} \Sym_{\mu_1 \dots \mu_j}
v_{\mu_1} \dots v_{\mu_j}
\bar N_v \tau^A_{\xi +} N_v
\nonumber\\
&+&\!\!\!\ b^N_j (2 M_N)^{j - 1}
(- i \partial^\nu)
\Sym_{\mu_1 \dots \mu_j}
v_{\mu_1} \dots v_{\mu_{j - 1}}
\bar N_v \tau^A_{\xi +} \frac{[S_{\mu_j}, S_\nu]}{M} N_v
+ \dots
\, ,
\end{eqnarray}
where
\begin{equation}
\tau^A_{\xi +} \equiv \frac{1}{2}
\left(
\xi \tau^A \xi^\dagger + \xi^\dagger \tau^A \xi
\right) \, .
\end{equation}
As in the pion case, the coefficients $a^N$ and $b^N$, matching partonic and
hadronic descriptions, are unknown, and have to be determined from experimental
data. Subleading operators are derived from this expression by replacing
the factors of the four-velocity $v^\mu$ in the Lorentz structure by derivatives.
Note, that, due to the time-reversal symmetry restrictions, one should always
replace an even number of velocities. In order to mimic the $D$-term structure,
one further introduces a subleading operator,
$$
\dots + c_2^N
\frac{1}{2 M}
\Sym_{\mu_1 \mu_2}
(- i \partial_{\mu_1}) (- i \partial_{\mu_2})
\bar N_v \tau^A_{\xi +} N_v
\, .
$$

\begin{figure}[t]
\begin{center}
\mbox{
\begin{picture}(0,55)(210,0)
\put(0,0){\insertfig{15}{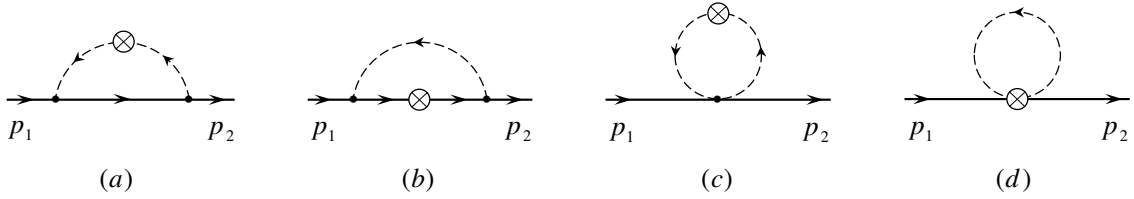}}
\end{picture}
}
\end{center}
\caption{\label{One-loop} One loop diagrams contributing to the nucleon
matrix elements of the twist-two operators. One should also add diagrams
with self-energy insertions into the external lines, which are not displayed
explicitly here.}
\end{figure}

Having constructed the operator basis, the leading-order chiral behavior is then
deduced from the Feynman graphs displayed in Fig.\ \ref{One-loop}. The isoscalar
form factors receive contributions only from the diagrams \ref{One-loop} (a) and
(c), since (b) and (d) vanish identically. Their calculation is straightforward
making use of the leading order chiral lagrangian (\ref{LOchiralLagrangian}) and
yields \cite{BelJi02}
\begin{eqnarray}
\label{IsovectorFFs}
G^E_{2, 0} (\Delta^2)
\!\!\!&=&\!\!\!
a^N_2 +
3 a^{\pi}_2 \frac{g_A^2 \pi}{2 M_N (4 \pi f_\pi)^2}
\left\{
\left( 2 m^2_\pi - \Delta^2 \right)
\int_{-1}^1 d \sigma \, \sqrt{m^2_\pi (\sigma)}
+
\frac{4}{3} m^3_\pi
\right\}
\, ,\nonumber\\
G^M_{2, 0} (\Delta^2)
\!\!\!&=&\!\!\!
b^N_2
+
3 \frac{g_A^2}{(4 \pi f_\pi)^2}
\int_{- 1}^1 d \sigma \,
\Big\{
a^{\pi}_2 m^2_\pi (\sigma) - 2 b^N_2 m^2_\pi \delta (\sigma - 1)
\Big\}
\ln m^2_\pi (\sigma)
\, , \nonumber\\
D_{2} (\Delta^2)
\!\!\!&=&\!\!\!
c^N_2 +
3 a^{\pi}_2 \frac{g_A^2 \pi}{4 (4 \pi f_\pi)^2}
M_N \left( 2 m^2_\pi - \Delta^2 \right)
\int_{-1}^1 d\sigma \, \frac{1 - \sigma^2}{\sqrt{m^2_\pi (\sigma)}}
\, .
\end{eqnarray}
Notice the absence of the one-loop contributions to the form factors $G^E_{2,0}$
and $D_2$ from the nucleon operators. This is a consequence of a cancellation
of the one-loop contribution (c) by the self-energy insertions into the external
lines. We completely absorb all analytic contributions present in
(\ref{IsovectorFFs}) into counterterms. Another comment concerns $G^E_{2,0}$.
The pion operators generate $1/M_N$-suppressed contributions to this structure.
Therefore, by power counting we must add ${\cal O} (1/M_N)$ bilinear nucleon
operators constructed from one large and one small component of the nucleon
field, and also analogous terms stemming from the chiral Lagrangian. Lorentz
invariance unambiguously fixes  their coefficients,  and no new low energy
constants arise. However, the effect of these contributions in form factors
is $\Delta^2$-independent and analytic in the pion mass. Thus, we will not
compute  them, although they can be anticipated to generate an additional
term $- 5 a_2^N g_A^2 m_\pi^3 / \left( 32 \pi f_\pi^2 M_N \right)$ to the
right-hand side of $G^E_{2,0}$ in Eq.\ (\ref{IsovectorFFs}) required by the
momentum sum rules. Also, there arises a Foldy-like term $(b^N_j - a^N_j)
\Delta^2/(4 M^2_N)$ in $G^E_{j,0}$, just like in the form factor case
\cite{BerFeaHemMei98}, from contributions of  small nucleon field components
in the heavy-mass expansion of relativistic nucleon operators.

Sum rules for the total momentum and spin of the nucleon (pion) impose the
following constraints on the coefficients $a^N_2$ and $b^N_2$ ($a^\pi_2$),
\begin{equation}
a^{N, q}_2 + a^{N, g}_2 = 1
\, , \qquad
b^{N, q}_2 + b^{N, g}_2 = 1
\, , \qquad
a^{\pi, q}_2 + a^{\pi, g}_2 = 1
\, .
\end{equation}
The latter two equations imply that the total gravitomagnetic moment of the
nucleon vanishes \cite{BroHwaMaSch00} (cf.\ Eq.\ (\ref{TotalGravitoMagneticMoment})).

The leading $(j,0)$-structures of the higher $j$-moments ($j > 2$) do not receive
non-analytic contributions in the momentum transfer $\Delta^2$ at the next-to-leading
order in the chiral expansion due to the absence of relevant pion operators. Thus,
one gets \cite{BelJi02}
\begin{equation}
G^E_{j,0} (\Delta^2) = a_j^N + \dots \, ,
\qquad
G^M_{j,0} (\Delta^2) = b_j^N
\left(
1 - 6 \frac{g_A^2 m^2_\pi}{(4 \pi f_\pi)^2} \ln m^2_\pi + \dots
\right) \,  ,
\end{equation}
where the ellipsis stand for analytic contributions from one-loop diagrams
and counterterms, which are suppressed by at least $m^2_\pi/(4 \pi f_\pi)^2$.

Due to the absence of the pion cloud contributions in diagrams (a) and (b), no
non-analytic dependence on the momentum transfer arises at this order of $\chi$PT.
However, the diagrams (c) and (d) develop chiral logarithms in the pion mass of
the form
\begin{eqnarray}
G^F_{2,0} (\Delta^2)
\!\!\!&=&\!\!\!
f^N_2
\left\{
1
-
\frac{2 m^2_\pi}{(4 \pi f_\pi)^2} ( \chi_F g_A^2 + 1) \ln m^2_\pi
\right\}
\, ,
\label{IsoScalarFFs}
\end{eqnarray}
where the overall coefficients $f^N_2 = (a_2^N, b_2^N, c_2^N)$ depend on the form
factor $G^F_{2,0}$ in question and $\chi_F$ takes the values $\chi_F = 3, 2, 3$
for $G^F_{2,0} = G^E_{2,0}, G^M_{2,0}, D_2$, respectively. It is  implied that
one adds counterterms to the right-hand side of these equations linear in $m^2_\pi$
whose (unknown) coefficients absorb minimally the ultraviolet divergences resulting
from loop integrals and also the Foldy-like term, as discussed after Eq.\
(\ref{IsovectorFFs}). The leading structures of the higher moments $G^E_{j,0}$ and
$G^M_{j,0}$, apart from the change of an overall normalization $b_2^N \to b_j^N$
and $a_2^N \to a_j^N$, have the same dependence on the chiral logarithms as in Eq.\
(\ref{IsoScalarFFs}). Finally, let us note that one also can include the delta-isobar
as a dynamical degree of freedom in $\chi$PT. Such kind of analysis was performed
for the orbital angular momentum sum rule (\ref{SpinRule}) in Ref.\ \cite{CheJi01}.

A pragmatic application of the results outlined in the present section can be either
in evolution of experimental measurements for the angular momentum sum rule
(\ref{SpinRule}) from nonzero to zero recoil or in chiral extrapolations of lattice
data normally taken at very high ``$m_\pi$'' all the way down to the physical point.

\subsection{Modeling GPDs}
\label{GPDModels}

The studies in the previous sections of generic properties of GPDs and their
asymptotic behavior in different regions of the momentum-fraction space allow to
develop some  ``semi-realistic'' models for them. Of course, the final word about
their validity is reserved to experimental measurements. There exists a number
of estimates of the functional form of GPDs based on various models of the hadronic
structure, including the chiral soliton quark model \cite{Petetal97,PolWei99,%
PenPolGoe00,Goeke:2001tz,SchBofRad02,SchColBof03,OssPolSchUrbGoe04,PraRos03,BroArr03},
bag model \cite{JiMelSon97,AniBinMedNogVen01}, constituent quark models
\cite{BofPasTra03,ScoVen03,Sim04}, instanton techniques \cite{AniDorMakTom00},
light-cone frameworks \cite{MukVan02,MukMusPauRad03}, Bethe-Salpeter
\cite{TheNogVen02,TibMil01,ChoJiKis01} and Schwinger-Dyson \cite{Tiburzi:2002kr,TibDetMil02}
approaches. The first-principle lattice simulations
\cite{LHPC03,QCDSF03,Negele:2004iu,Hagler:2004er} provided estimates for several
off-forward matrix elements of local operators and constrained their functional
dependence on the momentum transfer $\Delta^2$. First transverse lattice results
were reported in Ref.\ \cite{BraDalSanWat04}. In this section,  we will discuss how
available information can be used for construction of ans{\"a}tze for GPDs
based on very general principles. This will be done in a few steps. Since GPDs
depend on two longitudinal variables, $x$ and $\eta$, and one transverse $\Delta^2$,
we address first the question of modeling  the longitudinal dynamics, setting
$\Delta^2 = 0$. Then we consider interplay of longitudinal Feynman momentum
and transverse momentum $\bit{\Delta}_\perp$ dependence. Finally, we combine
the two models together building a single three-variable function.

\subsubsection{Longitudinal dynamics}
\label{LongDynamics}

We will rely  here  upon the approach that  uses  double distributions in order
to constrain plausible functional forms of the dependence on the $s$- and
$t$-channel longitudinal momenta $x$ and $\eta$. As already announced, we
consider first the limiting case $\Delta^2 = 0$. First, let us make two
observations. The $\beta$-variable of a double distribution $f (\beta, \alpha)$
can be interpreted as a fraction of the momentum $p$ carried by the parton.
The reduction formula (\ref{eq:redfsym}) states that the integral over $\alpha$
gives the usual parton density $f (\beta)$. When combined together they suggest
that the profile of $f (\beta, \alpha)$ in the $\beta$-direction follows the
shape of parton density $f (\beta)$. Thus, it makes sense to write
\begin{equation}
f (\beta, \alpha) = \pi (\beta, \alpha) \, f (\beta)
\, ,
\label{65n}
\end{equation}
where the function $\pi (\beta, \alpha)$ is normalized by the condition
\begin{equation}
\int_{-1 + |\beta|}^{1 - |\beta|} d \alpha \, \pi (\beta, \alpha) = 1
\, .
\end{equation}
It characterizes the profile of $f (\beta, \alpha)$ in the $t$-channel
$\alpha$-direction. The profile function should be symmetric with respect
to $\alpha \to -\alpha$ as a consequence of the time-reversal invariance
(\ref{TimeReversalDDs}), see Section \ref{TimeReversalHermiticity}. For a
fixed $\beta$, the function $\pi (\beta, \alpha)$ describes how the longitudinal
momentum transfer $\Delta^+$ is shared between the two partons. Hence, the shape
of $\pi (\beta, \alpha)$ should look like a symmetric meson distribution
amplitude $\varphi (\alpha)$. Recalling that the support region for DDs is
restricted by $|\alpha| \leq 1 - |\beta|$, to get a more close analogy with
DAs, we rescale $\alpha$ as $\alpha = (1 - |\beta|) \gamma$ introducing  the
variable $\gamma$ with $\beta$-independent limits: $-1 \leq \gamma \leq 1$.
The simplest model is to assume that the $\gamma$-profile is a universal
function $g(\gamma)$ for all $\beta$'s. As we already  emphasized, in general
this function does not have to vanish at the boundary of its support
\cite{Tib04}.  To proceed,  we disregard this complication and assume that
the double distribution is zero on the boundary  of its support region. More
elaborated models  may require lifting this condition. As possible simple
choices for $g(\gamma)$  one can take $\delta(\gamma)$, with no spread in
$\gamma$-direction, $(1 - \gamma^2)^b$ which is a characteristic shape for
the asymptotic limit of the quark and gluon distribution amplitudes for
$b = 1$ and $b = 2$, respectively, as we will explain in the next chapter.
In terms of $(\beta,\alpha)$ variables, all of these models can be summarized
by a single formula
\begin{equation}
\label{DD-Ansatz}
\pi (\beta, \alpha; b)
= \frac{{\mit\Gamma} \left( b + \ft32 \right)}{\sqrt{\pi} {\mit\Gamma} (b + 1)}
\frac{\left[ (1 - \beta)^2 - \alpha^2 \right]^b}{(1 - \beta)^{2 b + 1}}
\, ,
\end{equation}
with a parameter $b$. In the limiting case $b = \infty$, we have $\pi (\beta,
\alpha; \infty) = \delta (\alpha)$.

Let us analyze the structure of GPDs obtained from these simple models. In
particular, taking $f (\beta, \alpha; \infty) = \delta (\alpha) f (\beta)$
gives the simplest model
$$
H (x, \eta, \Delta^2 = 0) = f(x)
$$
in which the GPDs at $\Delta^2 = 0$ have no skewness dependence at all and
coincide with the usual parton distributions. For some values of $b$, one can
find analytic expressions for  GPDs also, provided that the functional form
of the forward density is not very involved. Taking, for instance,
\begin{equation}
f (\beta)
=
\frac{{\mit\Gamma} (5 - n)}{{\mit\Gamma} (4) {\mit\Gamma} (1 - n)}
\frac{(1 - \beta)^3}{\beta^n}
\theta (\beta)
\, ,
\qquad
\pi (\beta, \alpha; 1)
=
\frac{3}{4} \frac{(1 - \beta)^2 - \alpha^2}{(1 - \beta)^3}
\, ,
\end{equation}
gives  a simple analytical representation for the corresponding GPD,
\begin{equation}
\label{SimpleModel}
H (x, \eta, \Delta^2 = 0)
=
\frac{\left( 1 - \frac{n}{4} \right)}{\eta^3}
\Bigg\{
\theta (x > - \eta)
\left( \frac{x + \eta}{1 + \eta} \right)^{2 - n}
\left( \eta^2 - x + (2 - n) \eta (1 - x)\right)
- (\eta \to - \eta)
\Bigg\} .
\end{equation}
It is evident that no odd powers of $\eta$ would appear in the $x^j$-th
moments of this function. Furthermore, this expression is explicitly
non-analytic for $x = \pm \eta$. This  is true even if $n$ is integer.
Discontinuity at $x = \pm \eta$, however, appears only in the second
derivative of $H (x, \eta, \Delta^2 = 0)$, while the function itself
and its first derivatives are continuous so that the model curves for
$H(x,\eta)$ are very smooth.

In the case of   singlet quark distributions, the DDs $f_{\rm singl}
(\beta, \alpha)$ should be odd functions of $\beta$. Still, we can use
the model like (\ref{SimpleModel}) for the $\beta > 0$ part, but take
$f_{\rm singl} (\beta, \alpha)|_{\beta \neq 0} = A \, f (|\beta|, \alpha)
\, {\rm sgn} (\beta)$. Note, that $n \geq  1$  is the usual situation
for standard parametrizations of singlet quark distributions. Then the
integral producing $H_{\rm singl} (x, \eta)$ in the $|x| \leq \eta$
region  diverges for $\alpha \to x /\eta$. However, due to the antisymmetry
of $f_{\rm singl} (\beta, \alpha)$ with respect to $\beta \to - \beta$ and
its symmetry with respect to  $\alpha \to -\alpha$, the singularity at
$\alpha = x /\eta$ can be integrated using the principal value prescription.

\subsubsection{Longitudinal-transverse interplay}
\label{LongTransInterplay}

Above, we have discussed hypothetical GPDs with no transverse dynamics, setting
$\Delta^2 = 0$ from the start. Now we  are going to discuss the dependence on
$\Delta^2 \neq 0$ in a simplified situation when $\eta = 0$. Even in this case,
a first principle, model independent way to do this is currently not available,
we have to rely on some assumptions. The simplest idea is to assume a complete
factorization of longitudinal and transverse dependence, like  $H^q (x, \eta = 0,
\Delta^2) = f^q (x) F^q (\Delta^2)$, with $f^q (x)$ being a forward parton density
and $F^q (\Delta^2)$ the relevant form factor. To be specific, for $H^q$ and $E^q$
GPDs of the proton the factorized ansatz implies that the $\Delta^2$ dependence
is accumulated in Dirac and Pauli form factors $F_1^q (\Delta^2), F_2^q (\Delta^2)$,
respectively. Since the latter are defined through the matrix element of the
electromagnetic (or, in general, vector) current, only the valence  parts of $H^q$
and $E^q$ can be modeled in terms of $F_1^q$ and $F_2^q$. Thus, for the flavor-$q$
valence quark contribution we would write
\begin{equation}
H^{u, \rm val} (x, \eta = 0, \Delta^2)
=
\ft12 F^u_1 (\Delta^2) \, f^{u, {\rm val}} (x)
\, , \qquad
H^{d, \rm val} (x, \eta = 0, \Delta^2)
=
F^d_1 (\Delta^2) \, f^{d, {\rm val}} (x)
\, ,
\end{equation}
and analogously for the valence part of the helicity-flip function $E^q$, with
obvious replacements everywhere of the Dirac $F_1^q$ by the Pauli $F_2^q$ form
factor. Note however that the $x$-dependence of $E^q$ is not constrained by
the forward parton density $f^q$ so this is an ad hoc assumption. The presence
of the factor $1/2$ in the $u$-quark distribution in the proton is a consequence
of corresponding normalization of its form factor $F^u_1 (0) = 2$ and parton
density $\int dx \, f^{q, {\rm val}} (x) = 2$. The expression of the quark form
factors in terms of the measured proton and nucleon electromagnetic form factors
was given earlier in Section \ref{Spin12FFs}.

This factorized form cannot, however, be even remotely correct since as we
established in the previous section, at large values of the Feynman momentum
$x \to 1$, the $\Delta^2$-dependence must  become weaker. In other words,
the higher the Mellin moment of GPDs, the flatter is the dependence on the
momentum transfer. Due to a limited kinematical coverage in the $t$-channel
momentum transfer $\Delta^2$ in experiments, theoretical estimates confronted
to data are currently insensitive to this feature, however.

A hint for a nontrivial and phenomenologically viable interplay between
the $x$ and $\Delta^2$ dependence arises from the Regge theory. Recall that
the high-energy dependence of, say, elastic $p_1 + p_2 \to p'_1 + p'_2$
scattering amplitudes is governed by linear Regge trajectories $\alpha_q
(\Delta^2) = \alpha_q + \alpha'_q \Delta^2$ with $\Delta = p_1 - p'_1$, so
that the physical cross section goes like $\sigma \sim s^{\alpha_q (\Delta^2)}$
as a function of the center-of-mass energy $s = (p_1 + p_2)^2$. In our
case the role of $s$ is taken by the (inverse) momentum fraction $x$ which
in the physical process is equal to the generalized Bjorken variable $x =
x_{\rm B} \sim s^{-1}$. Therefore, at small-$x$ one can take as a working
hypothesis for the non-forward parton distribution \cite{Goeke:2001tz,BelMul02}
$$
f^q (x, \Delta^2)
\equiv
H^q (x , \eta = 0, \Delta^2)
\stackrel{x \to 0}{\sim}
x^{- \alpha_q (\Delta^2)}
=
{\rm e}^{- \alpha_q (\Delta^2) \ln x}
\, .
$$
This behavior was recently refined in dedicated studies in Refs.\ \cite{Diehl:2004cx,GPRV}.
Note, that this form satisfies the general requirement that the $\Delta^2$-dependence
should disappear in the $x \to 1$ limit.  Furthermore, the relevant  form factor
is determined by the $x$-integral of $f^q (x, \Delta^2)$. The latter admits the form
\begin{equation}
\label{t-depPDF}
f^q (x, \Delta^2) \sim x^{- \alpha_q - \alpha'_q (1 - x)^p \Delta^2} (1 - x)^N
\, .
\end{equation}
The requirement of a controlled transverse size of the hadron and, therefore,
finitness of the spectator distance $\bit{r}^s_\perp$ (see Fig.\ \ref{SkewlessImpactPS})
at large-$x$ where
$$
f^q (x \to 1, \Delta^2)
\sim
{\rm e}^{\alpha_q (1 - x) + \alpha'_q (1 - x)^{p + 1}} (1 - x)^N
$$
requires $p \geq 1$ \cite{Burkardt:2004bv}. Notice, that phenomenologically,
the form with $p = 0$ and $\alpha' \sim 1 GeV^{- 2}$ works fine up to at least
$x \sim 0.5$ \cite{Diehl:1998kh}. The form factor asymptotics at large
$\Delta^2$ is governed by the large-$x$ behavior of $f^q (x, \Delta)$ and is
given by
$$
F (\Delta^2 \to \infty) \sim |\Delta^2|^{- (N + 1)/(p + 1)}
\, .
$$
The Drell-Yan relation requires $p = 1$ \cite{Drell:1969km} and is consistent
with the above observations. Thus Eq.\ (\ref{t-depPDF}) with $p=1$ provides a
function with a plausible interplay between longitudinal and transverse dynamics.

\subsubsection{D-term}

We have discussed so far only the first component of the two-component form
of GPDs which can be reconstructed from a non-singular double distribution
(\ref{RedFor-PW}). The second component---the $D$-term---is entirely concentrated
inside the central region $|x| \leq \eta$. Originally, it was inspired by the
chiral quark-soliton model analysis and was parametrized as \cite{PolWei99}
\begin{equation}
D (\alpha, \Delta^2 = 0)
=
(1 - \alpha^2) \sum_{n = 0}^\infty d_n \, C_{2 n + 1}^{3/2} (\alpha)
\, ,
\end{equation}
where the expansion goes in odd powers of Gegenbauer polynomials due to the
antisymmetry of the $D$-term, $D (- \alpha) = - D (\alpha)$. The first few
parameters were found within the model \cite{Petetal97,Goeke:2001tz} as well
as recently extracted from lattice simulations \cite{LHPC03,QCDSF03}. For the
lowest term, the results deduced from them are
\begin{equation}
d_0^{\chi\rm QSM} = - 4.0 \frac{1}{N_f}
\, , \qquad
d_0^{\rm latt} = d_0^u \approx d_0^d \approx - 0.5
\, ,
\end{equation}
where $N_f$ is the number of active flavors. Both predictions are plagued by
uncontrollable uncertainties. In the lattice case, the effect of disconnected
diagrams was not calculated. However they are known to produce a sizable
negative contribution \cite{GadJiJun02}. Once the latter are properly taken
into account the lattice result might approach the model calculation. For our
subsequent estimates we choose an intermediate value $d_0 = - 1.0$.

To account for the $\Delta^2$-dependence one can assume the simplistic form
\begin{equation}
D (\alpha, \Delta^2)
=
\left(1 - \frac{\Delta^2}{m_{\rm D}^2} \right)^{- 3}
D (\alpha, \Delta^2 = 0) \, ,
\end{equation}
with the mass scale $m_{\rm D}^2 = 0.6 \, {\rm GeV}^2$.

\subsubsection{Nonfactorizable GPD ans\"atze}
\label{NonfactorGPDModels}

Now, we are in a position to give semi-realistic models for GPDs. The $q$-flavor
double distribution will be decomposed into the valence and sea components, $f^q
= f^{q, {\rm val}} + f^{q, {\rm sea}}$. Both of them are related to the non-forward
quark distribution $f^q (\beta, \Delta^2)$ with a profile function $\pi$, as was
discussed in Section \ref{LongDynamics},
\begin{eqnarray}
\label{DDanzats}
&&
f^{q, {\rm val}} (\beta, \alpha, \Delta^2)
=
f^{q, \rm val} (\beta, \Delta^2) \theta (\beta)
\pi (|\beta|, \alpha; b_{\rm val})
\, , \\
&&
f^{q, {\rm sea}} (\beta, \alpha, \Delta^2)
=
\left(
\bar f^q (\beta, \Delta^2) \theta (\beta)
-
\bar f^q (- \beta, \Delta^2) \theta (- \beta)
\right) \pi (|\beta|, \alpha; b_{\rm sea})
\, ,
\end{eqnarray}
where at $\Delta^2 = 0$ the function $f^q (\beta, \Delta^2 = 0)$ reduces to the
conventional parton densities.

Below we give separately the parametrization for the unpolarized and polarized
non-forward parton distributions. They have to be plugged into Eq.\ (\ref{DDanzats})
and then used to reconstruct GPDs by means of Eq.\ (\ref{RedFor-wro}).

\begin{figure}[t]
\begin{center}
\mbox{
\begin{picture}(0,165)(60,0)
\put(0,0){\insertfig{5.3}{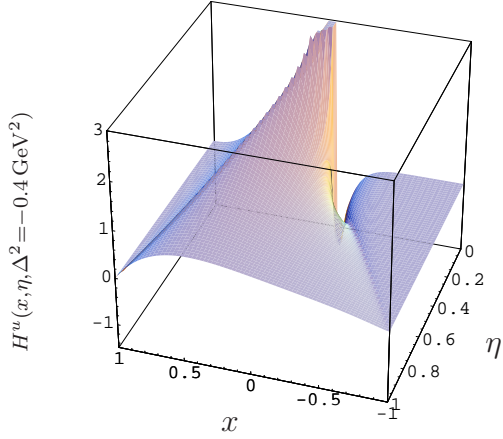}}
\put(50,0){$x$}
\put(150,30){$\eta$}
\put(-30,30){\rotate{$\scriptstyle H^u (x, \eta, \Delta^2 = - 0.4 \, {\rm GeV}^2)$}}
\end{picture}
}
\end{center}
\caption{\label{SampleUquarkGPD} The $u$-quark GPD at $\Delta^2 = - 0.4 \,
{\rm GeV}^2$ as a function of the Feynman momentum $x$ and skewness
$\eta$. Positive (negative) values of the distribution correspond to quark
(antiquarks) contributions to the GPD.}
\end{figure}

\begin{itemize}
\item Unpolarized GPDs. In our model we use the GRV leading order quark
distributions \cite{GluReyVog98} with discarded flavor asymmetry of the sea.
Then the nonforward distributions read
\begin{eqnarray}
\label{NonfactorGPD}
f^{u, \rm val} (\beta, \Delta^2)
\!\!\!&=&\!\!\!
1.239 \beta^{- \alpha_v - \alpha'_v (1 - \beta)^{1/2} \Delta^2}
\left( 1 - 1.8 \sqrt{\beta} + 9.5 \beta \right) (1 - \beta)^{2.72}
\, , \\
f^{d, \rm val} (\beta, \Delta^2)
\!\!\!&=&\!\!\!
0.761 \beta^{- \alpha_v - \alpha'_v (1 - \beta)^{1/2} \Delta^2}
\left( 1 - 1.8 \sqrt{\beta} + 9.5 \beta \right) (1 - \beta)^{3.62}
\, , \nonumber\\
\bar f^{u} (\beta, \Delta^2)
\!\!\!&=&\!\!\!
\bar f^{d} (\beta, \Delta^2)
=
0.76 \beta^{- \alpha_s - \alpha'_s (1 - \beta)^{3/2} \Delta^2}
\left( 1 - 3.6 \sqrt{\beta} + 7.8 \beta \right) (1 - \beta)^{9.1}
\, . \nonumber
\end{eqnarray}
They are assumed to be defined at the same renormalization scale as the forward quark
distributions used for their construction, i.e., $\mu^2_{\rm\scriptscriptstyle LO} =
0.26 \, {\rm GeV}^2$. These models closely reproduce the quark form factors with the
dipole parametrization of the proton and neutron Sachs form factors in the low
$|\Delta^2| < 3 \, {\rm GeV}^2$ region. The Regge intercepts and slope parameters
are taken to be
\begin{eqnarray}
&&\alpha_v = 0.52 \, , \qquad \alpha'_v = 1.1 \, {\rm GeV}^{-2}
\, , \\
&&\alpha_s = 0.85 \, , \qquad \alpha'_s = 0.3 \, {\rm GeV}^{-2}
\, . \nonumber
\end{eqnarray}
The Regge parameters for the valence quarks are numerically close to those of the
$\rho$-reggeons. Since the sea quarks are generated by the gluon radiation, their
Regge parameters are analogous to that of the pomeron. We use $p = 1/2$
exponentials instead of $p = 1$ dictated  by the Drell-Yan relation, since this
value fits better the form factor at small and moderate $\Delta^2$. For $p = 1$, one
can get a reasonable fit at moderate $\Delta^2$ with $\alpha'_u = 1.6 \, {\rm GeV}^2$.
For the $\alpha$-profile of the DDs we use in our estimates $b_{\rm val} = b_{\rm sea}
= 1$, though for the sea-quark distribution this value is not a consequence of
asymptotic considerations, rather $b_{\rm sea} = 2$ is a legitimate choice
\cite{Diehl:2003ny} (see Eqs.\ (136) and (255) there). The model of the $u$-quark
GPD at $\Delta^2 = - 0.4 \, {\rm GeV}^2$ is demonstrated in Fig.\ \ref{SampleUquarkGPD}.
\item Polarized GPDs. The model is quite analogous to the unpolarized case:
\begin{equation}
\label{NonfactorGPDpol}
{\mit\Delta} f^q (\beta, \Delta^2)
=
\eta_q A_q x^{a_q - \alpha'_q \Delta^2 (1 - \beta)}(1 - \beta)^{b_q}
\left( 1 + \gamma_q \beta + \rho_q \sqrt{\beta} \right)
\, .
\end{equation}
The parameters are fixed by the GSA forward densities \cite{GehSti96} at $\mu^2
= 4 \, {\rm GeV}^2$ in the $\Delta^2 = 0$ limit, and the   slopes  $\alpha'_u =
1.15 \, {\rm GeV}^{-2}$, $\alpha'_d = 1.0 \, {\rm GeV}^{-2}$ were chosen to fit
the dipole form of the axial form factor at low $|\Delta^2| < 3 \, {\rm GeV}^2$
with the effective mass $m_A^2 = 0.9 \, {\rm GeV}^2$ for the valence GPD. For the
sea-quark distribution, we take $\alpha'_s = 0.3 \, {\rm GeV}^{-2}$.

\end{itemize}

Let us emphasize that the models we have developed here serve solely an
illustration purpose. More realistic GPDs have to be developed by means of
combined fits to available data in different kinematical regions as was done
in recent works \cite{Diehl:2004cx,GPRV}. Presently, we address only quarks. The
spatial distribution of gluons was studied in Refs.\ \cite{StrWei03,FraStrWei04}.

\subsection{Visualizing proton via Wigner distributions}

\begin{figure}[t]
\begin{center}
\mbox{
\begin{picture}(0,270)(230,0)
\put(0,150){\insertfig{4.0}{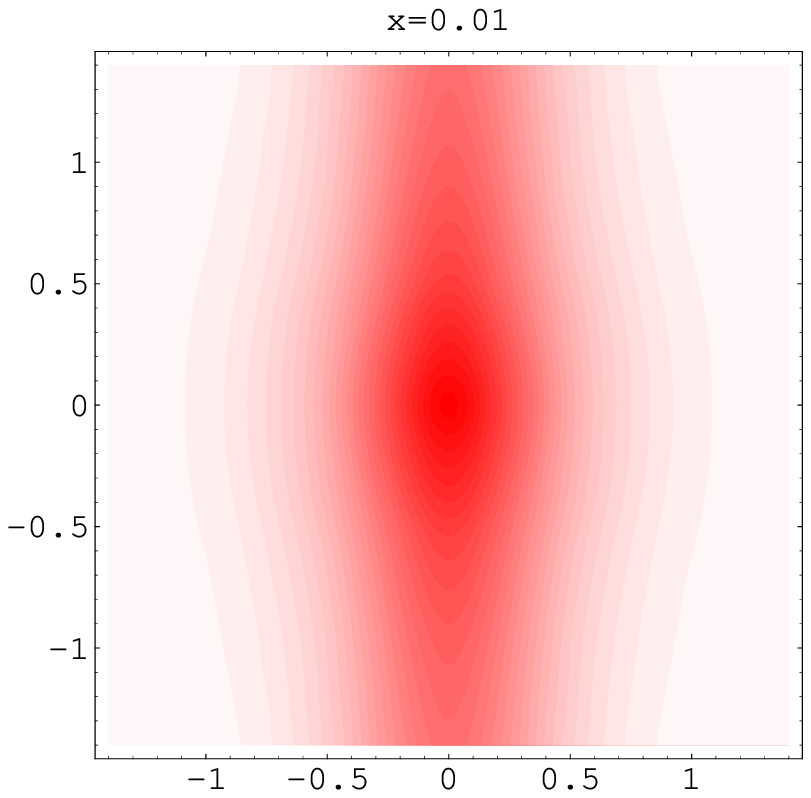}}
\put(170,150){\insertfig{4.0}{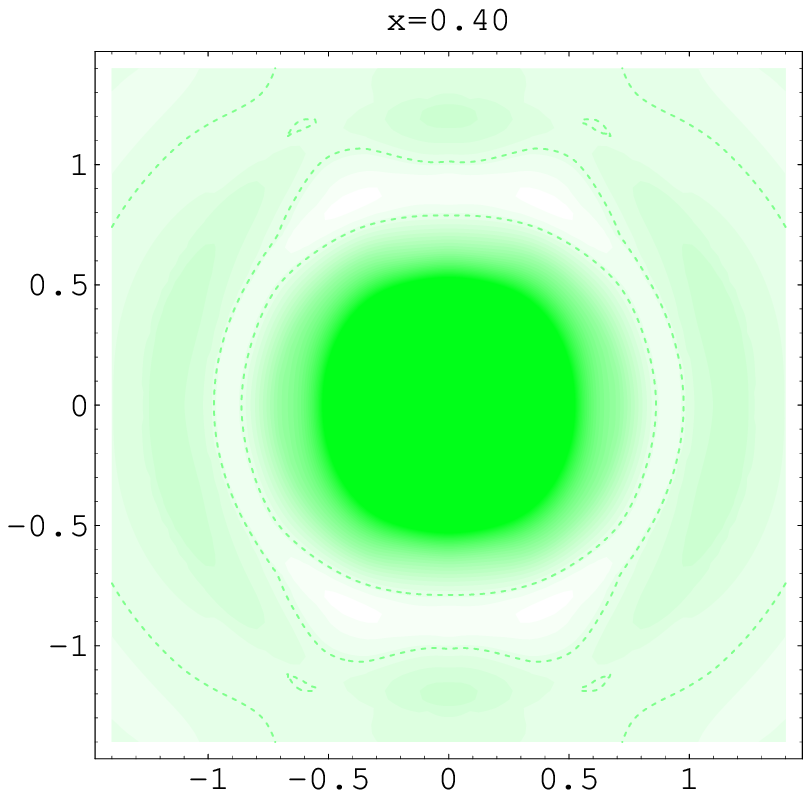}}
\put(340,150){\insertfig{4.0}{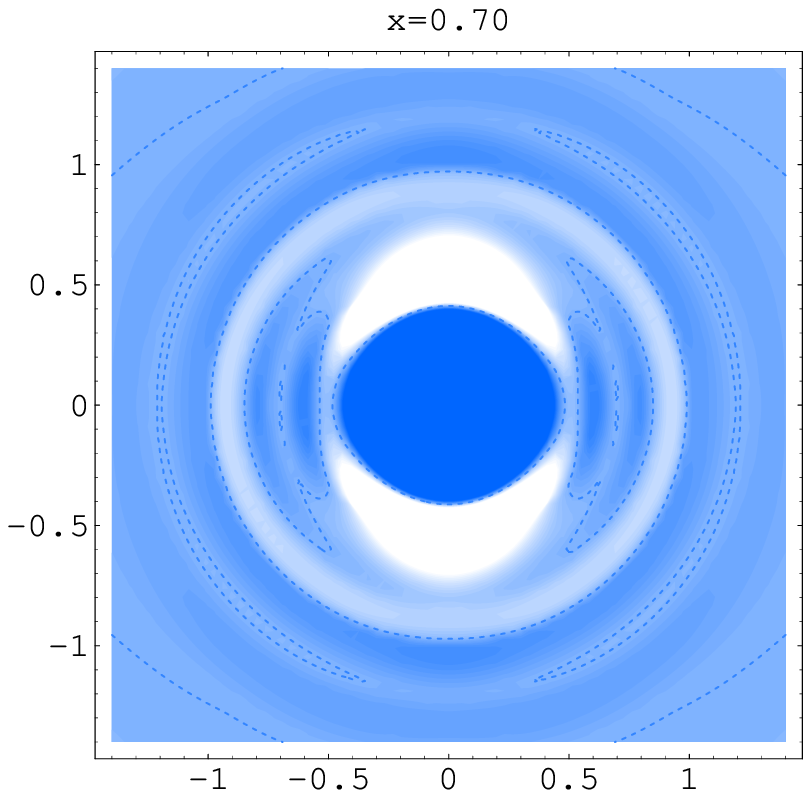}}
\put(0,0){\insertfig{4.0}{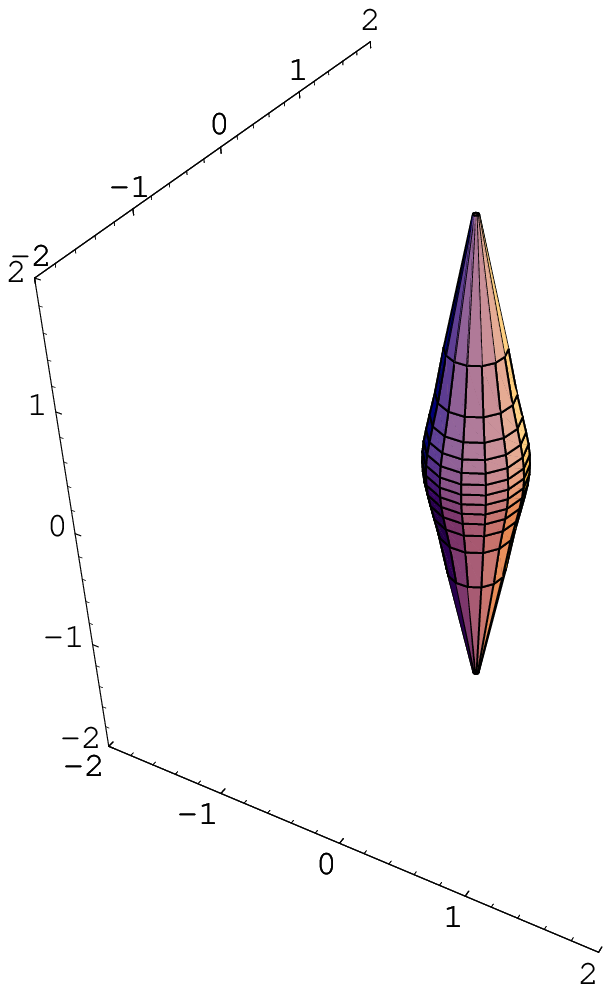}}
\put(170,0){\insertfig{4.0}{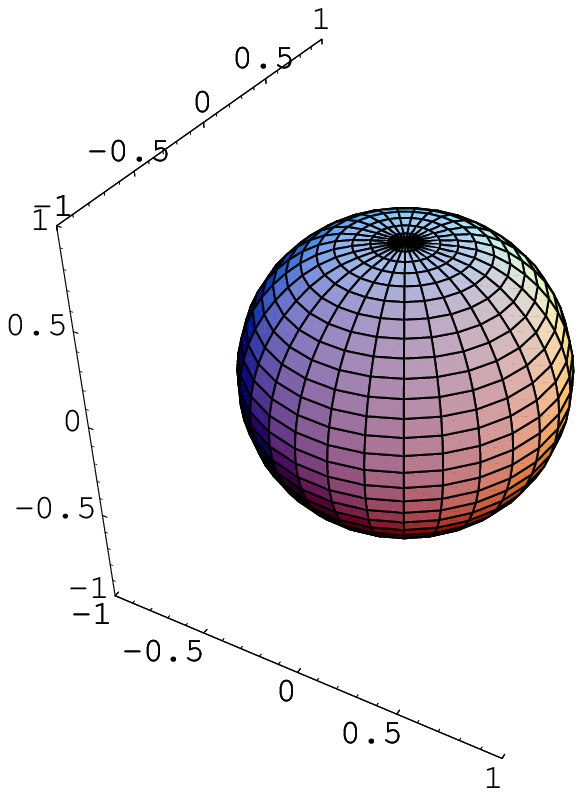}}
\put(340,0){\insertfig{4.0}{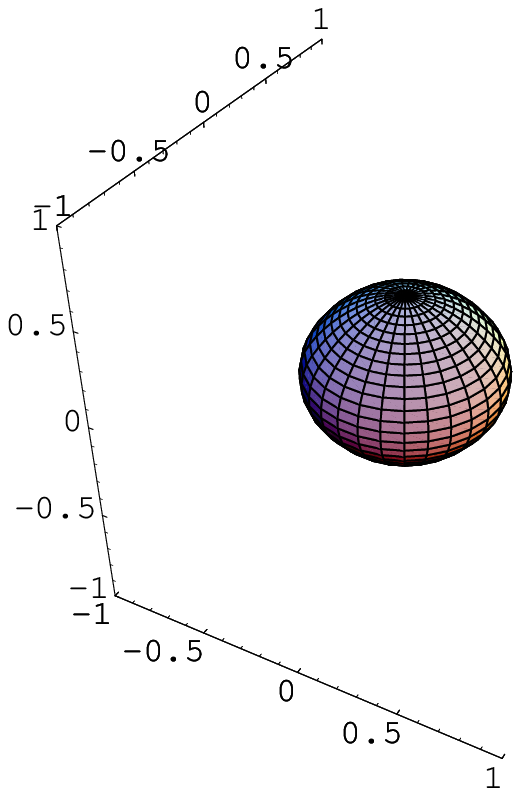}}
\end{picture}
}
\end{center}
\caption{\label{uQuarkNF} The $u$-quark Wigner distribution $W^u$ at different
values of the Feynman parameter for non-factorizable ansatz of generalized
parton distributions (\protect\ref{NonfactorGPD}). The vertical and horizontal
axis corresponds to $r^z$ and $|\bit{r}_\perp|$, respectively, measured in
femtometers. The [dashed] contours separate regions of positive [darker areas]
and negative [lighter areas] densities. Below each contour plot we presented the
shape of three-dimensional isodensity contours [$W^u = {\rm const}$].}
\end{figure}

Making use of the above models we can plot the phase-space quasi-probability
distribution of partons inside the proton, following Ref.\ \cite{BelJiYua03}.
According to our earlier discussion, the Wigner function of a quark $W^q
(x , \bit{r})$ is obtained from GPDs in the Breit frame through the
three-dimensional Fourier transformation with respect to the three-momentum
$\bit{\Delta}$ (see Section \ref{WignerDistrQuarks}):
$$
W^q (x, \bit{r})
=
\int \frac{d^3\bit{\Delta}}{(2\pi)^3}
{\rm e}^{i \bit{\scriptstyle \Delta} \cdot \bit{\scriptstyle r}}
H^q
\left(
x, \Delta_z/\sqrt{\bit{\Delta}{\!}^2 + 4 M^2_N}, - \bit{\Delta}^2
\right)
\, .
$$
We consider the quark Wigner quasi-probabilities resulting from the parametrization
for the GPDs given above.  In Fig.\ \ref{uQuarkNF} we show the up-quark Wigner
distributions calculated from $H^u(x, \eta, \Delta^2)$ for a few values of the
Feynman momentum $x = \{ 0.01, 0.4, 0.7 \}$. The intensity of the plots indicates
the magnitude of the positive density distribution. The lighter areas indicate
negative values, which are divided by the level zero, shown by dashed contours.
The plots exhibit significant dependence of the quark distribution on the
longitudinal momentum fraction $x$. The image is rotationally symmetric in the
transverse $\bit{r}_\perp$-plane. At small $x$, the distribution extends far
beyond the nominal nucleon size along the longitudinal $z$ direction. The physical
explanation for this is that the position space uncertainty of the quarks is large
when $x$ is small, and therefore the quarks are de-localized along the longitudinal
direction. This de-localization reflects the part of the nucleon wave function, and
shows long-range correlations as verified in high-energy scattering. On the other
hand, at larger $x$, the momentum along the $z$ direction is of order of nucleon
mass, and the quarks are localized to within $1/M_N$.

\subsection{Limitations of GPDs as Wigner distributions}
\label{GPDsAsWigner}

\begin{figure}[t]
\begin{center}
\mbox{
\begin{picture}(0,115)(100,0)
\put(0,-15){\insertfig{6}{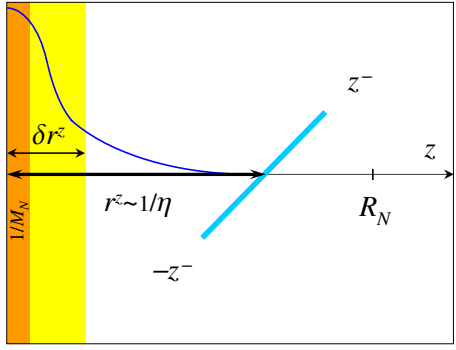}}
\end{picture}
}
\end{center}
\vspace{1cm}
\caption{\label{WignerMomentumScales} Distance scales for the Wigner
distribution in the Breit frame. The Compton wavelength and the radius of
the proton are $1/M_N$ and $R_N$, respectively. The partons are probed at
longitudinal distances $r^z$ in the proton localized by means of a wave
packet of the size $\delta r^z$ with a probe whose non-locality is $z^-$.}
\end{figure}

Let us analyze whether our interpretation of GPDs as Fourier transforms of
Wigner distributions receives any limitations from relativity and quantum
effects. To this end, consider the nucleon matrix element of the light-ray
operator
$$
\langle p_2 | {\cal O}^{qq} ( - z^- , z^- ) | p_1 \rangle
\, .
$$
Since we are interested in the Breit-frame interpretation of GPDs, we should
impose restrictions on momenta involved, similar to the ones used in discussion
of form factors, in order to avoid dangerous regions where relativistic and
quantum effects will spoil the interpretation. We construct the same wave packet
as in the analysis of the nucleon form factors in Section \ref{BreitFrameFFs}.
This allows one to find a sequence of inequalities between the longitudinal
$\Delta^z$ and transverse $\bit{\Delta}_\perp$ resolution scales of the probe
and typical momenta $p^z$, $\bit{p}_\perp$ of the wave packet as in Eq.\
(\ref{CombinedConstraint}). A straightforward consideration in one-to-one
correspondence with the one performed in Section \ref{BreitFrameFFs} yields
inequalities for the transverse momenta
\begin{equation}
1/R_N \ll |\bit{\Delta}_\perp| \ll |\bit{p}_\perp| \ll M_N
\, .
\end{equation}
The consideration for longitudinal variables is nontrivial, since the probe is
non-local in the light-like direction. Thus, the following three longitudinal
distance scales are involved (see Fig.\ \ref{WignerMomentumScales}). The
longitudinal position of partons $r^z$ is set in the Breit frame by the skewness
$r^z \sim 1/\Delta^z \sim 1/(\eta E)$ (see Eq.\ (\ref{SkewnessInBreit})). A typical
longitudinal momentum in the wave packet sets its intrinsic fuzziness $\delta r^z
\sim 1/p^z$. Finally, since the probe is non-local, we have an extra degree of
freedom---the Ioffe time $z^-$. The latter is a variable Fourier conjugate to the
the Feynman momentum $x = k^+/p^+$ (which has the meaning of the parton momentum
relative to that of the parent hadron in the infinite momentum frame as was
established in Section \ref{PDFsBjorkenFrame}). The non-locality grows with
decreasing parton momentum fraction $x$, $z^- \sim 1/(x E)$. This correlation
length is set by the distance which the quark-antiquark pair, originating from
the decay of the probe, propagates in the Breit frame from the space-time point
of its creation to recombination back into the off-shell gauge boson. So $z^-$
can exceed the nucleon size.

First, we have to analyze consequences of the uncertainty on the longitudinal
localization of the proton state. Obviously, to be insensitive to the structure
of the residual wave packet itself one has to set the condition that its fuzziness
$\delta r^z$ be much smaller that the resolving distances $r^z$ within the hadron,
$\delta r^z \ll r^z$. This translates into the condition in the reciprocal
(momentum) space
\begin{equation}
| \Delta^z | \ll | p^z | \ll M_N
\, ,
\end{equation}
with the right-most inequality being the limitation due to the finite Compton wavelength
of the nucleon. Next, the non-locality of the probe in the longitudinal direction $z^-$
cannot be larger than the distances $r^z$ it scans,
\begin{equation}
|z^-| \ll r^z
\, .
\end{equation}
When one makes the Fourier transformation of the above matrix element with respect
to the Ioffe time, one has to impose this restriction on the light-cone distances.
This immediately implies that the Feynman momentum $x$ probed in the proton cannot
be smaller than $1/r^z$, i.e., $x \gg 1/(E r^z) \sim \eta$. If this condition is
omitted, there will be a quantum uncertainty arising from the fuzziness of the picture
for very large Ioffe times. This consideration results in a limitation on the static
picture of GPDs (in the Breit frame) as Fourier transforms of Wigner distributions.
GPDs admit this interpretation only in a limited region of the parameter space, as
we have just established.

The unpolarized nucleon GPD in the Breit frame is given by \cite{BelJi02}
\begin{equation}
\label{GPD}
\frac{1}{2 M_N} F^q (x, \eta, \Delta^2) =
\left( H^q  - \frac{\bit{\Delta}^2}{4 M_N^2} E^q \right) (x, \eta, \Delta^2)
+
\frac{i}{M_N} [\bit{\Delta} \times \bit{s} ]^z
\left( H^q  + E^q \right) (x, \eta, \Delta^2)
\, ,
\end{equation}
where, for brevity, we set $w^\ast_2 w_1 = 1$ and $\bit{s}$ is defined as in Eq.\
(\ref{FFnormalization}). The physical content of the functions involved is
revealed by examining their spin structure. The matrix element in Eq. (\ref{GPD})
demonstrates that the first term is independent of the proton spin, and is
considered as the charge density
\begin{equation}
\rho (x, \bit{r})
=
\int
\frac{d^3 \bit{\Delta}}{(2\pi)^3} \,
{\rm e}^{i \bit{\scriptstyle \Delta} \cdot \bit{\scriptstyle r}}
\left( H - \frac{\bit{\Delta}^2}{4 M_N^2} E \right) (x, \eta, \Delta^2)
\, .
\end{equation}
The second term depends on the proton spin and can be regarded as
the third component of the vector current
\begin{eqnarray}
j^z (x, \bit{r})
=
\frac{i}{M_N}
\int
\frac{d^3 \bit{\Delta}}{(2\pi)^3} \,
{\rm e}^{i \bit{\scriptstyle \Delta} \cdot \bit{\scriptstyle r}} \,
[\bit{\Delta} \times \bit{s}]^z
\left( H + E \right) (x, \eta, \Delta^2)
\, .
\end{eqnarray}
The $E$-term generates a convection current due to the orbital angular momentum
of massless quarks and vanishes when all quarks are in the $s$-orbit. The physics
in separating $F^q$ into $\rho$ and $j^z$ can be seen from the Dirac matrix
$\gamma^+$ selected by the high-energy probes, which is a combination of time and
space components. Because the current distribution has no spherical symmetry, the
quark density seen in the infinite momentum frame, $\rho + j^z$, is deformed in
the impact parameter space \cite{Burkardt:2000za}. This is the kinematic effect
of Lorentz transformations.

The Wigner distributions introduced above allow one to calculate the single-quark
properties using the classical concept  of average in the phase space. If one
integrates over the momentum fraction $x$ in $\rho (x, \bit{r})$, one recovers
the spherical charge distribution. On the other hand,  integrating over $x$ in
$j^z (x, \bit{r})$  one obtains the electric current distribution. In the latter
case, if the integral is weighted with $x$, one gets  the mechanical momentum
density.

\subsection{Transition GPDs}
\label{FlavorChangingGPDs}

So far we have discussed GPDs involving the same hadron in the initial and final
state. There is, however, a number of reactions which change the flavor of partons
during the hard scattering.  This could be either due to weak interactions, or if
the original quark goes into a flavored meson, so that a quark with a different
flavor joins the outgoing baryon in the final state. In the present section, we
address these quantities and also show how one can relate them to the diagonal
ones making use of some approximations. Obviously, since there is a flavor
exchange between the incoming and outgoing baryons, there are no gluonic GPDs
associated with these transitions.

Contrary to the diagonal case, there are no positivity constraints on the
transition GPDs, since they do not become densities in any limit, like the
forward limit for the  diagonal GPDs. The transition GPDs rather describe
the correlations between the wave function of the resonance and that of the
nucleon.  Throughout our discussion, we will assume that resonances are stable
hadrons,  and thus the corresponding GPDs do not contain the final-state
strong-interaction phases. An alternative description would be in terms of
the hadrons actually  observed in the detector. The relevant GPD
$$
\langle H_1 H_2 | {\cal O}^{qq'} | p \rangle
\, ,
$$
describes the transition from the initial state nucleon (say, the proton $p$)
to the final state hadrons $H_1,H_2$ with a given invariant mass close to that
of the resonance $H$ that decays into them. In this case, the imaginary
phases have to be introduced.

\subsubsection{Baryon octet to octet transitions}

Let us start with discussion of matrix elements describing transitions between
hadrons within the baryon octet. They are very interesting phenomenologically
since they involve matrix elements of operators containing the strange quark.
For instance, matrix elements of the even and odd parity light-ray operators
for the transition $p \to {\mit\Sigma}^+$ can be  decomposed as
\begin{eqnarray}
\langle {\mit\Sigma}^+ (p_2) |
{\cal O}^{sd} (- z^-, z^-)
| p (p_1) \rangle
\!\!\!&=&\!\!\!
\int_{- 1}^{1} d x \ {\rm e}^{- i x p^+ z^- }
\left\{
h^+ H_{p {\mit\Sigma}^+} (x, \eta, \Delta^2)
+
e^+ E_{p {\mit\Sigma}^+} (x, \eta, \Delta^2)
\right\}
\, , \nonumber\\
\langle {\mit\Sigma}^+ (p_2) |
\widetilde {\cal O}^{sd} (- z^-, z^-)
| p (p_1) \rangle
\!\!\!&=&\!\!\!
\int_{- 1}^{1} d x \ {\rm e}^{- i x p^+ z^- }
\left\{
\tilde h^+ \widetilde H_{p {\mit\Sigma}^+} (x, \eta, \Delta^2)
+
\tilde e^+ \widetilde E_{p {\mit\Sigma}^+} (x, \eta, \Delta^2)
\right\}
\, , \nonumber\\
\end{eqnarray}
where the flavor structure of the operators should be understood as
${\cal O}^{sd} = \bar s \gamma^+ d$, etc. The first moments of these
functions are related to the vector and axial transition form factors
within the baryon octet \cite{Goeke:2001tz}
\begin{eqnarray}
\int_{-1}^1 dx \, H_{p {\mit\Sigma}^+} (x, \eta, \Delta^2)
\!\!\!&=&\!\!\!
F_1^{p {\mit\Sigma}^+} (\Delta^2)
-
\eta
F_3^{p {\mit\Sigma}^+} (\Delta^2)
\, , \\
\int_{-1}^1 dx \, E_{p {\mit\Sigma}^+} (x, \eta, \Delta^2)
\!\!\!&=&\!\!\!
F_2^{p {\mit\Sigma}^+} (\Delta^2)
+
\eta F_3^{p {\mit\Sigma}^+} (\Delta^2)
\, , \\
\int_{-1}^1 dx \, \widetilde H_{p {\mit\Sigma}^+} (x, \eta, \Delta^2)
\!\!\!&=&\!\!\!
G_A^{p {\mit\Sigma}^+} (\Delta^2)
+
\frac{M_{{\mit\Sigma}^+} - M_N}{M_{{\mit\Sigma}^+} + M_N}
G_{\rm eff}^{p {\mit\Sigma}^+} (\Delta^2)
\, , \\
\int_{-1}^1 dx \, \widetilde E_{p {\mit\Sigma}^+} (x, \eta, \Delta^2)
\!\!\!&=&\!\!\!
G_P^{p {\mit\Sigma}^+} (\Delta^2)
+
\frac{1}{\eta}
G_{\rm eff}^{p {\mit\Sigma}^+} (\Delta^2)
\, .
\end{eqnarray}
Compared to the parametrization of the matrix elements of the conserved vector
(\ref{vectorFFs}) and axial currents (\ref{axialFFs}), there are two extra
form factors, the weak electricity term $F_3$ in the vector current case
\begin{equation}
\langle {\mit\Sigma}^+ (p_2) | \bar s \gamma^\mu d (0) | p (p_1) \rangle
=
h^\mu F^{p {\mit\Sigma}^+}_1 ( \Delta^2)
+
e^\mu F^{p {\mit\Sigma}^+}_2 (\Delta^2)
-
\frac{b \Delta^\mu}{M_{{\mit\Sigma}^+} + M_N} F^{p {\mit\Sigma}^+}_3 (\Delta^2)
\, ,
\end{equation}
and the effective pseudoscalar term $G_{\rm eff}$ in the axial one,
\begin{equation}
\langle {\mit\Sigma}^+ (p_2) | \bar s \gamma^\mu \gamma^5 d (0) | p (p_1) \rangle
=
\widetilde h^\mu G^{p {\mit\Sigma}^+}_A ( \Delta^2)
+
\widetilde e^\mu G^{p {\mit\Sigma}^+}_P (\Delta^2)
+
\frac{
\widetilde t^{\nu\mu} \Delta_\nu
}{M_{{\mit\Sigma}^+} + M_N} G^{p {\mit\Sigma}^+}_{\rm eff} (\Delta^2)
\, .
\end{equation}
The parametrization is given in terms of the same bilinears as before, however,
with the outgoing wave function one corresponding to the hyperon. Both the
$F_3$ and $G_{\rm eff}$ terms appear due to the violation of the exact flavor
symmetry by the non-zero strange quark mass. They can be safely neglected
otherwise, of course, with a very high confidence for the strageness-free
transitions.

Now we are in a position to use the $SU(3)$ symmetry in order to relate all
transition GPDs within the octet to those of the diagonal generalized quark
distributions in the proton. The Lagrangian of the octet baryon coupling has
been given before in Eq.\ (\ref{EightToEightLagrange}). The only replacement
we have to make  there is to treat $M^a{}_b$ as an octet of non-local light-ray
quark operators,
\begin{equation}
\label{OctetLRoperators}
M^{q'}{}_q
=
{\cal O}^{q q'} (- z^- , z^-)
-
\frac{1}{3} \delta^{q q'}
\sum_{q''}
{\cal O}^{q''q''} (- z^- , z^-)
\, ,
\end{equation}
and similarly for the odd parity and maximal-helicity operators---the Dirac
structure is irrelevant. By the same token as in Section \ref{AxialCouplingSU3},
we can read off the strength of interaction in terms of the ``constants'' $D$
and $F$, see Eq.\ (\ref{ExampleUDtransition}). Removing them from the resulting
relations, we can re-express all the transition GPDs in terms of the $q$-flavor
quark distribution in the proton \cite{Goeke:2001tz}
\begin{eqnarray}
\label{EightToEightGPDs}
&&
H_{p n} = H^u - H^d
\, ,
\phantom{\frac{1}{\sqrt{2}}} \\
&&
H_{p {\mit\Lambda}} = - \frac{1}{\sqrt{6}} \left( 2 H^u - H^d - H^s \right)
\, , \qquad
H_{n {\mit\Lambda}} = - \frac{1}{\sqrt{6}} \left( 2 H^d - H^u - H^s \right)
\, , \nonumber\\
&&
H_{p {\mit\Sigma}^+} = - H^d + H^s
\, ,
\phantom{\frac{1}{\sqrt{2}}} \qquad\qquad\quad \ \ \,
H_{n {\mit\Sigma}^+} = - H^u + H^s
\, , \nonumber\\
&&
H_{p {\mit\Sigma}^0} = - \frac{1}{\sqrt{2}} \left( H^d - H^s \right)
\, , \qquad\qquad\ \ \,
H_{n {\mit\Sigma}^0} = - \frac{1}{\sqrt{2}} \left( H^u - H^s \right)
\, . \nonumber
\end{eqnarray}
The quark content of GPDs on the left-hand side of the equation are given
by flavor-changing quark operators ${\cal O}^{q' q}$. Obviously, these are
the same relations as those for the octet decay constants, addressed in
Section \ref{ProtonSpinPuzzle}. These relations are expected to hold to
reasonable accuracy, except for the distributions $\widetilde E$, where the
presence of the psedoscalar-meson pole induces a strong violation of the
flavor $SU (3)$ symmetry due to significant difference of the meson masses
within the octet.

\subsubsection{Baryon octet to decuplet transitions}

The transition amplitudes between the nucleon and the $\Delta$ isobar have
been under an intensive scrutiny for a long time because, among other things,
they allowed to address the question of the nucleon and its first resonance
``deformation''. The selection rules, for instance, for the electromagnetic
decay $\Delta \to N \gamma^\ast$ allow, apart from the magnetic, also for the
electric and the Coulomb quadrupole transitions. The latter two vanish in a
naive models with spherical symmetry, hence the term. In this sense, the
electric and Coulomb quadrupole  form factors quantify the amount of the
deformation and its dependence on the radial distance from the center-of-mass.

The ``nucleon-to-delta'' generalized parton distributions bring much more into
quantifying the precise partonic dynamics of these transitions, describing their
dependence with respect to a number of kinematical variables they depend upon.
According to our analysis in Section \ref{CountingGPDs}, there are eight
independent quark-helicity conserving transitions, hence, there will be eight
independent quark GPDs. Let us define the quark matrix elements for, say, the
transition $p \to \Delta^{++}$ and determine the rest by making use of the $SU(3)$
flavor symmetry,  as was done in the previous section for transitions within the
baryon octet. There are four GPDs for the even parity operators\footnote{Notice
that our parametrization differs from the one in Ref.\ \cite{Goeke:2001tz}
in the number of independent GPDs---we have more by one in the parity-even
operator matrix element to match the number of helicity amplitudes involved in
the transition---as well the basis of tensor structures used in the decomposition.}
\begin{eqnarray}
\langle \Delta^{++} (p_2) |
{\cal O}^{ud} (- z^- , z^-)
| p (p_1) \rangle
\!\!\!&=&\!\!\!
\int_{- 1}^1 dx \, {\rm e}^{- i x p^+ z^-}
\bar u_\nu (p_2)
\Bigg\{
\frac{ \Delta^\mu n^\nu - \Delta^\nu n^\mu }{M_N}
\bigg(
\gamma_\mu
G_1 (x, \eta, \Delta^2)
\nonumber\\
+
\frac{p_\mu}{M_N} G_2 (x, \eta, \Delta^2)
\!\!\!&-&\!\!\!
\frac{\Delta_\mu}{M_N} G_3 (x, \eta, \Delta^2)
\bigg)
+
\frac{\Delta^+ \Delta^\nu}{M_N^2} G_4 (x, \eta, \Delta^2)
\Bigg\}
\gamma^5 u (p_1)
\, ,
\end{eqnarray}
and four for the odd parity case
\begin{eqnarray}
\label{OddParityDelta}
\langle \Delta^{++} (p_2) |
\widetilde {\cal O}^{ud} (- z^- , z^-)
| p (p_1) \rangle
\!\!\!&=&\!\!\!
\int_{- 1}^1 dx \, {\rm e}^{- i x p^+ z^-}
\bar u_\nu (p_2)
\Bigg\{
\frac{ \Delta^\mu n^\nu - \Delta^\nu n^\mu }{M_N}
\bigg(
\gamma_\mu \widetilde G_1 (x, \eta, \Delta^2)
\nonumber\\
+
\frac{p_\mu}{M_N} \widetilde G_2 (x, \eta, \Delta^2)
\bigg)
\!\!\!&+&\!\!\!
n^\nu \widetilde G_3 (x, \eta, \Delta^2)
+
\frac{\Delta^+ \Delta^\nu}{M_N^2} \widetilde G_4 (x, \eta, \Delta^2)
\Bigg\}
u (p_1)
\, .
\end{eqnarray}
Here, $u^\mu$ is the Rarita-Schwinger spinor, whose properties are summarized
in Appendix \ref{RSchSpinors}. The $\gamma^5$  matrix  was added to match the
negative parity of the $\Delta^{++}$.  In the local limit,  this  parametrization
reduces
\begin{equation}
\int_{-1}^1 dx \, G_i (x, \eta, \Delta^2) = G_i (\Delta^2)
\, , \qquad
\int_{-1}^1 dx \, \widetilde G_i (x, \eta, \Delta^2) = \widetilde G_i (\Delta^2)
\, ,
\end{equation}
to the $p \to \Delta^{++}$ transition form factors \cite{Adl68,JonSca72,DevEisKor76},
\begin{eqnarray}
\langle \Delta^{++} (p_2) |
\bar u \gamma^\mu d (0)
| p (p_1) \rangle
\!\!\!&=&\!\!\!
u_\nu (p_2)
\Bigg\{
\frac{ \Delta^\rho g^{\mu\nu} - \Delta^\nu g^{\mu\rho} }{M_N}
\bigg(
\gamma_\rho
G_1 (\Delta^2)
+
\frac{p_\rho}{M_N} G_2 (\Delta^2)
\\
&&\qquad\qquad\qquad\qquad\qquad
-
\frac{\Delta_\rho}{M_N} G_3 (\Delta^2)
\bigg)
+
\frac{\Delta^\mu \Delta^\nu}{M_N^2} G_4 (\Delta^2)
\Bigg\}
\gamma^5 u (p_1)
\, , \nonumber\\
\langle {\mit\Delta}^{++} (p_2) |
\bar u \gamma^\mu \gamma^5 d (0)
| p (p_1) \rangle
\!\!\!&=&\!\!\!
u_\nu (p_2)
\Bigg\{
\frac{ \Delta^\rho g^{\mu\nu} - \Delta^\nu g^{\mu\rho} }{M_N}
\left(
\gamma_\rho \widetilde G_1 (\Delta^2)
+
\frac{p_\rho}{M_N} \widetilde G_2 (\Delta^2)
\right)
\nonumber\\
&&\qquad\qquad\qquad\qquad\qquad
+
g^{\mu\nu} \widetilde G_3 (\Delta^2)
+
\frac{\Delta^\mu \Delta^\nu}{M_N^2} \widetilde G_4 (\Delta^2)
\Bigg\}
u (p_1)
\, . \nonumber
\end{eqnarray}
The partial conservation of the vector and the axial currents implies that
\begin{equation}
G_4 \stackrel{\scriptscriptstyle\rm PCVC}{\approx} 0
\, , \qquad
\widetilde G_4
\stackrel{\scriptscriptstyle\rm PCAC}{\approx}
\frac{M_N^2}{m_\pi^2 - \Delta^2} \widetilde G_3
\, .
\end{equation}
However, for the  transitions in which $s$-quark is involved, the symmetry
is broken by its mass. This breaking is fully responsible for nonvanishing
$G_4^{(q \to s)} \sim m_s$.

In case of the electromagnetic current, the form factors $G_i$ are related
to the standard magnetic $M1$, electric $E2$ and Coulomb $C2$ form factors
by the following set of relations \cite{DevEisKor76}
\begin{eqnarray}
\label{MagneticGandG1}
&&\!\!\!\!\!\!\!\!\!\!\!
6 M_N M_{\mit\Delta}(M_{\mit\Delta} + M_N)
\left(
\begin{array}{l}
G_{M1} \\
G_{E2} \\
G_{C2}
\end{array}
\right)
\\
&&\!\!\!\!\!\!\!
=
\left(
\begin{array}{ccc}
(M_{\mit\Delta} + M_N )(3 M_{\mit\Delta} + M_N ) - \Delta^2
&
2 (M_{\mit\Delta}^2 - M_N^2)
&
2 \Delta^2
\\
M_{\mit\Delta}^2 - M_N^2+ \Delta^2
&
2 (M_{\mit\Delta}^2 - M_N^2)
&
2 \Delta^2
\\
4 M_{\mit\Delta}^2
&
2 (3 M_{\mit\Delta}^2 + M_N^2 - \Delta^2)
&
2 (M_{\mit\Delta}^2 - M_N^2 + \Delta^2)
\end{array}
\right)
\!\!
\left(
\begin{array}{l}
M_N G_1 \\
M_{\mit\Delta} G_2 \\
M_{\mit\Delta} G_3
\end{array}
\right)
. \nonumber
\end{eqnarray}
Some of these transition form factors have been accessed by  experimental
measurements. A very concise and convenient compendium of their parametrizations
can be found in Refs.\ \cite{SigVicOse98,Sig03}.

Let us now derive  $SU(3)$ relations between transition GPDs. To this end,
we have to write down the most general $SU(3)$ invariant couplings of the
non-local light-ray operators (\ref{OctetLRoperators}) to octet and decuplet
baryons. A simple symmetry argument shows that such a Lagragian contains one
term since $\bit{8}$ appears only once in the product $\bit{8} \otimes \bit{10}
= \bit{8} \oplus \bit{10} \oplus \bit{27} \oplus \bit{35}$. It  reads
\begin{equation}
{\cal L}_{\bit{\scriptstyle 8} \to \bit{\scriptstyle 10}}
=
E
\left(
T^{abc} \varepsilon_{cde} M^d{}_a \bar{B}^e{}_b
+
{\rm h.c.}
\right)
\, ,
\end{equation}
where the components of the totally symmetric tensor $T^{abc}$ describing
the decuplet are identified with the particles as follows
(see \cite{ButSav92,Leb94})
\begin{equation}
T^{ijk}
=
\parbox[c][70mm]{70mm}{\insertfig{7}{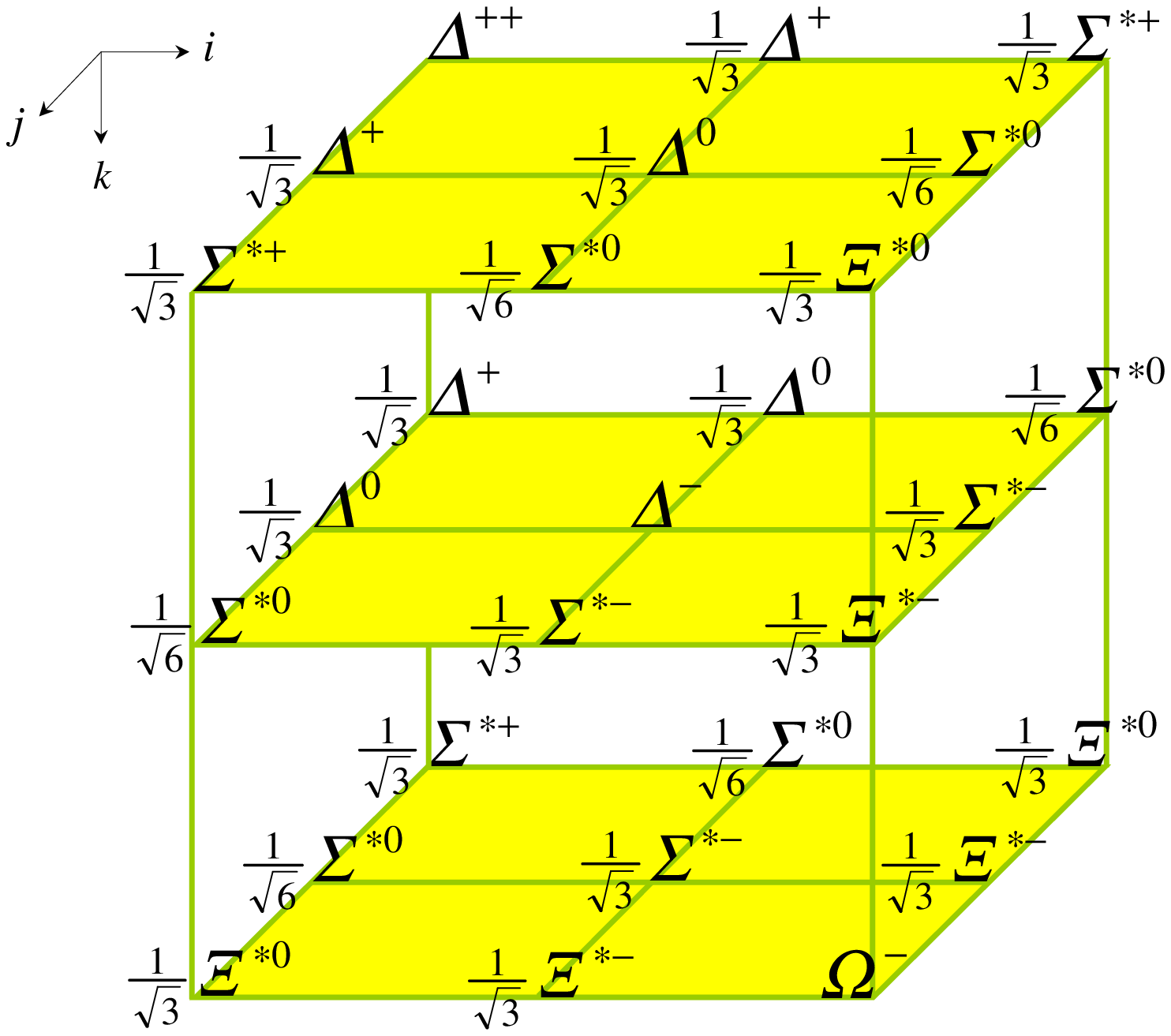}}
\, .
\end{equation}
For instance, $T^{111} = {\mit\Delta}^{++}$, $T^{122} =  \frac{1}{\sqrt{3}}
{\mit\Delta}^0$, etc. Expanding the invariant Lagrangian in the component
form, one finds for transitions involving the proton
\begin{equation}
{\cal L}_{\bit{\scriptstyle 8} \to \bit{\scriptstyle 10}}
=
E
\left(
{\mit\Delta}^{++} {\cal O}^{ud}
+
\frac{1}{\sqrt{3}} {\mit\Sigma}^{\ast +} {\cal O}^{sd}
-
\frac{1}{\sqrt{3}} {\mit\Delta}^0 {\cal O}^{du}
-
\frac{1}{\sqrt{6}} {\mit\Sigma}^{\ast 0} {\cal O}^{su}
-
\frac{1}{\sqrt{3}} {\mit\Delta}^+ ( {\cal O}^{uu} - {\cal O}^{dd} )
\right) \bar{p}
+
\, {\rm h.c.}
\, .
\end{equation}
Analogously,  one finds terms with the neutron. As usual,  the coefficients in
front of different interaction terms encode the strength of the transitions.
Following the same procedure as that outlined in Section \ref{OctetOctetOctet},
we find the following $SU(3)$ relations between matrix elements
\begin{eqnarray}
\langle {\mit\Delta}^{++} | {\cal O}^{ud} | p \rangle
\!\!\!&=&\!\!\!
- \sqrt{3}
\langle {\mit\Delta}^{0} | {\cal O}^{du} | p \rangle
= - \frac{\sqrt{3}}{2}
\langle {\mit\Delta}^+ | {\cal O}^{uu} - {\cal O}^{dd} | p \rangle
\nonumber\\
&=&\!\!\!
- \sqrt{6} \langle {\mit\Sigma}^{\ast 0} | {\cal O}^{su} | p \rangle
= \sqrt{3} \langle {\mit\Sigma}^{\ast +} | {\cal O}^{sd} | p \rangle
\, , \nonumber\\
&=&\!\!\!
- \langle {\mit\Delta}^- | {\cal O}^{du} | n \rangle
=
\sqrt{3}
\langle {\mit\Delta}^{+} | {\cal O}^{ud} | n \rangle
= - \frac{\sqrt{3}}{2}
\langle {\mit\Delta}^0 | {\cal O}^{uu} - {\cal O}^{dd} | n \rangle
\nonumber\\
&&\qquad\qquad\quad \ \
=
\sqrt{6} \langle {\mit\Sigma}^{\ast 0} | {\cal O}^{sd} | n \rangle
= - \sqrt{3} \langle {\mit\Sigma}^{\ast -} | {\cal O}^{su} | n \rangle
\, .
\end{eqnarray}
These relations hold irrespective to all other quantum numbers assigned to
bilocal quark operators,  and they provide  relations between GPDs. For instance,
the $p \to \Delta^{++, +, 0}$ transition GPDs obey the equations
\begin{equation}
\label{SU3DecupOctet}
G^{ud}_{p {\mit\Delta}^{++}}
=
-
\frac{\sqrt{3}}{2} G^{uu - dd}_{p {\mit\Delta}^{+}}
=
- \sqrt{3} G^{du}_{p {\mit\Delta}^0}
\, .
\end{equation}
Here, we did not display the index $i = 1, \dots, 4$ distinguishing different
species of GPDs for a given transition. Instead, we  exhibited their partonic
flavor content and the hadronic transition involved.

\subsubsection{Implications of the large-$N_c$ for decuplet-octet transitions}

It turns out that one can further reduce the GPDs for the $p \to \Delta^+$
transitions  to the diagonal $p \to p$ GPDs. This can be done by using the
multicolor limit, which proved to be extremely fruitful phenomenologically
over the last few decades since it captures many of the salient features
of the baryon physics. In this limit, the mesons and  baryons are described
by interpolating fields
$$
M = \sum_{i = 1}^{N_c} \bar\psi_i \psi^i
\, , \qquad
B = \varepsilon_{i_1 i_2 \dots i_{N_c}} \psi^{i_1} \psi^{i_2} \dots \psi^{i_{N_c}}
\, .
$$
Thus, a baryon is viewed as built from $N_c$ quarks,  so its mass grows linearly in
the number of colors $M \sim N_c$, while the meson mass is of order $N_c^0$. The
meson decay constants $M \to \bar\psi \psi$ scale as $N_c^{1/2}$, while the
triple-meson vertices vanish as $N_c^{-1/2}$. Therefore, the meson theory becomes
weakly coupled at large $N_c$,  and ultimately becomes free as $N_c \to \infty$,
so that mesons become stable in the multicolor world, and this approach cannot be
used  for the description of the physical meson spectrum.

For baryons, the situation is quite the opposite. The meson-baryon-baryon coupling
scales like  $N_c^{1/2}$,  so that the baryon theory becomes strongly coupled as
$N_c$ grows. There are a few ways for realization of  the large-$N_c$ limit for
baryons. It is either the large-$N_c$ quark model  (or the contracted $SU(4)$,
using the modern language) or the bosonic description in which the baryon arises
as a topologically stable soliton of a mesonic field theory. The simplest version
of the latter is the Skyrme model (see Appendix \ref{SkyrmeAppendix}). The nucleon
and the delta isobar represent different rotational modes of the same object---chiral
soliton. In the large-$N_c$ limit, the $\Delta$ isobar is degenerate in mass with
the nucleon, i.e., $M_{\mit\Delta} - M_N \sim \mathcal{O} (N_c^{- 1})$. A leading
order prediction in $1/N_c$ is a genuine property of QCD rather than a consequence
of model considerations, because the analysis of the leading large-$N_c$ order
contributions is purely kinematical. It relies on simple group theoretical arguments
only: no dynamical assumptions have to be made.

Before we explain the main steps in the derivation of the large-$N_c$ relations,
let us make it clear under what circumstances such relations may be valid. GPDs
depend on three independent kinematical variables, so let us see what constraints
and conditions are imposed on them by taking the multicolor limit.
\begin{itemize}
\item Since the (incoming and outgoing) baryon is heavy, $M \sim \mathcal{O} (N_c)$,
its four-momentum $p_i^\mu$ is dominated by its time component, and is of the same
order as the mass. Next observation is that the sum of all parton momenta---the
active and spectators---gives that of the parent hadron, $k + \sum_{\rm spectators}
k = p$. Then, because $p \sim {\cal O} (N_c)$, each of its $N_c$ partons carries
the momentum $k \sim {\cal O} (N_c^0)$. Hence, $x \equiv k^+/p^+ \approx \sqrt{2}
k^+/M \sim \mathcal{O}(N_c^{-1})$.
\item Since the baryons are considered as non-relativistic particles in the
infinite-mass limit, there should be no significant recoil transferred to the
outgoing particle. This suggests that the $t$-channel momentum $|\Delta^2| \ll
M^2$, thus at least $\Delta^2 \sim \mathcal{O} (N_c^0)$.
\item In the Breit frame, which is very suitable for discussing the non-relativistic
reduction of relativistic expressions, there is no energy exchange between the
incoming and outgoing hadrons,  and the momentum transfer is three-dimensional.
Thus, as we established earlier, the skewness in the heavy-baryon limit is $\eta =
\Delta^z/(2 M) \sim \mathcal{O} (N_c^{-1})$,  and it is parametrically of the same
order as the Feynman momentum fraction $x$.
\end{itemize}

The analysis shows \cite{AdkNapWit83}, that the only unsuppressed matrix elements
are mediated by scalar-isoscalar $\mathcal{O} = \sum_q \bar\psi_q \psi_q$ and
vector-isovector $\mathcal{O}^{ai} \sim \bar\psi_{q'} \gamma^i \tau^a_{q' q} \psi_q$
operators in the spin-flavor basis, with the spin of the state equal to its isospin,
\begin{equation}
\langle
{\scriptstyle S'} = {\scriptstyle I'}, {\scriptstyle S'_3}, {\scriptstyle I'_3}
|
{\cal O}
|
{\scriptstyle S} = {\scriptstyle I}, {\scriptstyle S_3}, {\scriptstyle I_3}
\rangle
\, , \qquad
\langle
{\scriptstyle S'} = {\scriptstyle I'}, {\scriptstyle S'_3}, {\scriptstyle I'_3}
| {\cal O}^{ai} |
{\scriptstyle S} = {\scriptstyle I}, {\scriptstyle S_3}, {\scriptstyle I_3} \rangle
\, .
\end{equation}
Here the state of a particle is parametrized in terms of its spin, which equals
to isospin, and their projections. The rest, i.e., scalar-isovector and
vector-isoscalar, transitions are subleading in the large $N_c$ counting.

Let us demonstrate how to relate GPDs for the proton-to-delta to the diagonal
proton-to-proton transitions. As an example, we consider matrix elements of
the odd parity quark operators. First, we need to understand what kind of
nucleon-to-nucleon GPDs  are matched into those for the nucleon-to-delta
transitions. In order to do this, we  perform the  non-relativistic reduction
of Dirac bilinears standing in front of the nucleon-nucleon GPDs. As we
advocated before, this can be consistently done by going to the Breit frame
first (see Section \ref{BreitFrameFFs}) and then taking the limit of the
heavy baryon mass $M_N = M_\Delta \equiv M \gg \bit{\Delta}^2$. As one can see
from Eqs. (\ref{ParityOddBreitFrame}), the resulting structures contain the
spin Pauli matrix sandwiched between two Weyl spinors. Thus,  all the structures
will be of leading order in the multi-color limit, provided that one takes
the isovector combination of quark operators
\begin{equation}
\label{IsovectorOperator}
\widetilde {\cal O}^{(3)} = \widetilde {\cal O}^{uu} - \widetilde {\cal O}^{dd}
\, .
\end{equation}
We will take the spin up states for incoming and outgoing nucleons. In this case
the $q$-flavor quark operator matrix elements yields
\begin{eqnarray}
\label{NtoNnonrelativistic}
\langle p_\uparrow (p_2) |
\widetilde {\cal O}^{qq} (- z^-, z^-)
| p_\uparrow (p_2) \rangle
\!\!\!&=&\!\!\!
\sqrt{2} M
\int_{- 1}^{1} d x \ {\rm e}^{- i x p^+ z^- }
\\
&\times&\!\!\!
\Big\{
\widetilde H^q (x, \eta, \Delta^2)
-
\frac{\Delta^2_z}{4 M^2} \left( \widetilde E^q (x, \eta,
\Delta^2) + \ft12 \widetilde H^q (x, \eta, \Delta^2) \right)
\Big\} \, , \nonumber
\end{eqnarray}
which has to be combined into the isovector combination (\ref{IsovectorOperator}).
For the analogous proton-to-delta transition, using  the Rarita-Schwinger
spin-vector (see Appendix \ref{RSchSpinors})  we find
\begin{eqnarray}
\label{NtoDeltanonrelativistic}
\langle \Delta_\uparrow{\!}^+ (p_2) |
\widetilde {\cal O}^{(3)} (- z^- , z^-)
| p_\uparrow  (p_1) \rangle
\!\!\!&=&\!\!\!
\frac{2 M}{\sqrt{3}} \int_{- 1}^1 dx \, {\rm e}^{- i x p^+ z^-}
\Bigg\{
\frac{\Delta_z}{M}
\left(
\widetilde G_1 (x, \eta, \Delta^2) + \widetilde G_2 (x, \eta, \Delta^2)
\right)
\nonumber\\
&+&\!\!\!
\widetilde G_3 (x, \eta, \Delta^2)
-
\frac{\Delta^2_z}{M^2} \widetilde G_4 (x, \eta, \Delta^2)
\Bigg\}
\, ,
\end{eqnarray}
where we do not indicate the hadronic transition involved, as was done in
the preceding section, $\widetilde G_i = \widetilde G_{i, p\Delta^+}^{uu-dd}$.
Thus, we observe that the isovector combination of proton-to-proton functions
$\widetilde H^u - \widetilde H^d$ is matched into $\widetilde G_3$, while the
combination $( \widetilde E^u - \widetilde E^d ) + \ft12 ( \widetilde H^u -
\widetilde H^d )$ into $\widetilde G_4$. Finally, we see that in the large-$N_c$
limit
\begin{equation}
\widetilde G_1 = - \widetilde G_2
\, .
\end{equation}

According to the large-$N_c$ limit, the nucleon and the delta-isobar are
eigenstates of the same object, the chiral soliton. In terms of the matrix
elements of the quark operators, we have the relation
\begin{equation}
\sqrt{2} \langle p_\uparrow | \widetilde{\cal O}^{(3)} | p_\uparrow \rangle
=
\langle  \Delta_\uparrow{\!\!}^+ | \widetilde{\cal O}^{(3)}| p_\uparrow \rangle
\, .
\end{equation}
Substituting Eqs.\ (\ref{NtoNnonrelativistic}) and (\ref{NtoDeltanonrelativistic})
into this relation and equating the coefficients in front of different powers of
the momentum transfer $\Delta_z$, we ultimately find
\begin{eqnarray}
\label{PolDPtoPP}
\widetilde G_3
\!\!\!&=&\!\!\!
\sqrt{3} \left( \widetilde H^u - \widetilde H^d \right)
\, , \\
\widetilde G_4
\!\!\!&=&\!\!\!
\frac{\sqrt{3}}{4}
\left\{
\left( \widetilde E^u - \widetilde E^d \right)
+
\ft12
\left( \widetilde H^u - \widetilde H^d \right)
\right\}
\, .
\end{eqnarray}

Analogous considerations apply to the vector operators. For the proton-to-proton
matrix element,  the non-relativistic reduction results in
\begin{eqnarray}
\langle p_\downarrow (p_2) | {\cal O}^{qq} (- z^-, z^-) | p_\uparrow (p_1) \rangle
=
- \frac{\Delta}{\sqrt{2}} \int_{- 1}^{1} d x \ {\rm e}^{- i x p^+ z^- }
\left( H^q (x, \eta, \Delta^2) + E^q (x, \eta, \Delta^2) \right)
\, ,
\end{eqnarray}
while for the transition to the delta isobar we have
\begin{eqnarray}
\langle \Delta^+{\!\!}_\downarrow (p_2) |
{\cal O}^{(3)} (- z^-, z^-)
| p_\uparrow (p_1) \rangle
=
- \frac{\Delta}{\sqrt{3}} \int_{- 1}^{1} d x \ {\rm e}^{- i x p^+ z^- }
G_1 (x, \eta, \Delta^2)
\, ,
\end{eqnarray}
where we kept only the leading contribution that  matches into the leading
contribution in the diagonal transition. Thus, one gets
\begin{equation}
\label{NonPolDPtoPP}
G_1
=
\sqrt{3} \left( H^u - H^d \right)
+
\sqrt{3} \left( E^u - E^d \right)
\, .
\end{equation}
The relations (\ref{PolDPtoPP}) -- (\ref{NonPolDPtoPP}) relate the ``unknown''
$p \to \Delta^+$ to ``known'' $p \to p$ GPDs. Using  $SU(3)$ relations of the
previous section,  one relates all other $\bit{8} \to \bit{10}$ matrix
elements to those of the flavor-diagonal proton transitions.

The relation (\ref{NonPolDPtoPP}) correctly reproduces the known large-$N_c$
prediction between the isovector combination of the proton and neutron magnetic
moments,
$$
\int_{-1}^1 dx \left( H^u - H^d + E^u - E^d \right) (x, \eta, \Delta^2 = 0)
=
\mu_p - \mu_n
\, ,
$$
and the $p \to \Delta^+$ transition magnetic moment $\mu_{\Delta^+ p}$.
The latter is determined by the zero-recoil limit of the magnetic form
factor associated with the matrix element of the electromagnetic current
$G_{M1}^{\rm em}$. This electromagnetic form factor is related the first
moment of the isovector GPD for the same transition $p \to \Delta^+$ via
the equation
$$
\int_{-1}^1 dx \, G_1 (x, \eta, \Delta^2 = 0)
=
2
G_1^{\rm em} (\Delta^2 = 0)
=
3
G^{\rm em}_{M1} (\Delta^2 = 0)
\equiv
\sqrt{6} \mu_{\Delta^+ p}
\, ,
$$
where in the first equality, we have used Eq.\ (\ref{EMtoIsovector}), while
in the second, the large-mass approximation of Eq.\ (\ref{MagneticGandG1})
for the $p \to \Delta^+$ electromagnetic form factors.

One can establish more precise scaling rules for the GPDs making use of known
results for large-$N_c$ scaling of nucleon form factors \cite{DasJenMan95,Man98,Jen98},
and also the parametric dependence of the kinematical variables on the number of
colors, as discussed above. As a result, one arrives at the following set of
scaling rules for the isosinglet \cite{Goeke:2001tz}
\begin{equation}
\begin{array}{cc}
H^u + H^d \sim \mathcal{O}(N_c^2)
\, , \qquad
& E^u + E^d \sim
\mathcal{O}(N_c^2)
\, , \\
\widetilde H^u + \widetilde H^d \sim \mathcal{O}(N_c)
\, , \qquad
& \widetilde E^u + \widetilde E^d \sim \mathcal{O}(N_c^3)
\, ,
\end{array}
\end{equation}
and isovector combinations
\begin{equation}
\begin{array}{cc}
H^u - H^d \sim \mathcal{O}(N_c^2)
\, , \qquad
&
E^u - E^d \sim \mathcal{O}(N_c^3)
\, , \\
\widetilde H^u - \widetilde H^d \sim \mathcal{O}(N_c)
\, , \qquad
& \widetilde E^u - \widetilde E^d \sim \mathcal{O}(N_c^4)
\, .
\end{array}
\end{equation}
These relation then allow to neglect the isovector combinations of $\widetilde H$
and $H$ GPDs compared to those of $\widetilde E$ and $E$ in Eqs.\ (\ref{PolDPtoPP})
and (\ref{NonPolDPtoPP}), respectively.

\section{Evolution equations}
\label{Evolution}

Having discussed nonperturbative characteristics of GPDs, we turn to the
exploration of their perturbative properties. Recall, that during a
deep-inelastic scattering event, the virtual photon of high virtuality
${\cal Q}^2$ exchanged between the lepton and the proton acts as an almost
local probe which, due to the uncertainty principle, selects distances
inversely proportional to its virtuality $|\delta \bit{z}_\perp| \sim
1/{\cal Q}$. The variation in the resolution scale changes the granularity
of the observed picture, or put it into the current context, the density of
the parton number $q(x)$ for a given momentum fraction $x$. Suppose that
the hard probe of a certain wavelength can resolve a parton having momentum
$x p^\mu$. By refining the probe's resolution, the same parton will be seen
as a system of several partons, each carrying a smaller momentum fraction
$x_i < x$ due to the momentum conservation $x = \sum_i x_i$. Thus, the
parton distribution $q(x)$ depends on the virtuality of the current $q(x)
\to q(x, {\cal Q}^2)$ and, as a result, the Bjorken scaling is violated.

\begin{figure}[t]
\begin{center}
\hspace{0cm} \mbox{
\begin{picture}(0,115)(180,0)
\put(0,0){\insertfig{12}{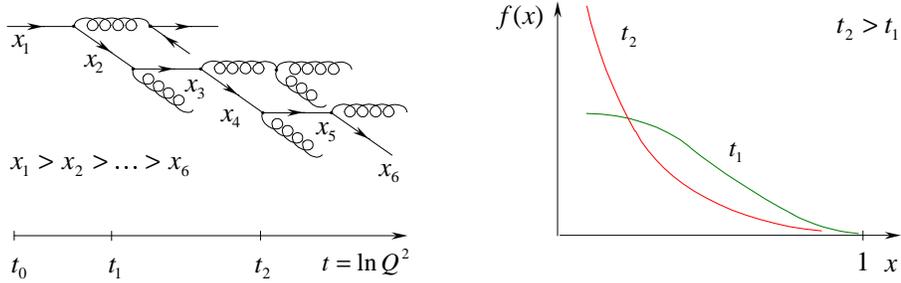}}
\end{picture}
}
\end{center}
\caption{\label{DisScalingViolation} Multiple emission of partons (left)
from the active quark diminishes its momentum via redistribution between
the particles it spawns. This shower decreases its momentum when measured
at later and later ``times'' $t = \ln {\cal Q}^2$ and results in scaling
violation in parton distributions.
}
\end{figure}

The  qualitative description  outlined above can be put in a rigorous theoretical
framework. In QCD, like in any theory with a dimensionless coupling constant, the
scaling violation is  logarithmic and, due to asymptotic freedom,  can be computed
to  arbitrary accuracy within  perturbation theory, the  limit being  set only by
the technical ability to actually perform the computation.

The general features of  the QCD evolution can be understood in the following
way. An active quark can emit a gluon (or several gluons) before the actual
interaction with the photon has occurred, as illustrated on  the left drawing in
Fig.\ \ref{DisScalingViolation}. When it radiates a gluon, the  quark looses a
portion of its original momentum. Though the loss  can be small for each particular
event, the quark,  being a relativistically moving particle tends to emit more
gluons. The bremsstrahlung of many gluons drifts the quark momentum into the low-$x$
region. Since the gluons, in their turn,  can dissociate  into  quark-antiquark
pairs, there is a proliferation of quarks with small momentum fractions and,
correspondingly, a decrease of the parton density in the large-$x$ domain. Thus,
the probability to find a quark with a small momentum fraction is higher at
larger ${\cal Q}^2$, as shown on the right part of Fig.\ \ref{DisScalingViolation}.
Therefore, one sees more and more quarks (or gluons) in the cloud forming the
parent ``fat'' quark.

The physical picture behind the evolution of generalized parton distributions
is more involved.  As we established in previous sections, in different regions
of the longitudinal momentum of the annihilated and created partons, these
functions resemble either parton densities or distribution amplitudes. The scaling
violation in parton distributions considered as an example above is especially
transparent and intuitive since one is  dealing with probabilities. The
distribution amplitudes do not have a classical interpretation, and the
evolution of DAs reflects their  quantum  nature. We will discuss the qualitative
features of the evolution of GPDs later in this chapter. Now we will concentrate
on the description of several calculational frameworks which allow one to quantify
the phenomenon.

\subsection{Divergences of perturbation theory and scale dependence}

\begin{figure}[t]
\begin{center}
\mbox{
\begin{picture}(0,80)(70,0)
\put(20,0){\insertfig{3}{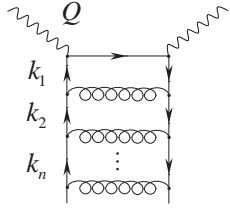}}
\end{picture}
}
\end{center}
\caption{\label{LadderDiag} Ladder diagrams producing the dominant
contribution in the leading logarithmic to the structure functions of
deeply inelastic scattering and resulting into logarithmic scaling
violation.}
\end{figure}

Formally, the  scaling violation arises due to the unrestricted growth of the
parton's transverse momentum in perturbation theory. This can be explicitly
demonstrated by computing the probability of the soft gluon emission\footnote{The
gluon is soft when all four components of its momentum are much smaller than
${\cal Q}$, i.e., $k^\mu \sim (\lambda, \lambda, \lambda, \lambda)$ and $\lambda
\ll {\cal Q}$. However, we can still apply perturbation theory if $\lambda  \gg
\Lambda_{\scriptscriptstyle\rm QCD}$.} from a quark. The amplitude in this
kinematic region takes the form (see, e.g., \cite{LanLif82})
\begin{equation}
i {\cal A}_{q \to qg} (p)
=
\
\parbox[c][15mm]{22mm}{
\insertfig{2}{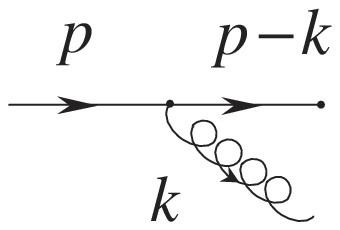}
}
=
i \frac{{\not\! p} - {\not\! k}}{(p - k)^2}
i g {\not\!\varepsilon} (k) u (p)
\simeq
g \frac{p \cdot \varepsilon (k)}{p\cdot k} u (p)
\, ,
\end{equation}
where we used the fact that the quark momentum is light-like $p^\mu = p^-
n^\mu$ with $p^- \sim {\cal Q}$, so that $p \cdot k \gg k^2 \sim p^2$. The
emission probability is then
\begin{equation}
\int
\frac{d^3 \bit{k}}{2 E_{\bit{\scriptstyle k}}(2 \pi)^3}
| {\cal A}_{q \to q g} (p) |^2
=
\frac{\alpha_s}{\pi^2}
\bar{u} (p) u (p)
\int \frac{d k^+}{k^+} \int \frac{d^2 \bit{k}_\perp}{\bit{k}_\perp^2}
\, ,
\end{equation}
with $p^\mu = n^{\ast\mu}$. Both the longitudinal and transverse integrals
here are logarithmically divergent. These divergences have different nature.
The longitudinal $k^+$ integral can be converted into the integral over the
fraction $x$, which cannot exceed one, so only the soft $x \to 0$ divergence
remains.  It does not induce the ${\cal Q}$-dependence,  hence this  is not
the phenomenon we are interested in now. It is rather an artifact of the
soft-gluon approximation. The scaling violation is related to the structure
of the transverse momentum integral. Its divergence at the lower limit,
when $\bit{k}_\perp \to 0$, corresponds to situation when the final quark
momentum $(p - k)$ is collinear to the initial momentum $p$. The {\sl collinear
divergence} converts into the $\ln p^2$ ({\sl mass singularity}) if no
approximation is made \cite{EllGeoMacPolRos78}. The divergence on the upper limit
is regulated by the kinematics of  deep inelastic scattering which restricts
$\bit{k}_\perp$ at a value of order ${\cal Q}$. Thus, for a one-loop diagram one
obtains a $\ln {\cal Q}^2/p^2$ contribution as a correction to the parton density,
and the naive scaling is violated. For a diagram with $n$ gluons, one can obtain
the $(\ln {\cal Q}^2/p^2)^n$ contribution. It comes from integration over the
region where the successive $\bit{k}_{i \perp}$'s are strongly ordered. This
means that the leading logarithm $(\ln {\cal Q}^2/p^2)^n$ for an $n$-rung
ladder diagram (see Fig.\ \ref{LadderDiag}) comes from the region
\begin{equation}
\label{StrongOderingDGLAP}
\begin{array}{ccccccccc}
{}| \bit{k}_{n \perp}|
&
\ll
&
\dots
&
\ll
&
{}|\bit{k}_{2 \perp}|
&
\ll
&
{}|\bit{k}_{1 \perp}|
&
\ll
&
{\cal Q}
\, , \\
k^-_n
&
\ll
&
\dots
&
\ll
&
k^-_2
&
\ll
&
k^-_1
\, ,
&
&
\\
k^+_n
&
\geq
&
\dots
&
\geq
&
k^+_2
&
\geq
&
k^+_1
\, .
&
&
\end{array}
\end{equation}
The last line here reflects the  decrease of the longitudinal quark momentum
due to the gluon emission (cf.\ Fig.\ \ref{DisScalingViolation} left). The
hierarchy in the second line is a consequence of the on-shellness of intermediate
states $k_i^2 = 2 k^+_i k^-_i - \bit{k}^2_{\perp i} = 0$ and the inequalities in
the other two lines. If one is interested in  contributions with non-leading
powers of $\ln {\cal Q}^2$ or higher-order non-logarithmic $\alpha_s$ corrections,
the strong ordering in $\bit{k}_{\perp}$  is lifted. It is worth mentioning here
that the resulting evolution equations in  momentum space can be interpreted in
terms of the kinetic equilibration of partons in the effective ``time'' $t \equiv
\ln {\cal Q}^2$ \cite{Lip72}.

\subsection{Factorization and evolution}

An important observation is that the parameter ${\cal Q}$ in this
calculation serves as an ultraviolet (UV) cut-off for the $\bit{k}_{\perp}$
integral. This allows one to treat the ${\cal Q}^2$-dependence of parton
densities as a particular case of the dependence on the UV regularization
parameter $\mu^2$ and to apply the renormalization group to organize the
summation of the $\ln {\cal Q}^2$  contributions.  Once the renormalization
group applicability is established, one can use the full power of this
formalism and compute the evolution effects to a given order of
perturbation theory.

\begin{figure}[t]
\unitlength1mm
\begin{center}
\mbox{
\begin{picture}(0,110)(65,0)
\put(0,10){\insertfig{4}{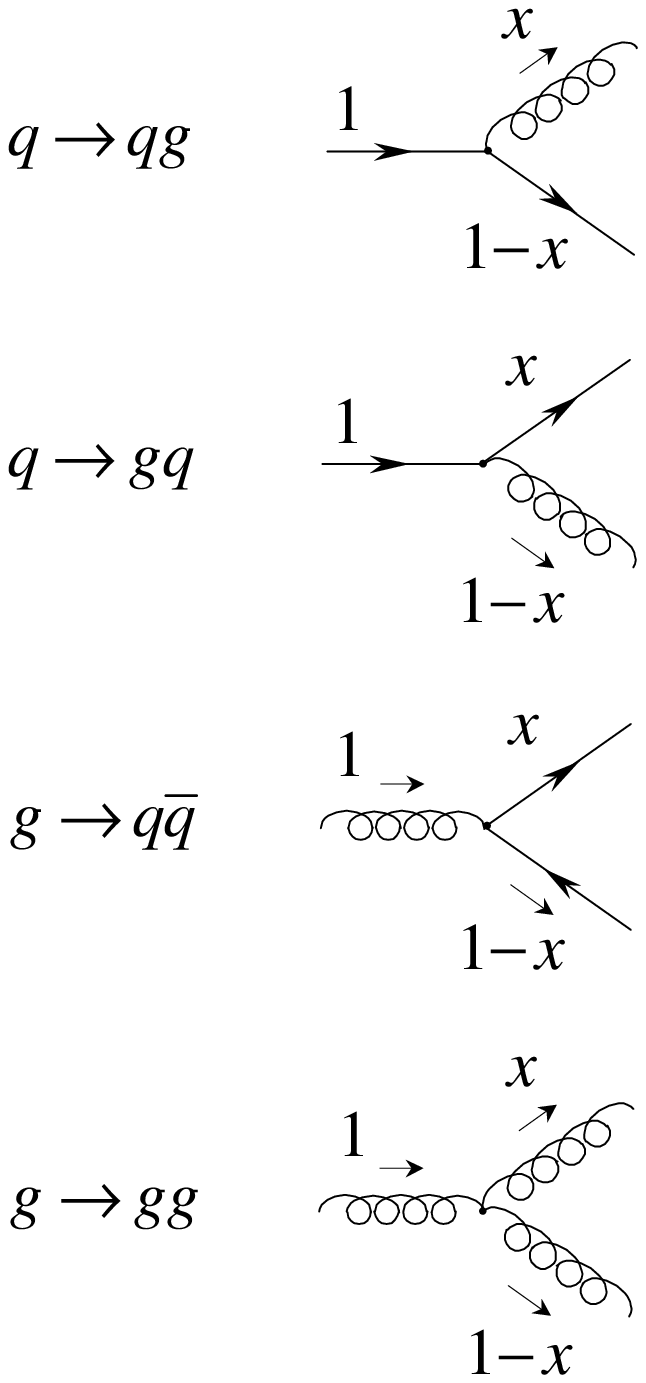}}
\put(60,0){\insertfig{7.2}{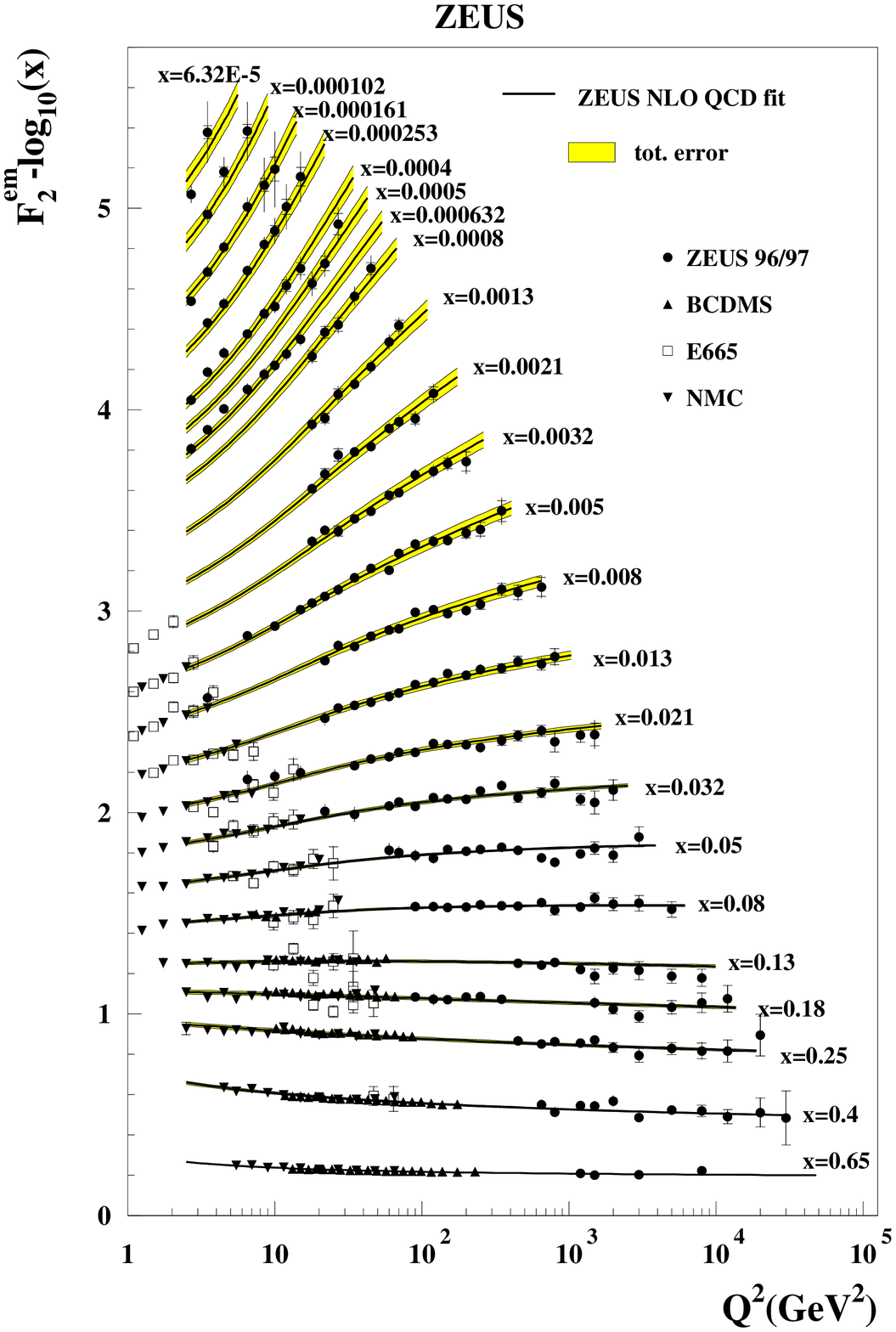}}
\end{picture}
}
\end{center}
\caption{\label{ExperimentalData} Elementary $a \to bc$ parton splitting
amplitudes encountered at each step of the evolution (left) and a fit to
experimental data \protect\cite{Che02} (right) implementing them (dressed
with next-to-leading order perturbative corrections) in the evolution
equation (\ref{evolmatrix}).}
\end{figure}

Within the QCD factorization approach,  the DIS structure functions are
represented as the convolution
\begin{equation}
F_i (x_{\rm B}, {\cal Q}^2)
=
\int_{x_{\rm B}}^1 \frac{d x}{x} \
C_i \left( x_{\rm B}/x, {\cal Q}^2/\mu^2 \right) q (x; \mu^2)
\end{equation}
of the coefficient function $C(x_{\rm B}/x, {\cal Q}^2/\mu^2)$ and
the parton distributions $q (x; \mu^2)$ given by the matrix elements
of composite operators. By construction, the factorization scale $\mu$
separates large virtualities from the small ones. For the coefficient
function $C$ determined by the large momentum integration, the scale
$\mu$ serves as a cut-off at the lower limit of the $\bit{k}_{\perp}$
integration, i.e., as the regularization parameter for the mass
singularities. At the same time, it serves as the upper limit for the
$\bit{k}_{\perp}$ integration (i.e., as an UV cut-off) in  matrix
elements defining the parton distributions. The crucial property is
the multiplicative renormalizability of the relevant composite
operators expressed by the renormalization group equation
\begin{eqnarray}
\label{RGgen}
\frac{d}{d \ln\mu^2} \, q (x , \mu^2)
\!\!\!&=&\!\!\!
\int_x^1 \frac{d y}{y} \,
P \left( x/y ; \alpha_s (\mu^2) \right) q(y ; \mu^2)
\, .
\end{eqnarray}
It is the well-known Dokshitzer-Gribov-Lipatov-Altarelli-Parisi (DGLAP)
equation \cite{Lip72,AltPar77,Dok77}. Since the structure functions $F_i
(x_B, {\cal Q}^2)$, being physical observables, do not depend on the
arbitrary momentum separation scale $\mu^2$
\begin{equation}
\frac{d F_i (x_{\rm B}, {\cal Q}^2)}{d \ln\mu^2} = 0
\, ,
\end{equation}
the same evolution kernels $P(z ; \alpha_s)$ appear in the evolution
equations for the coefficient functions
\begin{eqnarray}
\frac{d}{d \ln\mu^2} \,
C_{i}
\left(
x; {\cal Q}^2/{\mu^2}, \alpha_s (\mu^2)
\right)
=
- \int_x^1 \frac{d y}{y} \,
C_{i} \left( y; {\cal Q}^2/{\mu^2}, \alpha_s (\mu^2)
\right) P \left( x/y ; \alpha_s (\mu^2) \right)
\, .
\nonumber
\end{eqnarray}
Since the coefficient functions are given by integration over large momenta,
the evolution kernels are  computable in perturbation theory. To get rid of
complications due to gluons, one can consider a  specific combination of the
DIS cross sections where gluonic contributions drop out and which picks up a
specific flavor combination of quarks, the so-called non-singlet parton
distributions $q_{\scriptscriptstyle\rm NS}$. It is given, for instance, by
the difference of the scattering cross sections  on the proton and neutron
targets, the nonsinglet cross section  $\sigma^p - \sigma^n \equiv
\sigma^{\scriptscriptstyle\rm NS}$. Then the  evolution is governed by a
single kernel $P_{\scriptscriptstyle\rm NS} (x; \alpha_s(\mu^2))$. It is given
by an infinite series expansion in $\alpha_s$,
\begin{equation}
\label{PertExpansionDGLAPkernels}
P_{\scriptscriptstyle\rm NS} (x; \alpha_s)
=
\sum_{n = 0}^{\infty}
\left( \frac{\alpha_s}{2 \pi} \right)^{n + 1}
P^{(n)}_{\scriptscriptstyle\rm NS} (x)
\, .
\end{equation}
For the singlet quark distribution
\begin{equation}
q_{\scriptscriptstyle\rm S} = \sum_q \left( q + \bar q \right)
\, ,
\end{equation}
the evolution is more complicated since the quarks can mix with gluons
and, therefore, the evolution equation takes the form
\begin{equation}
\label{42}
\frac{d}{d \ln \mu^2}
\left(
\begin{array}{c}
q_{\scriptscriptstyle\rm S} \\
g
\end{array}
\right)
(x ; \mu^2)
= \int_x^1 \frac{d y}{y} \,
\left(
\begin{array}{cc}
P^{qq} & P^{qg} \\
P^{gq} & P^{gg}
\end{array}
\right)
\left( x/y ; \alpha_s (\mu^2) \right)
\left(
\begin{array}{c}
q_{\scriptscriptstyle\rm S} \\
g
\end{array}
\right)
(y ; \mu^2) \, .
\label{evolmatrix}
\end{equation}
The splitting functions are defined as decay probabilities,
\begin{equation}
P^{ab} (x ; \alpha_s (\mu^2)) = | A_{a \to bc} |^2
\, ,
\end{equation}
having the perturbative expansion like in Eq.\ (\ref{PertExpansionDGLAPkernels}).
The leading order amplitudes $A$ contributing to one-loop DGLAP kernels are
represented in Fig.\ \ref{ExperimentalData} (left). The solution of the DGLAP
evolution equations combined with higher order QCD radiative corrections allows
to obtain a perfect description of experimental data for the unpolarized deeply
inelastic cross section, shown in Fig.\ \ref{ExperimentalData} (right).

\subsection{Coordinate-space evolution equations}

As we know, the generalized parton distributions are determined by off-forward
matrix elements of bilocal operators living on the light cone. These are the
same operators that define the forward parton distribution. As long as the
separation of the field operators is neither zero $z^\mu = 0$ nor light-like
$z^2 = 0$, the renormalization of the product ${\mit\Phi}^*(0) {\mit\Phi}
(z)$ is trivially reduced to the familiar renormalization constants
${\cal Z}_{\mit\Phi}$ of the fundamental field operators ${\mit\Phi}$.
However, if $z^\mu = 0$ or $z^2=0$, additional divergences enter the game,
since the product of (at least) two field operators located at the same
space-time point (or on the light cone) produces an ill-defined quantity
from the point of view of the theory of distributions, and the corresponding
infinities have to be regularized and subtracted. This corresponds to the
ultraviolet divergencies discussed above within the context of the
momentum-space evolution equations.

Evolution equations for all ``momentum-fraction'' functions, i.e., parton
densities, distribution amplitudes and generalized parton distributions,
arise from the same set of  renormalization group equations for the light-ray
operators (\ref{QuarkLightRayOperators}) and (\ref{GluonLightRayOperators}).
The formalism was developed by several groups in early and mid eighties
\cite{BorRob82,Bal82,BraGeyRob85,BalBra89}.

In the following we will discuss the flavor singlet case, where the quark
and gluon operators mix with each other. For brevity we introduce the two
dimensional vector
\begin{equation}
\bit{\cal O} (z_1^- , z_2^-)
\equiv
\left(
\begin{array}{c}
\frac{1}{2} \sum_q
\left(
{\cal O}^{qq} (z_1^- , z_2^-)
\mp
{\cal O}^{qq} (z_2^- , z_1^-)
\right)
\\
{\cal O}^{gg} (z_1^- , z_2^-)
\end{array}
\right)
\, ,
\end{equation}
where the plus and minus signs stand for the axial $\widetilde {\cal O}$ and
vector channel ${\cal O}$, respectively. Note that due to the Bose symmetry,
the gluon operator also has definite symmetry with respect to the interchange of
$z^-_1 \leftrightarrow z^-_2$, i.e., it is (anti)symmetric in the case of
(axial)vector operator. The two-vector $\bit{\cal O}$ obeys the renormalization
group equation
\begin{equation}
\label{non-local-K}
\frac{d}{d\ln\mu^2}
\bit{\cal O} \left( z^-_1, z^-_2 \right)
=
\int_0^1 d u \int_0^1 d v \,
\bit{\cal K} (u, v) \,
\bit{\cal O}
\left( \bar u z^-_1 + u z^-_2 , v z^-_1 + \bar v z^-_2 \right)
\, ,
\end{equation}
(here and below $\bar{u} \equiv 1 - u$ and analogously for other variables)
with the perturbative kernel having an expansion in perturbation series
\begin{equation}
\bit{\cal K} (u, v)
=
\sum_{n = 1}^\infty
\left( \frac{\alpha_s}{2 \pi} \right)^n
\bit{\cal K}_{(n)} (u, v)
\, .
\end{equation}
The evolution kernel is a two-by-two matrix
\begin{equation}
\bit{\cal K} (u, v)
=
\left(
\begin{array}{cc}
{\cal K}^{qq}
&
\frac{z_2^- - z_1^-}{i} {\cal K}^{qg}
\\
\frac{i}{z_2^- - z_1^-} {\cal K}^{gq}
&
{\cal K}^{gg}
\end{array}
\right)
\, .
\end{equation}
which explicitly depends on the light-cone position $z_2^- - z_1^-$ in
the mixed channels.

Finally let us add that nonsinglet (or valence) combinations of even and odd parity
operators,
\begin{equation}
{\cal O}^{qq}_{\scriptscriptstyle\rm NS} (z_1^- , z_2^-)
=
\ft{1}{2}
\left(
{\cal O}^{qq} (z_1^- , z_2^-)
+
{\cal O}^{qq} (z_2^- , z_1^-)
\right)
\, , \quad
\widetilde{\cal O}^{qq}_{\scriptscriptstyle\rm NS} (z_1^- , z_2^-)
=
\ft{1}{2}
\left(
\widetilde{\cal O}^{qq} (z_1^- , z_2^-)
-
\widetilde{\cal O}^{qq} (z_2^- , z_1^-)
\right)
\, ,
\end{equation}
have an autonomous evolution, i.e., they do not mix with gluons.

\subsubsection{Symmetry properties}

\begin{figure}[t]
\unitlength1mm
\begin{center}
\mbox{
\begin{picture}(145,66)(0,0)
\put(0,0){\insertfig{14}{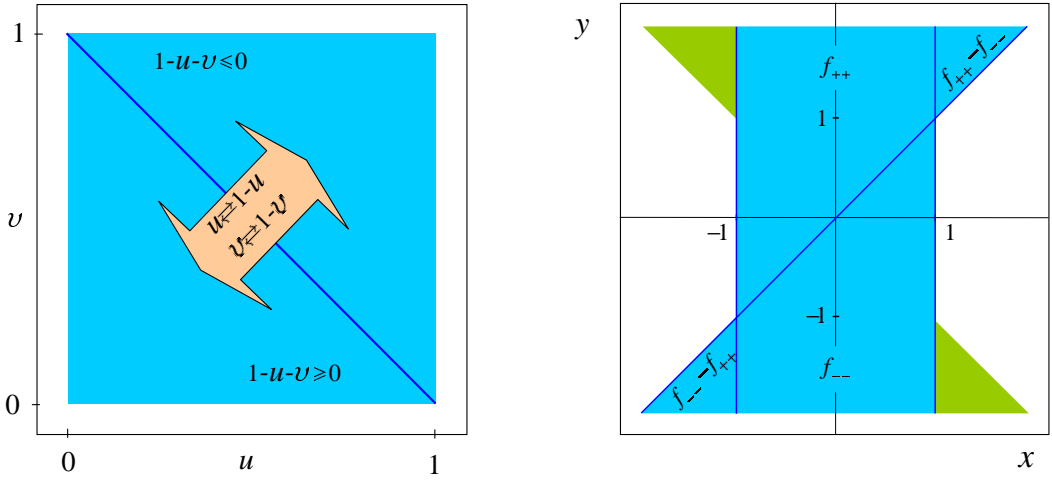}}
\end{picture}
}
\end{center}
\caption{\label{fig-sup-lr}
Support of the singlet anomalous dimensions $\bit{\cal K}$ in the light-cone
position (a) and fraction (b) representation. In (b) we show the support of
the momentum-space kernel which arises from $1 - u - v \ge 0$ of the light-cone
position representation (blue) and from $1 - u - v \leq 0$ (left-top and
right-bottom green corners). We use a short-hand notation in the graph
$f_{\pm \pm} = f (\pm x , \pm y)$ with $f$ defined by Eq.\ (\ref{FourierTransF}).
}
\end{figure}

The first important issue to understand is the anatomy of the evolution
kernels for GPDs, which arise from the support property of the kernels
$\bit{\cal K}$. This problem can be solved by means of the $\alpha$
or Feynman-parameter representation of Green functions with a non-local
operator insertion. It is sufficient to work in the light-cone gauge and
to formally generalize the $\alpha$-representation for the gluon propagator
\cite{MulDitRobGeyHor98}. From these studies one can deduce the support of
the kernels shown in Fig.\ \ref{fig-sup-lr} (a):

\begin{eqnarray}
\bit{\cal K} (u, v)
\neq 0 \, , \ \ \mbox{for} \ \ 0 \leq u, v \leq 1; \ \ 0, \ \ \mbox{otherwise}
\, .
\end{eqnarray}
Invariance under charge conjugation implies the following symmetry relation
\begin{eqnarray}
\bit{\cal K} (u, v)
=
\bit{\cal K} (v, u) \, .
\end{eqnarray}
It is also worth noting that the symmetry properties of the flavor singlet
operators with respect to the interchange of their light cone arguments,
i.e., $z^-_1 \leftrightarrow z^-_2$, can be used to map the region $u + v
\ge 1$ into $1 \ge u + v$ by the substitution $u \to 1 - v$ and $v \to
1 - u$. Here the region $1 \ge u + v$ corresponds in the forward case to
quark-quark mixing which occurs already at leading order, while $u + v \ge 1$
appears due to a quark-antiquark interaction.

\subsubsection{One-loop results}

The calculations of the one-loop kernels are rather straightforward. A
detailed exposition of the formalism on a simple example of the non-singlet
quark operator ${\cal O}^{qq}_{\scriptscriptstyle\rm NS}$ is given in
Appendix \ref{CoordinateEvolutionKernels}. The one-loop kernels have the
support only in the lower triangle of the diagram \ref{fig-sup-lr}, i.e.,
\begin{equation}
{\cal K}^{ab}_{(0)} ( u, v )
=
\theta (1 - u - v) \,
\kappa^{ab} (u, v)
\, .
\end{equation}
To avoid displaying the step-function in all formulas, we assume below that
$u + v \leq 1$, and merely list all the kernels for twist-two operators which
we classified in Section \ref{ClassificationTwistTwo}.

\noindent $\bullet\ $ Even parity sector
\cite{BraGeyRob87,BalBra89,Rad1996a,Rad1996b,Rad1997a,BluGeyRob97}:
\begin{eqnarray}
\label{QQV-kernel}
{\cal K}^{qq, V}_{(0)} ( u, v )
\!\!\!&=&\!\!\!
C_F
\left\{
1
+ \left[ \bar{u}/u \right]_+ \delta (v)
+ \left[ \bar{v}/v \right]_+ \delta (u)
- \ft{1}{2} \delta (u) \delta (v)
\right\}
\, , \\
{\cal K}^{qg, V}_{(0)} ( u, v )
\!\!\!&=&\!\!\!
2 T_F N_f
\left\{ 1 - u - v + 4 u v \right\}
\, , \\
{\cal K}^{gq, V}_{(0)} ( u, v )
\!\!\!&=&\!\!\!
C_F
\left\{
\delta (v) \delta (u) + 2
\right\}
\, , \\
\label{GGV-kernel}
{\cal K}^{gg, V}_{(0)} ( u, v )
\!\!\!&=&\!\!\!
C_A
\left\{
4 ( 1 - u - v + 3 u v )
+ \left[ \bar{u}^2/u \right]_+ \delta (v)
+ \left[ \bar{v}^2/v \right]_+ \delta (u)
\right\}
- \ft12 \left( \beta_0 + 6 C_A \right)\delta (u) \delta (v)
\, . \nonumber\\
\end{eqnarray}

\noindent $\bullet\ $ Odd parity sector \cite{BalRad97,BluGeyRob97}:
\begin{eqnarray}
{\cal K}^{qq, A}_{(0)} ( u, v )
\!\!\!&=&\!\!\!
C_F
\left\{
1
+ \left[ \bar{u}/u \right]_+ \delta (v)
+ \left[ \bar{v}/v \right]_+ \delta (u)
- \ft{1}{2} \delta (u) \delta (v)
\right\}
\, , \\
{\cal K}^{qg, A}_{(0)} ( u, v )
\!\!\!&=&\!\!\!
2 T_F N_f
\left\{ 1 - u - v \right\}
\, , \\
{\cal K}^{gq, A}_{(0)} ( u, v )
\!\!\!&=&\!\!\!
C_F
\left\{
\delta (v) \delta (u) - 2
\right\}
\, , \\
\label{GGA-kernel}
{\cal K}^{gg, A}_{(0)} ( u, v )
\!\!\!&=&\!\!\!
C_A
\left\{
4 ( 1 - u - v )
+ \left[ \bar{u}^2/u \right]_+ \delta (v)
+ \left[ \bar{v}^2/v \right]_+ \delta (u)
\right\}
- \ft12 \left( \beta_0 + 6 C_A \right)\delta (u) \delta (v)
\, . \nonumber\\
\end{eqnarray}

\noindent $\bullet\ $ Maximal-helicity sector \cite{BelMul97,BelMul00d}:
\begin{eqnarray}
{\cal K}^{qq, T}_{(0)} (u, v)
\!\!\!&=&\!\!\!
C_F
\left\{
\left[ \bar{v}/v \right]_+ \delta (u)
+
\left[ \bar{u}/u \right]_+ \delta (v)
-
\ft{1}{2} \delta (u) \delta (v)
\right\}
\, , \\
{\cal K}^{gg, T}_{(0)} ( u, v )
\!\!\!&=&\!\!\!
C_A \left\{
\left[ \bar{u}^2/u \right]_+ \delta (v)
+
\left[ \bar{v}^2/v \right]_+ \delta (u)
\right\}
-
\ft12 \left( \beta_0 + 6 C_A\right) \delta (u) \delta (v)
\, .
\label{MH-coord}
\end{eqnarray}
The plus-prescription here is  defined  as
\begin{equation}
\label{PlusPrescription}
\left[ \frac{f(u)}{u} \right]_+
\equiv
\frac{f(u)}{u}
-
\delta (u) \int_0^1 d v \, \frac{f(v)}{v}
\, .
\end{equation}

\subsection{Momentum-space evolution equations}

\begin{figure}[t]
\unitlength1mm
\begin{center}
\mbox{
\begin{picture}(0,23)(10,0)
\put(0,0){\insertfig{2}{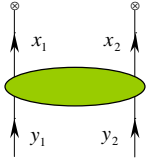}}
\end{picture}
}
\end{center}
\caption{\label{GenericKernel} Momentum flow in evolution kernels.}
\end{figure}

Let us discuss the evolution equation for the target independent GPDs.
In the singlet sector the quark and gluon GPDs can be organized into a
two-dimensional vector in complete analogy to the light-ray operators
discussed in the preceding section. Namely,
\begin{eqnarray}
\label{def-SPD-sing}
\bit{F} (x, \eta)
\equiv
\left(
{
\sum_q \left( F^q (x, \eta) \mp F^q (- x , \eta) \right)
\atop
\ft12 F^g (x, \eta)
}
\right)
\, ,
\end{eqnarray}
with $-$ and $+$ signs standing for parity-even and -odd sector. Here the
target-independent GPDs, entering this two-vector, are the Fourier transforms
of the light-ray operators
\begin{eqnarray}
\label{TargetIndepQuark}
\langle p_2 | {\cal O}^{qq} (z_1^- , z_2^-) | p_1 \rangle
\!\!\!&=&\!\!\!
p^+ \int_{-1}^1 dx \,
{\rm e}^{\frac{i}{2} p^+ \left( (\eta + x) z_1^- + (\eta - x) z_2^- \right)}
F^q (x, \eta)
\, , \\
\label{TargetIndepGluon}
\langle p_2 | {\cal O}^{gg} (z_1^- , z_2^-) | p_1 \rangle
\!\!\!&=&\!\!\!\frac{(p^+)^2}{4} \int_{-1}^1 dx \,
{\rm e}^{\frac{i}{2} p^+ \left( (\eta + x) z_1^- + (\eta - x) z_2^- \right)}
F^g (x, \eta)
\, .
\end{eqnarray}
Identical relations hold for the maximal helicity GPDs which carry open
Lorentz indices. The vector $\bit{F}$ satisfies a matrix evolution equation
\begin{eqnarray}
\label{MOMEE}
\frac{d}{d\ln\mu} \, \bit{F} (x, \eta)
\!\!\!&=&\!\!\!
-
\int_{-1}^1 d y \, \bit{K}
\left(
\left.
\frac{\eta + x}{2} , \frac{\eta - x}{2} \right|
\frac{\eta + y}{2} , \frac{\eta - y}{2}
\right)
\bit{F} (y, \eta)
\, ,
\end{eqnarray}
governed by the  two-by-two matrix kernel
\begin{equation}
\bit{K} ( x_1, x_2 \, |\,  y_1, y_2 )
=
\left(
\begin{array}{ll}
K^{qq}
&
K^{qg}
\\
K^{gq}
&
K^{gg}
\end{array}
\right)
\, ,
\end{equation}
having  an infinite expansion in perturbative series
\begin{equation}
\bit{K} ( x_1, x_2 \,|\, y_1, y_2 )
=
\sum_{n = 0}^\infty
\left( \frac{\alpha_s}{2 \pi} \right)^n
\bit{K}_{(n)} ( x_1, x_2 \,| \, y_1, y_2 )
\, .
\end{equation}

These kernels, via the Fourier transform incorporating Eqs.\
(\ref{TargetIndepQuark}) -- (\ref{TargetIndepGluon}), are related to the
light-cone position space kernels discussed in the preceding section
\cite{MulDitRobGeyHor98,Rad1996b,Rad1997a}
\begin{equation}
\label{FourierTransformDO}
\bit{K} ( x_1 , x_2 \,|\, y_1 , y_2 )
=
\int_0^1 du \int_0^1 dv \,
\left(
\begin{array}{cc}
- {\cal K}^{qq} (u, v)
&
{\cal K}^{qg} (u, v) \partial_{x_1}
\\
{\cal K}^{gq} (u, v) \partial^{- 1}_{x_1}
&
- {\cal K}^{gg} (u, v)
\end{array}
\right)
\delta \left( x_1 - y_1 (1 - u) - y_2 v \right)
\, , \nonumber
\end{equation}
where we have used the momentum conservation condition $\sum_{i = 1}^2 (x_i
- y_i) = 0$. The momentum fractions of incoming $y_i$ and outgoing $x_i$ legs
can be expressed in terms of the symmetric variables as
$$
x_1 = \frac{\eta + x}{2}
\, , \qquad
x_2 = \frac{\eta - x}{2}
\, , \qquad
y_1 = \frac{\eta + y}{2}
\, , \qquad
y_2 = \frac{\eta - y}{2}
\, .
$$
We denote  $\partial_{x_1} \equiv d/dx_1$ and $\partial_{x_1}^{- 1}
\equiv \int^{x_1} d x'_1$. The indefinite integration limits in the
$gq$ channel induces an ambiguity which, however,  affects only the
unphysical moments\footnote{\label{UnphysMoments} Those moments which
vanish by symmetry reasons, i.e., $j$ even (odd) in the singlet sector
of the vector (axial) operators.} and has to be fixed ``by hand'', e.g.,
by comparison of moments calculated in both representations. Note that
this ambiguity implicitly appears also in the diagrammatic calculation
of Feynman diagrams in the light-cone fraction representation and is
responsible for different results available in the literature.

The representation (\ref{FourierTransformDO}) implies a simple scaling property
of the $\bit{K}$ kernel, due to which  its components  are in fact two-variable
functions of the $x/\eta$ and $y/\eta$ ratios \cite{MulDitRobGeyHor98,Rad1997a},
\begin{eqnarray}
\label{scaling-rel}
\bit{K}
\left(
\left.
\frac{\eta + x}{2} , \frac{\eta - x}{2} \right|
\frac{\eta + y}{2} , \frac{\eta - y}{2}
\right)
=
\frac{1}{\eta} \left(
\begin{array}{rr}
k^{qq} \left( \frac{x}{\eta} , \frac{y}{\eta} \right)
&
\eta^{-1} k^{qg} \left( \frac{x}{\eta} , \frac{y}{\eta} \right)
\\
\eta k^{gq} \left( \frac{x}{\eta} , \frac{y}{\eta} \right)
&
k^{gg} \left( \frac{x}{\eta} , \frac{y}{\eta} \right)
\end{array}
\right)
\, .
\end{eqnarray}
The invariance under the charge conjugation implies now the symmetry for
the diagonal $k^{aa} (x, y) = k^{aa} (- x, - y)$ and off-diagonal ($a \neq b$)
$k^{ab} (x, y) = - k^{ab} (- x, - y)$ elements. The connection between the
parton-parton and parton-antiparton regions can be established by the
substitution $y \to - y$. As we already explained for the light-cone position
kernels, it is sufficient  to consider the $u + v \leq 1$ part of the total
support region $0 \le u, v \le 1$. The contribution that comes from the region
$u + v \ge 1$ can be formally  obtained from the first region by changing $y
\to - y$. Thus, the support of the kernels extends into two additional regions
in the upper-left and lower-right corners of the support diagram, see Fig.\
\ref{fig-sup-lr} (b). Then the integral representation (\ref{FourierTransformDO})
implies the support shown in Fig.\ \ref{fig-sup-lr} (b), or formally
\begin{equation}
\label{GAMTRE}
k^{ab} (x, y)
=
\vartheta (x, y) f^{ab} (x, y)
\pm
\vartheta (- x, - y) f^{ab} (- x, - y)
\, ,
\end{equation}
where the $(-)$ $+$ stands for (off-) diagonal entries. Here the step-function
is
\begin{equation}
\label{varThetaGeneric}
\vartheta (x, y)
=
\theta \left( \frac{y - x}{1 + y} \right)
\theta \left( \frac{1 + x}{1 + y} \right)
\mbox{sign} (1 + y)
=
\theta (y - x) \theta (1 + x) - \theta (x - y) \theta(- 1 - x)
\, ,
\end{equation}
and $f^{ab}$ are analytic functions related to integrals of the coordinate-space
evolution kernels as
\begin{equation}
\label{FourierTransF}
f^{aa} (x, y) = - \int_0^{\frac{1 + x}{1 + y}} d w \,
{\cal K}^{aa}
\left(
\frac{1 - x - (1 - y) w}{2} , \frac{1 + x - (1 + y) w}{2}
\right)
\, ,
\end{equation}
and analogously for off-diagonal elements with operators $- \partial_x$
and $- \partial^{- 1}_x$ acting on the right-hand side of the equality
for $qg$ and $gq$ kernels, respectively.

Looking  at Fig.\ \ref{fig-sup-lr} one gets  the impression that the whole
kernel can be obtained from the region $|x|,|y| \le 1$. Indeed, it was proved
in \cite{MulDitRobGeyHor98}  that the continuation is unique. For practical
purposes, it is sufficient to replace the $\theta$ structure:
\begin{eqnarray}
\theta(y - x)_{|x|,|y| \le 1} \to \vartheta(x, y)
\, .
\end{eqnarray}
Thus, the evolution kernels for GPDs can be considered as generalized
exclusive kernels, addressed below, and their restoration from the given
Efremov-Radyushkin-Brodsky-Lepage (ER-BL or exclusive) kernels is simple
and unique. We will elaborate more on these issues in Section \ref{app-LIM}.

\subsubsection{One-loop results}
\label{MomSpaceGerKernels}

A very simple computation method to calculate these kernels at one-loop
order is presented in Appendix \ref{CompTechMomentumKernels}. It was
originally suggested in Refs.\ \cite{BukKurLip83,BukFroLipKur85} and used
there to compute all leading order off-forward evolution kernels. Recently
they were recalculated by different means in a number of studies dedicated
specifically to analyses of GPDs \cite{Ji96b,FraFreGuzStr98,BelMul98a,Bel00}.
The results for sectors with different quantum numbers are summarized below.

\noindent $\bullet$ Even-parity sector:
\begin{eqnarray}
\label{PE-ext}
K^{qq; V}_{(0)} (x_1 , x_2 | y_1, y_2)
\!\!\!&=&\!\!\!
C_F
\Bigg[
\frac{x_1}{x_1 - y_1} \vartheta_{11}^0 (x_1 , x_1 - y_1)
+
\frac{x_2}{x_2 - y_2} \vartheta_{11}^0 (x_2, x_2 - y_2)
\\
&&
\qquad\qquad\qquad\qquad\qquad\qquad
+
\vartheta_{111}^0 ( x_1 , - x_2 , x_1 - y_1 )
\Bigg]_+
\, , \nonumber\\
K^{qg; V}_{(0)} (x_1 , x_2 | y_1, y_2)
\!\!\!&=&\!\!\!
2 T_F N_f
\left[
\vartheta_{112}^1 (x_1, - x_2 , x_1 - y_1)
+
2 \frac{x_1 - y_1}{y_1 y_2} \vartheta_{111}^0 (x_1, - x_2 , x_1 - y_1)
\right]
\, , \nonumber\\
K^{gq; V}_{(0)} (x_1 , x_2 | y_1, y_2)
\!\!\!&=&\!\!\!
C_F
\left[
( y_1 - y_2 ) \vartheta_{111}^0 (x_1, - x_2 , x_1 - y_1)
+
x_1 x_2
\vartheta_{111}^1 (x_1, - x_2 , x_1 - y_1)
\right]
\, , \nonumber\\
K^{gg; V}_{(0)} \left( x_1, x_2 | y_1, y_2 \right)
\!\!\!&=&\!\!\!
C_A
\Bigg[
\frac{x_1}{y_1}
\left[
\frac{x_1}{x_1 - y_1} \vartheta^0_{11} (x_1, x_1 - y_1)
\right]_+
+
\frac{x_2}{y_2}
\left[
\frac{x_2}{x_2 - y_2} \vartheta^0_{11} (x_2, x_2 - y_2)
\right]_+
\nonumber\\
&+&\!\!\!
2 \frac{x_1 x_2 + y_1 y_2}{y_1 y_2}
\vartheta^0_{111} (x_1, - x_2, x_1 - y_1)
+
2 \frac{x_1 x_2}{y_1 y_2} \frac{x_1 y_1 + x_2 y_2}{(x_1 + x_2)^2}
\vartheta^0_{11} (x_1, - x_2)
\nonumber\\
&&\qquad\qquad\qquad\qquad\qquad\qquad\qquad\quad\!\!
+
\left( \frac{1}{2} \frac{\beta_0}{C_A} + 2 \right) \delta (x_1 - y_1)
\Bigg]
\, . \nonumber
\end{eqnarray}

\noindent $\bullet$ Odd-parity sector:
\begin{eqnarray}
K^{qq; A}_{(0)} (x_1 , x_2 | y_1, y_2)
\!\!\!&=&\!\!\!
K^{qq; V}_{(0)} (x_1 , x_2 | y_1, y_2)
\, , \\
K^{qg; A}_{(0)} (x_1 , x_2 | y_1, y_2)
\!\!\!&=&\!\!\!
2 T_F N_f
\vartheta_{112}^1 (x_1, - x_2 , x_1 - y_1)
\, , \nonumber\\
K^{gq; A}_{(0)} (x_1 , x_2 | y_1, y_2)
\!\!\!&=&\!\!\!
C_F
\left[
( x_1 - x_2 ) \vartheta_{111}^0 (x_1, - x_2 , x_1 - y_1)
+
x_1 x_2 \vartheta_{111}^1 (x_1, - x_2 , x_1 - y_1)
\right]
\, , \nonumber\\
K^{gg; A}_{(0)} \left( x_1, x_2 | y_1, y_2 \right)
\!\!\!&=&\!\!\!
C_A
\Bigg[
\frac{x_1}{y_1}
\left[
\frac{x_1}{x_1 - y_1} \vartheta^0_{11} (x_1, x_1 - y_1)
\right]_+
+
\frac{x_2}{y_2}
\left[
\frac{x_2}{x_2 - y_2} \vartheta^0_{11} (x_2, x_2 - y_2)
\right]_+
\nonumber\\
&+&\!\!\!
2 \frac{x_1 y_2 + y_1 x_2}{y_1 y_2}
\vartheta^0_{111} (x_1, - x_2, x_1 - y_1)
+
2 \frac{x_1 x_2}{y_1 y_2} \vartheta^0_{11} (x_1, - x_2)
\nonumber\\
&&\qquad\qquad\qquad\qquad\qquad\qquad\qquad\quad\!\!
+
\left( \frac{1}{2} \frac{\beta_0}{C_A} + 2 \right)
\delta (x_1 - y_1)
\Bigg]
\, . \nonumber
\end{eqnarray}

\noindent $\bullet$ Maximal-spin (or transversity) sector:
\begin{eqnarray}
\label{MH-ext}
\!\!\!K^{qq; T}_{(0)} ( x_1, x_2 | y_1, y_2 )
\!\!\!&=&\!\!\!
C_F
\Bigg[
\left[
\frac{x_1}{x_1 - y_1} \vartheta^0_{11} (x_1, x_1 - y_1)
\right]_+
+
\left[
\frac{x_2}{x_2 - y_2} \vartheta^0_{11} (x_2, x_2 - y_2)
\right]_+
\\
&&\qquad\qquad\qquad\qquad\qquad\qquad\qquad\!\!
+
\frac{1}{2} \delta (x_1 - y_1)
\Bigg]
\, , \nonumber\\
\!\!\!K^{gg; T}_{(0)} ( x_1, x_2 | y_1, y_2 )
\!\!\!&=&\!\!\!
C_A
\left[
\frac{x_1}{y_1}
\left[
\frac{x_1}{x_1 - y_1} \vartheta^0_{11} (x_1, x_1 - y_1)
\right]_+
+
\frac{x_2}{y_2}
\left[
\frac{x_2}{x_2 - y_2} \vartheta^0_{11} (x_2, x_2 - y_2)
\right]_+ \right.
\nonumber\\
&&\left. \qquad\qquad\qquad\qquad\qquad\qquad\qquad\quad\!
+ \left( \frac{1}{2} \frac{\beta_0}{C_A} + 2 \right)
\delta (x_1 - y_1)
\right ]
\, . \nonumber
\end{eqnarray}
The regularization of the end-point behavior $y_i \to x_i$ is achieved via the
plus-prescription,
\begin{equation}
\left[
\frac{x_1}{x_1 - y_1} \vartheta^0_{11} (x_1, x_1 - y_1)
\right]_+
\equiv
\frac{x_1}{x_1 - y_1} \vartheta^0_{11} (x_1, x_1 - y_1)
-
\delta (x_1 - y_1)
\int d x'_1
\frac{x_1'}{x'_1 - y_1} \vartheta^0_{11} (x'_1, x'_1 - y_1)
\, .
\end{equation}
Note that the $\vartheta$-functions other than $\vartheta^{11}_0$ can
be reduced to the latter making use of Eqs.\ (\ref{ThetaFirst}) --
(\ref{ThetaLast}). Moreover, the step-function $\vartheta (x, y)$ from
the previous section is related to $\vartheta^{11}_0$ via
\begin{equation}
\label{RelationLipatovThetaGenericTheta}
\vartheta (x, y) = (1 + y) \vartheta^{11}_0 (1 + x, x - y)
\, ,
\end{equation}
as can be seen by matching their definitions.

The even parity kernels (\ref{PE-ext}) possess the off-diagonal pieces
in the non-physical sector as defined in the footnote \ref{UnphysMoments}.
To cure this problem in appropriate way, so that the forward limit would
not be affected, we should add certain terms to the kernels computed from
Feynman diagrams
\begin{equation}
K^{ab} \to K^{ab} + \delta K^{ab}
\, ,
\end{equation}
where \cite{BelMul98a}
\begin{eqnarray}
\delta K^{qg ; V}_{(0)} (x_1 , x_2 | y_1 , y_2 )
\!\!\!&=&\!\!\!
4 T_F N_f \frac{x_1 x_2}{y_1 y_2} (y_2 - y_1) \vartheta_{11}^0 (x_1 , - x_2)
\, , \\
\delta K^{gq ; V}_{(0)} (x_1 , x_2 | y_1 , y_2 )
\!\!\!&=&\!\!\!
2 C_F x_1 x_2 (x_2 - x_1) \vartheta_{11}^0 (x_1 , - x_2)
\, ,
\end{eqnarray}
In the forward limit $\vartheta_{11}^0 (x, - x) = 0$, so that these extra
terms die out.

\subsubsection{From coordinate to momentum-space kernels}

As we emphasized earlier, Eq. (\ref{FourierTransF}) connects the coordinate and
momentum space kernels. However, since the relation was given for the momentum
kernels (\ref{GAMTRE}) rather than those in Section \ref{MomSpaceGerKernels},
which depend on different variables, it is instructive to rewrite it. For the
diagonal kernels $K$ written in terms of $x_i$ and $y_i$ variables, Eq.\
(\ref{FourierTransF}) gives
\begin{eqnarray}
K^{aa} (x_1, x_2 | y_1, y_2)
=
\!\!\!&-&\!\!\!
\frac{y_1}{x_1 + x_2} \vartheta_{11}^0 (x_1 - y_1, x_1)
\int_0^{x_1/y_1} dw \,
{\cal K}^{aa}
\left(
\frac{x_2 - y_2 w}{x_1 + x_2} ,
\frac{x_1 - y_1 w}{x_1 + x_2}
\right)
\\
\!\!\!&-&\!\!\!
\frac{y_2}{x_1 + x_2} \vartheta_{11}^0 (x_2 - y_2, x_2)
\int_0^{x_2/y_2} dw \,
{\cal K}^{aa}
\left(
\frac{x_2 - y_2 w}{x_1 + x_2} ,
\frac{x_1 - y_1 w}{x_1 + x_2}
\right)
\, , \qquad \nonumber
\end{eqnarray}
where we assumed the momentum conservation $x_1 + x_2 = y_1 + y_2$. For the
off-diagonal elements, as it is evident from Eq.\ (\ref{FourierTransformDO}),
one should also apply the operators $- \partial_x$ and $- \partial^{- 1}_x$
on the right-hand side. Having these results, one can verify that the momentum
space evolution kernels (\ref{PE-ext}) -- (\ref{MH-ext}) arise from the coordinate
space ones (\ref{QQV-kernel}) -- (\ref{MH-coord}).

In the one-loop case, one can do the transformation in a more explicit way by
making use of Eq.\ (\ref{FourierTransformDO}).  The first step of the resuction
consists in  rescaling of the Feynman integration variables
\begin{equation}
\label{ShiftOfVariables}
\int_0^1 du \int_0^1 dv \, \theta (1 - u - v) f (u, v)
=
\int_{0}^1 du \int_0^1 dv \, \bar{u} f (u , \bar{u} v)
\, .
\end{equation}
The delta-function is eliminated then by using the general formula
\begin{equation}
\label{RedTheta}
\int_{0}^{1} d v f(v) \delta (x - v y)
=
f \left( \frac{x}{y} \right) {\vartheta}^0_{11} (x, x - y)
\, ,
\end{equation}
which brings in the  $\vartheta_{11}^0$-function of the momentum space
formulation. Notice that Eq.\ (\ref{scaling-rel}) is now an obvious
consequence of the rescaling property of the argument of the delta-function,
so that
$$
\vartheta_{11}^0 (x, x - y)
=
\frac{1}{\eta} \vartheta_{11}^0
\left(
\frac{x}{\eta} , \frac{x - y}{\eta}
\right)
\, .
$$
The second integration is accomplished making use of the result
\begin{eqnarray}
\int_{0}^{1}
d u \, \bar u^n \,
\vartheta_{11}^0 (x_1 - y_1 \bar{u} , x_1 - \eta \bar{u} )
\!\!\!&=&\!\!\!
\frac{1}{n}
\Bigg\{
\left[ 1 - \left( \frac{x_1}{\eta} \right)^n \right]
\vartheta_{11}^0 (x_1 - y_1 , - x_2)
\\
&-&\!\!\!
\frac{y_1}{y_2}
\left[
\left( \frac{x_1}{\eta} \right)^n - \left( \frac{x_1}{y_1} \right)^n
\right]
\vartheta_{11}^0 (x_1 - y_1 , x_1)
\Bigg\}
\, , \nonumber
\end{eqnarray}
where $\eta = x_1 + x_2 = y_1 + y_2$. As a simple demonstration of these
formulas, one finds
$$
{\cal K}^{qq}_{\scriptscriptstyle\rm test} (u, v) = 1
\qquad \Leftrightarrow \qquad
K^{qq}_{\scriptscriptstyle\rm test} (x_1, x_2 | y_1, y_2)
=
\vartheta_{111}^0 (x_1, - x_2, x_1 - y_1)
\, ,
$$
according to Eq.\ (\ref{FourierTransformDO}).

\subsubsection{Reduction to inclusive kernels}
\label{SectionDGLAPkernels}

The off-forward evolution kernels yield the usual DGLAP splitting functions
when the skewness parameter is set to zero, $\eta = 0$,
\begin{equation}
\label{ReductionToDGLAP}
K (x , - x | 1 , -1) \equiv - P (x)
\, ,
\end{equation}
where $0 \leq x \leq 1$ so that $\vartheta_{11}^0 (x - 1, x) = 1$. Then the
leading order kernels have the form, which coincide with the well-known results,
which present below for completeness.
\begin{itemize}
\item Even-parity sector:
\begin{eqnarray}
\label{LOvectorDGLAPqq}
P^{qq; V}_{(0)} (x)
\!\!\!&=&\!\!\!
C_F \left[ \frac{1 + x^2}{1 - x} \right]_+
\, , \\
P^{qg; V}_{(0)} (x)
\!\!\!&=&\!\!\!
2 T_F N_f \left\{ x^2 + (1 - x)^2 \right\}
\, , \\
P^{gq; V}_{(0)} (x)
\!\!\!&=&\!\!\!
C_F \frac{1}{x} \left\{ 1 + (1 - x)^2 \right\}
\, , \\
P^{gg; V}_{(0)} (x)
\!\!\!&=&\!\!\!
2 C_A
\left\{
\frac{1}{x} + \left[ \frac{1}{1 - x} \right]_+ - 2 + x (1 - x)
\right\}
-
\frac{\beta_0}{2} \delta(1 - x)
\, .
\end{eqnarray}
\item Odd-parity sector:
\begin{eqnarray}
P^{qq; A}_{(0)} (x)
\!\!\!&=&\!\!\!
C_F \left[ \frac{1 + x^2}{1 - x} \right]_+
\, , \\
P^{qg; A}_{(0)} (x)
\!\!\!&=&\!\!\!
2 T_F N_f
\left\{ x^2 - (1 - x)^2 \right\}
\, , \\
P^{gq; A}_{(0)} (x)
\!\!\!&=&\!\!\!
C_F \frac{1}{x} \left\{ 1 - (1 - x)^2 \right\}
\, , \\
P^{gg; A}_{(0)} (x)
\!\!\!&=&\!\!\!
2 C_A \left\{ 1 - 2 x + \left[ \frac{1}{1 - x} \right]_+ \right\}
-
\frac{\beta_0}{2} \delta(1 - x)
\, .
\end{eqnarray}
\item Maximal-helicity sector:
\begin{eqnarray}
P^{qq; T}_{(0)} (x)
\!\!\!&=&\!\!\!
C_F
\left\{
2 x \left[ \frac{1}{1 - x} \right]_+
+
\frac{3}{2} \delta (1 - z)
\right\}
\, , \\
P^{gg; T}_{(0)} (x)
\!\!\!&=&\!\!\!
2 C_A x^2
\left[ \frac{1}{1 - x} \right]_+
-
\frac{\beta_0}{2} \delta (1 - x)
\, . \\
\end{eqnarray}
\end{itemize}
We have used the standard form of the plus-regularization of the end-point
singularity,
\begin{equation}
\label{pluspres}
\left[ \frac{1}{1 - x} \right]_+ \equiv \frac{1}{1 - x}
-
\delta(1 - x) \int_{0}^1 \frac{dx'}{1 - x'}
\, .
\end{equation}

The Mellin moments of the evolution kernels define the anomalous dimensions
of the basic Wilson operators (i.e., those without total derivatives)
\begin{equation}
\label{ForwardADs}
\int_0^1 d x \, x^j P (x ; \alpha_s)
=
- \frac{1}{2} \gamma^{\scriptscriptstyle\rm fw}_j (\alpha_s) \, .
\end{equation}
As usual, the splitting functions are given by the  expansion in coupling constant
\begin{equation}
\label{ExpansionPDFeigenfunction}
\gamma^{\scriptscriptstyle\rm fw}_j (\alpha_s) = \sum_{n = 0}^{\infty}
\left( \frac{\alpha_s}{2 \pi} \right)^{n + 1}
\gamma^{\scriptscriptstyle\rm fw}_{(n)j}
\, .
\end{equation}
In the lowest order, the Mellin moments of the nonsinglet parton distributions
\begin{equation}
q_j (\mu^2) = \int_0^1 dx \, x^j \, q(x; \mu^2)
\end{equation}
have a simple evolution
\begin{equation}
q_j(\mu^2)
=
\left(
\frac{\alpha_s (\mu^2_0)}{\alpha_s (\mu^2)}
\right)^{\gamma^{qq; \scriptscriptstyle\rm fw}_{(0)}j / \beta_0}
q_j (\mu_0^2)
\, ,
\end{equation}
governed by the one-loop $qq$ anomalous dimension $\gamma^{qq;
{\scriptscriptstyle\rm fw}}_{(0)j}$ and the one-loop QCD beta-functions
$\beta_0$ defined in Eq.\ (\ref{BetaFunction}). To get the evolved
parton distribution $q (x; \mu^2)$, one can use the formal reconstruction
of a function  from  its Mellin moments
\begin{equation}
\label{deltaExpansion}
q (x; \mu^2) = \sum_{n=0}^{\infty} \frac{(-1)^j}{j!} \delta^{(j)} (x) q_j (\mu^2)
\, ,
\end{equation}
which gives
\begin{equation}
\label{PDFinDeltas}
q (x; \mu^2) = \sum_{j = 0}^\infty
\frac{(-1)^j}{j!} \delta^{(j)} (x)
\left(
\frac{\alpha_s (\mu^2_0)}{\alpha_s (\mu^2)}
\right)^{\gamma^{qq; \scriptscriptstyle\rm fw}_{(0)j} / \beta_0}
\int_0^1 dy \, y^j \, q (y; \mu_0^2)
\, .
\end{equation}
Using the fact that all $\gamma^{qq; \scriptscriptstyle\rm fw}_j > 0$ for $j > 0$,
while the lowest $j = 0$ moment of the parton distribution corresponds to a local
conserved axial/vector current with vanishing anomalous dimension, one finds
\begin{equation}
q (x; \mu^2 \to \infty) = \delta (x) \int_0^1 dy \, q (y; \mu_0^2)
\, ,
\end{equation}
Thus, we observe the qualitative feature of the inclusive evolution pointed
out at the beginning of this chapter. Namely, when the proton is probed with
infinitely fine resolution, one finds an infinite number of partons with tiny
(actually zero) momentum with respect to the momentum of the parent hadron.

The expression (\ref{deltaExpansion}), of course, has an obvious practical
difficulty: all the derivatives of the delta function are concentrated at a
single point $x = 0$, and the restoration of the correct support of the
function $q(x)$ is a result of the infinite summation of a  slow convergent
series.  In other words, Eq. (\ref{deltaExpansion})  should only be understood
in the sense of (mathematical) distributions, as a statement about its
integration with smooth functions.  A similar problem will be encountered
later in this chapter when we will attempt to solve the evolution equation
for GPDs using the eigenfunctions of the corresponding evolution equation.

\subsubsection{Reduction to exclusive kernels}
\label{ERBLleadingOrderEvKernels}

The meson and baryon distribution amplitudes discussed in Section
\ref{SectionDistributionAmplitudes} also evolve with the change of the resolution
scale $\mu^2$. The meson distribution amplitudes $\phi_M (x; \mu^2)$ obey the
ER-BL evolution equation \cite{EfrRad78,Chernyak:1977fk,Farrar:1979aw,BroLep79}
\begin{equation}
\label{ERBLevolEq}
\frac{d}{d \ln \mu^2} \, \phi_M (x; \mu^2)
=
\int_0^1 dy \, V (x,y) \phi_M (y; \mu^2)
\, .
\end{equation}
The functions $V ( x, y )$ correspond to the $\eta = 1$ limit of the general
off-forward evolution kernels:
\begin{equation}
\label{ReductionToERBL}
K ( x, 1 - x | y, 1 - y ) \equiv -  V ( x, y )
\, ,
\end{equation}
which holds for diagonal and off-diagonal elements of the mixing matrix.
Since $0 \leq x, y \leq 1$, we have $\vartheta_{11}^0 (x - y, x) = \frac{1}{y}
\theta (y - x)$, and  it is easy to derive that the leading order ER-BL kernels
have the following form\footnote{Note that the terms with $\delta$-function
accompanied by step-functions are understood in the following way $\delta(x - y)
\left[ \theta (y - x) + \theta (x - y) \right] = \delta(x - y)$.}
\begin{equation}
\label{DefLOERBLkernel}
V^{ab}_{(0)} ( x, y )
=
\theta( y - x ) F^{ab} ( x, y )
\pm
\theta( x - y ) \overline{F}^{ab} ( x, y ) ; \ \pm \,
\mbox{\ for\ }
\left\{ {\mbox{$a = b$} \atop \mbox{$a \not= b$}} \right.
\, ,
\end{equation}
where $\overline{F}^{ab} ( x, y ) \equiv F^{ab} ( \bar{x} , \bar{y} )$ with
$\bar{x} = 1 - x$ and $\bar{y} = 1 - y$. The $-$ and $+$ signs in Eq.\
(\ref{DefLOERBLkernel}) correspond off-diagonal $a \neq b$ and diagonal $a = b$
elements, respectively, of the $V^{ab}_{(0)}$ matrix. The use of Eqs.\
(\ref{ReductionToERBL}) reproduces the kernels obtained earlier in many papers.
The results are summarized below.
\begin{itemize}
\item Even-parity sector \cite{EfrRad78,BroLep79,Cha80}:
\begin{eqnarray}
\label{LOvectorERBLkernel}
F^{qq; V} ( x, y )
\!\!\!&=&\!\!\!
C_F \frac{x}{y}
\left\{
\frac{1}{[y - x]_+} + 1 + \frac{3}{2} \delta(x - y)
\right\}
\, , \\
F^{qg; V} ( x, y )
\!\!\!&=&\!\!\!
2 N_f T_F \frac{x}{y^2 \bar y} (2 x - y - 1)
\, , \\
F^{gq; V} ( x, y )
\!\!\!&=&\!\!\!
C_F \frac{x}{y} ( 2 y - x )
\, , \\
F^{gg; V} ( x, y )
\!\!\!&=&\!\!\!
C_A
\frac{x^2}{y^2}
\left\{
\frac{1}{[y - x]_+}
+
2 \bar x + 2 y ( 1 + 2 \bar x )
-
\frac{1}{2}\frac{\beta_0}{C_A} \delta (x - y)
\right\}
\, .
\end{eqnarray}

\item Odd-parity sector \cite{Ohr81,BaiGro81}:
\begin{eqnarray}
F^{qq; A} ( x, y )
\!\!\!&=&\!\!\!
C_F \frac{x}{y}
\left\{
\frac{1}{[y - x]_+} + 1 + \frac{3}{2} \delta(x - y)
\right\}
\, , \\
F^{qg; A} ( x, y )
\!\!\!&=&\!\!\!
- 2 N_f T_F \frac{x}{y^2}
\, , \\
F^{gq; A} ( x, y )
\!\!\!&=&\!\!\!
C_F \frac{x^2}{y}
\, , \\
F^{gg; A} ( x, y )
\!\!\!&=&\!\!\!
C_A
\frac{x^2}{y^2}
\left\{
\frac{1}{[y - x]_+}
+
2
-
\frac{1}{2}\frac{\beta_0}{C_A} \delta (x - y)
\right\}
\, .
\end{eqnarray}

\item Maximal-helicity sector \cite{BelMul97,JiHoo98}:
\begin{eqnarray}
F^{qq; T} ( x, y )
\!\!\!&=&\!\!\!
C_F \frac{x}{y}
\left\{
\frac{1}{[y - x]_+} + \frac{3}{2} \delta (x - y)
\right\}
\, , \\
F^{gg; T} ( x, y )
\!\!\!&=&\!\!\!
C_A \frac{x^2}{y^2}
\left\{
\frac{1}{[y - x]_+} - \frac{1}{2} \frac{\beta_0}{C_A} \delta (x - y)
\right\}
\, .
\label{LOtransvERBLkernel}
\end{eqnarray}
\end{itemize}
All of the singularities are regularized as follows
\begin{eqnarray*}
&&
\frac{\theta (y - x)}{[y - x]_+}
\equiv
\frac{\theta (y - x)}{y - x}
-
\delta (y - x) \int_0^1 dx' \frac{\theta (y - x')}{y - x'}
\, , \\
&&
\frac{\theta (x - y)}{[x - y]_+}
\equiv
\frac{\theta (x - y)}{x - y}
-
\delta (y - x) \int_0^1 dx' \frac{\theta (x' - y)}{x' - y}
\, .
\end{eqnarray*}

\subsubsection{From exclusive to inclusive kernels}
\label{app-LIM}

\begin{figure}[t]
\mbox{
\begin{picture}(0,100)(-150,0)
\put(0,0){\insertfig{7}{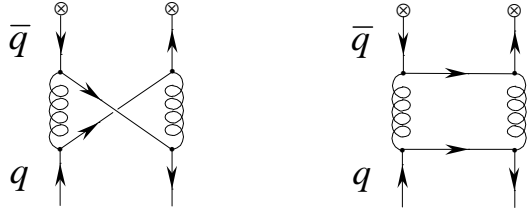}}
\end{picture}
}
\caption{\label{QuakrAntiquarkMixing} The typical contribution inducing
the parton-antiparton mixing in evolution kernels.}
\end{figure}

Both inclusive DGLAP splitting functions and exclusive ER-BL kernels for
mesons are particular  projections (reductions) of the same general off-forward
evolution kernels. In fact, there exist formulas providing a direct reduction
of the ER-BL kernels to the DGLAP ones. They can be most conveniently derived
if the ER-BL evolution is written in the matrix form, as an equation for the
moments of the distribution amplitude
\begin{equation}
a_j (\mu^2) = \int_0^1 d x \, x^j \phi_M (x; \mu^2)
\, .
\end{equation}
The coefficients $a_j$ are proportional to the matrix elements of local
operators
\begin{equation}
\label{MomDAlocOper}
\langle p | \bar{\psi} (0)
( - i\! \stackrel{{}_{\leftarrow}}{\partial}{\!}^+ )^j
\gamma^+ \psi (0) | 0 \rangle
= - i (p^+)^{j + 1} f_M \int_0^1 dx \, x^j \phi_M (x)
\, .
\end{equation}
Under evolution, these operators ${\mathcal O}_j$ mix with the operators
$(\partial^+)^{j-k} {\mathcal O}_k$ having the same total number $j$ of vector
indices but containing $j - k$ total derivatives acting on the operator
${\mathcal O}_k$ as a whole. Since a total derivative $\partial^{\mu}$ just
produces the overall factor $p^\mu$ in the matrix element (see footnote
\ref{TotalDerFoot}), the evolution equation for $a_j$'s has the form
\begin{equation}
\frac{d}{d \ln \mu^2} \, a_j(\mu^2)
=
- \frac12 \sum_{k = 0}^j \gamma_{jk} (g) a_k(\mu^2)
\, .
\end{equation}
The matrix elements $\gamma_{jk}$ are related to the kernel $V(x,y)$ by
\begin{equation}
\label{MellinMomERBL}
\int_0^1 dx \, x^j \, V(x,y) = - \frac12 \sum_{k = 0}^j \gamma_{jk} (g) y^k
\, .
\end{equation}
The appearance of a polynomial in $y$ on the right-hand side, rather than a single
power $y^j$, is a simple consequence of fact that the evolution kernel $V$ encodes
the mixing of operators with total derivatives. The local operator
$$
x^j
\leftrightarrow
\bar{\psi} (0)
( - i\! \stackrel{{}_{\leftarrow}}{\partial}{\!}^+ )^j
\gamma^+ \psi (0)
$$
is allowed to evolve with the ``probability'' $\gamma_{jk}$ into the operators with
less ``left derivatives'' $\stackrel{{}_\leftarrow}{\partial}{\!}^+$ and
more ``total derivatives'' $\partial^+$
$$
y^k
\leftrightarrow
\big( i \partial^+ \big)^k
\bar{\psi} (0)
( - i\! \stackrel{{}_{\leftarrow}}{\partial}{\!}^+ )^{j - k}
\gamma^+ \psi (0)
\, .
$$
In the forward case, the operators with total derivatives have zero matrix
elements, so only the diagonal term $\gamma_{jj}$ survives. The latter is
given by the $j$th moment of the DGLAP splitting function $P(z)$, and this
establishes its relationship to the ER-BL kernel. In the singlet case, the
relations are complicated by the quark-gluon mixing. So, we have
\begin{eqnarray}
\label{Mom-ERBL}
\int_0^1 dx\, \left( {x^j\atop  x^{j-1}} \right)
\left(
\begin{array}{rr}
V^{qq} & V^{qg}
\\
V^{gq} & V^{gg}
\end{array}
\right)(x,y)
=
-\frac{1}{2} \sum_{k = (0, 1)}^{j}
\left(
\begin{array}{rr}
\gamma^{qq}_{jk} & \gamma^{qg}_{jk}
\\
\gamma^{gq}_{jk} & \gamma^{gg}_{jk}
\end{array}
\right) \left( {y^k \atop  y^{k-1}} \right)
\, ,
\end{eqnarray}
for the ER-BL kernels, and
\begin{eqnarray}
\label{Mom-DGLAP}
\int_0^1 dx\, x^j
\left(
\begin{array}{rr}
P^{qq} & P^{qg}
\\
P^{gq} & P^{gg}
\end{array}
\right)(x)
=
- \frac{1}{2}
\left(
\begin{array}{rr}
\gamma^{qq; {\scriptscriptstyle\rm fw}}_j & \gamma^{qg; {\scriptscriptstyle\rm fw}}_j
\\
\gamma^{gq; {\scriptscriptstyle\rm fw}}_j & \gamma^{gg; {\scriptscriptstyle\rm fw}}_j
\end{array}
\right)
\, ,
\end{eqnarray}
for the splitting functions, such that $\gamma^{ab}_{jj} = \gamma^{ab;
{\scriptscriptstyle\rm fw}}_j$. To extract the diagonal elements  from Eq.\
(\ref{Mom-ERBL}), we substitute $y$ by $1/\eta$ and multiply each entry on both
sides with a sufficient power of $\eta$, found by inspection. As suggested by
Fig.\ \ref{fig-sup-lr}, the ER-BL kernels, having the support $0 \leq x, y \leq 1$,
can be extended to the entire $(x, y)$-plane according to the rules which we spell
out later. Rescaling the integration variable and taking the limit $\eta \to 0$,
we find
\begin{eqnarray}
\label{Mom-ERBL-Ext}
\lim_{\eta\to 0}
\int_{0}^{1} \frac{dx}{\eta}\, x^j
\left(
\begin{array}{rr}
k^{qq}
& \
\frac{1}{\eta} k^{qg}
\\
\frac{\eta}{x} k^{gq}
& \
\frac{1}{x} k^{gg}
\end{array}
\right)
\left( 2 \frac{x}{\eta} - 1 , \frac{2}{\eta} - 1 \right)
=
\frac{1}{2}
\left(
\begin{array}{rr}
\gamma^{qq}_{jj} & \gamma^{qg}_{jj}
\\
\gamma^{gq}_{jj} & \gamma^{gg}_{jj}
\end{array}
\right)
\, ,
\end{eqnarray}
where we used the extended kernels in the whole region. This is done by a
simple replacement of the $\theta$ functions in the ER-BL kernels by
\begin{eqnarray}
\label{Extension1}
\theta (y - x)
\!\!\!&\to&\!\!\!
\vartheta (2 x - 1 , 2 y - 1 )
=
\theta \left( 1 - \frac{x}{y} \right)
\theta \left( \frac{x}{y} \right)
{\rm sgn} (y)
\, , \\
\theta (x - y)
\!\!\!&\to&\!\!\!
\vartheta (1 - 2 x , 1 - 2 y )
=
\theta \left( 1 - \frac{\bar x}{\bar y} \right)
\theta \left( \frac{\bar x}{\bar y} \right)
{\rm sgn} (\bar y)
\, , \\
\theta (y - \bar x)
\!\!\!&\to&\!\!\!
\vartheta (1 - 2 x , 2 y - 1 )
=
\theta \left( 1 - \frac{\bar x}{y} \right)
\theta \left( \frac{\bar x}{y} \right)
{\rm sgn} (y)
\, , \\
\theta (\bar y - x)
\!\!\!&\to&\!\!\!
\vartheta ( 2 x - 1 , 1 - 2 y )
=
\theta \left( 1 - \frac{x}{\bar y} \right)
\theta \left( \frac{x}{\bar y} \right)
{\rm sgn} (\bar y)
\, ,
\label{Extension4}
\end{eqnarray}
where $0 \leq x, y \leq 1$ on the left-hand side of the substitution rules,
while on right-hand side they are unrestricted. The last two relations are
not required at leading order, since the corresponding momentum flow structures
start from two loops only. The latter originate from the region $u + v \geq 1$
of the coordinate space kernels and stem from the parton-antiparton mixing. Two
typical contributions inducing  the quark-antiquark mixing are demonstrated in
Fig.\ \ref{QuakrAntiquarkMixing}. Comparison with Eq.\ (\ref{Mom-DGLAP})
provides us with the desired formula:
\begin{eqnarray}
\label{DGLAPfromERBL}
P(x)
=
{\rm LIM}\, V (x,y)
\equiv
-
\lim_{\eta\to 0} \frac{1}{\eta}
\left(
\begin{array}{rr}
k^{qq}
& \
\frac{1}{\eta} k^{qg}
\\
\frac{\eta}{x} k^{gq}
& \
\frac{1}{x} k^{gg}
\end{array}
\right)
\left( 2 \frac{x}{\eta} - 1 , 2 \frac{1}{\eta} - 1 \right)
\, .
\end{eqnarray}
In particular, the step-function structures are reduced as follows
\begin{eqnarray}
\theta (y - x)
\!\!\!&\to&\!\!\!
\lim_{\eta \to 0^+} \vartheta \left( 2 \frac{x}{\eta} - 1 , 2 \frac{1}{\eta} - 1 \right)
=
\theta ( 1 - x ) \theta (x)
\, , \\
\theta (x - y)
\!\!\!&\to&\!\!\!
\lim_{\eta \to 0^+} \vartheta \left( 1 - 2 \frac{x}{\eta} , 1 - 2 \frac{1}{\eta} \right)
=
- \theta ( 1 - x ) \theta (x)
\, , \\
\theta (y - \bar x)
\!\!\!&\to&\!\!\!
\lim_{\eta \to 0^+} \vartheta \left( 1 - 2 \frac{x}{\eta} , 2 \frac{1}{\eta} - 1 \right)
=
\theta ( 1 + x ) \theta (- x)
\, , \\
\theta (\bar y - x)
\!\!\!&\to&\!\!\!
\lim_{\eta \to 0^+} \vartheta \left( 2 \frac{x}{\eta} - 1 , 1 - 2 \frac{1}{\eta} \right)
=
- \theta ( 1 + x ) \theta (- x)
\, .
\end{eqnarray}
For instance, the one-loop kernels $V^{ab}_{(0)}(x,y)$ contain only the two usual
$\theta$-structures, i.e., $\theta(y - x)$ and $\theta(x - y)$. Then the LIM
operation yields
\begin{eqnarray}
\label{def-LIM-mat}
{\rm LIM} \mbox{\boldmath$V$}_{(0)} (x,y)
&=&
\theta(1 - x) \theta(x)
\lim_{\eta\to 0} \frac{1}{\eta}
\left(
\begin{array}{rl}
( F^{qq} - \overline{F}^{qq} )
&
\frac{1}{\eta}( F^{qg} + \overline{F}^{qg} )
\\
\frac{\eta}{x} ( F^{gq} + \overline{F}^{gq} )
&
\frac{1}{x}( F^{gg} - \overline{F}^{gg} )
\end{array}
\right)
\left( \frac{x}{\eta} , \frac{1}{\eta} \right)
\, .
\nonumber
\end{eqnarray}
The plus-prescription involved in the diagonal kernels is reduced as follows
\begin{eqnarray}
{\rm LIM} \left[ f (x,y) \right]_+
= {\rm LIM} f (x,y)
- \delta(1 - x) \int_0^1 d x {\rm LIM}\,  f (x,y)
=
\left[ {\rm LIM}\,  f (x,y) \right]_+
\, .
\end{eqnarray}

\subsection{Evolution of double distributions}

Up to now we have addressed the scale dependence of GPDs and their
``off-springs''---parton densities and distribution amplitudes.
Before we turn to developing analytical tools for solving them, let
us briefly consider the evolution equations for double distributions
\cite{MulDitRobGeyHor98,Rad1996a,Rad1996b,Rad1997a}. Let us discuss
the non-singlet evolution of DDs $h^q_A$, introduced in section
\ref{SectionSingleCompDDnucleon}, as an example.

The generic form of the evolution equation for the double distribution
$h_A^q$ reads
\begin{equation}
\frac{d}{d \ln \mu^2} h_A^q (\beta_1, \alpha_1)
=
\int_{\mit\Omega} d \beta_1 \, d \alpha_2 \,
R^{qq} (\beta_1, \alpha_1; \beta_2, \alpha_2) h_A^q (\beta_2, \alpha_2)
, ,
\end{equation}
where the evolution kernel $R^{qq}$, having the standard perturbative expansion
\begin{equation}
R^{qq} (\beta_1, \alpha_1; \beta_2, \alpha_2)
=
\sum_{n = 0}^\infty
\left( \frac{\alpha_s}{2 \pi} \right)^n
R^{qq}_{(n)} (\beta_1, \alpha_1; \beta_2, \alpha_2)
\, ,
\end{equation}
can be found in terms of the light-cone evolution kernel for non-local operators.
To achieve this, we form the matrix element of both sides of Eq.\ (\ref{non-local-K})
and make use of the parametrization of the resulting operator matrix elements in
terms of DDs (\ref{SingleCompDDnucleonEven}). Then by performing the inverse Fourier
transform in $p^+$ and $\Delta^+$, one immediately finds
\begin{equation}
R^{qq} (\beta_1, \alpha_1; \beta_2, \alpha_2)
=
\int_0^1 d u \, d v \, \mathcal{K}^{qq} (u, v)
\, \delta ( \beta_1 - \beta_2 (1 - u - v) )
\, \delta ( \alpha_1 - \alpha_2 (1 - u - v) - v + u )
\, .
\end{equation}
Since the leading order coordinate kernel $\mathcal{K}_{(0)}^{qq}$ exists
only in the triangle $1 - u - v > 0$ there arises a spectral constraint in the
integrand. The calculation can be done by means of Eqs.\ (\ref{ShiftOfVariables})
-- (\ref{RedTheta}) and it yields
\begin{eqnarray}
R^{qq}_{(0)} (\beta_1, \alpha_1; \beta_2, \alpha_2)
\!\!\!&=&\!\!\!
\frac{1}{2 |\beta_2|}
\mathcal{K}^{qq}_{(0)}
\left(
\frac{1}{2}(1 - \alpha_1) - \frac{\beta_1}{2 \beta_2} (1 - \alpha_2) \, ,
\frac{\beta_1}{2 \beta_2} (1 + \alpha_1) - \frac{1}{2}(1 + \alpha_2)
\right)
\\
\times
\vartheta
\Bigg(
\frac{\beta_1}{\beta_2} (1 \!\!\!&-&\!\!\! \alpha_2)
- (1 - \alpha_1) \, ,
0
\Bigg)
\vartheta
\left(
(1 + \alpha_1) - \frac{\beta_1}{\beta_2} (1 + \alpha_2) \, ,
\frac{\beta_1}{\beta_2} (1 - \alpha_2) - (1 - \alpha_1)
\right)
\, . \quad\nonumber
\end{eqnarray}
Here the spectral constraints are expressed in terms of the generalized step
functions (\ref{varThetaGeneric}) via Eq.\ (\ref{RelationLipatovThetaGenericTheta}).
Other sectors are considered analogously. We refer the reader to Appendix
\ref{NonSymDDs} for a compendium of explicit results for one-loop evolution kernel,
albeit, for nonsymmetric DDs.

\subsection{Diagonalization of the mixing matrix}

Now we are in a position to construct a framework for the analytical solution of the
evolution equations for generalized parton distributions. To start with, let us recall
that the diagonalization of the leading order evolution equation (\ref{ERBLevolEq})
for the meson distribution amplitude (\ref{LeadingTwistPionDA})---the first nontrivial
encounter of the mixing pattern involving towers of operators containing total
derivatives---was found in Refs.\ \cite{EfrRad78,Farrar:1979aw,BroLep79}. It was
demonstrated there that the local operators
$$
\widetilde{\mathbb{O}}^{qq}_{jj}
\equiv
i^j \,
\bar\psi \gamma^+ \gamma^5
\left(
\stackrel{{}_{\rightarrow}}{\partial}{\!}^+
+
\stackrel{{}_{\leftarrow}}{\partial}{\!}^+
\right)^j
C_j^{3/2}
\left(
\frac{
\stackrel{{}_{\rightarrow}}{\partial}{\!}^+
-
\stackrel{{}_{\leftarrow}}{\partial}{\!}^+
}{
\stackrel{{}_{\rightarrow}}{\partial}{\!}^+
+
\stackrel{{}_{\leftarrow}}{\partial}{\!}^+
}
\right)
\psi
\, ,
$$
expressed through the conventional Gegenbauer polynomials $C_j^{3/2}$, do not
mix with other operators under renormalization for different $j$. In other words,
operators $\widetilde{\mathbb{O}}^{qq}_{jj}$ are multiplicatively renormalizable
for each values of $j$ or, equivalently, the matrix $\gamma_{jk}$ is diagonal in
the basis of Gegenbauer polynomials:  $\gamma_{jk} = \delta_{jk} \gamma_j$, where
$\gamma_j$ is the usual anomalous dimension of the forward matrix elements of the
quark operators with $j$ (left-right) derivatives. It was immediately recognized
\cite{EfrRad78} that the operators $\widetilde{\mathbb{O}}_{jj}^{qq}$ coincide with
conformal tensors of the free field theory \cite{Dobrev:1975ru}. In fact, it was
already anticipated in \cite{Migdal:1977nu} that the conformal operators renormalize
multiplicatively at one-loop order. The relationship between conformal invariance
and renormalization properties of composite operators was further studied in Refs.\
\cite{Mak81,BroFriLepSac80,Ohr82,CraDobTod85} and completely disentangled in higher
orders of the perturbation theory in the culminating studies of Refs.\
\cite{Mul92,Mul95,BelMul97,Mul98,BelMul98a,BelMul98b}.

These findings suggested that the anomalous dimensions of conformal operators are
enumerated by a single conserved quantum number $j$ identified with the conformal
spin of the composite operator. Therefore, one expects that this property is
reflected in QCD Lagrangian and it must be manifested as its space-time symmetry.
As is well-known the largest symmetry group of a massless classical four-dimensional
gauge theory is the fifteen-parameter group of conformal transformations $SO(4,2)$,
which contains dilatations $\mathbb{D}$ and special conformal transformations
$\mathbb{K}^\mu$ in addition to the generators of the conventional Poincar\'e group
with Lorentz transformations $\mathbb{M}^{\mu\nu}$ and translations $\mathbb{P}^\mu$.
The algebra as well as its representations are reviewed in Appendix
\ref{ConformalSymAppendix}. Below we will elaborate on this issues in greater
detail since these consideration allow one to understand the structure of higher
order corrections in a very efficient way.

\subsubsection{Representations of the collinear conformal algebra}

The discussion of twist two operators, to which we restrict our consideration,
allows to consider a subgroup of all confonformal transformations since the
former are built solely from ``good'' field components and ``live'' on the
light cone. This so-called collinear conformal subgroup contains the following
generators
\begin{equation}
\mathbb{L}^- \equiv \ft{i}{2} \mathbb{K}^-
\, , \qquad
\mathbb{L}^+ \equiv i \mathbb{P}^+
\, , \qquad
\mathbb{L}^0 \equiv \ft{i}{2} \left( \mathbb{D} + \mathbb{M}^{-+} \right)
\, ,
\end{equation}
projected on the light-cone with two tangent vectors $n^\mu$ and $n^{\ast\mu}$,
as specified in Appendix \ref{LCvectors}. There is yet another combination of
the dilatation and the light-cone projection of the Lorentz generator
\begin{equation}
\mathbb{E} = i \left( \mathbb{D} - \mathbb{M}^{-+} \right)
\, ,
\end{equation}
which commutes with all operators of the collinear conformal subgroup. It counts the
twist
\begin{equation}
\tau = d - s
\, ,
\end{equation}
defined in terms of the dimension $d$ and the spin $s$, of the operator it acts
on. The generators introduced above form a closed algebra and act non-trivially
on the ``good'' components of the quark and gluon operators living on the light-cone,
i.e., the elementary fields which enter the leading twist operators introduced in
Section \ref{ClassificationTwistTwo}. They obey the $su (1,1)$ commutation relations
\begin{equation}
[ \mathbb{L}^+ , \mathbb{L}^-] = - 2 \mathbb{L}^0
\, , \qquad
[ \mathbb{L}^0 , \mathbb{L}^\pm] = \pm \mathbb{L}^\pm
\end{equation}
From this commutator algebra one concludes that $\mathbb{L}^+$ is a step-up
operator and $\mathbb{L}^-$ is a step-down operator. The quadratic Casimir
operators is defined as
\begin{equation}
\mathbf{L}^2
=
\mathbb{L}^0 \left( \mathbb{L}^0 - 1 \right) - \mathbb{L}^+ \mathbb{L}^-
\, .
\end{equation}
The conformal spin is determined as an eigenvalue of the $\mathbb{L}^0$
\begin{equation}
[\mathbb{L}^0 , {\mit\Phi} (0)] = j {\mit\Phi} (0)
\, , \qquad
j \equiv \frac{1}{2} (d + s)
\, .
\end{equation}
It is composed from the dimension $d$ and the spin $s$ of the corresponding field
${\mit\Phi} (0)$. The values that the latter takes on elementary field operators
were given earlier in Eqs.\ (\ref{QuarkLCspin}) and (\ref{GluonLCspin}).

Since the canonical dimension of the fermion field is $d_q = 3/2$, one finds that the
``good'' $\psi_+$-component has the conformal spin $j_q = 1$. Similarly, for the ``bad''
$\psi_-$-component one finds $j_q = 1/2$. The canonical dimension of the field strength
tensor is $d_g=2$, the $F^{+ \mu}_{\phantom{+} \, \perp}$-component has the conformal
spin $j_g=3/2$. For $F^{+ -}$ and $F^{\mu\nu}_{\perp\perp}$-components one gets $j_g=1$,
while for the $F^{- \mu}_{\phantom{+} \, \perp}$-component one has $j_g = 1/2$. Thus,
the conformal spin of ``good'' QCD fields is
\begin{equation}
j_q = 1 \, , \qquad j_g = \ft32 \, .
\end{equation}
Now we are in a position to discuss the representation of the quantum operators
$\mathbb{L}^{\pm, 0}$ in the basis of quark and gluon fields.
\begin{itemize}
\item The representation of quantum generators acting on the quantum-field operators
in terms of differential operators is derived in Appendix \ref{ConformalSymAppendix}
using the method of induced representations. The action of quantum generators
$\mathbb{G}$ on a function ${\mit\Phi}$ living on the light-cone $z^-$ and
possessing the conformal spin $j$
\begin{equation}
\label{Rep1}
[ \mathbb{G} , {\mit\Phi} (z^-) ] \equiv \widehat{G} {\mit\Phi} (z^-)
\, ,
\end{equation}
results into the following differential representation for them
\begin{equation}
\label{HattedBasisGenerators}
\widehat{L}^- \equiv
2 j z^-
+
(z^-)^2 \frac{\partial}{\partial z^-}
\, , \qquad
\widehat{L}^+ \equiv \frac{\partial}{\partial z^-}
\, , \qquad
\widehat{L}^0 \equiv
j
+
z^- \frac{\partial}{\partial z^-}
\, .
\end{equation}
The quadratic Casimir operator in this representation is defined as
\begin{equation}
\widehat{\bit{L}}^2
=
\widehat{L}^0 ( \widehat{L}^0 - 1 ) - \widehat{L}^+ \widehat{L}^-
\, .
\end{equation}
Note that the commutation relations for the generators in this representation
differ from the generators acting on the Hilbert space by a minus sign,
\begin{equation}
[\widehat{L}^+ , \widehat{L}^- ] = 2 \widehat{L}^0
\, , \qquad
[\widehat{L}^0 , \widehat{L}^\pm ] = \mp \widehat{L}^\pm
\, .
\end{equation}
For a multiparticle operator, one defines the generators as
\begin{equation}
\widehat{L}^{\pm, 0}_{1 \dots N}
=
\sum_{\ell = 1}^{N} \widehat{L}^{\pm, 0}_\ell
\, .
\end{equation}
\item One can choose a different representation for quantum generators, which
is very handy in constructing the conformal operators \cite{DerKerMan97,BelDerKorMan05}.
Namely, one can realize the generators in the basis spanned by the elements
\cite{Ohr81}
\begin{equation}
Z^k \equiv \frac{\partial^{+ k} \, {\mit\Phi} (0)}{{\mit\Gamma} (k + 2 j)}
\, ,
\end{equation}
with the field possessing the conformal spin $j$. We construct a representation by
defining
\begin{equation}
[ \mathbb{G} , f (Z) ] \equiv G f (Z)
\, ,
\end{equation}
for any polynomial $f (Z)$ in the variable $Z$. The generators $G$ admit the
following differential representation\footnote{The derivation of generators
in this representation is very straightforward making use the available
representation (\ref{HattedBasisGenerators}):
$$
[\mathbb{G}, Z^k ]
=
\frac{\partial^{+ k} [ \mathbb{G} , {\mit\Phi} ]}{{\mit\Gamma} (2j + k)}
=
\frac{\partial^{+ k}  \widehat G {\mit\Phi}}{{\mit\Gamma} (2j + k)}
\equiv
G Z^k
\, .
$$
For instance, for $\mathbb{G} = \mathbb{L}^+$ one has according to these sequence
of transformations
$$
[\mathbb{L}^+, Z^k ]
=
( 2 j + k ) Z^{k + 1}
=
\left( 2 j Z + Z^2 \frac{\partial}{\partial Z} \right) Z^k
\, ,
$$
which gives $L^+$ in Eq.\ (\ref{UnhattedRepresentation}).
}
\begin{equation}
\label{UnhattedRepresentation}
L^- = \frac{\partial}{\partial Z}
\, , \qquad
L^+ = 2 j Z + Z^2 \frac{\partial}{\partial Z}
\, , \qquad
L^0 = j + Z \frac{\partial}{\partial Z}
\, .
\end{equation}
These generators satisfy the same commutation relations as the original
quantum generators
\begin{equation}
[ L^+ , L^- ] = - 2 L^0
\, , \qquad
[ L^0 , L^\pm ] = \pm L^\pm
\, .
\end{equation}
It is interesting to note that this representation differs from
(\ref{HattedBasisGenerators}) by a mere interchange of the functional
form of the raising and lowering operators. For multi-variable functions
$\chi ( Z_1, Z_2, \dots, Z_N )$ the generators are obviously generalized
by taking the sum of single-variable generators,
\begin{equation}
\label{MultiL}
L^{\pm, 0}_{1 \dots N} = \sum_{\ell = 1}^{N} L^{\pm, 0}_\ell
\, .
\end{equation}

The quadratic Casimir operator has the same form as before
$$
\bit{L}^2 = L^0 ( L^0 - 1 ) - L^+ L^-
\, .
$$
And its action on a two-argument function is determined by the following
differential operator
\begin{equation}
\label{2partCasimir}
\bit{L}^2_{12}
= - (Z_1 - Z_2)^2
\frac{\partial}{\partial Z_1}
\frac{\partial}{\partial Z_2}
+ 2 (Z_1 - Z_2)
\left(
j_2 \frac{\partial}{\partial Z_1}
-
j_1 \frac{\partial}{\partial Z_2}
\right)
+ (j_1 + j_2 ) \left( j_1 + j_2 - 1 \right)
\, .
\end{equation}

\end{itemize}

\subsubsection{Conformal operators}

In general, the Wilson operators (\ref{MaxSpinWilson}) do not transform
covariantly under the action of the $SO(4,2)$ conformal group and, as a
consequence, they do not have an autonomous renormalization. To identify
operators which are eigenfunctions of the renormalization group operator,
one uses the fact that at leading order of perturbation theory the latter
commutes with the generators of the conformal group. Thus they possess the
same eigenfunctions. Making use of the collinear $SL(2)$ subgroup of the
conformal group one can organize operators containing total derivatives
into conformally-covariant towers, the so-called conformal operators. They
are built as a linear superposition of bilinears
\begin{eqnarray*}
\left( \partial^{+ \, k } {\mit\Phi}_1 (0) \right)
\left( \partial^{+ \, n - k } {\mit\Phi}_2 (0) \right)
\, .
\end{eqnarray*}
Similarly to the construction of the ground state of the rotational
symmetry in quantum mechanics, we construct the ground---also called the
primary---state of the $SL(2)$ symmetry by annihilating a superposition of
the aforementioned bilinear states with the step-down operator $\mathbb{L}^-$.
The definite value of the conformal spin is assigned to them by the eigenvalues
of the quadratic Casimir operator (\ref{2partCasimir}). Since we construct
the state from derivatives acting on a field, to perform actual calculations
it is useful to use the representation (\ref{UnhattedRepresentation})
of the conformal generators. Thus, we build the operator as a linear
superposition\footnote{\label{TotalDerFoot}Notice that since the operators
are sandwiched between states with different momenta, one cannot ``integrate
by parts'' by moving derivatives from one field to another, like one does
in the forward matrix elements, and reduce the whole tower of operators for
$0 \leq k \leq n$ to a single one. In other words operators with total
derivatives do matter since $\langle p_2 | i \partial^+ \mathbb{O} | p_1
\rangle = \langle p_2 | [\mathbb{O} , \mathbb{P}^+] | p_1 \rangle = \Delta^+
\langle p_2 | \mathbb{O} | p_1 \rangle = \eta p^+ \langle p_2 | \mathbb{O} |
p_1 \rangle \neq 0$.}
\begin{equation}
\mathbb{O} (0) = \sum_{k = 0}^{n} c_{nk} \, Z_1^k Z_2^{n - k}
\, ,
\end{equation}
and impose two conditions.
\begin{itemize}
\item The lowest-weight vector condition
\begin{equation}
[ \mathbb{L}^- , \mathbb{O} (0)]
=
L_{12}^- \mathbb{O} (0) = 0
\, ,
\end{equation}
with multivariable generators (\ref{MultiL}) for $N = 2$.
\item Assignment of the eigenvalue $J$ of the conformal spin to the conformal
primary
\begin{equation}
\label{PrimaryField}
[ \bit{\mathbb{L}}^2 , \mathbb{O} (0) ]
=
\bit{L}^2_{12} \mathbb{O} (0)
=
J (J - 1) \mathbb{O} (0)
\, ,
\end{equation}
where the two-particle Casmir operators is given in Eq.\ (\ref{2partCasimir}).
\end{itemize}
The solution to the first equation is any translation-invariant function
$\mathbb{O} \sim f (Z_2 - Z_1)$ or, since we are interested in local operators,
a polynomial of an arbitrary order $\mathbb{O} \sim (Z_2 - Z_1)^n$, so that the
coefficients $c_{jk}$ are given by binomial coefficients. The second condition
immediately relates the order of the polynomial to the conformal spin
\begin{equation}
\label{ValueConfSpin}
J = n + j_1 + j_2
\, ,
\end{equation}
with $j_a$ for $a = 1, 2$ being the conformal spin of the elementary field
${\mit\Phi}_a$. Thus,  we come to the definition of conformal operators
\begin{equation}
\mathbb{O}_j (0)
=
i^{j - j_1 - j_2 + 2}
\frac{
{\mit\Gamma} (j + j_1 - j_2 + 2) {\mit\Gamma} (j + j_2 - j_1 + 2)
}{
{\mit\Gamma} (j - j_1 - j_2 + 3)
}
(Z_2 - Z_1)^{j - j_1 - j_2 + 2}
\, ,
\end{equation}
where we introduced $j = J - 2$. Here, the overall factor is chosen such that
in the basis spanned by elementary fields with derivatives, the operator will
be given in terms of the Jacobi polynomials,
\begin{equation}
P_j^{(\alpha, \beta)} (x)
=
\frac{1}{2^j}
\sum_{k = 0}^j
\left( {j + \alpha \atop k} \right) \left( {j + \beta \atop j - k} \right)
\, (x - 1)^{j - k} (x + 1)^k
\, .
\end{equation}
As usual,
$$
\left(
{j \atop k}
\right)
=
\frac{{\mit\Gamma} (j + 1)}{{\mit\Gamma} (k + 1) {\mit\Gamma} (j - k + 1)}
$$
is the binomial coefficient. In the explicit form, the conformal operator reads
\begin{equation}
\label{TwoParticleConformalOp}
\mathbb{O}_j (0) =
{\mit\Phi}_1 (0)
(i \partial^+ )^{j - j_1 - j_2 + 2}
P_{j - j_1 - j_2 + 2}^{(2 j_1 - 1 , 2 j_2 - 1)}
\left(
\stackrel{{}_{\leftrightarrow}}{\cal D}{\!}^+ \! / \partial^+
\right)
{\mit\Phi}_2 (0)
\, ,
\end{equation}
where we have used the notations
\begin{eqnarray*}
\partial_\mu
\equiv \
\stackrel{{}_\rightarrow}{\partial}_\mu + \stackrel{{}_\leftarrow}{\partial}_\mu
\, , \qquad
\stackrel{{}_\leftrightarrow}{\cal D}_\mu \
\equiv \
\stackrel{{}_\rightarrow}{\cal D}_\mu - \stackrel{{}_\leftarrow}{\cal D}_\mu
\, ,
\end{eqnarray*}
for the total and left-right derivatives, respectively. Then, the non-local
product ${\mit\Phi}_1(0) {\mit\Phi}_2(z^-)$ can be expanded over the conformal
operators:
\begin{eqnarray}
\label{NonlocalExpansionConformal}
{\mit\Phi}_1 (0) {\mit\Phi}_2 (z^-)
\!\!\!&=&\!\!\!
\sum_{n = 0}^\infty
\frac{
(2 n +  2 j_1 + 2 j_2 - 1)
{\mit\Gamma} (n + 2 j_1 + 2 j_2 - 1)
}{
{\mit\Gamma} (n + 2 j_1)
{\mit\Gamma} (n + 2 j_2)
}
\\
&\times&\!\!\!
(- i z^-)^n \int_0^1 d u \, u^{n + 2 j_1 - 1} \bar{u}^{n + 2 j_2 - 1}
\mathbb{O}_{j = n + j_1 + j_2 - 2} (u z^-)
\, . \nonumber
\end{eqnarray}
We implied the $A^+ = 0$ gauge here  to be able to neglect the path-ordered
exponentials.

\begin{figure}[t]
\mbox{
\begin{picture}(0,125)(-160,0)
\put(0,0){\insertfig{5}{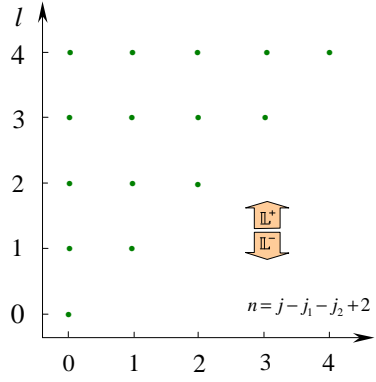}}
\end{picture}
}
\caption{\label{Tower} The tower of conformal operators is generated by the
step-up operator $\mathbb{L}^+$ from the lowest weights $\mathbb{O}_{jj}$
for different eigenvalues of the conformal spin $j$. The generator of the
conformal boost $\mathbb{L}^-$ serves as a step-down operator and decreases
the Lorentz spin $l$ of operators within a tower with the same conformal
spin $j$.}
\end{figure}

The conformal spin-$(j + 2)$ representation space is spanned by the conformal
operator $\mathbb{O}_{j} (0)$ and its descendants generated by applying the
step-up operator $\mathbb{L}^+$
\begin{equation}
\mathbb{O}_{jl}(0)
=
i^{l - j + j_1 + j_2 - 2}
[
\underbrace{
\mathbb{L}^+ ,
\ldots ,
[
\mathbb{L}^+ ,
[
\mathbb{L}^+
}_{l - j + j_1 + j_2 - 2}
, \mathbb{O}_j(0)]] \dots ]
=
( i \partial_+ )^{l - j + j_1 + j_2 - 2} \mathbb{O}_j (0)
\, ,
\end{equation}
to the vacuum state $\mathbb{O}_{jj}(0) \equiv \mathbb{O}_{j}(0)$. As it is
clear from the commutation relations, the momentum generator $\mathbb{P}^+$
is a raising operator, while the special conformal generator $\mathbb{K}^-$
is a lowering operator, see Fig.\ \ref{Tower}. The dilatation $\mathbb{D}$
and the Lorentz $\mathbb{M}^{-+}$ generators are diagonal operators. Then
the infinitesimal variations of the conformal operators under the light-cone
algebra are
\begin{eqnarray}
\begin{array}{lll}
\delta^P \, \mathbb{O}_{jl}
&\!\!\!=
i [ \mathbb{O}_{jl}, \mathbb{P}^+ ]
&\!\!\!=
i\, \mathbb{O}_{j l + 1}
\, , \\
\delta^M \, \mathbb{O}_{jl}
&\!\!\!=
i [ \mathbb{O}_{jl}, \mathbb{M}^{-+} ]
&\!\!\!=
-( l + s_1 + s_2 ) \mathbb{O}_{jl}
\, , \\
\label{confVaria-1}
\delta^D \, \mathbb{O}_{jl}
&\!\!\!=
i [\mathbb{O}_{jl}, \mathbb{D} ]
&\!\!\!=
- ( l + d_1 + d_2 ) \mathbb{O}_{jl}
\, , \\
\label{SpecialConformalBoosVar}
\delta^K \, \mathbb{O}_{jl}
&\!\!\!=
i [ \mathbb{O}_{jl}, \mathbb{K}^- ]
&\!\!\!=
i \, a (j,l) \mathbb{O}_{jl-1}
\, ,
\end{array}
\end{eqnarray}
with
\begin{equation}
a (j, l) = 2 (j - l)( j + l + 2 j_1 + 2 j_2 - 1 )
\, .
\end{equation}

Since the conformal spins of the elementary fields coincide $j_1 = j_2$ for
all phenomenologically interesting bilocal composite operators, the Jacobi
polynomials can be reduced to the Gegenbauer polynomials $C_k^\nu$:
\begin{eqnarray}
\label{JacPol-by-GegPol}
P_k^{\left(\nu - 1/2 , \nu - 1/2 \right)} (x)
= \frac{(\nu + 1/2)_k}{(2 \nu)_k} C_k^\nu (x)
\, ,
\end{eqnarray}
where $(a)_n \equiv {\mit\Gamma} (a + n)/{\mit\Gamma}(a)$ is the Pochhammer
symbol. The operators built from the same elementary fields and containing
Gegenbauer polynomials in light-cone derivatives will be referred to as conformal
operators, just like the operators given by Eq.\ (\ref{TwoParticleConformalOp}).
This will not lead to a confusion since we are not going to discuss the
mixed-field operators in this review. Thus, we introduce the quark and gluon
conformal operators,
\begin{equation}
\label{QuarkAndGluonConformalOperators}
\mathbb{O}^{qq}_{jl}
=
\bar\psi
(i \partial^+ )^l
\gamma^+
C_j^{3/2}
\left(
\stackrel{{}_\leftrightarrow}{\cal D}{\!}^+ \! / \partial^+
\right)
\psi
\, , \qquad
\mathbb{O}^{gg}_{jl}
=
F^+{}_\mu
(i \partial^+ )^{l - 1}
C_{j - 1}^{5/2}
\left(
\stackrel{{}_\leftrightarrow}{\cal D}{\!}^+ \! / \partial^+
\right)
F^{\mu +}
\, ,
\end{equation}
and analogously for other Dirac and Lorentz structures.

\subsubsection{Autonomous renormalization of conformal operators}
\label{AutonomousScaleDepConfOperators}

An operator which generates finite Green functions with products of elementary
field operators at different space-time positions is called the subtracted
operator and it is defined as
\begin{equation}
\bit{\mathbb{O}}^{\scriptscriptstyle\rm R}_{jl}
= \sum_{k = 0}^j {\cal Z}_{jk} \bit{\mathbb{O}}_{kl}
\, .
\end{equation}
From this definition, one immediately finds that it obeys the renormalization
group equation
\begin{equation}
\label{RGequation}
\frac{d}{d \ln\mu} {\mathbb{O}}^{\scriptscriptstyle\rm R}_{jl}
= -
\sum_{k = 0}^j {\gamma}_{jk} (g) \,
{\mathbb{O}}^{\scriptscriptstyle\rm R}_{kl}
\, .
\end{equation}
The matrix of anomalous dimensions\footnote{Notice that this matrix of anomalous
dimensions is different from the one introduced in Eq.\ (\ref{MellinMomERBL}). They
belong to different bases of states and only their diagonal elements $j = k$ are
related to each other via Eq.\ (\ref{relations}).} ${\gamma}_{jk} (g)$ possesses
a triangular form as a consequence of the Lorentz invariance, i.e., only operators
with the same number of derivatives can mix, so that $\gamma_{jk} \neq 0$ for $j
\geq k$, or explicitly
\begin{equation}
\label{DiagonalNonDiagonalADs}
{\gamma}_{jk}
=
\left(
\begin{array}{ccc}
\gamma_{jj} & \cdots & \gamma_{j0} \\
            & \ddots & \vdots \\
0           &        & \gamma_{00}
\end{array}
\right)
\equiv
\gamma_j \delta_{jk} + \gamma_{jk}^{\scriptscriptstyle\rm ND} \theta_{jk}
\, ,
\end{equation}
with the diagonal elements $\gamma_j \equiv \gamma_{jj}$ and non-diagonal elements
separated from them by means of the discrete step function $\theta_{jk} = \{ 1 ,
j > k ; 0 , j \leq k \}$. The anomalous dimension matrix ${\gamma}_{jk} (g)$
receives an infinite series expansion in perturbation theory;
\begin{equation}
\label{ADinPT}
{\gamma}_{jk} (g)
=
\sum_{n = 0}^\infty \left( \frac{\alpha_s}{2 \pi} \right)^{n + 1}
{\gamma}_{(n) jk}
\, .
\end{equation}

These anomalous dimensions induce a shift in the scaling transformation
of conformal operators (\ref{confVaria-1}), i.e.,
\begin{equation}
\delta^D \mathbb{O}_{jl}
= -
\sum_{k = 0}^j
\left\{
(l + d_1 + d_2) \delta_{jk} + \gamma_{jk}
\right\} \mathbb{O}_{kl}
\, .
\end{equation}
At leading order of perturbation theory, there are no counterterms for the
renormalization of the coupling constant, so there are no sources of special
conformal symmetry breaking in one-loop anomalous dimensions. Therefore,
in this approximation the tree form of the conformal boost is preserved. The
commutation relation between the dilatation and special conformal transformations
then immediately leads to the following condition
\begin{eqnarray*}
0
=
[\delta^D , \delta^K ] \mathbb{O}_{jl}
-
\delta^K \mathbb{O}_{jl}
=
- i \sum_{k = 1}^j \left\{ a (j, l) - a (k, l) \right\} \gamma_{(0)jk}
\mathbb{O}_{j l - 1}
=
- i \sum_{k = 1}^j a (j, k) \gamma_{(0)jk}
\mathbb{O}_{j l - 1}
\, .
\end{eqnarray*}
This imposes a constraint on the one-loop anomalous dimension matrix
\begin{equation}
\label{LOconstraint}
a (j,k) \gamma_{(0) jk}
=
0
\, .
\end{equation}
Since $a (j, k) \neq 0$ for $j \neq k$, the matrix $\gamma_{jk}$ does not possess
non-diagonal elements, but because $a (j, j) = 0$ vanishes, it can have the diagonal
ones: $\gamma_{(0) jk} = \gamma_{(0)j} \delta_{jk}$. This means  that conformal
operators with different conformal spins do not mix with each other at leading order
of perturbation theory supporting the observation made earlier. The constraint
(\ref{LOconstraint}) implies that the anomalous dimensions depend only on the
conformal spin,  but it gives no restrictions on the form of the diagonal elements.
As we already anticipate, the diagonal elements can be identified with the
anomalous dimensions of Wilson operators (without total derivatives) which appear
in the QCD description of the cross section of deeply inelastic scattering, which we
addressed in Section \ref{SectionDGLAPkernels}. A slight difference in their
normalization  will be addressed  in the subsequent section. This result also implies
that the perturbative expansion of non-diagonal anomalous dimensions starts at
two-loop order only
\begin{equation}
\label{NDgammaNLO}
\gamma_{jk}^{\scriptscriptstyle\rm ND}
=
\left( \frac{\alpha_s}{2 \pi} \right)^2
{\gamma}^{\scriptscriptstyle\rm ND}_{(1) jk}
+
{\cal O} (\alpha_s^3)
\, .
\end{equation}

So far we gave a generic discussion of the mixing pattern of conformal operators
$\mathbb{O}$ in perturbation theory. As we know in the singlet sector it is the
two-vector
\begin{equation}
\label{TwoVectorConfOperators}
\mathbf{O}_{jl}
\equiv
\left(
{
\mathbb{O}^{qq}_{jl}
\atop
\mathbb{O}^{gg}_{jl}
}
\right)
\, ,
\end{equation}
of the quark and gluon operators (\ref{QuarkAndGluonConformalOperators}), which
participates in the renormalization group evolution. It obeys the matrix equation
\begin{equation}
\label{MatrixRGequationConformalOp}
\frac{d}{d \ln \mu} \mathbf{O}_{jl} = - \sum_k \bit{\gamma}_{jk} (g) \mathbf{O}_{kl}
\, ,
\end{equation}
with the two-by-two matrix of anomalous dimensions
\begin{equation}
\label{TwoByTwoMatrixAnomalousDim}
\bit{\gamma}_{jk}
=
\left(
\begin{array}{cc}
\gamma^{qq}_{jk} & \gamma^{qg}_{jk} \\
\gamma^{gq}_{jk} & \gamma^{gg}_{jk}
\end{array}
\right)
\, .
\end{equation}
Each of its $\gamma^{ab}_{jk}$ four elements $ab = qq, qg, gq, gg$ is of the
triangular form (\ref{DiagonalNonDiagonalADs}).

\subsection{Diagonalization of generalized evolution kernels}
\label{DiagonalLOADs}

Our preceding discussion indicates that the generalized evolution kernels are
diagonalizable in the basis of Gegenbauer polynomials, i.e., the two-by-two
matrix of kernels can be cast in the form
\begin{equation}
\label{GenerKernelGegExp}
K_{(0)}^{ab} (x_1, x_2 | y_1, y_2)
=
\frac{1}{2}
\sum_{j = (0, 1)}^\infty
\widetilde P^{a}_j (x_1, x_2) \gamma_{(0)j}^{ab} \, P^{b}_j (y_1, y_2)
\, ,
\end{equation}
where the two-vectors $P^a$ ($a = q, g$) of Gegenbauer polynomials and their
conjugate are
\begin{equation}
P^a_j (x_1, x_2)
=
\left(
\begin{array}{c}
C_j^{3/2} \left( \displaystyle\frac{x_1 - x_2}{x_1 + x_2} \right)
\\ [4mm]
C_{j - 1}^{5/2} \left( \displaystyle\frac{x_1 - x_2}{x_1 + x_2} \right)
\end{array}
\right)
\, , \qquad
\widetilde P^a_j (x_1, x_2)
=
\left(
\begin{array}{c}
\displaystyle\frac{x_1 x_2}{n_j (\ft32)}
C_j^{3/2} \left( \displaystyle\frac{x_1 - x_2}{x_1 + x_2} \right)
\\ [4mm]
\displaystyle\frac{x_1^2 x_2^2}{n_{j - 1} (\ft52)}
C_{j - 1}^{5/2}
\left( \displaystyle\frac{x_1 - x_2}{x_1 + x_2} \right)
\end{array}
\right)
\, ,
\end{equation}
with the normalization factor being
\begin{equation}
\label{NormalizationX01}
n_j (\nu) = 2^{1 - 4 \nu}
\frac{ {\mit\Gamma}^2 (1/2) {\mit\Gamma} (2 \nu + j) }{
{\mit\Gamma}^2 (\nu) {\mit\Gamma} (j + 1) (\nu + j) }
\, .
\end{equation}
The low $j = 0$ limit in the series (\ref{GenerKernelGegExp}) stands for the
quark-quark sector and $j = 1$ for the rest. The one-loop anomalous dimensions,
deduced from the generalized evolution kernels from Section \ref{MomSpaceGerKernels},
are
\begin{itemize}
\item Even-parity sector \cite{Cha80,Ohr81,BelMul98a,BelMul98b}:
\begin{eqnarray}
\label{even-anomalous-dimensions}
\gamma^{qq; V}_{(0)j}
\!\!\!&=&\!\!\!
- C_F
\left(
- 4 \psi( j + 2 ) + 4 \psi(1) + \frac{2}{( j + 1 )( j + 2 )} + 3
\right)
\, , \\
\gamma^{qg; V}_{(0)j}
\!\!\!&=&\!\!\!
- 24 N_f T_F
\frac{j^2 + 3 j + 4}{j( j + 1 )( j + 2 )( j + 3 )}
\, , \\
\gamma^{gq; V}_{(0)j}
\!\!\!&=&\!\!\!
- C_F
\frac{j^2 + 3 j + 4}{3( j + 1 )( j + 2 )}
\, , \\
\label{gg-V-ad}
\gamma^{gg; V}_{(0)j}
\!\!\!&=&\!\!\!
- C_A \left(
- 4 \psi( j + 2 ) + 4 \psi(1)
+
8 \frac{j^2 + 3 j + 3}{j( j + 1 )( j + 2 )( j + 3 )}
- \frac{\beta_0}{C_A}
\right)
\, .
\end{eqnarray}
\item Odd-parity sector \cite{ShiVys81,BelMul98a,BelMul98b}:
\begin{eqnarray}
\label{odd-anomalous-dimensions}
\gamma^{qq; A}_{(0)j}
\!\!\!&=&\!\!\!
- C_F
\left(
- 4 \psi( j + 2 ) + 4 \psi(1) + \frac{2}{( j + 1 )( j + 2 )} + 3
\right)
\, , \\
\gamma^{qg; A}_{(0)j}
\!\!\!&=&\!\!\!
- 24 N_f T_F\frac{1}{( j + 1 )( j + 2 )}
\, , \\
\gamma^{gq; A}_{(0)j}
\!\!\!&=&\!\!\!
- C_F \frac{j ( j + 3 )}{3( j + 1 )( j + 2 )}
\, , \\
\gamma^{gg; A}_{(0)j}
\!\!\!&=&\!\!\!
- C_A \left(
- 4 \psi( j + 2 ) + 4 \psi(1)
+
\frac{8}{( j + 1 )( j + 2 )}
- \frac{\beta_0}{C_A}
\right)
\, .
\label{odd-anomalous-dimensions-last}
\end{eqnarray}
\item Maximal-helicity sector \cite{BelMul97,JiHoo98}:
\begin{eqnarray}
\gamma^{qq; T}_{(0)j}
\!\!\!&=&\!\!\!
- C_F \left( - 4 \psi( j + 2 ) + 4 \psi(1) + 3 \right)
\, , \\
\gamma^{gg; T}_{(0)j}
\!\!\!&=&\!\!\!
- C_A \left( - 4 \psi( j + 2 ) + 4 \psi(1) \right) + \beta_0
\, .
\label{maximal-helicity-anomalous-dimensions}
\end{eqnarray}
\end{itemize}
The anomalous dimensions depend on the Euler $\psi$-function which is expressed
in terms of the harmonic numbers
\begin{equation}
\psi (n)
=
\frac{d}{dn} \ln {\mit\Gamma} (n)
=
\psi (1)
+
\sum_{k = 1}^{n - 1} \frac{1}{k}
\, ,
\end{equation}
with the Euler constant $\gamma_{\scriptscriptstyle\rm E} = - \psi (1) \approx 0.577216$.

\subsubsection{Normalization of anomalous dimensions}
\label{NormalizationADs}

The difference of the off-diagonal entries $a \neq b$, displayed in Eqs.\
(\ref{even-anomalous-dimensions}) -- (\ref{maximal-helicity-anomalous-dimensions}),
from the forward anomalous dimensions (\ref{ForwardADs}) can be understood as
follows. When one sandwiches the conformal operators $\mathbb{O}_{jj}$ in
forward matrix elements, only the term which has no total derivatives $\partial^+$
is relevant. Thus, one needs the leading term of Gegenbauer polynomials at large
arguments, namely,
\begin{equation}
C_j^\nu (x) = c_j^\nu x^j + {\cal O} (x^{j - 1}) \, ,
\end{equation}
where the coefficient is
\begin{equation}
c_j^\nu =
\frac{2^{1 - 2 \nu - j} \sqrt{\pi}
{\mit\Gamma} (2 \nu + 2 j)}{
{\mit\Gamma} (j + 1) {\mit\Gamma} (\nu)
{\mit\Gamma} \left( \frac{1}{2} + \nu + j \right)}
\, .
\end{equation}
The ratio of these coefficients for quark and gluon sectors is therefore
\begin{equation}
\frac{c^{3/2}_j}{c^{5/2}_{j - 1}} = \frac{3}{j}
\, .
\end{equation}
Keeping only the contributions surviving in the forward limit, we thus get
for the quark and gluon (vector) conformal operators,
\begin{equation}
\label{WilsonToConformal}
\mathbb{O}^{qq}_{jj}
=
2^j c_j^{3/2}
\bar \psi \gamma^+ \left( i {\cal D}^+ \right)^j \psi
+
\cdots
\, , \qquad
\mathbb{O}^{gg}_{jj}
=
2^{j - 1} c_{j - 1}^{5/2}
F^{+ \mu} \left( i {\cal D}^+ \right)^{j - 1} F_\mu{}^+
+
\cdots
\, .
\end{equation}
Here the covariant derivative ${\cal D}^\mu$ acts on the right and the ellipses denote
terms containing total derivatives. Therefore, substituting (\ref{WilsonToConformal})
into Eq.\ (\ref{MatrixRGequationConformalOp}), keeping the leading term only and matching
it into the renormalization group equation for forward Wilson operators, we find that
the forward anomalous dimensions introduced in Eq.\ (\ref{ForwardADs}) are related to
the diagonal eigenvalues via
\begin{equation}
\gamma^{qg; {\rm fw}}_{j} = \frac{j}{6} \, \gamma^{qg}_{j}
\, , \qquad
\gamma^{gq; {\rm fw}}_{j} = \frac{6}{j} \, \gamma^{qg}_{j}
\, .
\end{equation}
This explains the difference in their normalization.

\subsubsection{Inclusive to exclusive reconstruction at leading order}
\label{ER-BLtoDGLAP}

In Section \ref{ERBLleadingOrderEvKernels}, we have already obtained the
one-loop ER-BL evolution kernels analytically by taking the appropriate limit
of generalized evolution kernels. As the formula (\ref{GenerKernelGegExp})
suggests the diagonal part (in the basis of the Gegenbauer polynomials) of
the exclusive kernels can be reconstructed directly from the corresponding
DGLAP analogues. The result of this reconstruction can be found analytically,
at least at leading order.

\begin{figure}[t]
\begin{center}
\hspace{0cm} \mbox{
\begin{picture}(0,150)(80,0)
\put(-10,-10){\insertfig{5.7}{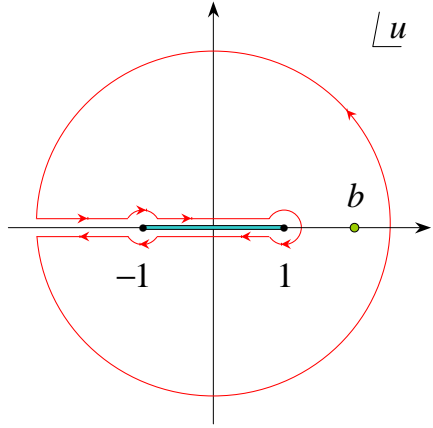}}
\end{picture}
}
\end{center}
\caption{\label{contour} The integration contour for the evaluation
of the integral (\protect\ref{integral}). There is a $n$-th order
pole in the point $u = b$ and the cut on the real axis from $-1$ to
$1$.}
\end{figure}

The formal expression for the diagonal part of the leading order exclusive
kernel admits the form identical to (\ref{GenerKernelGegExp}),
\begin{equation}
V_{(0)} ( x, y )
=
- \frac{1}{2} \sum_{j = 0}^{\infty} \frac{w(x | \nu)}{n_j(\nu)}
C^\nu_j (2x - 1) \gamma_{(0)j} C^\nu_j (2y - 1)
\, ,
\end{equation}
where $\gamma_{(0)j}$'s are the leading order anomalous dimensions of the
conformal operators, $w(x | \nu)$ is the weight factor
$$
w(x | \nu) = (x \bar x)^{\nu - 1/2}
\, ,
$$
and $n_j(\nu)$ is the normalization factor (\ref{NormalizationX01}).

Therefore, the main problem is to perform the infinite summation and
construct a generating function for the exclusive evolution kernels.
This goal can be  achieved in two straightforward steps.
\begin{itemize}
\item Use the Gegenbauer's summation theorem \cite{BatErd53},
\begin{eqnarray}
\label{SummationGegenbauer}
&&\frac{1}{n_j (\nu)} C_j^\nu (2x - 1) C_j^\nu (2y - 1)
\\
&&\qquad\qquad
= 2^{2 \nu} (\nu + j)
\int_0^{\pi} \frac{d \phi}{\pi} \, \left( \sin \phi \right)^{2 \nu - 1}
\, C_j^{\nu}
\left(
(2x - 1) (2 y - 1) + 4 \sqrt{x \bar{x} y \bar{y}} \cos \phi
\right)
\, . \nonumber
\end{eqnarray}
\item Multiply both sides of Eq.\ (\ref{SummationGegenbauer}) by
$w (x | \nu) z^j$ and sum over $j$. On the right-hand side of the
summation theorem, use the formula for the generating function of
the Gegenbauer polynomials
\begin{equation}
\label{GenerFunctGegenPolynom}
\sum_{j = 0}^\infty z^j C_j^\nu (a)
=
(1 - 2 a z + z^2)^{- \nu}
\, .
\end{equation}
To perform the summation, represent the ``contaminating factor'' $(\nu + j)$
by the differential operator $(\nu + z \partial_z)$.
\end{itemize}
Finally, changing the integration variable to $\cos \phi \equiv u$, we get
the generating function for reconstruction of the exclusive kernels from
the inclusive ones,
\begin{eqnarray}
\label{GeneratingFunctionERBL}
{\cal G} (z, \nu | x, y)
\!\!\!&\equiv&\!\!\!
\sum_{j = 0}^{\infty}
\frac{w(x | \nu)}{n_j(\nu)}
C^\nu_j (2x - 1) z^j C^{\nu}_j (2y - 1)
\\
&=&\!\!\!
2^{2\nu} \frac{\nu}{\pi} (x \bar x)^{\nu - \frac{1}{2}}
\int_{-1}^{1} d u (1 - u^2)^{\nu - 1}
\frac{(1 - z^2)}{[ 1 - 2a(u) z + z^2]^{\nu + 1}}
\, , \nonumber
\end{eqnarray}
where
\begin{equation}
a (u)
\equiv
(2x - 1)(2y - 1) + 4 u \sqrt{x \bar x y \bar y}
\, .
\end{equation}

In order to find an ER-BL evolution kernel, we have to convolute the equality
(\ref{GeneratingFunctionERBL}) with the corresponding DGLAP splitting function
and evaluate the integral on the right-hand side of this equation. Performing
the external integration first leads to the integrand which is an elementary
function of $u$. The remaining integral over $u$  can be  evaluated by considering
its analytical continuation to the whole complex plane and choosing the integration
contour as displayed in Fig.\ \ref{contour}. The calculation reduces to the
evaluation of the residues of the $n$-th order pole at  the point $u = b \equiv
(1 - (2x - 1)(2y - 1))/4 \sqrt{x \bar x y \bar y}$ and at the point $u = \infty$.
The final result is
\begin{equation}
\label{integral}
J_n
\equiv
\int_{-1}^{1} du \frac{\sqrt{1 - u^2}}{(u - b)^n}
=
\pi
\left\{ \res_{u = b} + \res_{u = \infty} \right\}
\frac{\sqrt{u^2 - 1}}{(u - b)^n}
\, .
\end{equation}
The first residue provides a contribution proportional to difference
of $\theta$-functions, while the second one to their sum. Namely, for
the $n = 1,2$ which appear in the transformation of the one-loop
splitting functions we get
\begin{equation}
\res_{u = b}
\frac{\sqrt{u^2 - 1}}{(u - b)^n}
=
\left\{
\frac{1}{2}\frac{y - x}{\sqrt{x \bar x y \bar y}}
\left[
\theta (y - x) - \theta (x - y)
\right], \qquad\qquad \mbox{for} \quad n = 1
\atop
\frac{1}{2}\frac{1 - (2x - 1)(2y - 1)}{y - x}
\left[
\theta (y - x) - \theta (x - y)
\right], \quad  \mbox{for} \quad n = 2
\right. ,
\end{equation}
and
\begin{equation}
\res_{u = \infty}
\frac{\sqrt{u^2 - 1}}{(u - b)^n}
=
\left\{
-
\frac{1}{4}
\frac{1 - (2x - 1)(2y - 1)}{\sqrt{x \bar x y \bar y}}
\left[
\theta (y - x) + \theta (x - y)
\right], \quad \mbox{for} \quad n = 1
\atop
-
\left[
\theta (y - x) + \theta (x - y)
\right], \qquad\qquad\qquad\quad \mbox{for} \quad n = 2
\right. .
\end{equation}
For instance, for the $qq$-kernel $P^{qq; V}_{(0)} (z) = (1 + z^2)/(1 - z)$
with $z > 0$, we have
\begin{equation}
\pi \int_{0}^{1} d z \ P^{qq; V}_{(0)} (z) \,
{\cal G} \left( z, \left. \ft32 \right| x, y  \right)
=
\frac{1}{2 y \bar y} J_2 - \sqrt{\frac{x \bar x}{y \bar y}} J_1
\, ,
\end{equation}
where
\begin{eqnarray}
J_1 &=& - \frac{\pi}{\sqrt{x \bar x y \bar y}}
\left\{
x \bar y \theta (y - x)
+
y \bar x \theta (x - y)
\right\} , \\
J_2 &=& 2 \pi
\left\{
\frac{x \bar y}{y - x} \theta (y - x)
+
\frac{y \bar x}{x - y} \theta (x - y)
\right\}
\, ,
\end{eqnarray}
which indeed coincides with the well-known result (\ref{LOvectorERBLkernel}).
Analogous calculations can be done for other channels. No additional subtleties
arise there.

\subsection{Asymptotic distribution amplitudes}
\begin{figure}[t]
\begin{center}
\hspace{0cm} \mbox{
\begin{picture}(0,115)(180,0)
\put(0,0){\insertfig{12}{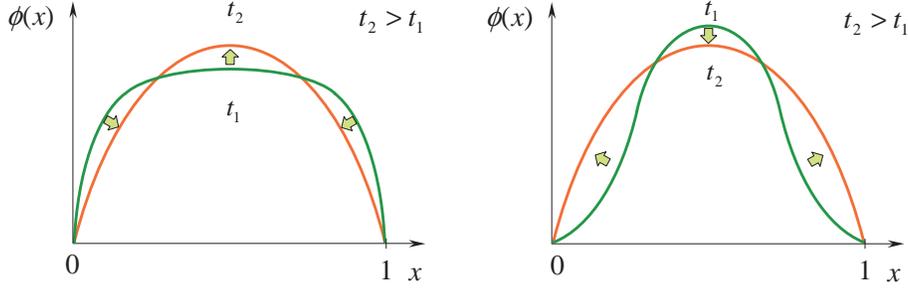}}
\end{picture}
}
\end{center}
\caption{\label{AsymptoticForm} At very large renormalization scales a distribution
amplitude is driven to its asymptotic form $(x\bar{x})^{2j - 1}$.
}
\end{figure}

The solution to the evolution equation of a two-particle distribution amplitude
can be written in the form of the conformal partial wave expansion
$$
\phi (x; \mu^2)
=
\sum_{j = (0, 1)}^\infty
\frac{w (x | \nu)}{n_j (\nu)}
C_j^{\nu} (2x - 1)
\left( \frac{\alpha (\mu_0^2)}{\alpha (\mu^2)} \right)^{\gamma_{(0)j} / \beta_0}
\int_0^1 dy \, C_j^{\nu} (2y - 1) \phi (y; \mu^2_0)
\, .
$$
The index of the Gegenbauer polynomial is related to the conformal spin of the
constituent fields $\nu_a = 2 j_a - 1$, it is either $\nu_q = 3/2$ for quarks
or $\nu_g = 5/2$ for gluons. The first term in the series is $j = 0 (1)$ for
quarks (gluons). Since the anomalous dimensions are positive, $\gamma_j > 0$
for $j > 0$, and growing with increasing $j$, only the lowest term of the
infinite series can be kept in the asymptotic limit $\mu^2 \to \infty$
$$
\phi (x; \mu^2 \to \infty) = a_0 (\mu^2) \left( x (1 - x) \right)^{\nu - 1/2}
\, .
$$
Actually, due to the conservation of the non-singlet quark current, $\gamma^{qq}_0
= 0$ and this equality is exact since the lowest term does not depend on the
renormalization point, $a_0 = \, {\rm const}$. Thus, the functional form of the
distribution amplitude is unambiguously fixed by applying the consequences of
the conformal symmetry. The only unknown parameter left is the nonperturbative
coefficient $c_0$. The general trend of the evolution of an input distribution
amplitude with increasing ``time'' $t = \ln \mu^2$ is demonstrated in Fig.\
\ref{AsymptoticForm}. Analogous considerations can be used  for multiparticle
distribution amplitudes, with the result
$$
\phi (x_1, x_2 , \dots , x_N ; \mu^2 \to \infty)
=
c (\mu^2) \prod_{k = 1}^N x_k^{2 j_k - 1}
\, ,
$$
where $j_k$ is the conformal spin of the $k$th constituent. However, in the
present circumstances there are no symmetry reasons prohibiting  the lowest
operator from developing a nonzero anomalous dimension.  Therefore, the
coefficient $c$ continues to evolve even in the asymptotic limit. The accuracy
of representing the large-$\mu^2$ behavior of  distribution amplitudes by the
lowest-order term only is determined by the ratio of the lowest and the
next-to-lowest anomalous dimensions.

\subsection{Conformal symmetry breaking}

The conformal symmetry is not a  quantum symmetry of QCD. Out of the
fifteen-parameter group only its Poincar\'e subgroup survives when quantum
corrections are taken into account. The currents associated with dilatation
and four conformal boosts are no longer conserved. This is a direct consequence
of the dimensional transmutation in QCD according to which the quantum theory
generates an intrinsic scale related to ultraviolet divergences. These
divergences cause the QCD coupling constant to depend on this scale and change
when the ultraviolet cutoff is varied. Therefore, since the conformal symmetry
ensured the multiplicative renormalizability of conformal operators at leading
order, in full quantum theory they will start to mix with each other. Thus, the
mixing arises beyond leading order of perturbation theory as we pointed out in
Eq.\ (\ref{NDgammaNLO}).

Moreover, there is yet another very subtle (scheme-dependent) source of the
conformal symmetry breaking due to the very necessity of a regularization
which makes the very calculation possible. The only regularization scheme
implemented in QCD perturbation theory\footnote{We exclude lattice regularization
because it is not very convenient  to use  in continuous perturbative computations.}
which preserves the non-abelian gauge invariance to all orders in coupling
constant is dimensional regularization and its off-springs like dimensional
reduction. In these approaches, the coupling constant becomes {\sl dimensionful}
when one goes away from four-dimensional space-time $d \neq 4$, and one can expect
the conformal symmetry breaking effects in renormalization of operators. It should
be emphasized that  the conformal invariance of the theory is destroyed even when
the four-dimensional beta-function vanishes identically, since the $d$-dimensional
function has a two-component form
$$
\beta_\varepsilon (g) = - \varepsilon g + \beta (g)
\, .
$$
The first source is nonvanishing for $d \neq 4$ even if $\beta (g) = 0$. This breaking
will be encoded below into the so-called special conformal anomaly matrix $\gamma^c$.

A very efficient formalism for computing higher order anomalous dimensions of
conformal operators consists of a few straightforward steps relying on the following
observations, some of which were established and spelled out earlier:
\begin{itemize}
\item The triangularity of the anomalous dimension matrix $\gamma_{jk}$ implies that
its eigenvalues are given by the diagonal elements and coincide with the well-known
forward anomalous dimensions.
\item Tree-level special conformal invariance implies that the leading order
mixing matrix is diagonal. One-loop violation of the symmetry induces
non-diagonal elements. Thus, the one-loop special conformal anomaly will
generate two-loop anomalous dimensions.
\item Scale Ward identity for the Green function with conformal operator insertion
is known to be with the Callan-Symanzik equation for the latter, and thus the
dilatation anomaly is the anomalous dimension of the composite operator
\begin{equation}
\label{AnomaliesFromProductThetaO}
\mathbb{O}^{\scriptscriptstyle\rm R}_j
\int\! d^d z \, {\mit\Theta}^\mu{}_\mu (z)
\sim
\sum_{k = 0}^{j} \gamma_{jk}
\mathbb{O}^{\scriptscriptstyle\rm R}_k
\, .
\end{equation}
\item The four-dimensional conformal algebra provides a relation between the
anomalies of dilatation and special conformal transformations via the commutator
$[ \mathbb{D}, \mathbb{K}^-] = i \mathbb{K}^-$. Using the latter, one can deduce
the form of the two-loop non-diagonal elements $\gamma_{jk}^{\scriptscriptstyle\rm ND}$
\cite{BelMul98b}.
\end{itemize}

\subsubsection{Anomaly in the trace of energy momentum tensor}

In order to write down conformal Ward identities, one has to know the
infinitesimal variation of the action in the regularized theory under the
conformal symmetry. Since the dilation and the special conformal symmetry
do not survive the regularization, the change of the action under these
transformations will be nonvanishing. To simplify the derivation it is
instructive to use the following generic variations of the Lagrangian
under the $\mathbb{D}$ and $\mathbb{K}^\mu$ transformations
\begin{eqnarray}
\label{D-variation}
\delta^D {\cal L} [{\mit\Phi}]
\!\!\!&=&\!\!\!
- \partial^\mu \big( z_\mu {\cal L} [{\mit\Phi}] \big)
+
d \, {\cal L} [{\mit\Phi}]
-
d_{\mit\Phi} \,
\frac{\partial{\cal L} [{\mit\Phi}]}{\partial{\mit\Phi}} {\mit\Phi}
-
(d_{\mit\Phi} + 1) \,
\frac{\partial{\cal L} [{\mit\Phi}]}{\partial(\partial_\mu {\mit\Phi})}
\partial_\mu {\mit\Phi}
\, , \\
\delta_\nu^K {\cal L} [{\mit\Phi}]
\!\!\!&=&\!\!\!
-
\partial^\mu
\left(
( 2 z_\mu z_\nu - z^2 g_{\mu\nu} ) {\cal L} [{\mit\Phi}]
\right)
+
2 x_\nu \left(
d \,
{\cal L} [{\mit\Phi}]
-
d_{\mit\Phi} \,
\frac{\partial{\cal L} [{\mit\Phi}]}{\partial{\mit\Phi}} {\mit\Phi}
-
(d_{\mit\Phi} + 1) \,
\frac{\partial{\cal L} [{\mit\Phi}]}{\partial(\partial_\mu {\mit\Phi})}
\partial_\mu {\mit\Phi} \right)
\nonumber\\
&-&\!\!\!
2 d_{\mit\Phi} \,
\frac{\partial{\cal L} [{\mit\Phi}]}{\partial(\partial_\nu {\mit\Phi})}
{\mit\Phi}
+
2 i \frac{\partial{\cal L} [{\mit\Phi}]}{\partial(\partial_\mu {\mit\Phi})}
{\mit\Sigma}_{\nu\mu} {\mit\Phi}
\, .
\label{C-variation}
\end{eqnarray}
Applying them to the dimensionally regularized $d = 4 - 2 \varepsilon$ QCD
Lagrangian in the covariant gauge (\ref{YM-renormalized}), with the following
assignment of scaling dimensions $d_g = 1$, $d_q = 3/2$, $d_{\bar{\omega}} =
d - 2$ and $d_\omega = 0$ to the gluon, quark and ghost fields, respectively,
one finds that the conformal variation of the action can be expressed by well
classified operators: gauge invariant operators $\mathcal{O}_A$, BRST exact
operators $\mathcal{O}_B$, and equation of motion (EOM) operators ${\mit\Omega}$.
A further advantage of the chosen set of scale dimensions is that it allows one
to use the Jackiw's conformal covariant transformation law for gauge fields
\cite{Jac78}. The final result of the calculation can be written as
\begin{eqnarray}
\label{VarAction}
\delta^D S_{\scriptscriptstyle \rm QCD} = \int d^d z \ \Delta (z)
\, , \quad
\delta^K_\mu S_{\scriptscriptstyle \rm QCD} = \int d^d z \ 2 z_\mu \, \Delta (z)
\, ,
\end{eqnarray}
where the integrand $\Delta (x)$ is expressed in terms of the regularized but
unrenormalized trace anomaly of the energy-momentum tensor ${\mit\Theta}_{\mu\nu}$
(\ref{SymTracelessEMtensorQCD}):
\begin{eqnarray}
\label{TraceAnomaly}
\Delta (z) &=& - {\mit\Theta}^\mu{}_\mu (z)
- d_g {\mit \Omega}_g (z)
- d_q {\mit \Omega}_{\bar{q}q} (z)
- d_{\bar\omega} {\mit \Omega}_{\bar\omega} (z) \nonumber\\
\!\!\!&=&\!\!\!
\varepsilon \,
\left\{
{\cal O}_A (z) + {\cal O}_B (z)
+ {\mit \Omega}_{\bar \omega} (z)
- {\mit \Omega}_{\bar{q}q} (z)
\right\}
+
(d - 2)\,\partial^\mu {\cal O}_{B\mu}(z)
\, .
\end{eqnarray}
Here we introduced the following set of gauge invariant and BRST-exact operators
\begin{eqnarray}
{\cal O}_A = \frac{\mathcal{Z}_3}{2} F^a_{\mu\nu} F^{a \mu\nu}
\, , \quad
{\cal O}_B = \frac{\delta^{\scriptscriptstyle\rm BRST}}{\delta\lambda}
\bar{\omega}^a \partial^\mu A_\mu^a
\, , \quad
{\cal O}_{B\mu} = \frac{\delta^{\scriptscriptstyle\rm BRST}}{\delta\lambda}
\bar{\omega}^a A^a_\mu
\, ,
\end{eqnarray}
as well as EOM operators
\begin{equation}
{\mit \Omega}_g
=
A^a_\mu \frac{\delta S}{\delta A^a_\mu},
\quad
{\mit \Omega}_{\bar{q}q}
=
\frac{\delta S}{\delta \psi} \psi
+
\bar{\psi} \frac{\delta S}{\delta \bar{\psi}}
\, , \quad
{\mit \Omega}_{\bar\omega}
=
\bar \omega^a \frac{\delta S}{\delta \bar \omega^a}
\, .
\end{equation}
It is desirable to express the variations of the action (\ref{VarAction}) in terms
of renormalized (via the modified minimal subtraction scheme) operators as this
allows one to neglect contributions proportional to the dimensional-regularization
parameter $\varepsilon $. For this reason, one should solve the renormalization
problem of the above mentioned operators \cite{JogLee75}. A straightforward analysis
yields \cite{BelMul98b}
\begin{eqnarray}
{\cal O}^{\scriptscriptstyle\rm R}_A (z)
\!\!\!&=&\!\!\!
\left( 1 + g \frac{\partial\ln \mathcal{Z}_{\rm c} }{\partial g} \right)
{\cal O}_A (z)
+ \left( g \frac{\partial\ln \mathcal{X}}{\partial g}
- 2 \xi \frac{\partial\ln \mathcal{X}}{\partial \xi} \right)
\Bigl\{ {\cal O}_B (z)
+ {\mit \Omega}_{\bar \omega} (x)
+ {\mit \Omega}_g (z) \Bigr\}\nonumber  \\
&&+ \left( g \frac{\partial\ln \widetilde{\mathcal{Z}}_3}{\partial g}
- 2\xi\frac{\partial\ln\widetilde{\mathcal{Z}}_3 }{\partial \xi}\right)
{\mit \Omega}_{\bar \omega} (z)
+ \frac{1}{2}
\left( g \frac{\partial\ln \widetilde{\mathcal{Z}}_2}{\partial g}
- 2\xi\frac{\partial\ln\widetilde{\mathcal{Z}}_2 }{\partial \xi}\right)
{\mit \Omega}_{\bar{q}q} (z)
\, , \\
{\cal O}^{\scriptscriptstyle\rm R}_B (z)
\!\!\!&=&\!\!\!
\left( 1 + 2 \xi\frac{\partial\ln \mathcal{X}}{\partial \xi} \right) {\cal O}_B (z)
+ 2 \xi\frac{\partial\ln \mathcal{X} }{\partial\xi}
\Bigl\{
{\mit \Omega}_g (z)
+ {\mit \Omega}_{\bar \omega} (z)
\Bigr\}
+ \xi\frac{\partial\ln\widetilde{\mathcal{Z}}_2 }{\partial \xi}
{\mit \Omega}_{\bar{q}q} (z)\nonumber \\
&&+ 2 \xi\frac{\partial\ln\widetilde{\mathcal{Z}}_3 }{\partial \xi}\
{\mit \Omega}_{\bar \omega} (z)
\, ,
\end{eqnarray}
where the charge renormalization constant is $\mathcal{Z}_{\rm c} =
\mathcal{X} \mathcal{Z}_3^{- 1/2}$ (see Eq.\ (\ref{BareToRenormalization})).
Thus, inserting these findings into Eq.\ (\ref{TraceAnomaly}) we obtain the
renormalized anomaly in the trace of the energy-momentum tensor in $d = 4 -
2 \varepsilon$ dimensions
\begin{eqnarray}
\label{trace-anomaly}
{\mit\Theta}^\mu{}_\mu (z)
\!\!\!&=&\!\!\!
\frac{\beta_\varepsilon}{g} {\cal O}^{\scriptscriptstyle\rm R}_A (z)
+
\left(
\frac{\beta_\varepsilon}{g} - \gamma_g
\right)
\Bigl\{
{\cal O}^{\scriptscriptstyle\rm R}_B (z) + {\mit \Omega}_{\bar\omega} (z)
\Bigr\}
-
\left( d_g + \gamma_g - \frac{\beta}{g} \right)
{\mit \Omega}_g (z)  \nonumber\\
&&\!\!\!
-
( d_q + \gamma_q - \varepsilon ) {\mit \Omega}_{\bar{q}q} (z)
-
( d_{\bar\omega} + 2 \gamma_\omega ) {\mit \Omega}_{\bar \omega} (z)
-
(d - 2) \partial^\mu {\cal O}^{\scriptscriptstyle\rm R}_{B\mu} (z)
\, .
\end{eqnarray}
We cannot set $\varepsilon = 0$ yet since ${\mit\Theta}^\mu{}_\mu$ will enter in a
product with the conformal operator into dilatation and special conformal Ward
identities and this product contains short distance singularities which once
regularized by means of dimensional regularization produce poles in $1/\varepsilon$.
The latter will compensate $\mathcal{O} (\varepsilon)$ terms in ${\mit\Theta}^\mu{}_\mu$
and result into finite contributions. After subtraction of the poles, the finite part
generates the dilatation (see Eq.\ (\ref{AnomaliesFromProductThetaO}) and Eq.\
(\ref{DilatationConfOperRenorm}) below) and special conformal anomalies (see Eq.\
(\ref{SpecialBooostConfOperRenorm}) below).

\subsubsection{Conformal Ward identities}
\label{SectConfWardIdentity}

Having found the form of the variations $\delta S_{\scriptscriptstyle\rm QCD}$,
we can now derive the dilatation and special conformal Ward identities.
It is well known that the Ward identity for the dilatation operator is
the Callan-Symanzik equation,
\begin{eqnarray}
&&\sum_k^L \left( d_k^{\rm can} + z_k \cdot \partial_k \right)
\langle
\mathbb{O}^{\scriptscriptstyle \rm R}_{jl} \
{\mit\Phi} (z_1) \dots {\mit\Phi} (z_L)
\rangle
\equiv
- \langle \mathbb{O}^{\scriptscriptstyle \rm R}_{jl} \
\delta^D \left( {\mit\Phi} (z_1) \dots {\mit\Phi} (z_L) \right)
\rangle
\\
&&\qquad\qquad\qquad\qquad
=
\langle \left( \delta^D \mathbb{O}^{\scriptscriptstyle \rm R}_{jl} \right)
{\mit\Phi} (z_1) \dots {\mit\Phi} (z_L)
\rangle
+
\langle \mathbb{O}^{\scriptscriptstyle \rm R}_{jl}
\left( i \delta^D S_{\scriptscriptstyle \rm QCD} \right)
{\mit\Phi} (z_1) \dots {\mit\Phi} (z_L)
\rangle
\, .
\nonumber
\end{eqnarray}
Here in the first line, we used the definition of the infinitesimal form
(\ref{DefInfinitsimalVariation}) of the scale transformation (\ref{RepConf})
and, in the second line, we integrated by parts in the functional integral
such that the variation is moved from the field monomial ${\mit\Phi} (z_1)
\dots {\mit\Phi} (z_L)$ on the conformal operator and the weight of the
functional measure $\exp (i S_{\scriptscriptstyle \rm QCD})$. In quantum
theory the Ward identity thus develops an extra term in its right-hand side
due to nonvanishing variation of the regularized QCD action $\delta^D
S_{\scriptscriptstyle \rm QCD}$. This generates an anomaly which arises from
the renormalization of the product of two composite operators: the conformal
operator and the trace anomaly in the energy-momentum tensor (\ref{VarAction})
-- (\ref{TraceAnomaly}),
\begin{equation}
\label{DilatationConfOperRenorm}
\mathbb{O}^{\scriptscriptstyle \rm R}_{jl}
\left( \delta^D S_{\scriptscriptstyle \rm QCD} \right)
=
i \sum_{k = 0}^{j} \gamma_{jk}
\mathbb{O}^{\scriptscriptstyle \rm R}_{k l}
+ \dots
\, .
\end{equation}
One can see that the canonical dimension $d^{\rm can}_j = j + 2 d_{\mit\Phi}$ of
the conformal operator $\mathbb{O}_j$ gets modified by the anomalous dimension,
$$
d^{\rm can}_j \to d^{\rm can}_j + \gamma_{jj}
\, ,
$$
however, it also induces the mixing of conformal operators starting at two-loop
order. On the other hand, in the Ward identity for the special conformal
transformation, there is an anomaly that appears from the renormalization of an
analogous product,
\begin{equation}
\label{SpecialBooostConfOperRenorm}
\mathbb{O}^{\scriptscriptstyle \rm R}_{k, l}
\left( \delta_-^K S_{\scriptscriptstyle \rm QCD} \right)
=
\sum_{k = 0}^{j} \gamma^c_{jk} (l)
\mathbb{O}^{\scriptscriptstyle \rm R}_{k, l - 1}
+ \dots
\, .
\end{equation}
Here $\gamma^c_{jk} (l)$ is the special conformal anomaly. It satisfies the following
condition
\begin{equation}
\gamma^c_{jj} (j) = 0
\, ,
\end{equation}
(no summation over $j$) which is an obvious consequence of the absence of the
counterterm in Eq.\ (\ref{SpecialBooostConfOperRenorm}) for $l = j$. Note, that
$\gamma^c \neq \gamma$, so that the dilatation and special conformal symmetries
are broken differently. However, their anomalies satisfy a constraint which we
are going to discuss next.

\subsubsection{Commutator constraints for conformal anomalies}

In order to establish relations between anomalous dimensions and special
conformal anomalies, one has to resort to the commutators between corresponding
generators from the conformal algebra. The symbolic form of the action of
the dilatation and conformal boost\footnote{They are understood as insertions
into a Green function with field monomials, i.e., $\langle [\dots] {\mit\Phi}
(z_1) \dots {\mit\Phi} (z_L) \rangle$.}, derived from the conformal Ward
identities, is given by\footnote{Here and below we assume the summation over
repeated indices.} \cite{BelMul98b}
\begin{eqnarray}
\label{Dilatation}
&&{}[ \mathbb{O}^{\scriptscriptstyle\rm R}_{jl} (0) , \mathbb{D}]
=
i \left[
\left( l + 2 d \right) \delta_{jk} + \gamma_{jk}
\right]
\mathbb{O}^{\scriptscriptstyle\rm R}_{kl} (0)
-
\left[
\mathbb{O}_{jl} (0) \int d^4 z \, {\mit\Theta}^\mu{}_\mu (z)
\right]^{\scriptscriptstyle\rm R}
+
\dots
\, , \\
\label{ConformalBoost}
&&{}[\mathbb{O}^{\scriptscriptstyle\rm R}_{jl} (0) , \mathbb{K}^-]
=
\left[
a (j, l) \delta_{jk} + \gamma^c_{jk} (l)
\right]
\mathbb{O}^{\scriptscriptstyle\rm R}_{k l - 1}
-
\left[
\mathbb{O}_{jl} (0) \int d^4 z \, 2 z^- {\mit\Theta}^\mu{}_\mu (z)
\right]^{\scriptscriptstyle\rm R} + \dots
\, ,
\end{eqnarray}
where the ellipses stand for the ``unpysical'' sector. Both of these transformations laws
are valid to all orders in perturbation theory and can be used for finding relations
between $\gamma$ and $\gamma^c$ \cite{BelMul98a,BelMul98b}.
\begin{itemize}
\item The first of these constraints stems from the commutator $[ \mathbb{D} ,
\mathbb{K}^\mu] = i \mathbb{K}^\mu$. It readily yields a matrix equation which
reads at the conformal point $\beta (g^\ast) = 0$
\begin{equation}
\label{GammaGammaC}
\left[ a (l) + \gamma^c (l) , \gamma \right] = 0
\, .
\end{equation}
It is an extension of Eq.\ (\ref{LOconstraint}) to all orders of perturbation
theory. Here the constant matrix $a_{jk} (l) = \delta_{jk} a (j, l)$ is a
remnant of the leading order conformal boost of the conformal operators
(\ref{SpecialConformalBoosVar}). This relation can be easily derived using
Eqs.\ (\ref{Dilatation}) and (\ref{ConformalBoost}). From this constraints,
one can draw very important conclusions. One immediately realizes that even at
this conformal point, the modified minimal subtraction scheme, used for
subtraction of UV divergences, breaks explicitly the conformal covariance, i.e.,
it leads to mixing of conformal operators with different conformal spins, so
that $\left. \gamma_{jk} \right|_{j \neq k} \neq 0$.

In QCD, the coupling is running and this commutator constraint is promoted to
\cite{Mul95,BelMul98b}
\begin{equation}
\label{GammaGammaCFull}
\left[ a (l) + \gamma^c (l) + 2 \frac{\beta}{g} b(l) , \gamma \right] = 0
\, ,
\end{equation}
with the four-dimensional $\beta$-function accompanied by a new matrix
\begin{equation}
b_{jk} (l)
=
\theta_{jk}
\left\{
2 (l + k + 3) \delta_{jk} - [1 + (-1)^{ j - k}] (2 k + 3)
\right\}
\, .
\end{equation}
\item Another equation between anomalies arises from the translation invariance
in the form of the commutator $[ \mathbb{K}^\mu , \mathbb{P}^\nu] = - 2 i \left(
g^{\mu\nu} \mathbb{D} + \mathbb{M}^{\mu\nu} \right)$ and implies a linear relation
between the same anomalies \cite{Mul98,BelMul98b}
\begin{equation}
\label{DiffCgamma}
\gamma^c_{jk} (l + 1) - \gamma^c_{jk} (l)
=
- 2 \gamma_{jk}
\, .
\end{equation}
It provides an identity between the diagonal elements of the special conformal
anomaly matrix and diagonal elements $\gamma_{jj} = \gamma_j$ of the anomalous
dimension matrix, namely,
\begin{equation}
\gamma^c_{jj} (l) = - 2 (l - j) \gamma_j
\, .
\end{equation}
Extension of Eq.\ (\ref{DiffCgamma}) to the case of the running strong coupling
is straightforward \cite{Mul98,BelMul98b}
\begin{equation}
\gamma^c_{jk} (l + 1) - \gamma^c_{jk} (l)
=
- 2 \left( \gamma_{jk} - 2 \frac{\beta}{g} \delta_{jk} \right)
\, .
\end{equation}
\end{itemize}

\subsubsection{NLO constraint and non-diagonal anomalous dimensions}
\label{NDanomalousDimensionsNLO}

Equation (\ref{GammaGammaCFull}) does not impose any constraints on the diagonal
elements of the anomalous dimension matrix $\gamma_{jj}$ since they merely cancel
in the commutator, however, it does immediately imply a relation between the
non-diagonal part of the elements of the two-loop anomalous dimension matrix
$\gamma_{(1)jk}^{{\scriptscriptstyle\rm ND}}$ in terms of the one-loop special
conformal anomaly, $\beta$-function and one-loop anomalous dimensions \cite{BelMul98b},
\begin{equation}
\label{GenericGammaND}
a (j,k) \gamma_{(1)jk}^{{\scriptscriptstyle\rm ND}}
=
\left( \gamma_{(0)j} - \gamma_{(0)k} \right)
\left( \gamma_{(0)jk}^c + \beta_0 b_{jk} \right)
\, .
\end{equation}
Here the one-loop special conformal anomaly was deduced to one-loop order
from Eqs.\ (\ref{SpecialBooostConfOperRenorm}) to be of the form
\begin{equation}
\gamma^{c}_{(0)jk}
=
- b_{jk} \gamma_{(0)k} + w_{jk}
=
a (j, k)
\left( - d_{jk} \gamma_{(0)k} + g_{jk} \right)
\, ,
\end{equation}
where the explicit form of the matrix $g_{jk}$ will be given below.

The generic expression (\ref{GenericGammaND}) can be used to find the
NLO non-diagonal anomalous dimensions for the singlet sector \cite{BelMul98b}
and represented as
\begin{eqnarray}
\label{andimND-QQ}
\gamma_{(1)jk}^{qq, {\scriptscriptstyle\rm ND}}
\!\!\!&=&\!\!\!
\left(
\gamma_{(0)j}^{qq} - \gamma_{(0)k}^{qq}
\right)
\left\{
d_{jk}
\left(
\beta_0 - \gamma_{(0)k}^{qq}
\right)
+
g^{qq}_{jk}
\right\}
-
\left(
\gamma_{(0)j}^{qg} - \gamma_{(0)k}^{qg}
\right) d_{jk} \,
\gamma_{(0)k}^{gq}
+
\gamma_{(0)j}^{qg} \, g^{gq}_{jk}
\, , \quad \\
\label{andimND-QG}
\gamma_{(1)jk}^{qg, {\scriptscriptstyle\rm ND}}
\!\!\!&=&\!\!\!
\left(
\gamma_{(0)j}^{qg} - \gamma_{(0)k}^{qg}
\right)
d_{jk}
\left( \beta_0 - \gamma_{(0)k}^{gg} \right)
- \left(
\gamma_{(0)j}^{qq} - \gamma_{(0)k}^{qq}
\right)
d_{jk} \gamma_{(0)k}^{qg}
+
\gamma_{(0)j}^{qg} g^{gg}_{jk}
-
g^{qq}_{jk} \gamma_{(0)k}^{qg}
\, , \nonumber\\
\label{andimND-GQ}
\gamma_{(1)jk}^{gq, {\scriptscriptstyle\rm ND}}
\!\!\!&=&\!\!\!
\left(
\gamma_{(0)j}^{gq} - \gamma_{(0)k}^{gq}
\right) d_{jk}
\left(
\beta_0 - \gamma^{qq}_{(0)k}
\right)
-
\left(
\gamma_{(0)j}^{gg} - \gamma_{(0)k}^{gg}
\right) d_{jk}
\gamma_{(0)k}^{gq}
+
\gamma_{(0)j}^{gq} g^{qq}_{jk}
-
g^{gg}_{jk} \gamma_{(0)k}^{gq}
\nonumber\\
&&\qquad\qquad\qquad\qquad\qquad\qquad\qquad\qquad\qquad\qquad\qquad\quad \ \
+
\left(
\gamma_{(0)j}^{gg} - \gamma_{(0)k}^{qq}
\right)
g^{gq}_{jk}
\, , \nonumber\\
\label{andimND-GG}
\gamma_{(1)jk}^{gg, {\scriptscriptstyle\rm ND}}
\!\!\!&=&\!\!\!
\left(
\gamma_{(0)j}^{gg} - \gamma_{(0)k}^{gg}
\right)
\left\{
d_{jk}
\left(
\beta_0 - \gamma_{(0)k}^{gg}
\right)
+
g^{gg}_{jk}
\right\}
-
\left(
\gamma_{(0)j}^{gq} - \gamma_{(0)k}^{gq}
\right) d_{jk}
\gamma_{(0)k}^{qg}
-
g^{gq}_{jk} \gamma_{(0)k}^{qg}
\, . \nonumber
\end{eqnarray}
The leading order diagonal anomalous dimensions has been given in Section
\ref{DiagonalLOADs}; the $d$-elements are
\begin{equation}
d_{jk}
= - \frac{1}{2}[ 1 + ( - 1)^{j - k} ]
\frac{(2k + 3)}{(j - k)(j + k + 3)}
\, .
\end{equation}
This matrix can be generated as a derivative of the Gegenbauer polynomial
with respect to its index,
\begin{equation}
\left. \frac{d}{d \nu} \right|_{\nu = 3/2} C_j^\nu (x)
=
- 2 \sum_{k = 0}^j d_{jk} C_k^{3/2} (x)
\, , \qquad
\left. \frac{d}{d \nu} \right|_{\nu = 5/2} C_{j - 1}^\nu (x)
=
- 2 \sum_{k = 1}^j d_{jk} C_{k - 1}^{3/2} (x)
\, .
\end{equation}
Finally, the renormalized special conformal anomalies $g$ are:
\begin{itemize}
\item Even and odd parity sectors:
\begin{eqnarray}
\label{g-QQ}
g^{qq, V}_{jk}
=
g^{qq, A}_{jk}
\!\!\!&=&\!\!\!
- C_F \left[ 1 + (-1)^{j-k} \right] \theta_{j-2,k}
\frac{( 3 + 2 k )}{(j - k)(j + k + 3)} \\
&\times&\!\!\!
\left\{
2 A_{jk} + ( A_{jk} - \psi( j + 2 ) + \psi(1) )
\frac{(j - k)(j + k + 3)}{( k + 1 )( k + 2 )}
\right\}
\, , \\
\label{g-GQ}
g^{gq, V}_{jk}
=
g^{gq, A}_{jk}
\!\!\!&=&\!\!\!
- C_F \left[ 1 + (-1)^{j-k} \right] \theta_{j-2,k}
\frac{1}{6}
\frac{( 3 + 2k )}{( k + 1 ) ( k + 2 )}
\, , \\
\label{g-GG}
g^{gg, V}_{jk}
=
g^{gg, A}_{jk}
\!\!\!&=&\!\!\!
- C_A [ 1 + ( - 1)^{j - k} ] \theta_{j - 2,k}
\frac{( 3 + 2 k )}{(j - k)(j + k + 3)} \\
&\times&\!\!\!
\Bigg\{
2 A_{jk} + ( A_{jk} - \psi (j+2) + \psi(1) )
\left[
\frac{{\mit\Gamma} (j + 4) {\mit\Gamma} (k)}{{\mit\Gamma} (j) {\mit\Gamma} (k + 4)} - 1
\right]
\nonumber\\
&&\qquad\qquad\qquad\qquad\qquad\ \
+ 2 (j - k)( j + k + 3 )
\frac{{\mit\Gamma} (k)}{{\mit\Gamma} (k + 4)}
\Bigg\}
\, . \nonumber
\end{eqnarray}
\item Maximal-helicity sector:
\begin{eqnarray}
g^{qq, T}_{jk}
\!\!\!&=&\!\!\!
g^{qq, V}_{jk}
\, , \\
g_{jk}^{gg, T}
\!\!\!&=&\!\!\!
-  2 C_A \sigma_{j - k} \theta_{j - 2,k}
\frac{( 3 + 2k )}{(j - k) (j + k + 3)} \\
&\times&\!\!\!
\left\{
2 A_{jk} + ( A_{jk} - \psi (j+2) + \psi(1) )
\left[
\frac{{\mit\Gamma} (j + 4) {\mit\Gamma} (k)}{{\mit\Gamma} (j) {\mit\Gamma} (k + 4)} - 1
\right]
\right\}\, . \nonumber
\end{eqnarray}
\end{itemize}
In all expressions we have introduced the matrix $A$ whose elements are defined by
\begin{equation}
A_{jk}
=
\psi\left( \frac{j + k + 4}{2} \right)
-
\psi\left( \frac{j - k}{2} \right)
+
2 \psi ( j - k ) - \psi ( j + 2 ) - \psi(1)
\, .
\end{equation}

\subsection{Solutions of renormalization group equation}

We have now all the elements of the anomalous dimension matrix of conformal
operators at NLO. In order to solve the renormalization group equation, it
remains find its eigenvectors (the eigenvalues are given by the diagonal
elements and coincide with known forward anomalous dimensions). This section
is devoted to the solution of this problem. It will be shown that the
eigefunctions are determined by the special conformal anomaly matrix,
determined above.

\subsubsection{Mixing of conformal operators}

Beyond leading order,  one has to perform an additional shuffling of
operators in order to get the eigenstates of the renormalization group
equation. Let us perform a transformation to the ``conformal''
renormalization scheme
\begin{equation}
\label{MStoCS}
{\mathbb{O}}^{\scriptscriptstyle\rm R}_{jl}
=
{\cal B}_{jk} \widehat {\mathbb{O}}^{\scriptscriptstyle\rm R}_{kl}
\, ,
\end{equation}
where the ${\cal B}$-matrix encodes the rotation to the diagonal basis,
in which the operators do not mix. This new set of operators satisfies
the renormalization group equation
\begin{equation}
\label{RGeqOhat}
\frac{d}{d \ln\mu} \widehat {\mathbb{O}}^{\scriptscriptstyle\rm R}_{jl}
=
-
\gamma_j
\widehat {\mathbb{O}}^{\scriptscriptstyle\rm R}_{jl}
\, ,
\end{equation}
with diagonal anomalous dimension matrix $\gamma_{jj} \equiv \gamma_j$. By
definition, they do not mix. The renormalization group equation is identical
to the dilatation operator, which is diagonal in this basis
\begin{equation}
{}[\widehat {\mathbb{O}}^{\scriptscriptstyle\rm R}_{jl} , \mathbb{D}]
=
i
\left( d^{\rm can} - \mu \frac{d}{d \mu} \right)
\widehat {\mathbb{O}}^{\scriptscriptstyle\rm R}_{jl}
=
i \left( l + 2 d + \gamma_j \right)
\widehat {\mathbb{O}}^{\scriptscriptstyle\rm R}_{jl}
\, .
\end{equation}
The solution to this equation is
\begin{equation}
\label{RGsolution}
\widehat {\mathbb{O}}^{\scriptscriptstyle\rm R}_{jl} (\mu^2)
= {\cal T} \exp
\left\{
-
\frac{1}{2} \int_{\mu_0^2}^{\mu^2}
\frac{d\tau}{\tau} \gamma_j (\alpha_s (\tau))
\right\}
\widehat {\mathbb{O}}^{\scriptscriptstyle\rm R}_{jl} (\mu_0^2)
\, .
\end{equation}
Since $\gamma_j$, which are two-by-two matrices in the singlet sector, do not commute
in general with each other for different scales of the coupling constant, we have
introduced the ${\cal T}$-ordered exponential along the ``proper time'' $\tau$.

From the renormalization group equations (\ref{RGequation}) and (\ref{RGeqOhat}), we
find that the ${\cal B}$-matrix satisfies the following differential equation:
\begin{equation}
\label{B-equation}
\beta (g) \frac{\partial}{\partial g} {\cal B}_{jk} (g)
+
\left( \gamma_j - \gamma_k \right) {\cal B}_{jk} (g)
+
\gamma^{\scriptscriptstyle\rm ND}_{jl} {\cal B}_{lk} (g)
=
0
\, .
\end{equation}
As will be shown in the following two sections, the formalism described above
allows to find the corrections to the eigenfunctions analytically, since the
${\cal B}$-matrix is defined in terms of the conformal anomaly $\gamma^c$
and the $\beta$-function. Below, we consider first the hypothetical conformal
limit and then address the realistic case with the running coupling constant.

\subsubsection{Restoration of conformal covariance in conformal limit}
\label{SectRestConfCov}

Our goal is to find the explicit form of the rotation matrix ${\cal B}$. It is
obvious that this matrix simultaneously diagonalizes both the dilatation and
conformal boosts given in Eqs.\ (\ref{Dilatation}) and (\ref{ConformalBoost}),
respectively. Let us discuss the nonsinglet case only, since the consideration
of the singlet sector does not bring any complications of principle. At the
conformal point $\beta = 0$, the rotation matrix $B$ satisfies a simplified
version of  Eq.\ (\ref{B-equation}), namely
\begin{equation}
{\cal B}^{- 1}_{j k'} \gamma_{k' k''} {\cal B}_{k'' k}
=
\gamma_j \delta_{jk}
\, ,
\end{equation}
(no summation on the right-hand side is implied). The last equation can be easily
solved recursively using the expansion
\begin{equation}
{\cal B}_{jk} = \sum_{N = 0}^\infty {\cal B}_{jk}^{(N)}
\, , \qquad
{\cal B}_{jk}^{(0)} = \delta_{jk}
\, .
\end{equation}
Note, that the terms ${\cal B}_{jk}^{(N)}$ with $N \geq 1$ are purely off-diagonal
matrices, i.e., their diagonal elements are identically zero. They are given in
terms of the $n$th power of a matrix $\bit{b} (k)$
\begin{equation}
{\cal B}_{jk}^{(N)} = \left\{ \bit{b}^N (k) \right\}_{jk}
\, ,
\end{equation}
which possesses the elements
\begin{equation}
b_{k' k''} (k)
=
- \frac{\gamma^{\scriptscriptstyle\rm ND}_{k' k''}}{\gamma_{k'} - \gamma_k}
\, .
\end{equation}
Let us demonstrate that the conformal boost is diagonal in this basis. The
transformation (\ref{MStoCS}) gives
\begin{equation}
{}[
\widehat {\mathbb{O}}^{\scriptscriptstyle\rm R}_{jl} , {\mathbb{K}}^-
]
=
{\cal B}^{-1}_{jk'}
\left\{
a (k', l) \delta_{k' k''} + \gamma^c_{k' k''} (l)
\right\}
{\cal B}_{k'' k}
\widehat {\mathbb{O}}^{\scriptscriptstyle\rm R}_{k l - 1}
\, .
\end{equation}
First we transform the right-hand side to the form
\begin{equation}
{\cal B}^{-1}_{jk'}
\left\{
a (k', l) \delta_{k' k''} + \gamma^c_{k' k''} (l)
\right\}
{\cal B}_{k'' k}
=
a (j, l) \delta_{jk}
+
{\cal B}^{-1}_{jk'} \gamma^c_{k' k''} (l) {\cal B}_{k'' k}
+
{\cal B}^{-1}_{jk'} a (k' , k) {\cal B}_{k' , k}
\, ,
\end{equation}
where we commuted ${\cal B}_{k''k}$ through the matrix $a_{k'k''}(l)$, first, and
then used the identity
\begin{equation}
\label{Bgamma}
a (k' , k) {\cal B}_{k' , k}
=
- \gamma^c_{k' k''} (k) {\cal B}_{k'' k}
\, ,
\end{equation}
proved in Appendix \ref{Bmatrix}. This yields
\begin{eqnarray}
{\cal B}^{-1}_{jk'}
\left\{
a (k', l) \delta_{k' k''} + \gamma^c_{k' k''} (l)
\right\}
{\cal B}_{k'' k}
\!\!\!&=&\!\!\!
a (j, l) \delta_{jk}
+
{\cal B}^{-1}_{jk'}
\left(
\gamma^c_{k' k''} (l)
-
\gamma^c_{k' k''} (k)
\right)
{\cal B}_{k'' k}
\nonumber\\
&=&\!\!\!
a (j, l) \delta_{jk}
+
2 (k - l)
{\cal B}^{-1}_{jk'} \gamma_{k' k''} {\cal B}_{k'' k}
\nonumber\\
&\equiv&\!\!\!
2 (j - l) (j + l + 2 \nu + \gamma_j)
\delta_{jk}
\, ,
\end{eqnarray}
where in the second step we used Eq.\ (\ref{DiffCgamma}) in the form
$\gamma^c_{kk'} (l) - \gamma^c_{k k'} (k) = 2 (k - l) \gamma_{kk'}$.
Therefore, for operators transformed according to Eq.\ (\ref{MStoCS})
one can restore the exact conformal covariance broken in the minimal
subtraction scheme,
\begin{equation}
[\widehat {\mathbb{O}}^{\scriptscriptstyle\rm R}_{jl} , \mathbb{K}^-]
=
2 (j - l) (j + l + 2 \nu + \gamma_j)
\widehat {\mathbb{O}}^{\scriptscriptstyle\rm R}_{j l - 1}
\, ,
\end{equation}
One observes that in the conformally covariant scheme the loop corrections to
the dilatation and conformal boost are determined solely by the anomalous
dimensions of conformal operators.

\subsubsection{Solution with running coupling}
\label{SectSolutRunCoupl}

In case of the running coupling constant,  the solution of the differential
equation (\ref{B-equation}) is more involved. First, we have to specify the
boundary condition for ${\cal B} (g)$. Note, that in the previous case of
the fixed coupling we are naturally led to a scheme with constant ${\cal B}$.
In the current case, the ``minimization'' of  radiative corrections corresponds
to the choice ${\cal B} (g_0) = 1$ with $g_0 = g (\mu_0^2)$ which implies the
absence of loop corrections at some initial scale $\mu_0$. The most important
advantage of this requirement is that, contrary to our previous discussion,
the initial condition for the solution of the renormalization group equation
(\ref{RGsolution}) can be easily determined by ordinary Gegenbauer operators
\begin{equation}
\label{NormalizationConfCovConf}
{\mathbb{O}}^{\scriptscriptstyle\rm R}_{jl} (\mu_0)
=
\widehat {\mathbb{O}}^{\scriptscriptstyle\rm R}_{jl} (\mu_0)
\, .
\end{equation}
To two-loop accuracy, we can write  equation for the first nontrivial iteration
of ${\cal B}_{jk} (g) = \1_{jk} + {\cal B}^{(1)}_{jk} (g)$,
\begin{equation}
\beta (g)
\frac{\partial}{\partial g} {\cal B}^{(1)}_{jk} (g)
+
\left( \gamma_j - \gamma_k \right) {\cal B}^{(1)}_{jk} (g)
+
{\gamma}^{\scriptscriptstyle\rm ND}_{jk}
=
0
\, ,
\end{equation}
the solution to which can be written in the form
\begin{equation}
\label{NLOsolutionBmatrix}
{\cal B}^{(1)}_{jk} (g)
=
- \int_{g_0}^{g} \frac{d g'}{\beta (g')}
\exp
\left(
- \int_{g'}^{g} \frac{d g''}{\beta (g'')}
\gamma_j (g'')
\right)
\gamma^{\scriptscriptstyle\rm ND}_{jk} (g')
\exp
\left(
- \int_{g}^{g'} \frac{d g''}{\beta (g'')}
\gamma_k (g'')
\right)
\, .
\end{equation}
Having constructed compact analytical solution to renormalization group equations
for conformal operators, we are ready to apply them to the evolution of GPDs.

\subsection{Solution of evolution equations for GPDs}

In the present section, we present an outline of methods used to solve
evolution equations for generalized parton distributions. We mostly focus
on the formalism of the orthogonal polynomial reconstruction which allows
for a fully analytical treatment of the problem. Analogous analytical
methods for coordinate-space functions, i.e., directly for non-local
light-ray operators, were designed in Refs.\ \cite{BalBra89,KivMan99}
for one-loop renormalization group equations. This formalism we will
briefly discuss next. We continue with a short discussion of the solution
of evolution equations for double distributions and the formalism of
``effective'' forward distributions. Then we turn to the machinery of
orthogonal polynomial reconstruction. Finally, we conclude this section
with a few remarks on the deficiency the truncation procedure of the
expansion in polynomials and advantages of a direct numerical integration
of evolution equations for GPDs.

\subsubsection{One-loop evolution in coordinate space}

The basis of local conformal operators, introduced and elaborated in great
detail in previous sections, is not the only choice. It turns out that one
can completely circumvent going through the local Taylor series expansion
and construction of conformal towers in order to solve analytically the
evolution equation for light-ray operators. In fact, such a procedure was
developed a long time ago and is based on the expansion in terms of non-local
conformal operators \cite{BalBra89}. It was recently applied to the study of
the evolution of the coordinate-space GPDs \cite{KivMan99}, which are the
partial Fourier transform of GPDs with respect to the Feynman momentum $x$,
$$
\mathcal{F}^a (z^-, \eta, \Delta^2)
\equiv
\int dx  \,{\rm e}^{- i x z^- p^+} F^a (x, \eta, \Delta^2)
\, .
$$
These functions are analogous to the Ioffe-time forward parton distributions
\cite{Iof69,Pestieau:1970in}.

In order to demonstrate the basic features of the formalism we discuss the
quark operator only and refer the interested reader to the original papers
\cite{BalBra89,KivMan99,BKM03,ManKirSch05} for an exhaustive treatment of
other aspects of the formalism and discussion of the singlet sector as well.
A non-local conformal operator is defined as a convolution of the light-ray
operator with a coefficient function \cite{BalBra89}
\begin{equation}
\label{CoherentState}
\mathbb{S}_j^{qq} (k)
=
\int_{- \infty}^\infty d r^- \int_0^\infty d z^- \, {\mit\Psi}_j (k; r^-, z^-)
\mathcal{O}^{qq} \left( \frac{r^- + z^-}{2} , \frac{r^- - z^-}{2} \right)
\, .
\end{equation}
The function ${\mit\Psi}_j (k; r^-, z^-)$ is found from the condition that this
coherent state is an eigenfunction of the conformal Casimir operator \cite{BKM03}
\begin{equation}
\label{CohConfStateConditions}
[\mathbb{L}^+ , \mathbb{S}_j^{qq} (k)] = - i k \, \mathbb{S}_j^{qq} (k)
\, , \qquad
[\mathbf{L}^2 , \mathbb{S}_j^{qq} (k)] = (j + 2)(j + 1) \, \mathbb{S}_j^{qq} (k)
\, ,
\end{equation}
where we adopted the second condition from Eq.\ (\ref{PrimaryField}) setting
there the conformal spin to $J = j + 2$ (see Eq.\ (\ref{ValueConfSpin})) with
the numerical value of the quark conformal spin $j_q = 1$. Taking into account
that the operator $\mathbb{L}^+$ acts as a total derivative on a non-local
operator, i.e., differentiation with respect to $r^-$, we can integrate by
parts in Eq.\ (\ref{CoherentState}) and get a simple equation for the wave
function
\begin{equation}
\frac{\partial}{\partial r^-} \, {\mit\Psi}_j (k; r^-, z^-)
=
i k \, {\mit\Psi}_j (k; r^-, z^-)
\, ,
\end{equation}
which admits a plane-wave solution in $r^-$
\begin{equation}
{\mit\Psi}_j (k; r^-, z^-) = {\rm e}^{i k r^-} \psi_j (k z^-)
\, .
\end{equation}
Since $r^-$ is the center of mass of the light-ray operator $\mathcal{O}^{qq}$,
the solution describes the propagation of the wave packet $\psi_j (k z^-)$ with
momentum $k$, while the $z^-$-dependent profile $\psi_j (k z^-)$ responsible for
the internal relative motion of quarks attached to the ends of the Wilson string.
From the second condition in Eq.\ (\ref{CohConfStateConditions}), one finds that
the wave function $\psi_j (k z^-)$ satisfies the Bessel equation,
\begin{equation}
\left(
\varrho^2 \frac{\partial^2}{\partial \varrho^2} + \varrho^2 - (j + 2) (j + 1)
\right)
\psi_j (\varrho)
= 0
\, ,
\end{equation}
and admits the solution
\begin{equation}
\psi_j (\varrho)
= \left( \frac{\varrho}{k} \right)^{1/2} J_{j + 3/2} (\varrho)
\, ,
\end{equation}
where the overall factor of $k^{-1/2}$ was chosen for normalization purposes.

By construction, the non-local conformal operator
\begin{equation}
\label{CoherentStateExplicit}
\mathbb{S}_j^{qq} (k)
=
\int_{- \infty}^\infty d r^- \, {\rm e}^{i k r^-} \int_0^\infty d z^- \,
(z^-)^{1/2} \,
J_{j + 3/2} (k z^-)
\mathcal{O}^{qq} \left( \frac{r^- + z^-}{2} , \frac{r^- - z^-}{2} \right)
\end{equation}
possesses an autonomous scale dependence at one-loop order,
\begin{equation}
\mathbb{S}^{qq}_j (k ; \mu^2)
=
\mathbb{S}^{qq}_j (k ; \mu_0^2)
\left(
\frac{\alpha_s (\mu_0^2)}{\alpha_s (\mu^2)}
\right)^{\gamma_{(0)j}^{qq}/\beta_0}
\, .
\end{equation}
This is equivalent to the multiplicative renormalization of local conformal
operators discussed in the preceding sections. Now, using the Neumann series
representation of the delta-function in the variable $z^-$ \cite{KivMan99}
\begin{equation}
\delta (z_1^- - z_2^-) = (z_2^-)^{1/2} (z_1^-)^{-3/2}
\sum_{j = 0}^\infty (3 + 2 j)
J_{j + 3/2} (k z_1^-) J_{j + 3/2} (k z_2^-)
\, ,
\end{equation}
and conventional Fourier representation for the delta-function in $r^-$
we can expand the non-local light-ray operator into the infinite series in
coherent states
\begin{eqnarray}
\label{ExpansionNLoperatorConformal}
\mathcal{O}^{qq} \left( \frac{r^- + z^-}{2} , \frac{r^- - z^-}{2} \right)
\!\!\!&=&\!\!\!
\int d r^{\prime -} \, d z^{\prime -}
\, \delta (r^- - r^{\prime -}) \, \delta (z^- - z^{\prime -})
\mathcal{O}^{qq}
\left(
\frac{r^{\prime -} + z^{\prime -}}{2} , \frac{r^{\prime -} - z^{\prime -}}{2}
\right)
\nonumber\\
&=&\!\!\!
(z^-)^{-3/2} \sum_{j = 0}^\infty (3 + 2 j)
\int \frac{d k}{2 \pi} \, {\rm e}^{i k r^-} J_{j + 3/2} (k z^-) \,
\mathbb{S}^{qq}_j (k)
\, .
\end{eqnarray}
Let us point out that the conformal operators constructed in Eq.\
(\ref{CoherentStateExplicit}) are defined for positive integer values of the
conformal spin $j$ only \cite{KivMan99}, such that non-local operators are
expanded in infinite series (\ref{ExpansionNLoperatorConformal}). On the
other hand, the decomposition found in the original work \cite{BalBra89}
uses the Kantorovich-Lebedev transformation \cite{KanLeb38} and is based on
an integral representation which requires analytical continuation of the
conformal spin $j$ into the complex plane, which thus uses a continuous
series representation of the collinear conformal group, rather than the
discrete series adopted in the local operator expansion. The equivalence
of the two solutions was demonstrated in Ref.\ \cite{KivMan99}, which
illustrates a relation between the Kantorovich-Lebedev transformation
and the Neumann series expansion.

\subsubsection{``Difficulties'' with eigenfunctions}
\label{SectDiffEigen}

Now we turn to a thorough discussion of constructing solutions to the evolution
equations for GPDs in the basis of local conformal operators. As we established
in Section \ref{DiagonalLOADs}, the one-loop evolution equation for GPDs is
diagonalized in the basis of Gegenbauer polynomials. Namely, taking the nonsinglet
quark sector as an example, one has the leading order equation\footnote{For brevity,
we do not display the superscript ${\rm NS}$ on quantities involved.}
\begin{equation}
\label{LOGenerEvEqforGPDs}
\frac{d}{d \ln \mu} F^q (x, \eta ; \mu^2)
=
- \frac{\alpha_s (\mu^2)}{2 \pi} \int_{- 1}^1 \frac{d y}{\eta}
k^{qq}_{(0)} \left( \frac{x}{\eta} , \frac{y}{\eta} \right)
F^q (y, \eta ; \mu^2) + {\cal O} (\alpha_s^2)
\, ,
\end{equation}
with the Gegenbauer polynomials $C_j^{3/2} (x)$---defined on the interval
$-1 \leq x \leq 1$---being its eigenfunctions,
\begin{equation}
\int_{- 1}^1 \frac{d x}{\eta} \, C_j^{3/2} \left( \frac{x}{\eta} \right)
k^{qq}_{(0)} \left( \frac{x}{\eta} , \frac{y}{\eta} \right)
=
\gamma_{(0)j}^{qq} C_j^{3/2} \left( \frac{y}{\eta} \right)
\, .
\end{equation}
One may try  to write the solution to Eq.\ (\ref{LOGenerEvEqforGPDs}) as an
expansion in terms of its eigenfunctions
\begin{equation}
\label{ExpansionGPDeigenfunctions}
F^q (x, \eta ; \mu^2)
=
\frac{1}{\eta} \sum_{j = 0}^\infty
\frac{w \big( \ft{x}{\eta} | \ft32 \big)}{\eta^j \, N_j ( \ft32)}
C_j^{3/2} \left( \frac{x}{\eta} \right)
\mathbb{F}^q_j (\eta ; \mu^2)
\, ,
\end{equation}
where the weight and normalization factors are
$$
w (x | \nu) = (1 - x^2)^{\nu - 1/2}
\, , \qquad
N_j (\nu) = 2^{1 - 2 \nu}
\frac{
{\mit\Gamma}^2 (1/2) {\mit\Gamma} (2 \nu + j)
}{
{\mit\Gamma}^2 (\nu) {\mit\Gamma} (j + 1) (\nu + j)
}
\, ,
$$
and the Gegenbauer moments of GPDs related to the conformal operators
(\ref{QuarkAndGluonConformalOperators}) via
\begin{equation}
\label{QuarkConfMoments}
\mathbb{F}^q_j (\eta ; \mu^2)
= \eta^j \int_{- 1}^1 dx \ C_j^{3/2} \left( \frac{x}{\eta} \right)
F^q (x, \eta ; \mu^2)
=
( p^+ )^{- j - 1}
\langle p_2 | \mathbb{O}^{qq}_{jj} (0) | p_1 \rangle
\, .
\end{equation}
However, one immediately faces a difficulty identical to the one we already
encountered before, when we studied the eigenfunctions of the inclusive DGLAP
equations. Namely, the Gegenbauer polynomials $C_j^\nu (x)$ are mutually orthogonal
on the interval $- 1 \leq x \leq 1$, which is their support region. Therefore,
each term on the right-hand side of  Eq.\ (\ref{ExpansionGPDeigenfunctions}),
vanishes outside  the interval $- \eta \leq x \leq \eta$, while the left-hand
side has the support in the entire region $- 1 \leq x \leq 1$. Of course, the
correct support on the right-hand side is restored due to the infinite summation
of the slowly converging series in Gegenbauer polynomials, in the very same
fashion as in Eq.\ (\ref{PDFinDeltas}). Thus, this form of the solution is not
useful in practical studies of the logarithmic scaling violation of GPDs. Still,
there is a feature which becomes extremely transparent in the representation
(\ref{ExpansionGPDeigenfunctions}). Due to fact that the
Gegenbauer moments evolve autonomously
\begin{equation}
\label{EvolutionConfMomGPDs}
\mathbb{F}^q_j (\eta ; \mu^2)
=
\mathbb{F}^q_j (\eta ; \mu_0^2)
\left(
\frac{\alpha_s (\mu_0^2)}{\alpha_s (\mu^2)}
\right)^{\gamma_{(0)j}^{qq}/\beta_0}
\end{equation}
with vanishing anomalous dimension for the lowest moment of GPDs in the
vector and axial-vector channels $\gamma_{(0)j = 0}^{qq, A, V} = 0$, at
asymptotically large scales only the lowest term in the expansion survives.
As a result, the GPD is driven at asymptotically large scales to the shape
\begin{equation}
\label{AsyQuarkGPDns}
F^q (x, \eta; \mu^2 \to \infty)
\equiv
F^q_{\rm asy} (x, \eta)
=
\frac{3}{4} \frac{1}{\eta}
\left( 1 - \frac{x^2}{\eta^2} \right) \theta (\eta - |x|) \, \mathbb{F}^q_0
\, ,
\end{equation}
where the lowest conformal moment $\mathbb{F}^q_0$ is related to the quark
electromagnetic form factors, i.e., it is the Dirac $\mathbb{F}^q_0 = F_1^q
(\Delta^2)$ and Pauli $\mathbb{F}^q_0 = F_2^q (\Delta^2)$ form factors for
parton helicity non-flip $F^q = H^q$ and flip $F^q = E^q$ GPDs. Note that
the GPD washes out completely from the ``inclusive'' region, i.e.,
\begin{equation}
F^q_{\rm asy} (x, \eta)
=
0
\, , \qquad \mbox{for} \qquad |x| > \eta
\, .
\end{equation}
The explicit asymptotic solutions for the singlet sector can be found below in
Eqs.\ (\ref{AsyFormSingletQuarkGPDs}) and (\ref{AsyFormGluonGPDs}).

Of course, the formal expansion (\ref{ExpansionGPDeigenfunctions}) should be
understood in a sense of the mathematical distribution in the same fashion
as the expansion in $\delta$-functions of the parton distribution
(\ref{deltaExpansion}). To make the analogy with the inclusive case more
explicit, it is easy to see that the Gegenbauer polynomials can be
represented in the form involving derivative of the $\delta$-function
making use of the Rodrigues' formula
\begin{equation}
\label{GegenbauerAsMathDistr}
\left( 1 - x^2 \right)^{\nu - 1/2} C_j^\nu (x)
=
2^{1 - 2 \nu - j}
\frac{
{\mit\Gamma} (\ft12) {\mit\Gamma} (j + 2 \nu)
}{
{\mit\Gamma} (\nu) {\mit\Gamma} (j + \nu + \ft12) {\mit\Gamma} (j + 1)
}
\int_{- 1}^1 dv \, (1 - v^2)^{j + \nu - \ft12}
\delta^{(j)} ( v - x )
\, ,
\end{equation}
where $\delta^{(j)} ( v - x ) = \partial_v^j \delta ( v - x )$. Hence, to get
a meaningful result, one should convolute both sides with a smooth function
first. This idea lies in the heart of the method which we will start discussing
in the next section using the same nonsinglet quark example.

\subsubsection{Orthogonal polynomial reconstruction}

\begin{figure}[t]
\unitlength1mm
\begin{center}
\mbox{
\begin{picture}(0,35)(89,0)
\put(0,0){\insertfig{4.3}{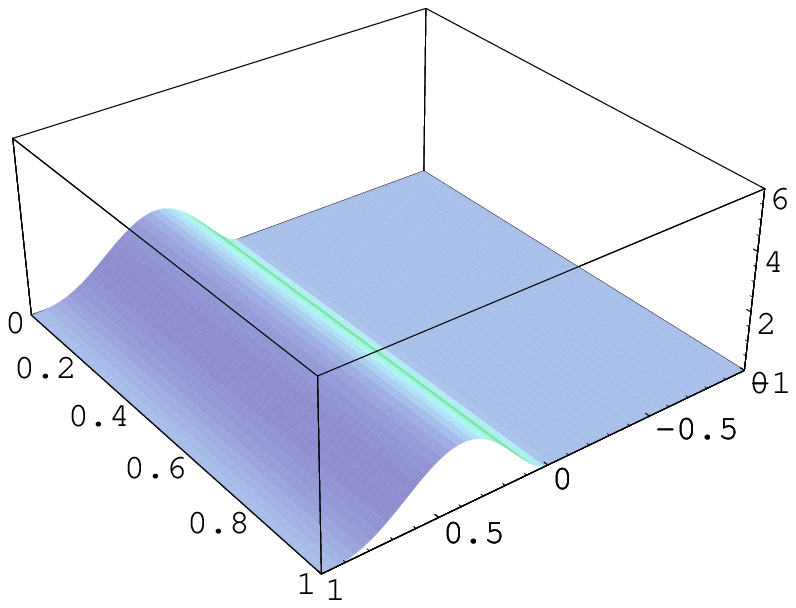}}
\put(45,0){\insertfig{4.3}{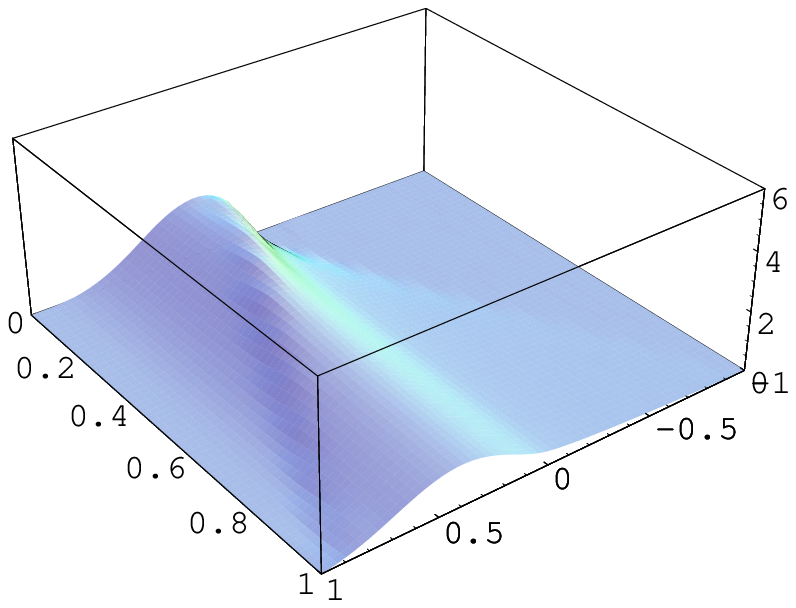}}
\put(90,0){\insertfig{4.3}{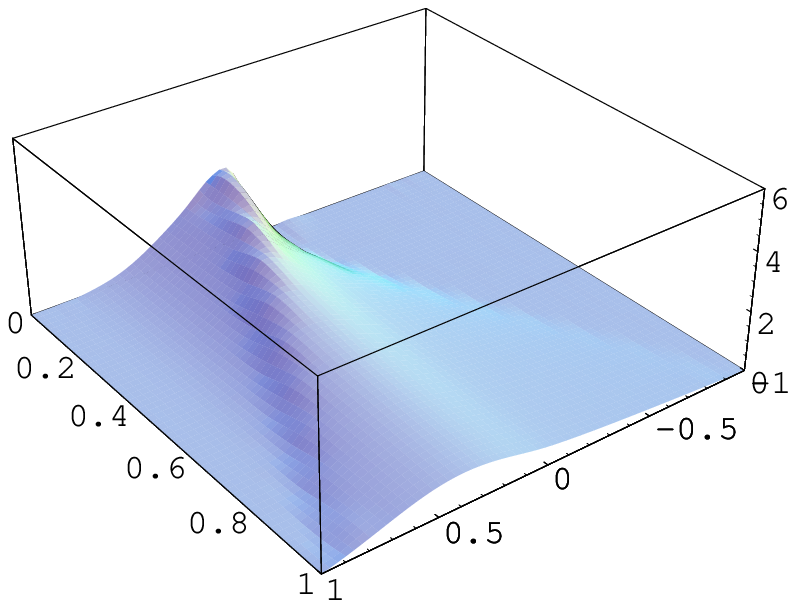}}
\put(135,0){\insertfig{4.3}{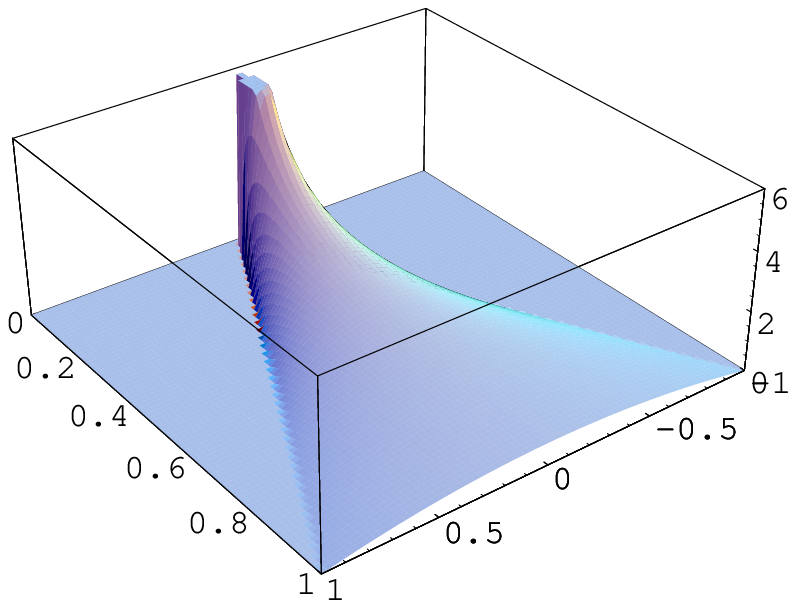}}
\put(4,3){${\scriptstyle \eta}$}
\put(32,3){${\scriptstyle x}$}
\end{picture}
}
\end{center}
\caption{\label{LOevolution} Evolution of a model distribution with the change of
the renormalization scale. Figures from left to right show the  skewless input
distribution $H (x ,\eta) = 60 x^2 (1 - x)^3 \theta (x)$ and its evolved shape at
$\mu^2 = 2 \, {\rm GeV}^2$, $\mu^2 = 200 \, {\rm GeV}^2$ and $\mu^2 = \infty$,
respectively. The qualitative features of generalized evolution are clearly seen:
Partons from the inclusive region $|x| > \eta$ migrate into the exclusive domain
$|x| < \eta$ and once they get there, they never come back. They are redistributed
in the exclusive region $|x| < \eta$ according to the usual ERBL evolution pattern
as examplified in Fig.\ \ref{AsymptoticForm}.}
\end{figure}

The restoration of the support properties of GPDs can be easily achieved
via their expansion in a series with respect to a complete set of orthogonal,
say, Gegenbauer, polynomials \cite{BelGeyMulSch97,ManPilWei97}, having the
same support as GPDs,
\begin{eqnarray}
\label{NonForEvo}
F^q \left( x, \eta ; \mu^2 \right)
=
\sum_{j = 0}^{\infty} \frac{w( x | \ft32)}{N_j(\ft32)}
C^{3/2}_j (x) F^q_j (\eta ; \mu^2)
\, .
\end{eqnarray}
It is straightforward to calculate the expansion coefficients $F^q_j$ in
terms of the conformal moments which evolve autonomously at leading order
(\ref{EvolutionConfMomGPDs}),
\begin{eqnarray}
\label{a-j}
F^q_j (\eta ; \mu^2) =
\sum_{k=0}^j E_{jk} ( \eta | \ft{3}{2} ) \mathbb{F}^q_k (\eta ; \mu^2)
\, .
\end{eqnarray}
Using Eq.\ (\ref{GegenbauerAsMathDistr}) and employing
the definition of hypergeometric functions,
we can write the expansion coefficients
$E_{jk}( \eta | \nu )$  in a very compact form
\begin{eqnarray}
\label{a-jk}
E_{jk}( \eta | \nu )
\!\!\!&=&\!\!\!
\int_{-1}^{1} dx \frac{w (x | \nu)}{\eta^k N_k (\nu) }
C^\nu_k ( x ) C^\nu_j ( \eta x )
\nonumber\\
&=&\!\!\!
\frac{1}{2} \theta_{jk} \left[ 1 + (-1)^{j - k} \right]
\frac{
(-1)^{\frac{j - k}{2}}
{\mit\Gamma} \left( \nu + \frac{j + k}{2} \right)
}{
{\mit\Gamma} \left( \nu + k \right)
{\mit\Gamma} \left( 1 + \frac{j - k}{2} \right)
}
\,
{_2F_1}
\left( \left. {\frac{k - j}{2}, \nu + \frac{j + k}{2} \atop \nu + k + 1}
\right| \eta^2 \right)
\, .
\end{eqnarray}
Since ${_2F_1}( - n, a + n; a + 1; 1 ) = \delta_{ n 0 }$, we immediately obtain
for $\eta = 1$ that $F_j^q (1 ; \mu^2)= \mathbb{F}^q_j (1 ; \mu^2)$,  which matches
to the solution to ER-BL equation for the quark component of meson distribution
amplitudes. The qualitative features of the evolution are demonstrated in Fig.\
\ref{LOevolution} for a simple GPD.

\subsubsection{Next-to-leading order: nonsinglet sector}

Let us now discuss the effects of mixing of conformal moments at next-to-leading
order, and how this modifies the solution to the evolution equations
\cite{BelMulNieSch98}. Again, we perform at first the expansion of GPDs in terms
of a set of orthogonal polynomials
\begin{eqnarray}
F^q \left( x, \eta ; \mu^2 \right)
=
\sum_{j = 0}^{\infty}
\frac{(1 - x)^\alpha (1 + x)^\beta}{n_j (\alpha, \beta)}
P_j^{(\alpha, \beta)} (x) F^q_j (\eta ; \mu^2)
\, .
\end{eqnarray}
Here we use the opportunity and illustrate the use of the Jacobi polynomials,
``living'' on the interval $- 1 \leq x \leq 1$. The normalization coefficients
are
$$
n_j (\alpha, \beta)
=
\frac{
{\mit\Gamma} (j + \alpha + 1) {\mit\Gamma} (j + \beta + 1)
}{
(2j + \alpha + \beta + 1) {\mit\Gamma} (j + 1) {\mit\Gamma} (j + \alpha + \beta + 1)
}
\, ,
$$
with parameters $\alpha$, $\beta$ which can be chosen from the condition of the
fastest convergence of the series, which is difficult to implement in practice.
The coefficients in the re-expansion in terms of conformal moments of GPDs
$\mathbb{F}^q_j (\eta ; \mu^2)$
\begin{eqnarray}
F^q_j (\eta ; \mu^2) =
\sum_{k=0}^j E_{jk} ( \eta | \alpha, \beta ) \mathbb{F}^q_k (\eta ; \mu^2)
\, ,
\end{eqnarray}
are now determined by the overlap integral
\begin{equation}
E_{jk} ( \eta | \alpha, \beta )
=
\int_{- 1}^1 dx \, \frac{w (x | \frac{3}{2})}{\eta^k N_k (\frac{3}{2})}
C_k^{3/2} (x) P_j^{(\alpha , \beta)} (\eta x)
\, .
\end{equation}
The results for all other classic orthogonal polynomials immediately follow
from this expression. For special values of the parameters $\alpha$ and
$\beta$, the Jacobi polynomials coincide \cite{AbrSte65} either with
Gegenbauer,
$$
P_j^{( \lambda - 1/2, \lambda - 1/2)} (x)
=
\frac{
{\mit\Gamma} (2 \lambda) {\mit\Gamma} (j + \lambda + \frac{1}{2})
}{
{\mit\Gamma} (j + 2 \lambda) {\mit\Gamma} (\lambda + \frac{1}{2})
} C_j^\lambda (x)
\, ,
$$
or Legendre,
$$
P_j^{(0,0)} (x) = P_j (x) \, ,
$$
or Chebyshev polynomials of the first,
$$
P_j^{(- 1/2, - 1/2)} (x)
=
\frac{{\mit\Gamma} (j + \frac{1}{2})}{\sqrt{\pi} j!} T_j (x)
\, ,
$$
and the second kind,
\begin{equation}
\label{ChebyshevPolynomial2}
P_j^{(1/2, 1/2)} (x)
=
\frac{2 {\mit\Gamma} (j + \frac{3}{2})}{\sqrt{\pi} (j + 1)!} U_j (x)
\, .
\end{equation}

Starting from two-loop order, the conformal operators start to mix and so do the
Gegenbauer moments of GPDs. The latter have to be re-expanded in terms of
multiplicatively renormalizable moments $\widehat{\mathbb{F}}^q_j (\eta ; \mu^2)$
as
\begin{equation}
{\mathbb{F}}^q_j (\eta ; \mu^2)
=
\sum_{k = 0}^j
\eta^{j - k}
{\cal B}_{jk} (\mu^2, \mu_0^2)
{\cal E}_k (\mu^2, \mu_0^2)
\widehat{\mathbb{F}}^q_k (\eta ; \mu^2_0)
\, .
\end{equation}
We have chosen the normalization condition for the two evolution operators
\begin{equation}
{\cal E}_j (\mu^2_0, \mu_0^2)
= 1
\, , \qquad
{\cal B}_{jk} (\mu^2_0, \mu_0^2)
= \delta_{jk}
\, ,
\end{equation}
which implies
\begin{equation}
\widehat{\mathbb{F}}^q_j (\eta ; \mu^2_0)
=
{\mathbb{F}}^q_j (\eta ; \mu^2_0)
\, ,
\end{equation}
so that at certain normalization point $\mu_0$ the multiplicatively
renormalizable operators coincide with conformal operators, as discussed in
Section \ref{SectSolutRunCoupl}. The two evolution operators satisfy the
following evolution equations with next-to-leading order solutions.
\begin{itemize}
\item The diagonal evolution operator ${\cal E}$ obeys the equation
\begin{equation}
\frac{d}{d\ln \mu} {\cal E}_j (\mu^2, \mu_0^2)
+
\gamma^{qq}_j {\cal E}_j (\mu^2, \mu_0^2)
=
0
\, ,
\end{equation}
with the two-loop solution
\begin{eqnarray}
{\cal E}_j (\mu^2 , \mu_0^2)
\!\!\!&=&\!\!\!
\left(
\frac{\alpha_s (\mu_0^2)}{\alpha_s (\mu^2)}
\right)^{ \gamma^{qq}_{(0)j} / \beta_0 }
\left(
\frac{
\beta_0 + \beta_1 \frac{\alpha_s (\mu^2)}{4 \pi}
}{
\beta_0 + \beta_1 \frac{\alpha_s (\mu^2_0)}{4 \pi}
}
\right)^{
\left(
\gamma^{qq}_{(0)j} / \beta_0
-
2 \gamma^{qq}_{(1)j} / \beta_1
\right)}
\nonumber\\
&\simeq&\!\!\!
\left(
\frac{\alpha_s (\mu_0^2)}{\alpha_s (\mu^2)}
\right)^{ \gamma^{qq}_{(0)j} / \beta_0 }
\left\{
1
+
\left(
\frac{\beta_1 \, \gamma^{qq}_{(0)j} }{ 2 \beta_0^2}
-
\frac{\gamma^{qq}_{(1)j}}{\beta_0}
\right)
\frac{\alpha_s(\mu^2) - \alpha_s(\mu_0^2) }{2 \pi}
\right\}
\, ,
\end{eqnarray}
where $\beta_0$ and $\beta_1$ are the first two coefficients of the
perturbative expansion of the QCD $\beta$-function, given in Appendix
\ref{CovGaugeRenormalization}.
\item The non-diagonal evolution operator ${\cal B}$ fulfills the equation
\begin{equation}
\frac{d}{d\ln \mu} {\cal B}_{jk} (\mu^2, \mu_0^2)
+
\left( \gamma^{qq}_j - \gamma^{qq}_k \right) {\cal B}_{jk} (\mu^2, \mu_0^2)
+
{\gamma}^{qq, \scriptscriptstyle\rm ND}_{jl} {\cal B}_{lk} (\mu^2, \mu_0^2)
=
0
\, .
\end{equation}
The solution to it in the two-loop approximation was given before in Eq.\
(\ref{NLOsolutionBmatrix}).  Keeping the first nontrivial terms in the
perturbative expansion for all functions involved, it reads
\begin{equation}
{\cal B}_{jk} (\mu^2, \mu_0^2)
=
\delta_{jk} - \frac{\alpha_s (\mu^2)}{2 \pi}
\frac{
\gamma^{qq, \scriptscriptstyle\rm ND}_{(1)jk}
}{
\gamma^{qq}_{(0)j}
-
\gamma^{qq}_{(0)k}
+
\beta_0
}
\left(
1
-
\left(
\frac{\alpha_s (\mu_0^2)}{ \alpha_s (\mu^2)}
\right)^{
( \gamma^{qq}_{(0)j} - \gamma^{qq}_{(0)k} + \beta_0 )/\beta_0}
\right)
\, .
\end{equation}
We remind that, compared to all other renormalization-group functions,
$\gamma^{qq, \scriptscriptstyle\rm ND}_{jk}$ starts at two-loop order
according to Eq.\ (\ref{NDgammaNLO}).
\end{itemize}

Finally, the coupling constant at two-loop order is given by the following
inverse-log expansion
\begin{equation}
\label{RunningCouplingConstantTwo}
\alpha_s (\mu^2)
=
- \frac{4\pi}{\beta_0 \ln ( \mu^2 / \Lambda^2_{\overline{\rm MS}})}
\left(
1 + \frac{\beta_1}{\beta_0^2}
\frac{
\ln \, \ln ( \mu^2 / \Lambda^2_{\overline{\rm MS}} )
}{
\ln ( \mu^2 / \Lambda^2_{\overline{\rm MS}})
}
\right) \ .
\end{equation}
If the input scale $\mu_0$ at which the initial conditions for renormalization
group evolution are defined is chosen very low, it is more reliable and accurate
to obtain $\alpha_s$ in two-loop approximation by solving the exact transcendental
equation
\begin{equation}
\label{ExactTwoLoopCoupling}
- \beta_0 \ln \frac{\mu^2}{\Lambda^2_{\overline{\rm MS}}}
= \frac{4 \pi}{\alpha_s (\mu^2)}
- \frac{\beta_1}{\beta_0}
\ln \left( - \frac{4 \pi}{\beta_0 \alpha_s (\mu^2)}
- \frac{\beta_1}{\beta_0^2} \right)
\, .
\end{equation}
The two formulas, Eqs.\ (\ref{RunningCouplingConstantTwo}) and
(\ref{ExactTwoLoopCoupling}), differ by $\sim 15$\% for $\mu^2 = 0.6\
{\rm GeV}^2$, and the deviation decreases to 3\%  for $\mu^2 = 4\ {\rm GeV}^2$.
However, this source of theoretical uncertainty is  conceptually irrelevant
for our study.

\subsubsection{Next-to-leading order: singlet sector}

Finally, let us turn to the singlet sector \cite{BelMulNieSch99}. The
``two-component vector'' of quark and gluon GPDs (\ref{def-SPD-sing}) is
expanded into a series
\begin{equation}
\label{PolySeries}
\bit{F} ( x, \eta ; \mu^2 )
= \sum_{j = 0}^{\infty} \widetilde{\bit{\cal P}}_j (x)
\bit{F}_j ( \eta ; \mu^2 )
\, ,
\end{equation}
involving  a matrix of orthogonal polynomials
\begin{equation}
\label{PolyMatrix}
\widetilde{\bit{\cal P}}_j ( x )
=
\left(
\begin{array}{cc}
\frac{w (x | \alpha_p )}{n_j ( \alpha_p )}
{\cal P}_j^{( \alpha_p )} (x)
& 0 \\
0 &
\frac{w (x | \alpha_p^\prime )}{
n_j ( \alpha_p^\prime )}
{\cal P}_j^{( \alpha_p^\prime )} (x)
\end{array}
\right) \  ,
\end{equation}
with $w (x | \alpha_p )$ and $n_j ( \alpha_p )$ being the weight and
normalization factors, respectively. Re-expanding the moments $\bit{F}_j
( \eta ; \mu^2 )$ in terms of the conformal moments $\mathbf{F}_k (\eta ;
\mu^2)$, one finds analogously to the nonsinglet sector
\begin{equation}
\label{JacobiMom}
\bit{F}_j ( \eta ; \mu^2 )
=
\sum_{k = 0}^{\infty} \bit{E}_{jk} (\eta) \mathbf{F}_k (\eta ; \mu^2)
\, .
\end{equation}
The matrix of coefficients
\begin{equation}
\label{E-matrix}
{\mbox{\boldmath$E$}}_{jk} (\eta)
=
\left(
\begin{array}{cc}
E_{jk} ( \eta | \frac{3}{2}; \alpha_p ) & 0 \\
0 & E_{j\,k-1} ( \eta | \frac{5}{2}; \alpha_p^\prime )
\end{array}
\right)
\, ,
\end{equation}
consists of elements
\begin{equation}
E_{jk} ( \eta | \nu ; \alpha_p )
=
\int_{- 1}^1 dx \, \frac{w (x | \nu)}{\eta^k N_k (\nu)}
C_k^\nu (x) {\cal P}_j^{(\alpha_p)} (\eta x)
\, ,
\end{equation}
while the matrix of conformal moments $\mathbf{F}_k (\eta ; \mu^2)$ is
defined by
\begin{equation}
\label{SingletGegenMomentsGPDs}
\mathbf{F}_j (\eta ; \mu^2)
=
\int_{- 1}^1 dx \,
\left(
\begin{array}{cc}
\eta^j \, C_j^{3/2} \left( x / \eta \right) & 0 \\
0 & \eta^{j - 1} \, C_{j - 1}^{5/2} \left( x / \eta \right)
\end{array}
\right)
\,
\bit{F} ( x, \eta ; \mu^2 )
\, ,
\end{equation}
where the two vector of the singlet quark and gluon GPDs is given in Eq.\
(\ref{def-SPD-sing}). These conformal moments are expressed in terms of the
non-forward matrix element of the two-vector of the conformal operators
(\ref{TwoVectorConfOperators}) as follows
\begin{equation}
\mathbf{F}_j (\eta ; \mu^2)
=
2 (p^+)^{- j - 1} \langle p_2 | \mathbf{O}_{jj} (0)| p_1 \rangle
\end{equation}
The latter evolve according to the formula
\begin{equation}
\mathbf{F}_k (\eta ; \mu^2)
=
\sum_{k = 0}^j
\eta^{j - k}
\bit{{\cal B}}_{jk} (\mu^2, \mu_0^2)
\bit{{\cal E}}_k (\mu^2, \mu_0^2)
\widehat{\mathbf{F}}_k (\eta ; \mu^2_0)
\, .
\end{equation}
involving two evolution operators as in the non-singlet case.

In NLO, the singlet diagonal evolution operator $\bit{\cal E}_j$ satisfies the
equation
\begin{equation}
\label{DiagEvolOperMatrix}
\frac{d}{d\ln \alpha_s (\mu^2)}
\bit{\cal E}_j \left( \mu^2, \mu_0^2 \right)
= - \frac{1}{\beta_0}
\left\{ \bit{\gamma}_{(0)j} + \frac{\alpha_s (\mu^2)}{2 \pi} \bit{R}_j \right\}
\bit{\cal E}_j \left( \mu^2, \mu_0^2 \right)
\, ,
\end{equation}
with the boundary condition
$$
\bit{\cal E}_j \left( \mu_0^2 , \mu_0^2 \right)
=
\1
=
\left( {1 \ \, 0 \atop 0 \ \, 1} \right)
\, .
$$
Here
\begin{equation}
\bit{R}_j
=
\bit{\gamma}_{(1)j} - \frac{\beta_1}{2 \beta_0} \bit{\gamma}_{(0)j}
\,
\end{equation}
is expressed in terms of the diagonal elements $\bit{\gamma}_{jj} = \bit{\gamma}_j$
of the one- and two-loop expressions of the two-by-two matrix of anomalous dimensions
(\ref{TwoByTwoMatrixAnomalousDim}). The solution to Eq.\ (\ref{DiagEvolOperMatrix})
can be written as \cite{FurPet82,GluReyVog90}
\begin{eqnarray}
\label{SingletDiagonalEvOp}
\bit{\cal E}_j \left( \mu^2 , \mu_0^2 \right)
\!\!\!&=&\!\!\!
\left(
\bit{P}^+_j -
\frac{\alpha_s (\mu^2) - \alpha_s (\mu^2_0)}{2 \pi}
\frac{1}{\beta_0}
\bit{P}^+_j \bit{R}_j \bit{P}^+_j
\right)
\left(
\frac{\alpha_s (\mu^2_0)}{\alpha_s (\mu^2)}
\right)^{\gamma^+_j / \beta_0}
\nonumber\\
&-&\!\!\!
\frac{\alpha_s (\mu^2)}{2\pi}
\frac{
\bit{P}^-_j \bit{R}_j \bit{P}^+_j}{\gamma^-_j - \gamma^+_j + \beta_0}
\left(
1 - \left( \frac{\alpha_s (\mu^2_0)}{\alpha_s (\mu^2)}
\right)^{\left( \gamma^-_j - \gamma^+_j + \beta_0 \right)/\beta_0}
\right)
\left(
\frac{\alpha_s (\mu^2_0)}{\alpha_s (\mu^2)}
\right)^{\gamma^+_j / \beta_0} \nonumber\\
&&\qquad\qquad
+ (\gamma^+_j \leftrightarrow \gamma^-_j,  \bit{P}^+_j \leftrightarrow \bit{P}^-_j)
\, ,
\end{eqnarray}
in terms of the projection operators
\begin{equation}
\bit{P}^\pm_j
= \frac{\pm 1}{\gamma^+_j - \gamma^-_j}
\left(
\bit{\gamma}_{(0)j} - \gamma^\mp_j \1
\right)
\, ,
\end{equation}
built from the eigenvalues of the LO anomalous dimension matrix $\bit{\gamma}_{(0)j}$,
\begin{equation}
\label{LOsingletEigenvalues}
\gamma^{\pm}_j
= \frac{1}{2}
\left(
\gamma^{qq}_{(0)j}
+
\gamma^{gg}_{(0)j}
\pm
\sqrt{\left(
\gamma^{qq}_{(0)j}
-
\gamma^{gg}_{(0)j}
\right)^2
+
4
\gamma^{gq}_{(0)j}
\gamma^{qg}_{(0)j}
}
\right)
\, .
\end{equation}
The projectors obey the properties
\begin{equation}
( \bit{P}^\pm_j )^2 = \bit{P}^\pm_j
\, , \qquad
\bit{P}^+_j \bit{P}^-_j = 0
\, , \qquad
\bit{P}^+_j + \bit{P}^-_j = \1
\, ,
\end{equation}
such that the expansion of the LO anomalous dimension matrix in its
eigenfunctions and eigenvalues reads
\begin{equation}
\bit{\gamma}_{(0)j} = \gamma_j^+ \bit{P}^+_j + \gamma_j^- \bit{P}^-_j
\, .
\end{equation}

The other evolution operator $\bit{\cal B}$, which fixes the corrections to the
eigenfunctions of the NLO kernels, can be represented to the required order by
\begin{equation}
\bit{\cal B} = \1 + \bit{\cal B}^{(1)}
\, ,
\end{equation}
where only the first iteration $\bit{\cal B}^{(1)}$ was kept. The latter is determined
by the first order differential equation
\begin{equation}
\frac{d}{d\ln \alpha_s (\mu^2)}
\bit{\cal B}^{(1)}_{jk} \left( \mu^2, \mu_0^2 \right)
= - \frac{1}{\beta_0}
\left\{
\bit{\gamma}_{(0)j} \bit{\cal B}^{(1)}_{jk} \left( \mu^2, \mu_0^2 \right)
-
\bit{\cal B}^{(1)}_{jk} \left( \mu^2, \mu_0^2 \right) \bit{\gamma}_{(0)k}
+ \frac{\alpha_s (\mu^2)}{2 \pi}
\bit{\gamma}^{{\scriptscriptstyle\rm ND}}_{(1)jk}
\right\}
\, ,
\end{equation}
and reads \cite{BelMulNieSch99}
\begin{eqnarray}
\bit{\cal B}^{(1)}_{jk}
\!\!\!&=&\!\!\!
- \frac{\alpha_s (\mu^2)}{2\pi}
\left(
\frac{
\bit{P}^+_j \bit{\gamma}^{{\scriptscriptstyle\rm ND}}_{(1)jk} \bit{P}^+_k
}{\gamma^+_j - \gamma^+_k + \beta_0}
\left(
1 - \left( \frac{\alpha_s (Q^2_0)}{\alpha_s (Q^2)}
\right)^{\left( \gamma^+_j - \gamma^+_k + \beta_0 \right)/\beta_0}
\right) \right.
\nonumber\\
&&\qquad\qquad +
\left.\frac{
\bit{P}^+_j \bit{\gamma}^{{\scriptscriptstyle\rm ND}}_{(1)jk} \bit{P}^-_k
}{\gamma^+_j - \gamma^-_k + \beta_0}
\left(
1 - \left( \frac{\alpha_s (\mu^2_0)}{\alpha_s (\mu^2)}
\right)^{\left( \gamma^+_j - \gamma^-_k + \beta_0 \right)/\beta_0}
\right) \right) \nonumber\\
&&\qquad\qquad\qquad\qquad\quad
+ (\gamma^+_j \leftrightarrow \gamma^-_j,  \bit{P}^+_j \leftrightarrow \bit{P}^-_j)
\, ,
\end{eqnarray}
where $j > k$.

\subsubsection{Asymptotic GPDs in singlet sector}
\label{SectAsyGPDSing}

An immediate consequence of these considerations is the functional form of the
singlet GPDs at asymptotically large scales. The evolution operator $\bit{\cal
E}_j$ (\ref{SingletDiagonalEvOp}), truncated to one-loop order in coupling,
possesses two eigenfunctions, which are found with the help of the projection
operators $\bit{P}^\pm_j$ acting on the two-vector of Gegenbauer moments of
GPDs (\ref{SingletGegenMomentsGPDs}),
\begin{equation}
\bit{P}^\pm_j \bit{F}_j
=
\frac{\pm 2}{\gamma^+_j - \gamma^-_j}
\left(
\begin{array}{c}
\gamma^{qq}_{(0)j} - \gamma^\mp_j
\\
\gamma^{gq}_{(0)j}
\end{array}
\right)
F^\pm_j
\, .
\end{equation}
These multiplicatively combinations of conformal moments,
\begin{equation}
F^{\pm}_j = \sum_q \mathbb{F}^q_j
+
\frac{1}{4}
\frac{\gamma^{qg}_{(0)j}}{\gamma^{qq}_{(0)j} - \gamma^{\mp}_j} \mathbb{F}^g_j
\, ,
\end{equation}
evolve with the anomalous dimensions $\gamma^{\pm}_j$, respectively. The $F^{\pm}_j$
are built as linear superpositions of the conformal moments of the quark GPDs, defined
earlier in Eq.\ (\ref{QuarkConfMoments}), and the gluon GPDs,
\begin{equation}
\label{GluonConfMoments}
\mathbb{F}^g_j (\eta; \mu_0^2)
\equiv
\eta^{j - 1}
\int_{- 1}^1 d x \,
C_{j - 1}^{5/2} \left( \frac{x}{\eta} \right) F^g (x, \eta; \mu_0^2)
=
4 (p^+)^{- j - 1} \langle p_2 | \mathbb{O}^{gg}_{jj} (0)| p_1 \rangle
\, .
\end{equation}

In particular, one finds for $j = 1$ that $F^-_1$ has a vanishing anomalous
dimension, while $F^+_1$ has a positive nonzero anomalous dimension,
\begin{equation}
\begin{array}{ll}
\gamma^-_1 = 0
\, , &\qquad
F_1^- = \sum_q \mathbb{F}_1^q + \frac{3}{2} \mathbb{F}_1^g
\, , \\
& \\
\gamma^+_1 = \frac{8}{3} C_F + \frac{2}{3} N_f
\, , &\qquad
F_1^+ = \sum_q \mathbb{F}_1^q - \frac{3 N_f}{8 C_F} \mathbb{F}_1^g
\, .
\end{array}
\end{equation}
The former result is expected since $F_1^-$ corresponds to the matrix element of
the conserved total energy-momentum tensor, see Eqs.\ (\ref{QuarkEMtensor}) and
(\ref{GluonEMtensor}). Namely,
\begin{equation}
\sum_q \mathbb{F}_1^q + \frac{3}{2} \mathbb{F}_1^g
=
2 c_1^{3/2} (p^+)^{- 2}
\langle p_2 | {\mit\Theta}^{q, ++} + {\mit\Theta}^{g, ++} | p_1 \rangle
\, .
\end{equation}
The presence of the factor $3/2$ in front of $\mathbb{F}^g_1$ is an artifact of
the normalization of the Gegenbauer polynomials. It can be seen to be gone when
transforming to Wilson operators via Eq.\ (\ref{WilsonToConformal}). Thus at
asymptotically large scales $\mu^2 \to \infty$, $F_1^+$ vanishes while $F_1^-$
remains constant. Introducing $\mathbb{F}^a_{1, \rm asy} = \mathbb{F}^a_1 (\mu^2
\to \infty)$ and, temporarily, a unit normalization for asymptotic $F^-_1$, one
can solve the two asymptotic equations
\begin{equation}
\sum_q \mathbb{F}^q_{1, \rm asy}
+
\frac{3}{2} \mathbb{F}^g_{1, \rm asy}
= 1
\, , \qquad
\sum_q \mathbb{F}^q_{1, \rm asy}
-
\frac{3 N_f}{8 C_F}
\mathbb{F}^g_{1, \rm asy}
= 0
\,
\end{equation}
in order to find separate singlet quark and gluon components
\begin{equation}
\sum_q \mathbb{F}^q_{1, \rm asy}
=
\frac{N_f}{4 C_F + N_f}
\, , \qquad
\frac{3}{2} \mathbb{F}^g_{1, \rm asy}
=
\frac{4 C_F}{4 C_F + N_f}
\end{equation}
This translates into the following asymptotic form $F^a_{\rm asy} = F^a (\mu^2
\to \infty)$ of the singlet quark \cite{Goeke:2001tz,Diehl:2003ny}
\begin{eqnarray}
\label{AsyFormSingletQuarkGPDs}
\sum_q H^q_{\rm asy} (x, \eta)
\!\!\!&=&\!\!\!
\frac{15}{4} \frac{N_f}{4 C_F + N_f} \,
\frac{x}{\eta^3} \, \left( 1 - \frac{x^2}{\eta^2} \right) \,
\theta (\eta - |x|) \,
\sum_q
\bigg\{
H^q_{2,0} + \eta^2 D^q_2
\bigg\}
\, , \\
\sum_q E^q_{\rm asy} (x, \eta)
\!\!\!&=&\!\!\!
\frac{15}{4} \frac{N_f}{4 C_F + N_f} \,
\frac{x}{\eta^3} \, \left( 1 - \frac{x^2}{\eta^2} \right) \,
\theta (\eta - |x|) \,
\sum_q
\bigg\{
E^q_{2,0} - \eta^2 D^q_2
\bigg\}
\, ,
\end{eqnarray}
and gluon GPDs
\begin{eqnarray}
\label{AsyFormGluonGPDs}
H^g_{\rm asy} (x, \eta)
\!\!\!&=&\!\!\!
\frac{15}{8} \frac{4C_F}{4 C_F + N_f} \, \frac{1}{\eta}
\left ( 1 - \frac{x^2}{\eta^2} \right)^2 \, \theta (\eta - |x|) \,
\bigg\{
H^g_{2,0} + \eta^2 D^g_2
\bigg\}
\, , \\
E^g_{\rm asy} (x, \eta)
\!\!\!&=&\!\!\!
\frac{15}{8} \frac{4C_F}{4 C_F + N_f} \, \frac{1}{\eta}
\left ( 1 - \frac{x^2}{\eta^2} \right)^2 \, \theta (\eta - |x|) \,
\bigg\{
E^g_{2,0} - \eta^2 D^g_2
\bigg\}
\, .
\end{eqnarray}
Here we restored the polynomials in form factors (\ref{FormFactorsABvsHE})
(cf.\ Eq.\ (\ref{PolynomialityNucleon})). Note that the boundary values all
$F^a (\pm \eta, \eta)$ vanish in the $\mu^2 \to \infty$ limit.

\subsubsection{Two-loop evolution in momentum space}

Though the method of orthogonal polynomial reconstruction discussed above
is rather elegant as it allows for an explicit analytic solution, its practical
implementation for GPDs faces difficulties due to poor convergence of the
truncated series\footnote{This is not the case in applications of the Jacobi
polynomial reconstruction of forward parton distributions \cite{Sha99}.} in
certain regions of the momentum fractions space, namely, around $x = \eta$ and
low $x$ and $\eta$ domain \cite{BelGeyMulSch97,BelMulNieSch98,BelMulNieSch99}.
A robust method, free from these complications, was developed by means of a
direct numerical solution \cite{FraFreGuzStr98,FreGuz98} of the integro-differential
evolution equation for GPDs (\ref{MOMEE}). Since the formalism allows straightforward
generalizations to high-loop orders, the two-loop generalized evolution kernels were
reconstructed in Ref.\ \cite{BelMulFre99} bypassing direct evaluations of Feynman
graphs, from the available two-loop anomalous dimensions of conformal operators
presented in Section \ref{NDanomalousDimensionsNLO}, inclusive splitting functions,
known two-loop non-singlet evolution kernels in exclusive kinematics
\cite{Sar84,DitRad84,MikRad85} and relying on supersymmetric relations, discussed
in Appendix \ref{SUSYrelationsADs}, and well-understood limiting procedure outlined
in Section \ref{app-LIM}.  A thorough numerical analysis demonstrated a high
efficiency of the formalism and revealed that the two-loop effects are strongly
enhanced around $x = \eta$, while they do not exceed 25\% within a very large
range in other kinematical regions \cite{FreMacD01,FreMcD01}.

\subsubsection{Solution of evolution equations for double distributions}
\label{subsoperatorsol}

The multiplicative renormalization of conformal operators can be used
for an efficient solution of evolution equations for double distributions.
From the autonomous leading order evolution of conformal operators one
can establish the eigenfunctions of the renormalization group equations
in the basis of double distributions and draw interesting consequences
for the functional form of the latter. For simplicity, we consider the
the nucleon GPD $A (x, \eta)$ and corresponding symmetric DDs $h_A
(\beta, \alpha)$ introduced in Sections \ref{PolynomSpin12GPDs} and
\ref{SectionSingleCompDDnucleon}, respectively.

Let us discuss the nonsinglet case first. Switching from the quark GPD $A^q$
in their multiplicatively renormalizable conformal moments to the DDs $h^q_A$,
we can write making use of Eq.\ (\ref{RelAtohA})
\begin{equation}
\label{101}
\mathbb{A}^q_n (\eta)
=
\eta^n
\int_{-1}^1 d x \, C_n^{3/2} \left( \frac{x}{\eta} \right) \, A^q (x, \eta )
=
\eta^n
\int_{\mit\Omega} d \beta \, d \alpha \,
C_n^{3/2} \left( \frac{\beta}{\eta} + \alpha \right)
h^q_A (\beta, \alpha)
\, .
\end{equation}
Using the identity
\begin{equation}
C_n^{3/2} \left( \frac{\beta}{\eta} + \alpha \right)
=
\sum_{l=0}^{n} \frac{{\mit\Gamma} (n - l + 3/2)}{{\mit\Gamma} (3/2) (n - l)!}
\, \left( \frac{2 \beta}{\eta} \right)^{n - l} \,
C_l^{3/2 + n - l} (\alpha)
\,  ,
\end{equation}
one finds
\begin{eqnarray}
\mathbb{A}^q_{n} (\eta)
=
\sum_{k = 0}^{[n/2]}
2^{n - 2k}
\frac{{\mit\Gamma} (n - 2k + 3/2)}{{\mit\Gamma} (3/2) (n - 2k)!}
\, \eta^{2k}
\, \int_{\mit\Omega} d \beta \, d \alpha \,
\beta^{n - 2k} C_{2k}^{3/2 + n - 2k} (\alpha) \, h^q_A (\beta, \alpha)
\, .
\label{gedd}
\end{eqnarray}
Hence, each $\beta^m C_l^{3/2 + m}(\alpha)$ moment of the DD $h^q_A (\beta, \alpha)$
is multiplicatively renormalizable and its evolution is governed by the anomalous
dimension $\gamma^{qq; V}_{(0)m + l}$ \cite{Rad1997a,Radyushkin:1998es},
\begin{equation}
\int_{\mit\Omega} d \beta \, d \alpha \,
\beta^m C_l^{3/2 + m} (\alpha) \, h^q_A (\beta, \alpha; \mu^2)
=
\left(
\frac{\alpha_s (\mu_0^2)}{\alpha_s (\mu^2)}
\right)^{\gamma_{(0)m + l}^{qq; V}/\beta_0}
\int_{\mit\Omega} d \beta \, d \alpha \,
\beta^m C_l^{3/2 + m} (\alpha) \, h^q_A (\beta, \alpha; \mu_0^2)
\, .
\end{equation}
In Eq.\ (\ref{gedd}), we took into account that the DDs $h^q_A (\beta, \alpha)$
are always even in $\alpha$, which gives an expansion of the Gegenbauer moments in
powers of $\eta^2$. In the nonsinglet case, the Gegenbauer moments $\mathbb{A}^q_n
(\eta)$ are nonzero for even $n$ only.

Another simple case is the evolution of the gluon distributions in pure
gluodynamics. Then the multiplicatively renormalizable Gegenbauer moments
of the GPD $A^g (x, \eta)$
\begin{equation}
\mathbb{A}_n^g (\eta)
=
\eta^{n - 1}
\int_{-1}^1 d x \, C_{n - 1}^{5/2} \left( \frac{x}{\eta} \right) \,
A^g (x, \eta)
\, ,
\label{1010}
\end{equation}
which are vanishing only for odd $n$, can be rewritten in terms of the DD $h^g_A$:
\begin{eqnarray}
\mathbb{A}_n^g (\eta)
=
\sum_{k=0}^{[(n - 1)/2]}
2^{n - 2k - 1}
\frac{{\mit\Gamma} (n - 2k + 3/2)}{{\mit\Gamma} (5/2) (n - 2k - 1)!}
\, \eta^{2k}
\int_{\mit\Omega} d \beta  \, d \alpha \,
\beta^{n - 2k} C_{2k}^{3/2 + n - 2k} (\alpha) \, h^g_A (\beta, \alpha)
\, .
\label{geddglu}
\end{eqnarray}
Note, that the two shifts, $n \to n - 1$, and $3/2 \to 5/2$ have compensated each
other. Again, every combined $\beta^m C_l^{3/2 + m}(\alpha)$ moment of $h^g_A
(\beta, \alpha)$ is multiplicatively renormalizable and its evolution is governed
by the anomalous dimension $\gamma_{(0)m+l}^{gg; V}$ \cite{Rad1997a,Radyushkin:1998es}.

Since the Gegenbauer polynomials $C_l^{3/2 + m}(\alpha)$ are orthogonal with the
weight $(1- \alpha^2)^{m + 1}$, the evolution of the $\beta^m$-moments of DDs in
both cases is given by the formula \cite{Radyushkin:1998es}
\begin{eqnarray}
h^q_{A, m} (\alpha; \mu^2)
\!\!\!&\equiv&\!\!\!
\int_{-1}^1 d \beta \, \beta^m  \, h^q_A (\beta, \alpha; \mu^2)
\nonumber \\
\!\!\!&=&\!\!\!
(1- \alpha^2)^{m+1} \sum_{k=0}^{\infty}
h^q_{A, ml} (\mu_0^2) \, C^{m + 3/2}_l (\alpha)
\left(
\frac{\alpha_s (\mu_0^2)}{\alpha_s (\mu^2)}
\right)^{\gamma_{(0)m + l}^{qq; V}/\beta_0}
\, ,
\label{eq:fnnons2}
\end{eqnarray}
where the expansion coefficients $h^q_{A, ml}$ are proportional to $\beta^m C_l^{3/2 + m}
(\alpha)$ moments of DDs $h^q_A$. The anomalous dimensions $\gamma_{(0)j}^{qq}$ increase
with $j = m + l$, and, hence, the $m$-th $\beta$-moment of $h_A^q (\beta, \alpha; \mu^2)$
is asymptotically, i.e., $\mu^2 \to \infty$, dominated by the $\alpha$-profile $(1 -
\alpha^2)^{m + 1}$. Such a correlation between $\beta$- and $\alpha$-dependence of $h_A^q
(\beta, \alpha;\mu )$ is not something exotic. Take a DD which is constant in its support
region. Then its $\beta^m$-moment behaves like $(1 - |\alpha|)^{m+1}$, i.e., the  width of
the $\alpha$ profile decreases with increasing $j$. This result is easy to understand: due
to the spectral condition $ |\alpha| \leq 1 -|\beta|$, the $\beta^m$ moments with larger
$m$ are dominated by regions which are narrower in the $\alpha$-direction.

These observations suggests to try a model in which the moments $h_{A, m} (\alpha; \mu^2)$
have the asymptotic $(1 - \alpha^2)^{m + 1}$  profile even at non-asymptotic $\mu^2$. This
is equivalent to assuming that all the combined moments $\beta^m C_l^{3/2 + m}(\alpha)$
with $l > 0$ vanish. Note that this assumption is stable with respect to evolution. Since
integrating $h^q_{A,m} (\alpha; \mu^2) $ over $\alpha$ one should get the moments $f_m^q
(\mu^2)$ of the forward density $f^q (\beta; \mu^2)$, the DD moments  $h^q_{A, m} (\alpha;
\mu^2)$ in this model are given by
\begin{equation}
h^q_{A, m} (\alpha; \mu^2) = \rho_{m + 1} (\alpha) f^q_m (\mu^2)
\, ,
\end{equation}
where $\rho_{m + 1} (\alpha)$ is the normalized profile function
\begin{equation}
\label{rhon}
\rho_m (\alpha)
\equiv
\frac{{\mit\Gamma} (m + 3/2)}{{\mit\Gamma} (1/2) {\mit\Gamma} (m + 1)} (1 - \alpha^2)^m
\, .
\end{equation}
In the explicit form:
\begin{equation}
\label{10810}
\int_{-1}^{1} d \beta \, \beta^m h^q_A (\beta,\alpha; \mu^2)
=
\frac{{\mit\Gamma} (m + 5/2)}{{\mit\Gamma} (1/2) \, (m + 1)!}
(1 - \alpha^2)^{m + 1} \int_{-1}^{1} dx \, x^m \, f^q (x; \mu^2)
\, .
\end{equation}
In this relation, all the dependence on $\alpha$ can be trivially shifted to the
left-hand side of this equation, and we immediately see that $h^q_A (\beta,\alpha;
\mu^2)$ in this model is a function of $\beta/(1 - \alpha^2)$,
\begin{equation}
h^q_A (\beta, \alpha; \mu^2)
=
F^q \left( \frac{\beta}{1 - \alpha^2}; \mu^2 \right)
\,
\theta \left( 0 < \frac{\beta}{1 - \alpha^2} < 1 \right)
\, .
\end{equation}
A direct relation between $f^q (x; \mu^2)$ and $F^q (u; \mu^2)$ can be easily obtained
using the basic fact that integrating $h^q_A (\beta, \alpha; \mu^2)$ over $\alpha$
one should get the forward density $f^q (x, \mu^2)$; e.g., for positive $x$ we have
\begin{equation}
f^q (x) = x \int_x^1 d u \frac{F^q (u)}{u^{3/2} \sqrt{u - x}}
\, .
\end{equation}
This relation has the structure of the Abel equation. Solving it for $F(u)$ we get
\begin{equation}
F^q (u) = - \frac{u^{3/2}}{\pi} \int_u^1 d x
\frac{\left[ f^q (x)/x \right]^{\prime}}{\sqrt{x - u}}
\, .
\label{inabel}
\end{equation}
Thus, in this model, knowing the forward density $f^q (x)$ one can calculate the double
distribution function $h^q_A (\beta, \alpha) = F^q (\beta/(1 - \alpha^2))$.

Note, however, that the model derived above violates the DD support condition
$|\beta| + |\alpha| \leq 1$: the restriction $|\beta| \leq 1 - \alpha^2$ defines a
larger area. Hence, the model is only applicable in a situation when the difference
between two spectral conditions can be neglected. A practically important case is
the shape of $A^q (x, \eta)$ for small skewness $\eta$. Indeed, calculating $A^q
(x, \eta)$ for small $\eta$ one integrates the relevant DDs $h^q_A (\beta, \alpha)$
over practically vertical lines. If $x$ is also small, both the correct $|\alpha|
\leq 1 - |x|$ and model $\alpha^2 \leq 1- |x|$ conditions can be substituted by
$|\alpha| \leq 1$. Now, if $x \gg \eta$, a slight deviation of the integration line
from  the vertical direction can be neglected and $A^q (x, \eta)$ can be approximated
by the forward limit $f^q (x)$.

Specifying the ansatz for $f^q (x)$, one can get an explicit expression for the model
DD by calculating $F^q (u)$ from Eq.\ (\ref{inabel}). However, in the simplest case
when $f^q (x) = A x^{-a}$ for small $x$, the result is evident without any calculation:
the DD $h^q_A (\beta, \alpha)$ which is a function of the ratio $\beta/(1-\alpha^2)$
and reduces to $A \beta^{-a}$ after an integration over $\alpha$ must be given by
$h_A^q (\beta,\alpha) = \rho_{a} (\alpha) f^q (\beta)$ where $\rho_{a} (\alpha)$ is
the normalized profile function of Eq.\ (\ref{rhon}), i.e.,
\begin{equation}
h^q_A (\beta, \alpha)
= A \, \frac{{\mit\Gamma} (a + 5/2)}{{\mit\Gamma} (1/2) \, {\mit\Gamma} (a + 2)}
(1 - \alpha^2)^{a}  \, \beta^{-a}
\, .
\end{equation}
This DD is a particular case of the general factorized ansatz $h_A^q (\beta,\alpha) =
\rho_m (\alpha) f^q (\beta)$ considered in Section \ref{LongDynamics}. The most
nontrivial feature of this model is the correlation $m = a$ between the profile
function parameter $n$ and the power $a$ characterizing the small-$\beta$ behavior
of the forward parton distribution.

Knowing the DDs $h^q_A (\beta, \alpha)$, the relevant GPDs $A^q (x, \eta)$ can be
obtained in a standard way. In particular, the GPD enhancement factor ${\cal R} (\eta)$
for small $\eta$ in this model is given by
\begin{equation}
{\cal R}^q \equiv \frac{ A^q (\eta, \eta) }{f^q (\eta)}
= \frac{{\mit\Gamma} (2a + 2)}{{\mit\Gamma} (a + 2) {\mit\Gamma} (a + 1)}
\, .
\end{equation}
The use of the asymptotic profiles for DD moments $h^q_{A, n} (\alpha)$ is the basic
assumption of the model described above. However, if one is interested in GPDs for
small $\eta$, the impact of deviations of $h^q_{A, n} (\alpha)$ from the asymptotic
profile is suppressed. Even if the higher harmonics are present in $h^q_{A, n} (\alpha)$,
i.e., if the $\beta^{n - 2k} C_{2k}^{3/2 + n - 2k} (\alpha)$ moments of $h^q_A (\beta,
\alpha)$ are nonzero for $k \geq 1$ values, their contribution into the Gegenbauer
moments $\mathbb{A}^q_n (\eta; \mu^2)$ is strongly suppressed by $\eta^{2k}$ factors
(see Eq.\ (\ref{gedd})). Hence, for small $\eta$, the shape of $A^q (\eta, \eta)$ for
a wide variety of model $\alpha$-profiles is very close to that based on the asymptotic
profile model.

Absence of higher harmonics in $h^q_{A, n} (\alpha)$ is equivalent to absence of
the $\eta$-dependence in the Gegenbauer moments $\mathbb{A}^q_n (\eta; \mu^2)$. The
assumption that the moments $\mathbb{A}^q_n (\eta; \mu^2)$ do not depend on $\eta$
is the starting point for the model of GPDs $A^q (x, \eta )$ constructed in Ref.\
\cite{ShuGolMarRys99}. Though the formalism of DDs is not used there, both
approaches lead to identical results: the final result of Ref.\ \cite{ShuGolMarRys99}
has the form of a DD representation for $A^q (x, \eta)$. The approach of Ref.\
\cite{ShuGolMarRys99}  is based on the concept of effective forward distributions,
which we discuss in the next subsection.

\subsubsection{Effective forward distributions}

Since the Gegenbauer moments $\mathbb{A}^a_n (\eta; \mu^2)$ evolve like $x^n$
Mellin moments of the usual forward parton densities $f^q (x; \mu^2)$, it was
proposed \cite{Shuvaev:1999fm} to introduce effective forward distributions
(EFDs) $q_{\eta} (x; \mu^2)$ and $g_{\eta} (x; \mu^2)$ as functions whose
$x^n$ Mellin moments are proportional to the Gegenbauer moments of GPDs, and
which coincide with the usual parton densities in the $\eta = 0$ limit. Since
\begin{eqnarray}
\label{qumom}
\mathbb{A}_n^q (\eta = 0; \mu^2)
\!\!\!&=&\!\!\!
\frac{2^{n} {\mit\Gamma} (n + 3/2)}{{\mit\Gamma} (3/2)\, n!}
\int_{-1}^1 d x \, x^n \, f^q (x; \mu^2)
\, , \\
\label{glumom}
\mathbb{A}_n^g (\eta = 0, \mu^2)
\!\!\!&=&\!\!\!
\frac{2^n {\mit\Gamma} (n + 3/2)}{3 \, {\mit\Gamma} (3/2) \, (n - 1)!}
\int_{-1}^1 d x \, x^n \, f^g (x; \mu^2)
\, ,
\end{eqnarray}
EFDs are defined \cite{ShuGolMarRys99} by
\begin{eqnarray}
\label{defq}
\int_{-1}^1 dz \, z^n \, q_{\eta}(z; \mu^2)
\!\!\!&=&\!\!\!
\frac{{\mit\Gamma} (3/2)\, n!}{2^n {\mit\Gamma} (n + 3/2)} \,
\mathbb{A}_n^q (\eta; \mu^2)
\, , \\
\label{defg}
\int_{-1}^1 dz \, z^n \, g_{\eta}(z;\mu)
\!\!\!&=&\!\!\!
\frac{3 \, {\mit\Gamma} (3/2)\, (n - 1)!}{2^n {\mit\Gamma} (n + 3/2)} \,
\mathbb{A}_n^g (\eta, \mu^2)
\, .
\end{eqnarray}
To proceed further, one can use the expansion (\ref{NonlocalExpansionConformal})
of bilocal light-cone operators ${\cal O}^{aa} (- z^-, z^-)$ over the
multiplicatively renormalizable local operators $\mathbb{O}^{aa}_n$. For quarks,
it reads
\begin{equation}
\label{108}
{\cal O}^{qq} (0, z^-)
=
\sum \limits_{n = 0}^{\infty}
\frac{2 n + 3}{2^{2n + 2}(n + 1)!} (- i z^-)^n
\int_{-1}^1 \, d\alpha \,
(1 - \alpha^2)^{n + 1} \,
\mathbb{O}_n^{qq} \left( \ft{1 - \alpha}{2} z^- \right)
\, .
\end{equation}
Inserting it into the nonforward matrix element gives
\begin{equation}
\label{1080}
A^q (x, \eta)
=
\sum \limits_{n=0}^{\infty}
\frac{2 n + 3}{2^{n + 2}(n + 1)!} \int_{-1}^1 \, d\alpha \,
(1 - \alpha^2)^{n + 1}
\delta^{(n)} (x- \eta\alpha ) \, \mathbb{A}_n^q (\eta)
\, ,
\end{equation}
where $\delta^{(n)} (x - \eta \alpha) = \partial^n_x \delta (x - \eta \alpha)$.
This formula has a typical structure of a formal inversion of Mellin moments,
\begin{equation}
\label{delty}
f (x)
=
\sum_{n=0}^{\infty}\frac{(-1)^n}{n!} \, \delta^{(n)} (x)
\int_{- 1}^1 d x' \, (x')^n f (x')
\, .
\end{equation}
In Ref.\ \cite{Shuvaev:1999fm}, it was proposed to use a  form more convenient
for summation over $n$
\begin{equation}
f(x) =
- \frac{1}{\pi} \, {\rm disc}_x \sum_{n = 0}^{\infty}\frac{1}{x^{n+1}}
\int_{- 1}^1 d x' \, (x')^n f (x')
\, ,
\end{equation}
based on treating $\delta^{(n)} (x)$ as discontinuities
\begin{equation}
{\rm disc}_x \ \Phi (x)
\equiv
\frac{1}{2i} \,
\biggl[
\Phi (x + i \varepsilon ) - \Phi (x - i \varepsilon)
\biggr] \, .
\end{equation}
of $1/x^{n + 1}$. Using Eq.\ (\ref{defq}), it gives
\begin{equation}
\label{hqsum}
A^q (x, \eta) =  - \, \frac{1}{\pi} \ {\rm disc}_x \int_{-1}^1
d \alpha \,\sum \limits_{n = 0}^{\infty}
\frac{{\mit\Gamma} (n + 5/2)}{{\mit\Gamma} (3/2) (n + 1)!}
\left( \frac{1 - \alpha^2}{x - \eta \alpha } \right)^{n + 1}
\int_{-1}^1 dz \, z^n \, q_{\eta} (z)
\, .
\end{equation}
Now one can perform summation over $n$ and take discontinuity. The formal
result, however, would contain  a rather singular integrand $[z (1 - \alpha^2)
/(x - \eta \alpha) - 1 ]^{-3/2}$ producing an end-point  divergence. To get a
correct result, one needs to soften the singularity by using the relation
\begin{equation}
\int_{0}^{1} d z \, z^n \, q_{\eta} (z)
=
- \frac{1}{n + 2} \int_{0}^{1} dz \, z^{n+2} \, \frac{d}{dz}
\left ( \frac{q_{\eta} (z)}{z} \right )
\, ,
\label{trick}
\end{equation}
obtained through integration by parts assuming that the surface terms vanish.
This gives
\begin{eqnarray}
A^q (x, \eta)
= - \frac1{\pi} \int_{-1}^1  d \alpha \int_{-1}^1 d z \,
\frac{x - \eta \, \alpha}{1 - \alpha^2}
\left[z (1 - \alpha^2)/(x - \eta \, \alpha) - 1 \right]^{-1/2}
\, \frac{d}{dz} \left( \frac{q_{\eta} (z)}{z} \right)
\, .  \label{Hq}
\end{eqnarray}
If one takes a model in which the effective forward distribution does not depend
on $\eta$, $q_\eta (z) = q (z) $, then this expression corresponds to a double
distribution representation for GPD, with the DD $h^q_A$ given by
\begin{eqnarray}
\label{fq}
h^q_A (\beta, \alpha) = - \frac{\beta}{\pi(1 - \alpha^2)}
\int_{-1}^1 dz\
\frac{ \theta(z / \beta\geq 1/(1-\alpha^2))}{\sqrt {z(1-\alpha^2)/\beta -1 } }
\, \frac{d}{dz}
\left ( \frac{ q \,(z)}{z} \right )
\, .
\end{eqnarray}
Note, that the model with $\eta$-independent ansatz for EFDs corresponds to DDs
which have a nontrivial $\alpha$-profile. As we already discussed, the $\beta^n$
moment of such a DD has the asymptotic shape $(1-\alpha^2)^{n+1}$ for all scales
$\mu^2$. The resulting model GPD has a nontrivial $\eta$-dependence, the fact also
evident from Eq.\ (\ref{Hq}). In other words, this model, though based solely on
input from forward distributions, is not a model in which GPD $H(x,\eta)$ coincides
with its forward limit $q(x)$ for all $\eta$. Such a purely forward model would
require $h^q_A (\beta, \alpha) = \delta (\alpha) \, f^q (\beta)$.

Equation (\ref{fq}) coincides with Eq.\ (\ref{inabel}), and, as pointed out at
the end of the previous subsection, the model for $H^q(x,\eta;\mu)$ based on
$\eta$-independent EFD was originally developed in Ref.\ \cite{ShuGolMarRys99}. The
spectral condition $\beta /z \leq 1 - \alpha^2$ relating the ``original'' fraction
$z$ of EFD and the ``produced''  fraction $\beta$ of DD resembles  the momentum
ordering $x/z \leq 1$ in the DGLAP equation: the produced fraction\footnote{For
positive fractions.} cannot be larger than the original one. In the present case,
if the parton also takes some nonzero fraction $\alpha$ of the momentum transfer,
the allowed values of $\beta$ cannot exceed $z (1 - \alpha^2)$. However, as we
already remarked, the model based on $\eta$-independent Gegenbauer moments gives
DD with incorrect support $|\beta| \leq 1 - \alpha^2$. This means that EFDs
{\sl must} have a nontrivial $\eta$-dependence needed to secure the correct
support $|\beta| \leq 1 - |\alpha|$.

For gluons, one can combine the expansion for the bilocal operator
\begin{equation}
\label{1080g}
\mathcal{O}^{gg} (0, z^-)
=
\sum \limits_{n = 0}^{\infty}
\frac{3 (2 n + 5)}{2^{2 n + 3}(n + 2)!}
\, (-i z^-)^n
\int_{-1}^1
d \alpha \, (1 - \alpha^2)^{n + 2} \,
\mathbb{O}_{n + 1}^{gg} \left( \ft{1 - \alpha}{2} z^- \right)
\end{equation}
and the EFD definition (\ref{defg}). The resulting expression for $A^g (x, \eta)$ has
the form close to Eq.\ (\ref{1080}) derived in the quark case:
\begin{eqnarray}
\label{Hg}
A^g (x,\eta)
=
- \frac{1}{\pi} \int_{-1}^1 d \alpha \int_{-1}^1 d z \,
\frac{(x - \eta \, \alpha)^2}{1 - \alpha^2}
\left[ z (1 - \alpha^2)/(x - \eta \, \alpha) - 1 \right]^{-1/2}
\,
\frac{d}{dz}
\left( \frac{g_{\eta} (z)}{z} \right)
\, .
\end{eqnarray}
The extra factor $(x - \eta \, \alpha)$ reflects the convention that $A^g (x, \eta)$
reduces to $x f^g (x)$ in the forward limit.

To get the inverse transformation, i.e., to express EFD $q_{\eta} (z; \mu^2)$ in terms
of GPD  $A^q (x, \eta; \mu^2)$, one can start again with EFD definition (\ref{defq}).
Using expression for the Gegenbauer moments and the beta-function representation for
the proportionality coefficients, one obtains\footnote{In this construction, we follow
the line of reasoning presented in Ref.\ \cite{Noritzsch:2000pr}, modifying without
notice some  formulas and statements of that paper where we deem it necessary.}
\begin{equation}
q_\eta (z) = - \ \frac{1}{\pi} \ {\rm disc}_z
\int_0^1 d s \, \frac{1}{2 z \sqrt{1 - s}} \,
\sum_{n = 0}^{\infty} \left( \frac{s \eta}{2 z} \right)^n
\,
\int_{-1}^1 d x \, C_n^{3/2} \left( \frac{x}{\eta} \right) \,
A^q (x, \eta)
\, ,
\end{equation}
and
\begin{equation}
g_\eta (z) = - \ \frac{1}{\pi} \ {\rm disc}_z
\int_0^1 d s \, \frac{3 \sqrt{1 - s}}{2 z^2} \,
\sum_{n = 0}^{\infty} \left( \frac{s \eta}{2 z} \right)^n
\,
\int_{-1}^1 d x \, C_n^{5/2} \left( \frac{x}{\eta} \right) \,
A^g (x, \eta)
\, .
\end{equation}
Incorporating the generating function representation for the Gegenbauer polynomials
(\ref{GenerFunctGegenPolynom}), these equations yield
\begin{equation}
q_\eta (z)
=
- \frac{1}{\pi} \ {\rm disc}_z \,
\int_0^1 \frac{d s}{2 z \sqrt{1 - s}} \, \int_{-1}^1 d x \,
\frac{A^q (x, \eta)}{\left[ 1 - s x/z + s^2 \eta^2/(4 z^2) \right]^{3/2} }
\end{equation}
and
\begin{equation}
g_\eta (z)
=
- \frac1{\pi} \ {\rm disc}_z \,
\int_0^1 d s \, \frac{3 \sqrt{1 - s}}{2 z^2} \, \int_{-1}^1 d x \,
\frac{A^g (x, \eta)}{\left[ 1 - s x/z + s^2 \eta^2/(4 z^2) \right]^{5/2} }
\, .
\end{equation}
There are evident discontinuities in the integrands of these representations in
the region
\begin{equation}
1 - \frac{s \, x}{z} + \frac{s^2 \eta^2}{4 z^2} \leq 0
\, ,
\label{ineq}
\end{equation}
where the arguments of the $z$-dependent square roots are negative. In this region
\begin{equation}
|x| \geq \frac{|z|}{s} + \frac{s \, \eta^2}{4 |z|}
\ .
\end{equation}
The function on the right hand side has minimum at $s = 2 |z|/\eta$ and is equal
there $\eta$. If $|z| \leq \eta/2$, the value of $s$ at the minimum is within the
range allowed for $s$, and integration over $x$ goes over the $\eta \leq |x| \leq
1$ limits. Note, that the restriction $\eta \leq |x|$ means that taking these
discontinuities we would include only the $\eta \leq |x|$ part of the Gegenbauer
moments $\mathbb{A}^q_n (\eta)$. The $|x| \leq \eta$ part of these moments, call
them $\mathbb{A}^{q, <}_n (\eta)$, can be added using Eq.\ (\ref{delty}), i.e.,
in the form of a sum of terms proportional to $\mathbb{A}^{q, <}_n (\eta) \,
\delta^{(n)}(z)$. Whether this part can also be written in a more closed form, is
an open question. An alternative way to write EFDs in analytic form is based on
using DDs and is discussed at the end of this section.

Integrating by parts over $x$ to reduce the singularity of the $3/2$ and $5/2$ roots
to the integrable power $1/2$, and then taking the discontinuity of the inverse square
root, one obtains
\begin{equation}
q_\eta (z)
= - \frac{1}{\pi}
\int_0^1 \frac{d s}{s \sqrt{1 - s}} \, \int_{-1}^1 d x \,
\frac{\theta ({\rm root})}{\sqrt{-1 + s x/z - s^2 \eta^2/(4 z^2)}}
\frac{\partial}{\partial x} \, A^q (x,\eta)
+
\delta q_{\eta} (z)
\, ,
\end{equation}
where
\begin{eqnarray}
\delta q_{\eta}(z)
\equiv \sum_{n = 0}^{\infty} \,
\frac{(- 1)^n {\mit\Gamma} (3/2)}{2^n {\mit\Gamma} (n + 3/2)}
\, \delta^{(n)} (z) \,
\eta^n \int_{- \eta}^{\eta} d x \,
C_n^{3/2} \left( \frac{x}{\eta} \right) \, A^q (x, \eta) \,
\, ,
\end{eqnarray}
and
\begin{eqnarray}
g_{\eta}(z)
\!\!\!&=&\!\!\!
\frac{2}{\pi}
\int_0^1 \frac{\sqrt{1 - s}}{s^2} \, d s \,  \int_{-1}^1 d x \,
\frac{\theta ({\rm root})}{\sqrt{-1 + s x/z - s^2 \eta^2/(4 z^2)}}
\frac{\partial^2 }{\partial x^2} \, A^g (x,\eta)
\\
&&\qquad\qquad\qquad\qquad\qquad\qquad\qquad\qquad
+
\sum_{n = 0}^{\infty}
\frac{3 (-1)^n {\mit\Gamma} (3/2)}{2^n \, n \, {\mit\Gamma} (n + 3/2)}
\,
\mathbb{A}^{g, <}_{n + 1} (\eta) \, \delta^{(n)} (z)
\, . \nonumber
\end{eqnarray}
We intentionally wrote the $|x| \leq |\eta|$ parts in two forms: in explicit form
for quarks (it will be used below) and in a shorter form for gluons. The notation
$\theta ({\rm root})$ implies that integration over $s,x$ is in the region specified
by Eq.\ (\ref{ineq}), where the argument of the square root is positive. If $|z|
\geq |\eta|/2$, the minimum value for $|x|$ in this region is achieved at the
boundary $s = 1$, and is given by a $z$-dependent expression $|x|_{\rm min} = |z|
+ \eta^2/4|z|$. The support region for the first term of EFDs is
\begin{equation}
0 \leq |z| \leq \frac{1}{2} \, \left( 1 + \sqrt{1 - \eta^2} \right)
\, , \qquad
0 \leq |\eta|\leq 1
\, .
\label{qsupp}
\end{equation}
It is smaller than the square $0 \leq |z|, \eta \leq 1$. Concerning the term coming
from the $|x| \leq \eta$ integration, one may be worried that, in principle, an
infinite sum of $\delta^{(n)} (z)$ terms can produce a function with any support.
However, using the relation
\begin{eqnarray}
\int_{-1}^1 d z \left( \frac{z}{\eta} \right)^n \delta q_\eta (z)
=
\frac{{\mit\Gamma} (3/2) \, n!}{2^n {\mit\Gamma} (n + 3/2)}
\,
\int_{- \eta}^{\eta} d x \, C_n^{3/2} \left( \frac{x}{\eta} \right) \, A^q (x, \eta)
\, ,
\end{eqnarray}
and taking into account that $C_n^{3/2} (x/\eta)$ are polynomials in $x/\eta$, one
can see that the support region of $\delta q_{\eta}(z)$ is $|z| \leq \eta$. The same
is true for the  $|x| \leq \eta$ gluonic term.

All these complications, related to different nature of GPDs $A (x, \eta)$ in $|x|
\geq \eta$ and $|x| \leq \eta$ regions, can be avoided  if one writes EFDs in terms
of DDs. The main point is that ``DDs do not know  about $\eta$''. In quark case,
using (\ref{gedd}), one has
\begin{equation}
\label{gedd2}
\int_{-1}^1 d z \, z^n \, q_\eta (z)
=
\sum_{k = 0}^{[n/2]} \left( \frac{\eta}{2} \right)^{2k}
\,
\frac{{\mit\Gamma} (n - 2k + 3/2) n!}{{\mit\Gamma} (n + 3/2) (n - 2k)!}
\int_{\mit\Omega} d \beta d \alpha
\, \beta^{n - 2k}\, C_{2k}^{3/2 + n - 2k} (\alpha) \, h^q_A (\beta, \alpha)
\, .
\end{equation}
The first terms of the $\eta^2$ expansion can be easily written:
\begin{eqnarray}
\label{highereta}
\int_{-1}^1 dz \, z^n \, q_{\eta}(z)
\!\!\!&=&\!\!\!
\int_{\mit\Omega} d \beta d \alpha \,
\biggl\{ \beta^n  h^q_A (\beta, \alpha)
\\
&+&\!\!\!
\frac{\eta^2}{2} \ \theta (n - 2) \, n (n - 1) \,
\biggl[ \alpha^2 - \frac{1}{2n + 1} \biggr] \, \beta^{n - 2}
h^q_A (\beta, \alpha) \biggr\} + \mathcal{O} (\eta^4)
\, . \nonumber
\end{eqnarray}
Inverting the Mellin transform, we obtain for positive\footnote{To reconstruct the
function $q_{\eta}(z)$ for negative $z$, one should use its $z$-symmetry.} $z$
\begin{eqnarray}
q_{\eta}(z)|_{z \geq 0}
=
f^q (z) + \frac{\eta^2}{2}
\, \left\{
\frac{d^2}{dz^2}
\left[
h^q_{A, 2} (z) - \frac{z^{3/2}}{2}
\int_z^1 d \beta
\, \frac{f^q (\beta)}{\beta^{5/2}} \,
\right]
\right \}_{+\, \{ 0,1 \}}
+
\mathcal{O} (\eta^4)
\, ,
\end{eqnarray}
where $f^q(z)$ is the forward distribution, i.e., the zeroth $\alpha$-moment of the
DD $h^q_A (\beta, \alpha)$, and $h^q_{A, 2} (\beta)$ is the second $\alpha$-moment
of the DD:
\begin{equation}
h^q_{A, 2} (\beta)|_{\beta \geq 0}
=
\int_{- 1 + \beta}^{1 - \beta} d \alpha \, \alpha^2 \, h^q_A (\beta, \alpha)
\, .
\end{equation}
The $\{ \ldots \}_{+\, \{ 0,1 \}}$ prescription generalizes the standard ``plus''
distribution: it  means that the function in the brackets should be accompanied by
a subtraction procedure which guarantees that its  $z^0$ and $z^1$ moments vanish
(even if taken in the $z \geq 0$ region only), to comply with the structure of the
generating  equation (\ref{highereta}). For gluons, using Eq.\ (\ref{geddglu}), one
obtains
\begin{eqnarray}
\label{higheretaglu}
\int_{-1}^1 dz \, z^n \, g_{\eta}(z)
\!\!\!&=&\!\!\!
\int_{\mit\Omega} d \beta d \alpha \,
\biggl\{
\beta^n h^g_A (\beta, \alpha)
\\
&+&\!\!\!
\frac{\eta^2}{2} \ \theta (n - 3) \, (n - 1)(n - 2) \,
\biggl[
\alpha^2 - \frac{1}{2n + 1} \biggr] \, \beta^{n - 2}
h^g_A (\beta, \alpha) \biggr\}
+
\mathcal{O} (\eta^4)
\, , \nonumber
\end{eqnarray}
and
\begin{equation}
g_{\eta}(z)|_{z \geq 0}
=
f^g (z)
+
\frac{\eta^2}{2} \,
\left\{
z \, \frac{d^2}{dz^2}
\left[
\frac{h^g_{A, 2} (z)}{z}
-
\frac{z^{1/2}}{2}
\int_z^1 d \beta \,
\frac{f^g (\beta)}{\beta^{5/2}} \,
\right]
\right\}_{+ \, \{ 0,1,2 \}}
+
\mathcal{O} (\eta^4)
\, ,
\end{equation}
with $\{ \ldots \}_{+\, \{ 0,1,2 \}}$ now meaning that $z^0,z^1$ and $z^2$ moments of
the function in the brackets should vanish. It is straightforward to include higher
terms of the $\eta^2$ expansion.

At present, the only practical way to construct GPDs consistent with the polynomiality
condition is to use factorized models ``forward density $(\beta)$ $ \otimes$ profile
function$(\beta,\alpha)$'' for DDs. In these  models, EFDs  $q_{\eta}(z)$,  $g_{\eta}(z)$
can be expressed in terms of forward  distributions through an expansion in $\eta^2$,
and one can study evolution of GPDs by existing codes for  numerical evolution of parton
densities.

\section{Compton scattering beyond leading order and power}
\label{Compton}

\begin{figure}[t]
\begin{center}
\hspace{0cm}
\mbox{
\begin{picture}(0,70)(170,0)
\put(-40,0){\insertfig{15}{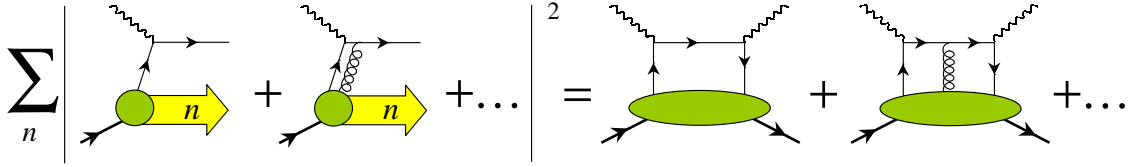}}
\end{picture}
}
\end{center}
\caption{\label{hard-interf} Hadronic structure functions of deeply inelastic
scattering as a series in a number of partons participating in a hard scattering,
truncated here to the lowest two terms in the Fock expansion. The first term on
the right-hand side of the equality stands for the leading twist-two contribution.
As was demonstrated in Section \ref{QCDPartonDistributions}, it is an absolute
value squared of a single-parton scattering amplitude. The second term is an
interference of single- and two-particle amplitudes and corresponds to
twist-three effects. The latter are power suppressed in the hard scale compared
to the leading term.}
\end{figure}

So far we have discussed the properties of the leading-twist GPDs irrespective
of specific high-energy processes where they emerge and can be ultimately
measured. The present section is a first step toward phenomenology of GPDs
in various scattering experiments. The subject of our analysis here is the
virtual Compton amplitude, a building block of the electron-nucleon Compton
scattering process. We present a thorough study of the Compton amplitude
including its gauge-invariant tensor decomposition, one-loop corrections to
the  short-distance coefficient function and power suppressed contributions
including the consideration of twist-three and target mass effects.

\subsection{Compton scattering in Bjorken limit}

As argued in the introduction, the scattering of a virtual photon on a single
parton scales at asymptotically large virtualities of the probe. The simultaneous
scattering on two partons is suppressed by a power of the hard scale (see Eq.\
(\ref{MultiPartScatDIS})). Since experimentally accessed processes are probed at
finite values of the momentum transfer, one has to know {\sl a priori} the
magnitude of power-suppressed contributions, which also go under the name of
power corrections or higher twist effects. A typical lepton-hadron cross section
at high momentum transfer ${\cal Q}$ is given by a series in $1/{\cal Q}$
\begin{equation}
\label{twist-exp}
\sigma (x_{\rm B} , {\cal Q}^2)
=
\sum_{\tau = 2}^{\infty}
\left( \frac{\Lambda}{{\cal Q}} \right)^{\tau - 2}
\int \{ dx_\tau \} \, C \! \left( x_{\rm B}, \{ x_\tau \} ; \alpha_s \right)
f \left( \{ x_\tau \} ; {\cal Q}^2 \right)
\, ,
\end{equation}
where $\tau$ stands for the twist of contributing operators, which parametrize
physics at soft scales. The first term in the expansion (\ref{twist-exp})
corresponds to the Feynman parton model picture. Subsequent terms, standing
for multi-parton correlations in hadron, reveal QCD dynamics not encoded in
conventional parton densities. They manifest genuine quantum mechanical effects,
since they involve interference of hadron wave functions with different numbers
of constituents, see Fig.\ \ref{hard-interf}. Twist-three correlations are unique
among all other higher-twist effects because they  appear as a leading effect in
certain hadronic spin asymmetries. The most familiar example is the transverse
spin structure function $g_2$ (see Eq.\ (\ref{DISstructurefunctions})) measured
in deeply inelastic scattering. Though this function is formally defined as a
Fourier transform of a two-quark bilocal operator---similarly to the twist-two
parton distributions, extensively discussed above---the presence of the transverse
Dirac matrix $\bit{\gamma}_\perp$ in its vertex results in the ``good-bad'' field
structure, in terminology of the formalism of light-cone quantization, as we
explain in Appendix \ref{LightConeQCD}. The elimination of the ``bad'' fields
in favor of the ``good'' ones (\ref{BadInTermsOfGood}) yields contributions with
an additional parton in the QCD light-cone operator compared to its original
two-particle counterpart. This is exactly the manifestation of the interference
nature of the transverse spin structure function, which makes it very attractive
for studies of genuine QCD interaction effects beyond the naive parton model.
Since deeply inelastic scattering is expressed in terms of the forward Compton
amplitude, the effects we have just outlined will arise in a more prominent and
enhanced way in the off-forward Compton amplitude, which we will address presently.
As a consequence of QCD factorization theorems, proved currently to leading twist
accuracy, the latter will be expressed in terms of GPDs.

\begin{figure}[t]
\begin{center}
\hspace{0cm}
\mbox{
\begin{picture}(0,80)(70,0)
\put(0,0){\insertfig{5}{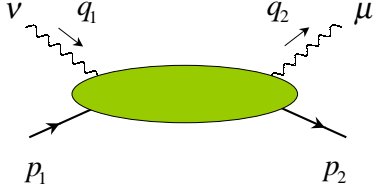}}
\end{picture}
}
\end{center}
\caption{\label{VirtualComptonAmplitude} Virtual Compton amplitude.}
\end{figure}

\subsubsection{Structure of the Compton amplitude in  generalized Bjorken limit}
\label{ComptonFactorization}

Several versions of the leading twist factorization theorems were discussed
in the literature \cite{Rad1997a,JiOsb98,ColFre99,BauFlePirRotSte02} for
the amplitude of the virtual Compton scattering on a hadron, which is defined
through the off-forward matrix element of the time-ordered product of two
quark electromagnetic currents (\ref{QuarkEMcurrent}),
\begin{equation}
\label{ComptonAmplitude}
T_{\mu\nu}
=
i \int d^4 z \, {\rm e}^{i q \cdot z}
\langle p_2 |
T \left\{ j_\mu (z/2) j_\nu (- z/2) \right\}
| p_1 \rangle
\, .
\end{equation}
To determine regions responsible for contributions with a powerlike
$\mathcal{Q}^{-N}$ behavior, it is sufficient to analyze the singularity
structure due to denominators of particle propagators. The numerators,
present in the QCD case, only change particular powers $N$, which are
not  important for our analysis at this stage. Furthermore, complications
due to gauge invariance in the realistic case of QCD can be treated in a rather
straightforward manner. Hence, to understand basic points of the factorization
approach of Ref. \cite{Rad1997a}, it is instructive to address the VCS
amplitude using the example of a cubic scalar model\footnote{The absence of
a stable vacuum state in this model is irrelevant for the demonstration of
generic perturbative properties of scattering amplitudes.}.
We will concentrate on the contribution due to the $s$-channel diagrams, since
the analysis of  $u$-channel diagrams does not bring any new insights into the
factorization property of the hadronic tensor.

\begin{figure}[t]
\begin{center}
\mbox{
\begin{picture}(0,100)(150,0)
\put(0,0){\insertfig{10}{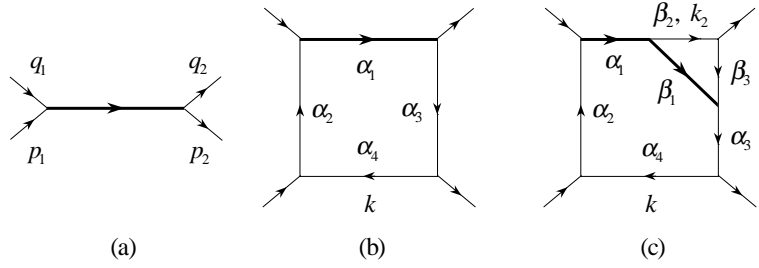}}
\end{picture}
}
\end{center}
\caption{\label{BoxDiagrams} Lowest order perturbative diagrams exhibiting
possible short-distance regimes contributing to the asymptotics of Compton
scattering amplitude. Thick lines correspond to the highly virtual
propagators.}
\end{figure}

The tree-level Compton amplitude, see Fig.\ \ref{BoxDiagrams} (a), is given by
\begin{equation}
T_{(0)}
=
\frac{1}{(p_1 + q_1)^2 + i 0}
=
\frac{1}{p_1^2 + q_1^2 + 2 (p_1 \cdot q_1)+ i 0}
\ .
\end{equation}
The scaling property is exhibited by considering, say, the kinematics when
both ${\cal Q}^2 \equiv -q_1^2$ and $(p_1 \cdot q_1)$ are large and $q_2^2
= 0$, while keeping the Bjorken variable $x_{\rm B}$ (\ref{QcalAndBjorken})
fixed. Neglecting the target mass $p_1^2$, one obtains
\begin{equation}
T_{(0)}
=
\frac{1}{2(p_1 \cdot q_1)} \frac{1}{1 - x_{\rm B} + i 0}
+
\mathcal{O} \left( \frac{p_1^2}{p_1 \cdot q_1} \right) \, .
\end{equation}

The analysis of loop amplitudes is conveniently performed making use of
the $\alpha$-representation of particle's propagators (\ref{ExpProp}).
The one-loop expression for the Compton amplitude displayed in Fig.\
\ref{BoxDiagrams} (b) can then be rewritten performing the momentum
integration with the help of Eq.\ (\ref{MomentumIntAlpha}) as
\begin{equation}
\label{treeC}
T_{(1)}
=
i g^2 \int \frac{d^4 k}{(2 \pi)^4}
\frac{1}{k^2 (k + p_1)^2 (k + p_2)^2 (k + p_1 + q_1)^2}
=
\frac{g^2}{(4 \pi)^2}\int_0^{\infty} \prod_{j = 1}^4
\frac{d \alpha_j}{\varrho^2} \,
{\rm e}^{i E_1 + i 0_\varrho}
\, ,
\end{equation}
($0_\varrho = 0 \cdot \varrho$) with the integrand depending on the exponent
\begin{eqnarray}
E_1
=
q_1^2 \, \frac{\alpha_1 \alpha_2}{\varrho}
+
q_2^2 \, \frac{\alpha_1\alpha_3 }{\varrho}
+
s \, \frac{\alpha_1 \alpha_4 }{\varrho}
+
t \, \frac{\alpha_2 \alpha_3 }{\varrho}
+
p_1^2 \frac{\alpha_2 \alpha_4 }{\varrho}
+
p_2^2 \frac{\alpha_3 \alpha_4 }{\varrho}
\, .
\label{tst1}
\end{eqnarray}
and $\varrho \equiv \sum_{j = 1}^4 \alpha_j$. In general, a two-to-two
scattering amplitude depends on seven kinematical invariants: the usual
Mandelstam variables
$$
s = (p_1 + q_1)^2 \, , \qquad t = (p_1 - p_2)^2 \, , \qquad u = (p_2 -  q_1)^2
$$
and external virtualities $q_1^2$, $q_2^2$, $p_1^2$, $p_2^2$ obeying the
constraint $s + t + u = q_1^2 + q_2^2 + p_1^2 + p_2^2$. It also depends
on the masses $m_\sigma$ of the internal lines $\sigma$ (which, for brevity,
were omitted in Eqs.\ (\ref{treeC}) and (\ref{tst1}) above). There exists
a simple rule to derive the coefficients accompanying the momentum invariants
in the exponential of the $\alpha$-representation. In particular, cutting the
lines 1 and 2 in Fig.\ \ref{BoxDiagrams} (b) corresponding to  $\alpha$-parameters
$\alpha_1$ and $\alpha_2$, one divides the diagram into two components with
$q_1^2$ being the total momentum squared entering into each of them. Cutting
the lines 1 and 3 gives the components corresponding to $q_2^2$, for lines
1 and 4 one gets $s$, and so on. In general, any diagram contributing to the
Compton amplitude can be written as
\begin{eqnarray}
&&\!\!\!\!\!\!\!\!
T (p_1, p_2, q_1 , q_2)
=
\int_0^{\infty} \prod_{\sigma = 1}^L d \alpha_{\sigma}
\frac{G( \alpha, \{p_i\},  m_\sigma )}{ D^2 (\alpha)}
\\
&&\times
\exp
\left\{
i q_1^2 \frac{ A_1( \alpha )}{D(\alpha) }
+ i q_2^2 \frac{ A_2( \alpha )}{D(\alpha) }
+ i s \frac{ A_s(\alpha)}{D(\alpha) }
+ i u \frac{ A_{u}(\alpha)}{D(\alpha)}
+ i t \frac{ A_{t}(\alpha)}{D(\alpha) }
- i  \sum_{\sigma} \alpha_{\sigma} (m_{\sigma}^2- i\epsilon)
\right\}
\, , \nonumber
\end{eqnarray}
where $L$ is the number of its internal lines. The pre-exponential factor
$G( \alpha, \{p_i\}, m_\sigma)$ results from the numerator structure of
the QCD diagram, and $D$, $A_s$, $A_u$, $A_t$, $A_1$, $A_2$ are functions
of the $\alpha$-parameters $\{\alpha_{\sigma}\}$ uniquely determined for
each diagram. The $D$ function (the ``determinant''
of a diagram) is given by a sum of products
of the $\alpha$-parameters of each set of lines the elimination
of which from the diagram
converts the diagram into a tree diagram (i.e., containing no loops).
The $A$ functions  are given by sums of products of the $\alpha$-parameters
corresponding to all possible cuts of the diagram into two components each
having a tree structure. A particular  product of the
$\alpha$-parameters is multiplied by the kinematical invariant corresponding
to a   specified cut.

Our objective is to consider the Compton  amplitude in the asymptotic
limit when some of the invariants are large and some are small,  and to
identify the powerlike scaling contributions. The hadron masses $p_1^2
= p_2^2 = M^2$ will always be treated as small variables. The main
point to realize is that if, say, $s$ is a large invariant, the result
of integration over a region where $A_s/D > \rho$ will be exponentially
$\sim \exp [-s\rho\, ]$ suppressed at large $s$. Only provided that
$A_s/D $ vanishes somewhere in the integration region, one can get
a power-behaved contribution. This may happen either when $A_s$ vanishes,
or when $D$ goes to infinity. In its  turn, $A_s$ can vanish only when
some $\alpha$-parameters vanish, and $D$ can tend to infinity only when
some $\alpha$-parameters tend to infinity. In particular, for the one-loop
diagram in Fig.\ \ref{BoxDiagrams} (b), there are two evident possibilities
$\alpha_1 = 0$ and $\alpha_4 = 0$ that make the coefficient in front of
$s$ equal to zero in Eq.\ (\ref{tst1}). These two choices do not exclude
each other, so one can take $\alpha_1 = 0$ and/or $\alpha_4 = 0$. Less
evident possibilities are $\alpha_2 \to \infty $ and/or $\alpha_3\to
\infty$. Note, that $\alpha_1 \to \infty $ or $\alpha_4 \to \infty $ do
not automatically make the $s$-coefficient $\alpha_1 \alpha_4 / \varrho$
vanish, since these parameters are present in the denominator factor
$\varrho$. In other words, there is a complication  that when $D \to \infty$,
it is quite possible that then also $A_s\to \infty $. Moreover, in the
opposite situation, when $A_s \to 0$, it is also possible that $D \to 0$,
and to get zero for the ratio $A_s/D$, the numerator $A_s$ should vanish
faster than the denominator $D$. The solution of the problem of finding
such sets of $\alpha$-parameters for which the ratio $A_s/D$ vanishes,
is facilitated by the fact that $\alpha_\sigma \to 0$ is analogous to
contraction of the corresponding line $\sigma$ into a point, while
$\alpha_\sigma \to \infty$ corresponds to removal of the line $\sigma$
from the diagram. Hence, we should find such sets of lines, whose
contraction into a point (``short-distance or SD  regime'') or their
removal from the diagram (``infrared or IR regime'') makes the diagram
independent of $s$, or, in general, of all  large variables.
The combined SD-IR regimes are also possible.
In this case,   the dependence on  large variables
is killed  by contracting some lines into points and eliminating some other lines.

\begin{figure}[t]
\begin{center}
\hspace{0cm}
\mbox{
\begin{picture}(0,150)(230,0)
\put(0,0){\insertfig{16}{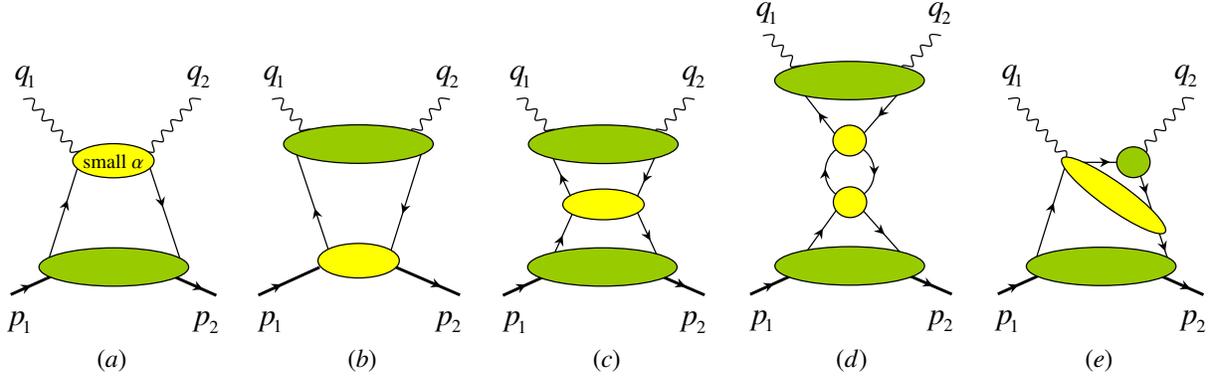}}
\end{picture}
}
\end{center}
\caption{\label{sdRegimesFig} Short-distance regimes in the generalized
Compton amplitude: the contractions (i), (ii), (iii) and (iv), discussed in the
text, are represented in (a), (b), (c) and (e), respectively. A multiperipheral
chain is shown in (d).}
\end{figure}

The $\alpha$-representation has the advantage that the dependence of the
amplitudes on the momentum invariants is displayed in the most direct way.
This makes it very convenient for a general analysis of the asymptotic
behavior of Feynman diagrams. Proceeding by analogy with the one-loop
example, one can easily find three evident possibilities to eliminate the
dependence of a four-point function on the large invariant $s$: (i) to
contract into a point a subgraph containing the external lines with
momenta $q_1$ and $q_2$, see Fig.\ \ref{sdRegimesFig} (a), (the $\alpha_1
= 0$ regime in the one-loop case); (ii) to contract into a point a subgraph
containing the external lines with momenta $p_1$ and $p_2$, see Fig.\
\ref{sdRegimesFig} (b), (the $\alpha_4=0$ regime); and (iii) to contract
into point a central subgraph so as to separate the sets $q_1$, $q_2$ and
$p_1$, $p_2$ from each other, see Fig.\ \ref{sdRegimesFig} (c). Since
small $\alpha_\sigma$'s correspond to large virtualities, the possibility (i)
corresponds  to large virtualities between the $q_1$ and $q_2$ vertices,
while the possibility (ii) implies that there is a large virtuality momentum flowing
between the $p_1$ and $p_2$ vertices. The latter are
  imitating  hadrons, and  since the
hadronic wave functions are expected to rapidly vanish at large virtualities,
the contraction (ii) should be excluded from possible contributing regions.
Note, that the contraction corresponding to the possibility (i) gives a
``reduced'' diagram that depends only on $t$ and $p_1^2$, $p_2^2$, but
does not depend on $q_1^2$ and $q_2^2$. If $t$ is small, all dependence on
large variables is eliminated. On the other hand, after the contraction of
type (iii), the upper part still depends on $q_1^2$ and $q_2^2$, and when
one of them or both are large, one would still need to apply the type (i)
contraction. When both $q_1^2$ and $q_2^2$ are small, we are not required
to contract a subgraph containing the $q_1, q_2$ vertices, and the large-$s$
(or $u$) behavior may be  governed by  central contractions (there may be several
of them,
forming a ``multiperipheral'' chain, see Fig.\ \ref{sdRegimesFig} (d)),
with the upper nonperturbative object similar to a photon-to-photon GPD.
In general, type (i) contraction eliminates the dependence on $q_1^2$,
$q_1^2$, $s$ and $u$. If  both of $q_1^2$, $q_2^2$ are large,
the type (i) contraction is the only possibility, since other  contractions
would involve the hadronic vertices.   If $t$ is small, the
lines outside the contracted subgraph do not need to have large virtualities.
 One
can parametrize the function describing the small-virtuality part by a
nonperturbative parton distribution, and calculate the large  virtuality
part using perturbative QCD and asymptotic freedom.
If only one of the photon virtualities, say $q_1^2$ is large,
there is also a possibility (iv), see Fig. \
\ref{sdRegimesFig} (e).
The reduced diagram in this case has an extra   small-virtuality part
corresponding to the distribution amplitude of the  $q_2$-photon.
 At this stage, the analysis should be supplemented by power counting.
 In QCD, the possibility (iv) is suppressed by $1/q_1^2$
 compared to possibility (i) \cite{Rad1996a,Ji96b}, thus if $q_1^2$ is large,
 the type (i) contraction is the only possibility to get the leading
 power contribution.

 The kinematics when
at least one of the invariants $q_1^2$, $q_2^2$ is large and $s$ is also
large, but the ratios $q_1^2/s$, $q_2^2/s$ are fixed, can be called the
generalized Bjorken limit.

\subsubsection{Compton amplitude in symmetric variables}
\label{ComptonVaribales}

In many cases, it is convenient to use symmetric combinations of particle
momenta
\begin{equation}
q = \frac{1}{2} (q_1 + q_2)
\, , \qquad
p = p_1 + p_2
\, , \qquad
\Delta = p_1 - p_2 = q_2 - q_1 \, .
\end{equation}
The three Mandelstam kinematical invariants can be traded for the
symmetric ones, which consist of the averaged photon virtuality and
two scaling variables,
\begin{equation}
\label{SymmetricVariables}
q^2 = - Q^2
\, , \qquad
\xi = \frac{Q^2}{p \cdot q}
\, , \qquad
\eta = \frac{\Delta \cdot q}{p \cdot q}
\, .
\end{equation}
The latter two are called  the  generalized Bjorken and skewness
variables, respectively. They can be re-expressed in terms of the
``experimental'' ones, i.e., masses of incoming and outgoing photons,
and the conventional Bjorken variable
\begin{equation}
\label{ExperVariables}
{\cal Q}^2 = - q_1^2
\, , \qquad
M_{\ell \bar\ell}^2 = q_2^2
\, , \qquad
x_{\rm B} = \frac{{\cal Q}^2}{2 p_1 \cdot q_1}
\, ,
\end{equation}
respectively, via the equations:
\begin{equation}
\label{ResolutionScale}
Q^2 = \frac{1}{2}
\left(
{\cal Q}^2 - M_{\ell \bar\ell}^2 + \frac{\Delta^2}{2}
\right) \, ,
\end{equation}
for the averaged photon virtuality, and
\begin{equation}
\label{XitoEta}
\xi =
\frac{
{\cal Q}^2 - M_{\ell \bar\ell}^2 + \Delta^2/2
}{
2{\cal Q}^2/ x_{\rm B} - {\cal Q}^2 - M_{\ell \bar\ell}^2 + \Delta^2
}
\, , \qquad
\eta =
\frac{
{\cal Q}^2 + M_{\ell \bar\ell}^2
}{
2{\cal Q}^2/ x_{\rm B} - {\cal Q}^2 - M_{\ell \bar\ell}^2 + \Delta^2
}
\, ,
\end{equation}
for the scaling variables. Note that $Q^2$ and $\xi$ can take both positive
(${\cal Q}^2 > M_{\ell \bar\ell}^2$) and negative (${\cal Q}^2 <
M_{\ell \bar\ell}^2$) values depending on the relative magnitude of spacelike
and timelike photon virtualities, while $\eta > 0$ and $\eta \pm \xi > 0$.
To complete the set of formulas, we give also the inverse transformations
\begin{equation}
M_{\ell \bar\ell}^2
= - \left( 1 - \frac{\eta}{\xi} \right) Q^2 + \frac{\Delta^2}{4}
\, , \qquad
{\cal Q}^2
=
\left( 1 + \frac{\eta}{\xi} \right) Q^2 - \frac{\Delta^2}{4}
\, ,
\end{equation}
and
\begin{equation}
x_{\rm B} =
\frac{(\xi + \eta) Q^2 - \xi \Delta^2/4}{(1 + \eta) Q^2 - \xi \Delta^2/2}
\, .
\end{equation}
The generalized Bjorken limit, where the perturbative QCD approach to the
general Compton amplitude makes sense and proves to be very fruitful, is
characterized by sending the Mandelstam variable $s$ and one or both photon
virtualities to infinity, while keeping $\xi$ and $\eta$ finite. The most
important limiting cases follow.
\begin{itemize}
\item Deeply inelastic scattering:
$$
\Delta = 0 \, , \qquad \eta = 0 \, , \qquad \xi = x_{\rm B}
\, ,
$$
with Bjorken kinematics
$$
s \sim - q_1^2 = - q_2^2 \to \infty
\, .
$$
\item Deeply virtual Compton scattering:
$$
q_2^2 = 0, \qquad \eta \simeq \xi
\, ,
$$
and
$$
s \sim - q_1^2 \to \infty \, , \qquad - \Delta^2 \ll s
\, .
$$
\item Timelike Compton scattering:
$$
q_1^2 = 0 \, , \qquad \eta \simeq - \xi
\, ,
$$
such that
$$
s \sim q_2^2 \to \infty \, , \qquad - \Delta^2 \ll s
\, .
$$

\item Doubly virtual Compton scattering:
$$
q_1^2 < 0 \, , \qquad q_2^2 > 0 \, , \qquad \eta > \xi
\, .
$$
with scaling limit
$$
s \sim - q_1^2 \sim q_2^2 \to \infty \, , \qquad
- \Delta^2 \ll s
\, .
$$
\end{itemize}
The  region $\eta < \xi$ is accessible only when both of the photon
virtualities are space-like (like in DIS). However, the Compton amplitude in
such a kinematics does not correspond to a leading contribution of any known
high-energy process. It arises, however, as a part of  the two-photon contribution
in exclusive electron-proton scattering, which measures electromagnetic form
factors \cite{AfaBroCarVan04}.

\begin{figure}[t]
\begin{center}
\mbox{
\begin{picture}(0,120)(175,0)
\put(0,0){\insertfig{12}{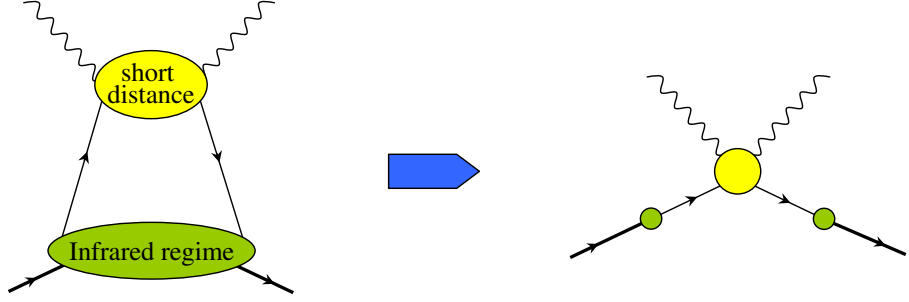}}
\end{picture}
}
\end{center}
\caption{\label{WAcomptonRegime} The diagram for the wide-angle Compton
scattering and its reduced graph.}
\end{figure}

Another process that attracted recently a lot of attention
\cite{Radyushkin:1998rt,Diehl:1998kh} is the wide-angle Compton scattering,
when both photons are real and $s \sim -t \sim -u$ is large. The contraction
of a subgraph containing the real-photon vertices ($\alpha_1$ in the one-loop
example) eliminates dependence on $s$ and $u$, but the resulting object still
has dependence on large variable $t$. For the one-loop diagram, this dependence
can be eliminated by the infrared regime $\alpha_4 \to \infty$. In general
case, the analog is the situation when the $\alpha$-parameters of a ``soft''
subgraph, connecting the incoming and outgoing nucleon vertices, tend to
infinity. This corresponds to Feynman mechanism, i.e., these lines can be
treated as  ``wee'' particles carrying decreasing fractions $x < \lambda
/\sqrt{-t}$ of the hadron momentum. As a result, large virtualities are not
induced in the lower object, and one  might talk about separation of hard and
soft contributions. The factorization properties of this process need further
study. In particular, they are also complicated by the possibility of
combining a central type (iii) contraction with Feynman mechanism for the upper and
lower components of the reduced diagram.

\subsubsection{One-loop factorization in symmetric variables}
\label{OneLoopFactorization}

Let us continue the discussion of the cubic scalar model and illustrate
the mechanism of factorization of the Compton amplitude into GPDs. We will use symmetric
variables which allow simultaneous treatment of different kinematical
situations. The tree-level Compton amplitude, see Fig.\ \ref{BoxDiagrams}
(a), is given now by
\begin{equation}
T_{(0)} = \frac{1}{(p_1 + q_1)^2 + i 0}
=
\frac{1}{(p \cdot q)}
\frac{1}{1 - \xi + \ft12 \epsilon + i 0}
\, ,
\end{equation}
where the small correction to the scaling contribution goes in powers of
$$
\epsilon \equiv \frac{4 M_N^2 - \Delta^2}{2 (p \cdot q)}
\, .
$$
Thus, the virtuality of the propagator of the tree diagram is proportional
to $(p \cdot q) = Q^2 /\xi $ rather than to  $Q^2=-q^2$ as one might expect
from the definition of the Compton amplitude (\ref{ComptonAmplitude}). In
this connection, we would like to note that the  variable
$Q^2$ related to the symmetric combination
$q_1 + q_2$ need not be mandatorily large in the
generalized Bjorken limit.
Actually, this scale becomes small  when the
virtualities of incoming and outgoing photons
cancel each other: $q_1^2+q_2^2=0$.
As we discussed above, to assure that the small-$\alpha$ regime (or short
``distances'' between the photon vertices) is the only possibility to get a
power-law contribution for large $s$, one or both external photon virtualities
$q_1^2$, $q_2^2$ need to be large. The momentum $q_1+q_2$ cannot be
arranged  as  a total momentum entering into one of the components
of a Compton diagram cut into two parts, so   its value is not directly relevant
to  the analysis of  asymptotic behavior.
In the generalized Bjorken
limit, it is convenient to treat $(p \cdot q)$ as  the basic large variable,
expressing  photon virtualities,  $Q^2$, etc. as $(p  \cdot q)$
multiplied by an appropriate dimensionless coefficient.
Since the variable $Q^2$ contains  ${\cal Q}^2 - M_{\ell \bar\ell}^2$
it  is  a large variable,
except for a region where ${\cal Q}^2 \approx  M_{\ell \bar\ell}^2$.
However, as far as ${\cal Q}^2 $ and   $  M_{\ell \bar\ell}^2$
are large, this
accidental cancellation  has no  significance.

When $p \cdot q \to \infty$ and $\xi$ is fixed, one recovers the scaling
coefficient function $1/(x - \xi + i 0)$ convoluted with the ``perturbative"
GPD which, to this order, is simply
\begin{equation}
F^\phi_{(0)} (x, \eta, \Delta^2) = \delta (1 - x)
\, .
\end{equation}
At one-loop order, the exponential factor of the $\alpha$-representation
in Eq.\ (\ref{tst1}) written  in symmetric variables is
\begin{eqnarray}
E_1
\!\!\!&=&\!\!\!
\alpha_1
\left\{
(1 - \xi + \ft12 \epsilon) \left(1 - \frac{\alpha_1}{\varrho}\right)
-
(1 + \eta + \epsilon) \frac{\alpha_2}{\varrho}
-
(1 - \eta + \epsilon) \frac{\alpha_3}{\varrho}
\right\}
(p \cdot q)
+
\frac{\alpha_2 \alpha_3}{\varrho} \Delta^2
\nonumber\\
&&+
(\alpha_2 + \alpha_3) \left( 1 - \frac{\alpha_2 + \alpha_3}{\varrho} \right) M_N^2
\, .
\end{eqnarray}
As we already discussed,   integration in the vicinity of $\alpha_1 \to 0$ corresponds to
large virtuality of the corresponding line in the Feynman diagram, the propagator
between the photon vertices. To leading order in $1/(p \cdot q)$, such an
integration gives
\begin{eqnarray}
\label{FactorizedForm}
T^{\mbox{\tiny SD}_1}_{(1)}
=
\frac{1}{(p \cdot q)}
\int_{-1}^1 d x \frac{F^\phi_{(1)} (x, \eta, \Delta^2)}{x - \xi + i 0}
\, ,
\end{eqnarray}
where we introduced the one-loop GPD, which absorbs all contributions which
diverge when  $M_N$ goes to zero, i.e., the so-called mass singularities,
\begin{equation}
F^\phi_{(1)} (x, \eta, \Delta^2)
=
\frac{i g^2}{(4 \pi)^2}
\int_0^\infty \prod_{j = 2}^4 d \alpha_j \,
\delta
\left(
x - 1
+
(1 + \eta) \frac{\alpha_2}{\tilde\varrho}
+
(1 - \eta) \frac{\alpha_3}{\tilde\varrho}
\right)
\frac{{\rm e}^{i \left(\widetilde E_1 + i 0_{\tilde\varrho} \right)}}{\tilde\varrho^2}
\, .
\end{equation}
Here,  $\tilde \varrho = \alpha_2 + \alpha_3 + \alpha_4$. Note, that   $\widetilde E_1 =
E_1 [\alpha_1 = 0]$ does not depend on large scales. Summing the tree and one-loop
contributions, one gets a factorized expression of the form (\ref{FactorizedForm})
with $F^\phi_{(1)}$ being replaced by $F^\phi = F^\phi_{(0)} + F^\phi_{(1)} +
\dots$. One can easily convince oneself that the perturbative expansion of the
GPD $F^\phi (x, \eta, \Delta^2)$ arises from the matrix element of the light-cone
operator $\mathcal{O}^{\phi\phi}$ from Eq.\ (\ref{NonlocalScalarOper}):
\begin{equation}
\label{ScalarGPD}
F^\phi (x, \eta, \Delta^2)
=
p^+ \int \frac{d z^-}{2 \pi}
{\rm e}^{i x z^- p^+}
\langle p_2 | \mathcal{O}^{\phi\phi} ( - z^-, z^- ) | p_1 \rangle
\, .
\end{equation}

Other short-distance and infrared regimes lead to contributions which are
power-suppressed compared to the leading one (\ref{FactorizedForm}), see Ref.\
\cite{Rad1997a}. When a larger number of lines is connecting the hadronic part
with hard propagators, this usually leads to the suppression of amplitudes by
powers of $1/(p \cdot q)$, merely because the number of hard propagators increases.
An important exception is provided by the longitudinally polarized $\varepsilon^\mu
\sim p^{\mu}$ vector particles in gauge field theories. Namely, in QCD the quark
propagator $\sim  {\not\!q} /(p\cdot q)$ combines with the adjacent gluon vertex
factor ${\not\!\varepsilon}\sim {\not\!p} $ to produce an unsuppressed ${\cal O}
(1)$ factor. Hence, one should take this contributions into account and sum
over all possible insertions of gluon lines into the hard quark propagator. The
summation is identical to that performed in Section \ref{QCDPartonDistributions}
for conventional parton densities and yields a path-ordered exponential stretched
between the elementary fields in the bilocal operator defining the GPD.
The analysis of the most general handbag diagram, i.e., with a single propagator
between the photon vertices and any possible configuration of lines forming the
lower part, does not bring in new complications compared to the already discussed
one-loop example, and can be treated in an analogous manner. In QCD case, the
numerator $\sim {\not\!q}$ of the hard propagator combines with the Dirac spinors
of the incoming and outgoing quark lines $u \bar u \sim \not\!p$ and cancels the
power of $p \cdot q$ coming from the denominator, so that $T^{\mbox{\tiny SD}_1}
\sim {\cal O} ((p \cdot q)^0)$.

\begin{figure}[t]
\begin{center}
\mbox{
\begin{picture}(0,110)(100,0)
\put(0,0){\insertfig{7}{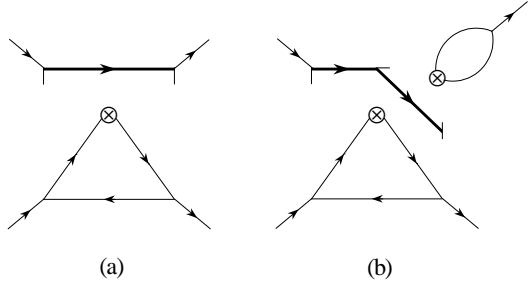}}
\end{picture}
}
\end{center}
\caption{\label{ShortDistanceRegimes} Short-distance regimes corresponding
to loop diagrams in Fig.\ \protect\ref{BoxDiagrams}.}
\end{figure}

\subsubsection{Hadronization corrections of the final-state photon}

When the final-state photon is timelike, the quark-antiquark pair can form
an on-shell intermediate hadronic state before annihilating into the heavy
photon. This corresponds to the hard momentum being  re-routed around the
photon vertex as shown in the diagram (c) of Fig.\ \ref{BoxDiagrams}. This
configuration can potentially generate an asymptotically leading  contribution
when the photon virtuality is low. Let us demonstrate that it is actually
suppressed. The two-loop diagram in Fig.\ \ref{BoxDiagrams} (c) has the form
\begin{eqnarray}
T_{(3)}
\!\!\!&=&\!\!\!
- g^4
\int \frac{d^4 k_1}{(2 \pi)^4}
\frac{1}{k_1^2 (k_1 + p_1)^2 (k_1 + p_2)^2 (k_1 + p_1 + p_2)^2}
\\
&\times&\!\!\!
\int \frac{d^4 k_2}{(2 \pi)^4}
\frac{1}{k_2^2 (k_2 - q_2)^2 (k_2 - k_1 - q_1 - p_1)^2}
=
i \frac{g^4}{(4 \pi)^4}
\int_0^\infty \prod_{i = 1}^4 d \alpha_i \prod_{j = 1}^3 d \beta_j
\frac{
{\rm e}^{i (E_2 + i 0_{\alpha + \beta})}
}{
[\alpha \beta + \beta_1 (\beta - \beta_1)]^2
}
\nonumber
\end{eqnarray}
with $\alpha \equiv \sum_{i = 1}^4 \alpha_i$ and $\beta \equiv \sum_{i = 1}^3
\beta_i$. The exponential here is given by
\begin{eqnarray}
E_2 \!\!\!&=&\!\!\!
\alpha
\frac{
\alpha_1 + \beta_1 \left( 1 - \frac{\beta_1 + \beta_3}{\beta} \right)
}{
\alpha + \beta_1 \left( 1 - \frac{\beta_1}{\beta} \right)
}
\nonumber\\
&&\times
\Bigg\{
(1 - \xi + \ft12 \epsilon)
\left(
1 - \frac{\alpha_1}{\alpha} + \frac{\beta_1 \beta_3}{\alpha \beta}
\right)
-
(1 + \eta + \epsilon) \frac{\alpha_2}{\alpha}
-
(1 - \eta + \epsilon)
\left(
\frac{\alpha_3}{\alpha} + \frac{\beta_1 \beta_3}{\alpha \beta}
\right)
\Bigg\} (p \cdot q)
\nonumber\\
&&+
\frac{
\alpha_2
\left(
\alpha_3 + \frac{\beta_1 \beta_3}{\beta}
\right)
}{
\alpha + \beta_1 \left(1 - \frac{\beta_1}{\beta} \right)
} \Delta^2
+
\left(
\alpha_2 + \alpha_3 + \frac{\beta_1 \beta_3}{\beta}
\right)
\left(
1
-
\frac{
\alpha_2 + \alpha_3 + \frac{\beta_1 \beta_3}{\beta}
}{
\alpha + \beta_1 \left(1 - \frac{\beta_1}{\beta} \right)
}
\right)
M_N^2
\nonumber\\
&&+
\beta_3 \left( 1 - \frac{\beta_1 + \beta_3}{\beta} \right) M_{\ell\bar\ell}^2
\, .
\end{eqnarray}
In the short-distance regime, i.e., $\alpha_1 \to 0$, $\beta_1 \to 0$,
defining  $\tilde \alpha = \alpha_2 + \alpha_3 + \alpha_4$ and $\tilde \beta
= \beta_2 + \beta_3$, we get, see Fig.\ \ref{ShortDistanceRegimes} (b),
\begin{equation}
\label{SD2}
T^{\mbox{\tiny SD}_2}_{(3)}
=
- \frac{g^2}{(p \cdot q)^2}
\int_{-1}^1 d x \int_0^1 d u
\frac{
F^\phi_{(1)} (x, \eta, \Delta^2) {\mit\Pi}_{(1)} (u, M_{\ell\bar\ell}^2)
}{
(1 - u)(x - \xi + i 0)^2
}
\, ,
\end{equation}
where $F^\phi_{(1)}$ was given above, since $\widetilde E_1 = E_2 [\alpha_1 =
\beta_i = 0]$, while
\begin{equation}
{\mit\Pi}_{(1)} (u, M_{\ell\bar\ell}^2)
=
\frac{1}{(4 \pi)^2}
\int_0^\infty
\prod_{j = 2}^3 d \beta_j \, \delta \left( u - \frac{\beta_3}{\tilde\beta} \right)
\exp
\left\{
i \beta_3 \left( 1 - \frac{\beta_3}{\tilde\beta} \right) M_{\ell\bar\ell}^2
\right\}
\end{equation}
is the first term in the perturbative expansion of the correlation function
\begin{equation}
{\mit\Pi} (u, M_{\ell\bar\ell}^2)
=
i q_2^- \int d^4 z \, {\rm e}^{i q_2 \cdot z}
\int \frac{d y^+}{2 \pi} {\rm e}^{- i u y^+ q_2^-}
\langle 0 |
T \left\{ \mathcal{O}^{\phi\phi} (0, y^+) , \, j (z) \right\}
| 0 \rangle
\, ,
\end{equation}
where the ``electromagnetic" current is $j (z) = \frac{1}{2} \phi^2 (z)$. Therefore,
we observe that this short-distance regime is suppressed compared to the leading
one, Eq.\ (\ref{FactorizedForm}).

In the QCD case, the suppression of contributions due to the hadronic component of
the photon is milder than in the scalar example---it is only $(p \cdot q)^{-1/2}$
compared to the handbag diagram---but still persists. The structure of the reduced
amplitude is the same as in Eq.\ (\ref{SD2}), but with only one power of the
hard-scattering coefficient $1/(x - \xi + i 0)$ involved, and the vacuum correlator
taking the following form
\begin{eqnarray}
\label{Def-VacCor}
{\mit\Pi}^\mu (u, q_2)
\!\!\!&\equiv&\!\!\!
\left(
q_2^\mu q_2^\nu
-
M_{\ell\bar\ell}^2 \ g^{\mu\nu}
\right) n^\ast_\nu
{\mit\Pi} (u, M_{\ell\bar\ell}^2)
\\
&=&\!\!\!
i \int d^4 z {\rm e}^{i z \cdot q_2}
\int \frac{d y^+}{2 \pi} {\rm e}^{- i u y^+ q_2^-}
\langle 0 |
T
\left\{
\bar\psi (0^-, 0^+, \bit{0}_\perp) \gamma^- \psi (0^-, y^+, \bit{0}_\perp) ,
j^\mu (z)
\right\}
|0 \rangle
\, .\nonumber
\end{eqnarray}
This correlation function can be saturated, using the jargon of QCD sum rules, by
the $\rho$-meson \cite{ShiVanZak79} and reads
\begin{equation}
\label{Mod-VacCor}
{\mit\Pi} (u, M_{\ell\bar\ell}^2)
=
- \frac{m_\rho^2}{g_\rho^2}
\frac{\varphi_\rho (u)}
{
M_{\ell\bar\ell}^2 - m_\rho^2 + i m_\rho {\mit\Gamma}_\rho
}
+
\frac{3}{4 \pi^2} u (1 - u)
\int_{s_0}^\infty \frac{d s}{s - M_{\ell\bar\ell}^2 - i 0}
\, ,
\end{equation}
where $\phi_\rho (u)$ is the $\rho$-meson distribution amplitude (\ref{RhoMesonDA}),
while $g_\rho^2/(4 \pi) = 2.36 \pm 0.18$, $m_\rho = 770 \, {\rm MeV}$ and
${\mit\Gamma}_\rho = 150 \, {\rm MeV}$ are the $\rho$-meson decay constant,
mass and width, respectively. The second term on the right-hand side comes
from the perturbative contribution to the correlator, known to two-loop order
\cite{ShiVanZak79}. A  part of it  is dual to the $\rho$-meson in the interval
$s \in [0, s_0]$ and is absorbed  in the $\rho$-pole contribution given by the
first term. The parameter $s_0 \approx 1.5 \, {\rm GeV}^2$ is the effective
continuum threshold\footnote{The approximation of the continuum contribution by
a step-function threshold in the spectral density causes a divergence in the
real part at $M_{\ell\bar\ell}^2 = s_0$, which is  spurious.}. Due to divergence
in the correlation function, one has to use a renormalized expression stemming
from the subtracted dispersion relation, ${\mit\Pi}_{\scriptscriptstyle\rm R}
(u, M_{\ell\bar\ell}^2) = {\mit\Pi} (u, M_{\ell\bar\ell}^2) - {\mit\Pi} (u, 0)$
in the Compton amplitude (\ref{SD2}).

{}From this result, it is apparent that, besides the usual $s$-channel discontinuity,
the VCS amplitude has  an extra imaginary part associated with the conversion  of the
quark-antiquark pair into an on-shell intermediate hadronic state.  Due to the current
conservation, which implies $\bar u(\ell_-) {\not\!q_2} u (- \ell_+) = 0$, only one
Lorentz structure contributes to the leptoproduction amplitude, and the result is
\begin{equation}
\label{Con-Lep-VacCor}
{\mit\Pi}_\mu (u, q_2)
\frac{g_{\mu\nu}}{M_{\ell\bar\ell}^2}
\bar u(\ell_-) \gamma_\nu u (- \ell_+)
=
- {\mit\Pi} (u, M_{\ell\bar\ell}^2)
\bar u(\ell_-) \gamma_- u (- \ell_+)
\, .
\end{equation}

It should be noted, that since the skewness and the generalized Bjorken variable
are not equal in general, $\eta \neq \pm \xi$, possible complications due to the
singular structure of the hard coefficient function do not arise because  its
poles $1/(x \pm\xi)$ are away from the ``turning" points $x=\pm \eta$ of GPDs
corresponding to the situation  when one of the  partons has zero momentum fraction.
Moreover, as we demonstrated in Section \ref{LongDynamics}, GPDs are continuous at
this point, i.e., have no jumps, so that this region does not present a problem even
in the case $\eta = \pm \xi$  \cite{Rad1997a,Radyushkin:1998bz,ColFre99}.

\begin{figure}[t]
\begin{center}
\mbox{
\begin{picture}(0,125)(170,0)
\put(0,0){\insertfig{12}{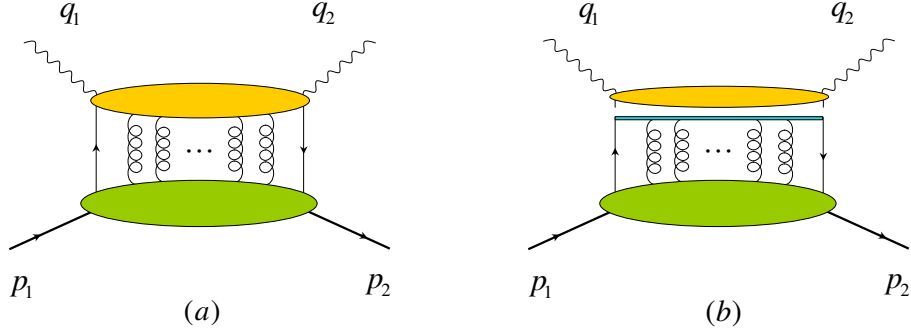}}
\end{picture}
}
\end{center}
\caption{\label{LeadingRegion} The leading asymptotic region in the Compton
scattering amplitude (a) and the factorization of longitudinal gluons into
the path-ordered exponential (b).}
\end{figure}

Summarizing, the leading region for  the Compton amplitude in QCD has the
structure depicted in Fig.\ \ref{LeadingRegion} (a).  In addition to the
simplest handbag of the scalar theory, one is allowed to attach to the hard
part an infinite number of zero-twist longitudinally polarized gluons $A^+$.
As we discussed above, their insertion does not result in power suppressed
contributions, and the net effect can be factorized into the Wilson line along
the trajectory of motion of the struck quark in the hard subprocess, as was
explained in Section \ref{QCDPartonDistributions}. It restores the color gauge
invariance of the light-ray operator involved in the correlation function
defining GPDs. Due to the unitary cancellation of the eikonal lines going
beyond the photon absorption and emission points, the path-ordered exponential
extends only between the quark fields, see Fig.\ \ref{LeadingRegion} (b).

\subsection{Restoration of electromagnetic gauge invariance}
\label{Amplitude}

Let us illustrate the peculiarities of the off-forward kinematics on a simple
example of a free Dirac fermion theory. As we will see, even and odd parity
structures start to ``talk'' to each other in this case. The reason is that,
in the off-forward kinematics, the operator product expansion should be used
in its full form, including operators with total derivatives because their
matrix elements  are nonzero.  Moreover, it is important to keep such operators
in order to restore the electromagnetic gauge invariance of the two-photon
amplitude defined by the chronological product $T \left\{ j_\mu (z_1) j_\nu (z_2)
\right\}$ of two currents $j_\mu (z) = \bar\psi (z) \gamma_\mu \psi (z)$. The
leading light-cone singularity $(z_1 - z_2)^2 \to 0$ arises from the handbag
diagram, as we established in the preceding section, and reads
\begin{eqnarray}
\label{OPE-naive}
T \left\{ j^\mu (z_1) j^\nu (z_2) \right\}
=
i \bar\psi (z_1) \gamma^\mu S (z_1 - z_2) \gamma^\nu \psi (z_2)
+
i \bar\psi (z_2) \gamma^\nu S (z_2 - z_1) \gamma^\mu \psi (z_1)
\, ,
\end{eqnarray}
where $S (z)$ is the coordinate-space free quark
propagator
$$S (z) = \frac{{\mit\Gamma} (d/2)}{2
\pi^{d/2}} \frac{{\not\! z}}{[- z^2]^{d/2}} \  .
$$
Taking into account the free
equations of motion
$$
{\not\!\partial} \psi = 0 \, , \qquad {\not\!\partial} S (z) = - i \delta^{(4)} (z)
\, ,
$$
it is straightforward to show that the handbag diagram respects current conservation.
After performing the decomposition of the Dirac structure in Eq.\ (\ref{OPE-naive})
with the help  of Eq.\ (\ref{ThreeToOne}) we obtain
\begin{eqnarray}
\label{OPE-naive-decom}
T \left\{ j^\mu (z_1) j^\nu (z_2) \right\}
\!\!\!&=&\!\!\!
S^{\mu \nu; \rho \sigma} i S_\rho (z_1 - z_2)
\left\{
\bar\psi (z_1) \gamma_\sigma \psi (z_2)
-
\bar\psi (z_2) \gamma_\sigma \psi (z_1)
\right\} \\
&-&\!\!\!
i \varepsilon^{\mu\nu\rho\sigma} i S_\rho (z_1 - z_2)
\left\{
\bar\psi (z_1) \gamma_\sigma \gamma^5 \psi (z_2)
+
\bar\psi (z_2) \gamma_\sigma \gamma^5 \psi (z_1)
\right\}
\, , \nonumber
\end{eqnarray}
where
\begin{equation}
S^{\mu \nu; \rho \sigma}
\equiv
g^{\mu\rho} g^{\nu\sigma}
+
g^{\mu\sigma} g^{\nu\rho}
-
g^{\mu\nu} g^{\rho\sigma}
\, .
\end{equation}
One easily finds that the current conservation does not hold  separately in the
parity even and odd terms. Rather, it is fulfilled due to the cancellation between
terms with different parity on the right-hand side of Eq.\ (\ref{OPE-naive-decom}).
In other words, if one takes only  the vector bilocal operator from  the right-hand
side of Eq. (\ref{OPE-naive-decom}) and parametrizes it through GPDs $H$, $E$, the
result will not be gauge-invariant. Apparently, the gauge invariance should be restored
if the contribution of the axial vector bilocal operator is added. However, the latter
is parametrized at the leading twist by another set of GPDs, $\widetilde H$,
$\widetilde E $, which, in general,  are dynamically independent of $H$, $E$. From
this point of view, the possibility of cancellations looks miraculous. The resolution
of the puzzle lies in the fact that in the twist decomposition of $\bar\psi (z_1)
\gamma_\mu (1, \gamma_5) \psi (z_2)$, one should  take into account operators with
total derivatives which formally have twist-three, but are ``kinematically'' related
to twist-two operators. After such terms are included, one obtains gauge invariant
expressions  for contributions associated both with  $H$, $E $ and $\tilde H$,
$\tilde E$ GPDs \cite{AniPirTer00,PenPolShuStr00,BelMul00,KivPolSchTer00,RadWei00,%
KivPol00,Radyushkin:2000ap,Radyushkin:2001fc,Whi01,AniTer01,BalLaz01,KipPol02}.
At twist-three, one should also include the interaction-dependent three-particle
contributions (see Fig.\ \ref{handbagdiagram}), but this does not represent any
difficulty of principle, as will be demonstrated in the next section.

\subsubsection{Tree-level twist-three Compton amplitude}
\label{TreeLevelCompton}

\begin{figure}[t]
\begin{center}
\hspace{0cm}
\mbox{
\begin{picture}(0,85)(100,0)
\put(-50,-14){\insertfig{10}{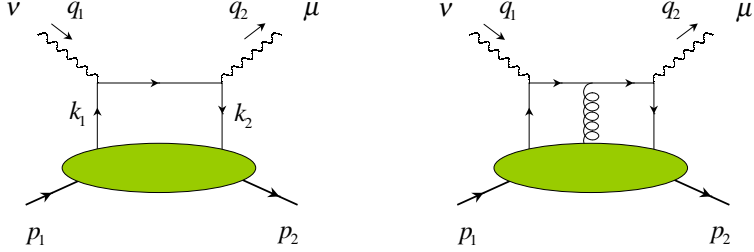}}
\end{picture}
}
\end{center}
\caption{\label{handbagdiagram} Two- and three-leg coefficient functions
for the virtual Compton process which build up a gauge invariant amplitude to
the twist-three accuracy. Addition of the $u$-channel diagrams is implied.}
\end{figure}

Let us use the ideas, we have just spelled out, and perform the calculation of
the Compton amplitudes in the Bjorken region in the light-cone gauge, in which
only the physical degrees of freedom propagate. Evaluating the handbag diagram
in the Bjorken limit,  in addition to the leading contributions, we will keep
also non-leading terms linear in partons' transverse momenta. For the $s$-channel
diagram, we perform the reduction of the Dirac structure as follows
\begin{eqnarray*}
\gamma^\mu S (k_1 + q_1) \gamma^\nu
\!\!\!&\simeq&\!\!\!
\frac{1}{(p \cdot q) (x - \xi + i 0)}
\Big\{
S^{\mu\nu;\rho\sigma}
\left(
\ft12 (p \cdot q) n_\sigma + k^\perp_\sigma + \ft12 (x - \xi) p_\sigma
\right) \gamma_\rho
\\
&&\qquad\qquad\qquad\qquad
+
i \varepsilon^{\mu\nu\rho\sigma}
\left( \ft12 x p_\sigma + q_\sigma + k^\perp_\sigma \right) \gamma_\rho \gamma^5
\Big\}
\, .
\end{eqnarray*}
Adding the $u$-channel diagram  gives  the amplitude in the twist-three
approximation
\begin{eqnarray}
\label{2Part}
T^{\mu\nu}_{(2)}
&=&\!\!\!
\frac{1}{2 p \cdot q}
\int d x \,
\sum_q
C^{q[-]}_{(0)} (x, \xi)
\left\{
(p \cdot q) S^{\mu \nu; \rho \sigma} n_\sigma F^q_\rho (x, \eta)
+
i \varepsilon^{\mu \nu \rho \sigma} p_\sigma
\left(
x \, \widetilde F^q_\rho (x, \eta)
-
\widetilde K^q_\rho (x, \eta)
\right)
\right\}
\nonumber
\\
\!\!\!&+&\!\!\!
\frac{1}{2 p \cdot q}
\int d x \,
\sum_q
C^{q[+]}_{(0)} (x, \xi)
\left\{2 i \varepsilon^{\mu \nu \rho \sigma} q_\sigma
\widetilde F^q_\rho (x, \eta) +
S^{\mu \nu; \rho -} \, p_\sigma K^q_\rho (x, \eta)
\right\}
\, ,
\end{eqnarray}
where the subscript $(2)$ on the amplitude means that it comes from the
two-particle contributions in the $t$-channel. The tree-level coefficient
functions are
\begin{equation}
\label{LOquarkCoeffFunct}
C^{q[\pm]}_{(0)} (x, \xi)
=
\frac{Q_q^2}{\xi - x - i 0} \pm \frac{Q_q^2}{\xi + x - i 0}
\, .
\end{equation}
In Eq.\ (\ref{2Part}) above we used the following boost-invariant parity-even
\begin{eqnarray}
F^q_\rho (x, \eta)
\!\!\!&=&\!\!\!
p^+ \int \frac{d z^-}{2 \pi} {\rm e}^{i x z^- p^+}
\langle p_2 | {\cal O}^{qq}_\rho ( - z^- , z^- ) | p_1 \rangle
\, , \\
K^q_\rho (x, \eta)
\!\!\!&=&\!\!\!
\int \frac{d z^-}{2 \pi} {\rm e}^{i x z^- p^+}
\langle p_2 | {\cal K}^{qq}_\rho ( - z^- , z^- ) | p_1 \rangle
\, ,
\label{EvenOpenIndex}
\end{eqnarray}
and parity-odd
\begin{eqnarray}
\widetilde F^q_\rho (x, \eta)
\!\!\!&=&\!\!\!
p^+ \int \frac{d z^-}{2 \pi} {\rm e}^{i x z^- p^+}
\langle p_2 | \widetilde {\cal O}^{qq}_\rho ( - z^- , z^- ) | p_1 \rangle
\, , \\
\widetilde K^q_\rho (x, \eta)
\!\!\!&=&\!\!\!
\int \frac{d z^-}{2 \pi} {\rm e}^{i x z^- p^+}
\langle p_2 | \widetilde {\cal K}^{qq}_\rho ( - z^- , z^- ) | p_1 \rangle
\label{OddOpenIndex}
\end{eqnarray}
functions
with an open Lorentz index of the Dirac matrix. They   are Fourier transforms
of the two-quark operators
\begin{eqnarray}
\label{OperatorOmu}
&&
{\cal O}^{qq}_\mu ( - z^- , z^- )
=
\bar\psi_q (- z^- ) \gamma_\mu \psi_q (z^-)
\, , \qquad\qquad
{\cal K}^{qq}_\mu ( - z^- , z^- )
=
\bar\psi_q ( - z^- )
\,
i
\!
\stackrel{{}_\leftrightarrow}{\partial}\!\!{}^\perp_\mu
\gamma^+
\,
\psi_q ( z^- )
\, , \nonumber\\
&&
\widetilde {\cal O}^{qq}_\mu ( - z^- , z^- )
=
\bar\psi_q (- z^- ) \gamma_\mu \gamma^5 \psi_q (z^-)
\, , \quad\qquad
\widetilde {\cal K}^{qq}_\mu ( - z^- , z^- )
=
\bar\psi_q ( - z^- )
\,
i
\!
\stackrel{{}_\leftrightarrow}{\partial}\!\!{}^\perp_\mu
\gamma^+ \gamma^5
\,
\psi_q ( z^- )
\, . \nonumber\\
\end{eqnarray}
In this notation, we do not intend to perform the form factor decomposition
for generalized parton distributions, so that the expression for the amplitude
is valid for a target of an arbitrary spin.

The first operator in Eq.\ (\ref{OperatorOmu}) has an open Lorentz index and
contains both the twist-two and twist-three contributions. The twist-two part
can be projected out by contracting the operators with the light-like vector
$n^\mu$. The operators ${\cal K}$ and $\widetilde{\cal K}$ reflect the
transverse momentum of partons in a hadron moving along the light-cone.
Neglecting them, one gets the conventional leading twist approximation to the
VCS amplitude \cite{MulDitRobGeyHor98,Ji96a,Rad1996a,BluRob00,BluGeyRob99}.

The functions introduced above are not independent, in fact,  they are
related to each other by means of the quark Heisenberg equation of motion,
${\not\!\!{\cal D}} \psi = 0$. Obviously, in the interacting theory they
are related to three-particle quark-gluon-quark operators. Namely, a
simple decomposition of Lorentz products and indices in terms of the light-cone
Sudakov components gives  the following relations
\begin{eqnarray*}
&&
\frac{\partial}{\partial z^-}
{\cal O}^{qq}{}^{, \perp}_\mu ( - z^-, z^- )
-
i \varepsilon^\perp_{\mu \nu} \partial^+
\widetilde {\cal O}^{qq}{}^{, \nu}_\perp ( - z^- , z^- )
+
i {\cal K}^{qq}_\mu ( - z^- , z^- )
+
i \varepsilon^\perp_{\mu \nu} \partial_\perp^\nu
\widetilde {\cal O}^{qq} ( - z^-  , z^- )
\\
&&\qquad\qquad
+ \int d z'^-
\left\{
w (z'^- - z^-) {\cal S}^{qgq}_{[-] \mu} ( - z^- , z'^- , z^- )
+
w (z'^- + z^-) {\cal S}^{qgq}_{[+] \mu} ( - z^- , z'^- , z^- )
\right\} = 0
\, , \nonumber\\
&&
\partial^+
{\cal O}^{qq}{}^{, \perp}_\mu ( - z^- , z^- )
-
i \varepsilon^\perp_{\mu \nu} \frac{\partial}{\partial z^-}
\widetilde {\cal O}^{qq}{}^{, \nu}_\perp ( - z^- , z^- )
+
\varepsilon^\perp_{\mu \nu}
\widetilde {\cal K}^{qq, \nu} (- z^- , z^- )
-
\partial^\perp_\mu {\cal O}^{qq} ( - z^- , z^- )
\\
&&\qquad\qquad
+ \int d z'^-
\left\{
w (z'^- - z^-) {\cal S}^{qgq}_{[-] \mu} ( - z^- , z'^- , z^- )
-
w (z'^- + z^-) {\cal S}^{qgq}_{[+] \mu} ( - z^- , z'^- , z^- )
\right\} = 0
\, , \nonumber
\end{eqnarray*}
where the weight  $w(z^-)$ depends on the boundary conditions assumed for the
gauge field, see Eq.\ (\ref{WeightForBoundaryCond}) in Appendix \ref{LightConeQCD}.
Obviously, the total contribution will not depend on this prescription. Notice that
$\partial^+$ is a total light-cone derivative, i.e.,
\begin{equation}
\label{TotalDerivative}
\partial_\mu {\cal O}^{\phi\phi} (- z , z)
\equiv
\left.
\frac{\partial}{\partial a^\mu}
\right|_{a = 0}
{\cal O}^{\phi\phi} (- z + a, z + a)
\, .
\end{equation}
The three-particle antiquark-gluon-quark operators ${\cal S}^{qgq}$ are given by
\begin{eqnarray}
\label{QGQoperator}
{\cal S}^{qgq}_{[\pm] \mu} (z^-_1, z^-_2, z^-_3)
\!\!\!&=&\!\!\!
i g \bar\psi_q (z^-_1)
\left\{
\gamma^+ F^+{}_\mu^\perp (z^-_2)
\pm
i \gamma^+ \gamma^5 \widetilde F^+{}_\mu^\perp (z^-_2)
\right\}
\psi_q (z^-_1)
\, , \nonumber\\
\widetilde {\cal S}^{qgq}_{[\pm] \mu} (z^-_1, z^-_2, z^-_3)
\!\!\!&=&\!\!\!
i g \bar\psi_q (z^-_1)
\left\{
\gamma^+ \gamma^5 F^+{}_\mu^\perp (z^-_2)
\pm
i \gamma^+ \widetilde F^+{}_\mu^\perp (z^-_2)
\right\}
\psi_q (z^-_1)
\, .
\end{eqnarray}
Here we adopted a notation for the $N_c \times N_c$ matrix of the gauge field
strength, $F_{\mu\nu} \equiv t^a F^a_{\mu\nu}$. This notation and a similar
notation for the gauge potential, $A_\mu \equiv t^a A_\mu^a$, will be implied
from now on whenever an adjoint $SU(3)$ index is not shown explicitly. The two
functions in Eq.\ (\ref{QGQoperator}) are related to each other by a ``duality''
equation
\begin{equation}
\label{Pro-Dual}
i \varepsilon^\perp_{\mu\nu} \, \widetilde {\cal S}^{qgq, \nu}_{[\pm]}
=
\pm \, {\cal S}^{qgq}_{[\pm] \mu}
\, .
\end{equation}
In terms of the  functions introduced in Eqs.\
(\ref{EvenOpenIndex}) -- (\ref{OperatorOmu}), these relations result in
\begin{eqnarray}
\label{EOM1}
x \, F^q{}^{, \perp}_\mu (x, \eta)
-
\eta \, i \varepsilon^\perp_{\mu \nu}
\widetilde F^q{}^{, \nu}_\perp (x, \eta)
\!\!\!&+&\!\!\!
i \varepsilon^\perp_{\mu \nu} \Delta^\nu_\perp \,
\widetilde F^q (x, \eta)
-
K^q_\mu (x, \eta)
\\
&-&\!\!\!
\int \frac{d \tau}{\tau}
\left\{
S^{qgq}_{[-] \mu} (x - \tau, x)
+
S^{qgq}_{[+] \mu} (x + \tau, x)
\right\}
= 0
\, , \nonumber\\
\label{EOM2}
x \, i \varepsilon^\perp_{\mu \nu}
\widetilde F^q{}^{, \nu}_\perp (x, \eta)
-
\eta \, F^q{}^{, \perp}_\mu (x, \eta)
\!\!\!&+&\!\!\!
\Delta^\perp_\mu \, F^q (x, \eta)
-
i \varepsilon^\perp_{\mu \nu} \widetilde K^{q, \nu} (x, \eta)
\\
&+&\!\!\!
\int \frac{d \tau}{\tau}
\left\{
S^{qgq}_{[-] \mu} (x - \tau, x)
-
S^{qgq}_{[+] \mu} (x + \tau, x)
\right\} = 0
\, , \nonumber
\end{eqnarray}
where the apparent singularity in $\tau$ is regularized in accordance with the
residual gauge fixing procedure, see  Eqs.\  (\ref{AinFaboundary}) and
(\ref{AinFConditions}). The boost-invariant twist-two functions are determined by
\begin{equation}
F^q \equiv (p^+)^{- 1}n^\mu F^q_\mu
\, , \qquad
\widetilde F^q \equiv (p^+)^{- 1} n^\mu \widetilde F^q_\mu
\, .
\end{equation}
The three-particle correlations $S^{qgq}_{[\pm] \rho} (x_1, x_2)$ are the functions
of two independent momentum fractions
\begin{equation}
S^{qgq}_{[\pm] \rho} (x_1, x_2)
= \int \frac{d z^-_1}{2 \pi} \frac{d z^-_2}{2 \pi}
{\rm e}^{i x_1 z^-_1 p^+ + i x_2 z^-_2 p^+} \
\langle p_2 |
{\cal S}^{qgq}_{[\pm] \rho}
\left( - z^-_1 , z^-_2, z_1^- \right)
| p_1 \rangle
\, .
\end{equation}

Since the two-particle contribution is related to the three-particle ones by QCD
equations of motion, to get the complete result we have to evaluate the contribution
from the Compton scattering on a quark-gluon-quark system. This situation is
analogous to the treatment of the transverse spin structure function $g_2$ in
deeply inelastic scattering \cite{BukKurLip83}. This observation was first made
in Ref.\ \cite{AniPirTer00} and applied there to off-forward Compton scattering
on a scalar target. The contribution corresponding  to the diagram in Fig.\
\ref{handbagdiagram} (b) is
\begin{eqnarray}
\label{3Part}
T^{\mu\nu}_{(3)}
\!\!\!&=&\!\!\!
\frac{1}{2 p \cdot q}
\int d x \int \frac{d\tau}{\tau}
\sum_q
C_{(0)}^{q[+]} (x, \xi) S^{\mu \nu; \rho \sigma} p_\sigma
\left\{
S^{qgq}_{[-] \rho} (x - \tau, x)
+
S^{qgq}_{[+] \rho} (x + \tau, x)
\right\}
\nonumber\\
&+&\!\!\!
\frac{1}{2 p \cdot q}
\int d x \int \frac{d\tau}{\tau}
\sum_q
C_{(0)}^{q[-]} (x, \xi) S^{\mu \sigma; \nu \rho} p_\sigma
\left\{
S^{qgq}_{[-] \rho} (x - \tau, x)
-
S^{qgq}_{[+] \rho} (x + \tau, x)
\right\}
\, .
\end{eqnarray}

Summing the two-particle (\ref{2Part}) and three-particle (\ref{3Part}) Compton
amplitudes and using Eq.\ (\ref{EOM1}) and (\ref{EOM2}) to remove the GPDs
$K^q_\rho$ and $\widetilde{K}^q_\rho$, we find that the quark-gluon-quark
correlation cancels, and  the result is simply
\begin{eqnarray}
\label{GaugeInvAmpl}
T^{\mu\nu}
=
T^{\mu\nu}_{(2)} + T^{\mu\nu}_{(3)}
\!\!\!&=&\!\!\!
\int_{- 1}^1 dx \,
\sum_q C^{q[-]}_{(0)} (x, \xi)
\left\{
{\cal T}_{(1)}^{\mu\nu} \,
F^q (x, \eta)
+
{\cal T}_{(2)}^{\mu\nu; \rho} \,
F^q{}^{, \perp}_\rho (x, \eta)
\right\}
\nonumber\\
&+&\!\!\!
\int_{- 1}^1 dx \,
\sum_q C^{q[+]}_{(0)} (x, \xi)
\left\{
\widetilde{\cal T}_{(1)}^{\mu\nu} \,
\widetilde F^{q} (x, \eta)
+
\widetilde{\cal T}_{(2)}^{\mu\nu; \rho} \,
\widetilde F^{q}{}^{, \perp}_\rho (x, \eta)
\right\}
\, .
\end{eqnarray}
The range of integration over the momentum fraction variable $x$ is
dictated by the support properties of GPDs, established in Section
\ref{TimeOrderingSupport}. The Lorentz tensors accompanying GPDs are
\begin{eqnarray}
\label{LorentzOfTwistThree}
&&
{\cal T}_{(1)}^{\mu\nu}
=
- g^{\mu\nu}
+
\frac{1}{p \cdot q}
p^\mu \left( 2 q^\nu + \xi p^\nu + \Delta_\perp^\nu \right)
+
\frac{1}{p \cdot q}
p^\nu \left( 2 q^\mu + \xi p^\mu - \Delta_\perp^\mu \right)
\, ,
\nonumber\\
&&
\widetilde {\cal T}_{(1)}^{\mu\nu}
=
\frac{i}{p \cdot q}
\left(
\varepsilon^{\mu \nu \rho \sigma} p_\rho q_\sigma
+
\ft{1}2 \varepsilon^\perp_{\rho \sigma} \Delta^\perp_\sigma
\left( p^\mu g^{\rho\nu}_\perp + p^\nu g^{\rho\mu}_\perp \right)
\right)
\, ,
\nonumber\\
&&
{\cal T}_{(2)}^{\mu\nu;\rho}
=
\frac{1}{p \cdot q}
\left( q^\nu + \left( \xi + \ft12 \eta \right) p^\nu \right)
g^{\mu \rho}_\perp
+
\frac{1}{p \cdot q}
\left( q^\mu + \left( \xi - \ft12 \eta \right) p^\mu \right)
g^{\nu \rho}_\perp
\, ,
\nonumber\\
&&
\widetilde {\cal T}_{(2)}^{\mu\nu;\rho}
=
\frac{i}{p \cdot q}
\left(
\varepsilon_{\mu \nu \rho \sigma} q_\sigma
+
\ft12 \eta \varepsilon^\perp_{\rho \sigma}
\left( p^\mu \, g^{\sigma\nu}_\perp + p^\nu \, g^{\sigma\mu}_\perp \right)
\right)
\, .
\end{eqnarray}
These expressions are obviously target independent. They were found here by means of a
rigorous field theoretical analysis \cite{AniPirTer00,PenPolShuStr00,BelMul00,RadWei00},
which confirms an earlier ``rule of the thumb'' restoration procedure \cite{GuiVan98}
suggested for the leading twist Lorentz tensor ${\cal T}_{(1)}^{\mu\nu}$. Furthermore,
it is easy to check that, after the twist-three corrections have been taken into account,
the result is gauge invariant up to ${\cal O} (\Delta_\perp^2)$, i.e., up to twist-four
effects:
$$
q_{1 \nu} {\cal T}_{(i)}^{\mu\nu}
={\cal O} (\Delta_\perp^2)
\qquad  \mbox{and} \qquad
q_{2 \mu} {\cal T}_{(i)}^{\mu\nu}
=
{\cal O} (\Delta_\perp^2)
\, .
$$
To prove this, we used the Sudakov decomposition of the photon momenta given in
Appendix \ref{DISframe}. It is interesting to note that the twist-three structures
${\cal T}_{\mu \nu; \rho}^{(i)}$ give exact zero when they are simultaneously
contracted both with incoming and outgoing photon momenta:
$$
q_{1 \nu} q_{2 \mu} {\cal T}^{(i)}_{\mu\nu; \rho} = 0
\, .
$$

The approximately gauge-invariant Lorentz structures introduced above can be
restored to exact ones. To this end, we define two projectors
\begin{equation}
{\cal P}^{\mu\nu}
\equiv
g^{\mu\nu}
-
\frac{q_1^\mu \ q_2^\nu}{q_1 \cdot q_2}
\, , \qquad
{\cal E}^{\mu \nu \rho \sigma}
\equiv
\varepsilon^{\alpha \beta \rho \sigma}
\left(
g^\mu{}_\alpha - \frac{p^\mu \, q_{2 \alpha}}{p \cdot q_2}
\right)
\left(
g^\nu{}_\beta - \frac{p^\nu \, q_{1 \beta}}{p \cdot q_1}
\right)
\, .
\end{equation}
They satisfy  the gauge invariance  conditions exactly
$$
{\cal P}^{\mu\nu} q_{2\mu}
=
{\cal P}^{\mu\nu} q_{1\nu}
=
{\cal E}^{\mu\nu\rho\sigma} q_{2\mu}
=
{\cal E}^{\mu\nu\rho\sigma} q_{1\nu}
= 0
\, .
$$
One can easily see that the tensors given in Eq. (\ref{LorentzOfTwistThree})
are the first terms in the $\Delta_\perp$ expansion of the following Lorentz
structures
\begin{equation}
\begin{array}{ll}
{\cal T}_{(1)}^{\mu\nu}
=
-
{\cal P}^{\mu\rho} g_{\rho\sigma} {\cal P}^{\sigma\nu}
+
\displaystyle\frac{\xi}{p \cdot q}
{\cal P}^{\mu\rho} p_\rho p_\sigma {\cal P}^{\sigma\nu}
\, , \qquad
&
\widetilde{\cal T}_{(1)}^{\mu\nu}
=
\displaystyle\frac{i}{p \cdot q} {\cal E}^{\mu\nu\rho\sigma} p_\rho q_\sigma
\, , \\
{\cal T}_{(2)}^{\mu\nu; \rho}
=
\displaystyle\frac{\xi}{p \cdot q}
\left(
{\cal P}^{\mu\rho} p_\sigma {\cal P}^{\sigma\nu}
+
{\cal P}^{\mu\sigma} p_\sigma {\cal P}^{\rho\nu}
\right)
\, , \qquad
&
\widetilde {\cal T}_{(2)}^{\mu\nu; \rho}
=
\displaystyle\frac{i}{p \cdot q} {\cal E}^{\mu\nu\rho\sigma} q_\sigma
\, ,
\end{array}
\end{equation}
which exactly satisfy the requirements of the gauge invariance.

\subsubsection{Lorentz structure of the Compton amplitude}

The analysis of  twist-three contributions given above allows one to find the
(electromagnetically) gauge-invariant tensor decomposition of the off-forward
Compton amplitude:
\begin{eqnarray}
\label{VCSamplitude}
T^{\mu\nu}
=
\!\!\!&-&\!\!\!
{\cal P}^{\mu\rho} g_{\rho\sigma} {\cal P}^{\sigma\nu}
{\cal F}_1 (\xi, \eta, \Delta^2; Q^2)
+
\frac{1}{p \cdot q}
{\cal P}^{\mu\rho} p_\rho p_\sigma {\cal P}^{\sigma\nu}
{\cal F}_2 (\xi, \eta, \Delta^2; Q^2)
\\
&+&\!\!\!
\frac{i}{p \cdot q}
{\cal E}^{\mu \nu \rho \sigma} p_\rho \, q_\sigma
\widetilde {\cal F}_1 (\xi, \eta, \Delta^2, Q^2)
+
{\cal P}^{\mu\rho} {\cal P}^{\sigma\nu}
{\cal T}_{\rho\sigma} (\xi, \eta, \Delta^2; Q^2)
\nonumber\\
&+&\!\!\!
\frac{\xi}{p \cdot q}
\left(
{\cal P}^{\mu\rho} p_\sigma {\cal P}^{\sigma\nu}
+
{\cal P}^{\mu\sigma} p_\sigma {\cal P}^{\rho\nu}
\right)
{\cal F}^\perp_{3 \rho} (\xi, \eta, \Delta^2; Q^2)
+
\frac{i}{p \cdot q}
{\cal E}^{\mu \nu \rho \sigma} \, q_\sigma
\widetilde {\cal F}^\perp_{3 \rho} (\xi, \eta, \Delta^2; Q^2)
\, . \nonumber
\end{eqnarray}
By analogy with terminology used in studies of deeply inelastic scattering, the
functions that appear in the decomposition (\ref{VCSamplitude}) will be called
the Compton form factors (CFFs). Four of these functions, ${\cal F}_1$,
${\cal F}_2$, $\widetilde{\cal F}_1$, ${\cal T}$, receive twist-two contributions
as their leading term, while the other two, ${\cal F}^\perp_{3 \rho}$ and
$\widetilde{\cal F}^\perp_{3 \rho}$, start at twist three. Similarly to the case
of the forward Compton amplitude, one can introduce the longitudinal-longitudinal
helicity\footnote{Referring to incoming-outgoing virtual photons.} amplitude
\begin{eqnarray}
{\cal F}_{L}
\equiv
\frac{1}{\xi} {\cal F}_2 - {\cal F}_1
\, .
\end{eqnarray}
In the preasymptotic Bjorken limit, the CFFs can be written as a power expansion in
the hard scale, with   each term, in its turn, being  a power series in the strong
coupling constant. In particular, the functions whose leading term is of twist two
will be written as
\begin{eqnarray}
{\cal F}_i (\xi, \eta, \Delta^2; Q^2)
\!\!\!&=&\!\!\!
\sum_{a = q, g} \int_{- 1}^1 dx \, C_i^{a[-]} \left( x, \xi, \eta ; Q^2/\mu^2 \right)
F^a (x, \eta, \Delta^2; \mu^2)
+
{\cal O} (Q^{- 2})
\, , \\
\widetilde {\cal F}_1 (\xi, \eta, \Delta^2; Q^2)
\!\!\!&=&\!\!\!
\sum_{a = q, g} \int_{- 1}^1 dx \, C_1^{a[+]} \left( x, \xi, \eta ; Q^2/\mu^2 \right)
\widetilde F^a (x, \eta, \Delta^2; \mu^2)
+
{\cal O} (Q^{- 2})
\, ,
\end{eqnarray}
where the index $i$ runs over $i = 1, L$. We have explicitly introduced the
factorization scale $\mu^2$ which separates short and long distance physics.
The perturbative expansion of coefficient functions in the strong coupling
constant reads to the lowest two orders
\begin{equation}
\label{CoefFuncPertExpansion}
C_i^{a[\pm]} (x, \xi, \eta ; Q^2/\mu^2)
=
C^{a[\pm]}_{i(0)} (x, \xi)
+
\frac{\alpha_s}{2 \pi} C_{i(1)}^{a[\pm]} (x, \xi, \eta ; Q^2/\mu^2)
+
{\cal O} (\alpha_s^2)
\, .
\end{equation}
The leading order quark coefficient functions for CFFs ${\cal F}_1$ and
$\widetilde{\cal F}_1$ are given in Eq.\ (\ref{LOquarkCoeffFunct}), i.e.,
$C_{1(0)}^{q[\pm]} = C_{(0)}^{q[\pm]}$. The function ${\cal F}_{L}$ vanishes
at leading order of perturbation theory,
\begin{equation}
\label{GeneralCallanGross}
{\cal F}_2
\stackrel{\rm LO}{=}
\xi {\cal F}_1
\, .
\end{equation}
This a generalization of the Callan-Gross relation (\ref{ForwardCallanGrossRel}),
just like in the forward Compton amplitude case, it is a consequence of the
spin-one-half nature of quarks. Thus, we have
$$
C^{q[\pm]}_{L(0)} = 0
\, .
$$
The photons can interact with gluons only through  quarks, hence the contribution
of gluons starts from one-loop order,
$$
C^{g[\pm]}_{i(0)} (x, \xi)  = 0
\, .
$$
The one-loop coefficient functions for all twist-two functions will be presented
later in Section \ref{OneLoopCoeffFunctions}.

The appearance of the function ${\cal T}$ in the  decomposition (\ref{VCSamplitude})
is a consequence of the gluon helicity-flip by two units. It arises due to nonzero
orbital angular momentum in the off-forward scattering. The twist-two photon
helicity-flip amplitude is absent in the handbag diagram because of the conservation
of the angular momentum along the photon-parton collision axis. Since  photons are
vector particles, to  flip its helicity one needs to compensate  two units of the
angular momentum. For the collinear twist-two partonic amplitude, this is only
possible by a simultaneous flip of gluon helicities. Since quarks have spin one-half,
their helicity flip can provide at most one unit of the angular momentum. As a
consequence, twist-two photon helicity-flip amplitude is sensitive to the
helicity-flip gluon distribution in a nucleon. However, a similar angular momentum
conservation argument shows that such distribution is forbidden in the forward limit,
i.e., in DIS on a spin-one-half target. In the off-forward case, the transverse
component of the momentum transfer $\Delta_\perp$ can provide one unit of the
angular momentum, so the off-forward virtual Compton scattering offers a unique
opportunity to investigate the helicity-flip gluon distribution in the nucleon. The
factorized form of the corresponding CFF is
\begin{eqnarray}
\label{GluonTransCCF}
{\cal T}_{\mu\nu} (\xi, \eta, \Delta^2; Q^2)
\!\!\!&=&\!\!\!
\int_{- 1}^1 dx \, C_T^g (x, \xi, \eta; Q^2/\mu^2)
F^g_{T \mu\nu} (x, \eta, \Delta^2; \mu^2)
+
{\cal O} (Q^{- 2})
\, , \qquad
\end{eqnarray}
where the boost-invariant maximal-helicity GPD $F^g_{T \mu\nu}$ is defined as
\begin{equation}
\langle p_2 | {\cal T}^{gg}_{\mu\nu} (- z^-, z^-) | p_1 \rangle
=
\frac{1}{4} (p^+)^2
\int_{- 1}^{1} d x \ {\rm e}^{- i x p^+ z^- } F^g_{T \mu\nu} (x, \eta, \Delta^2)
\, ,
\end{equation}
with the operator ${\cal T}^{gg}_{\mu\nu}$ determined in Eq.\
(\ref{GluonLightRayOperatorMaximalHelicity}). By the same token as before,
the CCF (\ref{GluonTransCCF}) has vanishing leading order coefficient
function $C^{g}_{T(0)} (x, \xi) = 0$.

Finally, the functions which start from twist three are displayed in the last line
of Eq.\ (\ref{VCSamplitude}). Of course, the collinear expansion of the corresponding
Lorentz structures (not taking into account the gluon helicity-flip contributions)
reproduces Eq.\ (\ref{GaugeInvAmpl}) of the preceding section up to terms linear in
the transverse momentum transfer $\Delta_\perp$. The twist-three functions at leading
order in strong coupling yield
\begin{eqnarray}
\label{TransvGPDeven}
{\cal F}^\perp_{3 \mu} (\xi, \eta, \Delta^2; Q^2)
\!\!\!&=&\!\!\!
\sum_{q} \int_{- 1}^1 dx \, C_{(0)}^{q[-]} (x, \xi)
F^{q \perp}_\mu (x, \eta, \Delta^2; Q^2)
\, , \\
\label{TransvGPDodd}
\widetilde {\cal F}^\perp_{3 \mu} (\xi, \eta, \Delta^2; Q^2)
\!\!\!&=&\!\!\!
\sum_{q} \int_{- 1}^1 dx \, C_{(0)}^{q[+]} (x, \xi)
\widetilde F^{q \perp}_\mu (x, \eta, \Delta^2; Q^2)
\, .
\end{eqnarray}
At one-loop order, different components of the twist-three functions receive different
radiative corrections, so that we have to decompose them first into different irreducible
components. As was demonstrated in Ref.\ \cite{ManPilSteVanWei97}, the coefficient
functions related to  separate components for the leading twist even-parity Lorentz
structure ${\cal T}^{\mu\nu}_{(1)}$  are no longer  equal starting from one-loop order.
This is due to a nonvanishing contribution of the longitudinally polarized photon
scattering. The coefficient functions of twist-three Compton form factors are also
different for different components \cite{KivMan03}.

In the forward limit the discontinuity of CFFs is related to the structure functions
of deeply inelastic scattering (\ref{DISstructurefunctions}) via
\begin{equation}
\frac{1}{2 \pi} \Im{\rm m} {\cal F}_1 (x_{\rm B}, 0 , 0, {\cal Q}^2)
=
F_1 (x_{\rm B}, {\cal Q}^2)
\, , \qquad
\frac{1}{2 \pi} \Im{\rm m} \widetilde {\cal F}_1 (x_{\rm B}, 0 , 0, {\cal Q}^2)
=
g_1 (x_{\rm B}, {\cal Q}^2)
\, .
\end{equation}

\subsection{Twist-three GPDs}
\label{TwistThreeGPDs}

Having introduced twist-three GPDs, we are in a position to explore their
properties. In this section, we define twist-three skewed parton distributions
and their relation to the twist-two ones.

\subsubsection{Geometric twist decomposition of operators}
\label{TwistDecomp}

Now we are going to present a decomposition of the two-quark operators
into separate twist components to the twist-three accuracy
\cite{BalBra89,BalBraKoiTan98,BelMul00,KivPolSchTer00,RadWei00,KivPol00}.
An essential ingredient of our presentation is the treatment of operators
with total derivatives. Since the group-theoretical notion of twist as
dimension minus spin of an operator is well defined for local operators,
we will adhere to the following strategy: first, we do the Taylor expansion
of a light cone operator into an infinite series of local operators; second,
we extract of definite twist components; and as a final step, we resum the
result back into a nonlocal form. An alternative approach---directly in terms
of the light-ray operators---is described in Appendix \ref{LightRayTwistThree}.
In the following we will not take care of the trace terms proportional to
$n^\rho$, since they only contribute at the twist-four level. They will be
considered in Section \ref{SectionTargetMassCorrections}.

We start with the Taylor expansion of Eq.\ (\ref{OperatorOmu}) in terms of
local operators
\begin{equation}
{\cal O}^{qq}_\rho (- z^-, z^-)
= \sum_{j = 0}^{\infty} \frac{(- i z^-)^j}{j!}
n^{\mu_1} \dots n^{\mu_j} {\cal O}^{qq}_{\rho \mu_1 \dots \mu_j}
\, ,
\end{equation}
with
\begin{equation}
{\cal O}^{qq}_{\rho \mu_1 \dots \mu_j}
=
\Sym_{\mu_1 \dots \mu_j}
\bar\psi_q \gamma_\rho \,
i\!\stackrel{{}_\leftrightarrow}{\cal D}_{\mu_1}
\dots
i\!\stackrel{{}_\leftrightarrow}{\cal D}_{\mu_j}
\psi_q
\, .
\end{equation}
To have a well-defined decomposition into twist-two and twist-three contributions,
we separate the tensors ${\cal R}^{\tau, qq}_{\rho \mu_1 \dots \mu_j}$ with $(j + 1)$
indices into components transforming irreducibly with respect to the Lorentz group.
Since the indices $\mu_1 \dots \mu_j$ already form an irreducible Lorentz tensor
$\left( \ft{j}{2} , \ft{j}{2} \right)$ of spin $j$,  we have to decompose its product
with the free Lorentz index $\rho$ in the representation $(\ft12, \ft12)$, which
yields
$$
\left( \ft{j}{2} , \ft{j}{2} \right) \otimes (\ft12, \ft12)
=
\left( \ft{j + 1}{2} , \ft{j + 1}{2} \right)
\oplus
\left( \ft{j}{2} , \ft{j}{2} \right)
\, ,
$$
or,  using  Young tableaux
\begin{eqnarray*}
\unitlength0.4cm
\begin{picture}(32,1)
\linethickness{0.06mm}
\put(1,0){\framebox(1,1){$\scriptstyle \mu_1$}}
\put(2,0){\framebox(1,1){$\scriptstyle \mu_2$}}
\put(3,0){\framebox(1,1){$\scriptstyle \mu_3$}}
\put(4,0){\framebox(3,1){$\ldots$}}
\put(7,0){\framebox(1,1){$\scriptstyle \mu_j$}}
\put(9.3,0.15){$\otimes$}
\put(11,0){\framebox(1,1){$\scriptstyle \rho$}}
\put(13.2,0.15){$=$}
\put(15,0){\framebox(1,1){$\scriptstyle \rho$}}
\put(16,0){\framebox(1,1){$\scriptstyle \mu_1$}}
\put(17,0){\framebox(1,1){$\scriptstyle \mu_2$}}
\put(18,0){\framebox(3,1){$\ldots$}}
\put(21,0){\framebox(1,1){$\scriptstyle \mu_j$}}
\put(23.2,0.15){$\oplus$}
\put(25,-1){\framebox(1,1){$\scriptstyle \rho$}}
\put(25,0){\framebox(1,1){$\scriptstyle \mu_1$}}
\put(26,0){\framebox(1,1){$\scriptstyle \mu_2$}}
\put(27,0){\framebox(3,1){$\ldots$}}
\put(30,0){\framebox(1,1){$\scriptstyle \mu_j$}}
\put(32,0.35){$.$}
\end{picture}
\end{eqnarray*}
In terms of operators, we perform  the decomposition of the non-local operator
${\cal O}^{qq}_\rho$ into twist-two and twist-three components. This can easily
be done by merely symmetrizing and antisymmetrizing the index $\rho$ with the
remaining indices,
\begin{eqnarray}
\label{NonLoc-Loc-Ope}
{\cal O}^{qq}_\rho (- z^-, z^-)
\!\!\!&=&\!\!\!
{\cal R}^{2, qq}_\rho (- z^- , z^-)
+
{\cal R}^{3, qq}_\rho (- z^- , z^-)
\nonumber\\
&=& \sum_{j = 0}^{\infty} \frac{(- i z^-)^j}{j!}
n^{\mu_1} \dots n^{\mu_j}
\left\{ {\cal R}^{2,qq}_{\rho \mu_1 \dots \mu_j}
+
\frac{2 j}{j + 1} {\cal R}^{3, qq}_{\rho \mu_1 \dots \mu_j}
\right\}
\, ,
\end{eqnarray}
where the operator with geometric twist $\tau = 2$ was discussed before
(see Eqs.\ (\ref{QuarkLocalOperator}) and (\ref{GluonLocalOperator})),
while the $\tau = 3$ operator is given by
\begin{equation}
\label{TwistThreeLocalOperators}
{\cal R}^{3,qq}_{\rho \mu_1 \dots \mu_j}
=
\Sym_{\mu_1 \dots \mu_j} \Asym_{\rho \mu_1} \Sym_{\mu_1 \dots \mu_j}
\bar\psi_q \gamma_{\rho} \,
i\!\stackrel{{}_\leftrightarrow}{\cal D}_{\mu_1}
\dots
i\!\stackrel{{}_\leftrightarrow}{\cal D}_{\mu_j}
\psi_q
\, .
\end{equation}
The antisymmetrization is defined as
$$
\Asym_{\mu_1 \mu_2} t_{\mu_1\mu_2}
=
\ft12
\left( t_{\mu_1\mu_2} - t_{\mu_2\mu_1} \right)
\, .
$$

The twist-two light-cone operator was found in Eq.\ (\ref{TwistTwoOpenIndex}),
while the twist-three light-ray operator contains two distinct contributions,
evaluated in Appendix \ref{LightRayTwistThree}. The first contribution is due
to the twist-two operators with total derivatives \cite{BelMul00,KivPolSchTer00,%
Radyushkin:2000ap,Geyer:2001qf,Eilers:2004mp,Geyer:2004bx,Geyer:2004by}, the
second one comes from genuine three-particle twist-three operators \cite{BelMul00}.
Combining them together, one gets
\begin{eqnarray}
\label{R3-part}
{\cal R}^{3,qq}_\rho (- z^-, z^-)
=
\frac{z^-}{2} \int_{0}^{1} du
\Bigg\{\!
-
u \, i \varepsilon_\rho{}^{+ \mu \nu} \partial_\mu
\left[
\widetilde {\cal R}^{2,qq}_\nu \left( (\bar u - u) z^-, z^- \right)
+
\widetilde {\cal R}^{2,qq}_\nu \left( - z^- , (u - \bar u) z^- \right)
\right]
&&
\nonumber\\
-
u \,
\left( \partial_\rho n_\sigma - g_{\rho\sigma} \partial_+ \right) \,
\left[
{\cal R}^{2,qq}_\sigma \left( (\bar u - u) z^-, z^- \right)
-
{\cal R}^{2,qq}_\sigma \left( - z^- , (u - \bar u) z^- \right)
\right]
&&
\nonumber\\
+
z^- \int_{- u}^{u} d \tau
\Big[
(u - \tau) {\cal S}^{qgq}_{[+]\rho}
\left( (\bar u - u) z^- , (\tau + \bar u) z^-, z^- \right)
\qquad\qquad
&& \nonumber\\
-
(u + \tau) {\cal S}^{qgq}_{[-] \rho}
\left( - z^- , (\tau - \bar u) z^-, (u - \bar u) z^- \right)
\Big]
\Bigg\}
\, .
\qquad\!\!\!
&&
\end{eqnarray}
An analogous equation holds for the parity-odd case, with obvious replacements
${\cal R} \leftrightarrow \widetilde {\cal R}$ and ${\cal S} \leftrightarrow
\widetilde {\cal S}$. Compared to the forward scattering case, the expression for
${\cal R}^{3,qq}$ contains the opposite parity terms $\widetilde {\cal R}^{2,qq}$.
Notice also that the ``center-of-mass'' of two- and three-particle operators gets
shifted by a total translation $\exp \left( \pm i \bar u z^- \partial^+ \right)$.
This equation gives a relation between generalized parton distributions of different
``twists'' when sandwiched between hadronic states.

In terms of local operators, the Eq.\ (\ref{R3-part}) reads
\begin{eqnarray}
\label{Sol-R3}
{\cal R}^{3, qq}_{\rho; j}
\!\!\! &=&\!\!\! \frac{1}{2 j}
\sum_{l = 0}^{j - 1} (j - l) \, (i \partial^+)^l
\left\{
- \sigma_{l + 1} \, i \varepsilon_\rho{}^{+ \mu \nu} \, i\partial_\mu \,
\widetilde {\cal R}^{2, qq}_{\nu; j - l - 1}
-
\sigma_l \,
\left( i\partial_\rho n^\sigma - g_\rho{}^\sigma i \partial^+ \right) \,
{\cal R}^{2, qq}_{\sigma; j - l - 1}
\right\} \nonumber\\
&-&\!\!\! \frac{1}{j}
\sum_{l = 0}^{j - 2} \sum_{k = 1}^{j - l - 1}
(i \partial^+)^l
\left\{
( j - k - l ) {\cal S}^{qgq}_{[+]\rho; j - l, k}
-
(-1)^l k \, {\cal S}^{qgq}_{[-]\rho; j - l, k}
\right\}
\, ,
\end{eqnarray}
with
$$
\sigma_l = \frac{1}{2} [1 - (- 1)^l]
\, .
$$
Similar equation holds for  odd parity. It is obtained by   replacing
$\gamma^+ \to \gamma^+ \gamma^5$ and removing or dressing functions
with a tilde. In Eq.\ (\ref{Sol-R3}) we used the convention
$$
{\cal O}_{\rho; j} = n^{\mu_1} \dots n^{\mu_j}
{\cal O}_{\rho \mu_1 \dots \mu_j} \ ,
$$
and also introduced three-particle local operators
\begin{equation}
{\cal S}^{qgq}_{[\pm]\rho; j, k}
=
i g
\bar\psi_q
\,
(i \! \stackrel{{}_\leftrightarrow}{\partial}{\!}^+ )^{k - 1}
\left\{
\gamma^+ F^+{}^\perp_\rho
\pm
i \gamma^+ \gamma^5 \widetilde F^+{}^\perp_\rho
\right\}
(i \! \stackrel{{}_\leftrightarrow}{\partial}{\!}^+ )^{j - k - 1}
\psi_q
\, .
\end{equation}

\subsubsection{Twist-three GPDs for spin-zero target}
\label{Spin0}

Turning to the analysis of twist-three GPDs, it is instructive to consider first
the case of a spinless target \cite{AniPirTer00,BelMul00,KivPolSchTer00,%
Radyushkin:2000ap,BelKirMulSch00c,AniTer01,AniBinMedNogVen01,BisCudCugLanSta03}.
First, we construct the decomposition of GPDs with an open Lorentz index up to
the twist-three contributions \cite{AniPirTer00}
\begin{equation}
F^q_\mu = p_\mu H^q + \Delta^\perp_\mu H_3^q
\, , \qquad
\widetilde F^q_\mu = \widetilde \Delta^\perp_\mu \widetilde H_3^q
\, ,
\end{equation}
where
\begin{equation}
\label{DefDeltaTilde}
\widetilde\Delta^\rho_\perp
\equiv i \varepsilon_\perp^{\rho \sigma} \Delta_\rho^\perp
\, .
\end{equation}
The time-reversal invariance implies that
\begin{equation}
H^q_3 (x, - \eta, \Delta^2) =  - H^q_3 (x, \eta, \Delta^2)
\, , \qquad
\widetilde H^q_3 (x, - \eta, \Delta^2) =  \widetilde H^q_3 (x, \eta, \Delta^2)
\, .
\end{equation}

The matrix elements of the twist-two local operators have been defined in Section
\ref{Polynomiality} in Eqs.\ (\ref{LorentzVscal}). Next we define the reduced
matrix elements of the antiquark-gluon-quark operators. Since these operators
are partially antisymmetrized, we have obviously two vectors: $\Delta_{\rho}^\perp$
and its ``dual'' $\widetilde\Delta_{\rho}^\perp$ as possible tensor structures.
Thus, the general decomposition of the reduced matrix elements is
\begin{eqnarray}
\label{Def-SV-Loc-Red}
\langle p_2| {\cal S}^{qgq}_{[\pm] \rho; j, k} | p_1 \rangle
=
\Delta_{\rho}^\perp S^{qgq}_{[\pm] j, k} ( p^+ )^j
\, ,
\qquad
\langle p_2|  \widetilde {\cal S}^{qgq}_{[\pm] \rho; j, k}  | p_1 \rangle
=
\widetilde \Delta_{\rho}^\perp \widetilde{S}^{qgq}_{[\pm] j, k} (p^+)^j
\, .
\end{eqnarray}
The duality relation (\ref{Pro-Dual}) gives $\widetilde{S}^{qgq}_{[\pm] j, k}
= S^{qgq}_{[\pm] j, k}$ and reduces the number of independent contributions to
two instead of four. It proves convenient to introduce a mixed representation
for corresponding GPDs, such that the transformed functions that depend on the
position of the gluon field and a Fourier conjugate variable with respect to
the distance between the two quark fields:
\begin{eqnarray}
\langle p_2 |
\left\{
{ {\cal S}^{qgq}_{[\pm]\rho} ( - z^- , u z^- , z^- )
\atop
\widetilde {\cal S}^{qgq}_{[\pm]\rho} ( - z^- , u z^- , z^- ) }
\right\}
| p_1 \rangle
\!\!\!&=&\!\!\! ( p^+ )^2
\left\{{\Delta_{\rho}^\perp \atop \widetilde\Delta_{\rho}^\perp } \right\}
\int dx \, {\rm e}^{- i x p^+ z^-}
S^{qgq}_{[\pm]}( x, u, \eta )
\, .
\end{eqnarray}
Taking the moments with respect to the momentum fraction $x$ yields the
polynomials of order $j - 2$ in the variable $u$:
\begin{eqnarray}
S^{qgq}_{[\pm] j} (u, \eta)
\equiv
\int_{-1}^1 dx \, x^{j-2}
S^{qgq}_{[\pm]} (x, u, \eta)
= \sum_{k = 1}^{j - 1} \left( {j - 2 \atop k - 1 } \right)
\left( \frac{1 + u}{2} \right)^{k - 1}
\left( \frac{1 - u}{2} \right)^{j - k - 1} S^{qgq}_{[\pm] j, k} (\eta)
\, .
\end{eqnarray}
The functions $S^{qgq}_{[\pm] j, k}(\eta)$ are defined in Eq.\ (\ref{Def-SV-Loc-Red}).
They are polynomials in $\eta$ of order $j$. Finally, to obtain  the matrix element
of the operator ${\cal O}^{qq}_{\rho; j}$, we  insert our findings (\ref{Def-Tw2-Sca})
into the solution (\ref{Sol-R3}):
\begin{eqnarray}
&&
( p^+ )^{- j} \langle p_2| {\cal O}^{qq}_{\rho; j} | p_1 \rangle
= \left( p_{\rho} + \frac{\Delta_{\rho}^\perp}{\eta} \right)
H^q_{j + 1} (\eta)
+
\Delta_{\rho}^\perp
\left\{
\sum_{k = 0}^{j} \frac{\sigma_{j + 1 - k}}{j + 1} \eta^{j - k}
\left( \frac{d}{d\eta} - \frac{k + 1}{\eta} \right)
H^q_{k + 1}(\eta)
\right\}
\nonumber\\
&&\qquad
-
\Delta_{\rho}^\perp
\sum_{k = 2}^{j} \frac{k! \eta^{j - k}}{(j + 1)(k - 2)!}
\int_{-1}^1 du
\left\{
\frac{1 - u}{2} S^{qgq}_{[+] k} (u, \eta)
- (-1)^{j - k} \frac{1 + u}{2} S^{qgq}_{[-] k} (u, \eta)
\right\}
\, ,
\nonumber\\
\end{eqnarray}
where we have used the relations
\begin{eqnarray*}
\int_{-1}^1 du \frac{1 + u}{2} S^{qgq}_{[\pm] j} (u, \eta)
\!\!\!&=&\!\!\!
2
\sum_{k = 1}^{j - 1} \frac{k}{j (j - 1)} S^{qgq}_{[\pm] j,k} (\eta),
\\
\int_{-1}^1 du \frac{1 - u}{2} S^{qgq}_{[\pm] j} (u, \eta)
\!\!\!&=&\!\!\!
2
\sum_{k = 1}^{j - 1} \frac{j - k}{j (j - 1)} S^{qgq}_{[\pm] j,k} (\eta)
\, .
\end{eqnarray*}

The final step is a summation of local operators, see Eq.\ (\ref{NonLoc-Loc-Ope}),
that yields an expression in terms of the usual twist-two GPDs $H^q (x, \eta)$ and
the mixed representation three-particle functions $S^{qgq}_{[\pm]} (y, u, \eta)$,
\begin{eqnarray}
\label{Res-Spi-O}
F^q_\mu (x, \eta)
\!\!\!&=&\!\!\!
\left( p_\mu + \frac{\Delta^\perp_\mu}{\eta} \right) H^q (x, \eta)
+
\frac{\Delta_\mu^\perp}{\eta}  \int_{-1}^{1}\frac{dy}{\eta} \,
W_+ \! \left( \frac{x}{\eta}, \frac{y}{\eta} \right)
\left( \eta \frac{\partial}{\partial\eta} + y \frac{\partial}{\partial y} \right)
H^q (y,\eta)
\!\!\!\!\!\\
&-&\!\!\!
\Delta_{\rho}^\perp
\int_{-1}^1 \frac{dy}{\eta} \int_{-1}^1 du
\bigg\{ \frac{1 - u}{2}
W^{\prime\prime} \! \left( \frac{x}{\eta}, \frac{y}{\eta} \right)
S^{qgq}_{[+]} (y, u, \eta)
+ \frac{1 + u}{2}
W^{\prime\prime} \! \left( - \frac{x}{\eta}, - \frac{y}{\eta} \right)
S^{qgq}_{[-]} (y, u, \eta) \bigg\}
\, . \!\!\!\!\nonumber
\end{eqnarray}
Here we have introduced the kernel
\begin{eqnarray}
W (x, y)
\!\!\!&\equiv&\!\!\!
\vartheta_{11}^0 (1 - x, y - x)
\\
&=&\!\!\!
\frac{\theta (1 - x) \theta (x - y) - \theta (x - 1) \theta(y - x)}{1 - y}
\, , \nonumber
\end{eqnarray}
which enters in symmetric and antisymmetric combinations
\begin{equation}
\label{Wpm}
W_\pm (x, y) = \frac{1}{2} \left\{ W (x, y) \pm W(- x, -y) \right\}
\, .
\end{equation}
Notice that we encounter the same generalized step function as in the section
devoted to the evolution of GPDs (see  Appendix \ref{GeneralStepFunctions}
for a summary of formulas involving these functions). We also used the notation
$$
W^{\prime\prime} \left( \pm \frac{x}{\eta}, \pm \frac{y}{\eta} \right)
\equiv
\frac{d^2}{dy^2}
W\!
\left( \pm \frac{x}{\eta}, \pm \frac{y}{\eta} \right)
$$
and incorporated the following result for the Mellin moments of the $W$-kernels
\begin{equation}
\label{MellinMomentsOfW}
\int_{-1}^1 \frac{dx}{\eta}
x^j W \! \left( \frac{x}{\eta}, \frac{y}{\eta} \right)
= \sum_{k = 0}^{j} \frac{\eta^{j - k} y^k}{j + 1}
\end{equation}
performing the restoration of the non-local form.

The only difference in the axial-vector case is that the corresponding operator
matrix element does not possess the twist-two part for a spinless target. However,
a non-vanishing twist-two contribution of the vector operator induces a kinematical
term due to the Levi-Civita tensor in Eq.\ (\ref{Sol-R3}). Thus, the twist-three
functions contain two parts, namely, the kinematical piece, expressed in terms of
the twist-two function, and the term given as a correlation function of
antiquark-gluon-quark operators \cite{BelMul00}:
\begin{eqnarray}
H^q_3 (x, \eta)
\!\!\!&=&\!\!\!
\frac{1}{\eta}
\int_{-1}^{1}\frac{dy}{\eta}
W_+ \! \left( \frac{x}{\eta}, \frac{y}{\eta} \right)
\left(
\eta \frac{\partial}{\partial\eta}
+ y \frac{\partial}{\partial y}
\right) H^q (y, \eta)
+
\frac{1}{\eta} \ H^q (x,\eta)
- S^{qgq}_{[+]} (x, \eta) , \nonumber\\
{\widetilde H}^q_3 (x, \eta)
\!\!\!&=&\!\!\!
\frac{1}{\eta}
\int_{-1}^{1} \frac{dy}{\eta}
W_-\! \left( \frac{x}{\eta}, \frac{y}{\eta} \right)
\left(
\eta \frac{\partial}{\partial\eta}
+
y \frac{\partial}{\partial y}
\right) H^q (y, \eta) - S^{qgq}_{[-]} (x, \eta)
\, .
\end{eqnarray}
Note that the first moment of these functions
with respect to $x$ vanishes. The  new dynamical
information is contained in the antiquark-gluon-quark GPDs \cite{BelMul00}
\begin{equation}
S^{qgq}_{\pm} (x, \eta)
= \int_{-1}^1 \frac{dy}{\eta} \int_{-1}^1 du
\left\{
\frac{1 - u}{2}
W^{\prime\prime} \! \left( \frac{x}{\eta},\frac{y}{\eta} \right)
S^{qgq}_{[+]} (y, u, \eta)
\pm \frac{1 + u}{2}
W^{\prime\prime} \! \left( - \frac{x}{\eta}, - \frac{y}{\eta} \right)
S^{qgq}_{[-]} (y, u, \eta)
\right\}
\, .
\end{equation}

\subsubsection{Twist-three GPDs for spin-one-half target}

Let us address the case of realistic spin-one-half target. For the vector
channel, the analysis repeating the one we outlined in the previous section
yields
\begin{eqnarray}
\label{VectorTw3Nucleon}
F^q_\mu (x, \eta)
\!\!\!&=&\!\!\!
\left( p_\mu + \frac{\Delta^\perp_\mu}{\eta} \right)
\left( \frac{h^+}{p^+} H^q (x, \eta) + \frac{e^+}{p^+} E^q (x, \eta) \right)
\\
&+&\!\!\!
\int_{-1}^{1} \frac{d y}{\eta}
\left\{
W_+ \! \left( \frac{x}{\eta}, \frac{y}{\eta} \right)
G^q_\mu (y, \eta)
+
W_- \! \left( \frac{x}{\eta}, \frac{y}{\eta} \right)
i \varepsilon^\perp_{\mu\nu} \widetilde{G}^{q, \nu} (y, \eta)
\right\}
\nonumber\\
&-&\!\!\!
\int_{-1}^{1} \frac{d y}{\eta}
\int_{-1}^1 du
\bigg\{
\frac{1 - u}{2}
W^{\prime\prime} \! \left( \frac{x}{\eta}, \frac{y}{\eta} \right)
S^{qgq}_{[+]\mu} (y, u, \eta)
+
\frac{1 + u}{2}
W^{\prime\prime} \! \left( - \frac{x}{\eta}, - \frac{y}{\eta} \right)
S^{qgq}_{[-]\mu} (y, u, \eta)
\bigg\}
\, ,
\nonumber
\end{eqnarray}
and similarly for the axial sector,
\begin{eqnarray}
\label{AxialTw3Nucleon}
\widetilde{F}^q_\mu (x, \eta)
\!\!\!&=&\!\!\!
\left( p_\mu + \frac{\Delta^\perp_\mu}{\eta} \right)
\left(
\frac{\widetilde{h}^+}{p^+} \widetilde{H}^q (x, \eta)
+
\frac{\widetilde{e}^+}{p^+} \widetilde{E}^q (x, \eta) \right)
\\
&+&\!\!\!
\int_{-1}^{1} \frac{d y}{\eta}
\left\{
W_+ \! \left( \frac{x}{\eta}, \frac{y}{\eta} \right)
\widetilde{G}^q_\mu (y, \eta)
+
W_- \! \left( \frac{x}{\eta}, \frac{y}{\eta} \right)
i \varepsilon^\perp_{\mu\nu} G^{q, \nu} (y, \eta)
\right\}
\nonumber\\
&-&\!\!\!
\int_{-1}^{1} \frac{d y}{\eta}
\int_{-1}^1 du
\bigg\{
\frac{1 - u}{2}
W^{\prime\prime} \! \left( \frac{x}{\eta}, \frac{y}{\eta} \right)
\widetilde{S}^{qgq}_{[+]\mu} (y, u, \eta)
+
\frac{1 + u}{2}
W^{\prime\prime} \! \left( - \frac{x}{\eta}, - \frac{y}{\eta} \right)
\widetilde{S}^{qgq}_{[-]\mu} (y, u, \eta)
\bigg\}
\, .
\nonumber
\end{eqnarray}
Here we have introduced a shorthand notation for the combinations
\begin{eqnarray}
G^{q, \mu} (x, \eta)
\!\!\!&=&\!\!\!
\left( h^\mu - p^\mu \frac{h^+}{p^+} \right)
\left( H^q (x, \eta) + E^q (x, \eta) \right)
\\
&+&\!\!\!
\frac{\Delta^\mu_\perp}{\eta}
\left\{
\frac{h^+}{p^+}
\left( \eta \frac{\partial}{\partial\eta} + x \frac{\partial}{\partial x} \right)
\big( H^q (x, \eta) + E^q (x, \eta) \big)
-
\frac{b}{2 M_N}
\left( \eta \frac{\partial}{\partial\eta} + x \frac{\partial}{\partial x} \right)
E^q (x, \eta)
\right\}
\, , \nonumber\\
\widetilde{G}^{q, \mu} (x, \eta)
\!\!\!&=&\!\!\!
\left( \widetilde{h}^\mu - p^\mu \frac{\widetilde{h}^+}{p^+} \right)
\widetilde{H} (x, \eta)
\\
&+&\!\!\!
\frac{\Delta^\mu_\perp}{\eta}
\left\{
\frac{\widetilde{h}^+}{p^+}
\left( \eta \frac{\partial}{\partial\eta} + x \frac{\partial}{\partial x} \right)
\widetilde{H} (x, \eta)
-
\frac{\widetilde{b}}{2 M_N}
\left( \eta \frac{\partial}{\partial\eta} + x \frac{\partial}{\partial x} \right)
\eta \widetilde{E} (x, \eta)
\right\}
\, . \nonumber
\end{eqnarray}
Next, for the time being we introduce three-particle GPDs without their explicit
decomposition in spinor bilinears,
\begin{eqnarray}
\langle p_2| \left\{
{
{\cal S}^{qgq}_{[\pm] \mu} (- z^-, u z^-, z^-)
\atop
\widetilde{\cal S}^{qgq}_{[\pm] \mu} (- z^-, u z^-, z^-)
}
\right\} | p_1 \rangle
=
(p^+)^2 \int dx \, {\rm e}^{- i x p^+ z^-}
\left\{
{
S^{qgq}_{[\pm] \mu} (x, u, \eta)
\atop
\widetilde{S}^{qgq}_{[\pm] \mu} (x, u, \eta)
}
\right\}
\, .
\end{eqnarray}

A general decomposition of the vector and axial-vector GPDs, in a complete basis
of structures to twist-three accuracy, reads
\begin{eqnarray}
\label{Def-V1}
F^q_\mu
\!\!\!&=&\!\!\!
p_\mu \frac{h^+}{p^+} H^q
+
p_\mu \frac{e^+}{p^+} E^q
+
\Delta_\mu^\perp \frac{h^+}{p^+} H^q_{3 +}
+
\Delta_\mu^\perp \frac{e^+}{p^+} E^q_{3 +}
+
\widetilde{\Delta}_\mu^\perp
\frac{\widetilde h^+}{p^+} \widetilde{H}^q_{3 -}
+
\widetilde{\Delta}_\mu^\perp
\frac{\widetilde e^+}{p^+} \widetilde{E}^q_{3 -}
\, ,\\
\label{Def-A1}
\widetilde F^q_\mu
\!\!\!&=&\!\!\!
p_\mu \frac{\widetilde h^+}{p^+} \widetilde H^q
+
p_\mu \frac{\widetilde e^+}{p^+} \widetilde E^q
+
\Delta_\mu^\perp \frac{\widetilde h^+}{p^+} \widetilde H^q_{3 +}
+
\Delta_\mu^\perp \frac{\widetilde e^+}{p^+} \widetilde E^q_{3 +}
+
\widetilde{\Delta}_\mu^\perp \frac{h^+}{p^+} H^q_{3 -}
+
\widetilde{\Delta}_\mu^\perp \frac{e^+}{p^+} E^q_{3 -}
\, .
\end{eqnarray}
The parametrization of the correlation functions $S$ and $\widetilde S$ then
can be cast in an analogous form,
\begin{eqnarray}
\label{TreeParticleDecomposition}
S^{qgq}_{[\pm] \mu} (x, u, \eta)
\!\!\!&=&\!\!\! \Delta_\mu^\perp \frac{b}{2 M}\ S_1^{\pm}
+
\Delta_\mu^\perp \frac{h^+}{p^+} S_2^{\pm}
+
\widetilde\Delta_\mu^\perp \frac{\widetilde b}{2 M} S_3^{\pm}
+
\widetilde\Delta_\mu^\perp \frac{\widetilde h^+}{p^+} S_4^{\pm}
,\nonumber\\
\widetilde S^{qgq}_{[\pm] \mu} (x,u,\eta)
\!\!\!&=&\!\!\!
\Delta_\mu^\perp \frac{\widetilde b}{2 M} \widetilde{S}_1^{\pm}
+ \Delta_\mu^\perp \frac{\widetilde h^+}{p^+}  \widetilde{S}_2^{\pm}
+ \widetilde\Delta_\mu^\perp \frac{b}{2 M} \widetilde{S}_3^{\pm}
+ \widetilde \Delta_\mu^\perp \frac{h^+}{p^+} \widetilde{S}_4^{\pm} ,
\end{eqnarray}
where duality (\ref{Pro-Dual}) requires $S_i^\pm = \widetilde{S}_i^\pm$.

In order to reduce the above GPDs to the form of Eqs.\ (\ref{Def-V1}) and
(\ref{Def-A1}), we decompose the (axial-) vector Dirac bilinears $h_\mu$
($\widetilde h_\mu$) in its twist-two and -three components making use of
the Dirac equation
\begin{eqnarray}
\label{Dec-FouVec}
h^\mu
\!\!\!&=&\!\!\!
p^\mu \frac{h^+}{p^+}
-
\eta \frac{\Delta^\mu_\perp}{\Delta^2_\perp}
\left\{
\Delta^2 \frac{h^+}{p^+} - 4 M_N^2 \frac{e^+}{p^+}
\right\}
-
\frac{\widetilde\Delta^\mu_\perp}{\Delta^2_\perp}
\left\{
\Delta^2 \frac{\widetilde{h}^+}{p^+} - 4 M_N^2 \frac{\widetilde{e}^+}{p^+}
\right\}
\, ,
\nonumber\\
\widetilde h^\mu
\!\!\!&=&\!\!\!
p^\mu \frac{\widetilde h^+}{p^+}
-
\eta \frac{\Delta^\mu_\perp}{\Delta^2_\perp}
\left\{
(\Delta^2 - 4 M_N^2) \frac{\widetilde h^+}{p^+}
-
4 M_N^2 \frac{1}{\eta^2} \frac{\widetilde e^+}{p^+}
\right\}
-
\frac{\widetilde\Delta^\mu_\perp}{\Delta^2_\perp}
\left\{
\Delta^2 \frac{h^+}{p^+} - 4 M_N^2 \frac{e^+}{p^+}
\right\}
\, .
\nonumber
\end{eqnarray}
Here twist-four terms, proportional to $n^\ast_\mu$, have been neglected. The
transverse momentum squared is re-expressed in terms of the momentum difference,
\begin{eqnarray*}
\Delta_\perp^2
\simeq (1 - \eta^2) \left( \Delta^2 - \Delta^2_{\rm min} \right)
\, , \qquad
\Delta^2_{\rm min} \simeq - 4 M_N^2 \frac{\eta^2}{1 - \eta^2}
\, .
\end{eqnarray*}
Note, that later on when computing physical amplitudes contributing to various
exclusive pro\-cesses, one can safely replace the light-like vector $n^\mu$ by
$q^\mu$, such that to twist-four accuracy
$$
\frac{v^+}{p^+} \to \frac{(v \cdot q)}{(p \cdot q)}
\, .
$$

According to our discussion, all twist-three GPDs are decomposed into the
so-called Wandzura-Wilczek (WW) term $F^{\scriptscriptstyle\rm WW}_\pm$ and
a function $F^{qgq}_\pm $ that contains dynamical information arising from
antiquark-gluon-quark correlations:
\begin{eqnarray}
F^q_{3\pm} = F^q_{{\scriptscriptstyle\rm WW} \pm} + F^{qgq}_\pm
\, ,
\end{eqnarray}
where $F$ is a unified notation for GPDs of different species
$$
F = \{ H, E, \widetilde H, \eta \widetilde E \}
\, .
$$
Notice that the pion-pole function $\widetilde E$ is accompanied by the skewness
$\eta$. As we just discussed, the WW parts are expressed solely in terms of the
twist-two functions and have the following form
\begin{eqnarray}
\label{Def-Ftw3}
&& F_{{\scriptscriptstyle\rm WW} +} (x, \eta)
=
\frac{1}{\eta}
\int_{-1}^{1} \frac{d y}{\eta} \,
W_+ \left( \frac{x}{\eta}, \frac{y}{\eta} \right)
\left(
y \frac{\partial}{\partial y}
+
\eta \frac{\partial}{\partial \eta}
\right)
F (y, \eta)
+
\frac{1}{\eta} F (x, \eta)
-
\frac{4 M^2_N}{\Delta^2_\perp} F^\perp_+ (x, \eta)
\, ,
\nonumber\\
&& F_{{\scriptscriptstyle\rm WW} -} (x, \eta)
=
\frac{1}{\eta}
\int_{- 1}^{1} \frac{d y}{\eta} \,
W_- \left( \frac{x}{\eta}, \frac{y}{\eta} \right)
\left(
y \frac{\partial}{\partial y}
+
\eta \frac{\partial}{\partial \eta}
\right)
F (y, \eta)
-
\frac{4 M^2_N}{\Delta^2_\perp} F^\perp_- (x, \eta)
\, ,
\end{eqnarray}
for the `$+$' and `$-$' component of the twist-three GPDs in the WW-approximation,
respectively. The functions $F^\perp_\pm$ specifically appear for the spin-one-half
targets and read
\begin{eqnarray}
H^\perp_\pm (x, \eta)
\!\!\!&=&\!\!\! \frac{\Delta^2}{4 M^2_N} \int_{- 1}^{1} \frac{d y}{\eta}
\left\{
\eta W_\pm \left( \frac{x}{\eta}, \frac{y}{\eta} \right)
\left( H + E \right) (y, \eta)
+
W_\mp \left( \frac{x}{\eta}, \frac{y}{\eta} \right)
\widetilde H (y, \eta)
\right\} \, ,
\nonumber\\
E^\perp_\pm (x, \eta)
\!\!\!&=&\!\!\! - \int_{- 1}^{1} \frac{d y}{\eta}
\left\{
\eta W_\pm \left( \frac{x}{\eta}, \frac{y}{\eta} \right)
\left( H + E \right) (y, \eta)
+
W_\mp \left( \frac{x}{\eta}, \frac{y}{\eta} \right)
\widetilde H (y, \eta) \right\},
\nonumber\\
\widetilde H^\perp_\pm (x, \eta)
\!\!\!&=&\!\!\!
\int_{- 1}^{1} \frac{d y}{\eta}
\left\{ \eta \left( \frac{\Delta^2}{4 M^2_N} - 1 \right)
W_\pm \left( \frac{x}{\eta}, \frac{y}{\eta} \right)
\widetilde H (y, \eta)
+
\frac{\Delta^2}{4 M^2_N} W_\mp \left( \frac{x}{\eta}, \frac{y}{\eta} \right)
\left( H + E \right) (y, \eta)
\right\}
\, ,
\nonumber\\
\eta \widetilde E^\perp_\pm (x, \eta)
\!\!\!&=&\!\!\! - \int_{- 1}^{1} \frac{d y}{\eta}
\left\{
W_\pm \left( \frac{x}{\eta}, \frac{y}{\eta} \right) \widetilde H (y, \eta)
+
\eta W_\mp \left( \frac{x}{\eta}, \frac{y}{\eta} \right)
\left( H + E \right) (y, \eta) \right\}
\, .
\end{eqnarray}
The antiquark-gluon-quark contributions can be read off from the parametrization
of the corresponding operators in Eq.\ (\ref{TreeParticleDecomposition}).

It is important to realize that the scale dependence of the twist-three GPDs
$F^q_\mu$ and $\widetilde F^q_\mu$ is not homogeneous, namely, their different
components have their own autonomous evolution in hard momentum transfer
${\cal Q}^2$. While in the WW-approximation the functions are expressed in
terms of standard twist-two GPDs, obeying the renormalization group equations
discussed at length in the previous chapter, the three-particle correlation
function satisfy an independent equation. The thorough analysis of the
latter reveals hidden symmetries of the problem which leads to its complete
integrability \cite{BDM98,Belitsky99b,DKM00,BKM00}. However, this is a subject
of separate reviews \cite{BKM03,BelBraGorKor04}.

\subsubsection{Properties of the kinematical kernels}

In Section \ref{Spin0} we introduced the notion of the WW-kernels (\ref{Wpm}).
Here, we consider their general properties \cite{KivPol00}. To make formulas
appear concise, let us adopt the following notations for the action of the
WW-kernels on a test function $\tau (x,\eta)$
\begin{eqnarray}
\label{conv}
{}[ W_\pm \otimes \tau ] (x,\eta)
\equiv
\int_{-1}^1 \frac{d y}{\eta}
W_\pm \left( \frac{x}{\eta} , \frac{y}{\eta} \right) \tau( y, \eta)
\, ,
\end{eqnarray}
The functions $[W_\pm \otimes \tau]$ resulting from the WW-transformation
will be referred to as the WW-transform.

It is instructive to analyze two limiting cases of the WW-transformation,
namely, the forward limit $\eta\to 0$ and the exclusive limit $\eta \to 1$,
as well as address the issue of polynomiality and singularity structure of
the WW kernels.

\noindent $\bullet$ In the forward limit, we easily obtain
\begin{eqnarray}
\label{fwd}
\lim_{\eta\to 0} [W_+ \otimes \tau] (x, \eta)
\!\!\!&=&\!\!\!
\theta(x)
\int_x^1 \frac{d y}{y} \, \tau (y, \eta = 0)
-
\theta(- x)
\int_{-1}^x \frac{d y}{y} \, \tau (y, \eta = 0)
\, , \nonumber\\
\lim_{\eta\to 0}
[W_- \otimes \tau] (x, \eta)
\!\!\!&=&\!\!\! 0
\, .
\end{eqnarray}
We can see that the action of the $W_+$ in the forward limit reproduces the
Wandzura--Wilczek relation for the spin structure function $g_T = g_1 + g_2$
\cite{WW}. The term with $\theta(x)$ corresponds to quark distributions
$$
\lim_{\Delta \to 0}
\widetilde{F}^{q\perp}_{\phantom{q}\mu} (x, \eta, \Delta^2)|_{x > 0}
=
h^\perp_\mu
\left\{
\Delta q (x) + \int_x^1 \frac{dy}{y} \Delta q (y)
\right\}
\, ,
$$
while the term with $\theta(- x)$ to antiquark ones. The $W_-$ kernel disappears
in the forward limit, so that this kernel is specific to off-forward kinematics.

\noindent $\bullet$ In the $\eta \to 1$ limit, the generalized parton distributions
have the properties of meson distribution amplitudes as we demonstrated in Section
\ref{PartonicContentGPDs}. In this limit, the WW-transforms have the form
\begin{eqnarray*}
\label{meson}
\lim_{\eta\to 1} [W_\pm \otimes \tau] (x, \eta)
\!\!\!&=&\!\!\!
\frac{1}{2}
\left\{
\int_{-1}^x \frac{d y}{1 - y}
\, \varphi (y)
\pm
\int_{x}^1
\frac{d y}{1 + y}
\, \varphi (y)
\right\}
\, ,
\end{eqnarray*}
where we used the notation $\tau (x, \eta = 1) = \varphi (x)$. This equation
corresponds to the WW-relations for meson distribution amplitudes derived
in Refs.\ \cite{BalBra96,BalBraKoiTan98}. So, we can say that the WW-transform
of GPDs interpolates between the WW-relations for parton distributions and
those for meson distribution amplitudes. Furthermore, the   general form of
the WW-kernels (\ref{Wpm}) allows one to derive WW relations for distribution
amplitudes of mesons with arbitrary spin.

\noindent $\bullet$ One can easily compute the Mellin moments of the WW-transform.
This was already used above in Eq.\ (\ref{MellinMomentsOfW}). The result for
WW-transform with the $W_\pm$ kernels is
\begin{eqnarray}
\label{Mellin}
\int_{-1}^1 d x \, x^j [W_\pm \otimes \tau] (x, \eta)
=
\frac{1}{j + 1}
\int_{-1}^1 dy
\left\{
\frac{y^{j + 1} - \eta^{j + 1}}{y - \eta}
\pm
\frac{y^{j + 1}- (-\eta)^{j + 1}}{y + \eta}
\right\} \tau (u,\eta)
\, .
\end{eqnarray}
Wherefrom we can see an important property of the WW-transformation, namely,
if the function $\tau (x, \eta)$ satisfies the polynomiality condition
\begin{eqnarray}
\int_{-1}^1 d x \, x^j \, \tau (x, \eta)
=
\sum_{k = 0}^{j + 1} \eta^k \tau_k
\, ,
\end{eqnarray}
i.e., it is a polynomial of order $j + 1$, so does its WW-transform.

Let us consider now two special cases of these Mellin moments, $j = 0,1$.
These particular moments of twist-three GPDs do not receive contribution
from the ``genuine twist-three'' quark-gluon operators, therefore, for
these particular cases the WW-approximation gives exact results. In the
forward limit this observation leads to Burkhard-Cottingham ($j = 0$)
\cite{BC} and Efremov-Teryaev-Leader ($j = 1$) \cite{ELT} sum rules for
polarized structure function $g_T$. Generalizations of these sum rules
for generalized parton distributions have been discussed in Refs.\
\cite{PenPolShuStr00,KivPolSchTer00}.

\noindent $\bullet$ In Ref.\ \cite{KivPolSchTer00} it was demonstrated that the
twist-three GPDs in the WW-approximation exhibit discontinuities at the points
$x=\pm\eta$. This feature stems from the analogous properties of the WW-kernels.
Namely, the discontinuities of a WW-transform at the points $x= \pm \eta$ yields
\begin{eqnarray}
\lim_{\varepsilon\to 0}
\Big[
{}[ W_\pm \otimes \tau ] (\eta + \varepsilon, \eta)
-
{}[ W_\pm \otimes \tau ] (\eta - \varepsilon, \eta)
\Big]
\!\!\!&=&\!\!\!
\frac{1}{2}
{\rm PV}
\int_{-1}^1 \frac{d y}{y - \eta} \tau (y, \eta)
\, , \nonumber\\
\lim_{\varepsilon\to 0}
\Big[
{}[W_\pm \otimes \tau] (- \eta + \varepsilon , \eta)
-
{}[W_\pm \otimes \tau] (- \eta - \varepsilon , \eta)
\Big]
\!\!\!&=&\!\!\!
\pm\frac{1}{2}
{\rm PV}
\int_{-1}^1 \frac{d y}{y + \eta} \tau (y, \eta)
\, .
\end{eqnarray}
We see that for a very wide class of functions $\tau (u,\eta)$ the discontinuity
of the corresponding WW-transforms is nonzero. This nonanalytic behavior of the
WW-transformation may result in violation of perturbative factorization for the
twist-three DVCS amplitude, since the singularity of the hard coefficient function
$x = \pm \xi$ overlaps the jump of the twist-three GPDs $x = \pm \eta = \pm \xi$.

\subsubsection{Discontinuities of twist-three GPDs}

Using the discontinuity structure of the WW-kernels one can find the behavior
of twist-three GPDs for spinless targets in the vicinity of the ``turning''
point $|x| = \eta$ \cite{KivPol00,RadWei00,Radyushkin:2000ap},
\begin{eqnarray}
\label{jumps}
&&H^q_3 (\eta + 0,\eta) - H^q_3 (\eta - 0,\eta)
\\
&&\hspace{2cm} =
\frac{1}{2}{\rm PV}\int_{-1}^1 \frac{dy}{\eta} \frac{1}{y - \eta}
\left\{
\left(
\eta \frac{\partial}{\partial\eta}
+
y \frac{\partial}{\partial y}
\right) H (y, \eta)
-
2 \frac{\partial^2}{\partial y^2}
\int_{-1}^1 du (1 + u) S^{qgq}_{[-]} (y, u, \eta)
\right\}
\, .
\nonumber
\end{eqnarray}
Here we have tacitly assumed that GPDs vanish at the boundary $x = \pm 1$ (see
Eq.\ \ref{VanishGPDoutside})). Obviously, these discontinuities are not an
artifact of the WW-approximation. Rather, they are intrinsic to the procedure
of separation of twist-two and -three contributions.

Notice however, that the combination
\begin{equation}
F^q_\mu (x, \eta, \Delta^2)
- i \varepsilon_{\mu\nu}^\perp
\widetilde{F}^{q \nu} (x, \eta, \Delta^2)
\, ,
\end{equation}
is free from discontinuities at $x = \eta$, while
\begin{equation}
F^q_\mu (x, \eta, \Delta^2)
+ i \varepsilon_{\mu\nu}^\perp
\widetilde{F}^{q \nu} (x, \eta, \Delta^2)
\, ,
\end{equation}
is continuous at $x = - \eta$. Fortunately only these combinations contribute
to the VCS amplitude to twist-four accuracy as we will discuss below.

\subsubsection{Factorization of the Compton amplitude at twist-three}
\label{FactorizationComptonAmplitude}

An important observation is that the observed divergences of the twist-three
DVCS amplitudes cancel in certain combinations of helicity amplitudes
\cite{KivPol00,RadWei00,Radyushkin:2000ap}. This is possible because, in
the Wandzura-Wilczek approximation, the discontinuities of contributing
twist-three GPDs are related to each other. Since the divergences occur
only at the points $x = \pm \xi$, it is sufficient that the jumps cancel
at these specific values to save factorizability. As we will see, though
divergent terms are formally present in $T^{\mu\nu}$, they do not contribute
to the physical Compton amplitude with a real, transverse photon in the
final state.

In particular, there is no singularity problem for the amplitudes with
longitudinal polarization of the virtual photon. Let us demonstrate this.
We can choose a frame in which the four momentum $q_1^\mu$ of the incoming
photon has no transverse components, i.e., $q_{1\perp}^\mu = 0$, and the
final-state photon carries the total transverse momentum transfer $q_{2 \perp}^\mu
= \Delta^\mu_\perp$. Then we construct the explicit form of the longitudinal
polarization vector of the incoming virtual photon. In this frame the tensor
structures (\ref{LorentzOfTwistThree}) of the Compton amplitude in terms
of the light-cone vectors loose their symmetric form, for instance,
$\widetilde{\cal T}^{\mu\nu}_{(1)} = i (\varepsilon^{\mu\nu\rho\sigma}
p_\rho q_\sigma + n^{\ast\mu} \varepsilon_\perp^{\nu\rho} \Delta^\perp_\rho)
/(p \cdot q)$. Imposing the usual normalization and orthogonality conditions,
we find for the incoming polarization vector
\begin{equation}
\label{LongPolVector}
\varepsilon^\mu_L (q_1)
=
\frac{1}{Q}
\left(
1 + \frac{\eta}{\xi}
+ \delta^2 \left( 2 \xi + \eta \right)^2
\right)^{-1/2}
\left(
q^\mu + \ft12 (2 \xi + \eta) p^\mu
\right)
\, ,
\end{equation}
with $\delta^2$ determined in Eq.\ (\ref{DefinitionOfDeltaEtAl}). The projection
of the Compton scattering amplitude (\ref{VCSamplitude}), with imposed on-shell
$q_2^2 = 0$ condition for  the outgoing photon, gives
\begin{equation}
T_{\mu\nu} \varepsilon^\nu_L (q_1)
\simeq
\frac{Q}{\xi^2 \sqrt{1 + \xi}} n_\mu \left( {\cal F}_2 - \xi {\cal F}_1 \right)
+
\frac{2 \xi}{Q \sqrt{1 + \xi}}
\left(
{\cal F}^\perp_{3 \mu}
-
i \varepsilon_{\mu\nu}^\perp \widetilde{\cal F}^\nu_{3 \perp}
\right)
\, .
\end{equation}
The first term does not represent any danger. Moreover, it vanishes at leading
order of perturbation theory due to the generalized Callan-Gross relation
(\ref{GeneralCallanGross}). It is the second term which is the structure of
interest. A closer inspection for a scalar target immediately suggests that,
it reduces to a combination of functions $H_3$ and $\widetilde{H}_3$,
\begin{equation}
\label{Tlong}
{\cal H}^3 - \widetilde{\cal H}^3
=
\sum_q \int_{-1}^1 \frac{dx}{x} C^{q[-]}_{(0)} (x, \xi)
\left( x H^q_3 (x, \xi) + \xi \widetilde H^q_3 (x, \xi) \right)
\, ,
\end{equation}
which is singularity-free according to Eq.\ (\ref{jumps}) and similar equation
for $\widetilde{H}_3$.

Analogous cancellations occur for the spin-one-half target. On the other hand,
a further analysis demonstrates that the DVCS amplitude on the nucleon with
transversely polarized virtual photon contains a different combination of
twist-three GPDs, namely,
$$
{\cal F}^\perp_{3 \mu}
+
i \varepsilon_{\mu\nu}^\perp \widetilde{\cal F}^\nu_{3 \perp}
\, ,
$$
and it does possess uncancelled discontinuities. However, in cross sections
for physical observables, the contribution of this ``problematic" part of the
twist-three DVCS amplitude is accompanied by a contracted polarization vector
of the emitted real photon. Thus is turns out to be suppressed by two powers
of the hard scale, relative to the leading order result. Thus, it does not
contribute to observables at the level of ${\cal O} (1/Q)$ power corrections
and is beyond the limits of approximations involved. In other words, the DVCS
differential cross section to the twist-three accuracy gets contributions only
from the longitudinal part of twist-three amplitudes which is free from
divergences, which could threaten the validity of the factorization.

\subsubsection{Sum rules for twist-three GPDs}

As we demonstrated in our previous discussion, the leading-twist GPDs
are related to hadronic form factor via integral relations---sum rules.
Let us analyze in the same fashion a few lowest Mellin moments of twist-three
GPDs. We established in the preceding section that the twist-three GPDs are
expressible in terms of ``known'' twist-two ones and multiparticle correlation
functions. Since the moments of GPDs are related to matrix elements of local
Wilson operators, thus a Mellin moment of a twist-three GPD will be a sum of
two- and three-particle operators. In order to compensate for the deficit in
the mass dimension of a local two-particle compared to a quark-gluon-quark
operator, which has an extra gluon field, the former will be inevitably dressed
with derivatives. A simple dimensional analysis suggests then that the lowest
moment of a twist-three GPD, expressed in terms of a local $\bar\psi (0)
{\mit\Gamma} \psi (0)$ operator, will not have contributions from three-particle
correlation, since otherwise it should possesses inverse derivatives, which
would lead to a non-locality and violation of the local operator product
expansion. Higher moments will contain multiparticle correlations, unless it
is prohibited by symmetry reasons. Let us discuss a few cases starting with the
spin-zero target and then addressing the nucleon.

\noindent$\bullet$ For spin-zero target, one immediately finds \cite{KivPolSchTer00}
the analogue of the Burkhardt-Cottingham sum rule
\begin{equation}
\int_{-1}^1 d x \, H^q_3 (x , \eta)
=
- \frac{1}{2} \frac{\partial}{\partial\eta}
\int_{-1}^1 d x \, H^q (x, \eta)
=
0
\, .
\end{equation}
There is also an analogue of the Efremov-Leader-Teryaev sum rule \cite{ELT} for
the same distribution, which takes the form
\begin{equation}
\int_{-1}^1 d x \, x \, H^q_3 (x , \eta)
=
- \frac{1}{4} \frac{\partial}{\partial\eta}
\int_{-1}^1 d x \, x \, H^q (x, \eta)
=
\frac{\eta}{2} P^q
\, .
\end{equation}
Here, as before, $P^q$ is the momentum fraction of the flavor-$q$ parton in a
spinless hadron, see Eq.\ (\ref{PionGPD}). As just pointed out, in both cases
the quark-gluon-quark operators do not contribute.

\noindent$\bullet$ For spin-one-half target, the first moment of the functions
$F^q_\mu$ and $\widetilde{F}^q_\mu$, introduced in Eqs.\ (\ref{VectorTw3Nucleon})
and (\ref{AxialTw3Nucleon}), respectively, yields the usual electromagnetic and
weak quark form factors
$$
\int_{- 1}^1 dx \, F^q_\mu (x, \eta, \Delta^2)
=
\langle p_2 | j^q_\mu (0) | p_1 \rangle
\, , \qquad
\int_{- 1}^1 dx \, \widetilde{F}^q_\mu (x, \eta, \Delta^2)
=
\langle p_2 | j^{5q}_\mu (0) | p_1 \rangle
\, .
$$
This means that the twist-three functions $F^q_{3 \pm}$ in Eqs.\ (\ref{Def-V1})
and (\ref{Def-A1}) have vanishing first moments,
\begin{equation}
\int_{- 1}^1 dx \, F^q_{3 \pm} (x, \eta, \Delta^2) = 0
\, .
\end{equation}
The most interesting information from the viewpoint of the proton spin crisis
comes from the first moment of the transverse part of the parity-even function,
\begin{eqnarray}
\label{SecondMomentVectorGPDs}
\int_{- 1}^1 dx \, x \, F^{q \perp}_{\phantom{q}\mu} (x, \eta, \Delta^2)
=
\ft12 h^\perp_\mu
\bigg\{
G_A^q (\Delta^2)
\!\!\!&+&\!\!\!
\int_{- 1}^1 dx \, x \,
\left( H^q (x, \eta, \Delta^2) + E^q (x, \eta, \Delta^2) \right)
\bigg\}
\nonumber\\
&-&\!\!\! \Delta^\perp_\mu \frac{b}{4 M_N}
\frac{\partial}{\partial\eta}
\int_{- 1}^1 dx \, x \,
E^q (x, \eta, \Delta^2)
\, .
\end{eqnarray}
It provides complimentary knowledge on the magnitude of the angular orbital
motion of quarks in building up the spin of the proton, since
\begin{equation}
\lim_{\Delta^2 \to 0}
\left\{
G_A^q (\Delta^2)
+
\int_{- 1}^1 dx \, x \,
\left( H^q (x, \eta, \Delta^2) + E^q (x, \eta, \Delta^2) \right)
\right\}
=
\Delta q + 2 J^q
\, .
\end{equation}
The term in Eq.\ (\ref{SecondMomentVectorGPDs}) proportional to $\Delta^\mu_\perp b$
measures the distribution of shear in the nucleon (see Eqs.\ (\ref{ShearForces}) and
(\ref{SecondMomentsHE})). Starting from the third moment, the sum rules receive
contributions from the quark-gluon-quark operators \cite{KipPol02}.

\subsection{Compton form factors of the nucleon}

Since factorization theorems have not been proven for the $1/{\cal Q}$
suppressed contributions, in contrast to the leading twist situation
\cite{Rad1997a,JiOsb98,ColFre99}, it may happen that, at this level of
accuracy, the results are plagued by singularities, and that the higher
twist distributions are not universal. Though, the first next-to-leading
order analysis in the Wandzura-Wilczek approximation \cite{KivMan03}
demonstrated the absence of these divergences and thus favors factorization
beyond leading power in hard momentum. For survival of factorization in
exclusive processes it is often vital that non-perturbative functions do
have specific analytic properties. For instance, in case of the pion transition
form factor measured in the process $\gamma^\ast \gamma^\ast \to \pi^0$,
the meson distribution amplitude should vanish at the end-points, and
there are some arguments \cite{BroLep79} that this is really the case.
For DVCS an analogous requirement amounts to continuity of generalized
parton distributions at $|x| = \eta$. The tree level analysis has been
given in the previous section, and as we have seen, for the transverse
polarization of the initial photon, one actually encounters a situation
with divergent expressions, however, they do not show up when the Compton
amplitude is projected on the real photon final state satisfying $\varepsilon^\ast
\cdot q_2 = 0$ \cite{RadWei00}.

We emphasize again that, in the absence of three-particle contributions, the
twist-three functions are completely known in terms of the twist-two ones, just
like in the naive parton model. It is an interesting non-perturbative problem
to estimate the size of the antiquark-gluon-quark matrix elements in comparison
with the Wandzura-Wilczek term. For forward kinematics, recent experimental
data \cite{SLAC00} and lattice results \cite{Gockeler:2000ja} suggest that the
dynamical twist-three effects encoded in three-particle operators are small.
Extrapolating these findings to off-forward scattering, one can hope to use the
Wandzura-Wilczek part of the relation (\ref{VectorTw3Nucleon}) as a reliable
model for the ``transverse'' twist-three GPDs $F^q_\mu$ and $\widetilde{F}^q_\mu$.

\subsubsection{Twist-three CFFs}

The Compton form factors introduced in (\ref{TransvGPDeven}) and
(\ref{TransvGPDodd}) are given as convolutions of perturbatively
calculable coefficient functions $C^{[\pm]}$ with a set of twist-two
and twist-three GPDs (\ref{Def-V1},\ref{Def-A1}) via
\begin{eqnarray}
\label{DefTw3}
&&\left\{
{\cal H}^q
,
{\cal E}^q
,
{\cal H}_{3+}^q
,
{\cal E}_{3+}^q
,
\widetilde {\cal H}_{3-}^q
,
\widetilde {\cal E}_{3-}^q
\right\}
(\xi, \eta)
=
\int_{- 1}^{1} \! dx \, C^{q[-]}_{(0)} (\xi, x)
\left\{
H^q
,
E^q
,
H^q_{3+}
,
E^q_{3+}
,
\widetilde H^q_{3-}
,
\widetilde E^q_{3-}
\right\}
(x, \eta)
\, , \nonumber\\
&& \\
&&\left\{
\widetilde {\cal H}^q
,
\widetilde {\cal E}^q
,
\widetilde {\cal H}_{3+}^q
,
\widetilde {\cal E}_{3+}^q
,
{\cal H}_{3-}^q
,
{\cal E}_{3-}^q
\right\}
(\xi, \eta)
=
\int_{- 1}^{1} \! dx \, C^{q[+]}_{(0)} (\xi, x)
\left\{
\widetilde H^q
,
\widetilde E^q
,
\widetilde H^q_{3+}
,
\widetilde E^q_{3+}
,
H^q_{3-}
,
E^q_{3-}
\right\}
(x, \eta)
\, .
\nonumber
\end{eqnarray}
As we already established in previous sections, only the singularity-free
combination of CFFs ${\cal F}_{3+} - {\cal F}_{3-}$ can show up in physical
amplitudes \cite{KivPolSchTer00,Radyushkin:2000ap,BelKirMulSch00c,BelKirMulSch01b}.
The absence of divergences can be immediately seen by computing the difference
of Compton form factors for a generic situation $\eta \neq \xi$, evaluating the
convolutions with the coefficient functions (\ref{LOquarkCoeffFunct}) and making
use of the identity
\begin{eqnarray*}
\int_{-1}^1 \frac{d x}{\eta}
\frac{1}{\xi - x} W \left( \frac{x}{\eta}, \frac{y}{\eta} \right)
=
\frac{1}{y - \eta} \ln \left( \frac{\xi - \eta}{\xi - y} \right)
\, .
\end{eqnarray*}
For instance, for unpolarized CFFs we find in the WW-approximation
\begin{eqnarray}
{\cal H}_{{\scriptscriptstyle\rm WW}+}^q (\xi, \eta)
-
{\cal H}_{{\scriptscriptstyle\rm WW}-}^q (\xi, \eta)
\!\!\!&=&\!\!\!
\frac{1}{\eta} {\cal H}^q (\xi, \eta)
+
\int_{- 1}^1 \frac{dy}{\eta} C_{3(0)}^{q[-]} (\xi, \eta, x)
\left( x \frac{\partial}{\partial x} + \eta \frac{\partial}{\partial \eta} \right)
H^q (x, \eta)
\!\!\! \\
&-&\!\!\!
\frac{\Delta^2}{\Delta^2_\perp}
\int_{- 1}^1 dx
\left\{
\eta C_{3(0)}^{q[-]} (\xi, \eta, x) (H^q + E^q) (x, \eta)
-
C_{3(0)}^{q[+]} (\xi, \eta, x) \widetilde H^q (x, \eta)
\right\}
, \nonumber
\end{eqnarray}
where the coefficient function
\begin{equation}
C_{3(0)}^{q[\pm]} (\xi, \eta, x)
=
\frac{Q^2_q}{\xi + x} \ln \frac{\xi + \eta}{\xi - x}
\pm
\frac{Q^2_q}{\xi - x} \ln \frac{\xi + \eta}{\xi + x}
\, ,
\end{equation}
is finite for $\eta = \xi$. Analogous relations hold for other CFFs.

In the real-photon limit $\eta = \xi$, relevant for the deeply virtual Compton
scattering,  a specific ``effective'' combination of Compton form factors
contributes to all amplitudes, namely,
\begin{eqnarray}
\label{Tw3Eff}
{\cal F}_{\rm eff}^q
\equiv
- \frac{2 \xi}{1 + \xi} {\cal F}^q
+
2 \xi \left( {\cal F}_{3 +}^q - {\cal F}_{3 -}^q \right)
\, ,
\end{eqnarray}
where ${\cal F}^q_{3 \pm}$ are defined in Eqs.\ (\ref{DefTw3}). Here, the CFF ${\cal F}$
in Eq.\ (\ref{Tw3Eff}) stands for ${\cal H}$, ${\cal E}$, $\widetilde {\cal H}$, and
$\widetilde {\cal E}$ functions. Therefore, only four new GPDs (corresponding to GPDs
$F = H$, $E$, $\widetilde H$, and $\widetilde E$) remain at twist-three level. The
explicit form for the effective twist-three functions yields
\begin{eqnarray}
\label{Res-tw3eff}
{\cal F}^q_{\rm eff} (\xi)
\!\!\!&=&\!\!\!
\frac{2}{1 + \xi} {\cal F}^q (\xi)
+ 2 \xi
\frac{\partial}{\partial \xi}
\int_{- 1}^{1} d x \,
C^{q[\mp]}_{3(0)} (\xi, \xi, x) F^q (x, \xi)
+
\frac{8 M^2 \xi}{\Delta_\perp^2}
{\cal F}^q_\perp (\xi)
\\
&&\!\!\! - 2 \xi \int_{-1}^{1} d u \int_{-1}^{1} d x \,
C^{qgq}_{(0)} (\xi, x, u)
\left( S^{qgq}_{[+] F} (-x, -u, -\xi) - S^{qgq}_{[-] F} (x, u, -\xi) \right)
\Bigg\}
\, , \nonumber
\end{eqnarray}
for ${\cal F} = {\cal H}, \ {\cal E}, \ \widetilde{\cal H}$ with $C^{q[-]}_{3(0)}$
coefficient function for the first two and $C^{q[+]}_{3(0)}$ for the latter. In
case of the function $\widetilde{\cal E}$, the formula takes a slightly different
form due to the accompanying factor of skewness $\eta$ in the helicity-flip
functions $\widetilde E^q$, namely,
\begin{eqnarray}
\label{Res-tw3effE-tilde}
\widetilde {\cal E}^q_{\rm eff} (\xi)
\!\!\!&=&\!\!\!
\frac{2}{1 + \xi} \widetilde {\cal E}^q (\xi)
+ 2 \frac{\partial}{\partial \xi}
\xi \int_{- 1}^{1} d x \,
C^{q[+]}_{3(0)} (\xi, \xi, x) \widetilde{E}^q (x, \xi)
+
\frac{8 M^2 \xi}{\Delta_\perp^2}
\widetilde {\cal E}^q_\perp (\xi)
\\
&&\!\!\! - 2 \xi \int_{-1}^{1} d u \int_{-1}^{1} d x \,
C^{qgq}_{(0)} (\xi, x, u)
\left( S^{qgq}_{[+] \widetilde{E}} (-x, -u, -\xi)
-
S^{qgq}_{[-]  \widetilde{E}} (x, u, -\xi) \right)
\Bigg\}
\, .
\nonumber
\end{eqnarray}
Here we have used a new notation for the three-particle coefficient functions
\begin{equation}
C_{(0)}^{qgq} (\xi, x, u)
= Q^2_q
\frac{{\partial^2}}{\partial x^2} \frac{1 + u}{\xi + x}
\ln \left( \frac{2 \xi}{\xi - x} \right)
\, .
\end{equation}
Finally, the functions ${\cal F}^q_\perp$ contributing to the effective CFFs
are given by
\begin{eqnarray}
\label{Def-Fperb}
{\cal H}^q_\perp (\xi)
\!\!\!&=&\!\!\!\
- \frac{\Delta^2}{4 M^2}
\int_{- 1}^{1} d x
\left\{
\xi \, C^{q[-]}_{3(0)} (x, \xi) \left( H^q + E^q \right) (x, \xi)
-
C^{q[+]}_{3(0)} (x, \xi) \widetilde H^q (x, \xi)
\right\} \, ,
\\
{\cal E}^q_\perp (\xi)
\!\!\!&=&\!\!\!\
\int_{- 1}^{1} d x
\left\{
\xi \, C^{q[-]}_{3(0)} (x, \xi) \left( H^q + E^q \right) (x, \xi)
-
C^{q[+]}_{3(0)} (x, \xi) \widetilde H^q (x, \xi)
\right\}
\, ,
\nonumber\\
\widetilde {\cal H}^q_\perp (\xi)
\!\!\!&=&\!\!\!
\int_{- 1}^{1} d x
\left\{
\xi \left(1 - \frac{\Delta^2}{4 M^2} \right)
C^{q[+]}_{3(0)} (x, \xi) \widetilde H^q (x, \xi)
+
\frac{\Delta^2}{4 M^2} C^{q[-]}_{3(0)} (x, \xi)
\left( H^q + E^q \right) (x, \xi)
\right\}
\, ,
\nonumber\\
\widetilde {\cal E}^q_\perp (\xi)
\!\!\!&=&\!\!\!
\frac{1}{\xi}
\int_{- 1}^{1} d x
\left\{
C^{q[+]}_{3(0)} (x, \xi) \widetilde H^q (x, \xi)
-
\xi \, C^{q[-]}_{3(0)} (x, \xi) \left( H^q + E^q \right) (x, \xi)
\right\} \, .
\nonumber
\end{eqnarray}
We should note that the kinematical factor $\Delta^{- 2}_\perp$ in Eqs.\
(\ref{Res-tw3eff}) and (\ref{Res-tw3effE-tilde}) cancels out in final results
for angular harmonics of the cross section (\ref{AmplitudesSquared}), calculated
later in Sections \ref{SubSec-AziAngDep-DVCS} and \ref{SubSec-AziAngDep-INT},
and therefore, it does not lead to spurious kinematical poles.

\subsubsection{Small-$x_{\rm B}$ behavior of Compton form factors}

Let us now study the properties of the DVCS CFFs for small $x_{\rm B}$
in the leading order approximation. At intermediate momentum transfer
$\Delta^2$, this kinematics was discussed within the BFKL approach in
Refs.\ \cite{ErmOlnShu98,BalKuc00} and in Ref.\ \cite{FavMac04} within
the dipole model. The small-$\xi$ behavior of the CFFs is governed by
the small momentum fraction asymptotics of parton densities. If the
latter behave like $x^{- \alpha_q}$ with $\alpha_q > 0$, then the real
and imaginary parts of ${\cal H}^q$ go like $\xi^{-\alpha_q}$, and the
ratio of the real to imaginary part is given by $\tan\left((\alpha_q - 1)
\pi/2 \right) = - \cot\left( \alpha_q \pi/2\right)$. For polarized parton
densities, the small-$x$ behavior of the Regge type $x^{-\widetilde\alpha_q}$
induces an equivalent growth of $\widetilde{\cal H}^q$ in $\xi$. However,
the ratio of the real to imaginary part is now $\tan\left( \widetilde\alpha_q
\pi/2\right)$ instead.

This behavior can be understood in a model-independent way. The essential
assumption is that double distributions can be represented in the following
form
\begin{eqnarray}
f^q (\beta, \alpha)
=
\beta^{- \alpha_q} \left\{ d_q (0, \alpha) + \dots \right\}
\, ,
\end{eqnarray}
in the vicinity of the point $\beta = 0$, where the parameter $\alpha_q$ may
depend on $\Delta^2$ and ${\cal Q}^2$ \cite{Goeke:2001tz,Vin05}, as we already
addressed in Section \ref{LongTransInterplay}. Double distributions can also
have $\delta$-like singularities at $\beta = 0$. Such contributions generate
GPDs entirely concentrated in the exclusive region $|x| < \eta$ and can be
interpreted as isolated mesonic-like states and treated separately. Moreover,
we assume that $f (\beta, \alpha)$ and, consequently, also $d (0, \alpha)$
vanishes fast enough when it approaches the support boundary $|\beta|  +
|\alpha| = 1$. By ``fast enough'' we mean that, when $\alpha \to \pm 1$, we
have $|d (0, \alpha)| < |1 \mp \alpha|^{{\rm Max} (0, \alpha_q - 1) +
\varepsilon}$ with $\varepsilon > 0$. Due to the symmetry properties of
corresponding DDs with respect to $\beta$, it is sufficient to discuss the
region $\beta > 0$ only. We also assume that $\alpha_q < 2$ in the vector
and $\widetilde \alpha_q < 1$ in the axial-vector sector to ensure the
existence of certain integrals. The latter inequalities are fulfilled
phenomenologically.

The usual parton densities $q (\beta)$ result from DDs after integration over
$\alpha$ (see Eq.\ (\ref{ForwardPDFfromDDs})). Thus, at small momentum fractions
$\beta$, the parton densities are given by $\beta^{- \alpha_q} \int_{-1}^1 d
\alpha \; d_q (0, \alpha)$. Then a straightforward calculation shows that the
imaginary and real parts of the twist-two CFFs at leading order behave like
\cite{ShuGolMarRys99,HebTeu01,BelKirMulSch00c,BelMulKir01}
\begin{eqnarray}
\label{Res-ReImVec}
\Im{\rm m} \, {\cal H}^q (\xi, \Delta^2; {\cal Q}^2)
\!\!\!&=&\!\!\!
Q_q^2 \, \pi \, \xi^{- \alpha_q}
\int_{-1}^{1} d \alpha \; (1 - \alpha)^{- \alpha_q}
d_q (0, \alpha, \Delta^2; {\cal Q}^2)
\, , \\
\Re{\rm e} \, {\cal H}^q (\xi, \Delta^2; {\cal Q}^2)
\!\!\!&=&\!\!\! \tan \left( (\alpha_q - 1) \frac{\pi}{2} \right)
\Im{\rm m} \, {\cal H}^q (\xi, \Delta^2; {\cal Q}^2)
\, .
\nonumber
\end{eqnarray}
An analogous formula holds for the spin-flip CFF ${\cal E}$. Note, that the D-term
is not important in this limit. However, the small-$\xi$ behavior of CFFs can, in
principle, be altered by other terms concentrated at $\beta = 0$, i.e.,
$\frac{d^n}{d\beta^n} \delta(\beta)$. Such dependence is not excluded so far by
sum rules. These $\delta^{(n)}(\beta)$-like singularities, convoluted with the
hard scattering part yield contributions proportional to $\xi^{-n-1}$. Due to
a definite cross symmetry of the hard scattering coefficient function, the
exponent $n$ of this dependence takes odd (even) values for ${\cal E}$
($\widetilde{\cal E}$), respectively. Implementing these considerations, the
complete set of asymptotic formulas for small-$\xi$ reads
\begin{eqnarray}
\label{Res-ParSmaBx}
\left\{
{
\Re{\rm e}
\atop
\Im{\rm m}
}
\right\}
{\cal H}^q (\xi, \Delta^2)
\!\!\!&=&\!\!\!
\left\{
{
- \cot\left( \alpha_q \, \pi/2 \right)
\atop
1
}
\right\}
N^q_{\cal H} (\Delta^2) \, \xi^{-\alpha_q(\Delta^2)}
\, , \\
\left\{
{
\Re{\rm e}
\atop
\Im{\rm m}
}
\right\}
{\cal E}^q (\xi, \Delta^2)
\!\!\!&=&\!\!\!
\left\{
{
- \cot\left( \beta_q \, \pi/2 \right)
\atop
1
}
\right\}
N^q_{\cal E} (\Delta^2) \, \xi^{- \beta_q (\Delta^2)}
+
\left\{ {1 \atop 0} \right\} M^q_{\cal E} (\Delta^2) \, \xi^{-2 p}
\, , \nonumber\\
\left\{
{
\Re{\rm e}
\atop
\Im{\rm m}
}
\right\}
\widetilde{\cal H}^q (\xi, \Delta^2)
\!\!\!&=&\!\!\!
\left\{
{
\tan\left( \widetilde\alpha_q \, \pi/2 \right)
\atop
1
}
\right\}
N^q_{\widetilde {\cal H}} (\Delta^2)
\xi^{-\widetilde\alpha_q (\Delta^2)}
\, , \nonumber\\
\left\{
{
\Re{\rm e}
\atop
\Im{\rm m}
}
\right\} \widetilde{\cal E}^q (\xi, \Delta^2)
\!\!\!&=&\!\!\!
\left\{
{
\tan ( \widetilde\beta_q \pi/2 )
\atop
1
}
\right\}
N^q_{\widetilde {\cal E}} (\Delta^2)
\xi^{-\widetilde\beta_q (\Delta^2)}
+ \left\{ {1 \atop 0} \right\} \frac{F^q_\pi (\Delta^2)}{2\xi}
+
\left\{ {1 \atop 0} \right\} M^q_{\widetilde{\cal E}} (\Delta^2)
\, \xi^{- 2 \widetilde p - 1}
\, , \nonumber
\end{eqnarray}
where at zero recoil $\alpha_q (\Delta^2 = 0) = \alpha_q$ and $\widetilde\alpha_q
(\Delta^2 = 0) = \widetilde\alpha_q = a_q$ coincide with the ``Regge intercepts''
of the forward parton distributions (\ref{NonfactorGPD}) and (\ref{NonfactorGPDpol}),
respectively. The constants $p, \widetilde p \ge 1$ are positive integers that
reflect the appearance of isolated meson-like states. They provide the dominant
contributions to the spin-flip CFFs and could overwhelm the small-$x_{\rm B}$
behavior of the spin non-flip CFFs. Note also, that because of the well-constrained
dependence at small and large $\xi$, and a simple shape of the (partonic) CFFs,
one can extend the parametrization (\ref{Res-ParSmaBx}) in the whole kinematical
region of the generalized Bjorken variable by allowing for a weak $\xi$-dependence
of the normalization $N$-factors and phases, keeping however the $M$-factors
$\xi$-independent. Equations (\ref{Res-ParSmaBx}) for unpolarized ${\cal H}$,
${\cal E}$ and polarized $\widetilde{\cal H}$, $\widetilde{\cal E}$ GPDs are in
agreement with relations stemming from derivative analyticity for corresponding
scattering amplitudes $A$ and $\widetilde{A}$,
$$
\Re{\rm e} \frac{A}{s}
=
\tan \left( \frac{\pi}{2} \frac{d}{d \ln s} \right)
\Im{\rm m} \frac{A}{s}
\, , \qquad
\Re{\rm e} \widetilde{A}
=
\tan \left( \frac{\pi}{2} \frac{d}{d \ln s} \right)
\Im{\rm m} \widetilde{A}
\, ,
$$
---even and odd, respectively, under crossing symmetry $s \leftrightarrow u$
\cite{GriMig68,BroKanSuk74}. These are obtained from dispersion relations by
approximating the amplitude with a simple Regge-pole form, $A \sim s^{\alpha_M}$,
and resumming the Taylor expansion in logarithms of the energy $s$, see, e.g.,
\cite{Diehl:2003ny}.

\subsubsection{One-loop corrections to twist-two CFFs}
\label{OneLoopCoeffFunctions}

\begin{figure}[t]
\begin{center}
\mbox{
\begin{picture}(0,60)(220,0)
\put(0,0){\insertfig{15}{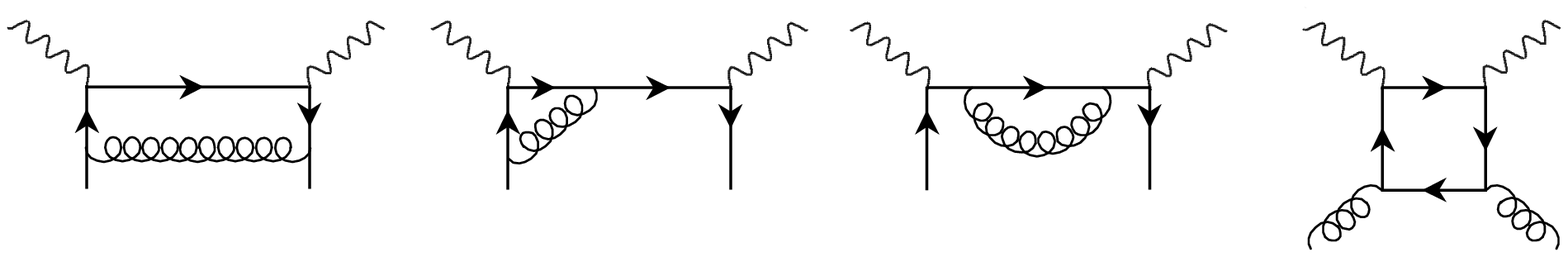}}
\end{picture}
}
\end{center}
\caption{\label{OneLoopCoeff} One-loop coefficient functions for quark and
and gluon GPDs in the Compton amplitude.}
\end{figure}

So far the analysis of the Compton amplitudes was carried out in Born approximation.
For reliable phenomenological analyses of experimental observables, one has to know
the magnitude of higher order effects in coupling constant. To match the two-loop
evolution of GPDs, analyzed in full in Section \ref{Evolution}, one has to complement
it with one-loop corrections to hard coefficient functions.

The next-to-leading coefficient functions $C_{i(1)}^{a[\pm]}$ (\ref{CoefFuncPertExpansion})
have been found in a number of studies by computing the one-loop Feynman diagrams shown
in Fig.\ \ref{OneLoopCoeff}. To simplify the presentation, let us extract the color
factors and quark charges from them and split the coefficient functions into two terms,
one of which entirely absorbs the renormalization group logarithm of the hard scale $Q^2$,
\begin{eqnarray}
C_{i(1)}^{q[\pm]} (x, \xi, \eta; Q^2/\mu^2)
\!\!\!&\equiv&\!\!\!
C_F Q_q^2
\left[
c_{i(1)}^{q[\pm]} (x, \xi, \eta)
+
\kappa_{i(1)}^{q[\pm]} (x, \xi, \eta)
\ln \left( Q^2/\mu^2 \right)
\right]
\, , \\
C_{i(1)}^{g[\pm]} (x, \xi, \eta; Q^2/\mu^2)
\!\!\!&\equiv&\!\!\!
2 T_F \sum_q Q_q^2
\left[
c_{i(1)}^{g[\pm]} (x, \xi, \eta)
+
\kappa_{i(1)}^{g[\pm]} (x, \xi, \eta)
\ln \left( Q^2/\mu^2 \right)
\right]
\, .
\end{eqnarray}
The index $i$ runs here over two values in the even-parity sector $i = 1, L$ and
it is $i = 1$ in the odd-parity one. One more value is obviously possible: $i = T$
corresponding to the gluon transversity, i.e., the maximal-helicity sector. The
parameter $\mu^2 = \mu^2_{\overline{\scriptscriptstyle\rm MS}}$ is the renormalization
scale in the $\overline{\rm MS}$ scheme. Varying it---normally within the range
$Q^2/2 \leq \mu^2 \leq 2 Q^2$---one can achieve minimization of the one-loop
corrections by choosing the optimal value. The reduced coefficient functions are
summarized below for different sectors.

\noindent $\bullet$ Parity-even sector \cite{JiOsb97,BelMul97,ManPilSteVanWei97,JiOsb98}
\begin{eqnarray}
c_{1(1)}^{q[-]} (x, \eta, \xi)
\!\!\!&=&\!\!\!
-
\frac{9}{2} \frac{1}{\xi - x}
-
\frac{
3 \left(2 x \, \xi - x^2 - \eta^2 \right)
}{
2 \left( \xi - x \right) \left( x^2 - \eta^2 \right)
}
\ln \left( 1 - \frac{x}{\xi} \right)
+
\frac{
\xi^2 + x^2 - 2 \eta^2
}{
2 \left( \xi - x \right) \left( x^2 - \eta^2 \right)
}
\ln^2 \left( 1 - \frac{x}{\xi} \right)
\nonumber \\
&+&\!\!\!
\frac{
3 x \left( \xi^2 - \eta^2 \right)
}{
\left( \xi^2 - x^2 \right) \left( x^2 - \eta^2 \right)
}
\ln \left(1 - \frac{\eta}{\xi} \right)
+
\frac{
x \left( \xi - \eta \right)
\left( x^2 - 2 \eta \xi - 2 \eta^2 - \xi^2 \right)
}{
2 \left( \xi^2 - x^2 \right) \eta \left( x^2 - \eta^2 \right)
}
\ln^2 \left( 1 - \frac{\eta}{\xi} \right)
\nonumber \\
&& +  (  \xi \leftrightarrow - \xi )
\, , \nonumber \\
c_{1(1)}^{g[-]}(x, \eta, \xi)
\!\!\!&=&\!\!\!
\frac{
4 \xi^2 - 4 x \, \xi + x^2 - \eta^2
}{
2 \left( x^2 - \eta^2 \right)^2
}
\ln \left( 1 - \frac{x}{\xi} \right)
-
\frac{
2 \xi^2 - 2 x \, \xi + x^2 - \eta^2
}{
4 \left( x^2 - \eta^2 \right)^2
}
\ln^2 \left( 1 - \frac{x}{\xi} \right)
\nonumber\\
&+&\!\!\!
\frac{
\left( \xi - \eta  \right) \left( x^2 - 4 \eta \xi - \eta^2 \right)
}{
2 \eta \left( x^2 - \eta^2 \right)^2
}
\ln \left( 1 - \frac{\eta}{\xi} \right)
-
\frac{
\left( \xi - \eta \right) \left( x^2 - 2 \eta \xi - \eta^2 \right)
}{
4 \eta \left( x^2 - \eta^2 \right)^2
}
\ln^2 \left( 1 - \frac{\eta}{\xi} \right)
\nonumber \\
&&
+ ( \xi \leftrightarrow - \xi)
\, . \nonumber
\end{eqnarray}
\noindent $\bullet$ Parity-odd sector \cite{BelMul97,ManPilSteVanWei97,JiOsb98}
\begin{eqnarray}
c_{1(1)}^{q[+]} (x, \eta, \xi)
\!\!\!&=&\!\!\!
-
\frac{9}{2} \frac{1}{\xi - x}
+
\frac{
x^2 + 3 \eta^2 - 2 x \, \xi - 2 \xi^2
}{
\left( \xi - x \right) \left( x^2 - \eta^2 \right)
}
\ln \left( 1 - \frac{x}{\xi} \right)
+
\frac{\xi^2 + x^2 - 2 \eta^2
}{
2 \left( \xi - x \right) \left( x^2 - \eta^2 \right)
}
\ln^2 \left(1 - \frac{x}{\xi}\right)
\nonumber \\
&+&\!\!\!
\frac{\left( \xi - \eta \right) \left( \xi^2 + 2 x^2 + 3 \eta \xi \right)
}{
\left( \xi^2 - x^2 \right) \left( x^2 - \eta^2 \right)
}
\ln \left(1 - \frac{\eta}{\xi}\right)
-
\frac{\left( \xi - \eta \right) \left( \xi^2 + x^2 + 2 \eta \xi \right)
}{
2 \left( \xi^2 - x^2 \right) \left( x^2 - \eta^2 \right)
}
\ln^2 \left(1 - \frac{\eta}{\xi}\right)
\nonumber\\
&&\!\!\!-
( \xi \leftrightarrow - \xi )
\, , \nonumber\\
\label{NLOPolGluon}
c_{1(1)}^{[+]g} (x, \eta, \xi)
\!\!\!&=&\!\!\!
- \frac{
3 x^2 + \eta^2 - 4 x \, \xi
}{
2 \left( x^2 - \eta^2 \right)^2
}
\ln \left( 1 - \frac{x}{\xi} \right)
-
\frac{
2 x \, \xi - x^2 - \eta^2
}{
4 \left( x^2 - \eta^2 \right)^2
}
\ln^2 \left(1 - \frac{x}{\xi}\right)
\\
&-&\!\!\!
\frac{
2 x \left( \xi - \eta  \right)
}{
\left( x^2 - \eta^2 \right)^2
}
\ln \left(1 - \frac{\eta}{\xi} \right)
+
\frac{
x \left( \xi - \eta \right)
}{
2 \left( x^2 - \eta^2 \right)^2
}
\ln^2 \left( 1 - \frac{\eta}{\xi} \right)
-
(\xi \leftrightarrow - \xi)
\, . \nonumber
\end{eqnarray}
\noindent $\bullet$ Longitudinal sector \cite{ManPilSteVanWei97}
\begin{eqnarray}
c_{L (1)}^{q [-]} (x, \eta, \xi)
\!\!\!&=&\!\!\!
-
\frac{
2 \xi
}{
\left( x^2 - \eta^2 \right)
}
\ln \left(1 - \frac{x}{\xi}\right)
+
\frac{
2 x \, \xi
}{
\eta (x^2 - \eta^2)
}
\ln \left(1 - \frac{\eta}{\xi} \right)
+ ( \xi \leftrightarrow - \xi)
\, , \nonumber \\
c_{L(1)}^{g[-]} (x, \eta, \xi)
\!\!\!& = &\!\!\!
\frac{
2 \xi \left( \xi - x \right)
}{
\left( x^2 - \eta^2 \right)^2
}
\ln \left(1 - \frac{x}{\xi} \right)
+
\frac{
\xi \left( x^2 - 2 \eta \xi + \eta^2 \right)
}{
\eta \left( x^2 - \eta^2 \right)^2
}
\ln \left( 1 - \frac{\eta}{\xi} \right)
+
( \xi \leftrightarrow - \xi  )
\, . \nonumber
\end{eqnarray}
\noindent $\bullet$ Maximal-helicity sector \cite{JiHoo98,BelMul00d}
\begin{eqnarray}
c_{T(1)}^{g} (x, \eta, \xi)
=
-
\frac{1}{x^2 - \eta^2}
\left\{
1 + \frac{\xi^2 - \eta^2}{\omega^2 (x^2 - \eta^2)}
\ln \frac{\xi^2 - x^2}{\xi^2 - \eta^2}
\right\}
\, .
\end{eqnarray}
The functions accompanying the logarithm of the ratio of the hard-to-renormalization
scale are given by the convolution\footnote{See the following Section \ref{ConformalOPE}
for details.} of the leading order quark coefficient functions (\ref{LOquarkCoeffFunct})
with corresponding one-loop evolution kernels (\ref{scaling-rel}) from Section
\ref{MomSpaceGerKernels},
\begin{equation}
\kappa_{1(1)}^{a[\mp]} (x, \xi, \eta)
\sim
Q_q^{-2}
\int_{-1}^1 d z \, C_{(0)}^{q[\mp]} (z, \xi) \,
k^{qa, V/A} \left( \frac{z}{\eta} , \frac{x}{\eta} \right)
\, ,
\quad\mbox{with}\quad a = q, g
\, ,
\end{equation}
if the mixing is allowed, and have the following explicit form
\begin{eqnarray}
\kappa_{1(1)}^{q[\pm]} (x, \xi, \eta)
\!\!\!&=&\!\!\!
\frac{3}{2} \frac{1}{\xi - x}
+
\frac{
\xi^2 + x^2 - 2 \eta^2
}{
\left( \xi - x \right) \left( x^2 - \eta^2 \right)
}
\ln \left( 1 - \frac{x}{\xi} \right)
\\
&&\qquad\quad
+
\frac{
x \left( \xi - \eta \right)
\left( x^2 - 2 \eta \xi - 2 \eta^2 - \xi^2 \right)
}{
\left( \xi^2 - x^2 \right) \eta \left( x^2 - \eta^2 \right)
}
\ln \left( 1 - \frac{\eta}{\xi} \right)
\pm (  \xi \leftrightarrow - \xi )
\, , \nonumber\\
\kappa_{1(1)}^{g[-]} (x, \xi, \eta)
\!\!\!&=&\!\!\!
-
\frac{
2 \xi^2 - 2 x \, \xi + x^2 - \eta^2
}{
2 \left( x^2 - \eta^2 \right)^2
}
\ln \left( 1 - \frac{x}{\xi} \right)
\\
&&\qquad\quad
-
\frac{
\left( \xi - \eta \right) \left( x^2 - 2 \eta \xi - \eta^2 \right)
}{
2 \eta \left( x^2 - \eta^2 \right)^2
}
\ln \left( 1 - \frac{\eta}{\xi} \right)
+
( \xi \leftrightarrow - \xi )
\, , \nonumber\\
\kappa_{1(1)}^{g[+]} (x, \xi, \eta)
\!\!\!&=&\!\!\!
-
\frac{
2 x \, \xi - x^2 - \eta^2
}{
2 \left( x^2 - \eta^2 \right)^2
}
\ln \left(1 - \frac{x}{\xi}\right)
+
\frac{
x \left( \xi - \eta \right)
}{
\left( x^2 - \eta^2 \right)^2
}
\ln \left( 1 - \frac{\eta}{\xi} \right)
-
(\xi \leftrightarrow - \xi)
\, .
\end{eqnarray}
Other functions are zero
\begin{equation}
\kappa_{L(1)}^{q[-]} (x, \xi, \eta)
=
\kappa_{L(1)}^{g[-]} (x, \xi, \eta)
=
\kappa_{T(1)}^{g} (x, \xi, \eta)
= 0
\, .
\end{equation}

Notice that in the real-photon limit $q_2^2 = 0$, relevant for DVCS, all singular
logarithmic contributions vanish at least linearly in $(\xi - \eta)$, supporting
the factorizability of the DVCS amplitude. Therefore, the DVCS coefficient
functions are easily obtainable from the above expressions by setting $\eta = \xi$
there. Let us point out that recently one-loop coefficient functions due to heavy
flavors propagating in loops were reported in Ref.\ \cite{Nor03}.

In the following section, we will demonstrate a formalism to reconstruct the one-loop
coefficient functions by means of conformal operator product expansion. It avoids
explicit loop calculations and is based on known next-to-leading forward coefficient
functions. The presentation will be rather technical and a practitioner who is
interested in final results rather than the use of the QCD machinery to deduce them,
can skip entirely the following Section \ref{ConformalOPE}.

\subsection{Application of conformal operator product expansion}
\label{ConformalOPE}

The next-to-leading order coefficient functions for off-forward Compton amplitudes
were deduced in the previous section by an explicit calculation of perturbative
one-loop diagrams. However, instead one could have used implications of the conformal
symmetry and avoid any computations of loop integrals completely. To this end, let us
note that in Section \ref{Evolution}, we found that the off-forward evolution kernels
at leading order of perturbation theory are defined entirely by the forward splitting
functions. Or in other words, the off-forward kernels are diagonalized by conformal
partial waves---conformal operators---and their eigenvalues are determined solely
by the forward anomalous dimensions of operators without total derivatives. Thus, as
we demonstrated in Section \ref{ER-BLtoDGLAP} one can unambiguously restore the
off-forward evolution kernels from the forward ones. By the same token, as we will
demonstrate in this section, one can reconstruct the off-forward coefficient functions.

As we explained before, the redefinition of conformal representations by shifting
the scale dimensions of fields, given originally in terms of their canonical mass
dimensions, by the anomalous ones, makes the theory respect conformal covariance.
However, the effect of the running of the gauge coupling inevitably breaks the
conformal symmetry. Therefore, supposing the existence of a nontrivial zero $g^*$
of the $\beta$-function ($\beta (g^*) = 0$) a conformaly covariant OPE can be
proven to exist even for interacting theory. Below we will shortly outline some of
the points which are of relevance for our further discussion.

Instead of constructing the operator product expansion in terms of local
Wilson operators, we can automatically take care of operators with total
derivatives by expanding over the basis of conformal operators. Keeping
the Lorentz indices open, the latter read for quarks
\begin{equation}
{\mathbb{O}}^{qq}_{\mu_0 \mu_1 \dots \mu_k}
=
\sum_{\ell = 0}^k e_{k\ell} (j_q) \Sym_{\mu_0 \dots \mu_k}
\bar\psi \, \gamma_{\mu_0}
\, i\!\stackrel{{}_\rightarrow}{\partial}_{\mu_1}
\dots
\, i\!\stackrel{{}_\rightarrow}{\partial}_{\mu_\ell}
\, i\!\stackrel{{}_\leftarrow}{\partial}_{\mu_{\ell + 1}}
\dots
\, i\!\stackrel{{}_\leftarrow}{\partial}_{\mu_{k - \ell}}
\psi
\, ,
\end{equation}
where the expansion coefficient for a generic conformal spin $j$ field is
\begin{equation}
e_{k\ell} (j)
\equiv
\frac{
(- 1)^{k - \ell} (4 j + k - 2)! (2 j - 1)! (2j + k - 1)!
}{
(4j - 2)! \ell ! (k - \ell)! (k - \ell + 2j - 1)! (\ell + 2 j - 1)!
}
\end{equation}
The descendants are obtained by acting with the step-up operators on the above
operator
\begin{equation}
{\mathbb{O}}^{qq}_{k ; \mu_0 \mu_1 \dots \mu_l}
\equiv
{}[
\dots
{}[
{}[
{\mathbb{O}}^{qq}_{\mu_0 \mu_1 \dots \mu_k} , \mathbb{P}_{\mu_{k + 1}}
{}] ,
\mathbb{P}_{\mu_{k + 2}} ] , \dots \mathbb{P}_{\mu_\ell}
{}]
\, ,
\end{equation}
with $l$ Lorentz indices, $(l - k)$ of which are are attached to total derivatives.
Contracting the open Lorentz indices with the light-like vectors $n^\mu$ we get the
conventional definition of the quark conformal operator
\begin{equation}
\mathbb{O}^{qq}_{kl} \equiv n^{\mu_0} n^{\mu_1} \dots n^{\mu_k}
{\mathbb{O}}^{qq}_{k; \mu_0 \mu_1 \dots \mu_l}
\, .
\end{equation}
We will apply the expansion in terms of the conformal operators to the product
of two electromagnetic currents. As can be seen from our results in Section
\ref{TreeLevelCompton}, for the leading twist contribution only the transverse
components of the Lorentz indices do contribute. The other tensor structures are
suppressed by at least one inverse power of the hard scale. Thus the expansion of
a product of two electromagnetic currents takes the form
\begin{equation}
\label{COPEnotreduced}
j_\perp (z) j_\perp (0)
=
\sum_{k = 0}^\infty
\left( \frac{1}{z^2} \right)^{d_j - t_k/2}
\sum_{l = k}^\infty c_{kl} \, (z^-)^{l + 2 s_q} \mathbb{O}^{qq}_{kl} (0)
\, ,
\end{equation}
where $d_j = 3$ is the mass dimensional of the electromagnetic current, $t_k =
d_k - k - 2 s_q = 2$ is the twist of the composite operator $\mathbb{O}^{qq}_{kk}$
and $s_q = 1/2$ is the spin of the quark. Actually, one can resum the infinite
series with respect to $l$. To this end one has to deduce a recursion relation
for the expansion coefficients $c_{kl}$. Acting on the left-hand side of Eq.\
(\ref{COPEnotreduced}) with the step-down conformal boost operator, we get
\begin{eqnarray}
[ j_\perp (z) j_\perp (0) , \mathbb{K}^-]
\!\!\!&=&\!\!\!
i \left( 2 d_j z^- - z^2 \partial^- + 2 z^- z^\mu \partial_\mu \right)
j_\perp (z) j_\perp (0)
\nonumber\\
&\simeq&\!\!\! 2 i \sum_{k = 0}^\infty
\left( \frac{1}{z^2} \right)^{d_j - t_k/2}
\sum_{l = k}^\infty (j_k + l - k) c_{kl} \, (z^-)^{l + 2 s_q + 1} \mathbb{O}^{qq}_{kl}
\, .
\end{eqnarray}
While applying $\mathbb{K}^-$ on the right-hand side of the equality (\ref{COPEnotreduced}),
we find
\begin{equation}
[ j_\perp (z) j_\perp (0) , \mathbb{K}^-]
\simeq
- 2 \sum_{k = 0}^\infty \left( \frac{1}{z^2} \right)^{d_j - t_k/2}
\sum_{l = k}^\infty (l - k + 1) (2 j_k + l - k) c_{k \, l + 1} \,
(z^-)^{l + 2 s_q + 1} \mathbb{O}^{qq}_{kl}
\, ,
\end{equation}
where we have used the covariant transformation law for conformal operators
established in Eq.\ (\ref{confVaria-1}). Here $j_k = 2j_q + k$ is the conformal
spin of the conformal operator $\mathbb{O}^{qq}_{kl}$. By comparing both sides
of the equation, we can get the following recursion relation
$$
c_{k \, l + 1} = - i \frac{(j_k + l - k)}{(l - k + 1)(2 j_k + l - k)} c_{kl}
\, ,
$$
the solution to which is given by
\begin{equation}
c_{kl} = (- i)^{l - k}
\frac{{\mit\Gamma} (j_k + l - k) {\mit\Gamma} (2j_k)
}{(l - k)! {\mit\Gamma} (2j_k + l - k) {\mit\Gamma} (j_k)}
c_{kk}
\, .
\end{equation}
Substituting these coefficients into (\ref{COPEnotreduced}) one can resum
contributions of total derivatives, $\mathbb{O}^{qq}_{kl} = (i \partial^+)^{l - k}
\mathbb{O}^{qq}_{kk}$, with the result
\begin{equation}
\label{COPE}
j_\perp (z) j_\perp (0)
=
\sum_{k = 0}^\infty c_{kk} \frac{{\mit\Gamma} (2 j_k)}{{\mit\Gamma}^2 (j_k)}
\left( \frac{1}{z^2} \right)^{d_j - t_k/2} (z^-)^{k + 2 s_q}
\int_0^1 du \, (u \bar{u})^{j_k - 1}
\mathbb{O}^{qq}_{kk} (u z^-)
\, ,
\end{equation}
which constitutes the conformal operator product expansion \cite{FerGatGri73}. Note
that the right-hand side can be expressed in terms of the confluent hypergeometric
function ${_1 F_1}$. Now sandwiching this expression between off-forward hadronic
states and transforming it into the momentum space, we find the Compton amplitudes
\begin{eqnarray}
T_{\perp\perp}
\!\!\!&=&\!\!\! i \int d^4 z \, {\rm e}^{i q \cdot z + i \Delta \cdot z/2 }
\langle p_2 | T \, \{ j_\perp (z) j_\perp (0) \} | p_1 \rangle
\\
&=&\!\!\!
\sum_{k = 0}^\infty c^\prime_k
\left( \frac{\mu^2}{Q^2} \right)^{2 - d_j + t_k/2}
\int_0^1 du \,
(u \bar{u})^{j_k - 1}
\left( 1 - \frac{\eta}{\xi} (1 - 2u) \right)^{d_j - k - t_k/2 - 3}
\xi^{- k - 1} \, \mathbb{F}^q_k (\eta)
\, , \nonumber
\end{eqnarray}
as an expansion in terms of the Gegenbauer moments of the quark GPDs
(\ref{QuarkConfMoments}). Here $\mu^2$ is the normalization scale. So
far the analysis was performed in noninteracting theory, i.e., in tree
approximation. By switching on strong interactions, the conformal spin
and the canonical dimensions get shifted by anomalous dimensions, since
the canonical dimension of conformal operators $\mathbb{O}^{qq}_{kk}$
changes accordingly (cf.\ Sections \ref{SectConfWardIdentity} and
\ref{SectRestConfCov})
$$
d_k^{\rm can}
\to
d_k = k + 2 d_q + \gamma_k^{qq} (\alpha_s)
\, .
$$
The expansion coefficients $c^\prime_k$ are promoted to functions of the coupling
constant $\alpha_s$ which are related to the Wilson coefficients of deeply inelastic
scattering. Apart from already present at tree level quark operator matrix elements,
the amplitude also acquires gluon operators starting from one-loop order. From the
above result for the Compton amplitude we can read off the expression for the
Compton form factors. For instance, for the quark contribution to the Compton
form factor ${\cal F}_1 (\xi, \eta)$ one immediately finds
\begin{eqnarray}
\label{LOOPE}
{\cal F}_1 (\xi, \eta)
\!\!\!&=&\!\!\! \sum_k
\sigma_k c_k^{\rm\scriptscriptstyle DIS} (\alpha_s)
\frac{2^{k + 2} B (k + 1, k + 2)}{B (k + 2 + \gamma_k^{qq}/2, k + 2 + \gamma_k^{qq}/2)}
\left( \frac{\mu^2}{Q^2} \right)^{\gamma_k/2}
\\
&\times&\!\!\!
\int_0^1 d u \, (u \bar{u})^{k + 1 + \gamma_k^{qq}/2}
\left( 1 - \frac{\eta}{\xi}(1 - 2u) \right)^{- k - 1 - \gamma_k^{qq}/2}
\xi^{- k - 1} \, \widehat{\mathbb{F}}^q_k (\eta)
\, , \nonumber
\end{eqnarray}
where $\sigma_k \equiv \ft12 [1 + (- 1)^{k + 1}]$, $B (a, b) = {\mit\Gamma} (a)
{\mit\Gamma} (b)/{\mit\Gamma} (a + b)$ is the Euler beta function and
$\widehat{\mathbb{F}}^q_k$ define the conformally covariant moments (\ref{MStoCS}),
which renormalize autonomously at the conformal point $\beta (g^\ast) = 0$. Due
to the normalization condition (\ref{NormalizationConfCovConf}), these conformally
covariant operators coincide with ordinary conformal operators at low normalization
point $\mu_0$, i.e., $\widehat{\mathbb{F}}^q_k (\mu_0) = {\mathbb{F}}^q_k (\mu_0)$.
Therefore, below we will display all equations at $\mu_0$ so that to avoid the
hat-symbol on top of moments. Analogously in the parity-odd Compton form factor
$\widetilde{\cal F}_1$, one replaces $\sigma_k$ by $\sigma_{k + 1}$ and $\mathbb{F}$
by $\widetilde{\mathbb{F}}$. Also one trades the DIS unpolarized Wilson coefficient
$c_k^{\rm\scriptscriptstyle DIS}$ by the one of polarized scattering. Obviously, at
leading order, $\gamma_k^{qq} = 0$ and $c_k = 1$. This reproduces the normalization
accepted before. Obviously, in the forward case it reduces to the conventional
expansion of the structure function in the unphysical region $x_{\rm B} > 1$,
\begin{equation}
{\cal F}_1 (\xi = x_{\rm B} , \eta = 0)
= F_1 (x_{\rm B})
=
\sum_k \frac{\sigma_k}{x_{\rm B}^{k + 1}}
\langle p | \bar\psi \gamma^+ (i \partial^+)^k \psi (0) | p \rangle
\, .
\end{equation}
where we limited the perturbative expansion to leading order only.

Let us demonstrate how to use the conformal operator product expansion in
order to find one-loop off-forward coefficient functions \cite{Mul98,BelMul97},
which were computed above making use of the standard Feynman rule formalism.
For demonstration purposes, let us use the odd-parity gluon sector. By expanding
the prediction of the conformal operator product expansion (\ref{LOOPE}) to
first order in the strong coupling constant
\begin{equation}
\widetilde{\cal F}_1 (\xi, \eta)
=
\widetilde{\cal F}_1^{(0)} (\xi, \eta)
+
\frac{\alpha_s}{2 \pi} \widetilde{\cal F}_1^{(1)} (\xi, \eta)
+
{\cal O} (\alpha_s^2)
\, ,
\end{equation}
we obtain for the next-to-leading correction in this sector
\begin{eqnarray}
\label{ConformalOPEnlo}
\widetilde{\cal F}_1^{(1)} (\xi, \eta)
\!\!\!&=&\!\!\!
\sum_q Q_q^2
\sum_k \frac{\sigma_{k + 1}}{N_f}
\left.
\Bigg\{
\widetilde{c}_{(1)k}^{qg}
-
\ft12 \gamma_{(0)k}^{qg; A} \ln \left( \frac{Q^2}{\mu^2} \right)
+
\ft12 \gamma_{(0)k}^{qg; A}
\frac{d}{d \varrho}
\Bigg\}
\right|_{\varrho = 0}
\\
&&
\times
\frac{2^k B (k + 1, k + 2)}{B (k + 2 + \varrho, k + 2 + \varrho)}
\int_0^1 d u (u \bar{u})^{k + 1 + \varrho}
\left( 1 - \frac{\eta}{\xi} (1 - 2 u) \right)^{- k - 1 - \varrho}
\xi^{- k - 1} \, \widetilde{\mathbb{F}}^g_k (\eta)
\, , \nonumber
\end{eqnarray}
with conformally covariant moments of the gluon GPD coinciding with conventional
conformal moments at low normalization point were defined in Eq.\ (\ref{GluonConfMoments}).
Note that the factor $4$ in $\widetilde{\mathbb{F}}^g_k$ (\ref{GluonConfMoments})
explains the relative normalization of Eqs.\ (\ref{LOOPE}) and (\ref{ConformalOPEnlo}).
The forward Wilson coefficient function in Eq.\ (\ref{ConformalOPEnlo}) is
\cite{AltRos88,BodQiu90}
\begin{equation}
\widetilde{c}_{(1)k}^{qg} = \frac{6}{k} \int_0^1 d x \, x^j \,
2 T_F N_f
\left[
(2x - 1) \ln \frac{1 - x}{x} + 3 - 4 x
\right]
=
\ft12 \gamma_{(0)k}^{qg; A}
\left[ \psi (k + 1) - \psi (1) + 1 \right]
\, .
\end{equation}
Here we have multiplied the forward coefficient function with the factor $6/k$ taking
into account the difference in the normalization of Wilson and conformal operators
(see Eq.\ (\ref{WilsonToConformal})). The above prediction from the operator product
expansion consists of two terms,
\begin{equation}
\label{TotalEvCF}
\widetilde{\cal F}_1^{(1)} (\xi, \eta)
=
\widetilde{\cal F}_1^{(1), {\rm ev}} (\xi, \eta)
+
\widetilde{\cal F}_1^{(1), {\rm cf}} (\xi, \eta)
\, ,
\end{equation}
---the correction due to renormalization group evolution of conformal operators
$\widetilde{\cal F}_1^{(1), {\rm ev}}$ and the next-to-leading order correction
to the coefficient function $\widetilde{\cal F}_1^{(1), {\rm cf}}$. While the
second one is the quantity in question, the former was already found in our
earlier discussion of the evolution of GPDs. It reads
\begin{equation}
\label{CorrEigenfunction}
\widetilde{\cal F}_1^{(1), {\rm ev}} (\xi, \eta)
=
\sum_q C_{(0)}^{q [+]} (x, \xi) \otimes \frac{1}{8 N_f}
\left\{
{\mit\Phi}^{qg; A}_{(1)} (x, y, \eta)
-
K^{qg; A}_{(0)} (x, y, \eta) \ln \left( \frac{Q^2}{\mu^2} \right)
\right\}
\otimes \widetilde{F}^g (y, \eta)
\, .
\end{equation}
Here to simplify formulas, we used the convention for the convolution
$$
\tau_1 (x, y) \otimes \tau_2 (y, z) \equiv \int d y \, \tau_1 (x, y) \tau_2 (y, z)
\, .
$$
The correction to the eigenfunction of the next-to-leading evolution equation
reads
\begin{equation}
{\mit\Phi}^{qg; A}_{(1)} (x, y, \eta)
=
(1 - {\rm diag})
S (x, z) \otimes K^{qg; A}_{(0)} (z, y, \eta)
\, .
\end{equation}
It reproduces the result for conformal operators when one form Gegenbauer moments
with it (see Eqs.\ (\ref{GammaGammaCFull}) and (\ref{Bgamma})). The shift operator
is defined as
$$
S (x, y) \otimes \, w (y | \nu) C_k^\nu (x)
\equiv
\left.
\frac{d}{d \varrho} \right|_{\varrho = 0} w (y | \nu + \varrho) C_k^{\nu + \varrho} (x)
\, ,
$$
while the subtraction of the diagonal contribution in the basis of conformal
operators yields for a test function $\tau (x, u)$
$$
C_k^\nu (x) \otimes (1 - {\rm diag}) \tau (x, y)
=
\sum_{l = 0}^k \left( \tau_{kl} - \tau_{kk} \delta_{kl} \right) C_k^\nu (y)
\, .
$$
Expanding Eq.\ (\ref{CorrEigenfunction}) in conformal partial waves, we find
\begin{eqnarray}
\widetilde{\cal F}_1^{(1), {\rm ev}} (\xi, \eta)
\!\!\!&=&\!\!\!
\sum_q Q_q^2
\sum_k \frac{\sigma_{k + 1}}{2 N_f}
\gamma_{(0)k}^{qg; A}
\left.
\Bigg\{ \frac{d}{d \varrho} - \ln \left( \frac{Q^2}{\mu^2} \right) \Bigg\}
\right|_{\varrho = 0}
\\
&&
\times
\frac{2^k B (k + 1, k + 2)}{B (k + 2 + \varrho, k + 2 + \varrho)}
\int_0^1 d u (u \bar{u})^{k + 1 + \varrho}
\left( 1 - \frac{\eta}{\xi} (1 - 2 u) \right)^{- k - 1}
\xi^{- k - 1} \, \widetilde{\mathbb{F}}^g_k (\eta)
\, . \nonumber
\end{eqnarray}
Using Eq.\ (\ref{TotalEvCF}) to solve for ${\cal F}_1^{(1), {\rm cf}}$ one finds from
the difference of the conformal operator product expansion (\ref{ConformalOPEnlo}) and
next-to-leading evolution corrections ${\cal F}_1^{(1), {\rm ev}}$ the one-loop
correction to the coefficient function
\begin{eqnarray}
\widetilde{\cal F}_1^{(1), {\rm cf}} (\xi, \eta)
\!\!\!&=&\!\!\!
C_{1 (1)}^{g [+]} (x, \xi, \eta; 1) \otimes \widetilde{F}^g (x, \eta)
=
\sum_q Q_q^2
\sum_k \frac{\sigma_{k + 1}}{N_f}
\left.
\Bigg\{
\widetilde{c}^{qg}_{(1)k} + \ft12 \gamma_{(0)k}^{qg; A} \frac{d}{d \varrho}
\Bigg\}
\right|_{\varrho = 0}
\\
&&
\times
\frac{2^k B (k + 1, k + 2)}{B (k + 2, k + 2)}
\int_0^1 d u (u \bar{u})^{k + 1}
\left( 1 - \frac{\eta}{\xi} (1 - 2 u) \right)^{- k - 1 - \varrho}
\xi^{- k - 1} \, \widetilde{\mathbb{F}}^g_k
\, . \nonumber
\end{eqnarray}
The resummation of this series is straightforward and gives the next-to-leading
coefficient function in the form of a convolution\footnote{The presence of the
logarithm in the coefficient function can be easily understood since $\frac{\ln x}{x}
= - \frac{d}{d \varrho}|_{\varrho = 0} x^{- 1 - \varrho}$.},
\begin{equation}
C_{1 (1)}^{g [+]} (y, \xi, \eta; 1)
=
\sum_q Q_q^2
\left\{
\frac{1 - \ln \left( 1 - \frac{x}{\xi} \right)}{\xi - x - i 0}
+
\frac{1 - \ln \left( 1 + \frac{x}{\xi} \right)}{\xi + x - i 0}
\right\}
\otimes
\frac{1}{8 N_f} K^{qg; A}_{(0)} (x, y, \eta)
\, .
\end{equation}
Evaluating the integrals we indeed reproduce the result for the off-forward
coefficient function (\ref{NLOPolGluon}), computed before by means of Feynman
diagrams. Presently we deduced it without any loop-integrals calculation but
merely relying on the conformal symmetry.

\subsection{Target mass corrections}
\label{SectionTargetMassCorrections}

The leading-twist approximation to hard processes in QCD is affected both by
radiative corrections in the strong coupling constant and higher twist power
suppressed contributions. The former were addressed in preceding sections. In
case when the energy scale of the reaction turns out to be rather low,
${\cal Q}^2 \sim 2 - 4\ {\rm GeV}^2$, one has to take care of power suppressed
effects since they modify significantly the scaling behavior of corresponding
cross sections. Power corrections can be divided into two classes according to
their origin: dynamical and kinematical. The dynamical corrections reflect
multiparton correlations inside hadrons \cite{Politzer:1980me}, i.e., new
information not contained in the leading-twist distributions. The kinematical
corrections include the power-suppressed contributions which can be expressed
in terms of the twist-two functions. We already discussed the Wandzura-Wilczek
type contributions \cite{WW} arising from separation of composite operators into
components with definite symmetry properties with respect to the Lorentz group.
Another prominent class accommodates target mass corrections \cite{GeoPol76}.
They stem from the subtraction of traces in Wilson operators possessing a well
defined geometrical twist.

In this section, we describe a formalism for resummation of the target mass
corrections for the virtual Compton scattering amplitude. Since the present day
studies of DVCS involve momentum transfers of the order of a few GeV, the ratio
of the nucleon mass to the momentum transfer can produce corrections as large
as 30\% to the leading twist massless results. Despite being suppressed by an
extra power of the hard momentum scale, the target mass corrections can still
compete at available energies with twist-three effects. There exists a standard
formalism \cite{GeoPol76} for resumming target mass corrections in deeply
inelastic scattering (see also \cite{Nac73} for an earlier  first discussion of
the topic). However, it is not readily applicable to off-forward processes since
new towers of Lorentz structures develop in matrix elements of local operators
and they cannot be handled in a fashion proposed in \cite{GeoPol76}. A discussion
of mass effects in exclusive processes with simpler kinematics of distribution
amplitudes can be found in Ref.\ \cite{Ball:1998ff}.

To resum infinite series of mass corrections, it is again more efficient to
use double distributions $h^q (\beta,\alpha,\Delta^2)$ from Section
\ref{SectionSingleCompDDnucleon} rather than GPDs $F^q (x,\eta, \Delta^2)$
\cite{BelMul01b}. This allows one to represent all-order results in a compact
fashion. Once this is accomplished, one can use the inverse Radon transformation
(\ref{Res-Inv}) to express DDs in terms of GPDs. Regretfully, this step can hardly
be used for successful phenomenological applications. However, each term in the
expansion of the resummed expressions in powers of $\left( M^2/Q^2 \right)^j$
can be straightforwardly cast in the conventional language of GPDs. The formalism
is described in the following two sections, and as a demonstration, we resum
target mass effects resulting from trace terms of twist-two operators. The
restriction to leading twist approximation in the Compton amplitude leads to the
violation of electromagnetic gauge invariance as we described at length in
Section \ref{Amplitude}. The problem of resumming mass effects stemming from
higher twist operator matrix elements was not addressed in the literature so far.
Therefore, it might be premature to use presently derived results for numerical
evaluation of the magnitude of hadronic mass effects in physical cross sections.

\subsubsection{Twist decomposition and harmonic polynomials}
\label{HarmonicPolynomials}

The goal of the present section is to construct the operator product expansion
for the off-forward Compton amplitude (\ref{ComptonAmplitude}) taking the
entire tower of target mass corrections into account. Since the latter originate
from the trace subtraction in the definition of the local Wilson operators
(\ref{QuarkLocalOperator}), the effects will be thus entirely kinematical in
nature. Hence, we can neglect dynamical higher twists stemming from multiparticle
operators and restrict our attention to two-particle contributions only. In this
approximation, the Fourier transformed two-photon amplitude is given by
\begin{eqnarray}
\label{AmplitudeT}
T^{\mu \nu}
\!\!\!&=&\!\!\!
\frac{1}{\pi^{2}}
\sum_q Q_q^2
\int d^4 z \ \frac{z_\sigma}{[- z^2 + i 0]^2}
\langle p_2 |
\cos \left( z \cdot q \right) S^{\mu \nu; \rho \sigma}
{\cal O}^{qq}_\rho \! \left( - z/2, z/2 \right)
\nonumber \\
&&\qquad\qquad\qquad\qquad\qquad\ +
\sin \left( z \cdot q \right) \varepsilon^{\mu \nu \rho \sigma}
\widetilde {\cal O}^{qq}_\rho \! \left( - z/2 , z/2 \right)
| p_1 \rangle
\, ,
\end{eqnarray}
where $\mathcal{O}^{qq}_\rho$ and $\widetilde{\mathcal{O}}^{qq}_\rho$ are defined
in Eqs.\ (\ref{QuarkLightRayOperators}) and (\ref{QuarkLightRayOperatorsOdd}),
respectively. Presently, we will keep the separation $z^\mu$ between the quark
fields non-light-like.

To solve the problem, we have to extract the increasing-twist traceless components
from the contributing operators. Such a procedure was designed in Section
\ref{TwistDecomp} for light-like interquark separations $z^2 = 0$. Since the
latter condition is lifted now, the twist separation has to be modified. The
general strategy is the following. Since a transparent formalism can be devised
in terms of local Wilson operators, we expand first the non-local operators
$\mathcal{O}^{qq}_\rho$ and $\widetilde{\mathcal{O}}^{qq}_\rho$ in Taylor series,
e.g., for the even parity sector,
\begin{equation}
\label{Def-Ope}
{\cal O}^{qq}_\rho (- z , z)
=
\sum_{j = 0}^{\infty} \frac{(-i)^j}{j!}
z^{\mu_1} \dots z^{\mu_j} {\cal O}^{qq}_{\rho \mu_1 \dots \mu_j} (0)
\equiv
\sum_{j = 0}^{\infty} \frac{(-i)^j}{j!} {\cal O}^{qq}_{\rho; j} (0)
\, ,
\end{equation}
with
\begin{equation}
\label{LocalWilsonOperatorsEven}
{\cal O}^{qq}_{\rho \mu_1 \dots \mu_j} (0) =
\bar\psi_q \gamma_\rho \,
i\!\stackrel{{}_\leftrightarrow}{\cal D}_{\mu_1}
\dots
i\!\stackrel{{}_\leftrightarrow}{\cal D}_{\mu_j}
\psi_q
\, .
\end{equation}
In terms of local operators, it is straightforward to symmetrize and antisymmetrize
the open Lorentz index $\rho$ with the rest, as we already demonstrated in Section
\ref{TwistDecomp}. Therefore, we can decompose the original operator $
{\cal O}^{qq}_\rho (- z , z)$ into its symmetric and antisymmetric components,
\begin{equation}
\label{DecomposionBody}
{\cal O}^{qq}_\rho (- z , z)
=
{\cal O}^{qq, \rm sym}_\rho (- z , z)
+
{\cal O}^{qq, \rm asym}_\rho (- z , z)
\, .
\end{equation}
When expanded in Taylor series of the type (\ref{Def-Ope}), they generate a tower of
Wilson operators (cf.\ (\ref{QuarkLocalOperator}) and (\ref{TwistThreeLocalOperators}))
\begin{eqnarray}
{\cal O}^{qq, \rm sym}_{\rho; j}
\!\!\!&=&\!\!\!
z^{\mu_1} \dots z^{\mu_j}
\bar\psi_q \gamma_{\{ \rho} i \stackrel{{}_\leftrightarrow}{\cal D}_{\mu_1}
\dots i \stackrel{{}_\leftrightarrow}{\cal D}_{\mu_j \}} \psi_q
\, , \\
\label{WilsonAntiSymmOper}
{\cal O}^{qq, \rm asym}_{\rho; j}
\!\!\!&=&\!\!\!
\frac{2j}{j + 1}
z^{\mu_1} \dots z^{\mu_j}
\bar\psi \gamma_{[ \rho} i \stackrel{{}_\leftrightarrow}{\cal D}_{\{ \mu_1]}
\dots i \stackrel{{}_\leftrightarrow}{\cal D}_{\mu_j \}} \psi_q
\, .
\end{eqnarray}
The curly $\{ \cdots \}$ and square $[ \cdots ]$ brackets denote the symmetrization
and antisymmetrization of the corresponding Lorentz indices (including the combinatoric
factors), respectively. Notice that the symmetrization is performed without the trace
subtraction. Obviously, if one  neglects  dynamical twist-four and higher effects, one
has  $z^\rho {\cal O}^{qq}_{\rho} (-z, z) = z^\rho {\cal O}^{qq, \rm sym}_{\rho} (- z, z)$
and the same holds true for local operators. The symmetric Wilson operators can be concisely
represented as
\begin{equation}
{\cal O}^{qq, \rm sym}_{\rho; j}
=
\frac{1}{j + 1} \frac{\partial}{\partial z^\rho} \, z^\sigma
{\cal O}^{qq}_{\sigma; j}
\, , \qquad
{\cal O}^{qq}_{\rho; j}
\equiv
z^{\mu_1} \dots z^{\mu_j} {\cal O}^{qq}_{\rho \mu_1 \dots \mu_j}
\, .
\end{equation}
The same formulas are valid in the axial sector $\widetilde{O}^{qq}_\rho$.

As a next step, we subtract traces from ${\cal O}^{qq, \rm sym}_{\rho}$
and ${\cal O}^{qq, \rm asym}_{\rho}$, thus rewriting the decomposition
(\ref{DecomposionBody}) in the form
\begin{equation}
{\cal O}^{qq}_\rho \! \left( - z , z \right)
=
{\cal R}^{2,qq}_\rho \! \left( - z , z \right)
+
{\cal R}^{3,qq}_\rho \! \left( - z , z \right)
+
{\cal R}^{r,qq}_\rho \! \left( - z , z \right)
\, ,
\end{equation}
which then gives definite twist-two ${\cal R}^{2, qq}_\rho$ and twist-three
${\cal R}^{3, qq}_\rho$ operators. Even higher twist operators are encoded
in ${\cal R}^{r, qq}_\rho$. A systematic construction of traceless operators
can be effectively achieved making use of the concept of the so-called harmonic
polynomials \cite{Vil68}. The construction of harmonic polynomials goes as
follows. A rank-$j$ (not necessarily symmetric and definitely not traceless)
tensor $T_{\mu_1 \dots \mu_j}$ is contracted first with $j$ four-vectors $z^\mu$,
\begin{equation}
\label{ContrTensor}
T_j (z) \equiv z^{\mu_1} \dots z^{\mu_j} T_{\mu_1 \dots \mu_j}
\, .
\end{equation}
The traceless tensor $\bar{T}_{\mu_1 \dots \mu_j}$
\begin{equation}
\label{TracelessConditionGeneral}
\bar{T}_j (z)
=
z^{\mu_1} \dots z^{\mu_j}
\bar{T}_{\mu_1 \dots \mu_j}
=
z^{\mu_1} \dots z^{\mu_j}
\Sym_{\mu_1 \dots \mu_j} T_{\mu_1 \dots \mu_j}
\, ,
\end{equation}
is built from (\ref{ContrTensor}) by means of the operation $\Sym$ introduced in
Section \ref{ClassificationTwistTwo}. However, so far we did not have its algebraic
definition. This is accomplished by noticing that $\bar{T}_j (z)$ satisfies the
condition \cite{Vil68}
\begin{equation}
\Box \bar{T}_j (z) = 0
\, , \qquad
\Box \equiv \frac{\partial}{\partial z^\mu} \frac{\partial}{\partial z_\mu}
\, .
\end{equation}
Then, the traceless and symmetric tensor $\bar{T}_{\mu_1 \dots \mu_j}$ can be
represented as a polynomial $H_j$ of the $j$-th order in the variable $z^2 \Box$
acting on $T_j (z)$
\begin{equation}
\label{TracelessnessCondition}
\bar{T}_j (z) = H^{j} \left( z^2, \Box \right) T_j (z)
\, .
\end{equation}
Substituting (\ref{TracelessnessCondition}) into (\ref{TracelessConditionGeneral})
one finds the explicit expression for the polynomials
\cite{Vil68,BalBra89,RobGeyLaz99}
\begin{eqnarray}
H^{j} \left( z^2, \Box \right)
=
\sum_{k=0}^{[j/2]}
\frac{
{\mit\Gamma} \left(j - k + 1 \right)
}{
k! {\mit\Gamma} \left( j + 1 \right)
}
\left( - \frac{z^2 \Box}{4} \right)^k
\, ,
\end{eqnarray}
which are dubbed harmonic polynomials.

For the case at hand, the tensor $T_{\mu_1 \dots \mu_j}$ is given by the local Wilson
operator (\ref{LocalWilsonOperatorsEven}). It is important to realize that the
derivatives $\partial/\partial z^\mu$ in Eq.\ (\ref{TracelessnessCondition}) act on
the coordinates $z^\mu$ which are contracted with the tensor indices of the operator
in question and not on the quark fields\footnote{This is, by the way, the reason that
we write the derivatives as $\partial/\partial z^\mu$ rather than $\partial_\mu$.}.
Compared to the example (\ref{ContrTensor}), the operators $\mathcal{O}_{\rho \mu_1
\dots \mu_j}^{qq}$ possess an open Lorentz index $\rho$. Therefore, the equation
defining the harmonic polynomials has to be complemented by yet another one. The
tracelessness condition now reads
\begin{equation}
\Box {\cal R}^{\tau,qq}_{\rho; j} = 0
\, , \qquad
\frac{\partial}{\partial z_\rho} {\cal R}^{\tau,qq}_{\rho; j} = 0
\, ,
\end{equation}
for each twist $\tau = 2, 3$. Straightforward algebra yields the decomposition
(cf.\ (\ref{NonLoc-Loc-Ope}))
\begin{equation}
{\cal O}^{qq}_{\rho; j}
=
{\cal R}^{2,qq}_{\rho; j}
+
\frac{2 j}{j + 1} {\cal R}^{3,qq}_{\rho; j}
+
{\cal R}^{r,qq}_{\rho; j}
\, ,
\end{equation}
with the traceless twist-two and -three operators \cite{RobGeyLaz99,BelMul01b}
\begin{eqnarray}
\label{R2traceless}
{\cal R}^{2,qq}_{\rho; j}
\!\!\!&=&\!\!\!
\frac{1}{j + 1}
\frac{\partial}{\partial z^\rho} H^{j + 1} \left( z^2, \Box \right)
z^\sigma {\cal O}^{qq}_{\sigma; j} \, ,
\qquad\qquad\qquad\qquad\qquad\qquad\quad\
j \geq 0
\, , \\
\label{R3traceless}
{\cal R}^{3,qq}_{\rho; j}
\!\!\!&=&\!\!\!
\frac{1}{2 j}
\left[
g_{\rho\sigma} z^\mu \frac{\partial}{\partial z^\mu}
-
z_\sigma \frac{\partial}{\partial z^\rho}
\right]
\left[
g^{\sigma\tau}  - \frac{1}{j + 1} z^\sigma \frac{\partial}{\partial z_\tau}
\right]
H^{j} \left( z^2, \Box \right) {\cal O}^{qq}_{\tau; j}
\, , \qquad
j \geq 1
\, ,
\end{eqnarray}
and the remainder being
\begin{eqnarray*}
{\cal R}^{r,qq}_{\rho; j} \!\!\!&=&\!\!\!
\frac{1}{j + 1}
\Bigg\{
\frac{\partial}{\partial z^\rho}
\left[ 1 - H^{j + 1} \left( z^2, \Box \right) \right] z^\sigma
+
\left[
g_\rho{}^\sigma z^\mu \frac{\partial}{\partial z^\mu}
-
z^\sigma \frac{\partial}{\partial z^\rho}
\right]
\left[ 1 - H^{j} \left( z^2, \Box \right) \right]
\nonumber\\
&&\qquad
+
\frac{1}{j + 1}
\left[ z_\rho z^\mu \frac{\partial}{\partial z^\mu}
-
z^2 \frac{\partial}{\partial z^\rho} \right]
\frac{\partial}{\partial z_\sigma}
H^{j} \left( z^2, \Box \right)
\Bigg\} {\cal O}^{qq}_{\sigma; j}
\, .
\end{eqnarray*}
It is an easy task to check that the operators ${\cal R}^{\tau, qq}_{\rho; j}$ ($\tau = 2,
3$) are indeed traceless. To this end, one uses
$$
\Box H^j \left( z^2, \Box \right) {\cal O}^{qq}_{\mu; j}
= 0
$$
and the Euler theorem
$$
z^\nu \frac{\partial}{\partial z^\nu} \frac{\partial}{\partial z_\mu} H^
j \left( z^2, \Box \right) {\cal O}^{qq}_{\mu; j}
=
(j - 1) \frac{\partial}{\partial z_\mu} H^j \left( z^2, \Box \right)
{\cal O}^{qq}_{\mu; j}
\, .
$$
When we set $z^2 = 0$ in these equations, we reproduce the earlier results given
in Eqs.\ (\ref{QuarkLocalOperator}) and (\ref{TwistThreeLocalOperators}). The
operator ${\cal R}^{r,qq}$ generates effects proportional to the total derivative
squared $\partial^2$ as well as multiparticle operators
$$
{\cal R}^{r, qq}_{\rho;j} \sim - \partial^2 {\cal R}^{2, qq}_{\rho; j}
+
\, {\rm multiparticle \ operators}
\, .
$$
The latter are not included into our consideration as we pointed out at the
beginning of this section. When ${\cal R}^{r, qq}_{\rho;j}$ is sandwiched
between states with different momenta, the total derivative term produces
contributions proportional to the momentum transfer squared $\Delta^2$.
Limiting our present analysis to the situation $M^2 \gg |\Delta^2|$, with
$M^2$ being the hadron mass, we will also neglect the latter in what follows
(see, however, Appendix \ref{LightRayTwistThree}).

\subsubsection{Spin-zero target}
\label{Sec-Res}

After these preliminary remarks, let us turn to the discussion of the general
framework for the resummation of kinematical mass corrections. Up to now, the
problem was addressed at leading twist only \cite{BelMul00}. Unfortunately, as
we described in detail in Section \ref{TreeLevelCompton}, this approximation
violates the electromagnetic gauge invariance. Since the target mass corrections
stemming from higher twist terms are not available, we restrict our present
consideration to the leading twist sectors only. The major steps in performing
the resummation are: parametrization of off-forward matrix elements of symmetric
local operators in terms of moments of DDs, subtraction of traces using the
harmonic projectors $H^j$, subsequent Fourier transformation to the momentum
space and, finally, resummation of infinite series.

To illustrate the main features of the formalism, we consider a spinless target.
As before, to leading twist accuracy only matrix elements of the parity even
operators are relevant in this case. To evaluate them, we perform as a first step
the form factor decomposition of the symmetric quark operator as follows
\begin{equation}
\label{Def-RedMattw2}
\langle p_2|
{\cal O}^{qq, {\rm sym}}_{\rho \mu_1 \cdots \mu_j}
|p_1 \rangle
= p_{\{\rho} p_{\mu_1} \cdots p_{\mu_j\}} H^q_{j + 1, 0}
+ \cdots
+ \Delta_{\{\rho} \Delta_{\mu_1} \cdots \Delta_{\mu_j\}} H^q_{j + 1,j + 1 }
+ \dots \ ,
\end{equation}
where $\{ \cdots \}$ denotes as usual the symmetrization. The ellipsis stands
for terms containing the metric tensor, which disappear when we subsequently
apply the harmonic polynomial $H^j$ to this expression. After the projection
via Eq.\ (\ref{R2traceless}), one gets
\begin{eqnarray}
\label{FinSumTw2}
\langle p_2| {\cal R}^{2, qq}_{\rho;j} | p_1 \rangle
\!\!\!&=&\!\!\!
\frac{1}{j + 1} \partial_\rho H^{j + 1} \left( z^2, \Box \right)
\sum_{k = 0}^{j + 1} \left( z \cdot p \right)^{j + 1 - k}
\left( z \cdot \Delta \right)^{k} H^q_{j + 1,  k} \\
&=&\!\!\! z_{\mu_1} \dots z_{\mu_j}
\Sym_{\rho \mu_1 \dots \mu_j}
\left\{
p_\rho p_{\mu_1} \dots p_{\mu_j} H^q_{j + 1, 0}
+ \cdots
+ \Delta_\rho \Delta_{\mu_1} \dots \Delta_{\mu_j} H^q_{j+1, j+1}
\right\}
\, .
\nonumber
\end{eqnarray}
The first line gives a compact expression for the subtracted operator, which is
extremely convenient for further  derivations, and is the basis of the formalism.

On the other hand, we can project out  the  twist-two operator from Eq.\
(\ref{Def-RedMattw2}) by  contracting  all Lorentz indices with the light-like
vector $n_\mu$. After such a projection, the trace terms disappear. This
observation suggests that the coefficients $H^q_{j + 1, k}$ in front of the
Lorentz tensors are related to the reduced matrix elements of the light-cone
operators and can be represented in terms of the moments of conventional
leading twist GPDs or DDs via (cf.\ Eq.\ (\ref{eq:one-DD}))
\begin{equation}
\label{BversusDD}
H^q_{j, k} = \frac{1}{k!}
\left. \frac{\partial^k}{\partial\eta^k} \right|_{\eta = 0}
\int_{-1}^{1} dx\; x^{j - 1} H^q (x,\eta)
= \left({j \atop k}\right) \int_{{\mit\Omega}}
d\beta \, d\alpha \,  \beta^{j - k} \alpha^{k} h^q (\beta, \alpha)
\end{equation}
where $0 \le k \le j$, $1 \le j$. Consequently, the parametrizations\footnote{As
discussed in Section \ref{spin0DD}, this is an example of the DD parametrization
by a single function \cite{BelKirMulSch00c}.} of the matrix element of the non-local
symmetric operator ${\cal O}^{qq, {\rm sym}}_\rho$ and the light-ray operator
$\mathcal{R}^{2,qq}_\rho$ coincides up to terms proportional to $z_\rho$,
$\Delta_\rho^\perp$ or $z^2$:
\begin{equation}
\label{definitionDD}
\langle p_2 | {\cal O}^{qq, {\rm sym}}_\rho (- z, z) | p_1 \rangle
=
\int_{{\mit\Omega}} d\beta \, d\alpha \, h^q (\beta, \alpha) {\cal P}_\rho
\ {\rm e}^{- i z \cdot {\cal P}} + \dots
\, ,
\end{equation}
where ${\cal P}_\mu \equiv \beta p_\mu + \alpha \Delta_\mu$.

Now, it easy to perform the resummation of matrix elements of traceless local
operators. Substitution of Eq.\ (\ref{BversusDD}) into  Eq.\ (\ref{FinSumTw2})
yields
\begin{equation}
\label{LocalOperDD}
\langle p_2| {\cal R}^{2, qq}_{\rho;j} | p_1 \rangle
= \int_{{\mit\Omega}} d \beta \, d \alpha \, h^q (\beta, \alpha)
\frac{1}{j + 1} \partial_\rho H^{j + 1} \left( z^2, \Box \right)
\left( z \cdot {\cal P} \right)^{j + 1}
\, .
\end{equation}
The action of the harmonic polynomials on the scalar product standing to its right
generates conventional Chebyshev polynomials (\ref{ChebyshevPolynomial2}), which
form an irreducible representation of the orthogonal group $SO (4)$ \cite{Vil68},
i.e., the Euclidean Lorentz group,
\begin{equation}
\label{ChebyshevResumm}
H^j \left( z^2, \Box \right) \left( z \cdot {\cal P} \right)^j
=
\left( \frac{z^2 {\cal P}^2}{4} \right)^{j/2}
U_{j} \left( \frac{z \cdot {\cal P}}{\sqrt{z^2 {\cal P}^2}} \right)
\, .
\end{equation}

Since a typical contribution of twist-two Wilson operators to the Compton amplitude
(\ref{AmplitudeT}) has the form
$$
\int d^4 z \, {\rm e}^{i z \cdot q} \frac{z_\sigma}{[-z^2]^k}
\langle p_2| {\cal R}^{2, qq}_{\rho;j} | p_1 \rangle
=
\langle p_2| {\cal R}^{2, qq}_{\rho \mu_1 \dots \mu_j} | p_1 \rangle
\int d^4 z \, {\rm e}^{i z \cdot q} \frac{z_\sigma z_{\mu_1} \dots z_{\mu_j}}{[-z^2]^k}
\, ,
$$
the Fourier transformation can be done using the following formula
\begin{eqnarray*}
\int d^4 z {\rm e}^{i z \cdot q}
\frac{ z_{\mu_1} \dots z_{\mu_j}}{[- z^2]^k}
= (- i)^{j + 1} 2^{4 - 2 k + j} \pi^{2}
\frac{{\mit\Gamma} (2 - k + j)}{{\mit\Gamma} (k)}
\frac{ q_{\mu_1} \dots q_{\mu_j}}{[- q^2]^{2 - k + j}}
\, ,
\end{eqnarray*}
which is a generalization of Eq.\ (\ref{DdimenCoordIntegral}). Here on the
right-hand side of the equality, we dropped all terms involving the metric
tensor $g_{\mu_i \mu_k}$, since they vanish when contracted with the traceless
tensor ${\cal R}^{2, qq}_{\rho \mu_1 \dots \mu_j}$. This yields
\begin{eqnarray}
\label{Integral}
&&\int d^4 z {\rm e}^{i z \cdot q} \frac{z_\sigma}{[- z^2]^{2}}
\frac{z_{\mu_1}}{2} \dots \frac{z_{\mu_j}}{2}
\langle p_2 | {\cal R}^{2, qq}_{\rho \mu_1 \dots \mu_j} | p_1 \rangle
\nonumber\\
&&\qquad\qquad\qquad\qquad= i^{j + 2} \pi^{2}
{\mit\Gamma} (j + 1)
\frac{1}{q^2} {\mit\Pi}_{\sigma \mu_1}
\frac{q_{\mu_2}}{q^2} \dots \frac{q_{\mu_j}}{q^2}
\langle p_2 | {\cal R}^{2, qq}_{\rho \mu_1 \dots \mu_j} | p_1 \rangle
\, ,
\end{eqnarray}
where we introduced a tensor
\begin{equation}
{\mit\Pi}_{\mu\nu}
\equiv g_{\mu\nu} - \frac{2}{q^2} q_\mu q_\nu ,
\qquad
{\mit\Pi}_{\mu\rho} {\mit\Pi}_{\rho\nu} = g_{\mu\nu}
\, .
\end{equation}
Assembling all results together, we find the contribution of the twist-two operators
to the Compton amplitude (\ref{AmplitudeT})
\begin{eqnarray}
\label{Fourier}
&&\int d^4 x {\rm e}^{i z \cdot q} \frac{z_\sigma}{[- z^2]^2}
\langle p_2 |
{\cal R}^{2, qq}_\rho \left( - \ft{z}{2}, \ft{z}{2} \right)
| p_1 \rangle \nonumber\\
&&\qquad\qquad\qquad
= 2 \pi^2
\left\{
\frac{q_\sigma}{q^2} \sum_{j = 0}^{\infty}
\langle p_2 | {\cal R}^{2, qq}_{\rho [j]} | p_1 \rangle
- \frac{1}{2}
{\mit\Pi}_{\sigma\tau} \frac{\partial}{\partial q_\tau}
\sum_{j = 1}^{\infty} \frac{1}{j}
\langle p_2 | {\cal R}^{2, qq}_{\rho [j]} | p_1 \rangle
\right\} .
\end{eqnarray}
Here we have used a new convention
\begin{equation}
{\cal R}^{2, qq}_{\rho [j]}
=
\frac{q_{\mu_1}}{q^2} \dots \frac{q_{\mu_j}}{q^2}
{\cal R}^{2, qq}_{\rho \mu_1 \dots \mu_j}
\, ,
\end{equation}
for contraction of the twist-two local operators $\mathcal{R}^{2, qq}$ with the
four-vectors $q^\mu/q^2$ and also exploited the identity
\begin{eqnarray*}
\frac{\partial}{\partial \frac{q_\sigma}{q^2}}
= q^2 {\mit\Pi}_{\sigma\tau} \frac{\partial}{\partial q_\tau}
\, .
\end{eqnarray*}

To evaluate the sums in (\ref{Fourier}), we substitute (\ref{ChebyshevResumm})
in (\ref{LocalOperDD}) and replace $z_\mu$ by $q_\mu/q^2$. This gives the local
traceless matrix elements $\langle p_2 | {\cal R}^{2, qq}_{\rho [j]} | p_1 \rangle$
in terms of the Chebyshev polynomials. The summation in Eq.\ (\ref{Fourier}) can
now be done with the help of the generating function for the Chebyshev polynomials
(\ref{ChebyshevPolynomial2})
\begin{equation}
G_U (a, b)
\equiv
\sum_{j = 0}^{\infty} a^j U_j (b)
=
(1 - 2 a b + a^2)^{- 1}
\, .
\end{equation}
The  required $j$-dependent coefficient in the series is obtained by
integrating  both sides of the equation with appropriate weights. In
this way one obtains the sums needed for calculation of (\ref{Fourier}),
\begin{eqnarray}
\label{ChebyshevSum}
&&\sum_{j = 0}^{\infty} \frac{a^{j + 1}}{(j + 1)} U_{j + 1} (b)
= \int_0^{a} \frac{da'}{a'}
\left( G_U (a', b) - 1 \right)
\, , \\
&&\sum_{j = 1}^{\infty} \frac{a^{j + 1}}{j(j + 1)} U_{j + 1} (b)
= \int_0^a da' \int_0^{a'} \frac{da''}{\left( a'' \right)^2}
\left( G_U (a'', b) - 1 - 2 a'' b \right)
\, .
\nonumber
\end{eqnarray}
Here $a = \ft12 \sqrt{{\cal P}^2/q^2}$ and $b = q \cdot {\cal P}/\sqrt{q^2 {\cal P}^2}$.
This immediately leads to
\begin{eqnarray}
\label{FouTra1}
&&\sum_{j = 0}^{\infty}
\langle p_2 | {\cal R}^{2, qq}_{\rho [j]} | p_1 \rangle
=
q^2 {\mit\Pi}_{\rho\tau} \frac{\partial}{\partial q_\tau}
\int_{{\mit\Omega}} d \beta \, d\alpha\, h^q (\beta, \alpha) \\
&&\qquad\times
\left\{
\frac{1 - \sqrt{1 + {\cal M}^2}}{2 \sqrt{1 + {\cal M}^2}}
\ln\left(1 + \frac{1 - \sqrt{1 + {\cal M}^2}}{2{\mit\Xi}} \right)
-
\frac{1 + \sqrt{1 +{\cal M}^2}}{2 \sqrt{1 + {\cal M}^2}}
\ln\left(1 + \frac{1 + \sqrt{1 + {\cal M}^2}}{2{\mit\Xi}} \right)
\right\} , \nonumber\\
\label{FouTra2}
&&\sum_{j = 1}^{\infty} \frac{1}{j}
\langle p_2 | {\cal R}^{2, qq}_{\rho [j]} | p_1 \rangle
= - q^2 {\mit\Pi}_{\rho\tau} \frac{\partial}{\partial q_\tau}
\int_{{\mit\Omega}} d \beta \, d\alpha\, h^q (\beta, \alpha) \nonumber\\
&&\qquad \times
\Bigg\{ \frac{1}{{\mit\Xi}}
+ \frac{1 - \sqrt{1 + {\cal M}^2}}{2 \sqrt{1 + {\cal M}^2}}
\left( 1 + \frac{1 - \sqrt{1 + {\cal M}^2}}{2{\mit\Xi}} \right)
\ln\left(1 + \frac{1 - \sqrt{1 + {\cal M}^2}}{2{\mit\Xi}} \right) \nonumber\\
&&\qquad\qquad\ \ -
\frac{1 + \sqrt{1 + {\cal M}^2}}{2 \sqrt{1 + {\cal M}^2}}
\left( 1 + \frac{1 + \sqrt{1 + {\cal M}^2}}{2{\mit\Xi}} \right)
\ln\left(1 + \frac{1 + \sqrt{1 + {\cal M}^2}}{2{\mit\Xi}} \right)
\Bigg\} ,
\end{eqnarray}
where we introduced the variables
\begin{equation}
{\mit\Xi} \equiv \frac{Q^2}{q \cdot {\cal P}}
=
\frac{\xi}{\beta + \eta \alpha}
\
\, , \qquad
{\cal M}^2
\equiv \frac{Q^2 {\cal P}^2}{\left( q \cdot {\cal P} \right)^2}
=
{\mit\Xi}^2
\frac{4 M^2 \beta^2}{Q^2}
\left\{ 1 - \frac{\Delta^2}{4 M^2} \left(1 - \frac{\alpha^2}{\beta^2} \right) \right\}
\, .
\end{equation}

After inserting (\ref{FouTra1}) and (\ref{FouTra2}) into Eq.\ (\ref{Fourier})
one should perform  differentiation with respect to $q_\mu$, which is done by
means of the formula for a test function $\tau$
\begin{equation}
\label{DiffQ}
q^2 {\mit\Pi}_{\mu\nu} \frac{\partial}{\partial q_\nu}
\tau \left( {\cal M}^2, {\mit\Xi} \right)
= \left\{
2 {\cal M}^2 \left( q_\mu + {\mit\Xi} {\cal P}_\mu \right)
\frac{\partial}{\partial{\cal M}^2}
- {\cal P}_\mu \frac{\partial}{\partial{\mit\Xi}^{-1}}
\right\}
\tau \left( {\cal M}^2, {\mit\Xi} \right)
\, .
\end{equation}
Finally, one needs to contract the resulting equation with the tensor
$S^{\mu\nu;\rho\sigma}$ to get the hadronic tensor
\begin{equation}
\label{Tw2Amplitude}
T^2_{\mu \nu} =
\sum_q Q_q^2
\int_{{\mit\Omega}} d\beta \, d\alpha \,
h^q (\beta, \alpha)
\left\{
-
\left( g_{\mu\nu} - \frac{q_\mu q_\nu}{q^2} \right) {\cal C}_1
+
\frac{1}{q \cdot {\cal P}}
\left( {\cal P}_\mu + \frac{q_\mu}{{\mit\Xi}} \right)
\left( {\cal P}_\nu + \frac{q_\nu}{{\mit\Xi}} \right) {\cal C}_2
\right\} ,
\end{equation}
with mass-dependent coefficient functions
\begin{eqnarray}
\label{F1}
&&{\cal C}_1
=
\frac{4 {\mit\Xi} - {\cal M}^2 \left( 1 - {\mit\Xi} \right)}
{{\mit\Xi} \left[ 4 {\mit\Xi} (1 + {\mit\Xi}) - {\cal M}^2 \right]}
+ \frac{{\cal M}^2 \left( 2 {\mit\Xi} - {\cal M}^2 \right)}
{4 {\mit\Xi} \left( 1 + {\cal M}^2 \right)^{3/2}}
\ln \left(
\frac{1 - \sqrt{1 + {\cal M}^2} + 2 {\mit\Xi}}{1
+ \sqrt{1 + {\cal M}^2} + 2 {\mit\Xi}}
\right)
+ \left( {\mit\Xi} \to - {\mit\Xi} \right) ,
\nonumber\\
\label{F2}
&&{\cal C}_2
= \frac{4 {\mit\Xi} - {\cal M}^2 \left( 1 - {\mit\Xi} \right)}
{(1 + {\cal M}^2) \left[ 4 {\mit\Xi} (1 + {\mit\Xi}) - {\cal M}^2 \right]}
+
\frac{ 3 {\cal M}^2 \left( 2 {\mit\Xi} - {\cal M}^2 \right)}
{4 \left( 1 + {\cal M}^2 \right)^{5/2}}
\ln \left(
\frac{1 - \sqrt{1 + {\cal M}^2} + 2 {\mit\Xi}}{1
+ \sqrt{1 + {\cal M}^2} + 2 {\mit\Xi}}
\right) \\
&&\qquad\qquad\qquad\qquad\qquad\qquad\qquad
\qquad\qquad\qquad\qquad\qquad\qquad\qquad\qquad\quad \
- \left( {\mit\Xi} \to - {\mit\Xi} \right) .
\nonumber
\end{eqnarray}
To pick up the correct sheet of the Riemann surface for the logarithm, it is
necessary to restore the suppressed Feynman prescription $Q^2 \to Q^2 - i 0$
in (\ref{Integral}). In the forward limit, Eq.\ (\ref{Tw2Amplitude}) coincides
with the well-known result \cite{GeoPol76}. As expected the current conservation
is fulfilled in the forward case but violated for off-forward kinematics. The
leading order massless (generalized) Callan--Gross relation \cite{BelMulKirSch00}
${\cal C}_2 = {\mit\Xi} {\cal C}_1$ is also broken by target mass corrections.

\subsubsection{Spin-one-half target}

In case of a spin-one-half target, the matrix elements $\langle p_2|
{\cal R}^{2, qq}_{\rho; j} |p_1\rangle$ are represented in terms of
the vectors $p$, $\Delta$ and a set of (independent) Dirac bilinears,
characterizing the spin content of the target. It is convenient to use
the same Dirac structures $h_\rho, b$ and $\tilde{h}_\rho, \tilde{b}$
as previously in Section \ref{PolynomSpin12GPDs} in order to exploit
the scalar sector analyzed above. Thus, the matrix element of, e.g.,
vector operator is parametrized according to
\begin{eqnarray}
\label{LorentzVscalZ}
\langle p_2|
{\cal O}^{qq, {\rm sym}}_{\rho \mu_1 \dots \mu_j}
| p_1 \rangle
\!\!\!&=&\!\!\!
h_{\{\rho} p_{\mu_1} \cdots p_{\mu_j\}} A^q_{j+1, 0}
+ \cdots
+ h_{\{ \rho} \Delta_{\mu_1} \cdots \Delta_{\mu_j \}} A^q_{j+1, j}
\\
\!\!\!&+&\!\!\!
\frac{b}{2M_N}
\left\{
p_{\{ \rho} p_{\mu_1} \cdots p_{\mu_j \}} B^q_{j+1, 0}
+ \cdots
+ \Delta_{\{ \rho} \Delta_{\mu_1} \cdots \Delta_{\mu_j \}} B^q_{j+1, j+1}
\right\} + \dots,
\nonumber
\end{eqnarray}
where again the ellipses denote terms proportional to the metric tensor, which
do not contribute to harmonic projections. The non-local twist-two vector operator
is given by
\begin{eqnarray}
\langle p_2 | {\cal R}^{2, qq}_\rho (- z, z) | p_1 \rangle
\!\!\!&=&\!\!\!
\int_{{\mit\Omega}} d \beta \, d \alpha \,
\left\{
h_A^q (\beta, \alpha) \, h \cdot \partial^{\cal P}
+
h^q_B (\beta, \alpha) \, \frac{b}{2M_N} \, {\cal P} \cdot \partial^{\cal P}
\right\} \\
&&\qquad\qquad\qquad\times
\partial_\rho \sum_{j = 0}^{\infty} \frac{(-i)^{j}}{j!(j + 1)^2}
\left( \frac{z^2 {\cal P}^2}{4} \right)^{(j + 1)/2}
U_{j + 1} \left( \frac{z \cdot {\cal P}}{\sqrt{z^2 {\cal P}^2}} \right)
\, ,
\nonumber
\end{eqnarray}
in terms of DDs introduced in Section \ref{SectionSingleCompDDnucleon}.
Here we have generated the factor of $z$ in $h \cdot z$ by the differentiation
with respect to ${\cal P}$. The same trick was done with the $b$ form factor
for the purpose of a uniform representation. The same equation holds for the
axial operator.

The Fourier transform and the resummation are done in the same vein as for a
spinless target, and the result is
\begin{eqnarray}
T^2_{\mu\nu}
\!\!\!&=&\!\!\!
\sum_q Q_q^2
\int_{{\mit\Omega}} d \beta \, d \alpha \,
\Bigg\{
\left( \tilde h^q_A (\beta, \alpha) \, \tilde h \cdot \partial^{\cal P}
+
\tilde h^q_B (\beta, \alpha) \, \frac{\tilde b}{2M_N} \,
{\cal P} \cdot \partial^{\cal P}
\right)
\frac{i}{q \cdot {\cal P}} \varepsilon_{\mu\nu\rho\sigma} q_\rho {\cal P}_\sigma
\ \widetilde {\cal C}_1 \\
&&\qquad\qquad\qquad\ \ \,
+ \left( h^q_A (\beta, \alpha) \, h \cdot \partial^{\cal P}
+
h^q_B (\beta, \alpha) \, \frac{b}{2M_N} \, {\cal P} \cdot \partial^{\cal P}
\right) \nonumber\\
&&\quad\qquad\qquad\qquad\qquad\qquad\times
\left(
-
\left( g_{\mu\nu} - \frac{q_\mu q_\nu}{q^2} \right) {\cal C}_1
+
\frac{1}{q \cdot {\cal P}}
\left( {\cal P}_\mu + \frac{q_\mu}{{\mit\Xi}} \right)
\left( {\cal P}_\nu + \frac{q_\nu}{{\mit\Xi}} \right) {\cal C}_2
\right) \nonumber
\Bigg\}
\, ,
\end{eqnarray}
with massive coefficient functions
\begin{eqnarray}
\label{Fn1}
{\cal C}_1
\!\!\!&=&\!\!\!
\frac{[2 {\mit\Xi} \left( 1 + 2 {\cal M}^2 \right) - {\cal M}^4]L_-}
{4 {\mit\Xi} \left( 1 + {\cal M}^2 \right)^{3/2}}
- \frac{L_+}{2 \left( 1 + {\cal M}^2 \right)}
- \frac{{\cal M}^2 \, L}{2 \left( 1 + {\cal M}^2 \right)^{3/2}}
+ \left( {\mit\Xi} \to - {\mit\Xi} \right)
\, , \nonumber\\
\label{Fn2}
{\cal C}_2
\!\!\!&=&\!\!\!
\frac{[ 2 {\mit\Xi} \left( 1 + 4 {\cal M}^2 \right) - 3 {\cal M}^4 ] L_-}
{4 \left( 1 + {\cal M}^2 \right)^{5/2}}
- \frac{{\mit\Xi} (1 - 2 {\cal M}^2) L_+}{2 \left( 1 + {\cal M}^2 \right)^2}
- \frac{3 {\mit\Xi} {\cal M}^2 \, L}{2
\left( 1 + {\cal M}^2 \right)^{5/2}}
- \left( {\mit\Xi} \to - {\mit\Xi} \right)
\, , \nonumber\\
\label{Gn1}
\widetilde {\cal C}_1
\!\!\!&=&\!\!\!
\frac{L_-}
{2 \left( 1 + {\cal M}^2 \right)^{1/2}}
- \frac{L_+}{2 \left( 1 + {\cal M}^2 \right)}
- \frac{{\cal M}^2 \, L}{2 \left( 1 + {\cal M}^2 \right)^{3/2}}
+ \left( {\mit\Xi} \to - {\mit\Xi} \right)
\, ,
\end{eqnarray}
expressed via
\begin{eqnarray*}
L_\pm
\!\!\!&\equiv&\!\!\!
\ln \frac{1 - \sqrt{1 + {\cal M}^2} + 2 {\mit\Xi}}{2 {\mit\Xi}}
\pm
\ln \frac{1 + \sqrt{1 + {\cal M}^2} + 2 {\mit\Xi}}{2 {\mit\Xi}}, \\
L
\!\!\!&\equiv&\!\!\!
{\rm Li}_2 \left( - \frac{1 - \sqrt{1 + {\cal M}^2}}{2 {\mit\Xi}} \right)
-
{\rm Li}_2 \left( - \frac{1 + \sqrt{1 + {\cal M}^2}}{2 {\mit\Xi}} \right)
\end{eqnarray*}
logarithms and Euler dilogarithms ${\rm Li}_2 (x) = - \int_0^x \ft{d y}{y}
\ln (1 - y)$.

\subsubsection{Twist-four mass corrections}

As we mentioned earlier, the resummed mass corrections can be expressed
in terms of GPDs by means of the inverse Radon transform (\ref{Res-Inv}),
which however involves them in unphysical regions of the momentum space.
Hence, this step does not prove useful in phenomenological applications.
However, every term in the mass expansion, $M^2/Q^2$, can be easily converted
into the conventional representation with GPDs involved only in the region of
their support. Let us demonstrate this for the expanded Compton form factor
${\cal F}_1$ for a (pseudo) scalar target (\ref{F1}). To the first non-trivial
order, i.e., to ${\cal O} \left( M^4/Q^4 \right)$ accuracy, the latter reads
in the DD form
\begin{eqnarray}
\label{F1-expa}
{\cal F}_1 \!\!\!
&\approx&\!\!\!
\sum_q Q_q^2
\int_{{\mit\Omega}} d \beta \, d \alpha \, h^q (\beta, \alpha)
\Bigg\{
\Bigg(
C_1^{(0)}\! \left( {\mit\Xi}^{-1} - i 0 \right)
+
C_1^{(0)}\! \left( - {\mit\Xi}^{-1} - i 0 \right)
\Bigg) \nonumber\\
&&\qquad\qquad\qquad\qquad\qquad +
\frac{M^2}{Q^2} \beta^2 {\mit\Xi}^2
\Bigg(
C_1^{(1)}\! \left( {\mit\Xi}^{-1} - i 0 \right)
+
C_1^{(1)}\! \left( - {\mit\Xi}^{-1} - i 0 \right)
\Bigg)
\Bigg\}
\, .
\nonumber
\end{eqnarray}
Here in the approximation $M^2 \gg \Delta^2$, we  set
$$
{\cal M}^2 \approx 4 \frac{M^2}{Q^2} \beta^2 {\mit\Xi}^2
$$
and introduced the mass-independent coefficient functions
$$
C_1^{(0)}\! \left( {\mit\Xi}^{-1} \right) = - \left( 1 + {\mit\Xi} \right)^{-1}
\, , \qquad
C_1^{(1)}\! \left( {\mit\Xi}^{-1} \right)
=
\left( 1 + {\mit\Xi} \right)^{-2}
+
2 \ln \left( \frac{{\mit\Xi}}{1 + {\mit\Xi}} \right)
\, .
$$
To switch from the DD to the usual GPD representation, we need the following
general result
\begin{eqnarray}
\label{DDtoGPD1}
&&
\int_{{\mit\Omega}} d \beta \, d \alpha \,
\left\{
\!\!
\begin{array}{c}
\alpha^n \\
\beta^n
\end{array}
\!\!
\right\}
\, {\cal C} \left( {\mit\Xi}^{-1} \right) h^q (y, z)
\\
&&\qquad\qquad\qquad
=
\int d x \, {\cal C} \left( \frac{x}{\xi} \right) \int d y \,
V^{(n)}_1 (x, y)
\left\{
\!\!
\begin{array}{c}
\frac{\partial^n}{\partial \eta^n} \\
\prod_{k = 0}^{n - 1}
\left(
- y \frac{\partial}{\partial y} - \eta \frac{\partial}{\partial \eta} - k
\right)
\end{array}
\!\!
\right\}
H^q (y, \eta)
\, , \nonumber
\end{eqnarray}
where the kernels are expressed in terms of generalized step-functions
(\ref{Theta110Appendix}),
\begin{eqnarray}
V^{(n)}_1 (x, y)
=
\frac{(y - x)^{n - 1}}{(n - 1)!}
\vartheta^0_{11} (x, x - y)
\, .
\end{eqnarray}
Substituting these expressions back into Eq.\ (\ref{F1-expa}) yields
\begin{eqnarray}
{\cal F}_1
\!\!\!&\approx&\!\!\!
\sum_q Q_q^2
\int_{-1}^1\!\! dx \int_{-1}^1 \!\! d y \;
\Bigg\{
\Bigg(
C_1^{(0)}\! \left( \ft{x}{\xi} - i 0 \right)
+
C_1^{(0)}\! \left( - \ft{x}{\xi} - i 0 \right)
\Bigg)
\delta (x - y)
\\
&-&\!\!\!
\frac{M^2}{Q^2}
\Bigg(
C_1^{(1)}\! \left( \ft{x}{\xi} - i 0 \right)
+
C_1^{(1)}\! \left( - \ft{x}{\xi} - i 0 \right)
\Bigg)
V_1^{(2)} (x, y)
\left(
y \frac{\partial}{\partial y} + \eta \frac{\partial}{\partial \eta} + 1
\right)
\Bigg\}
H^q (y, \eta) , \nonumber
\end{eqnarray}
which has now the desired form. Completely analogous manipulations produce
the GPD representation for all other form factors.

\section{Phenomenology of GPDs}
\label{Phenomenology}

\begin{figure}[t]
\begin{center}
\mbox{
\begin{picture}(0,100)(80,0)
\put(0,-15){\insertfig{5}{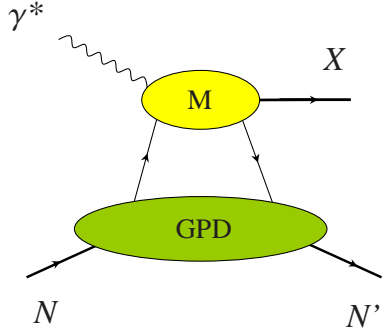}}
\end{picture}
}
\end{center}
\caption{\label{GenericExclusiveDiagram} A generic exclusive diagram for
virtual photoproduction of a gauge boson, hadronic system or jets, $X =
\gamma, \gamma^\ast, Z, \rho, \pi, J/\psi, {\rm jets}, \rho\rho, \dots$.
The process-initiating particle can be a hadron instead of the photon.}
\end{figure}

In this section, we give a comprehensive discussion of a few major exclusive
experiments where GPDs can be measured or accessed. All of the reactions involved
can be divided into a few categories according to the underlying mechanism $M$
of particle production, as demonstrated in Fig.\ \ref{GenericExclusiveDiagram}.
According to QCD factorization theorems, analogous to the one discussed in
Section \ref{ComptonFactorization} for the VCS, the scattering amplitudes for
the processes contain GPDs in convolution with a computable short-distance cross
section. This is the latter which assorts processes to the corresponding group,
which is either quark-photon Compton scattering, or hard re-scattering of quarks
via the exchange of a far off-shell gluon, or production of quark (gluon) jets
initiated by a photon or a meson from a quark/gluon in the target. Thus, we
classify the processes as follow.
\begin{itemize}
\item Compton-induced processes, which include
\begin{itemize}
\item leptoproduction of a real photon, $\ell N \to \ell' \gamma N'$:
\cite{Ji96a,Ji96b,GouDiePirRal97,Frankfurt:1998et,BelMulNieSch00,BelMulKirSch00,%
BelKirMulSch01b,BelMulKir01,Donnachie:2000rz,Donnachie:2000px,FreMcDStr02,GarSzcLon03,Fre03};
\item photoproduction of a lepton pair, $\gamma N \to ( \ell^+ \ell^- ) N'$:
\cite{BerDiePir01};
\item photoproduction of an electroweak boson, $\gamma N \to Z N'$:
\cite{BarLow82};
\item leptoproduction of a lepton pair, $\ell N \to \ell' ( \ell^+ \ell^- ) N'$:
\cite{GuiVan02,BelMul03,BelMul03a};
\item neutrino production of a real photon, $\nu_\ell N \to \ell N' \gamma$:
\cite{AmoCorGuz04,Psa04};
\end{itemize}
\item Hard rescattering processes, which include
\begin{itemize}
\item leptoproduction of a light meson, $\ell N \to \ell' M N'$:
\cite{Bro94,ColFraStr96,ManPilWei97,ManPilWei99,ManPilRad99,EidFraStr98,%
FraPobPolStr99,FraPolStrVan99,VanGuiGui99,CanLag02,DieVin04,GolKro05};
\item leptoproduction of a heavy meson, $\ell N \to \ell' M_h N'$:
\cite{Rys97,Hoo96,VanMan98a,FraKoeStr98};
\item photoproduction of a heavy meson, $\gamma N \to M_h N'$:
\cite{Rys93,Bro94,FraKoeStr98,VanMan98b};
\item mesoproduction of lepton pairs: $\pi N \to N' \ell \ell'$ \cite{BerDiePir01b};
\item hadron-antihadron annihilation into a lepton pair and a photon,
$h \bar{h} \to \ell^+ \ell^- \gamma$ \cite{PirSzy04};
\end{itemize}
\item Diffractive processes, which include
\begin{itemize}
\item photoproduction of two jets, $\gamma N \to (2\mbox{jets}) N'$:
\cite{GolKwiMar98,BraGotIvaSchSzy02};
\item pion dissociation into two jets, $\pi N \to (2\mbox{jets}) N'$:
\cite{BerBroGolGun81,FraMilStr93,NikSchSch00,BraIvaSchSzy01,Cher01};
\item neutrino production of heavy mesons, $\bar\nu_\mu N \to \mu D_s N'$ \cite{LehSch01};
\item leptoproduction of pion pairs, $\ell N \to \ell' \pi \pi N'$: \cite{LehPobPolSchGoe99};
\item leptoproduction of two vector mesons with large rapidity gap, $\ell N \to
\ell' \rho \rho N'$: \cite{IvaPirSzyTer02};
\item Higgs production, $pp \to p H p$: \cite{KhoMarRys00,RoeKhoMarOraRys02}.
\end{itemize}
\end{itemize}

In our subsequent presentation we will explore in detail three of the major
processes: leptoproduction of $\gamma$, $\ell^+ \ell^-$ pairs and light mesons.
The reason for this choice is that these reactions can be used to measure
GPDs to high accuracy. Moreover they have theoretically well-established and
rigorous status over some of the other processes whose description in terms of
GPDs is either not proven to all orders in strong coupling---like diffractive
production in $\gamma N$ collisions or production of two vector mesons with a
large rapidity gap---or is not even legitimate---for pion dissociation into jets
and diffractive Higgs production,---since factorization theorems are broken for
them. The light-cone dominance in virtual Compton scattering, on the other hand,
is a consequence of the external kinematical conditions on the process in the
same way as in deeply inelastic scattering. Therefore, one can expect the onset
of the precocious scaling as early as at ${\cal Q}^2 \sim \, 1 \, {\rm GeV}^2$.
It is not the case for hard exclusive meson production, giving access to GPDs as
well, where it is the dynamical behavior of the short-distance parton amplitude
confined to a small transverse volume near the light cone that drives the
applicability of the perturbative approach to the process. Here the reliability
of perturbative QCD predictions is postponed to larger momentum transfer. However,
the process of meson production is unique in its ability to potentially disentangle
GPDs into separate flavor components. No other experimentally feasible\footnote{Though
in principle, a deeply virtual Compton scattering with charged electroweak currents
will serve the purpose of flavor separation, its actual experimental implementation
faces gargantuan difficulties.} reactions sharing this property are available.

\subsection{Leptoproduction of a real photon}
\label{Sec-AziAngDep}

\begin{figure}[t]
\begin{center}
\mbox{
\begin{picture}(0,108)(155,0)
\put(0,0){\insertfig{10}{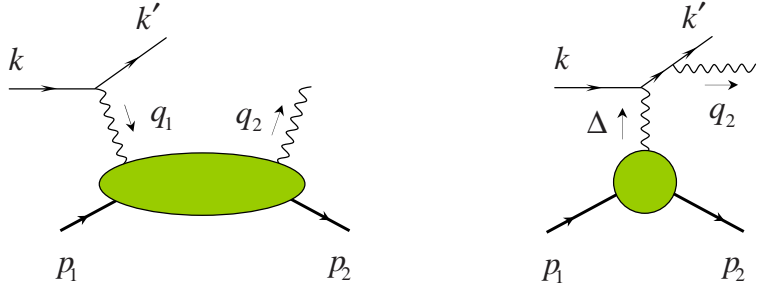}}
\end{picture}
}
\end{center}
\caption{\label{BHandDVCS} Subprocesses contributing to electroproduction
of a real photon: deeply virtual Compton scattering (left) and Bethe-Heitler
scattering (right).}
\end{figure}

\begin{figure}[t]
\unitlength1mm
\begin{center}
\mbox{
\begin{picture}(0,57)(85,0)
\put(30,0){\insertfig{11}{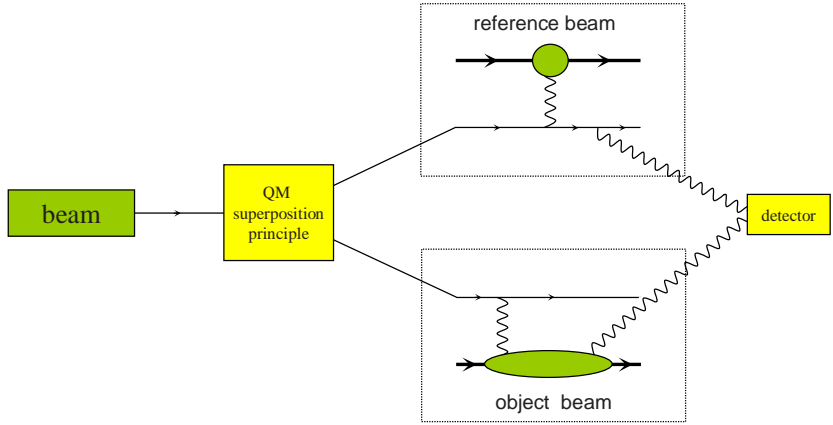}}
\end{picture}
}
\end{center}
\caption{\label{DVCSHolography} Proton ``holography'' with electroproduction
of a real photon \cite{BelMul02}.}
\end{figure}

Exclusive leptoproduction of a photon off the nucleon serves as a clean
electromagnetic probe of the nucleon's internal structure and currently is the
only measured reaction at present day experimental facilities in the required
energy range \cite{Aip01,Sau00,Adl01,Ste01}, which has a clear-cut legitimate
description in terms of leading twist GPDs. Since GPDs carry information on
both longitudinal and transverse degrees of freedom, their three-dimensional
experimental exploration requires a complete determination of the DVCS
amplitude, i.e., its magnitude and phase. For macroscopic objects viewed
with visible light, a way to measure the phase is known as holography,
which uses the interference of a source wave with the one diffracted off the
test body: The laser beam is split into two rays. One of them serves
as a reference source and the other reflects from the object's surface.
The reflected beam, which was in phase with the reference beam before
hitting the ``target", interferes with the reference beam and forms
fringes on the plate with varying intensity depending on the phase
difference of both. For microscopic quantum-mechanical states, as we
demonstrated in Section \ref{QMWignerExperiment}, one also resorts on
analogous interferometric schemes (see Fig.\ \ref{MachSender}). This
technology allows for a complete measurement of the corresponding Wigner
distribution. Unfortunately, the same method cannot be used for X-ray
holography of crystals or $\gamma$-ray interferomentry of hadrons via
scattering experiments due to the absence of practical ``splitters".
However, for leptoproduction of a photon from the nucleon there are
actually two contributing amplitudes (see Fig.\ \ref{BHandDVCS}):
the one due to deeply virtual Compton scattering off the nucleon
${\cal T}_{\scriptscriptstyle\rm DVCS}$, we are interested in, and
${\cal T}_{\scriptscriptstyle\rm BH}$ from the ``contaminating''
Bethe-Heitler (BH) process, in which the real photon spills off the
scattered lepton rather than the quark. The BH amplitude is completely
known since the only long-distance input turns out to be nucleon form
factors measured elsewhere. Thus the quantum-mechanical superposition
principle lead to the presence of an interference term
${\cal T}^\ast_{\scriptscriptstyle\rm DVCS} {\cal T}_{\scriptscriptstyle\rm BH}$
between these amplitudes in addition to the intensities of each of the
signals, $|{\cal T}_{\scriptscriptstyle\rm DVCS}|^2$ and
$|{\cal T}_{\scriptscriptstyle\rm BH}|^2$, where the phase information is
lost. Thus the BH process serves as a reference source for the measurement
of the DVCS signal---its magnitude and phase (see Fig.\ \ref{DVCSHolography}).
The most straightforward extraction of the interference term from the cross
section measurement is achieved by making use of the opposite lepton charge
conjugation properties of DVCS and BH amplitudes or their difference with
respect to the lepton and nucleon spin dependence.

\subsubsection{Kinematics of leptoproduction of the photon}
\label{DVCSkinematics}

We start our discussion of the process $e (k) N (p_1) \to e (k^\prime) N (p_2)
\gamma (q_2)$ with discussion of its kinematics. The reality of the final state
photon, immediately implies the equality of the skeweness $\eta$ and generalized
Bjorken $\xi$ variables, with the latter one expressed in terms of the standard
$x_{\rm B}$ via the equation
$$
\eta = \xi \approx \frac{x_{\rm B}}{2 - x_{\rm B}}
\, .
$$
The most transparent consideration arises in the target rest frame in which
the $z$-axis is directed opposite to the momentum of the spacelike virtual
photon\footnote{The transformation to the reference frame with the $z$-axis
directed along the lepton beam is addressed in Ref.\ \cite{DieSap05}.}, see
Fig.\ \ref{Fig-Kin}. In this frame, we obviously have for the initial proton
and virtual photon four-momenta
\begin{equation}
p_1^\mu = (M_N, \, 0, \, 0, \, 0)
\, , \qquad
q_1^\mu = (\omega_1, \, 0, \, 0, \, - q_1^z)
\, ,
\end{equation}
where the photon energy and the $z$-component of its three-momentum are
expressed in terms of the variables (\ref{ExperVariables}) as
\begin{equation}
\omega_1 = \frac{{\cal Q}}{\varepsilon}
\, , \qquad
q_1^z = \frac{{\cal Q}}{\varepsilon} \sqrt{1 + \varepsilon^2}
\, .
\end{equation}
Here and throughout our subsequent presentation we use the convention
\begin{eqnarray*}
\varepsilon \equiv 2 x_{\rm B} \frac{M_N}{{\cal Q}}
\, .
\end{eqnarray*}
The outgoing nucleon four-momentum in the target rest frame has the components
\begin{equation}
p_2^\mu = (E_2, \bit{p}_2)
\, , \qquad
E_2 = M_N - \frac{\Delta^2}{2 M_N}
\, , \qquad
|\bit{p}_2| = \sqrt{- \Delta^2 \left( 1 - \Delta^2/(4 M_N^2) \right)}
\, ,
\end{equation}
and the scattering angle of the recoiled nucleon is
\begin{equation}
\cos \theta_N =
-
\frac{
\varepsilon^2 \left( {\cal Q}^2 - \Delta^2 \right)
- 2 x_{\rm B} \Delta^2
}{
4 x_{\rm B} M_N |\bit{p}_2| \sqrt{1 + \varepsilon^2}
}
\, .
\end{equation}
The incoming electron four-momentum
\begin{equation}
k^\mu = (E, \, k^x, \, 0, \, k^z) = E (1, \, \sin \theta_e, \, 0, \, \cos \theta_e)
\, , \qquad
E = \frac{{\cal Q}}{y \varepsilon}
\, , \qquad
\cos \theta_e = - \frac{1 + y \varepsilon^2/2}{\sqrt{1 + \varepsilon^2}}
\, ,
\end{equation}
depends on the variable proportional to the lepton energy loss
\begin{equation}
y = \frac{p_1 \cdot q_1}{p_1\cdot k}
\, , \qquad
q_1 = k - k'
\, .
\end{equation}
Notice that from the relation
$$
\sin \theta_e =
\frac{\varepsilon \sqrt{1 - y - y^2 \varepsilon^2/4}}{\sqrt{1 + \varepsilon^2}}
\, ,
$$
it is clear that for large momentum transfer from the lepton to the target,
the lepton beam experiences forward scattering with almost no deflection as
$\theta_e \to 0$ for ${\cal Q}^2 \gg M_N^2$. Finally, the four-vector of the
real photon is given by
\begin{equation}
q_2^\mu = (\omega_2, \, \bit{v} \, \omega_2)
\, ,
\end{equation}
with momentum components and the scattering angle being
\begin{equation}
\omega_2 = \frac{{\cal Q}}{\varepsilon} + \frac{\Delta^2}{2 M_N}
\, , \qquad
|\bit{v}| = 1
\, , \qquad
\cos \theta_\gamma
= - \frac{
\varepsilon
\left(
{\cal Q}^2 + \Delta^2
\right)
+
2 {\cal Q} \omega_2
}{
2 {\cal Q} \omega_2 \sqrt{1 + \varepsilon^2}
}
\, .
\end{equation}

\subsubsection{Cross section for leptoproduction of the photon}

Now we are in a position to turn to physical observables, which give direct
access to GPDs in the measurement of the five-fold cross section for the
leptoproduction process
\begin{eqnarray}
\label{WQ}
d\sigma
=
\frac{\alpha^3_{\rm em}
x_{\rm B} y}{16 \, \pi^2 \, {\cal Q}^2 \sqrt{1 + \varepsilon^2}}
\left| \frac{\cal T}{e^3} \right|^2
d x_{\rm B} d y d ( - \Delta^2) d \phi d \varphi
\, .
\end{eqnarray}
Here the scattering amplitude ${\cal T}$ is a sum of the DVCS
${\cal T}_{\scriptscriptstyle\rm DVCS}$ and Bethe-Heitler (BH)
${\cal T}_{\scriptscriptstyle\rm BH}$ signals. The latter has no
absorptive part at lowest order in the QED fine structure constant
$\alpha_{\rm em} = e^2/(4 \pi)$, and is parametrized in terms of
the electromagnetic form factors (\ref{JintermsofF12}), which are
assumed to be known from other measurements. For the electron beam,
the separate contributions read
\begin{eqnarray}
\label{AmpDVCS}
{\cal T}_{\scriptscriptstyle\rm DVCS}
\!\!\!&=&\!\!\!
\frac{e^3}{q_1^2} \
\varepsilon^\ast_\mu (q_2) \
\bar u (k') \gamma_\nu u (k) \
T^{\mu\nu}
\, , \\
\label{AmpBH}
{\cal T}_{\scriptscriptstyle\rm BH}
\!\!\!&=&\!\!\!
\frac{e^3}{\Delta^2} \
\varepsilon^\ast_\mu (q_2) \
\bar u (k')
\left(
\gamma^\mu \frac{1}{ {\not\!k} - {\not\!\!\Delta} } \gamma^\nu
+
\gamma^\nu \frac{1}{ {\not\!k'} + {\not\!\!\Delta} } \gamma^\mu
\right)
u (k) \ J_\nu
\, .
\end{eqnarray}
They correspond to diagrams (a) and (b) in Fig.\ \ref{BHandDVCS},
respectively, including the crossed contributions in the BH cases. Here
the DVCS amplitude is expressed in terms of the tensor $T_{\mu\nu}$ from
Eq.\ (\ref{VCSamplitude}), while BH one via the nucleon electromagnetic
current
\begin{equation}
\label{EMcurrentMatrixEl}
J_\mu
\equiv
\langle p_2 | j_\mu (0) | p_1 \rangle
\, ,
\end{equation}
expressed in terms of the Dirac and Pauli form factors in Eq.\
(\ref{JintermsofF12}). It is very important to notice that the
DVCS amplitude changes its sign when one goes from the electron to
the positron beam,
\begin{equation}
\label{DVCSchargeConjug}
{\cal T}_{\scriptscriptstyle\rm DVCS}|_{e^+}
=
- {\cal T}_{\scriptscriptstyle\rm DVCS}|_{e^-}
\, ,
\end{equation}
while the BH process does not ${\cal T}_{\scriptscriptstyle\rm BH}|_{e^+}
= {\cal T}_{\scriptscriptstyle\rm BH}|_{e^-}$.

\begin{figure}[t]
\begin{center}
\mbox{
\begin{picture}(0,260)(160,0)
\put(-30,-15){\insertfig{12}{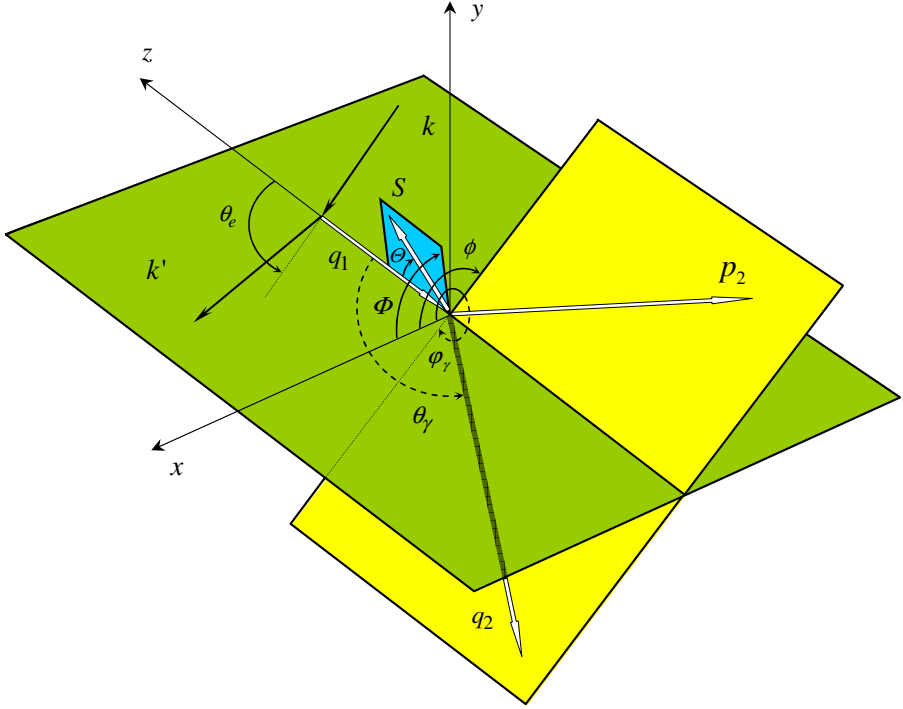}}
\end{picture}
}
\end{center}
\caption{\label{Fig-Kin} Kinematics of  leptoproduction in the target rest
frame. The $z$-direction is chosen counter-along the three-momentum of the
incoming virtual photon. The lepton three-momenta form the lepton scattering
plane, while the recoiled proton and outgoing real photon define the hadron
scattering plane. In this reference system the azimuthal angle between the
lepton plane and the recoiled proton momentum is $\phi$ and is related in
turn to the photon azimuthal angle $\varphi_\gamma = \phi + \pi$. The target
polarization three-vector is described by two spherical angles in this frame
$({\mit\Theta}, {\mit\Phi})$, so that the angle between the polarization
vector and the scattered nucleon is $\varphi = {\mit\Phi} - \phi$.}
\end{figure}

The dependence of the cross section (\ref{WQ}) on a number of kinematical
variables allows for a thorough exploration of the DVCS amplitude. Apart
from the Bjorken variable $x_{\rm B}$, the $t$-channel momentum transfer
$\Delta^2$ and the lepton energy loss $y$, it depends on the azimuthal
angle $\phi$ between the lepton and hadron scattering planes and the angle
between the nucleon polarization vector and the scattered nucleon $\varphi
= {\mit\Phi} - \phi$, as shown in Fig.\ \ref{Fig-Kin}. We parametrize the
polarization vector of the nucleon by polar and azimuthal angles ${\mit\Theta}$
and ${\mit\Phi}$, respectively,
\begin{equation}
\label{Spin}
S^\mu
=
(0, \sin{\mit\Theta} \cos{\mit\Phi}, \sin{\mit\Theta} \sin{\mit\Phi},
\cos{\mit\Theta})
\, ,
\end{equation}
so that it has solely longitudinal polarization in the target rest frame
when ${\mit\Theta} = 0, \pi$, and the transverse one for ${\mit\Theta} =
\pi/2$.

At this point, it is instructive to point out the relation of our angular variables
$\phi$ and ${\mit\Phi}$ to $\phi_h$ and $\phi_s$ of the ``Trento convention''
\cite{BacAleDieMil04}. They are simply related by
\begin{equation}
\phi + \phi_h = 2 \pi
\, , \qquad
{\mit\Phi} + \phi_s = 2 \pi
\, .
\end{equation}

As we emphasized in the previous section, the detailed experimental exploration
of the DVCS amplitude is possible via the extraction of the interference term.
One of the ways to achieve it is obvious from the charge-conjugation property
of the DVCS amplitude (\ref{DVCSchargeConjug}). By performing scattering
experiments with electron and positron beams and taking their difference, one
gets access to the interference
$$
d \sigma_{e^-} - d \sigma_{e^+}
\sim
\left(
{\cal T}_{\scriptscriptstyle\rm DVCS}^\ast
{\cal T}_{\scriptscriptstyle\rm BH}
+
{\cal T}_{\scriptscriptstyle\rm DVCS}
{\cal T}_{\scriptscriptstyle\rm BH}^\ast
\right)_{e^-}
\equiv {\cal I}_{e^-}
$$
Although this set up is a perfect filter of ${\cal I}$, there is currently
only one experimental facility possessing both types of the lepton beam,
Deutsches Elektronen-Synchrotron. Therefore, in order to be able to study
it with other machines, one has to resort to other techniques. As we will
demonstrate below they are based on lepton and nucleon polarization
observables and analysis of the azimuthal angle dependence of the outgoing
nucleon or photon. The successful completion of this program calls for a
detailed understanding of the spin and azimuthal angle dependence of the
differential cross section (\ref{WQ}). The azimuthal dependence of each of
the three terms in
\begin{equation}
\label{SquaredAmplitude}
{\cal T}^2
=
|{\cal T}_{\scriptscriptstyle\rm BH}|^2
+
|{\cal T}_{\scriptscriptstyle\rm DVCS}|^2
+
{\cal I}
\end{equation}
arises from the Lorentz products of leptonic and hadronic four-momenta
\cite{GouDiePirRal97,BelMulNieSch00,BelMulKir01}. In our frame, these
contractions yield finite sums of Fourier harmonics, whose maximal
frequencies are defined by the the rank $m$ of the corresponding leptonic
tensor\footnote{It is a product of two Dirac bilinear forms built from
leptonic bispinors entering the two scattering amplitudes (\ref{AmpDVCS})
and (\ref{AmpBH}), i.e., $\bar{u} (k') \Gamma_1 u (k) \bar{u} (k) \Gamma_2
u (k')$.} in the incoming lepton momentum $k_\mu$. Note, however, that in
the polarized part of the leptonic tensors, proportional to $\lambda$
times the $\varepsilon$-tensor, one has one four-vector $k_\mu$ less than
in the unpolarized part. Thus, the highest harmonic which is proportional
to $\lambda$ will be $\cos/\sin([m-1]\phi)$ instead of $\cos/\sin(m\phi)$.
The parity and time reversal invariance provide further constraints on the
Fourier coefficients.

There is an important difference between the interference term and the
squared DVCS amplitude. The former has a contaminating $\phi$-dependence
due to the lepton BH propagators,
\begin{equation}
\label{ExaBHpro}
{\cal Q}^2 {\cal P}_1 \equiv (k - q_2)^2 = {\cal Q}^2 + 2k\cdot \Delta
\, , \qquad
{\cal Q}^2 {\cal P}_2 \equiv (k - \Delta)^2 = - 2 k \cdot \Delta + \Delta^2
\, ,
\end{equation}
where
\begin{equation}
\label{kDelta}
k \cdot \Delta
= - \frac{{\cal Q}^2}{2y (1 + \varepsilon^2)}
\Bigg\{
1 + 2 K \cos{\phi} - \frac{\Delta^2}{{\cal Q}^2}
\left( 1 - x_{\rm B} (2 - y) + \frac{y \varepsilon^2}{2} \right)
+  \frac{y \varepsilon^2}{2}
\Bigg\} \, .
\end{equation}
The $1/{\cal Q}$-power suppressed kinematical factor $K$ appearing here
also shows up in the Fourier series
\begin{equation}
\label{DefKReal}
K
\equiv
\frac{1}{2 {\cal Q}^2}
\sqrt{
- \left( 1 - y - y^2 \varepsilon^2/4 \right)
\left(
4 x_{\rm B} (1 - x_{\rm B}) + \varepsilon^2
\right)
\left(
\Delta^2 - \Delta^2_{\rm min}
\right)
\left(
\Delta^2 - \Delta^2_{\rm max}
\right)
}
\, .
\end{equation}
We expressed $K$ in terms of the maximal and minimal momentum transfer in
the $t$-channel,
\begin{eqnarray}
\label{Def-Delta-minmax}
\Delta^2_{\rm min, max}
=
- \frac{1}{ 4 x_{\rm B} (1 - x_{\rm B}) + \varepsilon^2}
\Bigg\{
2
(1 - x_{\rm B}) {\cal Q}^2
+
\varepsilon^2 {\cal Q}^2
\mp 2 \sqrt{1 + \varepsilon^2} (1 - x_{\rm B}) {\cal Q}^2
\Bigg\}
\, ,
\end{eqnarray}
with $-$ ($+$) corresponding to $\Delta^2_{\rm min}$ ($\Delta^2_{\rm max}$).
It vanishes at the kinematical boundary $\Delta^2 = \Delta_{\rm min}^2$,
determined by the minimal value
\begin{eqnarray}
\label{Def-tmin}
- \Delta_{\rm min}^2
\approx
\frac{M^2 x_{\rm B}^2}{1 - x_{\rm B} + x_{\rm B} M^2/{{\cal Q}}^2}
\, ,
\end{eqnarray}
as well as at
\begin{eqnarray*}
y
\to
y_{\rm max}
\equiv
2 \frac{\sqrt{1 + \varepsilon^2} - 1}{\varepsilon^2}
\approx
1 - \frac{M^2_N x_{\rm B}^2}{{\cal Q}^2}
\, .
\end{eqnarray*}
As we see, the denominator of the $u$-channel lepton propagator ${\cal P}_1$
can be of order $1/{\cal Q}^2$ at large $y$. In the Bjorken limit it behaves
like $(1 - y)$. Moreover, if the outgoing photon is collinear to the incoming
lepton, ${\cal P}_1$ vanishes. Of course, the photon then lies in the lepton
scattering plane, i.e., $\varphi_\gamma = \phi + \pi = 0$, and both polar
angles coincide with each other. This condition is fulfilled if
\begin{eqnarray*}
y
\to
y_{\rm col}
\equiv
\frac{{\cal Q}^2 + \Delta^2}{{\cal Q}^2 + x_{\rm B} \Delta^2}
\approx
1 + (1 - x_{\rm B}) \frac{\Delta^2}{{\cal Q}^2} \, .
\end{eqnarray*}
Therefore, for large $y$, the squared BH and interference terms are enhanced
with respect to the squared DVCS one. Furthermore, the expansion of ${\cal P}_1$
in ${\cal Q}$ is not justified, and, thus, the Fourier analysis of experimental
data must be modified. For small $y$, it is legitimate to expand ${\cal P}_1$ and
${\cal P}_2$ in power series with respect to $1/{\cal Q}$. This generates higher
harmonics suppressed by powers of $K$.

\subsubsection{Angular dependence}

Each of the terms in the squared amplitude (\ref{SquaredAmplitude})
can be written as a finite Fourier sum of a few harmonic in the
azimuthal angle $\phi$,
\begin{eqnarray}
\label{Par-BH}
&&|{\cal T}_{\scriptscriptstyle\rm BH}|^2
=
\frac{
e^6
}{
x_{\rm B}^2 y^2 (1 + \varepsilon^2)^2 \Delta^2\,
{\cal P}_1 (\phi) {\cal P}_2 (\phi)
}
\left\{
c^{\scriptscriptstyle\rm BH}_0
+  \sum_{n = 1}^2
c^{\scriptscriptstyle\rm BH}_n \, \cos{(n\phi)}
+
s^{\scriptscriptstyle\rm BH}_1 \, \sin{(\phi)}
\right\} \, ,
\\
\label{AmplitudesSquared}
&& |{\cal T}_{\scriptscriptstyle\rm DVCS}|^2
=
\frac{e^6}{y^2 {\cal Q}^2}\left\{
c^{\scriptscriptstyle\rm DVCS}_0
+ \sum_{n=1}^2
\left[
c^{\scriptscriptstyle\rm DVCS}_n \cos (n\phi)
+
s^{\scriptscriptstyle\rm DVCS}_n \sin (n \phi)
\right]
\right\} \, ,
\\
\label{InterferenceTerm}
&&{\cal I}
=
\frac{
\pm e^6
}{
x_{\rm B} y^3 \Delta^2 {\cal P}_1 (\phi) {\cal P}_2 (\phi)
}
\left\{
c_0^{\scriptscriptstyle\rm INT}
+ \sum_{n = 1}^3
\left[
c_n^{\scriptscriptstyle\rm INT} \cos(n \phi)
+
s_n^{\scriptscriptstyle\rm INT} \sin(n \phi)
\right]
\right\} \, .
\end{eqnarray}
Here, the $+$ ($-$) sign in the interference stands for the negatively
(positively) charged lepton beam. The results for the Fourier coefficients,
presented below, show that the generation of new harmonics in the azimuthal
angular dependence is terminated at the twist-three level. The coefficients
$c^{\scriptscriptstyle\rm INT}_1$, $s^{\scriptscriptstyle\rm INT}_1$ as
well as $c^{\scriptscriptstyle\rm DVCS}_0$ arise at the twist-two level,
and their dependence on GPDs has been investigated in Refs.\
\cite{GouDiePirRal97,BelMulNieSch00}. The remaining coefficients provide
an additional angular dependence, with $c^{\scriptscriptstyle\rm DVCS}_1$,
$s^{\scriptscriptstyle\rm DVCS}_1$, $c^{\scriptscriptstyle\rm INT}_2$, and
$s^{\scriptscriptstyle\rm INT}_2$ involving twist-three GPDs
\cite{BelMulKirSch00,BelKirMulSch01b,BelMulKir01}. On the other hand, the
harmonics proportional to $\cos{(3\phi)}$ ($\cos{(2\phi)}$) or $\sin{(3\phi)}$
($\sin{(2\phi)}$) in the interference (squared DVCS) term originate  from
the twist-two double helicity-flip gluonic GPDs alone. They are not
contaminated by the twist-two quark amplitudes, however, they are affected by
twist-four power corrections \cite{KivMan01}. In what follows, we neglect
the effects of dynamical higher-twist (larger than three) contributions
since we expect Bjorken scaling to be effective in the Compton amplitude
starting with rather lower photon virtualities ${\cal Q}^2$.

To have a compact notation, we write the cross section for a polarized target
as
\begin{eqnarray}
\label{Def-DecCroSec}
d\sigma
=
d\sigma_{\rm unp}
+
\cos{\mit\Theta} \, d\sigma_{\scriptscriptstyle\rm LP}(\Lambda)
+
\sin{\mit\Theta} \, d\sigma_{\scriptscriptstyle\rm TP}(\varphi)
\, ,
\end{eqnarray}
where the polar angle ${\mit\Theta}$ appears in the decomposition of the nucleon
spin vector (\ref{Spin}), $S^\mu = \cos{\mit\Theta} S^\mu_{\scriptscriptstyle\rm LP}
(\Lambda) + \sin{\mit\Theta} S^\mu_\perp(\Phi)$ with the longitudinal polarization
$S^\mu_{\scriptscriptstyle\rm LP} = (0,0,0,\Lambda)$ expressed in terms of the
nucleon helicity. We will use the conventional definition of the lepton helicity,
i.e., $\lambda = 1$ if the spin is aligned with the direction of the lepton
three-momentum and $\lambda = - 1$, when they are opposite.

\subsubsection{Bethe-Heitler amplitude squared}
\label{BHcrosssection}

This part of the leptoproduction cross section is expressed solely in
terms of $F_1(\Delta^2)$ and $F_2(\Delta^2)$, the known Dirac and Pauli
form factors of the nucleon. There are several possibilities for the
target polarization, and the relevant Fourier coefficients are:
\begin{itemize}
\item Unpolarized target:
\end{itemize}
\begin{eqnarray}
\label{Def-FC-BH-unp0}
c^{\scriptscriptstyle\rm BH}_{0,{\rm unp}}
\!\!\!&=&\!\!\!
8 K^2
\left\{
\left( 2 + 3 \varepsilon^2 \right)
\frac{{\cal Q}^2}{\Delta^2}
\left( F_1^2 - \frac{\Delta^2}{4 M^2} F_2^2 \right)
+ 2 x_{\rm B}^2 \left( F_1 + F_2 \right)^2
\right\}
\\
&+&\!\!\! (2 - y)^2
\Bigg\{
\left( 2 + \varepsilon^2 \right)
\Bigg[
\frac{4 x_{\rm B}^2 M^2}{\Delta^2}
\left( 1 + \frac{\Delta^2}{{\cal Q}^2} \right)^2
+ 4 (1 - x_{\rm B})
\left( 1 + x_{\rm B}  \frac{\Delta^2}{{\cal Q}^2} \right)
\Bigg]
\left( F_1^2 - \frac{\Delta^2}{4 M^2} F_2^2 \right)
\nonumber\\
&+&\!\!\!
4 x_{\rm B}^2
\Bigg[
x_{\rm B} + \left(1 - x_{\rm B} + \frac{\varepsilon^2}{2} \right)
\left(1 -  \frac{\Delta^2}{Q^2} \right)^2
- x_{\rm B} (1 - 2 x_{\rm B}) \frac{\Delta^4}{Q^4}
\Bigg]
\left( F_1 + F_2 \right)^2
\Bigg\}
\nonumber\\
&+&\!\!\! 8 \left( 1 + \varepsilon^2 \right)
\left(1 - y - \frac{\varepsilon^2 y^2}{4} \right)
\Bigg\{
2 \varepsilon^2
\left( 1 - \frac{\Delta^2}{4 M^2} \right)
\left( F_1^2 - \frac{\Delta^2}{4 M^2} F_2^2 \right)
\nonumber\\
&&\qquad\qquad\qquad\qquad\qquad\qquad\qquad\qquad\qquad
- x_{\rm B}^2
\left( 1 -  \frac{\Delta^2}{Q^2} \right)^2
\left( F_1 + F_2 \right)^2
\Bigg\} \, ,
\nonumber\\
\label{Def-FC-BH-unp1}
c^{\scriptscriptstyle\rm BH}_{1,{\rm unp}}
\!\!\!&=&\!\!\!
8 K (2 - y)
\Bigg\{
\left(
\frac{4x_{\rm B}^2 M^2}{\Delta^2} - 2x_{\rm B} - \varepsilon^2
\right)
\left(
F_1^2 - \frac{\Delta^2}{4 M^2} F_2^2
\right)
\\
&&\qquad\qquad\qquad\qquad\qquad\qquad\qquad\qquad\qquad
+ \, 2 \, x_{\rm B}^2
\left( 1 - (1 - 2x_{\rm B}) \frac{\Delta^2}{{\cal Q}^2} \right)
\left( F_1 + F_2 \right)^2
\Bigg\},
\nonumber\\
\label{Def-FC-BH-unp2}
c^{\scriptscriptstyle\rm BH}_{2,{\rm unp}} \!\!\!&=&\!\!\!
8 x_{\rm B}^2 K^2  \left\{ \frac{4 M^2}{\Delta^2}
\left(F_1^2 - \frac{\Delta^2}{4 M^2} F_2^2\right) + 2 \left(F_1 +
F_2\right)^2 \right\} \, .
\end{eqnarray}

\begin{itemize}
\item Longitudinally polarized target:
\end{itemize}
\begin{eqnarray}
c^{\scriptscriptstyle\rm BH}_{0,{\scriptscriptstyle\rm LP}}
\!\!\!&=&\!\!\!
8 \lambda \Lambda x_{\rm B} (2 - y) y
\frac{\sqrt{1 + \varepsilon^2}}{1 - \frac{\Delta^2}{4 M^2}}
\left( F_1 +F _2 \right)
\Bigg\{
\frac{1}{2}
\left[ \frac{x_{\rm B}}{2} \left( 1 - \frac{\Delta^2}{{\cal Q}^2} \right)
- \frac{\Delta^2}{4 M^2}
\right]
\Bigg[ 2 - x_{\rm B}
\\
&-&\!\!\!
2(1 - x_{\rm B})^2  \frac{\Delta^2}{{\cal Q}^2}
+ \varepsilon^2
\left( 1 - \frac{\Delta^2}{{\cal Q}^2} \right)
- x_{\rm B} (1 - 2x_{\rm B}) \frac{\Delta^4}{{\cal Q}^4}
\Bigg]
\left( F_1 + F_2 \right)
\nonumber\\
&+&\!\!\!
\left(
1 - (1 - x_{\rm B})\frac{\Delta^2}{{\cal Q}^2}
\right)
\left[
\frac{x_{\rm B}^2 M^2}{\Delta^2}
\left( 1 + \frac{\Delta^2}{{\cal Q}^2} \right)^2
+
(1 - x_{\rm B})
\left( 1 + x_{\rm B} \frac{\Delta^2}{{\cal Q}^2} \right)
\right]
\left( F_1 + \frac{\Delta^2}{4 M^2} F_2 \right)
\Bigg\} \, ,
\nonumber\\
c^{\scriptscriptstyle\rm BH}_{1,{\scriptscriptstyle\rm LP}}
\!\!\!&=&\!\!\!
- 8 \lambda \Lambda x_{\rm B} y K
\frac{\sqrt{1 + \varepsilon^2}}{1 - \frac{\Delta^2}{4 M^2}}
\left( F_1 + F_2 \right)
\Bigg\{
\left[
\frac{\Delta^2}{2 M^2}
-
x_{\rm B} \left(1 - \frac{\Delta^2}{{\cal Q}^2} \right)
\right]
\left( 1 - x_{\rm B} + x_{\rm B} \frac{\Delta^2}{{\cal Q}^2} \right)
\left( F_1 + F_2 \right)
\nonumber\\
&+&\!\!\!
\left[
1 + x_{\rm B} - (3 - 2 x_{\rm B})
\left( 1 + x_{\rm B} \frac{\Delta^2}{{\cal Q}^2} \right)
-
\frac{4 x_{\rm B}^2 M^2}{\Delta^2}
\left( 1 + \frac{\Delta^4}{{\cal Q}^4} \right)
\right]
\left( F_1 + \frac{\Delta^2}{4 M^2} F_2 \right)
\Bigg\} \, .
\end{eqnarray}

\begin{itemize}
\item Transversely polarized target:
\end{itemize}
\begin{eqnarray}
c^{\scriptscriptstyle\rm BH}_{0,{\scriptscriptstyle\rm TP}}
\!\!\!&=&\!\!\!
- 8 \lambda \cos{(\varphi)} (2 - y) y \frac{{\cal Q}}{M}
\frac{\sqrt{1 + \varepsilon^2} K}{\sqrt{1 - y - \frac{\varepsilon^2 y^2}{4}}}
\left( F_1 + F_2 \right)
\Bigg\{
\frac{x_{\rm B}^3 M^2}{{\cal Q}^2}
\left( 1 - \frac{\Delta^2}{{\cal Q}^2} \right)
\left( F_1 + F_2 \right)
\\
&+&\!\!\!
\left(1 - (1 - x_{\rm B}) \frac{\Delta^2}{{\cal Q}^2} \right)
\left[
\frac{x_{\rm B}^2 M^2}{\Delta^2} \left(1 - \frac{\Delta^2}{{\cal Q}^2}\right) F_1
+
\frac{x_{\rm B}}{2} F_2
\right]
\Bigg\},
\nonumber\\
c^{\scriptscriptstyle\rm BH}_{1,{\scriptscriptstyle\rm TP}}
\!\!\!&=&\!\!\!
- 16 \lambda\cos{(\varphi)} x_{\rm B} y
\sqrt{1 - y - \frac{\varepsilon^2 y^2}{4}}\frac{M}{{\cal Q}}\sqrt{1 + \varepsilon^2}
\left( F_1 + F_2 \right)
\Bigg\{
\frac{2 K^2 {\cal Q}^2}{\Delta^2 \left( 1 - y - \frac{\varepsilon^2 y^2}{4} \right)}
\Bigg[
\frac{\Delta^2}{4 M^2} F_2
\nonumber\\
&+&\!\!\!
x_{\rm B} \left( 1 - \frac{\Delta^2}{{\cal Q}^2} \right) F_1 \Bigg]
+
(1 + \varepsilon^2) x_{\rm B} \left( 1 - \frac{\Delta^2}{{\cal Q}^2} \right)
\left( F_1 + \frac{\Delta^2}{4 M^2} F_2 \right)
\Bigg\} \, ,
\\
s^{\scriptscriptstyle\rm BH}_{1,{\scriptscriptstyle\rm TP}}
\!\!\!&=&\!\!\!
16 \lambda \sin{(\varphi)} y x_{\rm B}^2 \sqrt{1 - y -  \frac{\varepsilon^2 y^2}{4}}
\frac{M}{{\cal Q}}
\sqrt{(1 + \varepsilon^2)^3} \left( 1 - \frac{\Delta^2}{{\cal Q}^2} \right)
\left( F_1 + F_2\right)
\left( F_1 + \frac{\Delta^2}{4 M^2} F_2 \right) \, .
\end{eqnarray}

\subsubsection{DVCS amplitude squared}
\label{SubSec-AziAngDep-DVCS}

$|{\cal T}_{\scriptscriptstyle\rm DVCS}|^2$ is bilinear in the CFFs, and
its coefficients are given in terms of ${\cal C}^{\scriptscriptstyle\rm DVCS}$
functions, which are specified in Section \ref{App-CoeFun}:
\begin{itemize}
\item Unpolarized target:
\end{itemize}
\begin{eqnarray}
\label{Res-Mom-DVCS-UP}
c^{\scriptscriptstyle\rm DVCS}_{0,{\rm unp}}
\!\!\!&=&\!\!\!
2 ( 2 - 2 y + y^2 )
{\cal C}^{\scriptscriptstyle\rm DVCS}_{\rm unp}
\left(
{\cal F},{\cal F}^\ast\right) ,
\\
\label{Res-Mom-DVCS-UP-tw3}
\left\{
{c^{\scriptscriptstyle\rm DVCS}_{1,{\rm unp}}
\atop
s^{\scriptscriptstyle\rm DVCS}_{1,{\rm unp}}}
\right\}
\!\!\! &=&\!\!\!
\frac{8 K}{2 - x_{\rm B}} \left\{ {2-y \atop -\lambda y } \right\}
\left\{{ \Re{\rm e} \atop \Im{\rm m} }\right\} \,
{\cal C}^{\scriptscriptstyle\rm DVCS}_{\rm unp}
\left(
{\cal F}^{\rm eff},{\cal F}^\ast\right) \, ,
\\
c^{\scriptscriptstyle\rm DVCS}_{2,{\rm unp}}
\!\!\!&=&\!\!\!
- \frac{4 {\cal Q}^2 K^2}{M^2 (2 - x_{\rm B})}
\Re{\rm e}\,
{\cal C}^{\scriptscriptstyle\rm DVCS}_{T,{\rm unp}}
\left( {\cal F}_T, {\cal F}^\ast \right)
\, .
\end{eqnarray}

\begin{itemize}
\item Longitudinally polarized target:
\end{itemize}
\begin{eqnarray}
\label{Res-Mom-DVCS-LP}
c^{\scriptscriptstyle\rm DVCS}_{0,{\scriptscriptstyle\rm LP}}
\!\!\!&=&\!\!\!
2 \lambda \Lambda y(2-y)
{\cal C}^{\scriptscriptstyle\rm DVCS}_{\scriptscriptstyle\rm LP}
\left(
{\cal F},{\cal F}^\ast\right) ,
\\
\left\{ c^{\scriptscriptstyle\rm DVCS}_{1,{\scriptscriptstyle\rm LP}}
\atop
s^{\scriptscriptstyle\rm DVCS}_{1,{\scriptscriptstyle\rm LP}} \right\}
\!\!\! &=&\!\!\!
- \frac{8 \Lambda K}{2 - x_{\rm B}}
\left\{{-\lambda y \atop 2 - y}\right\}
\left\{{\Re{\rm e} \atop \Im{\rm m} }\right\}
{\cal C}^{\scriptscriptstyle\rm DVCS}_{\scriptscriptstyle\rm LP}
\left( {\cal F}^{\rm eff},{\cal F}^\ast \right) \, ,
\\
s^{\scriptscriptstyle\rm DVCS}_{2,{\scriptscriptstyle\rm LP}}
\!\!\!&=&\!\!\! - \frac{4 \Lambda {\cal Q}^2 K^2}{M^2 (2 - x_{\rm B})}
\Im{\rm m}\,
{\cal C}^{\scriptscriptstyle\rm DVCS}_{T,{\scriptscriptstyle\rm LP}}
\left({\cal F}_T,{\cal F}^\ast\right) \, .
\end{eqnarray}

\begin{itemize}
\item Transversely polarized target:
\end{itemize}
\begin{eqnarray}
\label{Res-Mom-DVCS-TP}
c^{\scriptscriptstyle\rm DVCS}_{0,{\scriptscriptstyle\rm TP}}
\!\!\!&=&\!\!\! - \frac{{\cal Q} K}{M\sqrt{1-y}}
\left[
- \lambda y (2 - y) \cos (\varphi)
{\cal C}^{\scriptscriptstyle\rm DVCS}_{\scriptscriptstyle\rm TP+}
+
(2 - 2 y + y^2) \sin (\varphi)
\Im{\rm m}\,
{\cal C}^{\scriptscriptstyle\rm DVCS}_{\scriptscriptstyle\rm TP-}
\right]
\left( {\cal F}, {\cal F}^\ast \right) \, ,
\nonumber\\
\\
\label{Res-Mom-DVCS-TP-Tw3}
\left\{
{ c^{\scriptscriptstyle\rm DVCS}_{1,{\scriptscriptstyle\rm TP}}
\atop
s^{\scriptscriptstyle\rm DVCS}_{1,{\scriptscriptstyle\rm TP}} }
\right\}
\!\!\! &=&\!\!\! -
\frac{4 {\cal Q} K^2}{M (2 - x_{\rm B}) \sqrt{1 - y}}
\\
&\times&\!\!\!
\left[
\cos (\varphi)
\left\{
{ - \lambda y \atop 2 - y}
\right\}
\left\{
{ \Re{\rm e} \atop \Im{\rm m} }
\right\} {\cal C}^{\scriptscriptstyle\rm DVCS}_{\scriptscriptstyle\rm TP +}
+
\sin (\varphi)
\left\{
{ 2 - y \atop \lambda y }
\right\}
\left\{
{ \Im{\rm m} \atop \Re{\rm e } }
\right\} {\cal C}^{\scriptscriptstyle\rm DVCS}_{\scriptscriptstyle\rm TP-}
\right]
\left( {\cal F}^{\rm eff}, {\cal F}^\ast \right) \, ,
\nonumber\\
\left\{
{ c^{\scriptscriptstyle\rm DVCS}_{2,{\scriptscriptstyle\rm TP}}
\atop
s^{\scriptscriptstyle\rm DVCS}_{2,{\scriptscriptstyle\rm TP}} }
\right\}
\!\!\!&=&\!\!\! - \frac{4 {\cal Q} \sqrt{1 - y} K}{M (2 - x_{\rm B})}
\Im{\rm m}
\left\{
{ \sin(\varphi)\,
{\cal C}^{\scriptscriptstyle\rm DVCS}_{T,{\scriptscriptstyle\rm TP}-}
\atop
\cos(\varphi)\,
{\cal C}^{\scriptscriptstyle\rm DVCS}_{T,{\scriptscriptstyle\rm TP}+ } }
\right\}
\left( {\cal F}_T, {\cal F}^\ast \right).
\end{eqnarray}

The coefficient $c^{\scriptscriptstyle\rm DVCS}_{0}$ is expressed in terms
of the twist-two CFFs ${\cal F}= \{{\cal H}, {\cal E}, \widetilde{\cal H},
\widetilde{\cal E} \}$, defined in Eq.\ (\ref{DefTw3}), while the
twist-three coefficients $c^{\scriptscriptstyle\rm DVCS}_{1}$ and
$s^{\scriptscriptstyle\rm DVCS}_{1}$ arise from the interference of twist-two CFFs
with the ``effective'' twist-three CFFs introduced in Eq.\ (\ref{Tw3Eff}). These
Fourier harmonics have the same functional dependence on CFFs as the leading
twist-two ones \cite{BelMulNieSch00}. However, this is not the case for the
Fourier coefficients $c^{\scriptscriptstyle\rm DVCS}_{2}$ and $s^{\rm DVCS}_{2}$,
induced by the gluon transversity (\ref{GluonTransCCF}).

\subsubsection{Interference of Bethe-Heitler and DVCS amplitudes}
\label{SubSec-AziAngDep-INT}

For the phenomenology of GPDs, the interference term ${\cal I}$ is the
most interesting quantity since it is linear in CFFs and thus simplifies
their extraction from experimental measurements. The Fourier harmonics
in this case have the following form:
\begin{itemize}
\item Unpolarized target:
\end{itemize}
\begin{eqnarray}
c^{\scriptscriptstyle\rm INT}_{0,{\rm unp}}
\!\!\!&=&\!\!\!
- 8  (2 - y)
\Re{\rm e}
\Bigg\{
\frac{(2 - y)^2}{1-y} K^2
{\cal C}^{\scriptscriptstyle\rm INT}_{\rm unp} \left({\cal F}\right)
+
\frac{\Delta^2}{{\cal Q}^2}  (1-y)(2 - x_{\rm B})
\left(
{\cal C}^{\scriptscriptstyle\rm INT}_{\rm unp}
+
\Delta {\cal C}^{\scriptscriptstyle\rm INT}_{\rm unp}
\right)
\left( {\cal F} \right)
\Bigg\} \, ,
\nonumber\\
\label{Res-IntTer-unp-c0}\\
\label{Res-IntTer-unp}
\left\{{c^{\scriptscriptstyle\rm INT}_{1, \rm unp}
\atop
s^{\scriptscriptstyle\rm INT}_{1, \rm unp}}\right\}
&\!\!\!=\!\!\!&
8 K
\left\{  {-(2 - 2y + y^2) \atop \lambda y (2-y)} \right\}
\left\{{\Re{\rm e} \atop \Im{\rm m} } \right\}
{\cal C}^{\scriptscriptstyle\rm INT}_{\rm unp}\left({\cal F} \right)
\, , \\
\label{Res-IntTer-unp-tw3}
\left\{ c^{\scriptscriptstyle\rm INT}_{2, \rm unp}
\atop
s^{\scriptscriptstyle\rm INT}_{2, \rm unp} \right\}
&\!\!\!=\!\!\!&
\frac{16 K^2}{2 - x_{\rm B}}  \left\{ { -(2 - y) \atop \lambda y} \right\}
\left\{{\Re{\rm e} \atop \Im{\rm m} } \right\}
{\cal C}^{\scriptscriptstyle\rm INT}_{\rm unp}
\left({\cal F}^{\rm eff}\right)
\, , \\
\label{Res-IntTer-Tra-unp}
c^{\scriptscriptstyle\rm INT}_{3, \rm unp}
\!\!\!&=&\!\!\! - \frac{8 {\cal Q}^2 K^3}{M^2 (2 - x_{\rm B})^2}
\Re{\rm e}\, {\cal C}^{\scriptscriptstyle\rm INT}_{T,{\rm unp}}
\left( {\cal F}_T \right)
\, .
\end{eqnarray}

\begin{itemize}
\item Longitudinally polarized target:
\end{itemize}
\begin{eqnarray}
c^{\scriptscriptstyle\rm INT}_{0, {\scriptscriptstyle\rm LP}}
\!\!\!&=&\!\!\!
-8 \lambda \Lambda y
\Re{\rm e}\Bigg\{
\left(\frac{(2 - y)^2}{1-y} +2 \right) K^2
{\cal C}^{\scriptscriptstyle\rm INT}_{\scriptscriptstyle\rm LP}
\left({\cal F}\right)
+
\frac{\Delta^2}{{\cal Q}^2} (1-y)(2 - x_{\rm B})
\left(
{\cal C}^{\scriptscriptstyle\rm INT}_{\scriptscriptstyle\rm LP}
+
\Delta {\cal C}^{\scriptscriptstyle\rm INT}_{\scriptscriptstyle\rm LP}
\right)
\left({\cal F}\right)
\Bigg\} \, ,
\nonumber\\
\label{Res-IntTer-LP-c0}\\
\label{Res-IntTer-LP}
\left\{{c^{\scriptscriptstyle\rm INT}_{1, \scriptscriptstyle\rm LP}
\atop
s^{\scriptscriptstyle\rm INT}_{1, \scriptscriptstyle\rm LP}}\right\}
&\!\!\!=\!\!\!&
8\Lambda K
\left\{  {-\lambda y (2 - y) \atop 2 - 2y + y^2 } \right\}
\left\{{\Re{\rm e} \atop \Im{\rm m} } \right\}
{\cal C}^{\scriptscriptstyle\rm INT}_{\scriptscriptstyle\rm LP}
\left({\cal F} \right) \, ,
\\
\left\{ c^{\scriptscriptstyle\rm INT}_{2, \scriptscriptstyle\rm LP}
\atop
s^{\scriptscriptstyle\rm INT}_{2, \scriptscriptstyle\rm LP} \right\}
&\!\!\!=\!\!\!&
\frac{16 \Lambda K^2}{2 - x_{\rm B}}
\left\{ { -\lambda y \atop  2 - y} \right\}
\left\{{\Re{\rm e} \atop \Im{\rm m} } \right\}
{\cal C}^{\scriptscriptstyle\rm INT}_{\scriptscriptstyle\rm LP}
\left(
{\cal F}^{\rm eff}\right) \, ,
\\
\label{Res-IntTer-Tra-pol}
s^{\scriptscriptstyle\rm INT}_{3, \scriptscriptstyle\rm LP}
\!\!\!&=&\!\!\!  \frac{8\Lambda {\cal Q}^2 K^3}{M^2 (2 - x_{\rm B})^2}
\Im{\rm m}\,
{\cal C}^{\scriptscriptstyle\rm INT}_{T,{\scriptscriptstyle\rm LP}}
\left( {\cal F}_T \right) \, .
\end{eqnarray}

\begin{itemize}
\item Transversely polarized target:
\end{itemize}
\begin{eqnarray}
\label{Res-IntTer-TP-c0}
c^{\scriptscriptstyle\rm INT}_{0,\scriptscriptstyle\rm TP}
\!\!\!&=&\!\!\!
\frac{8 M \sqrt{1-y}K}{\cal Q}\Bigg[
-\lambda y \cos(\varphi)\,
\Re{\rm e}
 \Bigg\{
\left(\frac{(2-y)^2}{1-y} +2\right)
{\cal C}^{\scriptscriptstyle\rm INT}_{{\scriptscriptstyle\rm TP}+}
\left({\cal F} \right) +
\Delta{\cal C}^{\scriptscriptstyle\rm INT}_{{\scriptscriptstyle\rm TP}+}
\left({\cal F} \right)
\Bigg\}
\\
&&\qquad\qquad\qquad
+ (2 - y)\sin(\varphi)\, \Im{\rm m}
\Bigg\{
\frac{(2-y)^2}{1-y}
{\cal C}^{\scriptscriptstyle\rm INT}_{{\scriptscriptstyle\rm TP}-}
\left({\cal F} \right)
+
\Delta{\cal C}^{\scriptscriptstyle\rm INT}_{{\scriptscriptstyle\rm TP}-}
\left({\cal F} \right)
\Bigg\}
\Bigg] \, ,
\nonumber\\
\label{Res-IntTer-TP}
\left\{
{c^{\scriptscriptstyle\rm INT}_{1, \scriptscriptstyle\rm TP}
\atop
s^{\scriptscriptstyle\rm INT}_{1, \scriptscriptstyle\rm TP}}
\right\}
\!\!\!&=&\!\!\!
\frac{8 M \sqrt{1-y}}{\cal Q}
\\
&\times&\!\!\!\left[
\cos(\varphi)
\left\{
{-\lambda y (2 - y) \atop 2 - 2y + y^2 }
\right\}
\left\{
{\Re{\rm e} \atop \Im{\rm m} }
\right\}
{\cal C}^{\scriptscriptstyle\rm INT}_{{\scriptscriptstyle\rm TP}+}
+
\sin(\varphi)
\left\{
{2 - 2y + y^2  \atop \lambda y (2 - y) }
\right\}
\left\{
{\Im{\rm m} \atop \Re{\rm e} }
\right\}
{\cal C}^{\scriptscriptstyle\rm INT}_{{\scriptscriptstyle\rm TP}-}
\right]
\left( {\cal F} \right) \, ,
\nonumber\\
\left\{
c^{\scriptscriptstyle\rm INT}_{2, \scriptscriptstyle\rm TP}
\atop
s^{\scriptscriptstyle\rm INT}_{2, \scriptscriptstyle\rm TP}
\right\}
\!\!\!&=&\!\!\!
\frac{16 M \sqrt{1-y}K}{{\cal Q}(2 - x_{\rm B})}
\\
&\times&\!\!\!
\left[
\cos(\varphi)
\left\{
{-\lambda y \atop 2 - y}
\right\}
\left\{
{\Re{\rm e} \atop \Im{\rm m}}
\right\}
{\cal C}^{\scriptscriptstyle\rm INT}_{{\scriptscriptstyle\rm TP}+}
+
\sin(\varphi)
\left\{
{2 - y  \atop \lambda y}
\right\}
\left\{
{\Im{\rm m}\atop \Re{\rm e}}
\right\}
{\cal C}^{\scriptscriptstyle\rm INT}_{{\scriptscriptstyle\rm TP}-}
\right]
\left( {\cal F}^{\rm eff} \right) \, ,
\nonumber\\
\label{Res-IntTer-Tra-Tp}
s^{\scriptscriptstyle\rm INT}_{3, \scriptscriptstyle\rm TP}
\!\!\!&=&\!\!\!
\frac{8 {\cal Q} \sqrt{1 - y} K^2}{M (2 - x_{\rm B})^2}
\cos(\varphi) \Im{\rm m}\,
{\cal C}^{\scriptscriptstyle\rm INT}_{T,{\scriptscriptstyle\rm TP}+}
\left( {\cal F}_T\right) \, ,
\\
c^{\scriptscriptstyle\rm INT}_{3, \scriptscriptstyle\rm TP}
\!\!\!&=&\!\!\!
\frac{8{\cal Q} \sqrt{1 - y} K^2}{M (2 - x_{\rm B})^2}
\sin(\varphi)
\Im{\rm m}\, {\cal C}^{\scriptscriptstyle\rm INT}_{T,{\scriptscriptstyle\rm TP}-}
\left({\cal F}_T\right) \, .
\end{eqnarray}

The twist-three coefficients, i.e., $ c^{\scriptscriptstyle\rm INT}_{2}$ and
$s^{\scriptscriptstyle\rm INT}_{2}$ have again the same functional dependence
as the twist-two ones. However, this is not the case for
$c^{\scriptscriptstyle\rm INT}_{0}$, which depends only on the twist-two
CFFs $\cal F$, and for  $ c^{\scriptscriptstyle\rm INT}_{3}$ and
$s^{\scriptscriptstyle\rm INT}_{3}$, induced by ${\cal F}_T$.

\subsubsection{Angular harmonics in terms of GPDs}
\label{App-CoeFun}

The Fourier coefficients displayed above are expressed in terms of the
coefficients ${\cal C}$. They depend on Compton form factors (\ref{DefTw3})
summed over quark flavors weighted with their charges squared,
$$
{\cal F} = \sum_q Q_q^2 {\cal F}^q
\, ,
$$
with ${\cal F}^q = \{ {\cal H}^q , {\cal E}^q , \widetilde{\cal H}^q ,
\widetilde{\cal E}^q , {\cal H}^q_{\rm eff} , {\cal E}^q_{\rm eff} ,
\widetilde{\cal H}^q_{\rm eff} , \widetilde{\cal E}^q_{\rm eff} \}$.
\begin{itemize}
\item Squared DVCS amplitude:
\end{itemize}
\begin{eqnarray}
\label{Def-C-DVCS-unp}
{\cal C}^{\scriptscriptstyle\rm DVCS}_{{\rm unp}}
\left(
{\cal F},{\cal F}^\ast
\right)
\!\!\!&=&\!\!\!
\frac{1}{(2 - x_{\rm B})^2}
\Bigg\{
4 (1 - x_{\rm B})
\left(
{\cal H} {\cal H}^\ast
+
\widetilde{\cal H} \widetilde {\cal H}^\ast
\right)- x_{\rm B}^2
\bigg(
{\cal H} {\cal E}^\ast
+ {\cal E} {\cal H}^\ast
+ \widetilde{{\cal H}} \widetilde{{\cal E}}^\ast
+ \widetilde{{\cal E}} \widetilde{{\cal H}}^\ast
\bigg)
\nonumber\\
&&\qquad\qquad\;
-
\left( x_{\rm B}^2 + (2 - x_{\rm B})^2 \frac{\Delta^2}{4M^2} \right)
{\cal E} {\cal E}^\ast
- x_{\rm B}^2 \frac{\Delta^2}{4M^2}
\widetilde{{\cal E}} \widetilde{{\cal E}}^\ast
\Bigg\},
\end{eqnarray}
\begin{eqnarray}
\label{Def-C-DVCS-LP}
{\cal C}^{\scriptscriptstyle\rm DVCS}_{{\rm LP}}
\left(
{\cal F},{\cal F}^\ast
\right)
\!\!\!&=&\!\!\!
\frac{1}{(2 - x_{\rm B})^2}
\Bigg\{
4 (1 - x_{\rm B})
\left(
{\cal H} \widetilde{\cal H}^\ast
+
\widetilde{\cal H} {\cal H}^\ast
\right)- x_{\rm B}^2
\bigg(
{\cal H} \widetilde{\cal E}^\ast
+ \widetilde{\cal E} {\cal H}^\ast
+ \widetilde{\cal H} {\cal E}^\ast
+ {\cal E} \widetilde{{\cal H}}^\ast
\bigg)
\nonumber\\
&&\qquad\qquad\;
-
x_{\rm B} \left( \frac{x_{\rm B}^2}{2} + (2 - x_{\rm B}) \frac{\Delta^2}{4M^2} \right)
\left(
{\cal E} \widetilde{{\cal E}}^\ast + \widetilde{\cal E} {\cal E}^\ast
\right)
\Bigg\},
\end{eqnarray}
\begin{eqnarray}
\label{Def-C-DVCS-TP}
{\cal C}^{\scriptscriptstyle\rm DVCS}_{{\rm TP}+}
\left(
{\cal F},{\cal F}^\ast
\right)
\!\!\!&=&\!\!\!
\frac{1}{(2 - x_{\rm B})^2}
\Bigg\{
2 x_{\rm B}(
{\cal H}\widetilde{\cal E}^\ast
+ \widetilde{\cal E} {\cal H}^\ast)
- 2 (2 - x_{\rm B}) (\widetilde {\cal H} {\cal E}^\ast
+ \widetilde {\cal H}^\ast {\cal E})
+ x_{\rm B}^2 ({\cal E}  \widetilde{\cal E}^\ast
+ \widetilde{\cal E} {\cal E}^\ast )
\Bigg\},
\nonumber\\
{\cal C}^{\scriptscriptstyle\rm DVCS}_{{\scriptscriptstyle\rm TP}-}
\left(
{\cal F},{\cal F}^\ast
\right)
\!\!\!&=&\!\!\!
\frac{2}{(2 - x_{\rm B})^2}
\Bigg\{
(2 - x_{\rm B})({\cal H} {\cal E}^\ast - {\cal E} {\cal H}^\ast)
- x_{\rm B} ( \widetilde{\cal H} \widetilde{\cal E}^\ast
- \widetilde{\cal E} \widetilde{\cal H}^\ast )
\Bigg\} \, .
\end{eqnarray}

\begin{itemize}
\item Interference of Bethe-Heitler and DVCS amplitudes:
\end{itemize}
In this case, a part of the twist-two level  result  is expressed in terms
of the functions, which appear  in the lowest twist approximation, and they
have the following form \cite{BelMulNieSch00}
\begin{eqnarray}
\label{Def-C-Int-unp}
{\cal C}^{\scriptscriptstyle\rm INT}_{\rm unp}
\!\!\!&=&\!\!\!
F_1 {\cal H} + \frac{x_{\rm B}}{2 - x_{\rm B}}
(F_1 + F_2) \widetilde {\cal H}
-
\frac{\Delta^2}{4M^2} F_2 {\cal E} \, ,
\\
\label{Def-C-Int-LP}
{\cal C}^{\scriptscriptstyle\rm INT}_{\scriptscriptstyle\rm LP}
\!\!\!&=&\!\!\!
\frac{x_{\rm B} }{2 - x_{\rm B}}(F_1 + F_2) \left({\cal H} +
\frac{x_{\rm B}}{2} {\cal E} \right)
+ F_1 \widetilde{\cal H}
- \frac{x_{\rm B}}{2 - x_{\rm B}}
\left(
\frac{x_{\rm B}}{2} F_1 + \frac{\Delta^2}{4 M^2} F_2
\right)
\widetilde{\cal E}
\, , \\
\label{Def-C-Int-TP}
{\cal C}^{\scriptscriptstyle\rm INT}_{{\scriptscriptstyle\rm TP}+}
\!\!\!&=&\!\!\!
(F_1 + F_2)
\left\{
\frac{x_{\rm B}^2}{2 - x_{\rm B}}
\left( {\cal H} + \frac{x_{\rm B}}{2} {\cal E} \right)
+  \frac{x_{\rm B} \Delta^2}{4 M^2}{\cal E}
\right\}
- \frac{x_{\rm B}^2}{2 - x_{\rm B}} F_1
\left( \widetilde{\cal H} + \frac{x_{\rm B}}{2} \widetilde{\cal E} \right)
\nonumber\\
&&
+ \frac{\Delta^2}{4M^2} \left\{4 \frac{1 - x_{\rm B}}{2 - x_{\rm B}} F_2
\widetilde{\cal H}
- \left( x_{\rm B} F_1 + \frac{x_{\rm B}^2}{2 - x_{\rm B}} F_2 \right)
\widetilde{\cal E} \right\}
\, , \\
{\cal C}^{\scriptscriptstyle\rm INT}_{{\scriptscriptstyle\rm TP}-}
\!\!\!&=&\!\!\!
\frac{1}{2 - x_{\rm B}}
\left( x_{\rm B}^2 F_1 - (1 - x_{\rm B}) \frac{\Delta^2}{M^2} F_2  \right)
{\cal H}
+ \Bigg\{
\frac{\Delta^2}{4 M^2}
\Bigg(
(2 - x_{\rm B}) F_1 + \frac{x_{\rm B}^2}{2 - x_{\rm B}} F_2
\Bigg)
\nonumber\\
&&
+ \frac{x_{\rm B}^2}{2 - x_{\rm B}} F_1 \Bigg\} {\cal E}
- \frac{x_{\rm B}^2}{2 - x_{\rm B}} (F_1 + F_2)
\left(
\widetilde{\cal H} + \frac{\Delta^2}{4 M^2}\widetilde{\cal E} \right) \, .
\nonumber
\end{eqnarray}
The additional terms that  appear  in the power-suppressed contributions are defined as
\begin{eqnarray}
\label{Def-C-IntAdd-unp}
\Delta {\cal C}^{\scriptscriptstyle\rm INT}_{{\rm unp}}
\!\!\!&=&\!\!\!
- \frac{x_{\rm B}}{2-x_{\rm B}}  (F_1 + F_2)
\left\{
\frac{x_{\rm B}}{2 - x_{\rm B}} ({\cal H} + {\cal E})
+ \widetilde {\cal H}
\right\} \, ,
\\
\label{Def-C-IntAdd-LP}
\Delta {\cal C}^{\scriptscriptstyle\rm INT}_{{\scriptscriptstyle\rm LP}}
\!\!\!&=&\!\!\! - \frac{x_{\rm B}}{2 - x_{\rm B}} (F_1 + F_2)
\left\{
{\cal H} + \frac{x_{\rm B}}{2}{\cal E} + \frac{x_{\rm B}}{2 - x_{\rm B}}
\left(
\widetilde{\cal H} + \frac{x_{\rm B}}{2}\widetilde{\cal E}
\right)
\right\} \, ,
\\
\label{Def-C-IntAdd-TP}
\Delta {\cal C}^{\scriptscriptstyle\rm INT}_{{\scriptscriptstyle\rm TP+}}
\!\!\!&=&\!\!\!
- \frac{\Delta^2}{M^2}
\left\{
F_2\widetilde {\cal H} - \frac{x_{\rm B}}{2 - x_{\rm B}}
\left( F_1 + \frac{x_{\rm B}}{2} F_2 \right) \widetilde{\cal E}
\right\} \, ,
\\
\Delta {\cal C}^{\scriptscriptstyle\rm INT}_{{\scriptscriptstyle\rm TP-}}
\!\!\!&=&\!\!\!
\frac{\Delta^2}{M^2}
\left( F_2 {\cal H} - F_1 {\cal E} \right) \, .
\end{eqnarray}

Let us now list the coefficients involving the gluon transversity:
\begin{itemize}
\item Squared DVCS amplitude:
\end{itemize}
\begin{eqnarray}
{\cal C}^{\scriptscriptstyle\rm DVCS}_{T,{\rm unp}}
\!\!\!&=&\!\!\!
\frac{1}{(2 -x_{\rm B})^2}
\Bigg\{
{\cal H}_T
\left[
(2 - x_{\rm B}) {\cal E}^\ast - x_{\rm B} \widetilde{\cal E}^\ast
\right]
- 2 (2 - x_{\rm B}) \widetilde{\cal H}_T
\left[
{\cal H}^\ast
+
\frac{\Delta^2}{4 M^2} {\cal E}^\ast
\right]
\\
&&\qquad\qquad- {\cal E}_T
\left[
(2 - x_{\rm B}) {\cal H}^\ast
- x_{\rm B} \widetilde{\cal H}^\ast
\right]
+ \widetilde{\cal E}_T
\left[
x_{\rm B} ( {\cal H}^\ast + {\cal E}^\ast )
- (2 - x_{\rm B}) \widetilde{\cal H}^\ast
\right]
\Bigg\} \, ,
\nonumber\\
{\cal C}^{\scriptscriptstyle\rm DVCS}_{T,{\scriptscriptstyle\rm LP}}
\!\!\!&=&\!\!\!
\frac{1}{(2 - x_{\rm B})^2}
\Bigg\{
{\cal H}_T
\left[
(2 - x_{\rm B}) {\cal E}^\ast - x_{\rm B} \widetilde{\cal E}^\ast
\right]
+ \widetilde{\cal H}_T
\left[
2 (2 - x_{\rm B}) \widetilde{\cal H}^\ast
-
x_{\rm B} \left( x_{\rm B} - \frac{\Delta^2}{2M^2} \right) \widetilde{\cal E}^\ast
\right]
\\
&&\qquad\qquad
- {\cal E}_T
\Bigg[
x_{\rm B} {\cal H}^\ast - (2 - x_{\rm B}) \widetilde{\cal H}^\ast
+ \frac{x_{\rm B}^2}{2}
\left( {\cal E}^\ast + \widetilde{\cal E}^\ast \right)
\Bigg]
\nonumber\\
&&\qquad\qquad
+ \widetilde{\cal E}_T
\left[
(2 - x_{\rm B})
\left( {\cal H}^\ast + \frac{x_{\rm B}}{2} {\cal E}^\ast \right)
- x_{\rm B}
\left(
\widetilde{\cal H}^\ast + \frac{x_{\rm B}}{2} \widetilde{\cal E}^\ast
\right)
\right]
\Bigg\} \, ,
\nonumber\\
{\cal C}^{\scriptscriptstyle\rm DVCS}_{T,{\scriptscriptstyle\rm TP+}}
\!\!\!&=&\!\!\!
\frac{1}{(2 - x_{\rm B})^2}
\Bigg\{
\left[
4 (1 - x_{\rm B}) {\cal H}_T
-
x_{\rm B}^2 {\cal E}_T + x_{\rm B} (2 - x_{\rm B}) \widetilde{\cal E}_T
\right]
\left(
{\cal H}^\ast + \widetilde{\cal H}^\ast
\right)
\\
&&\qquad\qquad
- \widetilde{\cal H}_T \left( x_{\rm B}^2 + (1 - x_{\rm B}) \frac{\Delta^2}{M^2} \right)
(2 \widetilde{\cal H}^\ast + x_{\rm B} \widetilde{\cal E}^\ast )
\nonumber\\
&&\qquad\qquad
- x_{\rm B}
\left[
x_{\rm B} {\cal H}_T
+ \left(
\frac{x_{\rm B}^2}{2} + (2 - x_{\rm B}) \frac{\Delta^2}{4 M^2}
\right) {\cal E}_T
\right]
( {\cal E}^\ast + \widetilde{\cal E}^\ast )
\nonumber\\
&&\qquad\qquad
+ \widetilde{\cal E}_T
\Bigg[ (2 - x_{\rm B})
\left(
\frac{x_{\rm B}^2}{2} + (2 - x_{\rm B}) \frac{\Delta^2}{4 M^2} \right)
{\cal E}^\ast
- x_{\rm B}^2 \left( \frac{x_{\rm B}}{2} - \frac{\Delta^2}{4 M^2} \right)
\widetilde{\cal E}^\ast
\Bigg]
\Bigg\} \, ,
\nonumber\\
{\cal C}^{\scriptscriptstyle\rm DVCS}_{T,{\scriptscriptstyle\rm TP-}}
\!\!\!&=&\!\!\!
\frac{1}{(2 - x_{\rm B})^2}
\Bigg\{
\left[
4 (1 - x_{\rm B}) {\cal H}_T
-
x_{\rm B}^2 {\cal E}_T + x_{\rm B} (2 - x_{\rm B}) \widetilde{\cal E}_T
\right]
\left( {\cal H}^\ast + \widetilde{\cal H}^\ast \right)
\\
&&\qquad\qquad
- 2 \widetilde{\cal H}_T
\left( x_{\rm B}^2 + (1 - x_{\rm B}) \frac{\Delta^2}{M^2} \right)
\left( {\cal H}^\ast + {\cal E}^\ast \right)
\nonumber\\
&&\qquad\qquad
-x_{\rm B}
\left(
x_{\rm B} {\cal H}_T - (2 - x_{\rm B}) \frac{\Delta^2}{4 M^2} \widetilde{\cal E}_T
\right)
\left( {\cal E}^\ast + \widetilde{\cal E}^\ast \right)
\nonumber\\
&&\qquad\qquad
- {\cal E}_T
\Bigg[
\left( x_{\rm B}^2 + (2 - x_{\rm B})^2 \frac{\Delta^2}{4 M^2} \right)
{\cal E}^\ast
 +
x_{\rm B}^2 \frac{\Delta^2}{4 M^2} \widetilde{\cal E}^\ast
\Bigg]
\Bigg\} \, .
\nonumber
\end{eqnarray}

\begin{itemize}
\item Interference of Bethe-Heitler and DVCS amplitudes:
\end{itemize}
\begin{eqnarray}
{\cal C}^{\scriptscriptstyle\rm INT}_{T,{\rm unp}}
\!\!\!&=&\!\!\!
- F_2 {\cal H}_T
+ 2 \left(F_1 + \frac{\Delta^2}{4 M^2} F_2\right) \widetilde{\cal H}_T
+ F_1 {\cal E}_T \, ,
\\
{\cal C}^{\scriptscriptstyle\rm INT}_{T,{\scriptscriptstyle\rm LP}}
\!\!\!&=&\!\!\!
F_2
\left\{
{\cal H}_T +
\frac{x_{\rm B}}{2}
( 2 \widetilde{\cal H}_T + {\cal E}_T  + \widetilde{\cal E}_T )
\right\}
+ F_1 ( x_{\rm B} \widetilde{\cal H}_T   + \widetilde{\cal E}_T ) \, ,
\\
{\cal C}^{\scriptscriptstyle\rm INT}_{T,{\scriptscriptstyle\rm TP}+}
\!\!\!&=&\!\!\!
(2 F_1 + x_{\rm B} F_2) {\cal H}_T
+ x_{\rm B} \left( \frac{x}{2} - \frac{\Delta^2}{4 M^2} \right)
\left[ 2 (F_1 + F_2) \widetilde{\cal H}_T + F_2 {\cal E}_T \right]
\\
&&+
\left\{
x_{\rm B} F_1
+
\left( \frac{x_{\rm B}^2}{2} + (2 - x_{\rm B}) \frac{\Delta^2}{4M^2} \right) F_2
\right\}
\widetilde{\cal E}_T \, ,
\nonumber\\
{\cal C}^{\scriptscriptstyle\rm INT}_{T,{\scriptscriptstyle\rm TP}-}
\!\!\!&=&\!\!\! (2 F_1 + x_{\rm B} F_2) {\cal H}_T
- (2 - x_{\rm B}) \frac{\Delta^2}{4 M^2}
\left[
2(F_1 + F_2) \widetilde{\cal H}_T  +  F_2 {\cal E}_T
\right]
+ x_{\rm B}
\left(
F_1 + F_2 \frac{\Delta^2}{4 M^2}
\right) \widetilde{\cal E}_T
\, . \nonumber\\
\end{eqnarray}
This set of formulas is the complete result for the real-photon
leptoproduction cross section in the twist-three approximation.
Below, presenting quantitative estimates, we will not discuss the
case of transversely polarized target, therefore, the integration
on the right-hand side of Eq.\ (\ref{WQ}) with respect to $\varphi$
gives $2 \pi$.

\subsubsection{Determination of Compton form factors from interference}

Restricting the consideration to four leading twist CFFs ${\cal F} = \{
{\cal H}, {\cal E}, \widetilde {\cal H}, \widetilde {\cal E} \}$, we have
eight observables given by the first harmonics $\cos(\phi)$ and $\sin(\phi)$
of the interference term, which are accessible away from the kinematical
boundaries in polarized beam and target experiments. Thus, experiments with
both longitudinally and transversely polarized target can measure all eight
Fourier coefficients $c_{1,{\mit\Lambda}}^{\scriptscriptstyle\rm INT}$ and
$s_{1,{\mit\Lambda}}^{\scriptscriptstyle\rm INT}$ and, thus, also
$\Re{\rm e}/\Im{\rm m}{\cal C}^{\scriptscriptstyle\rm INT}_{\mit\Lambda}$
with ${\mit\Lambda} = \{ {\rm unp}, {\rm LP}, {\rm TP}_x, {\rm TP}_y \}$.
Knowing these ${\cal C}$ functions, we can invert them to obtain the CFFs:
\begin{eqnarray}
\label{Inv-CFF-H}
{\cal H}
\!\!\!&=&\!\!\!
\frac{2 - x_{\rm B}}{(1 - x_{\rm B}) D}
\Bigg\{
\left[
\left(
2 - x_{\rm B} + \frac{4 x_{\rm B}^2 M^2}{(2 - x_{\rm B}) \Delta^2}
\right)
F_1 + \frac{x_{\rm B}^2}{2 - x_{\rm B}} F_2
\right] {\cal C}^{\scriptscriptstyle\rm INT}_{\rm unp}
\\
&&\hspace{2cm} - (F_1 + F_2)
\left[
x_{\rm B} {\cal C}^{\scriptscriptstyle\rm INT}_{{\scriptscriptstyle\rm LP}}
+
\frac{2 x_{\rm B}^2  M^2}{(2 - x_{\rm B})\Delta^2}
\left(
x_{\rm B} {\cal C}^{\scriptscriptstyle\rm INT}_{{\scriptscriptstyle\rm LP}}
-
{\cal C}^{\scriptscriptstyle\rm INT}_{{\scriptscriptstyle\rm TP}+}
\right)
\right]
+
F_2 {\cal C}^{\scriptscriptstyle\rm INT}_{{\scriptscriptstyle\rm TP}-}
\Bigg\}
\, , \nonumber\\
\label{Inv-CFF-E}
{\cal E}
\!\!\!&=&\!\!\!
\frac{2 - x_{\rm B}}{(1 - x_{\rm B}) D}
\Bigg\{
\left[
4 \frac{1 - x_{\rm B}}{2 - x_{\rm B}}F_2
-
\frac{4 M^2 x_{\rm B}^2}{(2 - x_{\rm B}) \Delta^2} F_1
\right]
{\cal C}^{\scriptscriptstyle\rm INT}_{\rm unp}
\\
&&\hspace{2cm}
+
\frac{4 x_{\rm B} M^2}{(2 - x_{\rm B}) \Delta^2}
\left( F_1 + F_2 \right)
\left(
x_{\rm B} {\cal C}^{\scriptscriptstyle\rm INT}_{{\scriptscriptstyle\rm LP}}
-
{\cal C}^{\scriptscriptstyle\rm INT}_{{\scriptscriptstyle\rm TP}+}
\right)
+
\frac{4 M^2}{\Delta^2} F_1
{\cal C}^{\scriptscriptstyle\rm INT}_{{\scriptscriptstyle\rm TP}-}
\Bigg\}
\, , \nonumber\\
\label{Inv-CFF-tH}
\widetilde{\cal H}
\!\!\!&=&\!\!\!
\frac{2 - x_{\rm B}}{(1 - x_{\rm B}) D}
\Bigg\{
(2 - x_{\rm B}) F_1
{\cal C}^{\scriptscriptstyle\rm INT}_{{\scriptscriptstyle\rm LP}}
\nonumber\\
&&\hspace{2cm}
- x_{\rm B} (F_1 + F_2) {\cal C}^{\scriptscriptstyle\rm INT}_{{\rm unp}}
+
\left[
\frac{2 x_{\rm B} M^2}{\Delta^2} F_1 + F_2
\right]
\left(
x_{\rm B} {\cal C}^{\scriptscriptstyle\rm INT}_{{\scriptscriptstyle\rm LP}}
-
{\cal C}^{\scriptscriptstyle\rm INT}_{{\scriptscriptstyle\rm TP}+}
\right)
\Bigg\}
\, , \\
\label{Inv-CFF-tE}
\widetilde{\cal E}
\!\!\!&=&\!\!\!
\frac{2 - x_{\rm B}}{(1 - x_{\rm B}) D}
\Bigg\{
\frac{4 M^2}{\Delta^2} \left( F_1 + F_2 \right)
\left(
x_{\rm B} {\cal C}^{\scriptscriptstyle\rm INT}_{{\rm unp}}
+
{\cal C}^{\scriptscriptstyle\rm INT}_{{\scriptscriptstyle\rm TP}-}
\right)
\\
&&\hspace{2cm}
+
\left[
4\frac{1 - x_{\rm B}}{x_{\rm B}} F_2 - \frac{4x_{\rm B} M^2}{\Delta^2} F_1
\right]
{\cal C}^{\scriptscriptstyle\rm INT}_{\scriptscriptstyle\rm LP}
-
\frac{4 (2 - x_{\rm B}) M^2}{x_{\rm B} \Delta^2} F_1
{\cal C}^{\scriptscriptstyle\rm INT}_{{\scriptscriptstyle\rm TP}+}
\Bigg\}
\, , \nonumber
\end{eqnarray}
where
\begin{eqnarray*}
D = 4 \left( F_1^2 - \frac{\Delta^2}{4 M^2} F_2^2 \right)
\left(
1 - \frac{\Delta^2_{\rm min}}{\Delta^2}
\right) \, .
\end{eqnarray*}
Consequently, the measurement of all four Fourier coefficients
$c_{0,\mit\Lambda}^{\scriptscriptstyle\rm INT}$, as well as the four
twist-two DVCS coefficients $c_{0,\mit\Lambda}^{\scriptscriptstyle\rm DVCS}$
can serve as experimental consistency checks. Alternatively, they can be used
to extract CFFs. Thus, experiments with longitudinally polarized target have
the potential to extract the real part of all four CFFs as well as two linear
combinations of their imaginary parts from the interference term alone. The
missing two imaginary parts could then, in principle, be obtained from the DVCS
cross section, i.e., by measuring $c_{0,{\rm unp}}^{\scriptscriptstyle\rm DVCS}$
and $c_{0,{\scriptscriptstyle\rm LP}}^{\scriptscriptstyle\rm DVCS}$.

\subsubsection{Physical observables and access to GPDs}
\label{Sec-PhyObs}

\begin{table}[t]
\begin{center}
\begin{tabular}{|l|c||c|c|c|c|c|c|c|c|}
\hline
\multicolumn{2}{|c||}{sector}
&
\multicolumn{4}{c|}{harmonics in $\cal I$}
&
\multicolumn{1}{c|}{$p$ of}
&
\multicolumn{2}{c|}{$\Delta_\perp^{l}$ behavior}
\\
\multicolumn{1}{|c|}{twist}
&
\multicolumn{1}{c||}{$\cal C$'s}
&
\multicolumn{1}{c|}{unp}
&
\multicolumn{1}{c|}{LP}
&
\multicolumn{1}{c|}{TP$_x$}
&
\multicolumn{1}{c|}{TP$_y$}
&
\multicolumn{1}{c|}{${\cal Q}^{-p}$}
&
\multicolumn{1}{c|}{unp, LP}&\multicolumn{1}{c|}{TP}
\\
\hline\hline
two
&
$\Re{\rm e} \, {\cal C}({\cal F}),\ \Delta{\cal C}({\cal F}) $
&
$c_1$, $c_0$
&
$c_1$, $c_0$
&
$c_1$, $c_0$
&
$s_1$, -
&
1,2
&
1,0
&
0,1
\\
\hline
&
$\Im{\rm m} \, {\cal C}({\cal F}),\ \Delta{\cal C}({\cal F})$
&
$s_1$, -
&
$s_1$, -
&
$s_1$, -
&
$c_1$, $c_0$
&
1,2
&
1,0
&
0,1
\\
\hline\hline
three
&
$\Re{\rm e} \, {\cal C}({\cal F}^{\rm eff})$
&
$c_2$
&
$c_2$
&
$c_2$
&
$s_2$
&
2
&
2
&
1
\\
\hline
&
$\Im{\rm m} \, {\cal C}( {\cal F}^{\rm eff})$
&
$s_2$
&
$s_2$
&
$s_2$
&
$c_2$
&
2
&
2
&
1
\\
\hline\hline
two
&
$\Re{\rm e} \, {\cal C}_T({\cal F}_T)$
&
$c_3$
&
-
&
-
&
-
&
1
&
3
&
2
\\
\hline
&
$\Im{\rm m} \, {\cal C}_T({\cal F}_T)$
&
-
&
$s_3$
&
$s_3$
&
$c_3$
&
1
&
3
&
2
\\
\hline
\end{tabular}
\end{center}
\caption{\label{Tab-FCs-Int} Fourier coefficients $c_{i}^{\scriptscriptstyle\rm INT}$
and $s_{i}^{\scriptscriptstyle\rm INT}$ of the interference term defined in Section
\ref{SubSec-AziAngDep-INT}, while the corresponding ${\cal C}$ coefficients are given
in Section \ref{App-CoeFun}.}
\end{table}

As we have seen, the cross section for leptoproduction of the real
photon possesses very rich angular structure. The goal of experimental
measurements is to pin down the GPDs, and this requires a clean
disentanglement of different components of the cross section (\ref{WQ}).
To go along this line, we introduce appropriate asymmetries in
Section \ref{SubSec-PhyObs-Asy} and demonstrate how a Fourier transform
can distinguish between the interference and squared DVCS contributions.

So far we have introduced eight CFFs at the twist-two level, with four
of them from the gluonic transversity contribution. Four new CFFs appear
at the twist-three level. These sectors can be separated due to their
characteristic azimuthal dependence as summarized in Table \ref{Tab-FCs-Int}
for the interference and in Table \ref{Tab-FCs-DVCS} for the squared DVCS
amplitude, respectively.

Let us first examine the issue of the dominance of each of the three terms
(\ref{Par-BH}) -- (\ref{InterferenceTerm}) in the leptoproduction cross section
(\ref{WQ}) in different kinematical regions. To do this, we need to know
the functional dependence of the Fourier coefficients (\ref{Par-BH}) --
(\ref{InterferenceTerm}) on scaling variables and transferred momenta. Apart
from the explicit $x_{\rm B}$-dependence of the multiplicative prefactors
there is also $x_{\rm B}$-dependence hidden in CFFs. For instance, the
unpolarized form factors ${\cal H}$ and ${\cal E}$ behave approximately
like $x_{\rm B}^{-1}$ in the small-$x_{\rm B}$ region. Thus, for general
kinematical settings, we expect from Eqs.\ (\ref{Def-FC-BH-unp0}),
(\ref{Res-Mom-DVCS-UP}), and (\ref{Res-IntTer-unp}) that
$c_0^{\scriptscriptstyle\rm BH} \sim x_{\rm B}^2 c_0^{\scriptscriptstyle\rm DVCS}
\sim x_{\rm B} c_1^{\scriptscriptstyle\rm INT}/K$ for the scattering on the
unpolarized target. Taking now into account the kinematical prefactors in Eqs.\
(\ref{Par-BH}) -- (\ref{InterferenceTerm}) and the behavior of the BH-propagators
(\ref{ExaBHpro}), we realize that the ratio of the DVCS to BH amplitude behaves
like $\sqrt{- (1-y) \Delta^2/ y^2 {\cal Q}^2 }$. Obviously, for small (large) $y$
the DVCS (BH) term dominates. As compared to the squared amplitudes, the
interference term, after subtraction of $c_0^{\scriptscriptstyle\rm INT}$, has
an additional factor $\sqrt{\Delta^2_\perp/\Delta^2}$. Note, that the beam spin-flip
contributions provide always an additional damping by a power of the factor $y$.
For the unpolarized or longitudinally polarized target, higher harmonics in any of
the three terms are suppressed by  powers of $K$. However, in case of gluonic
transversity, this goes in parallel with the enhancement by ${\cal Q}^2/M^2_N$.
It is important that the lower harmonics in the interference term, i.e.,
$c_0^{\scriptscriptstyle\rm INT}$, appear at power suppressed level but expressed
in terms of twist-two Compton form factors. Since, unlike $c_1$, they are not
proportional to the factor $K$, they can be rather important and sizable in the
regions close to the kinematical boundaries. In case of the transversely polarized
target, we observe that both higher and lower twist-three harmonics are suppressed
by one power of $K$ in the interference term.

The analytical structure and simple counting rules given above, provide a
guideline on how to separate the three different parts in the leptoproduction
cross section. In single spin-flip experiments, which give access to the
imaginary part of CFFs, the BH cross section drops out, while in unpolarized or
double spin-flip experiments it does not and one needs to subtract it. This can
certainly be done for not very small values of $y$. The interference and
squared DVCS terms have different azimuthal angular dependence due to the
presence of the BH-propagators in the interference term. In principle, this
fact can be used to separate them by a Fourier analysis. However, this method
requires very high-statistics data, from the experimental side, and a better
understanding of twist-four contributions, from the theoretical side.

\begin{table}[t]
\begin{center}
\begin{tabular}{|l|c||c|c|c|c|c|c|c|c|}
\hline
\multicolumn{2}{|c||}{interference of}
&
\multicolumn{4}{c|}{harmonics in $|{\cal T}_{\scriptscriptstyle\rm DVCS}|^2$}
&
\multicolumn{1}{c|}{$p$ of}
&
\multicolumn{2}{c|}{$\Delta_\perp^{l}$ behavior}
\\
\multicolumn{1}{|c|}{twist}
&
\multicolumn{1}{c||}{$\cal C$'s}
&
\multicolumn{1}{c|}{unp}
&
\multicolumn{1}{c|}{LP}
&
\multicolumn{1}{c|}{TP$_x$}
&
\multicolumn{1}{c|}{TP$_y$}
&
\multicolumn{1}{c|}{${\cal Q}^{-p}$}
&
\multicolumn{1}{c|}{unp, LP}
&
\multicolumn{1}{c|}{TP}
\\
\hline\hline
two \& two
&
$\Re{\rm e} \, {\cal C}({\cal F},{\cal F}^\ast)$
&
$c_0$
&
$c_0$
&
$c_0$
&
-
&
2
&
0
&
1
\\
\hline
&
$\Im{\rm m} \, {\cal C}({\cal F},{\cal F}^\ast)$
&
-
&
-
&
-
&
$c_0$
&
2
&
-
&
1
\\
\hline\hline
two \& three
&
$\Re{\rm e} \, {\cal C}({\cal F}^{\rm eff},{\cal F}^\ast)$
&
$c_1$
&
$c_1$
&
$c_1$
&
$s_1$
&
3
&
1
&
0
\\
\hline
&
$\Im{\rm m} \, {\cal C}({\cal F}^{\rm eff},{\cal F}^\ast)$
&
$s_1$
&
$s_1$
&
$s_1$
&
$c_1$
&
3
&
1
&
0
\\
\hline\hline
two \& two
&
$\Re{\rm e} \, {\cal C}_T({\cal F}_T,{\cal F}^\ast)$
&
$c_2$
&
-
&
-
&
-
&
2
&
2
&
-
\\
\hline
&
$\Im{\rm m} \, {\cal C}_T({\cal F}_T,{\cal F}^\ast)$
&
-
&
$s_2$
&
$s_2$
&
$c_2$
&
2
&
2
&
1
\\
\hline
\end{tabular}
\end{center}
\caption{\label{Tab-FCs-DVCS} Fourier coefficients
$c_{i}^{\scriptscriptstyle\rm DVCS}$ and $s_{i}^{\scriptscriptstyle\rm DVCS}$
of the squared DVCS amplitude $|{\cal T}_{\scriptscriptstyle\rm DVCS}|^2$
defined in Section \ref{SubSec-AziAngDep-DVCS}, while the corresponding
$\cal C$ coefficients are given in Section \ref{App-CoeFun}.}
\end{table}

Due to different charge conjugation properties of individual components, it
is possible to use the charge asymmetry to separate the interference and
squared DVCS terms. The interference term is charge-odd and can be extracted
in facilities that possess both positively and negatively charged lepton beams
\cite{BroCloGun72,KroSchGui96}, i.e.,
\begin{eqnarray}
d\sigma^+ - d\sigma^-
\sim
\frac{1}{x_{\rm B} y^3 {\cal P}_1 (\phi) {\cal P}_2 (\phi) \Delta^2}
\left\{
c_0^{\scriptscriptstyle\rm INT}
+
\sum_{n = 1}^3
\left[
c_n^{\scriptscriptstyle\rm INT} \cos(n \phi)
+
s_n^{\scriptscriptstyle\rm INT} \sin(n \phi)
\right]
\right\}
\, ,
\end{eqnarray}
where we used the convention $d^{\pm} \sigma = d \sigma_{e^\pm}$. The
measurement of the charge asymmetry and consequent extraction of separate
harmonics provides the real (unpolarized or double spin-flip experiments)
and imaginary (single spin-flip experiments) parts of linear combinations
of twist-two and twist-three CFFs. The explicit projection procedure of
these harmonics will be discussed below in Section \ref{SubSec-PhyObs-ChaAsy}.
Moreover, the charge-even part is given by the sum of the BH and DVCS cross
section. The subtraction of the BH part gives then the Fourier coefficients
of the DVCS cross section:
\begin{eqnarray}
d^+ \sigma + d ^-\sigma - 2d^{\scriptscriptstyle\rm BH} \sigma  \sim
\frac{1}{y^2 {\cal Q}^2}\left\{
c^{\scriptscriptstyle\rm DVCS}_0
+ \sum_{n = 1}^2
\left[
c^{\scriptscriptstyle\rm DVCS}_n \cos (n\phi)
+
s^{\scriptscriptstyle\rm DVCS}_n \sin (n \phi)
\right]
\right\}
\, .
\end{eqnarray}

For a polarized beam and target with all polarization options, the
real and imaginary parts of all four CFFs in the twist-three sector
can be extracted from the interference term alone by projecting onto
the $\cos(2\phi)$ and $\sin(2\phi)$ harmonics and using Eqs.\
(\ref{Inv-CFF-H}) -- (\ref{Inv-CFF-tE}). Alternatively, knowing the
twist-two sector and having only a longitudinally polarized target,
one can employ, in addition, the squared DVCS term, i.e., its
$\cos(\phi)$ and $\sin(\phi)$ harmonics, to access the full twist-three
sector.

For gluonic transversity, the $\cos(3\phi)$ and $\sin(3\phi)$ harmonics in
the interference term can only provide  one imaginary and three real parts of
certain linear combinations of ${\cal F}_T$. Missing information, in
principle, can be obtained from the $\cos(2\phi)$ and $\sin(2\phi)$ harmonics
of the squared DVCS term. Note here that again a polarized beam and target
with all polarizations is necessary. Moreover, the gluonic transversity is
suppressed by $\alpha_s/\pi$, so one expects a stronger contamination by
twist-four effects \cite{KivMan01}.

As we discussed, a combination of the charge asymmetry with different
nucleon/lepton polarizations and projection of the corresponding harmonics
provides, at least in principle, a way to explore the real and imaginary
part of all CFFs. This gives maximal access to all GPDs, which enter in a
convolution with the real or imaginary part of the coefficient functions.
Since these formulas cannot be deconvoluted in practice \cite{Fre00}, one
has to rely on models with a set of free parameters, as discussed in Section
\ref{GPDModels}, which has to be adjusted to experimental data on CFFs.

\subsubsection{Asymmetries}
\label{SubSec-PhyObs-Asy}

The measurements of the cross section (\ref{WQ}) in different setups,
as discussed in the preceding section, would directly lead to determination
of the CFFs. However, from the experimental point of view, it is simpler to
measure asymmetries, thus avoiding the issue of the absolute normalization.
Let us now discuss the separation of twist-two and twist-three sectors in
terms of asymmetries.

The charge asymmetry
\begin{equation}
\label{Def-CA-old}
A_{\rm C}
= \left.
\left(
\int_{-\pi/2}^{\pi/2} d \phi
\frac{d^+\sigma^{\rm unp} - d^-\sigma^{\rm unp}}{d \phi}
- \int_{\pi/2}^{3\pi/2} d\phi
\frac{d^+ \sigma^{\rm unp} - d^-\sigma^{\rm unp}}{d \phi}
\right)
\right/
\int_{0}^{2\pi} d\phi
\frac{d{^{-}\!\sigma}^{\rm unp} +  d{^{+}\!\sigma}^{\rm unp}}{d\phi}
\, ,
\end{equation}
for unpolarized settings contains the contribution of all harmonics due
to the presence of the non-negligible $\phi$-dependence of BH propagators:
\begin{equation}
A_{\rm C} \sim \sum_{n = 0}^3 I^c_{1,n} c_n^{\scriptscriptstyle\rm INT}
\, ,
\end{equation}
with
\begin{equation}
I^c_{1,n}
\sim
\left(
\int_{-\pi/2}^{\pi/2} d \phi
\frac{\cos(n \phi)}{{\cal P}_1(\phi) {\cal P}_2(\phi)}
-
\int_{\pi/2}^{3\pi/2} d \phi
\frac{\cos(n \phi)}{{\cal P}_1(\phi) {\cal P}_2(\phi)}
\right)
\, ,
\end{equation}
while the normalization is not affected by twist-three corrections. If the
final photon is collinear to the incoming (massless) lepton, ${\cal P}_1
(\phi)$ is peaked around $\phi = \pi$. Thus, the ratio $I^c_{1,n}/I^c_{1,1}$
approaches plus or minus one, and all harmonics contribute on equal footing.
However, since $K$ is then of  order $\Delta^2/{\cal Q}^2$, only
$c_0^{\scriptscriptstyle\rm INT}$ and $c_1^{\scriptscriptstyle\rm INT}$
give essential contributions at the same order in $\Delta^2/{\cal Q}^2$.
Thus, for this asymmetry, $c_0^{\scriptscriptstyle\rm INT}$ can give an
essential effect for large $y$, since the twist-two part becomes small.
For $y \ll y_{\rm col}$, all twist-three harmonics  are suppressed in
addition by the $K$-factor. From the expansion of $I^c_{1,n}$ in powers
of $K$ we get
\begin{eqnarray}
A_{\rm C}
\sim
c_{1,{\rm unp}}^{\scriptscriptstyle\rm INT}
-
\frac{1}{3} c_{3,{\rm unp}}^{\scriptscriptstyle\rm INT}
-
\frac{2 (3 - 2 y)}{2 - y} \frac{K}{1 - y}
\left(
c_{0,{\rm unp}}^{\scriptscriptstyle\rm INT}
-
\frac{1}{3} c_{2,{\rm unp}}^{\scriptscriptstyle\rm INT}
\right)
+
\dots
\, .
\end{eqnarray}
So, one would expect the contamination of the leading twist-two prediction by a
$\sqrt{-\Delta^2/{\cal Q}^2}$-suppressed term, which, however, contains only
twist-two CFFs. Even if this $1/{\cal Q}$-contamination is small, $A_{\rm C}$
still fails to extract solely the $c^{\scriptscriptstyle\rm INT}_{1, {\rm unp}}$
coefficient of the interference term since it gets an additive correction from
$c^{\scriptscriptstyle\rm INT}_{3, {\rm unp}}$, induced by the gluon transversity
and higher twist effects. Note, however, that the gluon contribution is suppressed
by a power of the strong coupling $\alpha_s$, so it probably does not strongly
affect the twist-two coefficient $c^{\scriptscriptstyle\rm INT}_{1, {\rm unp}}$.
Therefore, as a first order approximation, the definition (\ref{Def-CA-old}) can
be used for an order of magnitude estimate of the effect.

The (definite charge) beam-spin asymmetry on the unpolarized target,
\begin{equation}
\label{Def-SSA-old}
A_{\rm SL}
=
\left.
\left(
\int_0^\pi
d \phi
\frac{
d \sigma^\uparrow - d \sigma^\downarrow
}{
d\phi
}
- \int_\pi^{2\pi}
d \phi
\frac{
d \sigma^\uparrow - d \sigma^\downarrow
}{
d \phi
}
\right)
\right/
\int_0^{2\pi} d \phi
\frac{
d \sigma^\uparrow + d \sigma^\downarrow
}{
d \phi
}
\, ,
\end{equation}
does not separate the interference term alone. It does contain
contributions from the squared DVCS amplitude:
\begin{equation}
A_{\rm SL}
\sim
\pm \sum_{n = 1}^2 I^s_{1,n} s_{n,{\rm unp}}^{\scriptscriptstyle\rm INT}
-
\frac{\Delta^2}{y {\cal Q}^2} x_{\rm B} s_{1,{\rm unp}}^{\scriptscriptstyle\rm DVCS}
\, ,
\end{equation}
with
\begin{equation}
I^s_{1,n}
= - \frac{1}{y^2}
\left(
\int_{0}^{\pi} d \phi
\frac{\sin(n \phi)}{{\cal P}_1(\phi) {\cal P}_2(\phi) }
-
\int_{\pi}^{2\pi} d \phi
\frac{\sin(n \phi)}{{\cal P}_1(\phi) {\cal P}_2(\phi)}
\right)
\, . \nonumber
\end{equation}
In the collinear limit, the asymmetry vanishes. On the other hand, for
$y \to 0$, it is determined by the twist-three coefficient
$s_{1,{\rm unp}}^{\scriptscriptstyle\rm DVCS}$. The expansion with
respect to $K$ reads:
\begin{eqnarray}
\label{ASLexp}
A_{\rm SL}
\sim
s_{1,{\rm unp}}^{\scriptscriptstyle\rm INT}
-
\frac{2 (3 - 2 y)}{3(2 - y)} \frac{K}{1 - y}
s_{2,{\rm unp}}^{\scriptscriptstyle\rm INT}
-
\frac{(1-y)(2-y)\Delta^2}{y {\cal Q}^2} x_{\rm B}
s_{1,{\rm unp}}^{\scriptscriptstyle\rm DVCS}
+
\dots
\, .
\end{eqnarray}
The $1/{\cal Q}$-suppressed effect generated by  the BH propagators induces
contamination due to  the second harmonic suppressed by $K/(1-y)$. Depending
on the kinematics and the size of multi-particle contributions, this
contamination together with $s_{1,{\rm unp}}^{\scriptscriptstyle\rm DVCS}$
may prevent clean access to the twist-two GPDs even from high-precision
measurements of this asymmetry. Note, that for general reasons,
$s_{3,{\rm unp}}^{\scriptscriptstyle\rm INT}$
($s_{2,{\rm unp}}^{\scriptscriptstyle\rm DVCS}$) is absent in the
unpolarized interference (squared DVCS) term. However, the normalization
of $A_{\rm SL}$ is affected by $1/{\cal Q}$-effects in the interference term,
mainly due to the coefficient $c^{\scriptscriptstyle\rm INT}_{0,{\rm unp}}$.

\subsubsection{Electron and positron beam options}
\label{SubSec-PhyObs-ChaAsy}

When lepton beams of both charges are available, this provides  clean
separation of twist-two and twist-three GPDs. In these settings, one
can discuss charge-odd and charge-even parts of the cross section (\ref{WQ}),
which extract the interference, and squared DVCS and BH amplitudes,
respectively. The integrated charge-even part of the cross section does
not contain any twist-three contributions---the interference term cancels
there. On the other hand, after azimuthal averaging, the $c_0$ coefficient
of the squared DVCS and all harmonics of squared BH amplitudes survive.
So, it is convenient to use $c_0$  as a unique normalization of the
asymmetries discussed below. Namely, independently  from the target
polarization we introduce
\begin{equation}
\label{Def-Nor-Asy}
{\cal N}^{- 1}_{+-}
\equiv
\int_0^{2\pi} d \phi \frac{d^+ \sigma^{\rm unp} + d^-\sigma^{\rm unp}}{d\phi}
=
2\int_0^{2\pi} d\phi \frac{d^{\scriptscriptstyle\rm BH} \sigma^{\rm unp} +
d^{\scriptscriptstyle\rm DVCS}\sigma^{\rm unp}}{d\phi}
\, .
\end{equation}

\begin{itemize}
\item Charge-odd part:
\end{itemize}
In this case, one  ends up with the interference term alone. However, because of
the $\phi$-dependence of the BH propagators we have to include an additional
weight factor and use the measure
\begin{eqnarray}
d w =  2 \pi
\frac{
{\cal P}_1 (\phi) {\cal P}_2 (\phi) d \phi
}{
\int_0^{2\pi} {\cal P}_1 (\phi') {\cal P}_2 (\phi') d \phi'
}
\, ,
\end{eqnarray}
for the azimuthal integration in order to compensate for the strong
$\phi$-dependence of the product of lepton propagators. The measure $d w$ has
the properties
\begin{eqnarray*}
\int_0^{2\pi} d w = 2 \pi
\, , \qquad
d w (\phi) = d w (- \phi) = d w (\phi + 2 \pi) \, .
\end{eqnarray*}
Now we can exactly separate the Fourier coefficients in Eq.\ (\ref{InterferenceTerm}).
Note that $d w$ has its minimum at $\phi = \pi$, when the outgoing photon lies in
the lepton scattering plane. In case when its momentum is collinear to the lepton
beam, this minimum approaches zero in the massless limit.

To project out different harmonics, one can either (i) do the azimuthal
averaging with appropriate weights, namely,
\begin{equation}
\cos (n \phi) d w
\, , \qquad\qquad
\sin (n \phi) d w
\, ,
\end{equation}
where $n = 0, \dots, 3$, or, (ii) use the fact that the Fourier sum for
the cross section has only a finite number of terms, and integrate over
different partitions of the azimuthal sphere. It is important to realize
that the statistical error acquired via the extraction of the Fourier
harmonics with the first weighting procedure is smaller compared to the
latter, where the coefficients are found by forming asymmetries\footnote{This
generic statement can be definitely made for a cross section of the form
$d \sigma/d \phi = \alpha_0 + \alpha_1 \cos \phi$. The weights matching the
above two methods are then $w_1 = \cos \phi$, and $w_2 = {\rm sgn} (\cos \phi)$
corresponding to the left-right asymmetry. The former generates a smaller
statistical error in the course of the extraction of $\alpha_1$ compared to
the latter.}. Let us present the charge odd asymmetries (CoA), which distinguish
the cosine, $\cos (n \phi)$, and sine, $\sin (n \phi)$, harmonics.

The $\cos$-harmonics, i.e., $c^{\scriptscriptstyle\rm INT}_n$ coefficients,
are projected out by means of the integrals
\begin{eqnarray}
\label{Def-CA-Mom-c-0dd}
\CoA^{\mit\Lambda}_{c(0)}
\!\!\!&=&\!\!\!
{\cal N}_{+-}
\int_0^{2\pi} d w
\frac{
d^+ \sigma^{\mit\Lambda} - d^- \sigma^{\mit\Lambda}
}{
d\phi
}
\, , \\
\CoA^{\mit\Lambda}_{c(1)}
\!\!\!&=&\!\!\!
{\cal N}_{+-}
\left(
\int_{-\pi/2}^{\pi/2} d w
\frac{
d^+ \sigma^{\mit\Lambda} - d^- \sigma^{\mit\Lambda}
}{
d\phi
}
-
\int_{\pi/2}^{3\pi/2} d w
\frac{
d^+ \sigma^{\mit\Lambda} - d^- \sigma^{\mit\Lambda}
}{
d\phi
}
\right)
+ \frac{1}{3} \CoA^{\mit\Lambda}_{c(3)}
\, , \\
\CoA^{\mit\Lambda}_{c(n)}
\!\!\!&=&\!\!\!
{\cal N}_{+-}
\sum_{k = 1}^{2 n} (-1)^{k + 1}
\int_{(2 k - 3) \pi / (2n)}^{( 2 k - 1)\pi / (2n)} d w
\frac{
d^+ \sigma^{\mit\Lambda} - d^- \sigma^{\mit\Lambda}
}{
d\phi
} \, ,
\end{eqnarray}
with $n$ running over $n = 2, 3$ in the last equation, and ${\mit\Lambda}$
stands for ${\mit\Lambda} = \{ {\rm unp}, {\rm LP}, {\rm TP} \}$. The projection
can be achieved by an appropriate flip of the target polarization vector. For the
transversely polarized target, asymmetries are given in terms of two different
combinations of CFFs. They are separable by the projection of the first odd and
even harmonics in $\varphi$, while the average $\int_0^{2\pi} d \varphi\,
{\CoA}^{\scriptscriptstyle\rm TP} = 0$ vanishes.

Next, the $\sin$-harmonics, $s^{\scriptscriptstyle\rm INT}_n$, can analogously
be separated with the help of the formulas
\begin{eqnarray}
\label{Def-CA-Mom-s-Odd}
{\CoA}^{\mit\Lambda}_{s(1)}
\!\!\!&=&\!\!\!
{\cal N}_{+-}
\left(
\int_{0}^{\pi} d w
\frac{
d^+ \sigma^{\mit\Lambda} - d^- \sigma^{\mit\Lambda}
}{
d \phi
}
-
\int_{\pi}^{2\pi} d w
\frac{
d^+ \sigma^{\mit\Lambda} - d^- \sigma^{\mit\Lambda}
}{
d \phi
}
\right)
- \frac{1}{3} {\CoA}^{\mit\Lambda}_{s(3)}
\, , \\
\label{CoAsn}
{\CoA}^{\mit\Lambda}_{s(n)}
\!\!\!&=&\!\!\!
{\cal N}_{+-}
\sum_{k = 1}^{2n} (-1)^{k + 1}
\int_{(k - 1)\pi / n}^{k \pi / n}
d w
\frac{
d^+ \sigma^{\mit\Lambda} - d^- \sigma^{\mit\Lambda}
}{
d \phi
} \, ,
\end{eqnarray}
where $n = 2, 3$.

\begin{itemize}
\item Charge-even part:
\end{itemize}
Let us now  define the azimuthal asymmetries of the charge-even part. To do
this in the cleanest way, we subtract  first the BH term. This might be
possible in practice in the analysis of experimental data, since the BH
cross section is known exactly (up to electromagnetic radiative corrections),
see Section \ref{BHcrosssection}. Of course, it is assumed that the nucleon
form factors are known from other measurements. This gives the squared DVCS
amplitude
\begin{eqnarray*}
2 d^{\scriptscriptstyle\rm DVCS} \sigma
=
d^+ \sigma + d ^-\sigma
-
2 d^{\scriptscriptstyle\rm BH} \sigma
\, .
\end{eqnarray*}
Then an appropriate azimuthal averaging, now with the conventional measure
$d \phi$, separates the $\cos$-harmonics, $c^{\scriptscriptstyle\rm DVCS}_n$,
via
\begin{eqnarray}
\label{Def-CA-Mom-c-Even}
{\CeA}^{\mit\Lambda}_{c(0)}
\!\!\!&=&\!\!\!
2 {\cal N}_{+-}
\int_0^{2\pi} d \phi
\frac{
d^{\scriptscriptstyle\rm DVCS} \sigma^{\mit\Lambda}
}{
d\phi
}
\, , \\
\CeA^{\mit\Lambda}_{c(n)}
\!\!\!&=&\!\!\!
2 {\cal N}_{+-}
\sum_{k = 1}^{2n} (-1)^{k + 1}
\int_{(2 k - 3) \pi / (2n)}^{(2 k - 1) \pi / (2n)} d \phi
\frac{
d^{\scriptscriptstyle\rm DVCS} \sigma^{\mit\Lambda}
}{
d\phi
}
\, ,
\end{eqnarray}
with $n = 1, 2$,
and $\sin$-dependent coefficients by means of
\begin{eqnarray}
\label{Def-CA-Mom-s-Even}
\CeA^{\mit\Lambda}_{s(n)}
= 2 {\cal N}_{+-}
\sum_{k = 1}^{2n} (-1)^{k + 1}
\int_{(k - 1)\pi / n}^{k \pi / n}
d \phi
\frac{
d^{\scriptscriptstyle\rm DVCS} \sigma^{\mit\Lambda}
}{
d\phi
} \, ,
\end{eqnarray}
with $n = 1, 2$.

To conclude, an experimental facility having electron and positron beams
is an ideal place to study GPDs.

\subsubsection{Electron or positron beam option}

Another situation is when only one kind of the lepton beam is available.
Then, the study of single (lepton or nucleon) spin asymmetries allows one
to remove the background BH cross section. Note, however, that when both the
beam and target are polarized, and one studies double-spin asymmetries,
one gets the contamination from the BH harmonics too. In the single
(lepton or hadron) spin experiments, still the twist-two coefficient
$s_1^{\scriptscriptstyle\rm INT}$ is contaminated by power-suppressed
effects, since both the interference and squared DVCS terms contribute.
The best one can do in these circumstances is to cancel completely the
twist-three part of the interference term in the numerator. However,
one will have still the power-suppressed DVCS cross section. For
instance, for the single lepton-spin experiment one is able to probe
$s^{\scriptscriptstyle\rm INT}_{1,{\rm unp}}$ plus $(1 - y)\Delta^2/y
{\cal Q}^2$ corrections from $|{\cal T}_{\rm DVCS}|^2$.  This can be done
using the formula
\begin{eqnarray}
\label{Def-SSA-Mom}
\SSA_{1} \!\!\!&=&\!\!\!
\Bigg(
\int_{0}^\pi d w
\frac{ d \sigma^\uparrow - d \sigma^\downarrow
}{
d\phi
}
-
\int_\pi^{2\pi} d w
\frac{
d \sigma^\uparrow - d \sigma^\downarrow
}{
d\phi
}
\\
&&\qquad\qquad\qquad\qquad
- \frac{1}{3}
\sum_{k = 1}^{6} (-1)^{k + 1}
\int_{(k - 1)\pi / 6}^{k \pi / 6}
d w
\frac{
d \sigma^\uparrow - d \sigma^\downarrow
}{
d \phi
}
\left.\Bigg) \right/
\int_0^{2\pi} d \phi
\frac{
d \sigma^\uparrow + d \sigma^\downarrow
}{
d \phi
}
\, .
\nonumber
\end{eqnarray}
Analogous extraction of the twist-two coefficient
$s^{\scriptscriptstyle\rm INT}_{1,{\scriptscriptstyle\rm LP}}$
is available for the nucleon-spin asymmetry (with the unpolarized
lepton beam). The projection of the same components can be achieved
by weighting the integral with $\sin(\phi) d w$.

For smaller value of $y$, the contamination from the squared DVCS term may
be large. Let us demonstrate that a separation of the interference and
squared DVCS terms can be achieved by a Fourier transform. The multiplication
of the cross section with $d w / d \phi$ induces new harmonics in the squared
DVCS term. Projection of all three harmonics, i.e., measuring the asymmetries
\begin{equation}
\SSA_{n}
=
\Bigg(
\sum_{k = 1}^{2n} (-1)^{k + 1}
\int_{(k - 1)\pi / n}^{k \pi / n}
d w
\frac{
d \sigma^\uparrow - d \sigma^\downarrow
}{
d \phi
}
\left.\Bigg) \right/
\int_0^{2\pi} d \phi
\frac{
d \sigma^\uparrow + d \sigma^\downarrow
}{
d \phi
}
\, ,
\end{equation}
(where $n = 2, 3$) provides the desired Fourier coefficients:
\begin{eqnarray}
s_{1,{\rm unp}}^{\scriptscriptstyle\rm INT}
\sim
\left(
\SSA_{1} -
\frac{w_{s,11}}{w_{s,13}} \SSA_{3}
\right)
\, , \quad
s_{2,{\rm unp}}^{\scriptscriptstyle\rm INT}
\sim
\left(
\SSA_{2} - \frac{w_{s,12}}{w_{s,13}} \SSA_{3}
\right)
\, , \quad
s_{1,{\rm unp}}^{\scriptscriptstyle\rm DVCS}
\sim
\SSA_{3}
\, ,
\end{eqnarray}
with $w_{s,km}= \int_{0}^{2\pi} d w \sin(k \phi) \sin(m \phi)$. Thus, the
extraction of twist-three harmonics is, in principle, possible, however, it
requires high precision data. Such a modified Fourier analysis can  be used
to separate the coefficients of the interference and squared DVCS term also
in the unpolarized and double spin-flip experiments. But one should realize
that the weighted cross section contains now four odd and five even harmonics,
and that the BH cross section must be removed.

\subsubsection{Remarks on state-of-the-art of DVCS}

To date, theoretical analyses of the deeply virtual Compton scattering were
performed to next-to-leading order accuracy for a number of twist-two observables
\cite{BelMueNieSch99,FreMacD01,FreMcD01} and in Born approximation for certain
twist-three asymmetries \cite{KivPolVan00,BelMulKir01}. These calculations agree
rather well with the available experimental data \cite{Aip01,Sau00,Adl01,Ste01}
(see Ref.\ \cite{Now05} for a recent comprehensive summary) even with current crude
knowledge of parameter dependence of GPDs, as we outlined in Section \ref{GPDModels}.
Models for power suppressed effects in DVCS amplitudes were suggested based on
renormalon analysis \cite{BelSch98,VanManSte98} and an approach with introduction of
the intrinsic transverse momentum dependence in GPDs \cite{VanGuiGui99} with the
framework developed in Ref.\ \cite{JakKroRau94} for light-cone distribution amplitudes.

The current experimental facilities do not possess complete exclusivity with respect to
the final state particles. The usual situation is that the recoiled nucleon remains
undetected and reconstructed using the missing mass technique. This represents an
uncertainty in interpretation of the experimental data. Namely, the currently used
technique cannot unambiguously establish whether it was a single nucleon or a nucleon
accompanied by soft pion(s) that struck the detector. This calls for consideration of
the associate soft-pion production along with DVCS. This problem has been addressed in
several papers \cite{GuiMosVan03,CheSav03,KivPolStr04,Bir05} to which we refer for
further reading (see also Ref.\ \cite{BluEilGeyRob02}).

\subsection{Leptoproduction of lepton pairs}
\label{LeptonPairProductionNonidentical}

In the present section, we will discuss a process which allows to lift the
kinematical restriction $\eta = \xi$ of the deeply virtual Compton scattering. As
we have thoroughly discussed in the preceding sections, this restriction limited
the measurement of GPDs via the imaginary part of the deeply virtual Compton
scattering amplitudes in Born approximation solely to the line $\eta = \xi$ in
the longitudinal momentum fraction space, $F^q (\xi, \xi, \Delta^2)$ (see Fig.\
\ref{SampleUquarkGPD}). The way to eliminate this constraint is to have both the
incoming and outgoing photons to be off-shell and carry unequal virtualities. This
calls for consideration of the lepton pair production in the lepton scattering off
the nucleon $e (k) N (p_1) \to e (k') N (p_2) \ell (\ell_-) \bar\ell (\ell_+)$.

\subsubsection{Mapping the surface of GPDs}

\begin{figure}[t]
\begin{center}
\mbox{
\begin{picture}(0,214)(90,0)
\put(0,0){\insertfig{7}{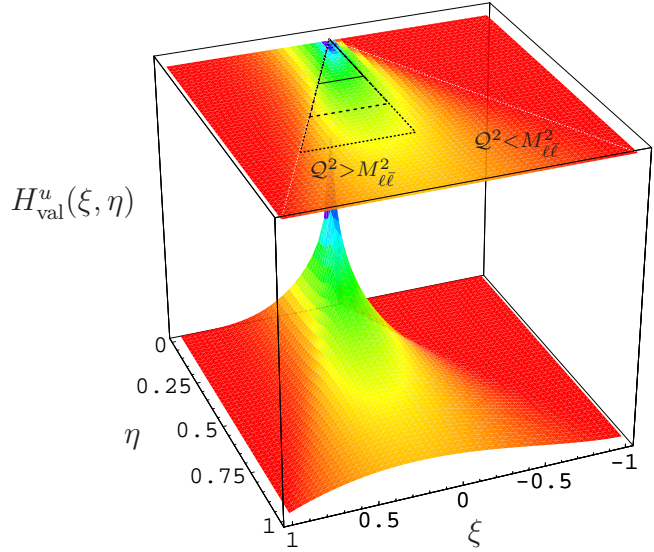}}
\put(128,5){$\xi$}
\put(-2,42){$\eta$}
\put(-45,130){$H^u_{\rm val} (\xi, \eta)$}
\put(68,143){{$\scriptstyle {\cal Q}^2 > M_{\ell\bar\ell}^2$}}
\put(130,154){$\scriptstyle {\cal Q}^2 < M_{\ell\bar\ell}^2$}
\end{picture}
}
\end{center}
\caption{\label{XiEtaPlane} The coverage of a sample GPD surface---the
valence component of the $u$-quark distribution---with electron beams of
different energies \cite{BelMul03,BelMul03a}. The three contours displayed on
the figure embrace the areas to be probed for different values of the electron
beam energy $E$, the lepton energy loss $y$, and the $t$-channel momentum
transfer $\Delta^2$: (i) solid contour corresponds to $E = 11 \, {\rm GeV}$,
$y = 0.5$, and $\Delta^2 = - 0.3 \, {\rm GeV}^2$ ($M_N^2 \leq {\cal Q}^2
\leq 10 \, {\rm GeV}^2$), (ii) dashed contour corresponds to $E = 25 \,
{\rm GeV}$, $y = 0.75$, and $\Delta^2 = - 1 \, {\rm GeV}^2$ ($- 4 \Delta^2
\leq {\cal Q}^2 \leq 20 \, {\rm GeV}^2$), (iii) dotted contour corresponds
to $E = 40 \, {\rm GeV}$, $y = 0.9$, and $\Delta^2 = - 3 \, {\rm GeV}^2$
($- 4 \Delta^2 \leq {\cal Q}^2 \leq 35 \, {\rm GeV}^2$). We discarded in
this plot the change of the GPD with $\Delta^2$ for different kinematical
settings.}
\end{figure}

The advantage of having varying masses of the incoming and outgoing photons
is that this allows to probe GPDs away from the diagonal $|\xi| = \eta$, the
only kinematics which is accessible in DVCS. The skewness variable $\eta$ is
proportional to the sum ${\cal Q}^2 + M_{\ell\bar\ell}^2$ (with $ {\cal Q}^2
= -q_1^2$  and $M_{\ell\bar\ell}^2 = q_2^2$) while $\xi$ is essentially the
difference ${\cal Q}^2 - M_{\ell\bar\ell}^2$ (cf.\ Eq.\ (\ref{XitoEta})). The
boundaries of the ($\xi$, $\eta$) region probed in the process are set by
the following kinematical constraints:
\begin{itemize}
\item The skewness parameter lies in the region $\eta_{\rm min} < \eta < 0$,
where the lower bound comes from the kinematical condition
$|\Delta^2| \geq |\Delta_{\rm min}^2|$:
\begin{equation}
\eta_{\rm min} \leq  \sqrt{- \Delta^2/(4 M_N^2 - \Delta^2)}
\, .
\end{equation}
\item The upper and lower value of $\xi$ is a consequence of the quasi-real
limit of the space- or timelike photon
\begin{equation}
- \eta < \xi < \eta
\, .
\end{equation}
\end{itemize}

It can be seen from Fig.\ \ref{XiEtaPlane} that the area of the surface
probed in the production of the lepton pair with electron beam of
increasing energy $E$ is quite extensive: (solid contour) $E = 11 \,
{\rm GeV}$, (dashed contour) $E = 25 \, {\rm GeV}$, (dotted contour)
$E = 40 \, {\rm GeV}$. It is obvious that the higher the energy of
the lepton beam, the higher $\Delta^2$ are allowed with observed
applicability of the perturbative analysis of the Compton amplitude,
and thus the higher values of $\eta$ are achieved.

Above,  only the case ${\cal Q}^2 > M_{\ell\bar\ell}^2$ was implied, which
probes $\xi > 0$ component of GPDs. For the reversed inequality sign, one
gets information on the region $\xi < 0$ and probes patches of the
two-dimensional surface analogous to the previous case. The positive mass
of the final-state photon allows to directly access only the ``exclusive''
component of GPDs, with $\eta > |\xi|$, where GPDs look like distribution
amplitudes. To access the  ``inclusive'' component characterized by $\eta
< |\xi|$, with GPDs following the behavior of parton distributions, requires
spacelike virtuality for the outgoing photon.  Such a configuration is possible
in two-photon exchange events in elastic electron-nucleon scattering. However,
since the hadronic tensor (\ref{ComptonAmplitude}) enters now via a loop
integral, the single spin asymmetry measurements cannot be used for a
direct extraction of GPDs.

\begin{figure}[t]
\begin{center}
\mbox{
\begin{picture}(0,108)(175,0)
\put(0,0){\insertfig{12}{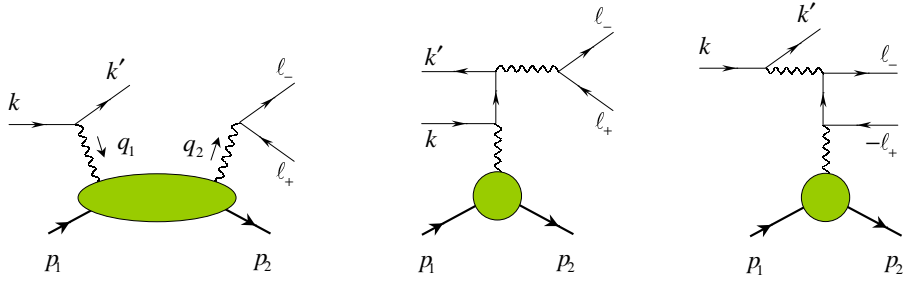}}
\end{picture}
}
\end{center}
\caption{\label{BHandDDVCS} Subprocesses contributing to electroproduction
of muon pairs.}
\end{figure}

\subsubsection{Cross section for lepton pair electroproduction}

\begin{figure}[t]
\begin{center}
\mbox{
\begin{picture}(0,310)(160,0)
\put(-30,-15){\insertfig{12}{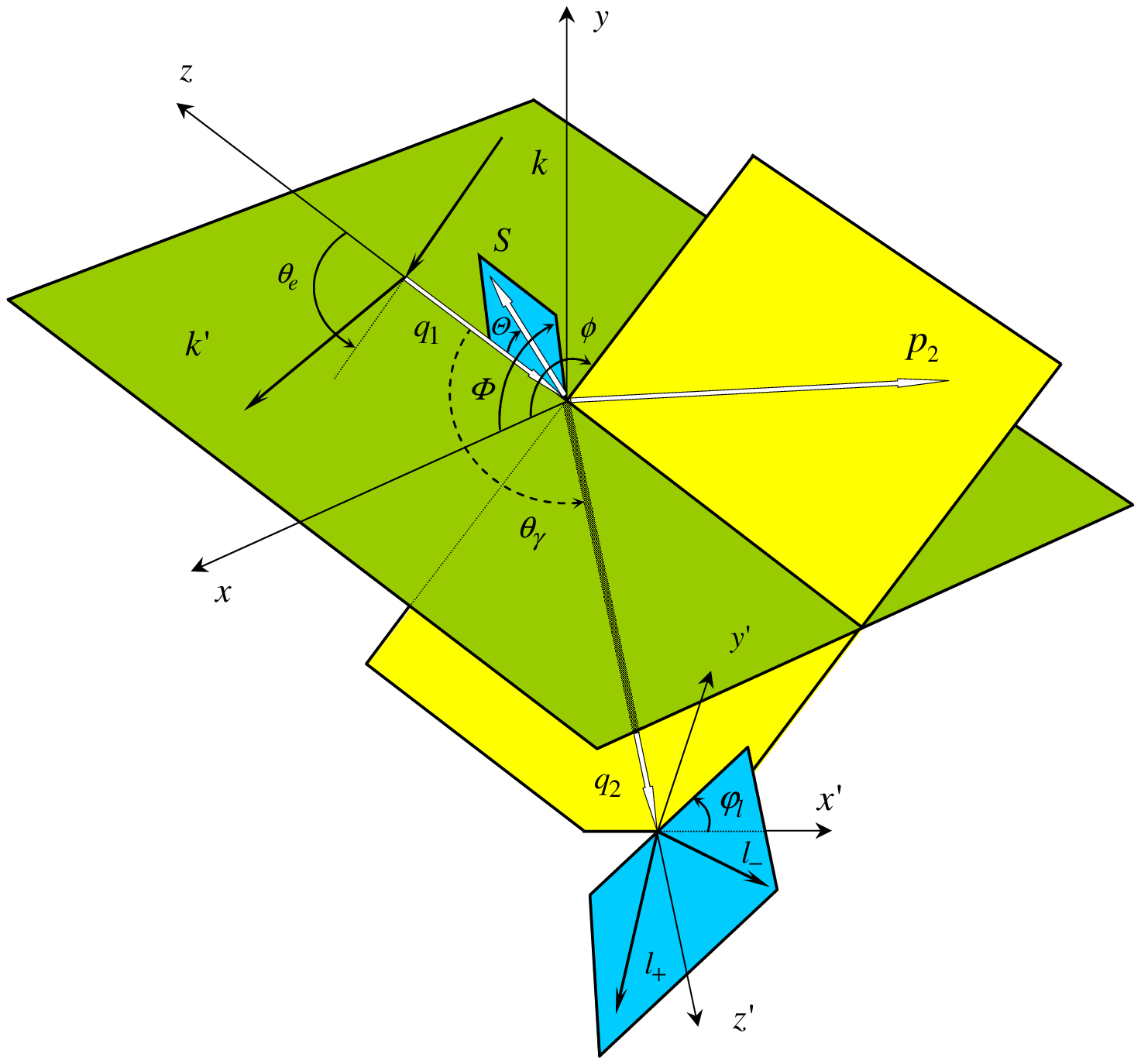}}
\end{picture}
}
\end{center}
\caption{\label{LeptonPairKinetic} The kinematics of the lepton pair production
in elastic electron-nucleon scattering. The coordinate system used in defining
the kinematical variables is determined by the $z$-axis counter-aligned to the
spacelike virtual photon. Another complimentary frame suited for evaluation
of scalar products is the one with $z'$ along the three-momentum of the timelike
photon.}
\end{figure}

The reaction we are going to discuss consists of three interfering processes,
depicted in Fig.\ \ref{BHandDDVCS} (the crossed contributions are implied).
However, only one of them is sensitive to the one-particle correlations in
the nucleon when at least one of the photon virtualities is large compared
to a typical hadronic scale. It arises from the virtual Compton scattering
amplitude, shown on the left hand side in Fig.\ \ref{BHandDDVCS}. The other
two amplitudes represent the Bethe-Heitler background. We will discuss the
simpler case of the production of the lepton pair of a different flavor compared
to that of the beam, i.e., the muon pair production. The consideration of
the electroproduction of electron pairs requires to add exchange contributions
due to the identical nature  of the electrons in the final state, i.e.,
$k' \to \ell_-$.

\begin{figure}[t]
\begin{center}
\mbox{
\begin{picture}(0,195)(80,0)
\put(0,-5){\insertfig{6}{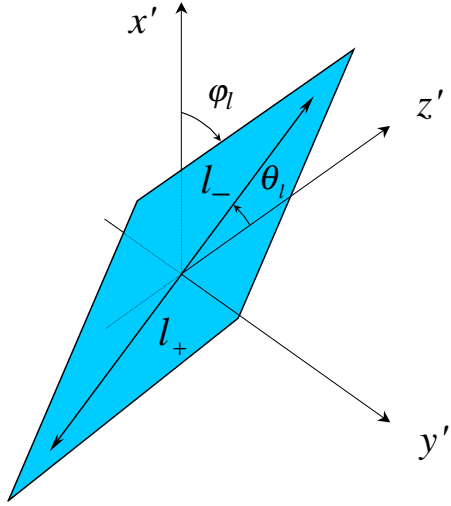}}
\end{picture}
}
\end{center}
\caption{\label{PhotonRestFrame} Azimuthal and polar angles of the lepton
pair in the rest frame of the timelike photon.}
\end{figure}

The generic form of the cross section of the exclusive electroproduction of
lepton pairs off the nucleon, $e (k) N (p_1) \to e (k') N (p_2) \ell (\ell_-)
\bar\ell (\ell_+)$, is
\begin{equation}
d \sigma = \frac{1}{4 p_1 \cdot k} |{\cal T}|^2 \, d {\rm LIPS}_4
\, ,
\end{equation}
where ${\cal T}$ is the sum of the amplitude of the virtual Compton scattering
and two Bethe-Heitler processes, ${\cal T} = {\cal T}_{\scriptscriptstyle\rm VCS}
+ {\cal T}_{{\scriptscriptstyle\rm BH}_1} + {\cal T}_{{\scriptscriptstyle\rm BH}_2}$,
shown in Fig.\ \ref{BHandDDVCS}. The four-particle Lorentz invariant phase space
\begin{equation}
\label{LIPS4}
d {\rm LIPS}_4
=
d M_{\ell \bar\ell}^2 \, d {\rm LIPS}_3 \, d {\mit \Phi}_{\ell \bar\ell}
\, ,
\end{equation}
is factorized, by introducing  integration over the invariant mass of the
lepton pair $M_{\ell \bar\ell}^2$, into Lorentz invariant phase-space factors
for the production of a heavy timelike photon off a nucleon
\begin{equation}
d {\rm LIPS}_3
=
(2 \pi)^4 \delta^{(4)} \left( k + p_1 - k' - p_2 - q_2 \right)
\frac{d^4 p_2}{(2 \pi)^3} \delta_+ \left( p_2^2 - M_N^2 \right)
\frac{d^4 k'}{(2 \pi)^3} \delta_+ \left( {k'}^2 \right)
\frac{d^4 q_2}{(2 \pi)^3} \delta_+ \left( q_2^2 - M_{\ell \bar\ell}^2 \right)
\, , \!
\end{equation}
and its subsequent decay into a lepton pair
\begin{equation}
d {\mit \Phi}_{\ell \bar\ell}
=
\frac{d^4 \ell_-}{(2 \pi)^3}
\delta_+ \left( \ell_-^2 - m_\ell^2 \right)
\delta_+ \left( (q_2 - \ell_-)^2 - m_\ell^2 \right)
\, .
\end{equation}
A simple calculation gives for them
\begin{equation}
\label{LIPS3-Phi}
d {\rm LIPS}_3
=
\frac{d x_{\rm B} d y d (- \Delta^2) d \phi}{16 (2 \pi)^4 \sqrt{1 + \varepsilon^2}}
\, , \qquad
d {\mit \Phi}_{\ell \bar\ell}
= \frac{\beta \, d {\mit \Omega}_\ell}{8 (2 \pi)^3}
\, ,
\end{equation}
respectively. The graphical definition of the angles involved is demonstrated
in Figs.\ \ref{LeptonPairKinetic} and  \ref{PhotonRestFrame}. Here the solid
angle and the velocity of the the final state lepton in the $\ell \bar\ell$
center-of-mass frame are
\begin{equation}
d {\mit\Omega}_\ell = \sin \theta_\ell d \theta_\ell d \varphi_\ell
\, , \qquad
\beta = \sqrt{1 - 4 m_\ell^2/M_{\ell \bar\ell}^2}
\, ,
\end{equation}
respectively. Introduction of the transverse polarization for the nucleon
spin-vector Eq.\ (\ref{Spin}) results in an extra integration variable in
the phase space given by Eqs.\ (\ref{LIPS4}) and (\ref{LIPS3-Phi}),
\begin{equation}
d {\rm LIPS}_4 \to  \frac{d {\mit\Phi}}{2 \pi} \times d {\rm LIPS}_4
\, .
\end{equation}
Extracting the lepton charge from the amplitudes, one gets for the cross
section, expressed in terms of experimentally measurable variables,
\begin{equation}
d \sigma
= \frac{\alpha_{\rm em}^4}{16 (2 \pi)^3}
\frac{x_{\rm B} y \beta}{{\cal Q}^2 \sqrt{1 + \varepsilon^2}}
\left| \frac{{\cal T}}{e^4} \right|^2
d x_{\rm B} d y d (- \Delta^2) d \phi d M_{\ell \bar\ell}^2 d {\mit \Omega}_\ell
\, .
\end{equation}

\subsubsection{Kinematics of the lepton pair electroproduction}
\label{ReferenceFrames}

There is a complementary rest frame to the conventional one discussed in Section
\ref{DVCSkinematics} with the $z$-axis directed opposite to the three-momentum
of the incoming virtual photon. It is very suitable for evaluation of scalar
products of four-momenta at intermediate stages. The new frame is defined
by rotating the $z$-axis in the hadronic plane such that it aligns with the
three-momentum $\bit{q}_2$ of the timelike virtual photon
\begin{equation}
\label{Velocity}
q_2 = (\omega_2, \, \bit{v} \, \omega_2)
\, , \qquad
\omega_2 = \frac{{\cal Q}}{\varepsilon} + \frac{\Delta^2}{2 M_N}
\, , \qquad
v \equiv |\bit{v}| = \sqrt{1 - M_{\ell \bar\ell}^2/\omega_2^2}
\, .
\end{equation}
Operationally, it is achieved by rotating the frame from Section \ref{DVCSkinematics}
in the hadron scattering plane by the scattering angle $\theta_\gamma$ of the
outgoing photon,
\begin{equation}
\cos \theta_\gamma
= - \frac{
\varepsilon
\left(
{\cal Q}^2 - M_{\ell \bar\ell}^2 + \Delta^2
\right)
+
2 {\cal Q} \omega_2
}{
2 {\cal Q} \omega_2 v \sqrt{1 + \varepsilon^2}
}
\, ,
\end{equation}
as shown in Fig.\ \ref{LeptonPairKinetic}. In the new frame with $z'$ axis directed
along $\vec{q}_2$, the particle four-momenta read
\begin{eqnarray}
q_2^\mu
\!\!\!&=&\!\!\!
(\omega_2, \, 0, \, 0, \, \omega_2 v)
\, , \qquad
q_1^\mu
=
(\omega_1, \, q_1^z \sin \theta_\gamma, \, 0, \, - q_1^z \cos \theta_\gamma)
\, , \qquad
p_1^\mu
=
(M_N, 0 , 0 , 0)
\, , \nonumber\\
k^\mu
\!\!\!&=&\!\!\!
E \left(
1
, \,
\sin \theta_e \cos \theta_\gamma \cos \varphi_\gamma
-
\cos \theta_e \sin \theta_\gamma
, \,
- \sin \theta_e \sin \varphi_\gamma
, \,
\sin \theta_e \sin \theta_\gamma \cos \varphi_\gamma
+
\cos \theta_e \cos \theta_\gamma
\right)
\, , \nonumber
\end{eqnarray}
with the azimuthal angle of the timelike photon $\varphi_\gamma = \pi + \phi$.
The outgoing proton and scattered lepton momentum can be found from these equation
using the momentum conservation laws. In these formulas, the sine of $\theta_\gamma$
can be expressed by means of invariants of scattering
\begin{equation}
\sin \theta_\gamma
=
\frac{
\sqrt{4 x_{\rm B} (1 - x_{\rm B}) + \varepsilon^2}
}{
2 {\cal Q} \omega_2 v \sqrt{1 + \varepsilon^2}
}
\sqrt{
- \left(
\Delta^2 - \Delta^2_{\rm min}
\right)
\left(
\Delta^2 - \Delta^2_{\rm max}
\right)
}
\, .
\end{equation}
Here the maximal and minimal momentum transfer in the $t$-channel is
\begin{eqnarray}
\label{Def-Delta-minmaxReal}
\Delta^2_{\rm min, max}
\!\!\!&=&\!\!\!
- \frac{1}{ 4 x_{\rm B} (1 - x_{\rm B}) + \varepsilon^2}
\Bigg\{
2
\left(
(1 - x_{\rm B}) {\cal Q}^2 - x_{\rm B} M_{\ell \bar\ell}^2
\right)
+ \varepsilon^2
\left(
{\cal Q}^2 - M_{\ell \bar\ell}^2
\right)
\nonumber\\
&&
\mp 2 \sqrt{1 + \varepsilon^2}
\sqrt{
\left(
(1 - x_{\rm B}) {\cal Q}^2 - x_{\rm B} M_{\ell \bar\ell}^2
\right)^2
-
\varepsilon^2
{\cal Q}^2 M_{\ell \bar\ell}^2
}
\Bigg\}
\, ,
\end{eqnarray}
with $-$ ($+$) corresponding to $\Delta^2_{\rm min}$ ($\Delta^2_{\rm max}$).
Their approximate form, i.e., neglecting corrections of order $(p \cdot q)^{- 1}$,
in symmetrical variables (\ref{SymmetricVariables}) is
$$
\Delta^2_{\rm max} \simeq - Q^2 \frac{1 - \eta^2}{\xi (1 - \xi)}
$$
and analogous result for $\Delta^2_{\rm min}$ given before in Eq.\
(\ref{MinMomDeltaSquared}). A boost from the timelike photon rest
frame to the $z'$ rest frame along the direction of motion of the
photon with velocity $\bit{v}$ (\ref{Velocity}) yields
\begin{equation}
\label{Def-FouVec-lm}
\ell_-
=
\left(
\frac{1}{2} \omega_2 \left( 1 + v \beta \cos \theta_\ell \right)
, \,
\frac{1}{2} M_{\ell \bar\ell} \, \beta \sin \theta_\ell \cos \varphi_\ell
, \,
\frac{1}{2} M_{\ell \bar\ell} \, \beta \sin \theta_\ell \sin \varphi_\ell
, \,
\frac{1}{2} \omega_2 \left( v + \beta \cos \theta_\ell \right)
\right) \, ,
\end{equation}
where $\theta_\ell$ and $\varphi_\ell$ are the solid angles of $\ell_-$ in
the $\ell \bar\ell$ center-of-mass frame. The vector $\ell_+$ is obtained
by the reflection, $\varphi_\ell \to \varphi_\ell + \pi $ and $\theta_\ell
\to \pi-\theta_\ell$, from $\ell_-$, or equivalently by the substitution
$\beta \to - \beta$.

For the bulk of our discussion it is enough to evaluate approximate forms
of scalar products which enter the cross section. Namely, keeping only the
leading and sub-leading terms in the $1/(p \cdot q)$-expansion, one arrives
at the following expressions
\begin{eqnarray}
\label{DefkdotDelta}
k \cdot \Delta
\!\!\!&\approx&\!\!\!
- \frac{Q^2}{y} \frac{\eta}{\xi}
\left( 1 - 2 K \cos \varphi_\gamma \right)
\, , \\
\ell_- \cdot \Delta
\!\!\!&\approx&\!\!\!
- \frac{Q^2}{\widetilde y} \frac{\eta}{\xi}
\left(
1 + 2 \widetilde K \cos \varphi_\ell
\right)
\, , \\
\label{Appldotk}
\ell_- \cdot k
\!\!\!&\approx&\!\!\!
\frac{Q^2}{y \widetilde y} \frac{1}{\xi}
\Bigg\{
\frac{1}{2} (\xi - \eta) ( 1 - \widetilde y )
+
\frac{1}{2} (\xi + \eta) (1 - y)
\\
&&\qquad
+
\sigma
\sqrt{(1 - y) (1 - \widetilde y) (\xi^2 - \eta^2)}
\cos (\varphi_\gamma - \varphi_\ell)
-
2 \eta K \cos \varphi_\gamma
-
2 \eta \widetilde K \cos \varphi_\ell
\Bigg\} \, . \nonumber
\end{eqnarray}
with
\begin{equation}
\label{Def-K}
\left\{
\begin{array}{c}
K
\\
\widetilde K
\end{array}
\right\}
\approx \frac{1}{2 \eta}
\sqrt{- \xi \frac{\Delta^2}{Q^2}}
\sqrt{ 1 - \frac{\Delta^2_{\rm min}}{\Delta^2} }
\sqrt{\frac{1 - \eta}{1 + \eta}}
\times
\left\{
\begin{array}{c}
\sqrt{(1 - y) (\xi + \eta)}
\\
\sqrt{(1 - \widetilde y) (\xi - \eta)}
\end{array}
\right\}
\, .
\end{equation}
Note,  that the original square root
$$\sqrt{(1 - y) (1 - \widetilde y) (1 - \eta^2/\xi^2) Q^4}
$$
in front of $\cos(\varphi_\gamma - \varphi_\ell)$  in Eq.\ (\ref{Appldotk})
was replaced by
$$
\sigma Q^2/\xi \sqrt{(1 - y) (1 - \widetilde y) (\xi^2 - \eta^2)}
\, ,
$$
with
$$
\sigma \equiv {\rm sgn} (Q^2/\xi)
\, .
$$
Since $Q^2/\xi$ is positive for the kinematics we are considering, $\sigma$
can be set to be equal to $+1$. However, discussing below relations between
scattering amplitudes, we will make use of  symmetries that induce the interchange
$\xi \to - \xi$ and $Q^2 \to Q^2$ in the underlying expressions. Under these
substitutions, $\sigma$ changes the sign and takes the negative value $-1$.
Equation (\ref{DefkdotDelta}) reduces to the known result  \cite{BelMulKir01}
in the limit of the real final-state photon $M_{\ell \bar\ell} = 0$, discussed
in the previous section. To make the results look symmetric, we introduce the
variable $\widetilde{y}$
\begin{equation}
\frac{1}{\widetilde y}
\equiv
\frac{p_1\cdot \ell_-}{p_1\cdot q_2}
=
\frac{1 + v \beta\cos \theta_\ell}{2}
\simeq
\frac{1 + \cos \theta_\ell}{2}
\, ,
\end{equation}
which varies in the interval $1 \leq \widetilde y \leq \infty$ and enters
in a fashion analogous to the fraction of the lepton energy loss $y$.

\subsubsection{Generating function for angular dependence}
\label{Sec-GenFun}

In this section, we will discuss the electroproduction cross section involving
the leading-twist Compton scattering amplitude. First, we present a generic
form of the squared amplitude, where the hadronic part left intact, while the
leptonic part is fully worked out. The square of the total amplitude, ${\cal T}
= {\cal T}_{\scriptscriptstyle\rm VCS} + {\cal T}_{{\scriptscriptstyle\rm BH}_1}
+ {\cal T}_{{\scriptscriptstyle\rm BH}_2}$, involves three essentially different
contributions
\begin{eqnarray}
{\cal T}^2 \!\!\!&=&\!\!\! | {\cal T}_{\scriptscriptstyle\rm VCS} |^2
+
{\cal I}
+
| {\cal T}_{{\scriptscriptstyle\rm BH}_1} + {\cal T}_{{\scriptscriptstyle\rm BH}_2} |^2
\, ,
\end{eqnarray}
the square of the virtual Compton scattering amplitude, --- bilinear in
Compton form factors, --- the square of the Bethe-Heitler processes, ---
independent on GPDs and expressed solely in terms of elastic form factors,
and the interference term
\begin{equation}
{\cal I}
=
{\cal T}_{\scriptscriptstyle\rm VCS} {\cal T}_{{\scriptscriptstyle\rm BH}_1}^\dagger
+
{\cal T}_{\scriptscriptstyle\rm VCS} {\cal T}_{{\scriptscriptstyle\rm BH}_2}^\dagger
+
{\cal T}_{\scriptscriptstyle\rm VCS}^\dagger {\cal T}_{{\scriptscriptstyle\rm BH}_1}
+
{\cal T}_{\scriptscriptstyle\rm VCS}^\dagger {\cal T}_{{\scriptscriptstyle\rm BH}_2}
\, ,
\end{equation}
which is linear in Compton form factors. For the electron beam, the separate
contributions to the total amplitude are given by
\begin{eqnarray}
\label{Def-AmpVCS}
{\cal T}_{\scriptscriptstyle\rm VCS}
\!\!\!&=&\!\!\!
\frac{e^4}{q_1^2 q_2^2} \
\bar u (\ell_-) \gamma_\mu u (- \ell_+) \
\bar u (k') \gamma_\nu u (k) \
T_{\mu\nu}
\, , \\
\label{Def-AmpBH-1}
{\cal T}_{{\scriptscriptstyle\rm BH}_1}
\!\!\!&=&\!\!\!
\frac{e^4}{q_2^2 \Delta^2} \
\bar u (\ell_-) \gamma_\mu u (- \ell_+) \
\bar u (k')
\left(
\gamma_\mu \frac{1}{ {\not\!k} - {\not\!\!\Delta} } \gamma_\nu
+
\gamma_\nu \frac{1}{ {\not\!k'} + {\not\!\!\Delta} } \gamma_\mu
\right)
u (k) \ J_\nu
\, , \\
\label{Def-AmpBH-2}
{\cal T}_{{\scriptscriptstyle\rm BH}_2}
\!\!\!&=&\!\!\!
\frac{e^4}{q_1^2 \Delta^2} \
\bar u (k') \gamma_\mu u (k) \
\bar u (\ell_-)
\left(
\gamma_\mu \frac{1}{ -{\not\!\ell_+} - {\not\!\!\Delta} } \gamma_\nu
+
\gamma_\nu \frac{1}{ {\not\!\ell_-} + {\not\!\!\Delta} } \gamma_\mu
\right)
u (- \ell_+) \
J_\nu
\, .
\end{eqnarray}
They correspond to the diagrams (a), (b), and (c) in Fig.~\ref{BHandDDVCS},
respectively, including the crossed contributions in the latter two cases. The
VCS tensor $T_{\mu\nu}$ and the electromagnetic current $J_\mu$ were introduced
previously in Eqs.\ (\ref{VCSamplitude}) and (\ref{EMcurrentMatrixEl}),
respectively. The two amplitudes ${\cal T}_{\scriptscriptstyle\rm VCS}$ and
${\cal T}_{{\scriptscriptstyle\rm BH}_2}$ change the overall sign when one
switches from the electron to the positron beam, and so do the interference
terms involving them, while the term ${\cal T}_{{\scriptscriptstyle\rm BH}_1}$
does not change sign. Obviously, both BH amplitudes are related by the
interchange of the momenta $k^\prime \leftrightarrow \ell_-$ and $k
\leftrightarrow -\ell_+$. Moreover,  the VCS and the first BH amplitude are
even under the interchange of the produced leptons in the pair, while the
second BH amplitude is odd. This symmetry property in the timelike DVCS plays
an analogous role as the charge asymmetry in the spacelike case \cite{BerDiePir01}.

The evaluation of separate terms gives expressions which are represented
as a Fourier sum of a few harmonics in the difference of the azimuthal
angles $\varphi_l - \varphi_\gamma = \varphi_l - \phi - \pi$. Below, we
list particular contributions to the total amplitude squared.

\noindent $\bullet$ The square of the VCS amplitude
\begin{equation}
\label{Def-VCS2}
| {\cal T}_{\scriptscriptstyle\rm VCS} |^2
=
\frac{2 \xi^2 e^8}{Q^4 y^2 \widetilde y^2 (\eta^2 - \xi^2)}
\sum_{n = 0}^{2}
\left(
a^{\scriptscriptstyle\rm VCS}_{n} + \lambda b^{\scriptscriptstyle\rm VCS}_{n}
\right)
\cos \Big(n ( \varphi_l - \phi ) \Big) \, ,
\end{equation}
has the following expansion coefficients
\begin{eqnarray}
\label{a0VCS}
a^{\scriptscriptstyle\rm VCS}_0
\!\!\!&=&\!\!\!
\frac12 (2 - 2y + y^2) (2 - 2 \widetilde y + \widetilde y^2)
\left(
{\cal F}_1 {\cal F}_1^\dagger
+
\widetilde{\cal F}_1 \widetilde{\cal F}_1^\dagger
\right)
+
4(1 - y)(1 - \widetilde{y}) \frac{\xi^2 - \eta^2}{\xi^2}
{\cal F}_L {\cal F}_L^\dagger
\, , \qquad\\
a^{\scriptscriptstyle\rm VCS}_1
\!\!\!&=&\!\!\!
- \frac{\sigma}{\xi} \sqrt{(1 - y)(1 - \widetilde{y})(\xi^2 - \eta^2)}
(2 - y) (2 - \widetilde{y})
\left(
{\cal F}_1 {\cal F}_L^\dagger
+
{\cal F}_L {\cal F}_1^\dagger
\right)
\, , \\
a^{\scriptscriptstyle\rm VCS}_2
\!\!\!&=&\!\!\!
2 (1 - y) (1 - \widetilde y)
\left(
{\cal F}_1 {\cal F}_1^\dagger
-
\widetilde{\cal F}_1 \widetilde{\cal F}_1^\dagger
\right)
\, , \\
b^{\scriptscriptstyle\rm VCS}_0
\!\!\!&=&\!\!\!
\frac12 y (2 - y) (2 - 2 \widetilde y + \widetilde y^2)
\left(
{\cal F}_1 \widetilde{\cal F}_1^\dagger
+
\widetilde{\cal F}_1 {\cal F}_1^\dagger
\right)
\, , \\
b^{\scriptscriptstyle\rm VCS}_1
\!\!\!&=&\!\!\!
- \frac{\sigma}{\xi} \sqrt{(1 - y)(1 - \widetilde{y})(\xi^2 - \eta^2) }
y (2 - \widetilde{y})
\left(
{\cal F}_L \widetilde{\cal F}_1^\dagger
+
\widetilde{\cal F}_1 {\cal F}_L^\dagger
\right)
\, , \\
b^{\scriptscriptstyle\rm VCS}_2
\!\!\!&=&\!\!\!
0 \, .
\label{b2VCS}
\end{eqnarray}
Due to the virtuality of both the incoming and outgoing photons, the Lorentz
structure accompanying ${\cal F}_L$, in contrast to the DVCS case, does
contribute to the cross section and generates, e.g., the coefficient
$a_1^{\scriptscriptstyle\rm VCS}$.

The interference of the VCS and  BH amplitude will involve lepton propagators
from the latter which will bring conveniently rescaled factors in the denominator
\begin{eqnarray}
\label{Def-P1P2}
&&(k^\prime + \Delta)^2 \equiv 2 \eta \, p \cdot q \, {\cal P}_1 (k)
\, , \qquad\quad\ \,
(k - \Delta)^2 \equiv 2 \eta \, p \cdot q \, {\cal P}_2 (k)
\, , \\
\label{Def-P3P4}
&&(\ell_+ + \Delta )^2 \equiv - 2 \eta \, p \cdot q \, {\cal P}_3 (\ell_-)
\, , \qquad
(\ell_- + \Delta)^2 \equiv - 2 \eta \, p \cdot q \, {\cal P}_4 (\ell_-)
\, .
\end{eqnarray}
The expressions  are rather lengthy  and are obtained by substituting Eqs.\
(\ref{DefkdotDelta}) and (\ref{Def-K}) into the left-hand side of the above
definitions. We also introduce the following shorthand notations for the
structures involving the nucleon matrix element of the quark electromagnetic
current to make the formulas look as concise as possible,
\begin{eqnarray}
\label{Def-S1}
{\cal S}_1
\!\!\!&\equiv&\!\!\!
- \eta \left( k - \frac{1}{y} q_1 \right) \cdot J^\dagger
-
\frac{1}{p \cdot q} \left( k - \frac{1}{y} q_1 \right) \cdot \Delta
\ q_1 \cdot J^\dagger \, , \\
\label{Def-S2}
{\cal S}_2
\!\!\!&\equiv&\!\!\!
- \eta \left( \ell_- - \frac{1}{\widetilde y} q_2 \right) \cdot J^\dagger
-
\frac{1}{p \cdot q} \left( \ell_- - \frac{1}{\widetilde y} q_2 \right) \cdot \Delta
\ q_2 \cdot J^\dagger
\, , \\
\label{Def-R12}
\widetilde{\cal S}_1
\!\!\!&\equiv&\!\!\!
\frac{i}{p \cdot q}
\varepsilon_{\mu\nu\rho\sigma} q_\mu k_\nu \Delta_\rho J^\dagger_\sigma
\, , \quad
\widetilde{\cal S}_2
\equiv
\frac{i}{p \cdot q}
\varepsilon_{\mu\nu\rho\sigma} q_\mu \ell_{- \nu} \Delta_\rho J^\dagger_\sigma
\, .
\end{eqnarray}
These functions depend on the azimuthal angles $\phi$ and $\varphi_l$.
Combining these results, one finds, in full analogy with the previous
analysis of $|{\cal T}_{\scriptscriptstyle\rm VCS}|^2$, the interference
contributions from the VCS and BH amplitudes.

\noindent $\bullet$ The interference ${\cal T}_{\scriptscriptstyle\rm VCS}
{\cal T}_{{\scriptscriptstyle\rm BH}_1}^\dagger$
\begin{equation}
\label{Def-BHINT1}
{\cal T}_{\scriptscriptstyle\rm VCS} {\cal T}_{{\scriptscriptstyle\rm BH}_1}^\dagger
=
\frac{2 \xi^2 e^8}
{y^2 \widetilde y^2 \eta^2 (\eta^2 - \xi^2) Q^4 \Delta^2
{\cal P}_1 (k) {\cal P}_2 (k)
}
\sum_{n = 0}^{2}
\left( a^1_n + \lambda b^1_n \right)
\cos \Big( n ( \varphi_l - \phi ) \Big) \, ,
\end{equation}
is determined by
\begin{eqnarray}
a^1_0
\!\!\!&=&\!\!\!
- 4 (1 - y) (1 - \widetilde y)
\left(
\eta {\cal S}_1 {\cal F}_1
+
\xi \widetilde{\cal S}_1 \widetilde{\cal F}_1
-
2 \frac{\xi^2 - \eta^2}{\xi} {\cal S}_1 {\cal F}_L
\right)
\nonumber\\
&&\hspace{0cm}
-
(2 - 2 y + y^2) (2 - 2 \widetilde y + \widetilde y^2)
\Big(
\xi {\cal S}_1 {\cal F}_1
-
\eta \widetilde{\cal S}_1 \widetilde{\cal F}_1
\Big)
\\
&&
-
2 \frac{\widetilde y}{y} (1 - y)(2 - y)(2 - \widetilde y) (\xi + \eta)
\left(
\widetilde{\cal S}_2 \widetilde{\cal F}_1
-
\frac{\eta}{\xi} {\cal S}_2 {\cal F}_L
\right)
\, ,
\nonumber\\
a^1_1
\!\!\!&=&\!\!\!
2 \sigma \sqrt{(1 - y)(1 - \widetilde y)(\xi^2 - \eta^2)}
\\
&&\times
\Bigg\{
(2 - y) (2 - \widetilde y)
\left(
{\cal S}_1 {\cal F}_1
+
\widetilde{\cal S}_1 \widetilde{\cal F}_1
-
\frac{\xi + \eta}{\xi}{\cal S}_1 {\cal F}_L
\right)
+
4 \frac{\widetilde y}{y} \frac{1 - y}{\xi - \eta}
\Big(
\eta {\cal S}_2 {\cal F}_1
+
\xi \widetilde{\cal S}_2 \widetilde{\cal F}_1
\Big)
\Bigg\}
\, , \nonumber\\
a^1_2
\!\!\!&=&\!\!\!
- 4 (\xi + \eta) (1 - y) (1 - \widetilde y)
\Big( {\cal S}_1 {\cal F}_1 + \widetilde{\cal S}_1 \widetilde{\cal F}_1 \Big)
\, , \\
b^1_0
\!\!\!&=&\!\!\!
- y (2 - y) (2 - 2 \widetilde y + \widetilde y^2)
\Big( \xi {\cal S}_1 \widetilde{\cal F}_1 - \eta \widetilde{\cal S}_1 {\cal F}_1 \Big)
- 2 (\xi + \eta) (1 - y) \widetilde y (2 - \widetilde y)
\widetilde{\cal S}_2 \left({\cal F}_1+{\cal F}_L \right)
\, , \qquad\\
b^1_1
\!\!\!&=&\!\!\!
2 \sigma \sqrt{(1 - y) (1 - \widetilde y) (\xi^2 - \eta^2)}
y (2 - \widetilde y)
\left(
{\cal S}_1 \widetilde{\cal F}_1
+
\widetilde{\cal S}_1 {\cal F}_1
+
\frac{\xi + \eta}{\xi} \widetilde{\cal S}_1 {\cal F}_L
\right)
\, , \\
\label{Def-BHINT1-b2}
b^1_2
\!\!\!&=&\!\!\! 0
\, .
\end{eqnarray}

\noindent $\bullet$ The interference ${\cal T}_{\scriptscriptstyle\rm VCS}
{\cal T}_{{\scriptscriptstyle\rm BH}_2}^\dagger$
\begin{equation}
\label{Def-BHINT2}
{\cal T}_{\scriptscriptstyle\rm VCS} {\cal T}_{{\scriptscriptstyle\rm BH}_2}^\dagger
=
\frac{2 \xi^2 e^8}
{y^2 \widetilde y^2 \eta^2 (\eta^2 - \xi^2) Q^4 \Delta^2
{\cal P}_3 (\ell_-) {\cal P}_4 (\ell_-)
}
\sum_{n = 0}^{2}
\left( a^2_n + \lambda b^2_n \right)
\cos \Big( n ( \varphi_l - \phi ) \Big) \, ,
\end{equation}
is encoded into the following Fourier coefficients
\begin{eqnarray}
\label{Def-BHINT2-a0}
a^2_0
\!\!\!&=&\!\!\!
- 4 (1 - y) (1 - \widetilde y)
\left(
\eta {\cal S}_2 {\cal F}_1
+
\xi \widetilde{\cal S}_2 \widetilde{\cal F}_1
+
2 \frac{\xi^2 - \eta^2}{\xi} {\cal S}_2 {\cal F}_L
\right)
\nonumber\\
&& +
(2 - 2 y + y^2) (2 - 2 \widetilde y + \widetilde y^2)
\Big(
\xi {\cal S}_2 {\cal F}_1
-
\eta \widetilde{\cal S}_2 \widetilde{\cal F}_1
\Big)
\\
&&
- 2 \frac{y}{\widetilde y} (1 - \widetilde y) (2 - y)(2 - \widetilde y) (\xi - \eta)
\left(
\widetilde{\cal S}_1 \widetilde{\cal F}_1
-
\frac{\eta}{\xi} {\cal S}_1 {\cal F}_L
\right)
\, ,
\nonumber \\
a^2_1
\!\!\!&=&\!\!\!
-2 \sigma \sqrt{(1 - y)(1 - \widetilde y)(\xi^2 - \eta^2)}
\\
&&\times
\Bigg\{
(2 - y) (2 - \widetilde y)
\left(
{\cal S}_2 {\cal F}_1
-
\widetilde{\cal S}_2 \widetilde{\cal F}_1
-
\frac{\xi - \eta}{\xi} {\cal S}_{2} {\cal F}_L
\right)
- 4 \frac{y}{\widetilde y} \frac{1 - \widetilde y}{\xi + \eta}
\Big(
\eta {\cal S}_1 {\cal F}_1
+
\xi \widetilde{\cal S}_1 \widetilde{\cal F}_1
\Big)
\Bigg\}
\, , \nonumber\\
a^2_2
\!\!\!&=&\!\!\!
4 (\xi - \eta) (1 - y) (1 - \widetilde y)
\Big(
{\cal S}_2 {\cal F}_1
-
\widetilde{\cal S}_2 \widetilde{\cal F}_1
\Big)
\, , \\
b^2_0
\!\!\!&=&\!\!\!
y (2 - y) (2 - 2 \widetilde y + \widetilde y^2)
\Big(
\xi {\cal S}_2 \widetilde{\cal F}_1 - \eta \widetilde{\cal S}_2 {\cal F}_1
\Big)
-
2 (\xi - \eta)
\frac{y^2}{\widetilde y} (2 - \widetilde y) (1 - \widetilde y)
\widetilde{\cal S}_1 \left( {\cal F}_1 + {\cal F}_L \right)
\, , \\
b^2_1
\!\!\!&=&\!\!\!
- 2 \sigma \sqrt{(1 - y) (1 - \widetilde y) (\xi^2 - \eta^2)}
y (2 - \widetilde y)
\Big(
{\cal S}_2 \widetilde{\cal F}_1
-
\widetilde{\cal S}_2 {\cal F}_1
-
\frac{\xi - \eta}{\xi} \widetilde{\cal S}_2 {\cal F}_L
\Big)
\, , \\
\label{Def-BHINT2-b2}
b^2_2
\!\!\!&=&\!\!\! 0
\, .
\end{eqnarray}

The unpolarized parts of the two interference terms must obey a symmetry
relation, since the BH amplitudes (\ref{Def-AmpBH-1}) and (\ref{Def-AmpBH-2})
are related to each other by the substitutions of the lepton's momenta $k
\leftrightarrow - \ell_+$ and $k^\prime \leftrightarrow \ell_-$. Obviously,
$q_1 \leftrightarrow - q_2$ under it. The Bose symmetry ensures on the other
hand the invariance of the Compton amplitude (\ref{Def-AmpVCS}) with respect
to this interchange. As we mentioned above, all of the amplitudes possess
definite symmetry properties under the permutation of leptons in the pair
$\ell_- \leftrightarrow \ell_+$ and, thus, one can take the advantages of
combining both transformations together, i.e., $k \leftrightarrow -\ell_-$
and $k^\prime \leftrightarrow \ell_+$. From the definitions of the four-vectors
(\ref{Def-FouVec-lm}) one can derive, after some algebra, a complete set of
substitution rules reflecting them
\begin{eqnarray}
\label{SymRel}
Q^2 \rightarrow Q^2
\, , \quad
\xi \rightarrow - \xi
\, , \quad
\sigma \rightarrow - \sigma
\, , \quad
\Delta \rightarrow \Delta
\, , \quad
\eta \rightarrow \eta
\, , \quad
y \leftrightarrow \widetilde y
\, , \quad
\varphi_{\ell} \leftrightarrow \phi
\, .
\end{eqnarray}
The prefactors in the interference terms (\ref{Def-BHINT1}) and (\ref{Def-BHINT2})
are even under the transformation (\ref{SymRel}). Moreover, from the definitions
(\ref{Def-S1}), (\ref{Def-S2}), and (\ref{Def-R12}), one can conclude also that
${\cal S}_1 \leftrightarrow - {\cal S}_2; \widetilde{\cal S}_1 \leftrightarrow -
\widetilde{\cal S}_2$ with (\ref{SymRel}). Taking all  the results together, one
obtains that the Fourier coefficients satisfy the equalities
\begin{equation}
a_n^2
=
-
\left.
a_n^1
\right|{\textstyle{
{\cal S}_1 \leftrightarrow -{\cal S}_2;
\widetilde{\cal S}_1 \leftrightarrow - \widetilde{\cal S}_2
\atop
y \leftrightarrow \widetilde y;
\xi \leftrightarrow - \xi \hfill
}}
\, ,
\end{equation}
where ${\cal F}_1$ and $\widetilde{\cal F}_1$ are even and odd functions in
$\xi$, respectively. Finally, one observes that the product of the BH propagators
(\ref{Def-P1P2}) and (\ref{Def-P3P4}) obeys the symmetry relation
\begin{eqnarray}
{\cal P}_3 {\cal P}_4
(Q^2, \Delta^2, \xi, \eta, y, \widetilde y, \varphi_{\ell})
=
{\cal P}_1 {\cal P}_2
(Q^2, \Delta^2, - \xi, \eta, \widetilde y, y, \phi = \varphi_{\ell})
\, .
\end{eqnarray}

\subsubsection{Angular dependence of the cross section}
\label{SubSec-AngDep}

Having the generic expression for the total amplitude squared, one can
proceed with the hadronic part and present the result as a Fourier expansion
in terms of the azimuthal angle, $\phi$ of the recoiled nucleon and
$\varphi_\ell$ of a lepton in the lepton pair. The starting point is to
write down the general angular decomposition of the squared amplitudes,
which results from the Lorentz structure of the contracted leptonic and
hadronic tensors. The harmonics appearing here, can be classified with
respect to the underlying twist expansion of the hadronic tensor
(\ref{ComptonAmplitude}), and the Fourier coefficients are in one-to-one
correspondence with helicity amplitudes defined in the target rest frame.
Extracting certain kinematical factors in order to match the normalization
adopted for the leptoproduction cross section of the real photon in Ref.\
\cite{BelMulKir01}, the square of the VCS amplitude and its interference
with the BH amplitudes as well as the squared BH amplitudes has the following
expansion in these azimuthal angles
\begin{eqnarray}
\label{Dec-VCS2}
&&| {\cal T}_{\scriptscriptstyle\rm VCS} |^2
=
\frac{2 \xi^2 e^8}{Q^4 y^2 \widetilde y^2 (\eta^2 - \xi^2)}
\sum_{n = 0}^2
\left\{
{\rm c}^{\scriptscriptstyle\rm VCS}_n (\varphi_\ell) \cos (n \phi)
+
{\rm s}^{\scriptscriptstyle\rm VCS}_n (\varphi_\ell) \sin (n \phi)
\right\}
\, , \\
\label{Dec-BHVCS}
&&{\cal I}
=
\frac{2 \xi (1 + \eta) e^8}{y^3 \widetilde y^3 (\eta^2 - \xi^2) Q^2 \Delta^2}
\sum_{n = 0}^3
\Bigg\{
\pm
\frac{\widetilde y}{{\cal P}_1 {\cal P}_2 (\phi)}
\left(
{\rm c}^1_n (\varphi_\ell) \cos (n \phi)
+
{\rm s}^1_n (\varphi_\ell) \sin (n \phi)
\right)
\\
&&\qquad\qquad\qquad\qquad\qquad\qquad\ \ \
+
\frac{y}{{\cal P}_3 {\cal P}_4(\varphi_\ell)}
\left(
{\rm c}^2_n (\phi) \cos (n \varphi_\ell)
+
{\rm s}^2_n (\phi) \sin (n \varphi_\ell)
\right)
\Bigg\}
\, ,
\nonumber\\
\label{Dec-BHsq}
&&| {\cal T}_{\scriptscriptstyle\rm BH} |^2
=
\frac{\xi (1 + \eta )^2}{y^4 {\widetilde y}^4 \Delta^2 Q^2 \eta (\eta^2 - \xi^2)}
\left\{ \sum_{n = 0}^4
\Bigg\{ \frac{{\widetilde y}^2}{{\cal P}_1^2 {\cal P}_2^2 (\phi)}
\left(
{\rm c}^{11}_n (\varphi_\ell) \cos (n \phi)
+
{\rm s}^{11}_n (\varphi_\ell) \sin (n \phi)
\right)\right.
\\
&&\qquad\qquad\qquad\qquad\qquad\qquad\qquad\quad\,\,\,
+ \frac{y^2}{ {\cal P}_3^2 {\cal P}_4^2 (\varphi_\ell)}
\left(
{\rm c}^{22}_n (\phi) \cos (n \varphi_\ell)
+
{\rm s}^{\rm 22}_n (\phi ) \sin (n \varphi_\ell)
\right)\Bigg\}
\nonumber\\
&&\qquad\qquad\qquad\qquad\qquad\qquad\quad\,\,\,
\left.
\left.
\pm \sum_{n = 0}^3
\frac{y \widetilde y}{ {\cal P}_1 {\cal P}_2 {\cal P}_3 {\cal P}_4}
\left(
{\rm c}^{\rm 12}_n (\varphi_\ell) \cos (n \phi)
+
{\rm s}^{\rm 12}_n (\varphi_\ell) \sin (n \phi)
\right)
\right\}
\right\}
\, .
\nonumber
\end{eqnarray}
Here the $+$ ($-$) sign stands for the electron (positron) beam, and the
relation between the azimuthal angles $\varphi_\gamma = \pi + \phi$ has
been used in the expansion.  The Fourier coefficients for the squared
VCS term ($i = {\scriptscriptstyle\rm VCS}$), the interference term with
the first BH amplitude (\ref{Def-AmpBH-1}) ($i=1$), and the squared BH
amplitude (\ref{Def-AmpBH-1}) ($i = 11$) are expanded up to the second
order harmonics in the azimuthal angle $\varphi_\ell$ of the lepton pair
and read
\begin{eqnarray}
\label{Dec-FC-gen}
{\rm c}^i_n (\varphi_\ell)
\!\!\!&=&\!\!\!
\sum_{m = 0}^2
\left\{
{\rm cc}^i_{n m} \cos (m \varphi_\ell)
+
{\rm cs}^i_{n m} \sin (m \varphi_\ell)
\right\}
\, ,
\nonumber\\
{\rm s}^i_n (\varphi_\ell)
\!\!\!&=&\!\!\!
\sum_{m = 0}^2
\left\{
{\rm sc}^i_{n m} \cos (m \varphi_\ell)
+
{\rm ss}^i_{n m} \sin (m \varphi_\ell)
\right\}
\, .
\end{eqnarray}
The Fourier coefficients
${\rm ab}_{nm}^i = ({\rm cc}_{nm}^i, {\rm cs}_{nm}^i, {\rm sc}_{nm}^i,
{\rm ss}_{nm}^i)$ depend on the nucleon polarization vector (\ref{Spin})
and can be decomposed further as
\begin{equation}
\label{Def-DecFC}
{\rm ab}_{nm}^i
=
{\rm ab}_{nm, {\rm unp}}^i
+
\cos {\mit\Theta} \,
{\rm ab}_{nm, {\scriptscriptstyle\rm LP}}^i
+
\sin {\mit\Theta} \,
{\rm ab}_{nm, {\scriptscriptstyle\rm TP}}^i ({\mit\Phi})
\, .
\end{equation}

A similar expansion in terms of the azimuthal angle $\phi$ is performed for
the interference of the second BH amplitude (\ref{Def-AmpBH-2}) with the
VCS amplitude  ($i = 2$) and for the  square of the second BH amplitude
($i = 22$).  The interference of both BH amplitudes ($i = 12$) is  analogous
to Eq.\ (\ref{Dec-FC-gen}), however, it contains now   third order harmonics
in the azimuthal angle $\varphi_\ell$. The Fourier coefficients linearly depend
on the polarization vector of the nucleon, see Eq.\ (\ref{Def-DecFC}). At the
edge of the phase space,  the overall coefficient in the BH amplitude gets
corrected according to
\begin{equation}
\frac{
\left( 1 + \eta \right)^2
}{
\eta^2 - \xi^2
}
\to
\frac{
\left(1 + \eta + \frac{\xi \Delta^2}{2 Q^2}\right)^2
}{
\eta^2
-
\left( 1 - \frac{\Delta^2}{4 Q^2} \right)^2 \xi^2
}
\, .
\end{equation}
We emphasize that $1/(\eta^2-\xi^2)$ expressions in the squared VCS (\ref{Dec-VCS2})
and the interference (\ref{Dec-BHVCS}) terms are corrected in an analogous manner to
ensure their correct behavior in the limits ${\cal Q}^2 \to 0$ and $M_{\ell\bar\ell}^2
\to 0$.

{}Finally, we remark that all BH propagators, defined in Eqs.\
(\ref{Def-P1P2}) and (\ref{Def-P3P4}), are even functions of the
azimuthal angle $\varphi$:
\begin{eqnarray}
{\cal P}_i (\varphi)
=
{\cal P}_i (2 \pi - \varphi)
\quad \mbox{for} \quad i = \{1, \dots, 4\}
\, .
\end{eqnarray}
Thus, even and odd harmonics can be clearly separated from each other. It is
also worth  mentioning  that the product ${\cal P}_3 {\cal P}_4$ as a function
of the lepton-pair solid angles $\varphi_{\ell}$ and $\theta_{\ell}$ satisfies
the symmetry relation
\begin{eqnarray}
{\cal P}_3 {\cal P}_4 (\theta_{\ell},\varphi_{\ell})
=
{\cal P}_3 {\cal P}_4 (\pi-\theta_{\ell},\varphi_{\ell}+\pi)
\, .
\end{eqnarray}
As a  consequence of this symmetry, any (anti)symmetric moment in $\theta_{\ell}$,
after integration over $d\theta_\ell$ in a symmetric interval around the point
$\theta_{\ell} = \pi/2$,  has  the following characteristic $\cos$-Fourier expansion
(for any integer $r$)
\begin{eqnarray}
\int_{\pi/2 - \vartheta}^{\pi/2 + \vartheta} d \cos\theta_{\ell} \;
\frac{\tau(\theta_{\ell}) }{[ {\cal P}_3 {\cal P}_4]^r}
=
\sum_{n = 0,1,2,\cdots} \tau_n (\vartheta)
\left\{
\cos([2 n + 1] \varphi_{\ell})
\atop
\cos(2 n\varphi_{\ell})
\right\}
\ \mbox{for} \
\left\{
\tau(\theta_{\ell}) = - \tau(\pi - \theta_{\ell})
\atop
\tau(\theta_{\ell}) = \tau(\pi - \theta_{\ell})
\right\}  ,
\end{eqnarray}
where $\vartheta \le \frac{\pi}{2}$.

\subsubsection{Squared virtual Compton amplitude}

At leading twist, it turns out that $|{\cal T}_{\scriptscriptstyle\rm VCS}|^2$
depends only on the harmonics $\cos [ n (\varphi_\ell - \phi) ]$ with $n = 0, 1, 2$.
As a result, we find that in this approximation, there exist the following relations
between the Fourier coefficients:
\begin{eqnarray}
\label{Rel-FC-VCS2-Tw2}
{\rm ss}^{\scriptscriptstyle\rm VCS}_{nn}
\!\!\!&\simeq&\!\!\!
{\rm cc}^{\scriptscriptstyle\rm VCS}_{nn}
\, , \nonumber\\
{\rm sc}^{\scriptscriptstyle\rm VCS}_{nm}
\!\!\!&\simeq&\!\!\!
{\rm cs}^{\scriptscriptstyle\rm VCS}_{nm} \simeq 0
\, , \\
{\rm ss}^{\scriptscriptstyle\rm VCS}_{nm}
\!\!\!&\simeq&\!\!\!
{\rm cc}^{\scriptscriptstyle\rm VCS}_{nm} \simeq 0
\, , \qquad n \neq m
\, . \nonumber
\end{eqnarray}
The nonvanishing Fourier coefficients ${\rm cc}^{\scriptscriptstyle\rm VCS}_{nn}$
can be easily evaluated from  Eqs.\ (\ref{a0VCS})-(\ref{b2VCS}) and products of
Compton form factors. The general structure of products of Compton form factors
is given by
\begin{eqnarray}
\label{Def-VV-full}
&&\!\!\!\!\!\!\!\!\!\!\!\!\!\!\!\!\!\!\!\!\!\!
\frac{1}{4} {\cal F}_1 {\cal F}_1^\dagger
\equiv
{\cal C}^{\scriptscriptstyle\rm VCS}_{{\cal VV}, {\rm unp}}
+
{\cal C}^{\scriptscriptstyle\rm VCS}_{{\cal VV}, {\scriptscriptstyle\rm LP}}
\cos{\mit\Theta}
+
i
\sqrt{- \frac{\Delta^2}{4M^2_N}}
\sqrt{1 - \frac{\Delta^2_{\rm min}}{\Delta^2}}
\sqrt{ \frac{1 - \eta}{1 + \eta} } \,
{\cal C}^{\scriptscriptstyle\rm VCS}_{{\cal VV}, {\scriptscriptstyle\rm TP}}
\sin \left( {\mit\Phi} - \phi \right) \sin{\mit\Theta}
\, , \quad\\
\nonumber\\
\label{Def-VA-full}
&&\!\!\!\!\!\!\!\!\!\!\!\!\!\!\!\!\!\!\!\!\!\!
\frac{1}{4} {\cal F}_1 \widetilde{\cal F}_1^\dagger
\equiv
{\cal C}^{\scriptscriptstyle\rm VCS}_{{\cal VA}, {\rm unp}}
+
{\cal C}^{\scriptscriptstyle\rm VCS}_{{\cal VA}, {\scriptscriptstyle\rm LP}}
\cos{\mit\Theta}
+
\sqrt{- \frac{\Delta^2}{4M^2_N}}
\sqrt{1 - \frac{\Delta^2_{\rm min}}{\Delta^2}}
\sqrt{ \frac{1 - \eta}{1 + \eta} } \,
{\cal C}^{\scriptscriptstyle\rm VCS}_{{\cal VA}, {\scriptscriptstyle\rm TP}}
\cos \left( {\mit\Phi} - \phi \right) \sin{\mit\Theta}
\, ,
\end{eqnarray}
with $\widetilde{\cal F}_1 \widetilde{\cal F}_1^\dagger$ product having the
decomposition similar to Eq.\ (\ref{Def-VV-full}). The polar ${\mit\Theta}$
and azimuthal ${\mit\Phi}$ angles parametrize the nucleon polarization vector
$S$ according to Eq.\ (\ref{Spin}). The functions ${\cal C} ({\cal F},{\cal F}^\ast)$
depend on Compton form factors $\mathcal{F}$. For the case at hand, we find at
leading order in $(p \cdot q)^{- 1}$
\begin{eqnarray}
\label{Res-C-unpVV}
{\cal C}^{\scriptscriptstyle\rm VCS}_{{\cal VV}, {\rm unp}} ({\cal F}, {\cal F}^\ast)
\!\!\!&=&\!\!\!
(1 - \eta^2) {\cal H} {\cal H}^\ast
-
\eta^2 ({\cal H} {\cal E}^\ast
+
{\cal E} {\cal H}^\ast)
-
\left( \frac{\Delta^2}{4 M^2_N} + \eta^2 \right) {\cal E} {\cal E}^\ast
\, , \\
\label{Res-C-unpAA}
{\cal C}^{\scriptscriptstyle\rm VCS}_{{\cal AA}, {\rm unp}} ({\cal F}, {\cal F}^\ast)
\!\!\!&=&\!\!\!
(1 - \eta^2) \widetilde {\cal H} \widetilde {\cal H}^\ast
-
\eta^2
(
\widetilde  {\cal H} \widetilde {\cal E}^\ast
+
\widetilde  {\cal E} \widetilde {\cal H}^\ast
)
- \eta^2 \frac{\Delta^2}{4 M^2_N} \widetilde {\cal E} \widetilde {\cal E}^\ast
\, , \\
{\cal C}^{\scriptscriptstyle\rm VCS}_{{\cal VA}, {\rm unp}} ({\cal F}, {\cal F}^\ast)
\!\!\!&=&\!\!\!
0
\, .
\phantom{\left(\frac{\Delta^2}{4 M^2}\right)}
\end{eqnarray}
The results for spin-dependent cases, including both longitudinally and
transversely polarized target options, are presented below. The nonvanishing
${\cal C}$-coefficients in the squared VCS amplitude, defined in Eq.\
(\ref{Def-VV-full}) -- (\ref{Def-VA-full}), for the polarized nucleon
target are given by
\begin{eqnarray}
{\cal C}^{\scriptscriptstyle\rm VCS}_{{\cal VA}, {\scriptscriptstyle\rm LP}}
({\cal F}, {\cal F}^\ast)
\!\!\!&=&\!\!\!
(1 - \eta^2) {\cal H} \widetilde {\cal H}^\ast
-
\eta^2 ({\cal H} \widetilde {\cal E}^\ast
+
{\cal E} \widetilde {\cal H}^\ast)
-
\eta
\left(
\frac{\Delta^2}{4 M^2_N} + \frac{\eta^2}{1 + \eta}
\right)
{\cal E} \widetilde{\cal E}^\ast
\, , \\
{\cal C}^{\scriptscriptstyle\rm VCS}_{{\cal VV}, {\scriptscriptstyle\rm TP}}
({\cal F}, {\cal F}^\ast)
\!\!\!&=&\!\!\!
(1 + \eta)
\left(
{\cal H} {\cal E}^\ast
-
{\cal E} {\cal H}^\ast
\right)
\, , \\
{\cal C}^{\scriptscriptstyle\rm VCS}_{{\cal AA}, {\scriptscriptstyle\rm TP}}
({\cal F}, {\cal F}^\ast)
\!\!\!&=&\!\!\!
- (1 + \eta) \eta
\left(
\widetilde {\cal H} \widetilde {\cal E}^\ast
-
\widetilde  {\cal E} \widetilde {\cal H}^\ast
\right)
\, , \\
{\cal C}^{\scriptscriptstyle\rm VCS}_{{\cal VA}, {\scriptscriptstyle\rm TP}}
({\cal F}, {\cal F}^\ast)
\!\!\!&=&\!\!\!
(1 + \eta)
\left(
\eta\, {\cal H} \widetilde {\cal E}^\ast
-
{\cal E} \widetilde {\cal H}^\ast
\right)
+
\eta^2 {\cal E} \widetilde{\cal E}^\ast
\, .
\end{eqnarray}
Note, that in the (spacelike) DVCS limit, i.e., when one sets $\eta = \xi$,
we retrieve our previous result from Ref.\ \cite{BelMulKir01}
\begin{eqnarray*}
{\cal C}^{\scriptscriptstyle\rm DVCS}_{\rm unp}
\stackrel{\mbox{\tiny DVCS}}{=}
{\cal C}^{\scriptscriptstyle\rm VCS}_{{\cal VV},\rm unp}
+
{\cal C}^{\scriptscriptstyle\rm VCS}_{{\cal AA},\rm unp}
\, ,
\end{eqnarray*}
and analogous relations for the polarized case in the DVCS limit $\eta = \xi$,
\begin{eqnarray}
\label{Cor-DVCS2}
{\cal C}^{\scriptscriptstyle\rm DVCS}_{{\scriptscriptstyle\rm LP}}
({\cal F}, {\cal F}^\ast)
\!\!\!&\stackrel{\mbox{\tiny DVCS}}{=}&\!\!\!
{\cal C}^{\scriptscriptstyle\rm VCS}_{{\cal VA}, {\scriptscriptstyle\rm LP}}
({\cal F}, {\cal F}^\ast)
+
{\cal C}^{\scriptscriptstyle\rm VCS}_{{\cal VA}, {\scriptscriptstyle\rm LP}}
({\cal F}^\ast, {\cal F})
\, , \nonumber\\
{\cal C}^{\scriptscriptstyle\rm DVCS}_{{\scriptscriptstyle\rm TP}+}
({\cal F}, {\cal F}^\ast)
\!\!\!&\stackrel{\mbox{\tiny DVCS}}{=}&\!\!\!
{\cal C}^{\scriptscriptstyle\rm VCS}_{{\cal VA}, {\scriptscriptstyle\rm TP}}
({\cal F}, {\cal F}^\ast)
+
{\cal C}^{\scriptscriptstyle\rm VCS}_{{\cal VA}, {\scriptscriptstyle\rm TP}}
({\cal F}^\ast, {\cal F})
\, , \\
{\cal C}^{\scriptscriptstyle\rm DVCS}_{{\scriptscriptstyle\rm TP}-}
({\cal F}, {\cal F}^\ast)
\!\!\!&\stackrel{\mbox{\tiny DVCS}}{=}&\!\!\!
{\cal C}^{\scriptscriptstyle\rm VCS}_{{\cal VV}, {\scriptscriptstyle\rm TP}}
({\cal F}, {\cal F}^\ast)
+
{\cal C}^{\scriptscriptstyle\rm VCS}_{{\cal AA}, {\scriptscriptstyle\rm TP}}
({\cal F}, {\cal F}^\ast)
\, . \nonumber
\end{eqnarray}

In this way, we find for the unpolarized target the following nonvanishing
Fourier coefficients in the twist-two sector
\begin{eqnarray}
\label{Def-FC-unp}
{\rm cc}^{\scriptscriptstyle\rm VCS}_{00, {\rm unp}}
\!\!\!&=&\!\!\!
2 (2 - 2y + y^2) (2 - 2 \widetilde y + \widetilde y^2)
\left\{
{\cal C}^{\scriptscriptstyle\rm VCS}_{{\cal VV}, {\rm unp}} ({\cal F}, {\cal F}^\ast)
+
{\cal C}^{\scriptscriptstyle\rm VCS}_{{\cal AA}, {\rm unp}} ({\cal F}, {\cal F}^\ast)
\right\}   \nonumber\\
&&
+
\frac{16}{\xi^2} (1 - y) (1 - \widetilde y) ( \xi^2 - \eta^2 )
{\cal C}^{\scriptscriptstyle\rm VCS}_{{\cal VV}, {\rm unp}} ({\cal F}_L, {\cal F}^\ast_L)
\, , \\
{\rm cc}^{\scriptscriptstyle\rm VCS}_{11,{\rm unp}}
\!\!\!&=&\!\!\!
\frac{4 \sigma}{\xi}
\sqrt{(1 - y)(1 - \widetilde y)(\xi^2 - \eta^2)} (2 - y)(2 - \widetilde y)\!
\left\{
{\cal C}^{\scriptscriptstyle\rm VCS}_{{\cal VV}, {\rm unp}} ({\cal F}, {\cal F}^\ast_L)
+
{\cal C}^{\scriptscriptstyle\rm VCS}_{{\cal VV}, {\rm unp}} ({\cal F}_L, {\cal F}^\ast) \!
\right\}
, \qquad\quad\\
{\rm cc}^{\scriptscriptstyle\rm VCS}_{22,{\rm unp}}
\!\!\!&=&\!\!\!
8 (1 - y) (1 - \widetilde y)
\left\{
{\cal C}^{\scriptscriptstyle\rm VCS}_{{\cal VV}, {\rm unp}} ({\cal F}, {\cal F}^\ast)
-
{\cal C}^{\scriptscriptstyle\rm VCS}_{{\cal AA}, {\rm unp}} ({\cal F}, {\cal F}^\ast)
\right\}
\, .
\end{eqnarray}
The remaining Fourier coefficients  for longitudinally polarized target
are given  by
\begin{eqnarray}
{\rm cc}^{\scriptscriptstyle\rm VCS}_{00,{\scriptscriptstyle\rm LP}}
\!\!\!&=&\!\!\!
2 \lambda (2 - y) y (2 - 2 \widetilde y + \widetilde y^2)
\left\{
{\cal C}^{\scriptscriptstyle\rm VCS}_{{\cal VA}, {\scriptscriptstyle\rm LP}}
({\cal F}, {\cal F}^\ast)
+
{\cal C}^{\scriptscriptstyle\rm VCS}_{{\cal VA}, {\scriptscriptstyle\rm LP}}
({\cal F}^\ast, {\cal F})
\right\}
\, , \nonumber\\
{\rm cc}^{\scriptscriptstyle\rm VCS}_{11,{\scriptscriptstyle\rm LP}}
\!\!\!&=&\!\!\! \lambda
\frac{4  \sigma}{\xi}
y (2 - \widetilde y)
\sqrt{(1 - y)(1 - \widetilde y)(\xi^2 - \eta^2)}
\left\{
{\cal C}^{\scriptscriptstyle\rm VCS}_{{\cal VA}, {\scriptscriptstyle\rm LP}}
({\cal F}_L, {\cal F}^\ast)
+
{\cal C}^{\scriptscriptstyle\rm VCS}_{{\cal VA}, {\scriptscriptstyle\rm LP}}
({\cal F}_L^\ast, {\cal F})
\right\}
\, ,
\end{eqnarray}
and, for transversely polarized target, they read
\begin{eqnarray}
\label{Def-FC-TP}
{\rm cc}^{\scriptscriptstyle\rm VCS}_{00,{\scriptscriptstyle\rm TP}}
\!\!\!&=&\!\!\!
\sqrt{-\frac{\Delta^2}{M^2_N}}
\sqrt{ 1 - \frac{\Delta^2_{\rm min}}{\Delta^2} }
\sqrt{\frac{1 - \eta}{1 + \eta}}
\nonumber \\
&&\!\!\!
\times
\Bigg\{
\lambda
\cos \left( {\mit\Phi} - \phi \right)
(2 - y) y (2 - 2 \widetilde y + \widetilde y^2)
\left\{
{\cal C}^{\scriptscriptstyle\rm VCS}_{{\cal VA},{\scriptscriptstyle\rm TP}}
({\cal F},{\cal F}^\ast)
+
{\cal C}^{\scriptscriptstyle\rm VCS}_{{\cal VA},{\scriptscriptstyle\rm TP}}
({\cal F}^\ast,{\cal F})
\right\}
\nonumber\\
&&\quad + \
i \sin \left( {\mit\Phi} - \phi \right)
(2 - 2 y + y^2) (2 - 2 \widetilde y + \widetilde y^2)
\left\{
{\cal C}^{\scriptscriptstyle\rm VCS}_{{\cal VV},{\scriptscriptstyle\rm TP}}
({\cal F},{\cal F}^\ast)
+
{\cal C}^{\scriptscriptstyle\rm VCS}_{{\cal AA},{\scriptscriptstyle\rm TP}}
({\cal F},{\cal F}^\ast)
\right\}
\nonumber\\
&&\quad + \
8 i \sin \left( {\mit\Phi} - \phi \right)
(1 - y) (1 - \widetilde y) \frac{\xi^2 - \eta^2}{\xi^2}
{\cal C}^{\scriptscriptstyle\rm VCS}_{{\cal VV}, {\scriptscriptstyle\rm TP}}
({\cal F}_L, {\cal F}^\ast_L)
\Bigg\}
\, , \\
{\rm cc}^{\scriptscriptstyle\rm VCS}_{11,{\scriptscriptstyle\rm TP}}
\!\!\!&=&\!\!\!
\sqrt{-\frac{\Delta^2}{M^2_N}}
\sqrt{ 1 - \frac{\Delta^2_{\rm min}}{\Delta^2} }
\sqrt{\frac{1 - \eta}{1 + \eta}}\,
\frac{2 \sigma}{\xi}  \sqrt{(1 - y)(1 - \widetilde y)(\xi^2 - \eta^2)}
\nonumber\\
&&\!\!\!
\times
\Bigg\{
\lambda
\cos \left( {\mit\Phi} - \phi \right)
y (2 - \widetilde y)
\left\{
{\cal C}^{\scriptscriptstyle\rm VCS}_{{\cal VA}, {\scriptscriptstyle\rm TP}}
({\cal F}_L, {\cal F}^\ast)
+
{\cal C}^{\scriptscriptstyle\rm VCS}_{{\cal VA}, {\scriptscriptstyle\rm TP}}
({\cal F}_L^\ast, {\cal F})
\right\}
\nonumber\\
&&\;\;+
i
\sin \left( {\mit\Phi} - \phi \right)
(2 - y) (2 - \widetilde y)
\left\{
{\cal C}^{\scriptscriptstyle\rm VCS}_{{\cal VV}, {\scriptscriptstyle\rm TP}}
({\cal F}_L, {\cal F}^\ast)
+
{\cal C}^{\scriptscriptstyle\rm VCS}_{{\cal VV}, {\scriptscriptstyle\rm TP}}
({\cal F}_L^\ast, {\cal F})
\right\}
\Bigg\}
\, ,\\
{\rm cc}^{\scriptscriptstyle\rm VCS}_{22,{\scriptscriptstyle\rm TP}}
\!\!\!&=&\!\!\!
\sqrt{-\frac{\Delta^2}{M^2_N}}
\sqrt{ 1 - \frac{\Delta^2_{\rm min}}{\Delta^2} }
\sqrt{\frac{1 - \eta}{1 + \eta}}
\nonumber\\
&&\!\!\!\times 4 i \sin \left( {\mit\Phi} - \phi \right)
(1 - y) (1 - \widetilde y)
\left\{
{\cal C}^{\scriptscriptstyle\rm VCS}_{{\cal VV}, {\scriptscriptstyle\rm TP}}
({\cal F}, {\cal F}^\ast)
-
{\cal C}^{\scriptscriptstyle\rm VCS}_{{\cal AA}, {\scriptscriptstyle\rm TP}}
({\cal F}, {\cal F}^\ast)
\right\}
\, .
\end{eqnarray}
Here, according to the general twist-two relation (\ref{Rel-FC-VCS2-Tw2}),
\begin{equation}
{\rm ss}^{\scriptscriptstyle\rm VCS}_{11,{\scriptscriptstyle\rm LP}/{\scriptscriptstyle\rm TP}}
\simeq
{\rm cc}^{\scriptscriptstyle\rm VCS}_{11,{\scriptscriptstyle\rm LP}/{\scriptscriptstyle\rm TP}}
\, , \qquad
{\rm ss}^{\scriptscriptstyle\rm VCS}_{22,{\scriptscriptstyle\rm TP}}
\simeq
{\rm cc}^{\scriptscriptstyle\rm VCS}_{22,{\scriptscriptstyle\rm TP}}
\, .
\end{equation}
All other coefficients are expressed using  Eq.\ (\ref{Rel-FC-VCS2-Tw2}).
Note, however, that the tensor-gluon contribution induces further second order
harmonics, which are not displayed here since they are suppressed by a power
of the coupling constant $\alpha_s$.

\subsubsection{Interference of virtual Compton and Bethe-Heitler amplitudes}
\label{SubSec-Int}

In the leading-twist approximation, the following general relations can be
established between the Fourier coefficients of the interference term
\begin{eqnarray}
\label{Rel-FC-INT-Tw2}
{\rm ss}^{\scriptscriptstyle\rm INT}_{nm}
\!\!\! &\simeq\!\!\!&
{\rm cc}^{\scriptscriptstyle\rm INT}_{nm}
\quad\mbox{and}\quad
{\rm cs}^{\scriptscriptstyle\rm INT}_{nm}
\simeq
- {\rm sc}^{\scriptscriptstyle\rm INT}_{nm}
\quad\mbox{for} \quad  \{ nm \} = \{ 12, 21, 32 \}
\, , \nonumber\\
{\rm ss}^{\scriptscriptstyle\rm INT}_{nm}
\!\!\!&\simeq\!\!\!&
{\rm cc}^{\scriptscriptstyle\rm INT}_{nm}
\simeq
{\rm cs}^{\scriptscriptstyle\rm INT}_{nm}
\simeq
{\rm sc}^{\scriptscriptstyle\rm INT}_{nm}
\simeq
0
\quad\mbox{for} \quad n \neq m \pm 1
\, ,
\end{eqnarray}
where $n,m+1 \le 3$ for INT$=\{1,2\}$. For the unpolarized target, there are
five nontrivial entries which arise for the unpolarized lepton beam, namely,
${\rm cc}^{\scriptscriptstyle\rm INT}_{01,{\rm unp}}$,
${\rm cc}^{\scriptscriptstyle\rm INT}_{10,{\rm unp}}$,
${\rm cc}^{\scriptscriptstyle\rm INT}_{12,{\rm unp}}$,
${\rm cc}^{\scriptscriptstyle\rm INT}_{21,{\rm unp}}$,
and ${\rm cc}^{\scriptscriptstyle\rm INT}_{32,{\rm unp}}$. They are
supplemented by three additional Fourier coefficients for the polarized-beam
option: ${\rm cs}^{\scriptscriptstyle\rm INT}_{01,{\rm unp}}$,
${\rm sc}^{\scriptscriptstyle\rm INT}_{10,{\rm unp}}$,
and ${\rm sc}^{\scriptscriptstyle\rm INT}_{21,{\rm unp}}$,
while ${\rm sc}^{\scriptscriptstyle\rm INT}_{12,{\rm unp}}
\simeq {\rm sc}^{\scriptscriptstyle\rm INT}_{32,{\rm unp}} \simeq 0$.
To find the explicit form of these Fourier coefficients, one should evaluate
the products of Dirac bilinears in Eqs.\ (\ref{Def-BHINT1}) -- (\ref{Def-BHINT2}).
This yields
\begin{eqnarray}
\label{Def-trace-SV}
\left\{
\!\!\!
\begin{array}{c}
{\cal S}_1
\\
{\cal S}_2
\end{array}
\!\!\!
\right\}\!
{\cal F}_1
\!\!\!&\equiv&\!\!\!
4 Q^2 \frac{(1 + \eta) \eta}{y \widetilde y \xi}
\left(
\left\{
\!\!\!
\begin{array}{c}
\widetilde y K
\\
y \widetilde K
\end{array}
\!\!\!
\right\} \left[
{\cal C}_{{\cal V},{\rm unp}} ({\cal F})
\left\{
\!\!\!
\begin{array}{c}
\cos \phi
\\
\cos \varphi_\ell
\end{array}
\!\!\!
\right\}
+
\cos{\mit\Theta} \,
i {\cal C}_{{\cal V},{\scriptscriptstyle\rm LP}} ({\cal F})
\left\{
\!\!\!
\begin{array}{c}
\sin \phi
\\
\sin \varphi_\ell
\end{array}
\!\!\!
\right\}
\right] \right.
\\
&+&\!\!\!
\left. \sin{\mit\Theta}
\left\{
\!\!\!
\begin{array}{c}
\widetilde y L
\\
y \widetilde L
\end{array}
\!\!\!
\right\} \left[
i {\cal C}_{{\cal V},{\scriptscriptstyle\rm TP}+} ({\cal F})
\left\{
\!\!\!
\begin{array}{c}
\sin \phi
\\
\sin \varphi_\ell
\end{array}
\!\!\!
\right\}
\cos ({\mit\Phi} - \phi)
+
i {\cal C}_{{\cal V},{\scriptscriptstyle\rm TP}-} ({\cal F})
\left\{
\!\!\!
\begin{array}{c}
\cos \phi
\\
\cos \varphi_\ell
\end{array}
\!\!\!
\right\}
\sin ({\mit\Phi} - \phi)
\right] \right)
\, ,
\nonumber\\
\nonumber\\
\label{Def-trace-RV}
\left\{
\!\!\!
\begin{array}{c}
\widetilde{\cal S}_1
\\
\widetilde{\cal S}_2
\end{array}
\!\!\!
\right\}\!
{\cal F}_1
\!\!\!&\equiv&\!\!\!
4 Q^2 \frac{(1 + \eta) \eta}{y \widetilde y \xi}
\left(
\left\{
\!\!\!
\begin{array}{c}
\widetilde y K
\\
y \widetilde K
\end{array}
\!\!\!
\right\}
\left[
- i {\cal C}_{{\cal V}, {\rm unp}} (\cal F)
\left\{
\!\!\!
\begin{array}{c}
\sin \phi
\\
\sin \varphi_\ell
\end{array}
\!\!\!
\right\}
-
\cos{\mit\Theta} \,
{\cal C}_{{\cal V}, {\scriptscriptstyle\rm LP}} (\cal F)
\left\{
\!\!\!
\begin{array}{c}
\cos \phi
\\
\cos \varphi_\ell
\end{array}
\!\!\!
\right\}\right]\right.
\\
&-&\!\!\!
\left. \sin{\mit\Theta}
\left\{
\!\!\!
\begin{array}{c}
\widetilde y L
\\
y \widetilde L
\end{array}
\!\!\!
\right\} \left[
{\cal C}_{{\cal V},{\scriptscriptstyle\rm TP}+} ({\cal F})
\left\{
\!\!\!
\begin{array}{c}
\cos \phi
\\
\cos \varphi_\ell
\end{array}
\!\!\!
\right\}
\cos ({\mit\Phi} - \phi)
-
{\cal C}_{{\cal V},{\scriptscriptstyle\rm TP}-} ({\cal F})
\left\{
\!\!\!
\begin{array}{c}
\sin \phi
\\
\sin \varphi_\ell
\end{array}
\!\!\!
\right\}
\sin({\mit\Phi} - \phi)
\right] \right)
\, ,
\nonumber\\
\nonumber\\
\label{Def-trace-SA}
\left\{
\!\!\!
\begin{array}{c}
{\cal S}_1
\\
{\cal S}_2
\end{array}
\!\!\!
\right\}\!
\widetilde{\cal F}_1
\!\!\!&\equiv&\!\!\!
4  Q^2 \frac{(1 + \eta) \eta}{y \widetilde y \xi}
\left(\left\{
\!\!\!
\begin{array}{c}
\widetilde y K
\\
y \widetilde K
\end{array}
\!\!\!
\right\}
\left[
i {\cal C}_{{\cal A},{\rm unp}} ({\cal F})
\left\{
\!\!\!
\begin{array}{c}
\sin \phi
\\
\sin \varphi_\ell
\end{array}
\!\!\!
\right\}
+
\cos{\mit\Theta} \,
{\cal C}_{{\cal A},{\scriptscriptstyle\rm LP}} ({\cal F})
\left\{
\!\!\!
\begin{array}{c}
\cos \phi
\\
\cos \varphi_\ell
\end{array}
\!\!\!
\right\}
\right]\right.
\\
&+&\!\!\!
\left. \sin{\mit\Theta}
\left\{
\!\!\!
\begin{array}{c}
\widetilde y L
\\
y \widetilde L
\end{array}
\!\!\!
\right\} \left[
{\cal C}_{{\cal A},{\scriptscriptstyle\rm TP}+} ({\cal F})
\left\{
\!\!\!
\begin{array}{c}
\cos \phi
\\
\cos \varphi_\ell
\end{array}
\!\!\!
\right\}
\cos ({\mit\Phi} - \phi)
-{\cal C}_{{\cal A},{\scriptscriptstyle\rm TP}-} ({\cal F})
\left\{
\!\!\!
\begin{array}{c}
\sin \phi
\\
\sin \varphi_\ell
\end{array}
\!\!\!
\right\}
\sin ({\mit\Phi} - \phi)
\right] \right)
\, ,
\nonumber\\
\nonumber\\
\label{Def-trace-RA}
\left\{
\!\!\!
\begin{array}{c}
\widetilde{\cal S}_1
\\
\widetilde{\cal S}_2
\end{array}
\!\!\!
\right\}\!
\widetilde{\cal F}_1
\!\!\!&\equiv&\!\!\!
4 Q^2 \frac{(1 + \eta) \eta}{y \widetilde y \xi}
\left(
\left\{
\!\!\!
\begin{array}{c}
\widetilde y K
\\
y \widetilde K
\end{array}
\!\!\!
\right\} \left[
-
{\cal C}_{{\cal A},{\rm unp}} ({\cal F})
\left\{
\!\!\!
\begin{array}{c}
\cos \phi
\\
\cos \varphi_\ell
\end{array}
\!\!\!
\right\}
-
\cos {\mit\Theta} \,
i {\cal C}_{{\cal A},{\scriptscriptstyle\rm LP}} ({\cal F})
\left\{
\!\!\!
\begin{array}{c}
\sin \phi
\\
\sin \varphi_\ell
\end{array}
\!\!\!
\right\}
\right]\right.
\\
&-&\!\!\!
\left. \sin{\mit\Theta}
\left\{
\!\!\!
\begin{array}{c}
\widetilde y L
\\
y \widetilde L
\end{array}
\!\!\!
\right\} \left[
i {\cal C}_{{\cal A},{\scriptscriptstyle\rm TP}+} ({\cal F})
\left\{
\!\!\!
\begin{array}{c}
\sin \phi
\\
\sin \varphi_\ell
\end{array}
\!\!\!
\right\}
\cos ({\mit\Phi} - \phi)
+
i {\cal C}_{{\cal A},{\scriptscriptstyle\rm TP}-} ({\cal F})
\left\{
\!\!\!
\begin{array}{c}
\cos \phi
\\
\cos \varphi_\ell
\end{array}
\!\!\!
\right\}
\sin ({\mit\Phi} - \phi)
\right] \right)
\, ,
\nonumber
\end{eqnarray}
where the angles ${\mit\Theta}$ and ${\mit\Phi}$ exhibit the dependence on the
polarization of the target. We used, in analogy to the definitions (\ref{Def-K}),
the shorthand notation for
\begin{equation}
\label{Def-L}
\left\{
\begin{array}{c}
L
\\
\widetilde L
\end{array}
\right\}
\approx
\frac{1}{2\eta}  \sqrt{\frac{\xi M^2_N}{Q^2}}
\left\{
\begin{array}{c}
\sqrt{(1-y)(\xi + \eta)}
\\
\sqrt{(1-\widetilde y)(\xi - \eta)}
\end{array}
\right\}
\end{equation}
We also introduced universal electric- and magnetic-like combinations
of electromagnetic form factors intertwined with CFFs
\begin{equation}
\label{Int-dif-Com}
{\cal C}_{{\cal V}, {\rm unp}}
=
F_1 {\cal H} - \frac{\Delta^2}{4 M_N^2} F_2 {\cal E}
\, , \qquad
{\cal C}_{{\cal A}, {\rm unp}}
=
\eta (F_1 + F_2) \widetilde {\cal H}
\, .
\end{equation}
For the longitudinally polarized nucleon, we get the following combinations of
the electromagnetic and Compton form factors:
\begin{equation}
\label{Def-C-int-LP}
{\cal C}_{{\cal V}, {\scriptscriptstyle\rm LP}}
=
\eta (F_1 + F_2)
\left(
{\cal H} + \frac{\eta}{1 + \eta} {\cal E}
\right)
\, , \qquad
{\cal C}_{{\cal A}, {\scriptscriptstyle\rm LP}}
=
F_1 \widetilde {\cal H}
-
\eta
\left(
\frac{\eta}{1 + \eta} F_1 + \frac{\Delta^2}{4 M_N^2} F_2
\right)
\widetilde {\cal E}
\, ,
\end{equation}
while for the transversely polarized case we have four more combinations
\begin{eqnarray}
\label{Def-C-int-TP-V+}
{\cal C}_{{\cal V}, {\scriptscriptstyle\rm TP}+}
\!\!\!&=&\!\!\!
\frac{2 \eta}{1-\eta} (F_1 + F_2)
\left\{
\eta
\left(
{\cal H} - \frac{\eta}{1 + \eta} {\cal E}
\right)
+ \frac{\Delta^2}{4 M_N^2} {\cal E}
\right\}
\, , \\
\label{Def-C-int-TP-A+}
{\cal C}_{{\cal A}, {\scriptscriptstyle\rm TP}+}
\!\!\!&=&\!\!\!
-
\frac{2}{1 + \eta}
\left\{
\eta^2 F_1
\left(
\widetilde {\cal H} + \frac{\eta}{1 + \eta} \widetilde {\cal E}
\right)
- \frac{\Delta^2}{4 M_N^2}
\left(
(1 - \eta^2) F_2 \widetilde {\cal H}
-
\eta (F_1 + \eta F_2) \widetilde {\cal E}
\right)
\right\}
\, ,\\
\label{Def-C-int-TP-V-}
{\cal C}_{{\cal V}, {\scriptscriptstyle\rm TP}-}
\!\!\!&=&\!\!\!
\frac{2}{1 + \eta}
\left\{
\eta^2 F_1
\left(
{\cal H} + {\cal E}
\right)
-
\frac{\Delta^2}{4 M_N^2}
\left(
(1 - \eta^2) F_2 {\cal H} - (F_1 + \eta^2 F_2) {\cal E}
\right)
\right\}
\, ,\\
\label{Def-C-int-TP-A-}
{\cal C}_{{\cal A}, {\scriptscriptstyle\rm TP}-}
\!\!\!&=&\!\!\!
-
\frac{2\eta^2}{1 + \eta} (F_1 + F_2)
\left\{
\widetilde  {\cal H} + \frac{\Delta^2}{4 M_N^2} \widetilde {\cal E}
\right\}
\, .
\end{eqnarray}

Now, using  Eqs.\ (\ref{Def-BHINT1}) -- (\ref{Def-BHINT1-b2}) and (\ref{Dec-BHVCS})
it is straightforward to derive the following nonzero Fourier coefficients for the
first interference term:
\begin{eqnarray}
\label{cc-INT1-01}
{\rm cc}^1_{01,{\rm unp}}
\!\!\!&=&\!\!\!
8 \widetilde K (1 - y)(2 - y)  (2 - \widetilde y)
\frac{\xi + \eta}{\eta}
\Re{\rm e}\!
\left\{\!
{\cal C}_{{\cal V},{\rm unp}}({\cal F})
+
{\cal C}_{{\cal A},{\rm unp}}({\cal F})
-
\frac{\xi - \eta}{\xi} {\cal C}_{{\cal V},{\rm unp}}({\cal F}_L)
\!\right\}\!
, \\
{\rm cs}^1_{01,{\rm unp}}
\!\!\!&=&\!\!\!
8 \lambda \widetilde K y (1 - y) (2 - \widetilde y)
\frac{\xi + \eta}{\eta}
\Im{\rm m}
\left\{
{\cal C}_{{\cal V},{\rm unp}}({\cal F})
+
{\cal C}_{{\cal A},{\rm unp}}({\cal F})
+
\frac{\xi - \eta}{\xi} {\cal C}_{{\cal V},{\rm unp}}({\cal F}_L)
\right\}
, \\
{\rm cc}^1_{10,{\rm unp}}
\!\!\!&=&\!\!\!
8 K \;
\Re{\rm e}
\left\{
-
(2 - 2 y + y^2) (2 - 2 \widetilde y + \widetilde y^2)
\left(
\frac{\xi}{\eta} {\cal C}_{{\cal V},{\rm unp}} ({\cal F})
+
{\cal C}_{{\cal A},{\rm unp}}({\cal F})
\right)
\right.
\nonumber \\
&&\qquad\qquad
\left.
+ 8 (1 - y) (1 - \widetilde y)\frac{\xi^2 - \eta^2}{\eta \xi}
{\cal C}_{{\cal V},{\rm unp}}({\cal F}_L)
\right\}
, \\
{\rm cc}^1_{12,{\rm unp}}
\!\!\!&=&\!\!\!
16 K (1 - y)  (1 - \widetilde y)
\frac{\xi - \eta}{\xi}
\Re{\rm e}
\left\{
{\cal C}_{{\cal V},{\rm unp}}({\cal F})
+
{\cal C}_{{\cal A},{\rm unp}}({\cal F})
\right\}
, \\
{\rm sc}^1_{10,{\rm unp}}
\!\!\!&=&\!\!\!
- 8 \lambda K y   (2-y)(2 - 2\widetilde y + \widetilde y^2)
\Im{\rm m}
\left\{
{\cal C}_{{\cal V},{\rm unp}}({\cal F})
+
\frac{\xi}{\eta} {\cal C}_{{\cal A}, {\rm unp}}({\cal F})
\right\}
, \\
{\rm cc}^1_{21,{\rm unp}}
\!\!\!&=&\!\!\!
8 \widetilde K (1 - y) (2 - y)  (2 - \widetilde y) \frac{\xi + \eta}{\eta}
\Re{\rm e}\!
\left\{\!
{\cal C}_{{\cal V},{\rm unp}}({\cal F})
-
{\cal C}_{{\cal A},{\rm unp}}({\cal F})
-
\frac{\xi + \eta}{\xi} {\cal C}_{{\cal V},{\rm unp}}({\cal F}_L)
\right\}\!
, \\
{\rm sc}^1_{21,{\rm unp}}
\!\!\!&=&\!\!\!
- 8 \lambda\widetilde K y (1 - y) (2 - \widetilde y) \frac{\xi + \eta}{\eta}
\Im{\rm m}
\left\{
{\cal C}_{{\cal V},{\rm unp}}({\cal F})
-
{\cal C}_{{\cal A},{\rm unp}}({\cal F})
+
\frac{\xi + \eta}{\xi} {\cal C}_{{\cal V},{\rm unp}}({\cal F}_L)
\right\}
, \qquad\quad\\
\label{cc-INT1-32}
{\rm cc}^1_{32,{\rm unp}}
\!\!\!&=&\!\!\!
- 16 K (1 - y) (1 - \widetilde y) \frac{\xi + \eta}{\eta}
\Re{\rm e}
\left\{
{\cal C}_{{\cal V},{\rm unp}}({\cal F})
-
{\cal C}_{{\cal A},{\rm unp}}({\cal F})
\right\}
.
\end{eqnarray}
One should also add to this list ${\rm ss}^1_{12,{\rm unp}}$, ${\rm ss}^1_{21,{\rm unp}}$,
${\rm cs}^1_{21,{\rm unp}}$, and ${\rm ss}^1_{32,{\rm unp}}$, which  follow  from Eq.\
(\ref{Rel-FC-INT-Tw2}).

Next, we display the explicit form of the Fourier coefficients in the interference
term with the second BH process for an unpolarized nucleon target, which can be
obtained through symmetry considerations as will be explained below. These coefficients
are
\begin{eqnarray}
\label{cc-INT2-01}
{\rm cc}^2_{01,{\rm unp}}
\!\!\!&=&\!\!\!
- 8 K (2 - y) (1 - \widetilde y) (2 - \widetilde y)
\frac{\xi - \eta}{\eta} \Re{\rm e}\!
\left\{\!
{\cal C}_{{\cal V},{\rm unp}}({\cal F})
-
{\cal C}_{{\cal A},{\rm unp}}({\cal F})
-
\frac{\xi + \eta}{\xi} {\cal C}_{{\cal V},{\rm unp}} ({\cal F}_L)
\right\}\!
, \qquad\quad\\
{\rm cs}^2_{01,{\rm unp}}
\!\!\!&=&\!\!\!
8 \lambda K  y (1 - \widetilde y) (2 - \widetilde y)
\frac{\xi - \eta}{\eta}
\; \Im{\rm m}\!
\left\{\!
{\cal C}_{{\cal V},{\rm unp}}({\cal F})
-
{\cal C}_{{\cal A},{\rm unp}}({\cal F})
+
\frac{\xi + \eta}{\xi} {\cal C}_{{\cal V},{\rm unp}}({\cal F}_L)
\right\}\!
, \\
\label{cc-INT2-10}
{\rm cc}^2_{10,{\rm unp}}
\!\!\!&=&\!\!\!
8 \widetilde K \; \Re{\rm e}
\left\{
(2 - 2 y + y^2) (2 - 2 \widetilde y + \widetilde y^2)
\left(
\frac{\xi}{\eta} {\cal C}_{{\cal V}, {\rm unp}} ({\cal F})
+
{\cal C}_{{\cal A},{\rm unp}}({\cal F})
\right)
\right.
\nonumber \\
&&\qquad\qquad
\left.
- 8 (1 - y)(1 - \widetilde y)
\frac{\xi^2 - \eta^2}{\eta \xi} {\cal C}_{{\cal V},{\rm unp}} ({\cal F}_L)
\right\}
, \\
{\rm cc}^2_{12,{\rm unp}}
\!\!\!&=&\!\!\!
16 \widetilde K  (1 -  y)(1 - \widetilde y)
\frac{\xi + \eta}{\eta}
\Re{\rm e}
\left\{
{\cal C}_{{\cal V},{\rm unp}} ({\cal F})
-
{\cal C}_{{\cal A},{\rm unp}}({\cal F})
\right\}
, \\
{\rm sc}^2_{10,{\rm unp}}
\!\!\!&=&\!\!\!
8 \lambda \widetilde K y (2 - y) (2 - 2 \widetilde y + \widetilde y^2)
\Im{\rm m}
\left\{
{\cal C}_{{\cal V},{\rm unp}}({\cal F})
+
\frac{\xi}{\eta} {\cal C}_{{\cal A},{\rm unp}}({\cal F})
\right\}
, \\
{\rm cc}^2_{21,{\rm unp}}
\!\!\!&=&\!\!\!
- 8 K (2 - y) (1 - \widetilde y) (2 - \widetilde y)
\frac{\xi - \eta}{\eta} \Re{\rm e}\!
\left\{\!
{\cal C}_{{\cal V},{\rm unp}} ({\cal F})
+
{\cal C}_{{\cal A},{\rm unp}} ({\cal F})
-
\frac{\xi - \eta}{\xi} {\cal C}_{{\cal V},{\rm unp}} ({\cal F}_L)
\right\}\!
, \\
{\rm sc}^2_{21,{\rm unp}}
\!\!\!&=&\!\!\!
8 \lambda K  y (1 - \widetilde y) (2 - \widetilde y)
\frac{\xi - \eta}{\eta}
\; \Im{\rm m}\!
\left\{\!
{\cal C}_{{\cal V},{\rm unp}}({\cal F})
+
{\cal C}_{{\cal A},{\rm unp}}({\cal F})
+
\frac{\xi - \eta}{\xi} {\cal C}_{{\cal V},{\rm unp}}({\cal F}_L)
\right\},
\\
\label{cc-INT2-32}
{\rm cc}^2_{32,{\rm unp}}
\!\!\!&=&\!\!\!
16 \widetilde K  (1 -  y)  (1 - \widetilde y)
\frac{\xi - \eta}{\eta}
\Re{\rm e}
\left\{
{\cal C}_{{\cal V}, {\rm unp}}({\cal F})
+
{\cal C}_{{\cal A},{\rm unp}} ({\cal F})
\right\}
.
\end{eqnarray}
These expressions follow from Eqs.\ (\ref{Def-BHINT2}) -- (\ref{Def-BHINT2-b2})
and (\ref{Dec-BHVCS}). For the unpolarized lepton beam, they appear from the
application of the symmetry under the exchange $k\leftrightarrow -\ell_-$ and
$k^\prime \leftrightarrow \ell_+$, as discussed above in Section \ref{Sec-GenFun},
\begin{eqnarray}
\left\{
{\rm cc}^2_{01},
{\rm cc}^2_{10},
{\rm cc}^2_{12},
{\rm cc}^2_{21},
{\rm cc}^2_{32}
\right\}_{\rm unp}
=
\left\{
{\rm cc}^1_{01},
{\rm cc}^1_{10},
{\rm cc}^1_{12},
{\rm cc}^1_{21},
{\rm cc}^1_{32}
\right\}_{\rm unp}
\Big|
{\textstyle {y \leftrightarrow \widetilde y\hfill \atop \xi \to - \xi}}
\, .
\end{eqnarray}
One should  keep in mind that ${\cal C}_{\cal V}$ and ${\cal C}_{\cal A}$
are even and odd functions in $\xi-i 0$, respectively, and  $\widetilde K
(\xi, \widetilde y) = K(- \xi, y = \widetilde y)$. It turns out that the
remaining coefficients for the polarized lepton beam satisfy the following
symmetry relations
\begin{eqnarray}
\label{FC-Int-Sym}
\left\{  {\rm cs}^2_{01},{\rm sc}^2_{21} \right\}_{\rm unp}
=
\sqrt{\frac{\widetilde y-1}{1 - y}}
\left\{  {\rm cs}^1_{01},{\rm sc}^1_{21} \right\}_{\rm unp}
\Big|
{\textstyle { \atop \xi \to - \xi}}
\, , \qquad
{\rm sc}^2_{10,{\rm unp}}
=
- \sqrt{\frac{\widetilde y-1}{1 - y}}  {\rm sc}^1_{10,{\rm unp}}
\Big|
{\textstyle { \atop \xi \to - \xi}}
\, .
\end{eqnarray}

Next let us comment on the Fourier coefficients for the polarized target. From
the results (\ref{Def-trace-SV})-(\ref{Def-trace-RA}), we immediately deduce
several relations which allow one to obtain the Fourier coefficients from Eqs.\
(\ref{cc-INT1-01}) -- (\ref{cc-INT1-32}) and (\ref{cc-INT2-01}) -- (\ref{cc-INT2-32})
for the unpolarized case by using simple substitution rules:
\begin{eqnarray}
\label{Def-FC-INT-LP}
\left\{
{\rm cc}_{01},{\rm cc}_{10},{\rm cc}_{21}
\right\}_{\scriptscriptstyle\rm LP}^{\scriptscriptstyle\rm INT}
&\!\!\!=\!\!\!&
\left\{
{\rm cs}_{01}, {\rm sc}_{10},{\rm sc}_{21}
\right\}_{\rm unp}^{\scriptscriptstyle\rm INT}
\Big|_{
\Im{\rm m}{\cal C}_{\rm unp} \to \Re{\rm e}{\cal C}_{\scriptscriptstyle\rm LP}
}\, ,
\\
\left\{
{\rm cs}_{01},{\rm sc}_{10},{\rm cs}_{12},{\rm sc}_{21} ,{\rm sc}_{32}
\right\}_{\scriptscriptstyle\rm LP}^{\scriptscriptstyle\rm INT}
&\!\!\!=\!\!\!&
\left\{
{\rm cc}_{01}, {\rm cc}_{10},{\rm cc}_{12},{\rm cc}_{21},{\rm cc}_{32}
\right\}_{\rm unp}^{\scriptscriptstyle\rm INT}
\Big|_{
\Re{\rm e}{\cal C}_{\rm unp} \to \Im{\rm m}{\cal C}_{\scriptscriptstyle\rm LP}
}\, .
\nonumber
\end{eqnarray}
In case of the transversely polarized target, as before, one has an additional
decomposition in $\cos({\mit\Phi} - \phi) $ and $\sin({\mit\Phi} - \phi)$ and
has to replace $K$ ($\widetilde K$) by  $L$ ($\widetilde L$)
\begin{eqnarray}
\label{Def-FC-INT-TP}
\left\{
{\rm cc}_{01},{\rm cc}_{10},{\rm cc}_{21}
\right\}_{{\scriptscriptstyle\rm TP}+}^{\scriptscriptstyle\rm INT}
&\!\!\!=\!\!\!&
\cos({\mit\Phi} - \phi)
\left\{
{\rm cs}_{01},{\rm sc}_{10},{\rm sc}_{21}
\right\}_{\rm unp}^{\scriptscriptstyle\rm INT}
\Big|_{
\Im{\rm m}{\cal C}_{\rm unp} \to \Re{\rm e}{\cal C}_{{\scriptscriptstyle\rm TP}+}
\atop
K \to L\, , \widetilde K \to \widetilde L \hspace{0.7cm}
}\, ,
\\
\left\{
{\rm cs}_{01},{\rm sc}_{10},{\rm sc}_{21}
\right\}_{{\scriptscriptstyle\rm TP}-}^{\scriptscriptstyle\rm INT}
&\!\!\!=\!\!\!&
\sin({\mit\Phi} - \phi)
\left\{
{\rm cs}_{01}, {\rm sc}_{10},{\rm sc}_{21}
\right\}_{\rm unp}^{\scriptscriptstyle\rm INT} \Big|_
{ \Im{\rm m}{\cal C}_{\rm unp} \to -\Re{\rm e}{\cal C}_{{\scriptscriptstyle\rm TP}-}
\atop
K \to L\, , \widetilde K \to \widetilde L \hspace{0.7cm}
}\, ,
\nonumber\\
\left\{
{\rm cs}_{01},{\rm sc}_{10},{\rm cs}_{12},{\rm sc}_{21},{\rm sc}_{32}
\right\}_{{\scriptscriptstyle\rm TP}+}^{\scriptscriptstyle\rm INT}  &\!\!\! =\!\!\! &
\cos({\mit\Phi} - \phi)
\left\{
{\rm cc}_{01}, {\rm cc}_{10}, {\rm cc}_{12}, {\rm cc}_{21},{\rm cc}_{32}
\right\}_{\rm unp}^{\scriptscriptstyle\rm INT}
\Big|_{
\Re{\rm e}{\cal C}_{\rm unp} \to \Im{\rm m}{\cal C}_{{\scriptscriptstyle\rm TP}+}
\atop
K \to L\, , \widetilde K \to \widetilde L \hspace{0.7cm}
}\, ,
\nonumber\\
\left\{
{\rm cc}_{01},{\rm cc}_{10},{\rm cc}_{12},{\rm cc}_{23} ,{\rm cc}_{32}
\right\}_{{\scriptscriptstyle\rm TP}-}^{\scriptscriptstyle\rm INT}  &\!\!\! =\!\!\! &
\sin({\mit\Phi} - \phi)
\left\{
{\rm cc}_{01},{\rm cc}_{10}, {\rm cc}_{12}, {\rm cc}_{23} ,{\rm cc}_{32}
\right\}_{\rm unp}^{\scriptscriptstyle\rm INT} \Big|_
{ \Re{\rm e} {\cal C}_{\rm unp} \to \Im{\rm m}{\cal C}_{{\scriptscriptstyle\rm TP}-}
\atop
K \to L\, , \widetilde K \to \widetilde L \hspace{0.7cm}
}\, .
\nonumber
\end{eqnarray}
Applying the relation (\ref{Rel-FC-INT-Tw2}), the remaining
nonvanishing Fourier coefficients, i.e., ${\rm ss}_{21}^{\scriptscriptstyle\rm INT}$,
${\rm sc}_{12}^{\scriptscriptstyle\rm INT} $ ${\rm cs}_{21}^{\scriptscriptstyle\rm INT}$,
${\rm cs}_{23}^{\scriptscriptstyle\rm INT} $ for LP and ${\scriptscriptstyle\rm TP}+$ as
well as ${\rm ss}_{12}^{\scriptscriptstyle\rm INT}$,
${\rm ss}_{21}^{\scriptscriptstyle\rm INT} $, ${\rm ss}_{23}^{\rm INT}$
and ${\rm cs}_{21}^{\scriptscriptstyle\rm INT} $ for ${\scriptscriptstyle\rm TP}-$
are easily found. Note also that
$$
\widetilde K(\xi,\widetilde y)
\approx
\sqrt{(\widetilde y - 1)/(1 - y)} K(- \xi,y)
\, .
$$

\subsubsection{Squared Bethe-Heitler amplitude}

The exact expressions for the Fourier coefficients of the pure BH term
(\ref{Dec-BHsq}) are extremely lengthy and, therefore, will not be displayed
here in analytic form. In the following, we limit ourselves instead to leading
terms in the asymptotic expansion as $Q^2/\xi \to \infty$. Namely,
\begin{eqnarray}
{\rm cc}_{00,{\rm unp}}^{11}
\!\!\!&\approx&\!\!\!
- 2 \frac{1 - \eta}{1 + \eta} \left(1 + \frac{\xi}{\eta}\right)
(1 - y)
\Bigg\{
(2 - 2 y + y^2) (2 - 2 \widetilde y + \widetilde{y}^2)
\left(1 + \frac{\xi^2}{\eta^2}\right)
\\
\!\!\!&-&\!\!\!
8 (1 - y)(1 - \widetilde y)
\left(1 - \frac{\xi^2}{\eta^2}\right)
\Bigg\}
\left\{
\left( 1 - \frac{\Delta^2_{\rm min}}{\Delta^2} \right)
\left( F_1^2 - \frac{\Delta^2}{4M^2}F_2^2 \right)
+
\frac{2\eta^2}{1-\eta^2} (F_1 + F_2)^2
\right\}
\, , \nonumber\\
{\rm cc}_{02,{\rm unp}}^{11}
\!\!\!&\approx&\!\!\!
2 \frac{1 - \eta}{1 + \eta} \left( 1 + \frac{\xi}{\eta} \right)
\left( 1 - \frac{\xi^2}{\eta^2} \right)
\left(1 - \frac{\Delta^2_{\rm min}}{\Delta^2}\right)  (1 - y)
\Bigg\{(2 - 2 y + y^2) (2 - 2 \widetilde y + \widetilde{y}^2)
\\
&+&\!\!\!
8 (1 - y) (1 - \widetilde y)
\Bigg\}
\left(F_1^2 - \frac{\Delta^2}{4M^2}F_2^2 \right)
\, , \nonumber\\
{\rm cc}_{11,{\rm unp}}^{11}
\!\!\!&\approx&\!\!\!
- 4 \frac{1 - \eta}{1 + \eta} \left(1 + \frac{\xi}{\eta}\right)
\frac{\sigma}{\eta}
\sqrt{(1 - y)(1 - \widetilde y)\left(\xi^2 - \eta^2\right)}
\\
&\times&\!\!\!
(1 - y) (2 - y) (2 - \widetilde y)
\left\{ \left(1 - 3 \frac{\xi}{\eta} \right)
\left( 1 - \frac{\Delta^2_{\rm min}}{\Delta^2} \right)
\left( F_1^2 - \frac{\Delta^2}{4M^2} F_2^2 \right)
-
\frac{4\xi\eta}{1 - \eta^2} (F_1 + F_2)^2
\right\}
, \nonumber\\
{\rm cc}_{13,{\rm unp}}^{11}
\!\!\!&\approx&\!\!\!
4 \frac{1 - \eta}{1 + \eta} \left( 1 + \frac{\xi}{\eta} \right)^2
\left( 1 - \frac{\Delta^2_{\rm min}}{\Delta^2} \right)
\frac{\sigma}{\eta}
\sqrt{(1 - y)(1 - \widetilde y) \left( \xi^2 - \eta^2 \right)}
\, , \\
&\times&\!\!\!
(1 - y) (2 - y) (2 - \widetilde y)
\left( F_1^2 - \frac{\Delta^2}{4M^2}F_2^2 \right)
\, , \nonumber\\
{\rm cc}_{20,{\rm unp}}^{11}
\!\!\!&\approx&\!\!\!
- 4 \frac{1 - \eta}{1 + \eta}
\left(1 - \frac{\xi}{\eta}\right) \left(1 - \frac{\xi^2}{\eta^2}\right)
\left( 1 - \frac{\Delta^2_{\rm min}}{\Delta^2} \right)
(1 - y)^2 (1-\widetilde y)
\left(F_1^2 - \frac{\Delta^2}{4M^2}F_2^2 \right)
\, , \\
{\rm cc}_{22,{\rm unp}}^{11}
\!\!\!&\approx&\!\!\!
8 \frac{1 - \eta}{1 + \eta}
\left(1 + \frac{\xi}{\eta}\right) \left(1 - \frac{\xi^2}{\eta^2}\right)
(1 - y)^2 (1 - \widetilde y)
\\
&\times&\!\!\!
\Bigg\{
\left(1 - \frac{\Delta^2_{\rm min}}{\Delta^2}\right)
\left(F_1^2 - \frac{\Delta^2}{4M^2}F_2^2 \right)
+
\frac{2\eta^2}{1 - \eta^2} (F_1 + F_2)^2
\Bigg\}
\, . \nonumber
\end{eqnarray}
The remaining coefficients  are expressed via the already known ones
\begin{eqnarray}
{\rm cc}_{24,{\rm unp}}^{11}
\!\!\!&\approx&\!\!\!
\frac{(\eta + \xi)^2}{(\eta - \xi)^2} {\rm cc}_{20,{\rm unp}}^{11}
\, , \qquad
{\rm ss}_{11,{\rm unp}}^{11}
\approx
{\rm cc}_{11,{\rm unp}}^{11}
+
2 \frac{\eta - \xi}{\eta + \xi} {\rm cc}_{13,{\rm unp}}^{11}
\, , \\
{\rm ss}_{13,{\rm unp}}^{11}
\!\!\!&\approx&\!\!\!
{\rm cc}_{13,{\rm unp}}^{11}
\, , \qquad\qquad\quad\;\;
{\rm ss}_{22,{\rm unp}}^{11}
\approx
{\rm cc}_{22,{\rm unp}}^{11}
\, , \qquad\qquad\qquad\quad
{\rm ss}_{24,{\rm unp}}^{11}
\approx
{\rm cc}_{24,{\rm unp}}^{11}
\, . \nonumber
\end{eqnarray}
One should realize that this expansion is only valid if one stays away
from kinematical boundaries, e.g., $y \ll 1$ is required. The reason
for this is that the leading terms vanish with $(1 - y)$, and subleading
corrections become important as $y \to 1$. In contrast  to the DVCS case,
no cancellations occur between the numerator and the denominator of the BH
amplitude squared, so that, in general, the Fourier decomposition goes as
high as up to the forth-order harmonics.

We also note that the Fourier coefficients for the second BH-amplitude
squared simply follow from the symmetry under the exchange $k \leftrightarrow -
\ell_- $ and $k^\prime \leftrightarrow \ell_+ $. Since we extracted a power of
$\xi$ in front of the squared BH amplitude (\ref{Dec-BHsq}), we obtain the
substitution rule
\begin{eqnarray}
{\rm cc}_{nm}^{22} = - {\rm cc}_{nm}^{11}\Big|
{\textstyle {y \leftrightarrow \widetilde y \hfill \atop \xi \to - \xi}}
\quad\mbox{and}\quad
{\rm ss}_{nm}^{22} = - {\rm ss}_{nm}^{11}\Big|
{\textstyle {y \leftrightarrow \widetilde y \hfill \atop \xi \to - \xi}}
\, ,
\end{eqnarray}
while the remaining variables $\{\eta,\Delta^2,Q^2\}$ are kept unchanged.
For the interference term of the first and second BH amplitudes,  the
Fourier coefficients are
\begin{eqnarray}
{\rm cc}_{00,{\rm unp}}^{12}
\!\!\!&\approx&\!\!\!
8 \frac{1 - \eta}{1 + \eta} \frac{\xi}{\eta}
\left( 1 - \frac{\xi^2}{\eta^2} \right)
(1 - y) (2 - y) (1 - \widetilde y) (2 - \widetilde y)
\\
&\times&\!\!\!
\left\{
\left( 1 - \frac{\Delta^2_{\rm min}}{\Delta^2} \right)
\left( F_1^2 - \frac{\Delta^2}{4M^2}F_2^2 \right)
+
\frac{2\eta^2}{1 - \eta^2} (F_1 + F_2)^2
\right\}
\, , \nonumber
\\
{\rm cc}_{02,{\rm unp}}^{12}
\!\!\!&\approx&\!\!\!
8 \frac{1 - \eta}{1 + \eta}
\left(1 + \frac{\xi}{\eta}\right) \left(1 - \frac{\xi^2}{\eta^2}\right)
(1 - y) (2 - y) (1 - \widetilde y) (2 - \widetilde y)
\left(1 - \frac{\Delta^2_{\rm min}}{\Delta^2}\right)
\left(F_1^2 - \frac{\Delta^2}{4M^2}F_2^2 \right)
\! ,
\nonumber\\
\\
{\rm cc}_{11,{\rm unp}}^{12}
\!\!\!&\approx&\!\!\!
8 \frac{1 - \eta}{1 + \eta}
\frac{\sigma}{\eta}
\sqrt{(1-y)(1-\widetilde y)\left(\xi^2-\eta^2\right)}
\\
&\times&\!\!\!
\Bigg\{
(2 - 2 y + y^2) (2 - 2 \widetilde y + \widetilde y^2)
\Bigg[
\frac{\xi^2}{\eta^2}
\left( 1 - \frac{\Delta^2_{\rm min}}{\Delta^2} \right)
\left( F_1^2 - \frac{\Delta^2}{4M^2}F_2^2 \right)
+
\frac{\xi^2 + \eta^2}{1 - \eta^2} \left( F_1 + F_2 \right)^2
\Bigg]
\nonumber\\
&-&\!\!\!
(1 - y)(1 - \widetilde y) \left( 1 - \frac{\xi^2}{\eta^2} \right)
\Bigg[
9 \left( 1 - \frac{\Delta^2_{\rm min}}{\Delta^2} \right)
\left( F_1^2 - \frac{\Delta^2}{4M^2}F_2^2 \right)
+ 10 \frac{\eta^2}{1 - \eta^2} \left( F_1 + F_2 \right)^2
\Bigg]
\Bigg\}
\, , \nonumber\\
{\rm cc}_{20,{\rm unp}}^{12}
\!\!\!&\approx&\!\!\!\!
- 8 \frac{1 - \eta}{1 + \eta}
\left(1 - \frac{\xi}{\eta}\right)
\\
&\times&\!\!\!\left(1 - \frac{\xi^2}{\eta^2}\right)
\left(1 - \frac{\Delta^2_{\rm min}}{\Delta^2}\right)
(1 - y)(2 - y) (1-\widetilde y) (2 - \widetilde y)
\left( F_1^2 - \frac{\Delta^2}{4M^2}F_2^2 \right) \! ,
\nonumber\\
{\rm cc}_{22,{\rm unp}}^{12}
\!\!\!&\approx&\!\!\!
8 \frac{1 - \eta}{1 + \eta} \frac{\xi}{\eta}
\left(1 - \frac{\xi^2}{\eta^2}\right)
\\
&\times&\!\!\!
(1 - y)(2 - y) (1 - \widetilde y) (2 - \widetilde y)
\Bigg\{
\left(1 - \frac{\Delta^2_{\rm min}}{\Delta^2}\right)
\left(F_1^2 - \frac{\Delta^2}{4M^2}F_2^2 \right)
+
\frac{2\eta^2}{1 - \eta^2} (F_1 + F_2)^2
\Bigg\}
\, , \nonumber\\
{\rm ss}_{11,{\rm unp}}^{12}
\!\!\!&\approx&\!\!\!
{\rm cc}_{11,{\rm unp}}^{12}
+
8 \frac{1 - \eta}{1 + \eta}
\frac{\sigma}{\eta}
\sqrt{(1 - y)(1 - \widetilde y) \left(\xi^2 - \eta^2\right)}
\left( 1 - \frac{\xi^2}{\eta^2} \right)
\\
&\times&\!\!\!
\left\{
(2 - 2 y + y^2) (2 - 2 \widetilde y + \widetilde y^2) + 8 (1 - y)(1 - \widetilde y)
\right\}
\left( 1 - \frac{\Delta^2_{\rm min}}{\Delta^2} \right)
\left( F_1^2 - \frac{\Delta^2}{4M^2}F_2^2 \right)
\, ,
\nonumber\\
{\rm ss}_{22,{\rm unp}}^{12}
\!\!\!&\approx&\!\!\!
{\rm cc}_{22,{\rm unp}}^{12}
\, .
\end{eqnarray}
Again, under the interchanges $k \leftrightarrow - \ell_-$ and $k^\prime
\leftrightarrow \ell_+$,  both amplitudes are odd, i.e.,
${\cal T}_{\scriptscriptstyle\rm BH_1} \leftrightarrow -
{\cal T}_{\scriptscriptstyle\rm BH_2}$, and, hence,  their interference
term is invariant. Thus, the Fourier coefficients must satisfy the relation
\begin{eqnarray}
{\rm cc}_{nm}^{12} = - {\rm cc}_{mn}^{12}\Big|
{\textstyle {y \leftrightarrow \widetilde y \hfill \atop \xi \to - \xi}}
\quad\mbox{and}\quad
{\rm ss}_{nm}^{12} = - {\rm ss}_{mn}^{12}\Big|
{\textstyle {y \leftrightarrow \widetilde y \hfill \atop \xi \to - \xi}}
\, ,
\end{eqnarray}
which is indeed the case.

\subsubsection{Single-spin asymmetries}

Having introduced a complete angular dependence of the cross section, we will now
have a closer look on the single-spin asymmetries, in particular, the beam-spin
asymmetry for the proton
\begin{eqnarray}
d\sigma^{\uparrow} - d\sigma^{\downarrow}
\sim
\left(
\pm {\cal T}_{{\scriptscriptstyle\rm BH}_1}^\ast
+
{\cal T}^\ast_{{\scriptscriptstyle\rm BH}_2}
\right)
\Im{\rm m}{\cal T}_{\scriptscriptstyle\rm VCS}
+
\cdots
\, .
\end{eqnarray}
Potentially, the interference term could be contaminated by the
imaginary part $\Im{\rm m} {\cal T}_{\scriptscriptstyle\rm VCS}
{\cal T}_{\scriptscriptstyle\rm VCS}^\dagger$ arising from the interference
of twist-two and twist-three Compton form factors. Just like in the DVCS
case \cite{BelMulKir01}, we expect, that this contribution can be safely
neglected, assuming the smallness of three-particle correlations. As we
mentioned above, in the leading order of QCD perturbation theory, the single
spin asymmetries are directly proportional to the linear combination of GPDs
\begin{eqnarray}
\Im{\rm m} \, {\cal F}
=
\pi \sum_{q} Q_q^2
\left\{
F^q (\xi, \eta, \Delta^2) \mp F^q ( - \xi, \eta, \Delta^2)
\right\}
+ {\cal O} (\alpha_s)
\, .
\end{eqnarray}
where $(+)$ $-$ applies for the (axial-) vector-type GPDs. For instance, eight
leading-twist observables are measurable in the beam-spin asymmetry, which
come in pairs from the interference of the VCS with the first and second BH
amplitudes\footnote{For brevity, throughout this section, we omit in the
Fourier coefficients the subscript ``unp'', which refers to the unpolarized
target.}:
${\rm cs}^1_{01}$,
${\rm sc}^1_{10}$, ${\rm sc}^1_{21}$, ${\rm cs}^1_{12}$ as well as
${\rm cs}^2_{01}$, ${\rm sc}^2_{10}$, ${\rm sc}^2_{21}$, ${\rm cs}^2_{12}$,
see Eq.\ (\ref{FC-Int-Sym}). However, they depend only on two different
linear combinations (\ref{Int-dif-Com}) of GPDs:
\begin{eqnarray}
\label{LinComBSA}
F_1 ({\cal H} + {\cal H}_L)
-
\frac{\Delta^2}{4 M_N^2} F_2 ({\cal E} + {\cal E}_L)
\, , \qquad
\eta (F_1 + F_2) \widetilde {\cal H} - \frac{\eta}{\xi}
\left(
F_1 {\cal H}_L - \frac{\Delta^2}{4 M_N^2} F_2 {\cal E}_L
\right)
\, .
\end{eqnarray}
Consequently, there exist six constraints among the whole set of coefficients,
which can be expressed as
\begin{eqnarray}
\label{FouCoeRel}
&&{\rm sc}^1_{10}
\simeq
-
\frac{
(2 - y)(2 - 2 \widetilde  y + \widetilde  y^2)
}{
2(1 - y)(2 - \widetilde y)
}
\frac{
\sqrt{(1 - y)(\xi + \eta)}
}{
\sqrt{(1-\widetilde y) (\xi - \eta) }
}
\left\{
{\rm cs}^1_{01} + \frac{ \xi - \eta}{\xi + \eta} {\rm sc}^1_{21}
\right\}
\, , \nonumber\\
&&\frac{{\rm cs}^2_{01}}{{\rm sc}^1_{21}}
\simeq
\frac{{\rm sc}^2_{10}}{{\rm sc}^1_{10}}
\simeq
\frac{{\rm sc}^2_{21}}{{\rm cs}^1_{01}}
\simeq
-
\frac{\sqrt{(1 - \widetilde y) (\xi - \eta)}}{\sqrt{(1 - y)(\xi + \eta)}}
\, ,
\end{eqnarray}
supplemented by the relation (\ref{Rel-FC-INT-Tw2}), i.e., ${\rm cs}^{1}_{21}
\simeq - {\rm sc}^{1}_{21}$ and ${\rm cs}^{2}_{21} \simeq - {\rm sc}^{2}_{21}$.
Another consequence of Eq.\ (\ref{LinComBSA}) is that the beam-spin asymmetry
gives no handle on the Callan-Gross relation, i.e., the longitudinal Compton
form factors cannot be separated from the leading ones. The Fourier coefficients
are projected out by taking the following moments when integrated over the
scattering and azimuthal angles:
\begin{eqnarray}
\label{Def-BSA}
&&\!\!\!\!\!\!\!\!\!
\sin\phi \to {\rm sc}_{10}^1
\, , \
\cos\theta_\ell \sin{\varphi_\ell} \rightarrow {\rm cs}_{01}^1
\, , \
\cos\theta_\ell \cos{\varphi_\ell} \sin(2\phi) \rightarrow {\rm sc}^1_{21}
\, , \
\cos\theta_\ell \sin{\varphi_\ell}  \cos(2\phi) \rightarrow {\rm cs}^1_{21}
\, , \nonumber\\
&&\!\!\!\!\!\!\!\!\!
\sin\phi_\ell \rightarrow {\rm sc}_{10}^2
\, , \
\cos\theta_\ell \sin{\phi} \rightarrow {\rm cs}_{01}^2
\, , \
\cos\theta_\ell \sin(2\varphi_\ell) \cos{\phi}  \rightarrow {\rm sc}^2_{21}
\, , \
\cos\theta_\ell \cos(2\varphi_\ell) \sin{\phi}  \rightarrow {\rm cs}^2_{21}
\, , \nonumber
\end{eqnarray}
where the weight in the first (second) row is even (odd) under the
reflection $\theta_{\ell} \to \pi - \theta_{\ell}$ and $\varphi_{\ell}
\to \pi + \varphi_{\ell}$. In the same line of thinking, one can study
the Fourier coefficients for the single target-spin asymmetries. Because
of the substitution rules (\ref{Def-FC-INT-LP}) and (\ref{Def-FC-INT-TP}),
the number of the Compton form factors will be the same as for the case of
charge and angular asymmetries discussed below. Of course, single
target-spin asymmetries are given by the imaginary part of new linear
combinations of GPDs with a characteristic angular dependence.

From Eq.\ (\ref{FouCoeRel}), one can conclude that the same information
about GPDs is obtained by taking the appropriate moments in $\phi$ or
$\varphi_\ell$. However, the size of the complementary beam-spin
asymmetries can vary. Moreover, if one takes the asymmetry from
the interference with the second BH amplitude, the weight must be odd
and, thus, we have no contamination from the squared VCS amplitude. To
suppress the squared BH amplitude, one can  integrate over the region
$\pi/4 \le \theta_\ell \le 3\pi/4$ and form alternatively the $\sin\phi$
or $\sin \varphi_\ell$ moments. This procedure extracts  the coefficients
${\rm sc}^1_{10}$ and ${\rm sc}^2_{10}$, respectively, cf.\ Eqs.\
(\ref{Dec-BHVCS}), (\ref{Dec-FC-gen}), and (\ref{Rel-FC-INT-Tw2}). Thus,
the beam-spin asymmetry
\begin{eqnarray}
\left\{
A_{\rm LU}^{\sin\phi} \atop A_{\rm LU}^{\sin\varphi_{{\ell}}}
\right\}
&\!\!\!=\!\!\!&
\frac{1}{{\cal N}}
\int_{\pi/4}^{3\pi/4}\! d \theta_{\ell}
\int_{0}^{2\pi}\! d \varphi_\ell
\int_{0}^{2\pi}\! d \phi
\left\{ 2 \sin\phi \atop 2 \sin\varphi_\ell \right\}
\frac{d \sigma^\uparrow - d\sigma^\downarrow}{d {\mit\Omega}_{\ell}d\phi}
\nonumber\\
\!\!\!&\sim&\!\!\!
\Im{\rm m}
\left\{
F_1 {\cal H}
-
\frac{\Delta^2}{4 M_N^2} F_2 {\cal E}
+
\xi (F_1 + F_2) \widetilde {\cal H}
\right\} \, ,
\end{eqnarray}
with the normalization factor given by
\begin{eqnarray*}
{\cal N}
=
\int_{\pi/4}^{3\pi/4}\! d \theta_{\ell}
\int_{0}^{2\pi}\! d \varphi_\ell
\int_{0}^{2\pi}\! d \phi
\frac{
d \sigma^\uparrow + d \sigma^\downarrow
}{d{\mit\Omega}_{\ell}d\phi}
\, ,
\end{eqnarray*}
is analogous to that defined in the case of space- and timelike DVCS.
For the proton target,  it is mainly sensitive to the contribution $F_1
\Im{\rm m} {\cal H} $ and, therefore, it may not be useful for the estimate
of the other two Compton form factors. The magnitude of the asymmetry can be
as large as 20\% depending on assumed models of GPDs.

\subsubsection{Charge and angular asymmetries}

Let us now comment on charge and angular asymmetries, in which the
Fourier coefficients ${\rm cc}^1_{01}$, ${\rm cc}^1_{10}$,
${\rm cc}^1_{12} \simeq {\rm ss}^1_{12}$, ${\rm cc}^1_{21} \simeq
{\rm ss}^1_{21}$, and ${\rm cc}^1_{32} \simeq {\rm ss}^1_{32}$ as
well as the complementary set due to  the second interference term are
accessible.  The Fourier coefficients of both interference
terms are related by
\begin{eqnarray}
\label{FouCoeRel-cc}
\frac{{\rm cc}^2_{01}}{{\rm cc}^1_{21}} \simeq
\frac{{\rm cc}^2_{10}}{{\rm cc}^1_{10}} \simeq
\frac{{\rm cc}^2_{12}}{{\rm cc}^1_{32}} \simeq
\frac{{\rm cc}^2_{21}}{{\rm cc}^1_{01}} \simeq
\frac{{\rm cc}^2_{32}}{{\rm cc}^1_{12}} \simeq
- \frac{\sqrt{(1-\widetilde y) (\xi - \eta) }}{\sqrt{(1 - y)(\xi + \eta)}}
\, .
\end{eqnarray}
There are now three independent Compton form factors ${\cal C}_{\cal
V,{\rm unp}}({\cal F})$, ${\cal C}_{\cal V,{\rm unp}}({\cal F}_L)$, and
${\cal C}_{\cal A,{\rm unp}}({\cal F})$, introduced in Eq.\ (\ref{Int-dif-Com}).
Consequently, there exist two constraints among five nontrivial Fourier
coefficients. If the Callan-Gross relation is assumed to be fulfilled, this
number increases to three:
\begin{eqnarray}
\label{FouCoeRelCG}
{\rm cc}^1_{10}
\!\!\!&\simeq&\!\!\!
-
\frac{(2 - 2 y +  y^2)(2 - 2 \widetilde y
+
\widetilde  y^2)}{2(2 - y)(1 - \widetilde y)(2 - \widetilde y)}
\frac{\sqrt{(1-\widetilde y) (\xi - \eta) }}{\sqrt{(1 - y)(\xi + \eta)}}
\left\{
\frac{\xi + \eta}{\xi - \eta} {\rm cc}^1_{01} + {\rm cc}^1_{21}
\right\}
\, , \\
{\rm cc}^1_{12}
\!\!\!&\simeq&\!\!\!
\frac{- 2 (1 - y)}{(2 - y)(2 - \widetilde y)}
\frac{\sqrt{(1-\widetilde y) (\xi - \eta) }}{\sqrt{(1 - y)(\xi + \eta)}}
\, {\rm cc}^1_{01}
\, , \nonumber\\
{\rm cc}^1_{32}
\!\!\!&\simeq&\!\!\!
\frac{-2(1-\widetilde y)}{(2 - y)(2 - \widetilde y)}
\frac{\sqrt{(1 - y)(\xi + \eta)}}{\sqrt{(1-\widetilde y)(\xi - \eta)}}
\,
{\rm cc}^1_{21}
\, . \nonumber
\end{eqnarray}
The  measurements of charge and angular asymmetries can be combined with double
spin-flip experiments, which offer information on a new superposition of
GPDs. As in the case of the beam-spin asymmetry, the number of independent
Compton form factors is, however, reduced to two as a consequence of
Eqs. \ (\ref{Def-FC-INT-LP}) and (\ref{Def-FC-INT-TP}).

The charge odd part is given by the interference of the first BH amplitude
with the VCS amplitude as well as with the second BH amplitude. For unpolarized
settings, the charge asymmetry yields
\begin{eqnarray}
d\sigma^{+} - d\sigma^{-}
\sim
\Re{\rm e} \left(
{\cal T}^\ast_{{\scriptscriptstyle\rm BH}_1} {\cal T}_{{\scriptscriptstyle\rm BH}_2}
+
{\cal T}^\ast_{{\scriptscriptstyle\rm BH}_1} {\cal T}_{\scriptscriptstyle\rm VCS}
\right)
\, .
\end{eqnarray}
Taking now the moments with respect to the solid angle of the final state
lepton pair that are even under the reflection, e.g., by means of the weight
function
\begin{eqnarray}
w^{\rm even}(\phi_\ell,\theta_\ell)
=
\{
1 , \
\cos\phi_\ell \cos\theta_\ell , \
\cos(2\phi_\ell) , \
\sin\phi_\ell \cos\theta_\ell , \
\sin(2\phi_\ell) , \ \dots \,
\}
\, ,
\end{eqnarray}
the contaminating BH interference drops out and we find
\begin{eqnarray}
\int d {\mit\Omega}_{\ell}\, w^{\rm even} (\phi_\ell,\theta_\ell)
\frac{d \sigma^{+} - d \sigma^{-}}{d {\mit\Omega}_\ell}
\sim
\int d {\mit\Omega}_{\ell}\, w^{\rm even} (\phi_\ell,\theta_\ell)\,
\Re{\rm e} \left(  {\cal T}^\ast_{{\scriptscriptstyle\rm BH}_1}
{\cal T}_{\scriptscriptstyle\rm VCS} \right)
\, .
\end{eqnarray}
Depending on  the choice of the weight function, this average will
provide Fourier series in $\phi$, where the zeroth, first, second and
third harmonics provide access to all leading-twist coefficients of
the first interference term. In case when only the lepton beam of a
specific charge---positive or negative---is available, on can form
asymmetries with an odd weight
\begin{eqnarray}
&&w^{\rm odd}(\phi_\ell,\theta_\ell)
\\
&&\quad=
\{
\cos\theta_\ell , \
\cos\varphi_\ell , \
\cos(2\varphi_\ell) \cos\theta_\ell , \
\cos(3\varphi_\ell) , \
\sin\varphi_\ell , \
\sin(2\varphi_\ell) \cos\theta_\ell , \
\sin(3\varphi_\ell) , \
\dots \,
\}
\, , \nonumber
\end{eqnarray}
so that the squared amplitudes exactly drop out
\begin{eqnarray}
\int d {\mit\Omega}_{\ell}\, w^{\rm odd} (\phi_\ell,\theta_\ell)
\frac{d\sigma}{d {\mit\Omega}_{\ell}}
\sim
\int d {\mit\Omega}_{\ell} \, w^{\rm odd}(\phi_\ell,\theta_\ell) \,
\left\{
\pm {\cal T}^\ast_{{\scriptscriptstyle\rm BH}_1}
{\cal T}_{{\scriptscriptstyle\rm BH}_2}
+
\Re{\rm e} \left( {\cal T}^\ast_{{\scriptscriptstyle\rm BH}_2}
{\cal T}_{\scriptscriptstyle\rm VCS} \right)
\right\}
\, .
\end{eqnarray}
After the subtraction of the remaining BH interference is done, it allows to
measure the leading twist-two Fourier coefficients. The procedure we just outlined
may give a handle on the real part of the Compton form factors. If both types of
the lepton-beam charge are available, the BH contribution disappears  in the charge
even combination and yields
\begin{eqnarray}
\int d {\mit\Omega}_{\ell}\, w^{\rm odd}(\phi_\ell,\theta_\ell)
\frac{d\sigma^{+} + d\sigma^{-}}{d {\mit\Omega}_\ell}
\sim
\int d {\mit\Omega}_{\ell}\, w^{\rm odd}(\phi_\ell,\theta_\ell)\,
\Re{\rm e}
\left(
{\cal T}^\ast_{{\scriptscriptstyle\rm BH}_2}
{\cal T}_{\scriptscriptstyle\rm VCS}
\right)
\, .
\end{eqnarray}

To illustrate the feasibility of the subtraction procedure, consider the
following charge and angular asymmetries
\begin{eqnarray}
\left\{
A_{\rm CA}^{\cos\varphi_{{\ell}}}
\atop
A^{\cos\varphi_{{\ell}}}
\right\}
&\!\!\!=\!\!\!&
\frac{1}{{\cal N}}
\int_{\pi/4}^{3\pi/4}\! d \theta_{\ell}
\int_{0}^{2\pi}\! d \phi
\int_{0}^{2\pi}\! d \varphi_\ell \;
2 \cos\varphi_\ell
\left\{
\left( d \sigma^+ + d \sigma^- \right)/2 d {\mit\Omega}_{\ell}d\phi
\atop
d \sigma^-/d {\mit\Omega}_{\ell}d\phi
\right\}
\, ,
\end{eqnarray}
integrated with the weight $2 \cos\varphi_\ell$. In both cases, the normalization
factor is given by
\begin{eqnarray*}
{\cal N}
=
\int_{\pi/4}^{3\pi/4}\! d\theta_{\ell}
\int_{0}^{2\pi}\! d \phi
\int_{0}^{2\pi}\! d \varphi_\ell\,
\frac{d \sigma^-}{d {\mit\Omega}_{\ell} d \phi}
\, .
\end{eqnarray*}
These asymmetries project the Fourier coefficient ${\rm cc}_{10}^2$ of
the second BH-VCS interference term, which, in the absence of ${\cal F}_L$,
is proportional to
\begin{eqnarray}
\label{Exp-cc210}
{\rm cc}_{10}^2 \sim \Re{\rm e}\left\{
-
\frac{\xi}{\eta} F_1 {\cal H}
+
\frac{\xi}{\eta} \frac{\Delta^2}{4M_N^2} F_2 {\cal E}
-
\eta
\left(F_1  + F_2 \right) \widetilde {\cal H}
\right\}
\, ,
\end{eqnarray}
making use of Eq.\ (\ref{cc-INT2-10}). One realizes that ${\cal H}$ is now
suppressed by a factor $\xi/\eta$ and, thus, with decreasing $|\xi|$ the
contribution of $\widetilde {\cal H}$ starts to be important. We remark that
as in case of DVCS, the charge and angular asymmetries might be contaminated
stronger by twist-three effects than the beam-spin asymmetries. These coefficients
are mainly of kinematical origin, i.e., they are expressed by twist-two GPDs
and generate Fourier coefficients ${\rm cc}^{\scriptscriptstyle\rm INT}_{00},
\ {\rm ss}^{\scriptscriptstyle\rm INT}_{11} \simeq
{\rm cc}^{\scriptscriptstyle\rm INT}_{11}$, and ${\rm ss}^{\rm INT}_{22}
\simeq {\rm cc}^{\scriptscriptstyle\rm INT}_{22}$ that do not necessarily
vanish at the kinematical boundaries.

\subsubsection{Advantages of lepton pair electroproduction}

In the preceding few sections, we have studied the process $e N \to e' N' \ell
\bar\ell$. The structure of the cross section was obtained to the leading power
in the hard momentum. We did not discuss here the power suppressed twist-three
contributions, which  generate further harmonics in the cross section. For
instance, they induce off-diagonal elements in the coefficient matrix of the
squared VCS amplitude, e.g., ${\rm cc}_{nm}$ with $n \neq m$ etc. However, they
will not contaminate already existing Fourier harmonics, e.g., ${\rm cc}_{nn}$
etc. The process of the lepton pair production is the most favorable for
experimental exploration of GPDs for a number of reasons:
\begin{itemize}
\item It is a clean electromagnetic process which does not involve other
unknown non-perturbative function and, thus, has no contamination from other
uncontrollable sources.
\item The virtuality of the final state photon allows to disentangle the
dependence of GPDs on both scaling variables and thus constrain the angular
momentum sum rule.
\item Studies of the angular distribution  of the recoiled nucleon and of the
lepton pair are complementary and lead to a rich angular structure of the cross
section that can be used for separation of various combinations of GPDs.
\item The variation of the relative magnitude of space- and timelike photon
virtualities allows to access distributions of partons and anti-partons in the
``exclusive" domain $\xi > |\eta|$.
\item The higher one goes in skewness $\eta$, whose maximal value is limited
by the magnitude of the momentum transfer to be within the region of
applicability of QCD factorization $|\Delta^2| \ll p \cdot q$, the larger
surface in the exclusive domain ($\eta > |\xi|$) one measures in experiment.
This diminishes the uncertainty coming from the inaccessible inclusive sector
($\eta < |\xi|$). The exclusive domain might saturate the spin sum rule
(\ref{SpinSumRule}) even for moderate $\eta$ since the second moment of GPDs
defining it is not extremely sensitive to the large-$\xi$ behavior of GPDs
where they are expected to behave just like conventional parton distributions.
\item Another interesting feature of this process is that  zero value of
the generalized Bjorken variable can be exactly reached when the incoming
and outgoing photons have about the same absolute values of virtualities
${\cal Q}^2 \simeq M_{\ell\bar\ell}^2$.
\item The major complication in the experimental measurement of the process
is a rather small magnitude of its  cross section which is suppressed by two
powers of the fine structure constant $\alpha_{\rm em}$ compared to a typical
deeply inelastic event. The other obstacle is the contamination of the
heavy-photon events by the background of meson production. The latter can be
circumvented in a relatively straightforward manner by avoiding the regions
of $M_{\ell\bar\ell}^2$ close to meson-resonance thresholds. However, this
also restricts the phase space in the measurements of GPDs. We also note that
a numerical estimate of this contamination can be done by means of Eqs.\
(\ref{Def-VacCor}) -- (\ref{Mod-VacCor}). In fact, according to a perturbative
QCD estimate \cite{GuiVan02}, the $\rho$ meson contribution to the beam spin
asymmetry turns out to be small.
\item A clear study of GPDs can be performed in experiments in which the tagged
flavor of the lepton pair differs from that of the beam. When they are the same,
one should add contributions in which the final electrons are interchanged, i.e.,
$k^\prime \leftrightarrow \ell_-$. These will coherently interfere with each
other. None of these results are available yet. They obviously will yield an
essentially different dependence of the VCS amplitude on the external variables.
To use the process with identical leptons for the extraction of information on
GPDs, one needs to ensure that the momentum flow in the quark propagator of the
handbag diagram remains large. Indeed, the scalar product
\begin{eqnarray}
p \cdot q^\prime = - p \cdot q \frac{ 2 -  y - y \cos\theta_\ell}{2y}
\left\{
1
+
{\cal O}
\left(
\frac{\Delta_\perp}{\sqrt{p\cdot q}}
\right)
\right\}
\, ,
\end{eqnarray}
that sets the scale in the exchanged VCS amplitude remains large. Note that here
and in the following we denote the kinematical variables that enter the exchanged
amplitudes with a prime. The power-suppressed contributions depend on all kinematical
variables, in particular, on $y$, $\widetilde y$ and both azimuthal angels $\phi$
and $\varphi_{\ell}$. Besides the condition $(2-y)/y > \cos\theta_\ell$, which is
fulfilled by the usual kinematical restriction $y< 1$, no other {\sl kinematical cuts}
are required to ensure the applicability of perturbative QCD. Moreover, $\eta^\prime$
is given by $\eta$ in leading order
\begin{eqnarray}
\eta^\prime =
\eta + {\cal O}
\left(
\frac{\Delta_\perp}{\sqrt{p\cdot q}}
\right)
\, ,
\end{eqnarray}
while $\xi^\prime$ receives  strong dependence on the leptonic variables:
\begin{eqnarray}
\label{XiToXiprime}
\xi^\prime
 =
\xi
\frac{
2 \cos \theta_\ell - y(1 + \cos \theta_\ell)
}{
2 - y (1 + \cos \theta_\ell)
}
-
\frac{
2 \sqrt{1 - y} \sqrt{\eta^2 - \xi^2} \sin\theta_\ell
}{
2 - y (1 + \cos\theta_\ell)
}
\cos(\phi - \varphi_\ell)
+
{\cal O}
\left(
\frac{\Delta_\perp}{\sqrt{p\cdot q}}
\right)
\, . \qquad
\end{eqnarray}
It should be  pointed out that if $\theta_\ell$  approaches the edge of the phase
space, i.e., $\theta_\ell \to \{0, \pi\}$, the absolute
values of the scaling variables in Eq.\ (\ref{XiToXiprime}) become
identical $|\xi^\prime| \simeq |\xi|$. The conclusions we draw from our
kinematical considerations are as follows. There are no crucial
difficulties in the application of perturbative QCD as long as $p \cdot
q$ is large. However, the analytic evaluation of observables and
further studies are required to find an ``optimal'' method to extract  the
$(\xi, \eta)$ shape of GPDs from the measurements of the reaction $e^\mp p
\to e^\mp p \ e^+ e^-$.
\end{itemize}

We have discussed the most favorable observables, namely, various lepton-spin
and azimuthal asymmetries that are sensitive to the imaginary part of the
Compton form factors and, thus, directly to GPDs. We  have  not discussed in
full, however, phenomenological consequences of using polarized targets, though
we have presented the complete set of formulas with explicit angular dependence
which can be used to extract complementary combinations of Compton form factors
from experimental data. Longitudinal and transverse nucleon-spin asymmetries
combined with the Fourier analysis will serve this purpose just like  the
lepton-spin asymmetries. In the case of DVCS (i.e.,  electroproduction of the
real photon), the   complete analysis along this line was given in  Ref.\
\cite{BelMulKir01}. The process $e N \to e' N' \ell \bar\ell$ with both photons
being virtual provides a unique possibility to make the  skewness  parameter
$\eta$ independent from the generalized Bjorken variable $\xi$. Unfortunately,
this process has a very low cross sections, the drawback which, hopefully, will
be circumvented with future high-luminosity machines. The  analysis of available
events from the CLAS detector at Jefferson Laboratory is under way \cite{Gar03}.

A closely related process to the one we addressed above is the photoproduction of
lepton pairs \cite{BerDiePir01}. It resembles the inverted deeply virtual Compton
scattering. Namely, the virtual photon with negative helicity in timelike Compton
scattering corresponds to a light quantum  with positive helicity in DVCS. As
compared to DVCS, the skewness is equal to the generalized Bjorken variable with
the opposite sign, i.e., $\xi \simeq \eta$. The unpolarized and polarized photon
beams allow to access the real and imaginary parts of a combination of Compton form
factors ${\cal C}_{\cal I}^{\rm unp}$ identical to DVCS.

\subsection{Hard exclusive production of mesons}

\begin{figure}[t]
\begin{center}
\hspace{0cm} \mbox{
\begin{picture}(0,114)(220,0)
\put(-10,-10){\insertfig{16}{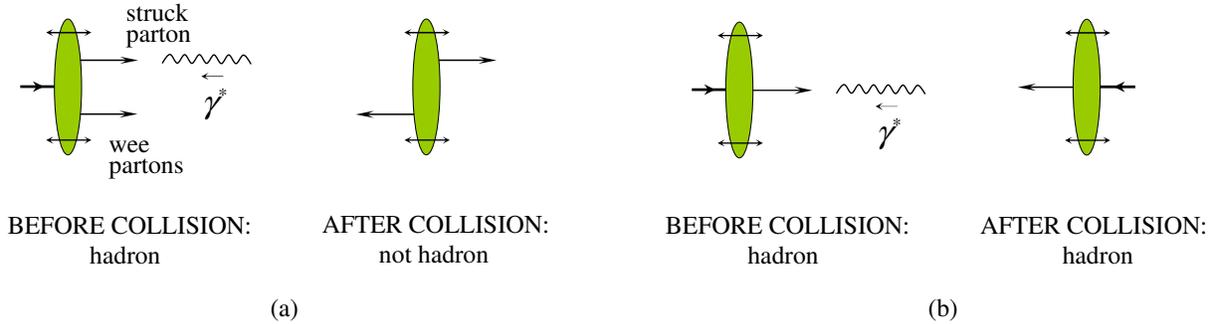}}
\end{picture}
}
\end{center}
\caption{\label{Fig03Chap01} Feynman (or soft) mechanism for electromagnetic
form factors. If more than one parton carries a large share of the parent's
hadron momentum (left diagram), after the absorption of the virtual the hadron
disintegrates since the active partons move back to back in the opposite
directions. However, provided a single parton carries the whole hadron's
momentum (right diagram), after the hard scattering the hadron as a whole
merely changes its direction of motion.}
\end{figure}

To conclude the  discussion of experimental probes of GPDs, we are going to
briefly address the exclusive electroproduction of light mesons $\ell (k) N (p_1)
\to \ell' (k') \gamma^\ast (q_1) N (p_1) \to \ell' (k') M (q_2) N (p_2)$. In
the large-${\cal Q}^2$ limit, the amplitude of this process is dominated by
the one-gluon-exchange mechanism, and can be calculated within perturbative
QCD in terms of GPDs and distribution amplitudes of produced mesons. The meson
electroproduction processes have one indisputable advantage over the
Compton-initiated processes: by changing the type of produced mesons one can
access different flavor combination of GPDs. This helps to perform the flavor
decomposition of GPDs. It should be emphasized, however, that the pQCD
diagrams for the  hard exclusive meson electroproduction are similar to
the one-gluon-exchange graphs determining the asymptotic behavior
of hadronic form factors, and they share the same problem about the range of
applicability of hard pQCD approach. The essence of the problem is that these
are the virtualities of the internal lines of the hard subprocess (in particular,
the virtuality of the exchanged gluon) which are much smaller than the
virtuality ${\cal Q}^2$ of the external probe, that determine whether the
use of pQCD is justified or not.

\subsubsection{Feynman versus hard mechanism in elastic form factors}
\label{FeynmanVsHard}

Let us briefly recall the microscopic description of hadronic electromagnetic
form factors. There are two basic mechanisms for their large-$\Delta^2$ behavior.
In general terms, the problem can be described in the following way. After the
large momentum is transferred from the probing lepton to the hadron through the
virtual photon interaction with a single quark, the latter radically changes the
direction of its motion. However, in an exclusive reaction, it should eventually
recombine together with the hadron's spectators to form the final state hadron.
Imagine a hadron possesses at least two partons that almost equally share the
bulk of its momentum, as shown in see Fig.\ \ref{Fig03Chap01} (a). If only one of
them absorbs the heavy photon, the active parton reverts its direction of motion,
while the other fast parton continues to move forward in the original direction
of the parent hadron. The probability for such momentum configuration to form an
outgoing hadron is vanishing and thus should be discarded. In the mechanism proposed
by Feynman \cite{Fey71}, only one quark has a large momentum fraction $x \sim 1$
of the original hadron's momentum, and after the collision it carries the bulk part
of the final hadron momentum. The ``wee'' spectators carry a very small $\sim
1/|\bit{\Delta}|$ fraction of the hadron momentum, and can be associated either
with initial or final hadron as in Fig.\ \ref{Fig03Chap01} (b). The more spectators
one has, the smaller is the phase space for keeping the spectators within the
$x_{\rm spect} \sim  1/|\bit{\Delta}|$ range. If the probability to have the momentum
fraction of the active quark close to one is proportional to $q (x) \sim (1 - x)^{\nu}$
where $\nu$ is proportional to the number of spectators, then the form factor drops
like $\sim |\bit{\Delta}|^{- \nu - 1}$. Thus, Feynman mechanism suggests spectator
counting rule: the larger is the number of quarks in a hadron, the faster its form
factor drops with increasing $|\bit{\Delta}|$.

\begin{figure}[t]
\begin{center}
\hspace{0cm} \mbox{
\begin{picture}(0,105)(90,0)
\put(-10,0){\insertfig{7.5}{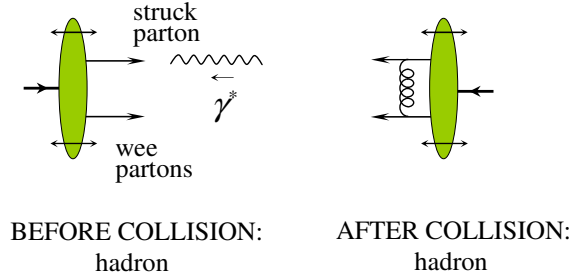}}
\end{picture}
}
\end{center}
\caption{\label{HardMechPictorial} Hard mechanism for electromagnetic
form factors.}
\end{figure}

In the hard perturbative QCD picture, all the valence quarks carry finite
fractions of the initial and final hadron momenta. The redistribution of
the momentum transfer among the quarks goes through exchange of hard gluons
(see Fig.\ \ref{HardMechPictorial}). The minimal number of exchanges is
$N - 1$ for a hadron composed of $N$ valence quarks. The perturbative QCD
prediction $F(- \bit{\Delta}^2) \sim (\bit{\Delta}^2)^{- N + 1}$ for the
spin-averaged form factors just counts the $1/\bit{\Delta}^2$ powers due
to the hard gluon propagators. The quark counting rules are the famous
prediction of perturbative QCD for exclusive processes.

\subsubsection{Quark counting rules}
\label{QuarkCountRules}

\begin{figure}[t]
\begin{center}
\mbox{
\begin{picture}(0,135)(220,0)
\put(0,0){\insertfig{6}{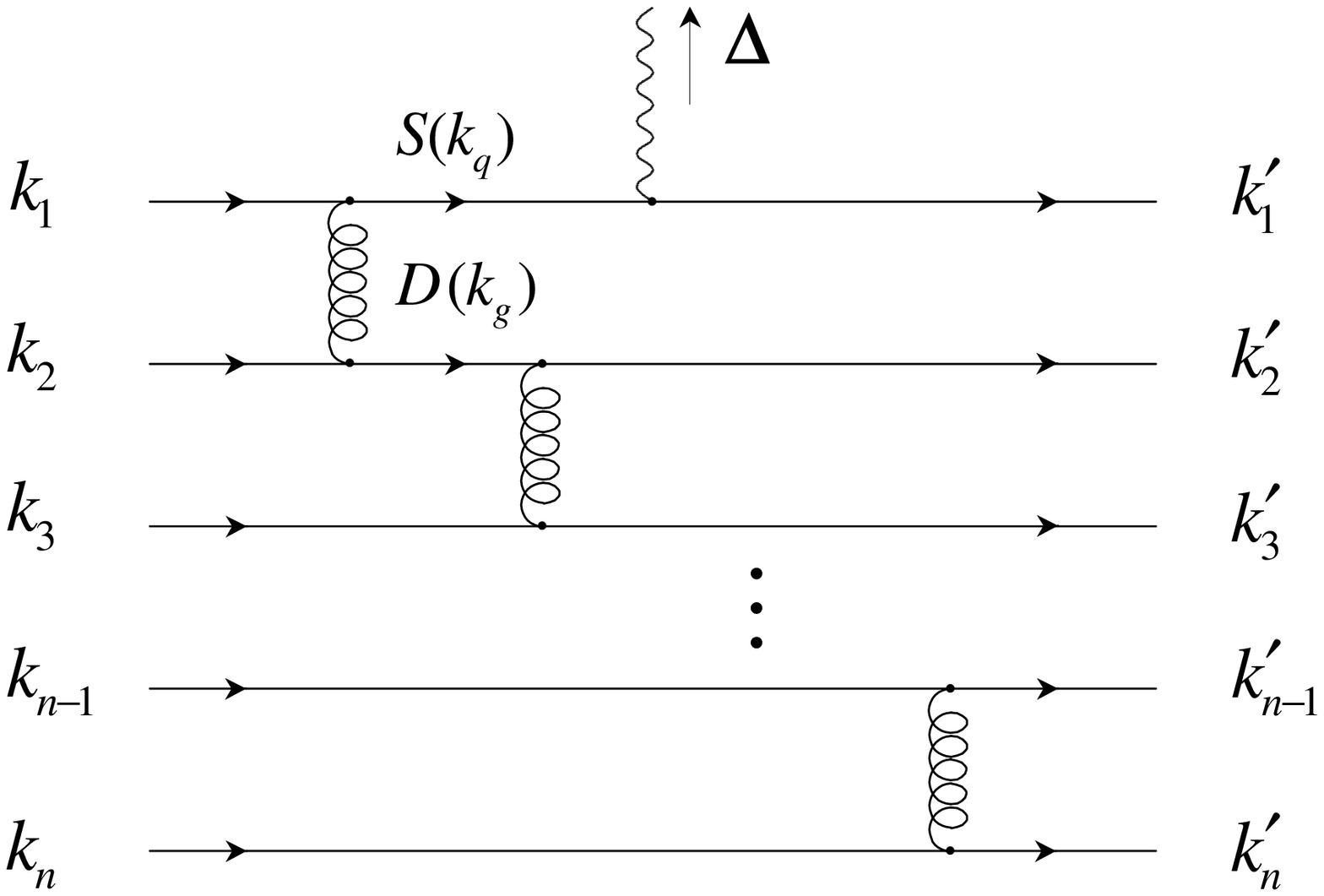}}
\put(240,0){\insertfig{6.7}{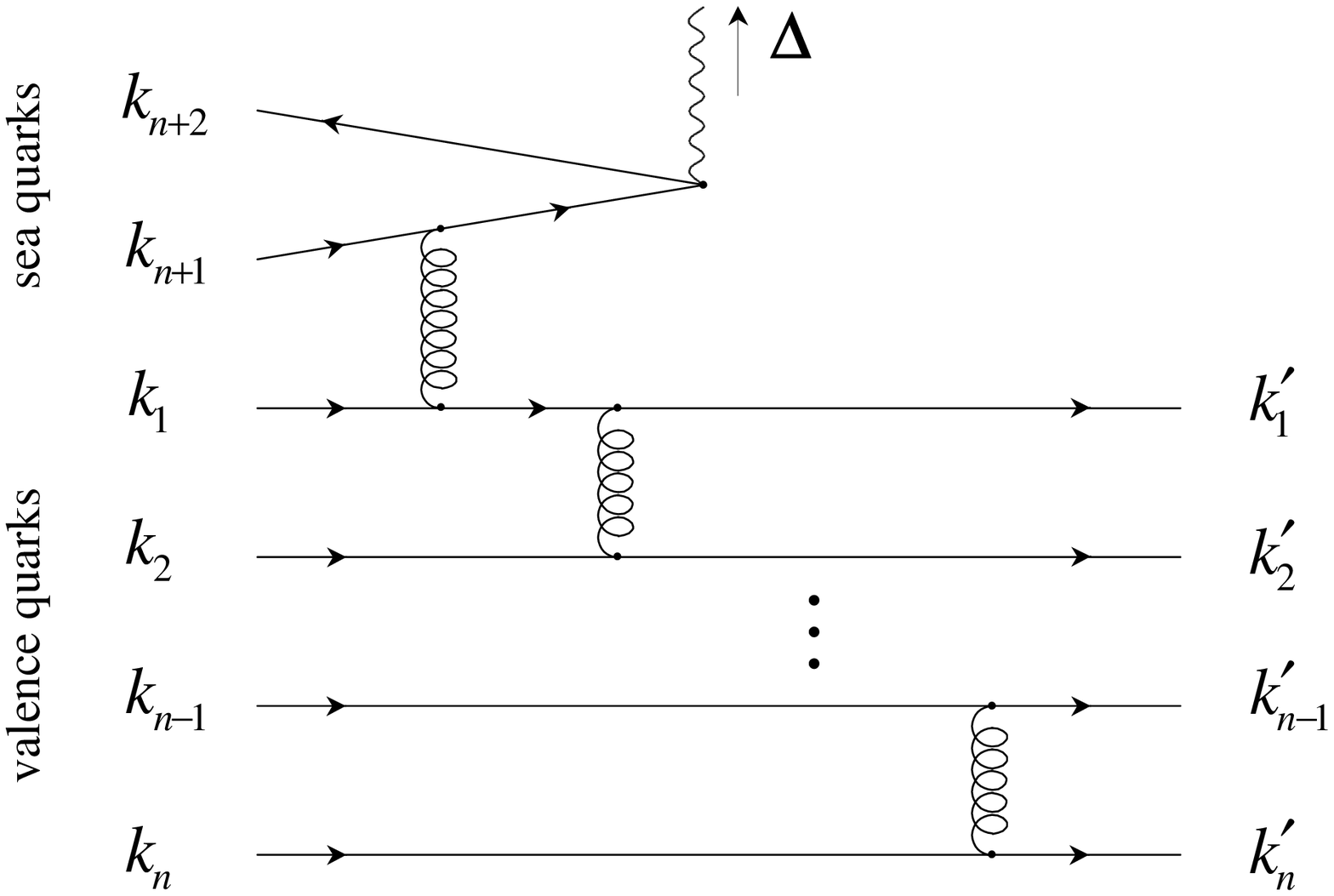}}
\end{picture}
}
\end{center}
\caption{\label{HardExchange} Origin of the quark counting rules from the
hard exchange mechanism for $n \to n$ and $n_i \to n_f$  components of form
factors.
}
\end{figure}

The derivation of the aforementioned counting rules is most transparent in
the Breit frame, where the  three-momenta of the incoming and outgoing
hadrons are given in terms of the momentum transfer $\bit{p}_1 = - \bit{p}_2
= \bit{\Delta}/2$. The partons momenta collinear to those of the incoming and
outgoing hadrons are then given by
$$
\bit{k}_i =  \frac{x_i}{2} \bit{\Delta}
\, , \qquad
\bit{k}'_i = -\frac{y_i}{2} \bit{\Delta}
\, ,
$$
respectively. The momentum fractions obey the obvious energy-momentum conservation
conditions
$$
\sum_{i = 1}^n x_i = \sum_{i = 1}^n y_i = 1
\, .
$$
The main observation is that there are no other scales except for the hard scale
$\bit{\Delta}$. Then the hard gluon exchange contribution to the form factor for
a hadron made of $n$-quarks (see Fig.\ \ref{HardExchange} for a typical diagram)
can be written as
\begin{eqnarray}
\label{MatrixElementCounting}
\langle p_2 | j^\mu_q (0) | p_1 \rangle
&=&
L^\mu F (\Delta^2) \nonumber \\
&\sim&
{\mit\Psi}^\ast_{q \cdots q} \,
\bigl \{\bar u (k_n) \dots \bar u (k_1) \bigr \} \otimes
\gamma^\mu D^{n - 1} S^{n - 1}\otimes
\bigl \{u (k'_1) \dots u (k'_n)\bigr \}
\, {\mit\Psi}_{q \cdots q}
\, ,
\end{eqnarray}
where $D$ stands for the exchanged gluons propagators, $S$ for the quark
propagators and $L^\mu = \bar{u} \gamma^\mu u$ is the quark current. Note,
that all of the momenta associated with the internal lines of the hard subprocess
are superpositions $[a(\{ x_j\})p_1 - b(\{y_k\})p_2]$  of the initial and final
hadron momenta, with the coefficients $a$ and  $b$ given by sums of particular
sets of fractions $\{x_j\}$ and $\{y_k \}$. Since $0 \leq a,b \leq 1$, the
virtualities $(- a b \bit{\Delta}{}^2)$ of the internal lines of the subprocess
are spacelike and of the order of $\bit{\Delta}{}^2$. In a situation when there
are no other scales, the combinations $ u (k_l) \bar u (k'_l)$ of Dirac spinors
can only  produce $\sim \bit{\Delta}$ factors. Thus, we get the counting rules
for building blocks of the amplitude
\begin{equation}
u, \bar u  \sim |\bit{\Delta}|^{1/2}
\, , \qquad
D \sim |\bit{\Delta}|^{-2}
\, , \qquad
S \sim |\bit{\Delta}|^{-1}
\, .
\end{equation}
Evidently, they could be guessed  on dimensional grounds alone.
For the form factor, we obtain
\begin{equation}
L^\mu F (\Delta^2)
\sim
\frac{|\bit{\Delta}|}{( \bit{\Delta}{}^2 )^{n - 1}}
\, .
\end{equation}
From the dimensional analysis, it also follows that the Lorentz structure
$L^\mu$ has the dimension of  mass, which gives the estimate $L \sim
|\bit{\Delta}|$. As a result, we find the scaling rule for the helicity
non-flip form factor
\begin{equation}
\label{QuarkCountFF}
F (\Delta^2)
\sim
\frac{1}{(\bit{\Delta}{}^2)^{n - 1}}
=
\frac{1}{( - \Delta{}^2)^{n - 1}}
\, .
\end{equation}

The counting rules can be generalized for the contributions involving Fock
components with different numbers $n_i, n_f$ of  quarks in the initial and
final state. In this case, one should change $n \to (n_i + n_f)/2$. In particular,
the annihilation-type $n + 2 \to n$ diagram  shown in Fig. \ref{HardExchange}
has an extra factor
$$
\bar u (k_{n + 2}) S D u (k_{n + 1}) \sim \frac{1}{\bit{\Delta}{}^2}
$$
resulting in an additional power of $1/\bit{\Delta}{}^2$ in the form factor
compared to (\ref{QuarkCountFF}).

It is rather popular to correlate the power behavior of form factors with the
behavior of inclusive structure functions $F_1 (x_{\rm B})$ of deeply inelastic
scattering at large Bjorken variable $x_{\rm B}$, when one approaches the
exclusive single-hadron pole. The invariant mass $W^2 = (p_1 + q_1)^2$ of the
produced hadronic system is related
to $x_{\rm B}$ by
\begin{equation}
\label{XbMass}
1 - x_{\rm B} = x_{\rm B} \frac{W^2 - M_h^2}{{\cal Q}^2}
\, ,
\end{equation}
where we shifted to the ${\cal Q}^2 = \bit{\Delta}{}^2$ notation. The single-hadron
contribution to cross section is given by the form factor squared multiplied by
$\delta (W^2 - m_h^2)$. According to Bloom and Gilman \cite{Bloom:1970xb}, the
$W^2$-integral of the hadron contribution is equal to the $x$-integral of the
structure function $F_1 (x)$ over a duality region with fixed boundaries in
the variable $W^2$. This gives a relation between the power $\nu$ specifying
the $(1 - x)^\nu$ behavior of the structure function $F_1 (x)$ in the $x \to 1$
region and the power-law behavior of the square elastic form factor: $F^2
({\cal Q}^2) \sim (1/{\cal Q}^2)^{\nu + 1}$, or $\nu = 2 n - 3$, if we use the
counting of Eq.\ (\ref{MatrixElementCounting}). In the proton case, with $n = 3$,
one obtains $\nu = 3$, or a dipole behavior for the Dirac $F_1 ({\cal Q}^2)$ form
factor.

It is worth emphasizing that one should not confuse the Bloom-Gilman duality with the
Drell-Yan relation \cite{Drell:1969km}, which relates the integral of a nonforward
parton density $f^q (x, {\cal Q}^2)$, given in terms of the skewless GPD via
$$
f^q (x, \mathcal{Q}^2) \equiv F^q (x, \eta = 0, \mathcal{Q}^2)
\, ,
$$
over the interval $x > 1- \lambda/{\cal Q}$ with the {\sl first power} of the form
factor. Also, the dominance of the region $x > 1- \lambda/{\cal Q}$ is a consequence
of a specific structure of  the density  $f^q (x, {\cal Q}^2)$. As we have seen before,
the Drell-Yan relation does not work for the Gaussian model, but holds for the
improved Regge-like model (\ref{t-depPDF}).

One should also realize that both relations were formulated in  nonperturbative
context. The shape of the structure function $F_1 (x)$ or nonforward parton density
$f^q (x, {\cal Q}^2)$ is not necessarily generated by pQCD dynamics. Knowing these
functions, or building some models for them, one can use Bloom-Gilman or Drell-Yan
relations to get predictions for form factors. Both relations have a common feature:
change of the power $\nu$ in the $(1 - x)^\nu$ behavior of the structure function
results in a change of the $1/{\cal Q}^{\nu+1}$ power behavior of the form factor.
Both relations give the same correlation between the two powers, and that is why
they are sometimes confused.

Perturbative QCD  predictions for asymptotic power behavior of form factors with
definite powers, e.g., $(\alpha_s/{\cal Q}^2)^{n-1}$ for a spin-averaged form factor
of $n$-quark hadron. It also predicts fixed powers $\alpha_s^{2n - 2} (1-x)^{2n - 3}$
for the $x \to 1$ behavior of the structure functions (see \cite{BroLep79}). Formally,
the powers of $(1 - x)$, $1/{\cal Q}^2$ and $\alpha_s$ are correlated in pQCD as in
Bloom-Gilman relation. However, a direct calculation of pQCD diagrams for $F_1 (x)$
gives expressions which have more complicated structure than the squares of form
factors (see, e.g., \cite{Yua03}, where the $x \to 1$ behavior of GPDs is also
discussed).

The Drell-Yan relation is invalid in pQCD, as emphasized in the pioneering paper
\cite{BroLep79}. In particular, the leading $(1 - x)^3$ term in $F_1 (x)$ is given
in pQCD by diagrams involving four hard gluon exchanges and is accompanied by the
$\alpha_s^4$ factor. Integrating it over the region $x> 1- \lambda/{\cal Q}$ one
would get a contribution $\sim \alpha_s^4/{\cal Q}^4$ that has the same $1/{\cal Q}^
4$ power as the pQCD prediction for the nucleon form factor, but is suppressed by
two powers of $\alpha_s$.

\subsubsection{GPDs and form factors}

The formalism of generalized parton distributions allows to combine the
two mechanisms---the soft and hard---in a single description. Form factors
result from integration of GPDs over the momentum fraction $x$ as was
demonstrated in Eq.\ (\ref{SumRule}),
$$
\int_{0}^{1} dx \, H^q (x, \eta, \Delta^2)
=
F^q (\Delta^2)
\, .
$$
The result does not depend on the skewness $\eta$, so it is convenient to
use here skewless generalized parton distributions, related via the Fourier
transform to the impact parameter parton densities as discussed in Section
\ref{SkewlessGPDsImpactParton}. The simplest case of the meson electromagnetic
form factor is also the most close diagram-wise to the meson electroproduction
process. The soft contribution to $f^q (x, \Delta^2)$ in this case can be
represented by the Drell-Yan formula
\begin{equation}
F_M (- \bit{\Delta}^2_\perp)
=
\int_0^1 \, dx \, \int \frac{d^2 \bit{k}_\perp}{(2 \pi)^3} \,
\psi^*_{qq} (x, \bit{k}_\perp - (1 - x) \bit{\Delta}_\perp) \,
\psi_{qq} (x, \bit{k}_\perp)
\, ,
\label{eq:DYW}
\end{equation}
corresponding to overlap of two light cone wave functions. Taking the Gaussian
ansatz
\begin{equation}
\psi_{qq} (x, \bit{k}_\perp)
\sim
\exp\left( - \frac{\bit{k}_\perp^2}{2 x (1 - x)\lambda^2} \right)
\end{equation}
one can easily perform the $\bit{k}_\perp$ integration and obtain
\begin{equation}
F^{\rm Gauss}_M (- \bit{\Delta}^2_\perp)
=
\int_0^1 dx \, q (x)
\exp \left( - \frac{\bit{\Delta}^2_\perp}{\lambda^2} \frac{1 - x}{4x}
\right)
\, ,
\label{FFBorel}
\end{equation}
where the integrand is a Gaussian model for the nonforward parton density. Its
specific feature is the structure of the exponential factor that has a nontrivial
non-factorized interplay between $\bit{\Delta}^2_\perp$ and $x$ dependence. The
pre-exponential factor may be interpreted as a forward quark density $q (x)$.
Similar expressions were obtained more than twenty years ago in QCD sum rules for
the soft contribution to hadronic form factors. In that approach, one considers a
three-point correlator of the electromagnetic current with two local currents
generating initial and final hadrons. It is convenient to work with the
so-called Borel transforms of the correlation functions (for details, see
\cite{Ioffe:1982qb,Nesterenko:1982gc}), and in the pion form factor case, the
double Borel transform of the basic triangle diagram has just a Gaussian form
\cite{Nesterenko:1982gc,Radyushkin:2004mt})
\begin{equation}
\Phi [\bit{\Delta}^2, \tau_1, \tau_2]
=
\frac{3}{2\pi^2(\tau_1 + \tau_2)} \int_0^1 dx \, x \bar{x} \,
\exp
\left(
- \frac{\bar{x}}{x} \frac{\tau_1 \tau_2}{\tau_1+ \tau_2} \bit{\Delta}^2
\right)
\, ,
\label{Borelpion}
\end{equation}
where as usual $\bar{x} = 1 - x$. It should be noted that the application of
the Borel transformation is equivalent to taking a Gaussian wave function with
$\lambda^2_i \sim 1/\tau_i$ for the vertices generating pions.

Turning back to the Gaussian ansatz (\ref{FFBorel}), it is evident that the
large-$\bit{\Delta}^2$ behavior of this expression is dominated by integration
over the region where $(1 - x) \bit{\Delta}^2 /\lambda^2 \sim 1$ or $1 - x \sim
\lambda^2/\bit{\Delta}^2$. Hence, the result of integration is completely
determined by the behavior of the pre-exponential factors (i.e., the model
forward densities) at $x$ close to 1. The ``Borel'' model (\ref{Borelpion})
implies  $\sim x(1-x)$ shape for $q_{\pi}(x)$, corresponding to the
$1/\bit{\Delta}^4$ asymptotic behavior. In a more general Gaussian model, if
$q (x) \sim (1 - x)^\nu$ for $x \to 1$, then the form factor asymptotically
drops like $(\bit{\Delta}^2)^{- \nu - 1}$ at large $\bit{\Delta}^2$.

\begin{figure}[t]
\begin{center}
\hspace{0cm} \mbox{
\begin{picture}(0,90)(180,0)
\put(0,0){\insertfig{12}{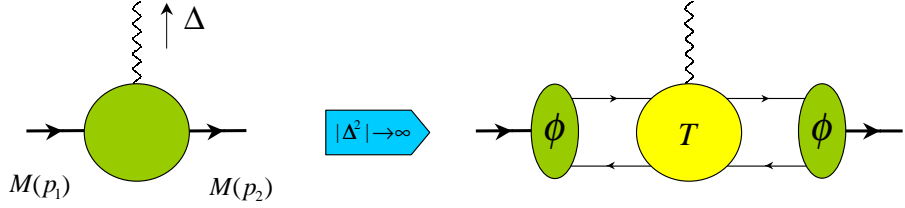}}
\end{picture}
}
\end{center}
\caption{\label{Fig04Chap01} Factorized expression for the meson form
factor at asymptotically large momentum transfer.}
\end{figure}

In the nucleon case, the analogue of Eq.\ (\ref{Borelpion}) for the $F_1$ form
factor is \cite{Nesterenko:1983gi,Radyushkin:2004mt}
\begin{eqnarray}
\label{Borelf1}
\Phi_1 [\bit{\Delta}^2,\tau_1,\tau_2]
=
\frac{1}{(2\pi)^4 (\tau_1 + \tau_2)^3}
\int_0^1 dx \, \left[ 3 \, Q_u \bar{x}^2 - (2 Q_u - Q_d) \bar{x}^3 \right]
\,
\exp
\left(
- \frac{\bar x}{x} \frac{\tau_1 \tau_2}{\tau_1 + \tau_2} \bit{\Delta}^2
\right)
\, ,
\end{eqnarray}
while in case of the magnetic form factor $G_M$ it is given by
\begin{eqnarray}
\label{Borelgm}
\Phi_M [\bit{\Delta}^2, \tau_1, \tau_2]
=
\frac{ 3 \, Q_u}{(2 \pi)^4 (\tau_1 + \tau_2)^3}
\int_0^1 dx \, \bar{x}^2 \,
\exp
\left(
- \frac{\bar{x}}{x} \frac{\tau_1 \tau_2}{\tau_1 + \tau_2} \bit{\Delta}^2
\right)
\, .
\end{eqnarray}
Since $G_M = F_1 + F_2$, one can derive the expression
\begin{eqnarray}
\Phi_2 [\bit{\Delta}^2, \tau_1, \tau_2]
= \frac{2 Q_u - Q_d}{(2\pi)^4(\tau_1 + \tau_2)^3}
\int_0^1 d x \, \bar{x}^3
\exp
\left(
- \frac{\bar x}{x} \frac{\tau_1 \tau_2}{\tau_1 + \tau_2} \bit{\Delta}^2
\right)
\, ,
\end{eqnarray}
for the function related to the $F_2$ form factor. Looking at the $x \to 1$ behavior
of the pre-exponential factors, we conclude that the Borel model, inspired by QCD
sum rules, gives $1/\bit{\Delta}^6$ for the large-$\bit{\Delta}^2$ behavior of
the nucleon form factors $F_1 (\bit{\Delta}^2)$ and $G_M (\bit{\Delta}^2)$, and
$1/\bit{\Delta}^8$ for $F_2 (\bit{\Delta}^2)$. In a general Gaussian model, taking
a more realistic $\sim (1-x)^3$ shape for the nucleon parton densities one would
obtain $1/\bit{\Delta}^8$ as the asymptotic behavior of $F_1(\bit{\Delta}^2)$.
Though this result seems to be in contradiction with the experimentally established
dipole behavior of this form factor, the Gaussian model for the proton form factor
$F_1^p (\bit{\Delta}^2)$ \cite{Radyushkin:1998rt,Diehl:1998kh} successfully describes
the data up to  $ \bit{\Delta}^2 \sim 10 \, {\rm GeV}^2$. The explanation of this
apparent ``paradox'' is very simple: the model $\bit{\Delta}^2$-dependence for this
and other form factors is given by more complicated functions than just pure powers
of $1/\bit{\Delta}^2$. Their nominal large-$\bit{\Delta}^2$ asymptotics is achieved
only at very large values of $\bit{\Delta}^2$, well beyond the accessible region.
Thus, conclusions made on the basis of asymptotic relations might be of little
importance in practice: a prediction with ``wrong'' large-$\bit{\Delta}^2$ behaviour
might be quite successful phenomenologically in a rather wide range of $\bit{\Delta}^2$.
In fact, the improved model
\begin{eqnarray}
f^q_{\rm Regge} (x, \Delta^2)
=
q (x) x^{- \alpha^{\prime} (1 - x) \Delta^2}
\, ,
\end{eqnarray}
which imposes the correct Regge behavior $x^{- \alpha (\Delta^2)}$ for small
momentum fractions $x$ by a minimal modification $(2 x \lambda^2)^{- 1}
\to - \alpha^\prime \ln x$ of the Gaussian ansatz, allows to obtain a rather
good description \cite{Diehl:2004cx,GPRV} of all four nucleon form factors
in the whole experimentally accessible region of spacelike momentum transfers.
Note, that the exponential factor of the improved model behaves like
$\exp[ -\alpha^\prime (1 - x)^2 \bit{\Delta}^2]$  for $x$ close to one.
As a result, if the parton density behaves like $(1 - x)^\nu$, then the
relevant form factor decreases as $(\bit{\Delta}^2)^{-(\nu + 1)/2}$ for
large $\bit{\Delta}^2$, which exactly corresponds to the Drell-Yan relation
\cite{Drell:1969km}. We already pointed this property out in Section
\ref{LongTransInterplay}.

\begin{figure}[t]
\begin{center}
\hspace{0cm} \mbox{
\begin{picture}(0,80)(150,0)
\put(-10,-10){\insertfig{12}{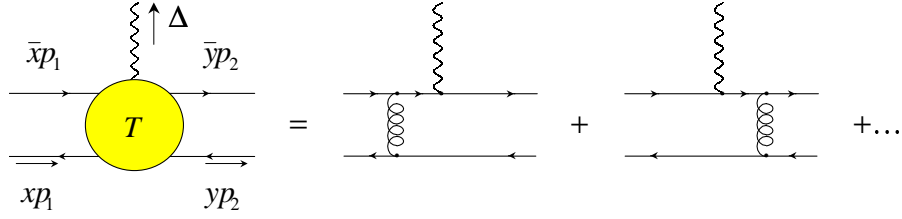}}
\end{picture}
}
\end{center}
\caption{\label{Fig05Chap01} Hard rescattering in the meson form factor.
The ellipses stand for the attachment of the photon to the lower quark
line in the one-gluon exchange coefficient functions as well as for the
higher order corrections.}
\end{figure}

The asymptotic behavior of the form factor has a factorized form within
perturbative QCD, see Fig.\ \ref{Fig04Chap01},
\begin{equation}
F^{\rm pQCD}_M (\Delta^2)
=
f_M^2
\int_0^1 d x \, d y \, \phi_M (x) \, T (x, y ; {\Delta}^2) \, \phi_M (y)
\, ,
\end{equation}
where $\phi_M (x)$ is the meson distribution amplitude and $T (x, y ; \Delta^2)$
is the amplitude of the hard gluon exchange subprocess shown in Fig.\
\ref{Fig05Chap01}. For the pion form factor, $M = \pi$,
$$
T (x, y ; \Delta^2)
=
\frac{
8 \pi \alpha_s
}{
9(1 - x)(1 - y) \bit{\Delta}^2
}
\, .
$$
Taking the asymptotic shape $\phi_\pi (x) = 6 x(1 - x)$, which apparently
is a good approximation even at low normalization points \cite{bakmik}, yields
\begin{equation}
\label{fpipQCD}
F_\pi^{\rm pQCD} ( - \bit{\Delta}^2)
=
\frac{8\pi \alpha_s f_\pi^2}{\bit{\Delta}^2} =
\frac{2\alpha_s}{\pi} \left( \frac{s_0}{\bit{\Delta}^2} \right )
\, ,
\end{equation}
where $s_0 = 4 \pi^2 f_{\pi}^2 \approx 0.7 \, {\rm GeV}^2$ is a typical hadronic
scale for the pion. Since $s_0 \approx m_\rho^2$, the hard contribution has an
obvious $2{\alpha_s}/{\pi}$ suppression compared to the vector meson dominance
expectation $F^{\scriptscriptstyle\rm VMD} (\Delta^2) \sim m_\rho^2/\Delta^2$.
It is well known that the ${\alpha_s}/{\pi}$ is the standard price for each extra
loop in a diagram.

\begin{figure}[t]
\begin{center}
\hspace{0cm} \mbox{
\begin{picture}(0,42)(50,0)
\put(-10,-10){\insertfig{4}{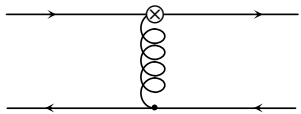}}
\end{picture}
}
\end{center}
\caption{\label{WilsonLineInGPDs} Additional hard rescattering diagram
contributing to the leading order large-$\Delta^2$ behavior of GPDs.}
\end{figure}

It is also possible to include the perturbative QCD contribution as an addition to
the nonforward parton density \cite{Diehl:1999ek,Vogt:2001if,Hoodbhoy:2003uu}
\begin{equation}
f^q_{\rm pQCD} (x, - \bit{\Delta}^2)
=
\frac{\alpha_s}{\pi} \left( \frac{s_0}{\bit{\Delta}^2} \right ) \, 4\, x
\left\{ 1 - 2 \left [1 + x \ln \left(\frac{1-x}{x}\right) \right] \right\}
\, .
\end{equation}
This functional form is computed from the sum of the covariant-gauge Feynman
diagrams in Fig.\ \ref{Fig05Chap01}, where the electromagnetic current is replaced
by the non-local light-ray operator (\ref{QuarkLightRayOperators}), and Fig.\
\ref{WilsonLineInGPDs} with the gluon line going out of the light-cone operator
vertex. Note, that the term in the square brackets arises from such a graph.
Therefore, it does not contribute to the form factor. The reason for vanishing
of the discussed contribution becomes obvious when one realizes that it stems
from the expansion of the Wilson line in the non-local light-cone operator
determining the GPD. For the local current relevant for the form factor, the
gauge link merely reduces to unity.

The comparison of the magnitude of the soft and hard contributions
can be made in the local quark-hadron duality model
\cite{Nesterenko:1982gc,Radyushkin:1990te,Radyushkin:1995pj},
where both are present simultaneously. Just like the Borel model,
this model is also inspired by QCD sum rules. However, instead of
the Borel transform
\begin{equation}
\label{eq:doublebor}
\Phi [\bit{\Delta}^2, \tau_1, \tau_2]
=
\frac{1}{\pi^2} \int_0^\infty d s_1 \int_0^\infty d s_2
\,
\varrho^{\rm pert} (s_1, s_2, \bit{\Delta}) \, {\rm e}^{- s_1 \tau_1 - s_2 \tau_2}
\, ,
\end{equation}
in which the spectral density $\varrho^{\rm pert} (s_1, s_2, \bit{\Delta}^2)$
(calculated from diagrams of perturbation theory for the relevant correlator of
three currents, e.g., electromagnetic and two axial ones, in case of the pion
form factor) is integrated with the exponential weights ${\rm e}^{-s_1 \tau_1}$,
${\rm e}^{-s_2 \tau_2}$, one takes the function
\begin{eqnarray}
\label{FLD}
T^{(3)} (\bit{\Delta}^2, s_0)
=
\frac{1}{\pi^2}
\int_0^{s_0} d s_1 \int_0^{s_0} d s_2\,
\varrho^{\rm pert} (s_1, s_2, \bit{\Delta}^2)
\, ,
\end{eqnarray}
in which $\varrho^{\rm pert} (s_1, s_2, \bit{\Delta}^2)$ is integrated over the
square $0 \leq s_1,s_2\leq s_0$. Then one assumes  that the pion contribution
$$
\varrho_{\pi} (s_1,s_2,\bit{\Delta}^2)
=
f_\pi^2 F_\pi (\bit{\Delta}^2) \delta (s_1 - m_\pi^2) \delta (s_2 - m_\pi^2)
$$
to the physical spectral density $\varrho^{\rm phys} (s_1, s_2, \bit{\Delta}^2)$
is dual to the ``perturbative'' spectral density $\varrho^{\rm pert} (s_1, s_2,
\bit{\Delta}^2)$ in the appropriate ``duality square'' $ [0 \leq s_1 \leq s_0]
\otimes [0 \leq s_2 \leq s_0]$. The size of the duality interval $s_0$ is fixed
from the local duality relation
\begin{eqnarray}
\label{fpiLD}
f_\pi^2
=
\frac{1}{\pi} \int_0^{s_0} d s \, \varrho^{\rm pert}(s)
\end{eqnarray}
for the correlator of two axial currents. Since
$$
\varrho^{\rm pert} (s)
=
\frac{1}{4\pi}
\left( 1 + \frac{\alpha_s}{\pi} \right) \, ,
$$
in the lowest order in $\alpha_s$ one obtains $s_0 = 4 \pi^2 f_\pi^2 \approx
0.7 \, {\rm GeV}^2$, the scale we just discussed above. Thus, the local
quark-hadron duality model corresponds to using the prescription $f_\pi^2 F_\pi
(\bit{\Delta}^2) = T^{(3)} (\bit{\Delta}^2, s_0)$ and a similar relation
Eq.\ (\ref{fpiLD}) for $ f_\pi^2$. The spectral density  $\varrho^{\rm pert}
(s_1, s_2, \Delta^2)$ is known up to $\alpha_s$ level. The lowest order term
found in Refs.\ \cite{Ioffe:1982qb,Nesterenko:1982gc} can be easily integrated
analytically over the duality square (see Ref.\ \cite{Nesterenko:1982gc}. The
${\cal O}(\alpha_s)$ term \cite{Braguta:2004ck} is much more complicated. However,
its large-$\bit{\Delta}^2$ asymptotic behavior is fixed by the pQCD result
(\ref{fpipQCD}), while  the  $\Delta^2 = 0$ value $\alpha_s/\pi$ is fixed by
the Ward identity. Assuming the simplest interpolation \cite{Radyushkin:1990te}
between these two values and combining the ${\cal O} (\alpha_s^0)$ and ${\cal O}
(\alpha_s)$ contributions gives
\begin{eqnarray}
\left( 1 + \frac{\alpha_s}{\pi} \right)
F_{\pi}^{\scriptscriptstyle\rm LD} (- \bit{\Delta}^2)
=
\left(
1 - \frac{1 + 6 s_0/\bit{\Delta}^2}{(1+4s_0/\bit{\Delta}^2)^{3/2}}
\right)
+
\frac{\alpha_s/\pi}{1+\bit{\Delta}^2/2s_0}
\, .
\label{LDpion}
\end{eqnarray}
Just like the Borel model is equivalent to a Gaussian wave function, the local
quark-hadron duality model corresponds to a step-like effective lightcone wave
function $\psi_{qq} (x, \bit{k}) \sim \theta [x (1 - x) s_0 - \bit{k}^2]$, and
the first term in (\ref{LDpion}) which is due  to the simplest  triangle
diagram, is given by an overlap of such  soft wave functions.  Hence, its
large-$\bit{\Delta}^2$ behavior is governed by the Feynman mechanism.

On the other hand, the ${\cal O}(\alpha_s)$ term includes the one-gluon exchange
contributions \cite{EfrRad78,BroLep79} responsible for  its leading asymptotic
behavior  at large $\bit{\Delta}^2$. One should realize, however, that to get the
``genuine  hard'' contribution, one should subtract from the ${\cal O}(\alpha_s)$
term of Eq.\ (\ref{LDpion}) the part produced by integration over a region where
the virtuality of the exchanged gluon is small, say, smaller than $s_0$. Such a
subtraction, as argued in Refs. \cite{Nesterenko:1983gi,Isgur:1984jm} would reduce
the hard part to a miniscule fraction of the  ${\cal O}(\alpha_s)$ term. But, even
without this subtraction, the purely soft term strongly exceeds the ${\cal O}
(\alpha_s)$ term of Eq.\ (\ref{LDpion}) for all accessible momentum transfers. The
two terms  become equal to each other only for $\bit{\Delta}^2 = 10 \, {\rm GeV}^2$.
The bulk of the ${\cal O}(\alpha_s)$  term even at that point should be treated as a
soft contribution. Within the local quark-hadron duality model, however, the question
of dividing the ${\cal O}(\alpha_s)$ term into soft and hard parts does not have
much of practical meaning. We only note, that with  $\alpha_s/\pi = 0.1$ and $s_0
= 4\pi^2 f_\pi^2$, the local duality  prediction is in perfect agreement with the
latest Jefferson Lab measurements of the pion form factor \cite{Volmer:2000ek}
at the momentum  transfer $0.6 \, {\rm GeV}^2 \leq \bit{\Delta}^2 \leq 1.6 \,
{\rm GeV}^2 $.

For the nucleon form factors, the asymptotically leading two-gluon exchange term
has an a priori $(\alpha_s/\pi)^2$ suppression resulting in a reduction by a
factor of hundred. It is further complemented by a huge reduction due to the
exclusion of the region of small gluon virtualities, so it is unlikely to be
relevant at any imaginable momentum transfers. Thus, we do not expect that hard
contributions will play a visible role in the nucleon case.

Comparing the $\Delta^2$-dependence of soft and hard parts of the nonforward
parton density, we observe that the soft part decreases exponentially at large
$\bit{\Delta}^2$, i.e., faster than any power of $1/\bit{\Delta}^2$. The power-law
behavior for the form factor is produced only after integration over $x$. The
resulting power depends on the $x \to 1$ behavior of $f^q (x, \Delta^2)$, in
agreement with the Feynman/Drell-Yan picture. In contrast, the perturbative QCD
term has a power-law behavior $(1/\Delta^2)^N$ before integration over $x$. As
already mentioned, the power $N = n-1$ is given by the number of gluon exchanges
necessary to transfer the large momentum  from the active quark to $n - 1$
spectators. Integration over $x$ does not affect this power: there is no
Drell-Yan/Feynman mechanism in perturbative QCD. In this connection, it is worth
recalling the pQCD  prediction  about the $x\to 1$ behavior of parton densities.
As we discussed, in the nucleon case, the leading $(1-x)^3$ term corresponds to
four-gluon-exchange diagrams, with inevitable  $(\alpha_s/\pi)^4 \sim 10^{-4}$
suppression. Furthermore, though the $(1 - x)^3$ behavior of this term and the
$1/\Delta^4$ behavior of the asymptotic prediction for the nucleon Dirac form
factor $F_1(\Delta^2)$ formally satisfy the Drell-Yan relation, they have nothing
to do with it, because these results have different power in $\alpha_s$. As noted
in the pioneering paper \cite{BroLep79}, the Drell-Yan relation in pQCD is violated
by two powers of $\alpha_s$.

There is extensive and growing evidence that perturbative QCD contributions are
just small $(\alpha_s/\pi)^n$ corrections incapable to explain the observed
absolute magnitude of hadronic form factors. The only exception is the $\gamma^*
\gamma \to \pi^0$ form factor for which the hard term has zero order in $\alpha_s$.
In this case, perturbative QCD should work, and it works starting with ${\cal Q}^2
\sim 2 \, {\rm GeV}^2$. Correspondingly, as we discussed earlier, Compton scattering
in which perturbative QCD also starts at $\alpha_s^0$ is an analogue of the $\gamma^*
\gamma \to \pi^0$ process---in a sense, $\gamma^* \gamma  \pi^0$ is a part of the
amplitude accounting for the most of the $\tilde E$ contribution---while the hard
meson electroproduction, where the leading hard term is ${\cal O} (\alpha_s)$, is
completely analogous to the pion form factor. In fact, the pion form factor is
extracted from the $\sigma_L$ part of the hard meson electroproduction cross section.
For this reason, one may expect that extraction of GPDs through perturbative QCD
formulas is justified for Compton-dominated processes starting at ${\cal Q}^2 \sim
2 \, {\rm GeV}^2$, while the meson electroproduction is expected to be strongly
contaminated by the soft contribution at least up to ${\cal Q}^2 \sim 10 \,
{\rm GeV}^2$. Such momentum transfers can be reached at upgraded Jefferson Lab
accelerator energies. Anticipating these future developments, we present below
the framework for perturbative description of hard meson electroproduction
processes.

\subsubsection{Exclusive meson production and QCD factorization}
\label{Sec-Int}

At sufficiently large ${\cal Q}^2$, the hard exclusive leptoproduction of a
meson $M$ from a nucleon target $N$,
\begin{equation}
\ell (k) N (p_1) \to \ell' (k') N' (p_2) M (q_2)
\end{equation}
is a promising process to test our understanding of perturbative QCD description
of exclusive reactions. It is also a unique tool to study  the properties of
nondiagonal transitions, $N \to N'$, with $N'$ being a baryon from an
$SU (3)$ multiplet, either octet or decuplet. The hadronic part of the
$\gamma_{\scriptscriptstyle L} N \to M N'$ process can be written as  the
Fourier transform of the matrix element of the electromagnetic current
Eq.\ (\ref{QuarkEMcurrent})
\begin{equation}
\label{CurrentToAmplitude}
\int d^4 z \, {\rm e}^{- i q_1 \cdot z}
\langle M (q_2) N' (p_2) | j^\mu (z) | N (p_1) \rangle
=
i (2 \pi)^4 \delta^{(4)} \left( q_1 + p_1 - q_2 - p_2 \right)
{\cal A}^\mu_M
\, ,
\end{equation}
with $q_1 = k - k'$ being  the momentum of the virtual photon, defined as
the difference of incoming and outgoing lepton momenta.

\begin{figure}[t]
\begin{center}
\mbox{
\begin{picture}(0,130)(90,0)
\put(0,0){\insertfig{5}{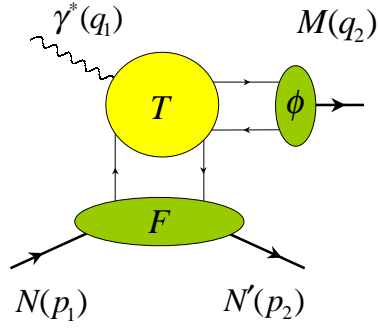}}
\end{picture}
}
\end{center}
\caption{\label{FactorizationHMP} Factorization of the meson leptoproduction at
$- q_1^2 \to \infty$ into hard rescattering amplitude $T$ and non-perturbative
functions $A$, the generalized flavour changing parton distribution, and $\phi$,
the distribution amplitude of the outgoing meson.}
\end{figure}

If the intermediate photon is longitudinally polarized and has a large virtuality
${\cal Q}^2 = - q_1^2$, the photoproduction amplitude $\gamma_{\scriptscriptstyle L}
N \to N' M$ can be rigorously treated within pQCD \cite{ColFraStr96}. A straightforward
leading twist calculation gives the following expression for the amplitude
\begin{equation}
\label{AmplitudeMeson}
{\cal A}^\mu_M
=
\frac{1}{{\cal Q}^2} \bigg( q_1^\mu + 2 x_{\rm B} p_1^\mu \bigg)
\frac{4 \pi f_M}{N_c}
{\cal F}^M_{NN'} (\eta, \Delta^2)
+
{\cal O} \left( 1/{\cal Q}^3 \right)
\, ,
\end{equation}
where $f_M$ stands for the meson decay constant, defined in Section
\ref{SectionDistributionAmplitudes}. The amplitude is explicitly
gauge invariant,
$$
q_1^\mu {\cal A}_\mu^M = 0
\, .
$$
Actually,  it was reconstructed from the structure $2 q^\mu + 3 \eta p^\mu$,
arising from the explicit calculation, by adding the transverse components
of the momentum transfer. The generalized structure function ${\cal F}$
depends on the skewness $\eta$,  the $t$-channel momentum transfer $\Delta^2$
and the resolution scale ${\cal Q}^2$. Neglecting the meson mass effects, the
skewness is equal to the generalized Bjorken variable $\eta \approx \xi$. In
the leading-twist approximation, $\mathcal{F}$ is expressed through  a convolution
of a meson distribution amplitude $\phi_M (u)$, a quark or gluon GPD $F^a$ and,
correspondingly, the quark or gluon coefficient function $T$ via \cite{ColFraStr96}
\begin{eqnarray}
\label{HMPCFFs}
&&
{\cal F}^M_{NN'} (\xi, \Delta^2)
\\
&&\qquad\equiv
\int_0^1 d u \int_{- 1}^1 d x \, \phi_M (u)
\left\{
T_M (u, x, \xi) F^M_{NN'} (x, \xi, \Delta^2)
+
T^g (u, x, \xi) F^g (x, \xi, \Delta^2)
\right\}
\, , \nonumber
\end{eqnarray}
(see Fig.\ \ref{FactorizationHMP}). Here, the quark $F^M_{NN'}$ and gluon $F^g$
GPDs provide a target-independent parametrization of the matrix elements of
light-ray operators between hadronic states, see Eqs.\ (\ref{TargetIndepQuarkFirst})
and (\ref{TargetIndepGluonFirst}). Choosing specific hadrons, one has to further
decompose them into Dirac/Lorentz structures, as was done in Section
\ref{SectionClassification}. In case of the flavor exchange in the $t$-channel,
the contribution of gluons is obviously absent, $T^g = 0$. It is also important
to realize that only chiral even GPDs enter the production amplitudes. As was
checked in \cite{Hoo02} by an explicit next-to-leading order calculation following
up an earlier all-order proof of Ref.\ \cite{ColDie00}, the chiral odd quark GPDs
do not contribute to Eq.\ (\ref{HMPCFFs}).

In the lowest order approximation, the function $T_M$ is given by the one-gluon
exchange mechanism shown in Fig.\ \ref{Fig05Chap01}. It encodes the short distance
dynamics of parton rescattering, and can be calculated in QCD perturbation theory
as a series in the strong coupling constant $\alpha_s$. The other two blocks, $\phi$
and $F$, are universal, i.e., process independent, and accumulate information about
the long-distance physics. The Bjorken limit implies  large invariant mass $W$ of
the hadronic final state, so that the produced baryon and meson are well separated
in the phase space.

The amplitude for the meson electroproduction process is very similar to the
pion form factor. Contributing diagrams can be decomposed into two sets. In the
first set, the photon is attached to the quark line. In the second set, it
is attached to the antiquark line. The momentum of the initial and final
(anti-) quark is given in the collinear approximation by $k_1 =  \frac{x +
\xi}{2\xi} \Delta$ $\left( k_2 =  \frac{x - \xi}{2\xi} \Delta \right)$ and
$k'_1 = u q_2$ $\left( k'_2 = (1 - u) q_2 \right)$, respectively. In the
leading twist approximation, both of these sets separately respect current
conservation. Obviously, if we formally replace $\frac{x + \xi}{2\xi} \to v$
and  $\frac{x - \xi}{2\xi} \to 1 - v$ with $0 < v < 1$, the kinematics reduces
to that  of the pion form factor. The essential difference is that we have to
keep track of the emerging imaginary part, which is absent in the pion form
factor case. The discontinuity develops in the region $|x| \ge \xi$ and can
easily be restored.

In case of flavor-exchange GPDs, one can use the $SU(3)$ symmetry to relate
them to the quark GPDs in the proton, as was discussed at length in Section
\ref{FlavorChangingGPDs}.

\subsubsection{Electroproduction of pseudoscalar mesons}
\label{HadronicAmplitude}

The production of pseudoscalar mesons is sensitive to the polarized
GPDs $\widetilde H$, $\widetilde E$. Therefore, in the present section,
sometimes we will use the symbol $F$ denoting either of them
$$
F^M_{NN'} = \widetilde H^M_{NN'} \, , \ \widetilde E^M_{NN'} \ .
$$
We recall that, in general, GPDs here may induce flavor-changing transitions.
For the function $\widetilde H^M_{NN'}$, one can  use the consequence of the
$SU(3)$ symmetry in the form of relations (\ref{EightToEightGPDs}). On the
other hand, for the helicity-flip functions $\widetilde E^M_{NN'}$, the effects
of the symmetry breaking are rather large and one cannot use the symmetry
arguments.

Let us present a compendium of leading order hard-scattering coefficient
function which enter the amplitude (\ref{HMPCFFs}).

\begin{itemize}
\item Charged pion production, $\gamma^\ast p \to n \pi^+$:
\begin{equation}
\label{ChargePionHardAmplitude}
T_{\pi^+} (u, x, \xi)
= C_F \alpha_s
\left\{
\frac{Q_u}{(1 - u) \left( \xi - x - i 0 \right)}
-
\frac{Q_d}{u \left( \xi + x - i 0 \right)}
\right\}
+
{\cal O} (\alpha_s^2)
\, .
\end{equation}
The function $\widetilde H^{\pi^+}_{p n}$ that accompanies this coefficient
function is the isovector combination of the proton ones
$$
\widetilde H^{\pi^+}_{pn}
=
\widetilde H_{p n}
=
\widetilde H^u - \widetilde H^d
\, ,
$$
see Eq.\ (\ref{EightToEightGPDs}).
\item Neutral pion production, $\gamma^\ast p \to p \pi^0$:
\begin{equation}
T_{\pi^0} (u, x, \xi)
= C_F \alpha_s
\left\{
\frac{1}{(1 - u) \left( \xi - x - i 0 \right)}
-
\frac{1}{u \left( \xi + x - i 0 \right)}
\right\}
+
{\cal O} (\alpha_s^2)
\, .
\end{equation}
The functions $F^{\pi^0}_{pp}$, both $\widetilde H^{\pi^0}_{pp}$ and
$\widetilde E^{\pi^0}_{pp}$  correspond to the combination
$$
F^{\pi^0}_{pp} = - \frac{1}{\sqrt{2}} \left( Q_u F^u - Q_d F^d \right)
\, .
$$
\item Charged kaon production, $\gamma^\ast p \to Y^0 K^+$, with the
neutral hyperon $Y^0 = {\mit\Lambda}, {\mit\Sigma}^0$ in the final state:
\begin{equation}
T_{K^+} (u, x, \xi)
= - C_F \alpha_s
\left\{
\frac{Q_u}{(1 - u) \left( \xi - x - i 0 \right)}
-
\frac{Q_s}{u \left( \xi + x - i 0 \right)}
\right\}
+
{\cal O} (\alpha_s^2)
\, .
\end{equation}
The transition GPDs $\widetilde H^{K^+}_{pY^0}$ with $Y = {\mit\Lambda},
{\mit\Sigma}^0$ are given by Eq.\ (\ref{EightToEightGPDs}),
$$
\widetilde H^{K^+}_{p {\mit\Lambda}}
=
\widetilde H_{p {\mit\Lambda}}
=
- \frac{1}{\sqrt{6}}
\left( 2
\widetilde H^u - \widetilde H^d - \widetilde H^s \right)
\, , \qquad
\widetilde H^{K^+}_{p {\mit\Sigma}^0}
=
\widetilde H_{p {\mit\Sigma}^0}
=
- \frac{1}{\sqrt{2}} \left( \widetilde H^d - \widetilde H^s \right)
\, .
$$
\item Neutral kaon production, $\gamma^\ast p \to K^0 {\mit\Sigma}^+$:
\begin{equation}
T_{K^0} (u, x, \xi)
= C_F \alpha_s
\left\{
\frac{Q_d}{(1 - u) \left( \xi - x - i 0 \right)}
-
\frac{Q_s}{u \left( \xi + x - i 0 \right)}
\right\}
+
{\cal O} (\alpha_s^2)
\, .
\end{equation}
The associated function $\widetilde H_{p {\mit\Sigma}^+}$ reads
$$
\widetilde H^{K^0}_{p {\mit\Sigma}^+}
=
\widetilde H_{p {\mit\Sigma}^+}
=
- \widetilde H^d + \widetilde H^s
\, .
$$
\end{itemize}

As we know, the GPD $\widetilde E$ possesses a pseudoscalar meson pole
(\ref{PionPoleInEtilde}). Since the mass difference between the pion and the
kaon are rather large, these functions  are more sensitive to the $SU (3)$ flavor
symmetry breaking. Thus, it will be more accurate to refrain from incorporating
the symmetry relations, and use instead the phenomenological values of low-energy
parameters. As we discussed in Section \ref{ChiPTGPDs}, in first approximation
one can just take the meson pole contribution only and use the following form for
the helicity-flip functions
\begin{equation}
\widetilde{E}_{p Y} (x , \eta , \Delta^2)
\simeq
\frac{\theta (\eta - |x|)}{\eta} \phi_K \left( \frac{x + \eta}{2 \eta} \right)
\frac{g_{KpY} M_N}{m_K^2 - \Delta^2}
\, ,
\end{equation}
with the couplings \cite{Goeke:2001tz}
$$
g_{Kp{\mit\Lambda}} \approx - 3.75 \sqrt{4 \pi}
\, , \qquad
g_{Kp{\mit\Sigma}} \approx 1.09 \sqrt{4 \pi}
\, .
$$

\subsubsection{Electroproduction of vector mesons}

The vector meson production process is sensitive to the unpolarized GPDs
$$
F = H , E
\, ,
$$
which enter in the spin sum rule (\ref{SpinRule}). In case of flavor-changing
transitions we will again use the $SU(3)$ relations since the symmetry breaking
effects in these quantities are very small compared to the meson-pole-dominated
function $\widetilde E$. We have the following set of results for perturbative
amplitudes and GPDs which contribute to cross sections.

\begin{itemize}
\item Neutral vector meson production, $\gamma^\ast p \to V^0_L p$:
\begin{eqnarray}
T_{V^0} (u, x, \xi)
\!\!\!&=&\!\!\!
C_F \alpha_s
\left\{
\frac{1}{(1 - u) \left( \xi - x - i 0 \right)}
-
\frac{1}{u \left( \xi + x - i 0 \right)}
\right\}
+
{\cal O} (\alpha_s^2)
\, , \\
T_g (u, x, \xi)
\!\!\!&=&\!\!\!
4 T_F \alpha_s
\frac{
\sum_q Q_q
}{
u (1 - u) ( \xi - x - i 0 )( \xi + x - i 0 )
}
+
{\cal O} (\alpha_s^2)
\, ,
\end{eqnarray}
with $V^0 = \rho^0 , \omega$. The flavor combinations of non-polarized
GPDs are
\begin{equation}
F^{\rho^0}_{pp} = Q_u F^u - Q_d F^d
\, , \qquad
F^{\omega}_{pp} = Q_u F^u + Q_d F^d
\, .
\end{equation}
The gluon GPD $F^g$ is given by the standard formula (\ref{GluonGPDnucleon}).
\item Charged vector meson production, $\gamma^\ast p \to \rho^+_L n$:
\begin{eqnarray}
T_{\rho^+} (u, x, \xi)
\!\!\!&=&\!\!\!
C_F \alpha_s
\left\{
\frac{Q_u}{(1 - u) \left( \xi - x - i 0 \right)}
-
\frac{Q_d}{u \left( \xi + x - i 0 \right)}
\right\}
+
{\cal O} (\alpha_s^2)
\, .
\end{eqnarray}
The flavor combination of non-polarized GPDs is
\begin{equation}
F^{\rho^+}_{pn} = F^u - F^d
\, .
\end{equation}
\end{itemize}

\subsubsection{Electroproduction of delta isobar}
\label{AmplitudesProtonToDelta}

Just like  in the production of pseudoscalar mesons with transitions within
the baryon octet, the pion production involving transitions from octet to
decuplet baryons is sensitive to the polarized GPDs, thus we identify $F^M_{NN'}$
in Eq.\ (\ref{HMPCFFs}) with
$$
F^M_{NN'} = \widetilde G^M_{i,NN'}
\, , \quad i = 1, \dots, 4
\, .
$$
The leading order perturbative amplitudes accompanying the corresponding
nucleon-to-delta transition GPDs read as follows.
\begin{itemize}
\item $\pi^+$ production, $\gamma^\ast p \to {\mit\Delta}^0 \pi^+$:
\begin{equation}
T_{\pi^+} (u, x, \xi)
= C_F \alpha_s
\left\{
\frac{Q_u}{(1 - u) \left( \xi - x - i 0 \right)}
-
\frac{Q_d}{u \left( \xi + x - i 0 \right)}
\right\}
+
{\cal O} (\alpha_s^2)
\, .
\end{equation}
The function $\widetilde F^{\pi^+}_{p {\mit\Delta}^0}$ which comes with this
coefficient function is the isovector combination of the proton GPDs
$$
\widetilde G^{\pi^+}_{p {\mit\Delta}^0}
=
\widetilde G_{p {\mit\Delta}^0}
=
- \frac{1}{\sqrt{3}} \widetilde G_{p {\mit\Delta}^{++}}
\, ,
$$
see Eqs.\ (\ref{OddParityDelta}) and (\ref{SU3DecupOctet}).
\item $\pi^-$ production, $\gamma^\ast p \to \Delta^{++} \pi^-$:
\begin{equation}
T_{\pi^-} (u, x, \xi)
= C_F \alpha_s
\left\{
\frac{Q_d}{(1 - u) \left( \xi - x - i 0 \right)}
-
\frac{Q_u}{u \left( \xi + x - i 0 \right)}
\right\}
+
{\cal O} (\alpha_s^2)
\, .
\end{equation}
The function $\widetilde F^{\pi^+}_{p {\mit\Delta}^{++}}$ accompanying $T_{\pi^-}$
is again the isovector combination of the proton ones
$$
\widetilde G^{\pi^-}_{p {\mit\Delta}^{++}}
=
\widetilde G_{p {\mit\Delta}^{++}}
\, ,
$$
see Eq.\ (\ref{OddParityDelta}).
\item $\pi^0$ production, $\gamma^\ast p \to \Delta^+ \pi^0$:
\begin{equation}
T_{\pi^-} (u, x, \xi)
= C_F \alpha_s
\frac{Q_u + Q_d}{2}
\left\{
\frac{1}{(1 - u) \left( \xi - x - i 0 \right)}
-
\frac{1}{u \left( \xi + x - i 0 \right)}
\right\}
+
{\cal O} (\alpha_s^2)
\, ,
\end{equation}
and $\widetilde F^{\pi^0}_{p {\mit\Delta}^{+}}$  is given by
$$
\widetilde G^{\pi^0}_{p {\mit\Delta}^+}
=
\widetilde G_{p {\mit\Delta}^+}
=
- \frac{2}{\sqrt{3}} \widetilde G_{p {\mit\Delta}^{++}}
\, ,
$$
see  Eqs.\ (\ref{OddParityDelta}) and (\ref{SU3DecupOctet}).
\end{itemize}

\subsubsection{Cross sections for electroproduction of mesons}

Having found the amplitudes, one is ready to discuss the event rates. The
cross section for exclusive electroproduction of mesons from the nucleon
target is given by
\begin{equation}
\label{XsectionInv}
d \sigma^M = \frac{1}{4 k \cdot p_1}
\left|
4 \pi \alpha_{\rm em} L_\mu {\cal A}^\mu_M
\right|^2 d {\rm LISP}_3 ,
\end{equation}
where the hadronic amplitude (\ref{AmplitudeMeson}) is contracted with the
leptonic current (which includes the photon propagator)
\begin{equation}
\label{LeptonCurrent}
L^\mu = \frac{i}{{\cal Q}^2} \bar u (k') \gamma^\mu u (k)
\, ,
\end{equation}
and the three-particle phase space volume has the standard form
\begin{equation}
\label{LISP}
d {\rm LISP}_3 = (2 \pi)^4 \delta^{(4)} (k + P_1 - k' - P_2 - q_2)
\frac{d^3 k'}{2 E' (2 \pi)^3}
\frac{d^3 P_2}{2 E_2 (2 \pi)^3}
\frac{d^3 q_2}{2 \varepsilon_2 (2 \pi)^3} .
\end{equation}
In the rest-frame of the target with the $z$-axis chosen counter-along the
momentum of the virtual photon, as shown in Fig. \ref{Fig-Kin}, one obtains
the following four-fold cross section
\begin{equation}
\label{4Xsection}
\frac{d \sigma^M}{d {\cal Q}^2 d x_{\rm B} d |\Delta^2| d \varphi_M}
=
\frac{\alpha_{\rm em}^2}{2 (4 \pi)^2} \frac{x_{\rm B} y^2}{{\cal Q}^4}
\left( 1 + 4 \frac{M_N^2 x_{\rm B}^2}{{\cal Q}^2} \right)^{-1/2}
\left|
L_\mu {\cal A}^\mu_M
\right|^2
\, .
\end{equation}
Here we use the standard variables as in VCS (\ref{ExperVariables}) and the
azimuthal angle of the outgoing meson with respect to the lepton scattering
plane is $\varphi_M = \varphi_\gamma$ with $\varphi_\gamma$ defined in Fig.\
\ref{Fig-Kin}.

The conversion from leptoproduction to photoproduction, $d
\sigma_{\scriptscriptstyle \! L}^M$, with longitudinally polarized
photons $\varepsilon^\mu_{\scriptscriptstyle \! L}$ is done by
multiplying the result (\ref{XsectionFinal}) with a kinematical
factor, namely,
\begin{equation}
\label{Conversion}
d \sigma_{\scriptscriptstyle \! L}^M = d \sigma^M
\left(
\frac{\left|
\varepsilon_{\scriptscriptstyle L} \cdot p_1
\right|^2}{q_1 \cdot p_1 }
\right)
\left(
\frac{\left| L \cdot p_1 \right|^2}{k \cdot p_1 }
\frac{d^3 k'}{2 E' (2 \pi)^3}
\right)^{-1}
=
d \sigma^M \frac{1}{\alpha_{\rm em}}
\frac{\pi}{1 - y} \frac{x_{\rm B}}{d x_{\rm B}} \frac{{\cal Q}^2}{d {\cal Q}^2}
\, .
\end{equation}
To get the last equality, we used
$$
\left|
\varepsilon_{\scriptscriptstyle \! L} \cdot p_1 \right|^2 = 4 {\cal Q}^2
\, ,
$$
where the  polarization vector of the longitudinal photon was defined in
Eq.\ (\ref{LongPolVector}), and
$$
\left| L \cdot p_1 \right|^2
=
16 \frac{1 - y}{y^2}
\, , \qquad
\frac{d^3 k'}{ 2 E' }
=
\frac{\pi y}{2} \frac{d x_{\rm B}}{x_{\rm B}} d {\cal Q}^2
\, .
$$
Thus, the cross section for scattering on a transversely polarized nucleon
target is given by
\begin{eqnarray}
\label{XsectionFinal}
\frac{d \sigma_{\scriptscriptstyle \! L}^{M}}{d |\Delta^2| d \varphi_M}
=
\frac{\alpha_{\rm em} \pi}{{\cal Q}^6}
\frac{f_M^2}{N_c^2} \frac{x_{\rm B}^2}{(2 - x_{\rm B})^2}
\left\{
\sigma_M + \sigma^\perp_M \sin{\mit\Theta} \sin ({\mit\Phi} - \varphi_M)
\right\}
\, .
\end{eqnarray}
This generic form of the cross section should be supplemented by the explicit
dependence of the unpolarized and transversely polarized components $\sigma_{M}$
and $\sigma_{M}^\perp$  on the corresponding GPDs. The calculations have been
performed in Refs.\ \cite{FraPobPolStr99,FraPolStrVan99,BelMul01,Goeke:2001tz}
for several types of outgoing baryons and mesons.
\begin{itemize}
\item Pion production:
\begin{eqnarray}
\sigma_{\pi}
\!\!\!&=&\!\!\!
8 (1 - x_{\rm B}) | \widetilde {\cal H}_{pN}^\pi |^2
-
x_{\rm B}^2 \frac{\Delta^2}{2 M_N^2}
| \widetilde {\cal E}_{pN}^\pi |^2
-
4 x_{\rm B}^2 \, \Re{\rm e}
\left(
\widetilde {\cal H}^{\pi \, \ast}_{pN} \widetilde {\cal E}_{pN}^\pi
\right)
\, , \\
\sigma^\perp_{\pi}
\!\!\!&=&\!\!\!
-
4 x_{\rm B} \sqrt{1 - x_{\rm B}} \sqrt{- \frac{\Delta^2}{M_N^2}}
\sqrt{1 - \frac{\Delta^2_{\rm min}}{\Delta^2}} \,
\Im{\rm m}
\left(
\widetilde {\cal H}^{\pi \, \ast}_{pN} \widetilde {\cal E}_{pN}^\pi
\right)
\, ,
\end{eqnarray}
where the final state baryon $N = n, p$ corresponds to the produced
$\pi = \pi^+, \pi^0$ meson, respectively. In this formula, we used the
isospin symmetry between the proton and the neutron and thus neglected
the mass difference between them, $M_p = M_n = M_N$.
\item Kaon production:
\begin{eqnarray}
\sigma_K
\!\!\!&=&\!\!\!
8 (1 - x_{\rm B}) | \widetilde {\cal H}_{pY}^K |^2
-
x_{\rm B}^2
\left\{
\frac{\Delta^2}{2 M_N^2}
-
\frac{1}{2} \left( 1 - \frac{M_Y}{M_N} \right)^2
\right\}
| \widetilde {\cal E}_{pY}^K |^2
\nonumber\\
&-&\!\!\!\
2 x_{\rm B}
\left\{
x_{\rm B} \left( 1 + \frac{M_Y}{M_N} \right)
-
(2 - x_{\rm B}) \left( 1 - \frac{M_Y}{M_N} \right)
\right\}
\, \Re{\rm e}
\left(
\widetilde {\cal H}^{K \, \ast}_{pY} \widetilde {\cal E}_{pY}^K
\right)
\, , \\
\sigma^\perp_K
\!\!\!&=&\!\!\!
-
4 x_{\rm B} \sqrt{1 - x_{\rm B}} \sqrt{- \frac{\Delta^2}{M_N^2}}
\sqrt{1 - \frac{\Delta^2_{\rm min}}{\Delta^2}} \,
\Im{\rm m}
\left(
\widetilde {\cal H}^{K \, \ast}_{pY} \widetilde {\cal E}_{pY}^K
\right)
\, ,
\end{eqnarray}
with the charged $K^+$ meson is produced with $Y = {\mit\Lambda}, {\mit\Sigma}^0$
and $K^0$ along with ${\mit\Sigma}^+$. Here, $M_Y$ is the mass of the hyperon
containing the strange quark.
\item Vector mesons:
\begin{eqnarray}
\sigma_V
\!\!\!&=&\!\!\!
8 (1 - x_{\rm B}) | {\cal H}_{pN}^V |^2
-
x_{\rm B}^2
\left(
2
+
(2 - x_{\rm B})^2 \frac{\Delta^2}{2 M_N^2}
\right)
| {\cal E}_{pN}^V |^2
-
4 x_{\rm B}^2 \, \Re{\rm e}
\left(
{\cal H}^{V \, \ast}_{pN} {\cal E}_{pN}^V
\right)
, \ \ \\
\sigma^\perp_V
\!\!\!&=&\!\!\!
4 (2 - x_{\rm B}) \sqrt{1 - x_{\rm B}} \sqrt{- \frac{\Delta^2}{M_N^2}}
\sqrt{1 - \frac{\Delta^2_{\rm min}}{\Delta^2}} \,
\Im{\rm m}
\left(
{\cal H}^{V \, \ast}_{pN} {\cal E}_{pN}^V
\right)
\, ,
\end{eqnarray}
with the outgoing nucleon being the proton $N = p$ for $V = \rho, \omega$
and the neutron $N = n$ for $V = \rho^+$.
\item Delta-isobar production:
\begin{eqnarray}
\sigma_{\Delta}
\!\!\!&=&\!\!\!
4 (1 - x_{\rm B})^2 | \widetilde{\cal G}^\pi_{3, p\Delta} |^2
+
x_{\rm B}^2
\left\{
\frac{\Delta^4}{M_N^4}
-
2 \frac{\Delta^2}{M_N^2}
\left( 1 + \frac{M_\Delta^2}{M_N^2} \right)
+
\left( 1 - \frac{M_\Delta^2}{M_N^2} \right)^2
\right\}
| \widetilde{\cal G}^\pi_{4, p\Delta} |^2
\nonumber\\
&+&\!\!\!
4 x_{\rm B}
\left\{
1 - \frac{M_\Delta^2 + \Delta^2}{M_N^2}
-
x_{\rm B}
\left(
1 + \frac{M_\Delta^2 - \Delta^2}{M_N^2}
\right)
\right\}
\Re{\rm e}
\left(
\widetilde{\cal G}_{3, p\Delta}^{\pi \, \ast} \widetilde{\cal G}^\pi_{4, p\Delta}
\right)
\, , \\
\sigma^\perp_\Delta
\!\!\!&=&\!\!\!
4 \sqrt{1 - x_{\rm B}} \sqrt{- \frac{\Delta^2}{M_N^2}}
\sqrt{1 - \frac{\Delta^2_{\rm min}}{\Delta^2}} \,
\frac{M_\Delta}{M_N}
\Im{\rm m}
\left(
\widetilde{\cal G}_{3, p\Delta}^{\pi \, \ast} \widetilde{\cal G}_{4, p\Delta}
\right)
\, ,
\end{eqnarray}
with the assignment of the reaction products specified in Section
\ref{AmplitudesProtonToDelta}.
\end{itemize}
As a concluding remark, let us point out that the electroproduction of exotics has
been discussed in Refs.\ \cite{DiPirSzy03,AniPirSzyTerWal04}.

\subsubsection{Perturbative corrections to meson production}

\begin{figure}[t]
\vspace{-1cm}
\begin{center}
\mbox{
\begin{picture}(0,155)(220,0)
\put(0,0){\insertfig{7}{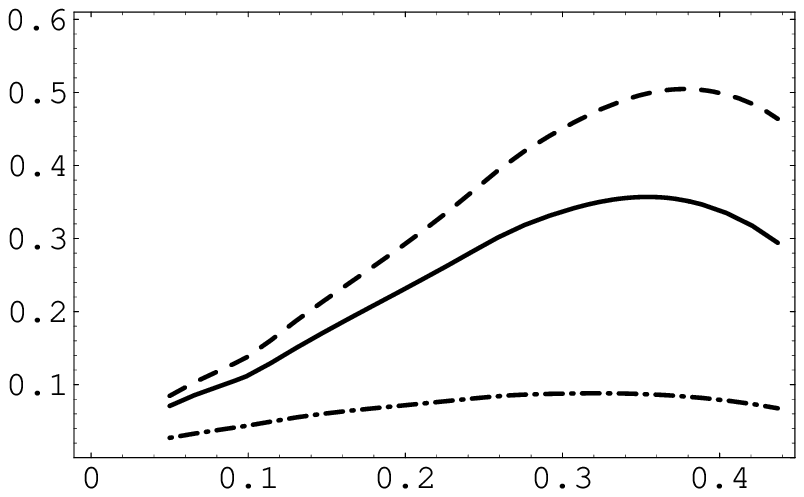}}
\put(30,100){$d \sigma_{\scriptscriptstyle L}^{\pi^+}/d|\Delta^2| \, \mbox{ vs. } x_{\rm B}$}
\put(230,0){\insertfig{7}{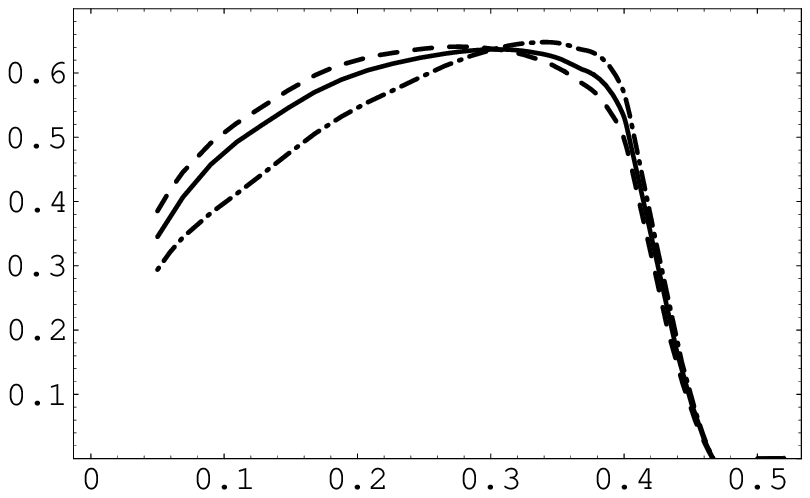}}
\put(260,20){$A^\perp_{\pi^+} \, \mbox{ vs. } x_{\rm B}$}
\end{picture}
}
\end{center}
\caption{\label{Fig-Pred} The leading twist predictions for the unpolarized
photoproduction cross section in units of nbarns $d\sigma_{\scriptscriptstyle
\! L}^{\pi^+}/d | \Delta^2 |$ are shown for $\Delta^2 = - 0.3 \ {\rm GeV}^2$
in the left panel. In the right panel, we display the transverse proton spin
asymmetry ${\cal A}^\perp_M$ for the same momentum transfer. The solid, dashed
and dash-dotted curves represent the leading and next-to-leading order with
naive and Brodsky-Lepage-Mackenzie scale setting, respectively, \cite{BelMul01}.}
\end{figure}

Let us shortly discuss the role of radiative corrections to the leading order amplitudes
taking the charged pion production as a case of study. At next-to-leading order, a
question arises about the value of the momentum scale in the argument of the strong
coupling constant. In estimates we will use two scale setting procedures: the naive
$\mu_{\scriptscriptstyle\rm R} = {\cal Q}$ in the running $\alpha_s ({\cal Q}^2)$ as
at leading order with $\Lambda^{\scriptscriptstyle\rm LO}_{\scriptscriptstyle\rm QCD}
= 220 \ {\rm MeV}$ and $N_f = 3$, in the first and the Brodsky-Lepage-Mackenzie scale
setting \cite{BroLepMak83,BelMul01,IvaSzy04,AniPirSzyTerWal04a} with a fixed coupling
at an ad hoc value $\alpha_s/\pi = 0.1$ below $1 \ {\rm GeV}^2$ in the second case.
The leading order predictions \cite{ManPilRad99,FraPobPolStr99,VanGuiGui99}
demonstrated in Fig.\ \ref{Fig-Pred} (left) are plagued by large uncertainties from
the higher order corrections in absolute cross sections, however, they largely cancel
in cross section ratios. For instance, the transverse target spin asymmetry defined
by \cite{FraPobPolStr99,BelMul01}
\begin{eqnarray}
\label{tSSA}
\left(
\int_{0}^{\pi}
d \varphi_M \frac{d \sigma_{\scriptscriptstyle \! L}^M}{d |\Delta^2| d \varphi_M}
-
\int_{\pi}^{2 \pi}
d \varphi_M \frac{d \sigma_{\scriptscriptstyle \! L}^M}{d |\Delta^2| d \varphi_M}
\right)
\left(
\int_{0}^{2 \pi}
d \varphi_M \frac{d \sigma_{\scriptscriptstyle \! L}^M}{d |\Delta^2| d \varphi_M}
\right)^{- 1}
\!\!\!&=&\!\!\!
- \frac{2}{\pi} \frac{\sigma^\perp_M}{\sigma_M} \sin {\mit\Theta} \cos {\mit\Phi}
\nonumber\\
&\equiv&\!\!\!
-
{\cal A}_M^\perp \sin {\mit\Theta} \cos {\mit\Phi}
\, . \nonumber\\
\end{eqnarray}
is shown in Fig.\ \ref{Fig-Pred} (right) for $\Delta^2 = - (0.1,\ 0.3) \
{\rm GeV}^2$ and exhibits very little sensitivity to next-to-leading order
corrections \cite{BelMul01}.

The next-to-leading order corrections are also available for heavy meson
\cite{IvaSzySchKra04} and light neutral vector meson \cite{IvaSzy04} production.

\subsubsection{Nonperturbative corrections}

The studies of higher-order perturbative corrections to the hard coefficient
function in many physical observables have demonstrated that ambiguities
generated by the perturbative resummation of fermion vacuum polarization
insertions were of the same order of magnitude as available non-perturbative
estimates of matrix elements of higher-twist operators \cite{Ben98}. The
development and sophistication of these ideas provided  some evidence that
infrared renormalons may explain the magnitude of higher-twist contributions
and even their functional dependence on scaling variables, and can thus be
used to get rough estimates of power-suppressed effects \cite{Ben98}. On
the practical side, to compute them one replaces the tree gluon propagator
in loop diagrams by the one with resummed fermion bubbles and restores
full $\beta$-function from the quark term---a procedure dubbed naive
nonabelianization. In this so-called single-chain approximation, the
propagator reads in the Landau gauge,
\begin{eqnarray*}
{\cal D}_{\mu\nu} (k) = - \frac{4 \pi}{\alpha_s \beta_0}
\int_0^\infty d \tau \,
{\rm e}^{4 \pi/(\alpha_s \beta_0) \tau}
\left( \frac{\mu^2 {\rm e}^C}{- k^2} \right)^\tau
\frac{1}{k^2} \left( g_{\mu\nu} - \frac{k_\mu k_\nu}{k^2}\right)
\, ,
\end{eqnarray*}
where $C_{\overline{\rm\scriptscriptstyle MS}} = \ft53$ in the $\overline{\rm MS}$
and $C_{\rm\scriptscriptstyle MS} = \ft53 - \gamma_{\rm E} + \ln 4 \pi$ in the
MS scheme. As usual, $\beta_0$ is the (negative) one-loop beta function
(\ref{BetaFunction}) and $\alpha_s = \alpha_s (\mu)$ is the running coupling
constant (\ref{RunningCouplingConstantTwo}). The quark coefficient function
with resummed renormalon chains for the $\pi^+$ production has the form
\cite{Bel03}
\begin{eqnarray}
T_{\pi^+} (u, x, \eta; {\cal Q}^2)
\!\!\!&=&\!\!\! - \frac{4 \pi C_F}{\beta_0}
\int_0^\infty \frac{d \tau}{\eta} \, {\rm e}^{4 \pi/(\alpha_s \beta_0) \tau}
\left( \frac{2 \mu^2 {\rm e}^C}{{\cal Q}^2} \right)^\tau
\nonumber\\
&\times&\!\!\!\Bigg\{
\frac{Q_u}{[ (1 - u) ( 1 - \frac{x}{\eta} - i 0 ) ]^{\tau + 1}}
-
\frac{Q_d}{[ u ( 1 + \frac{x}{\eta} - i 0 ) ]^{\tau + 1}}
\Bigg\}
\, ,
\end{eqnarray}
which is a straightforward modification of Eq.\ (\ref{ChargePionHardAmplitude}).

If we absorb the dependence on the momentum fraction into the argument of the
coupling constant, $\alpha_s ( \ft12 u (1 \pm \ft{x}{\eta}) {\cal Q}^2 {\rm e}^{- C})$,
it is easy to see that the end-point regions produce divergences. Infrared renormalons
are caused by the end-point singularities [Feynman mechanism] in exclusive hard-gluon
exchange amplitudes \cite{Aga96}, see also \cite{GodKiv98,KarSte01}. This can be viewed
as an estimate of the ambiguity in the resummation of higher-order perturbative
corrections or, taken to the extreme, as a model of higher-twist contributions
\cite{BenBraMag98}. Convolution of the coefficient function with the distribution
amplitude generates renormalon poles. For the asymptotic distribution amplitude
$\phi_{\rm asy} (u) = 6 u \bar u$, one gets two poles $\tau = 1$ and $\tau = 2$,
corresponding to ambiguities on the level of ${\cal Q}^{-2}$ and ${\cal Q}^{- 4}$
power corrections. Since the latter receives extra contributions from higher order
diagrams as well, it makes sense to rely only on $\tau = 1$ pole for  estimates of
the form of higher-twist corrections. Taking the imaginary part (divided by $\pi$)
arising from the contour deformation around the renormalon poles as a measure of
their magnitude, we get
\begin{equation}
\label{tilde-H}
\widetilde {\cal H}_{pn} (\eta, \Delta^2; {\cal Q}^2)
=
\widetilde {\cal H}^{\rm\scriptscriptstyle PV}_{pn} (\eta, \Delta^2; {\cal Q}^2)
+
\theta
\frac{{\mit\Lambda}^2_{\overline{\rm\scriptscriptstyle MS}}
\,
{\rm e}^{5/3}}{{\cal Q}^2}
\int_{- 1}^1 d x \, \Delta_{\widetilde H} (x, \eta)
\widetilde H_{pn} (x, \eta, \Delta^2)
\, ,
\end{equation}
where $\theta = \pm 1$ reflects the ambiguity of the contour going around the
renormalon pole in the Borel plane. Here, the one-loop expression for the QCD
coupling constant was used and
\begin{equation}
\Delta_{\widetilde H} (x, \eta)
=
- 48 \frac{\pi C_F}{\beta_0 \eta}
\Bigg\{
\frac{Q_u}{( 1 - \frac{x}{\eta} - i 0 )^2}
-
\frac{Q_d}{( 1 + \frac{x}{\eta} - i 0 )^2}
\Bigg\}
\, .
\end{equation}
Within the DD-based models for GPDs discussed in Sections \ref{GPDModels}, the
polarized GPDs and their first derivative are continuous functions at $x = \pm
\eta$ \cite{Radyushkin:1998bz}, and therefore the integral over $x$ is well-defined.
If we allow for meson exchange-like contributions in GPDs, as we discussed in
Section \ref{AnalyticPropertiesDDsAndGPDs}, this property would be lost. In the
first term of (\ref{tilde-H}) one should use the principal value prescription to
go around the poles in the Borel plane. Assuming the pion-pole dominated form for
$\widetilde E$ \cite{Goeke:2001tz} one obtains
\begin{equation}
\widetilde {\cal E}_{pn} (\eta, \Delta^2; {\cal Q}^2)
=
\widetilde {\cal E}^{\rm\scriptscriptstyle PV}_{pn} (\eta, \Delta^2; {\cal Q}^2)
-
\theta
\frac{{\mit\Lambda}^2_{\overline{\rm\scriptscriptstyle MS}}
\,
{\rm e}^{5/3}}{{\cal Q}^2} \Delta_{\widetilde E} (\eta, \Delta^2; {\cal Q}^2)
\, ,
\end{equation}
where only the single and double poles were kept at $\tau = 1$ in the second
term, so that
\begin{equation}
\Delta_{\widetilde E} (\eta, \Delta^2; {\cal Q}^2)
=
- 72 \frac{\pi C_F}{\beta_0 \eta}
F_\pi (\Delta^2)
\left(
2
+
\ln \frac{{\mit\Lambda}^2_{\overline{\rm\scriptscriptstyle MS}}
\,
{\rm e}^{5/3}}{{\cal Q}^2}
\right)
\, .
\end{equation}
Notice that the consideration of $\widetilde E$ completely parallels the
renormalon analysis of pion transition form factor \cite{Aga96,GodKiv98,KarSte01}.
In the vicinity of the pion pole one can approximate by $F_\pi (\Delta^2) = 4
g_A M_N/(m_\pi^2 - \Delta^2)$ according to Eq.\ (\ref{PionPoleInEtilde}).

\begin{figure}[t]
\mbox{
\begin{picture}(0,135)(-40,0)
\put(0,3){\insertfig{4.45}{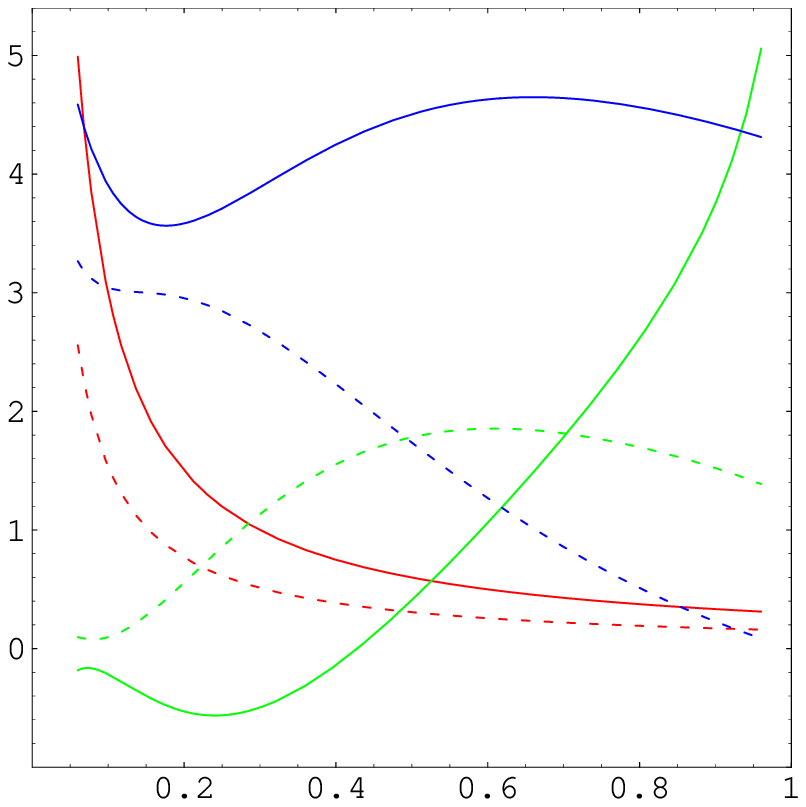}}
\put(23,87){${\scriptstyle 2}$}
\put(23,50.5){${\scriptstyle 3}$}
\put(23,24.5){${\scriptstyle 1}$}
\put(132,-4){\insertfig{4.88}{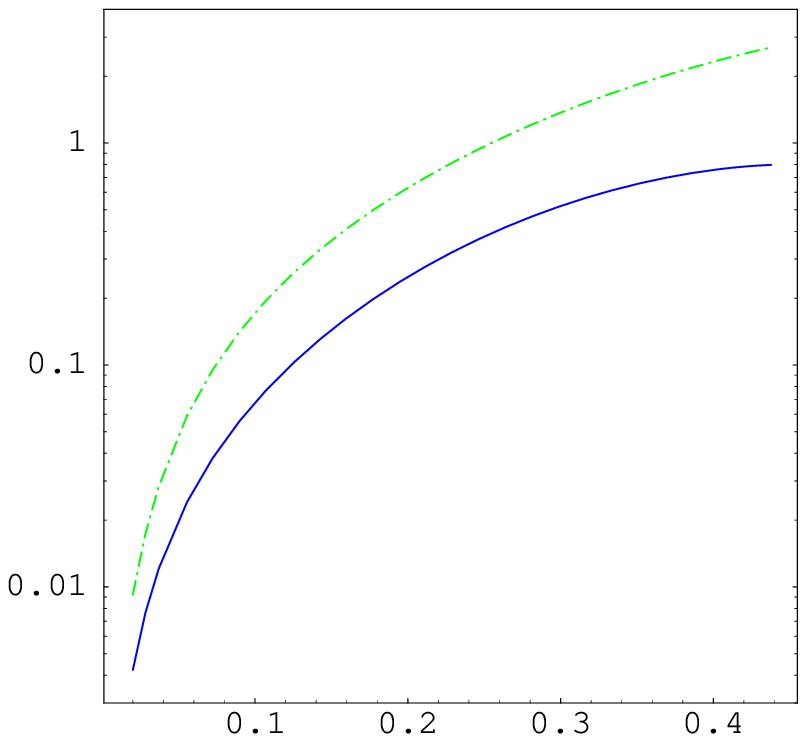}}
\put(168,20){$d \sigma_{\scriptscriptstyle L}^{\pi^+}/d|\Delta^2| \, \mbox{ vs. } x_{\rm B}$}
\put(285,0){\insertfig{4.65}{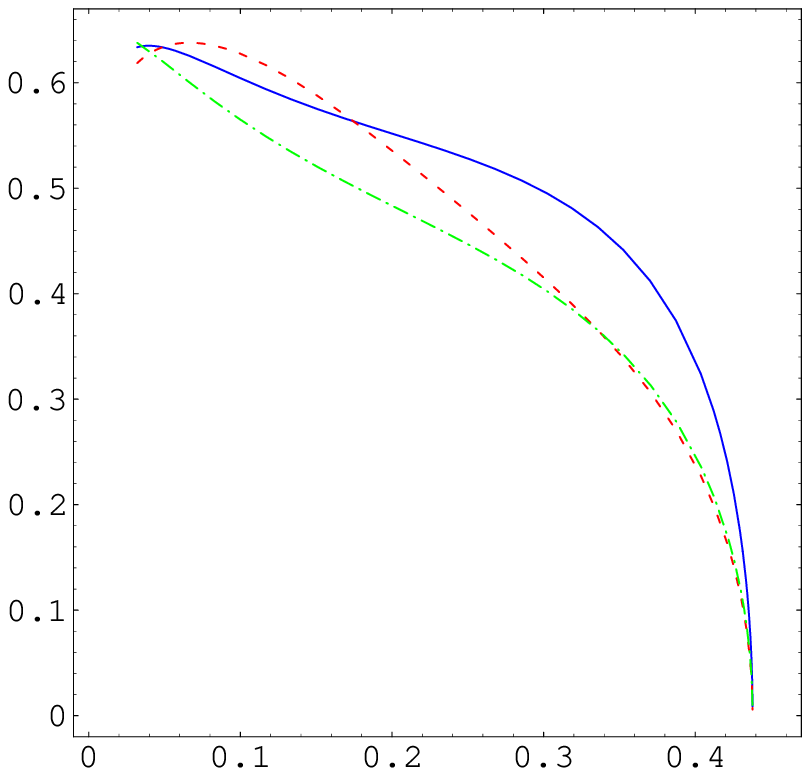}}
\put(330,20){${\cal A}^\perp_{\pi^+} \, \mbox{ vs. } x_{\rm B}$}
\end{picture}
}
\caption{\label{XsectionAndAsymmetry} Generalized structure functions (left)
in leading twist approximation (dashed) and including twist-four corrections
(solid) as a function of $x_{\rm B}$ for $\Delta^2 = - 0.3 \, {\rm GeV}^2$ and
${\cal Q}^2 = 10 \, {\rm GeV}^2$: (1) $\Re{\rm e} \widetilde {\cal H}$, (2)
$\Im{\rm m} \widetilde {\cal H}$, and (3) $10^{-2} \cdot \widetilde {\cal E}$.
The photoproduction cross section in units of nbarns (middle) without (solid)
and with (dash-dotted) power suppressed contributions for the same values of the
kinematical variables. The transverse spin asymmetry (right) at leading order
(solid) and with twist-four power effects taken into account for $\Delta^2
= - 0.3 \, {\rm GeV}^2$ and ${\cal Q}^2 = 4 \, {\rm GeV}^2$ (dashed) and
${\cal Q}^2 = 10 \, {\rm GeV}^2$ (dash-dotted). The maximal value of $x_{\rm B, max}$
is set by the kinematical constaint $|\Delta^2| > |\Delta^2_{\rm min}| =
M_N^2 x_{\rm B}^2/(1 - x_{\rm B})$.}
\end{figure}

The estimates illustrated in Fig.\ \ref{XsectionAndAsymmetry} rely on GPDs
discussed in Section \ref{NonfactorGPDModels}. The values  $\theta = 1$
and ${\mit\Lambda}_{\overline{\rm\scriptscriptstyle MS}} = 280 \, {\rm MeV}$
for $N_f = 4$ were used, and  the tree level result for
${\cal F}^{\rm\scriptscriptstyle PV} \to {\cal F}^{\rm\scriptscriptstyle LO}$.
Note, however, that to get the right magnitude of the higher-twist corrections
suggested  by  experimental data in deeply inelastic scattering, one should
take a larger value $|\theta| \approx 2 - 3$ \cite{MauSteSchMan97}.  The
extremely large size of power corrections to the absolute cross section of
the pion leptoproduction is in qualitative agreement with the earlier calculation
of Ref.\ \cite{VanManSte98}. It is interesting to note that the renormalon model
of higher-twist contributions only marginally affects the transverse target
spin asymmetry and thus leads to the apparent conclusion of {\sl the precocious
scaling} in ratios of observables,---a fact pointed out previously in various
circumstances \cite{FraPobPolStr99,BelMul01,BelJiYua02b}.

\section{Outlook}
\label{Outlook}

The last decade has witnessed substantial progress in unraveling the intriguing
puzzle of hadron's structure through the newly developed concept of generalized
parton distributions. Currently, theoretical understanding of various  aspects
of their physics and phenomenology have entered their early years of robust
maturity. Different frontiers have been explored so far and have reached profound
depth. As we discussed in this review, these developments include understanding of
precise microscopic information accumulated by these functions about the inner
content of hadrons and intuitive spatial picture encoded in GPDs; their symmetry
and partonic properties; a new emphasis on the  importance of the orbital angular
momentum of hadron's constituents in building up helicity-flip transitions and the
potential of GPDs to ultimately resolve the notorious proton spin crisis; the
renormalization group evolution of GPDs with the change of the resolution scale up
to the next-to-leading level of accuracy, just to name the most important ones.

The new functions, GPDs, are measurable in a number of exclusive reactions. The
latter are promising tools to disentangle, in a quantitative manner, the intricate
functional dependence of GPDs on intrinsic variables determining their shape. In
certain cases,  a precise theory has emerged from elaborate studies which includes
understanding of higher order and higher twist corrections. Despite these impressive
accomplishments, the subject is still in its early age of sophistication, and its
theory is far from being complete. Thus, instead of repeating the success story,
which we reviewed in the present work, let us give an outline of a few directions,
where significant improvements can be made.

The theory of exclusive processes for GPDs needs to be developed in a number of
directions, the major goal being  increasing its  predictive power. The leading
approximation for the amplitudes of  hard processes in QCD is affected by a number of
multiplicative and additive corrections. To the first class we attribute radiative
corrections in   strong coupling, while the second set  embraces higher-twist
contributions, which decay according to a power law in the hard scale. When the
given large scale ${\cal Q}$ (controlling  the applicability of factorization to
a reaction) becomes rather small, one must include power-suppressed effects since
they can significantly modify the scaling behavior and the magnitude of the
leading-twist predictions for corresponding cross sections. Power corrections are
divided, as we explained above, into two classes according to their origin:
dynamical and kinematical. The first consists of multiparton correlations inside
hadrons and gives rise to new non-perturbative functions. The second arises from
a separation of composite operators into components that have definite transformation
properties with respect to the Lorentz group, i.e., these components possess a
well-defined geometrical twist. This decomposition provides the so-called
Wandzura-Wilczek type contributions appearing from separation into components with
definite symmetry properties,  as well as target mass corrections stemming from
the subtraction of trace terms in the composite operators. All of these corrections
should  be under theoretical control to provide a reliable analysis of experimental
data at current facilities, which allow to probe only moderately hard scales.

We have reviewed a formalism based on harmonic polynomials, which span a
representation of the Lorentz group, to resum an infinite series of target
mass corrections associated with  twist-two operators. As we have seen, it
was not possible to stay within the framework of  single-variable functions
(i.e., GPDs), so we needed to employ the framework of double distributions.
The expansion in powers of $M^2$ can be cast in a conventional form of GPDs.
Without performing this mass expansion, the inverse Radon transform, of
course, directly converts the result into language of GPDs. However, a simpler
representation should be available. Another reason for extending the analysis
beyond the leading-twist level is the need to maintain electromagnetic gauge
invariance. Thus, inclusion of target mass corrections to twist-three operators
in off-forward Compton amplitudes and study of their implications for the
validity of phenomenology of leading twist approximation at available momentum
transfer is a natural next step. It would require a strong effort, since the
twist-three sector is trickier because it requires a thorough disentangling of
non-commuting procedures of twist separation and the use of Heisenberg equations
of motions for elementary fields.

Experimental and theoretical studies of quark-gluon  correlations represent
a further extension of the analysis of one-particle distributions. Studies of
the correlations between the ``QCD partons'' are interesting in two complimentary
aspects. First, they provide important quantitative information towards
understanding how the pointlike partons are bound together to produce hadrons.
The goal is, therefore, to progress  from the question ``Of what are the hadrons
built?'' to ``How they are built?'' Second, quark-gluon correlations, as we
already emphasized, are responsible for the  higher-twist corrections to hard
processes, the understanding of which is crucial in order to make QCD predictions
truly quantitative. This task is attracting growing attention recently, as it
has become clear that the discovery potential of the high-profile experimental
programs searching for  the physics beyond the standard model depends crucially
on having  effects of the strong interaction under control. A possible direction
for development of the theory of twist-four corrections in the virtual Compton
amplitude is a generalization of the Ellis-Furmanski-Petronzio formalism used
earlier for deeply inelastic scattering. A very outstanding and elaborate problem
is the construction of evolution equations for twist-four correlation functions.

We have outlined in this review the theory of the virtual Compton scattering
within the context of GPDs for spin-zero and spin-one-half targets by including
power-suppressed affects to twist-three accuracy. In order to explore the
structure of  the neutron, one must  resort to using an $A \! > \! 1$ nuclear
target, the lightest  being the the spin-1 deuteron. In order to understand how
precise one can determine neutron properties in hard reactions, one must understand
completely the structure of contributions in measurements using deuteron targets.
So far, the analysis has been performed only within the framework of the parton
model \cite{KirMul03,CanPir03} (see also \cite{FreStr03}). The consideration should
be extended at least to the same level of precision as in the DVCS on the nucleon
target, i.e., to twist-three accuracy.

A process unique in its ability to measure the surface of GPDs is the lepton-pair
production off the proton, $e p \to e' p' \ell \bar\ell$.  It has a doubly-virtual
Compton scattering (DDVCS) amplitude as a subprocess, encoding information on GPDs. An
immediate problem which has to be addressed is the calculation of the Bethe-Heitler
amplitudes for the final state leptons identical to the one in the initial state
$\ell = e$. This analysis requires the addition of exchange contributions and
revisions of all cross sections where they coherently interfere with already available
amplitudes, as discussed in Section \ref{LeptonPairProductionNonidentical}. This
will undoubtedly bring a vast variety of polarization and azimuthal observables which
will further help to unveil the complicated structure of GPDs. Yet another development
should include the study of twist-three effects in DDVCS in the same vein as it was
done for DVCS.

Currently the bulk of predictions for exclusive processes are available only to
zeroth order in strong coupling, excluding a few next-to-leading results for the
leading-twist off-forward Compton amplitude \cite{BelMul97,ManPilSteVanWei97,JiOsb98}
and certain final-state mesons in hard exclusive meson production
\cite{BelMul01,IvaSzySchKra04,IvaSzy04}. An important problem is the calculation
of one-loop corrections to twist-three Compton amplitudes with real and virtual
final-state photons and numerical studies of their magnitude. The first results
in the Wandzura-Wilczek approximation were reported in Ref.\ \cite{KivMan03}.
Continuous progress along these lines will allow one to extend the validity of
factorization theorems beyond leading order in hard scales. A promising direction
towards improvement of the leading-twist results is further development and use
of the conformal scheme for exclusive processes. A particular goal is to predict
two-loop corrections to the Compton scattering amplitude, starting from available
next-to-next-to-leading order results for deeply inelastic scattering. Another
project in higher order computation is to find one-loop corrections to the
twist-two exclusive production of tensor mesons and to the contribution of the
gluon transversity GPD in the double helicity flip production of vector mesons,
which arises at subleading twist-three level \cite{Kiv02}.

The importance of Sudakov effects in exclusive processes calls for understanding
an analogous mechanism in the leading-twist off-forward Compton scattering
amplitudes and hard exclusive meson production. The Sudakov effects are known
to reduce the sensitivity to edges of the phase space, which sometimes produce
uncontrollable contributions that ruin clean theoretical predictions. However, in
known circumstances this suppression happens away from the validity of perturbative
treatment, since the arguments of coupling run into the infrared region. This
issue must be understood for the presently discussed processes.

The description of the hard exclusive meson production in the framework of QCD
relies on  essentially the same framework as developed to address exclusive form
factors at large momentum transfer.  Therefore, all subtle issues related to the
description of helicity-flip form factors are immediately transferred to hard
meson production with a transversely-polarized photon \cite{ManPil00,AniTer03}.
In both cases, the endpoint contributions are uncontrollable in the naive form of
factorization, appearing as a convolution of a hard part with two distribution
amplitudes in case of form factors, or as a convolution of a hard part with a
meson distribution amplitude and a GPD for meson production.  And, in both cases,
one faces singularities arising from the endpoint behavior of distribution
amplitudes, which result in logarithmically divergent integrals.  This clearly
suggests that a part of the long-distance physics was erroneously included in the
hard coefficient function and calls for a thorough re-factorization.  A possible
line of investigations is to start with the $\rho$-meson form factor, where these
issues arise in the simplest circumstances, and analyze different regions of
Feynman integrals. The effective field theory that emerges from this consideration
is designed in terms of effective degrees of freedom parametrizing various regions
in the momentum space, and allows for a systematic treatment of more involved cases
like hard exclusive meson production or helicity-flip hadron form factors, including
the nucleon Pauli form factor. The latter has received intensive scrutiny from the
experimental side; it is one of the observables sensitive to the orbital motion
of constituents inside the proton and is intrinsically related to the proton's
spin content.

The study of hadron constituents that carry tiny momentum fractions allow to
probe an essentially new state of partonic matter: the color glass condensate
(for the most recent review see, e.g., \cite{Wei05}). This topic received a lot
of attention recently from the point of view of inclusive deeply inelastic
scattering. DVCS can contribute significantly toward advancement of the field.
At small $x$, a dipole picture can be applied to its description, with emphasis
on detailed understanding of the impact-parameter dependence of the two-dimensional
hadron profile (another outstanding feature made available by the use of GPDs),
which enters integrated in the analogous approach to deeply inelastic scattering.
An immediate theoretical problem is to design non-linear QCD evolution equations
to predict the small-$x$ behavior of GPDs from first principles.

The most prominent problem of GPDs concerns, of course, their reliable modeling.
Although we know their rough features in a limited range of parameter space, it
still remains a rather  weakly scrutinized area of the theory, and not because
of a lack of efforts.  A profound understanding can be achieved through elaborate
simultaneous fits to both lattice simulations and experimental data.

Transition GPDs, as described in Section \ref{FlavorChangingGPDs}, stand for matrix
elements of light-cone operators for which the hadron changes from the initial to the
final state.  Such contributions arise as a background to proton Compton scattering
and must be clearly identified in order to be subtracted from observed rates. However,
they are interesting on their own ground,  as one can use them to unveil many deep
facets of the structure of nucleon resonances, such as members of the baryon decuplet.
Questions like the intrinsic deformation of hadrons can be studied in a new environment,
distinct from the perspective of conventional form factor measurements. So far, a few
relations have been established between diagonal octet-decuplet transitions and those
of proton-proton GPDs by using large-$N_c$ QCD and SU(3) flavor symmetry. These
considerations, however, do not take into account  any deformation effects. An
interesting topic for further analysis is to study $1/N_c$ and SU(3) flavor violation
corrections to these relations.

The small-$\Delta^2$ dependence of GPDs can be revealed by making use of an effective
theory description at scales of the order of chiral symmetry breaking as was
demonstrated in Section \ref{ChiPTGPDs}. The conventional SU(2) chiral Lagrangian
can be used to compute corrections to all twist-two generalized parton distributions
to determine their dependence on momentum transfers up to ${\cal O} (m_\pi)$. This
analysis should be performed for diagonal as well as transition GPDs. So far, an
exploratory study has been done only for the parity-even GPDs as we reviewed above.
Improvements which allow to go to higher momentum transfers can be done by
systematically including the $\Delta$-isobar as a dynamical degree of freedom in
the microscopic Lagrangian.

Given the combined knowledge  gained in recent years and anticipating the
progress along the lines we outlined above, we  hope that the theory of GPDs
and related exclusive processes is capable to reach the level comparable to
that achieved in the theory of inclusive deeply inelastic processes, and to
provide access to new layers of information about hadronic structure.

\section*{Acknowledgements}

This work was supported by the U.S.\ National Science Foundation under grant no.\
PHY-0456520 (A.B.) and by the U.S.\ Department of Energy contract DE-AC05-84ER40150
under which the Southeastern Universities Research Association (SURA) operates the
Thomas Jefferson Accelerator Facility (A.R.). We are honored to dedicate this
paper to Anatoly Vasilievich Efremov, our Teacher, who taught us his approach
to physical phenomena and guided our first steps in physics. We were lucky to
work in the inspiring atmosphere he created in his group, and enjoyed collaboration
and discussions with A.V., and our friends, the members of his group, Alexander
Bakulev, Eduard Kuraev, Gregory Korchemsky, Sergei Mikhailov, Rusko Ruskov, and Oleg
Teryaev. Our work on generalized parton distributions was strongly influenced by our
contacts and collaboration with I.~Balitsky, V.~Braun, K.~Goeke, X.~Ji, L.~Mankiewicz,
I.~Musatov, D.~M\"uller, M.~Polyakov, A.~Sch\"afer, and C.~Weiss. One of us (A.B.)
expresses his sincere gratitude to Dieter M\"uller for years of fruitful and inspiring
collaboration. Over years, our GPD-effort was generously supported by the Alexander
von Humboldt Foundation, and we express our deep gratitude to the Foundation and
to our collaborators and colleagues with whom we worked and contacted in Germany:
J.~Bl\"umlein, D.~Diakonov, M.~Diehl, Th.~Feldmann, A.~Freund, B.~Geyer, V.~Guzey,
D.~Ivanov, R.~Jakob, A.~Kirchner, N.~Kivel, P.~Kroll, M.~Lazar, A.~Mukherjee,
N.~Nikolaev, H.-C.~Pauli, V.~Petrov, G.~Piller, P.~Pobylitsa, M.~Praszalowicz,
D.~Robaschik, J.~Speth, L.~Szymanowski, N.~Stefanis, W.~Schroers, M.~Vanderhaeghen,
and W.~Weise. We thank S.J.~Brodsky, M.~Burkardt, J.-W.~Chen, T.~Cohen, J.C.~Collins,
L.~Frankfurt, P.~Hoodbhoy, G.P.~Korchemsky, R.~Lebed, A.D.~Martin, G.~Miller,
B.~Pire, M.~Ryskin, M.~Savage, M.~Strikman, B.~Tiburzi, and F.~Yuan, for numerous
stimulating discussions. Last but not least, we are indebted to Markus Diehl, who
had generously agreed to be the first reader of this review, for his constructive
criticism, numerous suggestions and corrections, which lead to a significant
improvement of the manuscript.

\appendix

\setcounter{section}{0}
\setcounter{equation}{0}
\renewcommand{\theequation}{\Alph{section}.\arabic{equation}}

\section{Conventions}
\label{conventions}

We use the following convention for the metric tensor:
\begin{equation}
g_{\mu\nu} = g^{\mu\nu} = {\rm diag} (1, -1, -1, -1) \, .
\end{equation}
The contravariant and covariant vectors are defined as
\begin{equation}
v^\mu = (v_0, \mbox{\boldmath$v$})
\, , \qquad
v_\mu = (v_0, - \mbox{\boldmath$v$})
\, ,
\end{equation}
respectively. For the totally antisymmetric Levi-Civita tensor
$\varepsilon^{\mu\nu\rho\sigma}$, we adopt the normalization
\begin{equation}
\varepsilon^{0123} = 1
\, , \qquad
\varepsilon_{0123} = - 1
\, .
\end{equation}
It obeys the Schouten identity
$$
g^{\alpha\beta} \varepsilon^{\mu\nu\rho\sigma}
-
g^{\alpha\mu} \varepsilon^{\beta\nu\rho\sigma}
+
g^{\alpha\nu} \varepsilon^{\beta\mu\rho\sigma}
-
g^{\alpha\rho} \varepsilon^{\beta\mu\nu\sigma}
+
g^{\alpha\sigma} \varepsilon^{\beta\mu\nu\rho}
=
0
\, ,
$$
which stems from the absence of the totally antisymmetric rank-five tensor in
four dimensions.

\subsection{Representations of the Clifford algebra}

The Dirac $\gamma^\mu$ matrices obey the four-dimensional Clifford algebra:
\begin{equation}
\{ \gamma^\mu , \gamma^\nu \} = 2 g^{\mu\nu} \cdot \1_{[4] \times [4]} \, ,
\end{equation}
with $\1_{[4] \times [4]}$ being the $[4] \times [4]$ unit matrix. Let us
introduce two representations for this algebra used in the main text.

\noindent $\bullet$ {\sl Weyl representation:} In this representation, a bispinor
is constructed from the left $\bar\chi_\downarrow$ and right $\lambda_\uparrow$
Weyl spinors by  merely putting them in the same multiplet
\begin{equation}
\label{DiracWeyl}
\psi
=
\left(
\begin{array}{c}
\lambda_\uparrow
\\
\bar{\chi}_{\downarrow}
\end{array}
\right) \, .
\end{equation}
This is the Dirac spinor. The main reason to arrange these two-component spinors in
the same four-component vector is that the spatial parity is well defined for this
object, i.e., the left and right spinors jump their places $\bar{\chi}_\downarrow
\leftrightarrow \lambda_\uparrow$ under it. Recall that representations of the Lorentz
group $L_+^\uparrow = SO (3,1) = SO (4, \mathbb{C}) {}_{\downarrow R} \approx \left(
SL (2, \mathbb{C}) \otimes SL (2, \mathbb{C}) \right){}_{\downarrow R}$---a real form
of the complexified group of four-dimensional rotations \cite{BarRac86}---are labeled
by a pair $\left( j_1, j_2 \right)$ of numbers, which are the eigenvalues $j_i(j_i + 1)$
of the $SL (2)$ Casimir operators $\bit{J}{}^2_i$. The spinors $\bar{\chi}_\downarrow$
and $\lambda_\uparrow$ transform according to the $\left( \frac{1}{2}, 0 \right)$ and
$\left( 0, \frac{1}{2} \right)$ representations, respectively. The complex conjugate
of the $\bar{\chi}_\downarrow$ spinor transforms as $\lambda_\uparrow$, i.e.,
\begin{equation}
\bar{\chi}_\downarrow^\ast \sim i \sigma^y \lambda_\uparrow \, ,
\end{equation}
where the second Pauli $\sigma^y$ matrix serves as a charge conjugation matrix. The
Dirac matrices have the following form in the Weyl representation
\begin{equation}
\label{4Dmatrices}
\gamma^\mu
=
\left(
\begin{array}{cc}
0          & \bar\sigma^\mu \\
\sigma^\mu & 0
\end{array}
\right)
\, ,
\end{equation}
where the two-by-two matrices are
$$
\sigma^\mu = (1 , \bit{\sigma})
\, , \qquad
\bar\sigma^\mu = (1 , - \bit{\sigma})
$$
with $\bit{\sigma}$ being the vector of Pauli matrices $\bit{\sigma} =
(\sigma^x, \sigma^y, \sigma^z)$. The chirality and charge conjugation
matrices in this representation take the form
\begin{equation}
\label{Gamma5ChargeConj}
\gamma^5
=
i \gamma^0 \gamma^1 \gamma^2 \gamma^3
=
\left(
\begin{array}{rr}
1 & 0
\\
0 & - 1
\end{array}
\right)
\, , \qquad
C
=
i \gamma^2 \gamma^0
=
\left(
\begin{array}{cc}
- i \sigma_2 & 0          \\
0            & i \sigma_2
\end{array}
\right)
\, .
\end{equation}
The latter obeys the following representation-independent relations
\begin{equation}
C^T = - C
\, , \qquad
C^2 = - 1
\, , \qquad
C \gamma^\mu = - \left( \gamma^\mu \right){\!}^T C
\, .
\end{equation}

\noindent $\bullet$ {\sl Dirac representation} is obtained from the Weyl
representation by a unitary rotation
\begin{equation}
\gamma_{\scriptscriptstyle\rm D}
=
{\cal U}^\dagger \gamma_{\scriptscriptstyle\rm W} {\cal U}
\, , \qquad
{\cal U}
=
\frac{1}{\sqrt{2}}
\left(
\begin{array}{rr}
1 &   1 \\
1 & - 1
\end{array}
\right) \, .
\end{equation}
The Dirac matrices admit the form
\begin{equation}
\gamma^0
=
\left(
\begin{array}{rr}
1 &   0 \\
0 & - 1
\end{array}
\right)
\, , \qquad
\mbox{\boldmath$\gamma$}
=
\left(
\begin{array}{cc}
0 & \mbox{\boldmath$\sigma$} \\
- \mbox{\boldmath$\sigma$} & 0
\end{array}
\right)
\, , \qquad
\gamma^5
=
\left(
\begin{array}{rr}
0 & 1 \\
1 & 0
\end{array}
\right)
\, , \qquad
C = \left(
\begin{array}{cc}
0            & - i \sigma_2 \\
- i \sigma_2 & 0
\end{array}
\right) \, .
\end{equation}
The transposed matrices can be brought to the usual form with the action of the
charge conjugation matrices as follows
\begin{equation}
C \gamma_\mu^T C = \gamma_\mu
\, , \quad
C \gamma_5^T C = - \gamma_5
\, , \quad
C \left( \gamma_\mu \gamma_5 \right)^T C = - \gamma_\mu \gamma_5
\, , \quad
C \sigma^T_{\mu\nu} C = \sigma_{\mu\nu} \, ,
\end{equation}
where $\sigma_{\mu\nu} = \frac{i}{2} \left[ \gamma_\mu , \gamma_\nu \right]$
is an antisymmetric matrix, which has the property $\gamma^5 \sigma^{\mu\nu}
= \frac{i}{2} \varepsilon^{\mu\nu\rho\sigma} \sigma_{\rho\sigma}$. The product
of three $\gamma$-matrices can be reduced to one by means of the identity
\begin{equation}
\label{ThreeToOne}
\gamma^\mu \gamma^\nu \gamma^\rho
=
g^{\mu\nu} \gamma^\rho + g^{\nu\rho} \gamma^\nu - g^{\mu\rho} \gamma^\nu
+ i \varepsilon^{\mu\nu\rho\sigma} \gamma_\sigma \gamma_5
\, .
\end{equation}

\subsection{Spin-$1/2$ spinors}
\label{Spinors}

The amplitude of the plane-wave solution to the Dirac equation satisfies the equation
\begin{equation}
\label{DiracEquation}
\left( \not\! p - m \right) u (p) = 0 \, .
\end{equation}
Using the Dirac representation of the $\gamma$-matrices, the Dirac bispinor can be
written as
\begin{equation}
\label{bispinor}
u_\lambda (p)
=
\left(
\begin{array}{c}
\sqrt{E_{\bit{\scriptstyle p}} + m} \ w_\lambda
\\
\sqrt{E_{\bit{\scriptstyle p}} - m} \ ( \bit{n} \cdot \bit{\sigma} ) w_\lambda
\end{array}
\right)
\, .
\end{equation}
It defines a particle of the energy $E_{\bit{\scriptstyle p}}$ and three-momentum
$\bit{p}$, moving in the direction of the unit vector $\bit{n} \equiv \bit{p}/
\sqrt{\bit{p}^2}$. The Weyl spinors $w_\lambda$ are eigenfunctions of the helicity
operator. Recall, that the Pauli-Lubanski vector
$$
W^\mu = \ft12 \varepsilon^{\mu\nu\rho\sigma} P_\nu M_{\rho\sigma}
$$
in the rest-frame of the particle gives the helicity operator
\begin{equation}
\label{HelicityOperatorPLvector}
h = - \frac{i}{m} W^3 = i {\mit\Sigma}^{12}
=
\ft12
\left(
\begin{array}{cc}
\sigma^3 & 0        \\
0        & \sigma^3
\end{array}
\right)
\, ,
\end{equation}
where the relation ${\mit\Sigma}^{\mu\nu} = \ft14 [\gamma^\mu, \gamma^\nu]$ was
used to write the representation for the spin matrix. Since
\begin{equation}
u_\lambda (0)
=
\sqrt{2 m}
\left(
\begin{array}{c}
w_\lambda \\
0
\end{array}
\right)
\, ,
\end{equation}
in the particle's rest-frame, one finds two solutions of the eigenvalue equation
$h u_\lambda (0) = \lambda u_\lambda (0)$ with helicity $+ \ft12$ and $- \ft12$,
respectively,
\begin{equation}
w_\uparrow (0) =
\left(
\begin{array}{c}
1 \\
0
\end{array}
\right)
\, , \qquad
w_\downarrow (0) =
\left(
\begin{array}{c}
0 \\
1
\end{array}
\right)
\, .
\end{equation}
If one chooses a different direction to define polarization states, the two-component
spinor $w$ obeys the equation $\ft12 (\bit{n} \cdot \bit{\sigma}) w^\lambda = \lambda
w^\lambda$. The solutions to it are
\begin{eqnarray}
w_\uparrow
\!\!\!&=&\!\!\!
\left(
\begin{array}{c}
\cos \frac{\theta}{2}
\\
\sin \frac{\theta}{2} \, {\rm e}^{i \phi}
\end{array}
\!
\right)
\quad \ \
=
\frac{1}{\sqrt{2 |\bit{p}|( |\bit{p}| + p^3)}}
\left(
\begin{array}{c}
|\bit{p}| + p^3
\\
p_\perp
\end{array}
\right)
\, , \\
w_\downarrow
\!\!\!&=&\!\!\!
\left(
\begin{array}{c}
- \sin \frac{\theta}{2} \, {\rm e}^{- i \phi}
\\
\cos \frac{\theta}{2}
\end{array}
\right)
=
\frac{1}{\sqrt{2 |\bit{p}|( |\bit{p}| + p^3)}}
\left(
\begin{array}{c}
- \bar{p}_\perp
\\
|\bit{p}| + p^3
\end{array}
\right)
\, .
\end{eqnarray}
These Weyl spinors are normalized as $w^\ast_{\lambda'} w_\lambda =
\delta_{\lambda'\lambda}$. Here,  $(\theta, \phi)$ are the polar and
azimuthal angles specifying the direction of the vector $\bit{n}$. In
the second set of equalities, we use the holomorphic and anti-holomorphic
momenta, determined in terms of the components orthogonal to the $z$-axis,
\begin{equation}
p_\perp \equiv p^x + i p^y
\, , \qquad
\bar p_\perp \equiv p^x - i p^y
\, .
\end{equation}
The total Dirac bispinor is normalized as
\begin{equation}
\bar u_\lambda (p) u_{\lambda'} (p)
=
2 m \, \delta_{\lambda \lambda'}
\, .
\end{equation}
The density matrix is constructed from them as a direct product
\begin{equation}
u (p) \otimes \bar u (p)
=
\frac{1}{2} \left( \not\!p + m \right)
\left( 1 + \gamma_5 \frac{\not\! s}{m} \right) \, ,
\end{equation}
with the four-dimensional spin-vector
$$
s^\mu
=
2 m
\left(
\frac{(\bit{p} \cdot \bit{s})}{m}
,
\bit{s} + \frac{\bit{p} (\bit{p} \cdot \bit{s})}{m (E_{\bit{\scriptstyle p}} + m)}
\right)
\, , \qquad
s^2 = - m^2
\, .
$$
The latter is obtained from the spin three-vector $\bit{s}$ by a boost from the
rest frame along $\bit{p}$. Both relations are valid for general spin states. Using
the density matrix, we can find
\begin{equation}
\bar u (p) \gamma^\mu u (p) = 2 p^\mu
\, , \qquad
\bar u (p) \gamma^\mu \gamma^5 u (p) = 2 s^\mu
\, .
\end{equation}

The interpretation of the left and right spinors is transparent in Weyl
representation. Namely, taking the ultrarelativistic limit of the Dirac
equation,  one can neglect the mass $m = 0$ and set $|\mbox{\boldmath$p$}|
= E_{\bit{\scriptstyle p}}$, then Eq.\ (\ref{DiracEquation}) is reduced
to two uncoupled equations
\begin{equation}
\left( 1 + \mbox{\boldmath$n$} \cdot \mbox{\boldmath$\sigma$} \right)
\bar{\chi}_\downarrow = 0
\, , \qquad
\left( 1 - \mbox{\boldmath$n$} \cdot \mbox{\boldmath$\sigma$} \right)
\lambda_\uparrow = 0
\, ,
\end{equation}
for the left $\bar{\chi}_\downarrow$ and right $\lambda_\uparrow$ components of the
bispinor (\ref{DiracWeyl}). Here, $\bit{n} = \bit{p}/E_{\bit{\scriptstyle p}}$
defines the direction of the particle's motion. Choosing it along the $z$-axis,
$\bit{n} = (0, 0, 1)$, one obtains the solutions
\begin{equation}
\bar{\chi}_\downarrow
\sim
\left(
\begin{array}{c}
0 \\
1
\end{array}
\right)
\, , \qquad
\lambda_\uparrow
\sim
\left(
\begin{array}{c}
1 \\
0
\end{array}
\right)
\, .
\end{equation}
So that the left spinor $\bar{\chi}_\downarrow$ defines a particle with its
helicity pointed in the opposite direction to its motion, while the right
spinor $\lambda_\uparrow$ corresponds to the aligned orientation of its spin
and momentum.

Let us  give expressions for certain Dirac bilinears $\bar{u} (p_2) \Gamma u (p_1)$
which arise in parametrizations of operator matrix elements in different Lorentz
frames used throughout the paper.
\begin{itemize}
\item {\sl The Breit frame} is defined by the condition $\bit{p}_1 + \bit{p}_2 = 0$
and the absence of the energy exchange $\Delta^0 = E_{\bit{\scriptstyle p_1}} -
E_{\bit{\scriptstyle p_2}} = 0$, so that
\begin{equation}
E_{\bit{\scriptstyle p_1}} = E_{\bit{\scriptstyle p_2}}
=
m \sqrt{1 + \bit{\Delta}{}^2/(4 m^2)}
\, , \qquad
\bit{p}_1 = - \bit{p}_2 = \bit{\Delta}/2
\, .
\end{equation}
Thus, the parity-even Dirac bilinears read in terms of the two-component Weyl spinors
\begin{eqnarray}
\label{ParityEvenBreitFrame}
&&\bar u (p_2) \gamma^0 u (p_1)
=
2 m w_2^\ast w_1
\, , \qquad\qquad\quad\
\bar u (p_2) \gamma^k u (p_1)
=
i w_2^\ast [ \bit{\Delta} \times \bit{\sigma} ]^k w_1
\, , \\
&&\bar u (p_2) \frac{i \sigma^{\mu 0} \Delta_\mu}{2 m} u (p_1)
=
- \frac{\bit{\Delta}^2}{2 m} w_2^\ast w_1
\, , \qquad
\bar u (p_2) \frac{i \sigma^{\mu k} \Delta_\mu}{2 m} u (p_1)
=
i w_2^\ast [ \bit{\Delta} \times \bit{\sigma} ]^k w_1
\, , \nonumber
\end{eqnarray}
where $[ \bit{\Delta} \times  \bit{\sigma} ]^i \equiv \varepsilon^{ijk} \Delta^j
\sigma^k$. As it can be seen, the helicity-flip and non-flip transitions in the
Breit frame arise both from the $\gamma^\mu$ and $\sigma^{\mu\nu}$ vertices.
Analogously, the parity-odd Dirac bilinears yield
\begin{eqnarray}
\label{ParityOddBreitFrame}
&&
\bar u (p_2) \gamma^0 \gamma^5 u (p_1)
=
0
\, , \qquad\qquad\
\bar u (p_2) \gamma^k \gamma^5 u (p_1)
=
2 m \sqrt{1 + \bit{\Delta}{}^2/(4 m^2)} \, w_2^\ast \sigma^k w_1
\,  \nonumber\\
&&
\bar u (p_2) \gamma^ 5 (p_1)
=
w_2^\ast ( \bit{\sigma} \cdot \bit{\Delta} ) w_1
\, , \qquad\qquad\qquad\qquad\!
-
\frac{
w_2^\ast \Delta^k ( \bit{\sigma} \cdot \bit{\Delta} ) w_1
}{
2 m (1 + \sqrt{1 + \bit{\Delta}{}^2/(4 m^2)})
}
\, .
\end{eqnarray}
\item {\sl The Lorentz-covariant Breit frame} is a form of the Breit frame used
for applications involving heavy-baryon limit, see Section \ref{ChiPTGPDs}. In
the large-mass approximation the temporal component of the  incoming and outgoing
nucleons' four-momenta dominates over the rest and they can be decomposed as
\begin{equation}
p_{1,2}^\mu = m v^\mu + k^\mu \pm \Delta^\mu/2
\, ,
\end{equation}
with the four-velocity $v^\mu = (1, \mbox{\boldmath$0$})$ and the residual momentum
$$
k^\mu = \left( m (\sqrt{1 - \Delta{}^2/(4 m^2)} - 1), \mbox{\boldmath$0$}
\right)
\, .
$$
The large components of the velocity-dependent heavy baryon spinor are found from
the usual Eq.\ (\ref{bispinor}) with a projector
\begin{equation}
u_v (p_i) \equiv \frac{(1 + {\not\!v}) u (p_i)}{\sqrt{2 + 2 (v \cdot p_i )/ m}}
\, .
\end{equation}
They are normalized as $\bar u_v u_v = 2 m$. With these results, the Dirac bilinears
arising in Eq.\ (\ref{Decomposition}) are reduced via equations
\begin{eqnarray*}
&&\bar u (p_2) \gamma_\mu u (p_1)
= v_\mu \bar u_v (p_2) u_v (p_1)
+ \frac{1}{m} \bar u_v (p_2) [ S \cdot \Delta , S_\mu ] u_v (p_1)
\, , \\
&&\bar u (p_2) \frac{i \sigma_{\mu\nu} \Delta^\nu}{2 m} u (p_1)
= v_\mu \frac{\Delta^2}{4 m^2} \bar u_v (p_2) u_v (p_1)
+ \frac{1}{m} \bar u_v (p_2) [ S \cdot \Delta , S_\mu ] u_v (p_1)
\, .
\end{eqnarray*}
Here $S^\mu \equiv \frac{i}{2} \sigma^{\mu\nu} \gamma^5 v_\nu$ is the Pauli-Lubanski
spin vector. In the nucleon rest frame, it obviously coincides with the three-vector
of spin $S^\mu = (0, \bit{\mit\Sigma} /2 )$, where $\bit{\mit\Sigma} = {\rm diag}
(\bit{\sigma}, \bit{\sigma})$.
\item {\sl The light-cone frame} is another widely used frame, which has a unique
quantization axis defined by the boosted hadron. Thus, the helicity of all states
is defined with respect to this preferred direction. The spinors
$u^{\scriptscriptstyle\rm LC}$ in the light-cone helicity basis are related to the
conventional ones $u$, discussed above, via a unitary transformation  \cite{Diehl:2003ny}
\begin{eqnarray}
\left(
\begin{array}{c}
u_\uparrow
\\
u_\downarrow
\end{array}
\right)
\!\!\!&=&\!\!\!
\frac{1}{\sqrt[4]{8}\sqrt{p^+ |\bit{p}|
(|\bit{p}| + p^3)(E_{\bit{\scriptstyle p}} + |\bit{p}|)}}
\\
&\times&\!\!\!
\left(
\begin{array}{cc}
(|\bit{p}| + p^3)(E_{\bit{\scriptstyle p}} + |\bit{p}|)
&
m \, p_\perp
\\
- m \, \bar{p}_\perp
&
(|\bit{p}| + p^3)(E_{\bit{\scriptstyle p}} + |\bit{p}|)
\end{array}
\right)
\left(
\begin{array}{c}
u_\uparrow^{\scriptscriptstyle\rm LC}
\\
u_\downarrow^{\scriptscriptstyle\rm LC}
\end{array}
\right)
\, . \nonumber
\end{eqnarray}
The off-diagonal elements of this matrix  vanish when the particle is moving
fast and, therefore, the usual and light-cone helicities coincide. Below, we
will drop the superscript ${\scriptscriptstyle\rm LC}$, since we will talk
only about the light-cone spinors for the rest of this subsection. Explicit
form of bispinors with definite light-cone helicity can be found from the
above equation. The result for the Dirac representation is \cite{KogSop70}
\begin{equation}
\label{LightConeSpinors}
u_\uparrow (p)
= \frac{1}{\sqrt[4]{2} \sqrt{p^+}}
\left(
\begin{array}{c}
p^+ + m/\sqrt{2} \\
p_\perp/\sqrt{2} \\
p^+ - m/\sqrt{2} \\
p_\perp/\sqrt{2}
\end{array}
\right)
\, , \qquad
u_\downarrow (p)
= \frac{1}{\sqrt[4]{2} \sqrt{p^+}}
\left(
\begin{array}{c}
- \bar p_\perp/\sqrt{2} \\
p^+ + m/\sqrt{2}        \\
\bar p_\perp/\sqrt{2}   \\
- p^+ + m/\sqrt{2}
\end{array}
\right)
\, .
\end{equation}
Note, that the negative-energy spinors $u_\lambda (- p) = C \bar{u}^T_\lambda (p)$,
describing antiparticles, can be obtained from the positive-energy ones via
the simple substitution $u_{\uparrow \downarrow} (- p) = - u_{\downarrow \uparrow}
(p)|_{m \to -m}$. The bilinears which arise in the analysis of matrix elements
can be computed utilizing the explicit form of the light-cone helicity spinors:
\begin{equation}
\label{DiracBilinLCframe}
\begin{array}{ll}
\!\!
\bar u_\uparrow (p_2) \gamma^+ u_\uparrow (p_1)
=
2 \sqrt{p_1^+ p_2^+}
\, ,
&\quad\!\!\!
\bar u_\downarrow (p_2) \gamma^+ u_\uparrow (p_1)
=
0
\, , \\
\!\!
\bar u_\uparrow (p_2) \sigma^{+ k}_{\phantom{+} \perp} u_\uparrow (p_1)
=
0
\, ,
&\quad\!\!\!
\bar u_\downarrow (p_2) \sigma^{+ k}_{\phantom{+} \perp} u_\uparrow (p_1)
=
2 i \sqrt{p_1^+ p_2^+} \bit{e}^k_\perp
\, , \\
\!\!
\bar u_\uparrow (p_2) \sigma^{+-} u_\uparrow (p_1)
=
i m (p_2^+ - p_1^+)/\sqrt{p_1^+ p_2^+}
\, ,
&\quad\!\!\!
\bar u_\downarrow (p_2) \sigma^{+-} u_\uparrow (p_1)
=
i (p_1^+ p_{2 \perp} + p_2^+ p_{1 \perp})/\sqrt{p_1^+ p_2^+}
, \\
\!\!
\bar u_\uparrow (p_2) \gamma^+ \gamma^5 u_\uparrow (p_1)
=
2 \sqrt{p_1^+ p_2^+}
\, ,
&\quad\!\!\!
\bar u_\downarrow (p_2) \gamma^+ \gamma^5 u_\uparrow (p_1)
=
0
\, , \\
\!\!
\bar u_\uparrow (p_2) \gamma^5 u_\uparrow (p_1)
=
m (p_1^+ - p_2^+)/\sqrt{p_1^+ p_2^+}
\, ,
&\quad\!\!\!
\bar u_\downarrow (p_2) \gamma^5 u_\uparrow (p_1)
=
(p_2^+ p_{1 \perp} - p_1^+ p_{2 \perp})/\sqrt{p_1^+ p_2^+}
\, , \\
\end{array}
\end{equation}
where the two-dimensional vector is $\bit{e}^k = (1, i)$. Compared to the
Breit-frame bilinears in terms of conventional spinors, the helicity-flip
transitions in the present light-cone case arise from the chiral-odd Dirac
matrix $\sigma^{\mu\nu}$ and the chiral $\gamma^5$. Notice that, in general case,
the bilinears depend on each of the momenta $\bit{p}_{i \perp}$ separately.
Only when the plus-momenta are equal $p_1^+ = p_2^+$, do the bilinears depend
solely on the combination $\bit{\Delta}_\perp = \bit{p}_{1 \perp} - \bit{p}_{2 \perp}$
corresponding to the transverse momentum transfer. In the ``transverse'' Breit
frame $\bit{p}_{1 \perp} = - \bit{p}_{2 \perp} = \bit{\Delta}_\perp/2$ this also
occurs even if $p_1^+ \neq p_2^+$.
\end{itemize}

\subsection{Spin-$1$ vector field}

To describe a spin-one field, let us introduce a definite-helicity polarization
vectors $\bit{\varepsilon}$ in the particle's rest frame. The spin operator acts
on spin-one polarization vector as follows
\begin{equation}
{\mit\Sigma}^{\mu\nu} \varepsilon^\rho (0)
=
g^{\mu\rho} \varepsilon^\nu (0) - g^{\nu\rho} \varepsilon^\mu (0)
\, .
\end{equation}
Therefore, the eigenvalue equation for the helicity operator acting on the
three-vector $\bit{\varepsilon}$ of $\varepsilon^\mu = (0, \bit{\varepsilon})$
can be cast into the  matrix form
\begin{equation}
h \bit{\varepsilon}
=
\left(
\begin{array}{ccc}
0 & -i & 0 \\
i &  0 & 0 \\
0 &  0 & 0
\end{array}
\right)
\left(
\begin{array}{c}
\varepsilon^x \\
\varepsilon^y \\
\varepsilon^z
\end{array}
\right)
=
\lambda \bit{\varepsilon}
\, .
\end{equation}
One easily finds the eigenvectors of  this equation corresponding to helicity
$+1$ (right circular polarization), $-1$ (left circular polarization) and $0$
(longitudinal polarization),
\begin{equation}
\label{VectorPolarization}
\bit{\varepsilon}_\uparrow
=
- \frac{1}{\sqrt{2}}
\left(
\begin{array}{c}
1 \\
i \\
0
\end{array}
\right)
\, , \qquad
\bit{\varepsilon}_\downarrow
=
- \frac{1}{\sqrt{2}}
\left(
\begin{array}{c}
- 1 \\
i \\
0
\end{array}
\right)
\, , \qquad
\bit{\varepsilon}_0
=
\left(
\begin{array}{c}
0 \\
0 \\
1
\end{array}
\right)
\, ,
\end{equation}
respectively. By boosting these rest-frame polarization vectors into the system moving
with three-momentum $\bit{p}$, one gets the four-polarization vector in an arbitrary
frame
\begin{equation}
\label{SpinOneVector}
\varepsilon^\mu_s (p)
=
\left(
\frac{(\bit{p} \cdot \bit{\varepsilon}_s)}{m}
,
\bit{\varepsilon}_s
+
\frac{
\bit{p} (\bit{p} \cdot \bit{\varepsilon}_s)
}{
m (E_{\bit{\scriptstyle p}} + m)}
\right)
=
L^\mu{}_\nu (p) \varepsilon^\nu_s (0)
\, .
\end{equation}
To get  the last equation,  we used the representation in terms of the Lorentz
boost from the rest frame vector $\varepsilon^\mu_s (p) = (0, \bit{\varepsilon}_s)$,
with
\begin{equation}
L^\mu{}_\nu (p) = \frac{1}{m}
\left(
\begin{array}{cc}
E_{\bit{\scriptstyle p}}
&
p^j
\\
p^i
&
m \delta^{ij} + {\displaystyle\frac{p^i p^j}{E_{\bit{\scriptstyle p}} + m}}
\end{array}
\right)
\, .
\end{equation}
The polarization four-vector is normalized as
\begin{equation}
\varepsilon^\ast_{s_1} \cdot \varepsilon_{s_2} = - \delta_{s_1 s_2}
\, .
\end{equation}

\subsection{Spin-$3/2$ Rarita-Schwinger spin-vector}
\label{RSchSpinors}

The Rarita-Schwinger spin-vector $u^\mu (p)$ describes the wave function of
a spin-three-half particle \cite{RarSch46}, see Ref.\ \cite{HemHolKam98} for
a comprehensive summary. It satisfies the Dirac equation
$$
( {\not\! p} - m ) u^\mu (p) = 0
\, ,
$$
and two subsidiary conditions
$$
p_\mu u^\mu (p) = 0
\, , \qquad
\gamma_\mu u^\mu (p) = 0
\, .
$$
The second condition projects out the spin-one-half component contaminating the
spin-vector $u^\mu$. The spin-vector is constructed as a superposition of the
Dirac bispinor $u_\lambda (p)$ (\ref{bispinor}) and  the covariant spin-one
polarization vector (\ref{SpinOneVector})
\begin{equation}
\label{RSspinvector}
u^\mu_S (p)
=
\sum_{s_1, s_2}
\delta_{S, s_1 + s_2}
\left\langle \left. \ft12 , s_1 , 1 , s_2 \right| \ft32 , S \right\rangle
u_{s_1} (p) \varepsilon^\mu_{s_2} (p)
\, ,
\end{equation}
with the Clebsch-Gordan coefficients
$$
\left\langle \left. j_1 , m_1 , j_2 , m_2 \right| j , m \right\rangle
=
(- 1)^{j_1 - j_2 + m}
\sqrt{2 j_1 + 1}
\left(
\begin{array}{ccc}
j_1 & j_2 & j \\
m_1 & m_2 & - m
\end{array}
\right)
\, .
$$
Explicit form of different spin components of the Rarita-Schwinger spin-vector
(\ref{RSspinvector}) is given by
\begin{eqnarray}
u^\mu_{3/2} (p)
\!\!\!&=&\!\!\!
u_\uparrow (p) \varepsilon^\mu_\uparrow (p)
\, , \qquad
u^\mu_{1/2} (p)
=
\sqrt{\frac{1}{3}} u_\downarrow (p) \varepsilon^\mu_\uparrow (p)
+
\sqrt{\frac{2}{3}} u_\uparrow (p) \varepsilon^\mu_0 (p)
\, , \\
u^\mu_{- 3/2} (p)
\!\!\!&=&\!\!\!
u_\downarrow (p) \varepsilon^\mu_\downarrow (p)
\, , \qquad
u^\mu_{- 1/2} (p)
=
\sqrt{\frac{1}{3}} u_\uparrow (p) \varepsilon^\mu_\downarrow (p)
+
\sqrt{\frac{2}{3}} u_\downarrow (p) \varepsilon^\mu_0 (p)
\, .
\end{eqnarray}

Another widely used representation $u^\mu$ can be obtained from the above
expressions by introducing a (transposed) four-component spinor for
spin-three-half states,
$$
\chi_S^T = (\delta_{S,3/2}, \delta_{S,1/2}, \delta_{S,-1/2}, \delta_{S,- 3/2})
$$
and explicit summation over $s_1$ and $s_2$ in Eq.\ (\ref{RSspinvector}). This
yields
\begin{equation}
u^\mu_S (p) = L^\mu{}_\nu
\left(
\begin{array}{c}
\sqrt{E_{\bit{\scriptstyle p}} + m} \ \mathbb{R}^\nu \chi_S
\\
\sqrt{E_{\bit{\scriptstyle p}} - m} \ ( \bit{n} \cdot \bit{\sigma} )
\mathbb{R}^\nu \chi_S
\end{array}
\right)
\, ,
\end{equation}
where $\mathbb{R}^0 = 0$, while the remaining  $[2] \times [4]$ matrices
are given by
\begin{equation}
\mathbb{R}^1 =
\!
\left(
\!\!
\begin{array}{cccc}
- \ft{1}{\sqrt{2}}
&
0
&
\ft{1}{\sqrt{6}}
&
0
\\
0
&
- \ft{1}{\sqrt{6}}
&
0
&
\ft{1}{\sqrt{2}}
\end{array}
\!\!
\right)
\, , \ \
\mathbb{R}^2 =
\!
\left(
\!\!
\begin{array}{cccc}
- \ft{i}{\sqrt{2}}
&
0
&
- \ft{i}{\sqrt{6}}
&
0
\\
0
&
- \ft{i}{\sqrt{6}}
&
0
&
- \ft{i}{\sqrt{2}}
\end{array}
\!\!
\right)
\, , \ \
\mathbb{R}^3 =
\!\left(
\!\!
\begin{array}{cccc}
0
&
\sqrt{\ft{2}{3}}
&
0
&
0
\\
0
&
0
&
\sqrt{\ft{2}{3}}
&
0
\end{array}
\!\!
\right)
\, .
\end{equation}

\subsection{Particle states}

Throughout this paper, the vector of state is normalized in the following way. For a
one-particle state with three-momentum $\bit{p}$ and energy $E_{\bit{\scriptstyle p}}
= \sqrt{m^2 + \bit{p}^2}$, and other quantum numbers $\{ \alpha \}$, like spin, flavor
etc.,
\begin{equation}
\bigg|
h (\mbox{\boldmath$p$}, E_{\bit{\scriptstyle p}} , \{ \alpha \})
\bigg\rangle
\equiv
| p \rangle
\, ,
\end{equation}
we have
\begin{equation}
\langle p' | p \rangle
=
2 E_{\bit{\scriptstyle p}} (2 \pi)^3
\delta^{(3)}
\left( \bit{p}' - \bit{p} \right)
\delta_{\{ \alpha' \}\{ \alpha \}}
\, ,
\end{equation}
where $\delta_{\{ \alpha' \}\{ \alpha \}}$ is the Kronecker delta. For an $n$-particle
state, we have an obvious generalization,
\begin{eqnarray}
\langle
p_1 ,\dots p_N
|
p'_1 , \dots p'_N
\rangle
=
\prod_{k = 1}^{N}
2 E_{\bit{\scriptstyle p}_k}
(2 \pi)^3
\delta^{(3)}
\left(
\bit{p}'_k - \bit{p}_k
\right)
\delta_{\{ \alpha'_k \}\{ \alpha_k \}}
\, ,
\end{eqnarray}
where we assume that there are no identical particles. If they are  present,
one  simply  needs to  introduce symmetry factors. The completeness condition
written symbolically as
\begin{equation}
\label{CompletenessAppendix}
\sum_n \left| n \right\rangle \left\langle n \right| = 1 \, ,
\end{equation}
is a shorthand notation for
\begin{eqnarray}
\label{CompletenessExact}
\sum_{N}
\sum_{\{ \alpha_N \}}
\int \prod_{k = 1}^N \frac{d^3 \bit{p}_k}{2 E_{\bit{\scriptstyle p}_k } (2 \pi)^3}
| p_1 , \dots , p_k \rangle
\langle p_1 , \dots , p_k |
= 1 \, ,
\end{eqnarray}
where the element of the phase space can be rewritten in a relativistically-invariant
way:
\begin{equation}
\frac{d^3 \bit{p}_k}{2 E_{\bit{\scriptstyle p}_k } (2 \pi)^3}
=
\frac{d^4 p_k}{(2 \pi)^4} (2 \pi) \delta_+ (p^2_k - m^2_k)
\equiv
\frac{d^4 p_k}{(2 \pi)^4} (2 \pi)
\theta (p^0_k) \delta (p^2_k - m^2_k)
\, ,
\end{equation}
with the step function insuring the positivity of the energy flow.

\subsection{Color algebra}

The commutator algebra of the $SU (N_c)$ color group is
\begin{equation}
\left[ T^a , T^b \right] = i f^{abc} T^c \, .
\end{equation}
The generators in the fundamental $\left( T^a \right){}^i{}_j {\mit\Phi}^a = \left( t^a
\right){}^i{}_j {\mit\Phi}^a$ and adjoint $\left( T^c \right){}^{ab} {\mit\Phi}^c =
i f^{acb} {\mit\Phi}^c$ representations of the color group act on the quark and gluon
fields, respectively. The generators in the fundamental representation $t^a$ are
normalized by the condition
\begin{equation}
{\rm tr} \, t^a t^b = T_F \delta^{ab} \, .
\end{equation}
The color Casimir in the fundamental and adjoint representations are
\begin{equation}
\left( t^a \right)^i{}_j \left( t^a \right)^j{}_l
=
C_F \delta^i{}_l
\, , \qquad
f^{abc} f^{abd} = C_A \, \delta^{cd}
\, ,
\end{equation}
respectively. The numerical values of the constants involved are
\begin{equation}
T_F = \frac{1}{2}
\, , \qquad
C_F = \frac{N_c^2 - 1}{2 N_c}
\, , \qquad
C_A = N_c
\  .
\end{equation}

\section{Light-cone vectors and tensors}
\label{LCvectors}

In discussion of high-energy scattering, it is convenient to introduce a pair of
light-cone vectors such that $n^2 = n^{\ast 2} = 0$ and $n \cdot n^\ast \equiv
n^\mu n^\ast_\mu = 1$. They can be chosen in the following way
\begin{equation}
\label{LightLikeVectors}
n^\mu \equiv \ft{1}{\sqrt{2}} (1, 0, 0, -1)
\, , \qquad
n^{\ast \mu} \equiv \ft{1}{\sqrt{2}} (1, 0, 0, 1)
\, .
\end{equation}
Any four-vector $z^\mu$ can be decomposed into its light-cone components as
\begin{equation}
z^\mu = z^+ n^{\ast \mu} + z^- n^\mu + z^\mu_\perp
\, ,
\end{equation}
with
\begin{equation}
z^+ \equiv z \cdot n = \ft{1}{\sqrt{2}} (z^0 + z^z)
\, , \qquad
z^- \equiv z \cdot n^\ast = \ft{1}{\sqrt{2}} (z^0 - z^z)
\, .
\end{equation}
Scalar products are  written as
\begin{equation}
z \cdot y
\equiv
z_\mu y^\mu
=
z^+ y^- + z^- y^+ - \bit{z}_{\perp} \cdot \bit{y}_{\perp}
\, .
\end{equation}
The light-cone derivatives are conventionally defined as follows
\begin{equation}
\partial^+ = \frac{\partial}{\partial z^-}
\, , \qquad
\partial^- = \frac{\partial}{\partial z^+}
\, , \qquad
\partial^\mu_\perp = \frac{\partial}{\partial z_\mu^\perp}
\end{equation}
We use Euclidean notations for the transverse two-dimensional space with
the metric
$$
\delta_{\alpha\beta}
=
- g^\perp_{\alpha\beta}
\equiv - ( g_{\alpha\beta} - n_\alpha n^\ast_\beta - n_\alpha^\ast n_\beta )
= {\rm diag} (0, 1, 1, 0)
\, ,
$$
and the transverse projection of the totally antisymmetric tensor
$\varepsilon_\perp^{\mu \nu} \equiv \varepsilon^{\mu \nu - +}$, which has
the following nonzero components
$$
\varepsilon_\perp^{12} = - \varepsilon_\perp^{21}
=
\varepsilon^\perp_{12} = - \varepsilon^\perp_{21}
=
1
\, .
$$
Therefore, using the transverse Euclidean metric, we express $v_\perp \cdot
u_\perp = - \bit{v}_{\perp} \cdot \bit{u}_{\perp}$ in terms of the two-dimensional
vectors, $\bit{v} = (v^x, v^y)$. For contraction with transverse Dirac matrices,
we use the convention
\begin{equation}
\not\!\bit{k}_\perp
=
\bit{\gamma}_\perp \cdot \bit{k}_\perp
=
- \gamma_\perp \cdot k_\perp
\, .
\end{equation}
Let us note that, due to the boost invariance along the $z$-axis, one can rescale
the light-cone vectors,
$$
n^{\ast \mu} \to \varrho n^{\ast\mu}
\, , \qquad
n^\mu \to \varrho^{- 1} n^\mu
\, ,
$$
such that their product remains unchanged ($n \cdot n^\ast = 1$). One may choose
the value of $\varrho$ conveniently adjusting it to particular settings. Let us
consider a few choices used in practical applications.

\subsection{Compton frame}
\label{DISframe}

For most cases considered in this review, one can use the vectors defining the
kinematics of physical processes to construct the pair of the light-cone vectors
$n^\mu$ and $n^{\star \mu}$. In particular, take  the $2 \to 2$ particle kinematics,
$p_1 + q_1 = p_2 + q_2$, of the Compton amplitude. We can choose a reference frame
where the average momenta are collinear to the $z$-axis and are oppositely directed.
Then we find
\begin{eqnarray}
\label{LCvectorsComptonFrame}
n_\mu
\!\!\!&=&\!\!\!
\frac{
2 \xi
}{
Q^2 \sqrt{1 + 4 (\xi \delta)^2}
} \, q_\mu
-
\frac{
1 - \sqrt{1 + 4 (\xi \delta)^2}
}{
2 Q^2 \delta^2 \sqrt{1 + 4 (\xi \delta)^2}
} \, p_\mu
\, , \\
n^\star_\mu
\!\!\!&=&\!\!\!
- \frac{
\xi \delta^2
}{
\sqrt{1 + 4 (\xi \delta)^2}
} \, q_\mu
+
\frac{
1 + \sqrt{1 + 4 (\xi \delta)^2}
}{
4 \sqrt{1 + 4 (\xi \delta)^2}
} \, p_\mu \, ,
\nonumber
\end{eqnarray}
where we used the  momenta
\begin{equation}
\label{AverageMomenta}
p = p_1 + p_2
\, , \qquad
q = \ft12 (q_1 + q_2)
\, , \qquad
\Delta = p_1 - p_2 = q_2 - q_1
\, ,
\end{equation}
and the kinematical invariants
\begin{equation}
\label{DefinitionOfDeltaEtAl}
Q^2 = - q^2
\, , \qquad
\xi = \frac{Q^2}{p \cdot q}
\, , \qquad
\delta^2
\equiv
\frac{M_N^2 - \ft14 \Delta^2}{Q^2}
\, ,
\end{equation}
introduced in Section \ref{ComptonVaribales}. Using the reparametrization invariance,
we attributed a mass dimension to the vector $n^{\ast \mu}$ and correspondingly the
inverse mass dimension to $n^\mu$.

The light-cone decomposition of momenta for a generic $2 \to 2$ scattering then can
be found as an inverse transformation to Eq.\ (\ref{LCvectorsComptonFrame}),
\begin{eqnarray}
p_\mu
\!\!\!&=&\!\!\!
2 n^\star_\mu + Q^2 \delta^2 n_\mu
\, , \\
q_\mu
\!\!\!&=&\!\!\!
-
\frac{
2 \xi
}{
1 + \sqrt{1 + 4 ( \xi \delta )^2}
} n^\star_\mu
-
\frac{
\xi \, Q^2 \delta^2
}{
1 - \sqrt{1 + 4 ( \xi \delta )^2}
} n_\mu
\, . \nonumber
\end{eqnarray}

Expansion for  the momentum difference may have slightly different coefficients
depending on the way one introduces the skewness variable. It can be either defined
as a Lorentz invariant of measurable (external) momenta,
\begin{equation}
\eta = \frac{\Delta \cdot q}{p \cdot q}
\, ,
\end{equation}
or through  the light-cone coordinates,
\begin{equation}
\tilde\eta = \frac{\Delta^+}{p^+}
\, .
\end{equation}
Looking at the explicit expressions of the kinematical variables in terms of
light-cone vectors, one would expect a power-suppressed difference between these
two definitions. Specifically,
\begin{equation}
\tilde\eta =
\frac{
\eta
}{
\sqrt{1 + 4 ( \xi \delta )^2}
}
\, .
\end{equation}
The momentum transfer in terms of  these two  definitions is given by
\begin{equation}
\Delta_\mu
=
\frac{
2 \eta
}{
\sqrt{1 + 4 ( \xi \delta )^2}
} n^\star_\mu
-
\frac{
\eta \, Q^2 \delta^2
}{
\sqrt{1 + 4 ( \xi \delta )^2}
}
n_\mu
+
\Delta^\perp_\mu
\, , \qquad
\Delta_\mu
=
2 \tilde\eta n^\star_\mu
-
\tilde\eta Q^2 \delta^2 n_\mu
+
\Delta^\perp_\mu
\, .
\end{equation}
This difference is irrelevant for our presentation since
almost everywhere we are dealing  with accuracy
below the twist-four level, and
we can take $\eta \approx
\widetilde \eta$. The magnitude of the transverse momentum
component is expressed through  the
difference between the four-dimensional momentum squared and its minimal
accessible  value
\begin{equation}
\label{TranverseDeltaBreit}
\Delta_\perp^2
\equiv -\bit{\Delta}_\perp^2
=
(1 - \widetilde{\eta}^2) ( \Delta^2 - \Delta^2_{\rm min} )
\, , \qquad
\Delta^2_{\rm min}
=
- \frac{4 \widetilde{\eta}^2 M_N^2}{1 - \widetilde{\eta}^2}
\, .
\end{equation}
In many cases, one can neglect the corrections ${\cal O}(\delta^2)$
and use
 approximate expressions for  the momenta (\ref{AverageMomenta})
 in terms of the
light-cone
variables,
\begin{equation}
p_\mu
\simeq
2 n^\star_\mu
\, , \qquad
q_\mu
\simeq
\frac{Q^2}{2 \xi} n_\mu - \xi n^\star_\mu
\, , \qquad
\Delta^\mu
\simeq
2 \eta \, n^{\ast\mu} + \Delta_\perp^\mu
\, .
\end{equation}
There are similar forms for  the  incoming and outgoing momenta
\begin{eqnarray}
\label{MomentaComptonFrame}
&&
q_{1 \mu}
\simeq
\frac{Q^2}{2 \xi} n_\mu
-
( \xi + \eta ) n^\star_\mu
-
\ft12 \Delta^\perp_\mu
\, , \qquad
q_{2 \mu}
\simeq
\frac{Q^2}{2 \xi} n_\mu
-
( \xi - \eta ) n^\star_\mu
+
\ft12 \Delta^\perp_\mu
\, , \\
&&
p_{1 \mu}
\simeq
( 1 + \eta ) n^\star_\mu
+
\ft12 \Delta^\perp_\mu
\, , \qquad\qquad\quad \ \
p_{2 \mu}
\simeq
( 1 - \eta ) n^\star_\mu
-
\ft12 \Delta^\perp_\mu
\, .
\end{eqnarray}
As one can see, we have chosen the ``transverse'' Breit-frame assignment for
transverse components of the
 proton momenta, i.e., $p_{1 \perp}^\mu = -
p_{2 \perp}^\mu = \ft12 \Delta_\perp^\mu$. In certain cases,
 we will lift
this condition, in particular, in Sections \ref{GPDsInImpactParameterSpace} and
\ref{FactorizationComptonAmplitude}. Analogous relations apply to photons'
momenta. This approximation will be used throughout our analysis. Analogously,
for the light-cone vectors one takes
\begin{equation}
n_\mu \simeq \frac{\xi}{Q^2} (2 q_\mu + \xi p_\mu)
\, , \qquad
n^\ast_\mu \simeq \ft12 p_\mu
\, .
\end{equation}

\subsection{Breit frame}
\label{Breitframe}

Let us introduce the ``longitudinal'' Breit frame for the $2 \to 2$ hard elastic
scattering, which is useful in discussing the factorization theorems for exclusive
meson production \cite{ColFraStr96} addressed in Section \ref{Sec-Int}. This is a
frame where the incoming virtual photon has zero energy and moves along the
$z$-axis. The initial-state proton moves head-on into this ``brick-wall'' photon,
i.e., counter-along the $z$-axis, while the final-state hadron recoils backwards
and goes in the same direction as the virtual photon. The current fragmentation
system---the final state photon or meson---moves in the same direction as the
recoiled hadron. Neglecting the mass parameters and small transverse components,
the relevant vectors can be written in terms of dimensionless light-like vectors
(\ref{LightLikeVectors})
\begin{eqnarray}
q_1^\mu
\!\!\!&=&\!\!\!
\frac{{\cal Q}}{\sqrt{2}} n^{\ast \mu}
-
\frac{{\cal Q}}{\sqrt{2}} n^\mu
\, , \\
q_2^\mu
\!\!\!&=&\!\!\!
\frac{{\cal Q}}{\sqrt{2}} \frac{\xi - \eta}{\xi + \eta} n^{\ast \mu}
-
\frac{{\cal Q}}{\sqrt{2}} n^\mu
\, , \\
p_1^\mu
\!\!\!&=&\!\!\!
- \frac{{\cal Q}}{\sqrt{2}} \frac{1 + \eta}{\xi + \eta} n^{\ast \mu}
\, , \\
p_2^\mu
\!\!\!&=&\!\!\!
- \frac{{\cal Q}}{\sqrt{2}} \frac{1 - \eta}{\xi + \eta} n^{\ast \mu}
\, ,
\end{eqnarray}
 with ${\cal Q}^2= - q_1^2 $.

\subsection{Drell-Yan frame}
\label{DYframe}

There is yet another useful frame for discussion of the deeply inelastic scattering
of a virtual photon with momentum $q$ on a on-shell quark $l$ with production of a
quark jet $p_J = q + l$ in the final state, see Fig.\ \ref{LinkDIS}. It was used
in Section \ref{QCDPartonDistributions}. The frame is determined by transforming
into the ``center-of-mass'' frame of the incoming quark $l$ and the outgoing quark
jet $p_J$ and is dubbed the Drell-Yan frame \cite{MenOlnSop92,Col92}. Namely, the
four-momentum of the current jet $p_J$ is light-like, $p_J^2 = 0$, and can be
conveniently chosen as one of the light-like vectors, with the other one fixed by
the hadron momentum $p_\mu = n^\ast_\mu$. Thus, we define a normalized null-vector
$\tilde n_\mu$ tangent to $n^\mu$,
\begin{equation}
\tilde n_\mu \equiv \frac{p_{J\mu}}{p_J \cdot n^\ast} \, .
\end{equation}
Obviously, $\tilde n^2 = n^{\ast 2} = 0$ and $n \cdot \tilde n = 1$. The Sudakov
decomposition of all Lorentz vectors is straightforward
\begin{equation}
v_\mu = \tilde n_\mu v_- + n^\ast_\mu v_+ + v^\perp_\mu \, .
\end{equation}
We keep the same `$+$'-index notation for contractions with the vector $\tilde n$:
$v^+ \equiv \tilde n \cdot v$. In the Bjorken limit, $q^- \to \infty$, the difference
between $p_J^-$ and $q^-$ is negligible and, therefore, the Compton frame from
Appendix \ref{DISframe} and the present Drell-Yan frame \cite{MenOlnSop92,Col92}
coincide.

\section{Optical theorem}
\label{OpticalTheorem}

The optical theorem asserts  that
\begin{equation}
\label{OpticalTheoremEq}
\int d^4 z \ {\rm e}^{i q \cdot z}
\langle {\mit\Omega} |
\left[ j^\dagger_\mu (z) , j^\mu (0) \right]_-
| {\mit\Omega} \rangle
= 2 {\Im}{\rm m} \ i \!
\int d^4 x \ {\rm e}^{i q \cdot z}
\langle {\mit\Omega} |
T \left\{ j^\dagger_\mu (z) j^\mu (0) \right\}
| {\mit\Omega} \rangle \, ,
\end{equation}
where $| {\mit\Omega} \rangle$ stands for the vacuum state or any hadronic state.
Let us derive the theorem  for the polarization operator, i.e., when $| {\mit\Omega}
\rangle \to | 0 \rangle$ and the virtuality $q^2$ is timelike.

It is easy to derive  that
\begin{eqnarray}
\label{SpectralDensity}
\int d^4 z \ {\rm e}^{i q \cdot z}
\langle 0 |
\left[ j^\dagger_\mu (z) , j^\mu (0) \right]_-
| 0 \rangle
\!\!\!&=&\!\!\!
\sum_n \
\langle 0 | j^\dagger_\mu (0) | n \rangle
\langle n | j^\mu (0) | 0 \rangle
\ (2 \pi)^4
\delta^{(4)} \left( \sum_i^n p_i - q \right)
\nonumber\\
&\equiv&\!\!\! - \varrho (q^2)
\, .
\end{eqnarray}
To this end, one should use the completeness property of the intermediate states
(\ref{CompletenessAppendix}), translation symmetry
\begin{equation}
\label{TranslationInvariance}
j_\mu (z)
=
{\rm e}^{i z \cdot \mathbb{P}} j_\mu (0) {\rm e}^{- i z \cdot \mathbb{P}} \, ,
\end{equation}
with
\begin{equation}
\exp \left( i  z \cdot \mathbb{P} \right) | n \rangle
= \exp \left( i \sum_i^n p_i \cdot z \right) | n \rangle \, ,
\end{equation}
and the energy-momentum conservation $q = \sum_i^n p_i$. The latter is required,
in particular, to show that the second term in the commutator does not contribute
to the final result.

The function $\varrho (q^2)$ is a spectral density. Let us demonstrate this.
Due to the current conservation, $\partial_\mu j^\mu = 0$, we have
\begin{equation}
\label{MatrixElCurrentConserv}
\sum_i^n p_{i \mu} \langle n | j^\mu (0) | 0 \rangle = 0 \, ,
\end{equation}
with the timelike vector $\left( \sum_i^n p_i \right)^2 > 0$.
Hence,  to satisfy
Eq.\ (\ref{MatrixElCurrentConserv}), the vector $\langle n | j_\mu (0) | 0
\rangle$ has to be space-like. Therefore, one gets
\begin{equation}
\langle n | j_\mu (0) | 0 \rangle^\dagger
\langle n | j^\mu (0) | 0 \rangle < 0
\qquad
\Longrightarrow
\qquad
\varrho (q^2) > 0
\, ,
\end{equation}
which means that $\varrho (q^2)$ is positive definite.

Using Eq.\ (\ref{TranslationInvariance}),  one can reduce the chronological product
of currents in the integrand of the Fourier transform on the right-hand side of Eq.\
(\ref{OpticalTheoremEq}) to obtain
\begin{eqnarray}
\langle 0 |
T \left\{ j^\dagger_\mu (z) j^\mu (0) \right\}
| 0 \rangle
\!\!\!&=&\!\!\!
\sum_n
\langle 0 | j^\dagger_\mu (0) | n \rangle
\langle n | j^\mu (0) | 0 \rangle
\\
&\times&\!\!\!
\left\{
\exp \left( - i \sum_i^n p_i \cdot z \right) \theta (z^0)
+
\exp \left( i \sum_i^n p_i \cdot z \right) \theta (- z^0)
\right\} \, .
\nonumber
\end{eqnarray}
Inserting the unity
\begin{equation}
1 = \int d^4 q \ \delta^{(4)} \left( \sum_i^n p_i - q \right)
\end{equation}
into the sum, one can express this equation in terms of the spectral density
$\varrho (q^2)$ of Eq.\ (\ref{SpectralDensity}). Namely,
\begin{eqnarray}
\label{InterFormOT}
\langle 0 |
T \left\{ j^\dagger_\mu (z) j^\mu (0) \right\}
| 0 \rangle
\!\!\!&=&\!\!\!
-
\int \frac{d^4 q}{(2 \pi)^4} \ \varrho (q^2)
\left\{
{\rm e}^{- i q \cdot z } \theta (z^0)
+
{\rm e}^{ i q \cdot z } \theta (- z^0)
\right\}
\nonumber\\
&=&\!\!\!
- \frac{1}{2 \pi} \int_0^\infty d E
\int \frac{d^3 \mbox{\boldmath$q$}}{(2 \pi)^3} \
\varrho (q^2) \
{\rm e}^{
- i |z^0| E
+ i \mbox{\boldmath${\scriptstyle q}$}
\cdot \mbox{\boldmath${\scriptstyle z}$}
}
\, .
\nonumber
\end{eqnarray}
The final two steps involve the introduction of the spectral mass parameter $M^2$
via incorporating  the unity
\begin{equation}
\int_0^\infty d M^2 \ \delta \left( M^2 - q^2 \right) = 1
\end{equation}
into  the integrand of Eq.\ (\ref{InterFormOT}), and the use of the following
contour integral representation
\begin{equation}
\frac{{\rm e}^{- i |z^0| E}}{2 E}
= \frac{1}{2 \pi}
\int_{- \infty}^\infty \frac{d q_0}{(q^0)^2 - E^2 + i 0} \ {\rm e}^{- i z^0 q^0} \, .
\end{equation}
Integrating the result over $E$ and combining $q^0$ and $\mbox{\boldmath$q$}$
into one four-vector $q^\mu = (q^0 , \mbox{\boldmath$q$})$, we get
\begin{equation}
\label{SpectralRepresentCoord}
i \langle 0 |
T \left\{ j^\dagger_\mu (z) j^\mu (0) \right\}
| 0 \rangle
= \frac{1}{2 \pi}
\int \frac{d^4 q}{(2 \pi)^4}
{\rm e}^{- i q \cdot z} \int_0^\infty d M^2 \frac{\varrho (M^2)}{q^2 - M^2 + i 0}
\, .
\end{equation}
After the inverse Fourier transformation, rewriting the spectral density according
to its definition (\ref{SpectralDensity}), and taking the imaginary part of both
sides of Eq.\ (\ref{SpectralRepresentCoord}), one obtains the optical theorem
(\ref{OpticalTheoremEq}).

\section{Light-cone quantization of QCD}
\label{LightConeQCD}

Consider the action of quantum chromodynamics,
\begin{equation}
S_{\scriptscriptstyle\rm QCD} = \int d^4 z \, {\cal L}_{\scriptscriptstyle\rm QCD} (z)
\, ,
\end{equation}
with the Lagrangian (\ref{QCDLagrangian}). We will discuss it in the light-cone gauge;
therefore, it is convenient to cast the Lagrangian in terms of the light-cone components
of the elementary fields. Introducing the projectors for  fermions  ${\mit\Pi}^{\pm} =
\ft12 \gamma^{\mp} \gamma^{\pm}$, we can decompose the quark field into  ``good'' and
``bad'' components via
$$
\psi = {\mit\Pi}^+ \psi + {\mit\Pi}^- \psi
\equiv
\psi_+ + \psi_-
\, .
$$
The gauge potential can also be written in the light-cone coordinates as
\begin{equation}
A^\mu = (A^+ , A^- , \bit{A}^i_\perp)
\, , \qquad
A^\pm \equiv \ft{1}{\sqrt{2}} ( A^0 \pm A^3 )
\, ,
\end{equation}
where $\bit{A}^i_\perp = (A^x, A^y)$ is the transverse gauge field. The advantages
of this parametrization becomes transparent and well suited for the light-cone gauge
$A^+ = 0$. Note that the light-cone gauge can be implemented directly in the
Lagrangian, without introducing a gauge-fixing term as one does for implementation
of covariant gauges. Making use of the light-cone coordinates introduced in Appendix
\ref{LCvectors}, we immediately find that the QCD Lagrangian lacks terms with the
light-cone time derivatives $\partial^-$ acting on the bad field components $\psi_-$
and $A^-$. This implies that the latter do not evolve with time and therefore they
are not dynamical. Hence they can be integrated out in the functional integral without
the loss of any dynamical information. This integration out can be implemented with
a simple change of variables,
\begin{equation}
S_a
=
\partial^+ A_a^-
-
\frac{1}{\partial^+} \bit{\cal D}^i_{\perp ab}
\partial^+ \bit{A}^i_{\perp b}
-
g
\frac{1}{\partial^+} \bar\psi \gamma^+ t^a \psi
\, , \qquad
\chi
=
\psi_-
-
\frac{1}{2 \partial^+} \gamma^+ \bit{\gamma}^i_\perp \bit{\cal D}^i_\perp \psi_+
\, ,
\end{equation}
where $\bit{\cal D}^i_\perp = - \partial/\partial \bit{x}^i_\perp - i g
\bit{A}^{i a}_\perp t^a$ and $\bit{\cal D}^i_{\perp ab} = - \delta_{ab}
\partial/\partial \bit{x}^i_\perp + g f_{acb} \bit{A}^{i c}_\perp$ are
the fundamental and adjoint transverse covariant derivatives, respectively. This
yields a two-component structure of the QCD Lagrangian
\begin{eqnarray}
{\cal L}_{\scriptscriptstyle\rm QCD}
=
{\cal L}_{\scriptscriptstyle\rm LC} + \Delta {\cal L}
\, ,
\end{eqnarray}
where
\begin{eqnarray}
{\cal L}_{\scriptscriptstyle\rm LC}
\!\!\!&=&\!\!\!
i \bar\psi_+ \gamma^+ \partial^- \psi_+
-
\ft{i}{2}
\bar\psi_+ \bit{\gamma}^i_\perp \gamma^+ \bit{\gamma}^j_\perp
\bit{\cal D}^i_\perp \frac{1}{\partial^+} \bit{\cal D}^j_\perp \psi_+
\\
&+&\!\!\!
\left( \partial^+ \bit{A}^i_{\perp a} \right)
\left( \partial^- \bit{A}^i_{\perp a} \right)
-
\ft{1}{4} \bit{F}_{\perp ab}^i \bit{F}_{\perp ab}^i
-
\ft12
\left(
\frac{1}{\partial^+}
\bit{\cal D}^i_{\perp ab} \partial^+ \bit{A}^i_{\perp b}
+
g \frac{1}{\partial^+} \bar\psi_+ \gamma^+ t^a \psi_+
\right)^2
\, ,
\nonumber
\end{eqnarray}
and
\begin{equation}
\Delta {\cal L}
= \ft12 \left( S_a \right)^2
+
i \bar\chi \gamma^- \partial^+ \chi
\, .
\end{equation}
The latter non-dynamical part of the Lagrangian, i.e., $\Delta {\cal L}$, does not
depend on the light-cone time $z^+$ and can be completely absorbed into the
normalization of the functional integral measure. Equivalently, this boils down to
using the equations of motion for $S_a$ and $\chi$ and thus setting both fields to
zero,
\begin{equation}
\label{BadInTermsOfGood}
S_a = 0
\, , \qquad
\chi = 0
\, .
\end{equation}
Thus, at the expense of the manifest Lorentz invariance, the QCD Lagrangian can
be written in terms of the physical degrees of freedom only. At zero coupling,
the ``good'' fields can be identified as QCD partons. As we discussed in Section
\ref{PDFsBjorkenFrame}, at zero light-cone time $z^+ = 0$ the quarks and gluons
are free and one can decompose their fields in terms of creation and annihilation
operators as \cite{BroPauPin98}
\begin{eqnarray}
\label{FieldsCreatAnnihOps}
\psi_+ ( z^- , \bit{z}_\perp , z^+ = 0 )
\!\!\!&=&\!\!\!
\sum_{\lambda = \uparrow\downarrow}
\int \frac{d k^+ d^2 \bit{k}_\perp}{2 k^+ (2 \pi)^3}
\\
&\times&\!\!\!\left\{
b_\lambda (k^+ , \bit{k}_\perp) u_{+ \lambda} (k)
{\rm e}^{
- i (k^+ z^- - \bit{\scriptstyle k}_\perp \cdot \bit{\scriptstyle z}_\perp)
}
+
d^\dagger_\lambda (k^+ , \bit{k}_\perp) u_{+ \lambda} (- k)
{\rm e}^{
i (k^+ z^- - \bit{\scriptstyle k}_\perp \cdot \bit{\scriptstyle z}_\perp)
}
\right\}
\, , \nonumber\\
\bit{A}_\perp ( z^- , \bit{z}_\perp , z^+ = 0 )
\!\!\!&=&\!\!\!
\sum_{\lambda = \uparrow\downarrow}
\int \frac{d k^+ d^2 \bit{k}_\perp}{2 k^+ (2 \pi)^3}
\\
&\times&\!\!\!\left\{
a_\lambda (k^+ , \bit{k}_\perp) \bit{\varepsilon}^\lambda_\perp \,
{\rm e}^{
- i (k^+ z^- - \bit{\scriptstyle k}_\perp \cdot \bit{\scriptstyle z}_\perp)
}
+
a^\dagger_\lambda (k^+ , \bit{k}_\perp) \bit{\varepsilon}^{\lambda \ast}_\perp \,
{\rm e}^{
i (k^+ z^- - \bit{\scriptstyle k}_\perp \cdot \bit{\scriptstyle z}_\perp)
}
\right\}
\, . \nonumber
\end{eqnarray}
The bispinor $u_+$ is a ``good'' light-cone projection $u_+ = {\mit\Pi}^+ u$
of the light-cone bispinor (\ref{LightConeSpinors}). It possesses a definite
light-cone helicity \cite{KogSop70}, because it does not depend on the transverse
momentum $\bit{k}_\perp$. While the two-dimensional gluon polarization vectors are
\begin{equation}
\bit{\varepsilon}^\uparrow_\perp
=
- \frac{1}{\sqrt{2}}
\left(
\begin{array}{c}
1 \\
i
\end{array}
\right)
\, , \qquad
\bit{\varepsilon}^\downarrow_\perp
=
- \frac{1}{\sqrt{2}}
\left(
\begin{array}{c}
- 1 \\
i
\end{array}
\right)
\, ,
\end{equation}
for helicity up and down states, cf.\ Eq. (\ref{VectorPolarization}). The (anti-)
commutation relations between the creation and annihilation operators look as
follows,
\begin{eqnarray}
\{
b_\lambda (k^+ , \bit{k}_\perp) ,
b^\dagger_{\lambda'} (k^{\prime +}, \bit{k}^\prime_\perp )
\}
\!\!\!&=&\!\!\!
2 k^+ (2 \pi)^3
\delta (k^+ - k^{\prime +})
\delta^{(2)} (\bit{k}_\perp - \bit{k}'_\perp)
\delta_{\lambda \lambda^\prime}
\, , \\
{}[ \,
a_\lambda (k^+ , \bit{k}_\perp) ,
a^\dagger_{\lambda'} (k^{\prime +}, \bit{k}^\prime_\perp ) \,
{}]
\!\!\!&=&\!\!\!
2 k^+ (2 \pi)^3
\delta (k^+ - k^{\prime +})
\delta^{(2)} (\bit{k}_\perp - \bit{k}'_\perp)
\delta_{\lambda \lambda^\prime}
\, .
\end{eqnarray}

A drawback of the light-cone formalism is the loss of the Lorentz covariance,
since the non-dynamical degrees of freedom are integrated out in the functional
integral. If one keeps these degrees of freedom in the Lagrangian, then the
Feynman rules for the gluon propagator change, while the rest remains  the same
as in covariant gauges. Notice that the light-cone gauge does not fix the gauge
entirely, it allows for $z^-$-independent transformations $U(z^+, \bit{z})$, since
$A^+ (z) \to A^+ (z) + U^\dagger (z^+ , \bit{z}) \partial^+ U (z^+ , \bit{z}) = 0$.
The residual gauge freedom can be fixed by imposing a boundary condition on the
gauge potential or other means, as described below.

Since only the physical degrees of freedom propagate in the light-cone gauge,
the relation between the gauge potential and the field strength is linear,
$$
\partial^+ A_\mu^\perp = F^{+\perp}_{\phantom{+}\mu}
\, ,
$$
and can be easily inverted,
\begin{equation}
\label{FromPotentialToStrength}
A^\perp_\mu (z^-, z^+, \bit{z})
=
\int_{- \infty}^\infty dz'^- \, w (z'^- - z^-)
F^{+ \perp}_{\phantom{+}\,\mu} (z'^-, z^+, \bit{z})
\, .
\end{equation}
The weight $w$ depends on the boundary condition imposed on the potential.
It takes the following form
\begin{equation}
\label{WeightForBoundaryCond}
w (z^-)
=
\left\{
\begin{array}{ll}
- \theta (z^-) \, ,
& \quad \bit{A}_\perp (z^- = \infty) = 0   \\
\theta (- z^-) \, ,
& \quad \bit{A}_\perp (z^- = - \infty) = 0 \\
- \ft12 {\rm sgn} (z^-) \, ,
& \quad \bit{A}_\perp (z^- = \infty) + \bit{A}_\perp (z^- = - \infty) = 0
\end{array}
\right.
\, .
\end{equation}

Within perturbation theory, the incomplete gauge fixing elucidated above reveals
itself in a spurious $1/k^+$-singularity in the gluon Green function. The boundary
conditions, in turn, translate into a regularization of this pole. This  can be
seen from the following representation of the gauge potential in terms of the field
strength,
\begin{equation}
\label{AinFaboundary}
A^\perp_\mu (z^-, z^+, \bit{z})
=
i \int_{- \infty}^\infty \frac{d \tau}{[\tau]_{\rm\scriptstyle reg}} \,
{\rm e}^{- i \tau z^-}
\int_{- \infty}^\infty \frac{dz'^-}{2 \pi} \, {\rm e}^{i \tau z'^-}
F^{+ \perp}_{\phantom{+}\,\mu} (z'^-, z^+, \bit{z})
\, .
\end{equation}
The correspondence of the prescriptions with the boundary conditions is straightforward
\begin{equation}
\label{AinFConditions}
\frac{1}{[\tau]_{\rm\scriptstyle reg}}
=
\left\{
\begin{array}{ll}
\displaystyle\frac{1}{\tau - i0}
\, ,
& \quad \bit{A}_\perp (z^- = \infty) = 0   \\
\displaystyle\frac{1}{\tau + i0}
\, ,
& \quad \bit{A}_\perp (z^- = - \infty) = 0 \\
\displaystyle \frac{1}{2} {\rm PV }\frac{1}{\tau}
\, ,
& \quad \bit{A}_\perp (z^- = \infty) + \bit{A}_\perp (z^- = - \infty) = 0
\end{array}
\right.
\, .
\end{equation}
This issue will be further discussed in Appendix \ref{ComputationTechniques}
where we demonstrate some computational techniques which incorporate the
light-cone gauge.

\section{Improved Belinfant\'e energy-momentum tensor}
\label{SymmetricEMtensor}

One can construct a symmetric energy-momentum tensor directly from the
Lagrangian without going through the procedure of adding ad hoc extra
terms to the original non-symmetric energy-momentum tensor \cite{LanLif04}.
To this end, let us promote the Minkowskian space QCD action to that in a
curved background with the metric $g^{\mu\nu} (x)$,
\begin{equation}
S_{\rm\scriptscriptstyle QCD}
=
\int d^4 z \sqrt{- {\rm det} [ g_{\mu\nu} (z) ]} {\cal L}_{\rm\scriptscriptstyle QCD} (z)
\, ,
\end{equation}
where we introduced the inverse metric $g_{\mu\nu} = [ g^{\mu\nu} ]^{-1}$
satisfying  $g^{\mu\rho} (z) g_{\rho\nu} (z) = \delta^\mu_{\nu} = {\rm diag}
(1,1,1,1)$. The contraction of Lorentz indices in the QCD Lagrangian is
accomplished with the help of the metric tensor, i.e.,
\begin{equation}
{\cal L}_{\rm\scriptscriptstyle QCD} (z)
=
\ft{i}2
\left[
\bar\psi (z) \stackrel{{}_\leftrightarrow}{\cal D}_{ \{ \mu } \gamma_{ \nu \} }
\psi (z)
\right]
g^{\mu\nu} (z)
-
\ft14 F_{\mu\nu}^a (z) F_{\rho\sigma}^a (z) g^{\mu\rho} (z) g^{\nu\sigma} (z)
\, ,
\end{equation}
where the indices in the quark term are symmetrized $t_{\{ \mu\nu \}}
\equiv \ft12 \left( t_{\mu\nu} + t_{\nu\mu} \right)$.

Under a general coordinate transformation $z^\mu \to z'^\mu$, the metric
in two different coordinate systems transforms as a second-rank tensor
$$
g'^{\mu\nu} (z')
=
g^{\rho\sigma} (z)
\frac{\partial z'^\mu}{\partial z^\rho}
\frac{\partial z'^\nu}{\partial z^\sigma}
\, .
$$
For an infinitesimal transformation $z' = z + \zeta (z)$, the change
of the metric in different reference frames,  but at the same space-time
point $z$ is
\begin{equation}
\label{MetricVariation}
\delta g^{\mu\nu} (z)
\equiv
g'^{\mu\nu} (z) - g^{\mu\nu} (z)
=
g^{\mu\rho} (z) \partial_\rho \zeta^\nu (z)
+
g^{\nu\rho} (z) \partial_\rho \zeta^\mu (z)
-
\zeta^\rho (z) \partial_\rho g^{\mu\nu} (z)
\, .
\end{equation}
Neglecting the field variations $\delta {\mit\Phi}$ due to the above general
coordinate transformation, since they generate field equations of motion,
we write the variation of the action as
\begin{equation}
\label{ActionVariation}
\delta S_{\rm\scriptscriptstyle QCD}
=
\ft12 \int d^4 z \sqrt{- {\rm det} [ g_{\rho\sigma} (z) ]}
\,
{\mit\Theta}_{\mu\nu} (z) \delta g^{\mu\nu} (z)
+
\sum_{\mit\Phi} {\mit\Omega}_{\mit\Phi}
\, ,
\end{equation}
where we introduced the quantity
\begin{equation}
\label{SymEMtensorGravity}
\ft12 \sqrt{- {\rm det} [ g_{\rho\sigma} ]} \,
{\mit\Theta}_{\mu\nu}
=
\frac{\partial}{\partial g^{\mu\nu}}
\sqrt{- {\rm det} [ g_{\rho\sigma} ]} \ {\cal L}_{\rm\scriptscriptstyle QCD}
-
\partial_\rho
\frac{\partial}{\partial ( \partial_\rho g^{\mu\nu} )}
\sqrt{- {\rm det} [ g_{\rho\sigma} ]} \ {\cal L}_{\rm\scriptscriptstyle QCD}
\, ,
\end{equation}
after integrating by parts once in the second term and dropping the surface terms,
$\delta g^{\mu\nu} ({\rm surface}) = 0$.

The graviton is a spin-two particle, so it is a rank-two symmetric and traceless
tensor $h^{\mu\nu}$, which arises as a first term in the expansion of the metric
$g^{\mu\nu}$ around the flat Minkowskian space $\eta^{\mu\nu} = {\rm diag} (1, -1,
-1, -1)$, $g^{\mu\nu} = \eta^{\mu\nu} + h^{\mu\nu}$. Hence, the energy-momentum
tensor ${\mit\Theta}_{\mu\nu}$ coupled to it must be {\sl symmetric} and {\sl
traceless}. It is called the Belinfant\'e energy momentum tensor. Note, that on the
right-hand side of Eq.\ (\ref{SymEMtensorGravity}), one assumes that when
differentiating with respect to a component of $g^{\mu\nu}$, one does not use the
symmetry of the metric tensor in the differentiated function. For instance, while
$g^{\mu\nu}$ is differentiated, the transposed tensor $g^{\nu\mu}$ is not. Using
Eq.\ (\ref{MetricVariation}) in Eq.\ (\ref{ActionVariation}), it is easy to see
that the energy-momentum tensor is covariantly conserved in the curved background,
which translates into the usual conservation in the flat space, $\partial^\mu
{\mit\Theta}_{\mu\nu} = 0$.

To find the explicit form of the Belinfant\'e energy-momentum tensor for a given
Lagrangian, one needs the variation of the determinant of the curved metric, which
is easily to compute making use of its expansion in terms of minors $M$, and an
analogous representation for the elements of the inverse metric, i.e.,
$$
{\rm det} [ g_{\mu\nu} ]
=
\sum_{\nu = 0}^\mu (- 1)^{\mu + \nu} M_{\mu\nu} g_{\mu\nu}
\, , \qquad
[ g_{\mu\nu} ]^{-1}
=
\frac{
(- 1)^{\mu + \nu} M_{\nu\mu}
}{
{\rm det} [ g_{\mu\nu} ]
}
\, .
$$
Due to the fact that the metric is symmetric, one can interchange the indices
of the minor in the last equality. Then
$$
d \, {\rm det} [ g_{\mu\nu} ]
=
\sum_{\nu = 0}^\mu (- 1)^{\mu + \nu} M_{\mu\nu} d g_{\mu\nu}
=
{\rm det} [ g_{\mu\nu} ] g^{\mu\nu} d g_{\mu\nu}
=
-
{\rm det} [ g_{\mu\nu} ] g_{\mu\nu} d g^{\mu\nu}
\, ,
$$
where, at the final step, we used the property $g_{\mu\nu} d g^{\mu\nu} = -
g^{\mu\nu} d g_{\mu\nu}$ as a consequence of $g^{\mu\nu} g_{\mu\nu} = 4$. Since
the QCD Lagrangian does not depend on the derivatives of the metric, only the
first term in Eq.\ (\ref{SymEMtensorGravity}) gives a non-vanishing contribution
to the symmetric energy-momentum tensor. A calculation yields
\begin{equation}
\label{SymTracelessEMtensorQCD}
{\mit\Theta}^{\mu\nu}
=
- g^{\mu\nu} {\cal L}_{\rm\scriptscriptstyle QCD}
-
F^{\mu\rho}_a \, F^{\nu}_{a \, \rho}
+
\frac{i}{4}
\bar\psi
\left\{
\stackrel{{}_\leftrightarrow}{\cal D}{\!}^\mu \gamma^\nu
+
\stackrel{{}_\leftrightarrow}{\cal D}{\!}^\nu \gamma^\mu
\right\}
\psi
\, ,
\end{equation}
where we set the metric to the flat Minkowskian one.

Note, that although this technique produces a symmetric and also traceless (for
spin one and spin one-half fields) tensor, in contrast to the N\"other procedure,
it does not, however, generate automatically a traceless energy-momentum tensor
if scalar fields are present in the Lagrangian, e.g.,
$$
{\cal L}^{\rm sc}
=
\frac{1}{2}
\left( \partial^\mu \phi \right) \left( \partial_\mu \phi \right)
\, .
$$
To cure the problem one should add a superpotential $X^{\mu\nu\rho\sigma}$,
which results in the improved symmetric energy-momentum tensor for scalars
\cite{CalColJac70,Jac86}
$$
{\mit\Theta}^{\mu\nu}_{\rm sc}
=
\frac{1}{2}
\left( \partial^\mu \phi \right) \left( \partial^\nu \phi \right)
+
\frac{1}{2} \partial_\rho \partial_\sigma X^{\mu\nu\rho\sigma}
=
\frac{1}{2}
\left( \partial^\mu \phi \right) \left( \partial^\nu \phi \right)
-
\frac{1}{6} (\partial^\mu \partial^\nu - g^{\mu\nu} \Box) \phi^2
$$
with
$$
X^{\mu\nu\rho\sigma}
=
g^{\mu\nu} X^{\rho\sigma}
+
g^{\rho\sigma} X^{\mu\nu}
-
g^{\mu\rho} X^{\nu\sigma}
-
g^{\nu\rho} X^{\mu\sigma}
-
\frac{1}{3}
\left(
g^{\mu\nu} g^{\rho\sigma}
-
g^{\mu\rho} g^{\nu\sigma}
\right)
X^{\alpha\beta} g_{\alpha\beta}
\, ,
$$
expressed in terms of the symmetric tensor
$$
X^{\mu\nu} = \frac{1}{2} g^{\mu\nu} \phi^2
\, .
$$

\section{Basics of the Skyrme model}
\label{SkyrmeAppendix}

In the multicolor limit, QCD becomes equivalent to a theory of mesons \cite{tHo74}.
In the formal limit $N_c \to \infty$, this ``artificial'' world is populated with
stable and noninteracting mesons. For large but finite values of the numbers of
colors $N_c$, the mesons acquire nonvanishing width and two-particle cross sections
of order $N_c^{-1}$ and $N_c^{-2}$, respectively, they start to interact via the
exchange of single mesons rather than quarks and gluons. The low-energy limit of the
meson theory is given then by the chiral Lagrangian, which is a non-linear sigma
model of the spontaneously broken chiral symmetry. Restricting to the two-flavor
case, this effective theory is a theory of the pions as Goldstone bosons. The
baryons appear in it as topologically stable solitons \cite{Sky61,Wit79}, i.e.,
time-independent solutions to the classical equations of motion with conserved
energy. They have masses that diverge as the inverse coupling constant $m \sim N_c
= 1/(N_c^{-1})$. The solitons have the quantum numbers of the QCD baryons provided
one takes into account the Wess-Zumino term \cite{Wit79}.

The simplest chiral Lagrangian,
\begin{equation}
{\cal L} = \frac{f_\pi^2}{8} {\rm tr} \left[ \left( \partial_\mu {\mit\Sigma}
\right) \left( \partial^\mu {\mit\Sigma}^\dagger \right) \right] \, ,
\end{equation}
which we discussed in Section \ref{ChiPTGPDs}, with the $SU(2)$ matrix ${\mit\Sigma}$
of mesons, leads to unstable solitons. This drawback can be cured by adding a four
derivative ``Skyrme'' term
$$
\Delta {\cal L}
=
\frac{\varepsilon^2_{\rm\scriptscriptstyle B}}{32}
{\rm tr} [ ( \partial_\mu {\mit\Sigma} ) {\mit\Sigma}^\dagger ,
( \partial_\nu {\mit\Sigma} ) {\mit\Sigma}^\dagger ]
[ ( \partial^\mu {\mit\Sigma} ) {\mit\Sigma}^\dagger ,
( \partial^\nu {\mit\Sigma} ) {\mit\Sigma}^\dagger ]
\, ,
$$
where $\varepsilon_{\rm\scriptscriptstyle B}$ is a dimensionless parameter that
characterizes the size of finite energy configurations. The choice of this addendum
is unique since it provides a positive-definite Hamiltonian which is moreover
second order in time derivatives. The resulting chiral Lagrangian ${\cal L}
+ \Delta {\cal L}$ admits a hedgehog soliton solution  ${\mit\Sigma}_0 (\bit{r})
= \exp (i F(\bit{r}) \tau_i \hat{r}_i )$, due to Skyrme, with topologically
nontrivial boundary conditions $F (0) = - \pi$ and $F (\infty) = 0$ \cite{AdkNapWit83}
(for a review, see Ref.\ \cite{BroZah86}). Introducing the time-dependent matrix of
collective coordinates
$$
R (t) = a_0 (t) + i \tau_i a_i (t)
\qquad \mbox{with} \qquad
\sum_{i = 0}^3 a_i^2 = 1
\, ,
$$
and substituting the ansatz ${\mit\Sigma} (x) = R (t) {\mit\Sigma}_0 (\bit{r})
R^\dagger (t)$ into the original Lagrangian yields a Lagrangian, which is of
the second order in time derivatives of collective coordinates ${\cal L} \sim
\sum_{i = 0}^3 \dot{a}_i^2$. The canonical quantization procedure $\dot{a}_i
\sim \pi_i = - i \partial /\partial a_i$ then leads to a  Hamiltonian which
is a Laplace operator $\nabla^2 = \sum_{i = 0}^4 \partial^2/\partial a_i^2$
on a three-sphere. The eigenstates of this Laplacian are traceless symmetric
polynomials, in complete analogy with conventional three-dimensional spherical
harmionics. For instance, they satisfy $\nabla^2 (a_0 + i a_1)^L = - L (L + 2)
(a_0 + i a_1)^L$, and possess equal eigenvalues $S = I = \ft12 L$ of the spin
and isospin operators,
\begin{eqnarray}
S_k
\!\!\!&=&\!\!\!
\ft{i}{2}
\left(
a_k \frac{\partial}{\partial a_0}
-
a_0 \frac{\partial}{\partial a_k}
-
\varepsilon_{klm} a_l \frac{\partial}{\partial a_m}
\right)
\, , \\
I_k
\!\!\!&=&\!\!\!
\ft{i}{2}
\left(
a_0 \frac{\partial}{\partial a_k}
-
a_k \frac{\partial}{\partial a_0}
-
\varepsilon_{klm} a_l \frac{\partial}{\partial a_m}
\right)
\, ,
\end{eqnarray}
respectively.

Out of all hadronic states, we need for our purposes only the proton and
delta isobar spin-up wave functions. Assuming the unit normalization for the
Haar measure on the $SU(2)$ group, i.e.,
$$
\int dR
\equiv
\frac{1}{\pi^2} \int  d^4 a \,
\delta \left( \sum_i^4 a_i^2 - 1 \right)
=
1 \, ,
$$
one can write the hadron states normalized to unity as\footnote{Notice that
our normalization differs by a factor of $\sqrt{2}$ and a minus sign in the
$\Delta^+$ wave function compared to Ref.\ \cite{AdkNapWit83}. The first
difference is a consequence of the normalization of the Haar measure and
the requirement that $\langle H | H \rangle \equiv \int dR \, \langle H |
H \rangle = 1$. Since we are interested in the relative magnitude of $pp$
to $p \Delta^+$ transitions, this difference is totally irrelevant. The
minus sign in the delta wave function is introduced so as to get the usual
result for the magnetic moments (\ref{MagMomDeltaProton}).} \cite{AdkNapWit83}
\begin{equation}
| p_\uparrow \rangle = \sqrt{2} (a_1 + i a_2)
\, , \qquad
| \Delta_\uparrow{\!\!}^+ \rangle
=
2 (a_1 + i a_2) \left( 1 - 3 (a_0^2 + a_3^2) \right) \, .
\end{equation}
They have spin and isospin $S_3 = I_3 = 1/2$. Since only the vector-isovector
transitions survive in the large-$N_c$ limit, and are of interest for the
discussion in the main body of the paper, we write down the corresponding
matrix elements
\begin{eqnarray}
&&
\langle p_\uparrow |
{\rm tr} \left( \tau^z R \tau^z R^\dagger \right)
| p_\uparrow \rangle
\equiv
2
\int d R \, (a_1^2 + a_2^2) {\rm tr} \left( \tau^z R \tau^z R^\dagger \right)
=
-
\frac{2}{3} \, , \\
&&
\langle \Delta_\uparrow{\!\!}^+ |
{\rm tr} \left( \tau^z R \tau^z R^\dagger \right)
| p_\uparrow \rangle
=
- \frac{2 \sqrt{2}}{3}
\, .
\end{eqnarray}
From these relations, we find the result
\begin{equation}
\label{DeltaToProton}
\sqrt{2}
\langle p_\uparrow | {\rm tr} \left( \tau^z R \tau^z R^\dagger \right) | p_\uparrow
\rangle
=
\langle  \Delta_\uparrow{\!\!}^+ | {\rm tr} \left( \tau^z R \tau^z
R^\dagger \right) | p_\uparrow \rangle \, .
\end{equation}
we sought for. These matrix elements correspond to the following quark operator
matrix elements
$$
\langle B_2 | {\rm tr} \left( \tau^a R \tau^j R^\dagger \right) | B_1 \rangle
=
g_0
\langle B_2 | \bar{\psi}_{q'} \tau^a_{q' q} \gamma^j \psi_q | B_1 \rangle
\, ,
$$
up to an overall matching constant $g_0$.

Decomposing the two light-flavor electromagnetic current into the isovector and
isosinglet components $j_\mu = \ft12 (\bar{u} \gamma_\mu u - \bar{d} \gamma_\mu d)
+ \ft16 (\bar{u} \gamma_\mu u + \bar{d} \gamma_\mu d)$, one can neglect the
singlet component in the matrix element between the proton and the delta-isobar
and find
\begin{equation}
\label{EMtoIsovector}
\langle \Delta^+ | j^\mu | p \rangle
=
\ft12
\langle \Delta^+ | \bar{u} \gamma^\mu u - \bar{d} \gamma^\mu d | p \rangle
\, .
\end{equation}
This when combined with (\ref{DeltaToProton}) gives a relation between the spatial
components of the matrix elements of the quark electromagnetic current
\begin{equation}
\label{SkyrmeRelationDeltaProton}
\sqrt{2} \langle \Delta^+ | \bit{j} (0) | p \rangle
=
\langle p | \bit{j} (0) | p \rangle
-
\langle n | \bit{j} (0) | n \rangle \, .
\end{equation}
Here on the right-hand side, we have used the $SU(2)$ symmetry to relate the proton
matrix element of the isovector quark current to the difference of matrix elements
of the electromagnetic current between the proton and neutron states, i.e., $\langle p |
\bar{u} \gamma_\mu u - \bar{d} \gamma_\mu d | p \rangle = \langle p | j^\mu | p \rangle
- \langle n | j^\mu | n \rangle$. When Eq.\ (\ref{SkyrmeRelationDeltaProton}) is
re-expressed in terms of the magnetic moments (see Eq.\ (\ref{ParityEvenBreitFrame}))
we reproduce the well-known relation \cite{AdkNapWit83}
\begin{equation}
\label{MagMomDeltaProton}
\mu_{p \Delta} = \ft{1}{\sqrt{2}} (\mu_p - \mu_n)
\, .
\end{equation}

\section{Computation techniques for evolution equations}
\label{ComputationTechniques}

In this Appendix we present a number of techniques for evaluation of
one-loop evolution kernels in momentum and coordinate spaces.

\subsection{Feynman rules}
\label{FeynmanRules}

To start with, we give a comprehensive summary of Feynman rules for the
QCD Lagrangian. The quark and gluon propagators are determined by the
chronological product of two elementary fields,
\begin{eqnarray}
\langle 0 |
T
\left\{
\psi (z_1) \bar\psi (z_2)
\right\}
| 0 \rangle
\!\!\!&=&\!\!\!
\int \frac{d^4 k}{(2 \pi)^4}
{\rm e}^{- i k \cdot (z_1 - z_2)}
i {\cal S} (k)
\, , \\
\langle 0 |
T
\left\{
A_\mu^a (z_1) A^b_\nu (z_2)
\right\}
| 0 \rangle
\!\!\!&=&\!\!\!
\int \frac{d^4 k}{(2 \pi)^4}
{\rm e}^{- i k \cdot (z_1 - z_2)}
(- i) \delta^{ab} {\cal D}_{\mu\nu} (k)
\, ,
\end{eqnarray}
where the explicit gauge-dependent form of the momentum-space functions will be
established below.

\begin{itemize}
\item For every internal quark line we write
\begin{equation}
\parbox[c][12mm][t]{20mm}{
\insertfig{2}{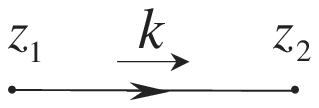}
}
\
=
\
i {\cal S} (k)
\, .
\end{equation}
\item For every internal gluon line we write
\begin{equation}
\parbox[c][12mm][t]{20mm}{
\insertfig{2}{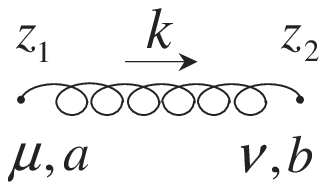}
}
\
=
\
- i \delta^{ab} {\cal D}^{\mu\nu} (k)
\, .
\end{equation}
\item Quark-gluon vertex is
\begin{equation}
\parbox[c][25mm]{30mm}{
\insertfig{2.5}{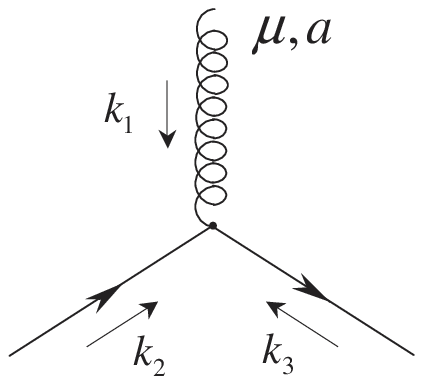}
}
\
=
\
{\cal V}^{qgq, a}_\mu (k_1 , k_2, k_3)
=
i g t^a \gamma_\mu
(2\pi)^4 \, \delta^{(4)} \left( k_1 + k_2 + k_3 \right)
\, .
\end{equation}
\item Three-gluon vertex is
\begin{eqnarray}
\parbox[c][25mm]{30mm}{
\insertfig{2.5}{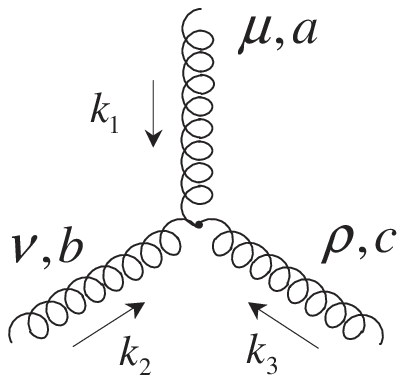} }
\!\!\!&=&\!\!\!
\
{\cal V}^{ggg, abc}_{\mu\nu\rho} (k_1 , k_1, k_3)
\\ [-1cm]
&&
=
g f^{abc}
\left\{
(k_1 - k_2)_\rho g_{\mu\nu}
+ (k_2 - k_3)_\mu g_{\nu\rho}
+ (k_3 - k_1)_\nu g_{\rho\mu}
\right\}
\nonumber\\
&&\qquad\qquad\qquad\qquad\qquad\qquad\qquad\times
(2\pi)^4 \, \delta^{(4)} \left( k_1 + k_2 + k_3 \right)
\, . \nonumber
\end{eqnarray}
\item Four-gluon vertex reads
\begin{eqnarray}
\parbox[c][30mm]{30mm}{
\insertfig{2.5}{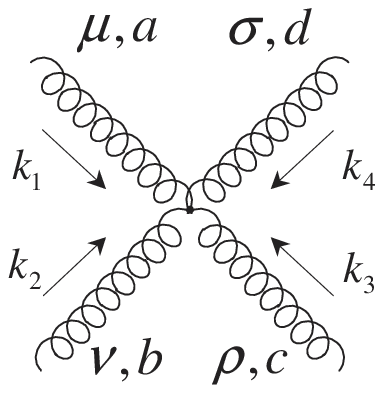} }
\!\!\!&=&\!\!\!
\
{\cal V}^{gggg, abcd}_{\mu\nu\rho\sigma} (k_1 , k_1, k_3, k_4)
\\ [-1cm]
&&
= - i g^2
\big\{
f^{abe}f^{cde}
\left(
g_{\mu\rho}g_{\nu\sigma} - g_{\mu\sigma}g_{\nu\rho}
\right)
\nonumber\\
&&\qquad\quad
+
f^{ace}f^{bde}
\left(
g_{\mu\nu}g_{\rho\sigma} - g_{\mu\sigma}g_{\nu\rho}
\right)
\nonumber\\
&&\qquad\quad
+
f^{ade}f^{cbe}
\left(
g_{\mu\rho}g_{\nu\sigma} - g_{\mu\nu}g_{\rho\sigma}
\right)
\big\}
(2\pi)^4 \delta^{(4)} \left( k_1 + k_2 + k_3 + k_4 \right) \nonumber
\, .
\end{eqnarray}
\item The resulting expression,constructed from propagators and vertices should
be integrated with respect to the momentum of every internal line, i.e., multiplied
by the factor
\begin{equation}
\int \prod_\ell \frac{d^dk_\ell}{(2\pi)^{\ell\cdot d}}
\, .
\end{equation}
\end{itemize}

We have not specified so far the explicit form of the propagators. It is the following.
The quark propagator we have
\begin{equation}
S (k) = \frac{\not\! k}{k^2 + i0} ,
\end{equation}
The form of gluon propagator depends on the gauge assumed. In the covariant gauge
$\partial \cdot A = 0$, it reads
\begin{equation}
{\cal D}^{\mu\nu} (k)
=
\frac{1}{k^2 + i0}
\left(
g^{\mu\nu} - (1 - \xi) \frac{k^\mu k^\nu}{k^2}
\right)
\, ,
\end{equation}
where $\xi$ is the gauge fixing parameter. The gluon propagator in the light-cone
gauge also depends on the fixing of residual gauge degrees of freedom. As we
discussed earlier in Appendix \ref{LightConeQCD}, the incomplete gauge fixing of
the $A^+ = 0$ condition exhibits itself in a spurious $1/k^+$-singularity in the
gluon Green function, which has to be regularized as well
\begin{equation}
\label{GluonPropagatorEq}
{\cal D}^{\mu\nu} (k)
=
\frac{d^{\mu\nu}_{\rm reg} (k)}{k^+ + i 0}
\, , \qquad
d^{\mu\nu} (k)
=
g^{\mu\nu} - \frac{k^\mu n^\nu + k^\nu n^\mu}{[k^+]_{\rm reg}}
\, .
\end{equation}
There are several possibilities to fix the gauge completely. They are (we put
the $k^+$ in square brackets to demonstrate that it is regularized via one of
the prescriptions):
\begin{itemize}
\item Principal value prescription
\begin{equation}
\label{PrincipalValue}
\frac{1}{[k^+]_{\scriptscriptstyle{\rm PV}}}
=
\frac{1}{2} \left( \frac{1}{k^+ + i 0} + \frac{1}{k^+ - i 0} \right)
\, .
\end{equation}
It corresponds to the anti-symmetric boundary conditions on the
gauge potential $\bit{A} (\infty) = - \bit{A} (- \infty)$.
\item Advanced prescription
\begin{equation}
d^{\mu\nu}_{\rm Adv} (k)
=
g^{\mu\nu} - \frac{k^\mu n^\nu}{k^+ + i 0} - \frac{k^\nu n^\mu}{k^+ - i 0}
\, ,
\end{equation}
arises from imposing the condition of vanishing of the gauge potential at
the light-cone future $\bit{A} (\infty) = 0$.
\item Retarded prescription
\begin{equation}
d^{\mu\nu}_{\rm Ret} (k)
=
g^{\mu\nu} - \frac{k^\mu n^\nu}{k^+ - i 0} - \frac{k^\nu n^\mu}{k^+ + i 0}
\, ,
\end{equation}
stems from the vanishing of the gauge potential in the past $\bit{A} (- \infty)
= 0$. The direction of momentum flow $k$ is very essential for the advanced and
the retarded prescriptions. We assume in Eq.\ (\ref{GluonPropagatorEq}) the flow
to be from the point $z_1$ to $z_2$.
\item Mandelstam-Leibbrandt prescription \cite{Man83,Lei94}
\begin{equation}
\label{MLpropagator}
\frac{1}{[k^+]_{\scriptscriptstyle{\rm ML}}}
=
\frac{1}{k^+ + i 0 \cdot k^-}
=
\frac{k^-}{k^+ k^- + i 0}
\, .
\end{equation}
This is prescription is not an obvious consequence of any boundary condition
on the gauge potential, see though \cite{SlaFro87}. It is however a consequence
of the equal-time (as opposed to the equal light-cone time) quantization
\cite{BasNarSol91}.
\end{itemize}

\subsection{Momentum integrals}

The basic integrals which appear in the course of calculations are listed below.
\begin{itemize}
\item  Minkowskian momentum integrals used in loop calculations
($d = 4 - 2 \varepsilon$)
\begin{eqnarray}
\label{MomentumIntAlpha}
\int \frac{d^dk}{(2\pi)^d} \,
{\rm e}^{A k^2 - i z \cdot k}
\!\!\!&=&\!\!\!
\frac{i}{(4\pi)^{d/2}}
\frac{1}{A^{d/2}} {\rm e}^{z^2/4A}
\, , \\
\int \frac{d^d k}{(2\pi)^d} \,
{\rm e}^{-i k \cdot z} \frac{1}{[-k^2]^m}
\!\!\!&=&\!\!\!
\frac{i}{2^{2m}{\pi}^{d/2}}
\frac{{\mit\Gamma} (d/2-m)}{{\mit\Gamma} (m)}
\frac{1}{[-z^2]^{d/2-m}}
\, .
\end{eqnarray}
\item Light-cone longitudinal integrals
\begin{equation}
\label{LongIntegral}
\int_{- \infty}^\infty d k^+ \, d k^- \,
{\rm e}^{i \left( A k^+ k^- + p^+ k^- + p^- k^+ + i 0\right)}
=
\frac{2 \pi}{A} {\rm e}^{- i p^+ p^- /A}
\, .
\end{equation}
\item Euclidean transverse momentum integrals ($d = 2 - 2 \varepsilon$)
\begin{eqnarray}
\label{Euclid2}
&&
\int \frac{d^d \bit{p}_\perp}{(2 \pi)^d} \,
{\rm e}^{- i \alpha \bit{\scriptstyle p}_\perp^2}
=
\frac{1}{(4 \pi)^{d/2}} \frac{1}{(i \alpha)^{d/2}}
\, .
\end{eqnarray}
\begin{eqnarray}
\label{Euclid1}
&&
\int \frac{d^d \bit{p}_\perp}{(2 \pi)^d}
\frac{
{\rm e}^{i \bit{\scriptstyle p}_\perp \cdot \bit{\scriptstyle z}_\perp }
}{
\bit{p}_\perp^{2m}}
=
\frac{1}{2^{2 m} \pi^{d/2}}
\frac{{\mit\Gamma} (d/2 - m)}{{\mit\Gamma} (m)}
\frac{1}{\bit{z}_\perp^{d - 2m}}
\, .
\end{eqnarray}
\item  Minkowskian space-time integrals
\begin{eqnarray}
\label{DdimenCoordIntegral}
\int d^d z {\rm e}^{i p \cdot z} \frac{1}{[- z^2]^k}
\!\!\!&=&\!\!\!
- i \pi^{d/2} 2^{d-2k} \frac{{\mit\Gamma} (d/2-k)}{{\mit\Gamma} (k)}
\frac{1}{[-p^2]^{d/2-k}}
\, , \\
\int d^d z {\rm e}^{A z^2 + i p \cdot z}
\!\!\!&=&\!\!\!
- i {\pi}^{d/2}\frac{1}{A^{d/2}} {\rm e}^{p^2/4A}
\, .
\end{eqnarray}
\end{itemize}

\subsection{Renormalization in covariant gauges}
\label{CovGaugeRenormalization}

First, let us recall some of the basic equations of QCD. The dimensionally
regularized and renormalized QCD Lagrangian in the covariant gauge is given by
\begin{eqnarray}
\label{YM-renormalized}
{\cal L}_{\scriptscriptstyle\rm QCD}
&=&
\mathcal{Z}_2 \bar\psi i \!\not\! \partial \psi
+
\bar \mathcal{Z}_1 \mu^\varepsilon g \bar\psi\!\not\!\!A^a t^a \psi
\nonumber\\
&-&
\frac{\mathcal{Z}_3}{4} \left(G^a_{\mu\nu}\right)^2
-
\frac{1}{2} \mu^\varepsilon g \mathcal{Z}_1 \ f^{abc} G^{a, \mu\nu} A^b_\mu A^c_\nu
-
\frac{\mathcal{Z}_4}{4} \mu^{2 \varepsilon} g^2 \left( f^{abc} A^b_\mu A^c_\nu \right)^2
-
\frac{1}{2\xi} \left( \partial^\mu A^a_\mu \right)^2
\nonumber\\
&+&
\widetilde{\mathcal{Z}}_3 \ \partial^\mu \bar\omega^a \partial_\mu \omega^a
+
\mu^\varepsilon g
\widetilde{\mathcal{Z}}_1 \ f^{abc} \partial^\mu \bar\omega^a A^b_\mu \omega^c
\, ,
\end{eqnarray}
where $d = 4 - 2 \varepsilon$ is the space-time dimension and $\mu$ is a mass
parameter, introduced in order to keep the coupling constant dimensionless. Here,
the renormalized quark, gluon and (anti-)ghost fields ${\mit\Phi} = \{ \psi ,
A^\mu , \bar\omega, \omega \}$, respectively, and the renormalized coupling are
expressed in terms of bare ones ${\mit\Phi}_{(0)}$ and $g^{(0)}$ via
\begin{equation}
\label{BareToRenormalization}
{\mit\Phi}_{(0)} = \sqrt{\mathcal{Z}_{\mit\Phi}} \, {\mit\Phi}
\, , \qquad
g^{(0)} = \mu^{\varepsilon} \mathcal{Z}_{\rm c} \, g
\, ,
\end{equation}
with corresponding renormalization constants $\mathcal{Z}_{\mit\Phi} =
\{ \mathcal{Z}_2, \ \mathcal{Z}_3 , \ \widetilde{\mathcal{Z}}_3 \}$, for the
quark, gluon and ghosts, respectively. The symbol $G^a_{\mu\nu}$ as a
shorthand notation for the Abelian part of the full QCD field strength tensor
$F^a_{\mu\nu} = G^a_{\mu\nu} + \mu^\varepsilon g \mathcal{X} f^{abc} A^b_{\mu}
A^c_{\nu}$. The canonical dimensions of elementary fields are $d^{\rm can}_q =
3/2 - \varepsilon$ for fermions, $d^{\rm can}_g = 1 - \varepsilon$ for gluons,
$d^{\rm can}_{\bar\omega} = d - 2$ and $d^{\rm can}_{\omega} = 0$ for the
anti-ghost and ghost fields, respectively. The Lagrangian (\ref{YM-renormalized})
is invariant under the following renormalized BRST-transformations:
\begin{eqnarray}
&&\delta^{\scriptscriptstyle\rm BRST} \psi
=
- i \mu^\varepsilon g \widetilde{\mathcal{Z}}_1 \omega^a t^a \psi \delta\lambda
\, , \quad
\delta^{\scriptscriptstyle\rm BRST} A^a_\mu
=
\widetilde{\mathcal{Z}}_3 \mathcal{D}_\mu \omega^a \delta\lambda
\, , \nonumber\\
&&\delta^{\scriptscriptstyle\rm BRST} \omega^a
=
\frac{1}{2} \mu^\varepsilon g \widetilde{\mathcal{Z}}_1 f^{abc}
\omega^b\omega^c \delta\lambda
\, , \quad
\delta^{\scriptscriptstyle\rm BRST} \bar{\omega}^a
=
\frac{1}{\xi} \partial_\mu A_\mu^a \delta\lambda
\, ,
\end{eqnarray}
where $\delta\lambda$ is a renormalized Grassman variable. The covariant
derivative is defined by  $\mathcal{D}_\mu = \partial_\mu -  i \mu^\varepsilon
g \mathcal{X} T^a A^a_\mu$, where $T^a$ is the generator in the fundamental ($T^a
\phi_i = t^a_{ij} \phi_j$) or adjoint ($T^b \phi^a = i f^{abc} \phi^c$) representations
depending on the object it is acting on. The Ward-Takahashi identities imply
the following relations between the renormalization constants $\mathcal{Z}_1
\mathcal{Z}^{- 3/2}_3 = \mathcal{Z}^{1/2}_4 \mathcal{Z}^{-1}_3 =
\widetilde{\mathcal{Z}}_1 \widetilde{\mathcal{Z}}^{-1}_3 \mathcal{Z}^{- 1/2}_3
= \bar{\mathcal{Z}}_1 \mathcal{Z}_2^{-1} \mathcal{Z}_3^{- 1/2}$. The charge
renormalization constant is expressed in terms of he vertex and wave function
renormalization constants via $\mathcal{Z}_{\rm c} = \mathcal{X} \mathcal{Z}_3^{-1/2}$.
In the minimal subtraction (MS) scheme, the $\mathcal{Z}$-factors are defined as
Laurent series in $\varepsilon$:
$$
\mathcal{Z} [g, \varepsilon]
=
1 + \sum_{n = 1}^\infty \frac{\mathcal{Z}^{[n]} [g]}{\varepsilon^n}
\, .
$$
For our consequent discussion,  we take the renormalization constants $\mathcal{Z}_2$,
$\mathcal{Z}_3$, $\widetilde{\mathcal{Z}}_3$ and $\mathcal{X} \equiv \mathcal{Z}_1
\mathcal{Z}^{-1}_3 = \widetilde{\mathcal{Z}}_1 \widetilde{\mathcal{Z}}^{-1}_3 =
\bar\mathcal{Z}_1 \mathcal{Z}_2^{-1}$ as independent ones.

The QCD Gell-Mann--Low function in $d$-dimensions is defined by
\begin{equation}
\beta_\varepsilon (g) = \frac{\partial g}{\partial\ln\mu}
= - \varepsilon g + \beta (g)
\, ,
\end{equation}
with $\beta(g)$ being the four-dimensional beta function. The anomalous
dimensions of the physical fields and the gauge fixing parameter are
\begin{equation}
\gamma_q (g) = \frac{1}{2} \frac{d}{d \ln\mu} \ln \mathcal{Z}_2
\, , \qquad
\gamma_g (g) = \frac{1}{2} \frac{d}{d \ln\mu} \ln \mathcal{Z}_3
\, , \qquad
\sigma = \frac{d}{d \ln\mu} \ln \xi = - 2 \gamma_g
\, ,
\end{equation}
respectively. Their perturbative expansions are determined by the series
\begin{equation}
\frac{\beta (g)}{g}
=
\sum_{n = 0}^\infty \left( \frac{\alpha_s}{4 \pi} \right)^{n + 1}
\beta_n
\, , \qquad
\gamma_a (g)
=
\sum_{n = 0}^\infty \left( \frac{\alpha_s}{2 \pi} \right)^{n + 1}
\gamma_{a(n)}
\, .
\end{equation}
The first two terms of the $\beta$-function, used in the present review, are
\begin{eqnarray}
\label{BetaFunction}
\beta_0
\!\!\!&=&\!\!\!
\frac{4}{3} T_F N_f - \frac{11}{3} C_A
\, , \\
\beta_1
\!\!\!&=&\!\!\!
\frac{10}{3} C_A N_f + 2 C_F N_f - \frac{34}{3} C_A^2
\, .
\end{eqnarray}
While the expressions for the one-loop anomalous dimensions of the elementary
fields read
\begin{equation}
\label{OneLoopADfieldsCovGauge}
\gamma_{q(0)} = \frac{\xi}{2} C_F
\, , \qquad
\gamma_{g(0)} = \frac{2}{3} T_F N_f
+ \frac{C_A}{4} \left( \xi - \frac{13}{3} \right)
\, .
\end{equation}
A full set of renormalization constants and anomalous dimensions in QCD up to
four-loop order can be found in Ref.\ \cite{Che04}.

\subsection{Renormalization of composite operators}

The subtracted operator\footnote{A vector of dimension $N$, composed from
different operators with the same quantum numbers, which mix under renormalization
group flow. For instance, in the flavor singlet case it is a two-dimensional object
(\ref{TwoVectorConfOperators}).} $\bit{{\cal O}}^{\scriptscriptstyle\rm R}
[{\mit\Phi}]$, constructed from the renormalized elementary fields ${\mit\Phi} =
\{ \psi , \bar\psi , A^\mu \}$, are expressed in terms of the bare ones
$\bit{{\cal O}} [{\mit\Phi}_{(0)}]$---built from the unrenormalized fields
${\mit\Phi}_{(0)} = \{ \psi_{(0)}, \bar\psi_{(0)}, A_{(0)}^\mu \}$---through the
multiplication with a renormalization factor
\begin{equation}
\bit{{\cal O}}^{\scriptscriptstyle\rm R}[{\mit\Phi}]
=
\bit{{\cal Z}} \bit{{\cal O}} [ \sqrt{{\cal Z}_{\mit\Phi}} {\mit\Phi}]
\, ,
\end{equation}
which admits Laurent expansion in the parameter $\varepsilon$ of dimensional
regularization,
\begin{equation}
\bit{{\cal Z}} [g , \varepsilon]
=
1
+
\sum_{n = 1}^\infty \frac{\bit{{\cal Z}}^{[n]} [g]}{\varepsilon^n}
\, .
\end{equation}
In turn, each term of this series $\bit{\cal Z}^{[n]} [g]$ has an infinite series
expansion in the strong coupling. The operators $\bit{{\cal O}}^{\scriptscriptstyle\rm R}
[{\mit\Phi}]$ generate finite contributions when inserted into Green functions with
elementary field operators:
\begin{equation}
\label{FiniteGreenFunctions}
\langle \bit{{\cal O}}^{\scriptscriptstyle\rm R} [{\mit\Phi}] \,
{\mit\Phi} (z_1) {\mit\Phi} (z_2) \dots {\mit\Phi} (z_N)
\rangle
= \, {\rm finite}
\, .
\end{equation}
The renormalization factors determine the anomalous dimensions of the corresponding
operators,
\begin{eqnarray}
\label{GammaZZ}
\bit{\gamma}
=
- \left( \frac{d}{d \ln \mu} \bit{{\cal Z}} \right) \bit{{\cal Z}}^{-1}
+
\frac{d}{d \ln \mu} \ln \bit{{\cal Z}}_{\mit\Phi}
\!\!\!&=&\!\!\!
g \frac{d}{d g}
\left(
\bit{{\cal Z}}^{[1]} - \bit{{\cal Z}}^{[1]}_{\mit\Phi}
\right)
\nonumber\\
&=&\!\!\!
g \frac{d}{d g}
\bit{{\cal Z}}^{[1]}
+
2 \gamma_{\mit\Phi}
\, ,
\end{eqnarray}
where at the second step we used the fact that only the single pole of the Laurent
series determine anomalous dimensions since
$$
\frac{d}{d \ln \mu} = - \varepsilon \frac{d}{d \ln g} + \dots
\, .
$$
All higher order terms in the Laurent expansion are expressed through it. Here, the
matrix $\bit{{\cal Z}}_{\mit\Phi}$ is diagonal and its entries are determined by the
renormalization constants of elementary fields forming a given bilocal composite
operator
\begin{equation}
\bit{{\cal Z}}_{\mit\Phi}
=
\left(
\begin{array}{cccc}
{\cal Z}_{\mit\Phi}
&
0
&
0
&
\dots
\\
0
&
{\cal Z}_{{\mit\Phi}'}
&
0
&
\dots
\\
0
&
0
&
{\cal Z}_{{\mit\Phi}''}
&
\dots
\\
\vdots
&
\vdots
&
\vdots
&
\ddots
\end{array}
\right)
\end{equation}
In order to calculate the renormalization factor, one computes the amputated%
\footnote{The amputated Green function is defined as a Green function
with removed propagators corresponding to the external lines $\langle
\bit{\cal O} [{\mit\Phi}'] \, {\mit\Phi} (z_1) \dots {\mit\Phi} (z_L)
\rangle_{\rm amp} = \langle \bit{\cal O} [{\mit\Phi}'] \, {\mit\Phi}
(z_1) \dots {\mit\Phi} (z_L) \rangle/ \prod_{k = 1}^L \langle
{\mit\Phi}' {\mit\Phi} (z_k) \rangle$.} Green function with insertion of
the bare composite operator, and uses the renormalized perturbation theory
to find the residue of the pole
\begin{equation}
\label{CountertermForO}
\langle
\bit{\cal O} [{\mit\Phi}] \, {\mit\Phi} (z_1) {\mit\Phi} (z_2)
\rangle_{\rm amp}
=
-
\frac{1}{\varepsilon} \bit{\cal R} \,
\langle
\bit{\cal O} [{\mit\Phi}] \, {\mit\Phi} (z_1) {\mit\Phi} (z_2)
\rangle_{\rm amp}^{\rm tree}
+
\dots
\, .
\end{equation}
Since the renormalized operator generates divergence-free Green functions
(\ref{FiniteGreenFunctions}), one immediately finds
\begin{equation}
\bit{\cal R} = \bit{\cal Z}^{[1]}
\, .
\end{equation}

\subsection{Construction of evolution equations in coordinate space}
\label{CoordinateEvolutionKernels}

In this appendix, we intend to review different approaches to the construction
of the renormalization group equations for non-local light-cone operators.
We will discuss two of them: one based on the use of covariant gauges, Ref.\
\cite{BraGeyRob85,BraGeyRob87}, and the other one relying on the light-cone
gauge with different prescriptions on the spurious pole in the gluon density
matrix. The machinery will be demonstrated on a simple example of the
unrenormalized quark string operator
\begin{equation}
\label{BareQuarkOperator}
{\cal O}^{qq} (z_1^- , z_2^-)
=
\bar\psi_{(0)} (z_1^-) [z_1^- , z_2^-] \gamma^+ \psi_{(0)} (z_2^-)
\, ,
\end{equation}
expressed in terms of the bare fields $\psi_{(0)}$. In the covariant gauge, one
needs the following Feynman rules for ${\cal O}^{qq}$ to one-loop accuracy,
\begin{eqnarray}
\label{FeynmanLRoperator1}
&&
\parbox[c][20mm][t]{25mm}{
\insertfig{2}{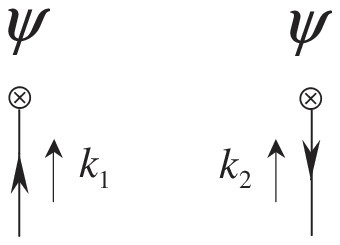}
}
\ \
=
{}[{\cal O}^{qq}_2 (k_1, k_2) ]^{\phantom{a}}
=
\gamma^+ \, {\rm e}^{ - i z^-_1 k_1^+ - i z^-_2 k_2^+ }
\, , \\
&&
\label{FeynmanLRoperator2}
\parbox[c][20mm][t]{25mm}{
\insertfig{2.5}{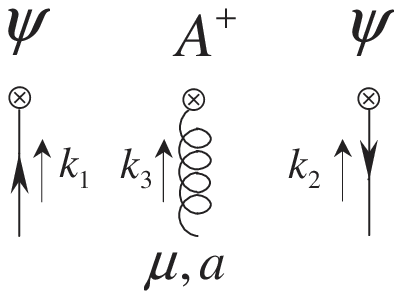}
}
\ \
=
{}[{\cal O}^{qq}_3 (k_1, k_2, k_3)]^a_\mu
=
g t^a n_\mu \gamma^+ \,
{\rm e}^{ - i z^-_1 k_1^+ - i z^-_2 k_2^+ }
\frac{ {\rm e}^{ - i z^-_1 k_3^+} - {\rm e}^{ - i z^-_2 k_3^+ } }{k_3^+}
\, .
\end{eqnarray}
Here the second contribution (\ref{FeynmanLRoperator2}) obviously arises from
the expansion of the path-ordered exponential in Eq.\ (\ref{BareQuarkOperator})
to the first non-trivial order in the coupling. In the light-cone gauge $A^+ = 0$,
only the first one, i.e., Eq.\ (\ref{FeynmanLRoperator1}), contributes. Let us
discuss both frameworks in turn.

\subsubsection{Covariant gauge formalism}

\begin{figure}[t]
\begin{center}
\mbox{
\begin{picture}(0,105)(140,0)
\put(20,0){\insertfig{8}{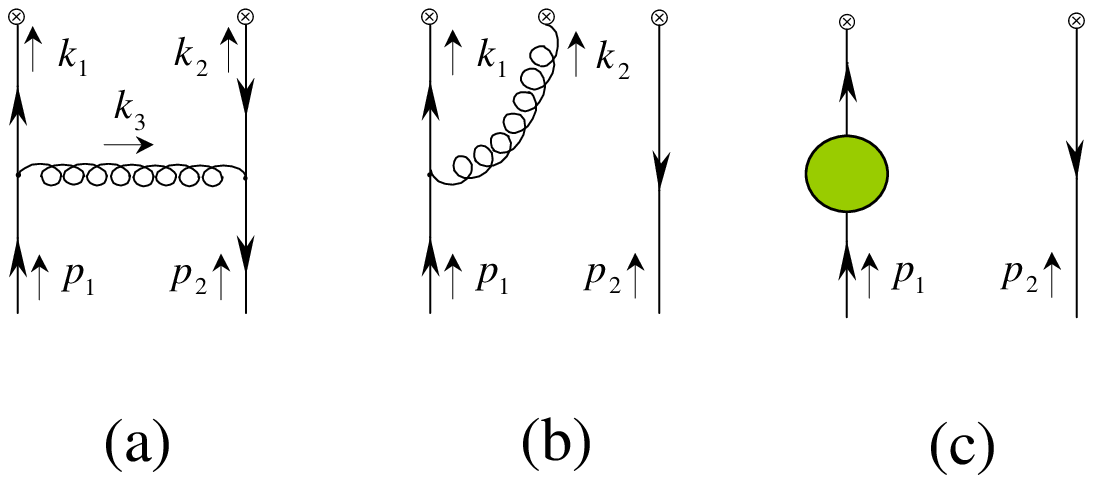}}
\end{picture}
}
\end{center}
\caption{\label{QQkernel} Feynman diagrams contributing to the
one-loop quark-quark mixing kernel.}
\end{figure}

In covariant gauges, we need to evaluate Feynman diagrams shown in Fig.\
\ref{QQkernel}. The diagrams (a) and (b) contribute to the renormalization
constant ${\cal R}$ in Eq.\ (\ref{CountertermForO}), while the diagram (c)
corresponds to the field renormalization which produces the addendum due to
the field anomalous dimension $\gamma_{\mit\Phi}$ defined by Eq.\ (\ref{GammaZZ}).

The most efficient method for computation of momentum integrals containing
Fourier exponents is to use the so-called $\alpha$-representation for
Feynman propagators,
\begin{equation}
\label{ExpProp}
\frac{1}{( k^2 + i 0 )^n}
=
\frac{(- i)^n}{{\mit\Gamma} (n)}
\int_0^\infty d \alpha \, \alpha^{n - 1} \, {\rm e}^{i \alpha (k^2 + i 0)}
\, ,
\end{equation}
taken here with an arbitrary power $n$ for generality. Exponentiating all
propagators, it is instructive the rescale the integrations variables,
\begin{equation}
\alpha_j  \equiv x_j \varrho
\, ,
\end{equation}
and switch to the integration over $\varrho$ and the Feynman parameters $x_j$
\begin{equation}
\int_0^\infty \prod_j^N d \alpha_j
=
\int_0^\infty d \varrho \, \varrho^{N - 1}
\int_0^1 \prod_j^N d x_j \, \delta \left( \sum_{k = 1}^N x_k - 1 \right)
\, .
\end{equation}
After integration over the loop momentum $k$, the subsequent integral over
the overall scale $\varrho$  results in poles in the parameter $\varepsilon$
of dimensional regularization. A typical integral reads
\begin{eqnarray}
\label{GenericIntegral}
&&\int \frac{d^d k}{(2 \pi)^d}
\frac{{\rm e}^{- i z^- \, k^+}}{k^2 \prod_{i = 1}^n (k - q_i)^2}
\\
&&\quad
=
\frac{i (- 1)^{n + 1}}{(4 \pi)^{d/2}}
{\mit\Gamma} \left( n + 1 - \ft{d}{2} \right)
\int_0^1 \prod_{i = 1}^{n + 1} dx_i \,
\delta \left( \sum_{i = 1}^{n + 1} x_i - 1 \right)
\frac{
{\rm e}^{- i z^- \sum_{i = 1}^n x_i q_i^+}
}{
\left[
- \sum_{i = 1}^n x_i q_i^2 + \left( \sum_{i = 1}^n x_i q_i \right)^2
\right]^{n + 1 - d/2}
}
. \nonumber
\end{eqnarray}

After these general preliminary remarks, we are now in a position to present
a step-by-step procedure of calculating the evolution kernel for the above
light-ray vector operator (\ref{BareQuarkOperator}) in the covariant Feynman
gauge $\xi = 1$.
\begin{itemize}
\item Use Feynman rules from Section \ref{FeynmanRules} and Eqs.\
(\ref{FeynmanLRoperator1}) and (\ref{FeynmanLRoperator1}) to
assign momentum intergals for the amputated Green function
$$
{\cal M}
=
\langle
{\cal O}^{qq} (z_1^-, z_2^-) \, \psi_{(0)} (k_1) \bar\psi_{(0)} (k_2)
\rangle_{\rm amp}
\equiv
\langle
{\cal O}^{qq} (z_1^-, z_2^-) \, \psi_{(0)} (k_1) \bar\psi_{(0)} (k_2)
\rangle
/
\prod_{j = 1}^2 i {\cal S} (k_j)
\, ,
$$
coming from each diagram in Fig.\ \ref{QQkernel}
\item Project out the color and Dirac (or Lorentz) structures of the corresponding
vertex
\begin{equation}
\label{QuarkProjection}
{\cal M}
=
\frac{1}{4 N_c} \gamma^+
{\rm tr}_{\rm c} {\rm tr} \left\{ \gamma^- {\cal M} \right\}
\, .
\end{equation}
The contributions of the diagrams  Fig.\ \ref{QQkernel} (a) and (b)
are then given by
\begin{eqnarray}
\label{MaGeneric}
{\cal M}^{(a)}_{\rm cov}
\!\!\!&=&\!\!\!
\frac{1}{4 N_c} \gamma^+
\int \prod_{j = 1}^3 \frac{d^d k_j}{(2 \pi)^4}
(-i) {\cal D}^{\mu\nu} (k_3)
\\
&\times&\!\!\!
{\rm tr}_{\rm c} {\rm tr}
\left\{
\gamma^- {\cal V}^a_\mu (k_3, - k_2, p_2)
i {\cal S} (- k_2)
{}[{\cal O}^{qq}_2 (k_1, k_2) ]
i {\cal S} (k_1)
{\cal V}^a_\nu (- k_3, p_1, - k_1) \right\}
\, , \nonumber\\
{\cal M}^{(b)}_{\rm cov}
\!\!\!&=&\!\!\!
\frac{1}{4 N_c} \gamma^+
\int \prod_{j = 1}^2 \frac{d^d k_j}{(2 \pi)^4}
(-i) {\cal D}^{\mu\nu} (k_2)
{\rm tr}_{\rm c} {\rm tr}
\left\{
\gamma^-
{}[{\cal O}^{qq}_3 (k_1, k_2, p_2)]^a_\mu
i {\cal S} (k_1)
{\cal V}^a_\nu (- k_2, p_1, - k_1) \right\}
\, . \nonumber
\end{eqnarray}
Perform integration over the redundant momenta using the energy-mom\-entum
delta functions leaving $k_3$ and $k_2$ as the loop momenta in diagrams (a)
and (b), respectively.
\item Compute traces. Use the fact that one can treat  external momenta
as light-like $p_i^\mu = p_i^+ n^{\ast\mu}$. We end up with momentum
integrals
\begin{eqnarray}
{\cal M}^{(a)}_{\rm cov}
\!\!\!&=&\!\!\!
2 i g^2 C_F \gamma^+ \int \frac{d^d k}{(2 \pi)^d}
\frac{\bit{k}^2_\perp}{k^2 (p_1 - k)^2 (p_2 + k)^2}
{\rm e}^{- i z_1^- (p_1 - k)^+ - i z_2^- (p_2 + k)^+}
\, , \\
{\cal M}^{(b)}_{\rm cov}
\!\!\!&=&\!\!\!
2 i g^2 C_F \gamma^+ \int \frac{d^d k}{(2 \pi)^d}
\frac{1}{k^2 (p_1 - k)^2} \frac{(p_1 - k)^+}{k^+}
\left(
{\rm e}^{- i z_1^- p_1^+ - i z_2^- p_2^+}
-
{\rm e}^{- i z_1^- (p_1 - k)^+ - i z_2^- (p_2 + k)^+}
\right)
\, . \nonumber
\end{eqnarray}
\item Compute the momentum integrals.

In the ${\cal M}^{(a)}$ amplitude, the lack of the $d$-dimensional covariance
of the integrand, due to the presence of the $d-2$ dimensional transverse
momentum $\bit{k}_\perp$, results in a slight complication in the use of the
generic integral (\ref{GenericIntegral}). However, with the help of Eqs.\
(\ref{LongIntegral}) and (\ref{Euclid2}), the integration is easily performed
in the light-cone (or Sudakov) coordinates
\begin{equation}
\label{SudakovMomentum}
k^\mu = k^+ n^{\ast\mu} + k^- n^\mu + k^\mu_\perp
\, ,
\end{equation}
with the volume element being
\begin{equation}
d^d k = d k^+ dk^- d^{d - 2} \bit{k}_\perp
\, .
\end{equation}
For the pole part\footnote{The PP is defined as ${\rm PP} \left\{
A {\mit\Gamma} (\varepsilon) \right\} = \frac{1}{\varepsilon} A$.} (PP)
of the dimensionally regularized integral,  we thus
find
\begin{equation}
{\rm PP} {\cal M}^{(a)}_{\rm cov}
=
\frac{1}{\varepsilon} \frac{\alpha_s}{2 \pi} C_F
\int_0^1 d x_1 \, d x_2 \, \theta (1 - x_1 - x_2)
{\rm e}^{
- i p_1^+ (\bar x_1 z_1^- + x_1 z_2^-)
- i p_2^+ (\bar x_2 z_1^- + x_2 z_2^-)
}
\, .
\end{equation}
Since we have set the external momenta to be light-like, we have no scale in
the integral. However, if introduced, it would simply add an exponential for
the integral with respect to the overall scale $\varrho$, and we can use the
substitution
$$
\int_0^\infty d \varrho \, \varrho^{1 - d/2}
\to
\int_0^\infty d \varrho \, \varrho^{1 - d/2}
{\rm e}^{- i m^2 \varrho}
=
\frac{{\mit\Gamma} (\varepsilon)}{(i m^2)^\varepsilon}
\simeq \frac{1}{\varepsilon}
+
\dots
\, .
$$

In the ${\cal M}^{(b)}$ amplitude, the integral with $1/k^+$ pole
can be computed making the trick
\begin{equation}
\frac{1}{k^+} \left( 1 - {\rm e}^{- i z^- k^+} \right)
=
i z^-  \int_0^1 d \tau \, {\rm e}^{- i \tau z^- k^+}
\, .
\end{equation}
Then, using the generic momentum integral (\ref{GenericIntegral}),  and
after that integrating over  $\tau$, we get
\begin{equation}
\label{MbFinal}
{\rm PP} {\cal M}^{(b)}_{\rm cov}
=
\frac{1}{\varepsilon} \frac{\alpha_s}{2 \pi} C_F
\int_0^1 d x_1
\left[
\frac{\bar x_1}{x_1}
\right]_+
{\rm e}^{
- i p_1^+ (\bar x_1 z_1^- + x_1 z_2^-)
- i p_2^+ z_2^-
}
\, ,
\end{equation}
with the plus-prescription defined in Eq.\ (\ref{PlusPrescription}).
\item Add the mirror symmetric contribution for the diagram (b)
${\cal M}^{(\tilde b)}$, which is obtained by the interchange
$p_1 \leftrightarrow p_2$ in the Eq.\ (\ref{MbFinal}), and notice
that, according to Eq.\ (\ref{CountertermForO}), the ${\cal Z}^{[1]}$
constant is determined to be
\begin{equation}
{\rm PP}
\left(
{\cal M}^{(a)} + {\cal M}^{(b)} + {\cal M}^{(\tilde b)}
\right)_{\rm cov}
=
- \frac{1}{\varepsilon} {\cal Z}^{[1]} \cdot [{\cal O}^{qq}_2 (p_1, p_2)]
\, ,
\end{equation}
with
\begin{eqnarray}
{\cal Z}^{[1]}_{qq} \cdot [{\cal O}^{qq}_2 (p_1, p_2)]
\equiv
-
\frac{\alpha_s}{2 \pi} C_F
\int_0^1 d x_1 \, d x_2 \, \theta (1 \!\!\!&-&\!\!\! x_1 - x_2)
\gamma^+
{\rm e}^{
- i p_1^+ (\bar x_1 z_1^- + x_1 z_2^-)
- i p_2^+ (\bar x_2 z_1^- + x_2 z_2^-)
}
\nonumber\\
&\times&\!\!\!
\left\{
1
+
\left[
\frac{\bar x_1}{x_1}
\right]_+
\delta (x_2)
+
\left[
\frac{\bar x_2}{x_2}
\right]_+
\delta (x_1)
\right\}
\, .
\end{eqnarray}
By comparing Eqs.\ (\ref{RGequation}) and (\ref{non-local-K}), we find
that the evolution kernel is
\begin{equation}
\label{LCkernelGeneric}
{\cal K} = - \alpha_s \frac{d}{d \alpha_s} {\cal Z}^{[1]} - \gamma_{\mit\Phi}
\, ,
\end{equation}
where  $\gamma_{\mit\Phi} = \gamma_q = \frac{\alpha_s}{2 \pi} \frac{C_F}{2} +
\cdots$ in the Feynman gauge $\xi = 1$ (see Eq.\ (\ref{OneLoopADfieldsCovGauge})).
Changing $x_1 = u$ and $x_2 = v$ we get the ${\cal K}^{qq, V}_{(0)}$ kernel in
Eq.\ (\ref{QQV-kernel}).
\end{itemize}
In order to study the singlet evolution involving mixing with gluonic operators,
we should consider  definite $C$-parity combinations of quark operators. The
summary of the corresponding Feynman rules, with the same  conventions as used here,
can be found in Appendices A.1 and A.2 of Ref.\ \cite{BelMul98b}. The calculation
of these evolution kernels goes through without any changes.

Yet another technique to evaluate the momentum integrals was developed  in Refs.\
\cite{BraGeyRob85,BraGeyRob87}. According to it, the propagators are combined using
the standard Feynman parametrization, and  the resulting momentum integrals take the
form
\begin{equation}
\int
\frac{d^d k}{(2 \pi)^d} {\rm e}^{-i k^+ (z^-_2 - z^-_1) }
\frac{P( p_i, x_i | k )}{[k^2 - L]^n}
\, ,
\end{equation}
where $P (p_i,x_i| k)$ is a polynomial function in $k$, $p_i$ and the Feynman
parameters $x_i$. The divergences can be evaluated by Taylor expanding the
exponential factor in the integrand: $e^{-i k^+ (z^-_2 - z^-_1)} = 1 - i k^+
(z^-_2 - z^-_1) + \dots$. Since the denominator depends on $k^2$ only, we can
average with respect to possible orientations of $k$. Due to the light-like
character of the vector $n$,  just  the first few terms survive after integration
(we maximally need to expand up to $(k^+)^3$). To present the result in the
conventional form, we have to remove terms proportional to $(z^-_2 - z^-_1)^m$.
This is achieved through the integration by parts with respect to the Feynman
parameters which play now the r\^ole of positions on the light-cone. A typical
integration for a test function $\tau(y,z)$ looks like
\begin{eqnarray}
i (z^-_2 - z^-_1) p^+_1 \, \mathbb{J} \cdot \tau (y, z)
= \mathbb{J} \cdot
\left\{
\delta(y) \tau (0 , z)
+
\frac{\partial \tau (y , z)}{\partial y}
-
\delta (1 - y - z) \tau (y , \bar y)
\right\}
\, , \nonumber
\end{eqnarray}
where we have introduced the following shorthand notation for the integral
\begin{equation}
\mathbb{J} \cdot \tau (u, v)
\equiv
\int_{0}^{1} d v \int_{0}^{\bar v} d u
{\rm e}^{
-i p^+_1 ( \bar u z^-_1 + u z^-_2)
-i p^+_2 ( \bar v z^-_2 + v z^-_1)
}
\tau (u, v)
\, .
\end{equation}
Similar equations hold for the $p^+_2$ contributions. When double and triple
integration by parts is required, it turns out that,  after the first
integration,   the dangerous terms proportional to $\delta (1 - u - v)$ that
might cause a problem vanish identically in all cases.

For diagrams which originate from the expansion of the phase factor
(\ref{FeynmanLRoperator2}), there appear terms in the integrand which
possess the structure
\begin{equation}
E (k, p_1, u)
\equiv
\frac{1}{( k^+ + u p^+_1 )}
\left[
1 - {\rm e}^{- i (z^-_2 - z^-_1) ( u p^+_1 + k^+ )}
\right]
\, ,
\end{equation}
with $k$ being the momentum of integration and $u$ one of the
Feynman parameters. The best way to treat them is to factorize
the $k$-dependence using the following identity utilizing the
translation invariance
\begin{equation}
E (k, p_1, u)
=
\frac{1}{u}
\exp
\left(
\frac{k^+}{u} \frac{\partial}{\partial p^+_1}
\right)
\frac{
1 - {\rm e}^{ - i u ( z^-_2 - z^-_1 ) p^+_1 }
}{p^+_1}
\, .
\end{equation}
As previously, expanding the exponential which depends on the differential
operator in power series of $(k^+)^n$, one can easily perform the final
momentum integration, since only a limited number of the lowest order terms
contribute as a consequence of the light-like character of the vector $n$.

However, this is not the end of the story in case of gluonic operators.
There, some contributions possess apart from the factor desired factor
$p^+_1 p^+_2$ accompanying the gluon operator vertex, also the factors with
$(p^+_i)^2$ dependence, which does not have the momentum structure of the
product of two field strength tensors. For completeness, let us outline a
few steps allowing to convert these structures into the required one, i.e.,
$(p^+_i)^2 \to p^+_1 p^+_2$. At first notice, that in the sum of all diagrams,
the coefficient in front of, e.g.,  $(p^+_1)^2$ has the following general
structure
\begin{equation}
\tau_1 (u) + \tau_2 (u) \delta (v) + \tau_3 (u) \delta ( 1 - u - v )
\, .
\end{equation}
The integration by parts reduces this expression to the desired form
up to contributions proportional to $\delta (1 - u - v)$. Namely,
\begin{eqnarray}
\label{reduction}
(p^+_1)^2 \,  \mathbb{J} \cdot \tau_1 (u)
\!\!\!&=&\!\!\!
(p^+_1)^2 \, \mathbb{J} \cdot
\left\{
\left[ \tau_1 (u) v + \tau_2 (u) \right] \delta (1 - u - v)
-
\tau_2 (u) \delta (v)
\right\}
\nonumber\\
&+&\!\!\! p^+_1 p^+_2 \, \mathbb{J} \cdot
\{
\left[ \tau_1 (u) v + \tau_2 (u) \right] \delta (1 - u - v)
\nonumber\\
&&\qquad\qquad
- \left[ \tau_1 (0) v + \tau_2 (0) \right] \delta (u)
- \left[ \tau_1'(u) v + \tau_2'(u) \right]
\}
\, .
\end{eqnarray}
Note also, that the resulting function
\begin{equation}
(p^+_1)^2
\left[
\tau_1 (u) v + \tau_2 (u) + \tau_3 (u)
\right] \delta (1 - u - v)
\end{equation}
can, in all known cases, be safely discarded on the grounds of the Bose symmetry
properties of the string operators it is convoluted with. Similar equations hold
for structures accompanying $(p^+_2)^2$. In particular, for the parity-even gluon
operator ${\cal O}^{gg} (z^-_1, z^-_2)$, the test functions involved above take
the following explicit form $\tau_1 (u) = 2 u (1 - u) - 1$ and $\tau_2 (u) =
\ft{1}{2} ( 1 - u ) ( 2 - u )$, and similarly, for the $u \to v$ contribution.

\subsubsection{Light-cone gauge formalism with ML prescription}

Turning to the calculation in the light-cone gauge, we will use the causal
prescription on the spurious light-cone pole in the gluon propagator, i.e.,
the Mandelstam-Leibbrandt (ML) prescription \cite{BasNarSol91,Lei94}. The most
important feature of this prescription is the presence of an additional
absorptive part of the vector boson Green function, namely
\begin{equation}
{\rm Disc}
\left\{
\frac{
d^{\mu\nu}_{\scriptscriptstyle{\rm ML}} (k)
}{
k^2 + i0
}
\right\}
= - 2 \pi i \theta(k_+)
\left\{
d^{\mu\nu}_{\scriptscriptstyle{\rm PV}} (k) \delta (k^2)
-
\frac{k^-}{k^2}
\left( k^\mu n^\nu + k^\nu n^\mu \right)
\delta (k^+ k^-)
\right\}
\, .
\end{equation}
The second contribution is of the ``ghost'' type,  since it has the opposite
sign compared  to the conventional one. As was proved in Ref.\ \cite{BasNarSol91},
the ML prescription is not an optional choice, rather, it is an unavoidable
consequence of the equal-time canonical quantization. The outcome of having this
additional term can be easily seen in the calculation of the one-loop evolution
kernels. Basically, it leads to the absence of spurious singularities in separate
diagrams, and to finiteness of individual contributions, in contrast to the
regularization by means of the principal value, which we will review next.

The momentum integrals with the Mandelstam-Leibbrandt prescription are most
easily computed with the help of the light-cone decomposition of momenta, Eq.\
(\ref{SudakovMomentum}). The longitudinal integrals are evaluated using the
Cauchy theorem, which yields
\begin{equation}
\int_{- \infty}^\infty
\frac{
d k^+ \, d k^-
}{
[k^+]^j_{\scriptscriptstyle{\rm ML}}
}
{\rm e}^{i \left( A k^+ k^- + k^+ p^- + k^- p^+ \right)}
=
- \frac{2 \pi}{p^+}
\left( - \frac{A}{p^+} \right)^{j - 1}
\left(
{\rm e}^{- i p^+ p^- / A}
-
\sum_{k = 0}^{j - 1} \left( - i \frac{p^+ p^-}{A} \right)^k
\right)
\, .
\end{equation}
For $j = 0$, it agrees with Eq.\ (\ref{LongIntegral}). The remaining Euclidean
integrals are evaluated via Eq.\ (\ref{Euclid2}). Combining two integrations ,
one gets the following result for a generic one-loop integral with a $1/k^+$-pole
regularized via the Mandelstam-Leibbrandt prescription
\begin{eqnarray}
\label{GenericLCintegralML}
&&\int \frac{d^d k}{(2 \pi)^d}
\frac{
{\rm e}^{- i z^- \, k^+}
}{
[k^+]_{\scriptscriptstyle{\rm ML}}
} \frac{1}{k^2 \prod_{i = 1}^n (k - q_i)^2}
\\
&&\qquad
=
\frac{i (- 1)^{n + 1}}{(4 \pi)^{d/2}}
{\mit\Gamma} \left( n + 1 - \ft{d}{2} \right)
\int_0^1 \prod_{i = 1}^{n + 1} dx_i \,
\delta \left( \sum_{i = 1}^{n + 1} x_i - 1 \right)
\frac{1}{\sum_{i = 1}^n x_i q_i^+}
\nonumber\\
&&\qquad\times
\left\{
\frac{
{\rm e}^{- i z^- \sum_{i = 1}^n x_i q_i^+}
}{
\left[
- \sum_{i = 1}^n x_i q_i^2 + \left( \sum_{i = 1}^n x_i q_i \right)^2
\right]^{n + 1 - d/2}
}
-
\frac{
1
}{
\left[
- \sum_{i = 1}^n x_i q_i^2 - \left( \sum_{i = 1}^n x_i \bit{q}_{i \perp} \right)^2
\right]^{n + 1 - d/2}
}
\right\}
\, , \nonumber
\end{eqnarray}
which agrees for $z^- = 0$ with results of Refs.\ \cite{Ogr86,CapDulLit83,HeiKuns98}.

The anomalous dimensions of quarks and gluons are, of course, gauge dependent
and their one-loop values for the present choice are given by
\begin{equation}
\label{OneLoopMLanomalousDim}
\gamma_{(0)q} = \frac{3}{2} C_F
\, , \qquad
\gamma_{(0)g} = \frac{\beta_0}{2}
\, .
\end{equation}
More generally, the anomalous dimension of the gluon field in the light-cone
gauge is equal to the renormalization group function of the running coupling,
$$
\gamma_g = \frac{\beta (g)}{g}
\, ,
$$
to all orders in the coupling constant.

In the light-cone gauge, only the diagrams (a) and (c) contribute. The Feynman
integral for the contribution (a) remains the same as in Eq.\ (\ref{MaGeneric}),
where we have to use now the gauge propagator (\ref{GluonPropagatorEq}) with
(\ref{MLpropagator}). Using the projection (\ref{QuarkProjection}) to evaluate
the trace, we find
\begin{eqnarray}
{\cal M}^{(a)}_{\rm lc}
\!\!\!&=&\!\!\!
2 i g^2 C_F \gamma^+ \int \frac{d^d k}{(2 \pi)^d}
\frac{
{\rm e}^{- i z_1^- (p_1 - k)^+ - i z_2^- (p_2 + k)^+}
}{k
^2 (p_1 - k)^2 (p_2 + k)^2
}
\\
&&\qquad\qquad
\times
\left\{
\bit{k}^2_\perp
+
\frac{(p_1 - k)^+}{[k^+]_{\scriptscriptstyle \rm ML}} (p_2 + k)^2
+
\frac{(p_2 + k)^+}{[k^+]_{\scriptscriptstyle \rm ML}} (p_1 - k)^2
\right\}
\, . \nonumber
\end{eqnarray}
The first term in the curly brackets produces the amplitude
${\cal M}^{(a)}_{\rm cov}$, while the integrals with the regularized pole
are given by Eq.\ (\ref{GenericLCintegralML}) adjusted for the present needs
\begin{equation}
\label{PPmomentumIntegral}
{\rm PP}
\int \frac{d^d k}{(2 \pi)^d}
\frac{{\rm e}^{- i k^+ z^-}}{[k^+]^n_{\scriptscriptstyle \rm ML} k^2 (p - k)^2}
=
\frac{i}{(4 \pi)^2 \varepsilon}
\int_0^1 \frac{d x}{x^n}
\left(
{\rm e}^{- i x p^+ z^-}
-
\sum_{l = 0}^{n - 1} \left( - i x p^+ z^- \right)^l
\right)
\, .
\end{equation}
Since $k^+/[k^+]_{\scriptscriptstyle \rm ML} = 1$, one can  use Eq.\
(\ref{GenericIntegral}) for other terms. Finally, adding the light
cone gauge quark anomalous dimensions to Eq.\ (\ref{LCkernelGeneric}),
one recovers the same result as obtained with the covariant gauge
fixing before, see Eq.\ (\ref{QQV-kernel}).

\subsection{Construction of evolution equations in momentum space}
\label{CompTechMomentumKernels}

Now we outline the momentum-space formalism for the calculation of
evolution kernels developed in Ref.\ \cite{BukKurLip83}. It relies
heavily on the use of the light-cone gauge with the principal value
prescription. To begin with, we outline the basics of the one-loop
renormalization of the QCD Lagrangian in this gauge.

\subsubsection{Renormalization in light-cone gauge with PV prescription}
\label{renormalization}

The principal value prescription on the unphysical pole at $k^+ = 0$
(\ref{PrincipalValue}) can be operationally formulated as a regularization
\be
\label{PrincipalValuePrescription}
\frac{1}{[k_+]_{\scriptscriptstyle\rm PV}}
=
\lim_{\delta \to 0}
\frac{k_+}{k_+^2 + \delta^2}
\, ,
\ee
with a parameter $\delta$. The introduction of this extra parameter allows
one to disentangle the spurious light-cone divergences from the conventional
ultraviolet and infrared poles in $\varepsilon$. They would be unseparable
otherwise if $\delta$ was set to zero from the onset.

As a first step, let us compute the one-loop quark self-energy due to the
gluon which is given by the momentum integral
\be
\Sigma (q)
=
\mu^{2 \varepsilon} \int \frac{d^d k}{(2 \pi)^d}
\gamma^\mu \frac{{\not\!q} - {\not\!k}}{(q - k)^2}
\gamma^\nu \frac{d_{\mu\nu} (k)}{k^2}
\, .
\ee
Reorganizing the numerator, we can cast it in the form
\be
\Sigma (q) = - 2 (1 - \varepsilon)  \gamma_\mu J^\mu_2 (q)
-
\left(
{\not\!q} \gamma_\mu \gamma_+ + \gamma_+ \gamma_\mu {\not\!q}
\right)
J_{2+}^\mu (q)
\, ,
\ee
where we set the dimensionally regularized tadpole $\int d^d k/k_+ k^2$ to
zero. The contributing integrals are
\ba
J^\mu_2 (q)
&=&
\mu^{2 \varepsilon}
\int \frac{d^d k}{(2 \pi)^d}
\frac{(q - k)^\mu}{k^2 (q - k)^2}
=
\frac{i {\mit\Gamma} (\varepsilon)}{(4 \pi)^2}
\left(
- \frac{4 \pi \mu^2}{q^2}
\right)^\varepsilon
\frac{q^\mu}{2}
\int_0^1
\frac{d u}{(u \bar{u})^\varepsilon}
\, , \\
\label{TwoPointAxial}
J_{2+}^\mu (q)
&=&
\mu^{2 \varepsilon}
\int \frac{d^d k}{(2 \pi)^d}
\frac{(q - k)^\mu}{k^2 (q - k)^2 [k_+]_{\scriptscriptstyle\rm PV}}
\nonumber\\
&=&
\frac{i {\mit\Gamma} (\varepsilon)}{(4 \pi)^2}
\left(
- \frac{4 \pi \mu^2}{q^2}
\right)^\varepsilon
\int_0^1
\frac{d u}{(u \bar{u})^\varepsilon [u q_+]_{\scriptscriptstyle\rm PV}}
\left\{
\bar\alpha q^\mu - (1 - 2 u) \frac{q^2}{2 q_+} n^\mu
\right\}
\, .
\ea
Let us add a comment on the calculation procedure involving the integrals with
$1/k_+$ pole in the integrand \cite{CapDulLit83}. Since its regularization
via the principal value prescription does not depend on $k_-$, after exponentiation
of propagators (\ref{ExpProp}), one easily performs the $d k_-$ integral resulting
in a delta-functions with the argument being an equality between the plus-components
of the loop and external momenta. Thus, the $k_+$ integration is trivial making use
of this delta-function. The transverse integral is performed using the Gaussian
integration (\ref{Euclid2}). Also, notice that if we would not use the regularization
with the parameter $\delta$ in Eq.\ (\ref{PrincipalValuePrescription}) this would
lead to an extra pole in the dimensionally regularized Feynman-parameter integral and
yield a double-pole in a one-loop diagram. Having distinct regularization parameters,
we disentangle the ultraviolet poles from the spurious poles due to the light-cone
singularity in the gauge propagator. Assembling everything together, we get
\be
\Sigma (q)
=
{\not\!q} \, \Sigma_1 (q) + \frac{q^2}{2 q_+} \gamma_+ \, \Sigma_2 (q)
\ee
where the functions in front of independent Dirac structures are
\be
\Sigma_1 (q)
=
- (1 - \varepsilon) \frac{i {\mit\Gamma} (\varepsilon)}{(4 \pi)^2}
\left(
- \frac{4 \pi \mu^2}{q^2}
\right)^\varepsilon
\int_0^1
\frac{d u}{(u \bar{u})^\varepsilon}
\, , \quad \
\Sigma_2 (q)
=
-
\frac{i {\mit\Gamma} (\varepsilon)}{(4 \pi)^2}
\left(
- \frac{4 \pi \mu^2}{q^2}
\right)^\varepsilon
\int_0^1
\frac{d u}{(u \bar{u})^\varepsilon}
\frac{4 \bar{u} q_+}{ [u q_+]_{\scriptscriptstyle\rm PV}}
\, .
\ee
The Feynman-parameter integral involving the principal-value regularized
denominator is computed as an expansion in the parameter of dimensional
regularization,
\be
\int_0^1
\frac{d \alpha}{(\alpha \bar\alpha)^\varepsilon}
\frac{\bar\alpha q_+}{ [\alpha q_+]_{\scriptscriptstyle\rm PV}}
=
- 1 - \ln \frac{\delta}{q_+}
- \varepsilon
\left(
2 - \ft12 \ln^2 \frac{\delta}{q_+} - \ft54 \zeta (2)
\right)
+
\mathcal{O} (\varepsilon^2)
\, .
\ee
The arising dependence on the plus-momentum is a well-known property of light cone
gauge regularized with principal value prescription, which propagates into the
anomalous dimensions of elementary field operators. Of course, the dependence on
$\delta$ (and $q^+$) cancels out in all gauge invariant quantities.

Keeping only the pole contributions in one-loop self-energies, the one-loop
gluon and quark propagators can be represented as a unitary rotations of the
tree propagators
\begin{equation}
{\cal D}^{\mu\nu} \to {\cal D}_{[1]}^{\mu\nu}
\, , \qquad
{\cal S} \to {\cal S}_{[1]}
\, .
\end{equation}
The renormalization constants are no longer numbers, but rather they are
matrices acting on the spinor (Lorentz) indices of the fermion (gluon) field
operators. Moreover as we emphasized above, the renormalization constants
depend on the plus components of the particles' momenta $k_i^+$. The origin
of this can be traced back to the lack of the rescaling invariance,
\begin{equation}
d^{\mu\nu}_{\scriptscriptstyle{\rm PV}} (\varrho k)
\neq
d^{\mu\nu}_{\scriptscriptstyle{\rm PV}} (k) ,
\end{equation}
which is otherwise obeyed by the unregularized propagator and the propagator
with Mandelstam-Leibbrandt prescription. We can represent the one-loop
propagators as follows.

\begin{itemize}
\item Quark propagator is
\begin{equation}
\label{QuarkPropOneLoopPV}
{\cal S}_{[1]} (k)
=
\bar{U} (k) {\not\! k} U (k) \frac{{\cal Z}_q (k)}{k^2 + i 0}
\, ,
\end{equation}
where
\begin{equation}
U (k)
=
1 - \frac{\widetilde{\cal Z}_q (k)}{2 k^+} {\not\! k} \gamma^+
\, , \qquad
\bar{U} (k)
\equiv
\gamma^0 U^\dagger (k) \gamma^0
=
1 - \frac{\widetilde{\cal Z}_q (k)}{2 k^+} \gamma^+ {\not\! k}
\, .
\end{equation}
The renormalization constants are
\begin{eqnarray}
\label{ZQConst}
{\cal Z}_q (k)
\!\!\!&=&\!\!\!
1 + \frac{\alpha_s}{\pi} C_F \frac{1}{\varepsilon}
\left\{
\int d q^+ \frac{k^+}{q^+ \! - k^+ \!} \,
\vartheta_{11}^0 (q^+ , q^+ \! - k^+)
+ \frac{3}{4}
\right\}
\, , \\
\label{ZQtildeConst}
\widetilde {\cal Z}_q (k)
\!\!\!&=&\!\!\!
\frac{\alpha_s}{2 \pi} C_F \frac{1}{\varepsilon}
\left\{
1
+
\int d q^+ \frac{k^+}{q^+ \! - k^+ \!} \,
\vartheta_{11}^0 (q^+ , q^+ \! - k^+)
\right\}
\, ,
\end{eqnarray}
\item Gluon propagator reads
\begin{equation}
{\cal D}_{[1]}^{\mu\nu} (k)
=
U^{\mu\rho} (k) d_{\rho\sigma} (k) U^{\sigma\nu} (k)
\frac{{\cal Z} (k)}{k^2 + i 0}
\, .
\end{equation}
The presence of the divergent tensor $U^{\mu\nu}$,
\begin{equation}
U^{\mu\nu} (k)
=
g^{\mu\nu}
-
\widetilde {\cal Z}_g (k) \frac{k^\mu n^\nu + k^\nu n^\mu}{k^+}
\, ,
\end{equation}
is an artifact of the Lorentz symmetry breaking effects by the gauge
fixing vector $n^\mu$ and of using  the principal value prescription
which leads to different renormalization constants for
``good'' and ``bad'' components of tensor fields. Namely,
\begin{eqnarray}
\label{ZGConst}
{\cal Z}_g (k)
\!\!\!&=&\!\!\!
1 + \frac{\alpha_s}{\pi} C_A \frac{1}{\varepsilon}
\left\{
\int d q^+ \frac{k^+}{q^+ \! - k^+ \!} \,
\vartheta_{11}^0 (q^+ , q^+ \! - k^+)
- \frac{1}{4} \frac{\beta_0}{C_A}
\right\}
\, , \\
\label{ZGtildeConst}
\widetilde {\cal Z}_g (k)
\!\!\!&=&\!\!\!
\frac{\alpha_s}{2 \pi} C_A \frac{1}{\varepsilon}
\left\{
1
+
\int d q^+ \frac{k^+}{q^+ \! - k^+ \!} \,
\vartheta_{11}^0 (q^+ , q^+ \! - k^+)
\right\}
\, .
\end{eqnarray}
\end{itemize}
Note that the ``good" components of the fields are renormalized by ${\cal Z}$
and are not affected by $\widetilde {\cal Z}$.

The appearing $\mathcal{Z}$-factors depend on the integral involving the
light-cone singularity
\begin{equation}
\int d q^+ \frac{k^+}{[\, q^+ \! - k^+ ]_{\scriptscriptstyle\rm PV}} \,
\vartheta_{11}^0 (q^+ , q^+ \! - k^+)
= \ln \frac{\delta}{k^+}
\, .
\end{equation}
However, for the consequent use we will not use the explicit form of this
integral.

Finally, notice that if one relies on the cut-off regularization with an
ultraviolet scale $\mu^2$ rather than dimensional regularization, then one
would be able to obtain one from the other by making the substitution
\begin{equation}
\frac{1}{\varepsilon} \to
\frac{\mu^{2 \varepsilon}}{\varepsilon}
\to
\ln \mu^2
\, ,
\end{equation}
where $\mu^2$ is the standard mass parameter of dimensional regularization.

\subsubsection{Evolution kernels in momentum space}

Computing the evolution kernels we have to isolate ultraviolet divergences
occurring in transverse-momentum integrals of partons interacting with a
bare  quark operator (\ref{BareQuarkOperator}). To extract this dependence
properly, it is sufficient to separate the perturbative loop from the
correlation function in question. To this end, the latter can be represented
in the form of the momentum integral in which the integration over the
fractional energies of the particles attached to the vertex is removed
\begin{equation}
\label{CorrelationFunctionGammaPlus}
{\cal O}^{qq} (x_1 , x_2)
=
\int \frac{d^4 k_1}{(2 \pi)^4} \frac{d^4 k_2}{(2 \pi)^4}
\delta ( k_1^+ - x_1 ) \delta ( k_2^+ - x_2 )
{\cal O}^{qq} (k_1 , k_2)
\, ,
\end{equation}
where the momentum-space operator is given here by
\begin{equation}
{\cal O}^{qq} (k_1 , k_2)
=
\int d^4 z_1 \, d^4 z_2 \,
{\rm e}^{i k_1 \cdot z_1 + i k_2 \cdot z_2}
\bar\psi (z_1) \gamma^+ \psi (z_2)
\equiv
\bar\psi (k_1) \gamma^+ \psi (k_2)
\, .
\end{equation}
Here in the last step  we introduced the momentum-space
quark fields.

The calculation of one-loop diagrams is extremely simple within the
framework of momentum-fraction space with the light-cone gauge
\cite{BukKurLip83,BukFroLipKur85,BelKur97,BelMul97a,Bel97}. Let us
describe again a step-by-step procedure for computing the kernel.
\begin{itemize}
\item Use Feynman rules from Section \ref{FeynmanRules} to write the one-loop
expression for the correlation function (\ref{CorrelationFunctionGammaPlus}).
\item Project the color and Dirac structure on the operator vertex
as in Eq.\ (\ref{QuarkProjection}). These two steps result in
\begin{eqnarray}
{\cal O}_{{\rm 1-loop} (a)}^{qq} (x_1 , x_2)
\!\!\!&=&\!\!\!
\frac{1}{4 N_c} \int \frac{d^4 p_1}{(2 \pi)^4} \frac{d^4 p_2}{(2 \pi)^4}
{\cal O}^{qq} (p_1 , p_2)
\int \prod_{i = 1}^3
\frac{d^4 k_i}{(2 \pi)^4}
\delta (k_1^+ - x_1) \delta (k_2^+ - x_2)
\\
&\times&\!\!\!
{\rm tr}_{\rm c} {\rm tr}
\left\{
\gamma^- {\cal V}^a_\mu (- k_3, p_1, - k_1)
i {\cal S} (- k_2) \gamma^+ i {\cal S} (k_1)
{\cal V}^a_\nu (k_3, - k_2, p_2) \right\}
(-i) {\cal D}^{\mu\nu} (k_3)
. \!\!\!\nonumber
\end{eqnarray}
\item Eliminate the momentum conservation delta-functions keeping $k_3^\mu$
as the loop momentum $k^\mu$.
\item Use the Sudakov decomposition for four-momenta in Eq. (\ref{SudakovMomentum}).
Compute the trace extracting only the piece which results in  the logarithmically
divergent transverse-momentum integral. Namely, since the loop-momentum volume
element produces the  factor $\bit{k}_\perp^2 d \bit{k}^2_\perp$,  and the denominators
of the three propagators give the  factor $(\bit{k}_\perp^2)^{- 3}$, one needs to extract
the term proportional to $\bit{k}_\perp^2$ from the trace, i.e.,
\begin{eqnarray}
&&\frac{1}{4} {\rm tr}
\left\{
\gamma^- \gamma_\mu ( {\not\! k}_1 - {\not\! p}_1 - {\not\! p}_2 )
\gamma^+ {\not\! k}_1 \gamma_\nu
\right\}
d^{\mu\nu} (k_1 - p_1)
\\
&&\qquad
\simeq
- 2 \bit{k}^2_{1\perp}
\left\{
1
+
\frac{x_1 - p_1^+ - p_2^+}{x_1 - p_1^+}
\left[ x_1 \beta_1 - 1 \right]
+
\frac{x_1}{x_1 - p_1^+}
\left[ (x_1 - p_1^+ - p_2^+) \beta_1 - 1 \right]
\right\}
\, . \nonumber
\end{eqnarray}
\item Use the strong-ordering approximation (\ref{StrongOderingDGLAP}),
$|\bit{k}_{1 \perp}| \gg |\bit{p}_{1 \perp}|$ and $k_1^- \gg p_1^-$.
For a typical propagator denominator this gives
\begin{equation}
(k_1 - p_1)^2 + i 0
\simeq
\bit{k}_{1 \perp}^2
\left[ (x_1 - p_1^+) \beta_1 - 1 + i 0 \right]
\, ,
\end{equation}
where one conveniently introduces a rescaled $k^-$ component
\begin{equation}
\beta \equiv \frac{2 k^-}{\bit{k}^2_\perp}
\, .
\end{equation}
The last three steps yield
\begin{eqnarray}
\label{RealOneLoop}
&&\!\!\!\!\!\!\!\!\!\!\!\!\!\!\!\!\!\!\!
{\cal O}^{qq}_{{\rm 1-loop} (a)} (x_1 , x_2)
=
- \frac{\alpha_s}{2 \pi^2} C_F
\int \frac{d^4 p_1}{(2 \pi)^4} \frac{d^4 p_2}{(2 \pi)^4}
{\cal O}^{qq} (p_1 , p_2)
\delta ( x_1 + x_2 - p_1^+ - p_2^+ )
\int \frac{d^2 \bit{k}_{1 \perp}}{\bit{k}^2_{1 \perp}}
\\
&&\!\!\!\!\!\!\!\!\!\!\!\!\!\!\!\quad\times
\frac{1}{x_1 - p_1^+}
\int \frac{d \beta_1}{2 \pi i}
\frac{
( x_1 - p_1^+ )
+
( x_1 - p_1^+ - p_2^+ )
\left[ x_1 \beta_1 - 1 \right]
+
x_1
\left[ (x_1 - p_1^+ - p_2^+) \beta_1 - 1 \right]
}{
[ x_1 \beta_1 - 1 + i 0]
[ (x_1 - p_1^+ - p_2^+) \beta_1 - 1 + i 0 ]
[ (x_1 - p_1^+) \beta_1 - 1 + i 0 ]
}
\, , \nonumber
\end{eqnarray}
\item Compute the transverse-momentum integral with a cut-off
\begin{equation}
\int^{\mu^2} \frac{d^2 \bit{k}_{1 \perp}}{\bit{k}^2_{1 \perp}}
= \pi \ln \mu^2
\, .
\end{equation}
\item Use the generalized step functions from Appendix \ref{GeneralStepFunctions}.

As can be seen, Eq.\ (\ref{RealOneLoop})  diverges at $x_1 =
p_1^+$. The  divergencies disappear after
one adds  the virtual radiative corrections
(renormalization of the field operators) discussed in the previous subsection.
\item Virtual corrections are computed using the one-loop propagator given
in Eq.\ (\ref{QuarkPropOneLoopPV}), namely, one gets
\begin{equation}
\label{VirtualOneLoop}
{\cal O}^{qq}_{{\rm 1-loop} (b)} (x_1 , x_2)
=
\int \frac{d^4 p_1}{(2 \pi)^4} \frac{d^4 p_2}{(2 \pi)^4}
\,
\sqrt{{\cal Z}_q (p_1) {\cal Z}_q (p_2)} \,
\bar\psi (p_1) U (p_1) \gamma^+ \bar U (p_2) \psi (p_2)
\, ,
\end{equation}
where the quark-field operators are renormalized as
\begin{equation}
\psi (k) = \sqrt{{\cal Z}_q (k)} U (k) \psi_{(0)} (k)
\, .
\end{equation}
Using the obvious fact that
$$
U (p) \gamma^+ = \gamma^+ \bar U (p) = 1
$$
and  explicit form of the one-loop renormalization constant
(\ref{ZQConst}), one finally gets  the contribution due to the  virtual corrections
\begin{eqnarray}
{\cal O}^{qq}_{{\rm 1-loop} (b)} (x_1 , x_2)
\!\!\!&=&\!\!\!
\frac{\alpha_s}{2 \pi} C_F \ln \mu^2
\int d x_1' \, d x_2' \,
\delta ( x_1 + x_2 - x_1' - x_2' )
\\
&\times&\!\!\!
\Bigg\{
\frac{3}{2} \delta (x_1 - x_1')
+
\delta (x_1 - x_1')
\int d x_1'' \frac{x_1'}{x_1'' - x_1'}
\vartheta_{11}^0 (x_1'' , x_1'' - x_1')
\nonumber\\
&&\qquad\qquad\quad \
+
\delta (x_2 - x_2')
\int d x_2'' \frac{x_2'}{x_2'' - x_2'}
\vartheta_{11}^0 (x_2'' , x_2'' - x_2')
\Bigg\}
{\cal O}^{qq} (x_1' , x_2')
\, , \nonumber
\end{eqnarray}
where $p_i^+ \equiv x_i^\prime$.
\item Adding Eqs.\ (\ref{RealOneLoop}) and (\ref{VirtualOneLoop}) yields
\begin{equation}
{\cal O}^{qq}_{\rm 1-loop} (x_1 , x_2)
=
- \frac{\alpha_s}{2 \pi} C_F \ln \mu^2
\int d x_1' \, d x_2' \,
\delta ( x_1 + x_2 - x_1' - x_2' )
K^{qq, V}_{(0)} (x_1 , x_2 | x_1', x_2')
{\cal O}^{qq} (x_1' , x_2')
\, ,
\end{equation}
with the kernel given in Eq.\ (\ref{PE-ext}), where the regularization of the
end-point behavior $y_i \to x_i$ results in the  plus-prescription,
\begin{equation}
\left[
\frac{y_1}{x_1 - y_1} \vartheta^0_{11} (x_1, x_1 - y_1)
\right]_+
\equiv
\frac{y_1}{x_1 - y_1} \vartheta^0_{11} (x_1, x_1 - y_1)
-
\delta (x_1 - y_1)
\int d y'_1
\frac{y_1}{y'_1 - y_1} \vartheta^0_{11} (y'_1, y'_1 - y_1)
\, . \!
\end{equation}
\end{itemize}

\subsection{Generalized step functions}
\label{GeneralStepFunctions}

The momentum space evolution kernels presented in Section \ref{MomSpaceGerKernels}
are expressed in terms of the generalized step-functions
\begin{equation}
\vartheta^{k}_{\alpha_1 \dots \alpha_j} (x_1, \dots , x_j)
\equiv
\int \frac{d \beta}{2 \pi i} \beta^k
\prod_{\ell = 1}^j \left( x_\ell \beta - 1 + i 0 \right)^{- \alpha_\ell}
\, .
\end{equation}
In general case, the reduction of the higher-rank functions is performed making
multiple use of the following identities
\begin{eqnarray}
\label{ReductionThetas}
\vartheta^n_{ i j k \cdots } (x_1, x_2, x_3, \dots)
\!\!\!&=&\!\!\!
\frac{1}{x_1 - x_2}
\left\{
\vartheta^{n - 1}_{ i - 1 j k \cdots } (x_1, x_2, x_3, \dots)
-
\vartheta^{n - 1}_{ i j - 1 k \cdots } (x_1, x_2, x_3, \dots)
\right\}
\\
&=&\!\!\! \frac{1}{x_1 - x_2}
\left\{
x_2 \vartheta^n_{ i - 1 j k \cdots} (x_1, x_2, x_3, \dots)
-
x_1 \vartheta^n_{ i j - 1 k \cdots} (x_1, x_2, x_3, \dots)
\right\}
\, . \nonumber
\end{eqnarray}
In case of two coinciding arguments, the function collapses to a lower rank,
$$
\vartheta^n_{ i j k \cdots } (x_1, x_1, x_3, \dots)
=
\vartheta^n_{ i + j, k \cdots } (x_1, x_3, \dots)
\, .
$$
Along this line of reasoning, all the functions can be reduced to the lowest
one $\vartheta^0_{11}$, which is given by
\begin{equation}
\label{Theta110Appendix}
\vartheta^0_{11} (x_1, x_2)
=
\frac{\theta (x_1) \theta (- x_2) - \theta(- x_1) \theta (x_2)}{x_1 - x_2}
=
\frac{\theta (x_1) - \theta (x_2)}{x_1 - x_2}
\, .
\end{equation}
It is defined in terms of the conventional step function $\theta (x) = \{ 1,
x \geq 0 \ ; \ 0, x < 0 \}$.

As a demonstration of how Eq.\ (\ref{ReductionThetas}) works, we give a list
of formulas for the generalized step functions that appear in the lowest order
calculations,
\begin{eqnarray}
\label{ThetaFirst}
&&
\
\vartheta^0_1 (x) = 0
\, , \\
&&
\vartheta^0_{21} (x_1,x_2)
=
\frac{x_2}{x_1-x_2}\vartheta^0_{11} (x_1,x_2)
\, , \\
&&
\vartheta^0_{22} (x_1,x_2)
=
- \frac{2x_1x_2}{(x_1-x_2)^2} \vartheta^0_{11} (x_1,x_2)
\, , \\
&&
\vartheta^1_{22} (x_1,x_2)
=
- \frac{x_1 + x_2}{(x_1 - x_2)^2} \vartheta^0_{11} (x_1,x_2)
\, , \\
&&
\vartheta^1_{21} (x_1,x_2)
=
\frac{1}{x_1-x_2}\vartheta^0_{11} (x_1,x_2)
-
\frac{1}{x_1-x_2}\vartheta^0_2 (x_1)
\, , \\
&&
\vartheta^0_{111} (x_1,x_2,x_3)
=
\frac{x_2}{x_1-x_2}\vartheta^0_{11} (x_2,x_3)
-
\frac{x_1}{x_1-x_2}\vartheta^0_{11} (x_1,x_3)
\, , \\
&&
\vartheta^1_{111} (x_1,x_2,x_3)
=
\frac{1}{x_1-x_2}\vartheta^0_{11} (x_2,x_3)
-
\frac{1}{x_1-x_2}\vartheta^0_{11} (x_1,x_3)
\, , \\
&&
\vartheta^1_{112} (x_1,x_2,x_3)
=
\frac{x_2}{(x_1 - x_2)(x_3 - x_2)}\vartheta^0_{11} (x_2, x_3)
+
\frac{x_1}{(x_1 - x_2)(x_1 - x_3)}\vartheta^0_{11} (x_1, x_3)
\, . \quad
\label{ThetaLast}
\end{eqnarray}

\section{Evolution of nonsymmetric double distributions}
\label{NonSymDDs}

Let us discuss the evolution of double distribution in nonsymmetric longitudinal
variables\footnote{These variables were denoted in the original work
\cite{Rad1996a,Rad1996b,Rad1997a} by $x$ and $y$, respectively. We have changed
the notation to $s$ and $r$ to avoid confusion with already used conventions in
Section \ref{ERBLleadingOrderEvKernels} and later.} $s$ and $r$ originally used
in Refs.\ \cite{Rad1996a,Rad1996b,Rad1997a} rather than symmetric $\alpha$ and
$\beta$ introduced in Section \ref{DDs}. These conventions will simplify somewhat
the explicit form of evolution kernels. The variables of the corresponding the
nonsymmetric double distribution $Q (s, r, \Delta^2)$ parametrize the four-momentum
of the initial parton
$$
k_1^\mu = s p_1^\mu + r \Delta^\mu
$$
in terms of the $s$-channels momentum of the incoming hadron $p_1$ and the
$t$-channel momentum transfer $\Delta$. The four-momentum of the final parton
is then $k_2^\mu = s p_1^\mu - (1 - r) \Delta^\mu$, i.e., the momentum transfer
$\Delta$ is shared by the partons in fractions $r$ and $1 - r \equiv \bar{r}$.
The relation between the parameters $s, r$ and the symmetric variables $\beta,
\alpha$ from Eq.\ (\ref{PartonMomentumDDsSymmetric}) is evidently given by
\begin{equation}
s = \beta
\, , \qquad
r = \frac{1}{2}(1 + \alpha - \beta) \ .
\end{equation}
Notice that by redefining the negative-$\beta$ part of as an antiquark distribution
(or using the $\beta$-symmetry property of the gluon distribution), the variable $s$
can always be taken positive, with the support region being $0 \leq s, r, s + r \leq 1$.
In particular, for quarks in the nucleon, the nonsymmetric DDs for the vector operator
are defined by the following representation \cite{Rad1996a}:
\begin{eqnarray}
\label{31}
\langle p_2 | \mathcal{O}^{qq} (0, z^-) | p_1 \rangle
\!\!\!&=&\!\!\!
\int_0^1 d s \, \int_0^{\bar{s}} d r
\Bigg\{
{\rm e}^{-i s z^- p_1^+ - i r z^- \Delta^+}
\left[
h^+ Q^q (s, r)
+
\frac{b}{M_N} (s p_1^+ + r \Delta^+) P^q (s, r)
\right]
\qquad\\
&&\qquad\qquad\quad \ \,
- \,
{\rm e}^{i s z^- p_1^+ - i \bar{r} z^- \Delta^+}
\left[
h^+ Q^{\bar q} (s, r)
+
\frac{b}{M_N} (s p_1^+ + r \Delta^+) P^{\bar q} (s, r)
\right]
\Bigg\}
\, , \nonumber
\end{eqnarray}
where we did not display the dependence of DDs $Q$ and $P$ on $\Delta^2$, since it
will be irrelevant throughout our discussion. At $\Delta^2 = 0$, the $r$-integral
of the combination $Q^{q, \bar{q}} (s, r) + s P^{q ,\bar{q}} (s, r)$ produces the
usual quark and antiquark densities $q (s)$ and $\bar{q} (s)$ as can be seen by
matching (\ref{31}) into Eq.\ (\ref{ForwardUnpolarizedPDF}). In a similar way, one
can introduce the nonsymmetric double distributions for gluons
\begin{eqnarray}
\langle p_2 | \mathcal{O}^{gg} (0, z^-) | p_1 \rangle
\!\!\!&=&\!\!\!
\frac{1}{2}
\int_0^1 d s \, \int_0^{\bar{s}} d r
\bigg\{
{\rm e}^{-i s z^- p_1^+ - i r z^- \Delta^+}
+
{\rm e}^{i s z^- p_1^+ - i \bar{r} z^- \Delta^+}
\bigg\}
\qquad
\\
&&\qquad\qquad\quad\times
\left[
h^+
p_1^+ \, s \, Q^g (s, r)
+
\frac{b}{M_N} (s p_1^+ + r \Delta^+)^2 P^g (s, r)
\right]
\, . \nonumber
\label{36}
\end{eqnarray}
Due to the factor $s$ included in this definition, the combination of gluon DDs
in the zero-recoil $\Delta^2 = 0$ limit $Q^g (s, r) + s P^g (s, r)$ reduces to
the usual gluon density $g (s)$ (\ref{NonpolGluonDensity}) after integration
over $r$. Analogous functions can be introduced in the odd parity sector, whose
definitions can be read from Eqs.\ (\ref{31}) and (\ref{36}) by dressing the
symbols on the right- and left-hand sides with tildes, i.e., $\mathcal{O}^{aa}
\to \widetilde{\mathcal{O}}^{aa}$ and $Q^a \to \widetilde{Q}^a$, $P^a \to
\widetilde{P}^a$, as well as changing the relative sign between the momentum
exponentials.

The use of $Q (s, r)$ DDs has some disadvantages when the symmetry properties
are discussed. In particular, the symmetry of the $f (\beta,\alpha)$ DDs with
respect to the change $\alpha \to -\alpha$ corresponds to the ``M{\" u}nchen
symmetry'' of $Q (s, r)$ with respect to the interchange $r \leftrightarrow
1 - s - r$ \cite{ManPilWei97}. However, evolution kernels for $Q (s, r)$ look
simpler than their symmetric analogues.

\subsection{Evolution kernels for nonsymmetric double distributions}
\label{SectEvolEqDDs}

The $\mu$-dependence of parton helicity-independent DDs $Q^a (s, r)$ is
governed by the evolution equation
\begin{equation}
\frac{d}{d\ln\mu^2} \,
Q^a (s_1, r_1; \mu^2)
=
\sum_b \,
\int_0^1 d s_2 \int_0^1 d r_2 \theta(1 - s_2 - r_2)
\, R^{ab;V} (s_1,r_1;s_2,r_2)
\, Q^b (s_2, r_2; \mu^2)
\, ,
\label{41}
\end{equation} where $a,b = q,g$. A similar set of equations, with the kernels
denoted by $R^{ab;A}(s_1,r_1;s_2,r_2)$ governs the evolution of the parton
helicity sensitive double distributions $\widetilde{Q}^a (s_1, r_1; \, \mu^2)$.

Integration over $r_1$ converts $Q^a(s_1, r_1; \mu^2)$ for $\Delta^2 = 0$ into
the parton distribution functions, whose evolution is described by the DGLAP
equations (\ref{42}). The kernels $R^{ab} (s_1,r_1; s_2, r_2)$ must satisfy
the reduction relation
\begin{equation}
\int_0^ {1 - s_1} d r_1 \, R^{ab} (s_1,r_1; s_2, r_2)
= \frac{1}{s_2} \, P^{ab} (s_1/s_2)
\, .
\label{43}
\end{equation}
Alternatively, integration over $s_1$ converts $Q^a (s_1,r_1; \mu^2)$ at the
point $\Delta^2 = 0$ into an object similar to a meson distribution amplitude,
so one may expect that the result of integration of $R^{ab} (s_1, r_1; s_2, r_2)$
over $s_1$ should be related to the ER-BL kernels governing the evolution of
distribution amplitudes \cite{EfrRad78,BroLep79} e.g., in case of the $qq$ kernel
\begin{equation}
\int_0^{1 - r_1} d s_1 \, R^{qq} (s_1,r_1; s_2, r_2) = V^{qq} (r_1, r_2)
\, .
\label{44}
\end{equation}
These reduction properties of the $R^{qq,V} (s_1,r_1;s_2, r_2)$ kernel can be
illustrated using its explicit one-loop form,
\begin{equation}
R^{ab} (s_1,r_1;s_2, r_2)
=
\frac{\alpha_s}{2 \pi}  R^{ab}_{(0)} (s_1,r_1;s_2, r_2)
+
\mathcal{O} (\alpha_s^2)
\, ,
\end{equation}
with \cite{Rad1996a}
\begin{eqnarray}
\label{46}
R^{qq; V}_{(0)} (s_1, r_1; s_2, r_2)
\!\!\!&=&\!\!\!
\frac{C_F}{s_2}
\Biggl \{
\frac{\theta (0 \leq s_1/s_2 \leq 1) s_1/s_2}{ (1-s_1/s_2)}
\left[
\, \frac1{r_2} \, \delta \left( \frac{s_1}{s_2} - \frac{r_1}{r_2} \right)
+
\frac{1}{\bar{r}_2} \, \delta \left( \frac{s_1}{s_2} - \frac{\bar{r}_1}{\bar{r}_2} \right)
\right]
\\
&+&\!\!\!
\theta
\left(
0 \leq \frac{s_1}{s_2} \leq {\rm min}
\left\{ \frac{r_1}{r_2}, \frac{\bar{r}_1}{\bar{r}_2} \right\}
\right)
-
\delta \left( 1 - \frac{s_1}{s_2} \right) \, \delta(r_1 - r_2)
\,
\left[
\frac12 + 2
\int_0^1 dz \, \frac{z}{\bar{z}}
\right]
\Biggr\}
. \nonumber
\end{eqnarray}
Here the last (formally divergent) term, as usual, provides the regularization
for the $1/(s_1 - s_2)$  singularities present in the kernel. This singularity
an be also written as $1/(r_2 -r_1)$  for the term containing $\delta(s_1/s_2
- r_1/r_2)$ and as $1/(r_1 - r_2)$ for the term accompanying $\delta(s_1/s_2
- \bar{r}_1/\bar{r}_2)$. In each particular case, incorporating the $1/(1 - z)$
term into a plus-type distribution, one should treat $z$ as $s_1/s_2$, $r_1/r_2$
or $\bar{r}_1/\bar{r}_2$, respectively. One can check that integrating $R^{qq;
V}_{(0)} (s_1,r_1;s_2, r_2)$ over $r_1$ or $s_1$ gives the DGLAP splitting
function (\ref{LOvectorDGLAPqq}) or the ER-BL kernel (\ref{DefLOERBLkernel})
with (\ref{LOvectorERBLkernel}).

A convenient way to get explicit expressions for $R^{ab} (s_1,r_1; s_2, r_2)$
is to extract them from the light-cone kernels ${\cal K}^{ab}_{(0)}(u, v)$
describing the evolution equations for the light-ray operators (\ref{non-local-K})).
Parametrizing nonforward matrix elements in terms of DDs, we can express $R^{ab}_{(0)}
(s_1,r_1;s_2,r_2)$ in terms of the universal ${\cal K}^{ab}_{(0)} (u, v)$ kernels.
Since the definitions of the gluon distributions $Q^g (s, r)$, $\widetilde{F}^g (s, r)$
contain an extra $p_1^+$ factor on the right-hand side, which results in the
differentiation $\partial/\partial s$ of the relevant kernel, it is convenient to
proceed in two steps. First, we introduce the auxiliary kernels $r^{ab}_{(0)} (s_1,r_1;
s_2,r_2)$ directly related by to the light-ray evolution kernels ${\cal K}^{ab}_{(0)}
(u, v)$,
\begin{eqnarray}
\label{410}
r^{ab}_{(0)} (s_1,r_1; s_2, r_2)
\!\!\!&=&\!\!\!
\int_0^1 d u \int_0^{\bar{u}} d v \, \delta (s_1 - s_2 (\bar{u} - v)) \,
\delta (r_1 - u - r_2 (\bar{u} - v)) \, {\cal K}^{ab}_{(0)} (u, v)
\\
&=&\!\!\!
\frac1{s_2} \,
{\cal K}^{ab}_{(0)}
\left(
r_1 - \frac{r_2 s_1}{s_2} , \bar r_1 - \frac{\bar{r}_2 s_1}{s_2}
\right)
\vartheta \left( \frac{r_2 s_1}{s_2} - r_1 , 0 \right)
\vartheta \left( \frac{s_1}{s_2} - 1, \frac{r_2 s_1}{s_2} - r_1 \right)
\, .
\nonumber
\end{eqnarray}
Here the spectral constraint arises as a result of the integration (\ref{RedTheta})
and expressing the appearing generalized step-functions by means of Eq.\
(\ref{RelationLipatovThetaGenericTheta}) in terms of $\vartheta (x, y)$,
Eq.\ (\ref{varThetaGeneric}).

The second step is to get the $R$-kernels using the relations
\begin{eqnarray}
R^{qq}_{(0)} (s_1, r_1; s_2, r_2)
\!\!\!&=&\!\!\!
r^{qq}_{(0)} (s_1, r_1; s_2, r_2)
\, , \\
R^{gg}_{(0)} (s_1, r_1; s_2, r_2)
\!\!\!&=&\!\!\!
\frac{s_2}{s_1} \, r^{gg}_{(0)} (s_1, r_1; s_2, r_2)
\label{78}
\, , \\
\frac{\partial}{\partial s_1}
\left[ s_1 R^{gq}_{(0)} (s_1, r_1; s_2, r_2) \right]
\!\!\!&=&\!\!\!
- r^{gq}_{(0)} (s_1, r_1; s_2, r_2)
\, , \\
R^{qg}_{(0)} (s_1, r_1; s_2, r_2)
\!\!\!&=&\!\!\! - s_2 \, \frac{\partial}{\partial s_1}
r^{qg}_{(0)} (s_1, r_1; s_2, r_2)
\, .
\label{420}
\end{eqnarray}
Hence, to obtain $R^{gq}_{(0)} (s_1, r_1; s_2, r_2)$, we should integrate $r^{gq}_{(0)}
(s_1, r_1; s_2, r_2)$ with respect to $s_1$. We fix the integration ambiguity  by the
requirement that $R^{gq}_{(0)} (s_1, r_1; s_2, r_2)$ vanishes for $s_1 > 1$.
Then
\begin{equation}
R^{gq}_{(0)} (s_1, r_1; s_2, r_2)
=
\frac{1}{s_1} \, \int_{s_1}^1 d s \, r^{gq}_{(0)} (s, r_1; s_2, r_2)
\,  \,  .
\label{421} \end{equation}
This convention guarantees a simple relation (\ref{43}) to the DGLAP kernels.

At one loop, $R^{qq; A}_{(0)} (s_1, r_1; s_2, r_2) = R^{qq; V}_{(0)} (s_1, r_1; s_2,
r_2)$, and this kernel was already displayed in Eq.\ (\ref{46}). Other kernels,
including the $R^{gg; V}_{(0)} (s_1, r_1; s_2, r_2)$ kernel originally obtained in
Ref.\ \cite{Rad1996b}, are given by
\begin{eqnarray}
\label{422dqg}
R^{qg; A}_{(0)} (s_1, r_1; s_2, r_2)
\!\!\!&=&\!\!\!
2 T_F N_f \frac{s_1}{s_2^2} \,
\Biggl\{
\delta \left( \frac{s_1}{s_2} - \frac{r_1}{r_2} \right) \, \theta(r_2 - r_1)
+
\delta \left( \frac{s_1}{s_2} - \frac{\bar{r}_1}{\bar{r}_2} \right)
\, \theta(r_1 - r_2)
\\
&&\qquad\qquad\qquad\qquad\qquad\qquad\
-
\frac{s_2}{s_1}
\theta \left( 0 \leq \frac{s_1}{s_2}
\leq
{\rm min} \left\{ \frac{r_1}{r_2}, \frac{\bar{r}_1}{\bar{r}_2}  \right\}
\right)
\Biggr\}
\, , \nonumber\\
\label{422dgq}
R^{gq; A}_{(0)} (s_1, r_1; s_2, r_2)
\!\!\!&=&\!\!\!
2 C_F \frac{1}{s_1} \,
\Biggl\{
\left( \frac{s_1}{s_2} - \frac{r_1}{r_2} \right)
\, \theta \left( \frac{s_1}{s_2} \leq \frac{r_1}{r_2} \leq 1 \right)
+
\left( \frac{s_1}{s_2} - \frac{\bar{r}_1}{\bar{r}_2} \right)
\, \theta \left( \frac{s_1}{s_2} \leq \frac{\bar{r}_1}{\bar{r}_2} \leq 1 \right)
\nonumber\\
&&\qquad\qquad\qquad\qquad\qquad\qquad\
+ \frac{1}{2} \delta(r_2 - r_1) \, \theta (0 \leq s_1 \leq  s_2 )
\Biggr\}
\, , \\
\label{422}
R^{gg; A}_{(0)} (s_1, r_1; s_2, r_2)
\!\!\!&=&\!\!\!
\frac{C_A}{s_2}
\,
\Biggl\{
\frac{\, \theta (0 \leq s_1/s_2 \leq 1) s_1/s_2}{(1 - s_1/s_2)}
\left[
\frac{1}{r_2} \, \delta \left( \frac{s_1}{s_2} - \frac{r_1}{r_2} \right)
+
\frac{1}{\bar{r}_2} \, \delta \left( \frac{s_1}{s_2} - \frac{\bar{r}_1}{\bar{r}_2} \right)
\right]
\\
&+&\!\!\!
4
\theta
\left(
0 \leq \frac{s_1}{s_2}
\leq
{\rm min} \left\{ \frac{r_1}{r_2}, \frac{\bar{r}_1}{\bar{r}_2} \right\}
\right)
-
\delta \left(1 - \frac{s_1}{s_2} \right) \delta(r_1 - r_2)
\left[
\frac{\beta_0}{2 C_A} \,
+
2 \int_0^1 \frac{d z}{\bar{z}}
\right]
\, \Biggr\}
, \nonumber
\end{eqnarray}
in the odd parity sector, and
\begin{eqnarray}
R^{qg; V}_{(0)} (s_1, r_1; s_2, r_2)
\!\!\!&=&\!\!\!
R^{qg; A}_{(0)} (s_1, r_1; s_2, r_2)
\label{422qg} \\
&+&\!\!\!
8 T_F N_f \frac{r_2 \bar{r}_2}{s_2}
\Biggl(
\frac{r_1}{r_2} + \frac{\bar{r}_1}{\bar{r}_2} - 2 \, \frac{s_1}{s_2}
\Biggr)
\,
\theta
\left(
0 \leq \frac{s_1}{s_2} \leq {\rm min}
\left\{ \frac{r_1}{r_2}, \frac{\bar{r}_1}{\bar{r}_2} \right\}
\right)
\, , \nonumber\\
R^{gq, V}_{(0)} (s_1, r_1; s_2, r_2)
\!\!\!&=&\!\!\! 2 C_F \frac{1}{s_1} \,
\biggl\{
\left( \frac{r_1}{r_2} - \frac{s_1}{ s_2} \right)
\, \theta \left( \frac{s_1}{s_2} \leq \frac{r_1}{r_2} \leq 1 \right)
+
\left( \frac{\bar{r}_1}{\bar{r}_2} - \frac{s_1}{ s_2} \right)
\, \theta \left( \frac{s_1}{s_2} \leq \frac{\bar{r}_1}{\bar{r}_2} \leq 1 \right)
\label{422gq}
\nonumber\\
&&\qquad\qquad\qquad\qquad\qquad\qquad\
+
\frac{1}{2}
\delta(r_2 - r_1) \, \theta (0 \leq s_1 \leq s_2 )  \biggr\}
\, , \\
\label{422gg}
R^{gg; V}_{(0)} (s_1, r_1; s_2, r_2)
\!\!\!&=&\!\!\!
R^{gg, A}_{(0)}(s_1, r_1; s_2, r_2)  \\
&+&\!\!\! 12 C_A \frac{1}{s_1}
\,
\left( r_1 - \frac{r_2 s_1}{s_2} \right)
\left( \bar{r}_1 - \frac{\bar{r}_2 s_1}{s_2} \right)
\,
\theta
\left(
0 \leq \frac{s_1}{s_2} \leq {\rm min}
\left\{ \frac{r_1}{r_2}, \frac{\bar{r}_1}{\bar{r}_2} \right\}
\right)
\, , \nonumber
\end{eqnarray}
for the even parity case. The equations (\ref{41}) with above evolution kernels
can be also solved analytically, with the result for the $s_1$-moments of DDs
given by an expansion in the Gegenbauer polynomials as we will explain in Section
\ref{subsoperatorsol}.

\subsection{Solution of evolution equation for double distributions}

Having discussed the evolution of GPDs, let us turn to the analogous
analysis of scale dependence of DDs, limiting our consideration however
to leading order in coupling constant. The analytic solution to the
evolution equations for double distributions (\ref{41}) introduced in
Section \ref{SectEvolEqDDs} was obtained in Refs.\ \cite{Rad1996a,Rad1996b}
by combining standard methods of solving evolution equations for parton
densities and distribution amplitudes. First, one should take moments
with respect to $s$. Utilizing the rescaling property $R^{ab} (s_1, r_1;
s_2, r_2) = R^{ab} (s_1/s_2, r_1; 1, r_2)/s_2$, one obtains
\begin{equation}
\frac{d}{d \ln\mu^2} Q_n^a (r_1; \mu^2)
= \sum_{b} \int_0^1 d r_2 \, R_n^{ab; V} (r_1, r_2) \, Q_n^b (r_2; \mu^2)
\, ,
\label{eq:fnev}
\end{equation}
where $Q_n^a (r; \mu^2)$ is the $n$-th $s$-moment of $Q^a (s, r; \mu^2)$
\begin{equation}
Q_n^a (r; \mu^2) = \int_0^{1} d s \, s^n  Q^a (s, r; \mu^2) \, .
\label{eq:fnmom}
\end{equation}
The one-loop kernels $R_{(0)n}^{ab; V} (r_1, r_2)$ and analogous kernels
$R_{(0)n}^{ab; A} (r_1, r_2)$ governing the evolution of parton helicity
dependent double distributions $\widetilde{Q}_n^a(y; \mu^2)$ are given by
\cite{BluGeyRob97,Radyushkin:1998es}
\begin{eqnarray}
\label{eq:rnkernelqq}
R^{qq; A}_{(0)n} (r_1, r_2)
\!\!\!&=&\!\!\!
C_F
\left\{
\left( \frac{r_1}{r_2} \right)^{n+1}
\left[ \frac{1}{n + 1} + \frac{1}{r_2 - r_1} \right]
\theta(r_2 - r_1)  \right.
\\
&+&\!\!\!
\left( \frac{\bar r_1}{\bar r_2} \right)^{n+1}
\left[ \frac1{n + 1} + \frac1{r_1 - r_2} \right]
\theta(r_1 - r_2)
\left.
- \delta(r_1 - r_2)
\left[ \frac{1}{2} + 2 \int_0^1 dz \, \frac{z}{\bar{z}} \right]
\right\}
, \nonumber\\
\label{56}
R_{(0)n}^{qg; A} (r_1, r_2)
\!\!\!&=&\!\!\!
2 T_F N_f \,
\frac{n}{n+1} \,
\left\{
\left( \frac{r_1}{r_2} \right)^{n+1} \theta(r_2 - r_1)
+
\left( \frac{\bar{r}_1}{\bar{r}_2} \right)^{n + 1} \theta(r_1 - r_2)
\right\}
, \\
\label{58}
R^{gq; A}_{(0)n} (r_1, r_2)
\!\!\!&=&\!\!\!
C_F
\frac{1}{n}
\left\{
\delta(r_1 - r_2)
-
\frac{2}{n + 1}
\left[
\left( \frac{r_1}{r_2} \right )^{n + 1} \theta(r_2 - r_1)
+
\left( \frac{\bar{r}_1}{\bar{r}_2} \right)^{n + 1} \theta(r_1 - r_2) \right]
\right\}
\, , \qquad\quad\\
\label{eq:rnkernelDgg}
R_{(0)n}^{gg; A} (r_1, r_2)
\!\!\!&=&\!\!\!
C_A
\left\{
\left( \frac{r_1}{r_2} \right)^{n+1}
\left[ \frac{4}{n + 1}
+
\frac{1}{r_2 - r_1} \right] \theta(r_2 - r_1)
\right.
\\
&+&\!\!\!
\left.
\left( \frac{\bar{r}_1}{\bar{r}_2} \right)^{n+1}
\left[ \frac{4}{n + 1}
+
\frac{1}{r_1 - r_2} \right] \theta(r_1 - r_2)
-
\delta(r_1 - r_2)
\left [\frac{\beta_0}{2 C_A} + 2 \int_0^1 \frac{dz}{\bar{z}} \,
\right]
\right\}
, \nonumber
 \end{eqnarray}
for odd parity sector, and
\begin{eqnarray}
R^{qq; V}_{(0)n} (r_1, r_2) \!\!\!&=&\!\!\! R^{qq, A}_{(0)n} (r_1,r_2)
\, , \\
\label{57}
R_{(0)n}^{qg; V} (r_1, r_2)
\!\!\!&=&\!\!\!
R_{(0)n}^{qg; A} (r_1, r_2)
+
8 T_F N_f \, \frac{n}{n+1}
\left\{
\left( \frac{r_1}{r_2} \right)^{n+1}
\left[ \frac{r_2 \bar{r}_1}{n} - \frac{r_1 \bar{r}_2} {n + 2} \right]
\theta(r_2 - r_1)
\right.
\nonumber\\
&+&\!\!\!
\left.
\left( \frac{\bar{r}_1}{\bar{r}_2} \right)^{n+1}
\left[ \frac{\bar{r}_2 r_1}{n} - \frac{\bar{r}_1 r_2} {n + 2} \right]
\theta(r_1 - r_2)
\right\}
\\
\label{59}
R_{(0)n}^{gq; V} (r_1, r_2)
\!\!\!&=&\!\!\!
C_F
\frac{1}{n}
\left\{
\delta(r_1 - r_2)
+
\frac{2}{n + 1}
\left[
\left( \frac{r_1}{r_2} \right)^{n + 1} \theta(r_2 - r_1)
+
\left( \frac{\bar{r}_1}{\bar{r}_2} \right)^{n + 1} \theta(r_1 - r_2) \right]
\right\}
, \qquad\quad \\
R_{(0)n}^{gg; V} (r_1, r_2)
\!\!\!&=&\!\!\!
R_{(0)n}^{gg; A} (r_1, r_2)
+
12 C_A \frac{1}{n + 1}
\left\{
\left( \frac{r_1}{r_2} \right)^{n + 1}
\left[ \frac{r_2 \bar{r}_1}{n} - \frac{r_1 \bar{r}_2}{n + 2} \right]
\theta(r_2 - r_1)
\right.
\nonumber\\
&+&\!\!\!
\left.
\left( \frac{\bar{r}_1}{\bar{r}_2} \right)^{n + 1}
\left[ \frac{\bar{r}_2 r_1}{n} - \frac{\bar{r}_1 r_2}{n + 2} \right]
\theta(r_1 - r_2)
\right\}
,
\label{eq:rnkernelgg}
 \end{eqnarray}
for even parity double distributions.

From Eqs.\ (\ref{56}), (\ref{58}) and (\ref{57}), (\ref{59}) one can derive the
following reduction formulas for the nondiagonal kernels:
\begin{eqnarray}
\frac{\partial}{\partial r_1} R_{(0)1}^{qg} (r_1,r_2)
\!\!\!&=&\!\!\!
- V^{qg}_{(0)} (r_1, r_2)
\, , \\
\lim_{n \to 0} n R_{(0)n}^{gq} (r_1, r_2)
\!\!\!&=&\!\!\!
-
\frac{\partial}{\partial r_1} V^{gq}_{(0)} (r_1, r_2)
\, ,
\end{eqnarray}
which hold for both even and odd parity sectors. To understand the structure of
these relations, one should realize that constructing nondiagonal $qg$ and $gq$
kernels, one faces mismatching $p_1^+$ factors which are converted in the ER-BL
limit into derivatives with respect to $r_1$.

It is straightforward to check that all the kernels $R_{(0)n}^{ab; V} (r_1,r_2)$
and $R_{(0)n}^{ab; A} (r_1, r_2)$ obey the property
$$
w_n (r_2) R^{ab}_{(0)n} (r_1,r_2) =  w_n (r_1) R^{ab}_{(0)n} (r_2,r_1)
\, ,
$$
where $w_n (r)= (r \bar{r})^{n + 1}$. Hence, the eigenfunctions of the evolution
equations are orthogonal with the weight $w_n (r)= (r \bar{r})^{n + 1}$, i.e., they
are proportional to the Gegenbauer polynomials $C^{n+3/2}_k (r - \bar{r})$, see Refs.\
\cite{BroLep79,MikRad85} and also the papers \cite{Cha80,ShiVys81,Ohr81,BaiGro81},
where the general algorithm was applied to the evolution of flavor-singlet distribution
amplitudes. Expanding  the moment functions  $Q_n^a (r; \mu^2)$ over the Gegenbauer
polynomials $C^{n+3/2}_k (r - \bar{r})$
\begin{equation}
Q_n^{a} (r; \mu^2) = (r \bar{r})^{n + 1}
\sum_{k=0}^{\infty} C^{n + 3/2}_k (r - \bar{r}) \, Q_{nk}^a(\mu)
\label{eq:rneq}
\end{equation}
we get the one-loop evolution equation for the expansion coefficients
\begin{equation}
\frac{d}{d \ln\mu^2} Q_{nk}^a (\mu^2)
=
\frac{\alpha_s}{2\pi} \sum_{b} {\mit\Gamma}_{(0)nk}^{ab; V} \, Q_{nk}^b (\mu^2)
\, ,
\label{a27}
\end{equation}
where ${\mit\Gamma}_{(0)nk}^{ab; V}$ are the eigenvalues of the kernels $R_{(0)}^{ab; V}
(r_1, r_2)$ related to the elements $\gamma^{ab; V}_{(0) j}$ of the usual flavor-singlet
anomalous dimension matrix given in Eq.\ (\ref{even-anomalous-dimensions})
\begin{equation}
{\mit\Gamma}_{(0)nk}^{qq; V}
=
- \frac{1}{2} \gamma_{(0)n+k}^{qq,V}
\, , \quad
{\mit\Gamma}_{(0)nk}^{qg; V} = - \frac{n}{2}  \gamma_{(0)n+k}^{qg,V}
\,  , \quad
{\mit\Gamma}_{(0)nk}^{gq; V} = - \frac{1}{2 n} \gamma_{(0)n+k}^{gq,V}
\, ,  \quad
{\mit\Gamma}_{(0)nk}^{gg; V} = - \frac{1}{2} \gamma_{(0)n+k}^{gg,V}
\, ;
\end{equation}
and similarly for the helicity-sensitive quantities ${\mit\Gamma}_{(0)nk}^{ab; A}$,
which are expressed through  $\gamma^{ab; A}_{(0)n + k}$ of Eqs.\
(\ref{odd-anomalous-dimensions}) -- (\ref{odd-anomalous-dimensions-last}).

In the approach outlined above, the outcome that the index $j$ of the anomalous dimensions
$\gamma_{(0)j}^{ab}$ is given just by the sum $n + k$ looks like an unexpected miracle.
But it can be easily explained within the framework of the lightcone operators
\cite{BalBra89}, where the solution of the evolution equation is known in the operator
form, with the $\mu$-dependence of multiplicatively renormalizable operators ${\mathbb O}_j$
governed by a single anomalous dimension $\gamma_j$. In subsection \ref{subsoperatorsol},
we discuss how this evolution on the operator level translates into the evolution of double
distributions.

\subsection{Asymptotic shape of double distributions}

It is instructive to consider first two simplified situations for evolution of
double distributions. In the quark nonsinglet case, the evolution is governed
by $\gamma_{n + k}^{qq}$ alone and the solution to Eq.\ (\ref{a27}) can be
written in the form (\ref{eq:rneq}) as
\begin{equation}
Q_n^{q, \rm\scriptscriptstyle NS} (r; \mu^2)
=
(r \bar{r})^{n+1} \sum_{k = 0}^{\infty}
C^{n + 3/2}_k (r - \bar{r}) \,
Q_{nk}^{q, \rm\scriptscriptstyle NS} (\mu_0^2)
\left(
\frac{\alpha_s (\mu_0^2)}{\alpha_s (\mu^2)}
\right)^{\gamma_{(0)n + k}^{qq; V}/\beta_0}
\, ,
\label{eq:fnnons}
\end{equation}
Since $\gamma_{(0)0}^{qq; V} = 0$ while all the anomalous dimensions $\gamma_{(0)j}^{qq; V}$
with $j \geq 1$ are positive, only $Q_0^{\rm\scriptscriptstyle NS} (r; \mu^2)$ survives in
the asymptotic limit $\mu^2 \to \infty$ while all the moments $Q_n^{q, \rm\scriptscriptstyle NS}
(r; \mu^2)$ with $n \geq 1$ evolve to zero values. Hence, in the formal $\mu^2 \to \infty$
limit, we have
\begin{equation}
Q^{q, \rm\scriptscriptstyle NS} (s, r; \mu^2 \to \infty)
\sim \delta(s) \, r \bar{r}
\, ,
\end{equation}
i.e., in each of its variables, the limiting function $Q^{\rm\scriptscriptstyle NS}
(s, r; \mu^2 \to \infty)$ acquires the characteristic asymptotic form dictated by the
nature of the variable: $\delta (s)$ is the limiting form of parton densities
\cite{GroWil73a,GeoPol74}, while the $r \bar{r}$-form is the asymptotic shape for the
lowest-twist two-body distribution amplitudes \cite{EfrRad78,BroLep79}.

Another example is the evolution of the gluon distribution in pure gluodynamics.
It is governed by $\gamma_{(0)n+k}^{gg;V}$ with $\beta_0 = - 11 N_c/3$. Note that
the lowest local operator in this case corresponds to $n = 1$. Furthermore, in pure
gluodynamics, $\gamma_{(0)1}^{gg;V}$ vanishes while $\gamma_{(0)j}^{gg;V} > 0$ if
$j \geq 2$. This means that in the $\mu^2 \to \infty$ limit we have
\begin{equation}
s \, Q^g (s, r; \mu^2 \to \infty) \sim \delta (s) \, (r \bar{r})^2
\, .
\end{equation}

Finally, in QCD one should take into account the effects due to the quark-gluon
mixing. Performing the analysis identical to the one in Section \ref{SectAsyGPDSing},
we find
\begin{equation}
\sum_q Q_{10}^q (\mu^2 \to \infty) \to \frac{N_f}{4 C_F + N_f}
\, , \qquad
Q_{10}^g (\mu^2 \to \infty) \to \frac{4C_F}{4 C_F + N_f}
\, .
\end{equation}
Since all the combinations of moments $Q_{nk}^a$ with $n + k \geq 2$ vanish in the
$\mu^2 \to \infty$ limit, we obtain
\begin{eqnarray}
s \, \sum_q Q^q (s, r ; \mu^2 \to \infty)
\!\!\!&\to&\!\!\!
30 \, \frac{N_f}{4 C_F + N_f} \, \delta(s) \, (r \bar{r})^2
\, , \\
s \, Q^g (s, r; \mu^2 \to \infty)
\!\!\!&\to&\!\!\!
30 \, \frac{4C_F}{4 C_F + N_f} \, \delta(s) \, (r \bar{r})^2
\, .
\end{eqnarray}
The former equation here can be rewritten in the form
\begin{equation}
\sum_q Q^q (s, r; \mu^2 \to \infty) \to
- 30 \, \frac{N_f}{4 C_F + N_f} \, \delta'(s)\, (r \bar{r})^2
\, .
\end{equation}
When reduced to GPDs, the asymptotic solutions found here reproduces the
results established in Sections \ref{SectDiffEigen} and \ref{SectAsyGPDSing}.

\section{Elements of the conformal group}
\label{ConformalSymAppendix}

\begin{figure}[t]
\begin{center}
\hspace{0cm} \mbox{
\begin{picture}(0,150)(185,0)
\put(0,0){\insertfig{5.25}{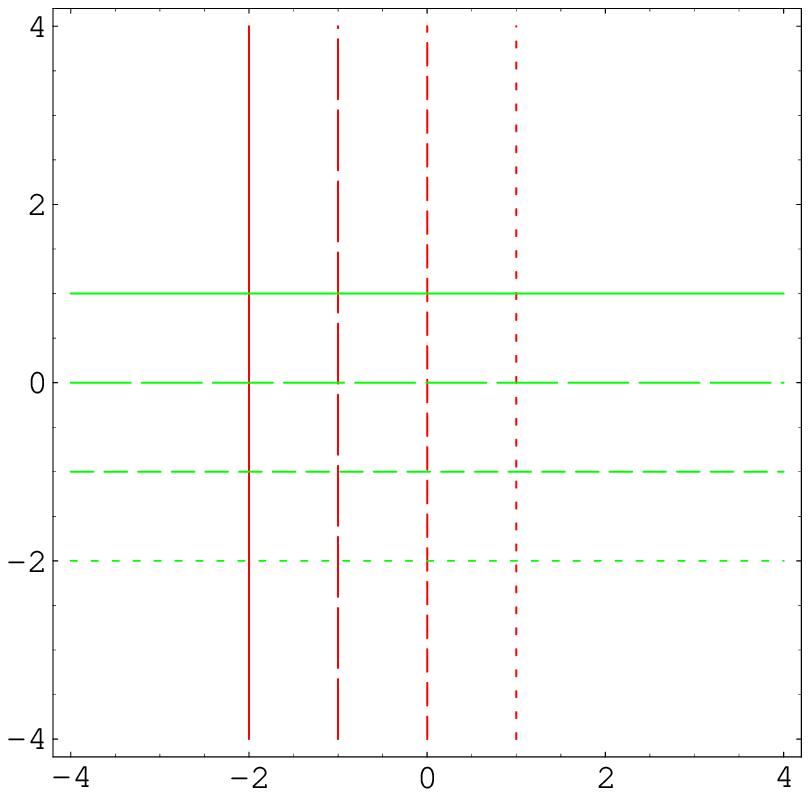}}
\put(200,-7){\insertfig{5.7}{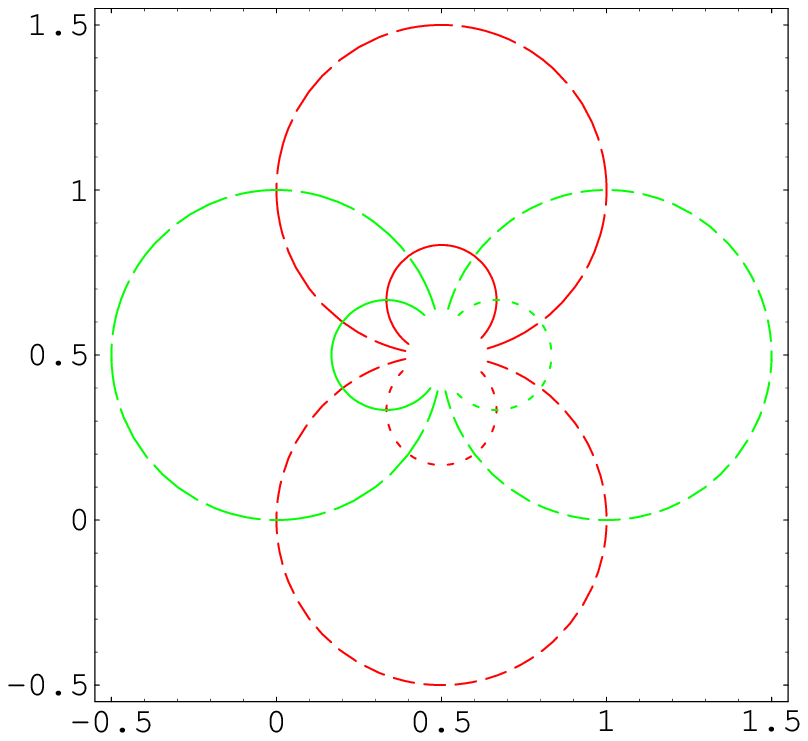}}
\put(75,-15){$x$}
\put(-10,75){$y$}
\put(273,-15){$x' (x, y)$}
\put(165,75){$y' (x, y)$}
\end{picture}
}
\end{center}
\caption{\label{ConformalTrans} Visualization of the finite conformal boost
(\ref{FiniteConformalBoost}) for $t = z = 0$ and transformation parameter
$d^\mu = (0, -1, -1, 0)$. The angles are preserved under the transformation.
}
\end{figure}

The conformal group is the maximal extension of the Poincar\'e group that leaves
the light cone invariant. It is the symmetry of the classical QCD Lagrangian and
is thus of physical interest for light-cone-dominated as well as high-energy
processes, where the asymptotic freedom of QCD implies that its approximation by
a non-interacting theory is good starting point.

Conformal transformations are coordinate transformation of the metric of
the $d$-dimensional space of the form
\begin{equation}
g^{\prime \, \mu\nu} (z') = {\mit\Omega} (z) g^{\mu\nu} (z)
\, .
\end{equation}
Let us consider the infinitesimal transformation,
\begin{equation}
{\mit\Omega} (z) \approx 1 + \omega (z) \, ,
\end{equation}
and count the number of independent parameters involved. The variation of the
metric at different space-time points is (cf.\ Eq.\ (\ref{MetricVariation}))
\begin{equation}
\label{MetricVariation2}
\delta g^{\mu\nu} (z) = g^{\prime \, \mu\nu} (z') - g^{\mu\nu} (z)
=
\omega (z) g^{\mu\nu} (z)
=
\partial^\mu \epsilon^\nu (z) + \partial^\nu \epsilon^\mu (z)
\, ,
\end{equation}
with $\delta z^\mu \equiv \epsilon^\mu (z)$. Contracting both sides with
the inverse metric tensor $[ g^{\mu\nu} ]^{- 1} = g_{\mu\nu}$, we get
\begin{equation}
\omega (z) = \frac{2}{d} \partial \cdot \epsilon (z)
\, .
\end{equation}
Substituting it back into (\ref{MetricVariation2}) we obtain the Killing
equation
\begin{equation}
\label{Killing}
\partial^\mu \epsilon^\nu (z)
-
\partial^\mu \epsilon^\nu (z)
-
\frac{2}{d}
g^{\mu\nu} (z) \partial \cdot \epsilon (z) = 0 \, .
\end{equation}
Differentiating this expression with respect to $z^\mu$ and $z^\nu$, i.e.,
applying $\partial^\mu$ and $\partial^\nu$, respectively, and summing up
the results, we use the Killing equation (\ref{Killing}) to find
\begin{equation}
\left\{
(d - 2) \partial^\mu \partial^\nu + g^{\mu\nu} (z) \partial^2
\right\}
\partial \cdot \epsilon (z) = 0 \, .
\end{equation}
Here we have assumed that $\partial \cdot \epsilon (z)$ is nonsingular. We
will not consider the case of $d = 2$ which leads to an infinite-dimensional
algebra. For other $d \neq 2$, the solution for the Killing vector reads
\begin{equation}
\label{GeneralSolutionKilling}
\epsilon^\mu (z)
=
d^\mu z^2 + (c \, \delta^{\mu\nu} + b^{\mu\nu}) z_\nu + a^\mu
\, .
\end{equation}
These infinitesimal transformations can be easily integrated to produce
finite transformations. As a result, one finds from the form of the general
solution (\ref{GeneralSolutionKilling}) that the well-known Poincar\'e
transformations, are augmented by dilatations
$$
z^\mu \stackrel{D}{\to} z^{\prime \mu} = {\rm e}^{c} z^\mu \  ,
$$
and special conformal boosts
\begin{equation}
\label{FiniteConformalBoost}
z^\mu
\stackrel{K}{\to}
z^{\prime \mu} = \frac{z^\mu + d^\mu z^2}{1 + 2 d \cdot z + d^2 z^2}
\, .
\end{equation}
The latter are composed of the sequence of an inversion $z^\mu \to z^\mu/z^2$,
a translation $z^\mu \to z^\mu + d^\mu$, and a further inversion. The finite
conformal boost is visualized in Fig.\ \ref{ConformalTrans}. It clearly
demonstrates the angle-preserving nature of the transformation: all angles
remain rectangular before and after the transformation is applied.

The conformal factor $[1 + 2 d \cdot z + d^2 z^2]^{-1}$ is singular on the
light-cone cone $z^2 = 0$. As a result,  the special conformal transformations
are not well defined as global transformations in the Minkowski space-time.
Moreover, it is possible to transform non-causally connected regions into one
another, which violates fundamental principles of nature. To apply the conformal
group to the quantum field theory in Minkowski space-time, it is sufficient in
our studies to restrict to infinitesimal special conformal transformations:
this eliminates both of the aforementioned problems.

\subsection{Conformal algebra}

The conformal transformations form a group. In four space-time dimensions
it is a fifteen-parameter group $SO(4,2)$. Below, we summarize specific
types of transformations and display the change of the space-time coordinates
which they induce:
\begin{center}
\begin{tabular}{lll}
$\bullet\ $ Lorentz transformation  & \quad ${\mathbb{M}}^{\mu\nu}$: & \quad
$\delta^M z^\mu = \omega^{\mu\nu} z_\nu$ \\
$\bullet\ $ Translation             & \quad ${\mathbb{P}}^\mu$:      & \quad
$\delta^P z^\mu = a^\mu$ \\
$\bullet\ $ Dilatation              & \quad ${\mathbb{D}}$:          & \quad
$\delta^D z^\mu = \lambda z^\mu$ \\
$\bullet\ $ Conformal boost         & \quad ${\mathbb{K}}^\mu$:      & \quad
$\delta^K z^\mu = c_\nu (2 z^\mu z^\nu - z^2 g^{\mu\nu})$ \\
\end{tabular}
\end{center}
The corresponding generators act in the Hilbert space of field operators and obey
the commutator algebra:
\begin{eqnarray}
\label{SO41ConformalAlgebra}
\begin{array}{ll}
{}i [{\mathbb{M}}^{\mu\nu} , {\mathbb{M}}^{\rho\sigma}]
&\!\!=
g^{\mu\rho} {\mathbb{M}}^{\nu\sigma}
+
g^{\nu\sigma} {\mathbb{M}}^{\mu\rho}
-
g^{\mu\sigma} {\mathbb{M}}^{\nu\rho}
-
g^{\nu\rho} {\mathbb{M}}^{\mu\sigma}
\, , \\
{}i [{\mathbb{M}}^{\mu\nu} , {\mathbb{P}}^\rho]
&\!\!=
g^{\mu\rho} {\mathbb{P}}^\nu - g^{\nu\rho} {\mathbb{P}}^\mu
\, , \\
{}i [{\mathbb{M}}^{\mu\nu} , {\mathbb{K}}^\rho]
&\!\!=
g^{\mu\rho} {\mathbb{K}}^\nu - g^{\nu\rho} {\mathbb{K}}^\mu
\, , \\
{}i [{\mathbb{P}}^\mu , {\mathbb{D}} ]
&\!\!=
- {\mathbb{P}}^\mu
\, , \\
{}i [{\mathbb{P}}^\mu , {\mathbb{K}}^\nu ]
&\!\!=
- 2 g^{\mu\nu} {\mathbb{D}}
+ 2 {\mathbb{M}}^{\mu\nu}
\, , \\
{}i [ {\mathbb{D}} , {\mathbb{K}}^\mu ]
&\!\!=
- {\mathbb{K}}^\mu
\, .
\end{array}
\end{eqnarray}
The remaining commutators vanish.

\subsection{Induced representations}

The infinitesimal conformal variation of a quantum field ${\mit\Phi} (x)$ is
determined by its commutator
\begin{equation}
\label{DefInfinitsimalVariation}
\delta^G {\mit\Phi} (z)
=
i [ {\mit\Phi} (z), {\mathbb{G}} ]
\, ,
\end{equation}
with one of the generators from the conformal algebra ${\mathbb{G}} = {\mathbb{P}}^\mu ,
{\mathbb{M}}^{\mu\nu} , \dots \ $. We are interested in the representation of these
generators, acting on the Hilbert space, in terms of differential operators,
\begin{equation}
\label{RepresentationSO42}
[ {\mathbb{G}} , {\mit\Phi} (z) ] \equiv \widehat{G} {\mit\Phi} (z)
\, .
\end{equation}
The differential representation of these generators can be easily derived with the
help of the standard technique of induced representations \cite{MacSal69} making use
of the coset parametrization of the Minkowski space
\begin{equation}
\label{MinkowskiCoset}
{\mit\Phi} (z)
=
{\rm e}^{i z \cdot {\mathbb{P}}}
{\mit\Phi} (0)
{\rm e}^{- i z \cdot {\mathbb{P}}}
\, .
\end{equation}

Let us consider the stability subgroup of the conformal group, i.e.,
transformations which leave the plane $z^\mu = 0$ unaffected. This
small group contains the generators $\{ {\mathbb{M}}^{\mu\nu}, {\mathbb{D}},
{\mathbb{K}}^\mu\}$,
\begin{eqnarray}
\label{StableDilatation}
{}[{\mathbb{M}}^{\mu\nu} , {\mit\Phi} (0)]
=
- i {\mit\Sigma}^{\mu\nu} {\mit\Phi} (0)
\, , \qquad
{}[{\mathbb{D}} , {\mit\Phi} (0))]
=
- i {\mit\Delta} {\mit\Phi} (0)
\, , \qquad
{}[ {\mathbb{K}}^\mu , {\mit\Phi} (0) ]
=
\kappa^\mu {\mit\Phi} (0)
\, .
\end{eqnarray}
The representation matrices ${\mit\Sigma}_{\mu\nu}$, ${\mit\Delta}$ and $\kappa_\mu$,
obey the algebra
\begin{equation}
\label{SmallRepressentation}
[ {\mit\Delta} , \kappa^\mu ] = \kappa^\mu
\, , \qquad
[ {\mit\Sigma}^{\mu\nu} , {\mit\Delta} ]_- = 0
\, , \qquad
[ {\mit\Delta} , {\mit\Delta} ] = 0
\, .
\end{equation}
We take ${\mit\Sigma}_{\mu\nu}$ that forms an irreducible representation of the
homogeneous Lorentz group, acting  on the fields as follows
\begin{equation}
{\mit\Sigma}^{\mu\nu} \psi
=
\ft{1}4 [\gamma^\mu , \gamma^\nu] \psi
\, , \qquad
{\mit\Sigma}^{\mu\nu} A^\rho
=
g^{\mu\rho} A^\nu - g^{\nu\rho} A^\sigma
\, , \qquad
{\mit\Sigma}^{\mu\nu} F^{\rho\sigma} = g^{\mu\rho} F^{\nu\sigma} -
g^{\nu\rho} F^{\mu\sigma} - (\rho \leftrightarrow \sigma)
\, .
\end{equation}
Since ${\mit\Delta}$ commutes with the generators of the small group, by the
Shur's lemma it must be multiple of unity,
\begin{equation}
{\mit\Delta} = d \cdot \1_{[n] \times [n]} \, ,
\end{equation}
with $\1_{[n] \times [n]}$ being an $[n] \times [n]$ unity matrix acting on the
$n$-dimensional vector ${\mit\Phi}$, which is a number for the scalar field, and
a four-component column for the Dirac fermion. The value of $d$  gives the mass
dimensions of fields
\begin{equation}
d_g = 1 \, , \qquad d_q = \ft32
\, .
\end{equation}
Then from (\ref{SmallRepressentation}) we have
\begin{equation}
\kappa^\mu = 0 \, .
\end{equation}

Our  goal now is to find the representation of the generators of scaling
${\mathbb{D}}$ and special conformal ${\mathbb{K}}^\mu$ transformations in
the basis of field operators depending on the space-time coordinate $z^\mu$.
Acting with ${\rm e}^{i z \cdot {\mathbb{P}}}$ and ${\rm e}^{- i z \cdot
{\mathbb{P}}}$ from the left and from the right on (\ref{StableDilatation}),
we get
\begin{equation}
\label{IntermedateRepresentation}
\left[ {\mit\Phi} (z) ,
{\rm e}^{i z \cdot {\mathbb{P}}}
{\mathbb{D}} \,
{\rm e}^{- i z \cdot {\mathbb{P}}}
\right]
= i d {\mit\Phi} (z) \, .
\end{equation}
A little algebra leads to the result
\begin{eqnarray}
{\rm e}^{i z \cdot {\mathbb{P}}} {\mathbb{D}} \, {\rm e}^{- i z \cdot {\mathbb{P}}}
\!\!\!&=&\!\!\!
\sum_{n = 0}^{\infty} \frac{i^n}{n!}
z_{\mu_1} \dots z_{\mu_n}
\left[ {\mathbb{P}}^{\mu_1} ,
\left[ \dots ,
\left[ {\mathbb{P}}^{\mu_n} , {\mathbb{D}} \right]
\right]
\right]
\nonumber\\
\!\!\!&=&\!\!\!
{\mathbb{D}} + i z_\mu \left[ {\mathbb{P}}^\mu , {\mathbb{D}} \right]
=
{\mathbb{D}} - z \cdot {\mathbb{P}}
\, ,
\end{eqnarray}
where only one commutator has survived in the sum, while all the other have
vanished by the use of the conformal algebra. Substituting this back into Eq.\
(\ref{IntermedateRepresentation}), we get finally the representation of the
dilatation in the basis of field operators
\begin{equation}
\left[ {\mit\Phi} (z) , {\mathbb{D}} \right]
= i \left( d + z \cdot \partial \right) {\mit\Phi} (z)
\, .
\end{equation}
Obviously, $d$ is the canonical dimension of the field.

In a similar way, we find the action of the special conformal boosts
\begin{equation}
\left[ {\mit\Phi} (z) ,
{\rm e}^{i z \cdot {\mathbb{P}}}
{\mathbb{K}}^\mu \,
{\rm e}^{- i z \cdot {\mathbb{P}}}
\right]
= 0
\, .
\end{equation}
With the first three terms surviving in the series, we can write
\begin{eqnarray}
{\rm e}^{i z \cdot {\mathbb{P}}}
{\mathbb{K}}^\mu \,
{\rm e}^{- i z \cdot {\mathbb{P}}}
\!\!\!&=&\!\!\!
\sum_{n = 0}^{\infty} \frac{i^n}{n!}
z_{\mu_1} \dots z_{\mu_n}
\left[
{\mathbb{P}}^{\mu_1} ,
\left[ \dots ,
\left[
{\mathbb{P}}^{\mu_n} , {\mathbb{K}}^\mu
\right]
\right]
\right]
\nonumber\\
\!\!\!&=&\!\!\!
{\mathbb{K}}^\mu
-
2 z_\nu \left( g^{\mu\nu} {\mathbb{D}} + {\mathbb{M}}^{\mu\nu} \right)
+
\left( 2 z^\mu z^\nu - z^2 g^{\mu\nu} \right) {\mathbb{P}}_\nu
\, .
\end{eqnarray}
Making use of the representation of $\mathbb{P}^\mu$, $\mathbb{M}^{\mu\nu}$ and
$\mathbb{D}$, we finally find
\begin{equation}
\left[ {\mit\Phi} (z) , {\mathbb{K}}^\mu \right]
=
i \left\{
( 2 z^\mu z^\nu - z^2 g^{\mu\nu} ) \partial_\nu
+
2 d z^\mu + 2 {\mit\Sigma}^{\mu\nu} z_\nu
\right\} {\mit\Phi} (z)
\, .
\end{equation}
To summarize, the representation of all generators from the conformal algebra reads
\begin{equation}
\label{RepConf}
\begin{array}{ll}
i \widehat{P}^\mu
&\!\!\! =
\partial^\mu
\, , \\
i \widehat{M}^{\mu\nu}
&\!\!\! =
z^\mu \partial^\nu
-
z^\nu \partial^\mu
+
{\mit\Sigma}^{\mu\nu}
\, , \\
i \widehat{D}
&\!\!\! =
d + z^\mu \partial_\mu
\, , \\
i \widehat{K}^\mu
&\!\!\! =
2 d z^\mu - z^2 \partial^\mu
+
2 z^\mu z^\nu \partial_\nu + 2 z_\nu {\mit\Sigma}^{\mu\nu}
\, .
\end{array}
\end{equation}

\section{Scheme transformation matrix}
\label{Bmatrix}

Let us prove Eq.\ (\ref{Bgamma}). Using  Eq.\ (\ref{GammaGammaC}),
written in the form
\begin{equation}
a (j, k) \gamma^{\scriptscriptstyle\rm ND}_{jk}
=
\left( \gamma_j - \gamma_k \right) \gamma^c_{jk} (l)
+
\left[ \bit{\gamma}^{\scriptscriptstyle\rm ND} , \bit{\gamma}^c (l) \right]_{jk}
\, ,
\end{equation}
we get
\begin{equation}
a (k' , k'') b_{k' k''} (k)
=
-
\gamma^c_{k' k''} (k)
-
\frac{1}{ \gamma_{k'} - \gamma_k }
\left[ \bit{\gamma}^{\scriptscriptstyle\rm ND} , \bit{\gamma}^c (l) \right]_{k' k''}
+
\frac{1}{\gamma_{k'} - \gamma_k}
\gamma^c_{k' k''} (l)
\left( \gamma_{k''} - \gamma_k \right)
\, .
\end{equation}
Note, that Eq.\ (\ref{DiffCgamma}) ensures that the right-hand side of this equation
does not depend on $l$, i.e.,
\begin{equation}
\left. a (k' , k'') b_{k' k''} (k) \right|_{l + n}
=
\left. a (k' , k'') b_{k' k''} (k) \right|_{l}
\, ,
\end{equation}
for any $n \in \mathbb{N}$.
Using the property
\begin{equation}
a (k_0 , k_N) = \sum_{i = 0}^{N - 1} a (k_i, k_{i + 1})
\, ,
\end{equation}
one proves by induction that
\begin{eqnarray}
\label{Recursion}
a (k', k'') \left\{ \bit{b}^N (k) \right\}_{k' k''}
=\!\!\!&-&\!\!\!
\gamma^c_{k' k'''} (l) \left\{ \bit{b}^{N - 1} (k) \right\}_{k''' k''}
\nonumber\\
&+&\!\!\!
\sum_{n = 0}^{N - 2}
\left\{
\bit{b}^n (k) \bit{M} (k) \bit{b}^{N - n - 2} (k)
\right\}_{k' k''}
\nonumber\\
&-&\!\!\!
\sum_{n = 0}^{N - 1}
\left\{
\bit{b}^n (k) \bit{M} (k) \bit{b}^{N - n - 1} (k)
\right\}_{k' k''}
\nonumber\\
&+&\!\!\!
\left\{ \bit{b}^{N - 1} (k) \right\}_{k' k'''}
\frac{1}{\gamma_{k'''} - \gamma_k}
\gamma^c_{k''' k''} (l)
\left( \gamma_{k''} - \gamma_k \right)
\, ,
\end{eqnarray}
where  the  matrix $\bit{M} (k)$ has  the following elements
\begin{equation}
M_{k' k''} (k) \equiv \frac{1}{\gamma_{k'} - \gamma_k}
\left[ \bit{\gamma}^{\scriptscriptstyle\rm ND} , \bit{\gamma}^c (l) \right]_{k' k''}
\, ,
\end{equation}
with the $l$-dependence made implicit in the notation for $\bit{M}$.

For the value $k = k''$, which enters (\ref{GammaGammaC}), the last term in
Eq.\ (\ref{Recursion}) is proportional to $\{ \bit{b} (k'') \}_{k' k''}
\gamma^c_{k'' k''} (l)$. Therefore, using the $l$-independence of the
right-hand side of (\ref{Recursion}),  one can get rid of this term by
setting $l = k''$. Thus,
\begin{eqnarray}
a (k' , k'') {\cal B}_{k' k''}
=
&-&\!\!\!
\gamma^c_{k' k'''} (k'')
\sum_{N = 1}^\infty
\left\{ \bit{b}^{N - 1} (k) \right\}_{k''' k''}
\\
&+&\!\!\!
\sum_{N = 1}^\infty
\left\{
\sum_{n = 0}^{N - 2}
\left\{
\bit{b}^n (k) \bit{M} (k) \bit{b}^{N - n - 2} (k)
\right\}_{k' k''}
-
\sum_{n = 0}^{N - 1}
\left\{
\bit{b}^n (k) \bit{M} (k) \bit{b}^{N - n - 1} (k)
\right\}_{k' k''}
\right\}
\, , \nonumber
\end{eqnarray}
so that the infinite sum of contributions in the curly brackets vanishes
identically. Therefore, we find that the transformation matrix from the
modified minimal subtraction scheme to the conformal scheme reads
\begin{equation}
\bit{\cal B}_{jk} = \left\{ \frac{1}{\1 + \bit{A}} \right\}_{jk}
\, , \qquad\mbox{with}\qquad
\bit{A}_{jk} = \frac{\gamma^c_{jk} (k)}{a (j, k)}
\, .
\end{equation}

\section{Two-loop anomalous dimensions}

In this appendix we give a concise summary of results about two-loop anomalous
dimensions: their explicit form for all twist-two sectors as well as equalities
among them arising in the minimal supersymmetric extension of QCD. Before we do
this, a few comments are in order.

First of all, we have to keep in mind the difference between the diagonal anomalous
dimensions of conformal operators, as defined in Eq.\ (\ref{DiagonalNonDiagonalADs}),
and the forward anomalous dimensions of the conventional Wilson operators
(\ref{ForwardADs}). They are related in the following way
\begin{equation}
\label{relations}
\gamma^{qq}_j
=
\gamma^{qq; {\scriptscriptstyle \rm fw}}_j
\, , \quad
\gamma^{qg}_j
=
\frac{6}{j}\ \gamma^{qg; {\scriptscriptstyle \rm fw}}_j
\, , \quad
\gamma^{gq}_j
=
\frac{j}{6}\ \gamma^{gq; {\scriptscriptstyle \rm fw}}_j
\, , \quad
\gamma^{gg}_j
=
\gamma^{gg; {\scriptscriptstyle \rm fw}}_j
\, .
\end{equation}
The pre-factors in the off-diagonal matrix elements of the two-by-two
anomalous dimension matrix come from the standard definition of the
Gegenbauer polynomials, as explained in Section \ref{NormalizationADs}.

The anomalous dimensions for even and odd moments of the forward parton
densities are not analytic functions of the spin. They depend on the
signature factor $\sigma = (-1)^{j + 1}$.

As it is well known, provided one is not restricted to particular flavor
combinations of quark operators, they start to mix with gluons, and the
resulting renormalization group equation takes a matrix form.

Let us now set up the terminology, which was implicitly used in the main text.
The combination of quark densities of the form
\begin{equation}
q^{(+)}_{\scriptscriptstyle\rm NS}
=
\left( u + \bar u \right) - \left( d + \bar d \right)
\, ,
\end{equation}
is called the nonsinglet combination with positive signature. It is distinct
from yet another nonsinglet combination
\begin{equation}
q^{(-)}_{\scriptscriptstyle\rm NS} = q - \bar q \equiv q_v
\, ,
\end{equation}
which is known as the valence quark distribution. It does not mix with gluons
either. Finally, the singlet quark distribution
\begin{equation}
q_{\scriptscriptstyle\rm S} = q + \bar q \, ,
\end{equation}
has the positive $\sigma = 1$ parity.

Without further ado, let us give the two-loop anomalous dimensions in all
twist-two sectors of QCD making use of the following harmonic sums
\begin{eqnarray*}
S_\ell (j) = \sum_{k = 1}^{j} \frac{1}{k^\ell}
\, , \qquad
S^\prime_\ell \left( j \right)
= 2^{\ell - 1} \sum_{k = 1}^{j}
[1 + (-1)^k] \frac{1}{k^\ell}
\, , \qquad
\widetilde{S} (j) = \sum_{k = 1}^{j} \frac{(-1)^k}{k^2} S_1 (k)
\, .
\end{eqnarray*}

\subsection{Even-parity sector}

The non-singlet anomalous dimension is \cite{FloKouLac81}
\begin{eqnarray}
\gamma^{{\rm \scriptscriptstyle NS, fw}}_{(1) j}
\!\!\!&=&\!\!\!
\left( C_F^2 - \frac{1}{2} C_F C_A \right)
\left\{
\frac{4 (2 j + 3)}{(j + 1)^2 (j + 2)^2} S(j + 1)
- 2 \frac{3 j^3 + 10 j^2 + 11 j + 3}{(j + 1)^3 (j + 2)^3}
\right.
\nonumber\\
&+&\!\!\! 4 \left( 2 S_1(j + 1) - \frac{1}{(j + 1)(j + 2)} \right)
\left(
S_2(j + 1) - S_2^{\prime} (j + 1)
\right) \nonumber\\
&+&\!\!\! \left.
16 \tilde{S}(j + 1) + 6 S_2(j + 1)
- \frac{3}{4} - 2 S_3^{\prime} (j + 1)
+ 4 (-1)^{j + 1}
\frac {2 j^2 + 6 j + 5}{(j + 1)^3 (j + 2)^3} \right\} \nonumber\\
&+& C_F C_A
\left\{ S_1(j + 1)
\left( \frac{134}{9} + \frac{2 (2 j + 3)}{(j + 1)^2 (j + 2)^2} \right)
\right. \nonumber\\
&-&\!\!\! 4 S_1 (j + 1) S_2(j + 1)
+ S_2(j + 1) \left( - \frac{13}{3} + \frac{2}{(j + 1) (j + 2)} \right)
\nonumber\\
&-&\!\!\!
\left.
\frac{43}{24} -
\frac{1}{9} \frac{151 j^4 + 867 j^3 + 1792 j^2 + 1590 j + 523
}{(j + 1)^3 (j + 2)^3} \right\} \nonumber\\
&+&\!\!\!
C_F T_F N_f
\left\{
- \frac{40}{9} S_1(j + 1) + \frac{8}{3} S_2(j + 1) + \frac{1}{3}
+ \frac{4}{9} \frac{ 11 j^2 + 27 j + 13}{(j + 1)^2 (j + 2)^2}
\right\} . \\
&& \nonumber
\end{eqnarray}
The singlet anomalous dimensions are \cite{Fur80}
\begin{eqnarray}
\gamma^{qq; {\scriptscriptstyle \rm fw}, V}_{(1)j}
\!\!\!&=&\!\!\!
\gamma^{{\rm \scriptscriptstyle NS, fw}}_{(1) j}
- 4 C_F T_F N_f
\frac{5 j^5 + 57 j^4 + 227 j^3 + 427 j^2 + 404 j + 160
}{j(j + 1)^3 (j + 2)^3 (j + 3)^2}
\, , \\
&& \nonumber\\
\gamma^{qg; {\scriptscriptstyle \rm fw}, V}_{(1)j}
\!\!\!&=&\!\!\!
- 2 C_A T_F N_f \left\{
\left( -2 S_1^2 (j + 1) + 2 S_2(j + 1)
- 2 S_2^{\prime} (j + 1) \right)
\frac{j^2 + 3 j + 4}{(j + 1) (j + 2) (j + 3)} \right.
\nonumber\\
&+&\!\!\!
\frac{960 + 2835 j + 4057 j^2 + 3983 j^3 + 3046 j^4 + 1777 j^5
+ 731 j^6 + 195 j^7 + 30 j^8 + 2 j^9}{j (j + 1)^3 (j + 2)^3 (j + 3)^3}
\nonumber\\
&+&\!\!\!
\left.
(- 1)^{j + 1}
\frac{141 + 165 j + 92 j^2 + 27 j^3 + 3 j^4}{(j + 1) (j + 2)^3 (j + 3)^3}
+
8 \frac{(2 j + 5)}{(j + 2)^2 (j + 3)^2} S_1 (j + 1)
\right\} \nonumber\\
&-&\!\!\!
2 C_F T_F N_f
\left\{ \left( 2 S_1^2(j + 1) - 2 S_2(j + 1) + 5 \right)
\frac{j^2 + 3 j + 4}{(j + 1)(j + 2)(j + 3)} \right. \nonumber\\
&-&\!\!\!
\left.
4 \frac{S_1(j + 1)}{(j + 1)^2}
+ \frac{11 j^4 + 70 j^3 + 159 j^2 + 160 j + 64}{(j + 1)^3(j + 2)^3(j + 3)}
\right\} \, ,\\
&& \nonumber\\
\gamma^{gq; {\scriptscriptstyle \rm fw}, V}_{(1)j}
\!\!\!&=&\!\!\!
- C_F^2
\left\{
\left( -2 S_1^2(j + 1) + 10 S_1(j + 1) - 2 S_2(j + 1) \right)
\frac{j^2 + 3 j + 4}{j(j + 1)(j + 2)}
\right.
\nonumber\\
&-&\!\!\!
\left.
4 \frac{S_1(j + 1)}{(j + 2)^2}
- \frac{12 j^6 + 102 j^5 + 373 j^4 + 740 j^3 + 821 j^2 + 464 j + 96
}{j(j + 1)^3(j + 2)^3} \right\}
\nonumber\\
&-&\!\!\!
2 C_A C_F
\left\{
\left( S_1^2(j + 1) + S_2(j + 1) - S_2^{\prime}(j + 1) \right)
\frac{j^2 + 3 j + 4}{j(j + 1)(j + 2)}
\right.
\nonumber\\
&+&\!\!\!
\frac{1296 + 10044 j + 30945 j^2 + 47954 j^3 + 42491 j^4 + 22902 j^5 + 7515 j^6
+ 1384 j^7 + 109 j^8}{9 j^2 (j + 1)^3 (j + 2)^2 (j + 3)^2}
\nonumber\\
&+&\!\!\!
\left.
(-1)^{j + 1}
\frac{8 + 9 j + 4 j^2 + j^3}{(j + 1)^3 (j + 2)^3}
- \frac{17 j^4 + 68 j^3 + 143 j^2 + 128 j + 24
}{3 j^2(j + 1)^2 (j + 2)} S_1(j + 1) \right\}
\nonumber\\
&-&\!\!\!
\frac{8}{3} C_F T_F N_f
\left\{
\left( S_1(j + 1)- \frac{8}{3} \right)
\frac{j^2 + 3 j + 4}{j(j + 1)(j + 2)} + \frac{1}{(j + 2)^2}
\right\} \, , \\
&&
\nonumber\\
\gamma^{gg; {\scriptscriptstyle \rm fw}, V}_{(1)j}
\!\!\!&=&\!\!\!
C_A T_F N_f \left\{
-\frac{40}{9} S_1(j + 1) + \frac{8}{3} + \frac{8}{9}
\frac{19 j^4 + 114 j^3 + 275 j^2 + 312 j + 138}{j(j + 1)^2(j + 2)^2(j + 3)}
\right\} \nonumber\\
&+&\!\!\!
C_F T_F N_f
\left\{ 2
+ 4 \frac{2 j^6 + 16 j^5 + 51 j^4 + 74 j^3 + 41 j^2 - 8 j - 16
}{j(j + 1)^3(j + 2)^3(j + 3)} \right\}
\nonumber\\
&+&\!\!\!
C_A^2
\left\{ \frac{134}{9} S_1(j + 1) + 16 S_1(j + 1)
\frac{
2 j^5 + 15 j^4 + 48 j^3 + 81 j^2 + 66 j + 18
}{
j^2(j + 1)^2(j + 2)^2(j + 3)^2
} \right.
\nonumber\\
&-&\!\!\! \frac{16}{3} + 8 S_2^{\prime}(j + 1)
\frac{j^2 + 3 j + 3}{j(j + 1)(j + 2)(j + 3)}
- 4 S_1(j + 1)S_2^{\prime}(j + 1)
\nonumber\\
&+&\!\!\!
8 \tilde{S}(j + 1) - S_3^{\prime}(j + 1)
-
\frac{1}{9}
\frac{457 j^9 + 6855 j^8 + 44428 j^7 + 163542 j^6
}{j^2(j + 1)^3(j + 2)^3(j + 3)^3} \nonumber\\
&-&\!\!\!
\left.
\frac{1}{9}
\frac{376129 j^5 + 557883 j^4 + 529962 j^3 + 308808 j^2 + 101088 j + 15552
}{j^2(j + 1)^3(j + 2)^3(j + 3)^3} \right\}
\, .
\end{eqnarray}

\subsection{Odd-parity sector}

The forward anomalous dimensions in the odd parity sector are given by
\cite{Nee96}:
\begin{eqnarray}
\gamma^{qq; {\rm \scriptscriptstyle fw}, A}_{(1) j}
\!\!\!&=&\!\!\!
\gamma^{{\rm \scriptscriptstyle NS, fw}}_{(1) j}
+
4 C_F T_F N_f
\frac{(j + 3) ( 4 + 5 j + 3 j^2 + j^3 )}{(j + 1)^3 (j + 2)^3}
\, , \\
\gamma^{qg; {\rm \scriptscriptstyle fw}, A}_{(1) j}
\!\!\!&=&\!\!\!
2 C_F T_F N_f
\Bigg\{
-
\frac{j ( 16 + 49 j + 60 j^2 + 30 j^3 + 5 j^4 )}{(j + 1)^3 (j + 2)^3}
+
\frac{4 j}{(j + 1)^2 (j +  2)} S_1 (j + 1)
\\
&-&\!\!\!
\frac{2 j}{(j + 1)(j + 2)} \left( S^2_1 (j + 1) - S_2 (j + 1) \right)
\Bigg\}
+
4 C_A T_F N_f
\Bigg\{
\frac{8 + 4 j - 7 j^2 - 10 j^3 - 6 j^4 - j^5}{(j + 1)^3 (j + 2)^3}
\nonumber\\
&-&\!\!\!
\frac{4 S_1 (j + 1)}{(j + 1)(j + 2)^2}
+
\frac{j}{(j + 1)(j + 2)}
\left( S_1^2 (j + 1) - S_2 (j + 1) + S^\prime_2 (j + 1) \right)
\Bigg\}
\, , \nonumber\\
\gamma^{gq; {\rm \scriptscriptstyle fw}, A}_{(1) j}
\!\!\!&=&\!\!\!
8 C_F T_F N_f
\left\{
\frac{(j + 3)(5j + 7)
}{
9 (j + 1) (j + 2)^2}
-
\frac{ (j + 3)}{3 (j + 1) (j + 2)} S_1 (j + 1)
\right\}
\\
&+&\!\!\!
C_F^2
\left\{
\frac{(j + 3)(3j + 4)(3 + 14 j + 12 j^2 + 3 j^3)}{(j + 1)^3 (j + 2)^3}
-
\frac{2 (j + 3)(3j + 4)}{(j + 1)(j + 2)^2} S_1 (j + 1)
\right.
\nonumber\\
&+&\!\!\!
\left.
\frac{2 (j + 3)}{(j + 1)(j + 2)} \left( S_1^2 (j + 1) + S_2 (j + 1) \right)
\right\}
\nonumber\\
&+&\!\!\!
2 C_A C_F
\left\{
-
\frac{750 + 2380 j + 3189 j^2 + 2098 j^3 + 651 j^4 + 76 j^5}{9 (j + 1)^3 (j + 2)^3}
\right.
\nonumber\\
&+&\!\!\!
\left.
\frac{45 + 44 j + 11 j^2}{3 (j + 1)^2 (j + 2)} S_1 (j + 1)
+
\frac{j + 3}{(j + 1) (j + 2)}
\left( - S_1^2 (j + 1) - S_2 (j + 1) + S^\prime_2 (j + 1) \right)
\right\}
\, , \nonumber\\
\gamma^{gg; {\rm \scriptscriptstyle fw}, A}_{(1) j}
\!\!\!&=&\!\!\!
2 C_F T_F N_f
\frac{8 + 30 j + 70 j^2 + 71 j^3 + 35 j^4 + 9 j^5 + j^6}{(j + 1)^3 (j + 2)^3}
\\
\!\!\!&+&\!\!\!
8 C_A T_F N_f
\left\{
\frac{35 + 75 j + 52 j^2 + 18 j^3 + 3 j^4}{9 (j + 1)^2 (j + 2)^2}
-
\frac{5}{9} S_1 (j + 1)
\right\}
\nonumber\\
\!\!\!&+&\!\!\!
C_A^2
\left(
-
\frac{1768 + 5250 j + 7075 j^2 + 4974 j^3 + 1909 j^4 + 432 j^5 + 48 j^6
}{
9 (j + 1)^3 (j + 2)^3}
\right.
\nonumber\\
\!\!\!&+&\!\!\!
\frac{2}{9}
\frac{484 + 948 j + 871 j^2 + 402 j^3 + 67 j^4}{(j + 1)^2 (j + 2)^2} S_1 (j + 1)
\nonumber\\
\!\!\!&+&\!\!\!
\left.
\frac{8 S^\prime_2 (j + 1)}{(j + 1)(j + 2)}
-
4 S_1 (j + 1) S^\prime_2 (j + 1) - S^\prime_3 (j + 1)
+
8 \widetilde{S} (j + 1)
\right)
\, . \nonumber
\end{eqnarray}

\subsection{Maximal-helicity sector}

Since the quark and gluon operators of the maximal helicity have different quantum
numbers with respect to the helicity operator (\ref{HelicityOperatorPLvector}), they
do not mix with each other under renormalization and thus have an autonomous scale
dependence. The anomalous dimension of the quark transversity sector is \cite{Vog97}
\begin{eqnarray}
\gamma^{qq; {\rm \scriptscriptstyle fw}, T}_{(1) j}
\!\!\!&=&\!\!\!
C_F^2
\left\{ - \frac{1}{4} - 2 S_1 (j + 1) + S_2 (j + 1) \right\}
+
\frac{8}{9} C_F T_F N_f
\left\{ \frac{3}{8} - 5 S_1(j + 1) + 3 S_2 (j + 1) \right\}
\nonumber\\
&+&\!\!\!
\frac{C_A C_F}{4}
\left\{
- \frac{20}{3} + \frac{572}{9} S_1 (j + 1)
- \frac{58}{3} S_2 (j + 1) - 16 S_1 (j + 1) S_2 (j + 1)
\right\}
\nonumber\\
&-&\!\!\!
2 C_F \left( C_F - \frac{C_A}{2} \right)
\Bigg\{
\frac{1}{4} + \frac{1 + (- 1)^j}{(j + 1) (j + 2)}
-
\frac{5}{2} S_2 (j + 1)
+
S^\prime_3 (j + 1) - 8  \widetilde{S} (j + 1)
\nonumber\\
&&\qquad\qquad\qquad\qquad\quad
-
S_1 (j + 1) \left( 1 + 4 S_2 (j + 1) - 4 S^\prime_2 (j + 1) \right)
\Bigg\}
\, ,
\end{eqnarray}
while the gluon it reads \cite{Vog98}
\begin{eqnarray}
\gamma^{gg; {\scriptscriptstyle \rm fw}, T}_{(1)j}
\!\!\!&=&\!\!\!
C_A^2 \left\{ S_1 (j + 1) \left( \frac{134}{9}
- 4 S_2^\prime \left( j + 1 \right)\right)
- S_3^\prime \left( j + 1 \right)
+ 8 \tilde{S}(j + 1) - \frac{1}{j (j + 3)} - \frac{16}{3}
\right\}
\nonumber\\
&+&\!\!\! C_A T_F N_f \left\{ \frac{8}{3}
- \frac{40}{9} S_1 (j + 1) - \frac{2}{j ( j + 3)} \right\}
+ C_F T_F N_f \frac{2 (j + 1)(j + 2)}{j (j + 3)}
\, .
\end{eqnarray}

\subsection{Supersymmetric relations between anomalous dimensions}
\label{SUSYrelationsADs}

The twist-two anomalous dimensions quoted in the previous appendix for
conventional Wilson operators and in Section \ref{NDanomalousDimensionsNLO}
for conformal operators are defined by rather lengthy expressions and an
independent check on their correctness is very instructive. In QCD per se,
there are no relations between the elements of the quark-gluon mixing matrix.
However, in its closest cousin---the $\mathcal{N} = 1$ super-Yang-Mills
theory which operates in terms of a single Majorana quark (gaugino)
coupled to gluodynamics---the presence of the space-time
supersymmetry, which transforms a gaugino into a gluon and then back,
allows one to establish a set of equations between the elements of the
mixing matrix. In perturbation theory the adjustments which have to be
made are minimal. Merely, one has to equate the color Casimir operators
in all QCD expression to the rank of the color group $N_c$, accounting for
the fact that contrary to QCD in $\mathcal{N} = 1$ super-Yang-Mills there
is just a single flavor of quarks and that it belongs to the adjoint
representation of the color group. Thus, the ``supersymmetric limit'' of
ordinary QCD is achieved via the identification $C_A = C_F = 2 T_F = N_c$
in all QCD anomalous dimensions. This is not the end of the story however,
since the conventional procedure of dimensional regularization, used to
operate with infinities in perturbation theory, explicitly breaks supersymmetry.
A way out was proposed a long time ago in the form of a regularization
procedure dubbed  dimensional reduction \cite{Sie79,CapJonNie80}.
According to it, the dimension of space-time is reduced from four down to
$d = 4 - 2 \varepsilon$, while all fields have the same number of components
as in four dimensions thus preserving the equality of bosonic and fermionic
degrees of freedom. Instead of redoing all calculations anew with the modified
regularization procedure, one can merely perform the scheme transformation via
a finite rotation of the two-vector (\ref{TwoVectorConfOperators}) of quark
and gluon conformal operators (\ref{QuarkAndGluonConformalOperators}) from
the standard dimensional regularization (DREG) to the dimensional reduction
(DRED) schemes
$$
\mathbf{O}^{\scriptscriptstyle\rm R}_{\scriptscriptstyle\rm DRED}
=
\bit{z}
\cdot
\mathbf{O}^{\scriptscriptstyle\rm R}_{\scriptscriptstyle\rm DREG}
\, .
$$
Thus, the quark-gluon anomalous dimension mixing matrix, $\mbox{\boldmath$\gamma$}$,
for the regularization with DRED is related to the one with DREG via
\begin{equation}
\bit{\gamma}^{\scriptscriptstyle\rm DRED}
= \bit{z}
\cdot
\bit{\gamma}^{\scriptscriptstyle\rm DREG}
\cdot
\bit{z}^{-1}
-
\beta (g) \frac{\partial}{\partial g}
\bit{z} \cdot \bit{z}^{-1}
\, .
\end{equation}
The transformation matrix $\bit{z}$ is deduced from the difference of counterterms
determining the anomalous dimensions of the quark and gluon conformal operators in
the above two schemes. It takes the following form to one-loop order in coupling
constant \cite{BelMulSch98}
\begin{equation}
\bit{z}_{jk} = \1_{[2] \times [2]} \delta_{jk}
+ \frac{\alpha_s}{2 \pi} N_c
\left\{
\bit{z}_j \delta_{jk}
+
\bit{z}^{\scriptscriptstyle\rm ND}_{jk}
\theta_{j - 2,k} [1 + (- 1)^{j - k}]
\right\} ,
\end{equation}
with the following matrices
\begin{equation}
\label{TransDiag}
\bit{z}^V_j
= \left(
\begin{array}{cc}
- \frac{j(j + 3)}{2(j + 1)(j + 2)}
&
\frac{12}{j(j + 2)(j + 3)}
\\
\frac{j}{6 (j + 2)}
&
-\frac{1}{6}
\end{array}
\right) , \quad
\bit{z}^A_j
= \left(
\begin{array}{cc}
- \frac{j(j + 3)}{2(j + 1)(j + 2)}
&
\frac{12}{j(j + 1)(j + 2)}
\\
- \frac{j}{3 (j + 1)(j + 2)}
&
-\frac{1}{6} - \frac{4}{(j + 1)(j + 2)}
\end{array}
\right) ,
\end{equation}
for the vector \cite{EllRosTer81,Blu98,BelMulSch98} and axial
\cite{Nee96,Blu98,BelMulSch98} channels, respectively, and a
universal non-diagonal (in the conformal spin) part \cite{BelMulSch98}
\begin{equation}
\label{TransNonDiag}
\bit{z}^{\scriptscriptstyle\rm ND}_{jk}
= \left(
\begin{array}{cc}
0
&
\frac{6 (2k + 3)}{k( k + 1 )( k + 2 )( k + 3 )} \\
- \frac{(2k + 3)}{6( k + 1 )( k + 2 )}
&
- \frac{(2k + 3)(j - k)(j + k + 3)}{k(k + 1)(k + 2)(k + 3)}
\end{array}
\right) \, .
\end{equation}
On the other hand, no rotation matrices are needed at the two-loop order for
the maximal-helicity operators, i.e., $z^T = 1$.

As a consequence of supersymmetric Ward identities and the commutator algebra of
the $\mathcal{N} = 1$ superconformal group \cite{Son85}, the transformed anomalous
dimensions in the DRED scheme obey a set of relations which hold to all orders in
perturbation theory \cite{BelMulSch98,BelMul02a}. They can be naturally classified
as follows.
\begin{itemize}
\item Autonomous relations for the vector and axial channels \cite{Dok77} (as
before, we set $\gamma_{jj} \equiv \gamma_j$ and we do not display the superscript
``DRED'' for brevity)
\begin{equation}
\label{DokSUSY}
\gamma^{qq; a}_j
+
\frac{6}{j} \gamma^{gq; a}_j
=
\frac{j}{6} \gamma^{qg; a}_j
+
\gamma^{gg; a}_j
\, ,
\quad a = V,A .
\end{equation}
\item Relations between the axial and vector channels
\cite{BukFroLipKur85,BukFroLipKur86,BelMulSch98,BelMul02a}
\begin{equation}
\label{SUSY2}
\gamma^{qq; V}_{j + 1}
+
\frac{6}{j + 1} \gamma^{gq; V}_{j + 1}
=
\gamma^{qq; A}_j
-
\frac{j}{6} \gamma^{qg; A}_j
\, , \qquad
\gamma^{qq; A}_{j + 1}
+
\frac{6}{j + 1} \gamma^{gq; A}_{j + 1}
=
\gamma^{qq; V}_j
-
\frac{j}{6} \gamma^{qg; V}_j
\, .
\end{equation}
\item Relations between the diagonal and non-diagonal elements of anomalous
dimensions in the axial and vector channels \cite{BelMulSch98,BelMul02a}
\begin{equation}
\label{for-nonfor}
\frac{6}{j} \gamma^{gq; V}_j
-
\frac{j + 3}{6} \gamma^{qg; V}_j
=
\frac{j + 1}{2j + 1} \Delta^{{\scriptscriptstyle\rm ND}, A}_{j + 1, j - 1}
\, ,
\quad
\frac{6}{j} \gamma^{gq; A}_j
-
\frac{j + 3}{6} \gamma^{qg; A}_j
=
\frac{j + 1}{2j + 1} \Delta^{{\scriptscriptstyle\rm ND}, V}_{j + 1, j - 1}
\, ,
\end{equation}
where
\begin{equation}
\Delta^{{\scriptscriptstyle\rm ND}, a}_{j + 1, j - 1}
\equiv
\frac{j - 1}{j + 1} \gamma^{gg; {\scriptscriptstyle\rm ND}, a}_{j + 1, j - 1}
+
\frac{j - 1}{6} \gamma^{qg; {\scriptscriptstyle\rm ND}, a}_{j + 1, j - 1}
-
\frac{6}{j + 1} \gamma^{gq; {\scriptscriptstyle\rm ND}, a}_{j + 1, j - 1}
-
\gamma^{qq; {\scriptscriptstyle\rm ND}, a}_{j + 1, j - 1}
\, .
\end{equation}
\item Relation between the anomalous dimensions of the maximal-helicity operators
\cite{Vog98,BelMulSch98}
\begin{equation}
\gamma^{qq; T}_j
=
\gamma^{gg; T}_j
\, .
\label{SUSYRelationMaxHelicity}
\end{equation}
\end{itemize}

As we established in Section \ref{AutonomousScaleDepConfOperators}, conformal operators
do not mix with each other at leading order in coupling constant. As a consequence, the
right-hand side of equations (\ref{for-nonfor}) is zero, i.e., the expansion starts from
$\mathcal{O} (\alpha_s^2)$. Beyond one loop, as we have seen in the main presentation,
the Eqs.\ (\ref{DokSUSY}) -- (\ref{SUSYRelationMaxHelicity}) provide a very non-trivial
check of the existing results. As a final remark, let point out that the above relations
are merely particular cases of more generic relations between the anomalous dimensions
of conformal operators $\gamma_{jk}$ for any $j \geq k$ \cite{BelMulSch98,BelMul02a}.

\section{Twist separation in light-ray operators}
\label{LightRayTwistThree}

The non-local form of the twist-two operators was introduced in the main
text in Eq.\ (\ref{TwistTwoOpenIndex}). An analogous representation for
the twist-three operators arises from the difference of ${\cal O}^{qq}_\rho$
and ${\cal R}^{2,qq}_\rho$. The result can be cast into the form \cite{BalBra89}
\begin{eqnarray}
\label{TwistThreeLRoperator}
{\cal R}^{3, qq}_\rho (- z, z)
\!\!\!&=&\!\!\!
{\cal O}^{qq}_\rho (- z, z) - {\cal R}^{2, qq}_\rho (- z, z)
\nonumber\\
&=&\!\!\!\
\int_0^1 du \, z^\mu
\left(
g_{\nu\rho} \frac{\partial}{\partial z^\mu}
-
g_{\nu\mu} \frac{\partial}{\partial z^\rho}
\right)
{\cal O}^{qq, \nu} (- u z, u z)
\, ,
\end{eqnarray}
(up to terms of higher order in $z^2$) making use of an obvious identity
$$
\int_0^1 du \, z^\mu \frac{\partial}{\partial z^\mu}
{\cal O}^{qq}_\rho (- u z, u z)
=
\int_0^1 du \, u \frac{\partial}{\partial u}
{\cal O}^{qq}_\rho (- u z, u z)
=
{\cal O}^{qq}_\rho (- z, z)
-
\int_0^1 du \, {\cal O}^{qq}_\rho (- u z, u z)
\, .
$$

Next, we have to differentiate the light-ray operator on the right-hand side
of Eq.\ (\ref{TwistThreeLRoperator}). This action   involves differentiation
of the path ordered exponential
$$
[z_2, z_1]
=
P \exp \left( i g \int_{z_1}^{z_2} dz^\mu \, A_\mu (z) \right)
\, .
$$
For a Wilson line (which is neither restricted to the light-cone here nor its
path is given by a straight line), a generic variation of the gauge link under
an infinitesimal deformation of the path/contour can be written as
\cite{GerNev79},
\begin{eqnarray}
\label{VariationPath}
\delta [z_2, z_1]
\!\!\!&=&\!\!\!
i g A_\mu (z_2) \delta z_2^\mu [z_2, z_1]
-
i g [z_2, z_1] A_\mu (z_1) \delta z_1^\mu
\nonumber\\
&+&\!\!\! i g \int_{- 1}^1 d \tau \,
[z_2, z(\tau)]
F_{\mu\nu} \left( z(\tau) \right)
\delta z^\mu (\tau) \dot z^\nu (\tau)
[z (\tau), z_1]
\, .
\end{eqnarray}
Here the dot on top of the path $z^\mu (\tau)$ from the initial to final
points stands for the differentiation with respect to the proper time $\tau$
on it, $\dot z^\nu (\tau) \equiv d z^\nu (\tau)/d \tau$. The field strength
appears here  because, in evaluating the infinitesimal displacement of the path, one
forms a closed loop, and thus induces a curvature, in it. The straight-line
parametrization of the path $z^\mu (\tau)$  from $z_1$ to $z_2$ is
$$
z_1^\mu = a^\mu + z^\mu
\to
z_2^\mu = a^\mu - z^\mu :
\qquad
z^\mu (\tau) = a^\mu - \tau z^\mu
\, .
$$
Therefore, depending on whether the variation in Eq.\ (\ref{VariationPath})
will be performed with respect to the total translation $a^\mu$ of the
path as a whole or, rather, with respect to the relative distance between
the ends $z^\mu$, one will not or will, respectively, acquire a factor of
the proper time $\tau$ under the integral. Namely:
\begin{itemize}
\item Differentiation with respect to the relative distance
$$
\delta z^\mu (\tau) = - \tau d z^\mu
\, , \qquad
\dot z^\nu (\tau) = - z^\mu
\, , \qquad
\delta z_1^\mu = d z^\mu
\, , \qquad
\delta z_2^\mu = - d z^\mu
\, ,
$$
results into the equation
\begin{equation}
\label{DiffWilsonDifference}
\frac{\partial}{\partial z^\mu} [ - z, z]
=
- i g A_\mu (- z) [- z , z]
- i g [- z , z] A_\mu (z)
- i g \int_{- 1}^1 d \tau \, \tau \,
{}[- z, \tau z] F_{\mu\nu} (\tau z) z^\nu [\tau z, z]
\, .
\end{equation}
\item Differentiation with respect to the total translation
$$
\delta z^\mu (\tau) = d a^\mu
\, , \qquad
\dot z^\nu (\tau) = - z^\mu
\, , \qquad
\delta z_1^\mu = d a^\mu
\, , \qquad
\delta z_2^\mu = d a^\mu
\, ,
$$
gives instead
\begin{eqnarray}
\label{DiffWilsonTranslation}
\frac{\partial}{\partial a^\mu} [ - z + a, z + a]
\!\!\!&=&\!\!\!
i g A_\mu (- z + a) [- z + a, z + a]
- i g [- z + a, z + a] A_\mu (z + a)
\nonumber\\
&-&\!\!\! i g \int_{- 1}^1 d \tau \,
{}[- z + a, \tau z + a] F_{\mu\nu} (\tau z + a) z^\nu [\tau z + a, z + a]
\, .
\end{eqnarray}
\end{itemize}
Thus, the differentiation in Eq.\ (\ref{TwistThreeLRoperator}) is performed
using Eq.\ (\ref{DiffWilsonDifference}) and yields
\begin{eqnarray}
\label{DiffOfNLoperator}
\frac{\partial}{\partial z^\mu} {\cal O}^{qq}_\nu (- z, z)
\!\!\!&=&\!\!\!
-
\bar\psi (- z) \stackrel{{}_\leftarrow}{\cal D}_\mu \! \gamma_\nu [- z, z] \psi (z)
+
\bar\psi (- z) [- z, z] \stackrel{{}_\rightarrow}{\cal D}_\mu \! \gamma_\nu  \psi (z)
\nonumber\\
&-&\!\!\!
i g \int_{- 1}^1 d\tau \, \tau \,
\bar\psi (- z) \gamma_\nu [- z, \tau z] F_{\mu\rho} (\tau z) z^\rho
[\tau z, z] \psi (z)
\, ,
\end{eqnarray}
where the left derivative acts on the argument $- z^\mu$ of the field $\bar\psi$
with the sign, i.e., $\bar\psi (- z) \! \stackrel{{}_\leftarrow}{\cal D}_\mu
= \bar\psi (- z) [ \partial/\partial (- z^\mu) + i g A_\mu (- z) ]$. We substitute
this expression back into Eq.\ (\ref{TwistThreeLRoperator}) and an analogous one
with changed indices. For further reduction we have to use the Heisenberg equation
of motion for the quark field ${\not\!\!{\cal D}} \psi = 0$ in the two-particle
operators containing total derivatives,
\begin{equation}
\left(
\gamma_\mu \stackrel{{}_\rightarrow}{\cal D}_\nu
-
\
\gamma_\nu \stackrel{{}_\rightarrow}{\cal D}_\mu
\right)
\psi
\stackrel{\rm EOM}{=}
\frac{i}{2}
\stackrel{{}_\rightarrow}{{\not\!\!{\cal D}}}
\sigma_{\mu\nu} \, \psi
\, , \qquad
\bar\psi
\left(
\gamma_\mu \stackrel{{}_\leftarrow}{\cal D}_\nu
-
\
\gamma_\nu \stackrel{{}_\leftarrow}{\cal D}_\mu
\right)
\stackrel{\rm EOM}{=}
\frac{i}{2}
\bar\psi \,
\sigma_{\mu\nu}
\stackrel{{}_\leftarrow}{{\not\!\!{\cal D}}}
\, .
\end{equation}
Assembling everything together, we get for the twist-three operator
\begin{eqnarray*}
{\cal R}^{3, qq}_\rho (- z, z)
= i z^\nu
\int_0^1 du
\bigg\{
\frac{u}{2} \bar\psi (- uz)
\Big(
\sigma_{\rho\nu}
\stackrel{{}_\leftarrow}{{\not\!\!{\cal D}}} [- uz, uz]
\!\!\!&+&\!\!\!
[- uz, uz]
\stackrel{{}_\rightarrow}{{\not\!\!{\cal D}}} \sigma_{\rho\nu}
\Big)
\psi (uz)
\nonumber\\
&+&\!\!\!
g \int_{- u}^u d\tau \, \tau
\bar\psi (- uz)
\gamma_\nu F_{\rho\mu} (\tau z) z^\mu
\psi (uz)
\bigg\}
\, .
\end{eqnarray*}
Now, the two-particle operators in round brackets can be further reduced to
operators with total derivatives and three-particle operators by means of Eq.\
(\ref{DiffWilsonTranslation}) and the use of the quark equation of motion,
namely,
\begin{eqnarray}
\left. \frac{\partial}{\partial a_\rho} \right|_{a = 0}
\bar\psi (a - z)
\left\{
\begin{array}{c}
\sigma_{\mu\nu} \gamma_\rho
\\
\gamma_\rho \sigma_{\mu\nu}
\end{array}
\right\}
\psi (a + z)
\!\!\!&=&\!\!\!
\bar\psi (- z)
\left\{
\begin{array}{c}
\sigma_{\mu\nu} \stackrel{{}_\leftarrow}{{\not\!\!{\cal D}}}
\\
\stackrel{{}_\rightarrow}{{\not\!\!{\cal D}}} \sigma_{\mu\nu}
\end{array}
\right\}
\psi (z)
\\
&-&\!\!\!
i g
\int_{- 1}^1 d\tau \,
\bar\psi (- z)
\left\{
\begin{array}{c}
\sigma_{\mu\nu} \gamma_\rho
\\
\gamma_\rho \sigma_{\mu\nu}
\end{array}
\right\}
F^{\rho\sigma} (\tau z) z_\sigma
\psi (z)
\, . \nonumber
\end{eqnarray}
Summing these contributions and reducing the product of three Dirac matrices
with Eq.\ (\ref{ThreeToOne}) to one, we find
\begin{eqnarray}
\bar\psi (- z)
\Big(
\sigma^{\mu\nu}
\stackrel{{}_\leftarrow}{{\not\!\!{\cal D}}} [- z, z]
+
[- z, z]
\stackrel{{}_\rightarrow}{{\not\!\!{\cal D}}} \sigma^{\mu\nu}
\Big)
\psi (z)
\!\!\!&=&\!\!\!
-
2 \varepsilon^{\mu\nu\rho\sigma} \partial_\rho
\widetilde {\cal O}^{qq}_\sigma (- z, z)
\\
&-&\!\!\!
2 i g \varepsilon^{\mu\nu\rho\sigma}
\int_{- 1}^1 d \tau \,
\bar\psi (- z)
\gamma_\sigma \gamma^5 F_{\rho\lambda} (\tau z) z^\lambda
\psi (z)
\, , \nonumber
\end{eqnarray}
with $\partial_\mu = \stackrel{{}_\rightarrow}{\partial}_\mu +
\stackrel{{}_\leftarrow}{\partial}_\mu$ being the total derivative as defined
in Eq.\ (\ref{TotalDerivative}). Thus, we express the twist-three parity-even
operator in terms of the total derivative of the parity-odd bilocal operator and
three-particle operators. Substituting the result into (\ref{TwistThreeLRoperator}),
we derive the following equation \cite{BalBra89}
\begin{eqnarray}
\label{recursion1}
{\cal O}^{qq}_\rho ( - z^- , z^- )
\!\!\!&=&\!\!\!
{\cal R}^{2, qq}_\rho ( - z^- , z^-)
+
z^- \int_{0}^{1} du
\Bigg\{
- i u \, \varepsilon_\rho{}^{+ \mu \nu} \partial_\mu
\widetilde {\cal O}^{qq}_\nu (- u z^-, u z^-)
\\
&&+ \frac{z^-}{2} \int_{- u}^{u} d\tau
\left[
(u - \tau)
{\cal S}^{qgq}_{[+]\rho} ( - u z^- , \tau z^- , u z^-)
-
(u + \tau)
{\cal S}^{qgq}_{[-]\rho} ( - u z^- , \tau z^- , u z^-)
\right]
\Bigg\}
\, , \nonumber
\end{eqnarray}
where we set $z^\mu = z^- n^\mu$ and use the three-particle operators
introduced in Eq.\ (\ref{QGQoperator}). An analogous equation holds for
the parity-odd twist-three operator with obvious dressing or removal of
tildes from the functions involved,
\begin{eqnarray}
\label{recursion2}
\widetilde {\cal O}^{qq}_\rho ( - z^- , z^- )
\!\!\!&=&\!\!\!
\widetilde {\cal R}^{2, qq}_\rho ( - z^- , z^-)
+
z^- \int_{0}^{1} du
\Bigg\{
- i u \, \varepsilon_\rho{}^{+ \mu \nu} \partial_\mu
{\cal O}^{qq}_\nu (- u z^-, u z^-)
\\
&&+ \frac{z^-}{2} \int_{- u}^{u} d\tau
\left[
(u - \tau)
\widetilde {\cal S}^{qgq}_{[+]\rho} ( - u z^- , \tau z^- , u z^-)
-
(u + \tau)
\widetilde {\cal S}^{qgq}_{[-]\rho} ( - u z^- , \tau z^- , u z^-)
\right]
\Bigg\}
\, . \nonumber
\end{eqnarray}

Notice that, on the right-hand side of this equation (\ref{recursion1}),
one gets the operator $\widetilde {\cal O}^{qq}_\rho$, which does not
possesses a definite twist. The same statement applies to (\ref{recursion2}).
Therefore, Eq.\ (\ref{recursion1}) combined with (\ref{recursion2}) forms a
coupled system of recursion relations, which should be solved in order
to determine the twist-three operators in terms of twist-two and twist-three
three-particle ones. These equations can be easily solved iteratively.
Namely, one substitutes $\widetilde {\cal O}^{qq}_\rho$ from (\ref{recursion2})
into (\ref{recursion1}) and so on. It is easier to perform this straightforward
calculation rather than describe it. The final result is given in Eq.\
(\ref{R3-part}) for $z^2 = z_\mu z^\mu = 0$ \cite{BelMul00}.

For certain applications---the resummation of target mass corrections
addressed in Section \ref{SectionTargetMassCorrections} at leading twist
level and accounting for twist-four effects due to $\Delta^2/Q^2$ effects
in amplitudes, just to name a few---one needs to keep non-light-like
inter-field separations, i.e., $z^2 \neq 0$. For the present discussion,
the multiparticle operators will be irrelevant and thus will be neglected.
We decompose a given non-local, say of even parity, operator ${\cal O}_\rho^{qq}
(-z, z)$ into its symmetric and antisymmetric parts
\begin{equation}
\label{Decomposion}
{\cal O}^{qq}_{\rho} (- z, z)
=
{\cal O}^{qq, \rm sym}_{\rho} (- z, z)
+
{\cal O}^{qq, \rm asym}_{\rho} (- z, z)
\, .
\end{equation}
The symmetric operator is easily constructed from ${\cal O}_\rho^{qq}$ as was
previously done in Eq.\ (\ref{TwistTwoOpenIndex}),
\begin{equation}
\label{SymmOperators}
{\cal O}^{qq, \rm sym}_{\rho} (- z, z)
=
\int_0^1 du \,
\frac{\partial}{\partial z^\rho} \, z^\sigma
{\cal O}^{qq}_\sigma (- uz, uz)
\, .
\end{equation}
When expanded in Taylor series it generates a tower of Wilson operators symmetric
in all Lorentz indices, however, since $z^2 \neq 0$ here, it is not automatically
traceless.

The antisymmetric operator is given by the difference between the ``full'' operator
and its symmetric component. After some algebra, which led to Eq.\ (\ref{recursion1}),
one finds the antisymmetric even-parity operator in terms of the odd-parity one
\begin{equation}
\label{AsymFromSym}
{\cal O}^{qq, \rm asym}_{\rho} (- z, z)
=
{\cal O}^{qq}_{\rho} (- z, z)
- {\cal O}^{qq, \rm sym}_{\rho} (- z, z)
=
- i \varepsilon_\rho{}^{\mu \nu \sigma} z_\mu \partial_\nu
\int_0^1 du \, u \
\widetilde {\cal O}^{qq}_{\sigma} (ux, - ux)
\, .
\end{equation}
Notice, that when we expand the second equality in Taylor series, we get Eq.\
(\ref{WilsonAntiSymmOper}). As we know, ${\cal O}^{qq, \rm asym}_{\rho}
(- z, z)$ can be expressed in terms of ${\cal O}^{qq, \rm sym}_{\rho} (- z, z)$
and $\widetilde{\cal O}^{qq, \rm sym}_{\rho} (- z, z)$ and higher twist
multi-particle operators, which we ignore. Thus, to find the antisymmetric
operator ${\cal O}^{\rm asym}$, we need to solve the system of equations given
by (\ref{AsymFromSym}) and a similar one for the axial vector operator.
Excluding, for instance, the axial operator we get an equation
\begin{eqnarray}
\label{EqForOV}
{\cal O}^{qq}_\rho (- z, z)
\!\!\!&=&\!\!\!
{\cal O}^{qq, \rm sym}_\rho (- z, z)
-
i \varepsilon_\rho{}^{\mu \nu \sigma} z_\mu \partial_\nu
\int_0^1 du \, u \, \widetilde
{\cal O}^{qq, \rm sym}_\sigma (- u z, u z)
\\
&+&\!\!\!
\left\{
\left[ (z \cdot \partial)^2 - z^2 \partial^2 \right] g_\rho{}^\sigma
-
\left[ (z \cdot \partial) \partial_\rho - \partial^2 z_\rho \right]
z^\sigma
\right\}
\int_0^1 du \, u \bar u \, {\cal O}^{qq}_\sigma (- u z, u z)
\, .
\nonumber
\end{eqnarray}
As before $\bar u \equiv 1 - u$. In the course of the calculation, the contraction of
two $\varepsilon$-tensors have produced six terms, four of them are displayed above,
and the other two are neglected since they are proportional to total derivatives
$\partial^\sigma {\cal O}_\sigma (- z, z)$. This contribution, my means of the
equations of motion, is expressed in terms of the twist-four antiquark-gluon-quark
operator (cf.\ (\ref{DiffOfNLoperator}))
\begin{equation}
\partial^\sigma {\cal O}^{qq}_\sigma (- z, z)
\stackrel{\rm EOM}{=}
- i g \int_{-1}^1 du \,
\bar\psi_q (- z) \gamma_\mu z_\nu F^{\mu \nu} (u z) \psi_q (z)
\, .
\end{equation}
Solving iteratively the recurrence relations (\ref{EqForOV}), we find to twist-four
accuracy
\begin{eqnarray}
\label{R3-partZ}
&&\!\!\!\!\!\!\!\!\!\!\!
{\cal O}^{qq, \rm asym}_{\rho} (- z, z)
=
- \frac{1}{2} \int_{0}^{1} du \, u
\Bigg\{
i \varepsilon_\rho{}^{\mu \nu \sigma} z_\mu \partial_\nu
\left[
\widetilde {\cal O}^{qq, \rm sym}_\sigma \left( (\bar u - u) z, z \right)
+
\widetilde {\cal O}^{qq, \rm sym}_\sigma \left( - z, (u - \bar u)z \right)
\right]
\nonumber\\
&&\qquad\qquad\qquad +
\left( \partial_\rho z_\sigma - g_{\rho \sigma} (z \cdot \partial) \right)
\,
\left[
{\cal O}^{qq, \rm sym}_\sigma \left( (\bar u - u) z, z \right)
-
{\cal O}^{qq, \rm sym}_\sigma \left( - z, (u - \bar u) z \right)
\right]
\Bigg\}
\nonumber\\
&&\!\!\!\!\!\!
+ \frac{1}{4} z^2 \partial^2
\int_{0}^{1} du \, u \bar u
\int_{0}^{1} dv \\
&&\ \times
\Bigg\{
i \bar u \, \varepsilon_\rho{}^{\mu \nu \sigma} z_\mu \partial_\nu
\left[
\widetilde {\cal O}^{qq, \rm sym}_\sigma
\left( (v \bar u - u) z , (v \bar u + u) z \right)
+
\widetilde {\cal O}^{qq, \rm sym}_\sigma
\left( - (v \bar u + u) z ,  - (v \bar u - u) z \right)
\right]
\nonumber\\
&&\
+
v \bar u \,
\left( \partial_\rho z^\sigma - g_\rho{}^\sigma (z \cdot \partial) \right) \,
\left[
{\cal O}^{qq, \rm sym}_\sigma
\left( (v \bar u - u) z , (v \bar u + u) z \right)
-
{\cal O}^{qq, \rm sym}_\sigma
\left(  - (v \bar u + u) z , - (v \bar u - u) z\right)
\right]
\nonumber\\
&&\
+ \frac{2}{z^2}
\left( z_\rho z^\sigma - g_\rho{}^\sigma z^2 \right) \,
\left[
{\cal O}^{qq, \rm sym}_\sigma
\left( (v \bar u - u) z , (v \bar u + u) z \right)
+
{\cal O}^{qq, \rm sym}_\sigma
\left( - (v \bar u + u) z , - (v \bar u - u) z \right)
\right]
\Bigg\}
\, . \nonumber
\end{eqnarray}
Here we dropped terms $\mathcal{O} (z^4)$. The corrections we accounted here
generate $\Delta^2/Q^2$ power suppressed effect in the Compton amplitude.



\end{document}